University of Patras
School of Natural Sciences
Department of Physics

# Signals for invisible matter from solar – terrestrial observations

A Thesis submitted by

**Marios R. Maroudas**

in fulfilment of the requirements for the degree of Doctor of Philosophy in Physics

Patras, Greece
July, 2022

Ευρωπαϊκή Ένωση
European Social Fund

Operational Programme
Human Resources Development, Education and Lifelong Learning
Co-financed by Greece and the European Union

ΕΣΠΑ 2014-2020
ανάπτυξη - εργασία - αλληλεγγύη

This research is co-financed by Greece and the European Union (European Social Fund- ESF) through the Operational Programme ≪Human Resources Development, Education and Lifelong Learning≫ in the context of the project "Strengthening Human Resources Research Potential via Doctorate Research – 2nd Cycle" (MIS-5000432), implemented by the State Scholarships Foundation (IKY).

ΠΑΝΕΠΙΣΤΗΜΙΟ ΠΑΤΡΩΝ
ΣΧΟΛΗ ΘΕΤΙΚΩΝ ΕΠΙΣΤΗΜΩΝ
ΤΜΗΜΑ ΦΥΣΙΚΗΣ

# Σήματα αόρατης ύλης από ηλιακές – επίγειες παρατηρήσεις

Διατριβή υποβληθείσα υπό του

Μάριου Ρ. Μαρούδα

προς απόκτηση του
τίτλου του Διδάκτωρος
στην Φυσική

Πάτρα
Ιούλιος, 2022

Επιχειρησιακό Πρόγραμμα
Ανάπτυξη Ανθρώπινου Δυναμικού,
Εκπαίδευση και Διά Βίου Μάθηση
Με τη συγχρηματοδότηση της Ελλάδας και της Ευρωπαϊκής Ένωσης

Ευρωπαϊκή Ένωση
Ευρωπαϊκό Κοινωνικό Ταμείο

ΕΣΠΑ 2014-2020
ανάπτυξη - εργασία - αλληλεγγύη

Το έργο συγχρηματοδοτείται από την Ελλάδα και την Ευρωπαϊκή Ένωση (Ευρωπαϊκό Κοινωνικό Ταμείο) μέσω του Επιχειρησιακού Προγράμματος «Ανάπτυξη Ανθρώπινου Δυναμικού, Εκπαίδευση και Διά Βίου Μάθηση», στο πλαίσιο της Πράξης «Ενίσχυση του ανθρώπινου ερευνητικού δυναμικού μέσω της υλοποίησης διδακτορικής έρευνας – 2ος Κύκλος» (MIS-5000432), που υλοποιεί το Ίδρυμα Κρατικών Υποτροφιών (ΙΚΥ).

**Advisory Committee:**

*Main supervisor:*

1. Professor Vassilis Anastassopoulos

*Co-supervisors:*

2. Professor emeritus Konstantin Zioutas
3. Professor Sergio Bertolucci

**Examining Committee:**

*President:*

1. Professor Vassilis Anastassopoulos (*Department of Physics, University of Patras*)

*Committee members:*

2. Professor emeritus Konstantin Zioutas (*Department of Physics, University of Patras*)
3. Professor Sergio Bertolucci (*Department of Physics, University of Bologna*)
4. Professor Yannis Semertizidis (*Department of Physics, Korea Advanced Institute of Science and Technology*)
5. Professor Athanassios Argiriou (*Department of Physics, University of Patras*)
6. Professor Giovanni Cantatore (*Department of Physics, University of Trieste*)
7. Professor Marin Karuza (*Department of Physics, University of Rijeka*)

*To my family and friends.*

*"Physics is like sex: sure, it may give some practical results, but that's not why we do it."*

—Richard P. Feynman

# CERTIFICATE OF ORIGINALITY

I certify that this thesis has not already been submitted for any other degree or diploma in any other university or other institute of higher learning, is not being submitted as part of candidature for any such degree or diploma, and does not contain any material which has been accepted as part of the requirements for any such degree or diploma.

I also certify that the thesis does not contain any material or research work previously published or written by another person, except where due acknowledgement is made. In addition, I certify that all information sources and literature used are indicated within the text.

Finally, I certify that the thesis has been written by me and that, to the best of my knowledge and belief, any help I have received in preparing the thesis, and all sources used, have been acknowledged in the thesis.

Patras, Greece
Wednesday 6th July, 2022

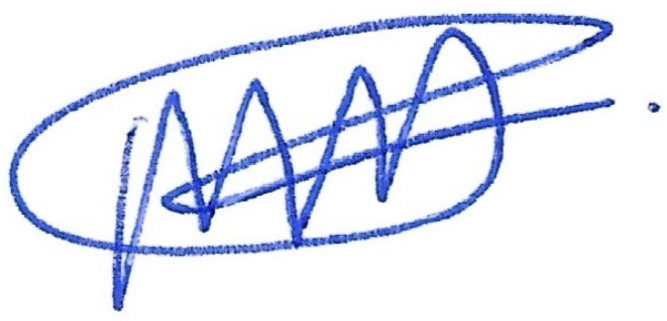

Marios R. Maroudas

# ACKNOWLEDGEMENTS

This thesis represents not only my work at the keyboard, it is a milestone in almost five years' work at the University of Patras, Greece and CERN in Geneva, Switzerland. Throughout these years I interacted with many friends, colleagues and mentors that have helped me, guided me, worked together with me and contributed directly or indirectly to the completion of this work. It is those remarkable individuals who I wish to acknowledge here.

First and foremost, I would like to express my sincerest gratitude to my thesis supervisor Prof. Konstantin Zioutas for his continuous support of my thesis research, experimental programme, data analysis and dissemination process. His numerous brilliant ideas along with his endless patience and his guidance during this long road are indeed noteworthy. Special thanks also to my supervisors Prof. Vasilis Anastassopoulos and Prof. Sergio Bertolucci for guiding and supporting me through all the phases of my thesis. I'm deeply indebted to all remaining members of my 7-member committee, Prof. Giovanni Cantatore, Prof. Marin Karuza, Prof. Athanassios Argiriou, and Prof. Yannis Semertzidis for their patience and attentiveness on reading through this huge work, their constructive comments, and for their general support throughout all the years of our collaboration.

Moreover, I owe my deepest gratitude to Fritz Caspers for teaching me all practical aspects of RF engineering, for his indispensable help on CAST-CAPP commissioning and for always being the advocate of the devil. I would like to express my appreciation to Horst Fischer and Wolfgang Funk for their extensive conversations on Dark Matter issues and their general support while being at CERN. I would like to also thank here explicitly Yannis Semertzidis for long fruitful discussions not only in the field of Dark Matter but also on the prospects of Physics in general. In addition, I would like to express my deepest appreciation to Kaan Ozbozduman, for his enormous work on all CAST-CAPP related software including DAQ, processing and analysis codes as well as for his patience with my tedious comments and never-ending requests. Thank also Theodoros Vafeiadis for sharing his CAST-related expertise and for his overall significant assistance in the assembly of the CAST-CAPP detector and for his help on any arisen issues. I have greatly benefited from Lino Miceli for his full-time commitment to the creation of the CAST-CAPP detector and for sharing his technical expertise and knowledge. Many thanks to Antonis Gardikiotis for his assistance with Fourier analysis as well as his long-term guidance. I am particularly grateful for the assistance given by Soohyung Lee in creating the first version of the CAST-CAPP data acquisition script and related python sub modules. Likewise, my sincerest gratitude extends to IKY foundation for the financial support. I would also like to express my gratitude to all fellow CAST shifters including Sergio Cuendis Arguedas, Justin Silvan Baier, Cristina Blasco Margalejo, Aydin Ozbey, Jessica Golm, Aleksandr Dermenev, Sebastian Michael Schmidt, and all the rest for long Dark Matter and stream discussions during early-morning CAST shifts as well as on their overall support. Thanks also to all my CERN colleagues, and services as well as my CERN friends for their material and moral support accordingly. Among them, I have to distinguish my friends Evangelos Motesnitsalis, Nikolas Papapetrou, and Vasilis Dimakopoulos for their constant encouragement and understanding throughout all the demanding years in Switzerland.

I can not thank enough Foteini Mitropoulou for her endless belief and psychological support throughout all the years without which this dissertation would not have materialised. A special thanks to Spyros Maroudas not only for his analytic and thorough advice but most importantly for providing a LaTeXtemplate as well as for creating several figures and schematics for this thesis and multiple scientific presentations. And finally, I am extremely grateful to my family and relatives, for standing by me and keep encouraging me to finish my thesis at my own pace. Without them, it would have taken me a lot longer to finish.

# TABLE OF CONTENTS

# VI Conclusions

# Appendices

# LIST OF FIGURES

## I Introduction

## II Theoretical Background

**7 EUV irradiance**

## 16 Stratospheric temperature

## 23 CAST-CAPP results

# VI Conclusions

# Appendices

## A Simulations

## B Dataset comparisons

# LIST OF TABLES

## II Theoretical Background

## III Solar Observations

# VI Conclusions

# Appendices

# PREFACE

This project has started unofficially around 2014 as a small-scale analysis of a few short solar-term datasets, such as solar EUV irradiance, to explore existing periodicities via manual handling of the data. In time, it evolved into a vast full-time Ph.D. project focused on investigating numerous diverse long-term observations for planetary dependencies. A specifically designed analysis procedure had to be developed from the ground up for an automatic sweeping of the thousands of rows of data. This document is the report of the "few (!)" most interesting and significant results of these analyses. Several publications and presentations have resulted from this research which are mentioned in Appendix Sect. E. It is noted that possible minimum differences in the figures of some of the published papers are due to more accurate planetary positions used in this work. Since this thesis contains the statistical analysis of various datasets, each chapter in Parts III and IV could also be read independently and separately from the rest. The main structure though remains the same for all Chap. 6 through 12 and Chap. 15 through 18.

The first Part II concentrates to introduce the reader to the theoretical background behind the various analyses performed for a wide range of solar and terrestrial datasets. Chap. 1 through 3 outlines the main characteristics of Dark Matter and streams as well as how gravitational lensing of non-relativistic particles works. Then, Chap. 4 proceeds with the combination of the aforementioned information into the formation of the hypotheses of this work, the main analysis procedures, and the research questions that arise.

The next Part III contains a considerable amount of different datasets from solar observations that are scrutinised for possible conventionally unexpected planetary signatures. The same evaluation techniques are performed in Part IV but this time for terrestrial observations. A first comparison between the several datasets for possible similarities and dissimilarities is also carried out. At the end of each Part as well as at the end of each Chapter the main conclusions, implications and future prospects are described.

Finally, the last Part V is dedicated to the experimental searches for Dark Matter axions with CAST-CAPP detector at CERN. This almost separate enormous project started around October 2018, and has transformed CAST experiment from an axion helioscope to an axion haloscope. Through this extensive procedure, a new data-taking strategy focused on streaming Dark Matter axions has been suggested that has the potential to redefine the field of direct

Dark Matter detection. Chap. 21 introduces axions as potential Dark Matter candidates, while Chap. 22 and 23 contain the technical details and the latest data-taking results of the CAST-CAPP detector which managed to set new world-class limits.

Consequently, the realisation of this PhD has hopefully provided new significant insights for long-standing mysteries via the retrieval of unexpected planetary signatures but also established a new strategy for “conventional” direct Dark Matter searches which have so far proven unsuccessful for all experiments. Notably, through the novel planetary analysis techniques and the unexpected statistically significant observations, which can extend even further to more datasets, this study has provided multiple indications that the dark universe may not be “invisible” after all.

# ABSTRACT

The composition of the dark universe although hypothesised, remains one of the biggest mysteries in modern physics. On smaller scales, there are various solar puzzling phenomena which known physics cannot explain like the coronal heating problem, the origin of sunspots, the trigger mechanism of solar flares, but also the open issue since the 1850's on the planetary impact of the active Sun. Interestingly, the 11-year solar cycle remains one of the oldest open questions in solar physics. At the same time, several terrestrial observations in the dynamic Earth's atmosphere such as the ionospheric ionisation around December are unexpected within conventional physics. Following this work, the suggested common solution of all these conventionally unexplained phenomena is based on an external triggering caused by low-speed streaming constituents from the dark sector, being gravitationally focused or deflected by the Sun and the orbiting planets. For this to happen, streams of invisible matter are assumed to exist which should interact with large cross-sections with baryonic matter. Existing favourable candidates from the dark sector include AntiQuark Nuggets, magnetic monopoles, and dark photons. Evidence in support of this hypothesis and on the existence of one or more streams or clusters has been provided based on a coincidence analysis of long-term astrophysical and planetary datasets. Of note, the planetary correlation is the novel key signature. By projecting the time of appearance of the measured observables on the orbital position of the various planets including the Moon, a striking clustering shows up. This statistically significant pronounced activity at certain planetary heliocentric longitudes points to preferred directions in the flow of the assumed streams, like probably one from the Galactic Centre. Notably, even stronger planetary correlations occur when the gravitational effect of two or more planets is combined. Some of the results are also supported by Fourier analyses. Additionally, the derived significant narrow periodicity of 27.32 d on most of the observables, which overlaps with the Moon's sidereal Month, strengthens the claim of a significant exo-solar influence. Finally, a redefined strategy for direct Dark Matter searches focused on streaming Dark Matter is proposed. This novel procedure has been successfully implemented in the CAST-CAPP detector at CERN searching for Dark Matter axions.

# ΠΕΡΙΛΗΨΗ

Η σύσταση του σκοτεινού σύμπαντος, παρόλες τις υποθέσεις παραμένει ένα από τα μεγαλύτερα μυστήρια της μοντέρνας φυσικής. Σε μικρότερες κλίμακες, υπάρχουν διάφορα ηλιακά φαινόμενα τα οποία η γνωστή φυσική δεν μπορεί να εξηγήσει, όπως το πρόβλημα της θέρμανσης του ηλιακού στέμματος, η προέλευση των ηλιακών κηλίδων, ο μηχανισμός πυροδότησης των ηλιακών εκλάμψεων, αλλά και το ανοιχτό ζήτημα από τη δεκαετία του 1850 σχετικά με τις πλανητικές επιπτώσεις του ενεργού Ήλιου. Σημειωτέον, ο 11-ετής ηλιακός κύκλος παραμένει ένα από τα παλαιότερα ανοικτά ερωτήματα της ηλιακής φυσικής. Ταυτόχρονα, αρκετές επίγειες παρατηρήσεις σχετικά με τη δυναμική γήινη ατμόσφαιρα, όπως ο ιονοσφαιρικός ιονισμός τον Δεκέμβριο, είναι απροσδόκητες στο πλαίσιο της συμβατικής φυσικής. Με βάση την συγκεκριμένη εργασία, η προτεινόμενη κοινή λύση για όλα αυτά τα συμβατικώς ανεξήγητα φαινόμενα βασίζεται σε μια εξωτερική πυροδότηση που προκαλείται από χαμηλής ταχύτητας συστατικά του σκοτεινού τομέα, τα οποία εστιάζονται ή εκτρέπονται βαρυτικά από τον Ήλιο και τους πλανήτες. Για να συμβεί αυτό, πρέπει να υπάρχουν ρεύματα αόρατης ύλης τα οποία θα πρέπει να αλληλεπιδρούν ισχυρά με την βαρυονική ύλη. Μερικοί ταιριαστοί υποψήφιοι από τον σκοτεινό τομέα περιλαμβάνουν τα AntiQuark Nuggets, τα μαγνητικά μονόπολα, και τα σκοτεινά φωτόνια. Για την υποστήριξη αυτής της υπόθεσης καθώς και για την ύπαρξη ενός ή περισσότερων ρευμάτων παρέχονται αποδείξεις βασισμένες στην ανάλυση συμπτώσεων μακροχρόνιων αστροφυσικών και πλανητικών βάσεων δεδομένων. Σημειώνεται πως η πλανητική συσχέτιση αποτελεί την βασική υπογραφή. Προβάλλοντας τον χρόνο εμφάνισης των παρατηρησιακών δεδομένων στην τροχιακή θέση των διαφόρων πλανητών, συμπεριλαμβανομένης της Σελήνης, εμφανίζεται μια εντυπωσιακή ομαδοποίηση. Αυτή η στατιστικώς σημαντική έντονη δραστηριότητα σε ορισμένα πλανητικά ηλιοκεντρικά μήκη υποδεικνύει προτιμώμενες κατευθύνσεις στη ροή των υποτιθέμενων ρευμάτων, όπως πιθανώς από το Γαλαξιακό μας κέντρο. Αξίζει να σημειωθεί ότι ακόμη ισχυρότερες πλανητικές συσχετίσεις εμφανίζονται όταν συνδυάζεται η βαρυτική επίδραση δύο ή περισσότερων πλανητών. Μερικά από τα αποτελέσματα υποστηρίζονται επίσης από αναλύσεις Fourier. Επιπροσθέτως, η προκύπτουσα στατιστικώς σημαντική περιοδικότητα 27.32 ημερών στις περισσότερες βάσεις δεδομένων, η οποία ταυτίζεται με τον αστρικό μήνα της Σελήνης, ενισχύει τον ισχυρισμό μιας σημαντικής εξω-ηλιακής επιρροής. Τέλος, προτείνεται μια νέα στρατηγική για άμεσες αναζητήσεις Σκοτεινής Ύλης που εστιάζουν σε σκοτεινές ροές. Αυτή η νέα διαδικασία εφαρμόστηκε με επιτυχία στον ανιχνευτή CAST-CAPP στο CERN ο οποίος ψάχνει για αξιόνια Σκοτεινής Ύλης.

# Part I: Introduction

*"Begin at the beginning," the King said gravely, "and go on till you come to the end: then stop."*

—Lewis Caroll, Alice in Wonderland,
*English author*

Since the mid-'70s when the formulation of the Standard Model (SM) of particle physics was actually completed, the confrontation with the experimental data, had been extremely successful in explaining with great accuracy all the fundamental interactions except gravity. However, despite its strengths, it was quickly realised that SM is also possessed by serious weaknesses. Some of the phenomena that it can not explain are the neutrino masses, the strong CP problem, and most importantly the composition of Dark Matter (DM) and Dark Energy (DE). The observations showed that almost 96% of the universe is "dark" and the usual observable baryonic matter constitutes only $\sim 4\%$, with DM being about 26.5% and DE about 68.5% of the total matter-energy density of the universe (Fig. 0.1). This constitutes a major gap in our understanding of the universe with only $\sim 4\%$ of it being currently explained. Therefore, one of the most fundamental open questions in physics is the identification of the constituents of DM in the universe.

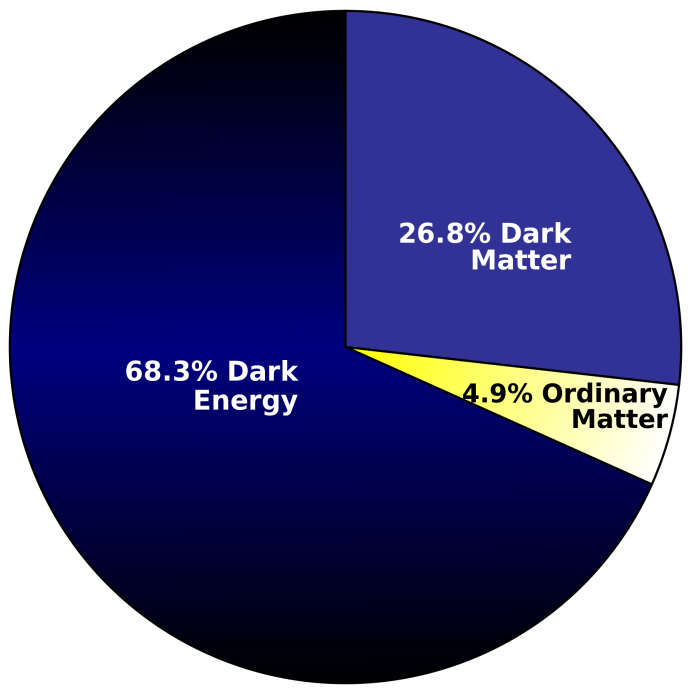

**Figure 0.1:** Estimated distribution of matter and energy in the universe extracted from Planck's high-precision CMB map (1).

The existence of DM was established through several large-scale observational evidence and got its name based on the fact that it was not interacting electromagnetically with normal matter and not emitting any light. A large number of experiments were then dedicated on direct and indirect searches, with the detection of DM constituents evolving into one of the most important challenges of modern physics. As we will see, all experimental efforts for the direct detection of DM have so far proven unsuccessful. An additional characteristic of DM which has been recently under investigation is its uniformity and especially around our solar neighbourhood. This, along with other conventional assumptions such as the assumed feeble interaction with normal matter may have to eventually be revisited.

Moving on to smaller scales, one of the most famous problems in solar physics concerns the solar activity and its $\sim 11\,\text{y}$ modulation. Even though several models have been suggested following conventional physics, still there are many unexplained issues regarding the triggering mechanism of solar phenomena such as the solar flares or the coronal heating problem (Fig. 0.2). Their origin remains unexplained within known physics since decades - centuries, while, by comparison, problems like the solar neutrino deficit problem were solved within a few decades.

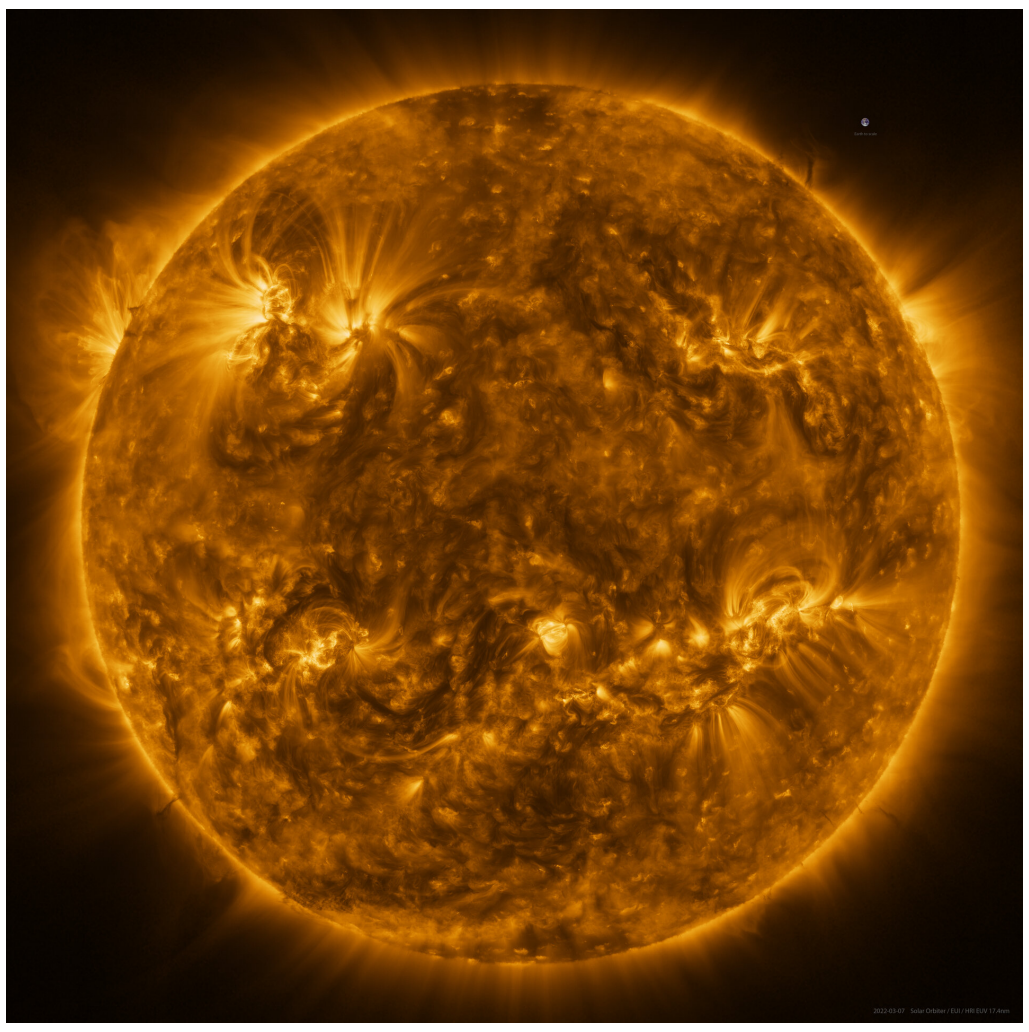

**Figure 0.2:** The highest ever resolution image of the Sun and its upper atmosphere, the corona, in EUV taken from the Solar Orbiter on 07/03/2022. An image of the Earth is also included for scale on the upper right part (2).

Similarly, beyond the Sun itself, there exist long-standing anomalies also in the Earth's atmosphere such the anomalously high degree of ionisation in December or other annual temperature anomalies in the Earth's stratosphere. Even in lower altitudes there exist phenomena such as Earthquakes (EQs) or even disease-related observations which are unexplained within conventional physics.

Historically, the study of such anomalous phenomena in physics has provided some interesting surprises (E.17). For example, the observation of an unexpected atmospheric ionisation in 1912 resulted in the discovery of cosmic rays [13]. Also, the discovery of DM by Zwicky in 1933 [14] was based on observed gravitational discrepancies in large cosmological systems like the Coma cluster containing over 1000 identified galaxies. Therefore, the analysis of the puzzling behaviour of solar and terrestrial phenomena although challenging, has the potential to eventually lead to a significant breakthrough. At the same time, given the long-lasting fruitless efforts on the direct detection of DM, some novel approaches seem to be required to decipher the nature and implications of the dark sector. In conclusion, there is an increasing need for a multidisciplinary approach on the aforementioned persisting physics problems which, through a much-needed unconventional phenomenological lens and a novel experimental approach, has the potential to advance our understanding of the universe that surrounds us.

# Part II:

# Theoretical Background

*"If the facts don't fit the theory, change the facts."*

—Albert Einstein,
*Theoretical Physicist*

1

# DARK MATTER

## 1.1 Observational evidence

### 1.1.1 Velocity dispersions

In 1933 Fritz Zwicky studied the redshifts of various galaxy clusters and noticed a large scatter in the apparent velocities of eight galaxies within the Coma cluster with differences that exceeded 2000 km/s [14, 15].

By applying the virial theorem, Zwicky estimated that the mean density of the Coma cluster would have to be 400 times greater than that which is derived from luminous matter. In other words, Zwicky found that the velocities of individual galaxies with respect to the cluster mean velocity are much larger than those expected from the estimated total mass of the cluster calculated from the masses of individual galaxies. The only way for the cluster not to expand rapidly is to assume that the cluster contains huge quantities of some invisible DM.

### 1.1.2 Galaxy rotation curves

The most convincing direct evidence for DM on galactic scales comes from the observations of the rotation curves of galaxies, namely the graph of circular velocities of stars and gas as a function of their distance from the galactic centre.

The rotation curves such as the one seen in Fig. 1.1 are usually obtained by combining observations of spectral lines with optical surface photometry by measuring the Doppler shifts of such characteristic spectral lines, such as the 21 cm emission line of neutral hydrogen. More specifically, in Newtonian dynamics, given any spherically symmetric mass distribution, a test particle (such as a star) in a circular orbit has a velocity:

$$v(r) = \sqrt{\frac{G_N M(r)}{r}} \tag{1.1}$$

where $M(r) = 4\pi \int \rho(r) r^2 dr$ is the total mass inwards of radius $r$ and $\rho(r)$ the density as a function of the radius. If we expect that $M(r)$ asymptotes to a constant $M$ (basically the total mass of all the stars in the galaxy) then the velocity is expected to be reduced with $v(r) \propto 1/\sqrt{r}$ beyond the optical disc.

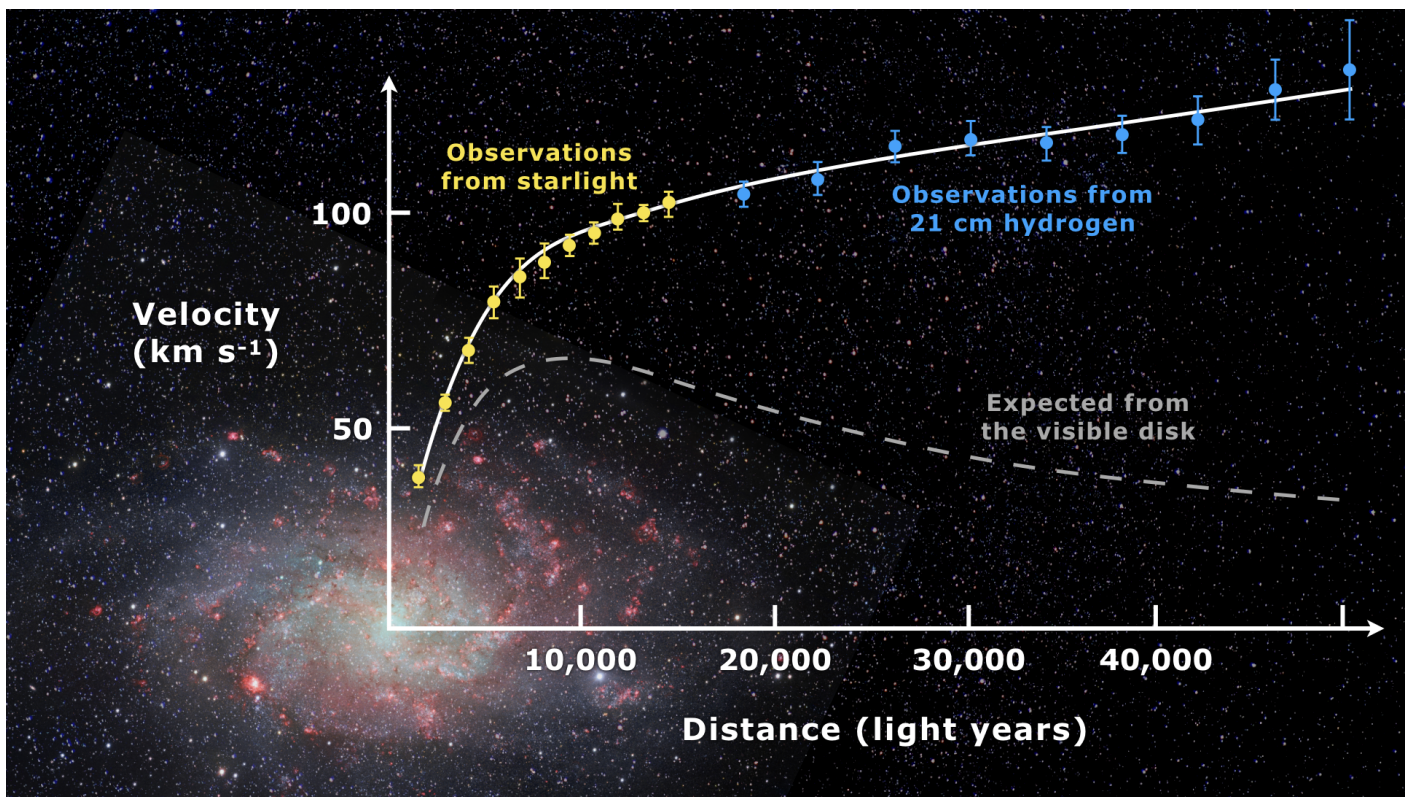

**Figure 1.1:** Observed rotation curve of spiral Messier 33 galaxy (yellow and blue points) and the predicted one from the visible matter distribution (grey line) (3).

However, in 1970 Ford and Rubin discovered that the rotation curves of galaxies are flat [16, 17] (see Fig. 1.1). The velocities of objects (stars or gas) orbiting the centres of the galaxies remain constant rather than decreasing as a function of the distance from the galactic centres as was expected. The simplest explanation is once more that galaxies contain much more mass than can be explained by the bright stellar objects residing in the galactic disks. Actually, the fact that $\upsilon(r)$ is approximately constant implies the existence of a DM halo with $M(r) \propto r$ and a mass density distribution of:

$$\rho(r) \propto \frac{M(r)}{r^3} \sim \frac{1}{r^2} \tag{1.2}$$

### 1.1.3 Cosmic microwave background

The existence of background radiation originates from the propagation of photons in the early universe, once they decoupled from matter. It was predicted in 1948 [18] and discovered in 1965 [19]. The Cosmic Microwave Background (CMB) is the thermal radiation that is left from the time of recombination in the universe history (about 400 000 y) when the universe had cooled down enough for atoms to be formed. This radiation is strongest in the microwave region. After many decades of experimental efforts, the CMB is known to be isotropic at the $10^{-5}$ level and follows the spectrum of a black body with temperature $T_0 = 2.7255 \pm 0.0006\,\mathrm{K}$ [20].

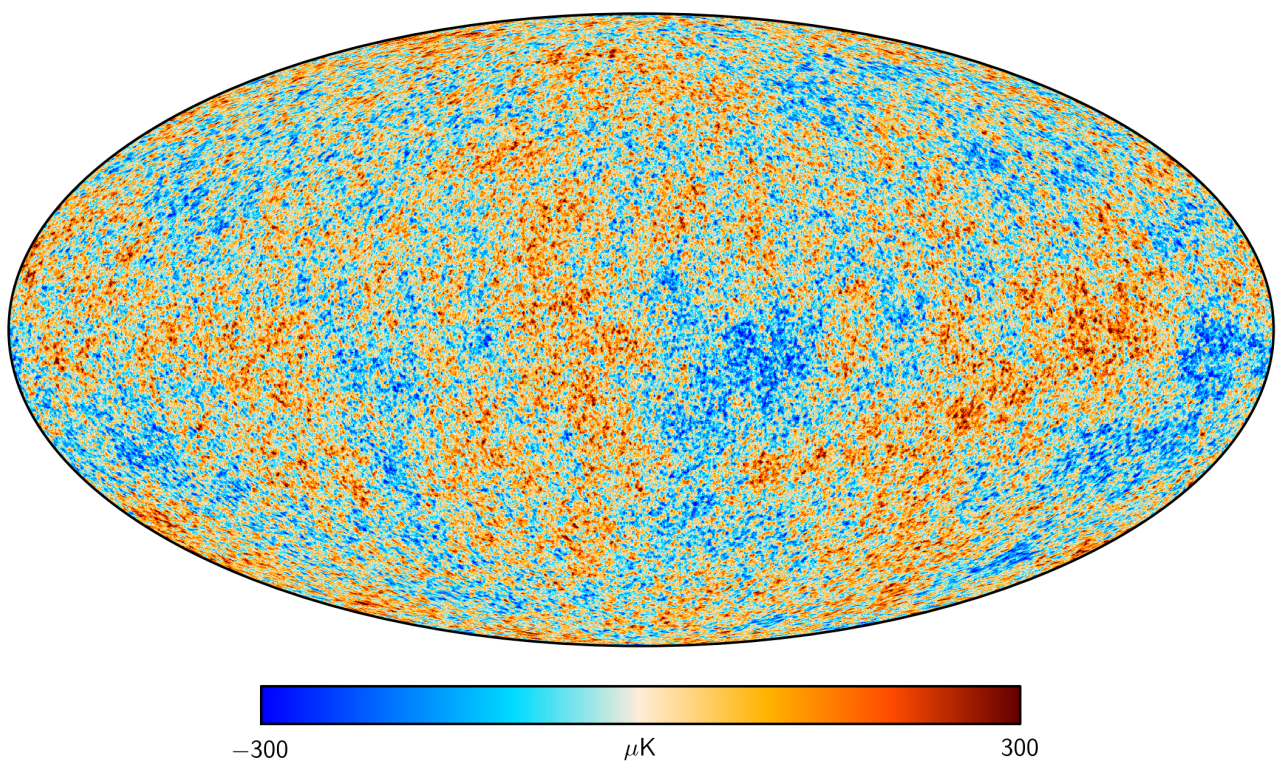

**Figure 1.2:** All-sky map of the CMB temperature anisotropies as obtained by Planck (4).

The tiny temperature anisotropies appearing in the CMB spectrum (see Fig. 1.2) enable an accurate testing of cosmological models and can put stringent constraints on cosmological parameters.

From the latest results of the Planck space telescope, the abundance of baryons ($b$) and matter ($m$) in the universe has been measured to be [21]:

$$\begin{aligned} \Omega_b {h_0}^2 &= 0.0224 \pm 0.0001 \\ \Omega_m &= 0.315 \pm 0.007 \end{aligned} \tag{1.3}$$

where $h_0 = 0.678 \pm 0.009$ is the *dimensionless Hubble parameter* (or *rescaled Hubble's constant*) defined by the present day Hubble constant $H_0 = 100 h_0$ km/sec/Mpc.

From the comparison of the cosmological baryon density $\Omega_b$ of the universe today, as well as the net cosmological matter density $\Omega_m$ the universe must contain matter which is not composed of protons and neutrons. This extra contribution from DM to the total matter of the universe ($\Omega_m = \Omega_b + \Omega_{DM}$) has also been estimated to be [21]:

$$\Omega_{DM} {h_0}^2 = 0.120 \pm 0.001 \tag{1.4}$$

As it has been measured, in order to have a total cosmological density of about:

$$\Omega_{\text{total}} = \frac{\rho}{\rho_{\text{cr}}} = 1.011 \pm 0.006 \tag{1.5}$$

where: $\rho$ : the observed density,

$\rho_{\text{cr}} = \frac{3{H_0}^2}{8\pi G_N}$ : the critical density of the Friedmann universe required to close the Universe.

we need an additional density contribution of about $\Omega_\Lambda = 0.685 \pm 0.007$ coming from DE. Therefore, the the total density of the universe is very close to the critical density of $\Omega_{\text{cr}} = 1$ indicating that the universe is flat with a composition of about 69.1% DE, 25.9% DM and 4.9% baryonic matter.

### 1.1.4 Primordial nucleosynthesis

From Big Bang theory, when the universe was a few hundred seconds old, at a temperature of about $10^{10}$ K, deuterium became stable: $p + n \rightarrow D + \gamma$. After the formation of deuterium, also helium and lithium could form. The formation of heavier elements such as $C$, $N$, and $O$, can only occur inside stars where there are densities high enough for triple interactions of three helium atoms into a single carbon atom. This means that their creation can only happen after about a billion years. The predictions provide a mass abundance of about 75% $^1$H, 25% $^4$He, $10^{-5}$ D and $10^{-10}$ $^7Li$ [22]. These predictions can match the data only if atoms are only 4% of the total constituents of the universe, as, if there are more baryons, then there should also be more helium, lithium and heavier elements formed during the Big Bang [23].

### 1.1.5 Gravitational lensing

As light is affected by the curvature of space-time, it provides a method of accurately determining the mass of massive astrophysical objects [15, 24]. More specifically, because the gravitational field of a massive object, such as a cluster, curves the space around it, light rays emitted from other astrophysical objects behind the cluster travel along paths which are curved rather than straight on their way to an observer at the line of sight [25] (see Fig. 1.3).

If the lensing is strong enough, there are multiple paths from the same object, past the cluster, that arrive at the observer which results in multiple images of the same object (see Fig. 1.3a). Furthermore, because the light from different sides of the same galaxy travels along slightly different paths, the images of strongly lensed sources are distorted into arcs. Moreover, when the lens, the source and the observer are in perfect alignment then the image produced is a perfect ring (see Fig. 1.3b), which is called *Einstein ring*. On the other hand, if the lensing is weak, then the images may become slightly elongated, even if they are not multiply imaged [26].

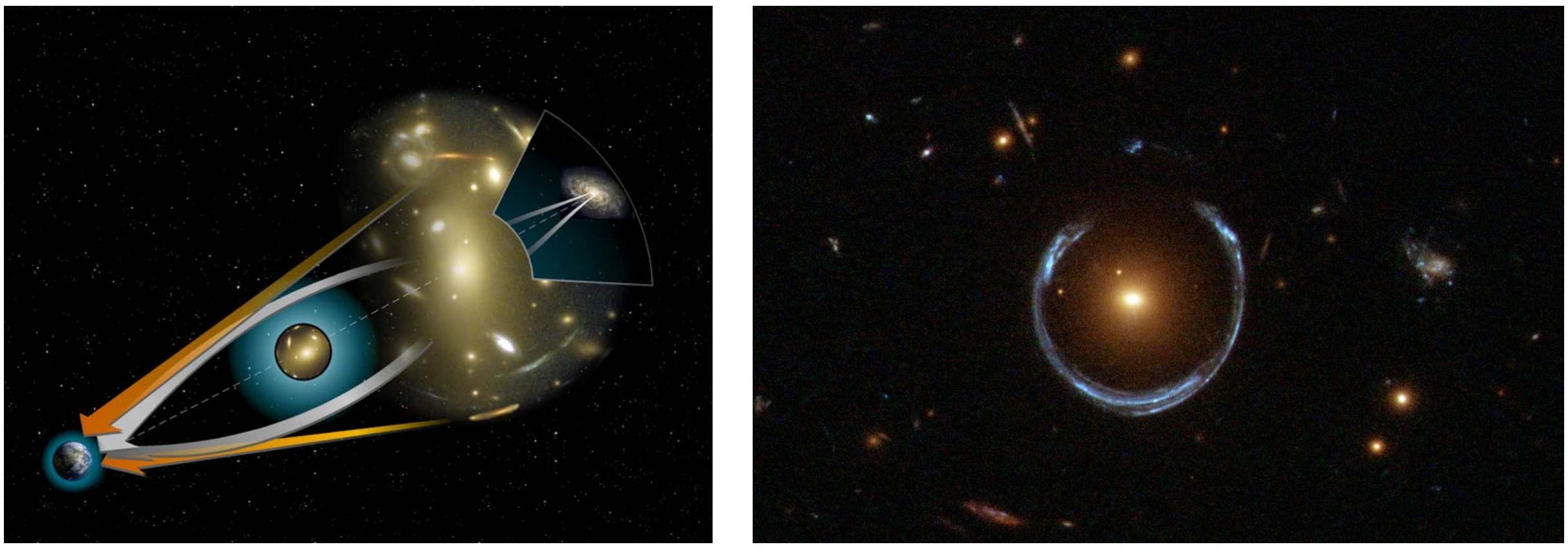

**(a)** Bending light around a massive object from a distant source. The orange arrows show the apparent position of the background source. The white arrows show the path of the light from the true position of the source. (52).

**(b)** The gravity of a luminous red galaxy has gravitationally distorted the light from a much more distant blue galaxy. The lens alignment is so precise that the background galaxy is distorted into a horseshoe – a nearly complete ring (53).

**Figure 1.3:** Images of gravitational lensing effects.

For a lensing cluster with total mass $M$ and impact parameter $b$, the deflection angle is:

$$a = 4\frac{G_N M}{bc^2} \tag{1.6}$$

Therefore, from measurements of the deflection angle and impact parameter, which can be calculated from the redshift to the lensing cluster and source, the total mass $M$ of the cluster can be calculated. Another calculation for strong lensing comes from the Einstein radius which is proportional to the square root of the projected mass inside it. Several galaxy clusters have been observed for the determination of their matter content using this technique. The advantage of this technique over rotation curves (Sect. 1.1.2) is the evidence that DM can be seen out to much larger distances, even 200 kpc from the centres of galaxies. As an example for studies of galaxy clusters such as the Sloan Digital Sky Survey (SDSS) J1004+4112 [27], it was concluded that no more than 10% of the matter in the cluster was luminous meaning that at least 90% of the matter was DM.

### 1.1.6 Structure formation

The *cosmological principle* asserts that the universe is homogeneous and isotropic on large scales. This is also supported by early studies [28] which revealed that the mean spatial density of galaxies is approximately independent of the distance and direction in the sky. However, on small scales (Mpc) the universe is exceedingly inhomogeneous and increasingly homogeneous on scales approaching the entire horizon. More specifically, in clusters, the density reaches values $10^3$ times larger than the average density, and in galaxies $10^5$ times larger [29].

The problem is that before the decoupling time, where matter and radiation decoupled, inhomogeneities in baryonic components could not grow because photos and baryons were strictly coupled. This is due to the radiation pressure of photons which pushes away any concentration of matter that may be created under the effect of gravity thus preventing any fluctuation in the distribution of ordinary matter which grows denser. However, these inhomogeneities are required to explain the observed transformation of a highly homogeneous universe at early times to a highly local non-homogeneous one. These small anisotropies in the universe were the ones that due to gravitational attraction, drew more matter from their surroundings, growing denser and more massive eventually giving rise to stars, galaxies and the larger structures that we observe today. In order to solve this problem the introduction of DM in the early universe is required [30]. As we will see in Sect. 1.2.2, the growth of primordial fluctuations in different types of DM gives rise to completely different distributions of cosmic structure.

Furthermore, from several large-scale $N$-body simulations [31–35] it was demonstrated that the large-scale structure of the luminous matter that we observe today could only have been formed in the presence of a substantial amount of DM.

## 1.2 Characteristics of Dark Matter

### 1.2.1 Basic properties

Based on the various aforementioned observations, there are several requirements that every DM particle candidate should satisfy [26]:

- DM must interact gravitationally and form stable halos around the galaxies.
- DM particles must be electrically neutral and interact very weakly with normal matter.
- DM must have no, or extremely weak, interactions with photons.
- Self-interactions between DM particles must be very small. The elastic-scattering cross section must be $\sigma_{\text{DM-DM}}/m_{\text{DM}} < 0.47\,\text{cm}^2/\text{g} \approx 0.84\,\text{b/GeV}$ [36].
- DM lifetime must be long ($> 160\,$Gy) compared to cosmological timescales, otherwise they would have decayed by today [37].

- Interactions of DM particles with baryons should also be weak, otherwise during the formation of galaxy a baryon-DM disk would be created in contradiction with the more diffuse and extended DM halos that are observed.
- DM should be everywhere in the universe with a bigger density in the halo of galaxies. However, it must have begun uniformly distributed throughout the universe with the same initial spectrum of density fluctuations that normal matter possessed.
- DM should be non-baryonic based on the CMB measurements and the Big Bang nucleosynthesis predictions.
- DM must be non-relativistic at the epoch of structure formation in order for structures such as galaxies to emerge by gradual accumulation of particles.

### 1.2.2 Classification

There are three different theories on DM named Cold Dark Matter (CDM), Hot Dark Matter (HDM) and Warm Dark Matter (WDM) [38, 39] which, contrary to their name, do not refer to the temperatures of matter itself, but the size of particles with respect to the size of a protogalaxy, from which dwarf galaxies are later formed. This results in the following classification of DM particles:

- CDM, where DM particles are much smaller than a protogalaxy.
- HDM, where DM particles are much larger than a protogalaxy.
- WDM where DM particles are similar to a protogalaxy.

In turn, the size of these particles determines the velocity that they can travel, which determines their thermodynamical properties and indicates how far they could have travelled (their *free streaming length*) before being slowed by cosmic expansion. Primordial density fluctuations smaller than this length get washed out as particles spread from over-dense to under-dense regions, while larger fluctuations remain unaffected [40]. Therefore, the free streaming length sets a minimum scale for later structure formation, and thus different types of DM result in different evolutionary histories for the universe (see Fig. 1.4).

#### 1.2.2.1 Cold Dark Matter

CDM consists of particles which were non-relativistic at the time they decoupled from the other components of the universe [41–43]. Specifically, in CDM models, the particles are relatively heavy, with a mass similar to that of protons or more. Because of their high mass, these DM particles would move relatively slowly, at similar speeds as the gas and dust in our galaxy. Examples of CDM candidates are Weakly Interacting Massive Particles (WIMPs) (see Sect. 1.3.3), axions (see Sect. 1.3.2) and MAssive Compact Halo Objects (MACHOs) (see Sect. 1.3.1.1).

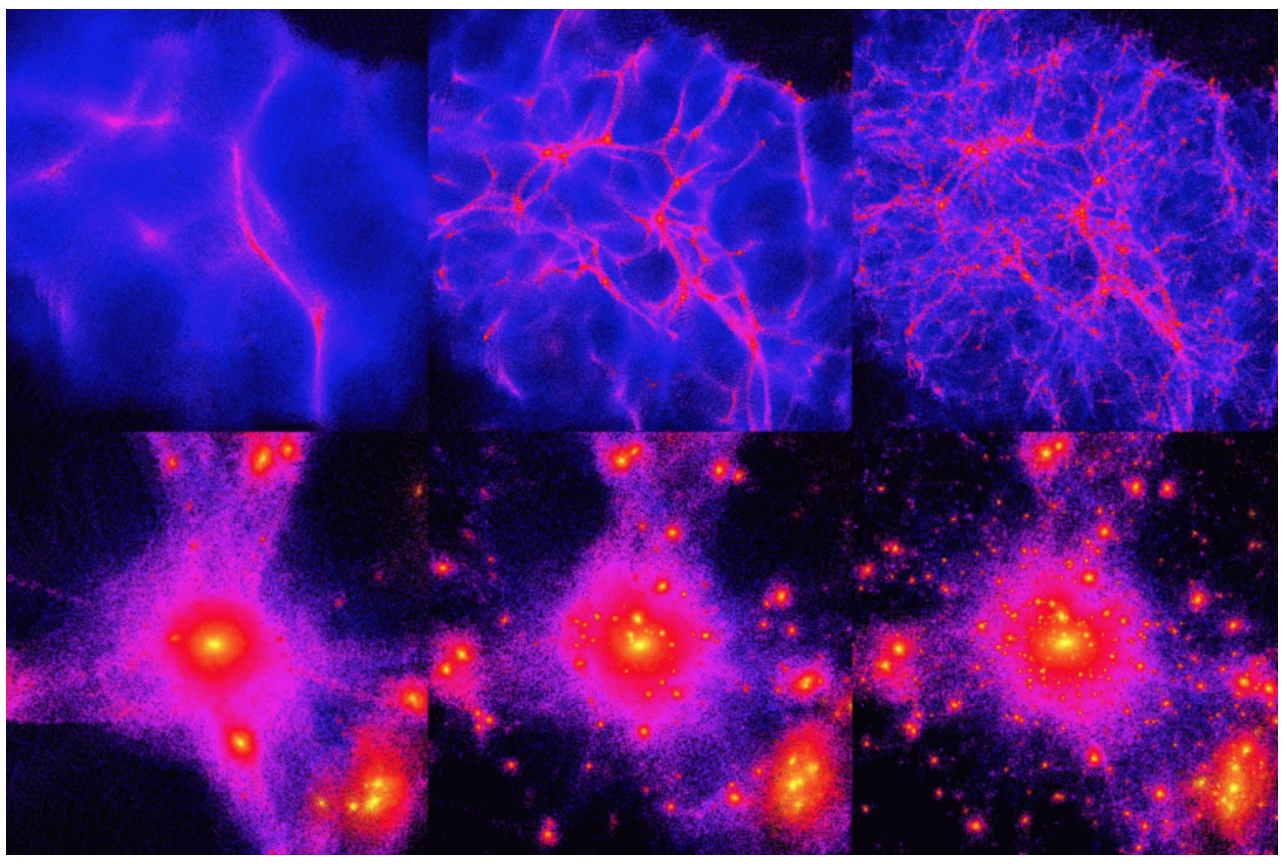

**Figure 1.4:** Simulations for structure formation in a universe dominated by HDM (left), WDM (middle) and CDM (right) at early times (high redshift) (top row) and at present time ($z = 0$) (bottom row) (5).

The non-gravitational interactions of CDM particles are much weaker than the weak interactions. In CDM small-scale structures such as galaxies can form at early times, and they can clump together into large-scale structures such as clusters and superclusters over time through their mutual gravitational attraction. This is called a *bottom-up* structure formation scenario and is in compliance with what we observe in deep field observations, therefore making this type of DM the most favourable one.

#### 1.2.2.2 Hot Dark Matter

HDM consists of particles that were relativistic at the time they decoupled from the other components of the universe and which remained relativistic until the mass contained within a Hubble volume (a sphere of proper radius $R_H$) was large compared to the mass of a galaxy [44]. HDM particles must also be still in thermal equilibrium after the last phase transition in the early hot universe which presumably took place at $T_{\rm QCD} \approx 10^2$ MeV. HDM particles have a cosmological number density almost comparable to that of the CMB photons which implies an upper bound to their mass of a few tens of eV [45].

The most natural candidate of HDM is the massive neutrino(s) (see Sect. 1.3.4), but other possibilities have also been suggested such as the *majoron* of non zero mass which is lighter than the lightest neutrino species and in which all neutrinos decay [46].

The problem with HDM is that its free streaming length is much larger than the size of a galaxy. Therefore, if all of DM were hot, all galaxy-sized fluctuations would get washed out due to free streaming. This means that the first structures that can form are huge supercluster-sized pancakes, which were then theorised to fragment and break up into smaller clusters and then further break up into galaxies. This is called a *top-down* structure formation scenario. However, deep field observations indicate the opposite, which means that any model dominated by HDM disagrees with the current observations [40].

#### 1.2.2.3 Warm Dark Matter

WDM refers to particles with a free-streaming length comparable to the size of a region, which subsequently grew to become a dwarf galaxy. WDM can also be seen as just HDM cooled down and thus it is usually discussed in comparison with HDM scenario [47].

WDM particles interact much more weakly than neutrinos and this is why the most common WDM candidate is sterile neutrino (see Sect. 1.3.4.2). Also, the gravitino (the Supersymmetry (SUSY) partner of graviton) [48] can have properties similar to WDM, although strictly speaking, it is a CDM particle candidate.

The predictions of WDM are similar to CDM on large scales (including the CMB, galaxy clustering, and large galaxies), but with fewer small-scale density perturbations. WDM reduces the number of dwarf galaxy abundance [49] giving a better agreement with observations [50]. Furthermore, the lower concentrations of WDM halos [51, 52] leads to lower densities in the inner regions of galaxies (where there are stars), producing a constant density core, and thus providing a solution to the core-cusp problem of CDM. Finally, $N$-body simulations of WDM haloes with $m_{\mathrm{WDM}} \sim 1 - 2\,\mathrm{keV}$, show that the too-big-to-fail problem does not exist [50].

## 1.3 Composition of Dark Matter

### 1.3.1 Baryonic matter

#### 1.3.1.1 MACHOs

A theory for the DM composition that was popular for a time was that of MACHOs [53]. MACHOs are normal astrophysical objects made of ordinary baryonic matter in the form of faint stars, planetary objects such as brown dwarfs or stellar remnants such as white dwarfs, neutron stars and black holes. They could also be Primordial Black Hole (PBH) or mirror matter [54].

These systems of condensed baryons can appear dark because they can be dim, but can also be effectively transparent to background light because their compact nature causes the effective cross-section for light absorption per unit mass across an entire halo to be very small. The only way to detect such objects is to look for micro-lensing effects, where a lensing object can pass near the line of sight towards a background star which would then slightly increase the star's luminosity [55].

The main problem of MACHOs, is that they can not be the dominant contributor to DM as there are not enough of stellar remnants to completely solve the DM issue [56, 57]. Furthermore, Big Bang could not have produced enough baryons and still be consistent with the observed elemental abundances [58] including the abundance of deuterium [59]. Together with observations from CMB, it is concluded that a large fraction of non-baryonic matter is required regardless of the existence of MACHOs.

#### 1.3.1.2 Primordial Black Holes

PBH are hypothetical types of black holes that were formed soon after the Big Bang [60]. These black holes could be small enough for Hawking radiation to be important [61], which however has not yet been confirmed experimentally. Black holes larger than $10^{15}$ g which are unaffected by Hawking radiation can have various astrophysical consequences and they could also be a part of DM. PBH should have formed before the era of Big-Bang nucleosynthesis, otherwise they would have been counted in Eq. 1.3 instead of Eq. 1.4. However, such an early creation of a large number of black holes is possible only in a few cosmological models [62]. This early creation can classify them also as non-baryonic and they could behave like any other form of CDM [63]. Finally, a large number of astrophysical observations constrain PBH allowing only a very narrow range of masses [64].

### 1.3.2 Axions

The existence of axions was first postulated to solve the strong $CP$ problem of QCD. They are pseudo Nambu-Goldstone bosons produced from the spontaneous breaking of a new global $U(1)$ symmetry, called *Peccei-Quinn symmetry*. However, the string theory also predicts such (pseudo-)scalar fields within a wide range of mass scales [65,66] called *Axion Like Particles (ALPs)*.

The axions are expected to be extremely weakly interacting with ordinary matter, which in turn implies that they were not in thermal equilibrium in the early universe. Although they are very light, axions can constitute CDM as they were produced non-thermally. The calculation of the axion relic density depends on the assumptions made on their production mechanism but there is still an acceptable range where axions satisfy all present-day constraints and at the same time represent a good DM candidate [67].

For relic axions to account for the right amount of CDM, their mass should be in the *classic axion window* of $10^{-6} - 10^{-3}$ eV. Much lower masses on the order of neV are still possible in fine-tuned models called *anthropic axion window*. An example is the ultralight axion with mass on the order of $10^{-22}$ eV, which can resolve the core-cusp problem on sub-galactic scales [68]. Finally, also masses above the classic window are possible, but do not constitute a solution to the DM problem.

More details about the axions will be provided in Chap. 21.

### 1.3.3 WIMPs

The most widely-studied DM candidates are the WIMPs ($\chi$). These particles interact with the SM fields only via the weak nuclear force, thus being non-baryonic and electrically neutral. WIMPs should carry a conserved quantum number for them to be stable on cosmological timescales. Usually, for this to happen, the WIMP have to be the lightest member of a matter sector which is charged under a discrete symmetry, and often this symmetry needs to be essentially imposed by hand on the underlying theory. WIMPs should naturally acquire masses $m_\chi$ within a few orders of magnitude of the $SU(2)_L \times U(1)_Y$ electroweak symmetry breaking scale (a few GeV or TeV) since they are members of $SU(2)_L$ multiplets and no other symmetries are present to force their masses down [69]. This makes them sufficiently heavy to constitute CDM even if they have been produced thermally in the early universe. Examples of WIMPs are the *lightest neutralino* in supersymmetry [70], the *lightest Kaluza-Klein particle* which appear in models of universal extra dimensions [71, 72] and an *additional inert Higgs boson* [73].

#### 1.3.3.1 Lightest Supersymmetric Particle

The currently best motivated WIMP candidate is the Lightest Supersymmetric Particle (LSP) with exact $R$-parity. This conserved discrete symmetry arises naturally in some Grand Unified Theories (GUTs) and makes the LSP stable. The fact that it does not interact electromagnetically is achieved by having the LSP be a neutral particle, such as the lightest neutralino or sneutrino (SUSY partner of SM neutrinos). The sneutrino is strongly constrained due to its large nuclear-scattering cross-section which disagrees with limits found by direct DM detection experiments [74] (see also Sect. 1.4.1.1). However the neutralino remains arguably the leading candidate for DM.

Other candidates include axinos (SUSY partner of axion) [75, 76] and gravitinos (SUSY partner of graviton) however they arise in only a subset of supersymmetric scenarios and have some unattractive properties (see [70] and references therein). For example, gravitinos can be overproduced in the early universe if the temperature of the reheating epoch is not sufficiently low. Also, their presence can destroy the abundance of primordial light elements in some scenarios. Despite existing solutions, axions and gravitinos have very weak interactions and would be practically impossible to detect [77].

### 1.3.4 Neutrinos

#### 1.3.4.1 SM neutrinos

Neutrinos were considered, until recently excellent DM candidates as they interact weakly with normal matter, they don't have an electric charge and they also have a small mass. They are also long-lived, abundant in the universe but most importantly they are known to exist within the SM. If a light $m_\nu \leq 100\,\text{eV}$ Dirac neutrino exists, its cosmological density would be:

$$\Omega_\nu {h_0}^2 \simeq \frac{\sum_i m_i}{94\,\text{eV}} \tag{1.7}$$

Using current cosmological upper bounds on the neutrino mass ($\sum m_\nu < 0.59 - 0.66\,\text{eV}$), the upper bound on the contribution of light neutrinos becomes [78]:

$$\Omega_\nu {h_0}^2 \leq 0.0062 \tag{1.8}$$

However, comparing with Eq. 1.4 this is a very small fraction of the total DM. In order to have the correct abundance, the sum of masses (calculated from Eq. 1.7) should have been on the order of $\sum m_\nu \sim 11.3\,\text{eV}$. This limit is called the *Gerstein-Zeldovich limit* [79], which however is in conflict with the experimental limits.

Finally, as the mass of neutrinos is much smaller than their decoupling temperature, they decouple relativistic and become non-relativistic only deeply in the matter-dominated epoch. But N-body simulations of structure formation in a universe dominated by such HDM candidates cannot reproduce the observed structure of the Universe today [80]. Therefore, massive SM neutrinos cannot be simultaneously astrophysical and cosmological DM meaning to account for the missing mass in the galaxies and to contribute to the cosmological expansion.

#### 1.3.4.2 Sterile neutrinos

A solution to the above problems is to postulate the existence of heavier *sterile neutrinos* $\nu_s$ with weaker interactions that fulfil the constraints from cosmic structure formation and phase-space densities. Their existence is also predicted by many theories as they would provide a very simple explanation of the observed neutrino oscillations via the seesaw mechanism. They are similar to the SM neutrinos, but they don't interact through the SM weak interactions [81–83]. However, they communicate with the rest of the neutrino sector through fermion mixing [84, 85].

Sterile neutrinos were first proposed as DM candidates in 1993 [81] and their most stringent cosmological and astrophysical constraints come from the analysis of their cosmological abundance and the study of their decay products [86]. The implications of the existence of such heavy neutrinos depend strongly on the magnitude of their mass [87, 88]. Sterile neutrinos of mass of a few keV and small mixing are stable particles (with $\tau_{\nu_s} \gg 10^{17}\,\text{s}$) and therefore are viable DM candidates [89].

### 1.3.5 Magnetic monopoles

#### 1.3.5.1 Dirac monopoles

In 1931 Dirac introduced the magnetic monopole in order to explain the quantisation of the electric charge, which follows from the existence of at least one free magnetic charge [90]. The relation between the elementary electric charge $e$ and the basic magnetic charge $g$ is:

$$g = \frac{n\hbar c}{2e} \tag{1.9}$$

where $n$ is an integer $n = 1, 2, 3, ....$ This is known as the *Dirac quantisation condition*. The magnetic charge is therefore $g = ng_D$, where $g_D = \hbar c/2e = 68.5e$ is called the *unit Dirac charge* [91]. The existence of magnetic charges and magnetic currents would provide a symmetry on Maxwell's equations but not a perfect one, since numerically $e \neq g$. The minimum mass of the Dirac monopole (for $n = 1$) can be derived from Eq. 1.9 and is in the order of the proton mass $m_p$. More specifically [92]:

$$m_M = \left(\frac{g}{e}\right)^2 m_e \approx 2.56 m_p \approx 2.4\,\text{GeV}/\text{c}^2 \tag{1.10}$$

So far no such Dirac monopoles have been discovered.

#### 1.3.5.2 GUT monopoles

Magnetic monopoles were also existent in GUTs of the strong and electroweak interaction. More specifically, in the broken phase of a GUT there are typically localised classical solutions carrying magnetic charge under an unbroken $U(1)$ symmetry [93,94]. These magnetic monopoles (also called *'t Hooft–Polyakov monopoles*) can be produced during a possible GUT phase transition in the early universe [95] and can have an enormous mass:

$$m_M > 10^{16} - 10^{17}\,\text{GeV} \tag{1.11}$$

These masses can not be produced at any man-made accelerator. Even larger masses are expected if gravity is brought into the unification picture as well as in some SUSY models. The velocity spectrum of these monopoles is usually in the range $4 \times 10^{-5} < \beta < 0.1$ (where $\beta = v/c$) with possible peaks corresponding to the escape velocities of the Earth, the Sun and the galaxy.

If the $U(1)$ group would appear in a later phase transition in the early universe, this could have lead to the production of intermediate mass monopoles which could be multiply

charged and have masses:

$$m_M \sim 10^5 - 10^{13}\,\text{GeV} \tag{1.12}$$

These monopoles may be accelerated to relativistic velocities $\beta \geq 0.1$ by the galactic magnetic field and by several astrophysical sites. Very energetic monopoles could also be behind the highest energy cosmic rays [96]. An important characteristic is that the lowest mass magnetic monopoles should be stable since the magnetic charge is conserved like the electric charge. Therefore, magnetic monopoles that are produced in the early universe should still exist as cosmic relics, whose kinetic energy has been affected by the expansion of the universe as well as their travel through galactic and intergalactic magnetic fields. This means that magnetic monopoles can be components of DM. However, the basic problem of monopoles is their predicted density today which is in contrast with the experimental evidence thus giving rise to the *cosmological monopole problem* [97, 98].

### 1.3.6 Alternative proposals

Other candidate proposals [70] for DM include particles [99–102] derived from *little Higgs* models [103–106], which pose an alternative mechanism to supersymmetry in order to stabilise the weak scale, *Q-balls* [107, 108], *mirror particles* [109–113], *CHArged Massive Particles (CHAMPs)* [114], *Self-Interacting Dark Matter (SIDM)* [115], *D-matter* [116], *cryptons* [117, 118], *super-weakly interacting DM* [119], *brane world DM* [120], *heavy fourth generation neutrinos* [121, 122].

Moreover, in addition to WIMPs, there are the Gravitationally-Interacting Massive Particles (GIMPs), which are singular structures in spacetime, embedded within a geometry whose average forms the DE. They do not couple to any of the standard elementary particles via the gauge fields of strong, electromagnetic or weak interactions but are subject only to gravitational interactions thus making them good DM candidates [123, 124].

An alternative also scenario for DM includes the generic term Strongly-Interacting Massive Particles (SIMPs) whose interactions with nucleons have large cross-sections [125, 126]. In several models within SIDM, such particles could also explain the astrophysical problem of small-scale structure formation [127].

Other hypothetical particle candidates from the hidden sector are *dark photons* [128] which are neutral light vector bosons that can kinetically mix with the ordinary photos. Therefore, they should be produced in any process in which a virtual or real photon is involved (see [129] for a review).

AntiQuark Nuggets (AQNs) are also possible DM particle candidates. The are massive objects which can still satisfy the astrophysical and cosmological constraints but can also explain the matter-antimatter asymmetry [130–132]. This kind of candidate belongs to a more generic class of antimatter macroscopic bound states which can account for DM. These candidates called *macros* interact with ordinary matter primary through annihilation with large cross-sections [133].

Finally, there is also a wide search on on the validity of the Newton law of gravity at very large distances such as Modified Newton Dynamics (MOND) [134, 135], which can reproduce many observations on galactic scales, such as galactic rotation curves without the need of DM. However, the basic problem with MOND is that it is a purely non-relativistic theory. There have been several attempts to embed it in a relativistic theory but this requires the existence of additional fields [136, 137].

## 1.4 Detection of Dark Matter particles

### 1.4.1 Direct detection

#### 1.4.1.1 WIMPs

Since WIMPs interact with SM particles via the weak force, they should have weak-scale scattering cross sections with normal nuclei. Therefore, one of the most promising ways to detect WIMPs is to look for nuclear recoils in large-volume target materials on Earth [138]. The expected WIMP masses are in the range of 10 GeV to 10 TeV, resulting to typical nuclear recoil energies on the order of 1 to 100 keV.

The expected interaction rates depend on the product of the local WIMP flux and the interaction cross-section. The local flux depends on the local density of DM, the mean WIMP velocity, the galactic escape velocity, and the mass of the WIMP. The expected interaction rate depends on the mass and the cross-section of the WIMP, both of which are unknown. The cross-section in turn depends on the nature of the couplings, which means that spin-dependent and spin-independent couplings are available for non-relativistic WIMPs. Cross-section calculations in Minimal Supersymmetric Standard Model (MSSM) usually provide rates of at most 1 evt/d/kg of detector, which is much lower than the usual radioactive backgrounds. This is why most laboratories are underground to shield against cosmic rays and select extremely radio-pure materials [78].

Due to the proper motion of the solar system within the galactic halo, there should be a mean net velocity between WIMPs and the Earth. This means that there are two different experimental signatures that are expected for WIMPs. The first one is a strong daily forward/backward asymmetry of the nuclear recoil direction due to the alternate sweeping of the WIMP cloud by the rotating Earth. This effect can be detected by gaseous detectors, anisotropic response scintillators and extremely fine grain solid-state detectors. The second effect is a few percent annual modulation of the recoil rate due to the non-perpendicularity of the solar ecliptic and galactic planes resulting in the Earth's speed ($\sim 30\,\mathrm{km/s}$) adding to or subtracting from the speed of the Sun ($\sim 220\,\mathrm{km/s}$) (see Fig. 1.5). This effect can only be detected with large masses, together with an identification of the nuclear recoil as the much larger background can also undertake seasonal modulation.

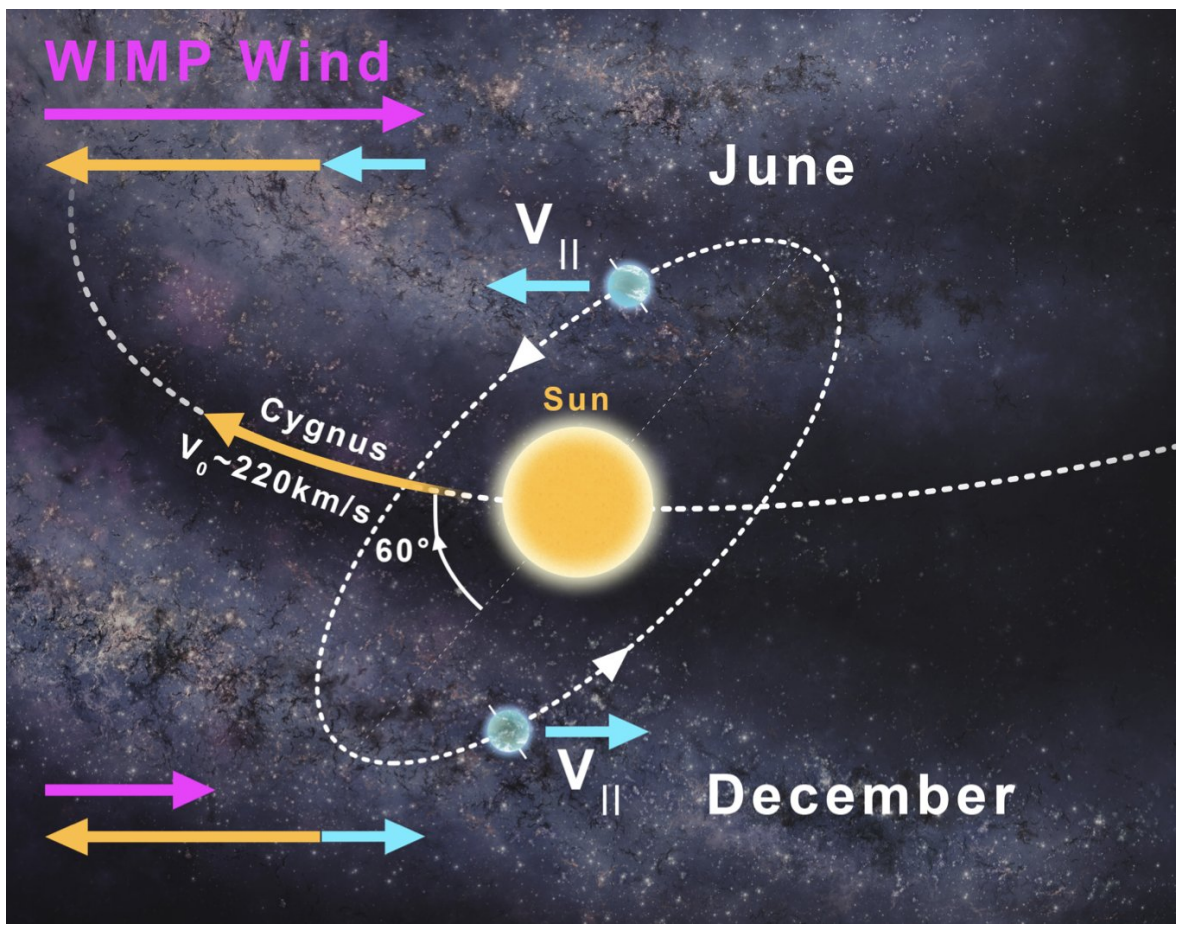

**Figure 1.5:** Illustration of the rotation of the Earth around the Sun as well as the Sun around the GC, resulting in a WIMP wind with a predictable annual variation of its intensity. (6).

In Fig. 1.6 the upper limits for WIMP scattering cross-sections, normalised to scattering on a single nucleon, for spin-independent (Fig. 1.6a) and spin-dependent (Fig. 1.6b) couplings are plotted as a function of the WIMP mass. The constraints from indirect observations and collider searches are also presented.

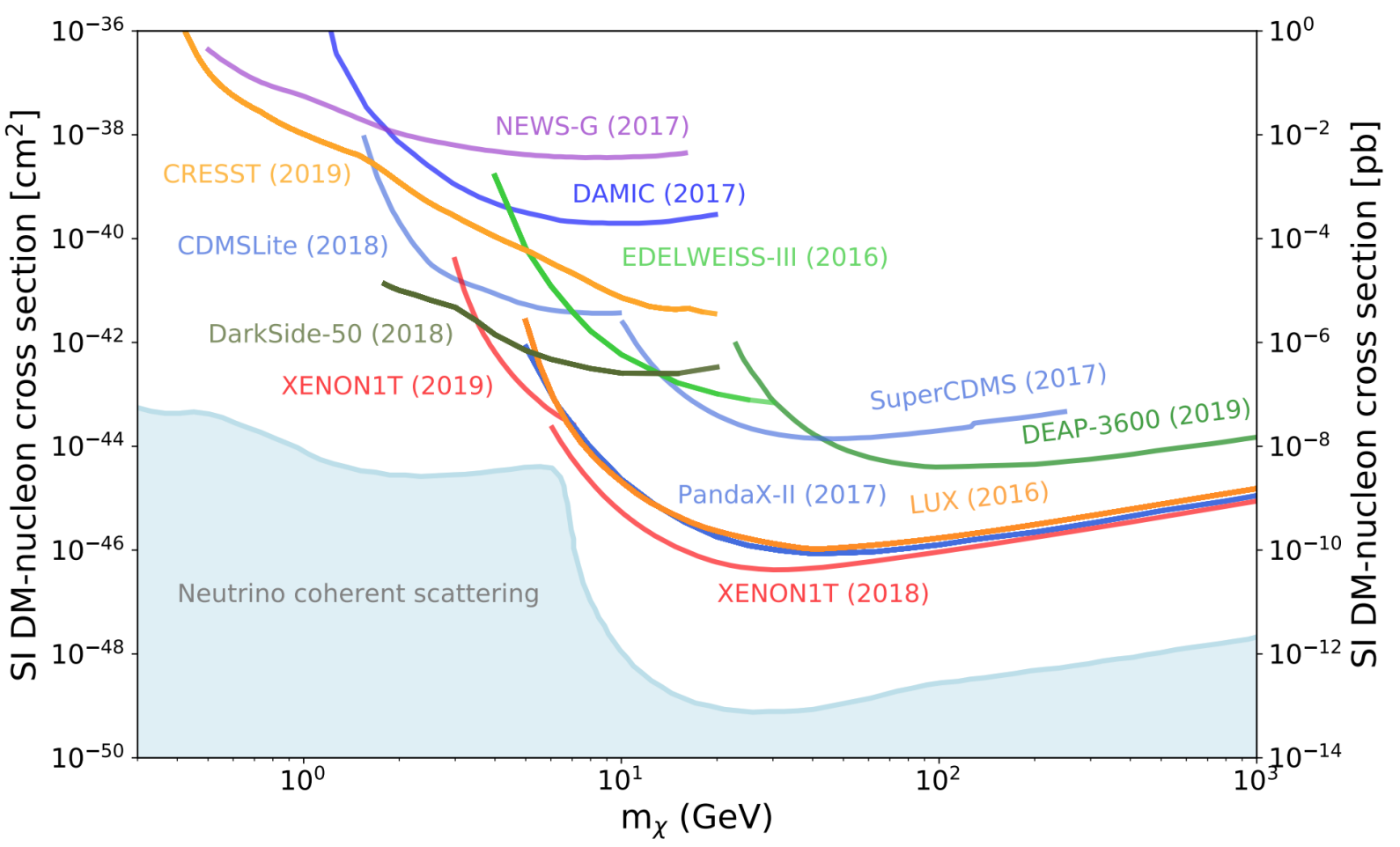

**(a)** WIMP cross-sections for spin-independent WIMP-nuclon coupling as a function of mass (45).

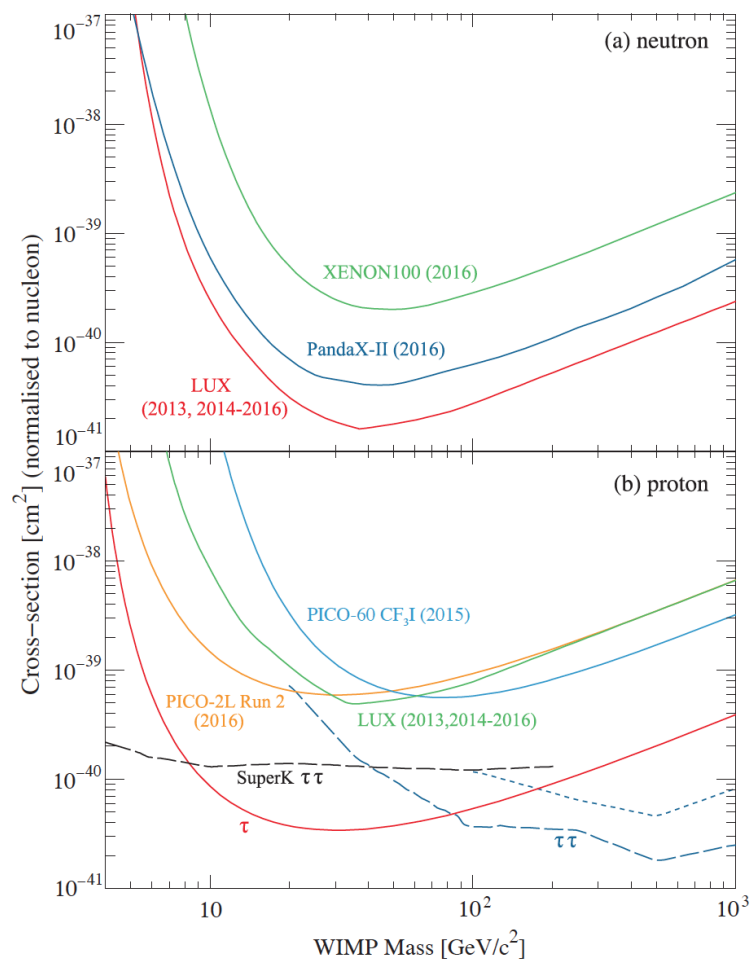

**(b)** WIMP cross-sections for spin-dependent coupling as a function of mass. (a) shows interactions with the neutron while (b) shows interactions with the proton (54).

**Figure 1.6:** Latest results from WIMP searches.

#### 1.4.1.2 Axions

Axion detection is based on the conversion of an axion into photon in a strong magnetic field (see Sect. 21.3). Experiments looking for DM axions are cavity experiments which attempt to detect halo axions, by tuning the frequency of a magnetic field to the axion mass and searching for microwaves from resonant conversion (see Sect. 21.3.2). Astrophysical limits can also be acquired by considering the maximum amount of energy that they could carry out of supernovae and stellar cores without exceeding the observed cooling rates.

#### 1.4.1.3 Magnetic monopoles

GUT and intermediate-mass monopoles, given their large expected masses, can only be searched for as relic particles from the early universe in the cosmic radiation. Magnetic monopoles of cosmic origin must have been formed shortly after the Big Bang, as topological defects when the universe expanded and cooled. However, the existence of the galactic magnetic field ($\sim 3\,\mathrm{mG}$) would accelerate such monopoles, thus draining energy from the magnetic field. Therefore, for the galactic magnetic field to be sustained, its dissipation must not exceed its regeneration, which implies an upper flux limit called *Parker bound* [139]:

$$\Phi \leq 10^{-15}\,\mathrm{cm}^{-2}\,\mathrm{s}^{-1}\,\mathrm{sr}^{-1} \tag{1.13}$$

The only experiment ever reporting a single magnetic monopole (with high energies $10^{15} - 10^{16}\,\mathrm{GeV}$) event, coming from cosmic rays, was the experiment of Blas Cabrera in 14/02/1982 which consisted of a four-turn, 5 cm-diameter loop with its axis vertically oriented, connected to the superconducting input coil of a Superconducting Quantum Interference Device (SQUID) [140]. The simple idea was that if a monopole passed through the loop it would induce a current whose amplitude is proportional to the magnetic charge. The magnetic flux of a monopole passing through the loop is given by $4\pi g = \frac{hc}{e}$. The result was not able to be repeated and confirmed by other experiments. However if the event is considered to be spurious, these data would set an upper limit of $6.1 \times 10^{-10}\,\mathrm{cm}^{-2}\,\mathrm{s}^{-1}\,\mathrm{sr}^{-1}$ which is much larger than the Parker bound.

The interaction of a GUT monopole core with a nucleon can lead to a reaction in which the nucleon decays. The cross-section for this process is on the order of magnitude of the core size, $\sigma \sim 10^{-56}\,\mathrm{cm}^2$ which is practically negligible. However, the catalysis process can proceed with a cross-section of the order of the strong interaction cross-section [141, 142]. The Monopole, Astrophysics and Cosmic Ray Observatory (MACRO) experiment, has set an upper limit to the flux of the local monopoles of $1.4 \times 10^{-16}\,\mathrm{cm}^{-2}\,\mathrm{s}^{-1}\,\mathrm{sr}^{-1}$ for $4 \times 10^{-5} < \beta < 1$, which is well below the Parker limit in almost all the $\beta$ range for GUT monopoles [143] (see Fig. 1.7).

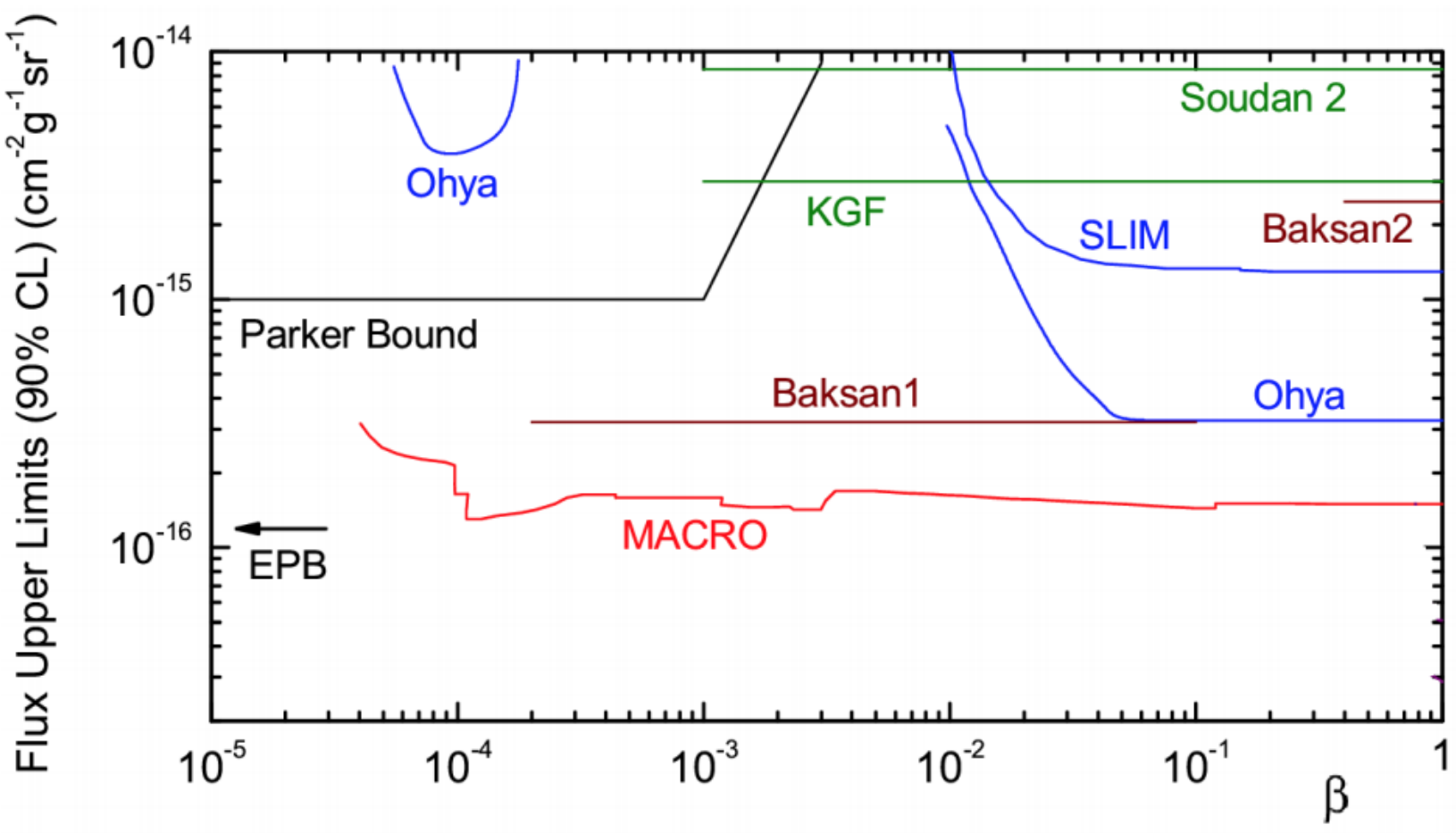

**Figure 1.7:** Upper limits versus velocity $\beta$ for a flux of cosmic GUT monopoles with magnetic charge $g = g_D$ (7).

## 1.4.2 Indirect detection

### 1.4.2.1 MACHOs

The only way to detect hidden galactic baryonic matter in the form of MACHOs is through microlensing effects. The main surveys for the detection of such events, by monitoring the luminosity of millions of stars in the Large and Small Magellanic Clouds for several years, were carried out by the MACHO [144], Expérience pour la Recherche d'Objets Sombre (EROS) [145], and Optical Gravitational Lensing Experiment (OGLE) [146] collaborations. MACHO results limited these events to about $f \sim 20\%$ of the halo fraction made of dark mass objects of $\sim 0.4 M_\odot$ but the issue is basically left open [147] as the EROS collaboration put an upper limit on the halo fraction of $f < 0.8$ for $\sim 0.4 M_\odot$. Future campaigns [148–150] are under way to further address this issue.

### 1.4.2.2 WIMPs

If WIMPs are thermal relics, they should by default have weak-scale self-annihilation cross-sections no matter if the particles are Majorana or Dirac. If DM was produced non-thermally it would also have a comparable or even higher self-annihilation cross-section. Therefore, WIMPs should annihilate at a non-vanishing rate today, except from the case of a non-thermally produced Dirac particle with an initial mater-antimatter asymmetry.

The primary or secondary annihilation products of WIMPs are the ones that can be detected indirectly. These products include neutrinos, gamma rays, positrons, antiprotons and antinuclei. Indirect searches can place constraints on annihilation cross-sections, in annihilating models and on the lifetime in decaying models [69].The most promising targets, whether looking for decays or annihilations, are the ones with large DM densities and/or low astrophysical backgrounds. Such typical targets are the Galactic Centre (GC), dwarf galaxies, and galaxy clusters.

Finally, WIMPs can be slowed down, captured or even trapped in celestial objects like the Earth or the Sun, thus enhancing their density and their probability of annihilation. This is a source of muon neutrinos that can interact in the Earth. Upward going muons can then be detected in large neutrino telescopes. However, for standard halo velocity profiles, only the limits from the Sun, which probe mostly spin-dependent couplings are competitive with limits from direct WIMP searches (see Sect. 1.4.1.1).

#### 1.4.2.3 Sterile neutrinos

Sterile neutrinos with $m_{\nu_s} \sim$ keV have a lifetime larger than the age of the universe but they do decay into a lighter neutrino state and a photon ($\nu_s \rightarrow \gamma\nu$). These mono-energetic photons will have an energy $E_\gamma = m_{\nu_s}/2$. Therefore, the most promising way to search for sterile neutrinos is with X-ray observatories. Furthermore, the flux of the X-rays depends on the sterile neutrino abundance [151].

An unidentified feature in the X-ray spectra of galaxy clusters as well as Andromeda and the Milky Way galaxies has been reported by several groups leading a signal which can be interpreted as the decay of a DM particle with $\sim 7\,\mathrm{keV}$ mass [152–154] (see Fig. 1.8a). This was suggested to be a sterile neutrino with a mixing angle $\sin^2 2\theta \simeq (0.2-2)\times 10^{10}$. Several discussions were made whether this 3.5 keV X-ray spectral line could be a statistical fluctuation, an unknown astrophysical line or an instrumental feature, but the results are not yet conclusive (see for example [89] and references therein). Moreover, the existence of this line was not confirmed by data from other missions [155, 156].

Finally, in addition to their impact on small-scale structures and the x-ray spectrum, sterile neutrinos also have astrophysical effects on the velocity distribution of pulsars and the formation of the first starts. The most recent combined exclusion limits if sterile neutrinos make up all of DM are shown on Fig. 1.8b.

#### 1.4.2.4 Magnetic monopoles

Indirect searches for magnetic monopoles include a variety of searches in matter as for example when analysing a piece of matter which was exposed at an accelerator beam or cosmic rays. Large neutrino telescopes also performed searches for very fast, relativistic, intermediate-mass magnetic monopoles searching for Cherenkov light being emitted in water or ice. The idea is that when traversing through a medium such as water or ice, relativistic monopoles would lose some of their energy due to Cherenkov radiation. When the monopole speed exceeds the group velocity of light in that particular medium, photons are emitted from excited atoms in the medium. Electrically charged particles also give rise to Cherenkov

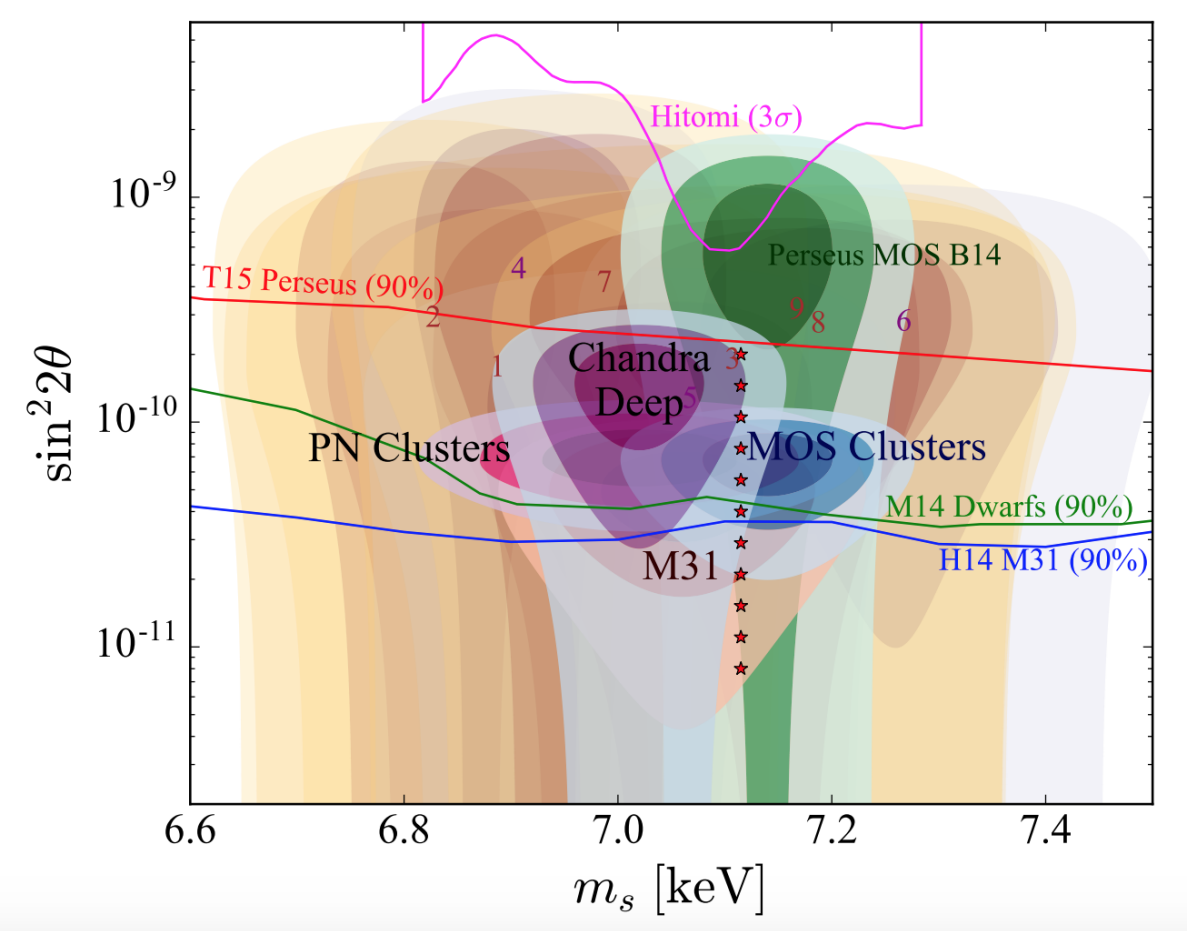

(a) X-ray detections consistent with sterile neutrino DM.

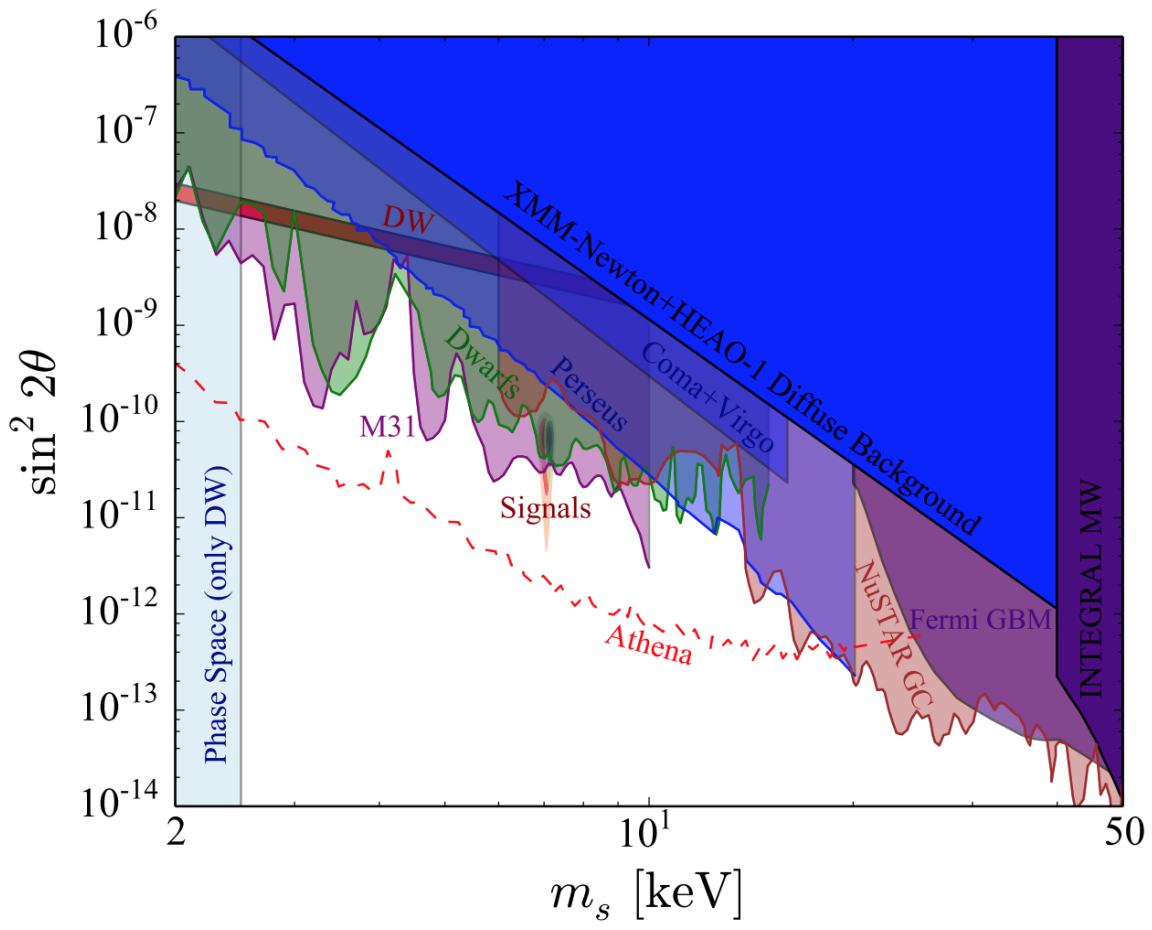

(b) Constraints in the full parameter space of neutrino DM. The red dashed line shows the forecast sensitivity of the planned Athena X-ray telescope.

**Figure 1.8:** Latest results from sterile neutrino searches (8).

radiation, but the number of photons emitted is much smaller in that case. For example, in water and ice, a monopole with $g = g_D$ generates 8300 more photons than a particle with one electric unit charge $e$ travelling at the same speed [157]. The most recent results from neutrino telescopes are shown in Fig. 1.9.

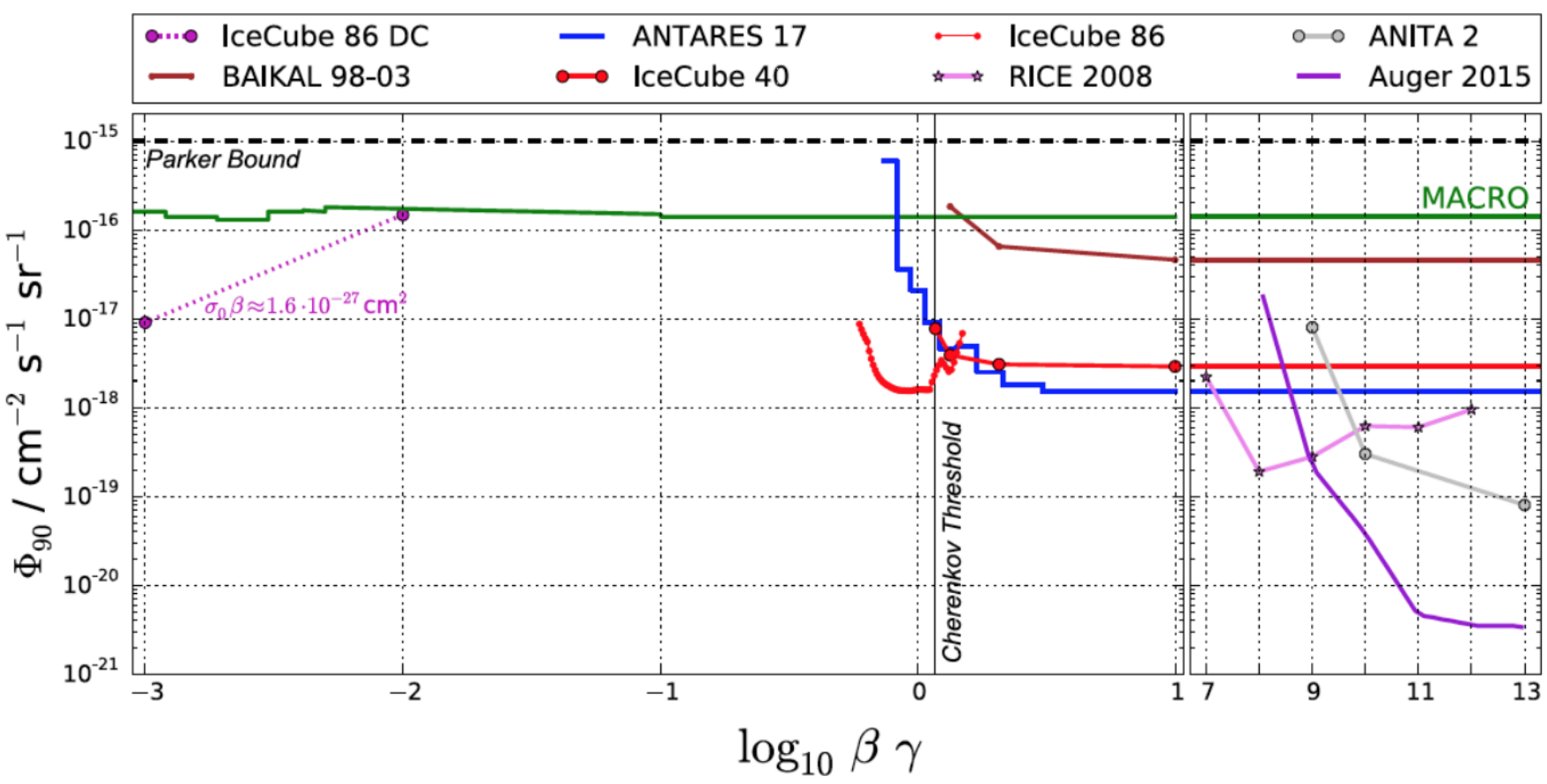

**Figure 1.9:** Current upper limits as a function of velocity $\beta$ of very energetic cosmic GUT monopoles with $g = g_D$ (9).

## 1.4.3 Collider searches

### 1.4.3.1 WIMPs

Due to their extremely weak interaction with SM particles, which constitute the detectors, a visible direct signal in collider experiments is impossible. Therefore, the usual indirect signature for WIMPs in accelerators is a missing transverse energy $E_T$ which refers to an apparent missing component of the total final-state momentum in the transverse to a collider beam direction. If the sum of the outgoing transverse momenta of a reaction is not zero, as they do in the original beam, this could be the signature of the production and escape of a massive particle with a very small interaction cross-section with the material of the detector. While the particle should be stable enough to go beyond the detector, the stability on cosmological scales will be unknown which means that their relic abundance can not be calculated [69].

In Fig. 1.10 the limits on spin-independent DM nucleon cross-section is shown from analysis of the A Toroidal LHC Apparatus (ATLAS) detector of Large Hadron Collider (LHC) compared with direct detection results.

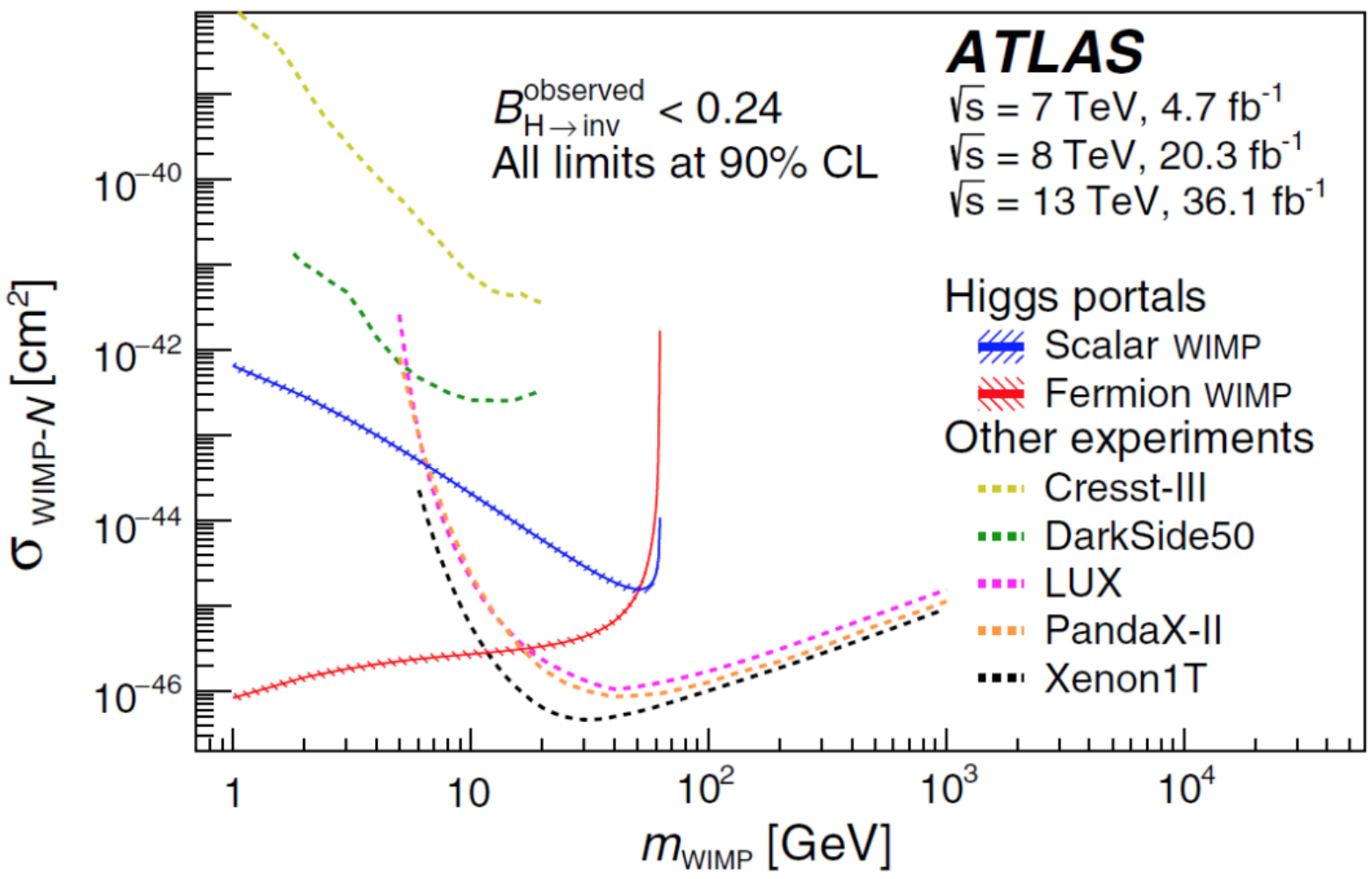

**Figure 1.10:** Comparison of the upper limits obtained by ATLAS compared with several direct detection results on WIMP-nucleon scattering cross-section. The regions above the contours are excluded (10).

#### 1.4.3.2 Magnetic monopoles

Several searches have been performed at hadron-hadron, electron-positron and lepton-hadron colliders mostly directly using scintillation counters, gas chambers and nuclear-track detectors taking advantage of the monopole high ionisation power (see [157] for a recent review). For example, magnetic monopoles formed in collisions in LHC, can go through the Monopole and Exotics Detector at the LHC (MoEDAL) detector, breaking long-chain molecules in the plastic nuclear-track detectors thus creating a minute trail of damage through all 10 sheets of plastic nuclear-track detectors. A clear indication of the path of a monopole would be an aligned set of holes with the trajectory pointing back to the collision point. In Fig. 1.11, an overview of the monopole mass limits set by collider experiments is presented.

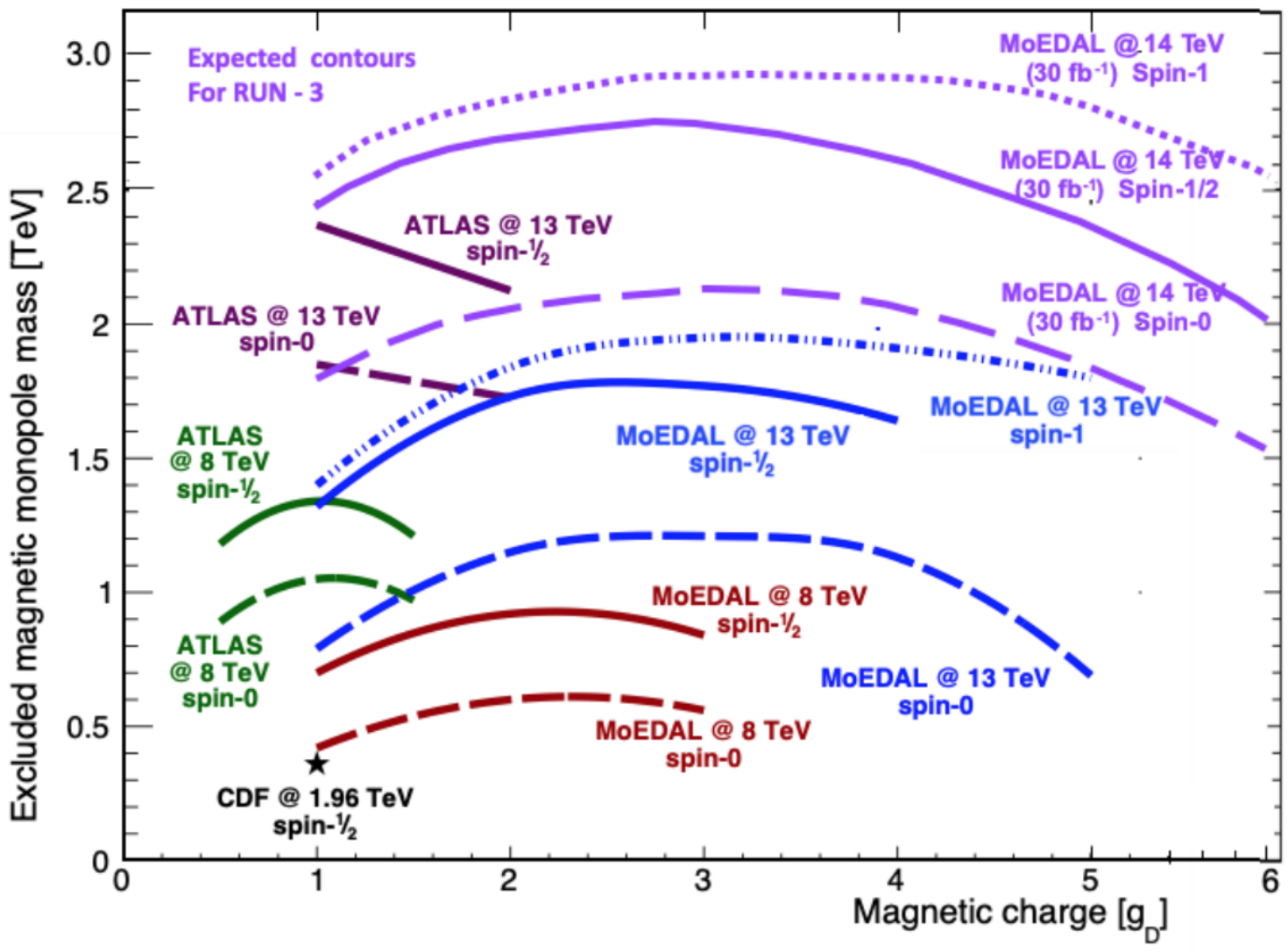

**Figure 1.11:** Excluded magnetic monopole masses for pair produced monopoles at collider experiments with MoEDAL latest result (blue) and expected results from RUN-3 (purple) (11).

# 2
# STREAMS

## 2.1 Introduction

As we have seen in Sect. 1.1 most of the matter in the universe is "dark" and assumed to be distributed in diffuse halos around galaxies. In the Standard Halo Model (SHM) it is also assumed that there is an isotropic halo distribution on our galaxy with a density of $\rho \sim 1/r^2$ and with the local DM density being around 0.2–0.6 $\mathrm{GeV/cm^3}$ [158]. However, as we will see, a much higher value could also be possible with the isotropy assumption also being debatable.

Multiple results for N-body and hydrodynamical simulations show deviations from the putative smooth and isotropic SHM [159] which point to the existence of tidal streams [160], a dark disk [161,162] and debris flows [163,164]. These could have a significant effect on the local DM density and velocity distribution, and therefore also to direct DM experiments. Other cases including miniclusters with masses $M \sim 10^{-12} M_\odot$ [165, 166] or caustics [167] could also enhance the local DM flux in our solar neighbourhood (see also Sect. 21.4.1).

It is noted, that a conservative upper bound on the possible DM density in the solar system can be set by high precision positional observations on planets and spacecrafts [168, 169]. More specifically, for the orbital distances of Earth, Mars and Saturn studies have shown that $\rho_{DM}$ is less than $1.4 \cdot 10^{-19}\,\mathrm{g/cm^3}$, $1.4 \cdot 10^{-20}\,\mathrm{g/cm^3}$ and $1.1 \cdot 10^{-20}\,\mathrm{g/cm^3}$ respectively. At the same time the total mass of DM in a sphere within the orbit of Saturn has to be less than $1.7 \cdot 10^{-10} M_\odot$ [169]. Finally, assuming a spherical DM halo centred on the Sun, the planetary motions can derive a maximum allowed DM density at the location of the Earth, on the order of $10^5\,\mathrm{GeV/cm^3}$ [168].

## 2.2 Tidal streams

### 2.2.1 Stellar streams

Stellar streams are associations of stars grouped in elongated filaments arcing around a host galaxy. These streams originate from the tidal disruption of globular clusters or satellite dwarf galaxies from their host galaxy (see Fig. 2.1). They have more or less a uniform stellar density along their length as well as small velocity dispersions. They are found throughout the galactic halos of the Milky Way, Andromeda and other nearby massive galaxies [170].

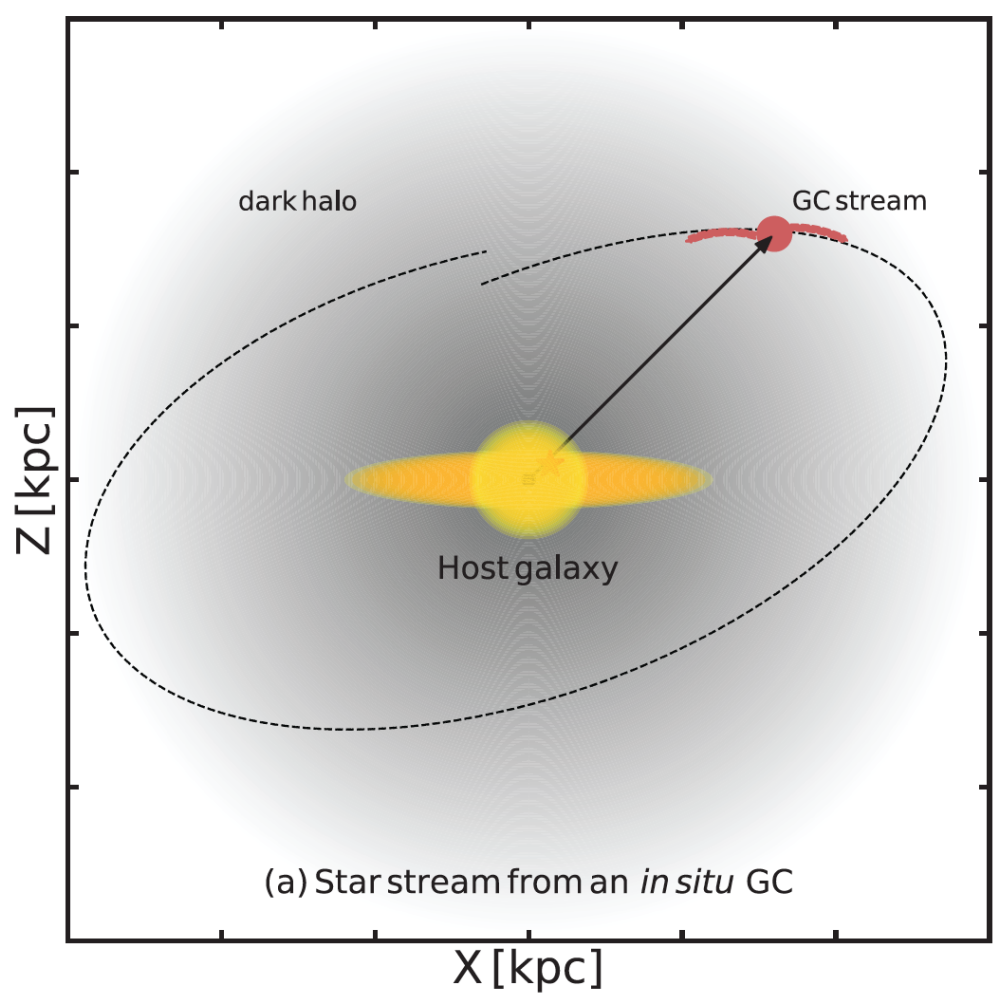

**Figure 2.1:** Schematic illustration of the formation of a stellar stream from a globular cluster orbiting a Milky Way type host galaxy (12).

When a stellar stream encounters a DM sub-halo, its structure is perturbed producing a gap in the otherwise relatively smooth stream density (see Fig. 2.2) [171]. An alternative explanation for the creation of these gaps is due to the interaction with a massive dark object on the order of $10^6 - 10^8 M_\odot$ [172].

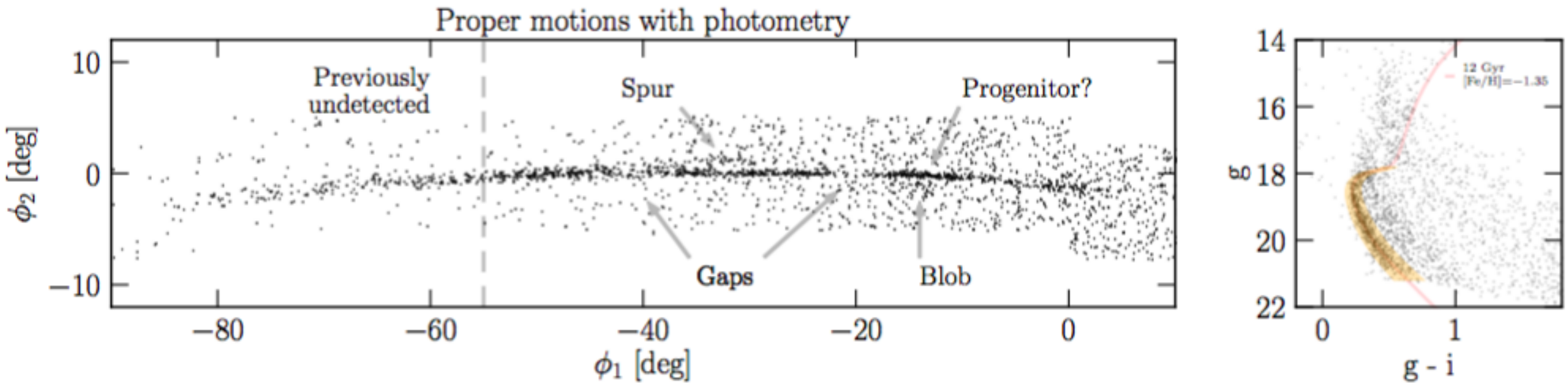

**Figure 2.2:** GD-1 stream containing high (blobs, spurs) or low (gaps) number of stars along the spatial distribution of the stream (13).

Based on the density variations on these streams, the abundance of DM halos can also be constrained and eventually the nature of DM itself can be inferred [172, 173]. As outlined in Sect. 1.2.2 different theoretical expectations exist, for example for the CDM and WDM models.

With the Global Astrometric Interferometer for Astrophysics (GAIA) data releases, several stellar streams have started to emerge and more are expected. The GAIA mission is the follow-up astrometric survey to the HIgh Precision PARallax COllecting Satellite (Hipparcos) mission (1989 – 1993) which provides measurements and full-space coordinates for over a billion stars aiming to chart a three-dimensional map of the Milky Way while revealing the composition, formation and evolution of our galaxy. Until today more than 60 long and thin stellar streams have been discovered in the Milky Way [174] (see Fig. 2.3).

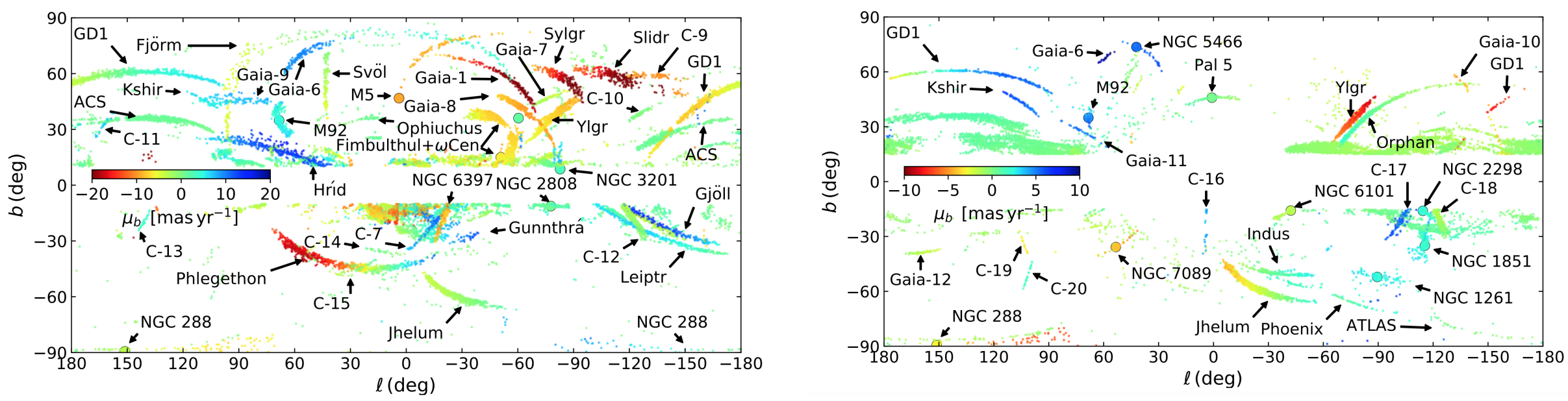

**(a)** Heliocentric distance range [3, 12]kpc.

**(b)** Heliocentric distance range [10, 30]kpc.

**Figure 2.3:** GAIA map from EDR3 data of the proper motion distribution. The likelihood detection threshold for stars being stream members is set at $> 10\sigma$ (14).

#### 2.2.1.1 Sagittarius stream and others

The most prominent and brightest known stellar structure on the Milky Way is the Sagittarius (Sgr) stream (see Fig. 2.4a). It is a complex structure made of tidally stripped stars as well as DM deriving from the Sgr dwarf galaxy due to the ongoing merging with Milky Way the last 5 Gy (see Fig. 2.4b). The observed orbit of the Sgr dwarf galaxy, which is one of the largest satellite galaxies orbiting the Milky Way, is nearly polar with the resulting tidal stream wrapping 360° across the sky [175]. Observations on the properties of the stars forming the Sgr stream can provide insights on the star formation history as well as the origin of the stream itself.

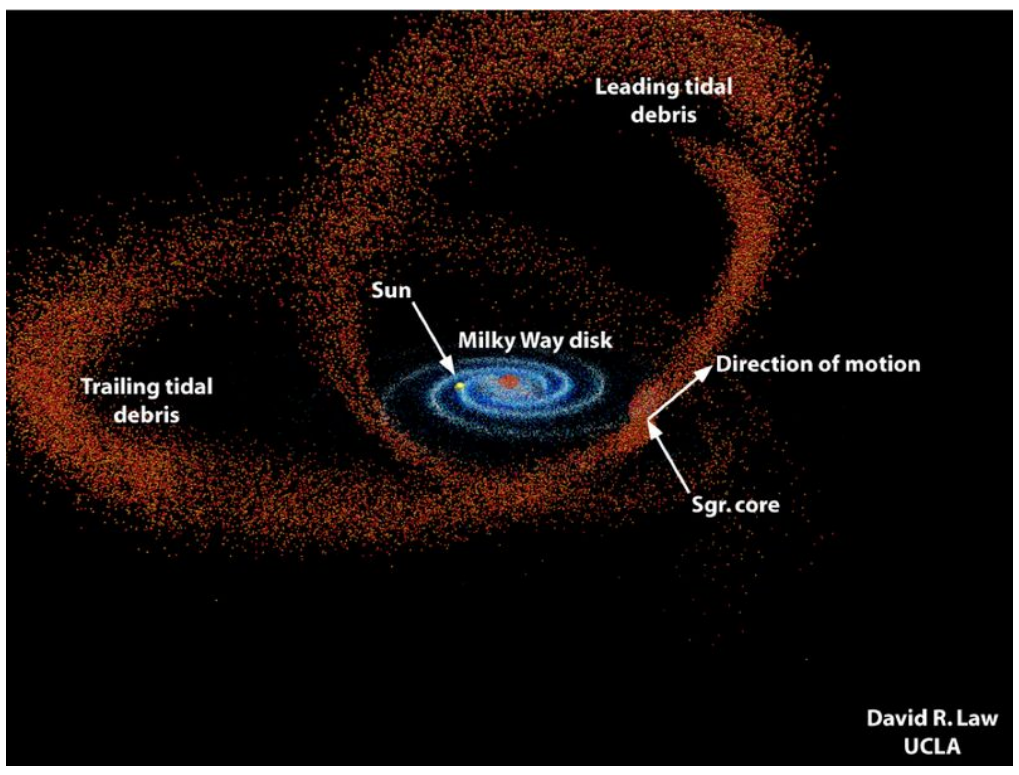

**(a)** The streams of stars left behind as the Sagittarius dwarf galaxy is torn apart by the gravitational potential of the Milky Way over a period of billions of years. Long tails extend in either direction from the parent galaxy (labeled Sgr core) and stretch across the northern and southern skies, wrapping in a full circle around the Milky Way (55).

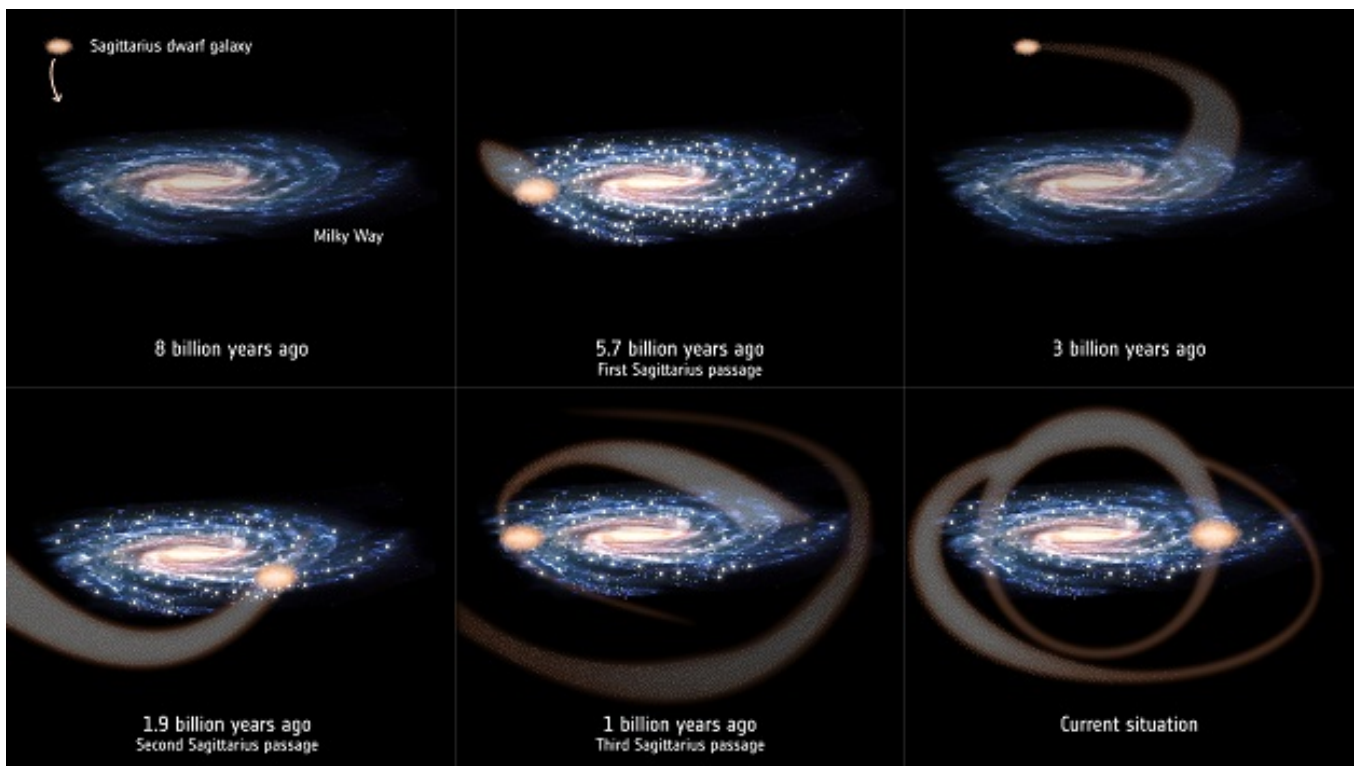

**(b)** Schematic of the subsequent passages of the Sgr dwarf galaxy as it orbits Milky Way and its impact on the Milky Way's star formation activity based on data from GAIA mission (56).

**Figure 2.4:** Sgr stream and its formation.

As already mentioned, there are several known examples of stellar tidal streams in the Milky Way halo except for the Sgr stream. Even before GAIA mission, the SDSS *field of streams* [176], contained a number of additional tidal stream candidates such as the *Monoceros* and *Orphan* streams. The Monoceros tidal stream is a low-latitude tidal stream in a ring-like structure whose progenitor, possibly the Canis Major dwarf galaxy, moves on a prograde nearly circular orbit in the Milky Way disk [177]. The Orphan stream, on the other hand, forms a smooth arc that is significantly longer, wider and farther from the GC than another of the commonly studied globular cluster streams [178].

Besides large over-densities such as the aforementioned ones, numerous thin streams have been discovered which cross the halo at various distances (see Fig. 2.3). These can provide additional information on the destruction of small dwarf galaxies and globular clusters which are useful for constraining the galactic potential and the distribution of mass at large radii [179].

Finally, an important relatively recent discovery, was that several halo streams exist also in our solar neighbourhood [180–182]. It is also mentioned that the existence of cosmic streams of normal matter in our solar system was already proven in the case of neutral Helium, which flows with a velocity peaking at about 25 km/s [183].

### 2.2.2 Dark Matter streams

Stellar streams are spatially aligned and have small velocity dispersions. Their presence strongly suggests that DM streams should also have formed with a similar mechanism, i.e. from material stripped from in-falling satellites. A DM stream is dynamically cold and has a one-dimensional morphology.

Even though stellar streams could be several kpc away from our solar neighbourhood, the associated DM stream could be significantly broader and thus have an influence on direct DM experiments. As an example, the DM content of a stream like Sgr is likely to be significantly more extended than its stellar content and could have an offset of as much as a few kpc [159]. Furthermore, the stellar and DM streams are not necessarily spatially coincident and can drift from co-axiality during evolution in the Milky Way's tidal field [184]. As a result, stellar and DM streams do not follow the same orbits (see Fig. 2.5) and therefore the proximity of the Sun to a stellar stream alone can not necessarily be used as an indicator of the stream contribution in the local DM density.

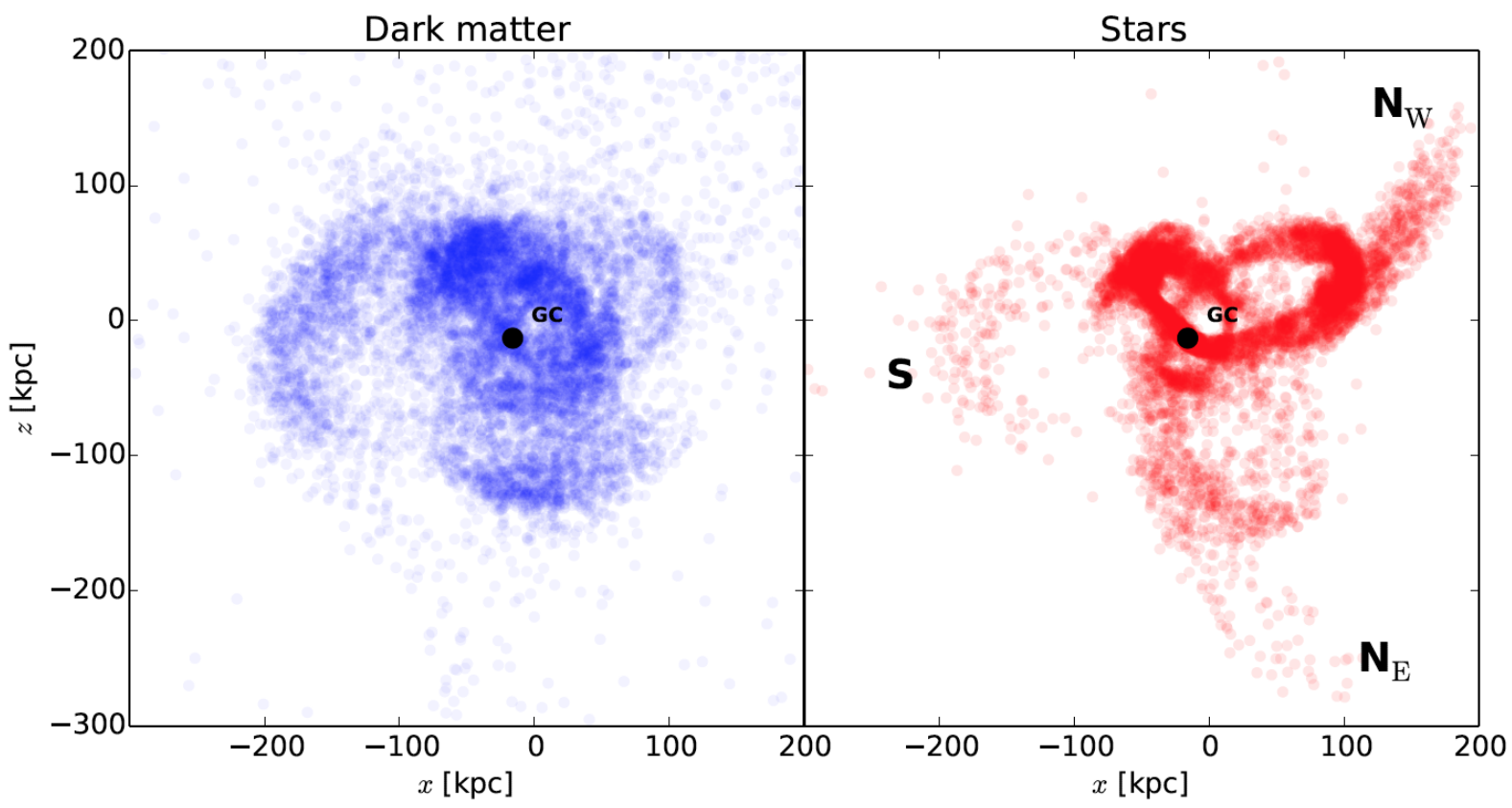

**Figure 2.5:** View of the Sgr debris stream where the projection plane is defined by its pole at Galactocentric coordinates $(l, b) = (275°, -14°)$. The locations of the Sun and the GC are marked by a filled black circle and the label GC, respectively (15).

The presence of such streams in the local neighbourhood can significantly affect the predictions of event rates and recoil spectra at direct detection experiments (presented in Sect. 1.4.1). Assuming a passage of a DM stream from the solar position, its density could be on the order of $\sim (0.3 - 23)$ % of the local density of our galactic halo [185, 186]. For the case of WIMPs this extra contribution would produce a steplike feature in the energy recoil spectrum.

While their existence is proved from N-body simulations, simulations show that the probability of a single stream dominating the local DM density is less than 1% [187]. Notably, other numerical simulations on the Sgr stream including specific structures, as the Sgr accretion and tidal stream indicate an increase in the high-energy recoil events of up to 45% depending on the local DM density contributed by the halo of Milky Way [184]. Finally, there are also other possibilities that could affect direct detection probabilities, such as the Earth lying within a sub-halo, or within a dense tidal stream created by disruption of an earlier sub-halo. These cases would also produce a spike in the velocity distribution of local DM particles [160].

## 2.3 Debris flows

Debris flow is an example of a kinematic substructure that is spatially mixed on large scales. It arises from the accretion of one or more older satellites that completed several orbital wraps. In this case, any structure in position space is washed out, while velocity-space features are preserved. Bellow their basic characteristics and formation will be outlined.

As mentioned in Sect. 1.1.6 the galactic DM halo forms through a process of hierarchical structure formation with smaller halos merging together to form a larger host in a period of billions of years [188]. While some of these DM halos merge completely and virialise, others, more recent mergers, do not and can continue to orbit the galaxy as satellites (see Fig. 2.6).

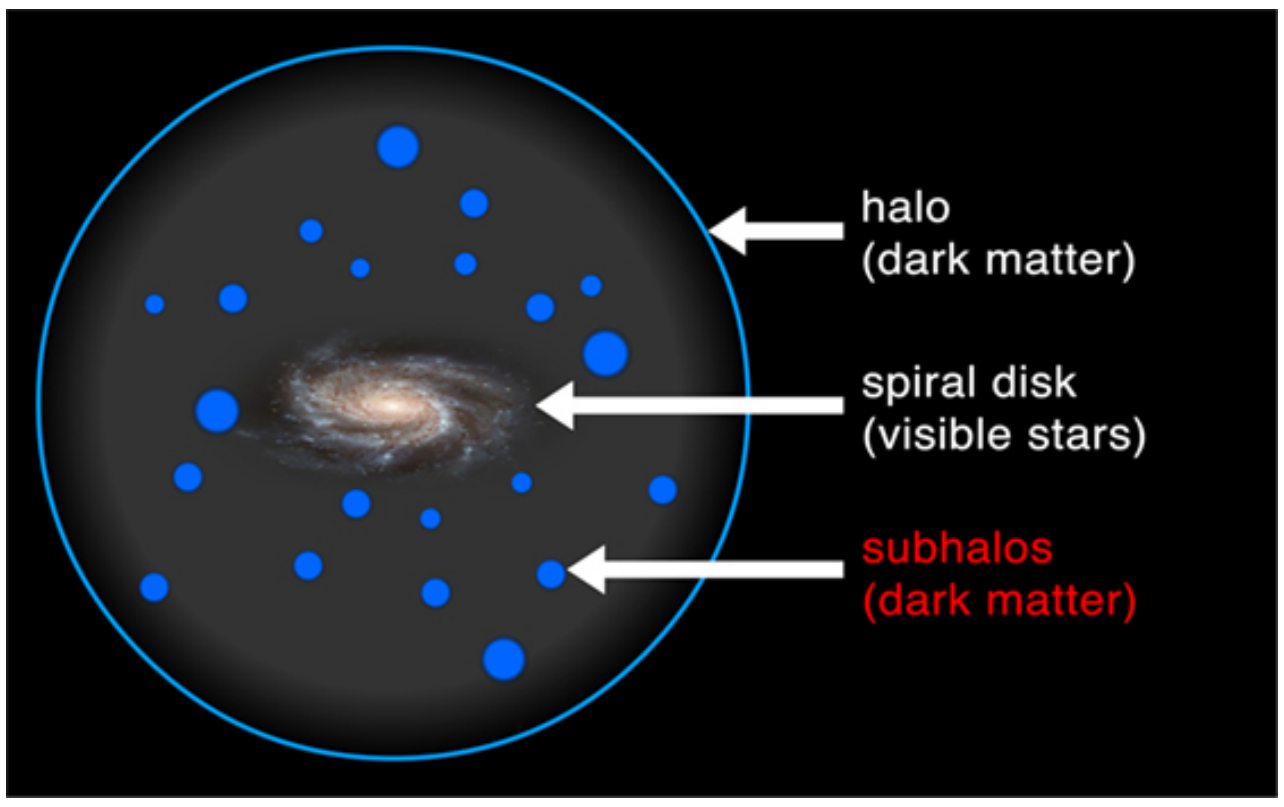

**Figure 2.6:** Illustration of a spiral galaxy in the center of its host halo with several sub-haloes around it. The picture is not to scale, as the halo is actually many times bigger than the galaxy (16).

These sub-halos experience tidal forces leading to substantial mass loss [189, 190]. The stripped particles attach to the Galactic halo and those with shared progenitors sometimes retain distinctive phase-space features. The stars formed near the cores of the sub-halos may also be stripped. As a result, these relic structures in the Milky Way can be observed as features in the spatial and velocity distribution of galactic halo stars [191–193].

The velocity structures that are spatially localised are identified as tidal streams, whereas the ones being spatially homogenised are called *debris flow*. Both of them arise from the disruption of satellites that fall into the Milky Way but are different in the relative amount of phase-mixing. The debris flow velocity substructure refers to the sum of all material stripped from in-falling sub-halos that has not been completely phase-mixed. Therefore, it comprises of dynamically cold and narrow tidal streams from recently in-falling sub-halos, older tidal streams that have been wrapped multiple times, and lost material from halos in the form of sheets and plumes due to gravitational effects at pericenter passages [163].

### 2.3.1 Dark Matter debris

The substructures in star distributions can point to potential structures also in the DM distribution the same way as stellar streams were associated with the disruption of dwarf galaxies. While DM is generally assumed to be smoothly distributed in space with a Maxwellian velocity distribution, simulations of the hierarchical formation predict large substructures throughout the halo. Gravitationally bound sub-halos, such as those to host the Milky Way dwarf satellite galaxies are examples of spatial phase-space substructure. Simulations also predict many thousands of dark sub-clumps within our galaxy's virial volume too small to host a star but potentially interesting DM annihilation sources. The substructure abundance compared to the smooth host halo mass distribution decreases towards the GC due to the much stronger tidal forces and shorter dynamical times closer to the centre of the gravitational potential.

An important property is that the same tidal disruption processes that render the local DM distribution spatially smooth are sources of velocity substructure. Most of the high-speed DM particles in our solar system have been recently accreted and are partially phase-mixed having not yet come into equilibrium with the rest of the halo, with the spatially homogenised velocity structures being the debris flow [163, 164].

### 2.3.2 Properties

The basic difference between an individual tidal stream and debris flow is that a tidal stream is dynamically cold and has a one-dimensional morphology while the latter is dynamically hot and is spatially ubiquitous in the central regions of the halo of the Milky Way [163] (see Fig. 2.7). Moreover, since debris flow is the superposition of tidal debris from many disrupting satellites, its velocity dispersion comes from the relative velocity of the material stripped from the sub-halos as well as from the intersections of a single halo's tidal stream with itself.

The identification of particles as debris is made when it is not bound to a sub-halo today. Based on Via Lactea-II N-body simulation it is found that the tidally-stripped particles are distributed within a 100 kpc radius of the GC, with a very high density within $\sim 10$ kpc from the center [164]. These debris particles constitute a few percent of the local density today and have a speed distribution in the solar neighbourhood that increases for radial shells closer to the GC and peaks at about 340 km/s in the galactic frame (see Fig. 2.8).

Finally, despite the distinct high-speed behaviour, tidal debris does not have characteristic spatial distribution and therefore can not be distinguished from the background halo.

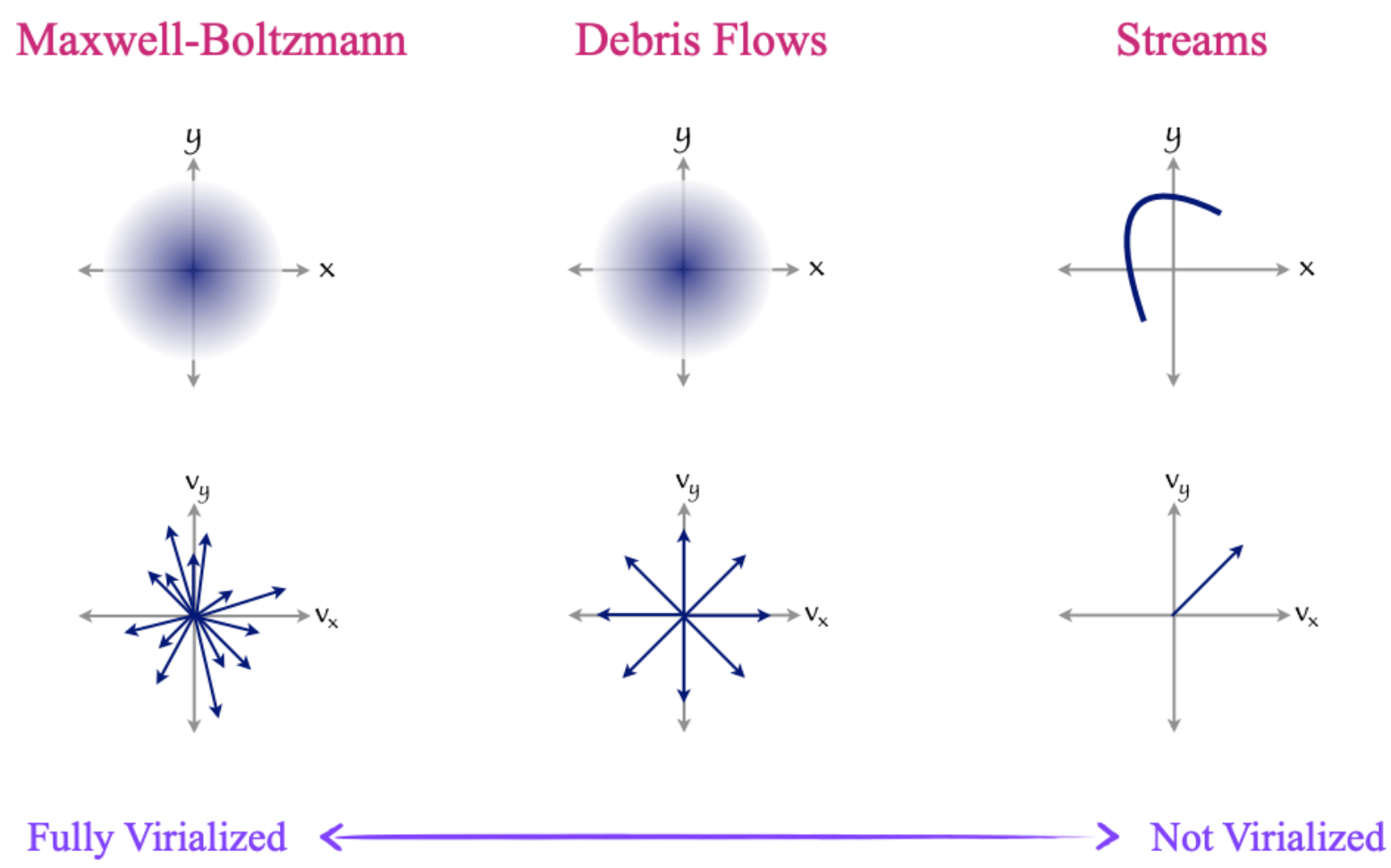

**Figure 2.7:** Differences between normal DM halo, debris flows and tidal streams (17).

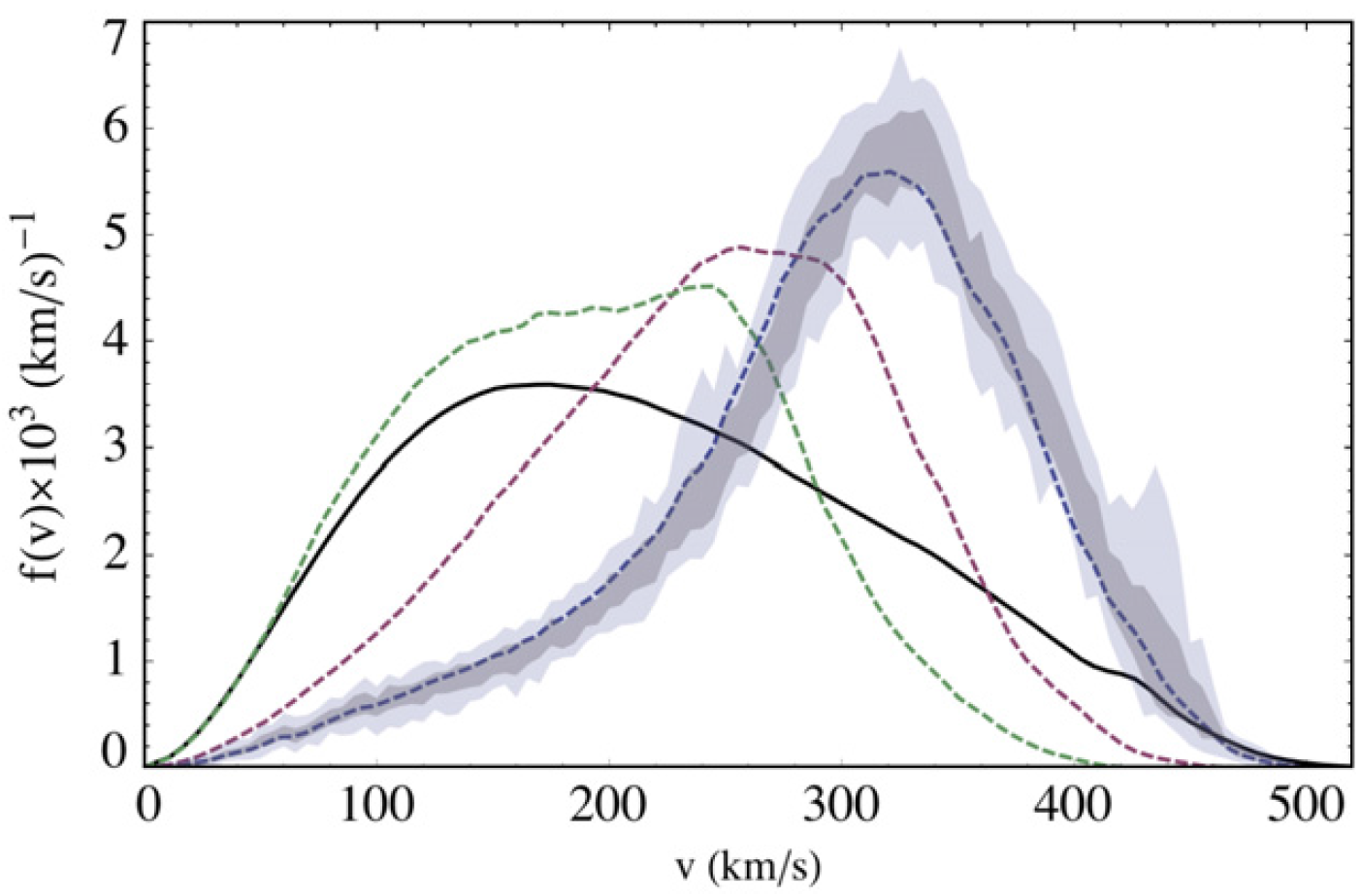

**Figure 2.8:** Normalised speed distributions for DM debris in the Milky Way. The dotted lines correspond to the distributions of debris particles located at 30 kpc to 45 kpc (green), 15 kpc to 30 kpc (pink) and 5 kpc to 15 kpc (blue) from the GC. The black solid line is the distribution for the particles from Via Lactea-II in a 5 kpc to 15 kpc shell (18).

### 2.3.3 Direct detection implications

Due to the aforementioned spatial homogeneity, the debris flow is present in the solar neighbourhood in a nearly spherical distribution, and therefore also has implications in experiments of direct DM detection. It is found that debris flow consist about 20% of all DM particles in the Via Lactea-II simulation between 7.5 kpc to 9.5 kpc, which increases to 50% for speeds greater than 450 km/s and to 80% for 600 km/s [163] (see Fig. 2.9).

If DM has a large minimum scattering threshold, this means that direct detection experiments will be sensitive to it. The non-Maxwellian distribution would result in a different recoil spectrum, with more scattering events at large nuclear recoil energies. In addition, due to the angle between the velocities of the flow particles in the galactic frame and the direction of Earth's motion, the appearance of an angular distribution of events in directional experiments

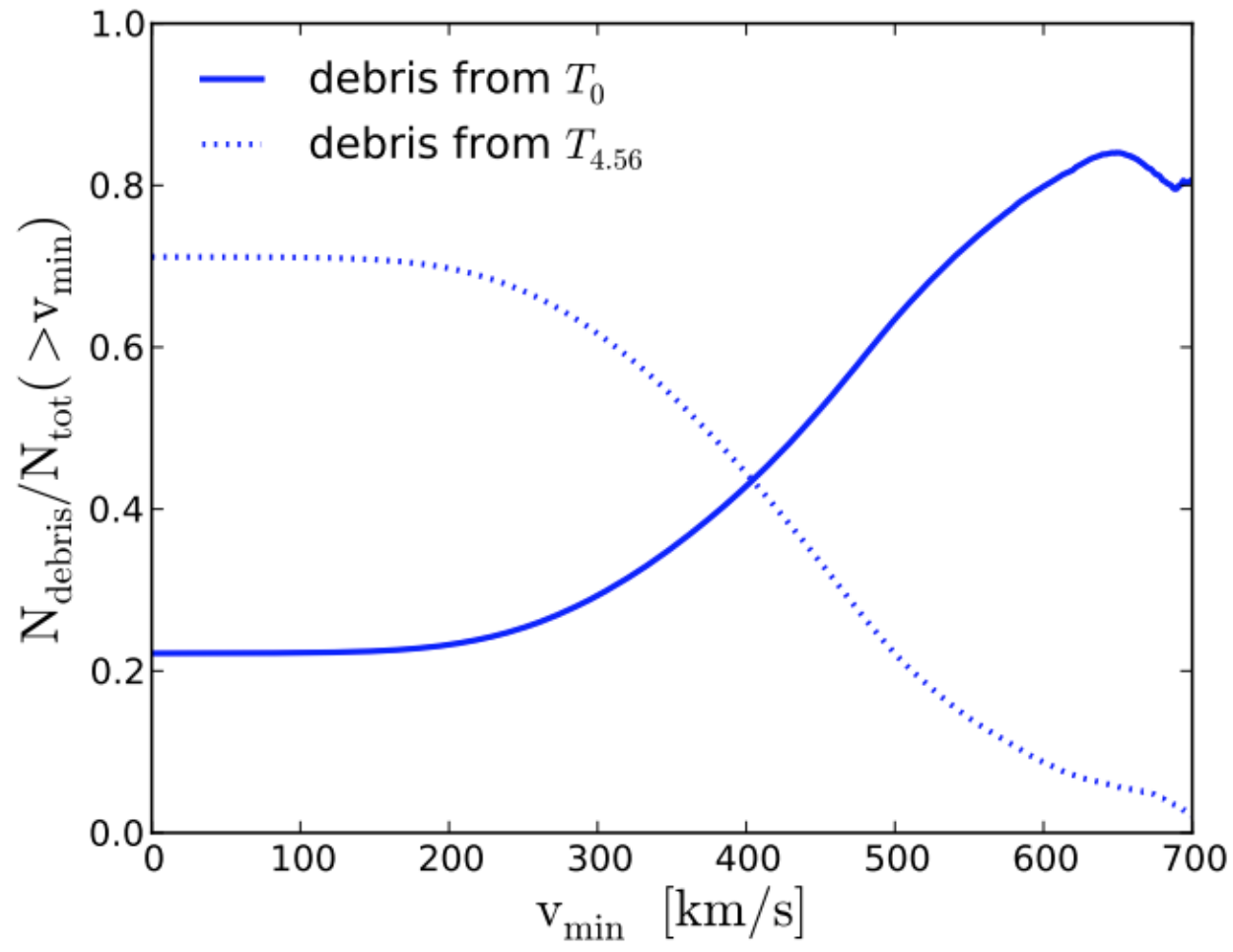

**Figure 2.9:** Fractional density of debris particles above a minimum speed $v_{min}$ in the Earth's rest frame (in June). Solid line corresponds to debris particles with a $z = 0$ remnant halo and the dotted line to high redshift debris from completely disrupted from $z = 0$ halos (19).

is also expected [163].

As for the experimental verification of the existence of debris flow, this can be realised through imprints of the flow in the local stellar distribution. For example, the observation of metal-poor stars which exhibit distinct velocity behaviour without spatial features over large areas of the sky is a distinct signature. From the latest GAIA results such signatures could be further exploited. Such a discovery would indicate that a significant amount of the local DM halo is not virialised and has distinctive phase-space features despite being spatially uniform [164].

## 2.4 Dark disk

The structure of our galaxy is based on interacting baryons that can cool by dissipating energy through photon emission as they collapse to form a structure. Cooling is a necessary property to baryonic structures occupying relatively small volumes and forming compact objects such as stars or planets. On a larger scale, cooling is necessary for the formation of disk galaxies. In contrast to baryons, DM is usually assumed to be cold and collision-less distributed through a large halo in a random way. This paradigm could be relaxed in the SIDM, or WDM scenarios. However, there is a possibility that the dark sector could have the complexity of the baryonic sector with a subdominant fraction of it being able to interact more strongly and maybe even cool like baryons. This is called *Partially Interacting Dark Matter (PIDM)* scenario and is in contrast with theories predicting a single cold collisionless particle such as WIMP or axion. The energy density of PIDM could be about as large as that of baryons and if its dynamics are dissipative it will cool and form a disk within galaxies like baryons. As a result, Milky Way could contain such structures from interacting DM analogous to the structures in the visible world, in an invisible disk parallel to the visible one. This possibility is called *Double-Disk Dark Matter (DDDM)* [194]. The difference of PIDM with other DM scenarios involving long-range forces and bound states such as mirror matter or SIDM (mentioned in Sect. 1.3.6) is that PIDM is a subdominant component of DM which is far less restrictive.

Another interesting possibility for the formation of dark disks is from ordinary DM accreting onto the stellar disk [161, 195–197]. In these cases, massive satellites are preferentially dragged into the baryonic disc plane by dynamical friction where they are disrupted by tides. When they dissolve they leave behind a thick DM disk, with its precise properties depending on the stochastic merger history and cosmology. Their spatial distribution is expected to be roughly coincident with the visible disk, co-rotating with it, but with a lagging angular velocity. The rotation lag is on the order of 40 km/s to 50 km/s with respect to the local circular velocity [195]. For Milky Way, the dark disk with $\Lambda CDM$ is of longer scale length and height than the accreted stellar thick disk whereas its density is in the range of $\rho_d/\rho_{halo} \sim 0.2 - 1$ within $|z| < 1.1$ kpc at the solar neighbourhood, where $\rho_{halo}$ is the ordinary local DM halo density.

### 2.4.1 Properties

#### 2.4.1.1 DDDM abundance

The fraction of the energy density in PIDM compared to ordinary DM is defined as:

$$\varepsilon \equiv \frac{\Omega_{PIDM}}{\Omega_{\text{DM}}} \tag{2.1}$$

In the case of DDDM the fraction of the Milky Way's mass localised in a dark disk is denoted as:

$$\varepsilon_{\text{disk}} = \frac{M^{\text{disk}}_{\text{DDDM}}}{M^{\text{gal}}_{\text{DM}}} \tag{2.2}$$

where: $M^{\text{gal}}_{\text{DM}}$ : the total mass of all DM in the galaxy.

Due to the fact that only about a third of baryons end up in the galactic disc this means that the total energy fraction in DDDM would be $\varepsilon_{\text{disk}} \approx \varepsilon/3$.

The current bounds on SIDM are derived from halo shapes and cluster interactions. These bounds assume that all DM is self-interacting. However, in PIDM a sufficiently small fraction of all matter could have extremely strong interactions without affecting observations. From the Bullet Cluster, it is implied that no more than 30% of DM is lost due to collisional effects i.e. $\varepsilon \leq 0.3$. In addition, when dissipation and therefore cooling happens with the consequent formation of a dark disk a strong bound can be derived, on the order of $\varepsilon_{disk} \leq 0.05$, which from Eq. 2.2 means that the mass of the DDDM disk can be on the order of 5% of the total mass of the galaxy. This means that the mass of the DDDM disk can be on the same order of magnitude as the mass of the baryonic disk, which in turn implies that DDDM can have comparable energy density with the ordinary baryonic matter [198].

#### 2.4.1.2 Disk formation

DDDM will have angular momentum from tidal torques while falling into a halo and will cool into a rotationally-supported disk as it also happens with baryonic matter. Supernova as well as other important evolution processes do not exist in DDDM as in the baryonic galactic disk. However, simulations that provide the required results for the disk formation point to a formation of the disk without the need of such supernova and stellar procedures, with the initial conditions favouring an approximate alignment of the baryonic and dark disk [194, 198].

An important parameter is the angle between the baryonic and the DDDM disk which has also an important effect on the detection possibilities outlined in Sect. 2.4.2. Gravity is expected to align these structures within a period of about $10^7$ y. Numerical simulations of the galaxy show that stellar and gaseous components of the baryonic disk are aligned within $\sim 7°$, whereas the angular momentum vector of DM in the inner halo is less aligned with the median angle to the angular momentum of the gaseous disk being 18° (see also Fig. 2.10). A similar gravitational alignment is expected also between the two disks in the DDDM model.

### 2.4.2 Detection possibilities

#### 2.4.2.1 Indirect detection

A thin dark disk would lead to a considerable increase of the local density of DDDM compared with ordinary DM in the plane of our galaxy, which could be detected via gravitational effects if DDDM is all bound into dark atoms. In general, depending on the dark disk height, a large boost factor from local density enhancement on the range of $10 - 1000$ is predicted [198].

Indirect detection signals, such as gamma rays, different from ordinary DM could be produced due to the annihilation of DM particles with their antiparticles. Because photons travel freely towards us, such a signal could provide a map of the dark disk in the sky resulting in a visual confirmation on the fact that DM has cooled into a structure different from a usual halo. The enhancement of local density compared to normal DM depends on the disk height, with the boost factor resulting not only from the compression of the disk in the vertical direction but also because the disk scale length in the radial direction is smaller than the radial spread in the ordinary DM distribution. The resulting spatial distribution of photons as a function of the distance from the GC is strikingly different between DDDM and ordinary DM due to the shape of the disk, as depicted in Fig. 2.10.

An additional indirect detection signal is derived from the smaller velocity dispersion of DDDM that could result in a larger Sommerfeld enhancement factor. The ordinary DM halo is approximately isothermal, with a Gaussian velocity distribution and a $10^{-3}\,c$ dispersion. On the other hand, DDDM travels in circular orbits around the GC and its velocity dispersion could be much smaller than $10^{-3}\,c$. This could give rise to an enhancement of the local density by a factor of 100 compared to non-dissipative DM as the Sommerfeld enhancement factor scales with $1/v$.

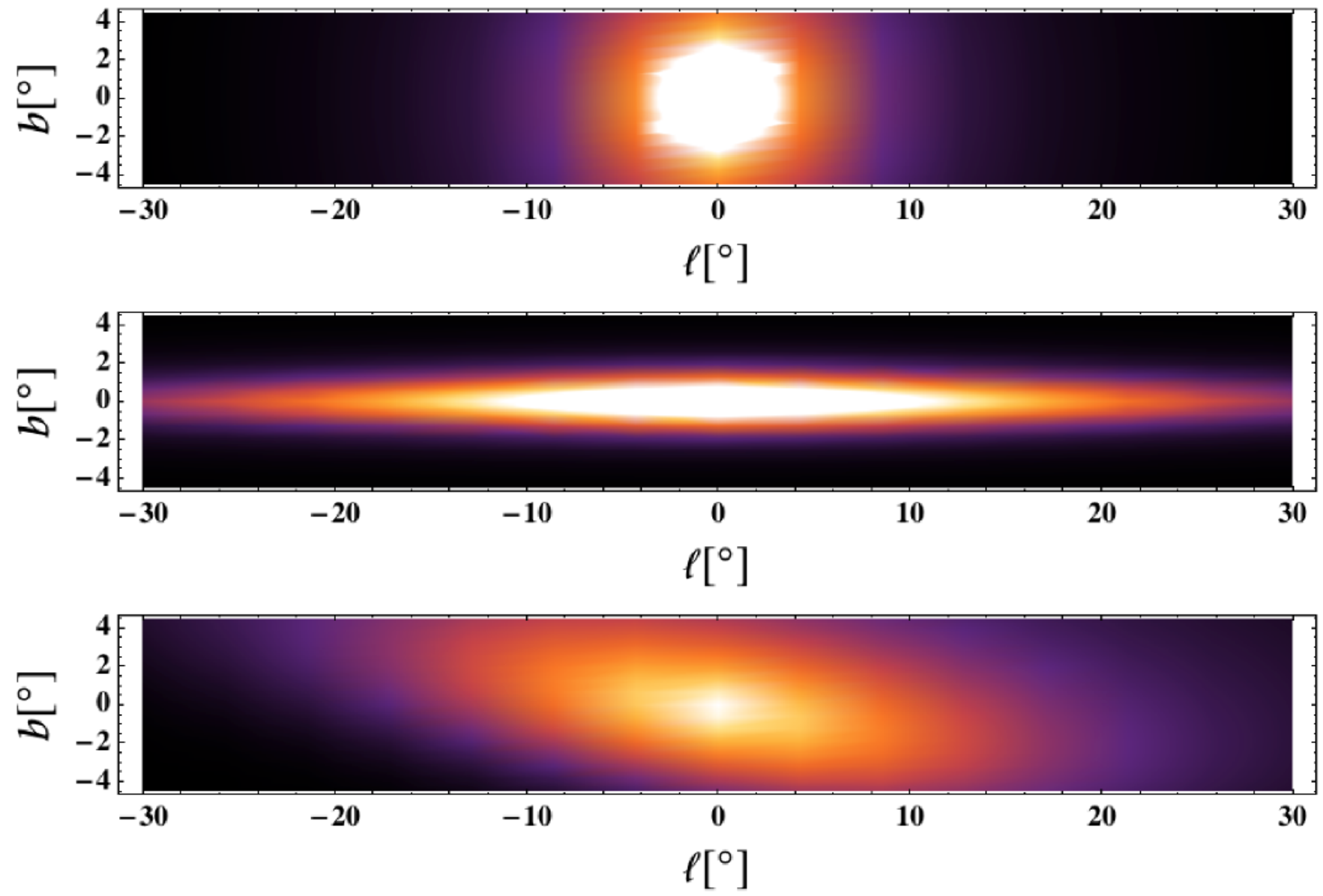

**Figure 2.10:** Sky maps in galactic coordinates of the photon flux shape for different DM profiles. Top: Ordinary DM. Middle: DDDM in a disk with height $z_d = 100\,\text{pc}$ aligned with the baryonic disk. Bottom: The same DDDM disk but misaligned with our galaxy by 18° (20).

#### 2.4.2.2 Direct detection

The most basic problem with direct detection is that the Earth might be placed outside the dark disk either due to misalignment between the dark and the baryonic disc or due to the thickness of the dark disk. In the optimal case where the Sun is inside the DDDM disk, the DDDM density at the position of the Sun could be about 20 times higher than ordinary DM i.e. $6\,\text{GeV/cm}^3$. The resulting spectrum from DDDM scattering off nucleons would be much different from a WIMP of a similar mass, but due to the low kinetic energy of DDDM, a measurable signal in direct detection experiments can not be produced.

This comes as a consequence of the formation of the DDDM disk. After cooling, DDDM is in a form of a rotationally supported disk, which means that a typical DDDM particle would move in a circular orbit around the GC. Therefore, in our solar neighbourhood, both the solar system and DDDM would be in approximately the same circular orbit, meaning that they would have the same average large radial velocity component. Only the deviations from this velocity could contribute to nucleus scattering, but the small relative velocity does not exceed the current experimental thresholds, of about a few keV in nuclear recoil.

It is noted that the ordinary dark disk model, which as mentioned forms from the accretion of massive satellites onto the galactic disk, can also enhance significantly the flux and thus detection rates in direct detection experiments at low energies. Also due to its low velocity with respect to the Earth it can also increase the annual modulation signal in DM detectors (mentioned in Sect. 1.4.1.1), with an energy-dependent phase shift that can point to the mass of the DM particle. More specifically, the relative change in the mean streaming velocity due to Earth's motion is about 19% larger compared to the SHM ($\sim 6\%$). The phase difference between the dark disk and the SHM is due to the misalignment of the Sun's motion relative to the dark disk [196, 197].

#### 2.4.2.3 Solar capture

The aforementioned small relative velocity between the two disks could enhance capturing in the Sun which can provide signatures in neutrino observations. As DM particles pass through the Sun, they could scatter off nuclei inside the Sun and become gravitationally bound. With subsequent scattering, they could eventually accumulate in the centre of the Sun. These captured particles and antiparticles could then annihilate into various SM particles, which would then decay into energetic neutrino pairs that could be observed with neutrino telescopes.

An important point is that if the Sun is inside the dark disk, then DDDM should have a local density $\sim 10$ times as large as the ordinary DM density near the Sun thus enhancing the capture rate. Another effect for the enhancement of capture rate could be by a larger gravitational Sommerfeld enhancement giving rise to another factor 10 compared to ordinary DM. The lower velocity giving rise to this enhancement also happens in a dark disk made of ordinary DM [196].

## 2.5 Fine-grained streams and caustics

### 2.5.1 Formation

Before the non-linear structure formation, DM was assumed to be evenly distributed with the particles occupying a thin 3-dimensional sheet filling all the full 6-dimensional phase-space (3 spatial and 3 velocity dimensions). The subsequent gravitational collisionless evolution stretched and folded this sheet resulting in the DM distribution at a typical point today to be a superposition of many fine-grained streams. Each of these streams has a very small velocity dispersion, and a smooth position-varying density and mean velocity defined by the initial linear conditions [199]. Therefore, the DM distribution at a typical point in a halo could be a superposition of many fine-grained streams.

The existence of such distinct streams is a direct consequence of the collisionless character and the coldness of CDM [200]. The folds in the fine-grained phase-space sheet produce catastrophes known as *caustics* [201, 202]. These are regions with a very high DM spatial density which is inversely proportional to the square root of the velocity dispersion. This means that they could provide an increased annihilation rate into photons. They are limited only by the small nonzero thickness of the phase-sheet [203].

### 2.5.2 Characteristics

If the local DM density consists of a few streams, then, its velocity distribution will comprise of a few discrete values, one for each stream. However, if the DM in our solar neighbourhood is dominated by a large number of streams, the velocity distribution will be smooth and the individual streams will be undetectable. The difficulty for the prediction of such a distribution comes from the inability of current N-body simulations to resolve the relevant cosmological scales [204].

Based on the fine-grained phase-space structure of DM throughout the galactic halo, the estimated stream density on the solar neighbourhood is on the order of $10^5$ [204] whereas more recent and realistic simulations for CDM indicate a number of $\sim 10^{14}$ [199]. Moreover, about half of the local DM density is expected to be comprised of the $10^6$ most massive streams, with the most massive individual stream contributing a 0.1% to the local DM density (see Fig. 2.11). This results to a smooth velocity distributions of DM particles. The simulations also showed that the probability to see a local DM signal is about 20% when a single stream accounts for 1% of the total signal [199].

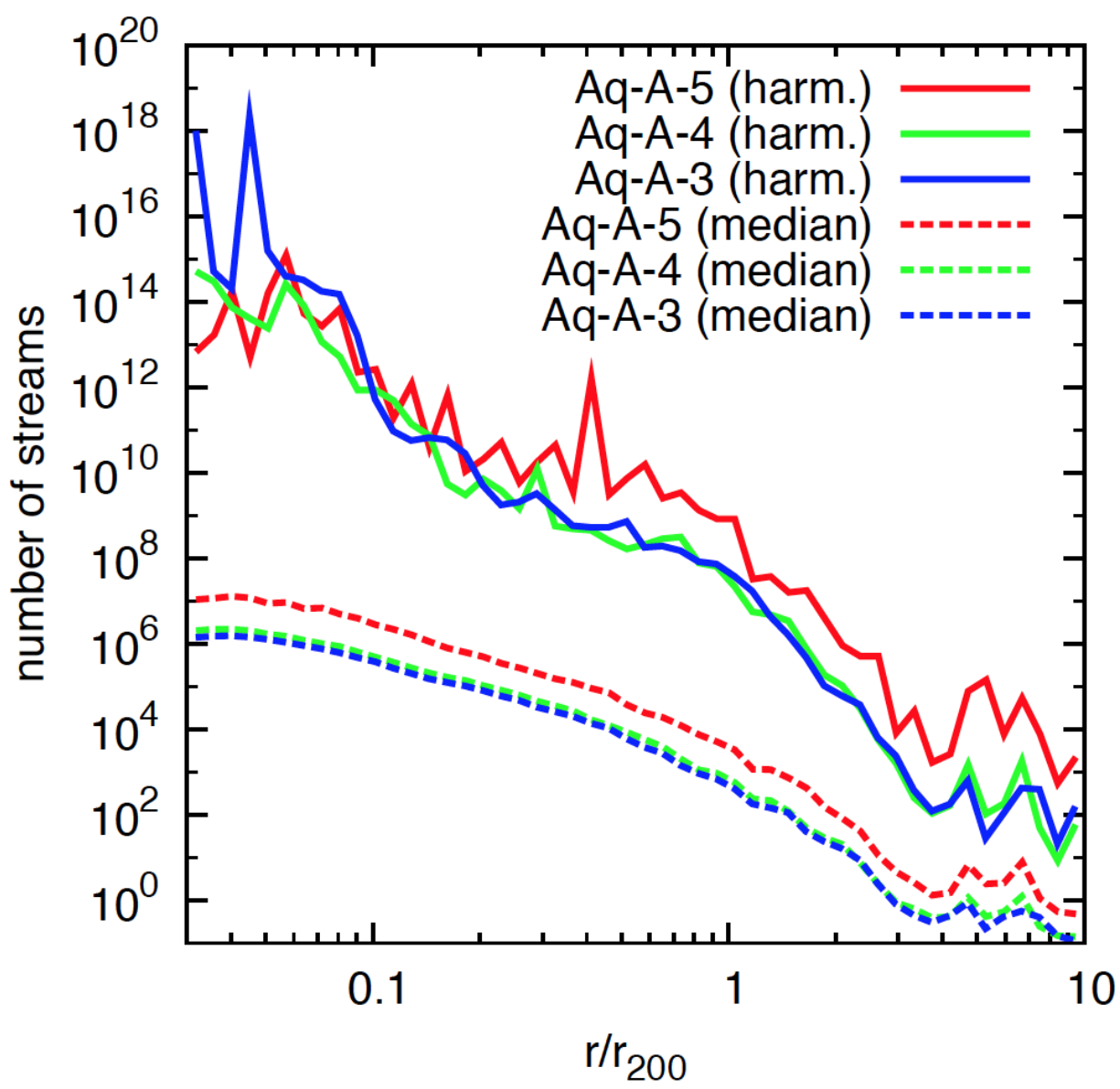

**Figure 2.11:** Number of fine-grained streams in the Aquarius (Aq-A) halo simulations as a function of radius. The number of streams are estimated by dividing the mean density at each radius by an estimate of the characteristic local density of individual streams. For this estimate, both harmonic mean (solid lines) and median (dashed lines) stream density for particles in each radial shell are used (21).

## 2.6 Summary

Even though DM is assumed to be smooth and isotropic, various observations and simulations have shown that streams of DM should also exist in various forms. These deviations from she SHM which include tidal streams, dark disk and debris flows, can enhance the local DM density and even have observable effects on DM experiments. In addition, following the fine-grained phase-space structure of our galactic halo, $10^{14}$ streams with a broad range of individual densities are expected to exist in our local neighbourhood. Therefore, as we will see in the following chapters, depending on the cross-section, the velocity distribution and the constituents of DM, a significant impact on various solar and terrestrial observables as well as on direct DM experiments could take place which would not happen for the case of the normal isotropic SHM.

3

# GRAVITATIONAL LENSING OF SLOW-MOVING PARTICLES

## 3.1 Introduction

In the SHM the DM particles are assumed to be in an isotropic, collisionless and isothermal halo with a density profile of $\rho(r) \propto r^{-2}$. They are also assumed to follow a Maxwell-Boltzmann velocity distribution withf a dispersion of $\sigma_v = v_c/\sqrt{2}$, where the local circular speed $v_c$ is calculated at around $(233 \pm 3)$ km/s. The cutoff of this distribution is at the escape velocity of the galaxy at $v_{\text{esc}} = 528^{+24}_{-25}$ km/s [158]. However, there is a significant uncertainty on the velocity distffribution as well as the parameters on which it depends [159], with different observations providing different results.

As an example, based on observations of metal-poor stars from the SDSS in the solar neighbourhood, the local DM velocities can be derived directly. As seen in Fig. 3.1, these measurements indicate a non-isotropic distribution with a lower peak and a smaller dispersion than the usual assumptions of the SHM [205]. Some simulations have also shown deviations from the Maxwell-Boltzmann distribution, with the appearance of tidal-stream induced narrow spikes above the dominant smooth distribution [206]. Several deviations from the SHM picture were outlined in Chap. 2.

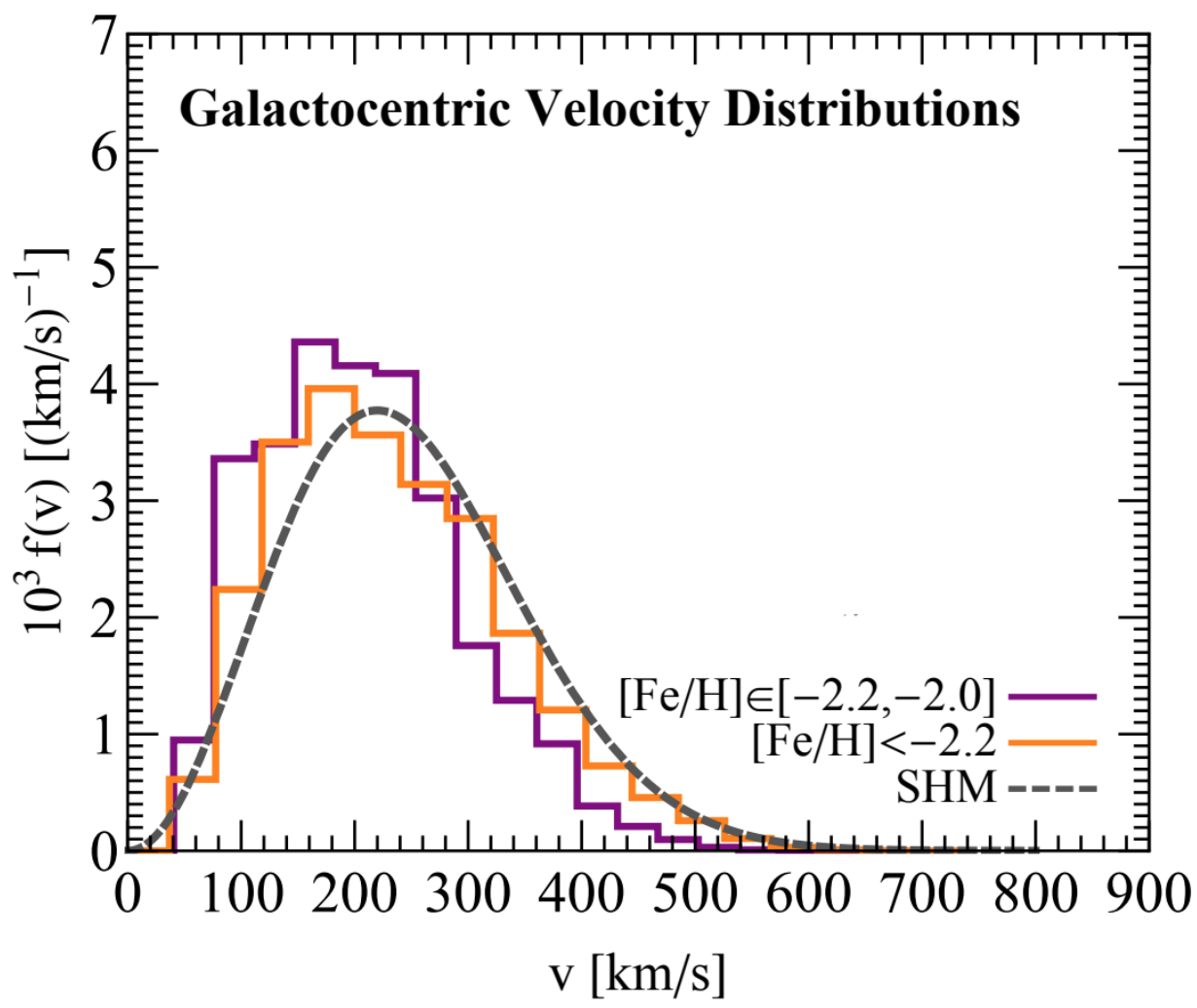

**Figure 3.1:** Galactocentric speed distributions for metal-poor stars from SDSS within 4 kpc from the Sun and and Galactocentric distances of $7 < r < 10$ kpc. In dashed grey, the SHM with 220 km/s is also shown for comparison (22).

As we will see also later in Sect. 23.2.1 and 23.3.4 the DM density and velocity distribution is of utmost important for direct DM searches as it affects the predicted signal in the detectors. An additional important characteristic is also the time variation or modulation in the lab frame of reference. As an example, annual [207], monthly [208], and daily [167, 209] modulations of the DM flux due to the Earth's motion and the gravitational effect of the entire solar system [210, 211] have been investigated in the literature (see also Sect. 1.4.1). Interestingly, the gravitational effects of the Solar system bodies can have a significant effect on the spectral DM density for the case of directional streaming DM.

An additional important characteristic of slow-moving CDM is that it can undergo gravitational focusing induced by the solar system bodies, which, can in turn result in a large flux amplification. Interestingly, as seen in Sect. 1.1.5, gravitational lensing played a key role in the indirect detection of DM, and as we will see in the next chapters, it could also be crucial for its direct detection.

## 3.2 Deflection angles

### 3.2.1 Classical approach

The general formula for the deflection in classical relativity for particles with an arbitrary velocity $\upsilon$, and for a mass $M$ enclosed within an impact parameter $b$ (radius of cylinder) is given by [212]:

$$a_c = \frac{2G_N M}{bc^2\beta^2}\left[1+\beta^2+bg\left(1+\frac{\beta^2}{4}\right)\frac{3\pi}{2}+ \right. \\ \left. +b^2g^2\left(5\beta^{-2}-\frac{1}{3}\beta^{-4}+\frac{5}{3}\beta^2\right)+\mathcal{O}\left(g^3\right)\right] \tag{3.1}$$

where $\beta = \upsilon/c$ and $g = G_N M/b^2$. If we keep only the first order terms from Eq. 3.1, then the deflection angle becomes:

$$a_c = \frac{4G_N M}{bc^2}\left(\frac{1+\beta^2}{2\beta^2}\right) \tag{3.2}$$

For the limit $\beta \to 1$ the classic formula for the deflection angle for a light ray in the weak field [24] is recovered as presented in Eq. 1.6, whereas for $\beta \ll 1$ the amount of deflection is reduced by a factor of two which corresponds to the Newtonian prediction [213].

### 3.2.2 Semiclassical approach

An alternative relation for the deflection can be derived with the semiclassical approach. In this case we include terms corresponding to the interaction of a massive photon field with a minimally coupled static gravitational field in the action integral. The deflection angle then becomes [214]:

$$a_{sc} = \frac{4G_N M}{bc^2}\left(\frac{3-\beta^2}{2\beta^2}\right) \tag{3.3}$$

Again, Eq. 3.3 gives Eq. 1.6 when $\beta \to 1$, but for $\beta \ll 1$ it becomes $a_{sc} = 1.5a = 3a_c$, which means that for very slow-moving particles the semiclassical results gives about 50% more deflection than the the one obtained with the standard formula of Eq. 1.6 [215]. The variation of the deflection angle as a function of particle speed is shown in Fig. 3.2. There, it is observed that for speeds around 0.01 c both approaches seem to fail to focus particles in a specific point.

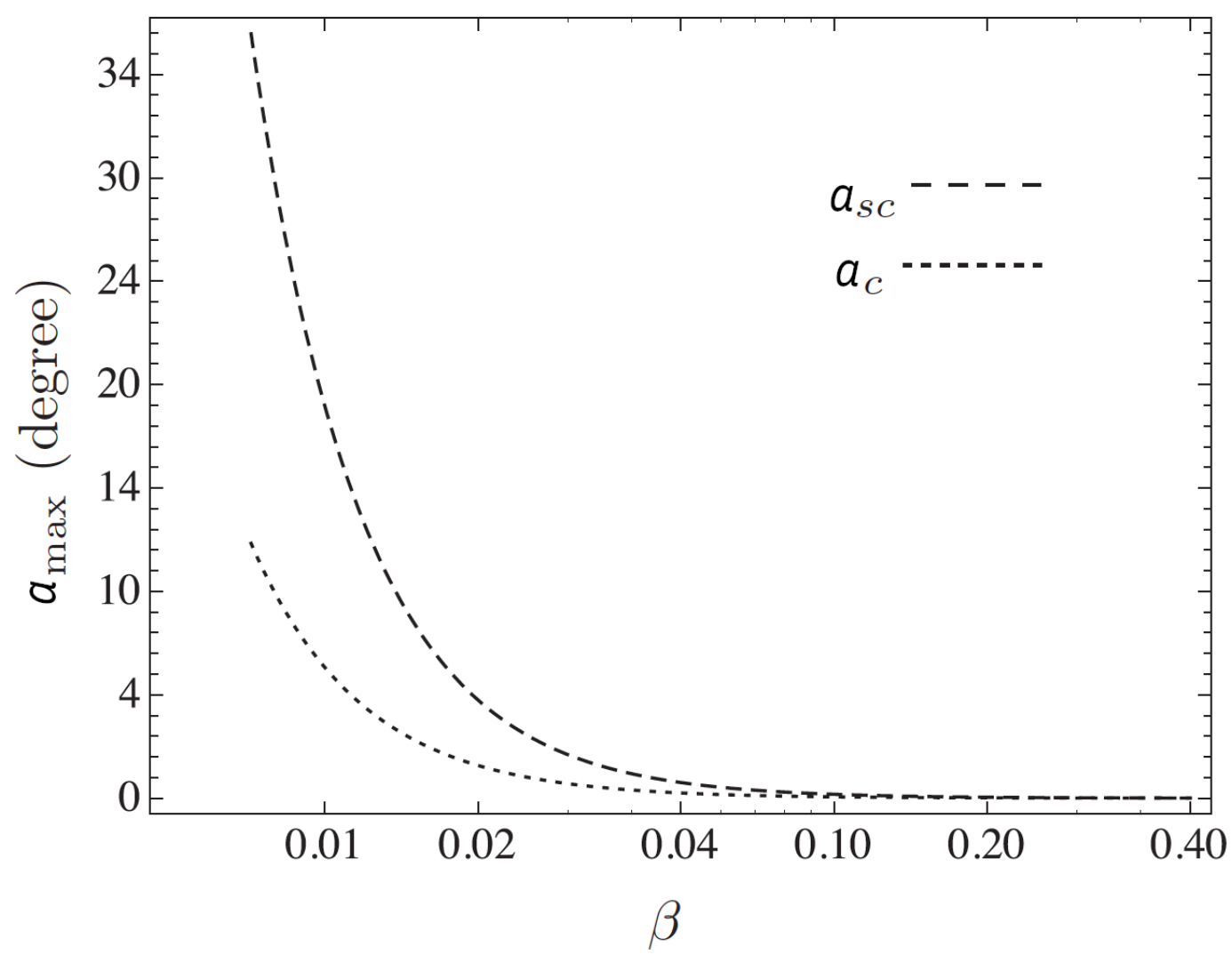

**Figure 3.2:** Maximum deflection angle as a function of particle speeds $\beta$ for the classical and semi-classical approach (23).

## 3.3 Focal length and particle velocities

### 3.3.1 Lensing by the Sun

For photons and neutrinos the minimum focal length $F_{\min}$ of the Sun is about 550 AU and 23.5 AU respectively [215]. However, for slow-moving particles the minimum focal length is shorter since the deflection angle is $a \propto \frac{1}{bv^2}$, as seen from Eq. 3.1 and 3.3.

#### 3.3.1.1 Opaque Sun

For the case of the Sun assumed to be opaque, assuming a small deflection angle, the minimal focal length is approximated by [216]:

$$F_{\text{min, op}} = \frac{c^2}{4g_\odot} = 548.30 \pm 0.01\,\text{AU} \tag{3.4}$$

where: $g_\odot = G_N M_\odot / {R_\odot}^2$ : constant.

The uncertainty in $F$ is due to the statistical errors on $R_\odot$ and $G_N M_\odot$.

#### 3.3.1.2 Transparent Sun

The transparent Sun can also be used as a gravitational lens where the minimum focal length is defined as “the minimum distance between the centre of the lens and the point on the optic axis where the deflected rays corresponding to different impact parameters converge” [216]. This will generate several focal lengths starting with the minimum of 23.5 AU for neutrinos or other relativistic particles. It is noted that this distance is between the orbits of Uranus and Neptune.

Since the Sun is in hydrostatic equilibrium, its pressure and density are simply proportional to their radius raised to a polytropic (power-law) index $n$. So, by considering the mass enclosed within the transition impact parameter, at which the index $n$ of the density changes from 0 to -2, the approximate minimum focal length is [215]:

$$F_{\text{min, tr}} \sim 2\left(\frac{1+n}{3+n}\right)\left(\frac{c^2}{4g_\odot}\right)\left(\frac{r_n^2}{m_n}\right) \tag{3.5}$$

where: $r_n = 0.075$ : the normalised radius corresponding to the transition of the power-law index,

$m_n = 0.1$ : the normalised mass enclosed within $r_n$.

Eq. 3.5 gives a value of $F_{\text{min, tr}} \sim 21$ AU, while when using Eq. 1.6 and considering again the mass $m_n$ within an impact radius $r_n = 0.075$, we get a higher value of $F_{\text{min, tr}} \sim 31$ AU.

From Eq. 3.2 and 3.3 we can derive the relationship between the approximate focal length and the velocities of slow-moving particles for the two cases of the classical and semiclassical formulae [215]:

$$D_{L,c} = F_{\text{min}}\left(\frac{2\beta^2}{1+\beta^2}\right) \tag{3.6}$$

$$D_{L,sc} = F_{\text{min}}\left(\frac{2\beta^2}{3-\beta^2}\right) \tag{3.7}$$

where $D_L$ is the distance between the lens and the observer and $F_{\text{min}}$ is the minimum focal length for particles with speed $\beta = 1$.

By demanding the approximate focal length to be $D_L \sim 1$ AU meaning to have particles focused by the Sun towards the Earth we can get lower and upper bounds for the speeds of the particles that are lensed by the Sun. From Eq. 3.6 and 3.7 accordingly we get for the minimum focal length of 21 AU $\beta \sim 0.15$ and $\beta \sim 0.26$. On the other hand for $F_{\text{min, tr}} = 31$ AU we get respectively the values $\beta \sim 0.13$ and $\beta \sim 0.22$.

As a result, in order for particles to be focused on the Earth by the lensing action of the transparent Sun, their velocities have to be on the order of $0.1 - 0.3$ c. This also means that particles with less than 0.1 c will be focused in less than 1 AU whereas particles with speeds greater than 0.3 c will be focused beyond Earth [216, 217]. From numerical calculations of Eq. 3.6 and 3.7, the exact values for the limiting particle speeds are $\upsilon \sim (0.145 \pm 0.001)$ c and $\upsilon \sim (0.247 \pm 0.001)$ c. This means that we cannot have a higher flux of particles with speeds higher than these values. It is also noted that weak lensing basically fails for particle speeds $\sim 0.01$ c, as seen in Fig. 3.2 [215].

Therefore, the limit for gravitational focusing of the transparent Sun towards the Earth is about $\upsilon$ ∼0.01 c to 0.14 c, whereas for the semiclassical case it becomes $\upsilon \sim 0.01$ c to 0.24 c.

### 3.3.2 Planetary lensing

Under the same reasoning, planets are also capable of gravitational lensing slow-moving particles. We use here Jupiter as an example which has a mass of $\sim 10^{-3}\,\mathrm{M}_\odot$ and a distance from the Earth which varies between 4.2 AU and 6.2 AU. Using these two minimum and maximum values for the distance of Jupiter from Earth, from Eq. 3.6 we get [215]:

$$\beta_c^{jup} = \begin{cases} 0.04, & \text{for } D_L = 4.2\,\mathrm{AU} \\ 0.05, & \text{for } D_L = 6.2\,\mathrm{AU} \end{cases} \tag{3.8}$$

whereas from Eq. 3.7 we get:

$$\beta_{sc}^{jup} = \begin{cases} 0.07, & \text{for } D_L = 4.2\,\mathrm{AU} \\ 0.08, & \text{for } D_L = 6.2\,\mathrm{AU} \end{cases} \tag{3.9}$$

By calculating again the maximum deflection angle for Jupiter, as in Fig. 3.2, we find that the lower limit for particle speed is one order of magnitude less than it was in the Sun's case. This means that for Jupiter a focusing of particles with speeds as low as 0.001 c is possible.

Other examples of objects capable for gravitational lensing of slow moving particles are the Earth and the Moon which can have a focal length of $\geq 1$ AU for particle speeds of $\upsilon \sim 10^{-2}$ c and $\upsilon \sim 3 \times 10^{-3}$ c respectively. In addition, the Moon itself can focus on the site of the Earth particles with speeds of $\upsilon \sim 2 \times 10^{-4}$ c (see also Publication E.1).

## 3.4 Amplifications and flux

### 3.4.1 Lens equation

If we assume $D_{LS} \approx D_S$ for a distant source, then the equation satisfied by the geometry of Fig. 3.3 is:

$$\vec{\beta} = \vec{\theta} - \vec{a} \tag{3.10}$$

where $\vec{a}$ is a measure of the deflection given by Eq. 3.2 and 3.3.

The numerical solution of Eq. 3.10 gives the image locations for a given source location. It is noted that the source and the image locations do not align on a straight line.

For an optical lens the correlation between the focal length, the impact parameter and

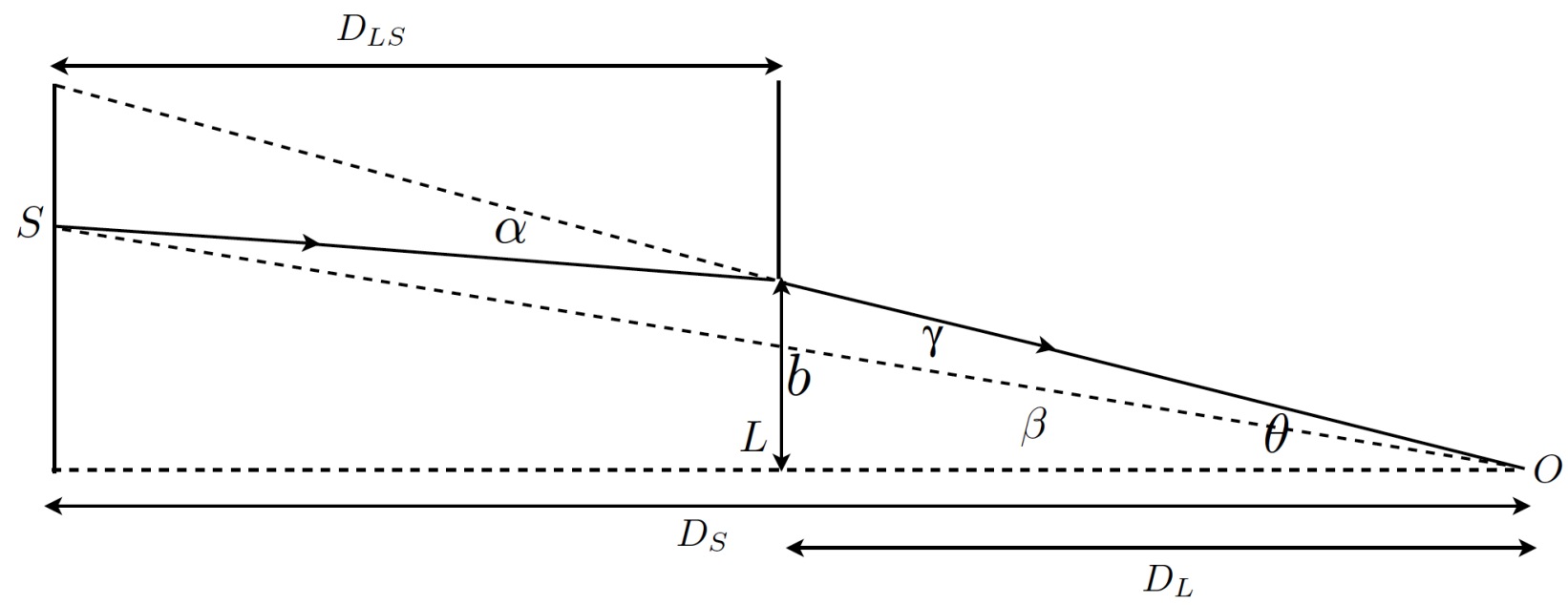

**Figure 3.3:** Schematic of the lens $L$ being at a distance $D_L$ from the observer $O$ and a distance $D_{LS}$ from the source S. $\beta$ and $\theta$ are the angles subtended by the unlensed source and its image as seen from the observer (24).

the deflection angle is:

$$b = D_L \tan a \tag{3.11}$$

where: $b$ : the impact parameter,

$D_L$ : the focal length,

$a$ : the deflection angle.

The focal length $D_L$ here is the distance between the center of the lens and the point at which the paraxial light rays converge or appear to diverge.

In contrast, a gravitational lens has a focal length for each impact parameter bounded by a minimum focal length. This is due to the deflection angle $a$ not being a linear function of the impact parameter $b$. As a result, paraxial light rays converge at different points along the optic axis [216].

## 3.4.2 Caustics and magnification

### 3.4.2.1 Caustics

The Sun is assumed to be a centrally condensed spherically symmetric lens. The lens plane is defined as the plane perpendicular to the line connecting the point source and the fixed observer whereas the source plane is the plane perpendicular to the line extending from the fixed observer through the lens centre. Some areas can exist in the source plane where a point source can be placed so that the fixed lens creates a specific odd number of images visible to the fixed observer. The boundaries between these areas of the source plane are called *caustics*. For spherical lenses the caustics are circles. The magnification is very high along the caustics in the source plane [215, 216]. For non-relativistic particles, the radii of the Einstein ring and the caustics is larger if the observer is at or near the minimum focal length.

#### 3.4.2.2 Magnification

In general, the magnification at a given location is defined as the ratio of the size of the image to that of the source, and for a point-like source is given by [216]:

$$\mu = \frac{\theta}{\beta}\frac{d\theta}{d\beta} \tag{3.12}$$

The magnification of an extended source is calculated by averaging all the individual points that make up the extended source. Assuming that all the sources are small compared to the lens, Eq. 3.12 holds for small impact parameter crossing of point sources across the lens.

The maximum possible magnification by the Sun can be calculated to be on the order of $\sim 10^6$. In Fig. 3.4 a specific case for particle speeds of 0.01 c is shown. It is noted that for particles with lower speeds than $\beta = 0.06$ the magnification is less whereas particles with speeds of $\sim 0.01$ c or less are deflected strongly and cannot be focused. Finally, using Eq. 3.3 the results are similar, with the particles speeds corresponding to maximum magnification being a bit higher, and the maximum magnifications being about one order of magnitude higher than the classical case [215].

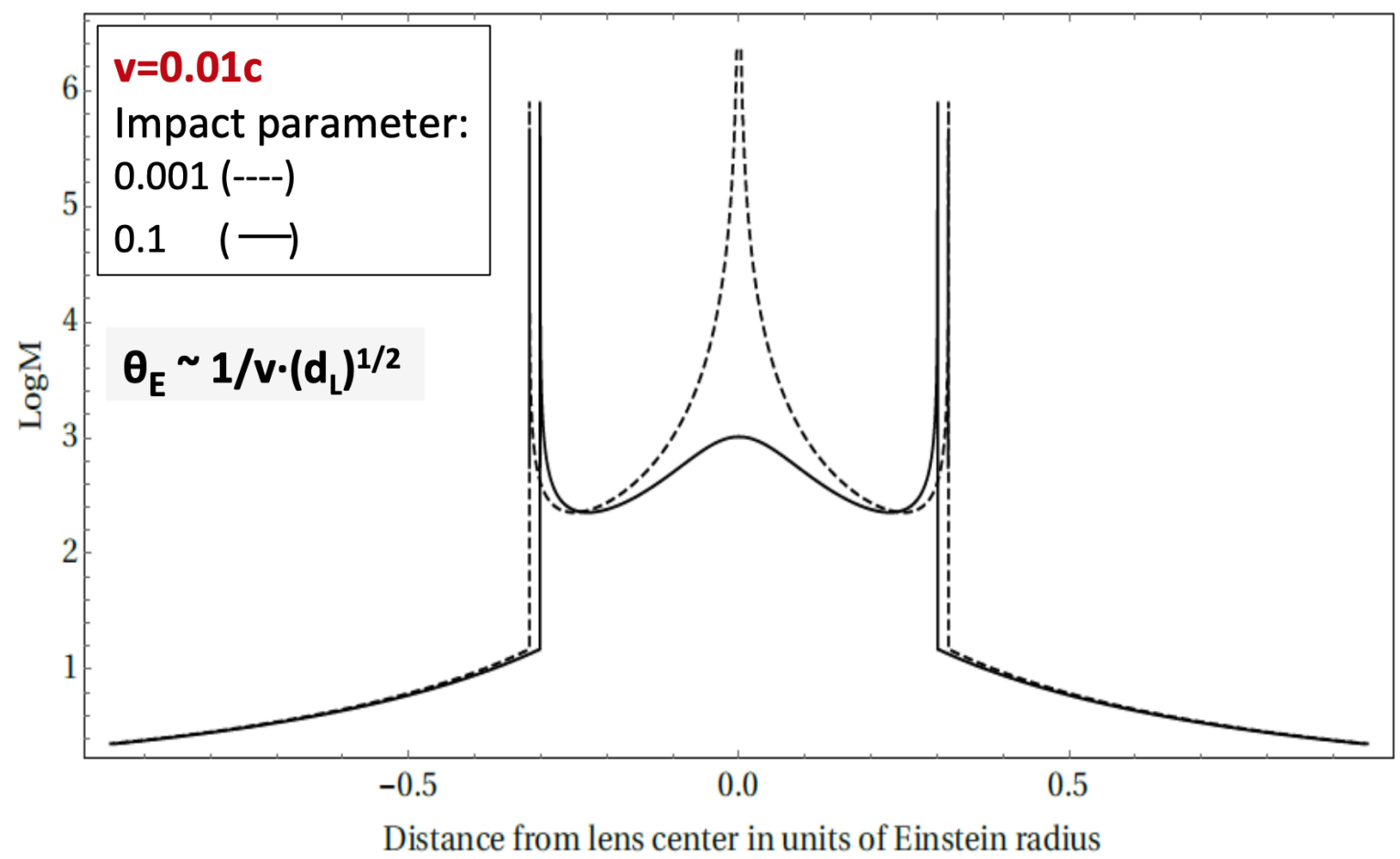

**Figure 3.4:** Magnification of the source by the Sun as a function of the angular separation of the source from the center of the lens for impact parameters $b = 0.001$ (dashed line) and $b = 0.1$ (continuous line). The maximum magnification of $\sim 10^6$ is at the crossing of the caustics (25).

#### 3.4.2.3 Planetary focusing

As we have outlined in Sect. 3.3.2, the rest of the solar system bodies are also capable of focusing slow moving particles. The planets could act as primary lenses but also as a secondary lens to further amplify the lensing action by the Sun as they transit and align along the Sun-Earth axis.

As an example, for the case of Jupiter, the magnifications near the caustics are expected to be similar as for the case of the Sun [215] which means that the amplification factors are significant not only for the Sun but also for large planets like Jupiter. For the calculation of the exact values, an accurate density profile of the used planet should be used.

For the case of the Earth's Moon, the flux enhancement at the position of the Earth can be about $10^4$ for particles of speeds of $\sim 10^{-4}$ c (see Fig. 20.1 and also Publication E.10).

### 3.4.3 Relativistic considerations

In the ideal case considering a wave-theoretical description of the solar gravitational lens, the amplification by the Sun to the brightness of light from distant faint sources is $\sim \frac{2G_N M}{c^2 \lambda}$. This corresponds to an amplification of $1.2 \times 10^{11}$ for the case of incident monochromatic coherent electromagnetic emission at the near-infrared wavelength of $\lambda = 1\,\mu\text{m}$ downstream at about 547 AU [218–220].

However, we have seen that the Einstein ring and the deflection angle (see Eq. 3.2 and 3.3) correlation with velocity is $\upsilon^{-1}$ and $\upsilon^{-2}$ respectively. This means that the situation is improved even further for an Earth observer searching for non-relativistic particles with $\upsilon \ll$ c [221].

## 3.5 Self-focusing

The intrinsic mass distribution of a solar body, like the Earth, can also perform gravitational focusing on incident directional streaming DM on its opposite side which can also result in a flux amplification depending on the incoming velocities [222]. Notably, any object with a sufficiently large solid angle and uniform density acts as a real convex gravitational collector having a focal point (see Publication E.9).

### 3.5.1 Earth

The trace of the motion of streaming monochromatic DM can be found by solving the equations of motion:

$$\frac{d^2\mathbf{r}}{dt^2} = -G_N \frac{M(r)}{r^2}\frac{\mathbf{r}}{r} \tag{3.13}$$

where: $\mathbf{r} = (x, y, z)$ : the geocentric coordinates,

$M(r) = \int_0^r 4\pi p^2 \rho(s)\, ds$ : the mass of the gravitating body.

For the gravitational field of the Earth, and assuming a density profile like the one in Fig. 3.5, an efficient focusing is found to occur on the Earth's surface on the opposite side of the incoming DM particles, when the initial velocity of the DM particles is close to $\upsilon_{\text{inj}} = 17\,\text{km/s}$. The various orbits for this case are depicted in Fig. 3.6.

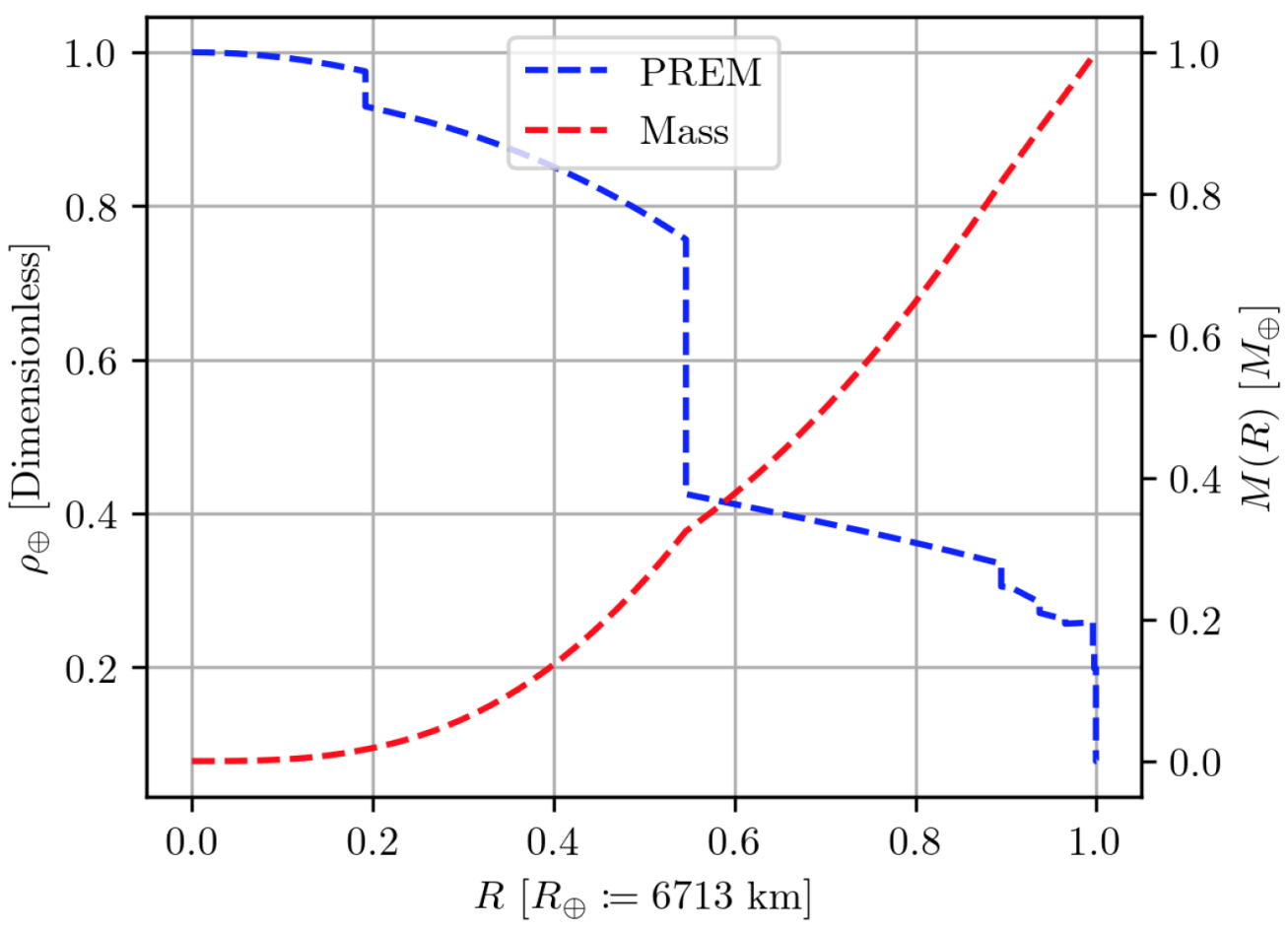

**Figure 3.5:** Earth's density profile and the enclosed mass in a sphere of radius $R$ from the geo-center (26).

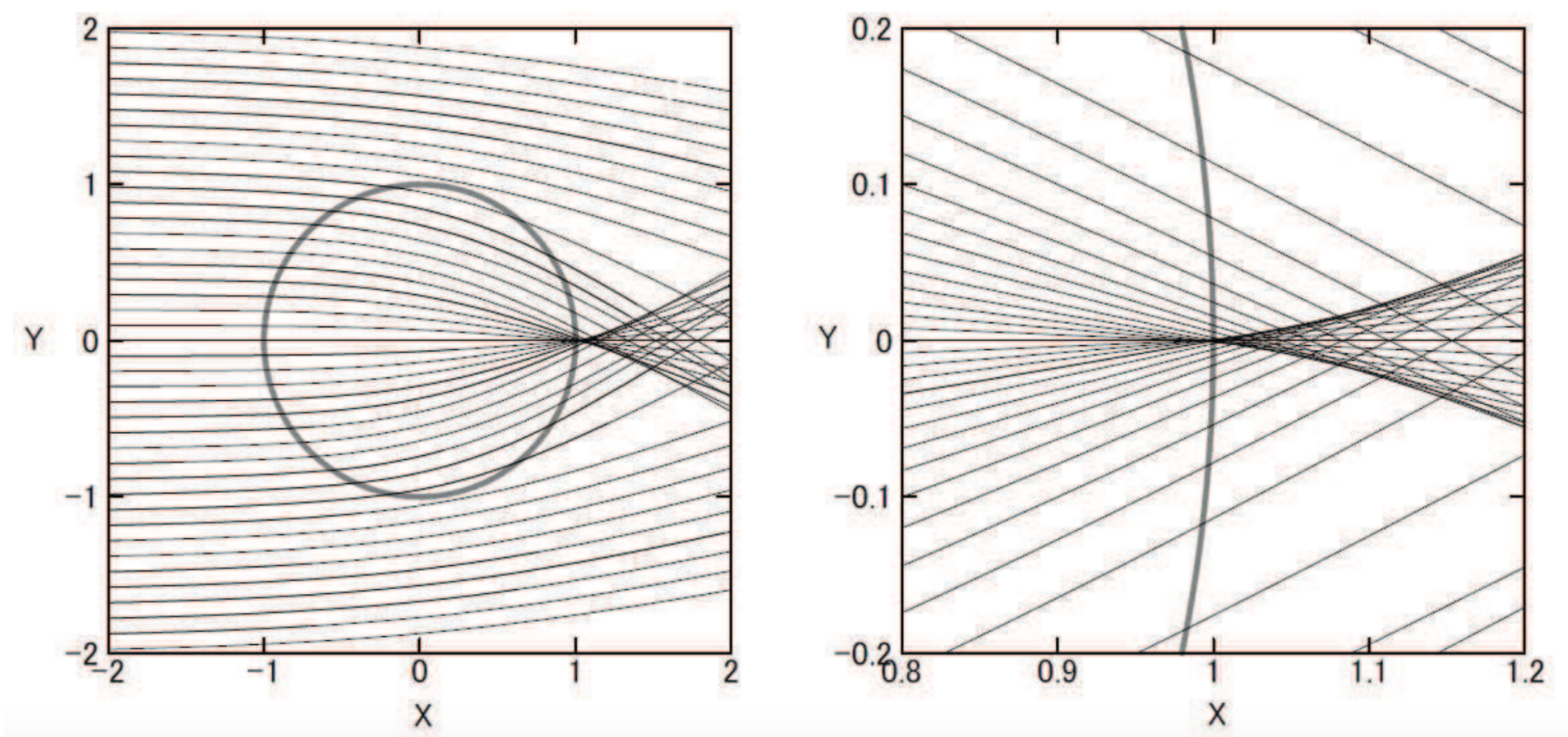

**Figure 3.6:** Gravitational self-focusing of a DM stream by the Earth's mass distribution on its surface with an aperture diameter of about 6000 km and an initial DM velocity of 17 km/s. The right figure is a zoom-in on the Earth's surface (27).

A slightly different velocity of 8.8 km/s has also been found in similar research, with the difference being attributed mainly to the use of different numerical techniques [223].

#### 3.5.1.1 Amplification

The amplification factor is defined as:

$$A(\upsilon) = \frac{S_{\text{aperture}}}{S_{\text{focus}}} = \left(\frac{r_{\text{aperture}}}{r_{\text{focus}}}\right)^2 \tag{3.14}$$

where: $S_{\text{aperture}} = \pi r_{\text{aperture}}^2$ : the collecting area (aperture) of the injection flux enclosed by a circle with impact parameter $r_{\text{inj}} = r_{\text{aperture}}$,

$S_{\text{focus}} = \pi r_{\text{focus}}^2$ : the focal area.

The various amplifications as a function of the distance from the geo-centre for various injection velocities are shown in Fig. 3.7. For the case of an injection velocity of 17 km/s, where the focusing takes place on the surface of the Earth, the flux amplification is on the order of $A \sim 10^9$ the initial flux. Obviously, the focal position moves with the injection velocity with smaller velocity values being focused on the Earth's sub-surface, and larger values towards the Earth's atmosphere. It is noted that in theory, the amplification factor can be infinite at the focal point according to the geometry of the trajectories of the particles.

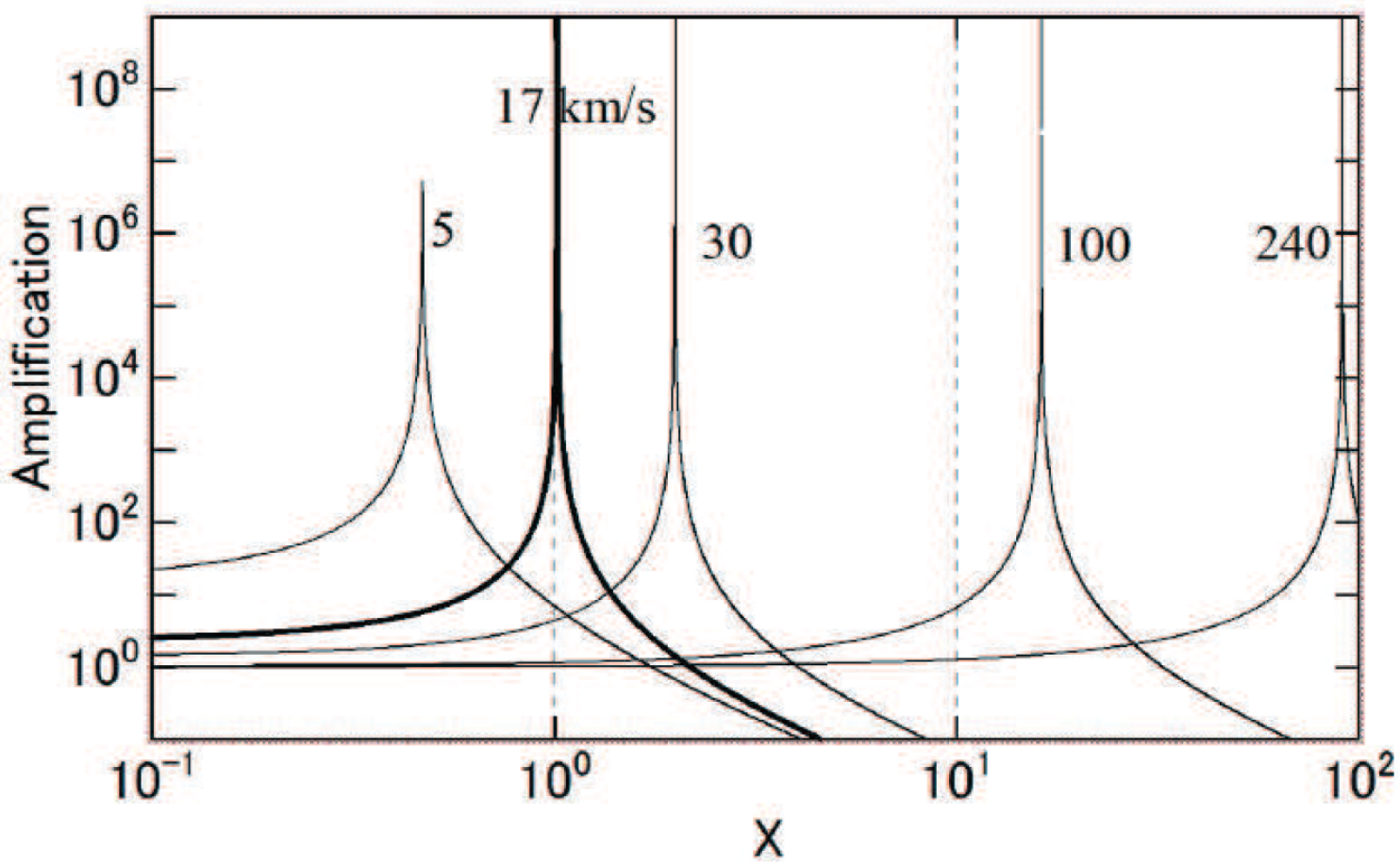

**Figure 3.7:** Flux amplification by the Earth as a function of the distance from the geo-center for DM flows with injection velocities 5, 17, 30, 100 and 240 km/s (27).

#### 3.5.1.2 Focusing velocity

A particle focused by the Earth is also accelerated by the Earth's gravity, which means that at the position of the focusing, the velocity is slightly larger than the injection velocity.

$$\upsilon_{\text{focus}} = \sqrt{\upsilon_{\text{inj}}^2 + \frac{2G_N M_\oplus}{R_\oplus}} \tag{3.15}$$

As an example a DM particle with $\upsilon_{\text{inj}} = 17$ km/s will be focused on the opposite side of the Earth's surface at $\upsilon_{\text{focus}} = 20$ km/s.

### 3.5.2 Different focusing objects

Similarly, with the Earth, the same procedure of gravitational focusing by taking into account the specific inner mass distribution can take place with other solar system bodies. An extended object with radius $r_{\text{aperture}}$ can cause convex-lens like convergence in a focal point with the amplification being proportional to the area of the aperture as seen in Eq. 3.14.

As an example, the Moon can act as a filled-aperture DM flow collector which can focus onto the Earth particles with injection velocities up to $\sim 1.3 \times 10^{-3}\,\text{c}$ (or about 400 km/s) [222].

Alternatively, any object can also be used as concave-lens like collector, where a cylindrical flow of DM creates an enhanced rung around the point mass [224]. For example, the Sun can focus on the position of the Earth particles with an injection velocity of $\sim 200\,\text{km/s}$ and impact parameter $\sim 60 R_\oplus$ (see Fig. 3.8a). The created DM ring will have a diameter of $2\theta \sim 25°$. In this case an Earth observer will see a typical Einstein ring at the position of the Sun with an enhanced DM density being the order of $\sim 3000$ for a zero dispersion stream (see Fig. 3.8b) [225].

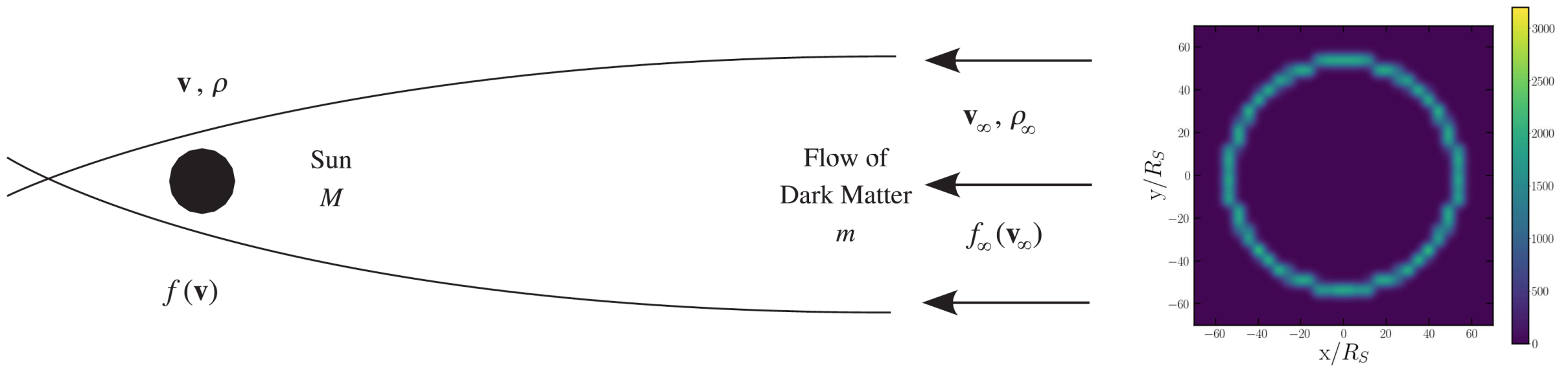

**(a)** Schematic illustration of a DM flow coming from infinity with a mean velocity **v** relative to the Sun and a velocity distribution function $f(\mathbf{v})$, going through the gravitational field of the Sun (57).

**(b)** Gravitational effect by the Sun on a stream with zero dispersion focused on the position of the Earth (41).

**Figure 3.8:** Gravitational focusing by the Sun towards the Earth.

## 3.6 Free fall

### 3.6.1 Cross-section

Considering the case of the Sun, its gravity can deflect incoming unbound particles, increasing their density and velocity. This basically reflects the free fall of slow-moving particles towards the Sun. In Fig. 3.9 the impact distribution on the solar surface for gravitationally focused particles with an example initial velocity is shown.

The cross-section for such a procedure is given by [226]:

$$\sigma(v_{\text{inj}}, r) = \pi r^2 \left[ 1 + \left( \frac{v_{\text{esc}}(r)}{v_{\text{inj}}} \right)^2 \right] \tag{3.16}$$

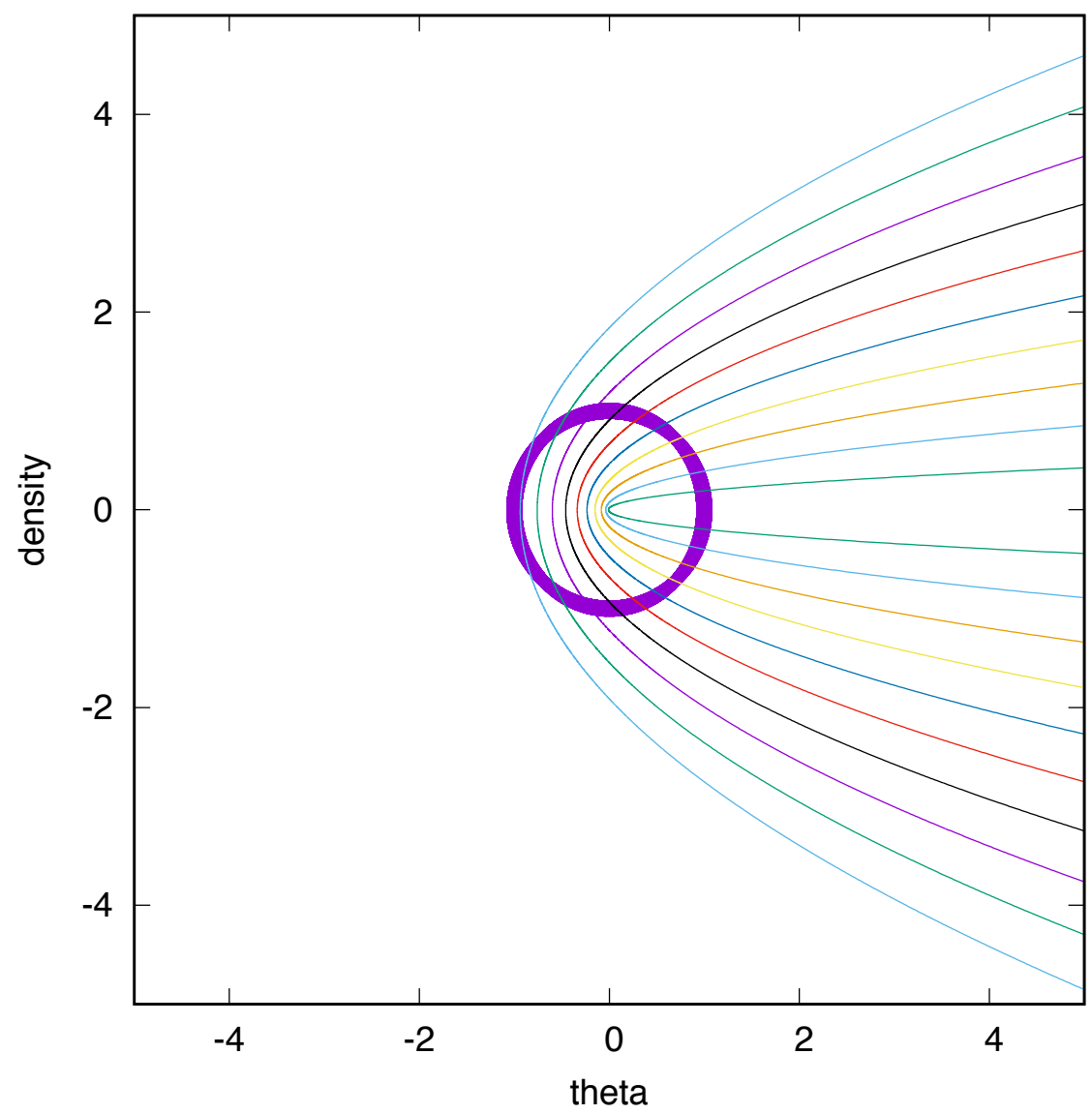

**Figure 3.9:** Free fall trajectories showing the density distribution as a function of the angle (in radian) for an incoming monochromatic uniform flow of particles towards the Sun, with a velocity of $\upsilon_{\text{inj}} = 6\,\text{km/s}$ (28).

where: $r$ : the approach distance,

$\upsilon_{\text{esc}}$ : the escape velocity in distance $r$,

$\upsilon_{\text{inj}}$ : the injection velocity of a particle at infinity.

The escape velocity $\upsilon_{\text{esc}}$, which is basically the critical velocity above which a body is no longer gravitationally bound to a larger mass, is given by [227]:

$$\upsilon_{\text{esc}}(r) = \sqrt{\frac{2G_N M}{r}} \tag{3.17}$$

which for the case of the surface of the Sun, it is $\upsilon_{\text{esc}} \sim 617.7\,\text{km/s}$. Therefore, for the Sun, the flux amplification will be on the order of:

$$\frac{(617.7\,[\text{km/s}])^2}{(\upsilon_{\text{inj}}\,[\text{km/s}])^2} \tag{3.18}$$

For the case of the Earth, the same phenomenon takes place with the corresponding escape velocity from its surface being about $\upsilon_{\text{esc}} \sim 11.2\,\text{km/s}$.

## 3.7 Summary

Slow-moving particles can be gravitationally lensed by the solar system bodies towards the Earth or the Sun because the deflection angle is inversely proportional to the square of the velocity of the particles ($a \sim \frac{G_N M}{b\upsilon^2}$). As an example, the Sun can focus on Earth particles with speeds between ∼0.01 c to 0.2 c, whereas Jupiter can do the same with the values being scaled down by an order of magnitude, i.e., $\sim 10^{-3} - 10^{-2}$ c. The flux amplification from the Sun can be up to a factor $10^{11}$. Similar numbers are expected at the caustics for point sources also for large planets like Jupiter. The Moon can also focus particles with velocities $\sim 10^{-4}$ c onto the Earth with a flux amplification of about $10^4$. On the opposite side, Earth itself can focus particles of $\sim 10^{-3} - 10^{-2}$ c towards the Sun.

Considering also the intrinsic mass distribution of the solar bodies we can also have self-focusing effects. As an example, the Earth can act as a high-efficiency gravitational collector for particles with an initial velocity of about 17 km/s with respect to the geo-centre. The increase of the flux, in this case, can be about $10^9$. With in the same scenario, the Moon can also focus towards the Earth particles with an injection velocity of about 400 km/s. Moreover, for an impact parameter of $\sim 60 R_{\oplus}$, the Sun can focus towards the Earth particles with lower velocities of about 200 km/s with the density enhancement on a zero dispersion stream being on the order of 3000 (see also Publication E.9).

Finally, there is also the case of free fall which can increase the flux of incoming slow-moving particles by $(\upsilon_{\rm esc}/\upsilon_{\rm inj})^2$, where $\upsilon_{\rm inj}$ the initial velocity of the focused particles far away from the attracting solar object.

Therefore, we see that slow-moving particles depending on their speed and the assumed impact parameter can be focused towards the Earth or the Sun and result in a significant flux amplification at the focal plane. As we will see also in the following chapters, this can also have implications in direct DM detection experiments.

4

# METHODOLOGY

## 4.1 Introduction

One of the biggest challenges in modern physics concerns the identification and detection of the constituents of DM. So far the strongest evidence of DM derives from large scale gravitational observations while direct and indirect searches have not provided so far convincing evidence of it. Large-scale observations suggest that the ordinary DM halo in the galaxy is rather uniform at least for the size of the solar system. However, as described in Chap. 2 the co-existence of dark streams or the galactic disk hypothesis is probable. At the same time, due to the non-relativistic characteristics of some DM candidates, planetary gravitational lensing becomes efficient which can result to a significant flux amplification. In the following chapters, a combination of these ideas will be performed while investigating the as yet unanswered question as to whether the motor of the active Sun as well as several unanswered terrestrial phenomena are entirely internal in origin, or if they are triggered by some external influence. The scenario that will be followed in this thesis is the latter, which assumes that the triggering mechanism is based on planetary gravitational focusing of some invisible massive matter stream(s). It is noted, that the generic dark candidate constituents that could trigger the aforementioned phenomena will be referred as *invisible massive matter*, to distinguish them from ordinary DM candidates such as WIMPs and axions which have very small interaction cross-sections with baryonic matter, at least following the already excluded parameter phase-space.

## 4.2 Hypotheses

There is a plethora of phenomena which remain unexplained within known physics. More specifically, despite several attempts and numerous proposed theories, some aspects of the solar activity, such as the origin of the solar flares or the solar emission in Extreme UltraViolet (EUV), sunspots, as well as several observations in the Earth's atmosphere, such as some anomalies on the stratospheric temperatures and on the ionisation of the ionosphere, contain unresolved mysteries.

The driving idea behind this thesis is based on the gravitational focusing and self-focusing effects by the Sun and its planets of low-speed invisible streaming massive matter. This slow-moving invisible matter of galactic or cosmic in origin could interact somehow with the Sun and the Earth and cause at least part of the observed anomalous behaviour. Apparently, similar gravitational lensing can take place between the planets and other celestial bodies (see Fig. 4.1).

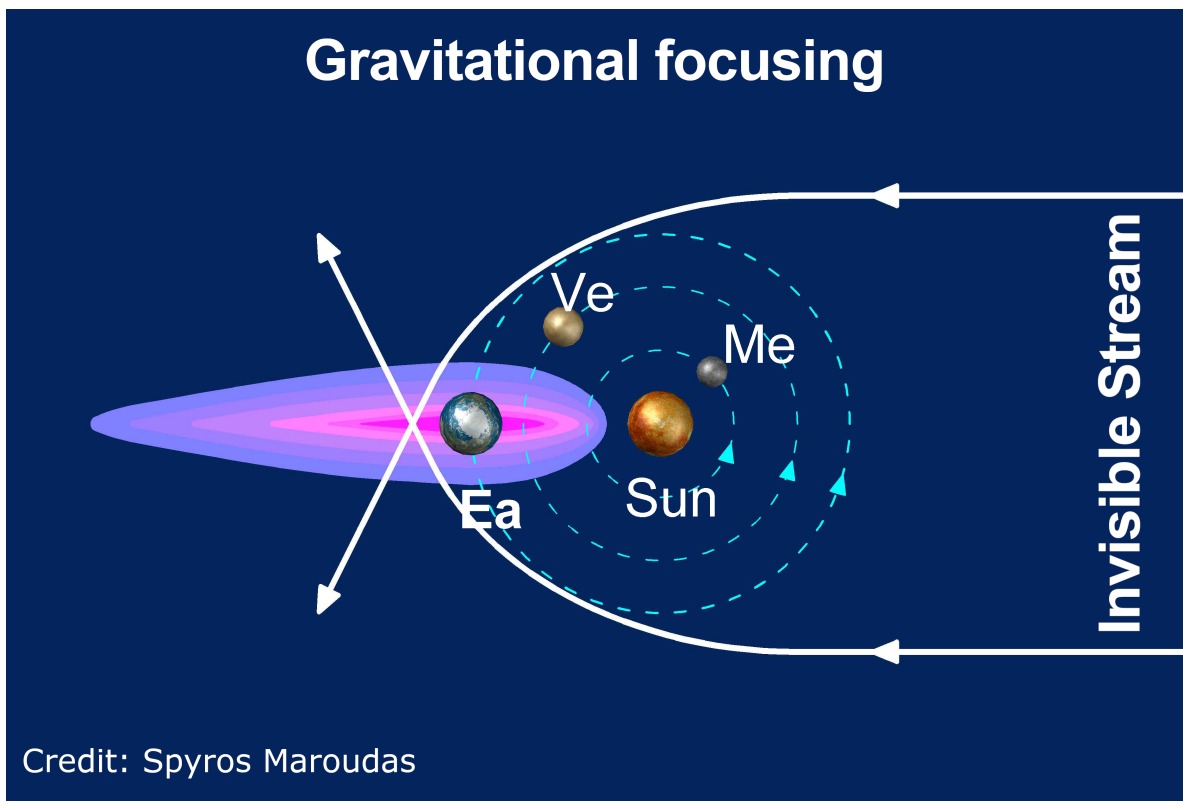

**(a)** Gravitational focusing effect by the Sun towards the Earth. The orbiting planets disturb periodically the dominating Sun's gravitational focusing effect (inspired by [183]).

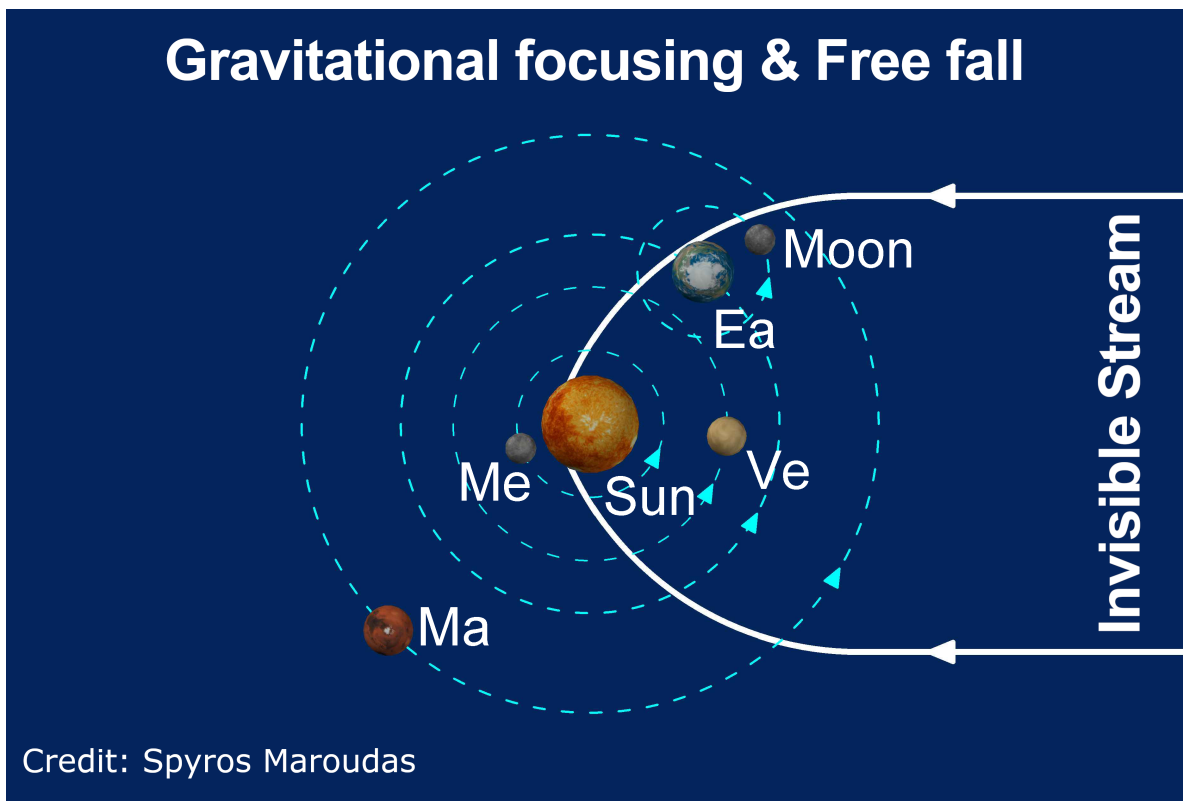

**(b)** Gravitational focusing and free fall of incident low-speed streams towards the Sun. The flux can be gravitationally modulated by an intervening planet resulting in a specific planetary dependence.

**Figure 4.1:** Schematic view of the flow of a putative slow-speed invisible matter stream, with gravitational (self)-focusing effects by the Sun, Earth, Venus, Mercury and/or the Moon.

For example, when a stream is aligned towards the Sun with an intervening planet there is a transitory marked flux enhancement at the site of the Sun, increasing its impact up to several orders of magnitude, for as long as the alignment lasts. The amplification factors, can even be up to $10^{11}$ in the ideal case for solar focusing. Furthermore, an additional contribution comes from the self-focusing effect of the solar system bodies due to their intrinsic mass distribution, causing an additional flux enhancement. For example, for the case of the Earth for particles with 17 km/s the flux enhancement is on the order of $10^9$ (see also Publication E.9). Finally, a large amplification takes place from the free fall towards the Sun of slow-moving particles (see Fig. 4.1b). Similarly, this free-fall effect also applies to the rest of the planets like the Earth with the corresponding escape velocities. If such a hypothesis is correct then an unexpected statistically significant planetary relationship of the selected observable should be derived based on the optimum alignment of the planets with the incoming stream(s). Therefore, an observed planetary dependence will be the decisive signature for the involvement of the dark sector.

In order for the above to happen, the following assumptions are made:

- The widely addressed DM candidates such as WIMPs and axions can not have such an impact as they are extremely weakly interacting. The invisible massive matter candidates should instead interact strongly with normal matter to produce macroscopically observable effects.
- The stream constituents should have a velocity distribution which will allow planetary and solar gravitational lensing to occur which will increase the flux impinging into the Sun and the Earth accordingly (see Fig. 4.1).
- The irradiation of the solar system by any kind of feeble interacting particles must not be isotropic and must be arriving preferentially along the ecliptic plane.
- The temporally increased influx can trigger solar activity as well as terrestrial observations.

There is currently no known force that can explain remote planetary effects, except the widely suspected gravitational tidal forces such as those acting on Earth by the Moon. However, the strength of this force is far too feeble to trigger for example solar phenomena, by a factor $10^{-12}$ [228, 229]. Nevertheless, the aforementioned planetary gravitational lensing effects of streaming invisible massive matter would result in a seemingly remote interaction within the solar system [230]. Therefore, following this mechanism, it is natural to expect a planetary dependence of various phenomena which however should not occur within conventional physics (see also Publications E.1, E.3, E.5, E.14, E.15, E.17, E.18, E.19, E.20, E.21, and E.23).

At this point, no assumption can be made about the exact nature of the streaming invisible matter nor its exact interaction cross-section with normal matter, although some hints and suggestions will be provided. The main goal of this thesis is the proof of the gravitational focusing scenario within our solar system and the existence of preferred direction(s) pointing to an external triggering mechanism of the made observations.

### 4.2.1 Particle candidates

The conventional DM candidates like the aforementioned WIMPs or axions, cannot be used to explain the observations made in this research due to their far too small interaction cross-sections with normal matter. Even though an assumption about the exact nature of the streaming invisible matter can not be made at this point, there are possible particle candidates that seem to be fitting in (see also Sect. 1.3.6).

A first example particle candidate is AQNs which are dust-like particles with nuclear density and they should interact strongly with the solar and the Earth's atmospheres which are both ionised from a certain height outwards and permeated with magnetic field [231–233] (see also Publication E.11). Another candidate that deserves further consideration are magnetic monopoles (see Sect. 1.3.5). Finally, dark photons, which are also DM candidates, are of potential interest with their rest mass being still unknown (see also Publication E.4).

Notably, there are several publications discussing potential constituents from the dark sector having a large cross-section with normal matter without contradicting cosmological arguments [234]. Examples are particles with a rest mass around $10^{-3}\,\mathrm{GeV}/c^2$ [235] and $10^{25}\,\mathrm{GeV}/c^2$ [236] which interact strongly with normal matter (see also Publication E.3). Discussed cross-sections are on the order of $\sigma = 10^{-23}\,\mathrm{cm}^2$ or even up to $\sigma = 10^{-18}\,\mathrm{cm}^2$ [235].

Some specific estimations on the order of magnitude of the potential energy range of probable candidates will be made in the following chapters.

### 4.2.2 Planetary configurations

The different planets according to their mass and distance from the Sun select different velocity ranges for an optimal focusing on the Sun. For this reason, instead of simple synodic alignments, a different planetary configuration can be associated with transient solar phenomena. As an example, particles with $\upsilon \sim 10^{-3}c$ have a time of flight of about 5 d for 1 AU. This means that if the Earth participates with the gravitational lensing, then it will appear about 5° advanced in heliocentric longitude $l$ when the solar activity is being triggered. This means that simple synodic alignments are not expected as it happens for particles with $\upsilon \sim c$. However, several planetary periodicities are expected which can provide more insight on the direction and the velocity distributions of the streams.

For orientation purposes, in the specific Earth - Sun configuration (see Fig. 4.1a) around the 18$^{\text{th}}$ of December the assumed invisible stream comes within 5.5° from the direction of the GC which corresponds to a heliocentric longitude of $\sim 266.5°$. Then, Earth in Fig. 4.1a is on a heliocentric longitude of $\sim 86.5°$ thus having the alignment GC → Sun → Earth.

## 4.3 Research design

A statistical analysis of time (=position) distributions of the various planets for each solar or terrestrial long-term dataset is required. For this purpose, the heliocentric longitude coordinate system of the solar system bodies is used. A peaking distribution at any planetary position would provide a strong piece of evidence that the occurrence of the analysed phenomena is modulated by the position of the planets. The observation of narrow (on the order of even days or weeks) and/or peaking distributions are very important for this kind of analysis as it excludes any planetary tidal-force inspired models as well as any intrinsic phenomena.

Due to the different revolution times around the Sun of the different planets, the appearance of a significant excess around the same longitudinal range will signal the presence of one (or more) stream(s), since each planet is passing through the same longitude at different times. Therefore, a search around specific longitudinal ranges for different datasets and a following comparison of the results is essential as it can pinpoint the possible direction(s) of the assumed stream(s).

### 4.3.1 Eccentricity correction

Some planets, such as Mercury have a significant orbital eccentricity. This effect is translated in the amount of time that the planets stay in a position, which is not stable and changes according to their eccentricity. For example, for Mercury the stay-time per longitude-bin varies in the course of one orbit by a factor $\sim 2.2$. Therefore, for this generic modulation to be factored-out, this distribution has to be subtracted from each dataset for the specific planet that is to be used. This way the resulting time (=position) distributions of each dataset will be clear from such eccentricity related effects (see also Fig. 4.4). For this subtraction of the eccentricity, the longitudinal distribution of the data is basically divided with the expected longitudinal distribution of the stay-time of the selected planet.

From another perspective, the eccentricity of a planetary orbit corresponds to the random distribution of a dataset that one would expect if no planetary relationship due to gravitational lensing is taking place. Therefore, if the dataset analysed was randomly distributed across the whole orbit of a planet, the resulting distribution from the subtraction of the data from the simulated eccentricity distribution would give a flat distribution.

### 4.3.2 Periodic distributions

Due to the planetary gravitational lensing of slow-moving flows of dark constituents towards the Sun and the Earth, a periodic behaviour is created. The solar system has in total about 30 orbital and synodic periods plus several multiples (see Tab. 4.1). Many of these have

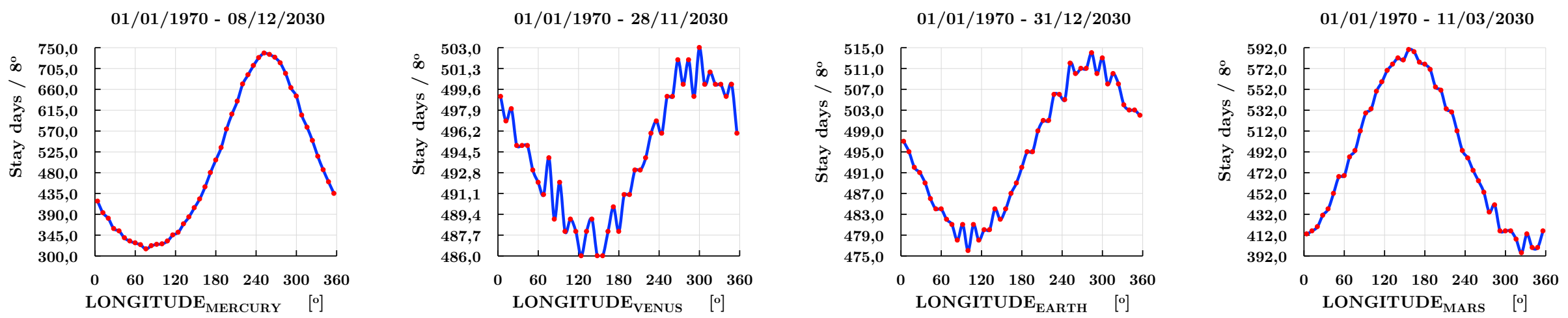

**(a)** Mercury's eccentricity for 253 orbits with 8° bin resulting in 57.32% modulation.

**(b)** Venus' eccentricity for 99 orbits with 8° bin resulting in 3.38% modulation.

**(c)** Earth's eccentricity for 61 orbits with 8° bin resulting in 7.39% modulation.

**(d)** Mars' eccentricity for 32 orbits with 8° bin resulting in 33.05% modulation.

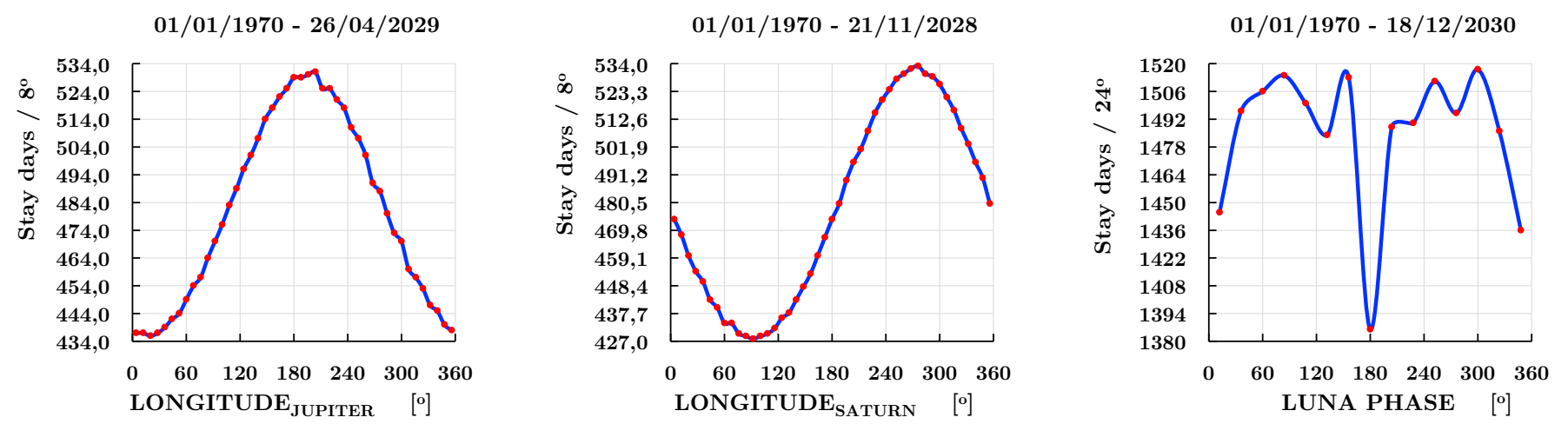

**(e)** Jupiter's eccentricity for 5 orbits with 8° bin resulting in 17.89% modulation.

**(f)** Saturn's eccentricity for 2 orbits with 8° bin resulting in 19.70% modulation.

**(g)** Moon's eccentricity for 815 orbits with 24° bin resulting in 8.64% modulation.

**Figure 4.2:** The *stay days* per bin of heliocentric longitude for the various planets including Earth's Moon. In each case the dates have been chosen to result in multiple orbits for each planet. These plots depict the eccentricity of each planet and at the same time the expected isotropic distribution for any non planetary dependent dataset.

already been observed in large datasets such as sunspots and solar flares and have not yet been understood through conventional physics. However, they could be studied and perhaps explained with the aforementioned reasoning.

As an example, the Jupiter's orbital revolution around the Sun of 11.8 y or the synod Jupiter-Earth-Venus of 11.01 y are probably not symptomatically close to the 11 y periodicity of the solar activity [237]. By performing a Fourier analysis of the selected dataset such periodicities can be measured which strengthen the planetary dependence claim (see Sect. 4.5.2).

**Table 4.1:** Synodical periods between all planets in days and years [12].

| | Mercury | Venus | Earth | Mars | Jupiter | Saturn | Uranus | Neptune | Pluto |
|---|---|---|---|---|---|---|---|---|---|
| Mercury | - | 145 d | 116 d | 101 d | 90 d | 89 d | 88 d | 88 d | 88 d |
| Venus | 145 d | - | 584 d | 334 d | 237 d | 229 d | 226 d | 226 d | 225 d |
| Earth | 116 d | 584 d | - | 780 d | 399 d | 378 d | 370 d | 367 d | 367 d |
| Mars | 101 d | 334 d | 780 d | - | 816 d | 734 d | 703 d | 695 d | 692 d |
| Jupiter | 90 d | 237 d | 399 d | 816 d | - | 19.8 y | 13.8 y | 12.8 y | 12.5 y |
| Saturn | 89 d | 229 d | 378 d | 734 d | 19.8 y | - | 45.5 y | 36.0 y | 33.5 d |
| Uranus | 88 d | 226 d | 370 d | 703 d | 13.8 y | 45.5 y | - | 171.9 y | 127.8 y |
| Neptune | 88 d | 226 d | 367 d | 695 d | 12.8 y | 36.0 y | 171.9 y | - | 498.2 y |
| Pluto | 88 d | 225 d | 367 d | 692 d | 12.5 y | 33.5 y | 127.8 y | 498.2 y | - |

# 4.4 Data collection

## 4.4.1 Planetary positions

The data projected on the planetary heliocentric longitude positions have been downloaded from Caltech/Jet Propulsion Laboratory (JPL)'s Horizons System of National Aeronautics and Space Administration (NASA) `https://ssd.jpl.nasa.gov/horizons.cgi`. The main bulk of the downloaded data correspond to daily binning but depending on the characteristics of the dataset, data with different time resolutions have also been used. The "Heliocentric Ecliptic Longitude" $l$ and "Latitude" $b$ (in °) as well as the "Heliocentric Range" $r$ have been primarily used for the various analyses (see Fig. 4.3).

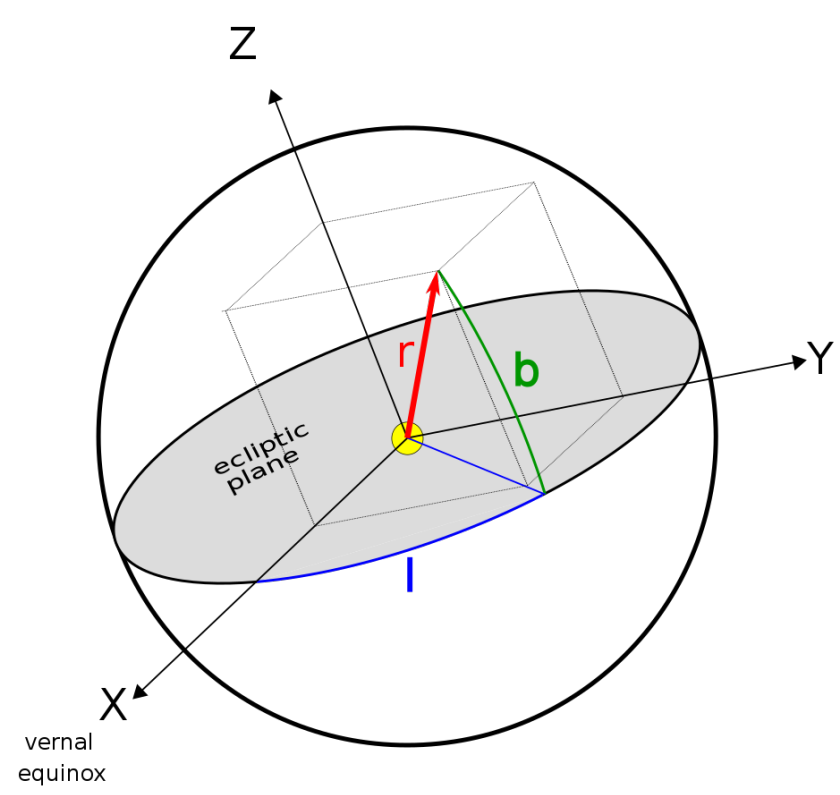

**Figure 4.3:** The heliocentric ecliptic coordinate system which describes the planets' orbital movement around the Sun. The system is centred on the center of the Sun, its plane of reference is the ecliptic plane, and the primary direction (in the x-axis) is the vernal equinox. The longitude is indicated by $l$ and the latitude by $b$ , whereas $r$ is the distance from the Sun's center (29).

### 4.4.2 Solar and terrestrial datasets

Due to the large number of datasets that have been analysed in this thesis, the specific source and details of each dataset will be mentioned directly in each of the following chapters of Parts III and IV. Most of the datasets have been downloaded from publicly available sources but others have been acquired from private communications with specialists.

In general though, and as a rule of thumb, any dataset that points to an 11 y periodicity is suggestible to a streaming invisible matter analysis which can validate the existence of invisible stream(s) as well as provide some hints on the nature and properties of the actual particle candidates and their interactions with baryonic matter.

## 4.5 Data analysis procedure

A variety of programming languages as well as computer software programs and online tools were used for the data analysis of the various datasets presented in Parts III and IV. The main programming languages used were Visual Basic for Applications (VBA), Fortran and Python (see Sect. D). In addition, scientific software such as Mathematica, Origin, Microsoft Excel and Veusz were used for additional numerical and statistical analysis as well as for graphing and curve-fitting purposes.

The very first step before the analysis of any dataset and the search for planetary dependence is the projection of the time distribution of the data-points over the full existing period which depicts directly any existing periodicities but also points to possible peculiarities or corrections that have to be performed before the application of the analysis algorithms. An additional visualisation is through the histogram presentation of the data per year, per month, per week and per day of a month.

### 4.5.1 Main analysis algorithm

After the graphing, the data correction steps and the first quality checks, the data are analysed searching for possible planetary relationships. The main idea is the calculation of the distribution of the data as a function of the heliocentric longitude of each planet from 0°–360°. Therefore, the first step is the longitudinal distribution of the data for each one of the planets by projecting the day of each data-point on the corresponding planetary longitude (see Fig. 4.4a). Depending on the total available time range of the data, the planets which perform the most orbits are usually selected, based on their revolution period to exclude systematics. The main analysis algorithm used for this procedure is described in more technical detail in Appendix Sect. D.1.

As already mentioned the eccentricity effect has to be removed from each distribution and therefore the eccentricity of each planet is also plotted in the same way and with the same settings and parameters as the data. This corresponds to a number of "Stay days" that the chosen planet(s) stay in each specific bin of heliocentric longitude (see Fig. 4.4b). The longitudinal distribution of the raw data is then divided with the longitudinal distribution of the "Stay days" to remove completely such kinematical effects (see Fig. 4.4c). In the rest of the plots that will follow in Parts III and IV with the analysis of the various datasets, the final "eccentricity corrected" plots will be presented unless stressed otherwise. The label "Stay days" is also removed to facilitate reading comprehension.

Finally, the label "No Planetary Constraints" that appear at the top of the plots, above the used period, indicates that no other planet has been constrained to propagate in a specific longitudinal range. It is noted that the line is a simple connection between the data points and is added for visual purposes.

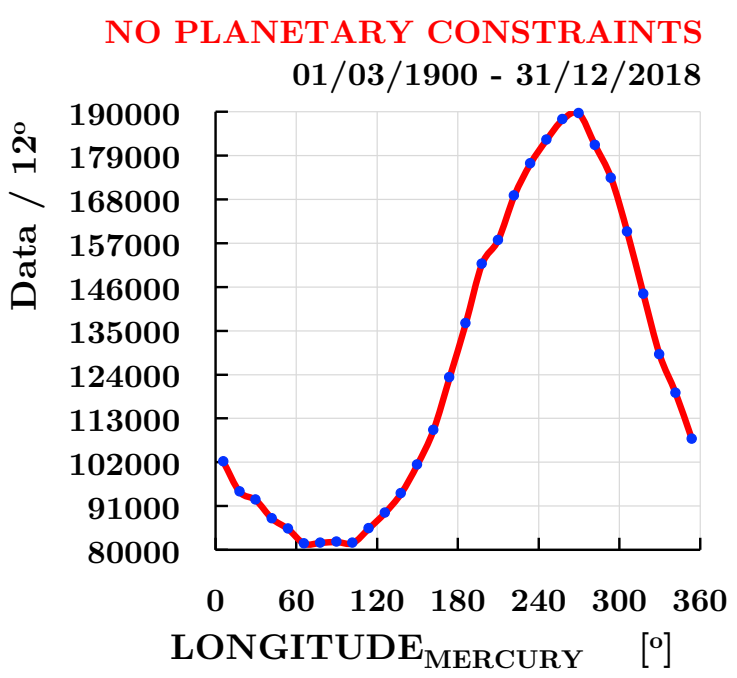

**(a)** Mercury's longitudinal distribution of the data.

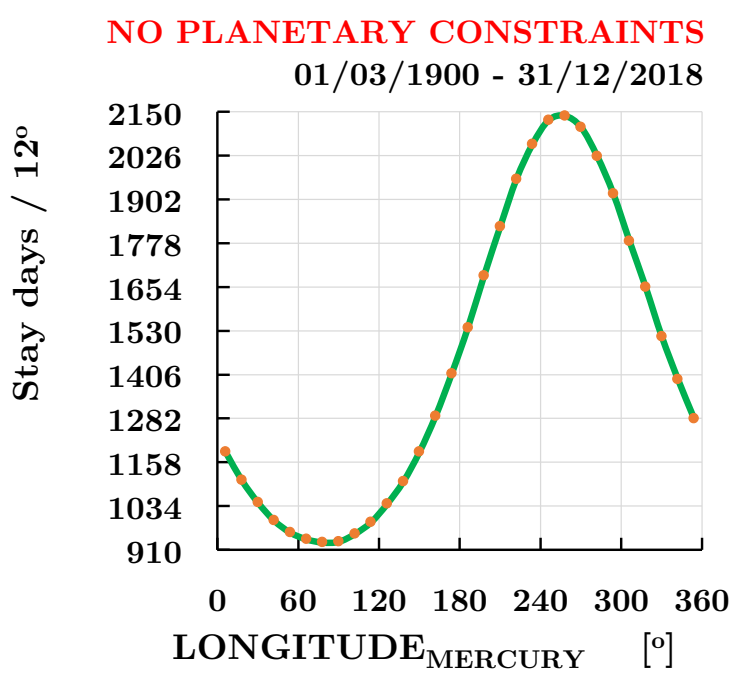

**(b)** Stay time of Mercury per bin of longitude due to its eccentricity.

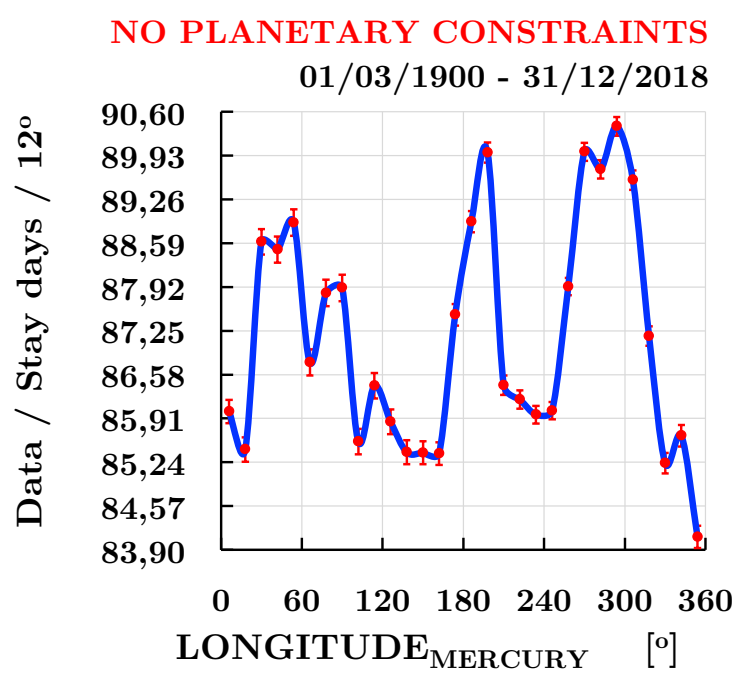

**(c)** Mercury's longitudinal distribution of the eccentricity corrected data by division of the data in Fig. 4.4a by the data in Fig. 4.4b. The error bars in each point are also shown.

**Figure 4.4:** Example of the main analysis procedure in a test dataset. The data are plotted as a function of the heliocentric longitude of Mercury for a 12° bin.

#### 4.5.1.1 Combination of planets

The aforementioned procedure considers the effect of each planet regardless of the position of the other planets. However, it is difficult to infer only from one plot a particularly significant clustering. Moreover, and whatever clustering is observed, a single planet will never be able to distinguish between a single inner solar clock mechanism from an external cause. This is why a variety of analyses has to be performed using single planets as well as combinations of them (see Fig. 4.5). The easiest and most logical approach is to look at whether the combined effect of two planets influences the observed distributions. Subsequently, even more planetary combinations can be selected to be analysed. Assuming Poisson statistics the observed excessed are compared to a randomly occurring rate and the significance of each peak can be derived.

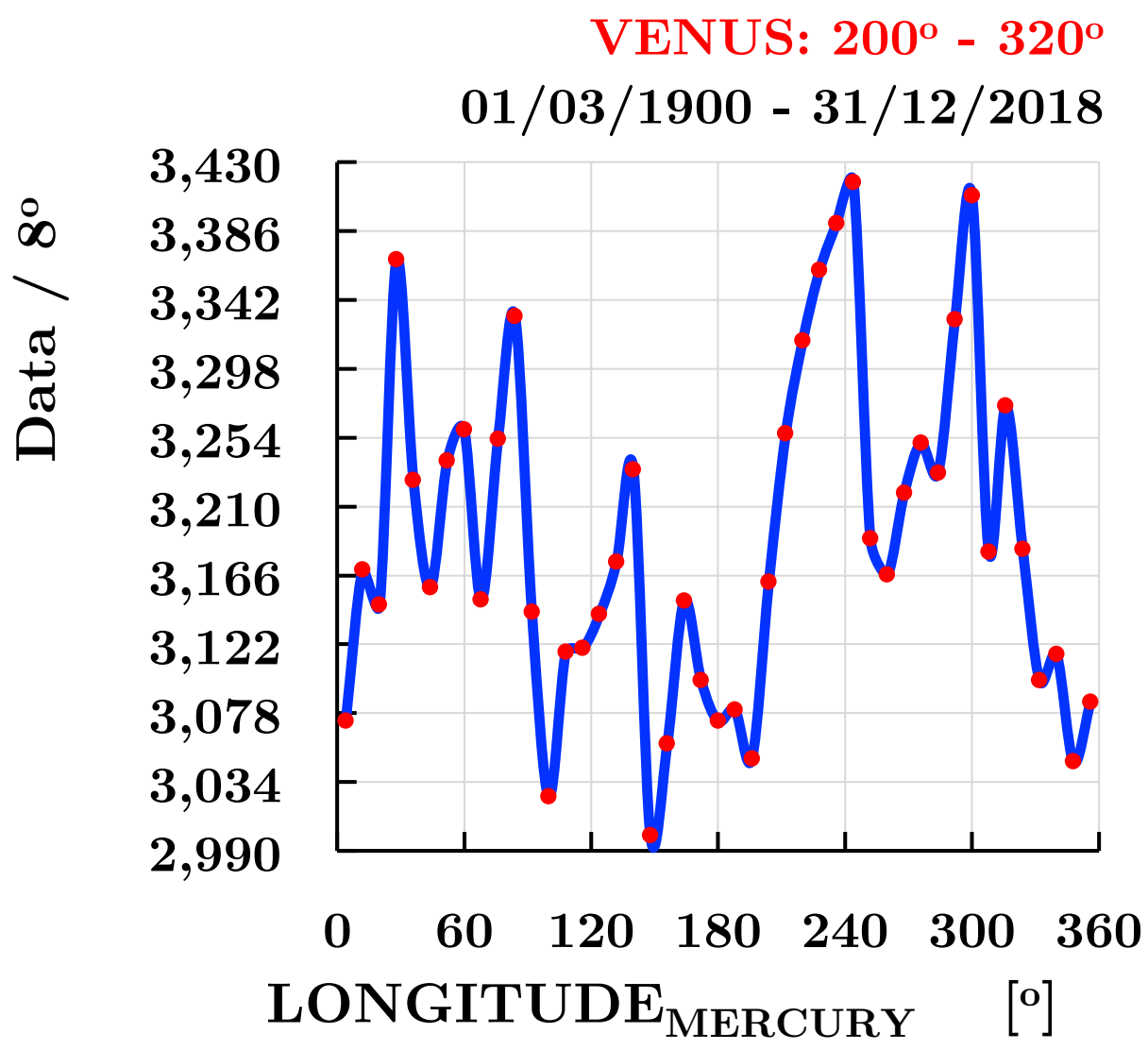

**Figure 4.5:** Example of a dataset distribution as a function of Mercury heliocentric longitude with the constraint of Venus being at longitude between 200°–320°.

An additional analysis tool that can provide further insight into the planetary dependence is the comparison of planetary longitudinal regions being 180° apart in the combined effect of two planets. For example, if the distribution of the data as a function of Mercury heliocentric longitude with the constraint of Venus being at a longitude between 200°–320° shows a significant peak then they can be compared with the distribution of the data as a function of Mercury heliocentric longitude when Venus is between 20°–140°. And if these two distributions show a clear difference or a complete absence of the peak in the second case, this means that the 200°–320° of Venus is a preferable region, emphasising also the strong influence of the relative position of Venus and Mercury.

### 4.5.2 Fourier analysis

A complementary technique for the retrieval of periodic frequencies over a large dataset is through the calculation of the Fourier periodogram, which estimates the spectral density of a signal. More specifically, the "online periodogram interface service" from NASA's "Exoplanet Archive" is used `https://exoplanetarchive.ipac.caltech.edu/cgi-bin/Pgram/nph-pgram`, where a Lomb-Scargle periodogram spectral analysis is performed.

However, this technique can not distinguish specific planetary distributions and structures but also effects from the combination of more than one planet as the specifically-designed algorithms described before, but can only point to existing periods associated with planetary orbits, synodic periodicities and multiples thereof and can provide further evidence in support of the principal hypotheses. The reason for this is that the whole solar system affects the trajectories of invisible matter constituents, even though occasionally only two or even one single celestial body can dominate a planetary signature.

## 4.6 Summary

An observed peaking relation on the positional planetary distribution of an observable, excludes on its own any conventional explanation, either from remote planetary interactions which are smooth over an orbit, or intrinsic. On the other hand, a statistically significant planetary relationship supports the working hypotheses of streams of unidentified slow-moving massive particles which are gravitationally focused by the planets and/or the Sun of our solar system. The huge flux amplification of these invisible massive particles which can take place together with the assumed large interaction cross-sections with normal matter could then trigger a variety of phenomena on the Sun and the dynamic atmosphere of the Earth which can not otherwise be explained.

Such a statistical analysis based on this new proposed mechanism can be performed over a wide range of solar and terrestrial observations with the results having the ability to test and/or constrain unconventional theoretical models for the invisible matter. As we will see also in the next chapters, the results of these analyses could have deep implications in current DM experiments and the potential to eventually redefine the strategy of the conventional direct searches for DM which has not yet provided any signature for all experiments.

In conclusion, this research aims to establish a totally different view of the dark sector and provide new tools in particle and astroparticle physics, which could constitute a major breakthrough in our quest for the exploration and explanation of our physical universe.

# Part III:

# Solar Observations

*"If I had to choose a religion, the sun as the universal giver of life would be my god."*

—Napoleon Bonaparte (1769–1821),
*Military Leader*

# 5

# INTRODUCTION

Solar activity is known to be modulated within a $\sim 11\,\mathrm{y}$ cycle, moving from its most quiet period called *solar minimum* to its most active one called solar *maximum* and back (see Fig. 5.1a). Although several theories about this periodicity have been proposed [238], the solar cycle remains as one of the biggest problems in solar physics [237, 239]. Increased solar activity is associated with an increase in the number of sunspots, solar flares and Coronal Mass Ejections (CMEs) (see Fig. 5.1b), However, the origin of these phenomena along with others, such as the solar corona heating mechanism [240] (see Chap. 7) or the coronal element abundances, remain unexplained with no consensus model currently being widely accepted.

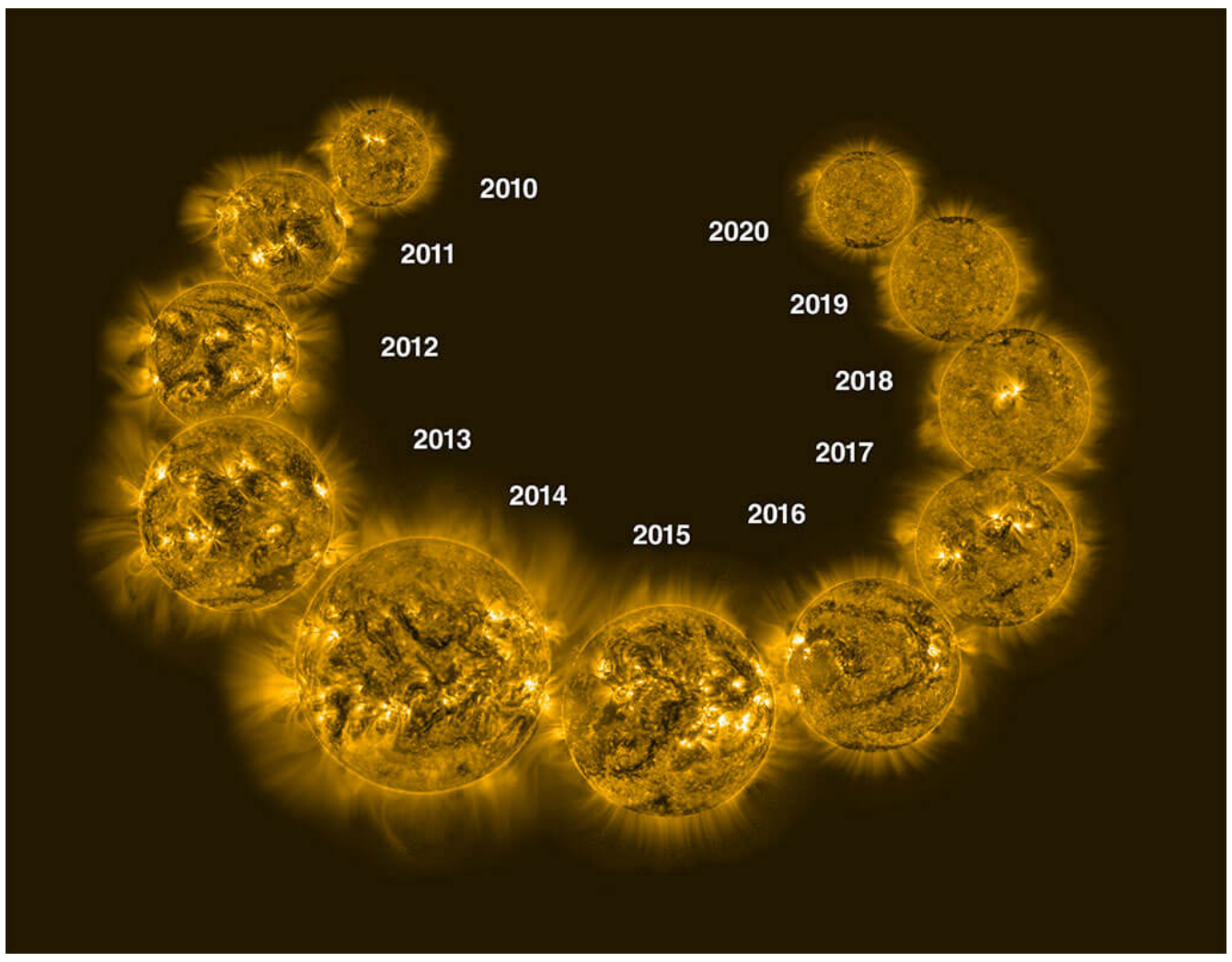

(a) Sun's evolution in EUV light from 2010 to 2020 (58).

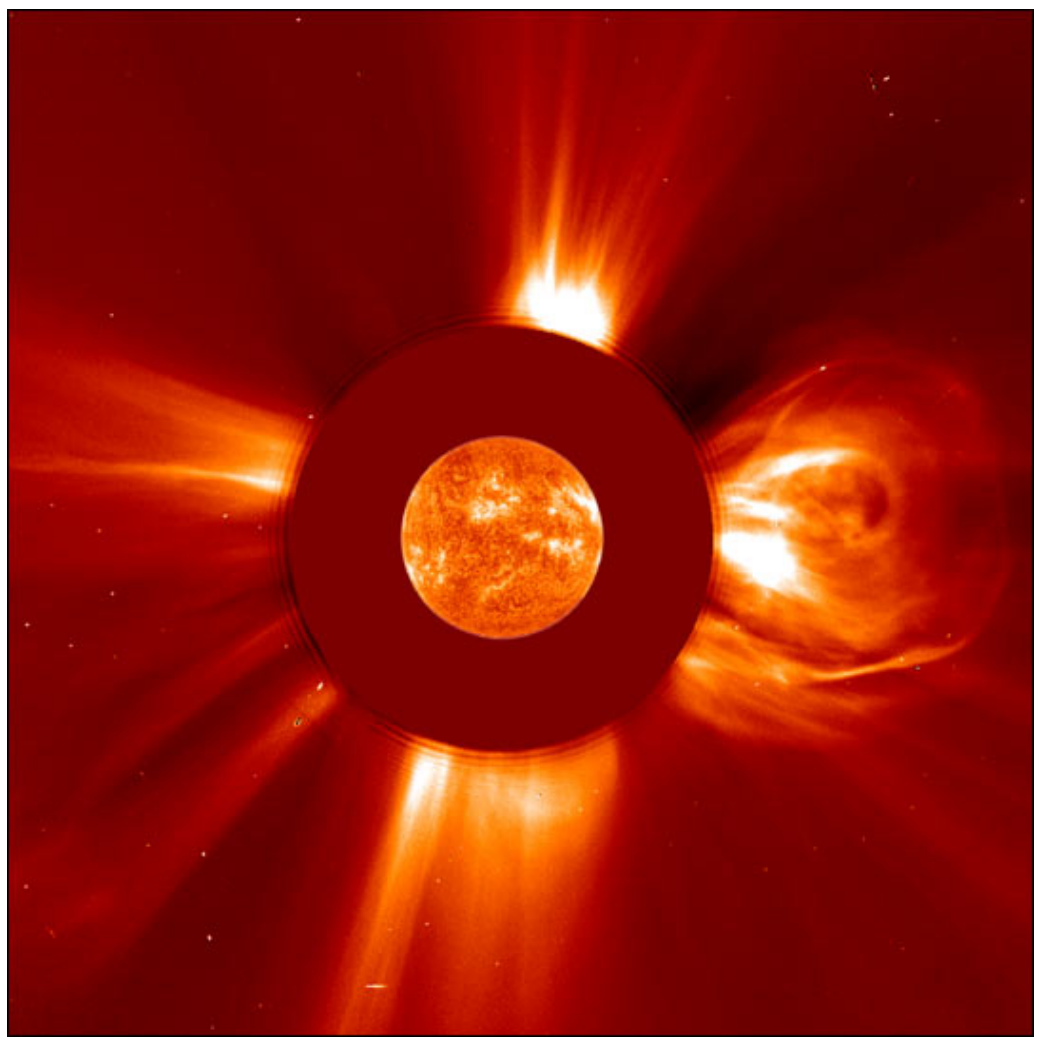

(b) Picture of a CME observed by SOHO in 2001 (59).

**Figure 5.1:** Solar cycle and associated phenomena.

The alternative solution proposed in this work is that behind the at-least-partly triggering of these observations is the gravitational focusing of non-relativistic streaming invisible massive matter by the Sun and its planets (see Fig. 5.2) [237]. The reason is that the gravitational deflection goes with $1/v^2$ and therefore streaming invisible matter particles with velocities around $10^{-3}$ c or less can get strongly influenced over the solar system distances. As an example, the Earth can focus invisible slow moving particles towards the Sun with speeds of about $10^{-3} - 10^{-2}$ c. Additionally, the Sun's gravity can attract particles with speeds less than $\upsilon_{esc} \sim 617.7\,\mathrm{km/s}$ reflecting free fall, with the flux enhancement increasing by $(617.7/\upsilon_{inj})^2$. If such mechanisms are at work, then the occurrence of the investigated phenomena would be strongly modulated by the position of the planets interacting with the invisible streams (see Publications E.1, E.5, E.14, E.15, E.17, and E.13).

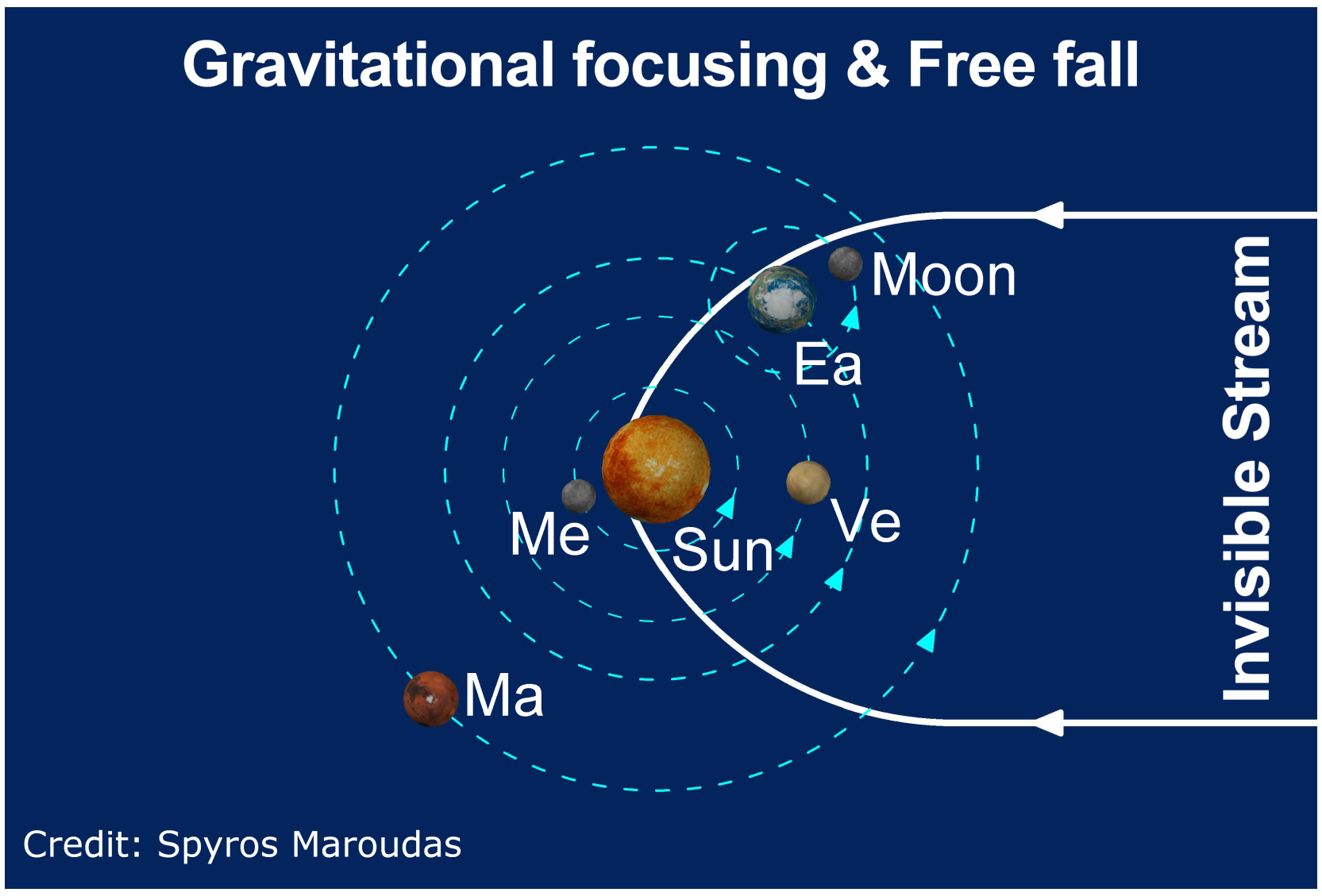

**Figure 5.2:** Schematic view of gravitational focusing of invisible slow-moving highly interacting particles towards the Sun (same as Fig. 4.1b). The dominating free fall effect of the Sun is indicated by the two white thick lines, with the flux being also gravitationally modulated by the intervening planets resulting to an observed planetary relationship.

The link of solar activity with planetary motions is not new [230]. Wolf was first to suggest on 1859 that solar dynamics is partially driven by planetary tides [241], and, since then, this claim has been supported by various empirical pieces of evidence (see references within [242]). The only known forces that could explain these observations were the widely suspected gravitational tidal forces. Since then, several calculations have been made to quantify the tidal effects of the planets on the solar surface, resulting in a value about $10^{-12}$ the gravity of the Sun [228, 229]. This means that the planetary tidal forces are too feeble to modulate solar activity [229, 243, 244], although more complex mechanisms can not be excluded [242, 245].

As a result, a discovered planetary relationship on small-scale solar observations would point to the alternative scenario of streaming invisible matter which undergoes planetary gravitational focusing towards the Sun. Notably, possible constituents from the dark sector with large cross-sections with normal matter [235], such as AQNs, magnetic monopoles or dark photons, can further amplify the impact of focused DM streams. It is mentioned that the derivation of short-lasting statistically significant peaks on the planetary longitudinal distributions of solar observables would further exclude tidal-force models, which should extend smoothly over an orbital period. In the following Chap. 6 through 12 some of the most significant planetary distributions will be presented.

# 6

# SOLAR FLARES

## 6.1 Introduction

One of the greatest solar mysteries is the origin and physical mechanism behind solar flares. Solar flares are a brief eruption of intense high-energy radiation from the Sun's surface towards outer space (see Fig. 6.1). High-energy particles, electron streams, hard X-rays and radio bursts are often emitted together with the occurrence of a shock wave when the solar flares interact with the interplanetary medium. They were first observed by Carrington [246] and Hodgson [247] in 1859 as localised visible brightening of small areas within a sunspot group. Flares are usually accompanied by a CME, which are large releases of plasma and magnetic field from the Sun's corona. Together with CMEs, solar flares are the most powerful explosions in our solar system. The produced radiation, which spans across the whole electromagnetic spectrum at all wavelengths from radio waves to gamma rays. The frequency of the occurrence of solar flares follows the 11 y solar cycle and varies from several per day when the Sun is particularly "active" to less than one per week when the Sun is "dormant". The large flares have a duration ranging from about 20 min up to several hours [248].

Although multiple mechanisms have been proposed [249–252] (see also [253] for a review), their physical mechanism as a well as a forecasting of their occurrence remains an open problem

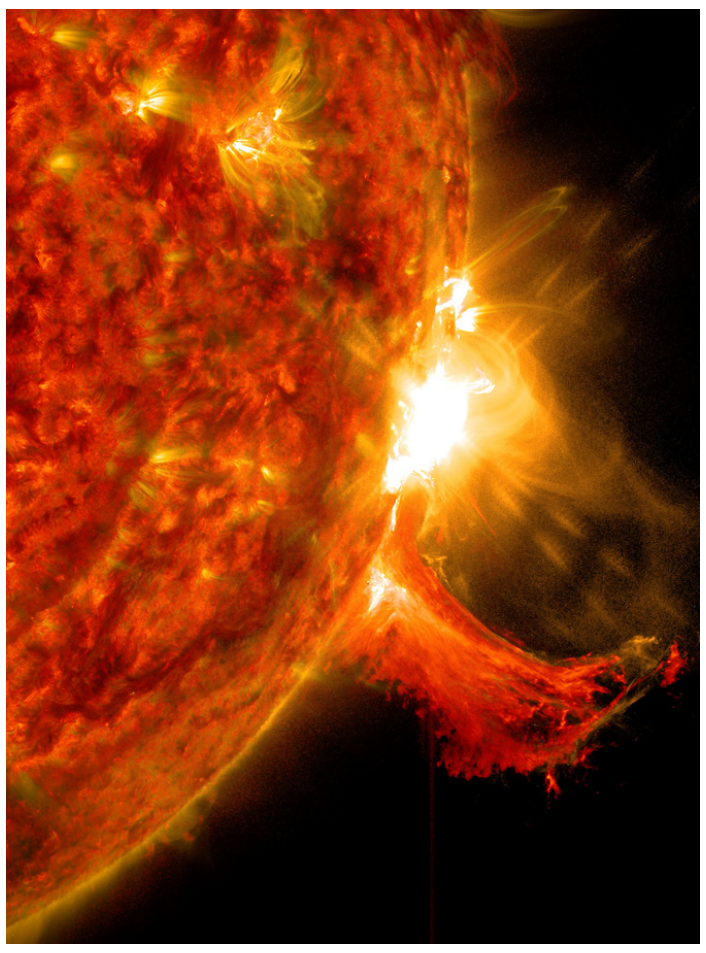

**(a)** Image of a M7.3 solar flare shown in the bright flash of light on the right limb of the Sun. A burst of solar material erupting into space is also seen below it. The image was captured by SDO on 02/10/2014 (60).

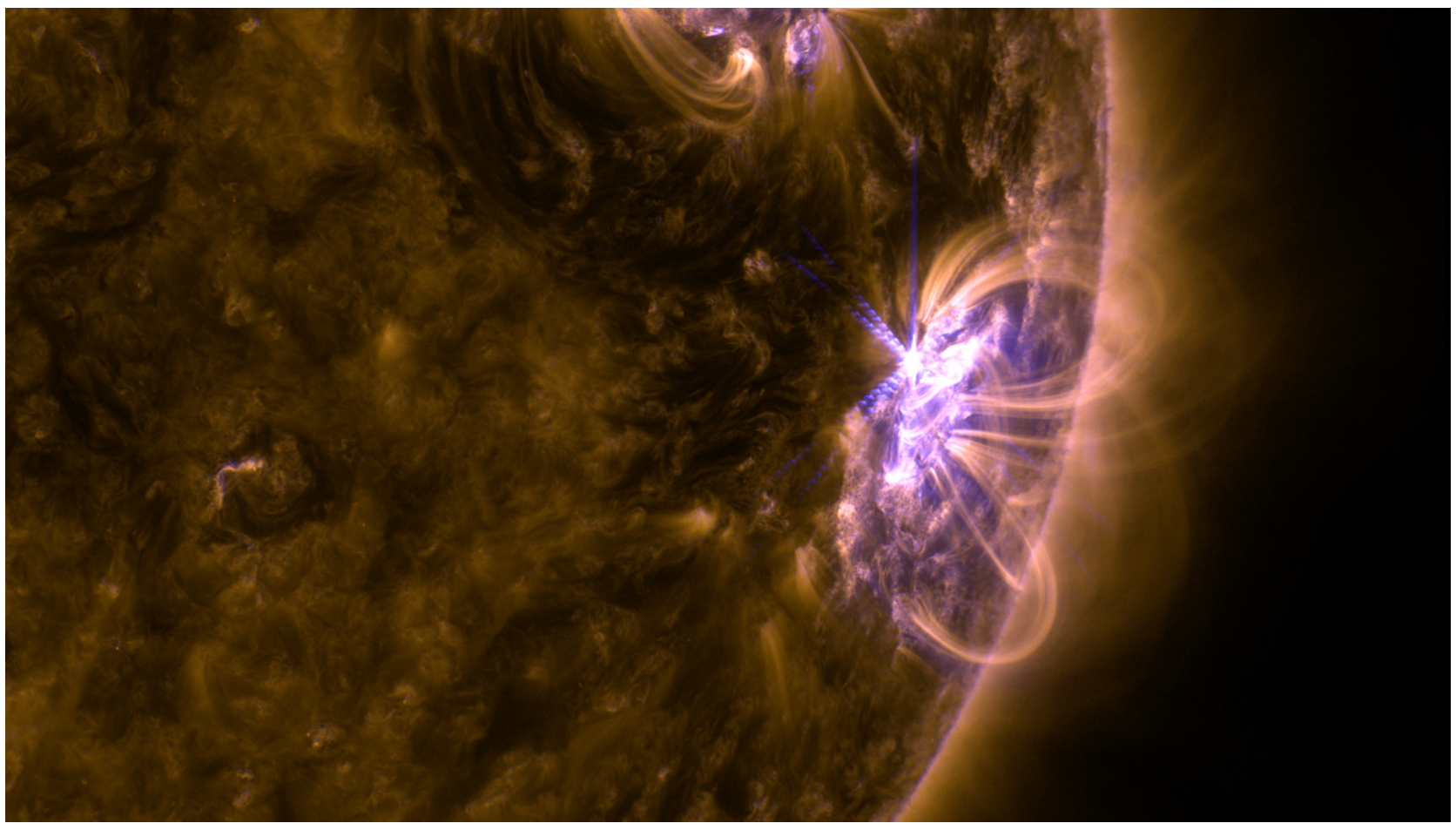

**(b)** Image of a M8.1 solar flare seen in the bright area on the right as capture by SDO on 08/09/2017. The image is a mixture of EUV light in 131 Å and 171 Å (61).

**Figure 6.1:** Solar flare images.

in solar physics [254, 255]. Therefore, based on the assumptions of this work it is possible that at least part of the solar activity is triggered by some incoming highly interacting, slow-moving invisible matter which is gravitationally focused towards the Sun by the planets. If this is the case, then the occurrence of flares would be strongly modulated by the position of the planets interacting with these invisible stream(s). In order to search for such a planetary relationship a statistical analysis of the temporal distribution of the number of flares has to be made. Therefore, a derived planetary relationship would be the key signature for the involvement of dark sector in the occurrence of solar flares (see also Publications E.1, E.14, E.15).

## 6.2 Solar flare data

### 6.2.1 Data origin

The X-ray solar flares used in this analysis have been acquired from National Oceanic and Atmospheric Administration (NOOA) National Geophysical Data Center (NGDC) [256]. The measurements belong to Geostationary Operational Environmental Satellites (GOES) program, from the two X-Ray Sensors (XRSs) instruments providing the solar X-ray fluxes for the wavelength bands from 0.5 Å to 4 Å and 1 Å to 8 Å. GOES are geosynchronous satellites in circular orbits around the Earth at an altitude of 35 790 km tracking the Sun with almost full-time coverage. They provide whole-Sun soft X-ray fluxes with an X-ray threshold energy of $\sim 1.5$ keV. The time resolution of the raw data from GOES is 3 s and they cover from 1975 until today.

The magnitude of the flares, called *flare index* is defined by Space Weather Prediction Center (SWPC) based on the 1 min average at the peak of the flare. More precisely, they are classified according to the order of magnitude of the peak burst intensity within the 0.1 nm, to 0.8 nm, from A- to X-class flares as shown in Tab. 6.1. With "A" the weakest flares are denoted, and with "X" the strongest, while classes "C", "M" and "X" stand for *Common*, *Medium* and *Xtreme*.

**Table 6.1:** Flare classification based on peak burst intensity (50, 51).

| Flare Class | Burst Intensity ($I$) [W/m$^2$] |
|---|---|
| A | $I < 10^{-7}$ |
| B | $10^{-7} \leq I < 10^{-6}$ |
| C | $10^{-6} \leq I < 10^{-5}$ |
| M | $10^{-5} \leq I < 10^{-4}$ |
| X | $I > 10^{-4}$ |

Moreover, since the NOOA list of open accessed data [256] go up to 28/06/2017, these data have been completed with the data from [257] which comprise with real-time data from GOES. Therefore, the full period of the downloaded data from both sources spans from 01/09/1975 to 12/03/2021 corresponding to 16 630 d or about 46 y.

### 6.2.2 Data curation

The acquired lists contain data belonging to flare class B to X. However, in this analysis, only the M and the X-flare data were used since they were the strongest ones. Therefore, the rest of the available flare data have not been analysed here. Moreover, the M and X-flare energy threshold is about 100 to 1000 times above the non-flaring Sun level, with the quiescent time noise being constant within less than 10% variation [258]. Therefore, this choice of flares avoids any significant noise-related effects.

More specifically, the sum of the number of flares per day based on their class was used. In the days where no M or X-flares have been observed, the number zero has been assigned. Therefore, a daily list of the number of M-flares and X-flares has been created (see Fig. 6.2). In total 6872 M- and X- class flares are contained in the selected period of $1975 - 2021$ (see also Appendix Sect. B.2.1).

In addition, an alternative analysis, searching for planetary relationship has also been performed with the daily sum of the Flare Index ($FI$) of M-flares and X-flares which roughly gives the total energy emitted by the flares. The observed correlations were similar to the daily number of flares and therefore are not presented here.

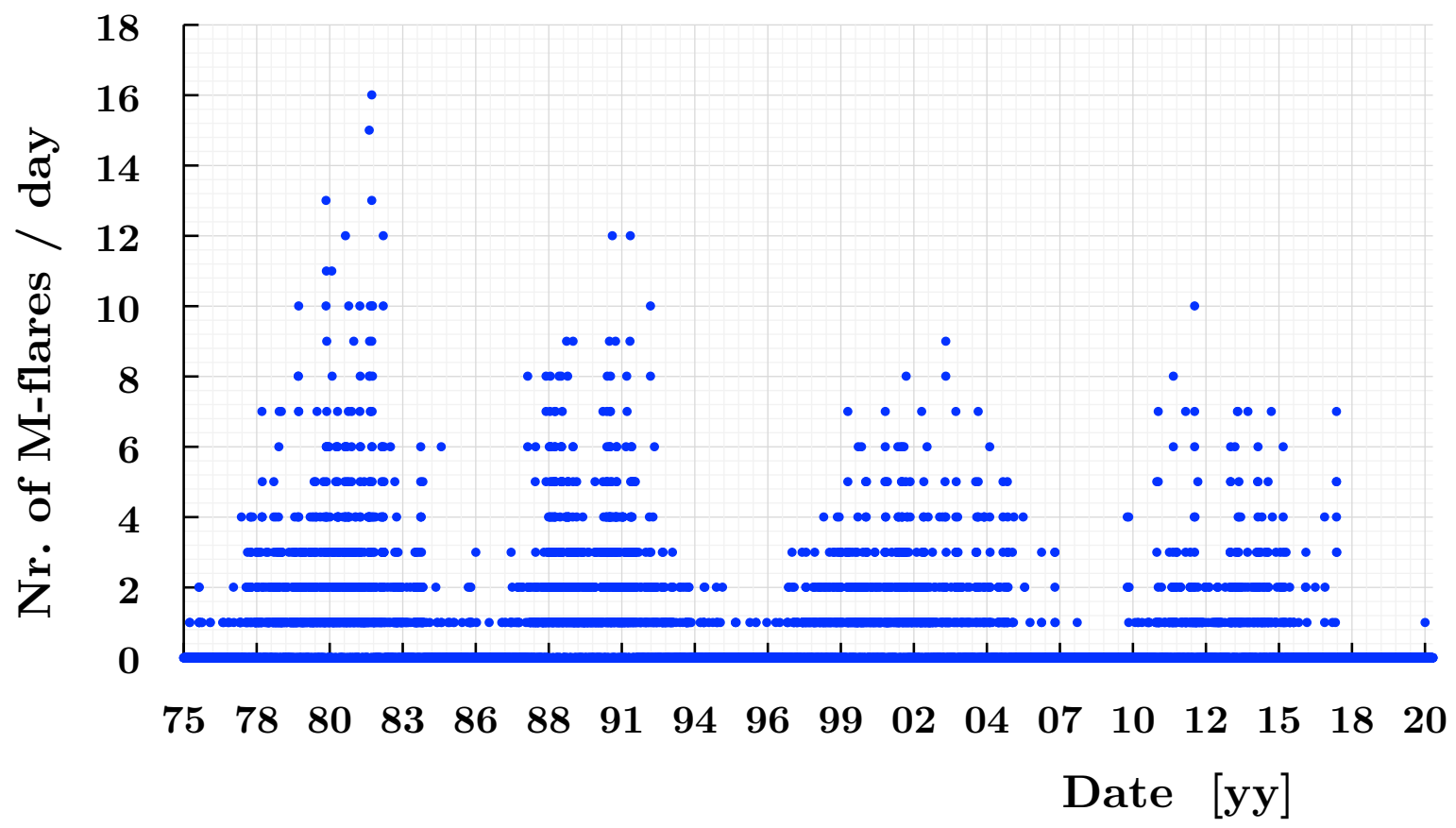

**Figure 6.2:** The daily Nr. of M-class solar flares for the period 01/09/1975 - 12/03/2021..

### 6.2.3 Data statistics

From the data seen at Fig. 6.2 various histograms have been created in Fig. 6.3 accordingly, that show the frequency distribution of the number of M-flares respectively for the period 01/09/1975 to 12/03/2021.

The total number of M-flares for the selected period is 6377 with an average of 0.38 M-flares per day and a maximum of 16 M-flares per day. The number of days without M-flares (= 0) is 13459 out of 16 630 d. This means that between 01/09/1975 and 12/03/2021 3171 d contained M-flares.

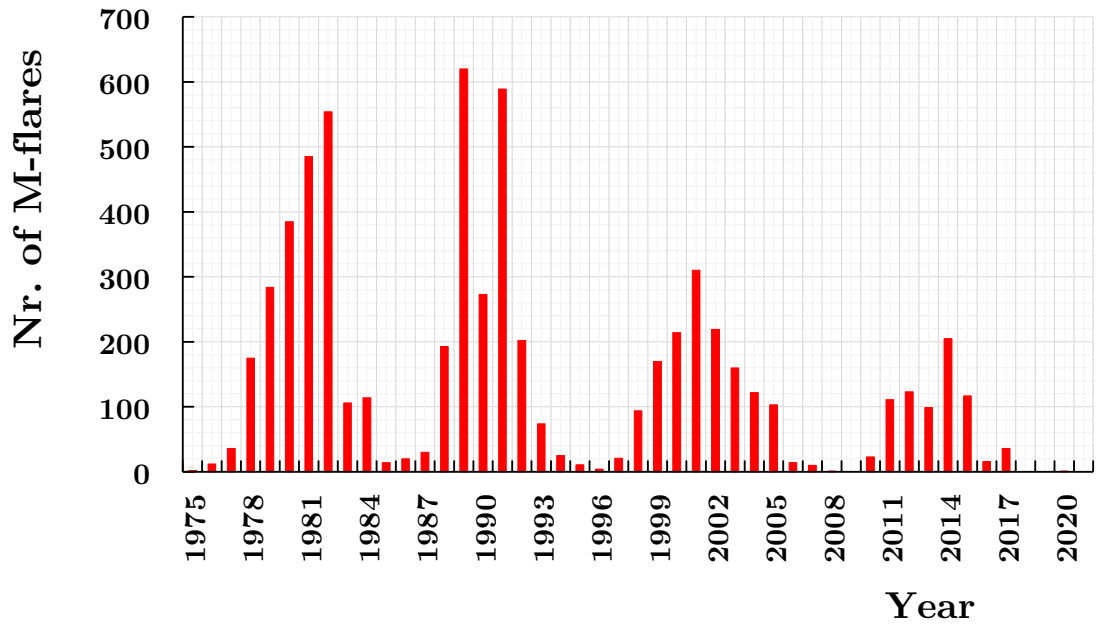

**(a)** Nr. of M-flares per year.

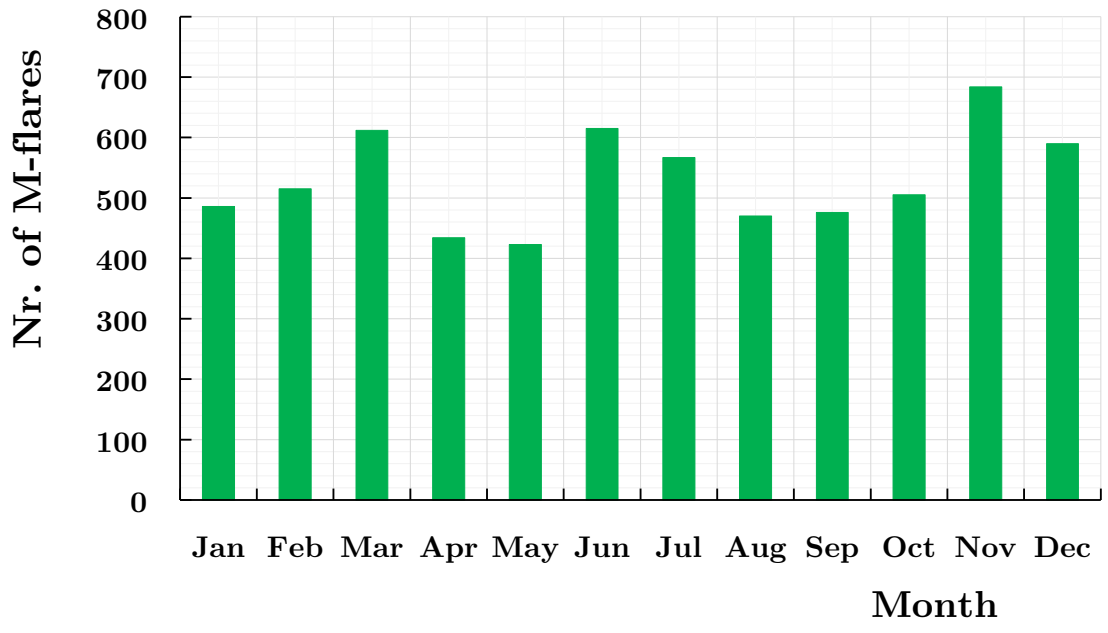

**(b)** Nr. of M-flares per month.

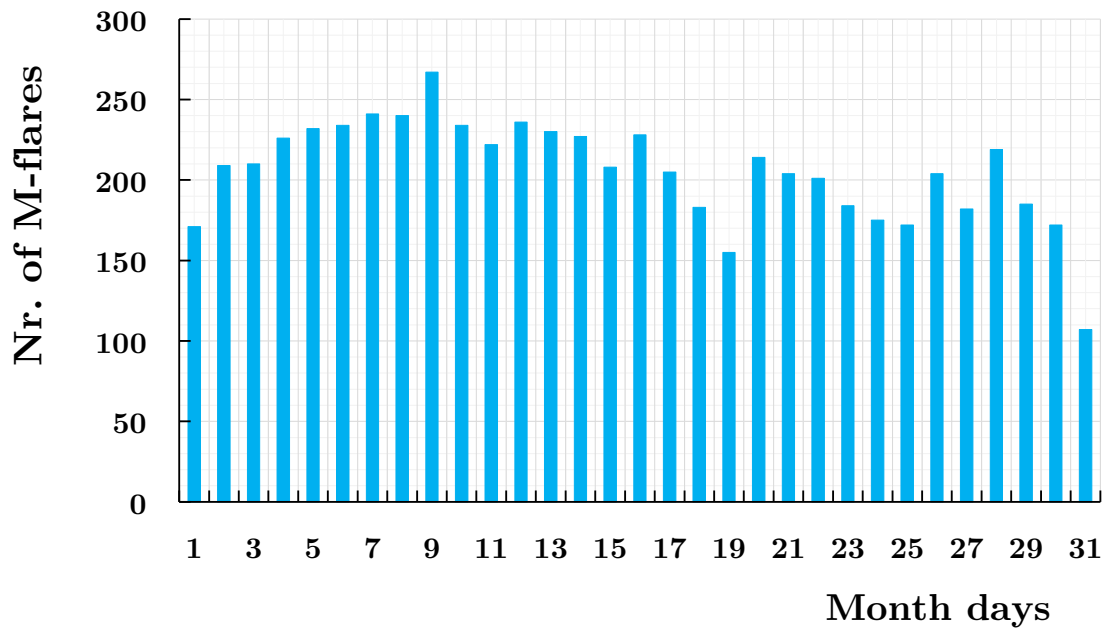

**(c)** Nr. of M-flares per day of month.

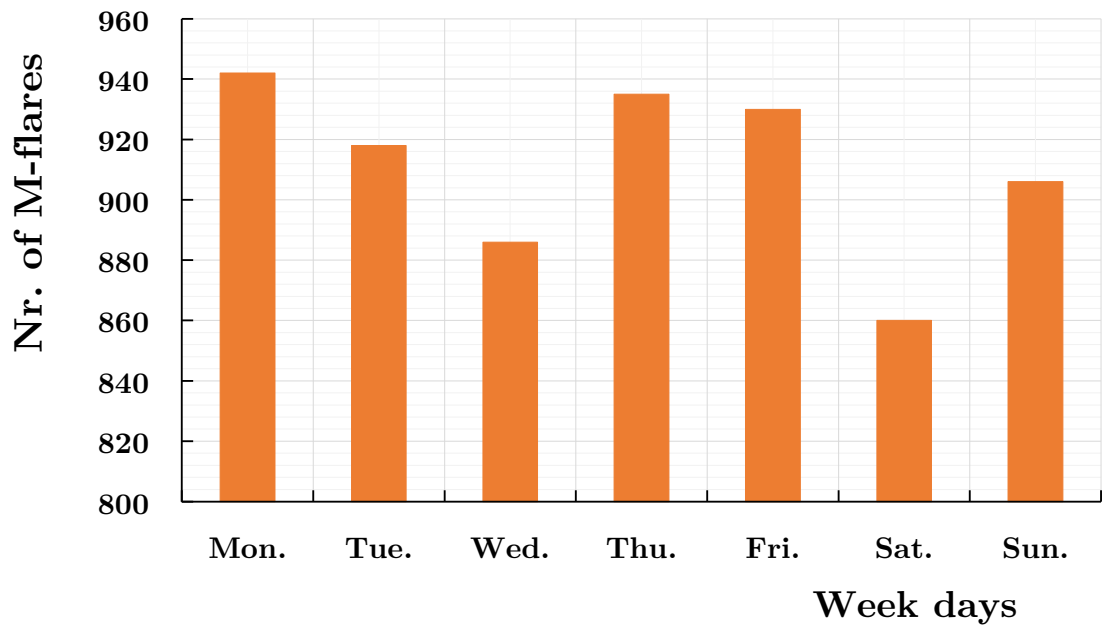

**(d)** Nr. of M-flares per day of week.

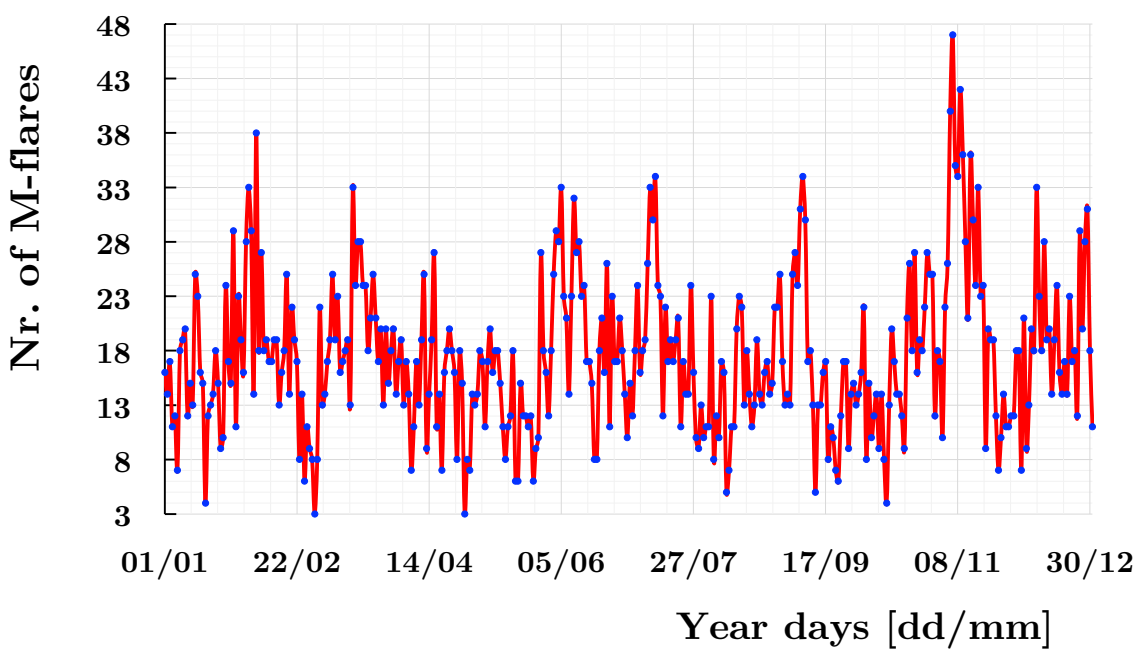

**(e)** Nr. of M-flares per day of year.

**Figure 6.3:** Histograms for the Nr. of M-flares for the period 01/09/1975 - 12/03/2021.

# 6.3 Data analysis and results

## 6.3.1 Planetary longitudinal distributions

### 6.3.1.1 Single planets

To search for a possible planetary relationship, the daily number of M-flares is projected on the corresponding planetary heliocentric longitudinal coordinates. The first step is to use single planets without any other constraint on the position of the rest of the planets. The three inner planets Mercury, Venus and Earth along with Mars are used due to their relatively short revolution time (88 d, 225 d, 365 d and 687 d) in comparison to the 46 y available period. As a result any randomly occurring event rate is expected to be averaged out after about 189, 74, 45 and 24 revolutions respectively.

In Fig. 6.4 the four longitudinal distributions for Mercury, Venus, Earth, and Mars are shown, depicting statistically significant peaks which are otherwise unexpected within conventional physics. It is noted that any eccentricity-related effect is already removed due to the implied time normalisation.

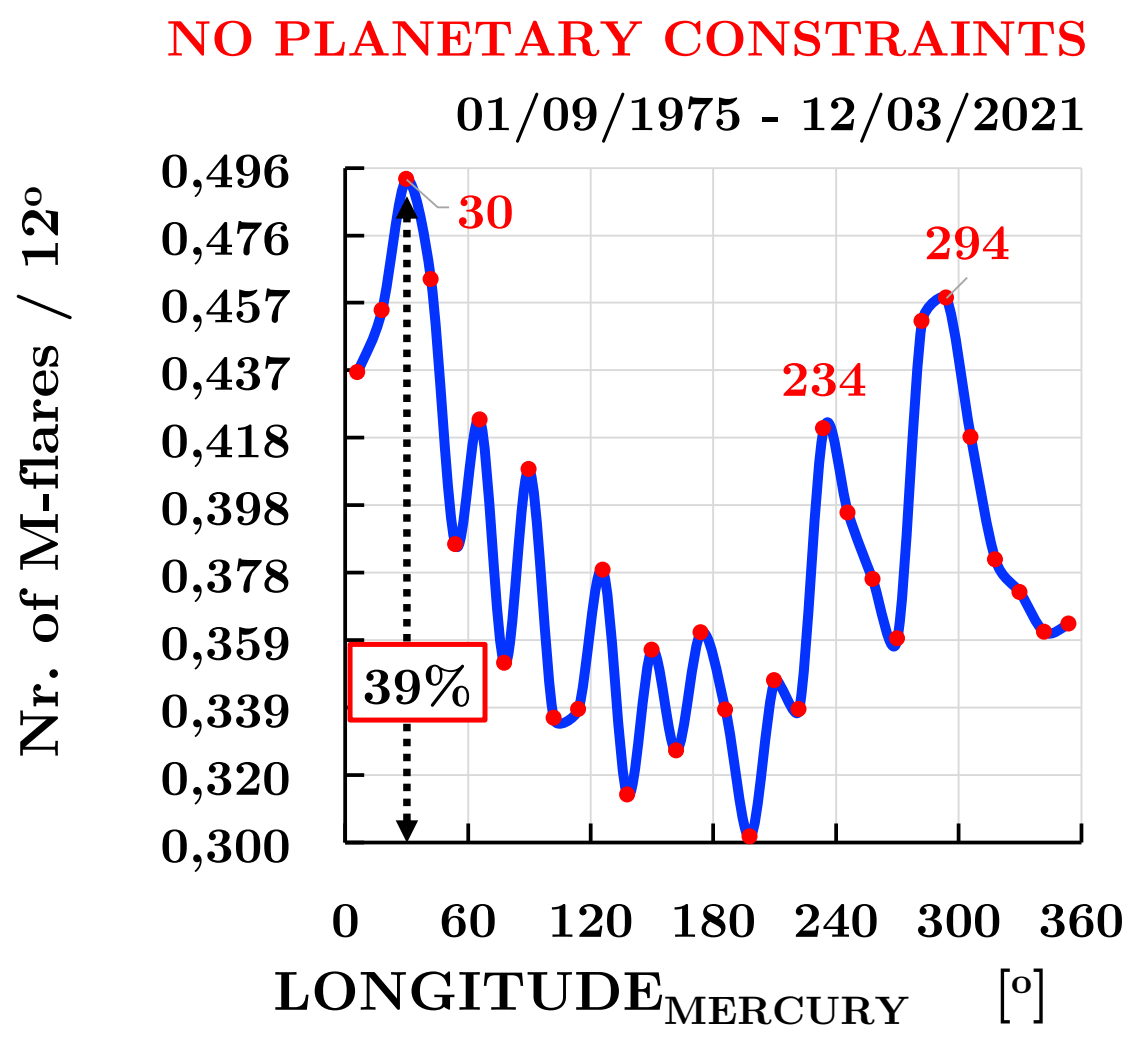

**(a)** Mercury for 12° bin. $\sigma \sim 6.9\%$.

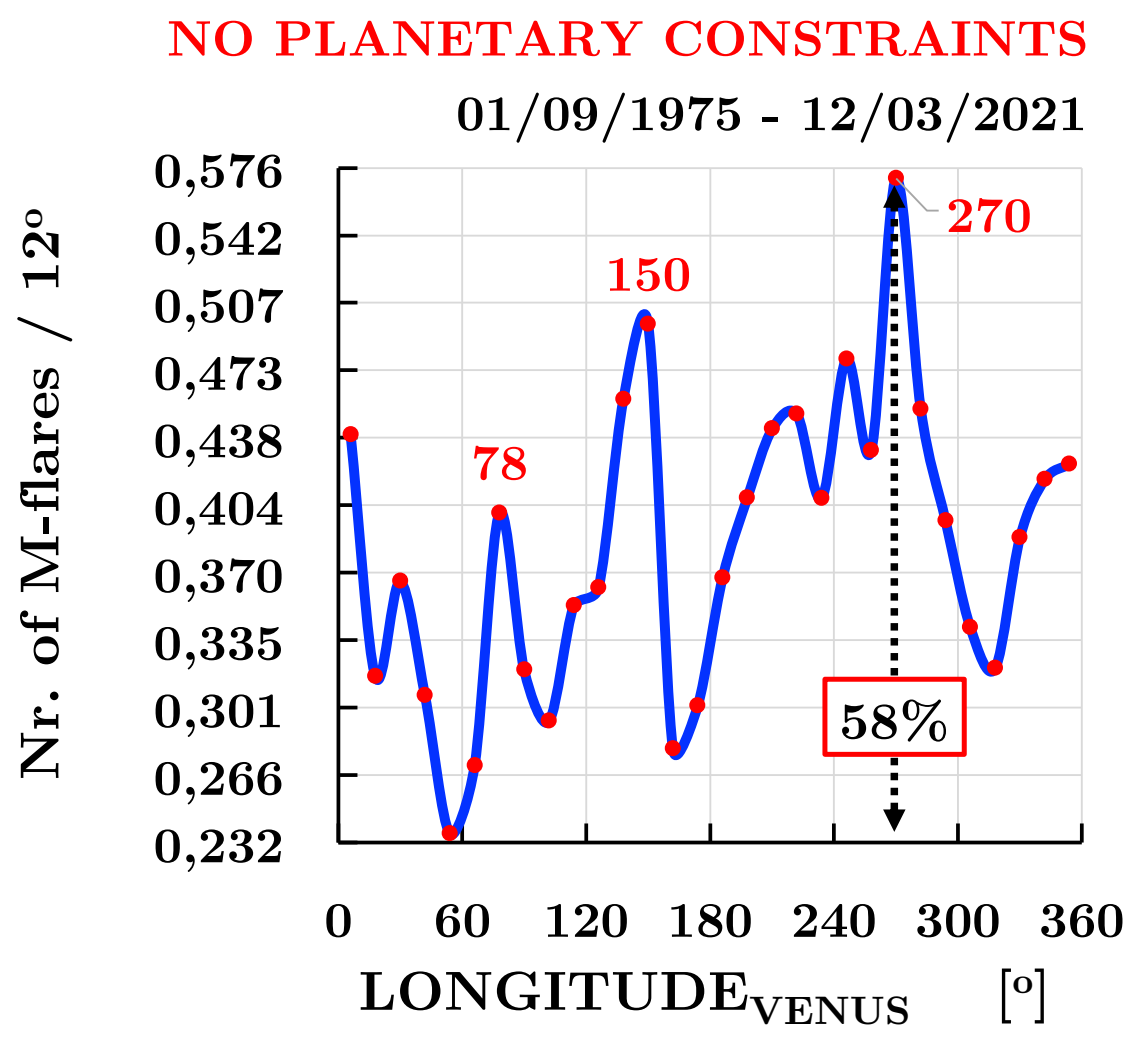

**(b)** Venus for 12° bin. $\sigma \sim 6.9\%$.

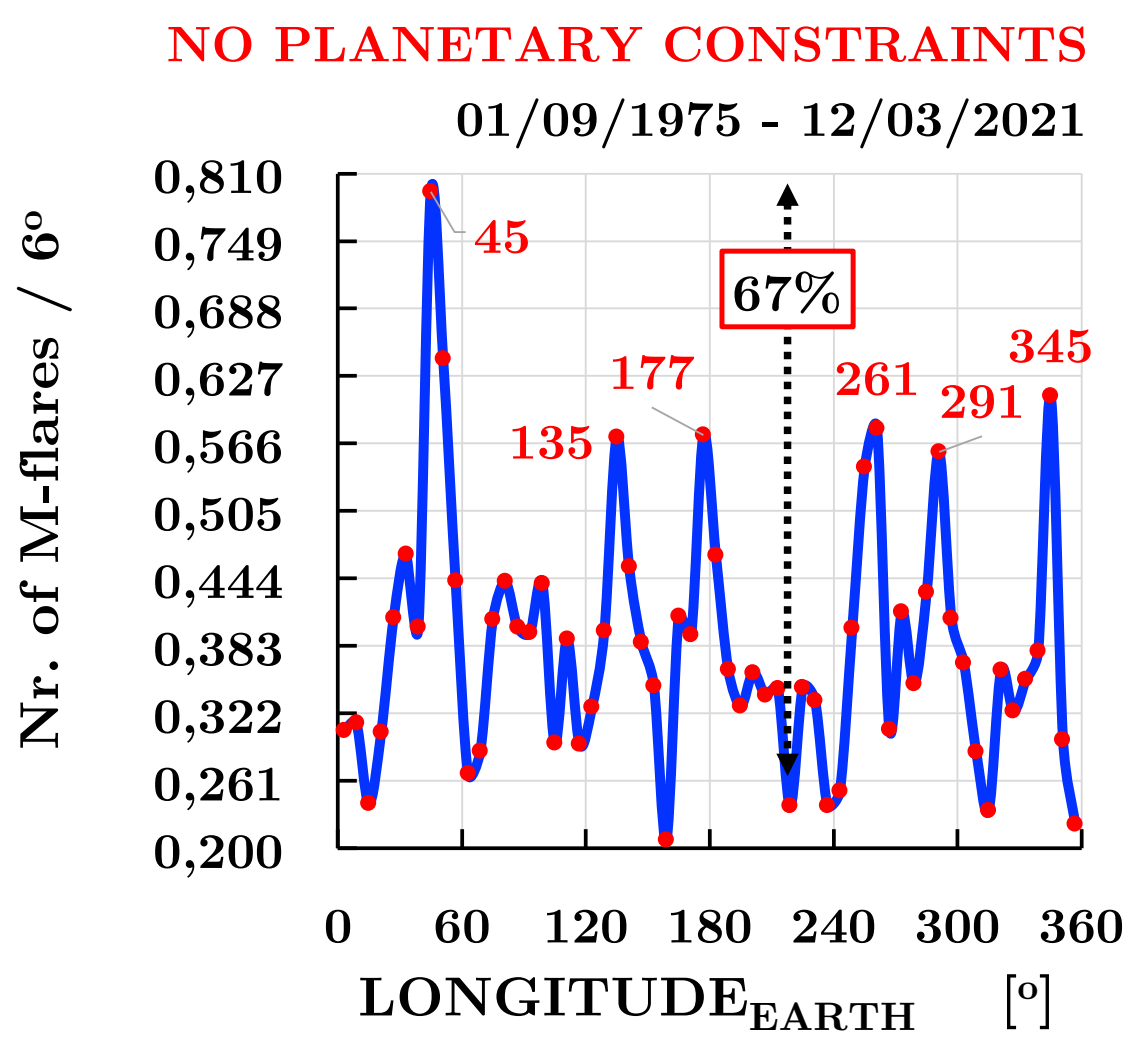

**(c)** Earth for 6° bin. $\sigma \sim 9.7\%$.

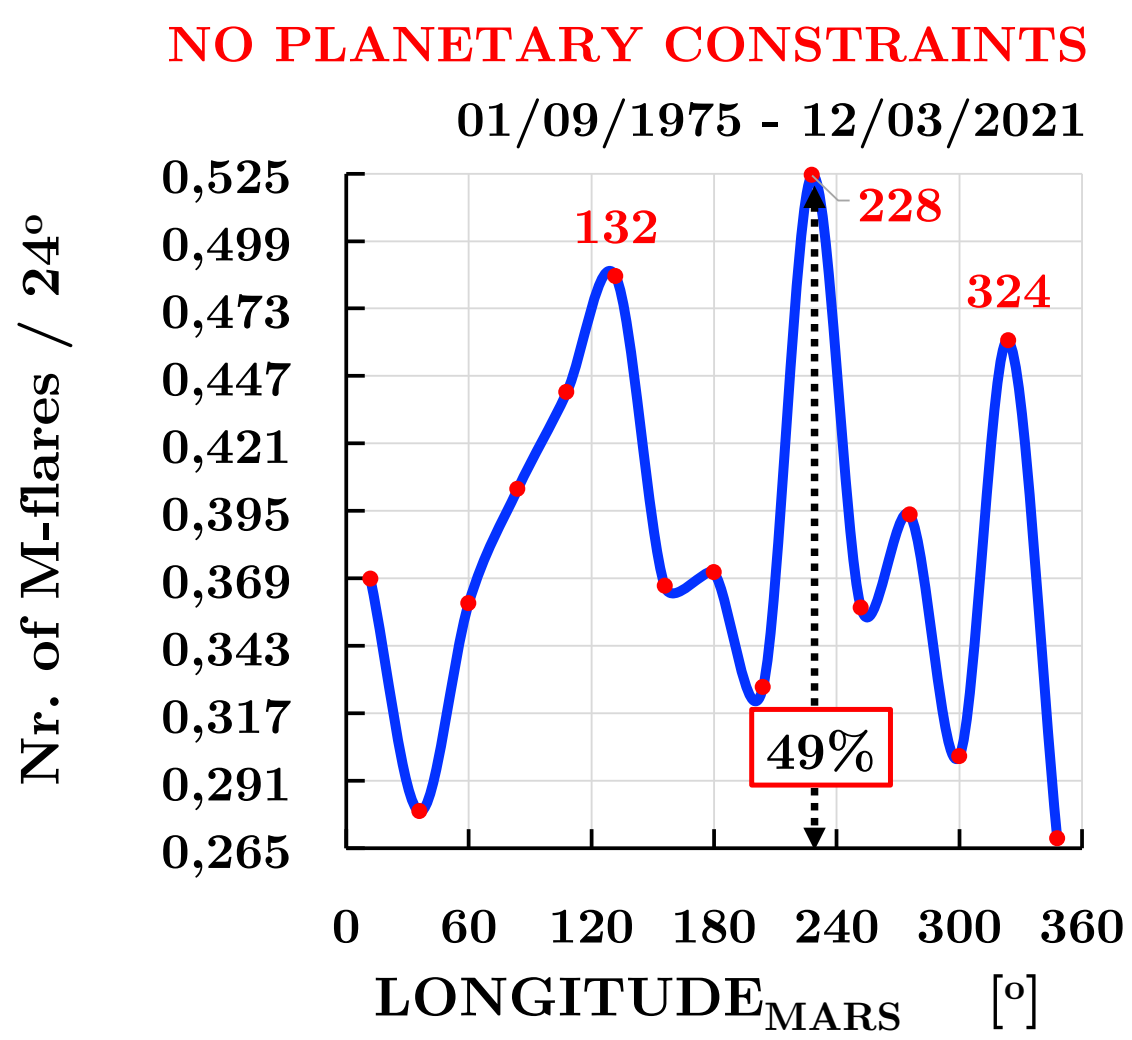

**(d)** Mars for 24° bin. $\sigma \sim 4.8\%$.

**Figure 6.4:** Planetary heliocentric longitude distributions, time normalised, of the Nr. of M-flares for the period 01/09/1975 - 12/03/2021. The total Nr. of M-flares for this period is 6377. The relative mean statistical errors per point are also given.

As already mentioned, the total number of M-flares for the full period 01/09/1975 - 12/03/2021 is 6377. Therefore, the relative mean statistical error per point for 12° bin is $\sigma \sim 6.9\%$. Accordingly for bin = 6° we have $\sigma \sim 9.7\%$, whereas for bin = 24° we have $\sigma \sim 4.8\%$.

For the case of Mercury in Fig. 6.4a the total observed difference for the bin of 12°

between the maximum and the minimum point is $\sim 38.8\%$. There are three clearly observable peaks around 30°, 234° and 394°. It is noted that the mean value of the time-normalized number of M-flares is 0.38. Similarly, in the reference frame of Venus in Fig. 6.4b we have two peaks around 150° and 270° with a total observed difference being 58.6%. Finally, for Earth in Fig. 6.4c the total maximum - minimum difference is 73.9% while for Mars in Fig. 6.4d it is 48.8%.

Assuming Poisson statistics most of the observed excesses are significant ($> 5\sigma$) compared to a randomly occurring rate. As an example, for Mercury (Fig. 6.4a) the peak around 294°, considering only one point has a significance of $\sim 2.4\sigma$ compared to the point around 342°. For Venus (Fig. 6.4b) the maximum observed point around 270° has a $\sim 6.6\sigma$ statistical significance compared to the point around 162°. For the case of the Earth (Fig. 6.4c), the peak around 45° compared to the single point around 15° has a significance of $\sim 6.9\sigma$. Lastly, for Mars (Fig. 6.4d) the peak around 132° has a significance of $\sim 7.7\sigma$ compared to the minimum point on the left around 36°. It s also noted that these estimations are conservative since the appearance of one or more large peaks in a spectrum increases the mean value, which results in a decrease of the calculated significance.

#### 6.3.1.2 Combining planets

Next, we introduce the combination of more than one planets into the distribution of the number of M-flares. This is because a possible combined focusing efficiency by several planets could enhance the possible effects. Therefore, for the case of Mercury and Venus, a combination of these two planets is performed searching for an increased or decreased influence on the observed distributions. In Fig. 6.5 the number of flares is projected on the reference frame of Mercury's heliocentric longitude adding the constraint of Venus being ±60° around 260° and ±60° around 80°. These positions define two opposite orbital arcs around the angular region where a large excess of M-flares has already been observed from Fig. 6.4a and 6.4b.

In the case of Venus being between 200° to 320° (see Fig. 6.5a) we have a total of 2400 flares in 5584 d fulfilling this constraint. The overall maximum - minimum difference observed is $\sim 73.5\%$ with the statistical accuracy per point being $\sigma \sim 12.9\%$. In this case we clearly see that the two peaks around 23° and 239° appear better resolved than in the case without any additional constraints (Fig. 6.4a). As an example the point around 239° compared to the minimum point in the right around 266° has a difference of $\sim 5.3\sigma$.

On the other hand, for the case of Venus propagating in the range of 20° to 140° (see Fig. 6.5b), the corresponding numbers are 1819 flares and 5510 d with the observed amplitude being 77.1% and the statistical error per point being $\sigma \sim 14.8\%$. Their difference is clear as shown in Fig. 6.5c and 6.5d. The statistical significance of the integrated excess between both the positions of Venus, i.e.

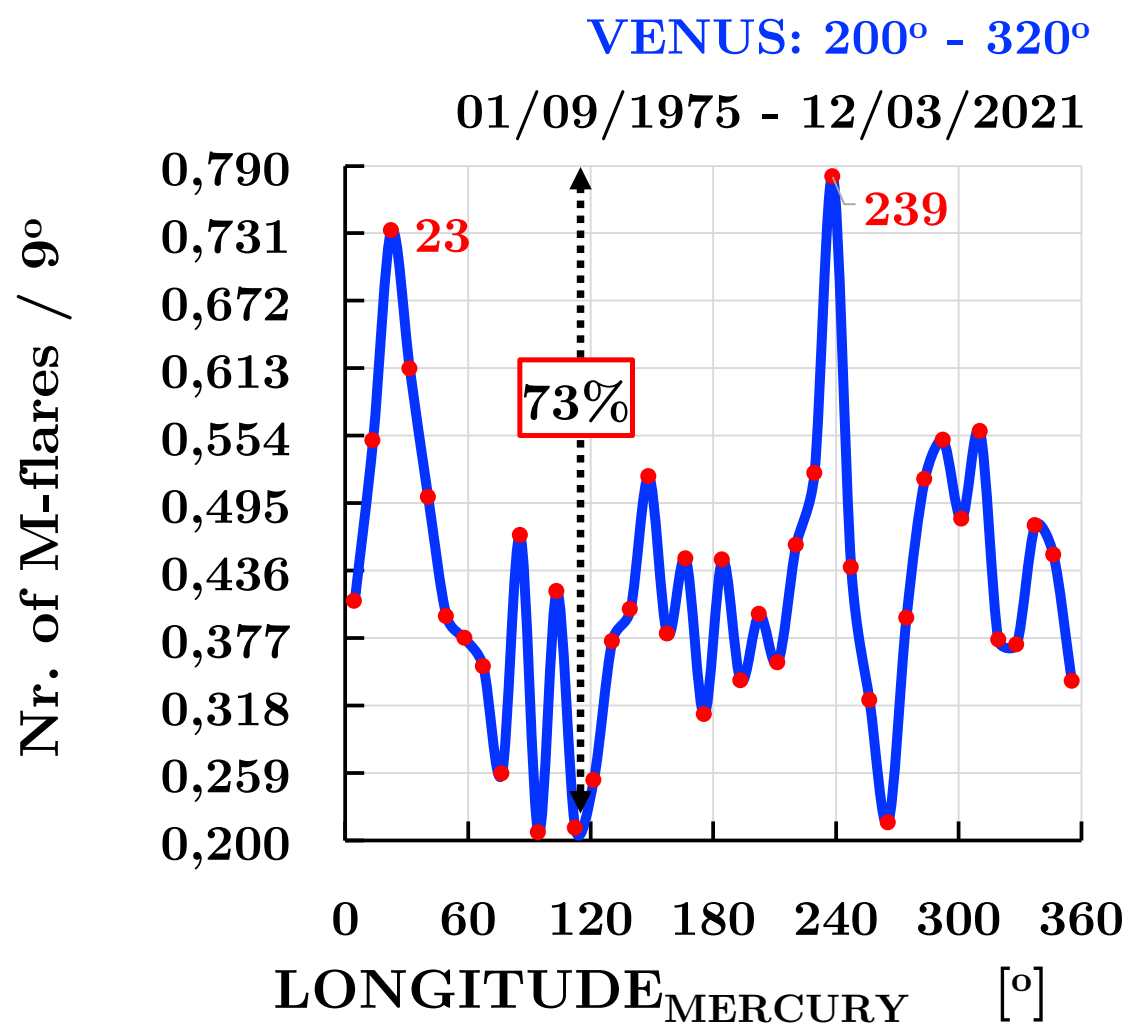

**(a)** Mercury when Venus is between 200° to 320°. $\sigma \sim 12.9\%$.

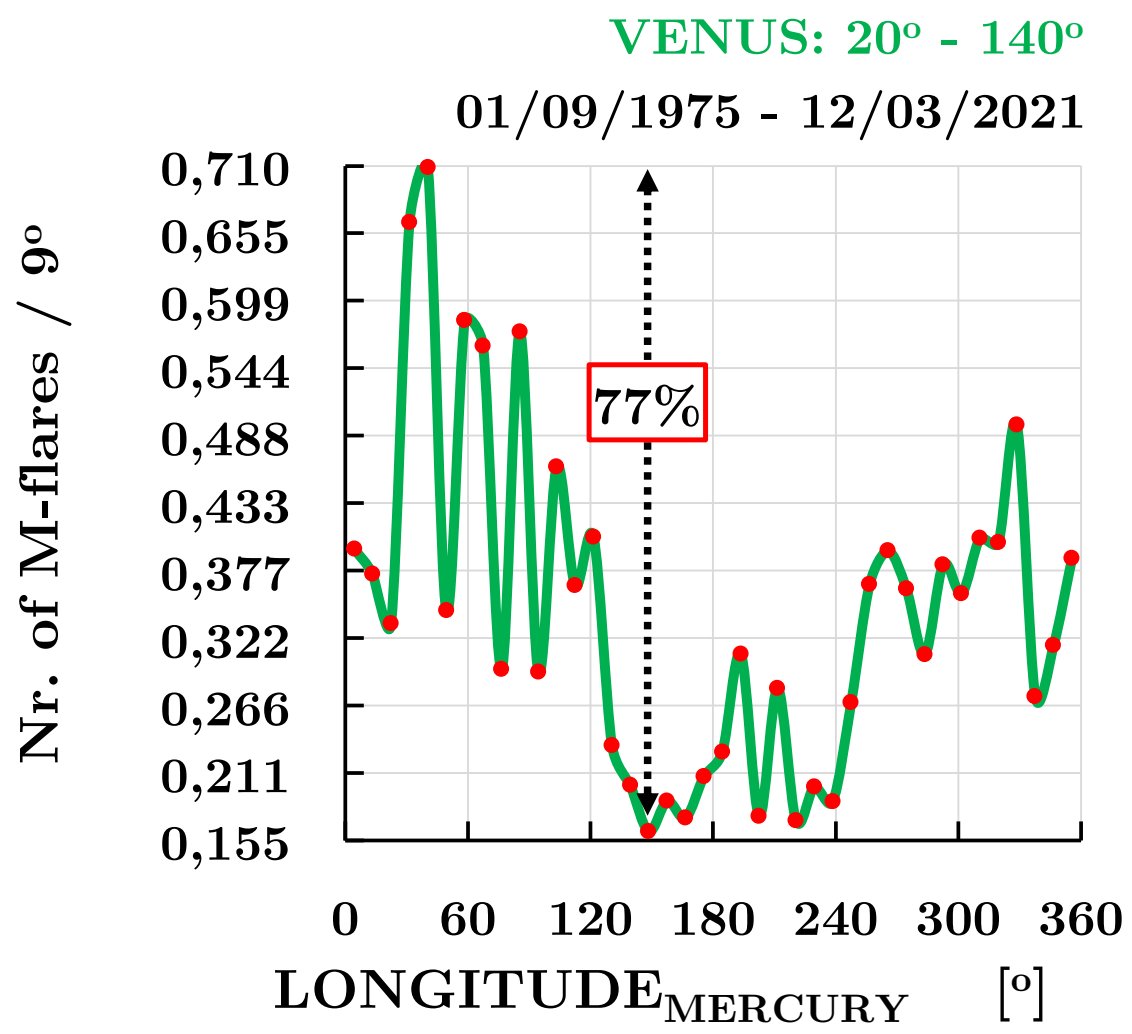

**(b)** Mercury when Venus is between 20° to 140°. $\sigma \sim 14.8\%$.

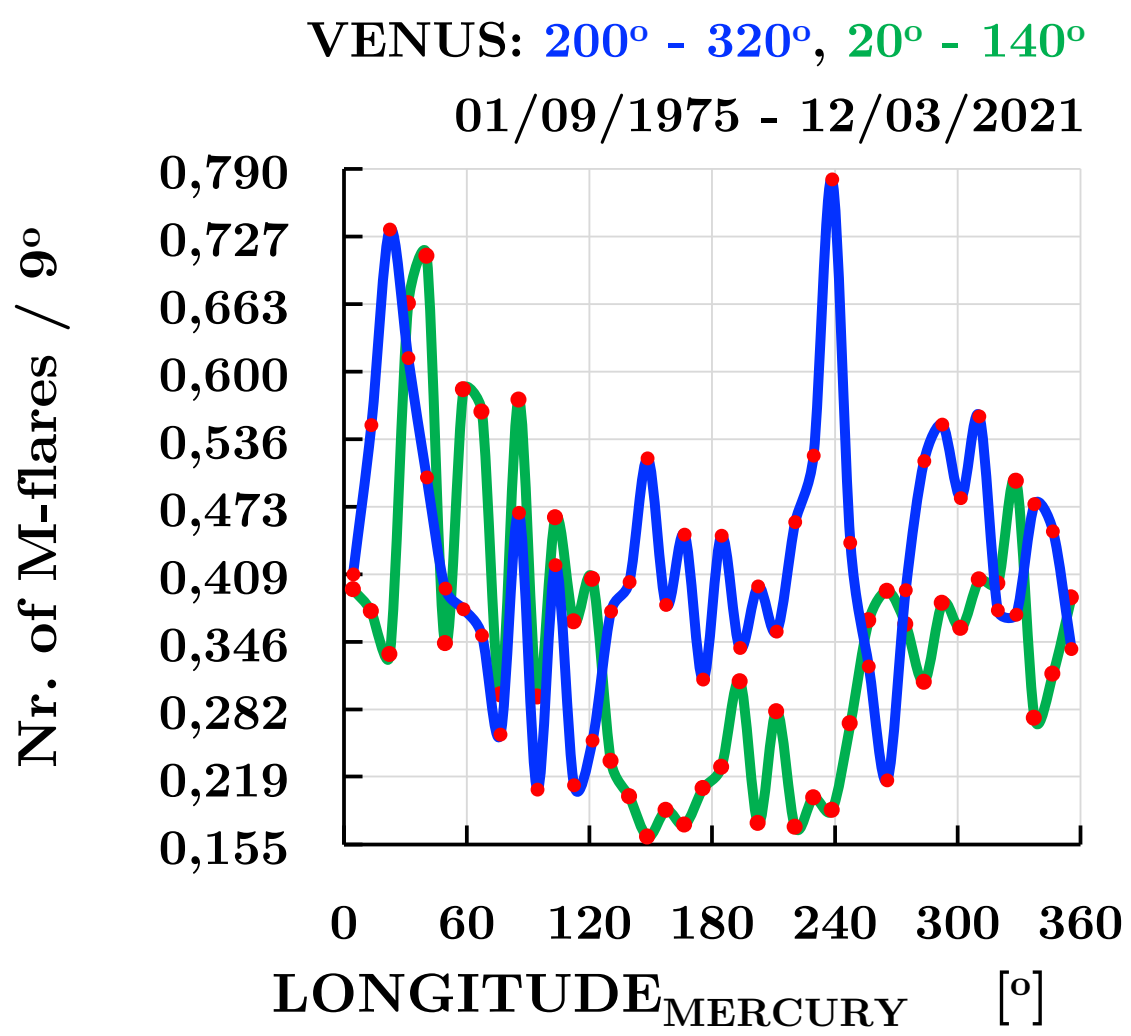

**(c)** Comparison between Fig. 6.5a and Fig. 6.5b.

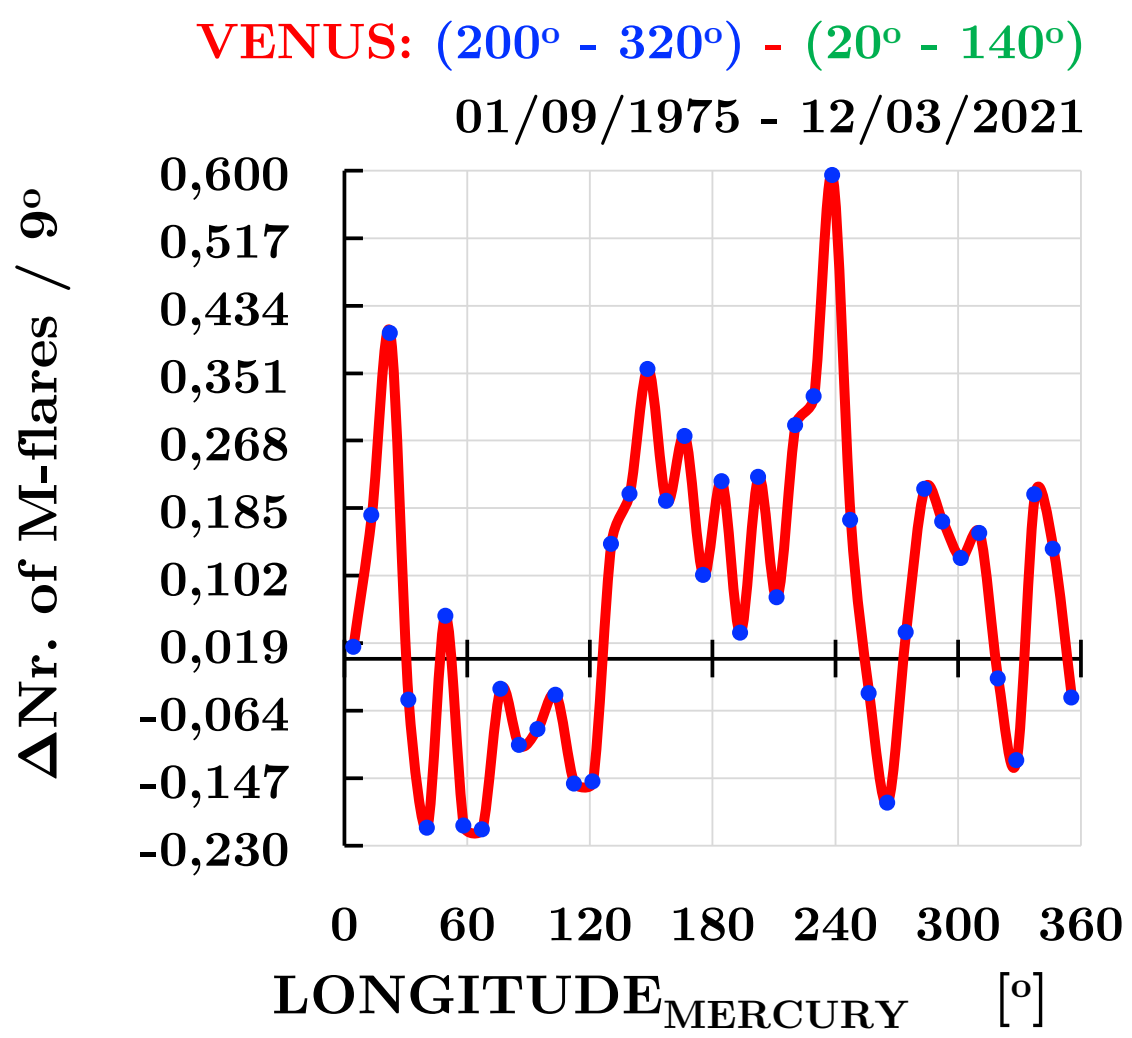

**(d)** Fig. 6.5a minus Fig. 6.5b.

**Figure 6.5:** Distribution of the Nr. of M-flares for the reference frame of Mercury when Venus propagates between 180° opposite orbital arcs with 120°

, the difference of the number of M-flares $2400 - 1819 = 581$ is $\frac{581}{\sqrt{2400+1819}} \sim 8.9\sigma$. This observation contradicts the predicted flat spectra in both cases without any significant difference if the observed excesses were due to randomly occurring events. Finally, it is mentioned that the range 200° to 320° which has an increased excess of the number of flares was chosen while searching also for a possible correlation with the GC (around 266°).

An additional case with the combined effect of two planets is shown in Fig. 6.6 where the M-fare distribution is depicted in the reference frame of Earth when Mercury is constrained on a 60°-wide orbital arc around 230°. The observed difference between the maximum and the minimum point for a duration of 3847 d and 1426 M-flares is 83%. The mean error per point in this case is 16.8% which means that assuming Poisson statistics, as an example the single point around 59° has a difference of $\sim 4.5\sigma$ compared to the point around 5°.

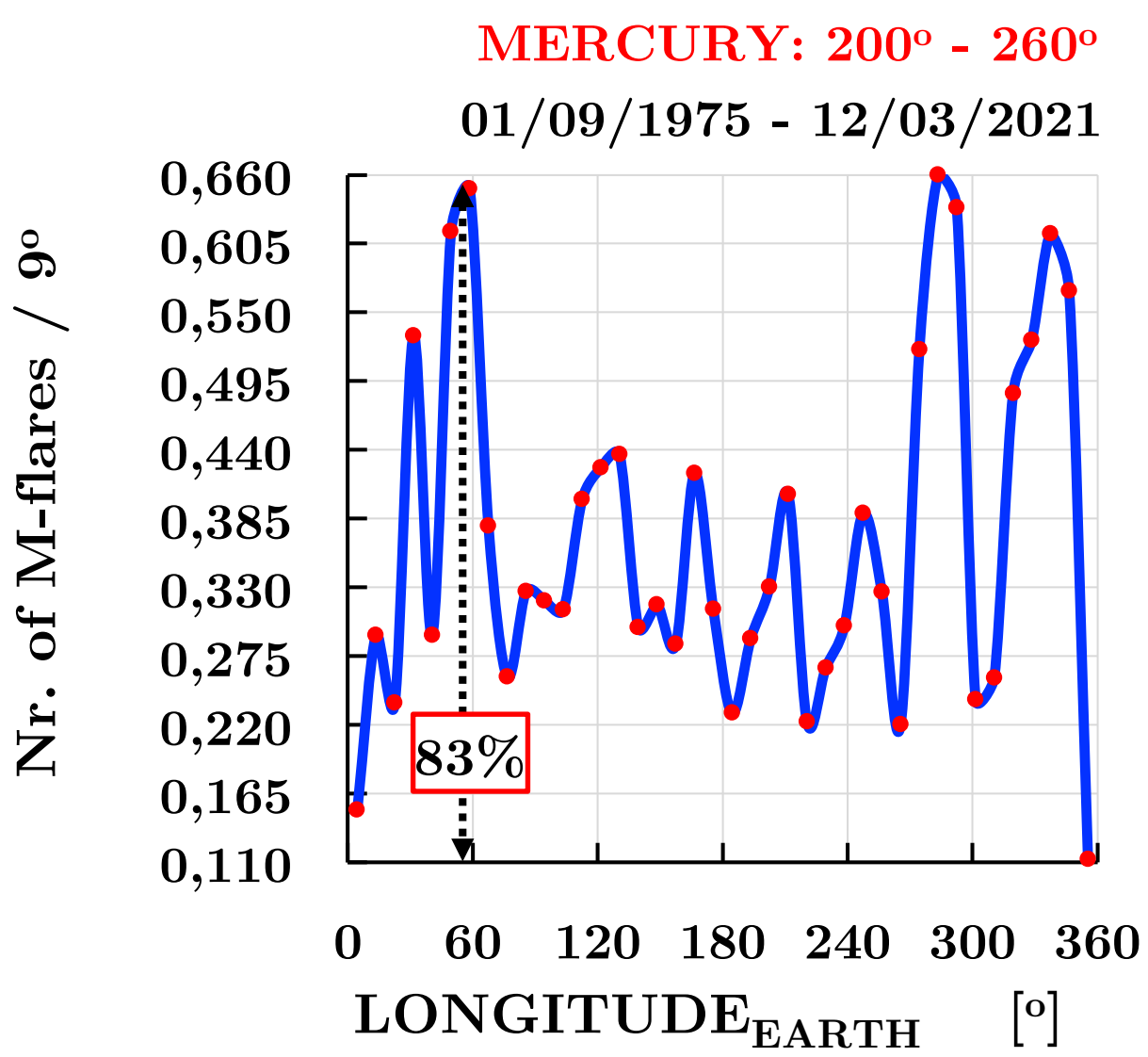

**Figure 6.6:** Nr. of M-flares distribution as a function of the heliocentric longitude of Earth when Mercury is allowed to propagate between 200° to 260°. The relative mean statistical error per point is $\sigma \sim 16.8\%$.

### 6.3.2 Multiplication spectra

To further exclude systematics and follow up the behaviour in time of the various distributions, the time series of the M-flares have been divided into four physical sub-periods defined by the four solar cycles. Then, the product between each bin of the four individual distributions has been calculated. The resulting spectrum is the so-called "multiplication spectrum". If some excess peaks are acquired with this procedure, then these peaks would have a much better Signal-to-Noise Ratio (SNR) than the same peaks in the individual periods. Therefore, such peaks can provide a clear indication that the observations are not randomly distributed, pointing to a planetary-related clustering. The analytical simple equation of the multiplication spectrum is:

$$Y_{tot}\left(J\right) = \prod_{SC=21}^{24} \Theta_{SC}\left(J\right) \tag{6.1}$$

where: $SC$ : the corresponding solar cycle,

$J$ : the bin number,

$\Theta$ : the heliocentric longitude.

The specific periods of the solar cycles which are used are shown in Tab. 6.2. The total number of M-flares and the corresponding number of days are also shown.

**Table 6.2:** Exact dates of the latest four solar cycles, used for the multiplication spectra and the corresponding Nr. of days and M-flares.

| Solar cycle Nr. [-] | Start date [UTC] | End date [UTC] | Nr. of Days [-] | Nr. of M-flares [-] |
|---|---|---|---|---|
| 21 | 01/03/1976 | 01/09/1986 | 3837 | 2184 |
| 22 | 02/09/1986 | 01/08/1996 | 3622 | 2021 |
| 23 | 02/08/1996 | 01/12/2008 | 4505 | 1439 |
| 24 | 02/12/2008 | 01/05/2020 | 4169 | 730 |

The three inner planets Mercury, Venus and Earth are used for the creation of each corresponding multiplication spectrum in Fig. 6.7. We observe that the significance of the various peaks is enhanced resulting in a better SNR as expected. This is a clear indication that these observations are not randomly distributed. This fact is supported by random Poisson simulations where the resulting Coefficient of Variation (CV) of the multiplication spectrum is twice as large as the corresponding one of each of the four individual spectra i.e.
, the multiplication spectrum has a higher statistical fluctuation than the individual spectra if it is derived from random Poisson distributions. On the other hand, as an example in Fig. 6.7c, we see a considerable peak around 45° which, when compared to the rather low level of the rest of the spectrum, is against the conventional expectation when no planetary correlation is at work. The same reasoning applies also to the smaller peak candidates, which therefore point to multiple clustering phenomena appearing in these "coincidence" spectra.

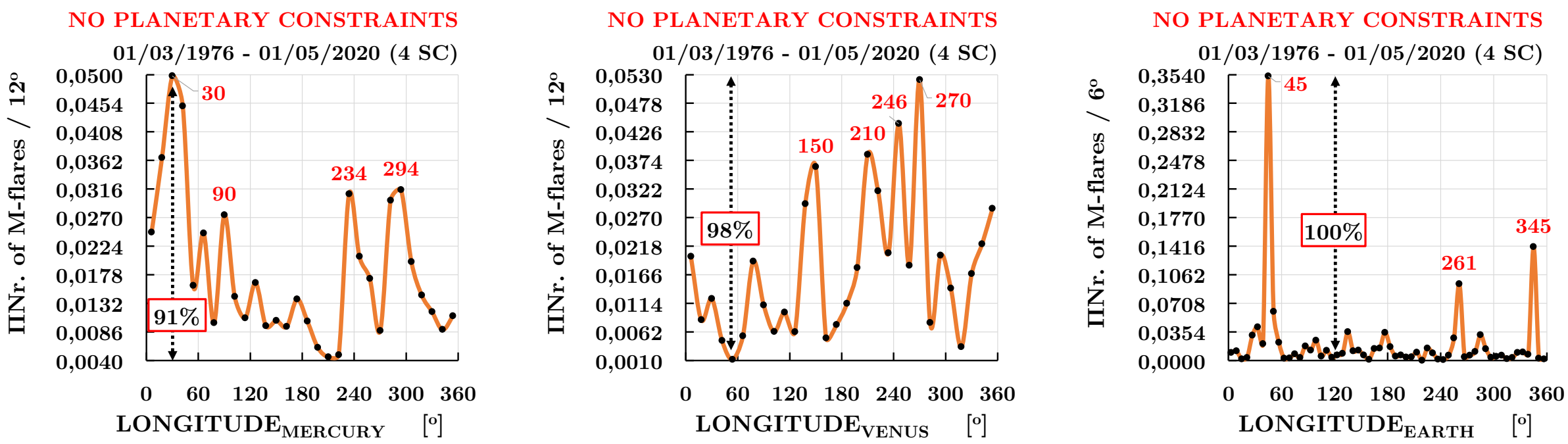

**(a)** Mercury for bin = 12°. **(b)** Venus for bin = 12°. **(c)** Earth for bin = 6°.

**Figure 6.7:** Multiplication spectra for the Nr. of M-flares using the last four solar cycles for Mercury, Venus and Earth.

### 6.3.3 Fourier analysis

As an alternative independent analysis searching for periodicities, a conventional Lomb-periodogram spectral analysis has been performed in all the data from the period 01/09/1975 - 12/03/2021. From the resulting periodogram, several peaks appear which are then compared with the known periods given by the planetary revolutions, as well as the planetary synods from the solar bodies.

In particular, the 8th biggest peak of the Fourier periodogram is located around 144.4 d with an amplitude of 27.6 dB (see Fig. 6.8a). Its Full Width at Half Maximum (FWHM) obtained by a Gaussian fit is $\sim 1.1$ d. This significant peak is very close to the synodic period of 145 d between Mercury and Venus. Furthermore, the 10th biggest peak has a period of about 336.34 d with an amplitude of 23.1 dB and a FWHM of 5.8 d. This means that it is within the $\sim \pm 2.4\sigma$, with the synod 334 d of Venus - Mars. Then, the 19th peak with the biggest amplitude of 14.1 dB corresponds to 224.2 d and has a $FWHM \sim 3.9$ d, which coincides with the 224.7 d revolution period of Venus. Finally, an important observation is that of a peak around 27.32 d $\pm$ 0.02 d with an amplitude of $\sim 9$ dB (see Fig. 6.8b). This period matches perfectly Moon's sidereal month, fixed to remote stars and points to an additional exo-solar influence on the number of flares. This fits this work hypothesis with the Moon modulating the assumed incoming invisible massive stream(s) towards the Sun.

Although these results cannot prove on their own a planetary relationship of the number of M-flares produced in the Sun, they further establish the significance of the peaks observed in the planetary longitudinal distributions.

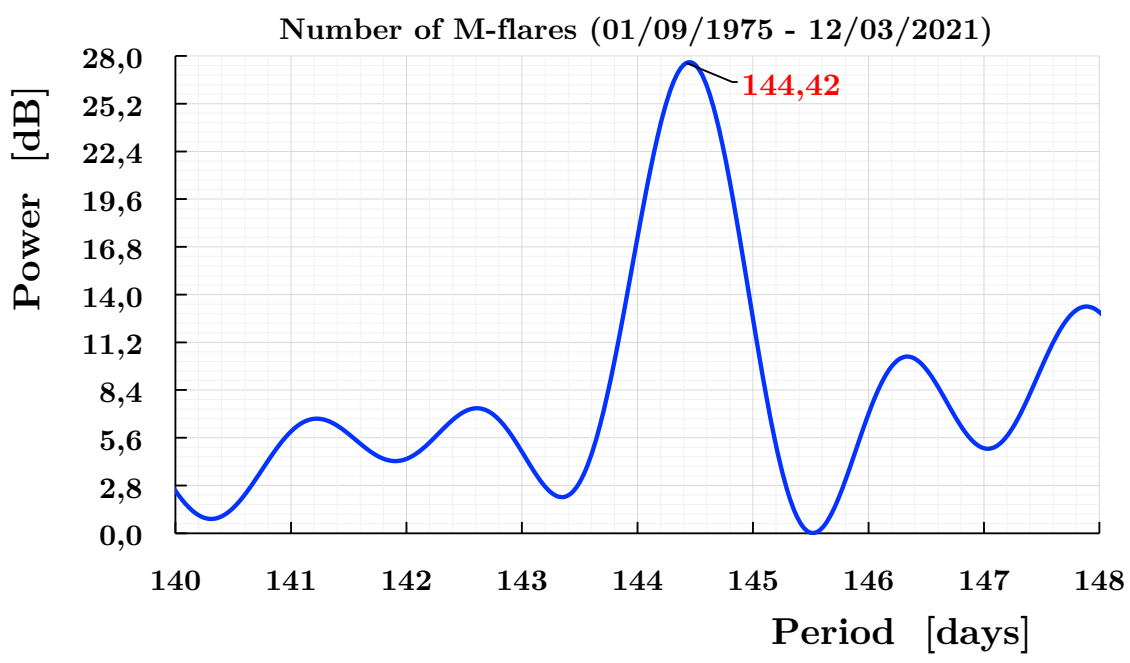

(a) Around 144 d.

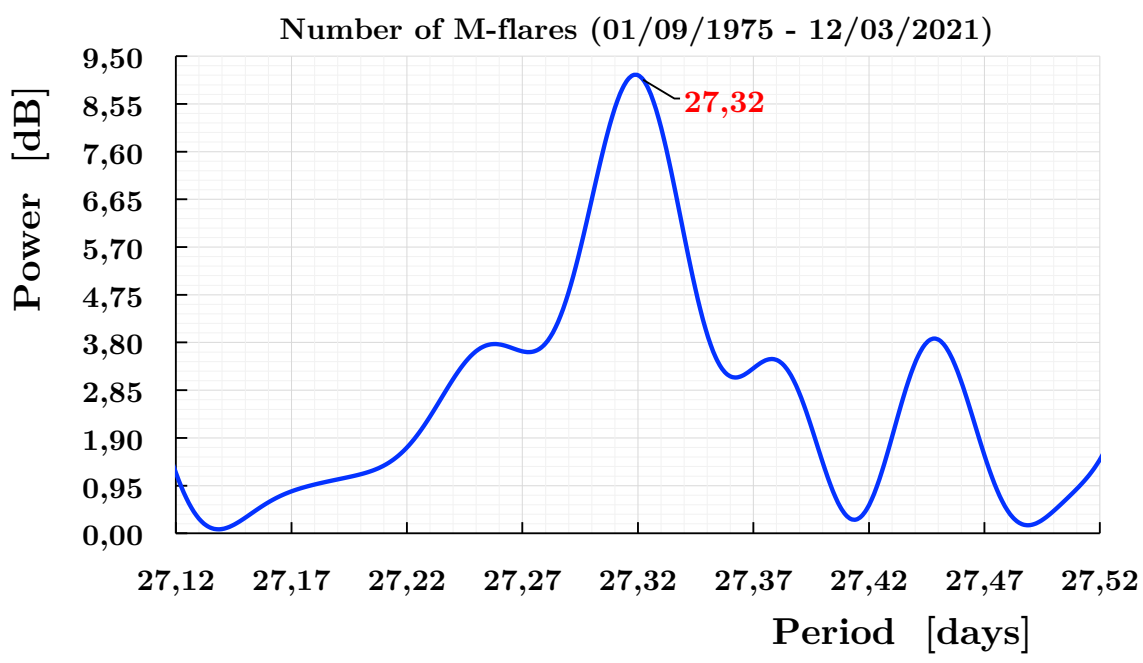

(b) Around 27.3 d.

**Figure 6.8:** Fourier periodogram of the Nr. of M-flares from 01/09/1975 - 12/03/2021.

## 6.4 Summary

From the analysis of the occurrence of X- and M-flares during the last four solar cycles, a clear evidence is found that their occurrence is strongly modulated by the position of the planets. This is based on the narrow peaks appearing in the heliocentric longitude distributions of the selected planets. This contradicts the expectation of a flat distribution if the data were randomly distributed. In fact, it is widely accepted that flares do not follow Poisson distribution as time series of highly energetic solar flares show Fourier peaks as the Rieger peak around 154 d [259], and the 11 y. Such observations point to an underlying non-random cause of the solar activity, with the assumed scenario of this work fitting in since the various observed spectral shapes cannot be produced by Poisson statistics alone. Moreover, even if there was any random distribution of the flares, then the large averaging taking place while analysing several orbits, would produce flat non-peaking distributions. Instead, striking peaks appear in all distributions of both M-flares and X-flares showing also the non-Poissonian character of the two solar activities (see Appendix Sect. B.2.1).

In addition, the statistically significant ($> 8\sigma$) difference observed between the two 180° opposite orbital arcs of Venus while looking at the distribution of Mercury, in Fig. 6.5, as well as the individual peaks in all inner planets around the same region of 200° to 320° point to a correlation with the direction of the GC which is located around 266.5° heliocentric longitude. This means that there is a significantly higher rate of the flaring Sun when the planets are pointing towards the side of the GC. This behaviour is in agreement with the working assumption of an influx of slow-moving massive invisible particles coming from the direction of the GC and being gravitationally focused by the planets and the Sun itself via free fall (see Publication E.1). Further, the multiplication spectra as well as the Fourier periodograms of both M and X-flares, which are part of the cross-checking procedure, provide additional verification of the significance of the observed peaks and the periodicities of the observed planetary relationships of the number of M- and X-flares.

Finally, an additional analysis (not shown here) has been made with the flare index as well as with flares derived from Lockheed Martin Solar and Astrophysics Laboratory (LMSAL) [260] originating from GOES but filtered with a different detection algorithm than NOOA catalog. In this investigation, various datasets have been used with all the flares but also with the 1020, 5895, 15615 and 50215 biggest flares, based on the *flare index (FI)* factor. The results are similar to the ones presented here however with even better SNR in some cases. The same procedure has also been performed with the 50248, and 48267 smallest flares (as well as the so-called *nano-flares*) with some peaks, like the ones in Mercury, appearing around the same heliocentric longitudes but with a much smaller overall effect and significance. In conclusion, the biggest flares as well as the days that contain a higher number of flares exhibit the biggest effects.

In the future, a more refined analysis in all flares including the smaller ones, but also analysis with better than 1 d time resolution involving all planets of our solar system, could provide even more accurately the preferred directions of the assumed stream(s) but also provide hints on the velocity spectrum of their constituents. It is noted that the available time resolution from the GOES data is 3 s and thus the already publicly available data allow for a much more precise analysis. Lastly, a comparison with more datasets of the solar activity could also reveal more about the nature of the possible constituents and their form of interaction with normal matter. An interesting candidate from the dark sector which deserves further attention is that of AQN since a correlation with solar flares has already been investigated [232].

# 7

# EUV IRRADIANCE

## 7.1 Introduction

One of the most fundamental open problems [240] in solar physics since 1939 [261, 262] is that of the solar corona which relates to the question of why the temperature of the Sun's outer atmosphere, i.e. corona, is millions of degrees higher than that of the photospheric surface. In fact, the photosphere lies at a temperature of about 5800 K while the solar corona at about 1 MK to 10 MK (see Fig. 7.1). This behaviour of the Sun's atmosphere is beyond the expected reduction of the temperature while moving further away from the core of the Sun. More specifically, the heat of the corona should originate below the photosphere, which implies a temperature inversion above the Sun's surface which rules out equilibrium thermal transport processes, since heat should be flowing from a hotter to a cooler body. As a result, one or more non-thermal processes are required to heat the corona, the chromosphere, and the transition region seen in Fig. 7.1b [263].

In Fig. 7.2 the analog spectrum of the quiet Sun is shown. At high energy $> 10^{15}$ Hz we see a distribution which is not expected within a $\sim 5800$ K black-body emission. This striking excess in EUV of the quiet Sun is the manifestation of the solar corona problem [264].

Solar EUV covers the wavelengths from 10 nm to 120 nm, and originates in the corona and chromosphere of the Sun's atmosphere, and varies by a factor of ten over the course of a typical

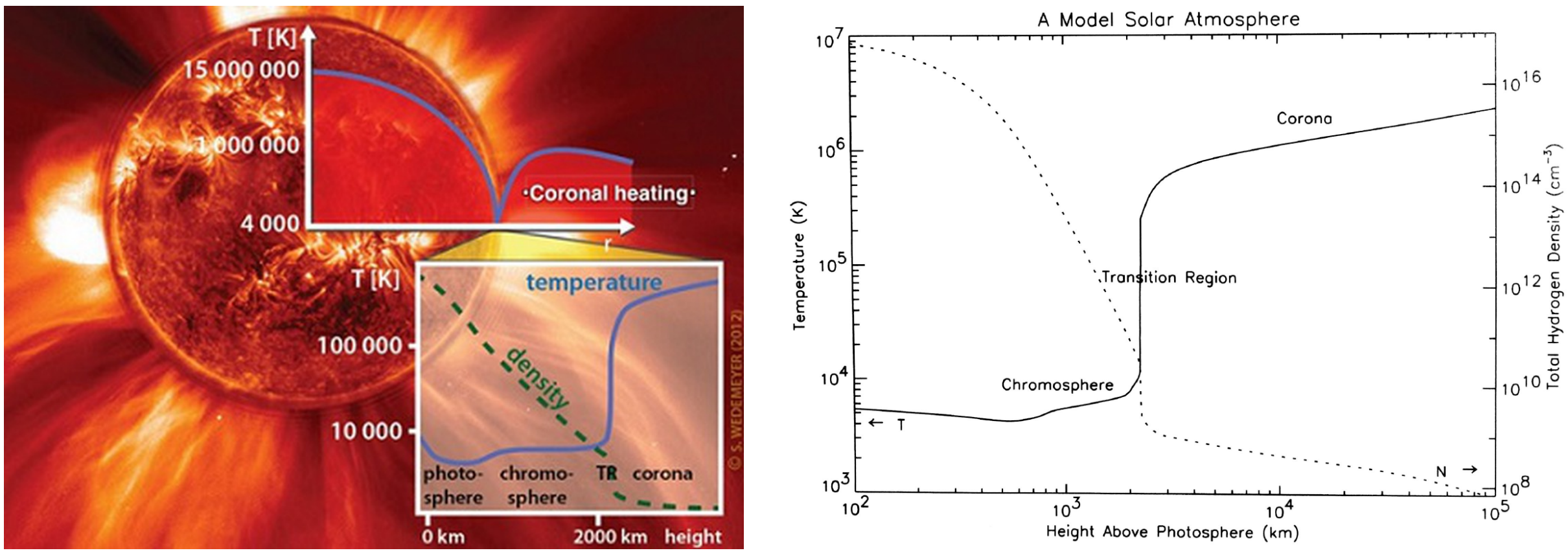

**(a)** (62). **(b)** (63).

**Figure 7.1:** Plasma temperature (solid line) and Hydrogen density (dotted line) distributions as a function of altitude above the optically thick solar surface.

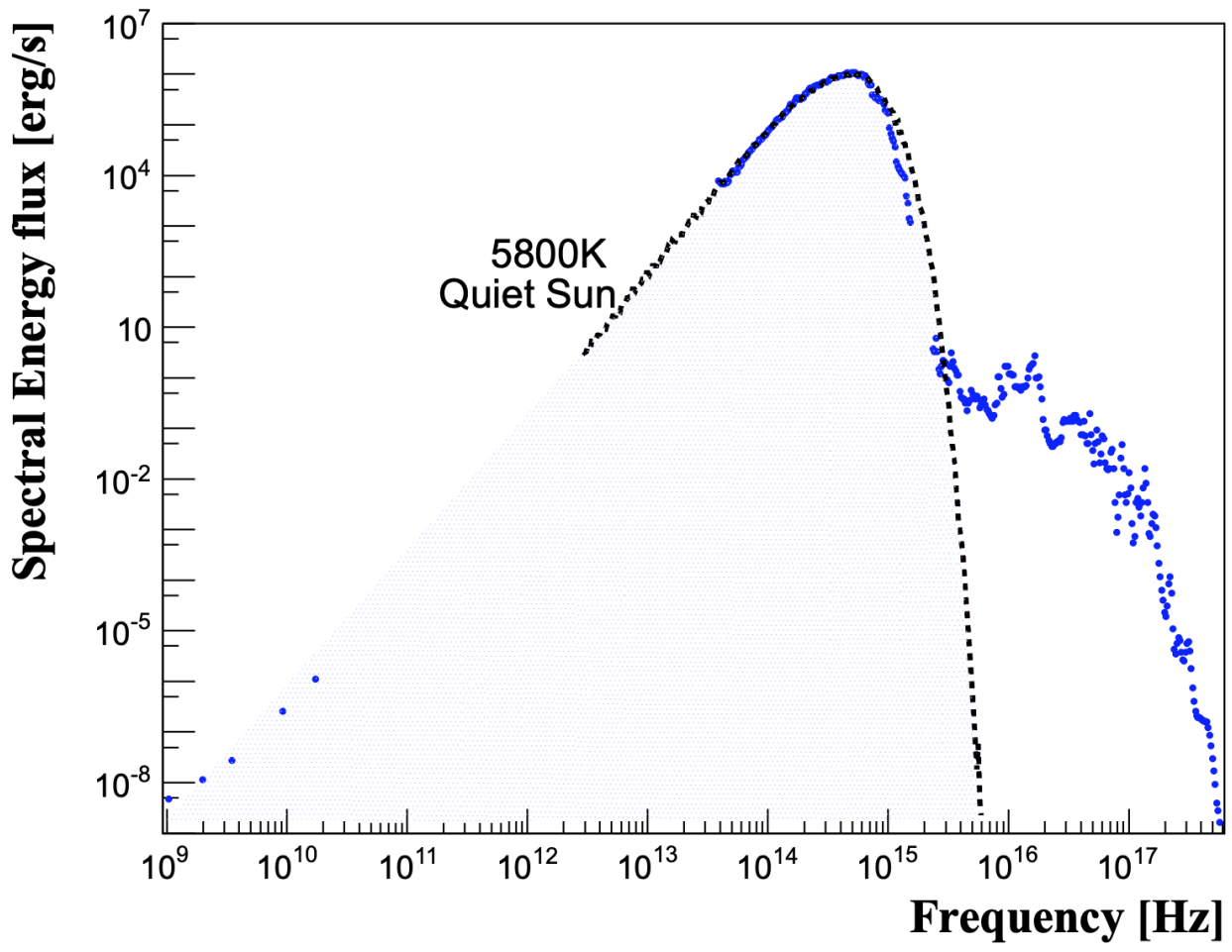

**Figure 7.2:** The spectral energy distribution of the quiet Sun having a near-black body shape at visible and near-IR. At EUV and far-IR wavelengths and beyond (blue points) there is an excess over the fitted blackbody function(in dashed line) (30, 31).

solar cycle [265]. It is noted that the EUV photons reaching the Earth are completely absorbed in the upper atmosphere above 80 km and thus the measurement of the EUV radiation is only possible through rockets and satellites. These absorbed highly energetic photons not only heat the upper atmosphere but also ionise it creating the ionosphere (see Chap. 15).

Therefore, based on the above it is worth performing an analysis looking for a planetary relationship of the solar EUV irradiance, as well as search for possible correlations and differences with other solar phenomena like the solar flares (see also Publications E.1, E.14, and E.15).

## 7.2 EUV data

### 7.2.1 Data origin

The EUV data used in this analysis have been obtained from the Solar EUV Monitor (SEM) on the Solar and Heliospheric Observatory (SOHO) spacecraft which is a joint European Space Agency (ESA) and NASA mission. The EUV is observed in three different channels [266]. From channel 1 to 3, the first-order flux from 26 nm to 34 nm is derived, whereas from channel 2, the zero-order flux at 0.1 nm to 50 nm is measured [267]. In this analysis, the SEM data that are used correspond to the daily averaged full solar disk central order flux at a range of 1 AU, extending from 0.1 nm to 50 nm. The units are in $10^{10}$ photons/cm$^2$/s.

The acquired data period spans from 01/01/1996 - 01/03/2021 which corresponds to 9192 days. The sum of the raw EUV measurements is $26\,014.8541 \times 10^{10}$ photons/cm$^2$ with a minimum of $0.7372 \times 10^{10}$ photons/cm$^2$/s, a maximum of $44.8740 \times 10^{10}$ photons/cm$^2$/s and an average of $2.8302 \times 10^{10}$ photons/cm$^2$/s (see Fig. 7.3).

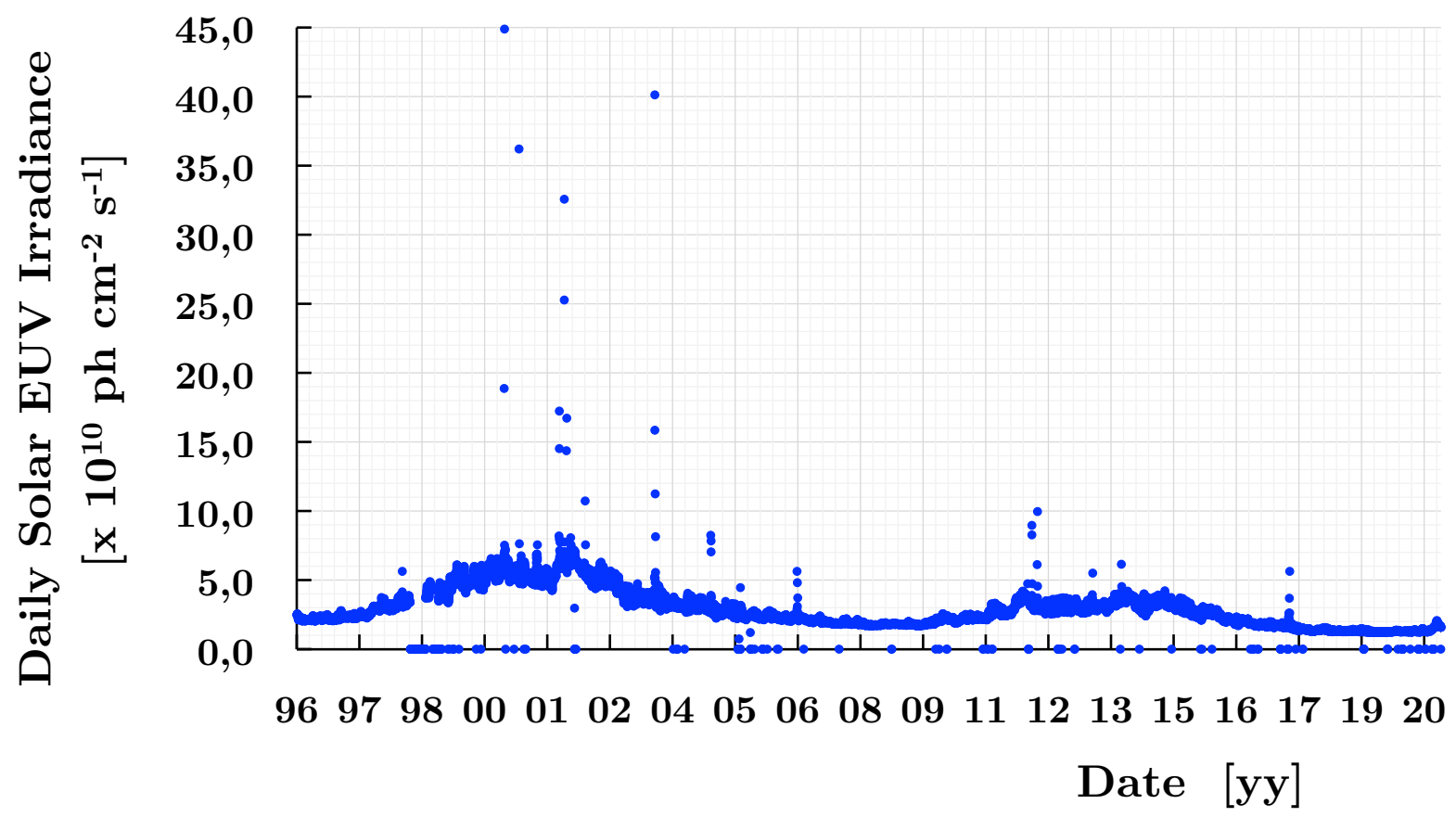

**Figure 7.3:** The daily average raw solar EUV data from 01/01/1996 - 01/03/2021.

### 7.2.2 Data curation

The raw data contain 314 zero values (out of 9192 measurements i.e. $\sim 3.4\%$) which correspond to no data being recorded during these days. The 165 of these "empty" dates are until 02/02/1999. Furthermore, as seen in Fig. 7.3 in several dates there are EUV values which are far above or far below the local average and seem to correspond to wrong measurements. These are in total 59 days out of 9192 (i.e. $\sim 0.64\%$).

In order for all the aforementioned irregular values (in total 373/9192 i.e $\sim 4.1\%$) to not interfere with the analysis, they have been corrected via linear interpolation, with the new corrected time series shown in Fig. 7.4. Even after correcting these values, to avoid even further possible systematics in our analysis, the uninterrupted high precision measurement period 03/02/1999 - 01/03/2021 will be used which skips the many continuous zero measurements of 1998 − 1999. However, it is noted that the results that will be presented here remain the same with the inclusion of the corrected data from the period 01/01/1996 - 02/02/1999.

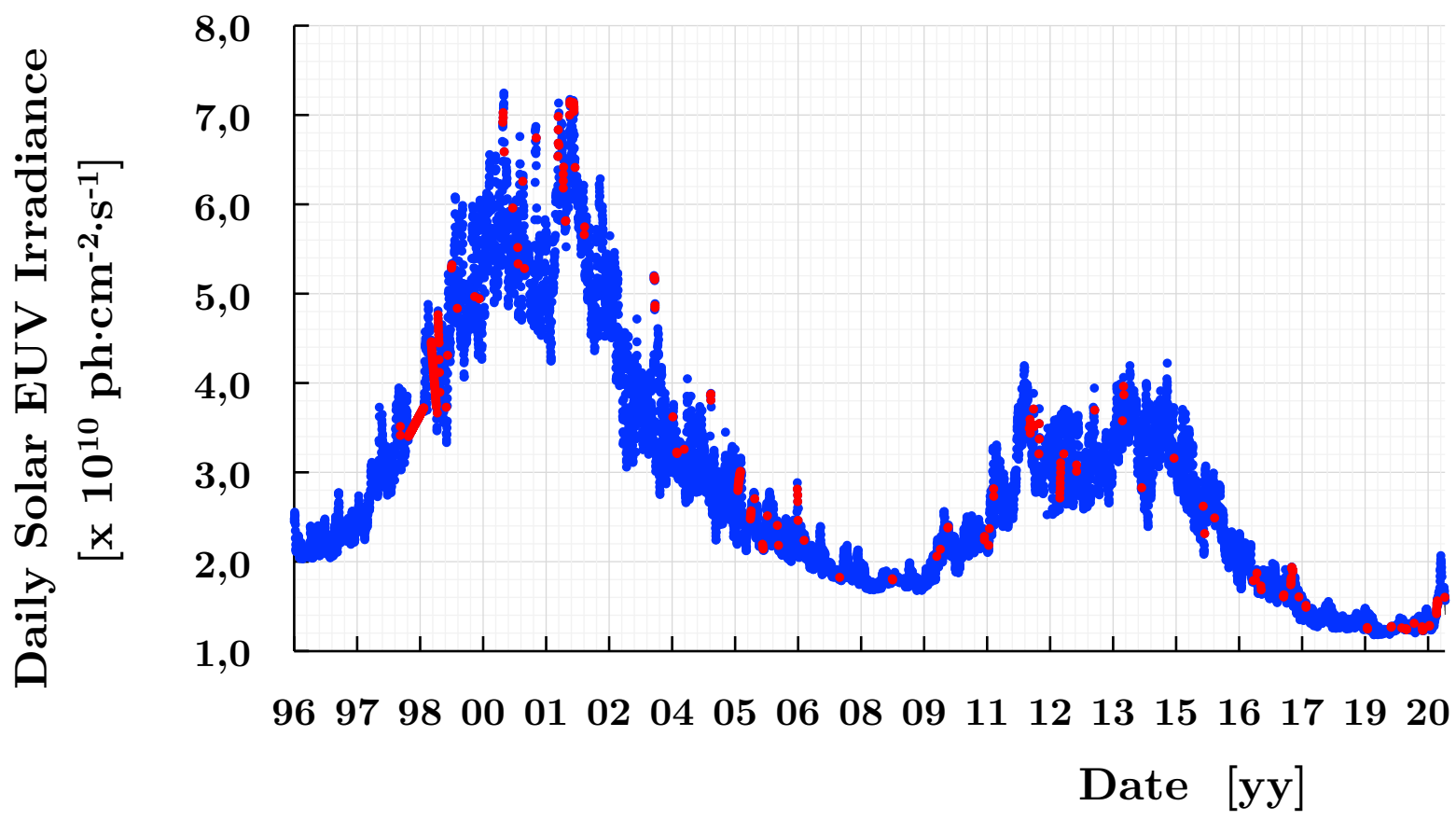

**Figure 7.4:** The daily average corrected solar EUV data. The corrected values are seen in red.

### 7.2.3 Data statistics

From the corrected data seen at Fig. 7.4, the various corresponding histograms have been created in Fig. 7.5 that show the frequency distribution of the average daily EUV for the period 01/01/1996 - 01/03/2021. Fig. 7.5a is an alternative representation of Fig. 7.4. In Fig. 7.5e we can already see an interesting variation with 13 peaks which could be related to the 27.32 d Moon's sidereal month and thus pointing to an exo-solar source. Even though in the literature the 27 d period is conventionally assumed to be associated to the solar rotation [268], however, as we will see also later the solar differential rotation ranges from 25 d to 36 d [269, 270] and thus should not be able to produce narrow peaks around 27.3 d. This period will be followed further in the next sections.

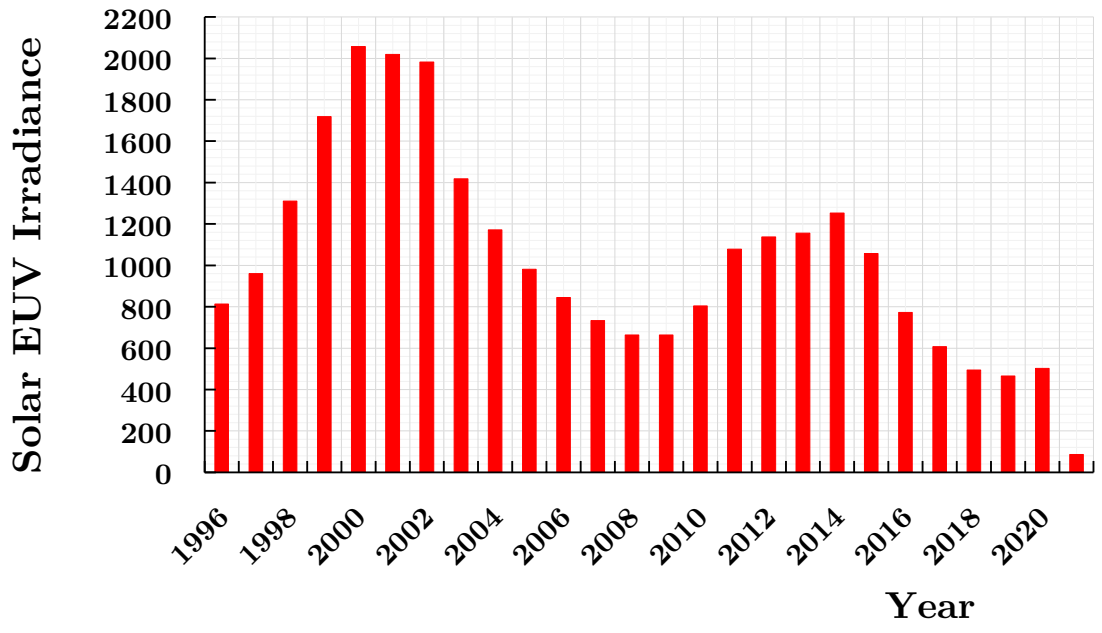

**(a)** Solar EUV irradiance per year.

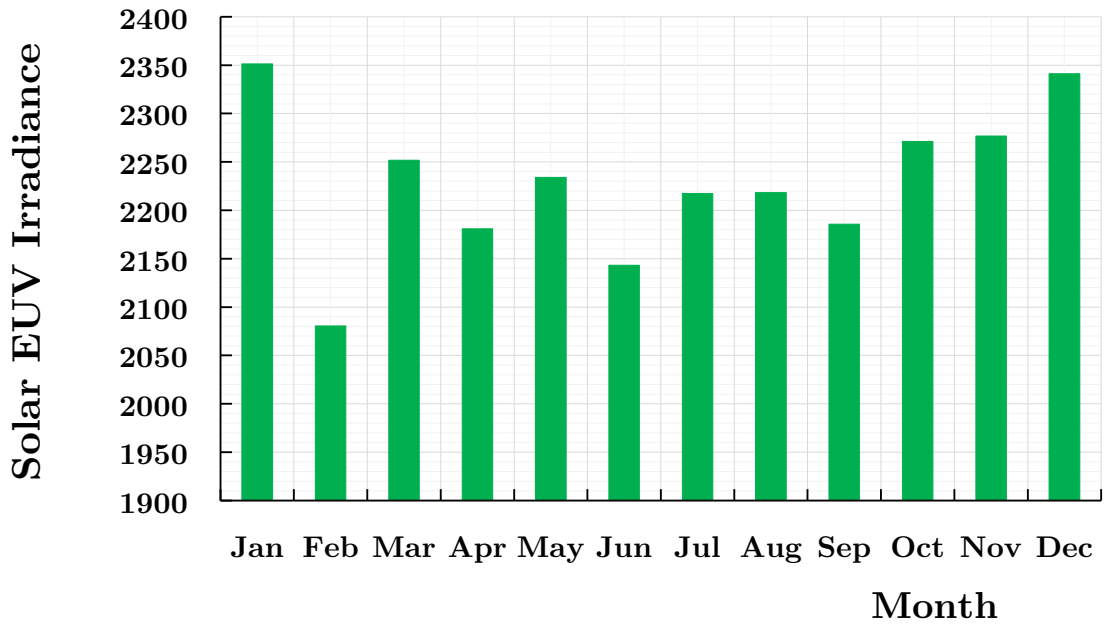

**(b)** Solar EUV irradiance per month.

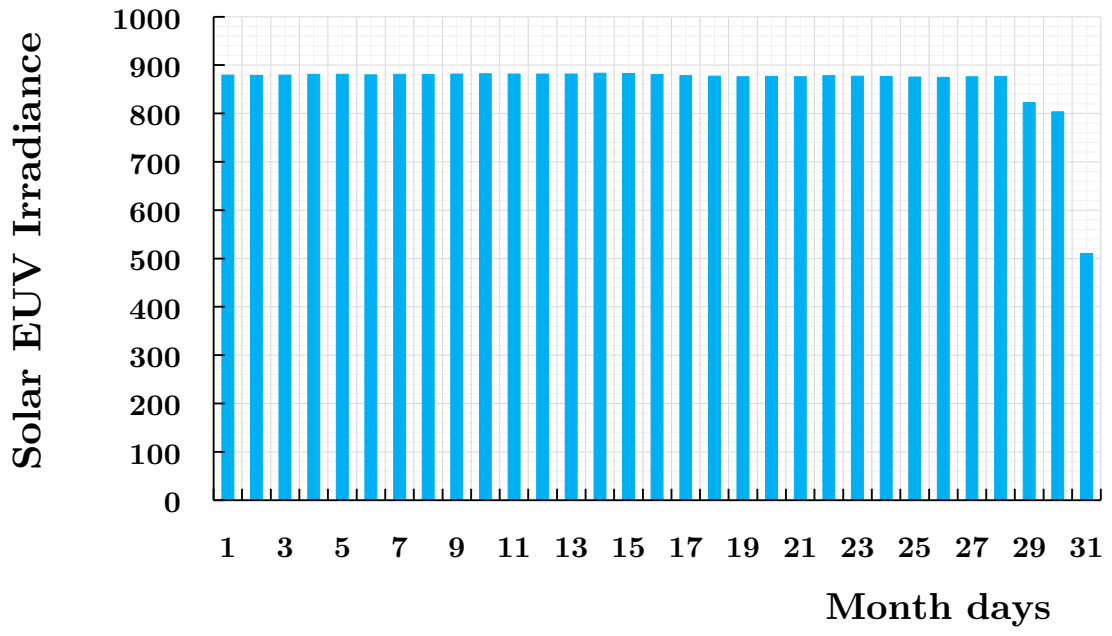

**(c)** Solar EUV irradiance per day of month.

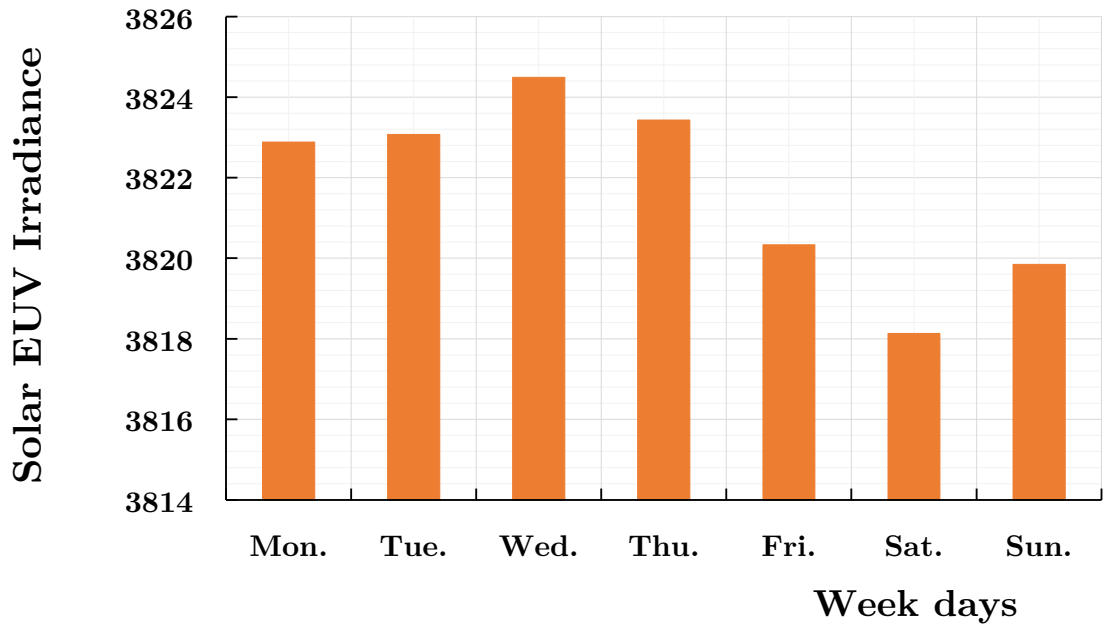

**(d)** Solar EUV irradiance per day of week.

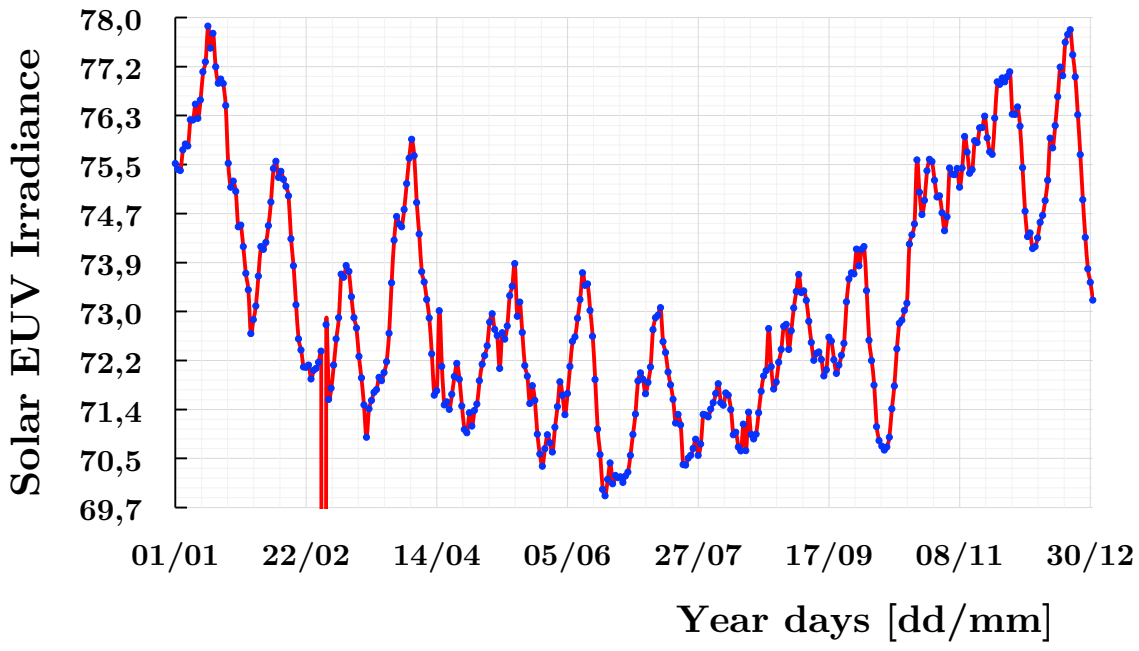

**(e)** Solar EUV irradiance per day of year. The observed point with the reduced amplitude corresponds to 29/02 which is a leap day added only to leap years and thus appears less frequently in the total period.

**Figure 7.5:** Histograms for the daily average solar EUV irradiance for 01/01/1996 - 01/03/2021.

# 7.3 Data analysis and results

## 7.3.1 Planetary longitudinal distributions

### 7.3.1.1 Single planets

To look for a planetary relationship, the daily data of EUV are projected on the corresponding planetary heliocentric longitudinal coordinates. The first step is to use single planets without any further constraint on the position of the rest of the planets. As seen in Fig. 7.6 the inner planets Mercury, Venus and Earth, as well as Mars, Jupiter and Moon, are selected since they perform multiple orbits during the given period 03/02/1999 - 01/03/2021 (8063 d). Moreover, it is noted that any eccentricity related effect is already removed. In all plots in Fig. 7.6 the sum of the EUV values for the selected time period is $23\,535.6 \times 10^{10}$ photons/cm$^2$.

During these $\sim 22$ y Mercury has performed about 92 revolutions and any random effect would be averaged out. Instead, in Fig. 7.6a, we observe a big minimum-maximum difference of 4.5%. Similarly we see some peaking distributions in Venus (Fig. 7.6b), Earth (Fig. 7.6c) and Mars (Fig. 7.6d) with the overall amplitude being 5.7%, 5.5% and 5.8% respectively. In Jupiter (Fig. 7.6e) a single wide peak around 80° is observed with an amplitude of 68.3%. Its FWHM as derived by a Gaussian fit is about $152.8° \pm 9.6°$. Finally, in Fig. 7.6f we can see a wide peak with an amplitude of 2.9% that could perhaps explain the observed variation of Fig. 7.5e.

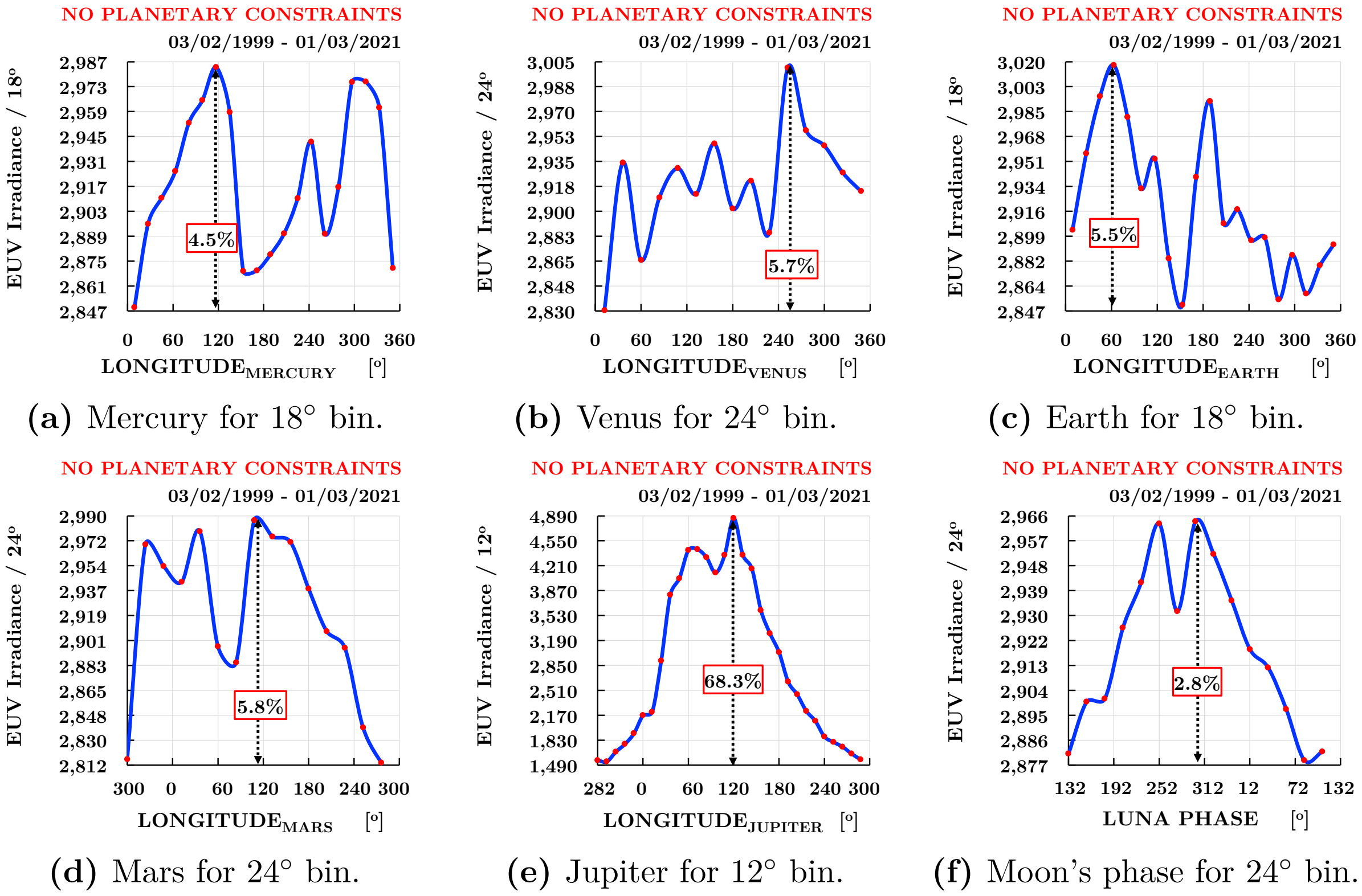

**(a)** Mercury for 18° bin. **(b)** Venus for 24° bin. **(c)** Earth for 18° bin.

**(d)** Mars for 24° bin. **(e)** Jupiter for 12° bin. **(f)** Moon's phase for 24° bin.

**Figure 7.6:** Planetary heliocentric longitude distributions of solar EUV irradiance for the period 03/02/1999 - 01/03/2021.

#### 7.3.1.2 Combining planets

The next step is to use a combination of planetary positions in order to strengthen further the planetary results. First we use the reference frame of Mercury for which an interesting result was obtained in Fig. 7.6a. However, this time Venus was allowed to propagate in a fixed heliocentric orbital arc during the selected time period. This is shown in Fig. 7.7 where the solar EUV irradiance is plotted in the reference frame of Mercury while requiring Venus to propagate in two opposite orbital arcs 120° wide. For the case of Venus being constrained to propagate between 200° to 320° (Fig. 7.7a) we have in total $7999.9 \times 10^{10}$ photons/cm$^2$ in 2718 d which results to a difference between the maximum and the minimum point of 11.2% and two peaks around 239° and 302°. On the other hand, for the case of Venus being between 20° to 140° we have in total $7801.3 \times 10^{10}$ photons/cm$^2$ in 2681 d giving an amplitude of 10.4%. As seen from Fig. 7.7c and 7.7d, from the difference between these two cases we can conclude on the strong correlation of the relative position of Venus and Mercury on EUV. Furthermore, it is noted that the longitudinal position of Venus in Fig. 7.7a (200° to 320°) is centred around the direction of the GC at 266°. This prominent direction derives from the alignment GC - Earth - Sun which takes place within $\sim 5.5°$ every June. As explained also before, if there was no planetary correlation, then not only the two spectra in Fig. 7.7c would be flat, but also no difference between the two would be expected. However, this is not the case as seen in Fig. 7.7d.

In Fig. 7.8 this time the EUV distribution is plotted as a function of Venus position with Earth being constrained to propagate in the orbital arc between 200° to 320°. Once more the distribution is far from random, with the corresponding time being 2747 d and the sum of the events being $7933.5 \times 10^{10}$ photons/cm$^2$ giving a maximum - minimum difference of 14.5%.

Finally, in Fig. 7.9, an additional plot for the distribution of EUV is made, but this time for the reference frame of Earth while Mercury's longitudinal position is vetoed between 0° to 180°. The resulting minimum-maximum difference is 29.9% and the number of days fulfilling this constraint are 3016. In this distribution, we see four clear significant peaks around 38°, 128°, 222° and 315° (with an accuracy of $\sim \pm 2°$) with a FWHM of about 47°, 45°, 53° and 55° respectively (with an accuracy of $\sim \pm 12°$), as obtained by a Gaussian fitting algorithm. The number of peaks can be associated with the ratio between the revolution period of Earth and Mercury which is about $365.256/87.969 \sim 4.1$.

### 7.3.2 Fourier analysis

In order to have an additional hint about the underlined periodicities, a Fourier analysis has been performed with all data (01/01/1996 - 01/03/2021). The most interesting derived periodicity is the 8$^{\text{th}}$ biggest peak at 27.32 d with an amplitude of 6.2 dB and a calculated FWHM, via a Gaussian fit, of 0.08 d (see Fig. 7.10). This value totally overlaps with Moon's sidereal month of 27.32 d fixed to remote stars. This observation by it's own points to an additional exo-solar influence on the solar activity, and verifies the assumed external influence on the Sun from the Earth with the intervening Moon derived from Fig. 7.6f and 7.5e.

As expected, in Fig. 7.10, there is also a less pronounced peak at 29.52 d $\pm$ 0.06 d (43$^{\text{rd}}$ biggest with amplitude of 1.4 dB) associated with Moon's synodic period of 29.53 d, i.e. fixed

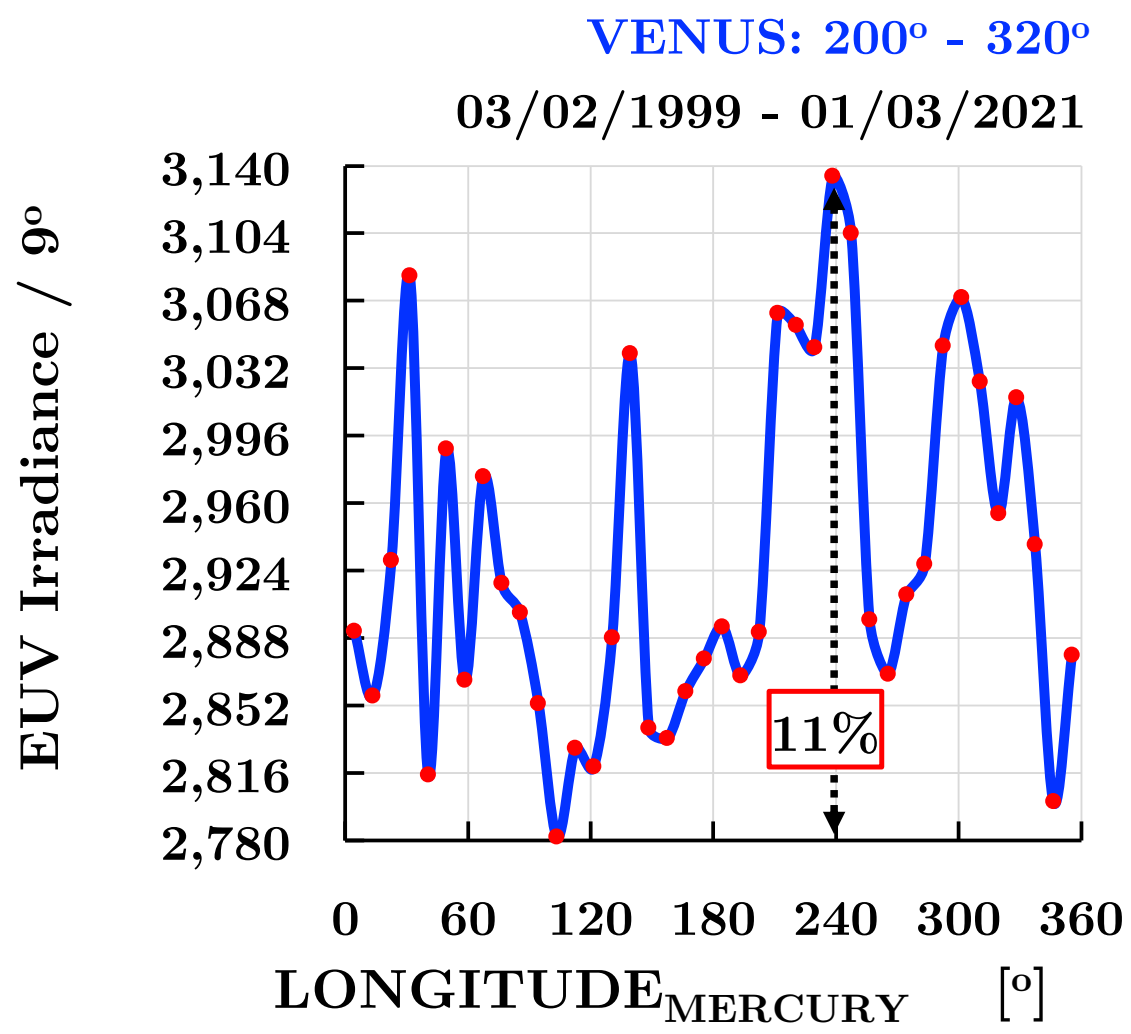

**(a)** Mercury when Venus is between 200° to 320°.

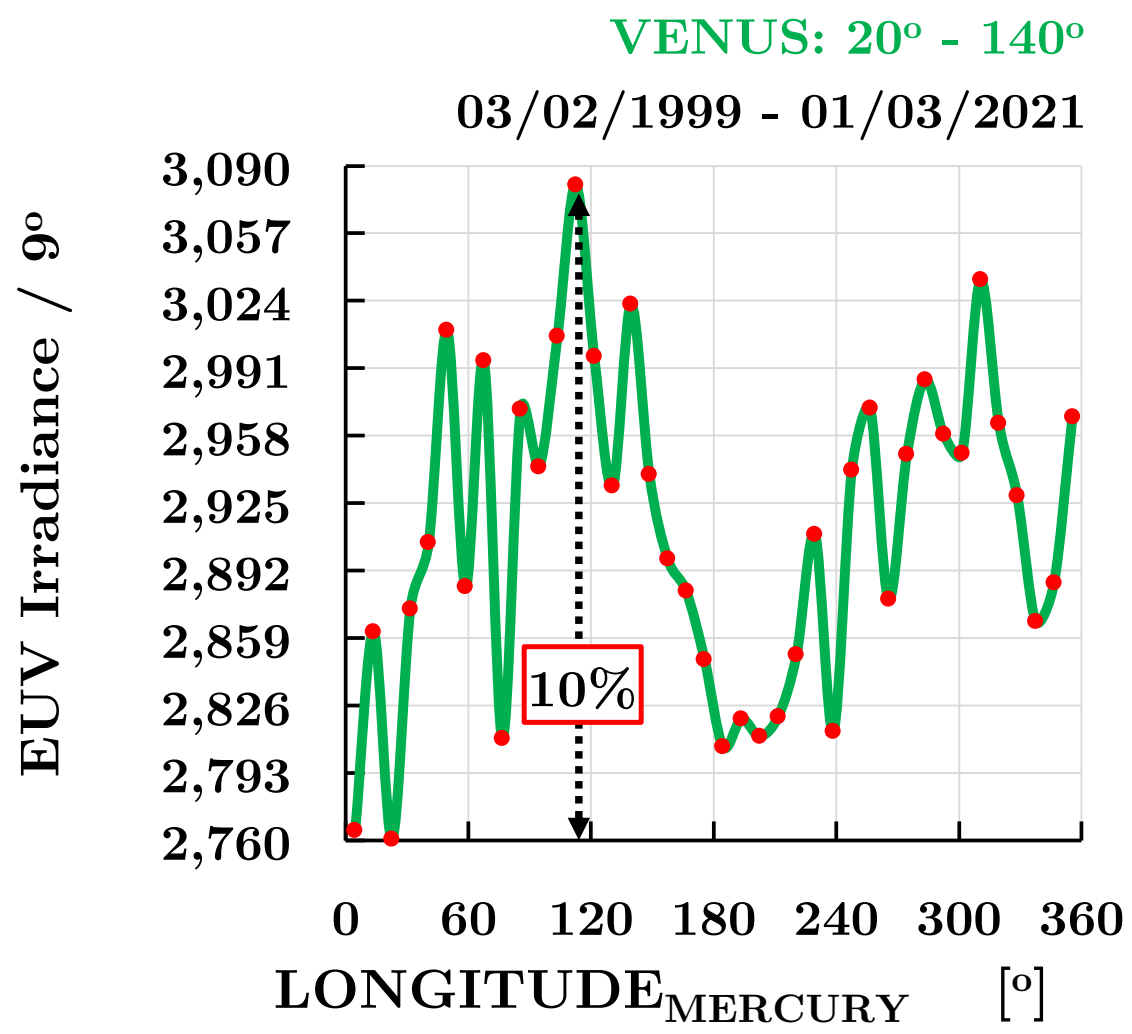

**(b)** Mercury when Venus is between 20° to 140°.

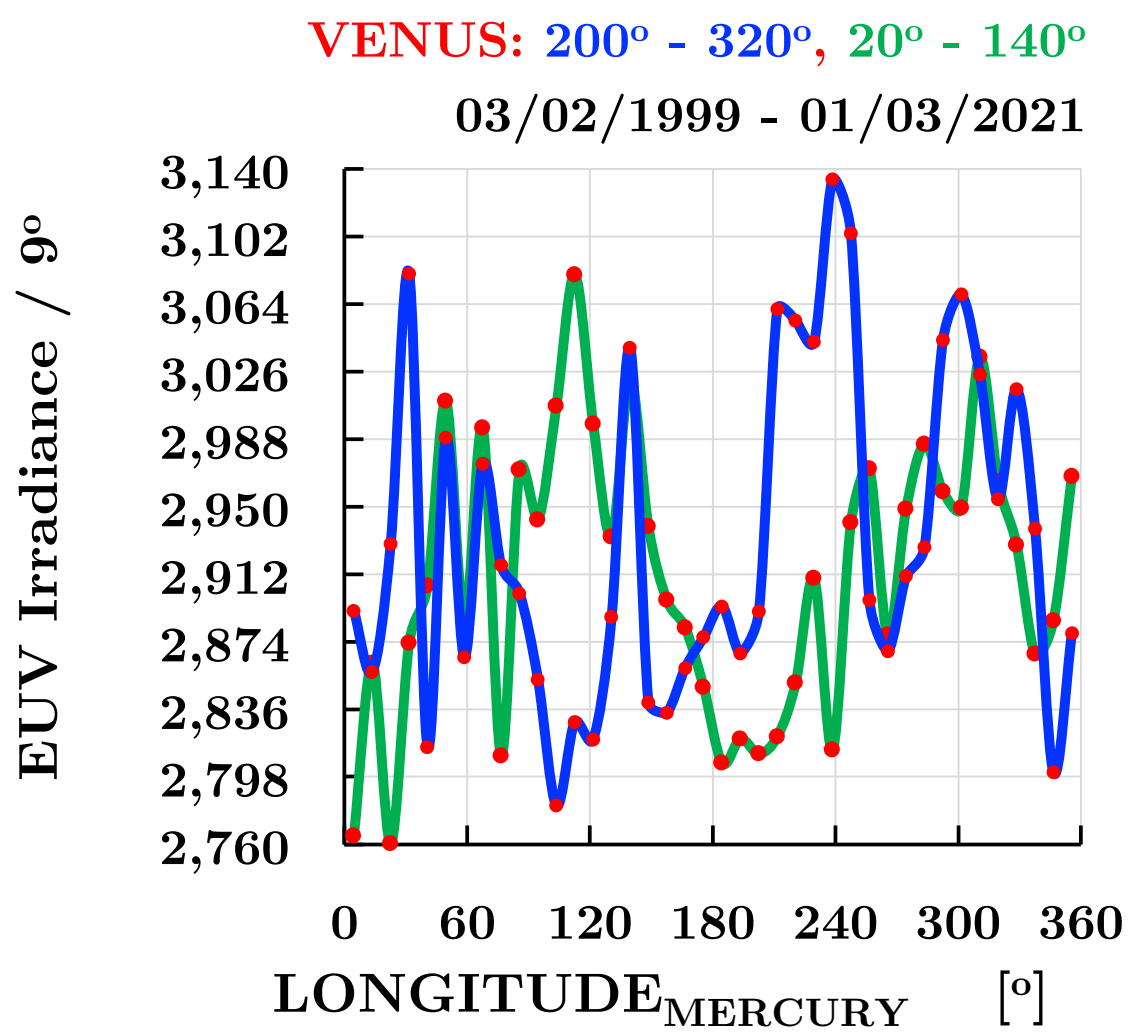

**(c)** Comparison between Fig. 7.7a and Fig. 7.7b.

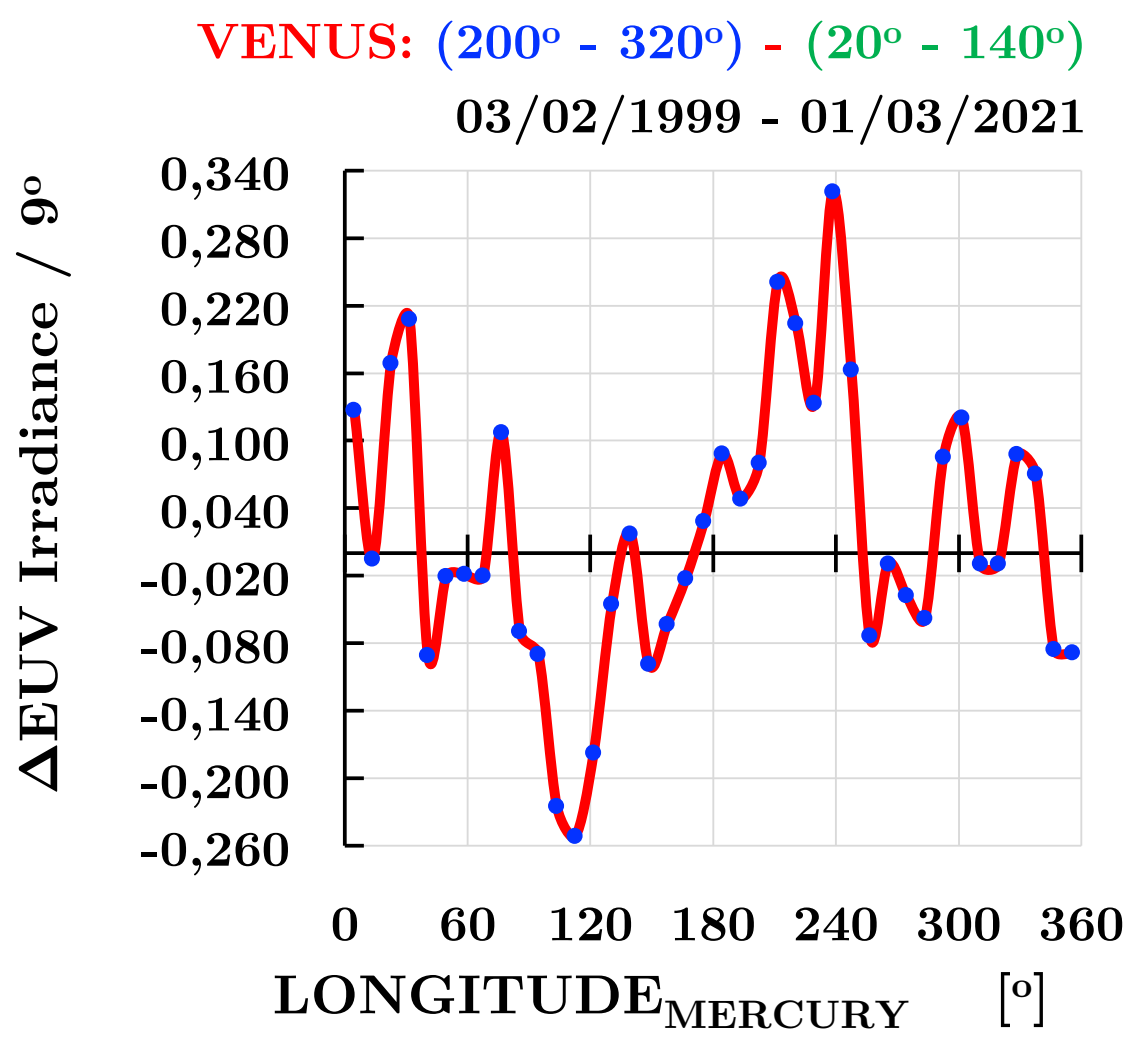

**(d)** Fig. 7.7a minus Fig. 7.7b.

**Figure 7.7:** Solar EUV intensity distribution for the reference frame of Mercury when Venus propagates between 180° opposite orbital arcs with 120° width, for 9° bin.

to the Sun.

Other derived interesting peaks from the Fourier periodogram are presented bellow.

- The 33rd biggest peak is around 145.2 d ± 0.8 d and has an amplitude of about 2 dB. This overlaps with the 145 d synodic period of Mercury-Venus and also verifies the Mercury-Venus correlation seen in Fig. 7.7.
- The 4th biggest peak with an amplitude of about 12.4 dB, is around 593.3 d ± 11.8 d which is within $\leq 1\sigma$ equal to the synodic period of Earth-Venus 584 d and supports the Earth-Venus correlation found in Fig. 7.8.
- The 17th biggest peak is located around 222.8 d ± 2.6 d and has an amplitude of 3.7 dB.

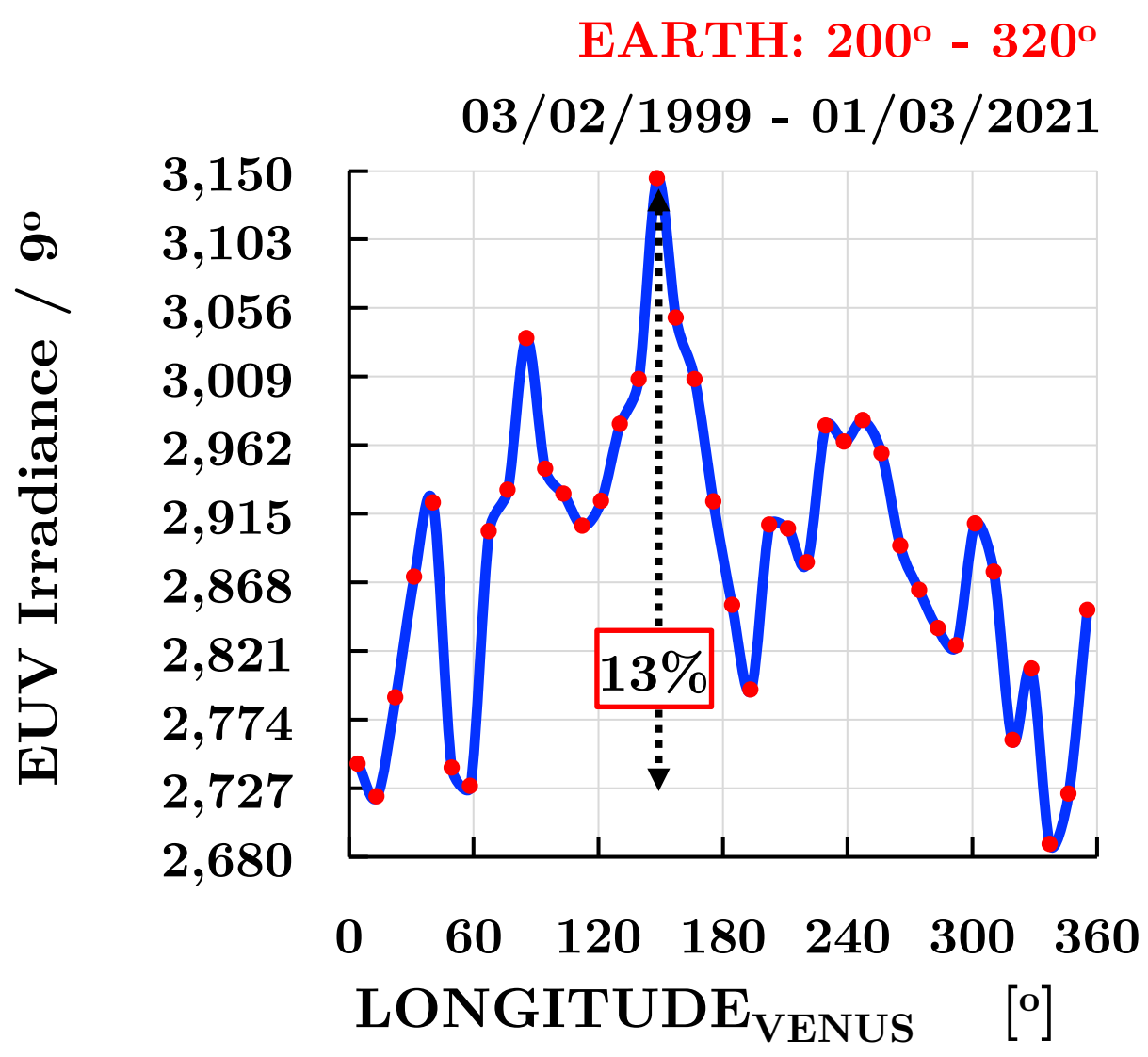

**Figure 7.8:** Solar EUV distribution in the reference frame of Venus when Earth is constrained between 200° to 320° for bin = 9°.

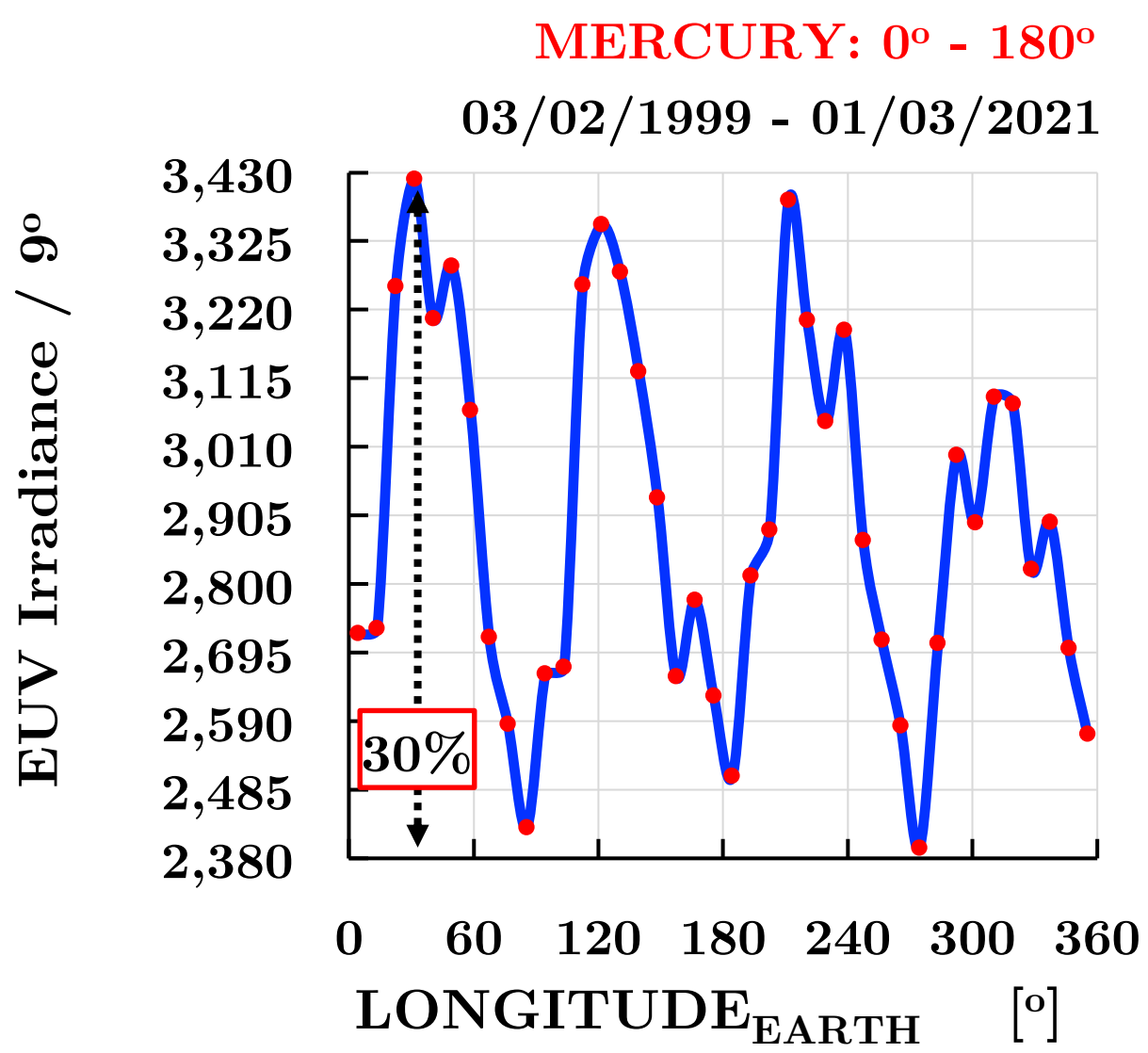

**Figure 7.9:** EUV as a function of Earth's heliocentric longitude while Mercury is restricted to propagate between 0° to 180° for bin = 9°.

This period is also within $\leq 1\sigma$ with the Venus revolution period of 224.7 d.

It is noted that the period 03/02/1999 - 01/03/2021 gives the same results as with the 01/01/1996 - 01/03/2021. As an example, the 8$^{th}$ biggest peak which coincides with the Moon's sidereal month remains at 27.32 d with an amplitude of 5.15 dB, whereas around 29.54 d the peak which can be identified as the Moon's synodic period has again a much smaller amplitude at about 1.53 dB.

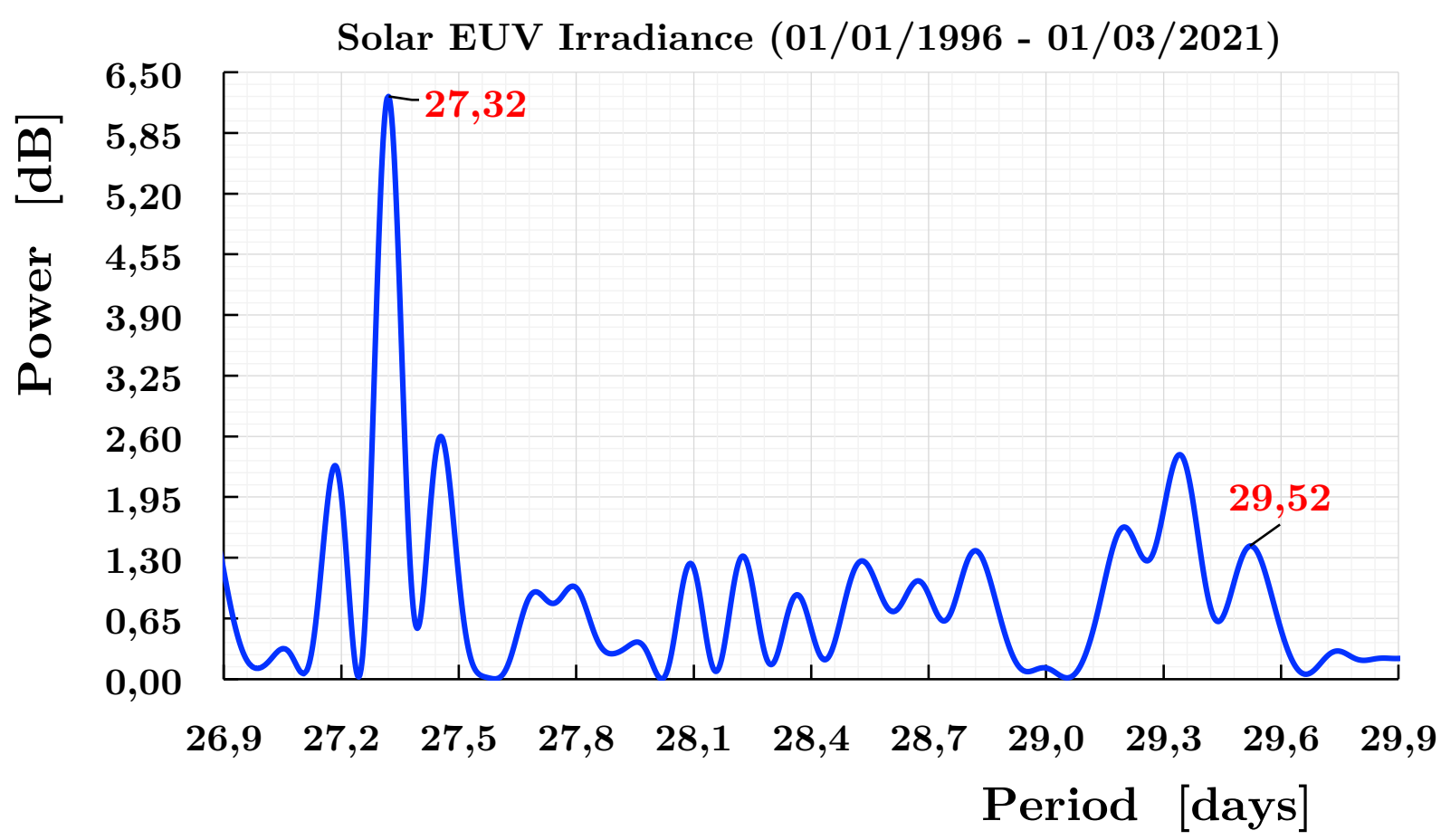

**Figure 7.10:** Fourier periodogram of the corrected solar EUV irradiance data from 01/01/1996 to 01/03/2021.

## 7.4 Summary

The analysed solar emission in EUV, which is the manifestation of the solar corona mystery, shows a significant planetary dependance when both single inner planets were used as well as when positional constraints of different planets were applied. It is also remarkable that a peaking EUV distribution takes place when the longitudinal position of Venus is on a 120° window around the direction of the GC similarly to what was observed with solar flares.

At the same time, a significant lunar relationship is found, which is supported by the Fourier analysis. In fact the 27.32 d periodicity associated with the Moon's sidereal month fixed to the stars points to an additional exo-solar origin of the EUV. An additional derived information is the observation of the absence of a statistically significant correlation of EUV with another solar activity manifestation, e.g. that of the M-flares in most of the small scale characteristics (see Appendix Sect. B.3.1). However, a depletion of the the EUV irradiation is observed $1-2$ months before the occurrence of big X-flares with a sharp increase of the emission $\sim 1$ week before (see Appendix Sect. B.3.2).

In summary, considering the observation that the occurrence of the full disk EUV irradiance of the Sun is modulated by the position of the planets, together with the deduction of the existence of an additional mechanism in the Sun before the appearance of the biggest flares, supports the working hypothesis that the solar activity is triggered by an influx of highly interacting invisible massive matter in form of streams which gets gravitationally lensed by the planets or the free fall effect of the Sun itself. The gravitational modulation by other solar system bodies, like the interposed Earth's Moon also agrees with the observations made in this work. Finally, particle candidates such as AQNs fit in the invisible streaming scenario since their impact on the solar coronal heating has already been argued [231].

# 8

# SUNSPOTS

## 8.1 Introduction

Sunspots are temporary phenomena manifested on the solar photosphere appearing as spots that are darker and cooler than the surrounding areas (Fig. 8.1a). Sunspots are characterised by a dark core, called *umbra*, and a less dark halo called *penumbra* (Fig. 8.1b). The umbra can be up to 20 000 km wide and has a temperature of about 3700 K while the surrounding photosphere has a temperature of around 5800 K. The magnetic field is strongest within the umbra, at about 2000 G to 3000 G on average, and weaker in the penumbra. Additionally, the magnetic field in the umbra is oriented rather vertically to the solar surface, becoming slightly inclined as we approach the umbral-penumbral boundary [271]. Sunspots are mainly seen in groups, and while they move across the surface of the Sun they can expand and contract.

The sunspot number is commonly accepted as a proxy of solar activity with the number of sunspots varying strongly over a solar cycle of about 11 y [272], from almost zero observable sunspots on the solar disc during solar minimum and 150 or more during a solar maximum (Fig. 8.2). The sunspot activity has also shown that the amplitude of the sunspot cycle varies in each cycle (see Fig. 8.3 and Fig. 8.4). Each sunspot typically lasts for several days, however, the very large ones may live for several weeks. With the solar rotation, each sunspot, depending on its location, comes back to its original position after about 27 d which is defined as the

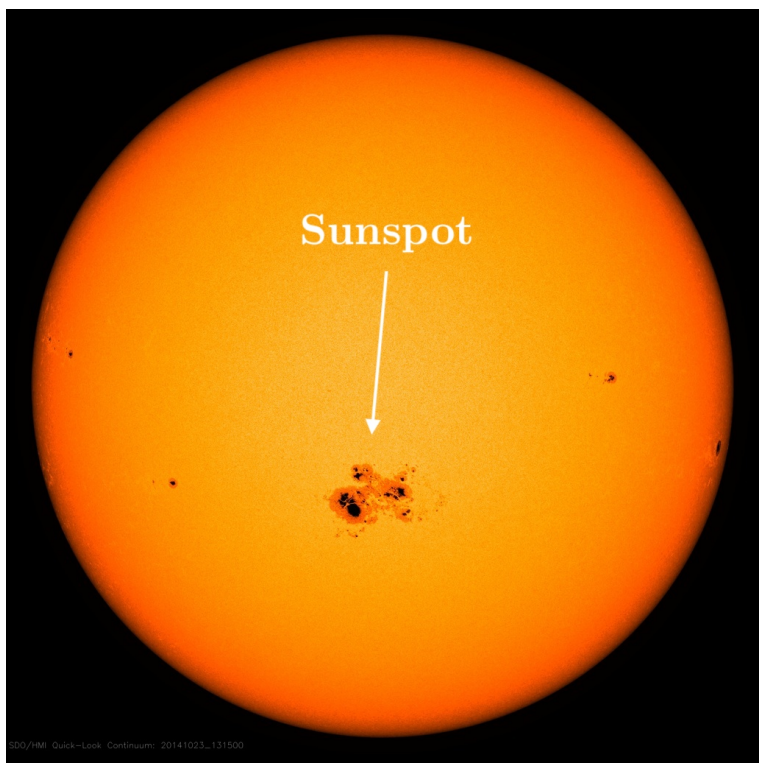

(a) A huge sunspot with of about 130 000 km diameter seen on the lower center of the Sun's image captured by SDO on 23/10/2014 (64).

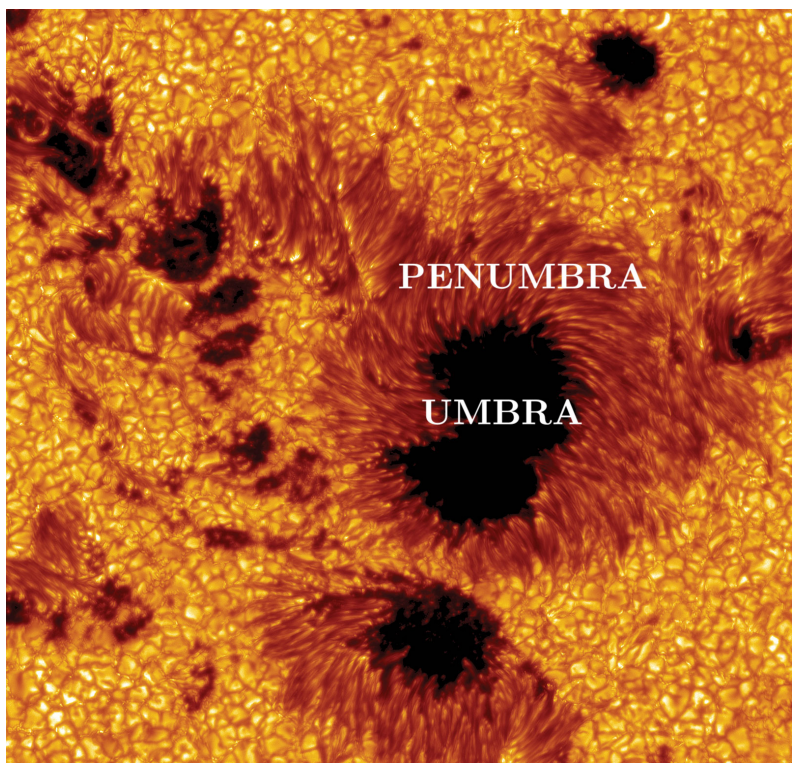

(b) Sunspot group in an active region observed on 15/07/2002 (65).

**Figure 8.1:** Pictures of sunspots.

synodic rotation period of the Sun as seen from the Earth. However, by correcting with the motion of the Earth, the derived mean value of the true rotation period is only 25.4 d [273]. Furthermore, sunspots usually come in groups with two sets of spots, which are known as *active regions*. One set has a positive or north magnetic field and another one with a negative or south magnetic field.

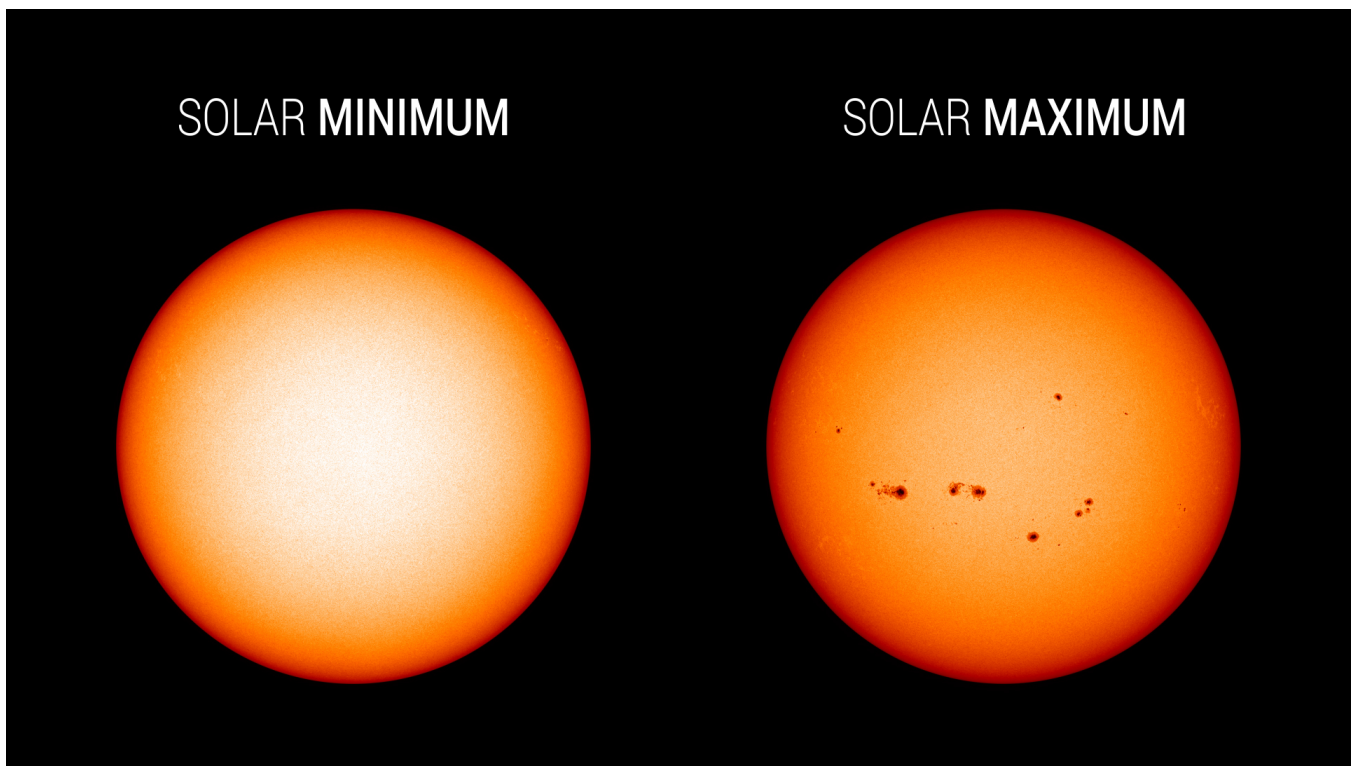

(a) Comparison of visible light images from SDO for the solar minimum in 12/2019 and the solar maximum in 04/2014 (66).

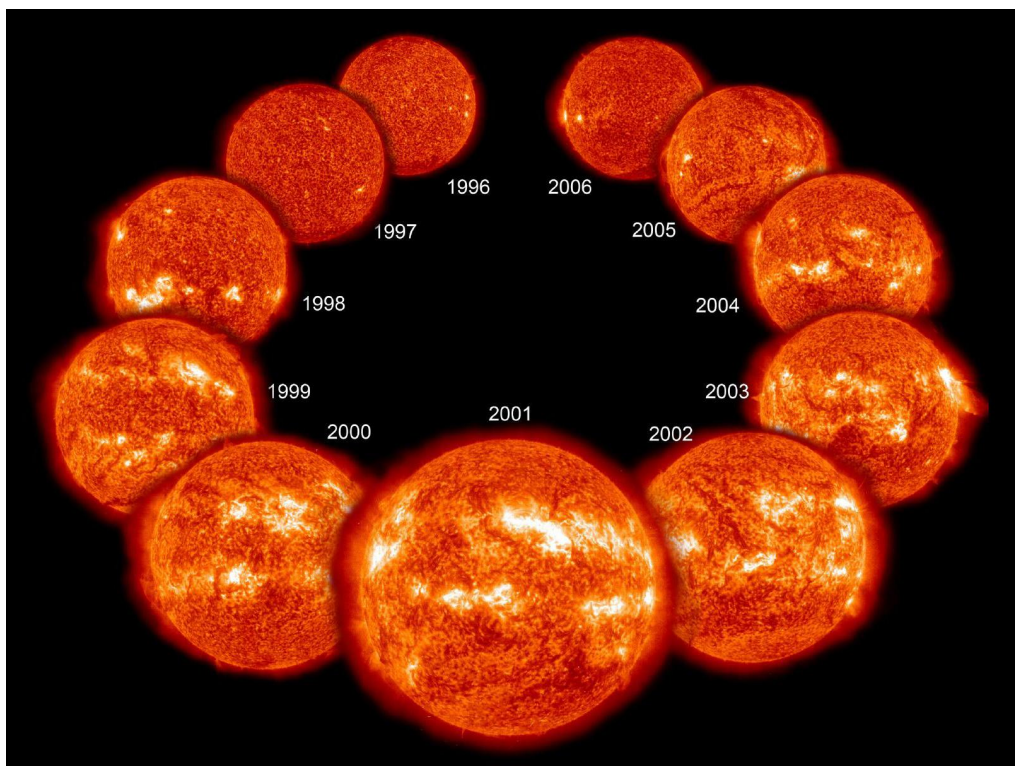

(b) Full-disc images of lower corona for the solar cycle 23 progressed from solar minimum (1996) to maximum (2001) and back to minimum (2006) (67).

**Figure 8.2:** Variation of the Nr. of sunspots during a $\sim 11$ y solar cycle.

The location of the sunspots on the Sun is not random, but are gathered in two latitude bands on either side of the equator, and also varies throughout the solar cycle. During solar minima, sunspots tend to form around 30° to 45° North and South of the Sun's equator, while during solar maxima they tend to appear closer to the equator around 15° latitude [274]. This behaviour when plotted on a graph where the latitudes are plotted against time, like Fig. 8.3, is called a *butterfly diagram* due to the shape of the distributions for both northern and southern hemispheres resembling the wings of a butterfly.

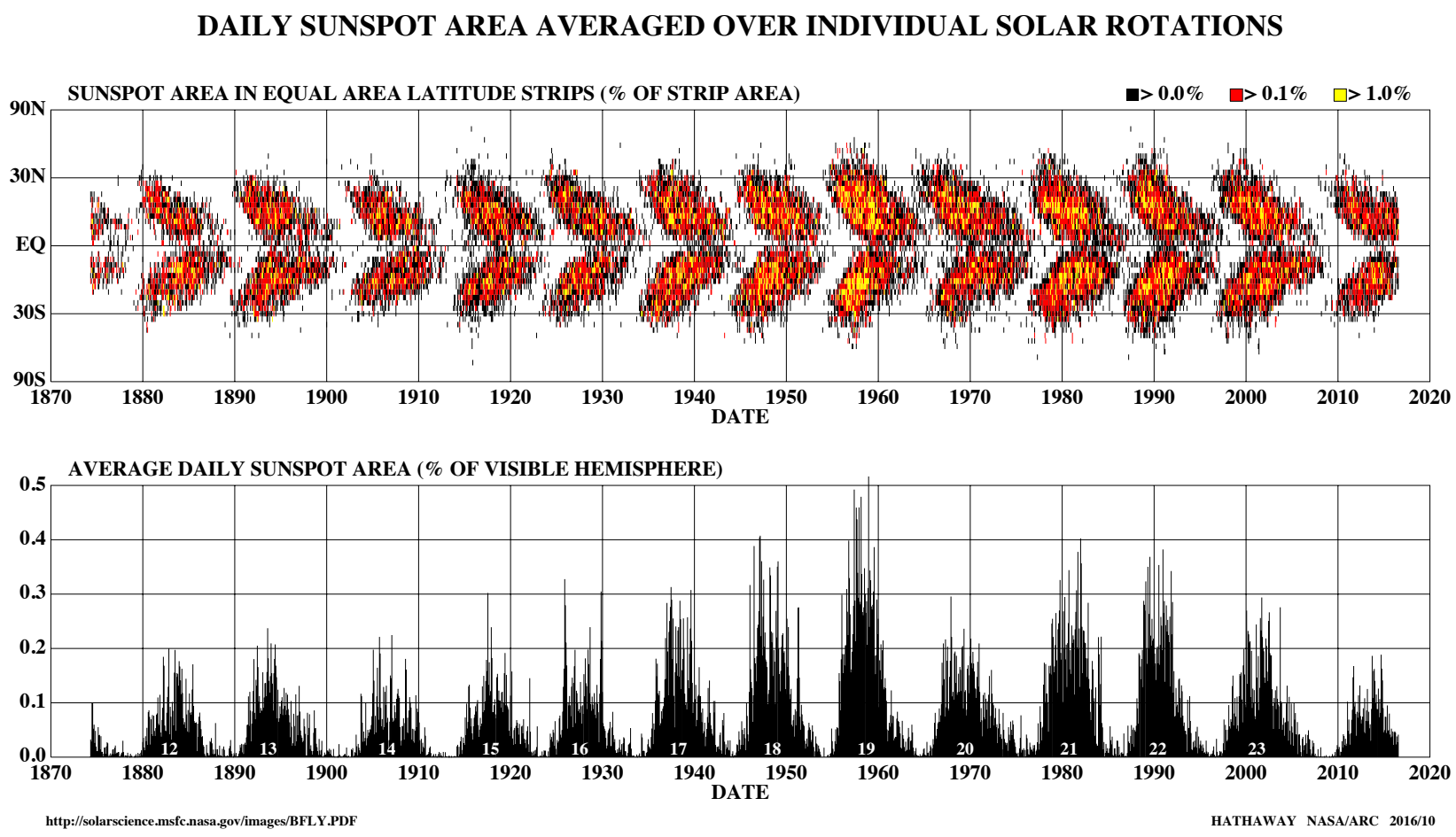

**Figure 8.3:** Butterfly diagram (upper panel) showing the positions of sunspots for each rotation of the Sun since 05/1874 and the relative solar surface area covered by the sunspots (lower panel) (32).

The solar magnetic field is considered as the root cause of the sunspot phenomenon, since they are assumed to be formed when a magnetic flux tube rises above the photosphere, forming a pair of positive and negative spots at the intersection of the solar surface with such a flux tube [275]. However, multiple long-standing problems regarding the conventional formation model remain unresolved [276]. Although a variety of theories have been proposed, the nature as well as the origin of the sunspot cycle, remains one of the biggest questions in solar physics [239, 255, 277]. As a result, in the following analysis, a correlation will be searched with planetary positions which following the hypothesis of this work will be the key signature for the involvement of the dark sector. If this is the case, then the only viable explanation for the external triggering of the sunspot cycle is through a highly interacting slow-moving invisible matter in the form of streams being focused by the planets towards the Sun's position but also from the Sun itself due to free fall.

## 8.2 Sunspot data

### 8.2.1 Data origin

The sunspot number series have been acquired by the open-access World Data Center (WDC)-Sunspot Index and Long-term Solar Observations (SILSO) database from the Royal Observatory of Belgium in Brussels, which is preserving the longest record of solar activity over the last four centuries [278]. The sunspot number time series document the variations of the last 35 solar cycles.

The daily total sunspot number ($R$) is derived from the total number of sunspots observed on the Sun ($N_s$) and the number of sunspot groups ($N_g$) counted over the entire solar disk. The formula used is:

$$R_i = 10N_g + N_s \tag{8.1}$$

The sunspot number which is based on all visual observations, was introduced by Rudolf Wolf in 1849. The acquired time series contain daily data from 01/01/1818 up to today. Before that, daily observations were too sparse and only monthly and yearly sets were compiled.

Regarding the error values, the dataset contains the standard deviation ($\sigma$) of the base counts included in the calculations as well as the number of observations for each measurement. This way the statistical uncertainty can be computed by:

$$\mathrm{SE} = \frac{\sigma}{\sqrt{\mathrm{N}}} \tag{8.2}$$

where: SE : the daily standard error,

$N$ : the number of observations for the day.

Before 1981, the number of observations is set to 1, as the Sunspot Number was essentially the raw number calculated by Wolf from the Zurich observatory. The errors for that period (1818-1981) are estimated from an auto-regressive model based on the Poissonian distribution of the actual sunspot numbers [278].

The acquired data which are shown in Fig. 8.4, and span from 01/01/1818 to 28/02/2021, a period which corresponds to 74 204 d. It is noted that 151 d i.e. 0.2% are marked as provisional meaning that the number provided is subject to a possible revision. These provisional numbers basically correspond to the latest dates between 01/10/2020 - 28/02/2021.

### 8.2.2 Data statistics

For the analyses of this work, the raw data have been used without any correction or modification. The standard daily error for each number has also been calculated based on Eq. 8.2.

The full downloaded dataset (01/01/1818 - 28/02/2021) contain 3247 (i.e. 4.4%) missing values. However, for the formal analysis algorithms to work properly the data from 1900 thereafter hare used. Therefore, the following analysis contains the more recent and more accurate data from 01/03/1900 - 28/02/2021 corresponding to 121 full years or 44 195 d. During this period no missing value was recorded. The sum of the total number of sunspots observed during this period is 3767690 with an average of 85 sunspots per day, a minimum of 0, and a maximum of 503 sunspots per day.

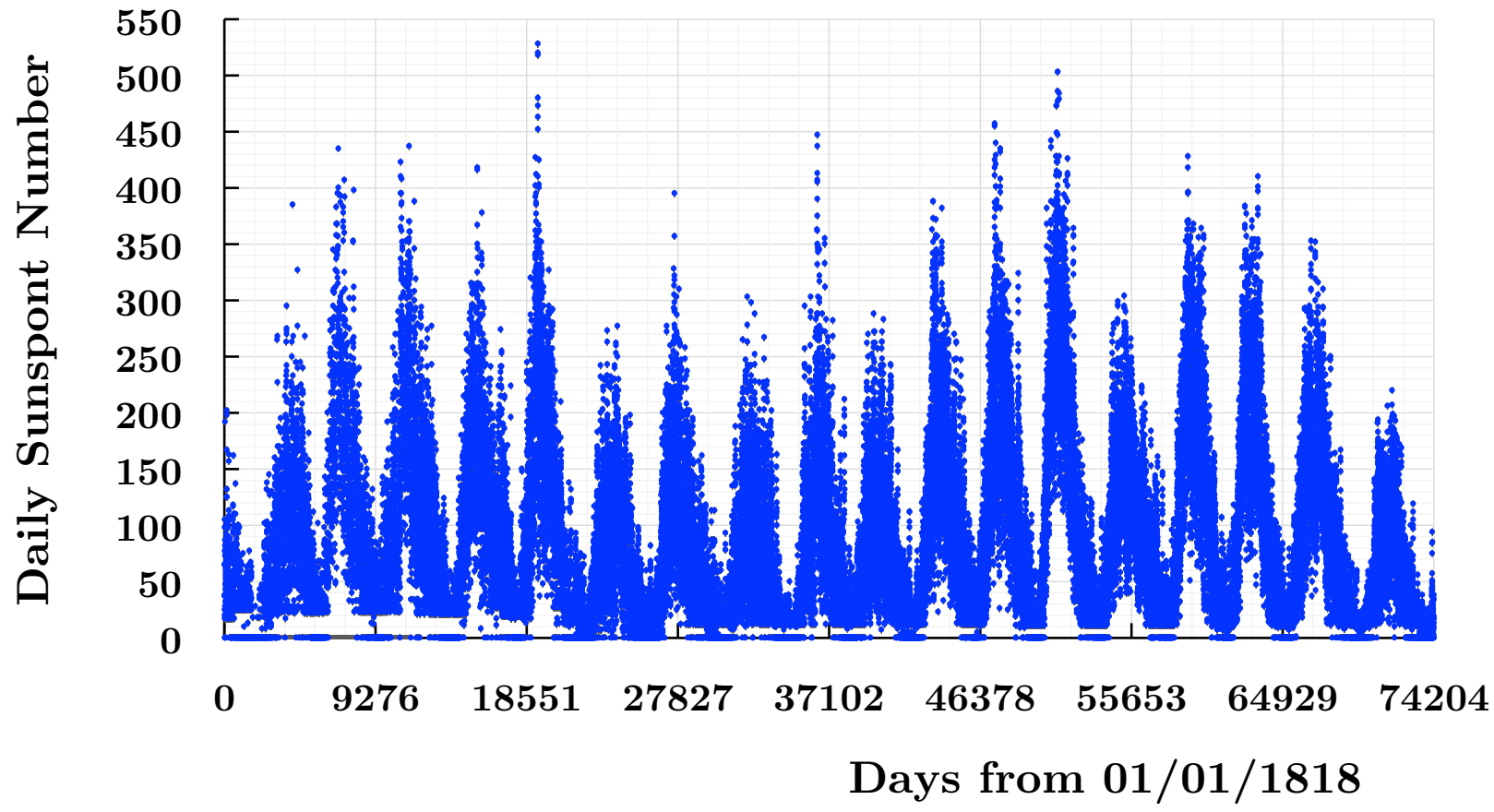

**Figure 8.4:** Daily sunspot Nr. for the full available period 01/01/1818 - 28/02/2021.

The corresponding histograms for the selected period which show the frequency distribution of the daily measured sunspot number are presented in Fig. 8.5.

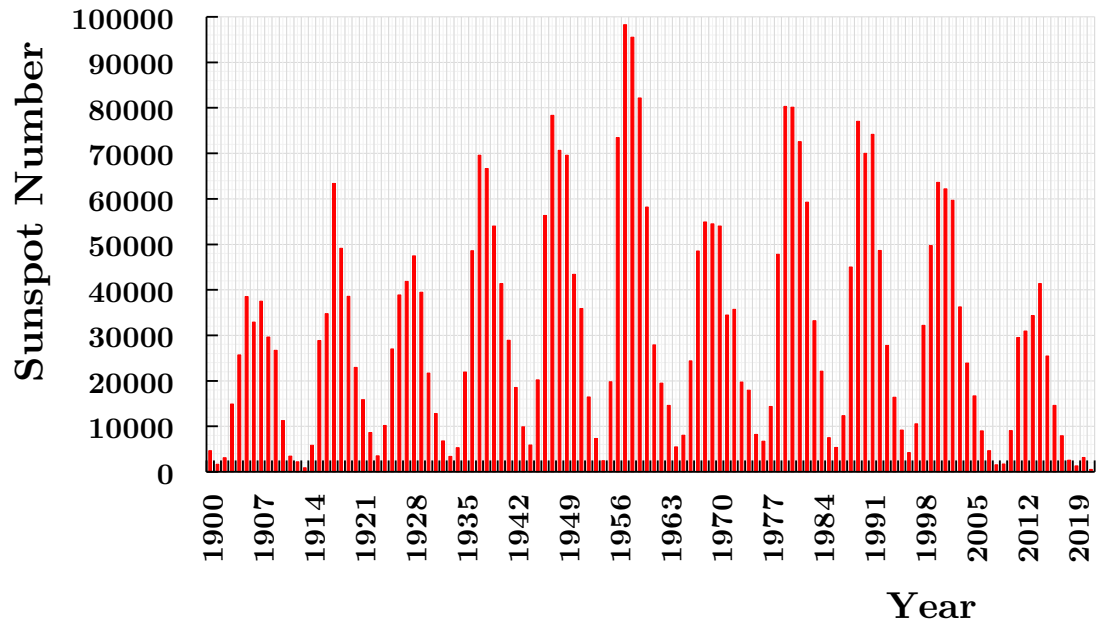

**(a)** Sunspot Nr. per year.

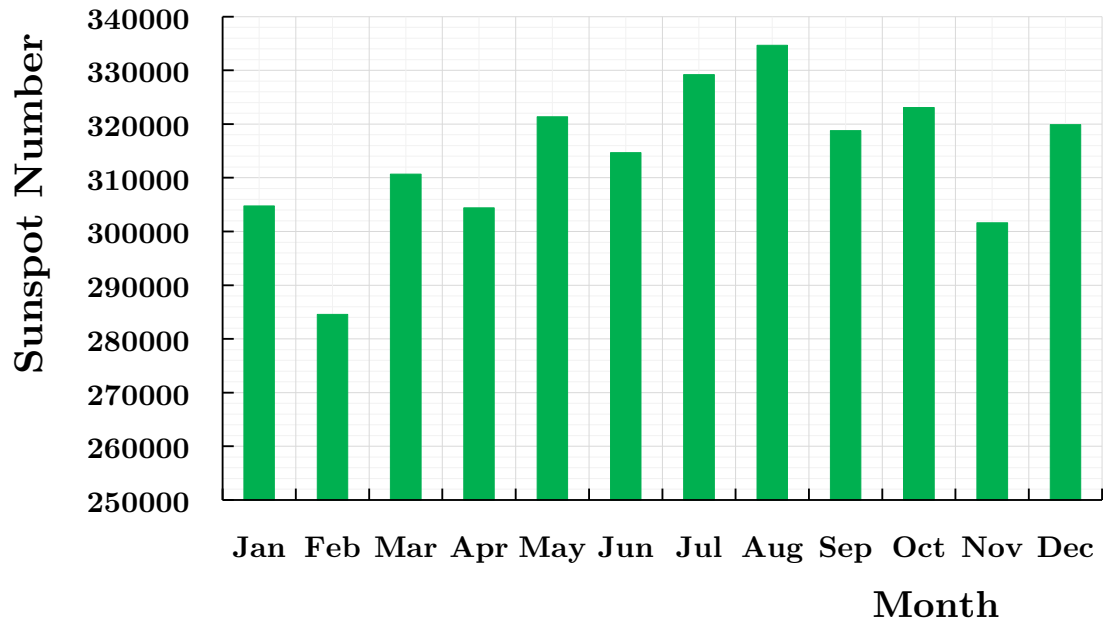

**(b)** Sunspot Nr. per month.

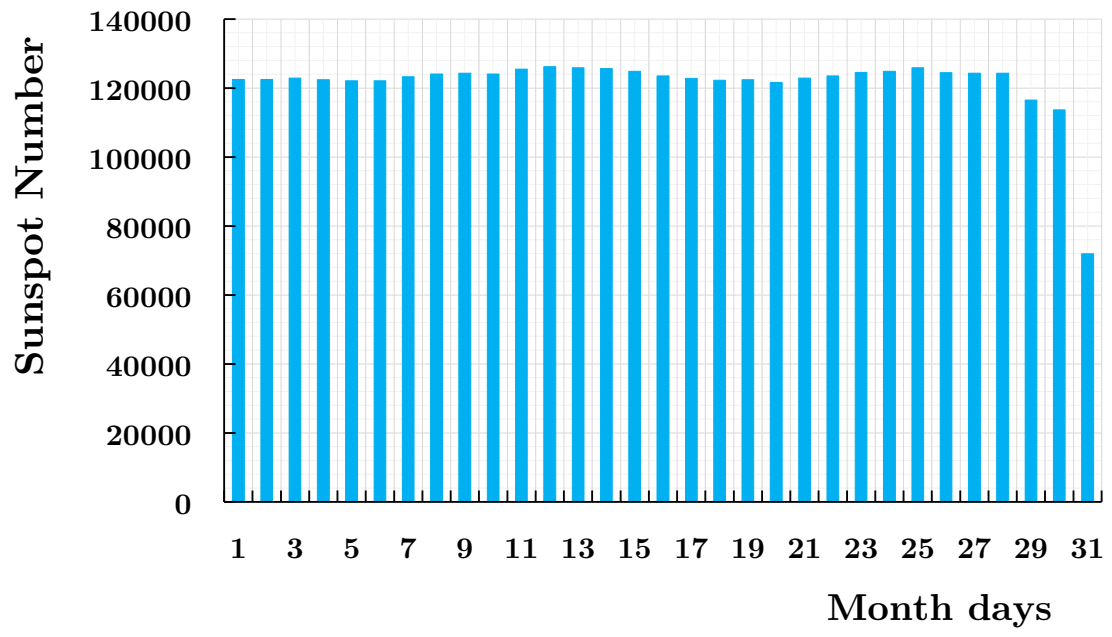

**(c)** Sunspot Nr. per day of month.

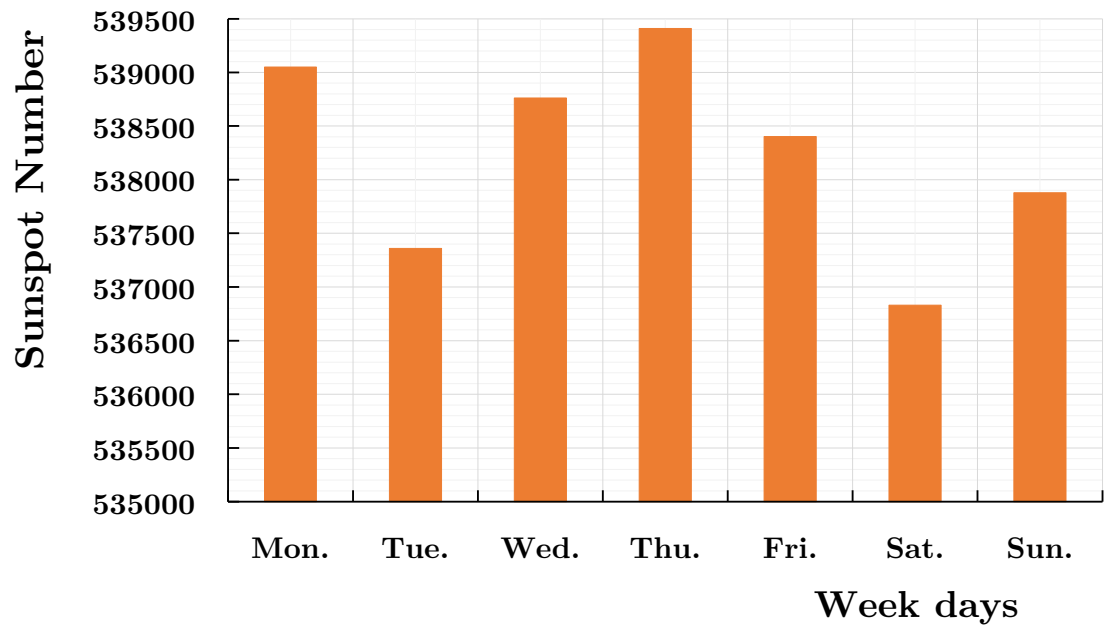

**(d)** Sunspot Nr. per day of week.

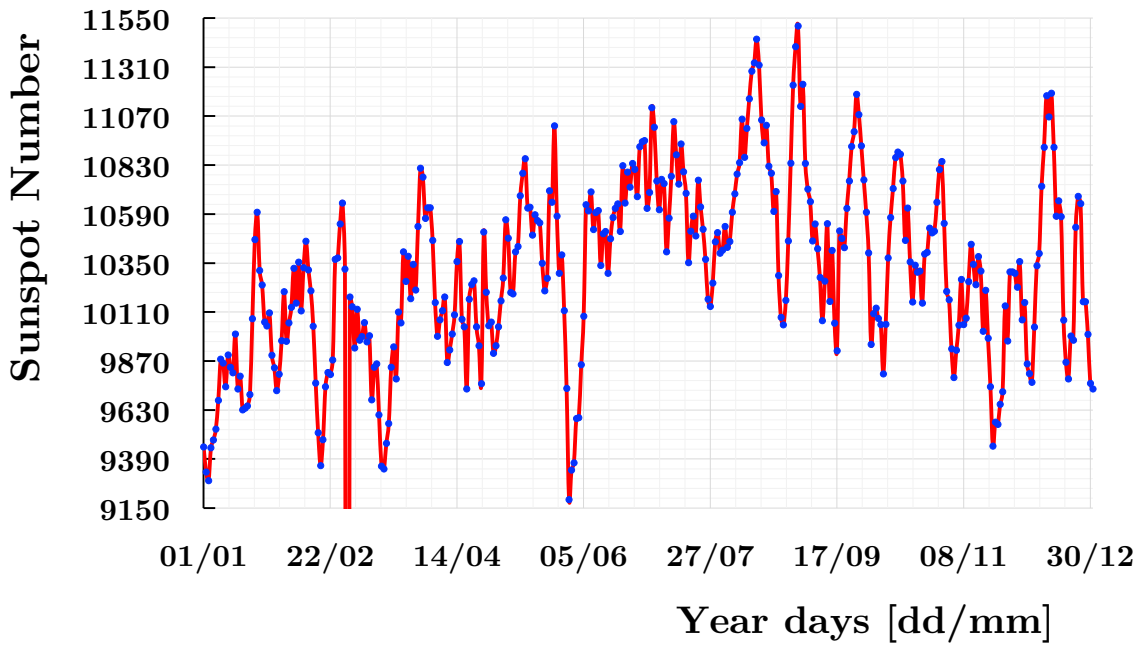

**(e)** Sunspot Nr. per day of year. The observed point with the reduced amplitude corresponds to 29/02 which is a leap day added only to leap years and thus appears less frequently in the total period.

**Figure 8.5:** Histograms for the daily Nr. of sunspots for the period 01/03/1900 - 28/02/2021.

## 8.3 Data analysis and results

### 8.3.1 Planetary longitudinal distributions

#### 8.3.1.1 Single planets

Like in the previous chapters, to look for a planetary relationship, the daily number of sunspots has to be projected on the corresponding planetary heliocentric longitudinal coordinates. A derived significant planetary correlation is the key signature for the involvement of the dark sector following the driving hypothesis of this work. The planetary heliocentric longitudes for the various planets that are used in this case are for 12:00 Coordinated Universal Time (UTC) and are acquired from NASA's Horizons system.

The planetary distributions from Mercury up to Saturn are shown in Fig. 8.6. In all these plots any eccentricity effects are already removed. The total number of sunspots from 01/03/1900 to 28/02/2021 is 3767690. In addition, since the standard errors are already provided for each daily measurement, all the following plots contain error bars. It is noted that in some cases the error bars are very small, and thus not visible. Regarding the statistical mean error, for a 12° bin we get $\sigma \sim 0.3\%$, whereas for a 24° bin we get $\sigma \sim 0.2\%$. As a result, we see that in all planets significant peaks $> 10\sigma$ arise. This is unexpected within conventional solar physics since any random effect is factored-out due the many revolutions of the planets, and a remote planetary influence on the number of sunspots does not exist. More specifically, during the selected $\sim 121$ y we have about 502.4, 196.7, 121, 64.3, 10.2, and 4.1 revolutions for Mercury, Venus, Earth, Mars, Jupiter and Saturn respectively which should factor out any randomly occurring excess.

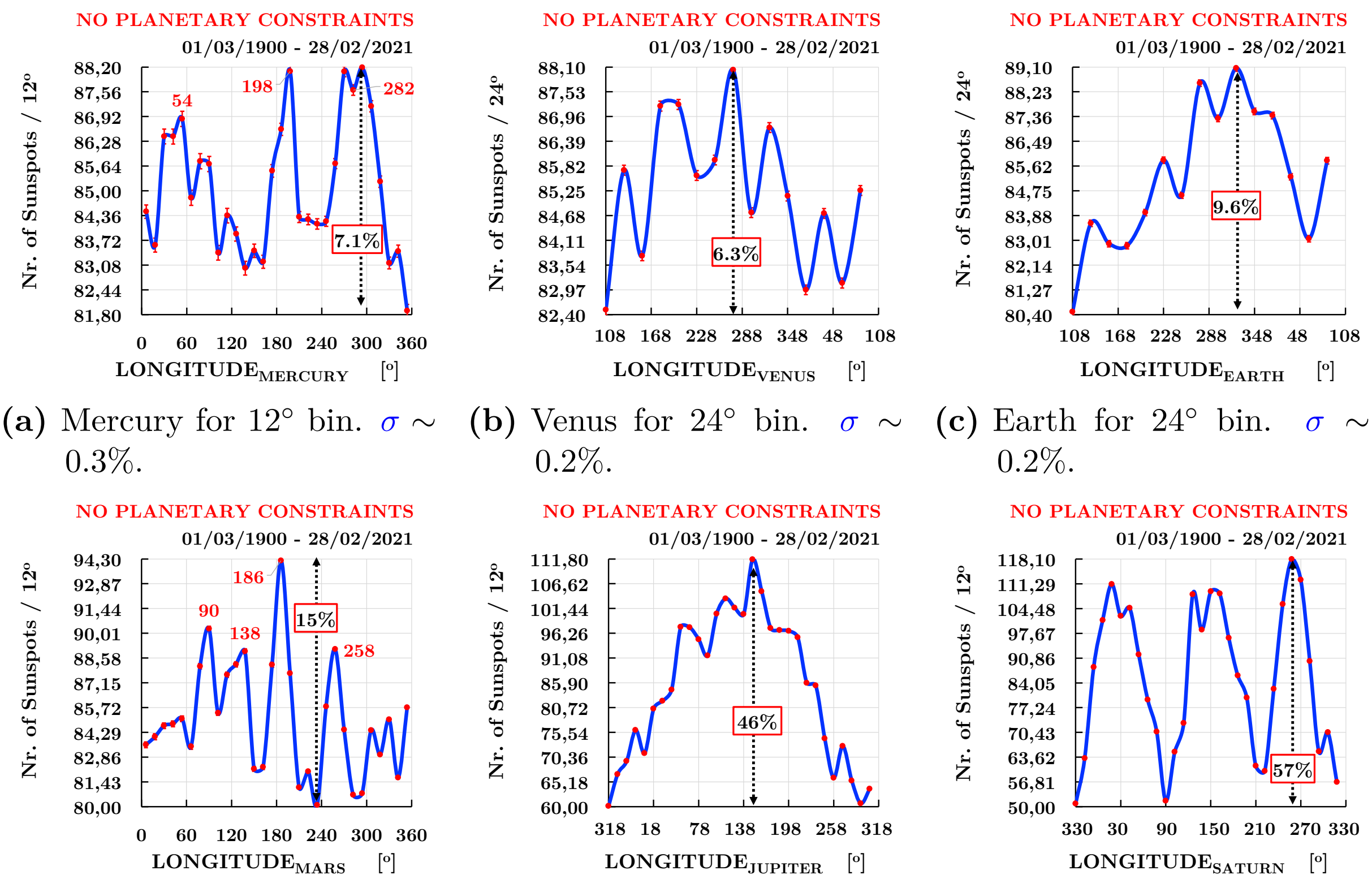

**(a)** Mercury for 12° bin. $\sigma \sim 0.3\%$.

**(b)** Venus for 24° bin. $\sigma \sim 0.2\%$.

**(c)** Earth for 24° bin. $\sigma \sim 0.2\%$.

**(d)** Mars for 12° bin. $\sigma \sim 0.3\%$.

**(e)** Jupiter for 12° bin. $\sigma \sim 0.3\%$.

**(f)** Saturn for 12° bin. $\sigma \sim 0.3\%$.

**Figure 8.6:** Planetary heliocentric longitude distributions of the daily sunspot Nr. for the period 01/03/1900 - 28/02/2021. The relative mean statistical errors per point are also given.

More specifically, for the individual plots of Fig. 8.6, the observed amplitude between the maximum and minimum points in the distributions of Mercury, Venus, Earth, Mars, Jupiter and Saturn is 7.1%, 6.3%, 9.6%, 15%, 46.2% and 56.9% respectively. In Mercury (Fig. 8.6a) we clearly see two narrow peaks around $\sim 198°$ and $\sim 282°$ with a FWHM derived by a Gaussian fit of 25.5° and 68.5° respectively. As an example the significance of the peak around 282° compared to the mean value of 85.1 is $\sim 21\sigma$. Similarly, a significant wide peak is observed in the case of Jupiter in Fig. 8.6e around $130.5° \pm 2.9°$ with a FWHM of $254° \pm 42.9°$. Finally, in Saturn three distinctive peaks are observed around $26.9° \pm 2.5°$, $153.1° \pm 2.6°$ and $261° \pm 1.9°$ with a corresponding FWHM of about $65.5° \pm 9.2°$, $70.4° \pm 9°$ and $47.1° \pm 6.7°$ as derived from a multiple Gaussian fit function. Finally, it is worth noticing that the wide peak of Jupiter in Fig. 8.6e can not cause the observed narrow peaks in Mercury, neither the three peaks in Saturn, as shown by a simulation in Appendix Sect. A.2.1.1.

#### 8.3.1.2 Combining planets

The next step is to search for a combined effect of more than one planets in order to search for a possible enhancement of the already observed highly significant effects. This is done by constraining one ore more planets to propagate in a specific orbital arc as explained in Sect. 4.5.1.1.

So, first in Fig. 8.7 Mercury's position is chosen to be observed while Venus is propagating around $210° \pm 50°$ in the first case and around $30° \pm 50°$ in the second case. For both cases the amplitude is bigger than the 7.1% observed when no constraints where applied in Fig. 8.6a, with the amplitude for Fig. 8.7a being 17.1% while for Fig. 8.7b 12.4%. The difference between the two 180° opposite orbital arcs which is shown in Fig. 8.7c and 8.7d points to an overall enhanced phenomenon when Venus is on the region 160° to 260°. The number of sunspots and days in the two cases is 1061234 and 12 268 d and 1029960 and 12 264 d respectively.

As next, in Fig. 8.8 more planets are constrained while the reference frame remains on Mercury. For the case of Mars being around $320° \pm 50°$, shown in Fig. 8.8a, we have a maximum - minimum difference of 20.5% while the number of corresponding days and sunspots is 10 429 d and 867110 accordingly. Again, the peak around 282° appears even better resolved than the case without any constraints in Fig. 8.6a. Then, in Fig. 8.8b, when the Moon is around $70° \pm 50°$ we have in total 1043978 sunspots for a total of 12 390 d fulfilling these conditions. The overall observed amplitude in this case is 49.6%. Like in the case of Saturn, in Fig. 8.6f, three clear peaks are observed around $40.8° \pm 1.4°$, $188.4° \pm 1°$ and $280° \pm 1°$ with a corresponding FWHM of about $82.7° \pm 7°$, $52.8° \pm 4°$ and $43° \pm 3.7°$ accordingly. It is noted that the ratio of the revolution periods of Mercury and Moon is about 2.98 which could explain the appearance of these three peaks. Therefore, we see that in all cases in Fig. 8.7 and 8.8, the initial 7.13% amplitude from Fig. 8.6a is greatly enlarged with the inclusion of positional constraints in more planets.

As next, we move to the reference frame of Venus while constraining Earth and/or Mars

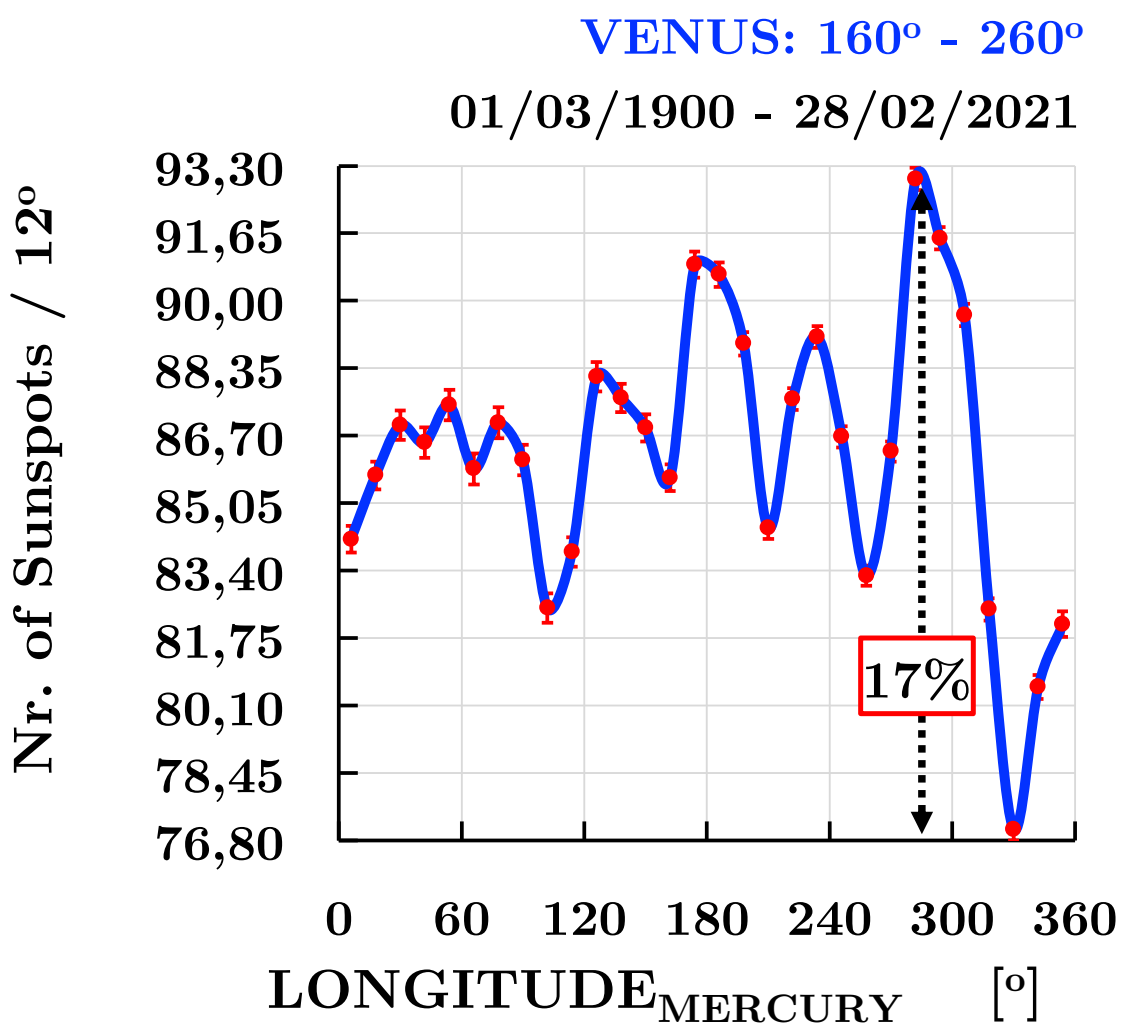

(a) Mercury when Venus is between 160° to 260°. $\sigma \sim 0.5\%$.

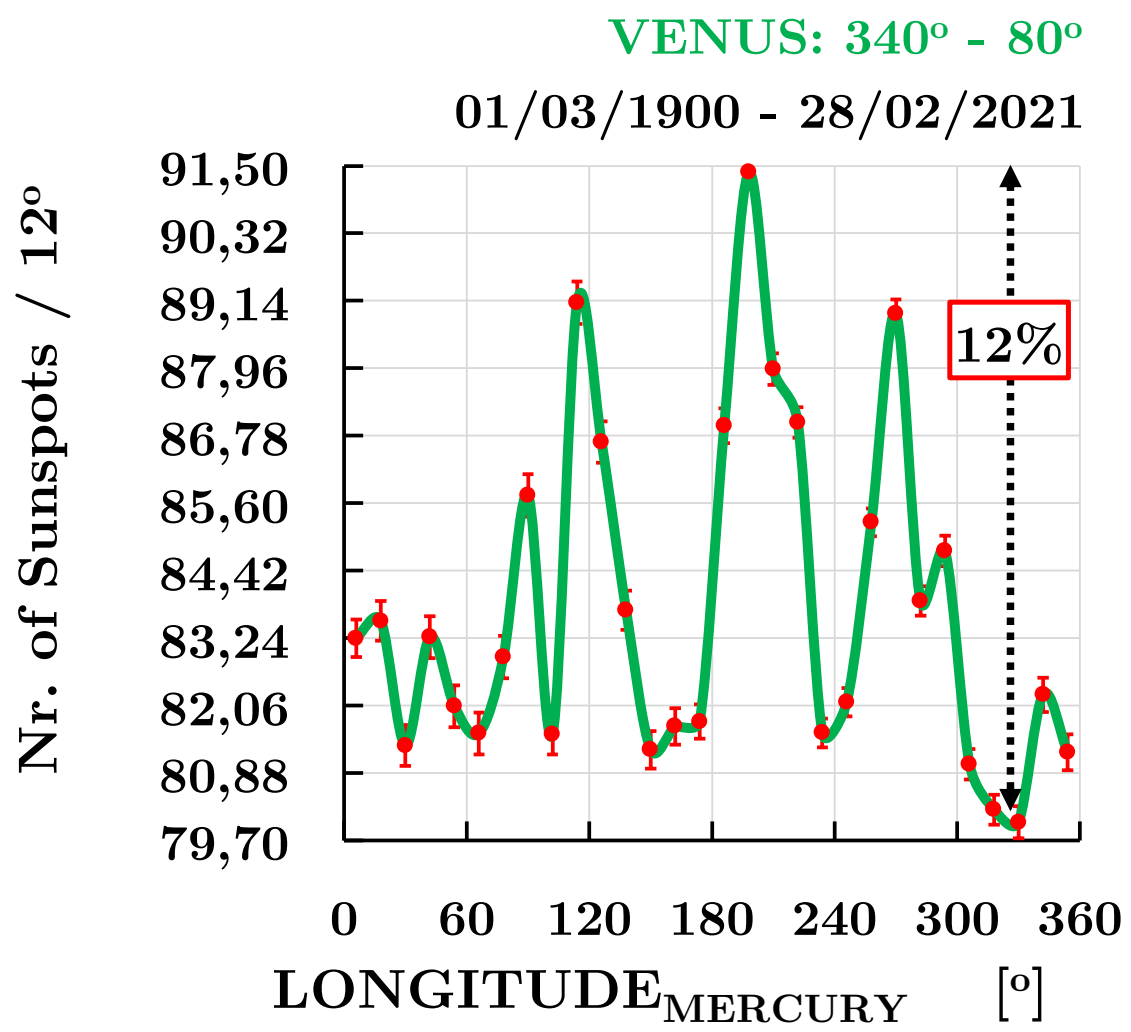

(b) Mercury when Venus is between 340° to 80°. $\sigma \sim 0.5\%$.

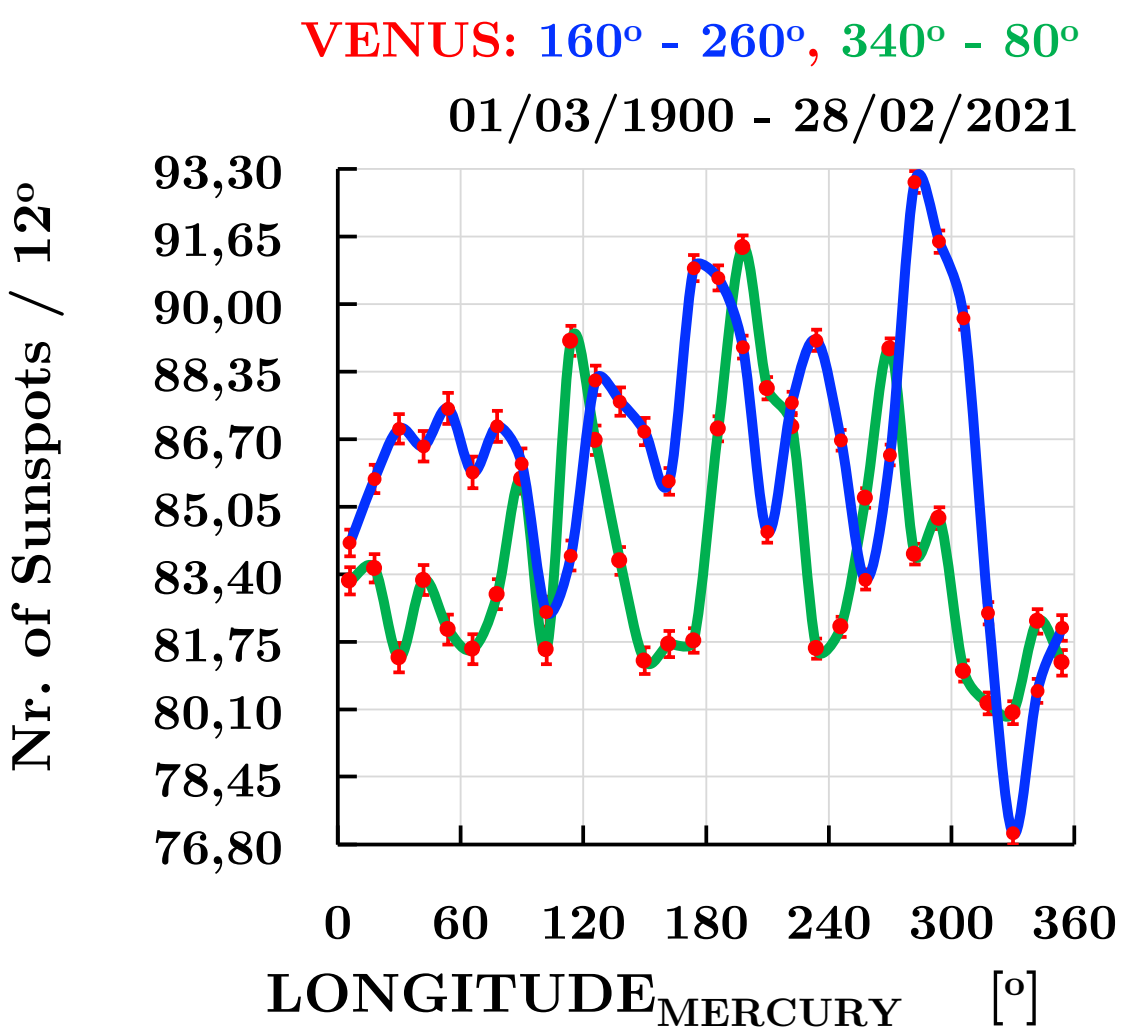

(c) Comparison between Fig. 8.7a and Fig. 8.7b.

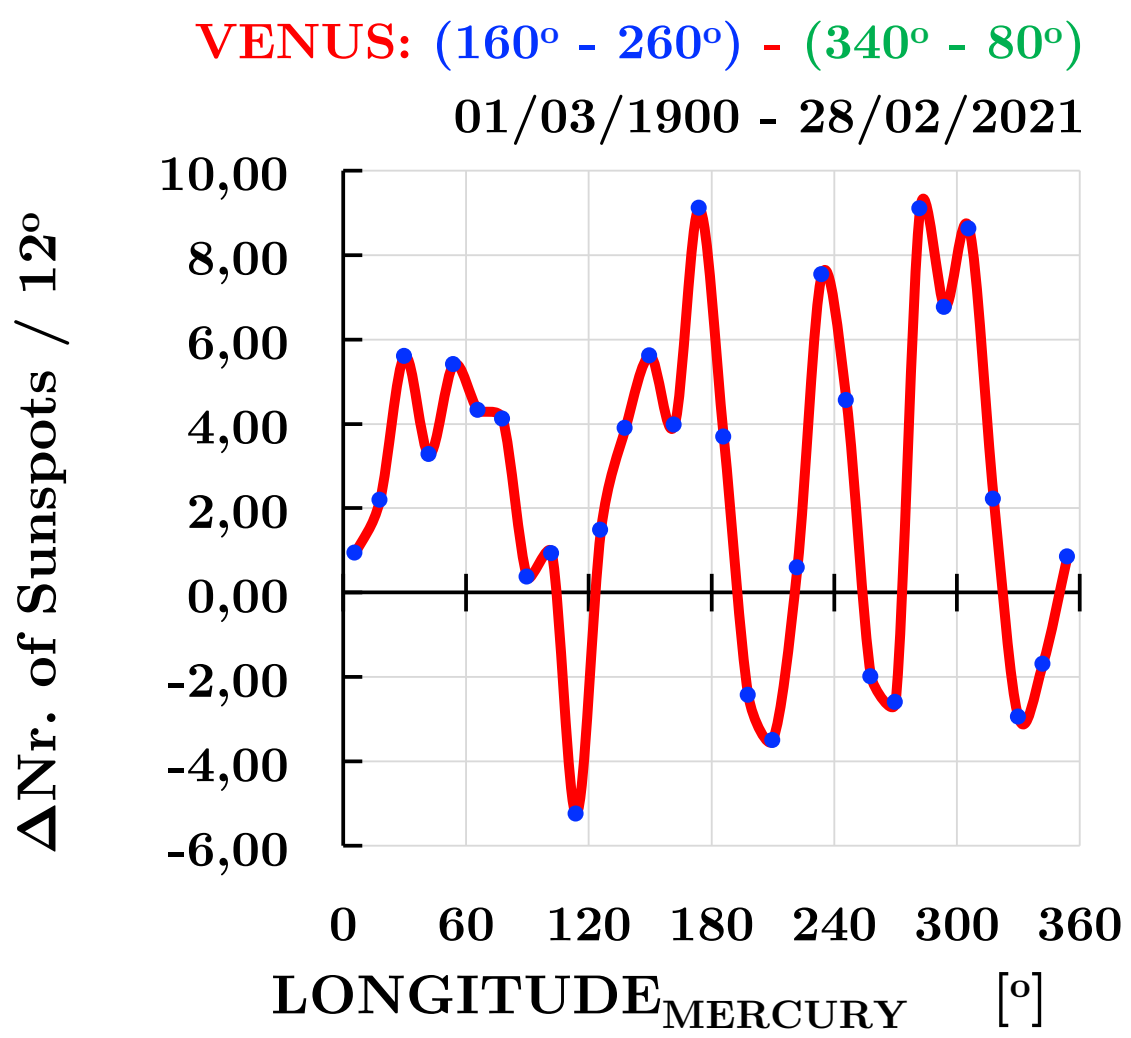

(d) Fig. 8.7a minus Fig. 8.7b.

**Figure 8.7:** Distribution of sunspot Nr. for the reference frame of Mercury while Venus propagates between two 180° opposite orbital arcs with 100° width, for 12° bin. The relative mean statistical errors per point are also given.

to propagate around $30° \pm 50°$. These positions were selected based on an optimum analysis search, since they were the ones with the biggest observed effects for the respective bin size (12°), selected period (01/03/1900 - 28/02/2021), width of the constrained region (100°) and step of search (10°). For Fig. 8.9a where Earth's orbital position is constrained between 340° to 80°, we have a maximum - minimum difference of 28.2% with a total of 1029057 sunspots in 12 168 d. We basically observe three peaks around $34.6° \pm 49.8°$, $154.4° \pm 123.7°$ and $268.7° \pm 48.3°$ where the FWHM is derived from a Gaussian fit in each individual peak. Then, for Fig. 8.9b where Mars is constrained between 340° to 80° the amplitude is 32.1% with 948280 sunspots in 11 189 d. In this case if we assume we have three peaks, then these peaks

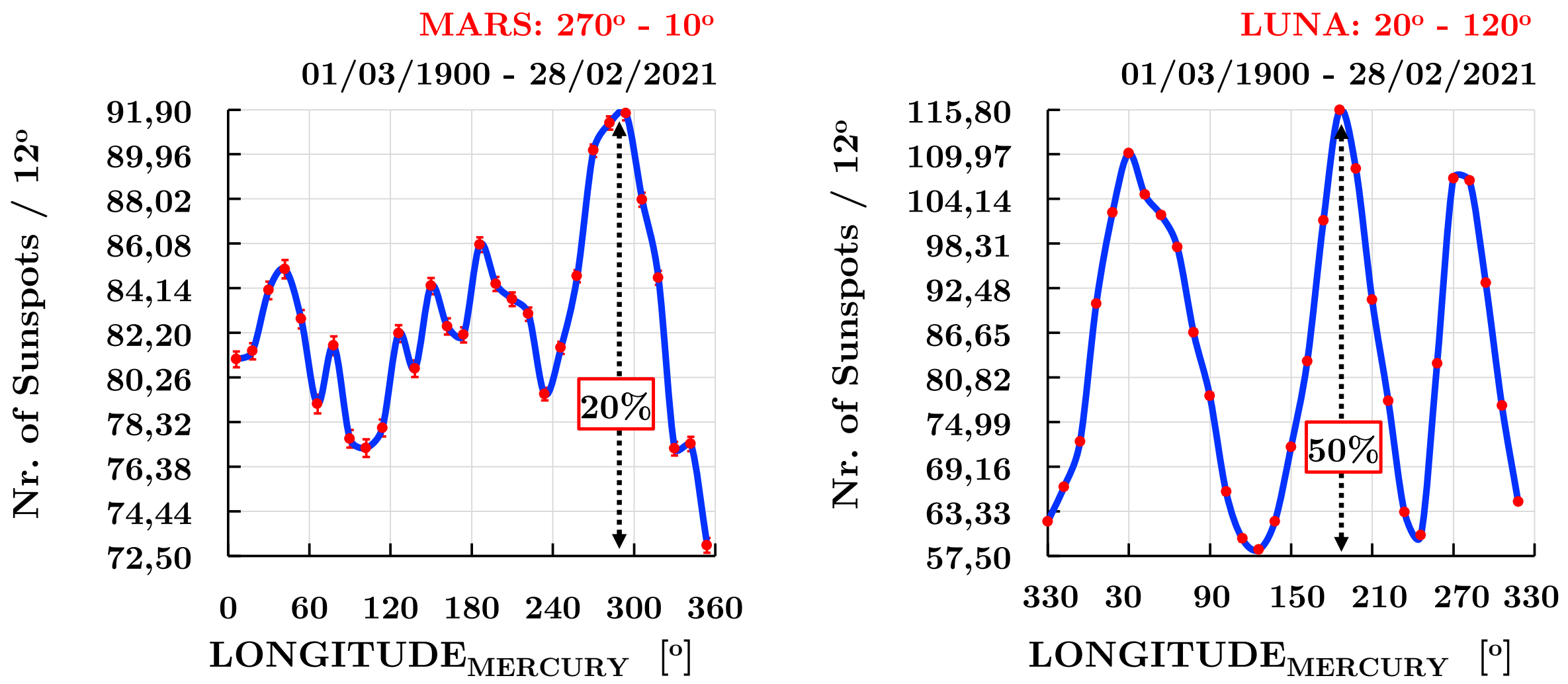

**(a)** Mercury when Mars is between 270° to 10°. $\sigma \sim 0.6\%$.

**(b)** Mercury when Moon is between 20° to 120°. $\sigma \sim 0.5\%$.

**Figure 8.8:** Distribution of sunspot Nr. for the reference frame of Mercury while Mars or Moon propagate in two 100°-wide regions, for 12° bin. The relative mean statistical errors per point are also given.

are located around $64.8° \pm 104.3°$, $181.1° \pm 67.4°$ and $297.2° \pm 154.5°$. Finally, in Fig. 8.9c when we combine both Venus and Earth to propagate between 340° to 80°, the total amplitude is 85.3% with 251233 sunspots in 3013 d. A wide peak is observed in this case around 213.4° with a FWHM = 302.3°.

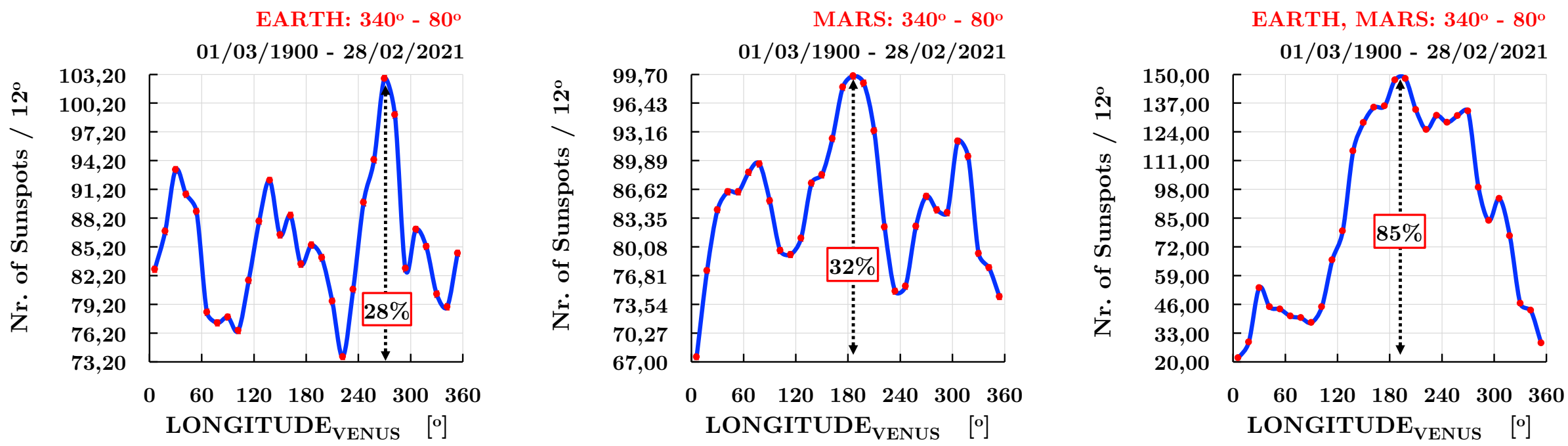

**(a)** Venus when Earth is between 340° to 80°. $\sigma \sim 0.5\%$.

**(b)** Venus when Mars is between 340° to 80°. $\sigma \sim 0.6\%$.

**(c)** Venus when Earth and Mars are both between 340° to 80°. $\sigma \sim 1.1\%$.

**Figure 8.9:** Distribution of sunspot Nr. for the reference frame of Venus while Earth or Mars are constrained in two 100°-wide regions, for 12° bin. The relative mean statistical errors per point are also given.

In Fig. 8.10 Mars' orbit is chosen to be observed, while Mercury's heliocentric longitude is constrained to be around $90° \pm 50°$. There are eight peaks observed with the difference between the maximum and minimum point being 56.8%. For this constraint, 8352 d and 710479 sunspots are selected. The number of peaks in this case could be associated with the ratio of the orbital periods of Mars and Mercury being 7.8 .The eight Gaussian fitted peaks observed in Fig. 8.10b have an average $FWHM \sim 26.3°$.

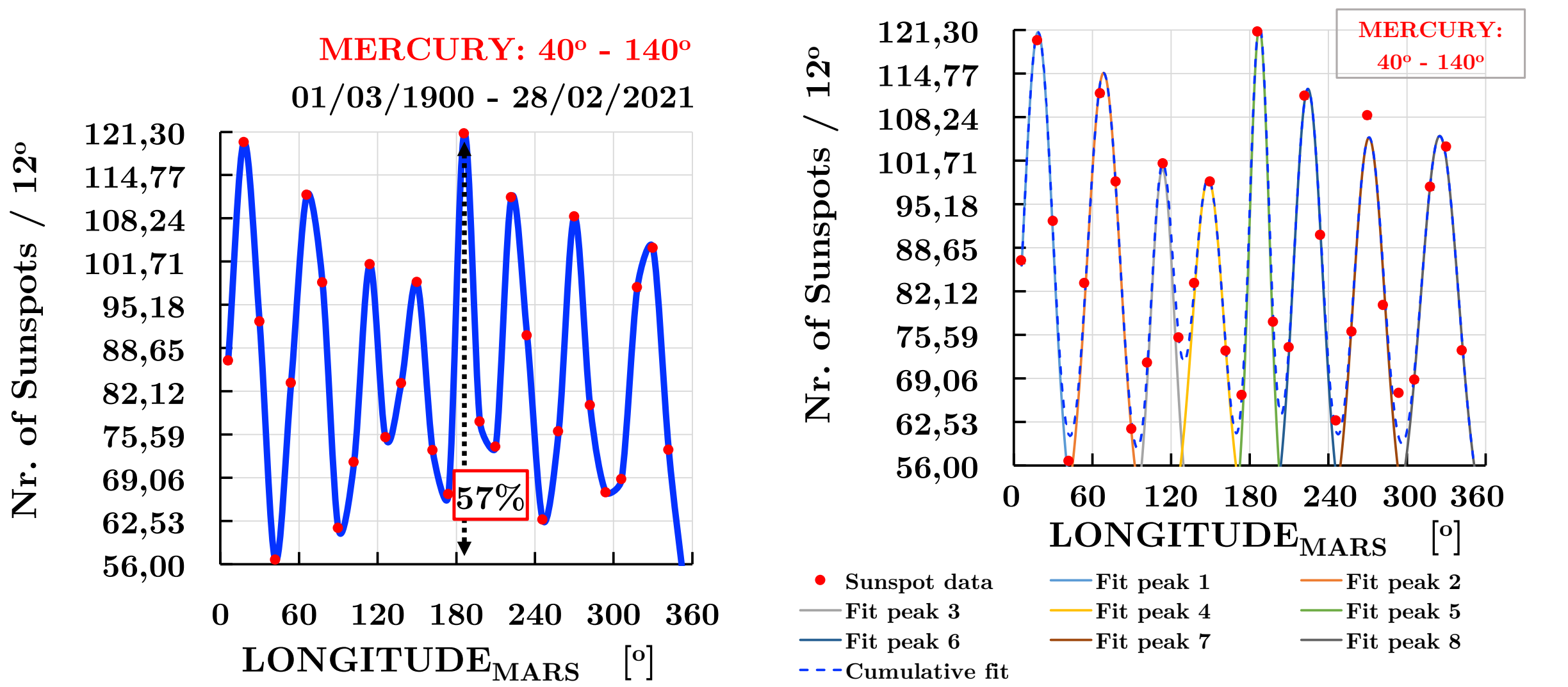

**(a)** Sunspot data. $\sigma \sim 0.6\%$.

**(b)** Sunspot and fitted data.

**Figure 8.10:** Distribution of sunspot Nr. as a function of Mars' heliocentric longitude when Mercury is allowed to propagate between 40° to 140° for 12° bin size. The relative mean statistical errors per point are also given.

Finally, in Fig. 8.11 we see the sunspot number distribution for the phase of Moon when no constrain is applied and when Mercury is constrained to be around $90° \pm 50°$. In the first case, in Fig. 8.11a, the observed amplitude is 6.5% whereas in the second case, in Fig. 8.11b, the amplitude is 35.7%. In the first case we have in total all 3767690 sunspots in 44 195 d whereas in the second case we have 710479 sunspots in 8352 d. The wide peak observed when constraining Mercury's position is around 314.9° and has a FWHM of 188.8°. The considerable difference between the two cases is better observed in Fig. 8.11c where the same scale has been selected on the y-axis. It is noted that for the 180° opposite case of Moon's phase while Mercury is around 220° to 320°, the overall amplitude is smaller at about 16% with 17 014 d and 1466301 sunspots corresponding in this constraint. This plot is not shown here due to space considerations.

### 8.3.2 Fourier analysis

An additional search strategy for underlining periodicities is through the conventional Fourier analysis. The Lomb periodogram produced here was based on the daily data from the period 01/03/1900 - 28/02/2021. Then the various derived periods are compared with the revolution periods of the solar system objects as well as with the planetary synods.

The 6$^{th}$ biggest in amplitude peak that is derived, has a period of $\sim 27.34$ d and an amplitude of 32.4 dB (see Fig. 8.12a). From a Gaussian fit, its FWHM is calculated at 0.015 d.

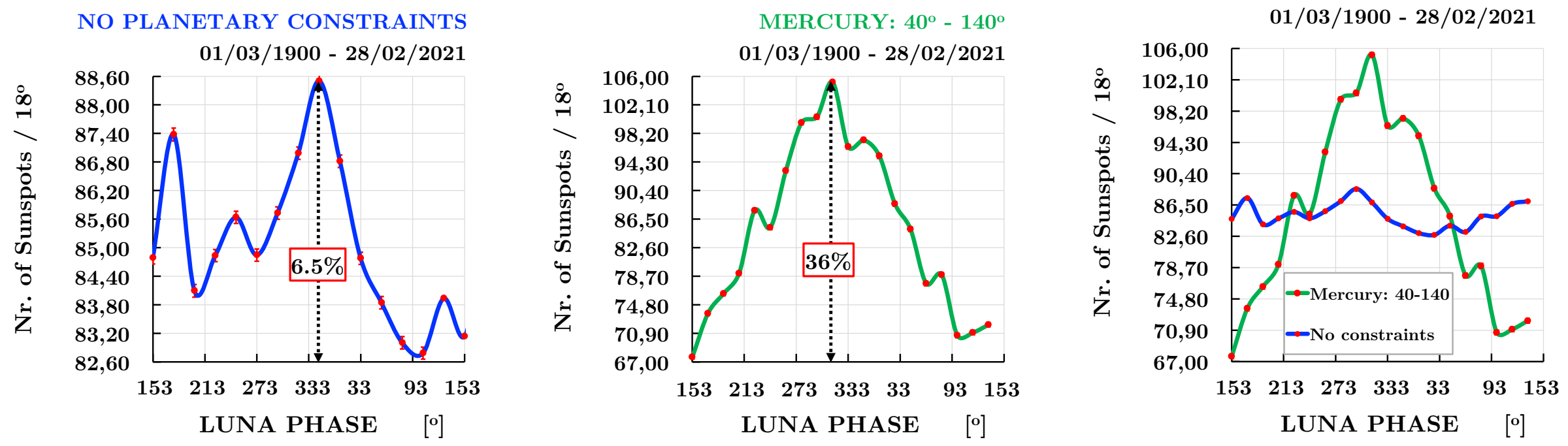

**(a)** Moon's phase without any planetary constraint. $\sigma \sim 0.2\%$.

**(b)** Moon's phase when Mercury propagates between 40° to 140°. $\sigma \sim 0.5\%$.

**(c)** Comparison between Fig. 8.11a and Fig. 8.11b.

**Figure 8.11:** Distribution of sunspot Nr. as a function of Moon's phase with and without additional planetary longitudinal constraints, for 18° bin. The relative mean statistical errors per point are also given.

This period is very close with Moon's sidereal month of 27.32 d fixed to remote stars, indicating an exo-solar impact on the number of sunspots. Furthermore, the 9$^{th}$ biggest peak is located around 399.3 d has an amplitude of $\sim$ 28 dB and a FWHM of 3.2 d (see Fig. 8.12b). The error for this period, is 1.3 d. This means that it overlaps with the synod between Earth and Jupiter of 399 d. There is also an additional peak around 335.8 d with 14.7 dB and FWHM = 3.2 d giving an error of $\pm$1.4 d which is very close to the 334 d synod of Venus - Mars (see Fig. 8.12c). This peak lists as 34$^{th}$ in the list of biggest peaks in amplitude. One more observed peak is around 225.1 d $\pm$ 0.47 d with an amplitude of $\sim$ 6.3 dB, which can be identified as the Venus revolution period of 224.7 d. Then, there is wide peak around 693.1 d $\pm$ 5.3 d with an amplitude of $\sim$ 12.3 dB (making it the 46$^{th}$ biggest peak) for which the synods of Mars - Pluto (692 d) and Mars - Neptune (695 d) are within the calculated error. Finally, and as expected, the biggest peak is around 3859.1 d $\sim$ 10.6 y with a power of 10 426.2 dB and a FWHM of 326.9 d.

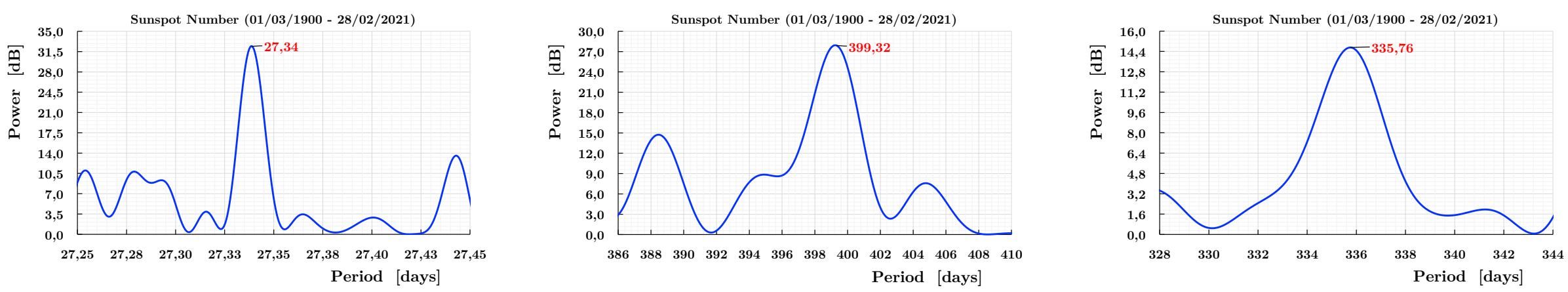

**(a)** Around 27.34 d.

**(b)** Around $\sim$ 399.3 d.

**(c)** Around $\sim$ 336 d.

**Figure 8.12:** Fourier periodogram of the sunspot data for the period 01/03/1900 to 28/02/2021 zoomed around a few interesting periods.

Additionally, and for cross-checking reasons a Fourier analysis has also been performed for the number of sunspots but this time for the full available period 01/01/1818 to 28/02/2021. In this case, the same peaks as with the period 01/03/1900 to 28/02/2021 were observed but with slightly different amplitudes (see Fig. 8.13).

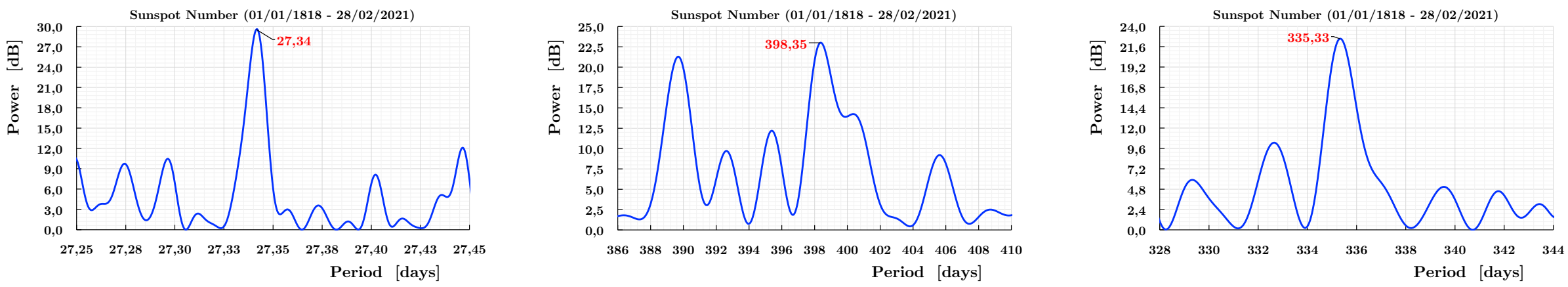

**(a)** Around 27.34 d. **(b)** Around 398.4 d. **(c)** Around 335.3 d.

**Figure 8.13:** Fourier periodogram of the sunspot data for the full available period 01/01/1818 - 28/02/2021 zoomed around a few interesting periods.

Lastly, for the full period 01/01/1818 to 28/02/2021, as seen in Fig. 8.14 around 11 y there are three peaks observed. Around 3679 d ∼ 10.1 y with an amplitude of 2939 dB, around 3965 d ∼ 10.9 y with an amplitude of 12 292 dB and around 4326 d ∼ 11.8 y with an amplitude of 2562 dB. Their corresponding FWHM are about 195 d, 196 d and 181 d respectively. It is noted that, these values are most probably not randomly very close [237] to the Jupiter - Saturn spring tidal period of 9.93 y, the Mercury - Venus orbital combination repeating every 11.1 y and the 11.07 y cycle for the alignment of Venus, Earth and Jupiter as well as Jupiter's revolution period of 11.87 y [242, 279].

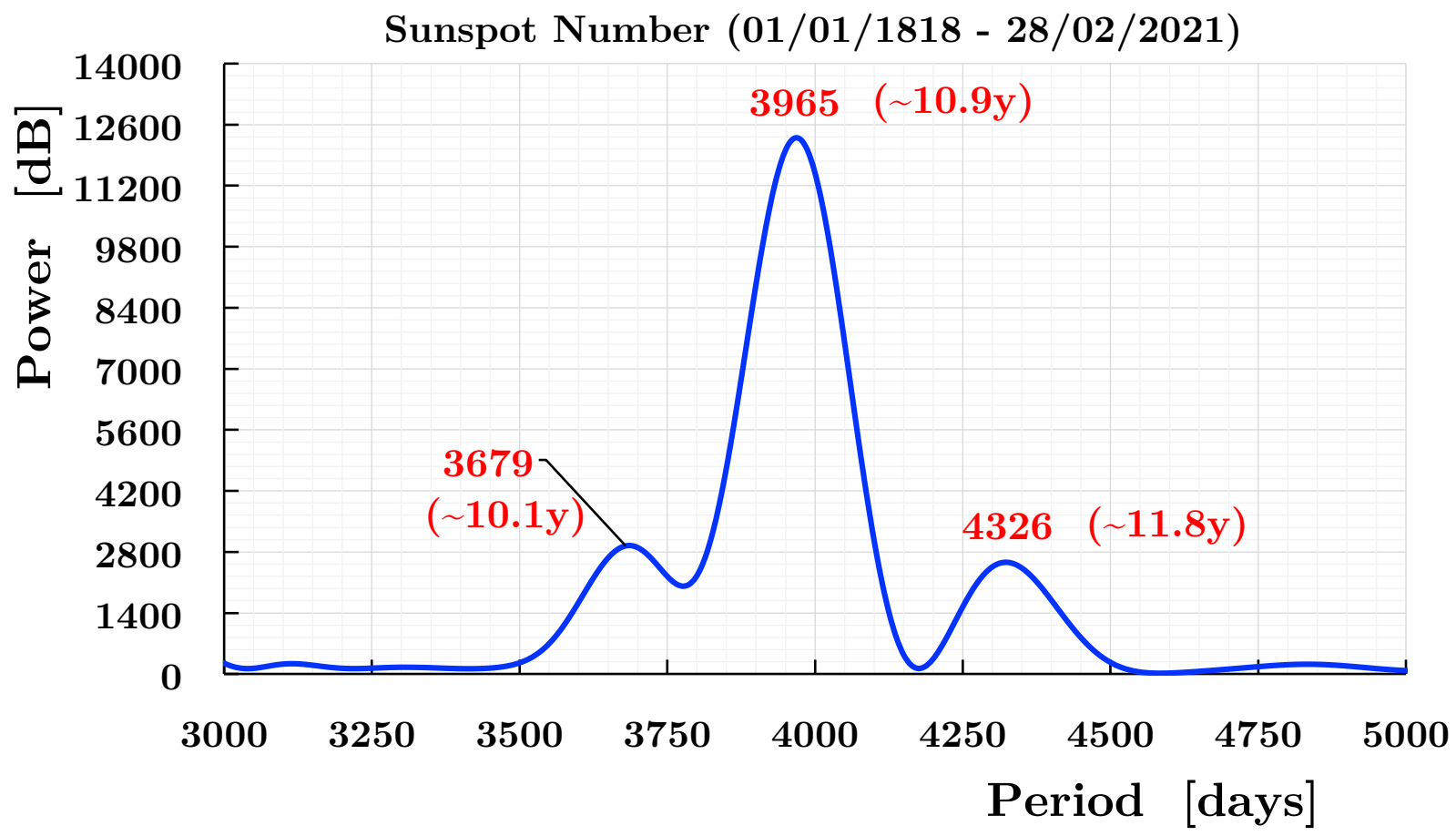

**Figure 8.14:** Fourier periodogram of the sunspot Nr. for the full available period 01/01/1818 - 28/02/2021 zoomed in around 11 y ∼ 4015 d.

## 8.4 Summary

The statistical analysis of the distribution of the number of sunspots as a function of the various planetary heliocentric longitude positions and configurations over a 121 y period, showed statistically significant peaks which are unexpected within conventional physics. These observations support the assumptions of this work, involving slow-moving massive invisible matter interacting "strongly" with the Sun after being gravitationally focused by the planets and eventually the Sun itself through free fall. As an example, the sunspot distributions on the reference frame of planets such as Mercury, Jupiter and Saturn exhibit a statistically significant peaking behaviour. These observations become even more enhanced when the combined effect of more planets is introduced.

In particular, based on the observed optimal heliocentric orbital arcs of the various planets an estimation of the direction(s) of the stream(s) could be envisaged, even though there could be multiple streams with a wide velocity spectrum being focused differently by the different planets. As an example, the orbital arc 40° to 140° for Mercury's position seems to coincide for both the reference frames of Mars and Moon's phase (see Fig. 8.10 and 8.11c). Similarly, both Earth and Mars seem to have a preferred longitudinal position around $30° \pm 50°$ when observed from the reference frame of Venus (see Fig. 8.9). Furthermore, the Fourier analysis is also supportive of the planetary periodicities exhibiting significant peaks around 27.34 d which is probably not coincidentally close with Moon's sidereal month, but also around 399.3 d, 335.8 d and 225.1 d coinciding with Earth - Jupiter synod, Venus - Mars synod and Venus revolution period respectively (see also Publication E.13).

Finally, the comparison of all the planetary distributions with the corresponding ones from EUV irradiation exhibits common characteristics between the two manifestations of solar activity which, however, in some degree, are already expected [280–282] (see Appendix Sect. B.4.1). In future, more refined analyses also for different time periods while using narrower longitudinal windows for all the planets could provide further hints towards the identification of the preferred direction(s) of the assumed streams of invisible massive matter.

9

# F10.7

## 9.1 Introduction

The various manifestations of solar activity are assumed to be related with the total amount of magnetic flux emerging through the photosphere to the chromosphere and corona as well as its temporal and spatial distributions. The 10.7 cm solar flux is a measurement of the integrated emission at a wavelength of 10.7 cm from all the sources that are present on the solar disc. It is conventionally assumed to be thermal in origin, and related to the total amount of plasma trapped in the magnetic field over the active regions of the Sun [283]. In Fig. 9.1 the flux density as a function of the wavelength $\lambda$ is shown on the radio frequencies where at wavelengths greater than 1 cm the spectral density is becoming greater than what is expected from a 6000 K blackbody radiator.

It is widely accepted that the solar radio flux at 10.7 cm (2800 MHz) is an excellent indicator proxy of solar activity [284, 285]. This is due to the fact that solar emissions at these wavelengths are very sensitive to the conditions in the upper chromosphere and the lower corona. F10.7 seem to correlate with the sunspot number [286–288] as well as with the UltraViolet (UV) [289, 290] and the visible solar irradiance. Furthermore, EUV emissions that are known to impact the ionosphere also correlate with the F10.7 index [291–295] even though some discrepancies are observed during certain short-term periods [296, 297].

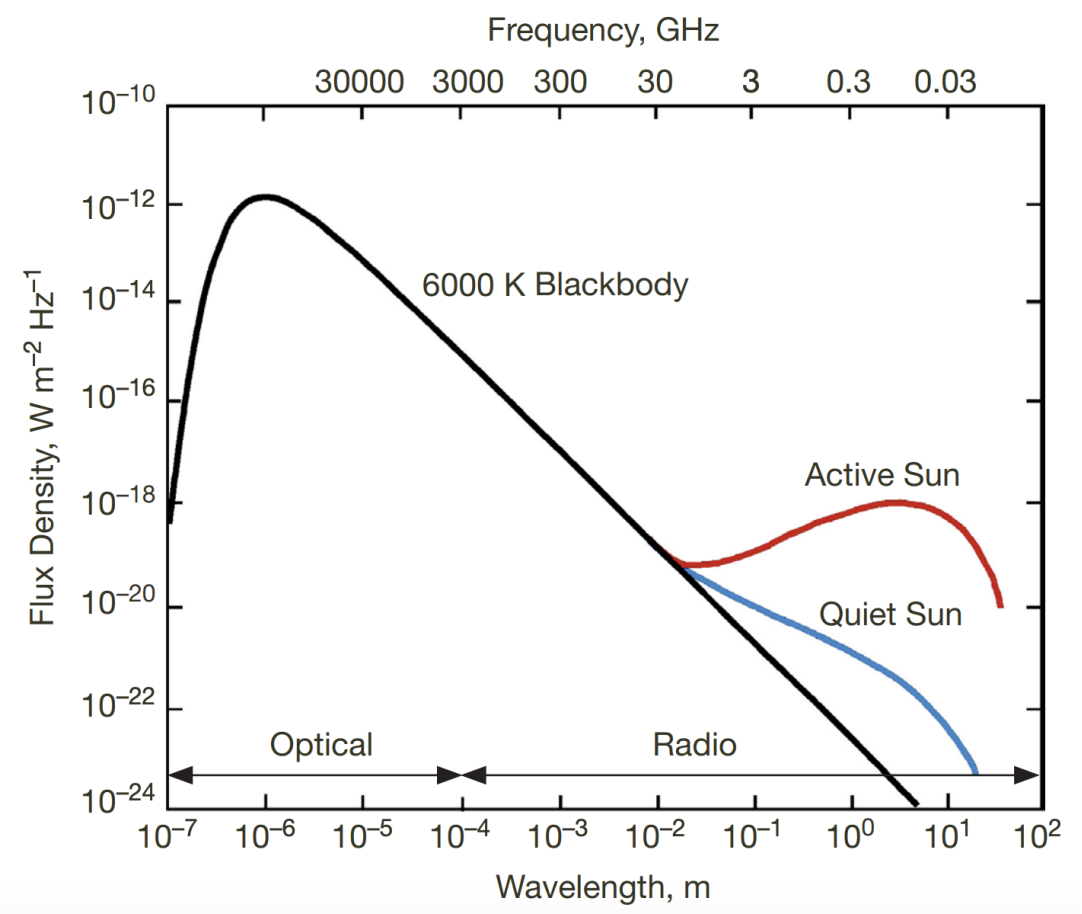

**(a)** Solar spectrum at optical and radio frequencies. For $\lambda > 1\,\text{cm}$ the active Sun emissions deviate, even more than the quiet Sun, from the expected blackbody radiation (68).

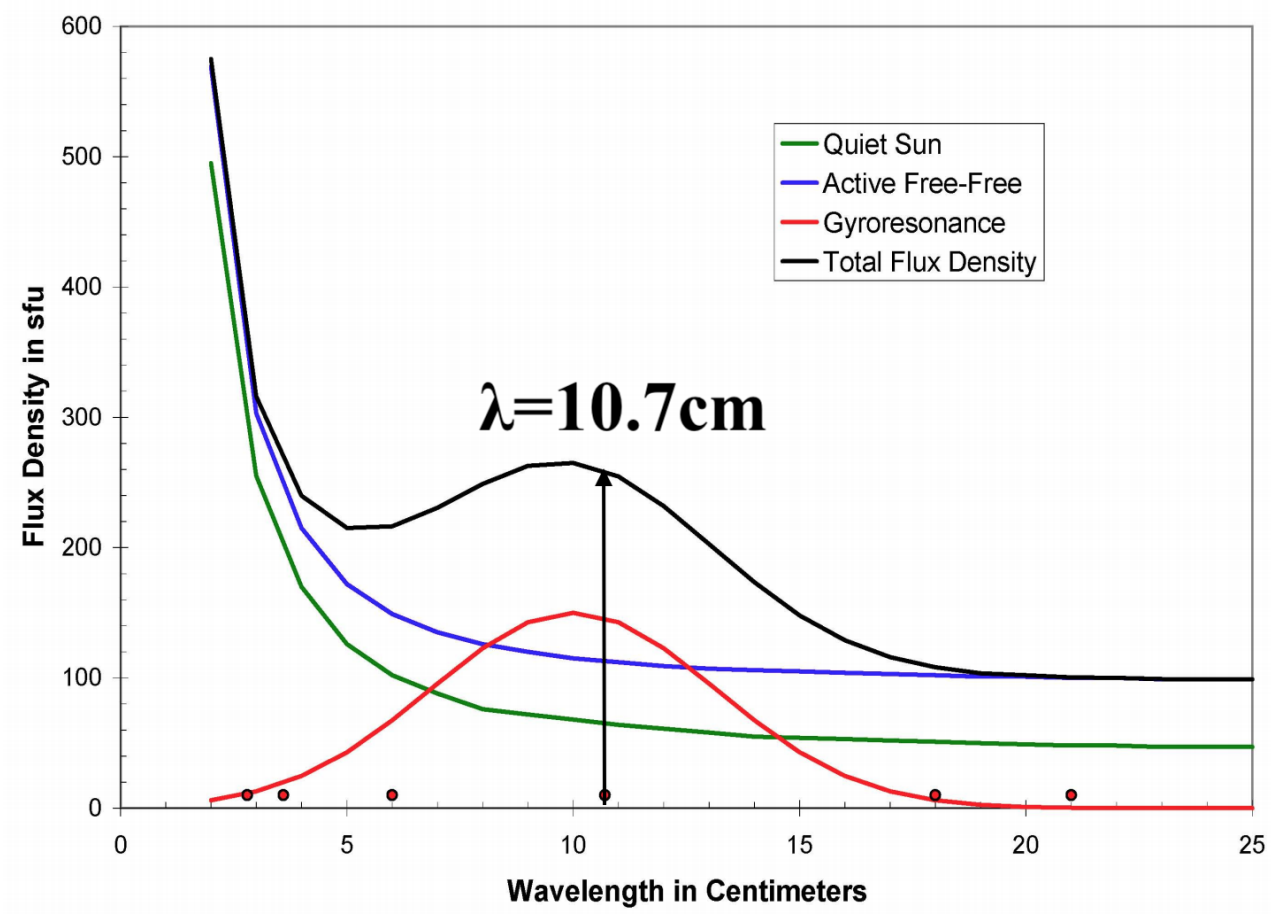

**(b)** Solar spectrum around the 10.7 cm wavelength. The quiet Sun emission is shown in green while the black curve represents the more active Sun (69).

**Figure 9.1:** Solar radiation spectrum.

Therefore, since a planetary relationship has already been found in the EUV as well as in the sunspot number, as seen in Chap. 7 and 8 respectively, it is natural to search also for a planetary correlation with the solar F10.7 index.

## 9.2 F10.7 data

### 9.2.1 Data origin

The provided daily F10.7 measurements are courtesy of the National Research Council Canada (NRC) in partnership with the Natural Resources Canada (NRCan) [283]. However, the acquired and used data in this work have been downloaded from the OMNIWeb interface [298] due to the direct availability of the full-time period.

The 10.7 cm solar flux, i.e. the solar flux density at 10.7 cm wavelength is measured by two radio telescopes located at the Dominion Radio Astrophysical Observatory. There are three different flux measurements made each day for the total emission from all sources present on the solar disc in a 100 MHz wide band centred around the frequency of 2800 mHz. Between March and October (summer) the measurements are made at 17:00, 20:00 and 23:00 UTC, whereas from November to February (winter) the corresponding times are 18:00, 20:00 and 22:00 UTC due to the location of the radio telescope and the high latitude. Each measurement lasts 1 h and contains the average of the solar emission within that hour [289, 299]. In the following analysis, the adjusted measurements made at 20:00 UTC are used, which correspond to local noon. The total acquired time range is 28/11/1963 - 03/03/2021.

The original measurements contain two different values. The "observed" ones and the "adjusted" ones. The observed values are the raw numbers measured by the solar radio telescope. However, these values are modulated by the level of solar activity but also by the varying distance between the Earth and Sun. Since this is a study of the Sun, the annual modulation due to the changing distance Earth-Sun has to be normalised. This is done with the adjusted values which are given for the average distance of 1 AU.

### 9.2.2 Data curation

The data contain 13 out of 20916 measurements with unspecified values (i.e. 0.06%). These dates, have been "corrected" by taking the average of four points before and four points after the problematic value.

The raw data along with the corrected values that are used for the analysis are shown in Fig. 9.2. Finally, the error associated with each measurement is 1 sfu or 1% depending on which is bigger in a given daily value. These errors have been assigned in each value so as the resulting longitudinal distributions to contain error bars. As a result, the statistical significance of a possible peak can easily be derived.

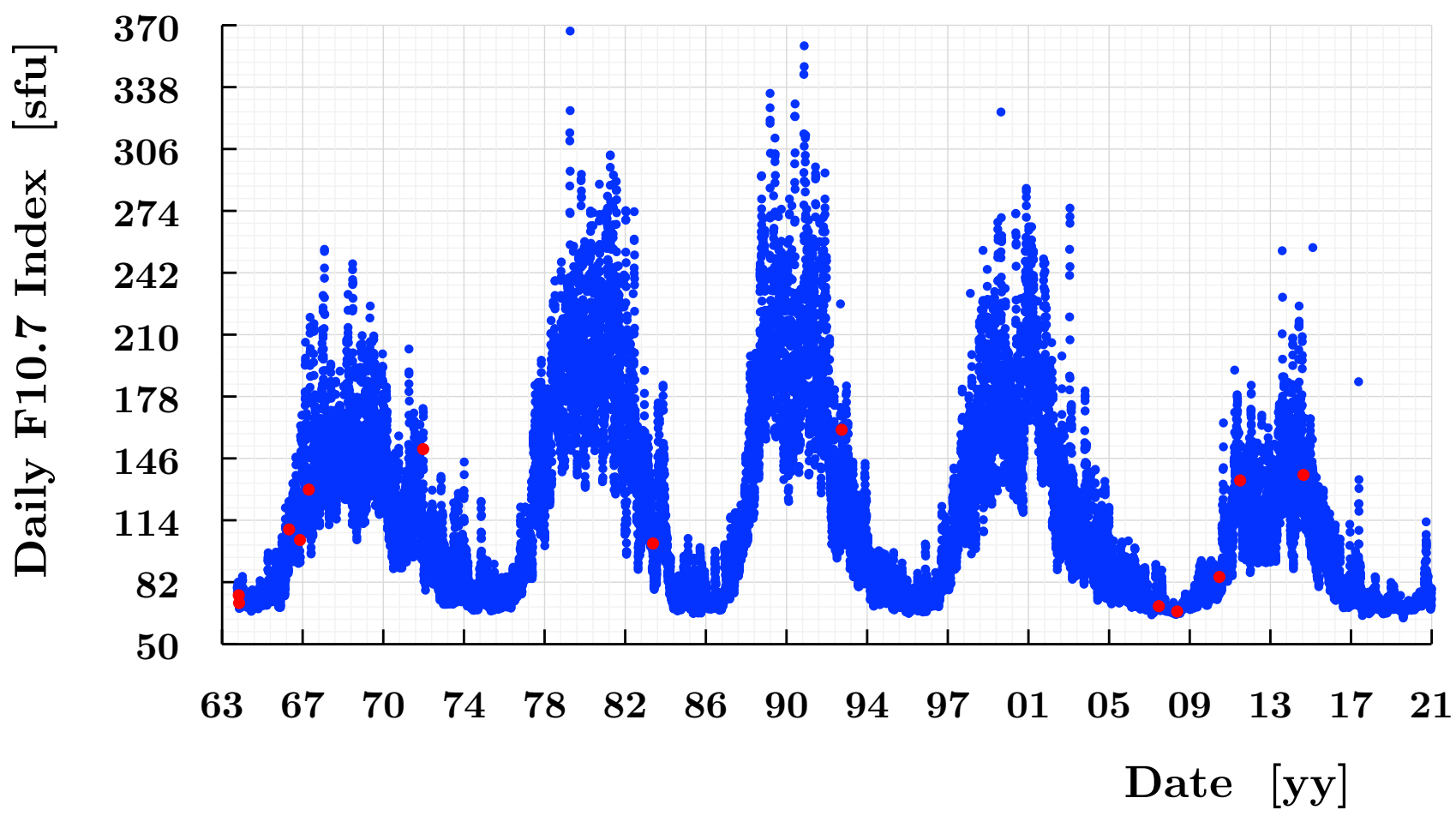

**Figure 9.2:** Temporal evolution of the daily F10.7 index for the period 28/11/1963 to 03/03/2021. The corrected values are also seen in red.

The total range of the data that are used spans from 28/11/1963 to 03/03/2021 which is in total 20 916 d. The minimum value is 63.4 sfu and the maximum 367 sfu with the average value being 117.5 sfu and the total sum 2 458 016.3 sfu.

## 9.3 Data analysis and results

### 9.3.1 Planetary longitudinal distributions

#### 9.3.1.1 Single planets

To search for planetary relationships, the daily data of F10.7 are projected on the planetary heliocentric longitudinal coordinates. Firstly, single planets are used without applying any constraint on the position of the rest of the planets. Eccentricity-related effects are already removed from all plots. In Fig. 9.3 the heliocentric longitude positions of the planets Mercury, Venus Earth, as well as Mars, Jupiter and Saturn are used for the distributions of the F10.7 data. The observed maximum to minimum differences are 4.9%, 5.0%, 5.9%, 7.1%, 54.2% and 51.7% for Mercury (Fig. 9.3a), Venus (Fig. 9.3b), Earth (Fig. 9.3c), Mars (Fig. 9.3d), Jupiter (Fig. 9.3e) and Saturn (Fig. 9.3f) respectively. In all plots in Fig. 9.3 the sum of the F10.7 solar radio flux for the selected time period of 20 916 d is 2 458 016.3 sfu. In addition, during these 20 916 d the planets starting from Mercury to Saturn have performed about 237.8, 93.1, 57.3, 30.4, 4.8 and 1.9 revolutions around the Sun accordingly. This means, especially for the inner planets which perform many revolutions on the selected time period, that any randomly occurring event would have eventually been factored out.

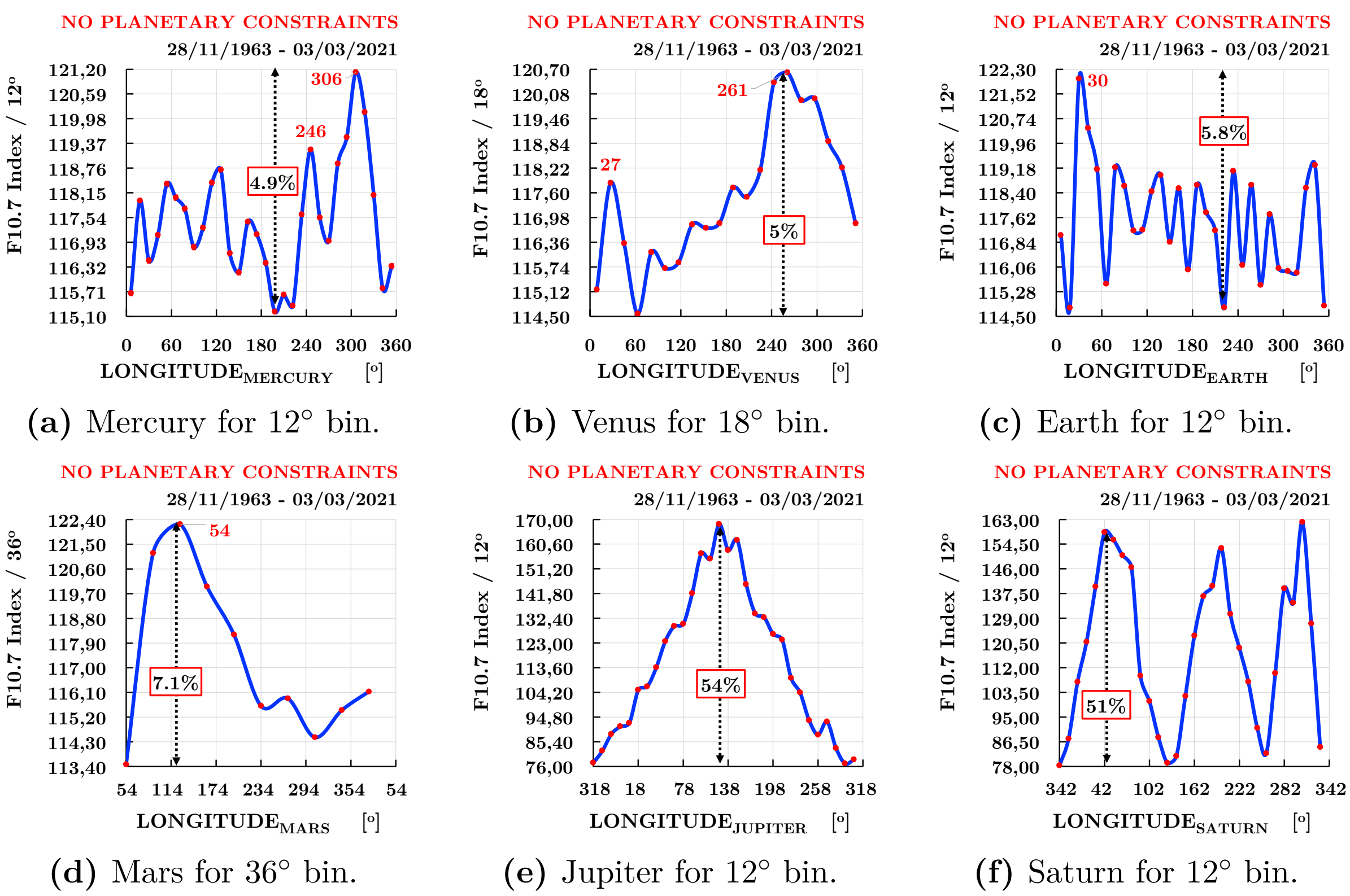

(a) Mercury for 12° bin. (b) Venus for 18° bin. (c) Earth for 12° bin.

(d) Mars for 36° bin. (e) Jupiter for 12° bin. (f) Saturn for 12° bin.

**Figure 9.3:** Planetary heliocentric longitude distributions of F10.7 index for the period 28/11/1963 - 03/03/2021.

All points in the plots contain error bars, however since they are very small they are not directly observable. We therefore see some significant peaks far above $10\sigma$ in all planetary distributions. As an example, in Mercury we have two peaks around 246° ($64\sigma$ compared to the point around 222°) and 306° ($77\sigma$ compared to the point around 342°), for Venus around 27° ($57\sigma$ compared to the point around 63°), and 261° ($53\sigma$ compared to the point around 207°), for Earth around 30° ($89\sigma$ compared to the point around 66°) and for Mars a wide peak around 54° ($164\sigma$ compared to the point around 234°). Furthermore, in Jupiter we observe a wide peak around 128.9° with a very big amplitude of 54%. From a Gaussian fit we find that the FWHM of this peak is about $162.8° \pm 8.5°$. Finally, in Saturn we see three clear peaks around $51.4° \pm 1.9°$, $193.1° \pm 2°$ and $299.1° \pm 1.8°$ with a corresponding FWHM of about $88.6° \pm 14.6°$, $81.2° \pm 10.8°$ and $52.9° \pm 7.7°$ accordingly, as derived from a multiple Gaussian fit function. This behaviour resembles the 11 y solar cycle. Interestingly, as it was found by a simulation, the considerable 54% amplitude of Jupiter cannot produce an additional peaking distribution to another planet, resembling the ones in Fig. 9.3, based on orbital kinematical effects (see Appendix Sect. A.2.1.1).

#### 9.3.1.2 Combining planets

As usual to search for an even bigger effect, a combination of more planets has to be included to the distribution of the F10.7 index. In Fig. 9.4 Mercury is kept as a reference frame and Venus is allowed to move between 200° to 320° and 20° to 140°. In the first case in Fig. 9.4a, the observed amplitude is 8% which is bigger than the 4.9% in Fig. 9.3a where no additional planetary constraint was made. Furthermore, the total number of days fulfilling this constraint is 7029 d with the integral of F10.7 being 839 230.9 sfu. On the other hand, in the case of Venus being between 20° to 140°, in Fig. 9.4b, we have a total maximum - minimum difference of 8.8% with the number of the applicable days being 6923 and the total F10.7 measurements being 804 460.78 sfu. As we see in Fig. 9.4c and 9.4d there is a significant difference between the two 180° opposite heliocentric longitude ranges with the peaks around 30° and 234° appearing only in the case of Venus being around 200° to 320°. This strengthens the case of a stream coming from the direction of the GC around 266°.

Then, in Fig. 9.5 Mercury is again selected, but this time with Mars propagating in 120°-wide opposite regions. For the case of Mars being around $270° \pm 60°$, in Fig. 9.5a, we have a 14.4% amplitude with a big peak around 300° which is in contrast with what we would expect from Fig. 9.3d. The number of selected days out of the total $\sim 58$ y in this case is 6499 with the total F10.7 being 751 005.1 sfu. For the opposite case when Mars is propagating around $90° \pm 60°$ we have an amplitude of 5.5% with 7409 d and 882 674.1 sfu in total. The conventionally unexpected big difference between the two cases is seen in Fig. 9.5c. This case verifies also the effect of Mars on the F10.7 with the biggest amplitude being again when being

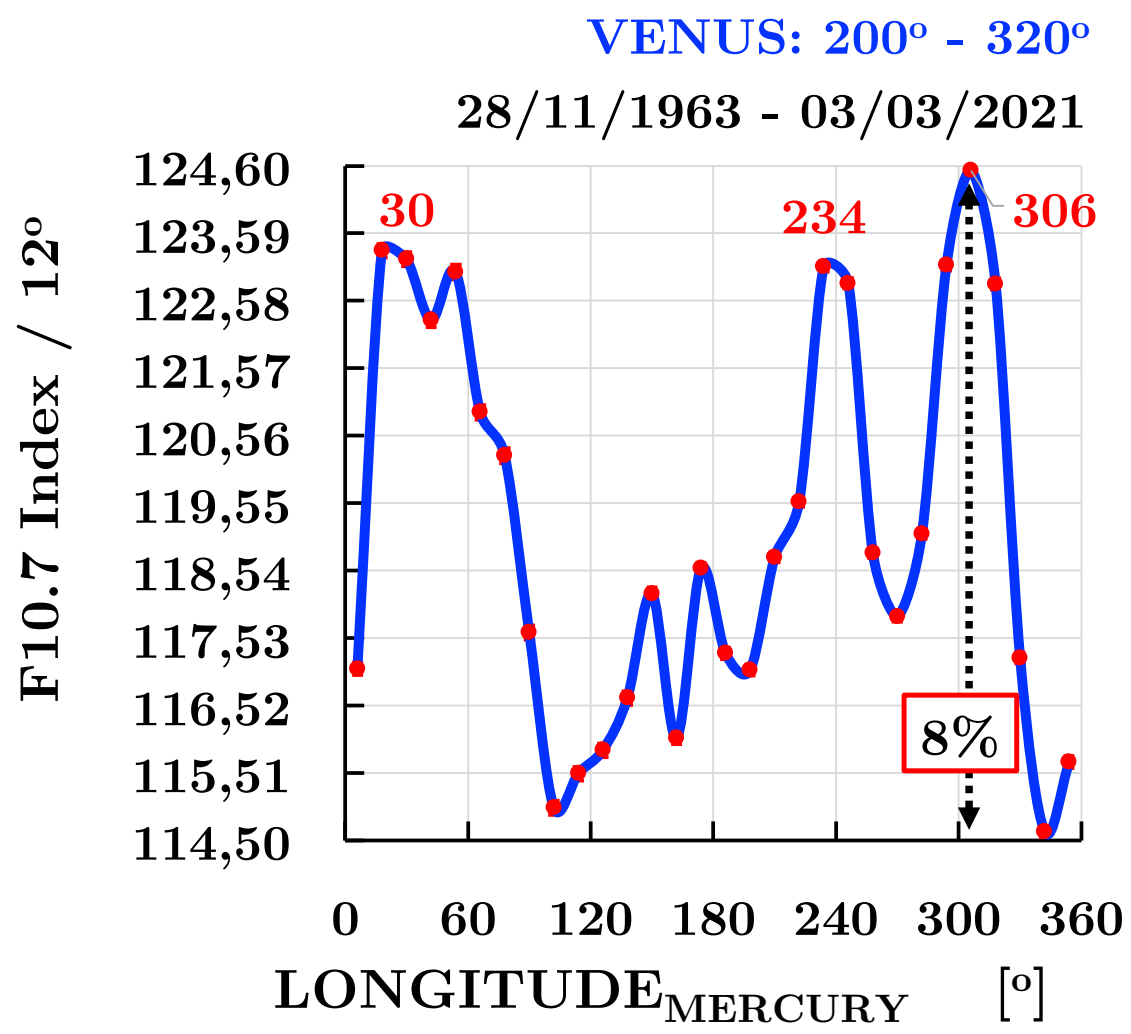

(a) Mercury when Venus is between 200° to 320°.

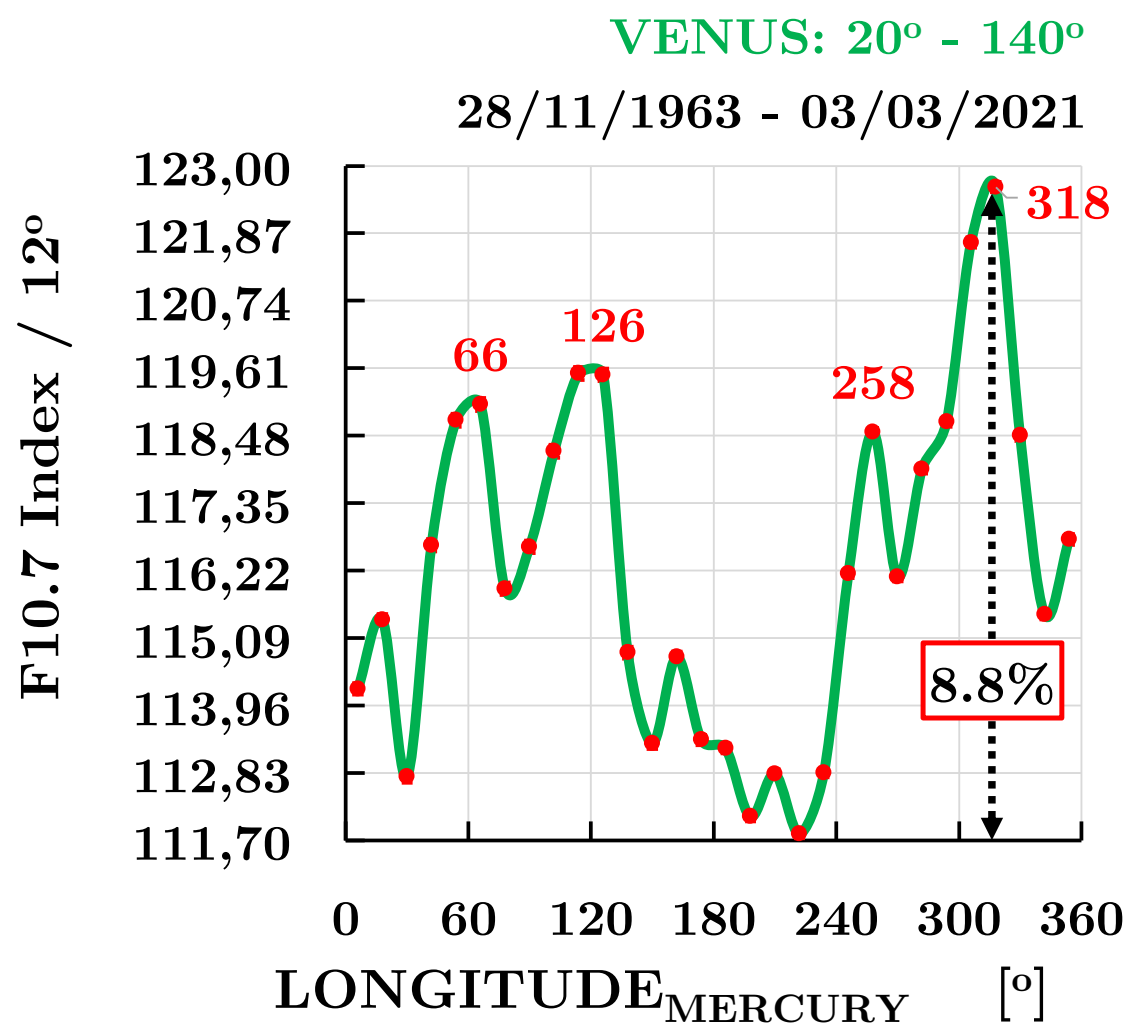

(b) Mercury when Venus is between 20° to 140°.

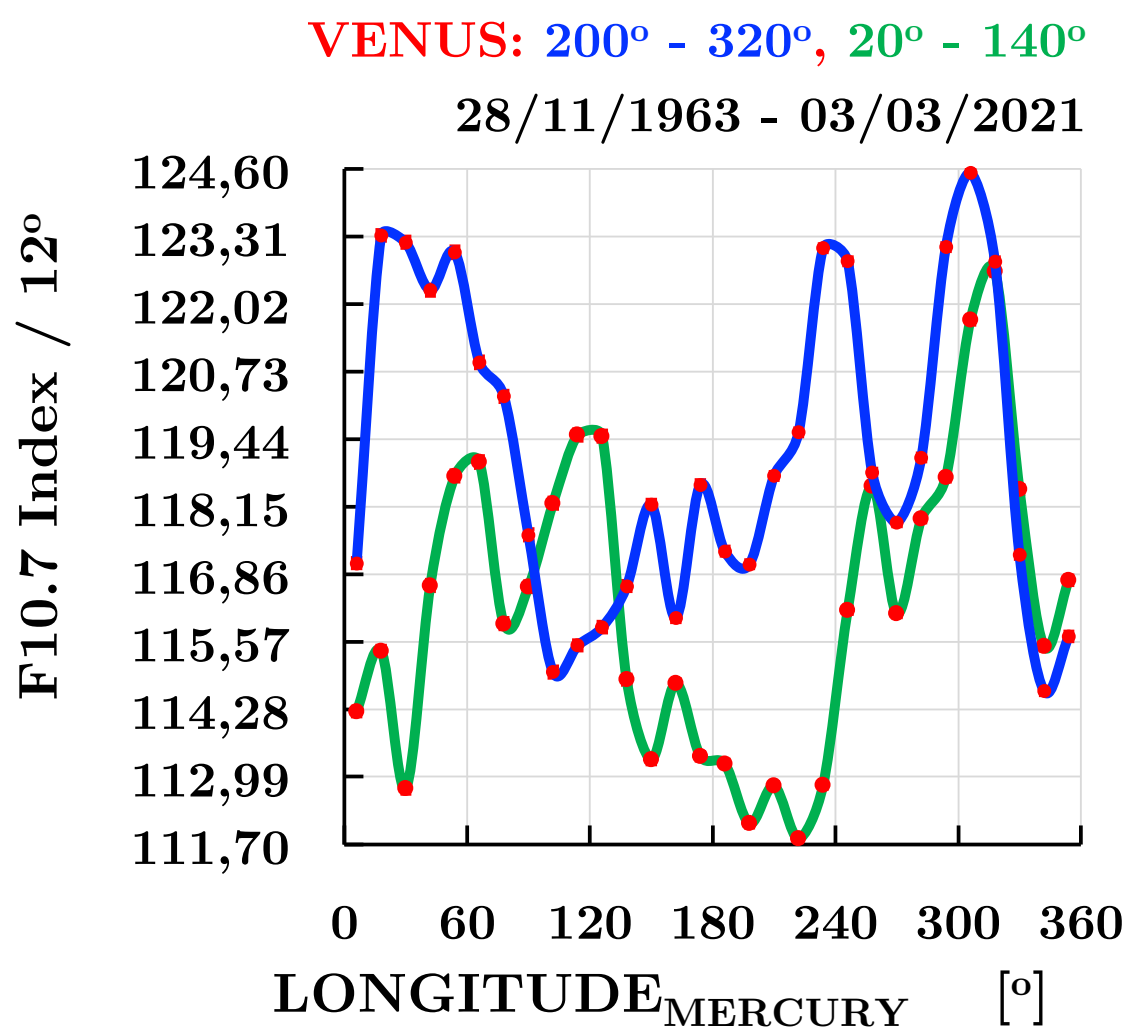

(c) Comparison between Fig. 9.4a and Fig. 9.4b.

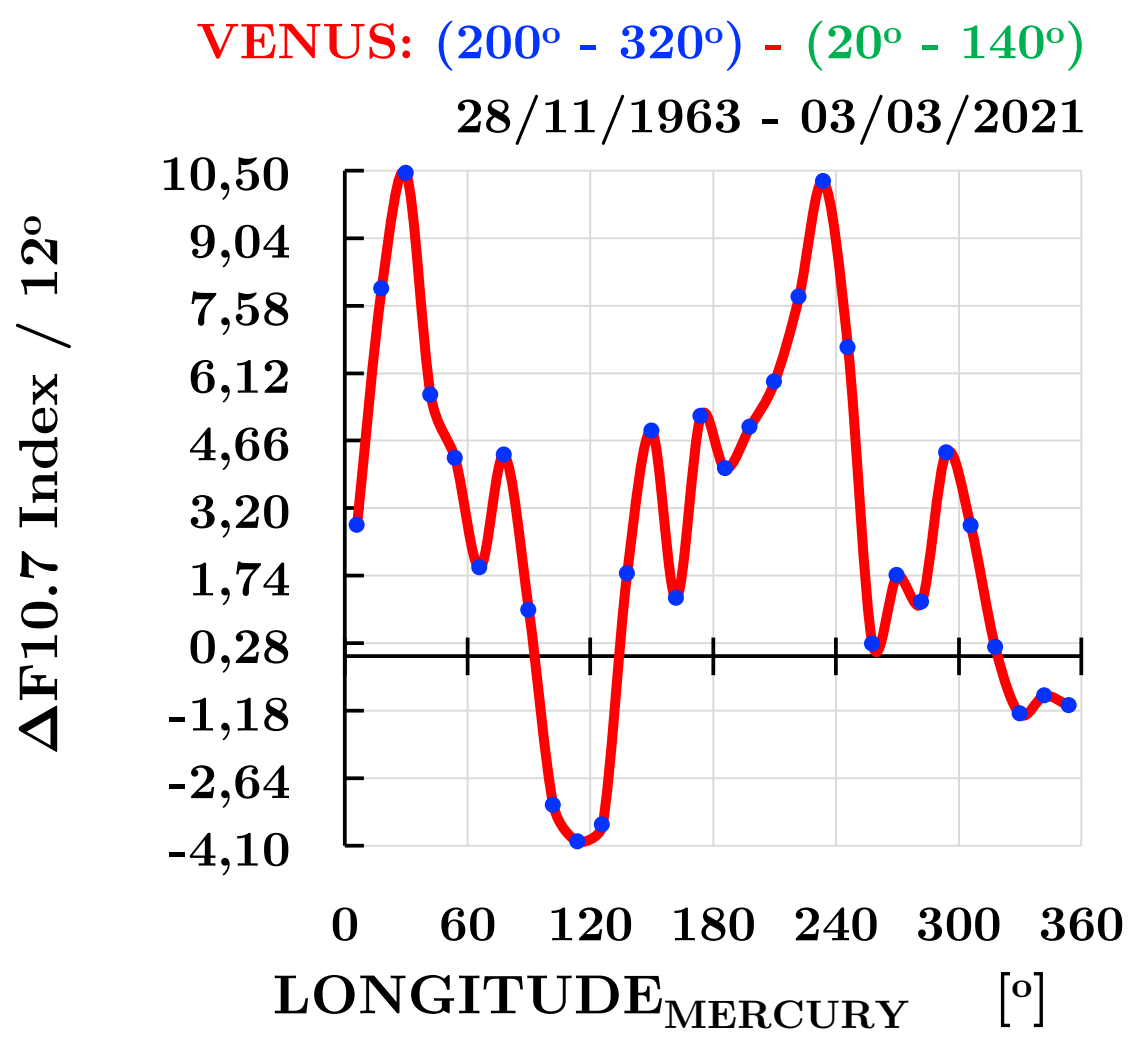

(d) Fig. 9.4a minus Fig. 9.4b.

**Figure 9.4:** Distribution of F10.7 for the reference frame of Mercury while Venus propagates between two 180° opposite orbital arcs with 120° width, for 12° bin.

around the direction of the GC.

In Fig. 9.6 we move on to the reference frame of Venus. In Fig. 9.6a Mercury is allowed to propagate around $50° \pm 60°$, with the total maximum - minimum difference being 16.3% in 4997 d, and with a total of 585 126.6 sfu. Four peaks are observed around $78.1° \pm 3.3°$, $148° \pm 3.4°$, $220.9° \pm 3°$ and $294.7° \pm 3.3°$ with a FWHM derived by a Gaussian fit of $32.7° \pm 8.6°$, $13.6° \pm 5.8°$, $33.6° \pm 7.7°$ and $34.7° \pm 8.6°$ respectively. Next, for the case of Jupiter propagating around $350° \pm 60°$ in Fig. 9.6b we have a 15.9% amplitude in 5954 d, and a sum of 548 991.5 sfu. The big peak observed is around 294°. It is noted that the region 290° to 50° is the one with the minimum effect on Jupiter as seen in Fig. 9.3e. Both distributions in Fig. 9.6 have a big amplitude of around 16% in contrast with the 5% amplitude when no planetary constraint was

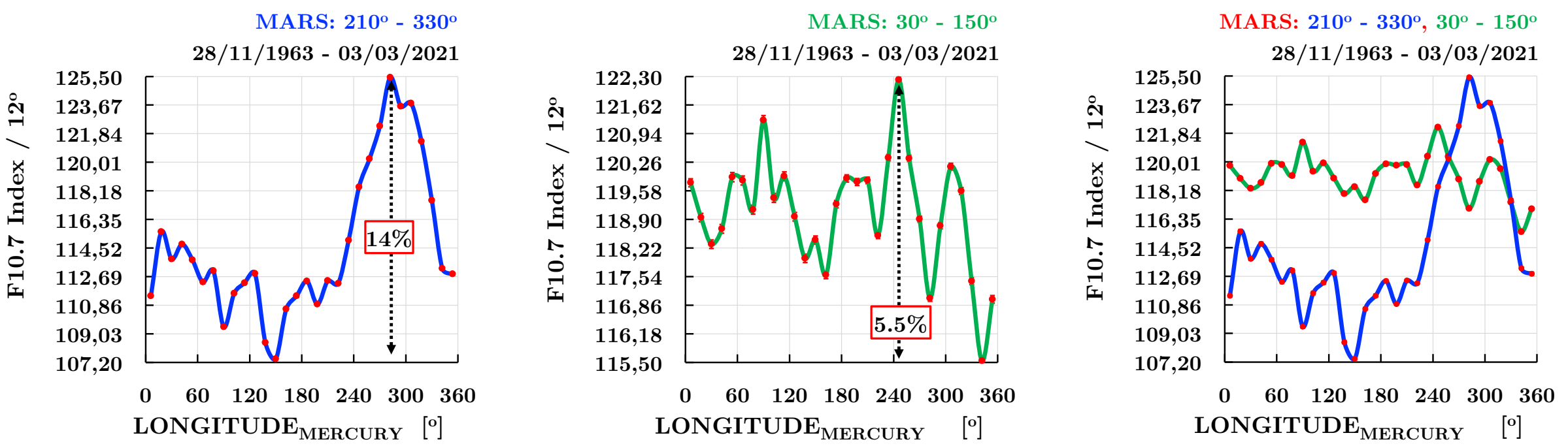

**(a)** Mercury when Mars is between 210° to 330°.

**(b)** Mercury when Mars is between 30° to 150°.

**(c)** Comparison between Fig. 9.5a and Fig. 9.5b.

**Figure 9.5:** Distribution of F10.7 for the reference frame of Mercury while Mars propagates between two 180° opposite orbital arcs with 120° width, for 12° bin.

in place (Fig. 9.3b).

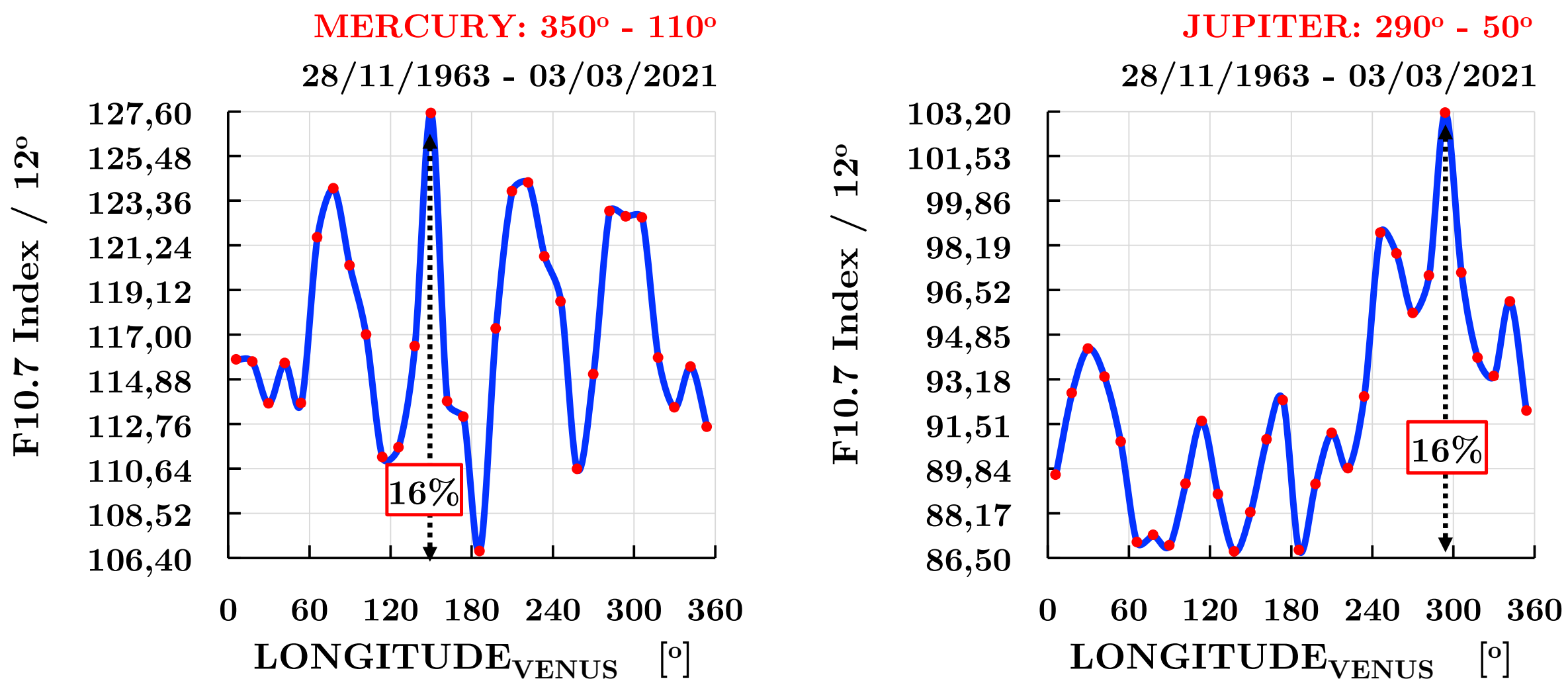

**(a)** Venus when Mercury is between 350° to 110°.

**(b)** Venus when Jupiter is between 290° to 50°.

**Figure 9.6:** Distribution of F10.7 for the reference frame of Venus while Mercury or Jupiter are constrained within 120° wide range, for 12° bin.

Moving on to Fig. 9.7, the F10.7 radio line is plotted as a function of Earth's heliocentric position while Mercury propagates around $90° \pm 60°$. For this case we have 4828 d out of 20 916 d, 567 776.8 sfu out of 2 458 016.3 sfu and a total difference between the maximum and minimum point of 25.8%. This shows a distinct difference with Fig. 9.3c where no positional constraint was applied with the observed amplitude in the first peak around 30° being 5%.

Finally, in Fig. 9.8 F10.7 is projected in the phase of the Moon both without any positional constraints and when Mercury is constrained around $85° \pm 45°$. In Fig. 9.8a the total amplitude is 1.8% while for Fig. 9.8b the amplitude becomes 42.7% showing the big advantage of using more than one planets in this kind of analysis. The number of days and the sum of F10.7 for Fig. 9.8b is 3512 d and 413 164.3 sfu accordingly, with the wide peak being around 293° having a FWHM of 189.3°. The comparison with the notable difference between these cases is seen more clearly in Fig. 9.8c where the same y-axis scale has been selected. It is noted that for the 180° opposite case when Mercury is between 220° to 310°, the maximum to minimum difference is much smaller at about 7.2% with 7360 d and 870 753 sfu being the integrals for the total number of selected days and the sum of the F10.7 values respectively. This means that the heliocentric longitude of $\sim 85°$ for Mercury seems to be preferred for the reference frame of the Moon.

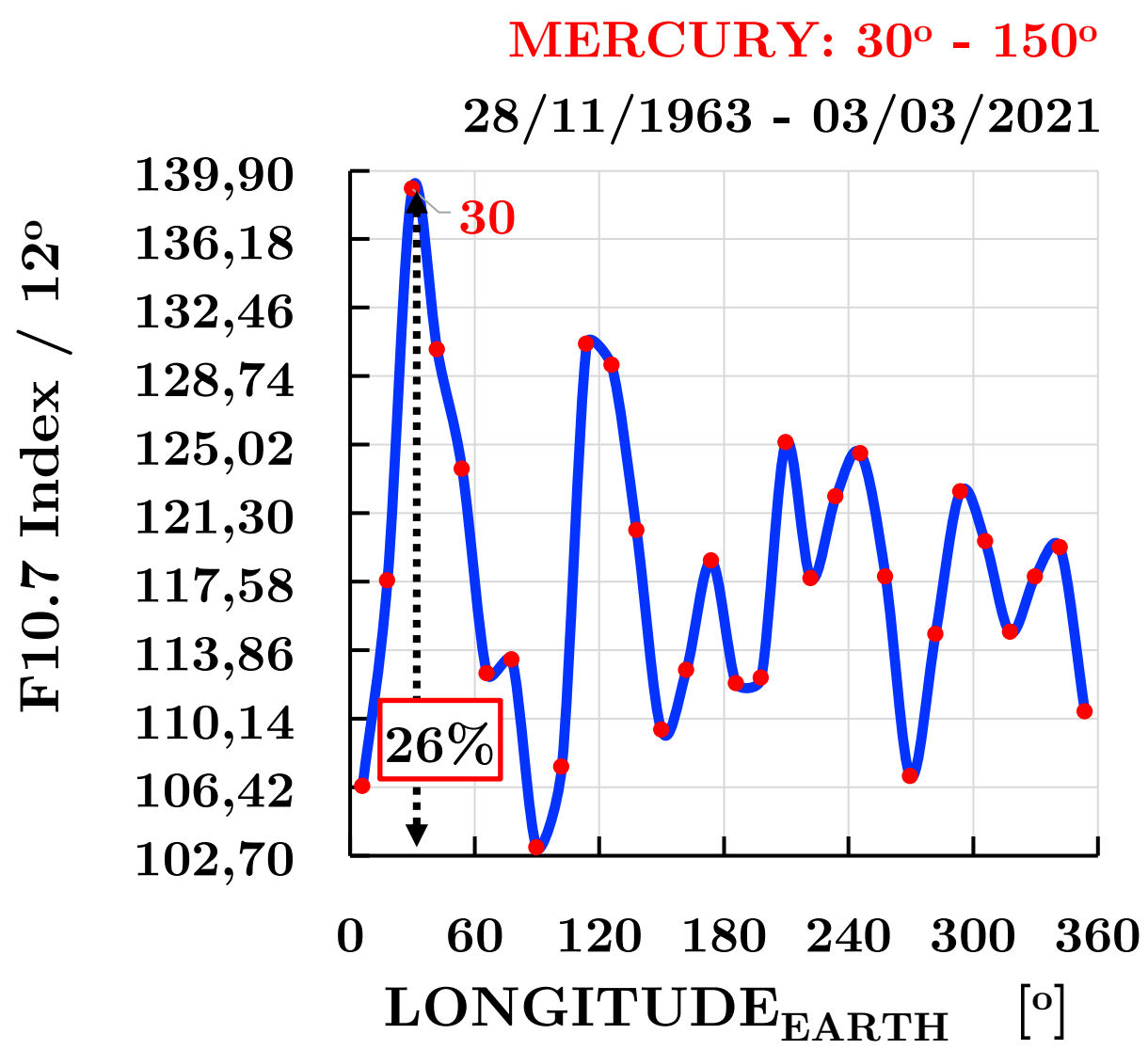

**Figure 9.7:** F10.7 distribution as a function of Earth's heliocentric longitude when Mercury lies between 30° to 150° for bin = 12°.

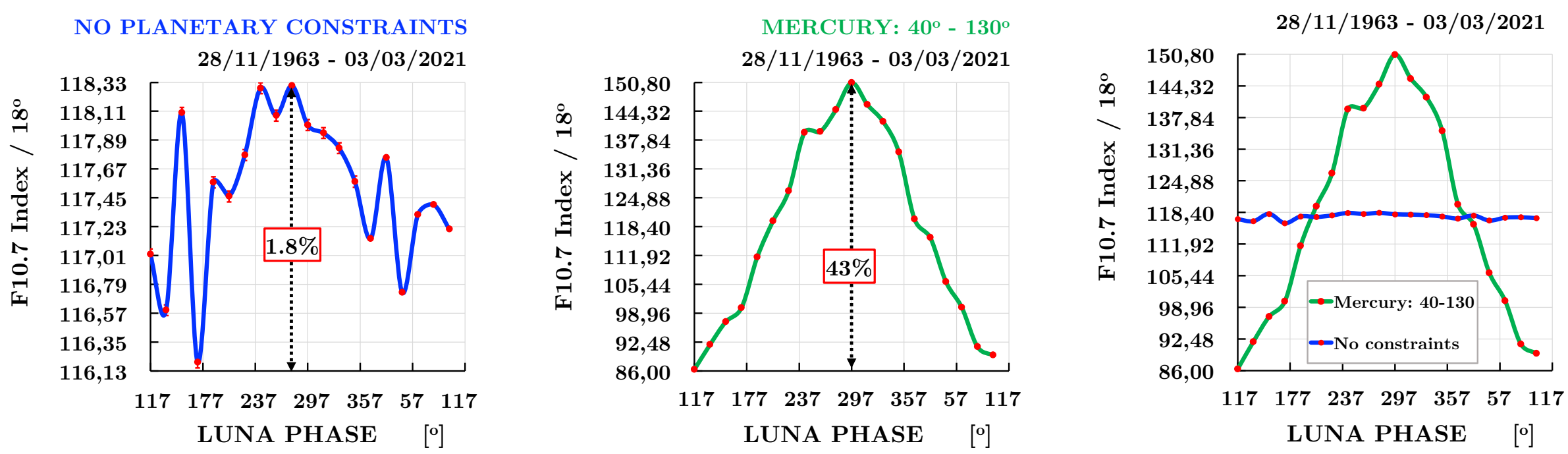

**(a)** Moon's phase without any planetary constraint.

**(b)** Moon's phase when Mercury propagates between 40° to 130°.

**(c)** Comparison between Fig. 9.8a and Fig. 9.8b.

**Figure 9.8:** Distribution of F10.7 as a function of Moon's phase with and without any additional longitudinal constraints in Mercury's heliocentric position, for 18° bin.

### 9.3.2 Fourier analysis

A Fourier analysis has been performed for the full period between 28/11/1963 to 03/03/2021. The results contain some interesting frequencies coinciding with the ones found in the planetary distributions. The 5$^{th}$ biggest peak is located around 684.3 d ± 9.3 d with an amplitude of 30 dB which coincides with the revolution period of Mars at 687 d (see Fig. 9.9a). In addition, the 24$^{th}$ biggest peak with an amplitude of 10.3 dB is located at 225 d ± 1.7 d which overlaps with Venus revolution period around the Sun of 224.7 d (Fig. 9.9b). Furthermore, the 19$^{th}$ biggest peak has an amplitude of 11.4 dB and a frequency of 27.32 d ± 0.02 d which is exactly Moon's sidereal month, fixed to remote stars (see Fig. 9.9c). At the same time, around 29.53 d, the period associated with Moon's synodic period i.e. the reference frame is fixed to the Sun, there is no peak observed (Fig. 9.10.) This observation points to an additional significant exo-solar influence on the solar activity with the Moon modulating the assumed incoming stream(s). Finally, there is also a less pronounced peak around 88 d ± 0.2 d with an amplitude of 2.7 dB coinciding with the revolution period of Mercury of 87.97 d.

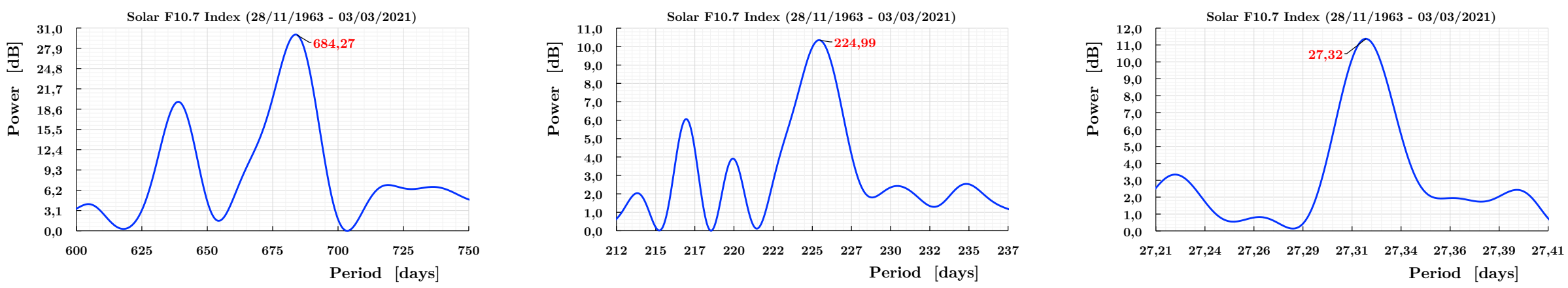

**(a)** Around ∼ 684 d. **(b)** Around ∼ 225 d. **(c)** Around ∼ 27.3 d.

**Figure 9.9:** Fourier periodogram of the F10.7 data from 28/11/1963 to 03/03/2021 zoomed around a few interesting periods.

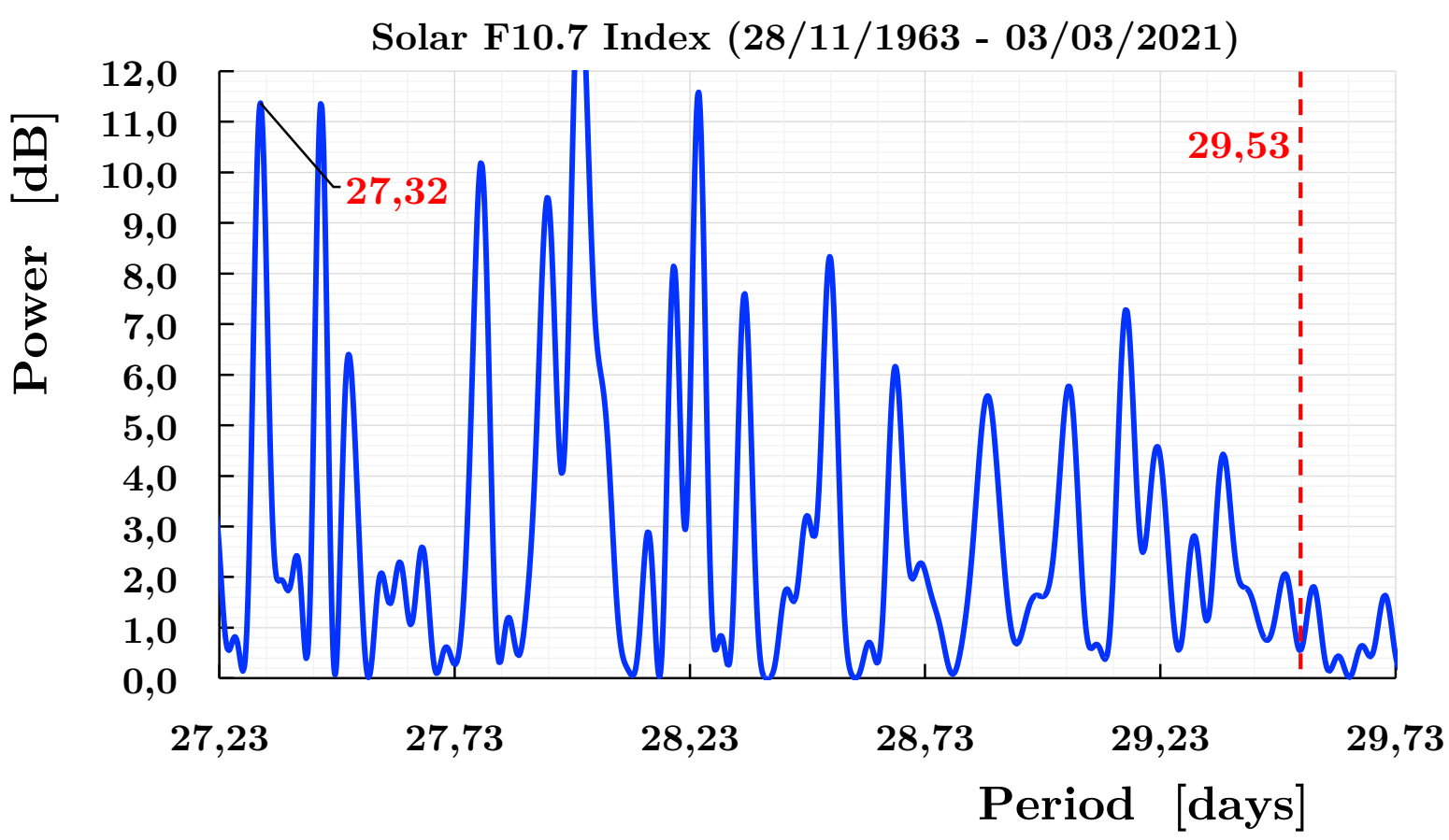

**Figure 9.10:** Fourier periodogram of the F10.7 data comparing 27.32 d and 29.35 d corresponding to Moon's sidereal month and synodic period respectively. The presence of a significant peak around the sidereal month, together with the absence of a peak around the synodic period strengthens the claim for an additional exo-solar impact on the Sun's activity modulated also by the Moon.

## 9.4 Summary

From the statistical analysis on the distribution of the F10.7 solar index as a function of the heliocentric longitudinal position of the planets, a highly significant planetary relationship has been found, which is unexpected within conventional solar physics. The only viable explanation seems to be through gravitational focusing of invisible slow-moving matter by the planets towards the Sun and eventually the free-fall effect of the Sun itself. Notably, statistically significant peaks were observed in all the selected planets, with all of them being far above the $5\sigma$ level. Notably, these peaks should not have existed after an averaging of about 58 y due to the several planetary revolutions around the Sun which smear-out any randomly occurring event. Noticeably, the observed already statistically significant effects become even more enhanced when a combination of more than one planet takes place. Finally, additional confirmation of the retrieved periodicities of single planets is found through a Fourier analysis. The most important one seems is through the retrieval of the Moon's sidereal month around 27.32 d which together with the absence of a peak in the synodic period around 29.53 d indicates an additional significant exo-solar impact on the F10.7.

A comparison of the derived dependencies, with the corresponding spectra of solar EUV irradiation via a statistical correlation analysis shows that the two datasets have similar characteristics. The same holds for the comparison with the number of sunspots where a statistically significant positive linear correlation is found for both the general trend as well as for the individual small-scale characteristics in the various planetary distributions (see Appendix Sect. B.5).

An additional future comparison could also be performed with the rest of the provided times from NRC like 22:00, 23:00, 17:00 and 18:00 UTC, as well as for different periods and using the combination of more planets. This way, possible discrepancies or similarities could show up which can provide more information on the direction(s) of the assumed stream(s) of invisible matter being focused by the planets towards the Sun as well as on their velocity distribution.

# 10

# SOLAR RADIUS

## 10.1 Introduction

The accurate measurement of the radius of the Sun and its variations is one of the oldest astronomical challenges. Its importance ranges from the fact that it can serve as an astronomical standard, up to the understanding of the physical processes behind its variation. However, a high-accuracy measurement of the diameter of the Sun remains a challenge even for modern techniques with the helioseismology providing an alternative measurement of the solar radius (*seismic radius*) which is different from the conventional photospheric radius. This way, a high precision radial displacement of the subsurface layer is possible through the analysis of the oscillation frequencies of surface gravity (f) modes.

Latest analyses for the determination of the variation of helioseismic radius proxy during the period from 1996 to 2017 show an anti-correlation with the 11 y solar activity as it is manifested by the solar sunspots number [300,301] (see Fig. 10.1). The anti-correlation between the solar radius and the solar activity is suggesting some kind of external impact-like pressure being exerted towards the whole Sun as the higher the solar activity the more compressed the solar size seems to get. Therefore, to study further the origin of this behaviour, a search for planetary correlations has been performed suggesting that streaming slow-moving invisible

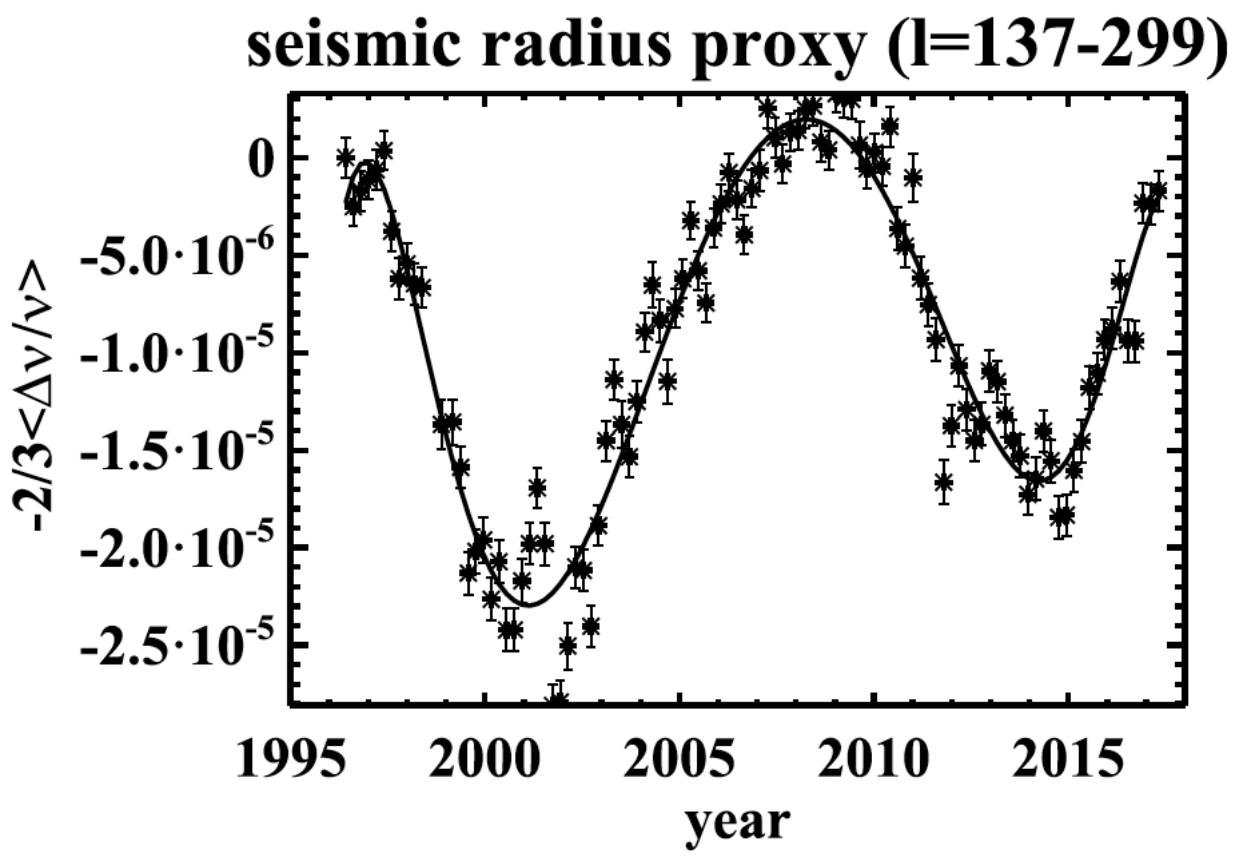

**(a)** Variation of the seismic radius proxy relative to the first measurement in 1996.

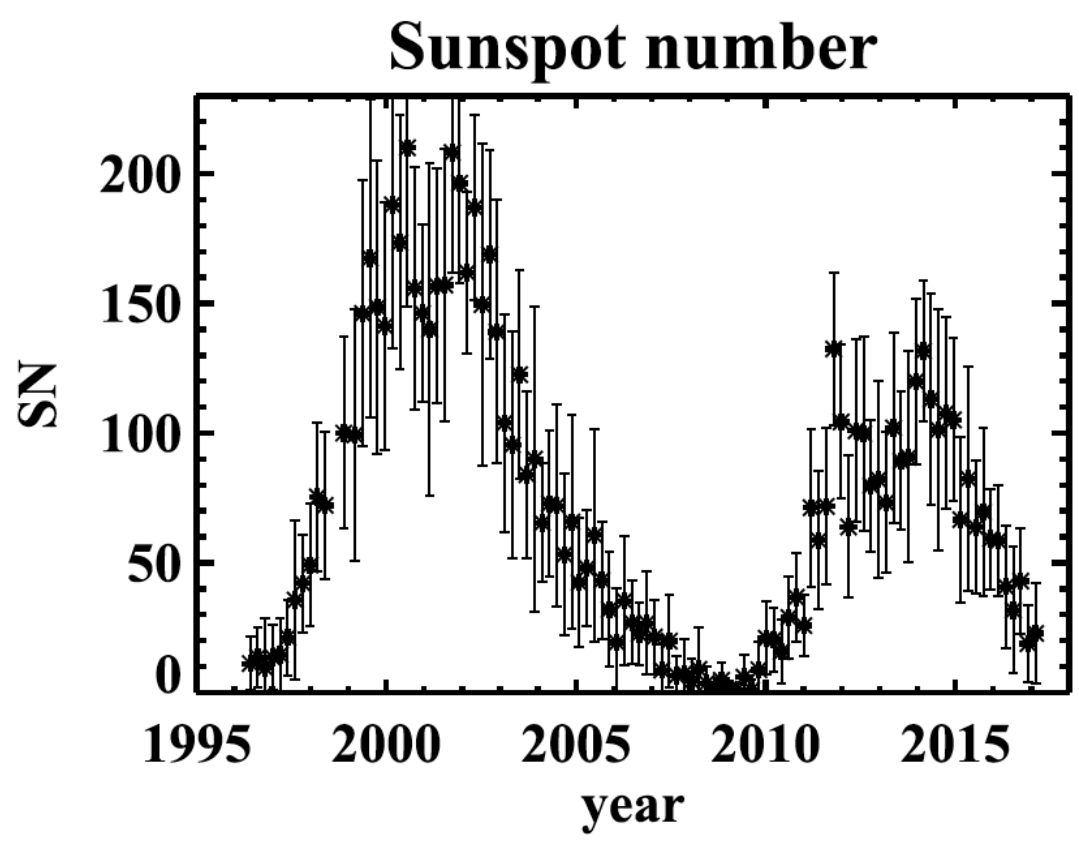

**(b)** The sunspot number, averaged for the 72-days period corresponding to the intervals of the helioseismic analysis.

**Figure 10.1:** Comparison of solar sunspot Nr. and seismic radius proxy shows an anti-phase between the two (33).

massive matter towards the Sun, enhanced by the gravitational focusing of the planets as well as the free fall effect Sun itself, could be the underlying reason (see also Publication E.5).

## 10.2 The solar radius data

### 10.2.1 Data origin

The time-series of the current helioseismology data are derived from the SOHO (1996 − 2010) [302] and the Solar Dynamics Observatory (SDO) (2010 − 2017) [303] space missions. More specifically, the variations of the helioseismic radius proxy $\Delta R$ has been deduced from the analysis of the f-mode frequencies extracted from the Michelson Doppler Imager (MDI) and Helioseismic and Magnetic Imager (HMI) data onboard the two missions. The f-mode frequencies are sensitive to the sharp density gradient in the near-surface layers and thus can be used for measurements of the seismic radius of the Sun [301].

$$\frac{\Delta R}{R} = \frac{2}{3}\left\langle \frac{\delta v}{v} \right\rangle_l \tag{10.1}$$

where the relative variations of the f-mode frequency $v$ are averaged over a range of angular degree $l$.

The raw seismic radius data that are used cover the period between 06/06/1996 and 01/12/2017. More specifically, the MDI data cover the initial period until 20/03/2011 while the HMI data cover the rest of the period. The total number of the measured that are used, combined from the two instruments, is 108. The measurements contain both positive and negative values ranging from −4.936 to 1.503, since they are normalised to the seismic radius measured in 2009 during the solar minimum between the solar cycles 23 and 24 [300]. These values are shown in Fig. 10.2. In the same figure, the solar radius data are compared with the solar radio line at 10.7 cm (2.8 GHz) which acts as a solar proxy (see also Chap. 9). We clearly see again here the anti-correlation with the sunspot number as in Fig. 10.1.

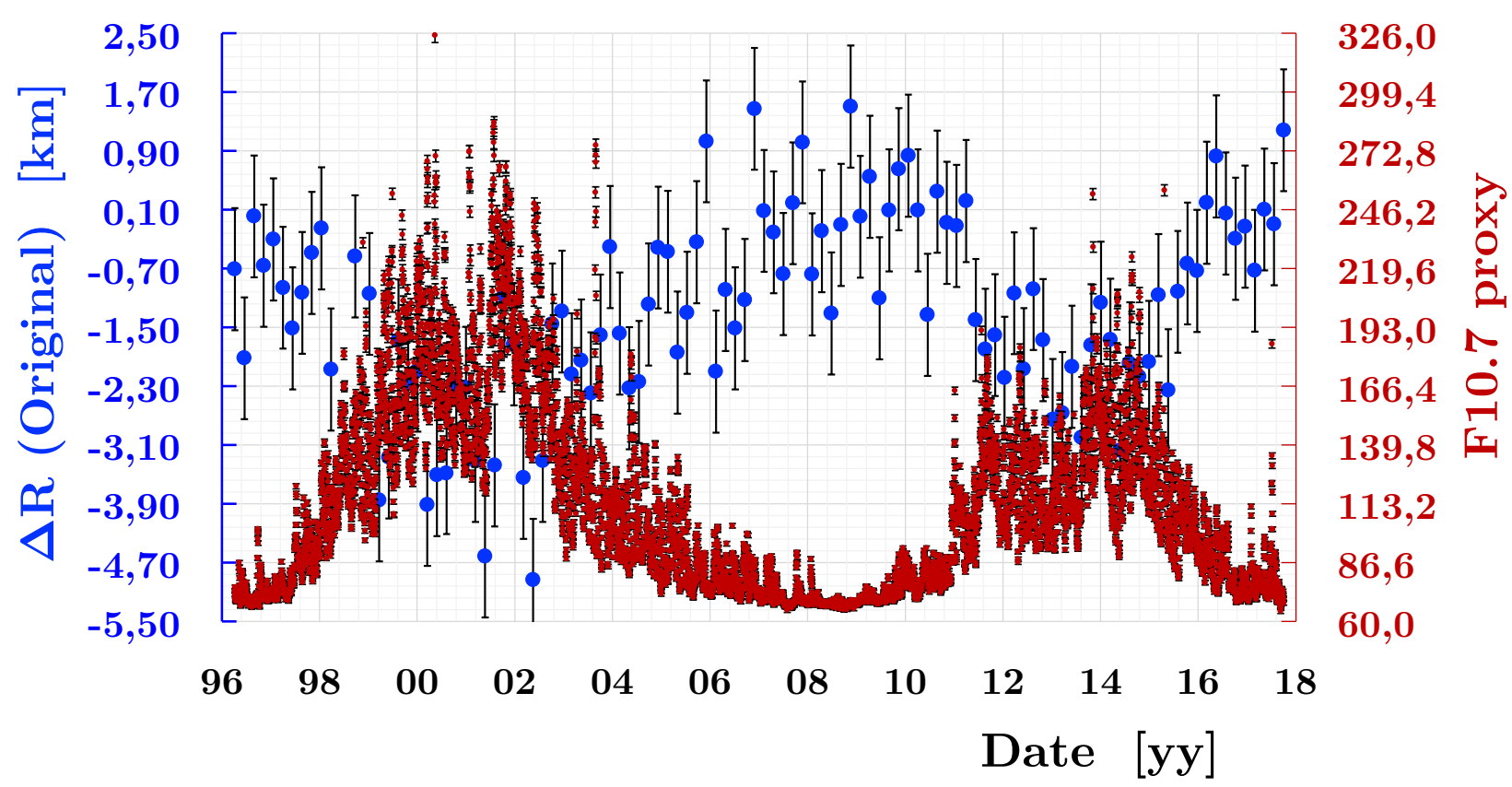

**Figure 10.2:** Comparison of solar radius with the F10.7 solar proxy. The errors from both datasets are also shown.

The provided error for each raw solar radius measurements is ±0.83 km.

### 10.2.2 Data curation

To be able to perform the planetary analysis procedure the raw positive and negative values of the solar radius are converted into positive ones. Thus, the first step is to subtract the minimum value (4.936) from each one of the raw values thus converting all of them into positive ones (see Fig. 10.3a). This procedure does not insert any kind of artefacts in the analysis since, as mentioned, the raw data are already normalised to the first measurement.

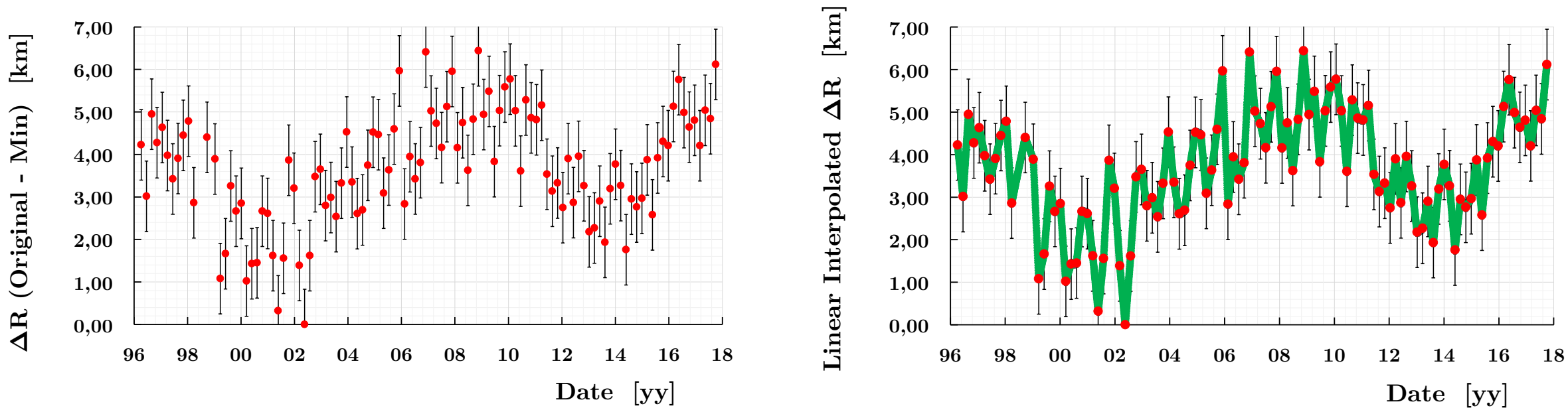

**(a)** Initial data treatment of solar radius data. Raw solar radius data minus minimum value. The error bars in each point are also shown.

**(b)** Linearly interpolated data (in green) between the 72 d binned values (in red) from Fig. 10.3a. The errors for the original 72 d binned values are also shown.

**Figure 10.3:** Data treatment on solar radius data.

#### 10.2.2.1 Linear Interpolation

The raw solar radius data are recorded every 72 d. To produce a heliocentric longitudinal distribution for the inner planets as well as for a combination of planetary positions, we have applied a linear interpolation between every two neighbouring 72 d positive values. This way, we can reproduce daily data allowing us to apply our analysis code and get credible results. The results from this technique are depicted in Fig. 10.3b. It is noted that even though the actual statistical error is given for the raw values, for the linearly interpolated data a calculation is not straightforward and therefore it has not been included in the following results with the daily analysis plots. The verification of the accuracy of this procedure and its effects on the various longitudinal distributions is shown in Appendix Sect. B.6.1.

## 10.3 Data analysis and results

### 10.3.1 Planetary longitudinal distributions

The derived daily data of solar radius are used to produce the various longitudinal distributions of each planet. However, there are minimum allowed values for the bins that are can be used due to the time resolution of the data of the angle that each planet covers during the selected cadence time. These are shown in Tab. 10.1.

**Table 10.1:** Minimum, maximum and average 72 d binned heliocentric longitude differences for each planet and some of their satellites over the period 01/01/1970 - 31/12/2030.

| | Minimum Longitude Difference [degrees] | Maximum Longitude Difference [degrees] | Average Longitude Difference [degrees] |
|---|---|---|---|
| Mercury | 263.3058 | 315.5962 | 294.2272 |
| Venus | 114.0638 | 116.7055 | 115.3592 |
| Earth | 68.7782 | 73.2216 | 70.9398 |
| Mars | 31.5126 | 45.5255 | 37.7623 |
| Jupiter | 5.4342 | 6.6061 | 5.9930 |
| Saturn | 2.1643 | 2.6873 | 2.4661 |

However, with the linear interpolation method (see Sect. 10.2.2.1) a smaller bin size can be used as the 72 d data have been "transformed" into daily ones. Mercury due to its $\sim 88$ d short revolution period it is practically useless to create a spectrum from the 72 d cadenced data. The same applies to the Earth's Moon. Therefore, the rest of the planets will be used for the analysis.

#### 10.3.1.1 Single planets

The linearly interpolated data are first plotted as a function of the heliocentric longitude positions of the various planets. In Fig. 10.4 Venus, Earth, Mars and Jupiter are used for the whole period of 06/06/1996 to 01/12/2017.

By fitting a Gaussian function, we find that the peak appearing in Venus distribution in Fig. 10.4a, is located around $107.2° \pm 2°$, has an amplitude of 6.8% and a FWHM of $248.5° \pm 28.1°$ which considering the revolution period of Venus corresponds to about $154\,\mathrm{d} \pm 17\,\mathrm{d}$.

Similarly, we find that in Earth (see Fig. 10.4b) the wide peak with an amplitude of 27.8% is centred around $121.3° \pm 2.05°$ and has a $FWHM \simeq 194.5° \pm 14.8° \simeq 193\,\mathrm{d} \pm 15\,\mathrm{d}$. This strong seasonal variation on the behaviour of the Earth's spectrum could be associated with the 8.5° inclination of the Sun's spin axis relative to the ecliptic as it is defined by the orbit of the Earth, and therefore at this stage is not considered a planetary dependence.

Moreover, in Fig. 10.4c, Mars' spectrum comprises of three peaks with the biggest one having an amplitude of 11.6%. From a Gaussian fitting algorithm, we can derive the location of the three peaks to be around $176.6° \pm 1.7°$, $274.8° \pm 3.2°$ and $24.4° \pm 4.6°$ with their FWHM being accordingly $44.8° \pm 7.7°$, $53° \pm 5.4°$ and $108.3° \pm 20°$. These three peaks in Mars, could be explained by the single peak in Venus in Fig. 10.4a, due to Mars orbit being about 3.05 times Venus orbit. This claim is also verified in Appendix Sect. A.2.2. For comparison, an Earth planetary relationship would result to two peaks instead of three. The time difference between the calculated centres of three peaks is found to be equal to 198 d ± 7 d, 178 d ± 12 d and 314 d ± 10.8 d which gives an average distance of about 230 d ± 5.8 d. Accordingly the full width of each peak is about 192 d, 196 d and 264 d accordingly, giving a mean value of 217.3 d. Therefore, the overall mean value is about $(230\,\text{d} + 217.3\,\text{d})/2 = 223.6\,\text{d}$, which is within $1\sigma$ with the orbital period of Venus around 224.7 d.

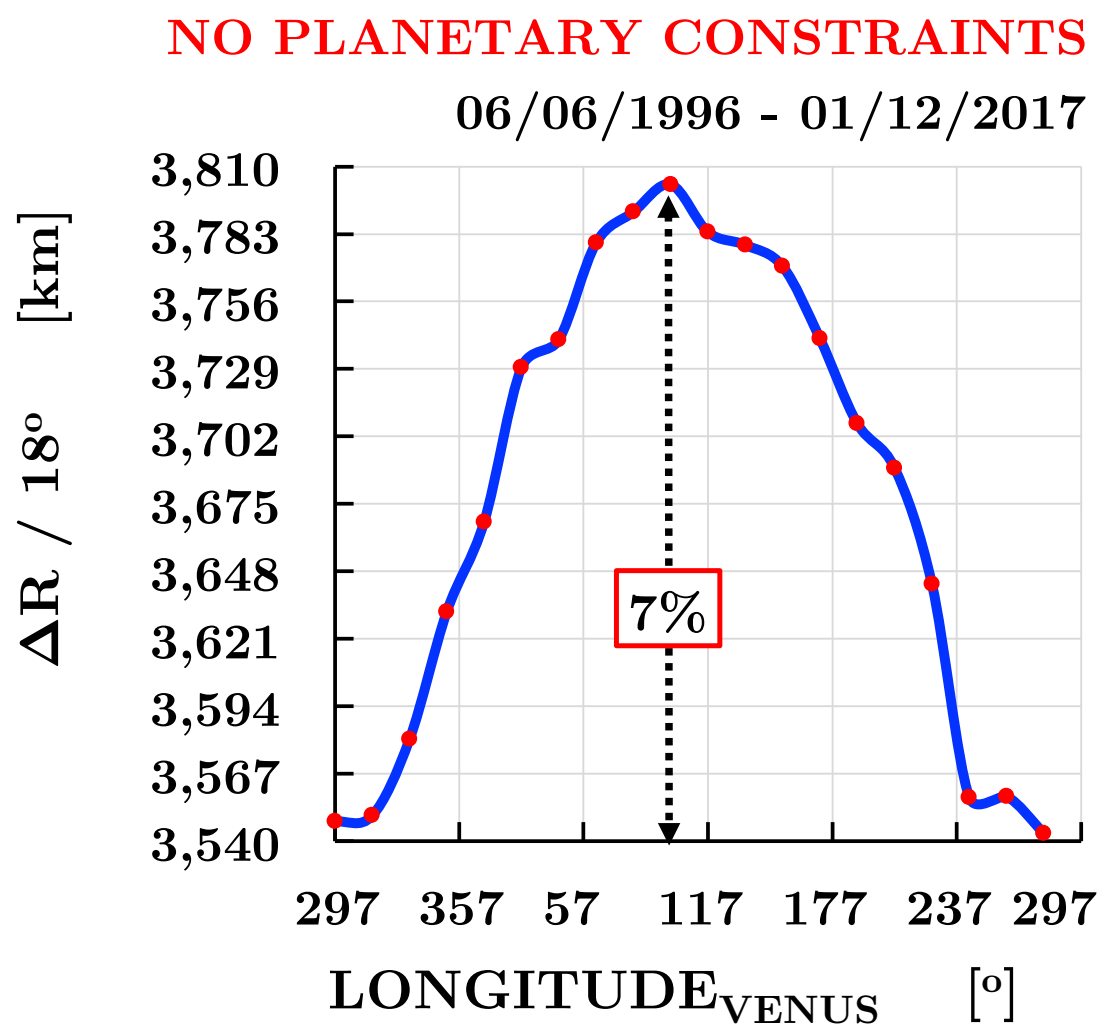

(a) Venus for bin = 6°.

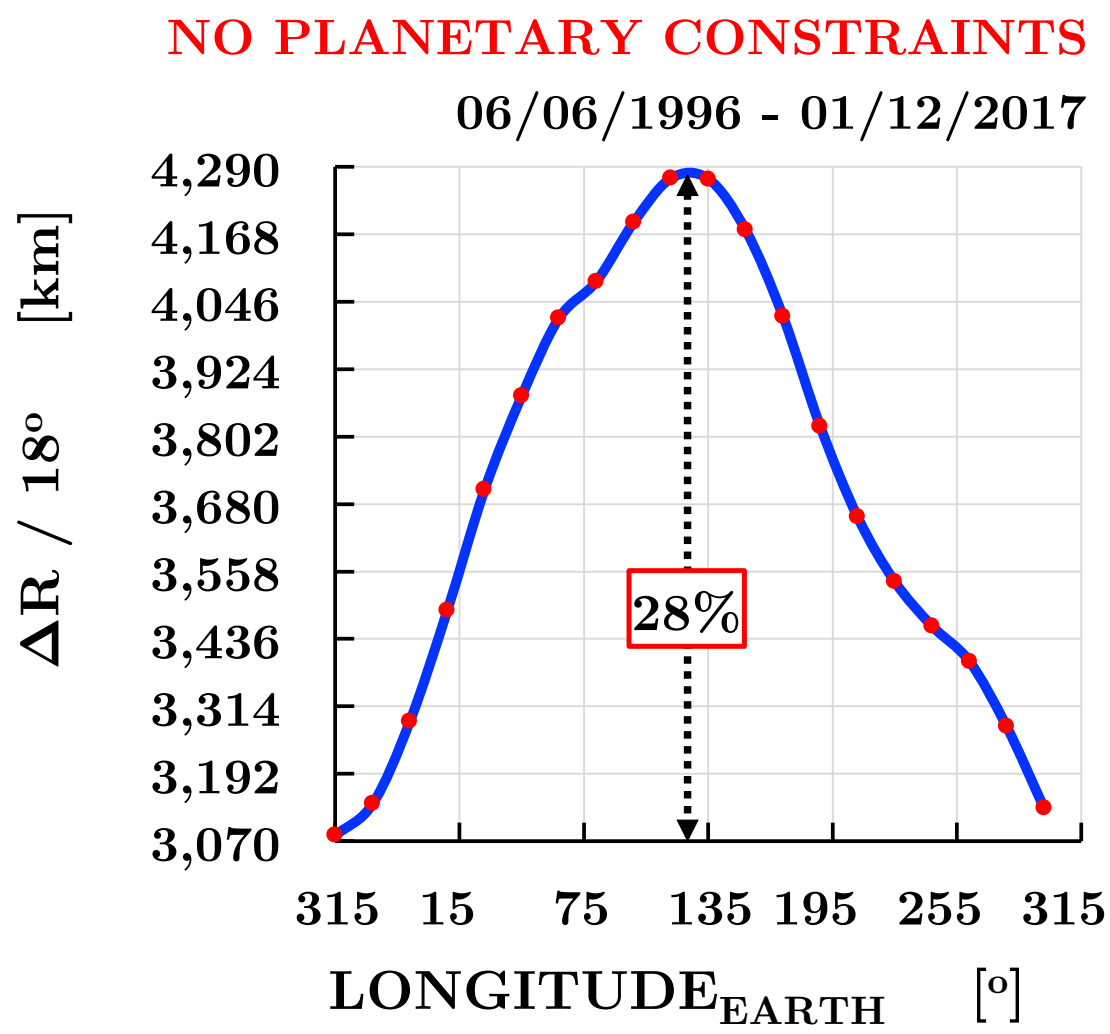

(b) Earth for bin = 18°.

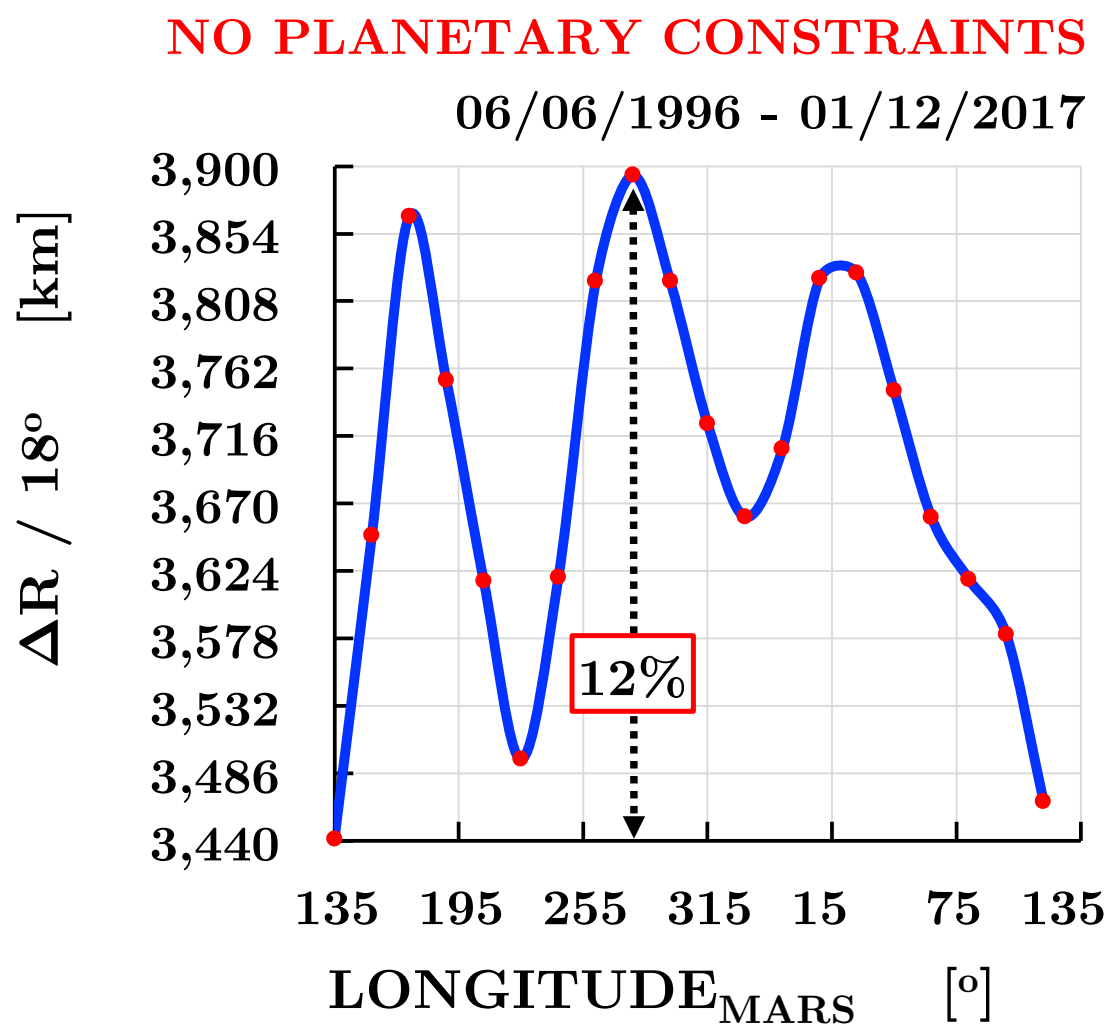

(c) Mars for bin = 18°.

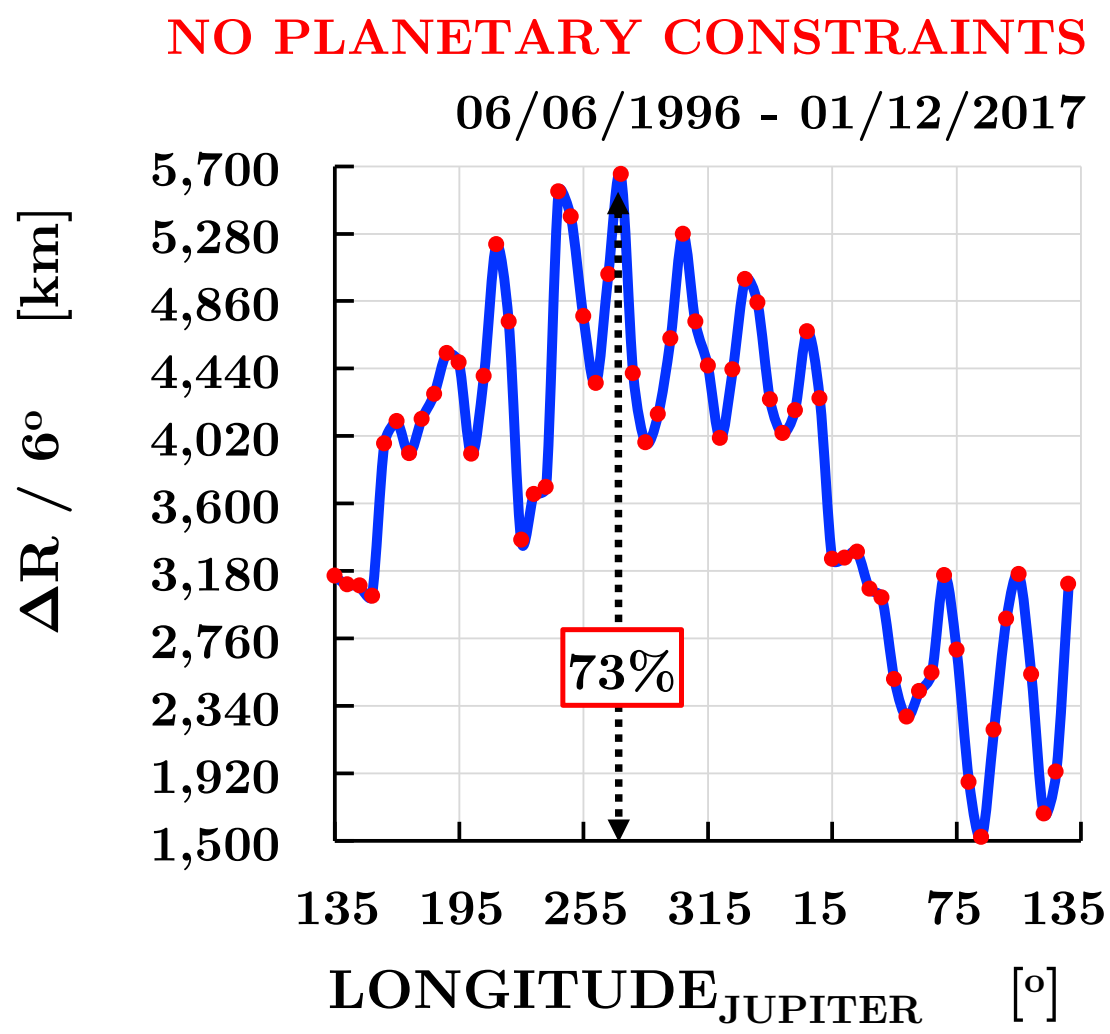

(d) Jupiter for bin = 6°.

**Figure 10.4:** Solar radius data as a function of the position of the various planets for 06/06/1996 – 01/12/2017 for various bins.

Finally, in Jupiter in Fig. 10.4d, the minimum-maximum difference observed is 73.1% and the whole spectrum seems to be comprised of 12 peaks which could be originating from the Earth's spectrum, as simulated in Appendix Sect. A.1.3.

#### 10.3.1.2 Combining planets

The main result from combinations of two planetary longitudinal positions comes from the distribution of Mars while constraining Jupiter to propagate between 20° to 200° compared with the 180° opposite case of 200° to 20° (see Fig. 10.5). Not many planetary combinations can be performed again due to the relatively high cadence of 72 d in the raw data. The maximum - minimum difference of the case of Jupiter being on 20° to 200° longitudinal range is about 36.7% while for the case of Jupiter being on the opposite 200° to 20° longitudinal range is about 34.1%. As for the number of days that fulfil each constraint, we have 4358 d for the former case and 3491 d for the latter case. The difference of the integrated $\Delta R$ excess between the two orbital positions of Jupiter shown in Fig. 10.5 is estimated to be on the order of 12.5$\sigma$

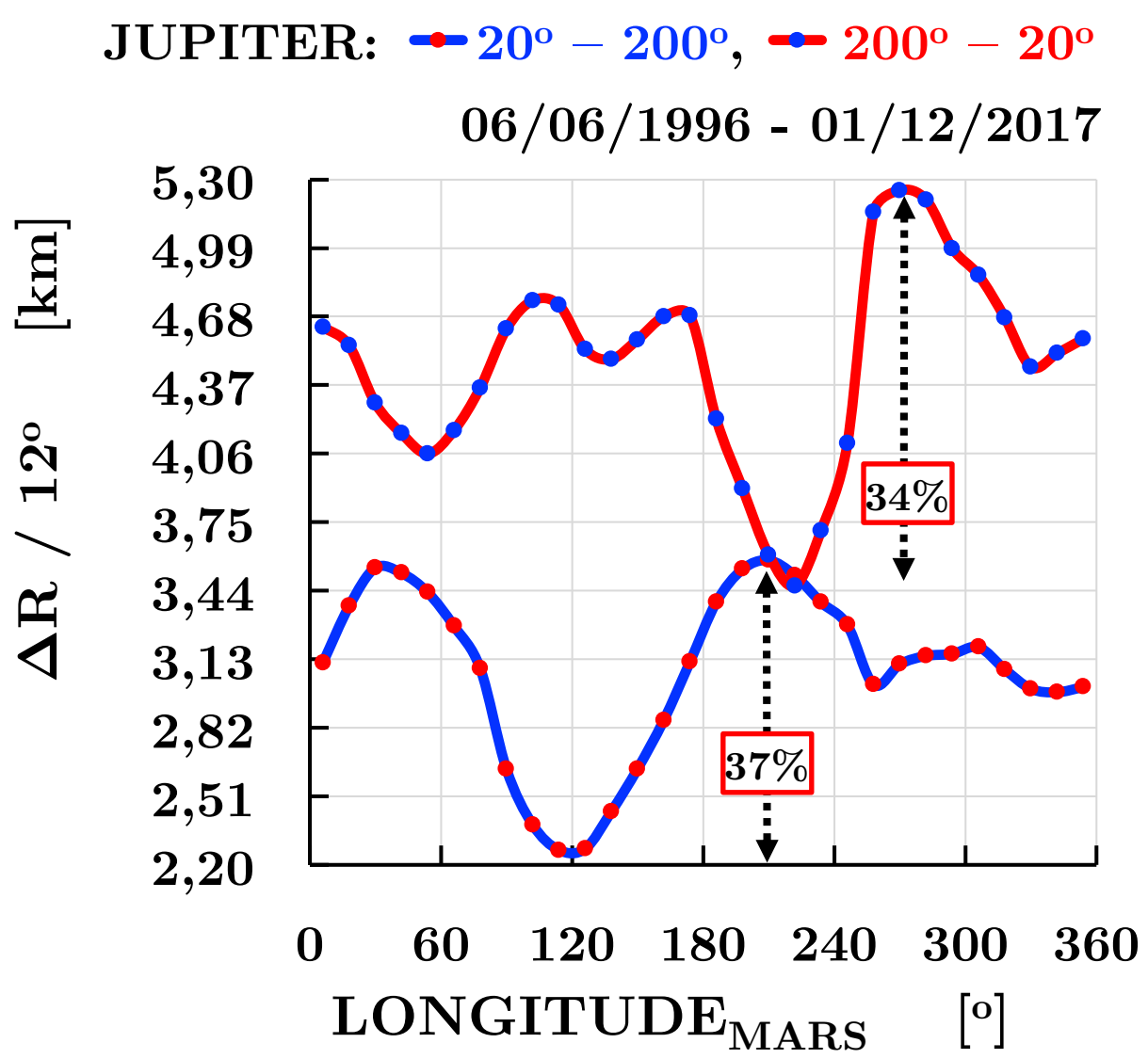

**Figure 10.5:** Mars longitudinal distribution while Jupiter is vetoed to be between 20° to 200° and 200° to 20°.

An additional constraint, this time with Mercury's position when the reference frame is fixed on Venus is presented in Fig. 10.6. For the case of Mercury being in the sector 0° to 180°, we have an integrated $\Delta R$ of 10 713.9 km while the corresponding number of days is 2920 giving an overall amplitude of 13.6%. On the other hand for the longitude sector 180° to 360°, the integrated $\Delta R$ is 18 163.1 km and the total number of days is 4929. The difference between the maximum and the minimum point is 14.6%. The difference between the two cases of Mercury in Fig. 10.6 appears quite strong, as an opposite behaviour is observed with the crests of one case corresponding to troughs of the other case and vice versa.

### 10.3.2 Planetary longitudinal distributions with raw data

To strengthen the derived first evidence on a planetary relationship behind the solar radius variation an additional analysis is performed wit the raw positive solar radius data with

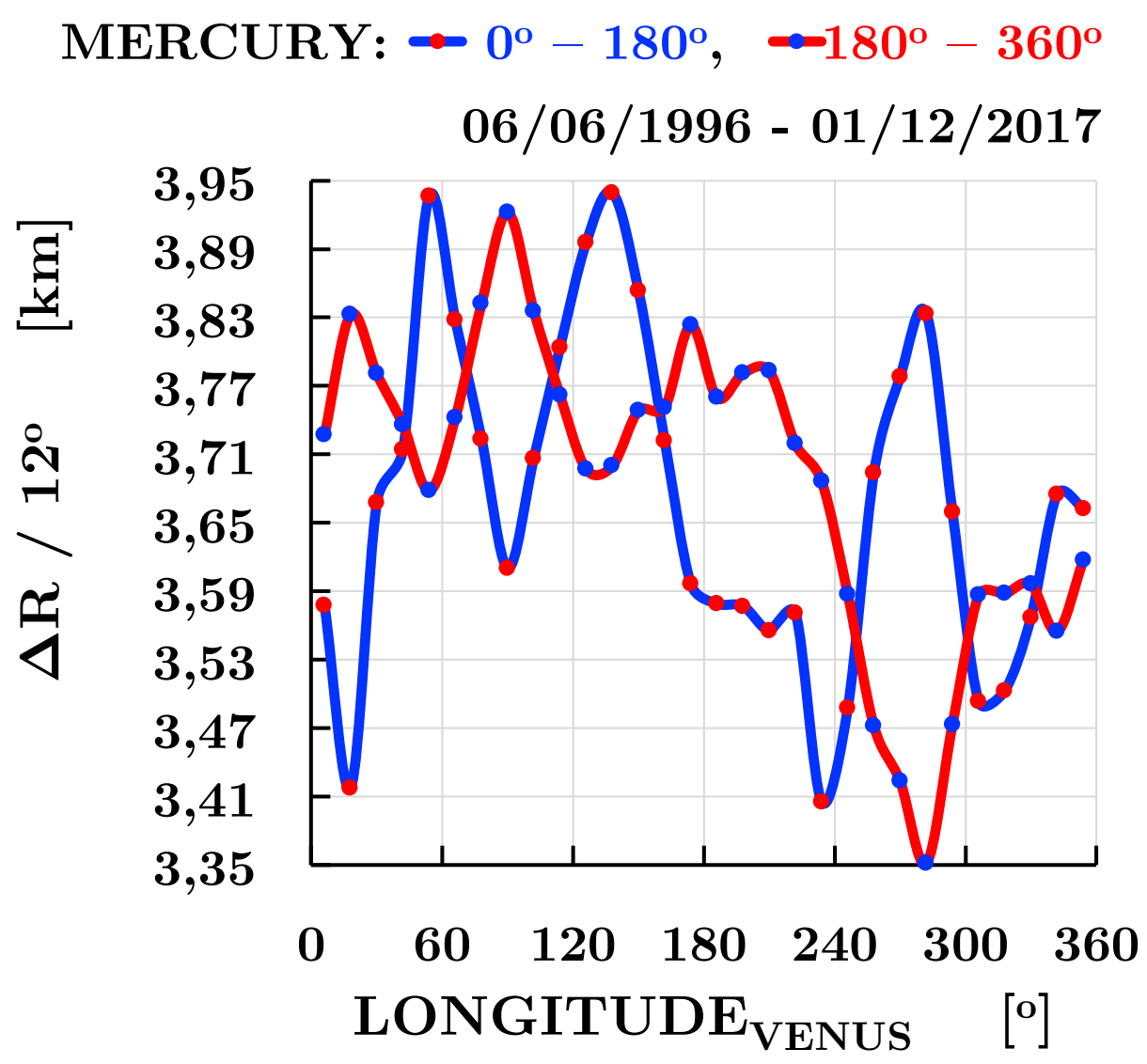

**Figure 10.6:** Solar radius as a function of Venus longitudinal position while Mercury is allowed propagate between 0° to 180° and 180° to 360°.

cadence = 72 d. In this case Saturn's reference of frame is used since the fast-orbiting inner planets are practically unusable given the large cadence. The distribution is shown in Fig. 10.7, with the clear double peaking distribution providing a striking planetary dependence. The observed large amplitude is about 70.9%, whereas the calculated significance between the four points around 162° and the four points around 54° is $\sim 11\sigma$.

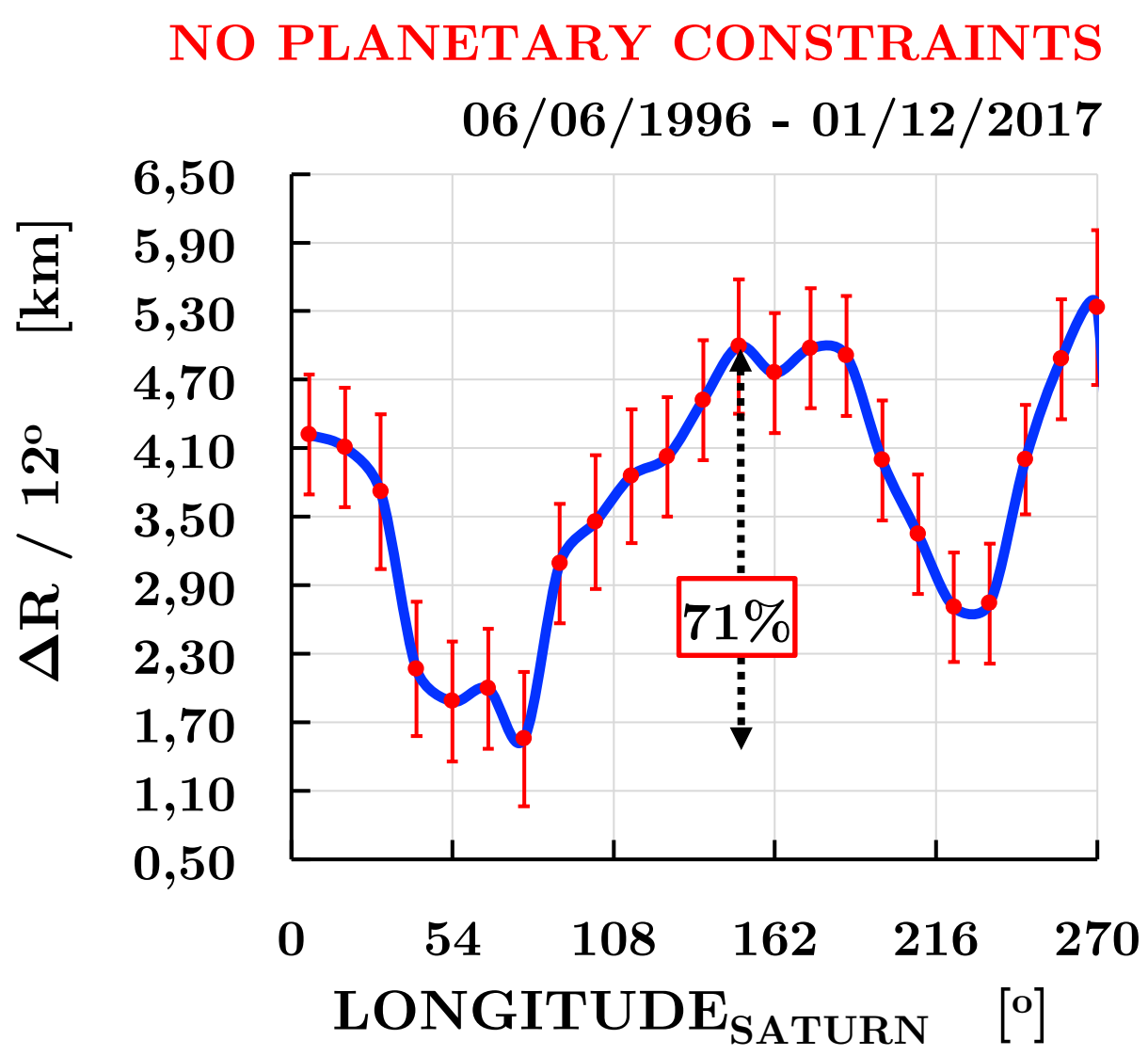

**Figure 10.7:** Distribution of the positive raw solar radius data as a function of Saturn's orbital position for bin = 12°.

Then, the same distribution is reproduced in Fig. 10.8 when imposing a wide planetary constraint for Mercury's longitudinal position when it is split into two halves, namely 0°-180° (Fig. 10.8a) and 180°-360° (Fig. 10.8b). When performing a statistical correlation analysis for the two cases compared with the distribution in Saturn's orbital frame of reference when no planetary positional constraints were applied (Fig. 10.7), we find both correlations to be statistically significant on the 0.05 level. More specifically, when Mercury is between 0°-180° we have ($r = 0.94$, $p = 1.1 \times 10^{-11}$), whereas when Mercury is between 180°-360° we have ($r = 0.89$, $p = 1.77 \times 10^{-8}$). The exact same observation is made for the case of Venus (see Publication E.5).

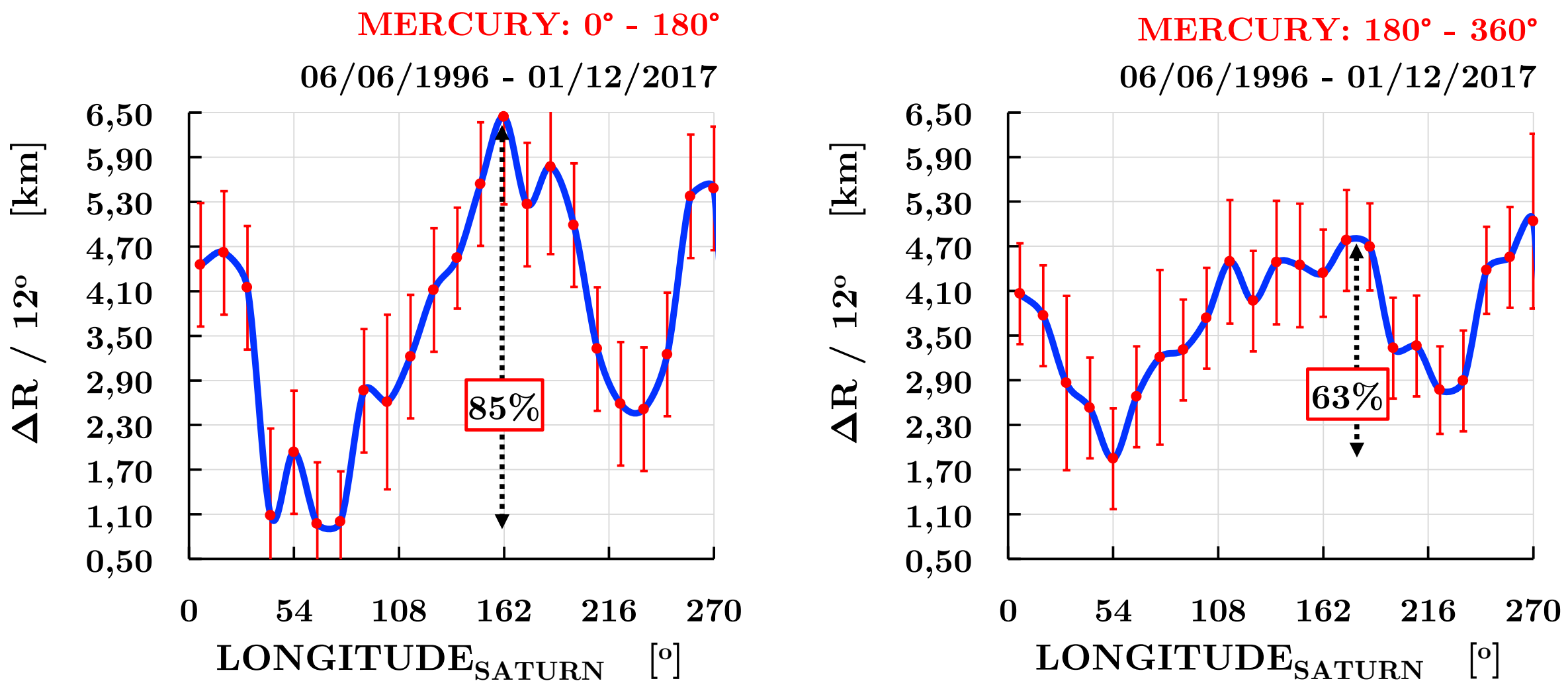

**(a)** Mercury between 0°-180°. **(b)** Mercury between 180°-360°.

**Figure 10.8:** Distribution of the positive raw solar radius data as a function of Saturn's orbital position when Mercury is assumed to propagate in a 180° orbital arc. The scale in both x and y axes is the same with Fig. 10.7 for comparison.

This result shows that even with the 72 d binned data there is a directional preference based on the aforementioned planetary configurations between 0°-180° for Mercury's orbital position. This strengthens the claim of a planetary relationship for the solar radius variation.

### 10.3.3 Fourier analysis

In order to enhance the already derived periodicities we also perform a Fourier analysis on the raw 72 d dataset of solar radius. The results from the normalised data from Fig. 10.3a, where the minimum value is subtracted from the raw values are the same as in the raw data, whereas a Fourier analysis on the interpolated data does not make sense. In Fig. 10.9 the results of the Fourier analysis are presented with the period in the x-axis presented in days. This is done by multiplying the 72 d cadenced results with the numerical value 72.

The various derived peak positions from the Lomb-periodogram are compared with the usual planetary revolution periods as well as the known synodical periods between all planets. Interestingly, as seen in Fig. 10.9 the 7$^{\text{th}}$ biggest peak is located around 227.6 d with 1.2 dB amplitude, and with a FWHM, resulting from a Gaussian fit, of about 5.3 d. This value is close to the 224.7 d revolution period of Venus. Similarly, the 3$^{\text{rd}}$ biggest peak corresponding to Earth's revolution period is located around 365.2 d with a FWHM of about 38.2 d and an amplitude of 4.6 dB.

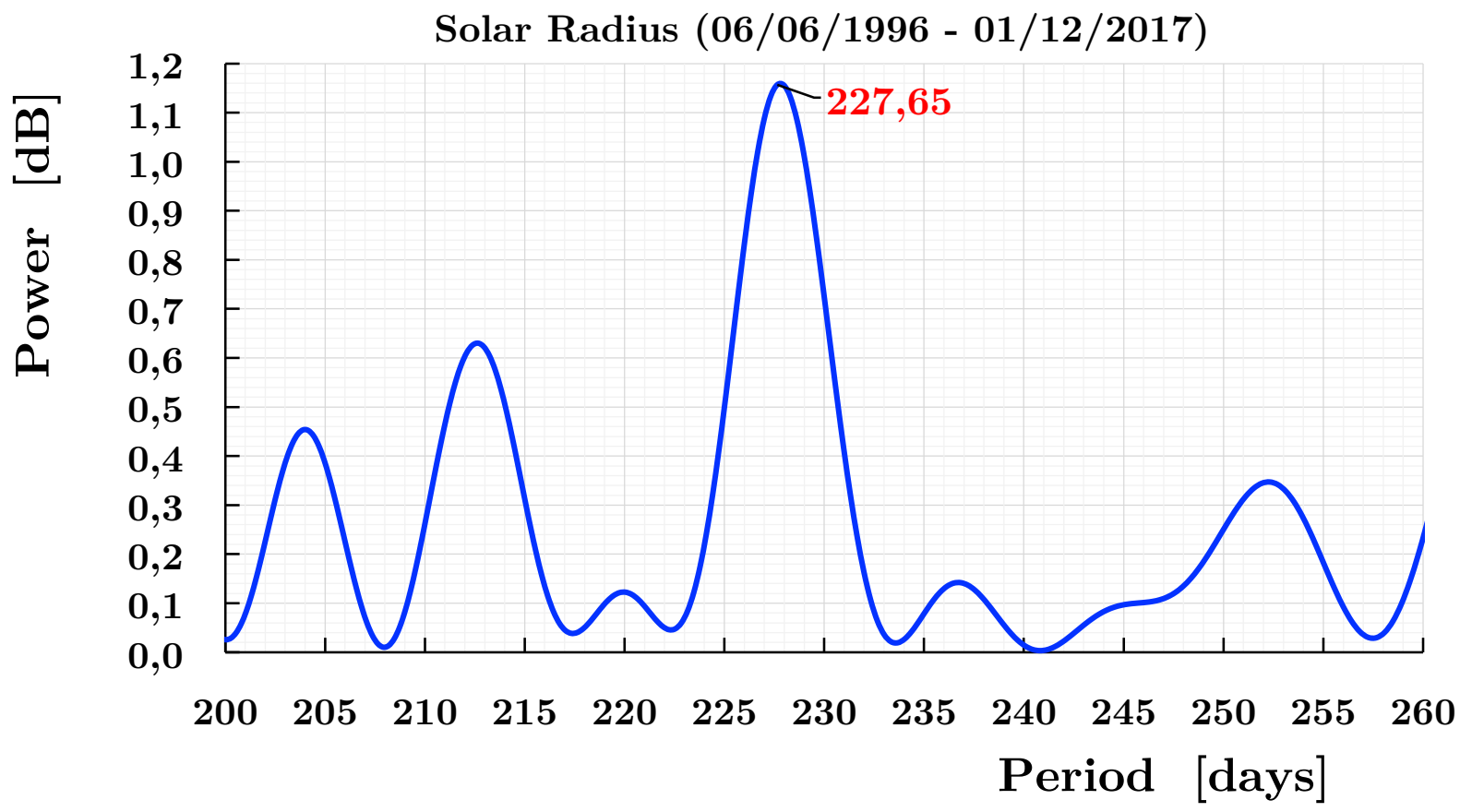

**Figure 10.9:** Fourier analysis of the raw solar radius data around 230 d.

## 10.4 Summary

Following recent helioseismology measurements, the observed 11 y rhythmic variation of Sun's seismic radius with time [300, 301] shows a planetary relationship, and more specifically, with Venus' 225 d orbital period. The results of a specifically designed simulation for Venus as well as for Earth validate the performed analysis and the derived claim of the striking planetary relationship of the solar radius with the orbital period of Venus since, as expected, three peaks appear in the distribution of Mars verifying their kinematical relationship (see Appendix Sect. A.2.2). In addition, by analysing also the raw 72 d cadenced solar radius data in the reference frame of Saturn when combined also with Mercury, a statistically significant planetary impact is found which is maximum in the orbital arc between 0° to 180°.

Independent cross-checks for the applied analysis and simulation have also been validated with other available daily datasets, which show planetary relationship such as the daily number of solar flares the total electron content of the atmosphere (see Appendix Sect. B.6.1). The analysis reconstructed the spectral features seen with the daily data, thus verifying the reliability of the applied simulation procedure. This improves the credibility of the simulation including the linear interpolation procedure using the 72 d cadence.

Finally, when comparing the solar radius data with the solar activity proxy given by the F10.7 data, the Venus related relationship for the solar radius data appears different from that of the solar activity proxy. Therefore the cause of the solar activity should be different from that behind the 11 y rhythmically varying solar radius at the level of $\sim 10^{-5}$ (= few km/solar radius). A similar conclusion comes from the comparison of the solar radius with the total tidal induced force by the eight planets. In this case, no significant association whatsoever is found for the two datasets, thus pointing to a different external triggering (see Appendix Sect. B.6.2 and B.6.3).

In summary, the rhythmic inflating and deflating of the whole Sun like a giant balloon throughout $\sim 11\,\mathrm{y}$ requires a large energy input, while its planetary relationship points at streaming invisible matter from the dark sector. DM might have some as yet overlooked streaming component(s), which in addition might interact effectively with Sun's ordinary matter. Notably, both effects combined can enormously enhance the DM impact increasing its flux above the cosmic basal value due to planetary focusing towards the Sun and the free fall towards the Sun itself (see also Publication E.5). Future analysis with a smaller cadence time could provide more accurate results on the exact planetary relationship of Venus as well as search for a possible relationship with other planets including also more combinations of them. This has the potential that the putative direction(s) of the incoming invisible massive stream(s) could be identified.

# 11

# CORONAL COMPOSITION

## 11.1 Introduction

As initially hinted by Tolstoy in 1889 [304] and much later from Pottasch in 1963 [305] the abundances of the elements in the solar corona such as Mg, Si and Fe should be significantly higher than the corresponding values in the underlying photosphere, an effect that can not be explained by the experimental uncertainties. Eventually, this resulted in a significant abundance anomaly in the solar corona [306].

This anomaly between the coronal and photospheric abundances seems to be related to the First Ionisation Potential (FIP) of the elements. More specifically, from Solar Wind (SW) and Solar Energetic Particle (SEP) measurements it has been observed that the average ratio of the elements with a low FIP ($< 10\,\mathrm{eV}$) such as Mg, Si, and Fe, to the elements with high FIP ($> 11\,\mathrm{eV}$) such as C, O, Ne, and He in corona is a factor of 3 to 4 higher than in the photosphere [307, 308] (see Fig. 11.1). This phenomenon is called the *FIP effect* with the exact mechanisms responsible for this elemental fractionation according to FIP not being yet understood [306, 309] even though it is nearly as old as that of coronal heating.

The degree of enhancement (fractionation) of the low FIP elements in the corona compared to the photosphere manifests itself also through spectroscopic measurements. More specifically, it has been shown that it can also be measured using the intensities of spectral lines from elements with different FIP such as Si (with low FIP) and S (with high FIP) [310]. The derived spectroscopy results of the coronal abundances are similar with the SW and SEP measurements but with slightly different numbers depending on the elements [311–313].

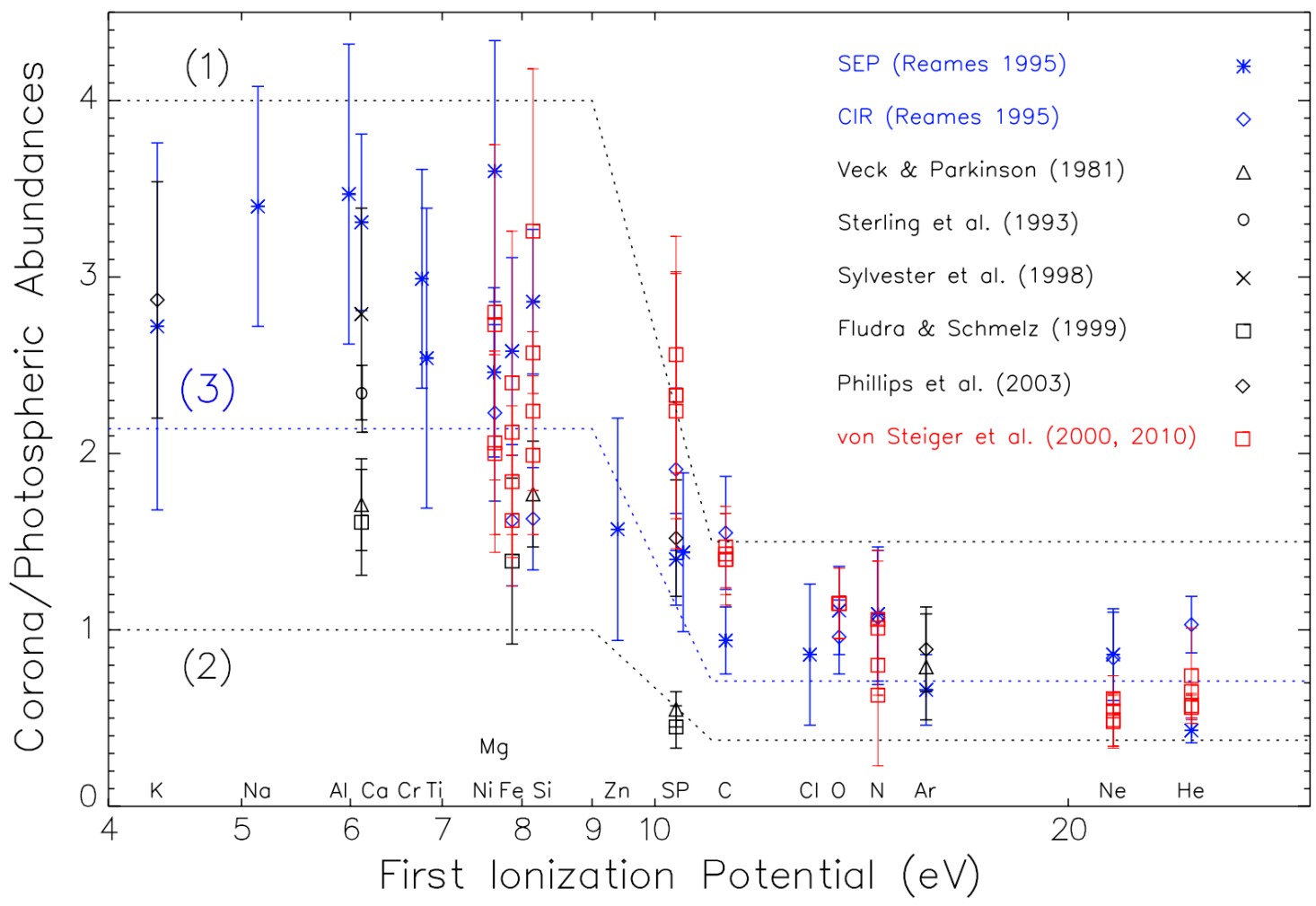

**Figure 11.1:** Coronal / photospheric elemental abundances as a function of FIP from SW (red), SEP (blue) and coronal spectroscopy (black) measurements. Each data point corresponds to 25 to several hundred events / measurements. The two dashed lines represent two empirical models where (1) is low-FIP elements enhanced by a factor of 4 with respect to the photospheric values and high-FIP elements the same as in the corona and the photosphere; and (2) is low-FIP elements the same in corona and photosphere and high-FIP elements depleted by a factor of 4 with respect to their photospheric values. The middle blue line (3) is the best fit to the data (34, 35).

It is also worth mentioning that the FIP effect appears also in other stars [314], and in some case it even appears reversed (*inverse FIP effect*) where the low FIP elements appear depleted relative to the high FIP elements [309, 315].

Last but not least, the solar coronal element composition is found to be related with solar cycle phase [316] and therefore an analysis on the temporal distribution of the ratio of element abundances between corona and photosphere, based on the assumptions of this work, as described in Sect. 4.2, could provide additional information on the origin of this anomaly (see also Publication E.17).

## 11.2 FIP-bias data

### 11.2.1 Data origin

The data that will be used in this analysis are obtained from [316]. They belong to spectroscopic measurements obtained by SDO which measures the solar EUV irradiance from 0.1 nm to 105 nm with 0.1 nm spectra resolution, 10 s temporal cadence and 20% accuracy [317]. From these measurements the average daily ratio of coronal to photospheric composition is calculated, which is called *FIP bias*:

$$\text{FIP-bias} = \frac{A_c}{A_p} \tag{11.1}$$

where: $A_c$ : the coronal abundance derived from Fe, Mg and Si,

$A_p$ : the photospheric abundance of Ne.

The selected period spans from 30/04/2010 to 11/05/2014 covering from the end of the extended solar minimum of cycle 23 to beyond the solar maximum of cycle 24.

Finally, the uncertainty in each daily average FIP bias measurement has been calculated to be $\sim 0.3$ [316].

### 11.2.2 Data curation

In the raw data, there are only a few days with unspecified values. More specifically, these correspond to the dates 19/05/2010 to 03/06/2010 (16 d) and 28/07/2010 (1 d). So in total we have 17 d out of the 1473 d i.e. 1.2% which have unspecified measurements. To correct them we have replaced these measurements with the mean value of five days before and five days after each gap. The resulting data are denoted as "corrected data" and are used in the rest of the analysis. The corrected along with the raw data are seen in Fig. 11.2.

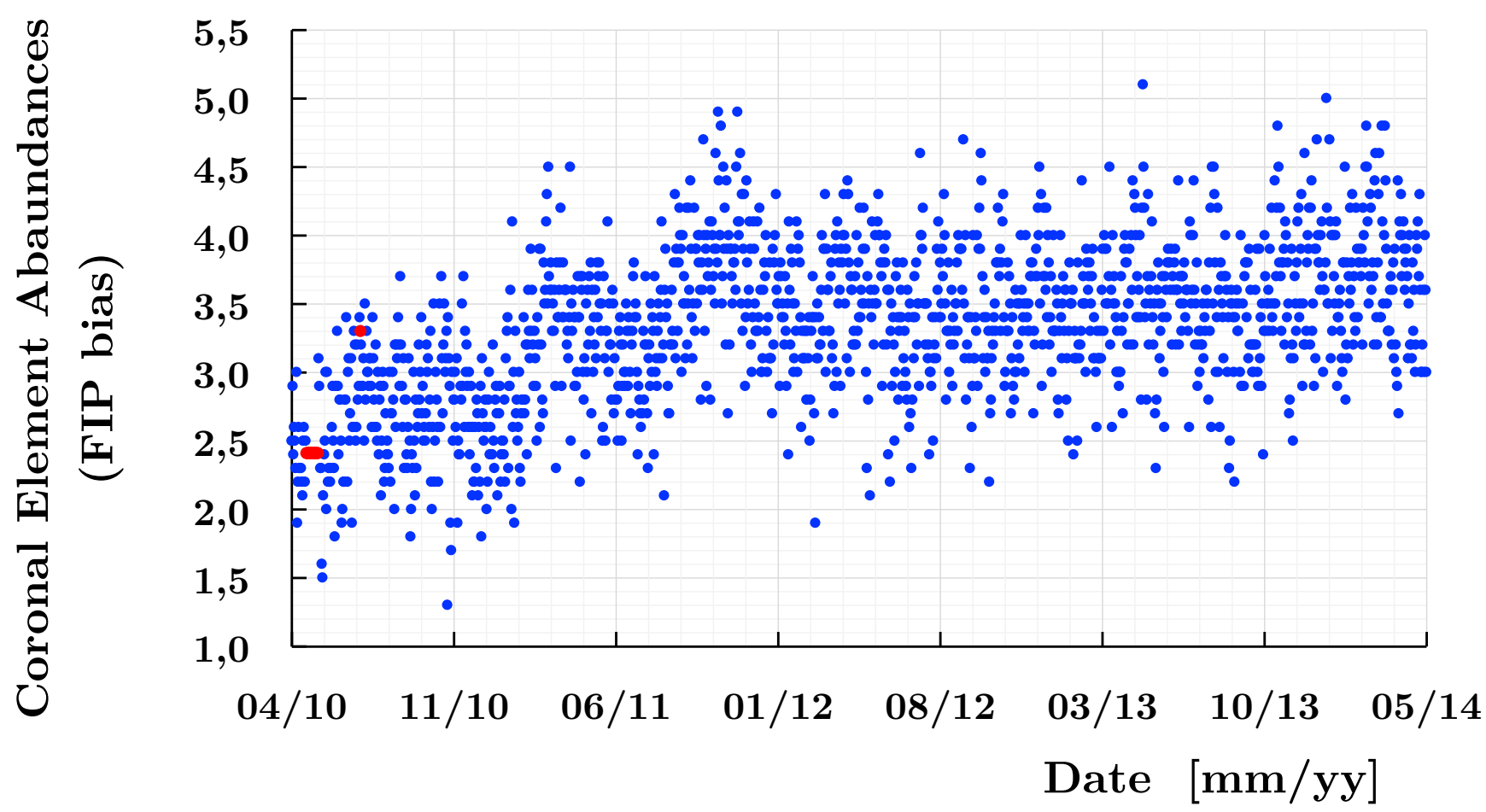

**Figure 11.2:** Temporal evolution of the daily average corrected FIP bias data. The 17 corrected values are seen in red.

It is noted that the analysis results that will be presented in the next sections do not change even if we use the raw data, but also if we only use the data from 04/06/2010 to exclude the 16 empty consecutive dates. The overall observed effects remain the same with just a small reduction of about 1.6% on the overall significance of some peaks.

In the raw FIP bias data the minimum value is 1.3 and the maximum 5.1. Also, for the total of 1456 d, we have a sum of 4852.3 giving an average of 3.3 per day. Accordingly, in the corrected data for a total of 1473 d, we have a sum of 4894.2 corresponding to an average of 3.3 FIP bias per day.

## 11.3 Data analysis and results

### 11.3.1 Planetary longitudinal distributions

#### 11.3.1.1 Single planets

To look for a possible planetary relationship on the daily FIP bias data, we have to assign the corresponding heliocentric longitude of each planet on each given day. Then, a projection has to be made on the full orbital cycle of each planet. For this reason, the corrected data are used for the whole period 30/04/2010 - 11/05/2014. The assigned daily planetary positions correspond to 12:00UTC.

In Fig. 11.3 the heliocentric longitude positions of Mercury, Venus, Earth, Mars as well as the Moon's phase have been used for the distribution of FIP bias data. The observed differences between the maximum - minimum points for the various plots are 5.8%, 15.3%, 13.6%, 18.7% and 5.5% for Mercury (Fig. 11.3a), Venus (Fig. 11.3b), Earth (Fig. 11.3c), Mars (Fig. 11.3d), and Moon (Fig. 11.3e) respectively. In all plots the calculated error bars are also visible. It is noted that the total number of days is 1473 and the sum of FIP bias values is 4894.2.

Based on the peaks observed, there seems to be a significant planetary relationship in Earth and Mars spectra. For example, in Earth shown in Fig. 11.3c, the statistical significance of the peak around 200° compared with the mean value of 3.3 is calculated to be $\sim 7.6\sigma$ whereas if in the calculation we exclude the maximum value of the peak, then the significance becomes $\sim 8\sigma$. A Gaussian fit on this peak gives a center value of about 188.2° with a FWHM of 92°. In the case of Mars in Fig. 11.3d the significance is even bigger, with the peak around 130° compared to the mean value giving a significance of $\sim 11.3\sigma$. The wide peak in the middle of Mars' spectrum is fitted with a Gaussian function and is found to be located around 122.5° with a $FWHM = 136.3°$.

#### 11.3.1.2 Combining planets

We then move on to different planetary distributions while using a combination of planets in order to search for an enhanced phenomenon in specific longitude regions.

The most interesting results are derived when constraining the longitudinal positions of Venus, Earth and Mars while the reference frame is fixed on Mercury's orbit (Fig. 11.4). The

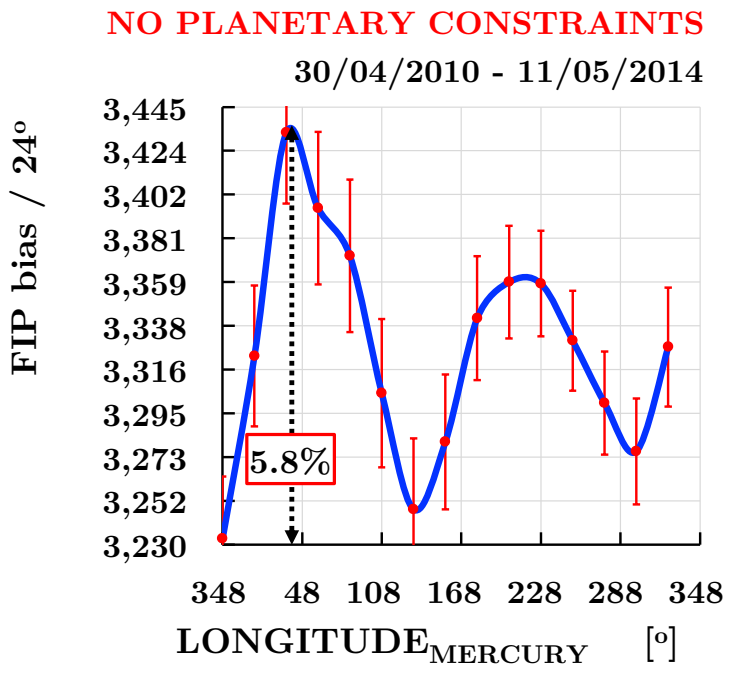

**(a)** Mercury for 24° bin.

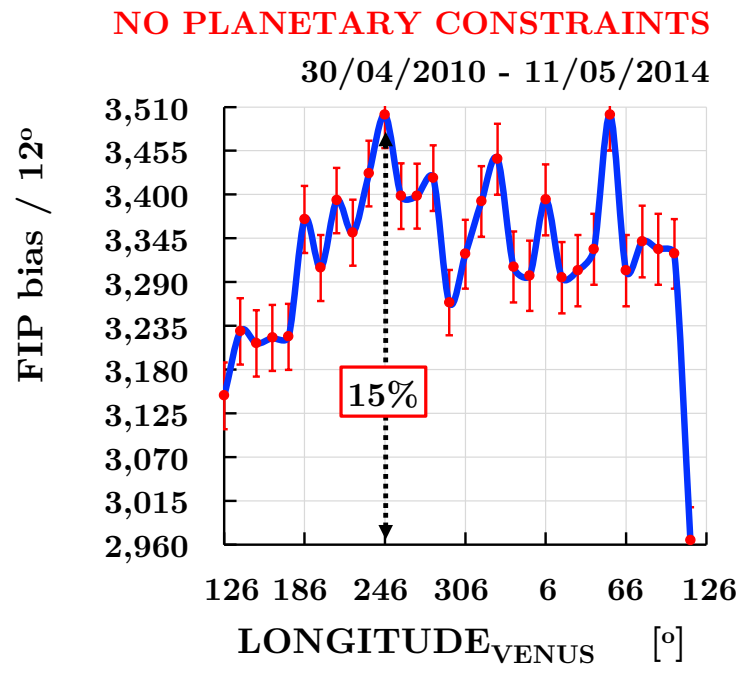

**(b)** Venus for 12° bin.

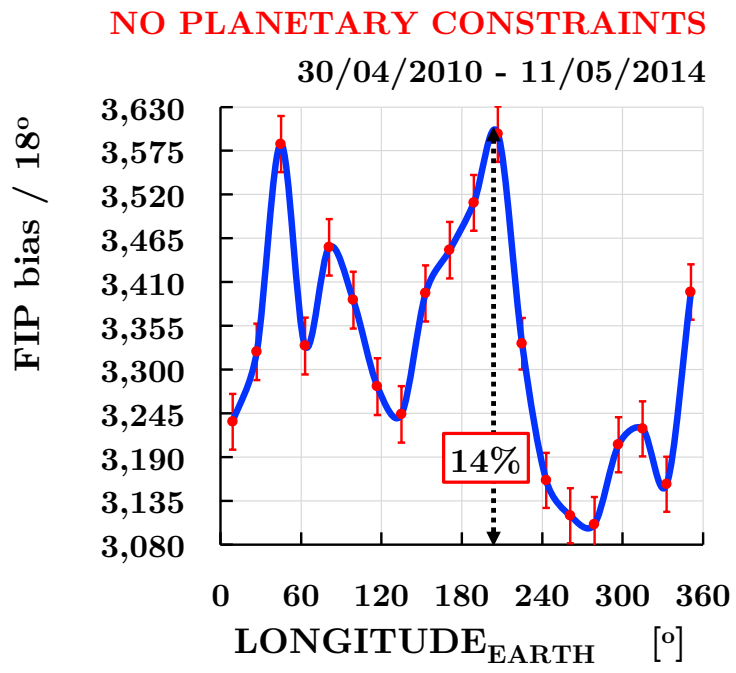

**(c)** Earth for 18° bin.

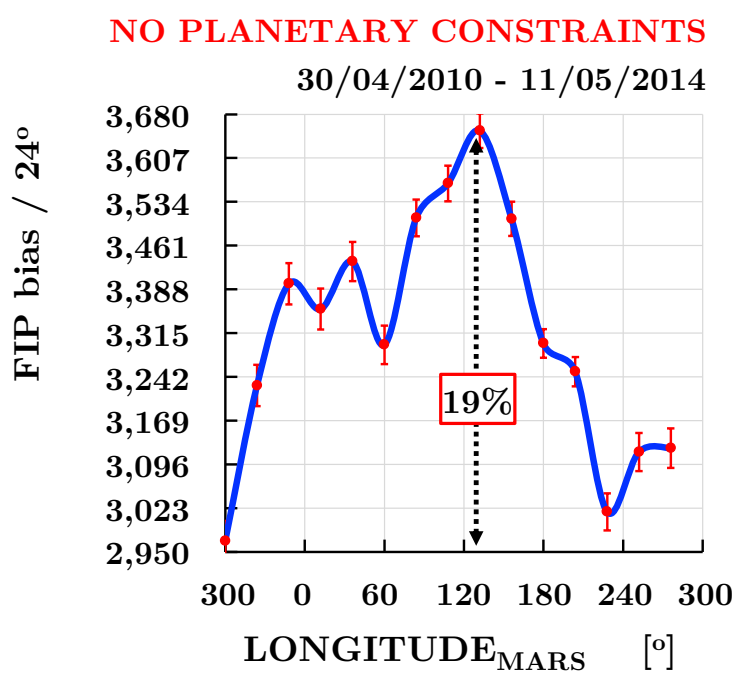

**(d)** Mars for 24° bin.

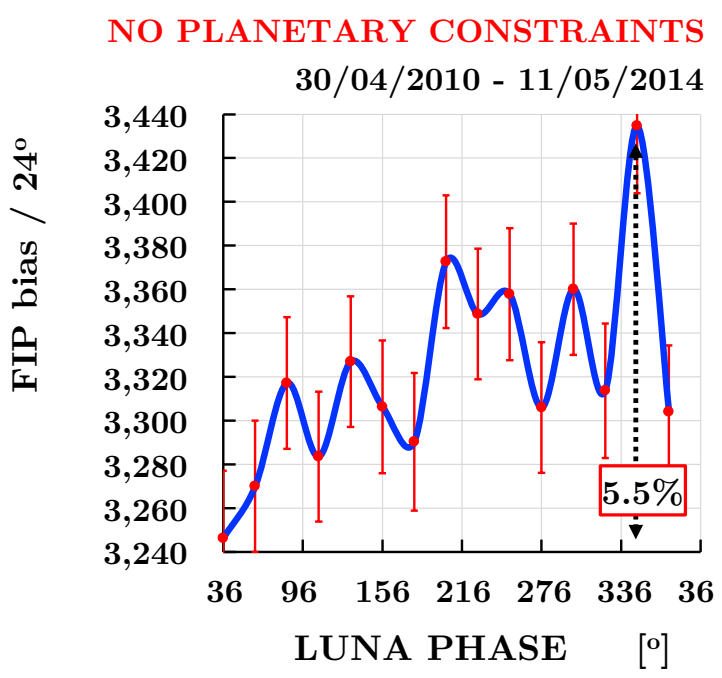

**(e)** Moon's phase for 24° bin.

**Figure 11.3:** Planetary heliocentric longitude distributions of FIP bias for the period 30/04/2010 - 11/05/2014.

derived numbers for the overall FIP bias and selected days for each case are 1256.3 and 376 d, 1345.2 and 398 d, 1496.8 and 473 d for Figures 11.4a, 11.4b and 11.4c respectively. For the case of Venus being around 30° ± 50° in Fig. 11.4a we have a total amplitude of 23.7%, with the peak around 36° having a 7.5$\sigma$ difference from the mean value. This is a major difference from the case where no constraints where applied in Fig. 11.3a where the significance of the same peak was at about 3$\sigma$. Thus, this observation confirms our initial expectations of even more significant results when using this procedure with the combined effect of more planets. Similarly, for the case of the Earth in Fig. 11.4b propagating around 50° ± 50° the observed amplitude in Mercury's distribution is 18.6% with the same peak in Mercury (around 36°) having this time a 5.4$\sigma$ difference from the mean value of 3.4. Lastly, when Mars in Fig. 11.4c is allowed to move in the heliocentric longitude window around 230° ± 50° the total derived maximum to minimum difference is 17% with the significance being this time somewhat smaller than the other two cases, at about 3.9$\sigma$. Due to the presence of multiple big peaks the mean value increases and therefore the calculated significance between peak and mean value is by default expected to be smaller. On the other hand, if a comparison is made with the minimum value on the base of the peaks the significance is above 5$\sigma$.

As next we move on to the reference frame of Venus where in Fig. 11.5 the longitudinal position of Earth has been constrained around 50° ± 50°. In this case we have an overall maximum - minimum difference of 23.6% while the total FIP bias is 1345.2 and the selected

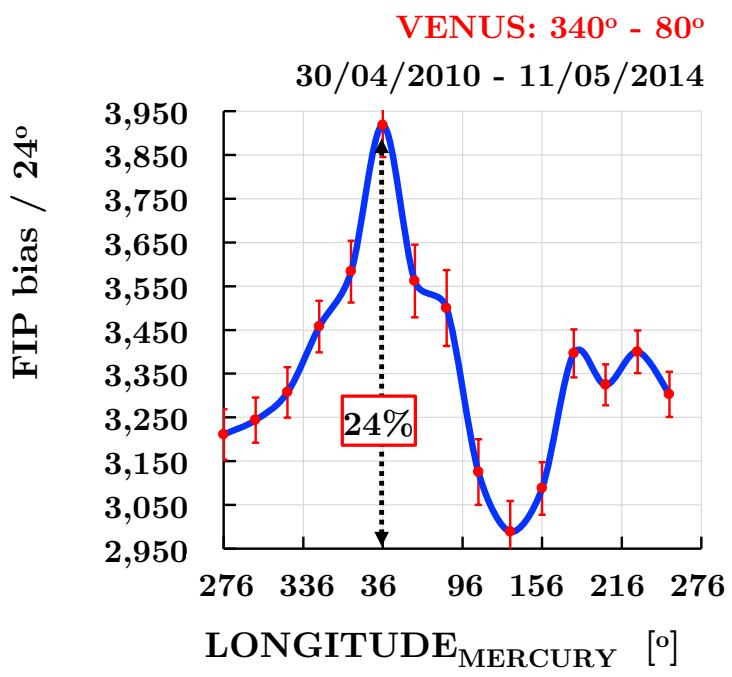

**(a)** Mercury when Venus is between 340° to 80°.

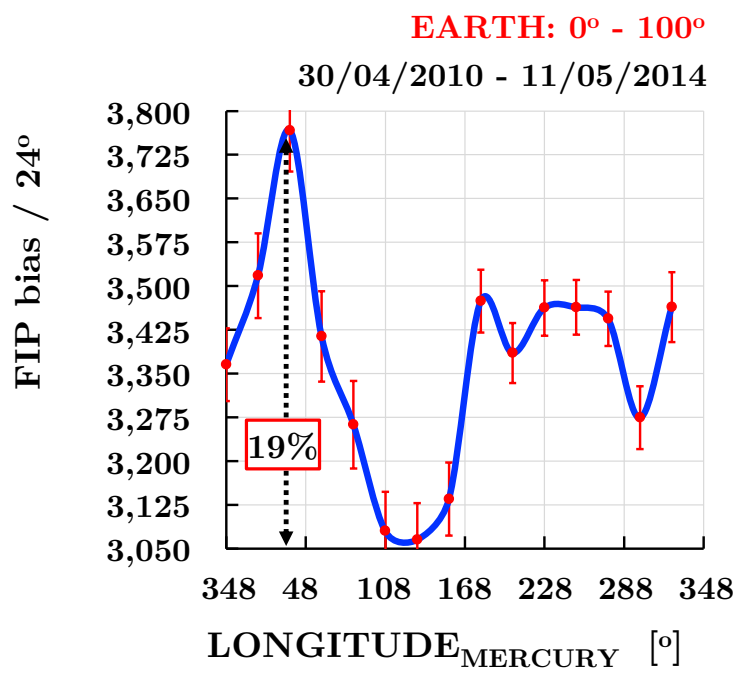

**(b)** Mercury when Earth is between 0° to 100°.

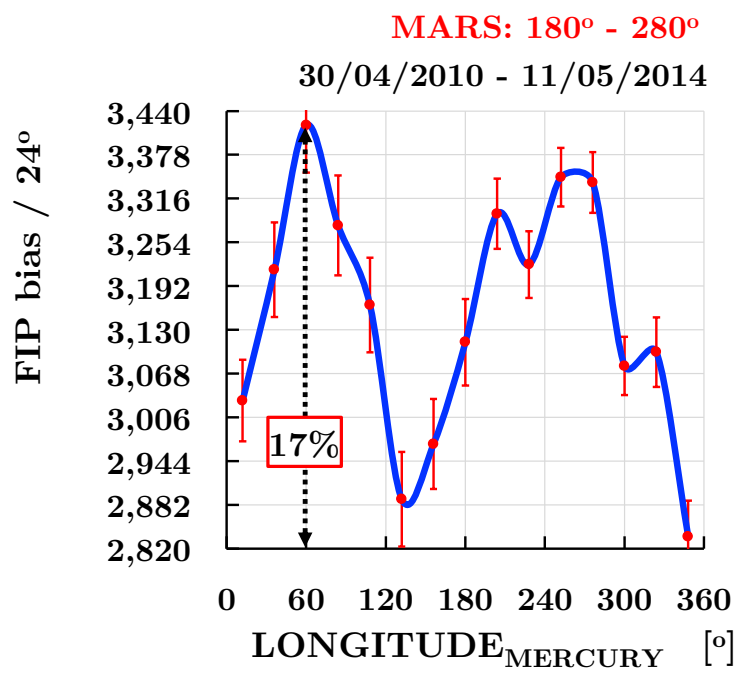

**(c)** Mercury when Mars is between 180° to 280°.

**Figure 11.4:** FIP bias distribution for the reference frame of Mercury when other planets are constrained to propagate in a specific longitude region for bin = 24°

days are 398. The distinct wide peak observed around 276° has significance of about 7.97$\sigma$ when compared to the mean value of 3.4 which, once more, is much better than Fig. 11.3b where no highly significant peak was observed.

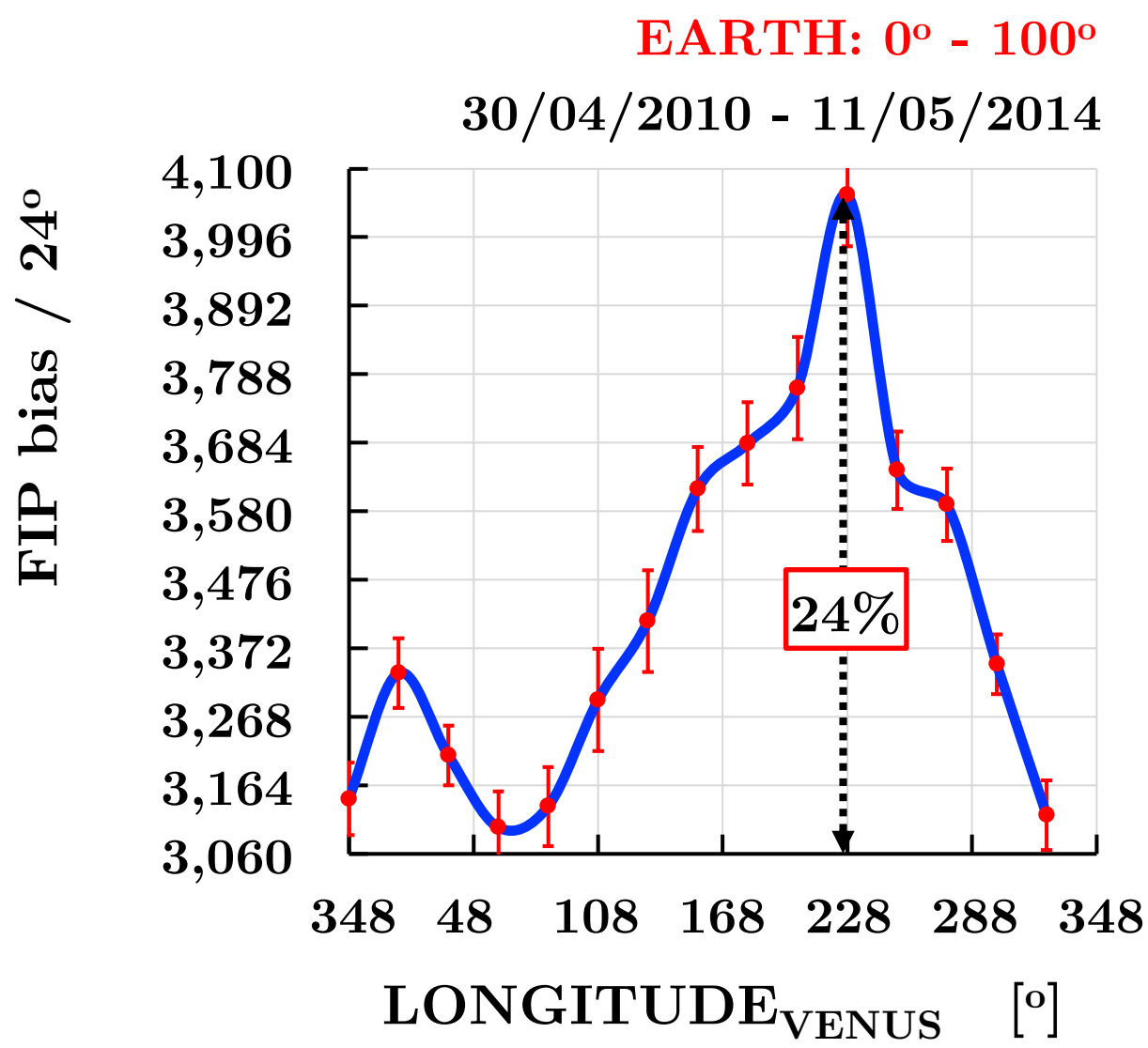

**Figure 11.5:** FIP bias distribution as a function of Venus' position while Earth is allowed to propagate between 0° to 100° for bin = 24°.

Finally, in Fig. 11.6 we look at the position of Earth while at the same time Venus is allowed to propagate in a 120 degree wide window around 310°. The number of days corresponding to this case is 489 having a sum of 1644.9 FIP bias. The difference between the maximum and minimum point is about 31.2%. The statistical significance of the maximum point on the peak around 45° compared to the mean value of 3.4 is 11$\sigma$ when in Fig. 11.3c it was calculated to be 7.2$\sigma$.

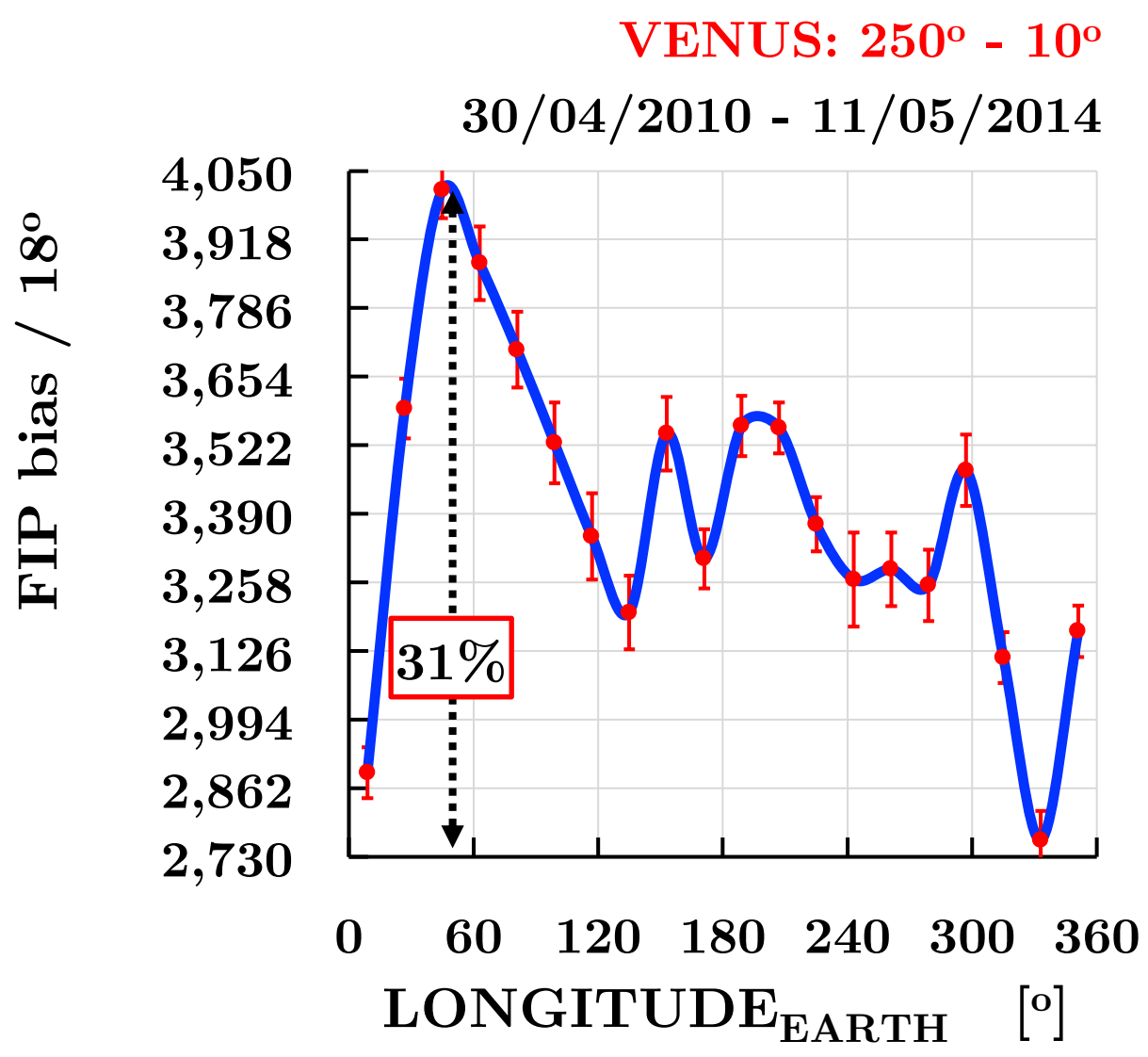

**Figure 11.6:** FIP bias distribution as a function of Earth's position while Venus propagates between 250° to 10° for bin = 18°.

### 11.3.2 Fourier analysis

An additional strategy to search for underlining periodicities is through the Lomb-periodogram spectral analysis. The corrected data have been used here for the full period 30/04/2010 - 11/05/2014. The procedure involves the comparison of the derived periods with the various revolution periods of the solar system objects and the planetary synods.

The biggest in amplitude peak that is derived, corresponds to 796.4 d and has a power of 105.4 dB (see Fig. 11.7a). From a Gaussian fit we find a FWHM of 311.8 d which gives an error of about ±132.4 d. This means that a variety of periods overlap with this peak, including Mars revolution period of 686.98 d and the synods Mars - Pluto (692 d), Mars - Neptune (695 d), Mars - Uranus (703 d), Mars - Saturn (734 d), Mars - Earth (780 d) and Mars - Jupiter (816 d).

Moreover, the 3$^{rd}$ biggest peak with a power of about 15.2 dB has a period of 373.2 d±23.1 d (see Fig. 11.7b) which is $\leq 1\sigma$ with the Earth's revolution period of 365.25 d.

Finally, the 9$^{th}$ biggest peak is located around 148.2 d ± 3.9 d and has an amplitude of 5.67 dB (see Fig. 11.7c). The synodic period of 145 d corresponding to Mercury - Venus is fitting in.

These results indicate that the already derived planetary correlations with Earth and Mars which show the most significant relationship, as well as the combination of Mercury and Venus, are supported also by the Fourier analysis. Thus, this result strengthens the hypotheses of this work indicating an involvement of the dark sector through gravitational focusing by the planets towards the Sun.

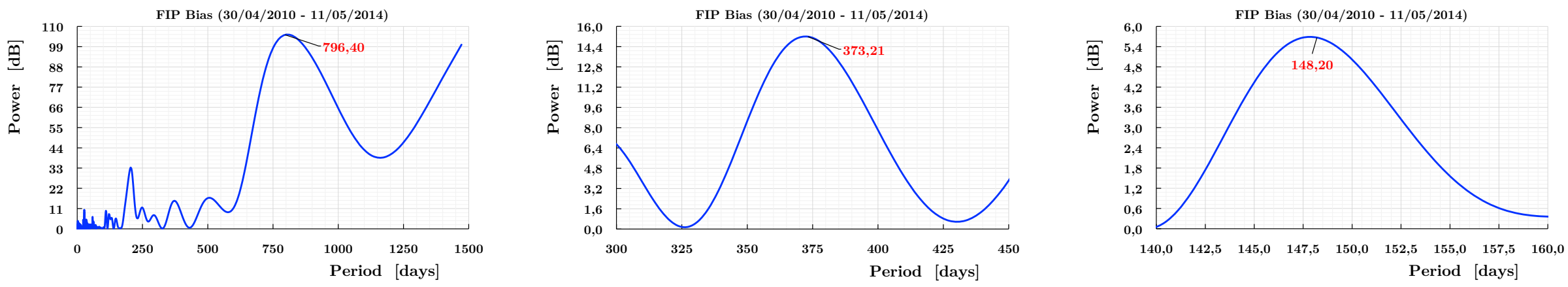

**(a)** Around 796.4 d. **(b)** Around 373.21 d. **(c)** Around 148.20 d.

**Figure 11.7:** Fourier periodogram of FIP bias for the period 30/04/2010 - 11/05/2014 zoomed in around a few interesting periods.

## 11.4 Summary

Based on the statistical analysis on the ratio of the coronal element abundances to the photospheric ones, a significant planetary relationship has been observed. This strengthens further the assumption of this work where invisible slow-moving massive matter interacts "strongly" with the Sun's atmosphere after it has been gravitationally focused by the planets and the Sun itself via free fall. This behaviour is depicted primarily in the distributions of Earth and Mars but also when combining multiple planets. Such a result is unexpected within conventional physics since any additional remote interactions are already excluded. The Lomb-periodogram spectral analysis provides an additional support on the claims of this work verifying the discovered periodicities.

The comparison with other solar observations such as with F10.7 solar radio proxy as well as with EUV solar irradiance exhibits several similarities in the distributions of Earth and Mars as well as in the combined analysis of Mercury-Earth and Earth-Venus. However, interesting differences are derived on the rest of the distributions as quantified also by statical correlation analysis (see Appendix Sect. B.7).

These striking unexpected results deserve further investigation with the addition of more data from more datasets as well as an analysis in a shorter time-cadence (see also Publication E.17). This has the possibility to provide better insights on the nature of the assumed interaction of invisible massive stream(s) with Sun, but also on the correlation of the coronal element abundances with other solar observables, therefore exploring alternative solutions to long-standing solar mysteries.

12

# LYMAN-ALPHA

## 12.1 Introduction

The hydrogen Ly-$\alpha$ line at 121.567 nm is the most intense solar vacuum UV emission line. This line results from the $2p - 1s$ transition in hydrogen (see Fig. 12.1) and is the main excitation source for atomic hydrogen resonant scattering in cool material in the solar system [318]. It is produced in the transition region of the Sun and radiated to the upper chromosphere where coherent scattering results in spectral broadening [319]. Due to a high abundance of neutral hydrogen in the solar chromosphere, the Ly-$\alpha$ line is optically thick and has a width of about 0.1 nm with a 30% central self-reversal dip [320]. Its variability over a solar cycle is about a factor of two [321].

It has already been shown that Ly-$\alpha$ can serve as a proxy measurement for solar activity [322] and, the other way around, as a proxy for other solar emissions [323, 324]. Lastly, a correlation has been made with the solar flares [325]. Therefore, based also on the significant planetary relationship found on a variety of solar observables such as the number of flares, EUV, the number of sunspots, and F10.7 radio flux presented in Chapters 6, 7, 8 and 9 respectively, the application of the analysis procedure also on the Ly-$\alpha$ irradiance is also expected to provide some interesting results.

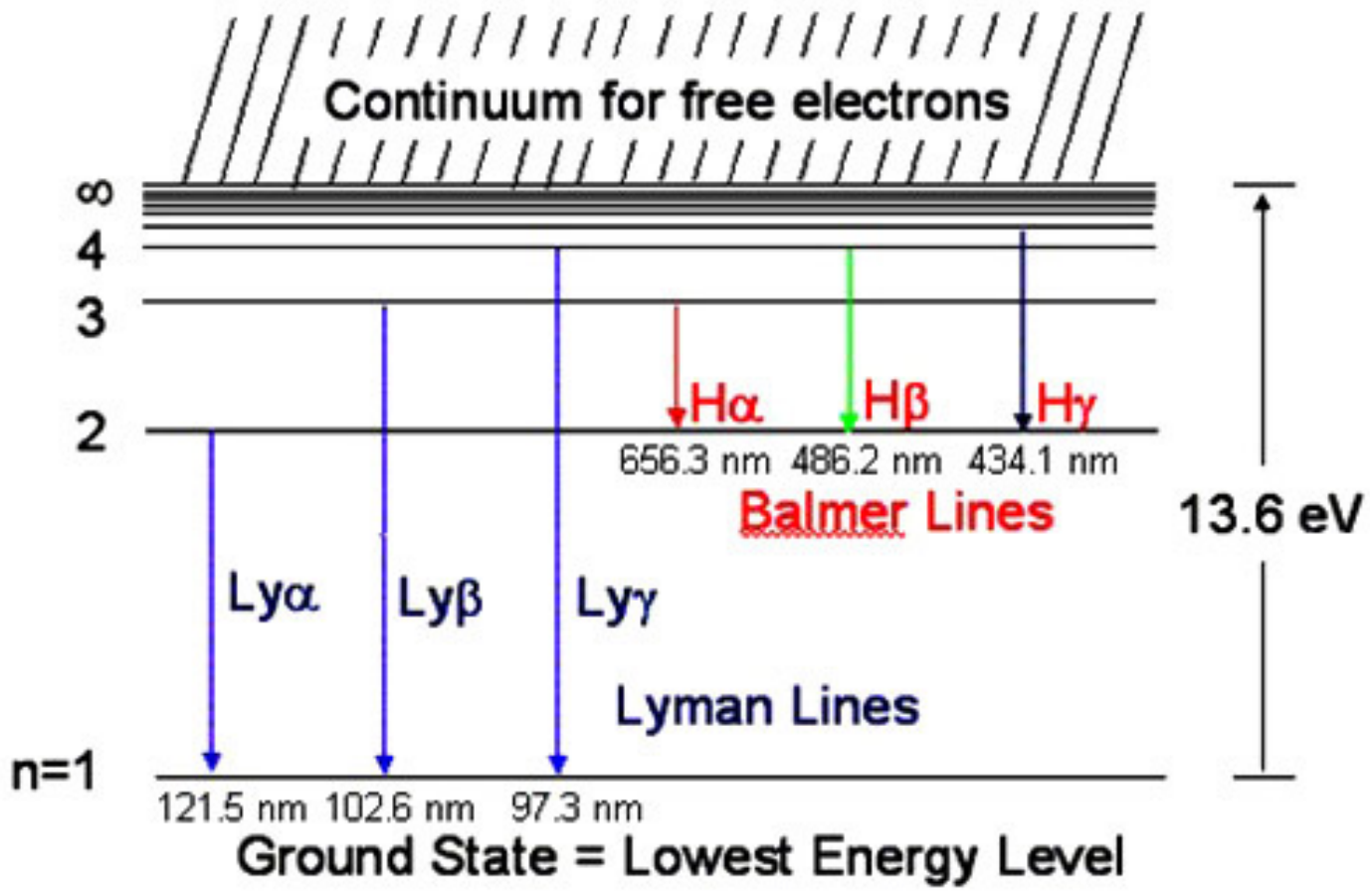

**Figure 12.1:** Energy levels of the hydrogen atom with some of the transitions between them that give rise to the indicated spectral lines. The Lyman series observed are in the UV range while the Balmer series are in the visible range (36).

## 12.2 Ly-$\alpha$ data

### 12.2.1 Data origin

The raw open-access Ly-$\alpha$ data have been downloaded from Lasp Interactive Solar Irradiance Datacenter (LISIRD) and belong to Laboratory for Atmospheric and Space Physics (LASP) of the University of Colorado [326]. The acquired version-4 data [318] have a cadence of 24 h, and they span from 14/02/1947 to 08/04/2021 which is in total 27 083 d. The acquired composite Ly-$\alpha$ time series are shown in Fig. 12.2.

The solar Ly-$\alpha$ irradiance data are scaled at 1 AU and include average daily measurements from multiple instruments and models that construct a long time series history of the full disk solar irradiance integrated from 121 nm to 122 nm which is dominated by the bright solar H I 121.6 nm emission. The corresponding units are $W/m^2$, and therefore the values are identical to the solar spectral irradiance in 1 nm centred at 121.5 nm with units of $W/m^2/nm$. Finally, each average daily value also contain a corresponding $1\sigma$ uncertainty value in units of $W/m^2$ [326].

The minimum daily irradiance is $5.6 \times 10^{-3}$ $W/m^2$ and the maximum $1.2 \times 10^{-2}$ $W/m^2$. Also, the average irradiance per day is $7.4 \times 10^{-3}$ $W/m^2$ with the total sum for all $\sim 74$ y being about $2 \times 10^2$ $W/m^2$.

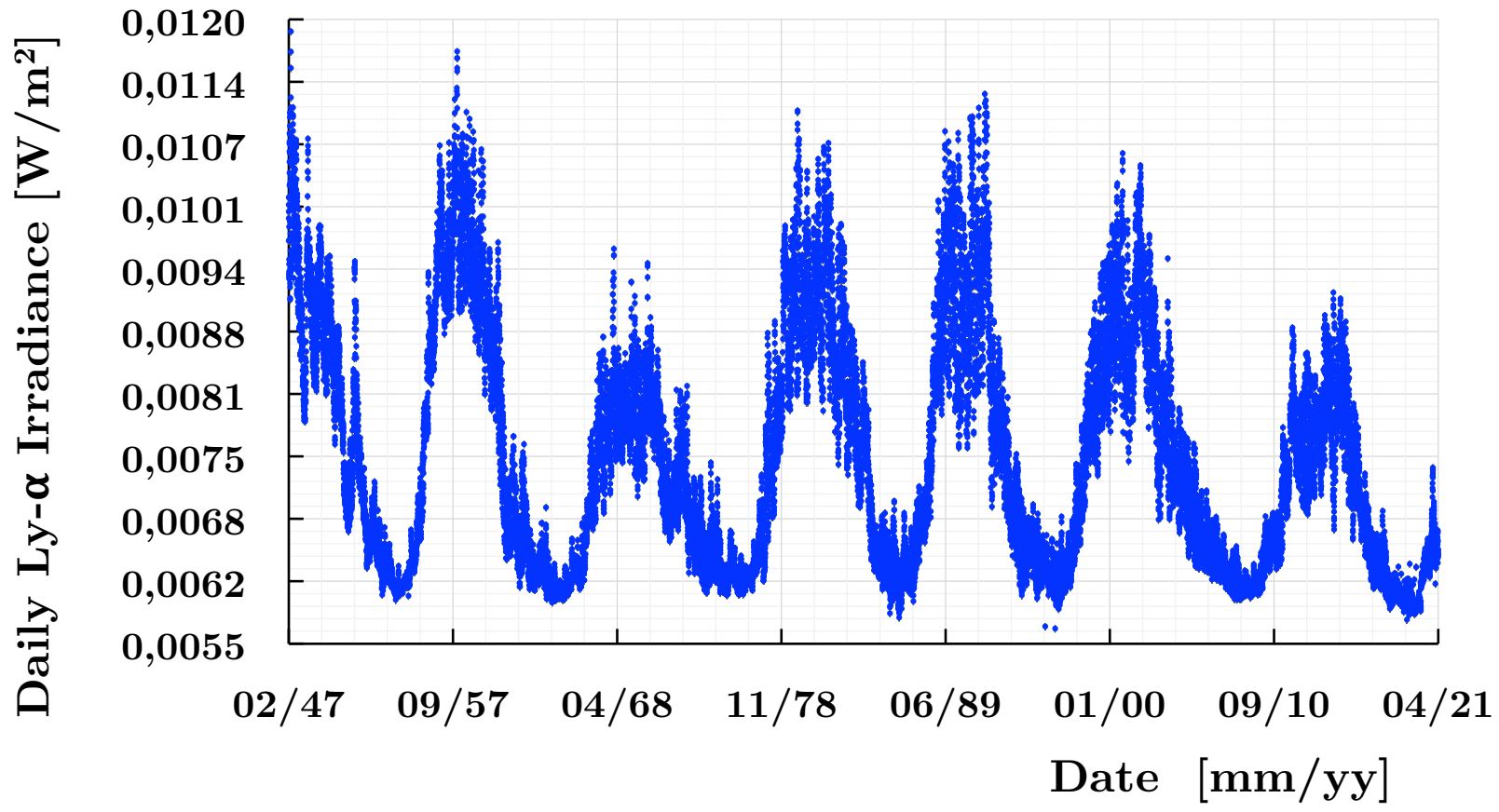

**Figure 12.2:** Temporal evolution of he daily average Ly-$\alpha$ solar irradiance for the period 14/02/1947 to 08/04/2021.

# 12.3 Data analysis and results

## 12.3.1 Planetary longitudinal distributions

To search for a possible planetary modulation of the distribution of Ly-$\alpha$, the daily data are projected on the planetary heliocentric longitude position of each planet. For the y-axes of the plots to be better readable, the raw data along with the errors have been multiplied with $10^3$. Therefore, the resulting units have become $10^{-3}$ W/m$^2$.

It is noted that during the $\sim 74$ y period, Mercury has performed about 307.9, Venus 120.5, Earth 74.2, Mars 39.4, Jupiter 6.2 and Saturn 2.5 revolutions around the Sun, while Moon has performed 919 revolutions around the Earth. This means that any randomly occurring effect should average out due to the many revolutions which can be viewed as a repeating measurement.

### 12.3.1.1 Single planets

In Fig. 12.3 individual planets without any additional constraint on the position of the rest of the planets are used for Ly-$\alpha$. The observed difference between the maximum to minimum points for the distributions of Mercury, Venus, Earth, Mars, Jupiter and Saturn are 0.8%, 1.3%, 0.7%, 2.4%, 21.8% and 23.5% respectively. It is noted, that for the case of Earth in Fig. 12.3c if the bin becomes smaller, in the order of 2° to 6° then a modulating behaviour is observed. The error bars in each point are also visible in the plots. Due to the relatively big errors, in the cases of Mercury (Fig. 12.3a), Venus (Fig. 12.3b) and Earth (Fig. 12.3c) the statistical significance of the peaks observed is relatively small on the order of 2 to 3$\sigma$. However, in Mars shown in Fig. 12.3d, the significance of the second peak around 130° compared to the mean value of 7.4 has a 5.3$\sigma$ significance, and if we exclude the maximum point of the peak the significance increases to 5.7$\sigma$. On the other hand, in Jupiter and Saturn the distinct peaks observed have a very highly statistical significance of about 54$\sigma$ and 26$\sigma$ respectively when compared to the mean value of 7.4. For the case of Jupiter in Fig. 12.3e, the wide peak around 159.1° in Fig. 12.3e has been fitted with a Gaussian function which gives a FWHM of $\sim$ 171.3°. For Saturn, shown in Fig. 12.3f, three peaks are observed around 50.7° $\pm$ 2°, 169.6° $\pm$ 4.3° and 276.4° $\pm$ 4.1° with their FWHM being accordingly 77.9° $\pm$ 10.7°, 83.1° $\pm$ 11.3° and 90.5° $\pm$ 15.5°. It is noted, that a similar behaviour for Jupiter and Saturn was observed also in F10.7 (Fig. 9.3) and sunspots (Fig. 8.6).

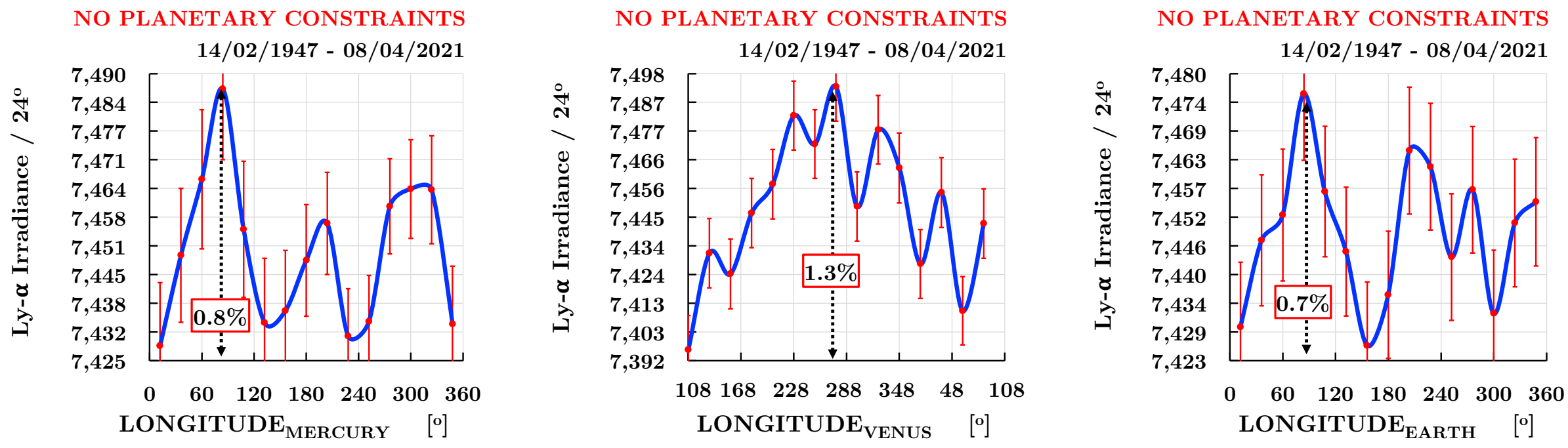

(a) Mercury for 24° bin. (b) Venus for 24° bin. (c) Earth for 24° bin.

(d) Mars for 24° bin. (e) Jupiter for 12° bin. (f) Saturn for 12° bin.

**Figure 12.3:** Planetary heliocentric longitude distributions of Ly-$\alpha$ irradiance for the period 14/02/1947 - 08/04/2021.

#### 12.3.1.2 Combining planets

The next step is to try search for an enhanced and more significant effect, especially on the inner planets, by constraining the longitudinal position of one of the remaining planets as explained in Sect. 4.5.1.1. In Fig. 12.4 Mercury's heliocentric longitude is selected as the reference frame while Mars is constrained to propagate between 310° $\pm$ 60 degree and the 180° opposite orbital arc around 130° $\pm$ 60 degree. For the first case in Fig. 12.4a the overall amplitude is about 3.7% with the peak in the middle around 270° being about 6.4$\sigma$ compared to the mean value of 7.4. The total number of days corresponding to this case is 7813 with the sum of the Ly-$\alpha$ irradiance being 57.9 W m$^{-2}$. On the other hand, for the second case in Fig. 12.4b the overall amplitude is very small at around 1.4% with no significant peaks observed. In this case we have 76.9 W m$^{-2}$ in 10 287 d. The distinct difference between the two cases supporting the assumptions of this work, pointing to a preferred direction of Mars, is shown in Fig. 12.4c. Even more importantly, in Fig. 12.4d the importance of the application of constraints on the longitudinal position of a second planets is depicted, which, through the comparison with the case when no constraints are applied from Fig. 12.3a, strengthens further the streaming scenario.

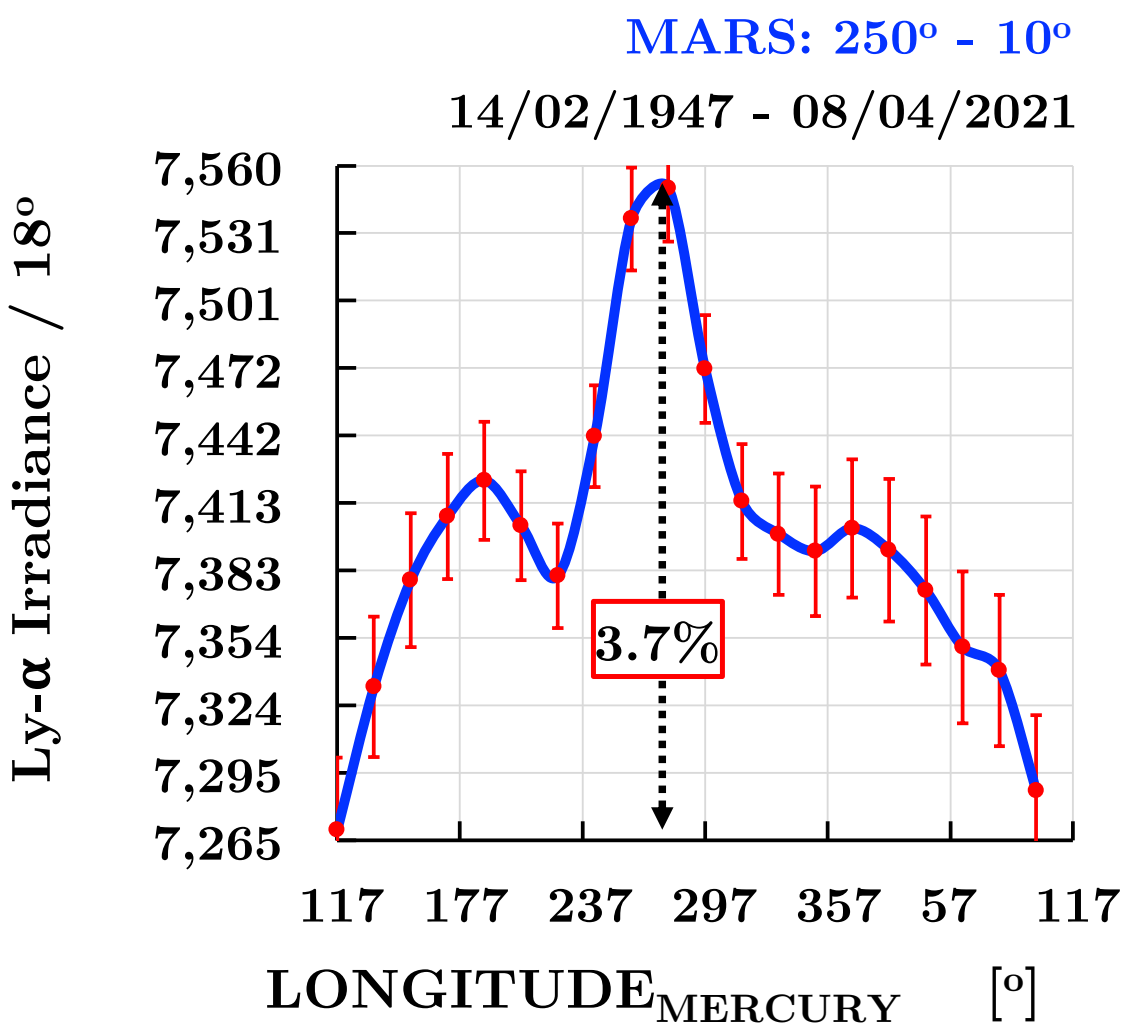

(a) Mercury when Mars is between 250° to 10°.

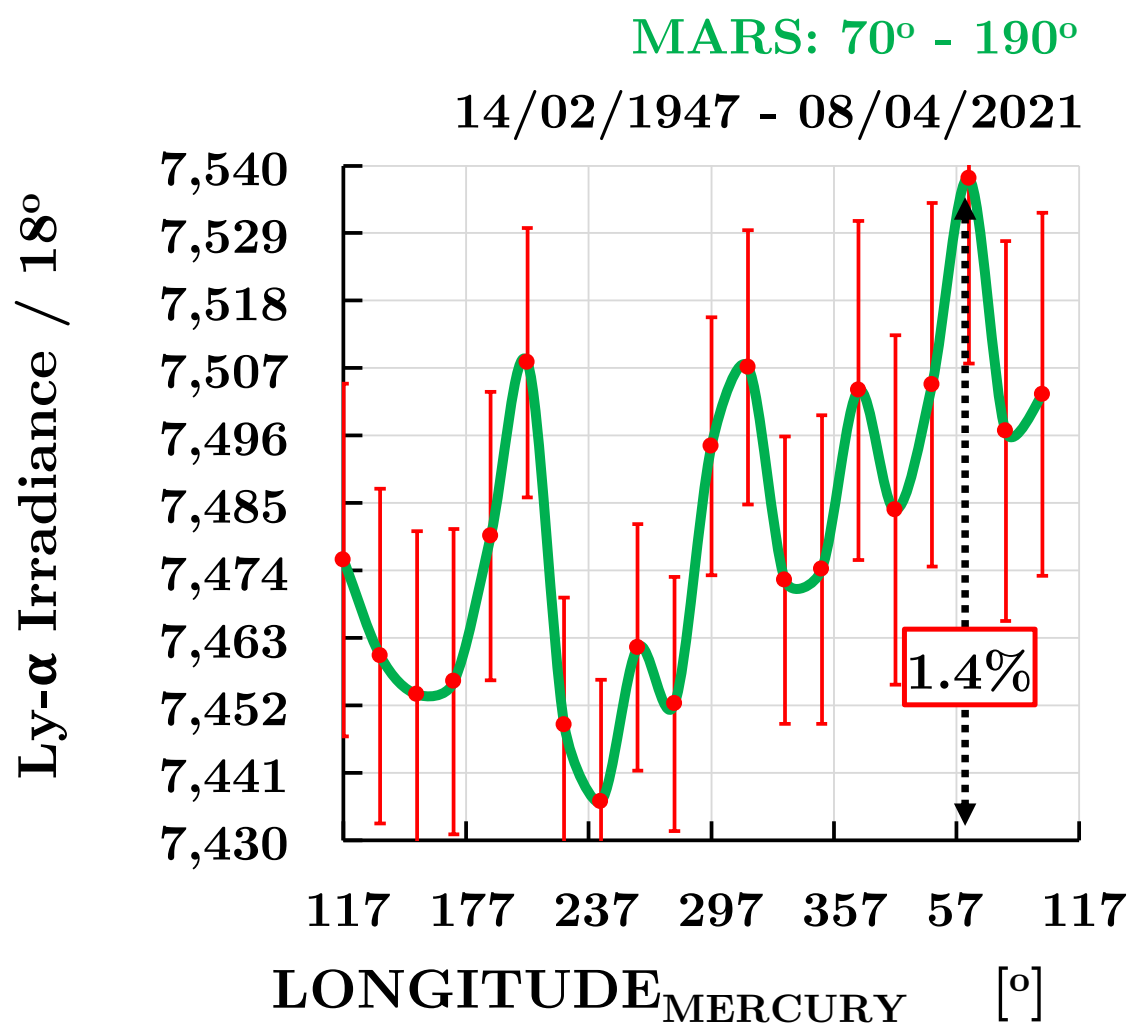

(b) Mercury when Mars is between 70° to 190°.

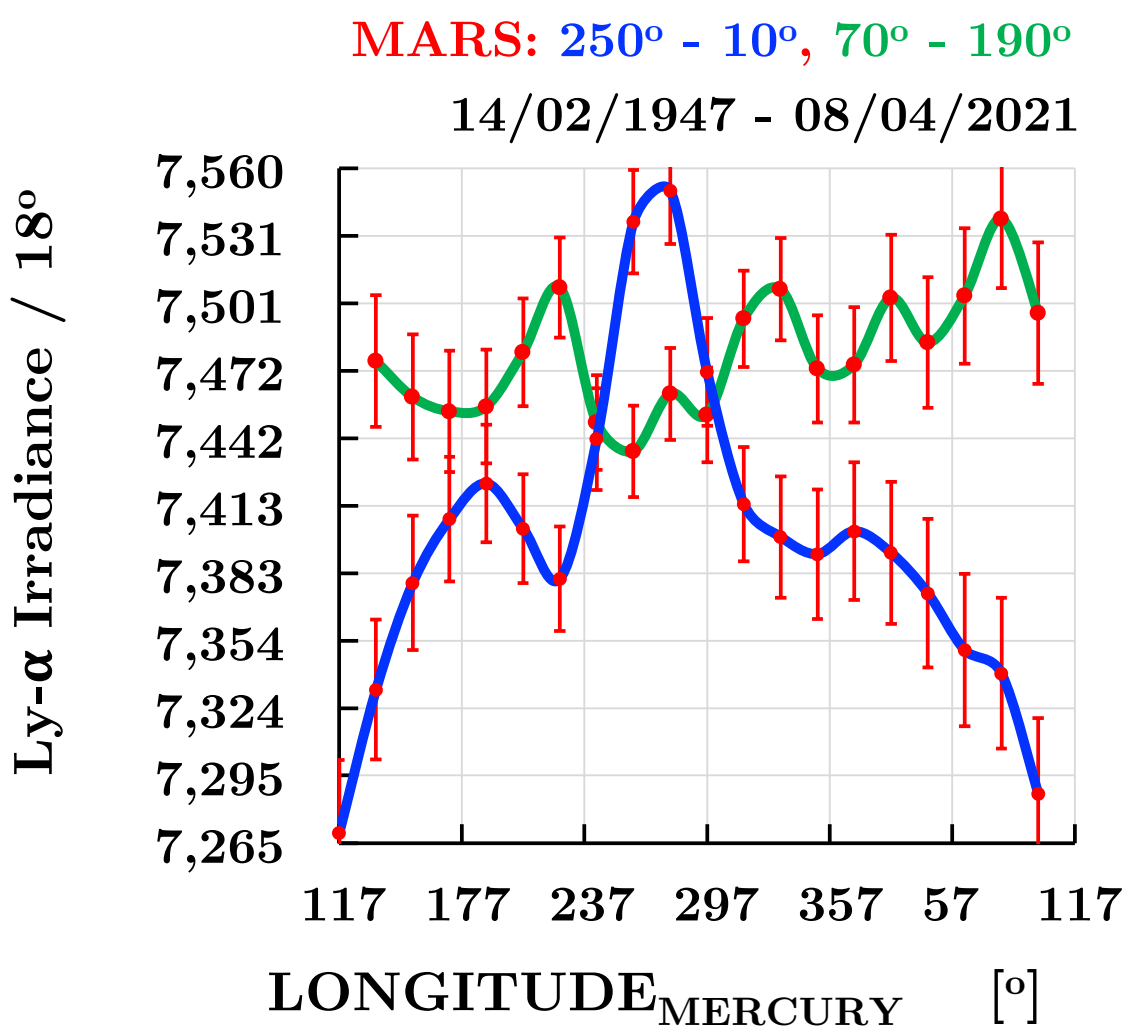

(c) Comparison between Fig. 12.4a and Fig. 12.4b.

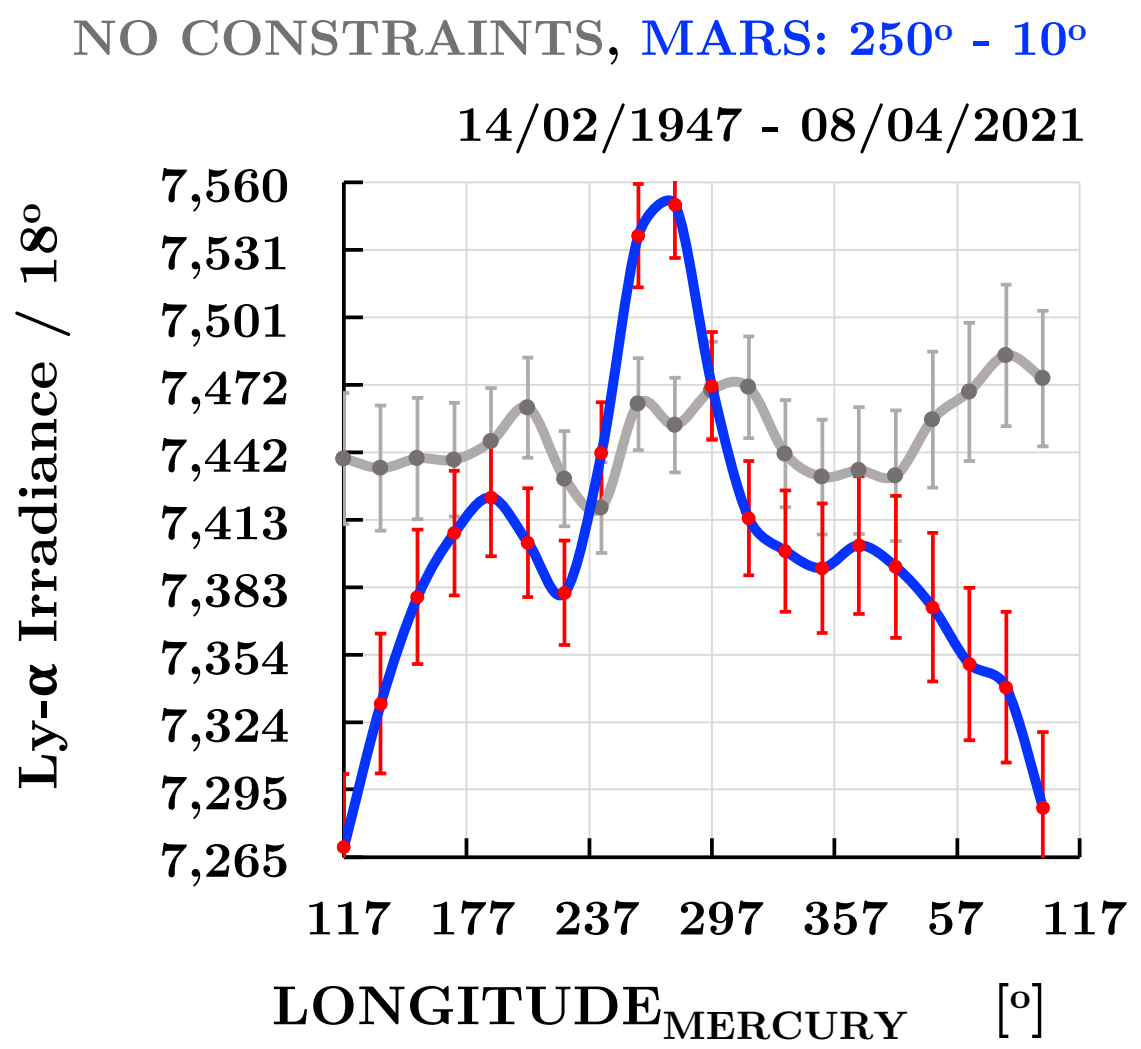

(d) Comparison between Fig. 12.4a and Fig. 12.3d where no constraints are applied.

**Figure 12.4:** Ly-$\alpha$ irradiance distribution for the reference frame of Mercury when Mars propagates between 180° opposite orbital arcs with 120° width, for 12° bin.

In Fig. 12.5, the reference frame is changed to Venus's position, while constraining the Earth and Mars to move in two 100°-wide orbital arcs. These distributions should once more be compared with the unconstrained case in Fig. 12.3b where no significant peaks were observed.

Instead, in Fig. 12.5 we see in all three cases some highly significant peaks. More specifically, when Earth is constrained to propagate around $150° \pm 50°$ in Fig. 12.5a we have 7413 d fulfilling the constraint with a total of $55.2\,\mathrm{W\,m^{-2}}$ giving a 7.2% amplitude and a $7.5\sigma$ significance of the middle peak compared to the mean value of 7.4. From a multiple fit Gaussian we find that the three peaks are centred around $94.6° \pm 4°$, $229.9° \pm 2.7°$ and $343.3° \pm 3.6°$ and have a FWHM of $65.4° \pm 21°$, $63.1° \pm 9.1°$ and $89.4° \pm 14°$ respectively.

For the second case of Mars being constrained around $280° \pm 50°$ in Fig. 12.5b we have 6771 d fulfilling the constraint with a total of $49.9\,\mathrm{W\,m^{-2}}$ giving a 8.7% maximum - minimum difference and a $10.5\sigma$ on the middle peak. The three observed large peaks are centred around $119.9° \pm 3.9°$, $233.6° \pm 2.6°$ and $356.1° \pm 4.1°$ and have a FWHM of $92.4° \pm 36.1°$, $78.5° \pm 7.9°$ and $104.4° \pm 44.6°$ respectively.

Then, when we combine the effect of Fig. 12.5a and 12.5b and we apply both of these positional constraints in Fig. 12.5c we end up with a single 154.3°-wide peak which is highly significant at about $24\sigma$. This peak is located around 219.4° and has a 33.9% amplitude, with $12.9\,\mathrm{W\,m^{-2}}$ in 1734 d.

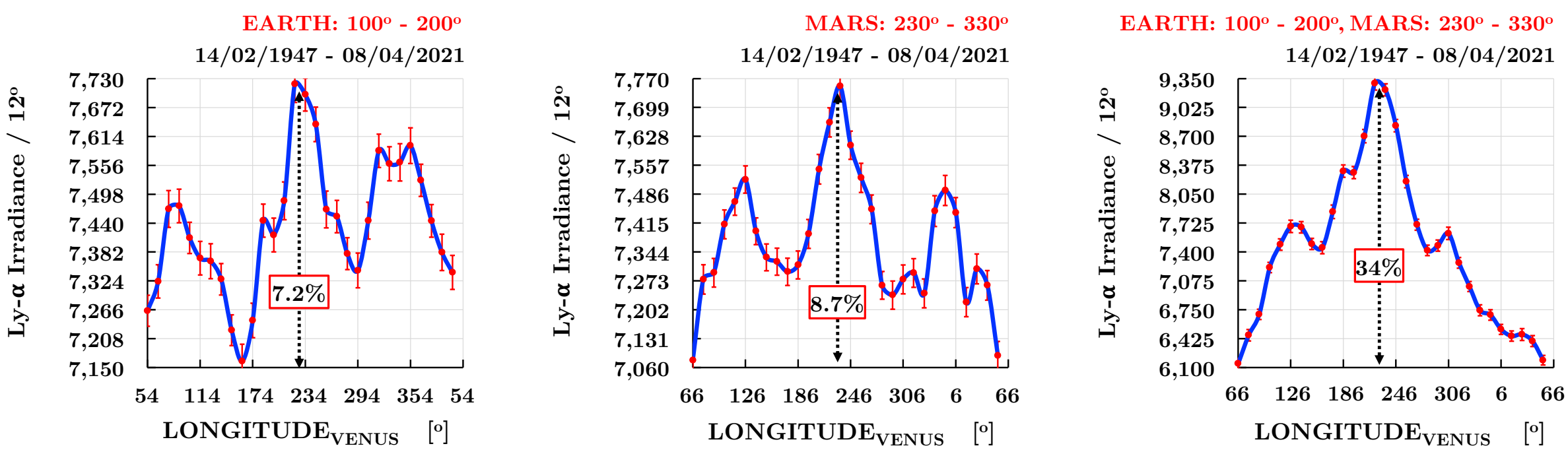

**(a)** Venus when Earth is between 100° to 200°.

**(b)** Venus when Mars is between 230° to 330°.

**(c)** Venus when Earth is between 100° to 200° and Mars between 230° to 330°.

**Figure 12.5:** Ly-$\alpha$ irradiance distribution for the reference frame of Venus when Earth and Mars are constrained in some 100°-wide regions, for 12° bin.

Likewise, in Fig. 12.6 we move to the Earth's reference frame. Here, a significant difference is observed between the case where no constraints are applied, as in Fig. 12.3c, and when Mars' position is constrained around $310° \pm 60°$. This orbital arc was chosen to be the same as in Fig. 12.4d. The total number of days for the applied conditions are 7813 and the total Ly-$\alpha$ irradiance is $57.9\,\mathrm{W\,m^{-2}}$ which lead to a 4.5% difference between the maximum and minimum points and a $6.6\sigma$ significance for the peak in the middle around 170° when compared to the mean value of 7.42.

Lastly, we investigate also the case of Moon's phase distribution for Ly-$\alpha$. Even though in the case where no constraints are applied in Fig. 12.7a where the amplitude is negligible at around 0.6% with no significant peak, in the case of Fig. 12.7b where we constrain Mercury to propagate around $85° \pm 45°$ the amplitude becomes 19.2% with the observed wide peak having a significance of 21.6$\sigma$. More specifically, for the case of Fig. 12.7b we have in total 6457 d and 47.7 W m$^{-2}$. From a Gaussian fit, the exact peak location is determined around 314° with its FWHM being 184°. The notable difference between the two cases is shown in Fig. 12.7c. As an additional comparison, the 180° opposite angle of Mercury at $265° \pm 45°$ gives only a 2.8% non-significant amplitude for a total of 71.1 W m$^{-2}$ in 9552 d. This strengthens even further the observation of Fig. 12.7c with the 40° to 130° region of Mercury providing the biggest result.

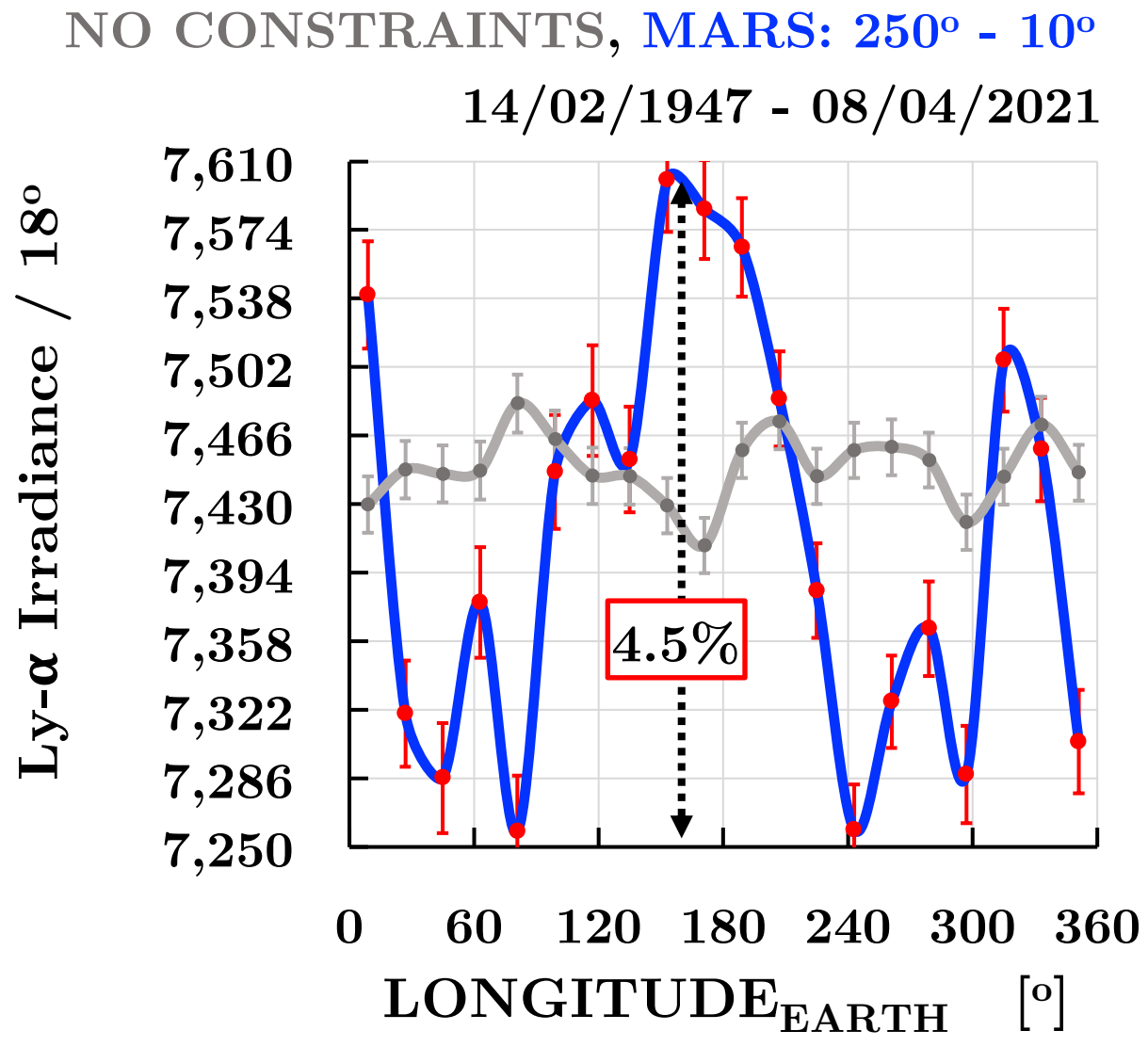

**Figure 12.6:** Ly-$\alpha$ as a function of Earth's heliocentric longitude position without constraints (in gray) and while Mars is constrained to propagate around 250° to 10° (in blue) for bin = 18°.

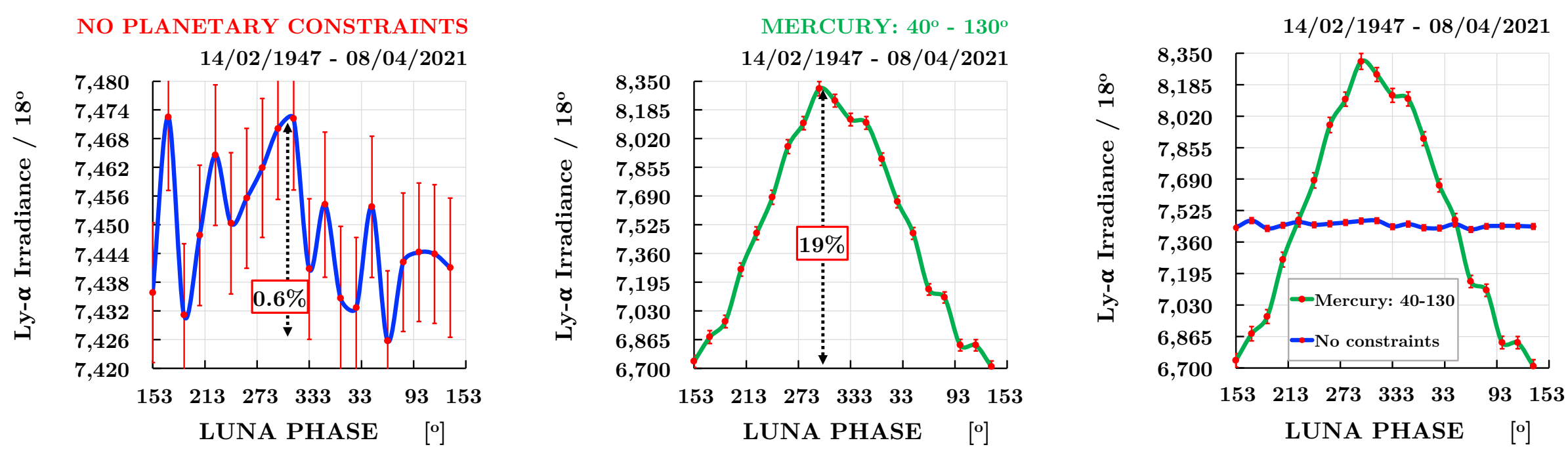

(a) Moon's phase without any planetary constraint.

(b) Moon's phase when Mercury propagates between 40° to 130°.

(c) Comparison between Fig. 12.7a and Fig. 12.7b.

**Figure 12.7:** Distribution of Ly-$\alpha$ irradiance as a function of Moon's phase with and without any additional planetary longitudinal constraints, for 18° bin.

### 12.3.2 Fourier analysis

To search for additional possible periodicities in the raw Ly-$\alpha$ data, as well as to establish further the ones already discovered, we perform a Lomb periodogram spectral analysis. From the resulting periods, a comparison is made with the planetary revolution periods as well as with the planetary synods.

From the resulting periods, the biggest observed peak with a power of 9127 dB is around 3982.08 d which corresponds to about 10.90 y (see Fig. 12.8a). Its Gaussian derived FWHM is $\sim 515.9$ d $\sim 1.4$ y. This peak was expected based on the 11 y solar cycle. Moving to smaller peaks, the 7$^{th}$ biggest peak, seen in Fig. 12.8b, is around 399 d $\pm$ 1 d with an amplitude of 37.1 dB. This means that it overlaps with the synod of Earth - Jupiter at 399 d. Furthermore, the 9$^{th}$ and 17$^{th}$ biggest peaks are centred around 693.1 d $\pm$ 5.8 d and 735.3 d $\pm$ 5.8 d and have amplitudes 25.7 dB and 13 dB respectively as seen in Fig. 12.8c. These peaks, are well within 1$\sigma$ with Mars revolution period (687 d) the synods of Mars - Pluto at 692 d, the synod of Mars - Neptune at 695 d and the synod of Mars - Saturn around 734 d. Last but not least, as seen in Fig. 12.8d, the 18$^{th}$ peak in amplitude is around 27.34 d $\pm$ 0.01 d and has a power of 12.4 dB. This is very close to Moon's sidereal month of 27.32 d, i.e. fixed to remote starts. Although a 27 d periodicity has been initially correlated with the Sun's rotation [321], a peak with 0.03 d width can not originate from the differential rotation of the Sun which spans from about 25 d to 36 d [269, 270]. Therefore, this unexpected observation strengthens the hypothesis of an exo-solar impact on the Sun's activity.

It is noted that a few more less significant periodicities are also found but are not shown here. For example around 220.4 day there is peak with an amplitude of 8.2 dB which by fitting a Gaussian function its center is determined around 224.7° with a FWHM of 1.5°. This can be identified as Venus' revolution period of 224.7 d. Finally, around 770 d, the 5$^{th}$ biggest peak with amplitude of 46.2 dB has a FWHM of 20 d and is close to Mars - Earth synod of 780 d.

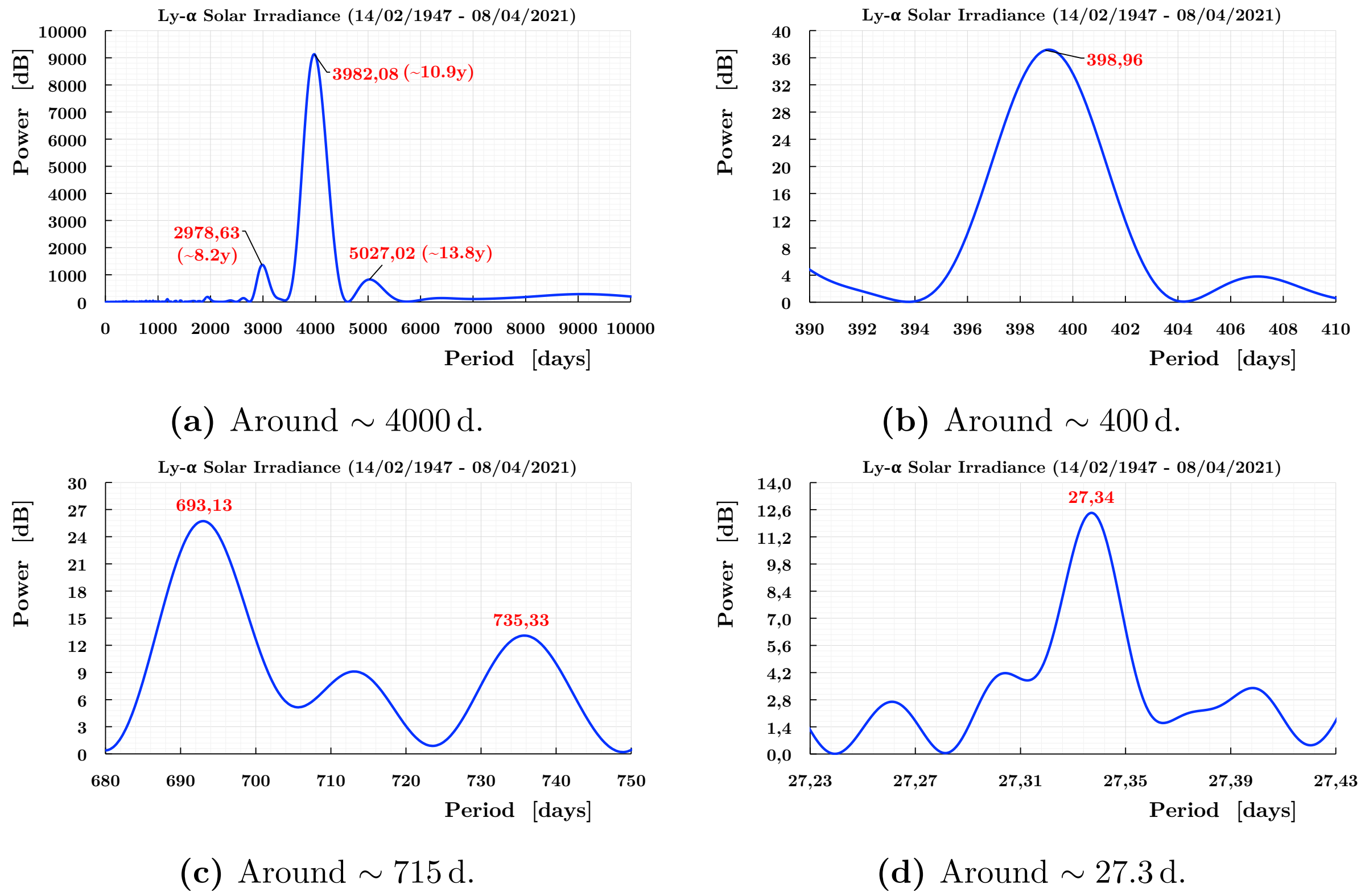

**(a)** Around $\sim 4000$ d. **(b)** Around $\sim 400$ d. **(c)** Around $\sim 715$ d. **(d)** Around $\sim 27.3$ d.

**Figure 12.8:** Fourier periodogram of the raw Ly-$\alpha$ data for the period 14/02/1947 to 08/04/2021 zoomed around a few interesting periods.

## 12.4 Summary

Based on the single and combined planetary distributions of the hydrogen Ly-$\alpha$ line some highly statistically significant peaks were observed of the inner planets as well as for Jupiter and Saturn. In addition, while the Moon's phase on its own does not show any effect on the Ly-$\alpha$ distribution, when Mercury is constrained to propagate on a 90°-wide orbital arc, a $> 20\sigma$ peak is observed (see Fig. 12.7). It is noted that the same behaviour was observed for the number of sunspots and the F10.7 solar index (see Fig. 8.11 and 9.8 respectively). These observations are unexpected within conventional physics since any remote gravitational planetary effect is extremely feeble. The common viable scenario is that of streaming invisible "strongly" interacting massive matter which undergoes planetary gravitational focusing towards the Sun enhancing enormously the incident flux.

The Fourier analysis performed on the full period of about 74 y strengthens the derived planetary periodicities and especially the one associated with Moon's sidereal month of $\sim 27.32$ d which points to an unexpected exo-solar impact on the Sun's activity.

A comparison with other solar observables like EUV and F10.7 solar radio flux via a statistical correlation analysis showed a significant strong positive correlation in the 0.05 level for almost all distributions, pointing to a common external origin on all three datasets (see Appendix Sect. B.8).

Future analyses including smaller cadence time as well as a selection of different periods, such as an analysis of each solar cycle independently, can provide more details on the direction(s) of the assumed stream(s) and the strength of the gravitational focusing by the different planets. At the same time the more thorough comparison with other datasets has the potential to provide a quantitate indication on the assumed “strong” interaction of the invisible massive streams with normal matter.

# 13
# DISCUSSION

The observation of statistically significant narrow peaks in the longitudinal planetary distributions of most solar observables studied in this Part III are not explained within known physics. Therefore, the only viable explanation, at the moment, is that of gravitational focusing of streams of non-relativistic highly interacting invisible massive matter by the various planets towards the Sun including eventually the free-fall of the Sun itself. At the same time the deficiency of the current solar models to explain and predict the solar cycles as well as the origin of phenomena such as the solar corona heating as it is manifested by EUV irradiance, the appearance of solar flares and sunspots, hints on the existence of an additional external influence. This is strongly supported by the the exo-solar observations made so far in this Part III, with the assumptions of this work, fitting-in.

For an accurate and statistically significant analysis, daily continuous long-term observations were used (see Tab. 13.1). The inner planets were mainly chosen due to their short revolution periods which results in a large number of revolutions throughout the selected periods that smears out any randomly-produced statistical fluctuation. The observed statistically significant peaking distributions which exclude conventional explanations such as tidal forces, were enhanced even further when more planets were combined. In these cases, where double or even triple planetary combinations were chosen, the preferred orbital positions of each planet were highlighted for each observable which in the long term can provide even more information on the direction(s) of the assumed invisible stream(s). As an example, a prominent derived direction in space was that of the GC, around 266.5° heliocentric longitude. This direction is supported by the distributions of most solar observations where an enhanced effect was observed for the orbital arc 200° to 300° over the 180° opposite one at 20° to 140°. Additionally, a common spectral shape was observed for the longitudinal distributions of Jupiter and Saturn for EUV (Fig. 7.6e), sunspots (Fig. 8.6e and 8.6f), F10.7 (Fig. 9.3e and 9.3f) as well as Ly-$\alpha$ (Fig. 12.3e and 12.3f). This is also supported by the corresponding statistically significant positive linear correlation found by the statistical correlation analysis performed in Appendix Sect. B. It is worth mentioning here, that the possible simultaneous presence of multiple streams having different velocity ranges makes the identification of the optimum planetary positions and thus stream directions even more challenging (see also Publication E.13).

**Table 13.1:** The various datasets that have been analysed in this Part III with their corresponding chapter reference and the used time-period.

| Dataset | Chapter | Time period |
|---|---|---|
| M-Flares | 6 | 01/09/1975 - 12/03/2021 |
| X-Flares | 6 | 01/09/1975 - 12/03/2021 |
| EUV | 7 | 01/01/1996 - 01/03/2021 |
| Sunspots | 8 | 01/03/1900 - 28/02/2021 |
| F10.7 | 9 | 28/11/1963 - 03/03/2021 |
| Solar Radius | 10 | 06/06/1996 - 01/12/2017 |
| FIP bias | 11 | 30/04/2010 - 11/05/2014 |
| Ly-$\alpha$ | 12 | 14/02/1947 - 08/04/2021 |

Additional supporting evidence for the existence of an external influence is derived from the Lomb periodogram analyses of the various datasets. As seen in Fig. 13.1 almost all observations show a significant narrow peak around 27.32 d overlapping with the Moon's sidereal month, i.e. fixed to remote stars. More specifically, the peaks are located at 27.32 d $\pm$ 0.02 d, 27.32 d $\pm$ 0.03 d, 27.34 d $\pm$ 0.01 d, 27.32 d $\pm$ 0.02 d and 27.34 d $\pm$ 0.01 d for the number of M-flares, EUV, number of sunspots, F10.7 and Ly-$\alpha$ solar irradiance accordingly. These exo-solar signatures are fitting-in the assumed scenario of incoming streams(s) that are gravitationally focused towards the Sun by Earth and are modulated also by the Moon's orbital position. The 27 d periodicity is conventionally attributed to the differential solar rotation. However, it can not explain such very narrow periodicities as it spans from about 25 d to 36 d [269, 270], with the wide range of these values not being able to produce sharp narrow peaks around 27.32 d. In addition, the 29.53 d period which is associated with the Moon's synodic period, i.e. fixed to the Sun, appears less pronounced in all cases, thus stressing the exo-solar character of the aforementioned observables, which has been widely overlooked in standard solar physics etc.

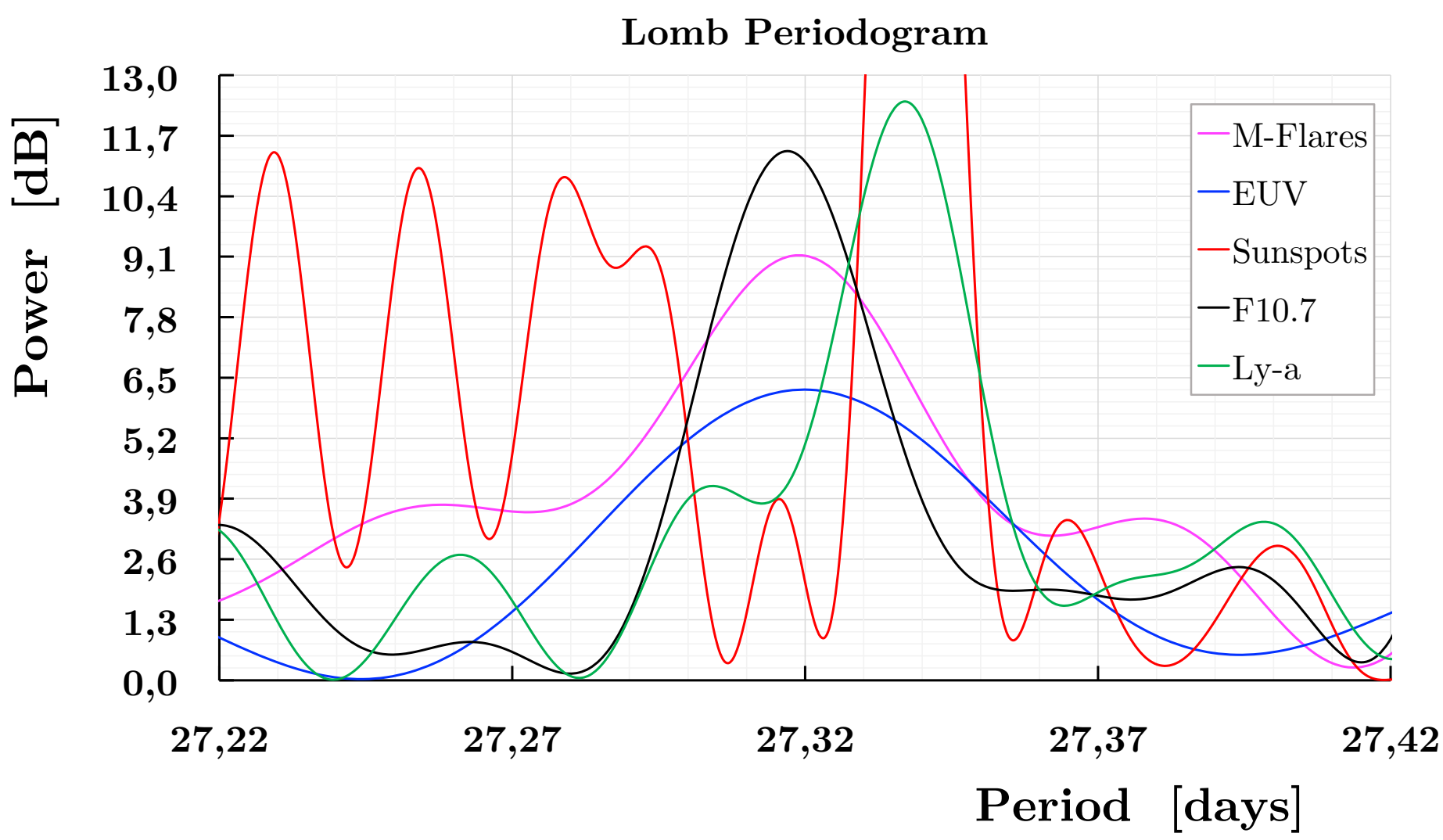

**Figure 13.1:** Fourier spectra for the various solar observables showing some significant peaks around 27.32 d corresponding to Moon's sidereal month.

Finally, a statistical correlation analysis has been performed both on the general trend (time distribution) of the various datasets but also on the specific planetary longitudinal distributions (see Appendix Sect. B). A summary of the Pearson's correlation coefficients calculated between all the datasets, for their respected time series, is shown in Tab. 13.2. The first significant observation is the negative linear correlation of solar radius with the rest of the solar observables. The remaining observables show a positive linear correlation between them with all but one, that between FIP and X-Flares, being statistically significant on the 0.05 level. It is noted that these smaller or bigger correlations refer to the general trend of the datasets and not to the specific planetary relationship of each case. For these small-scale comparisons, the reader can refer to Appendix Sect. B.

**Table 13.2:** Pearson's correlation matrix for the various datasets studied in this part. The coefficients showing a positive linear correlation are in blue while the ones showing negative correlation are in red. The statistically significant results on the 0.05 level are marked with *. The maximum available periods were used for these calculations.

| | M-Flares | X-Flares | EUV | Sunspots | F10.7 | $\Delta R$ | FIP | Ly-$\alpha$ |
|---|---|---|---|---|---|---|---|---|
| M-Flares | 1* | 0.29* | 0.34* | 0.44* | 0.50* | -0.23 | 0.10* | 0.41* |
| X-Flares | 0.29* | 1* | 0.13* | 0.18* | 0.22* | -0.09 | 0.02 | 0.17* |
| EUV | 0.34* | 0.13* | 1* | 0.90* | 0.93* | -0.75 | 0.54* | 0.94* |
| Sunspots | 0.44* | 0.18* | 0.90* | 1* | 0.95* | -0.69 | 0.47* | 0.91* |
| F10.7 | 0.50* | 0.22* | 0.93* | 0.95* | 1* | 0.70 | 0.48* | 0.94* |
| $\Delta R$ | -0.23* | -0.09* | -0.75* | -0.69 | -0.70* | 1* | -0.43* | -0.72* |
| FIP | 0.10* | 0.02 | 0.54* | 0.47* | 0.48* | -0.43 | 1* | 0.49* |
| Ly-$\alpha$ | 0.41* | 0.17* | 0.94* | 0.91* | 0.94* | -0.72 | 0.49* | 1* |

Notably, only daily, or larger, cadenced data were analysed in this work. However, most of the publicly available datasets contain even finer resolution data with hourly or even minute cadence. Therefore, the application of the specifically designed analysis procedure on these data has a potential for new results. A more refined analysis could also be elaborated introducing all planets and providing even more constraints both on the planetary positions but also on the amplitude of the selected observable. This can provide more hints towards the identification of the invisible stream(s), their preferred direction(s) but also to the nature of their constituents and their interactions. In addition it has the potential to also improve significantly the forecasting of the sunspot activity which is a key objective of space weather and space climate research. Furthermore, an interesting comparison of the derived directions of the invisible streams from these analyses could be performed with the ones from stellar halo streams. These are also expected to carry DM halos [327], and are currently being identified through the GAIA astrometry mission data [328, 329].

Additional solar datasets could also be investigated for planetary relationships, with the ones depicting an 11 y periodicity being the most favourable candidates [237]. Such an example, which is expected to produce interesting results, is that of Total Solar Irradiance (TSI) [324]. A similar search could also be carried out with the Sun's visible and infrared irradiance [330].

An additional interesting dataset is that of the mean magnetic field of the Sun. Data were initially downloaded for that purpose from Wilcox Solar Observatory (WSO) [331], but due to the many missing dates ($\sim 18.2\%$), no reliable results could be derived. It is mentioned that an intriguing observation for which however no long-term datasets exist is that of sunquakes [332] (see Fig. 13.2). They have been observed to occur during solar flares but their origin remains a mystery [333]. Interestingly they have been shown to have a downward propagating motion [334] which is in agreement with the scenario of in-falling gravitationally focused invisible massive streams.

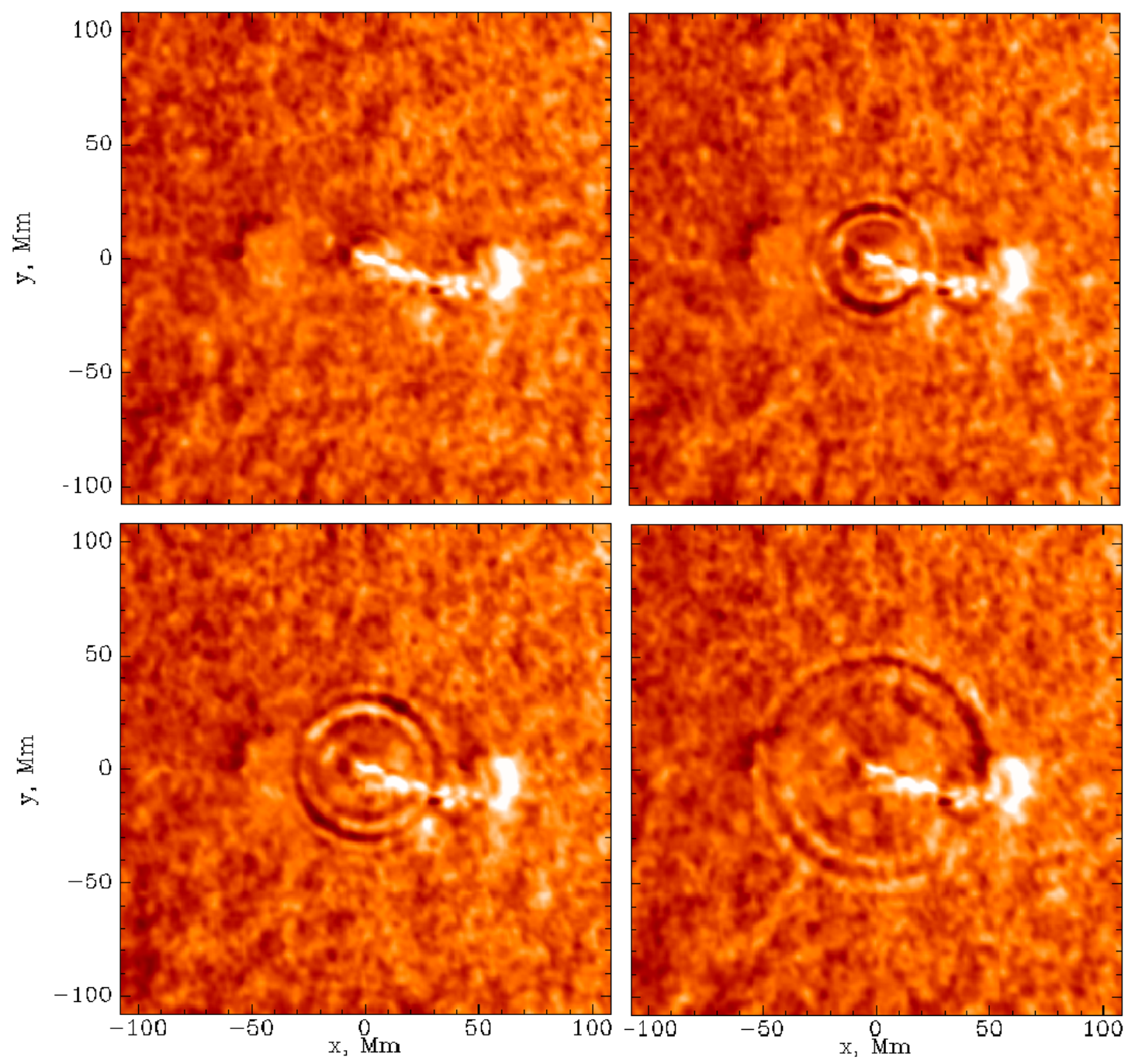

**Figure 13.2:** Sunquakes depicting a ripple pattern resembling the wave spreading from a rock dropped into a pool of water (37).

Moreover, exo-planetary systems could also exhibit similar planetary relationships since the gravitational focusing of invisible streams could also occur there [335]. An initial search could be made by investigating the associated stellar activity as a function of the exo-planetary orbital phases (see also Publication E.17). This could be searched in the GAIA latest catalogues [230, 336]. Finally, a re-analysis searching for planetary relationship could be performed in unexplained observations like that from Fermi-Large Area Telescope (LAT) studying the anomalous multi-GeV $\gamma$-ray emission from the solar disk [337].

# Part IV:

# Terrestrial Observations

*"O earth, what changes hast thou seen! "*

—Alfred Lord Tennyson (1809–1892),
*English Poet*

# 14

# INTRODUCTION

The Sun is capable to focus on Earth particles with speeds of about 0.01 c to 0.2 c with the flux amplification being up to $10^{11}$. Similarly, other solar system bodies like Jupiter can perform gravitational focusing towards the Earth. As an example, the Moon itself can focus towards the Earth particles with velocities around $10^{-4}$ c with a flux amplification of $\sim 10^4$. At the same time, self-focusing effects from the Earth itself for speeds on the order of 17 km/s, could also influence incoming slow-moving cosmic streams with the amplification in this case being up to $\sim 10^9$ depending on the stream dispersion [222].

Therefore, similarly to Part III, we can have a big flux enhancement of invisible massive streams, but this time at the location of the Earth, whose strong interaction with ordinary matter could trigger a variety of unexplained phenomena (see Fig. 14.1). Since the orbiting planets affect periodically the dominating Sun's gravitational field they should give rise to a planetary time-dependent imprint if the studied dataset is caused or affected by a gravitationally focused stream. This means that the key signature for the involvement invisible massive streams will once more be that of a statistically significant planetary dependence through peaking distributions. As with the solar observations, also with terrestrial observations, such a distinct signature can not be associated with any long-range forces such as tidal-force inspired models since they are too weak while their strength changes smoothly over an orbital period [229, 242–244].

The usual DM constituents such as WIMPs or axions can not cause any observable effect since their interaction probability is extremely small for the experimentally excluded parameter phase space. This assessment applies of course only to the parameter phase space accessible so far to experimentation. In addition, there are already theoretically expected candidates from the dark sector which have large cross-sections with normal matter, such as AQNs, magnetic monopoles or dark photons. Even though a direct DM signature has not been yet observed, this could be explained considering for example energy threshold effects [235]. However, when a low-speed stream is aligned towards the Earth with an intervening planet, there is a transitory flux enhancement at the site of the Earth increasing the interaction probability by several orders of magnitude and thus exceeding any threshold constraints. Such a scenario could be at the origin of various as yet unexplained phenomena in the Earth's upper and lower atmosphere (see also Publications E.1, E.2, E.3, E.4, E.14, E.15, and E.19).

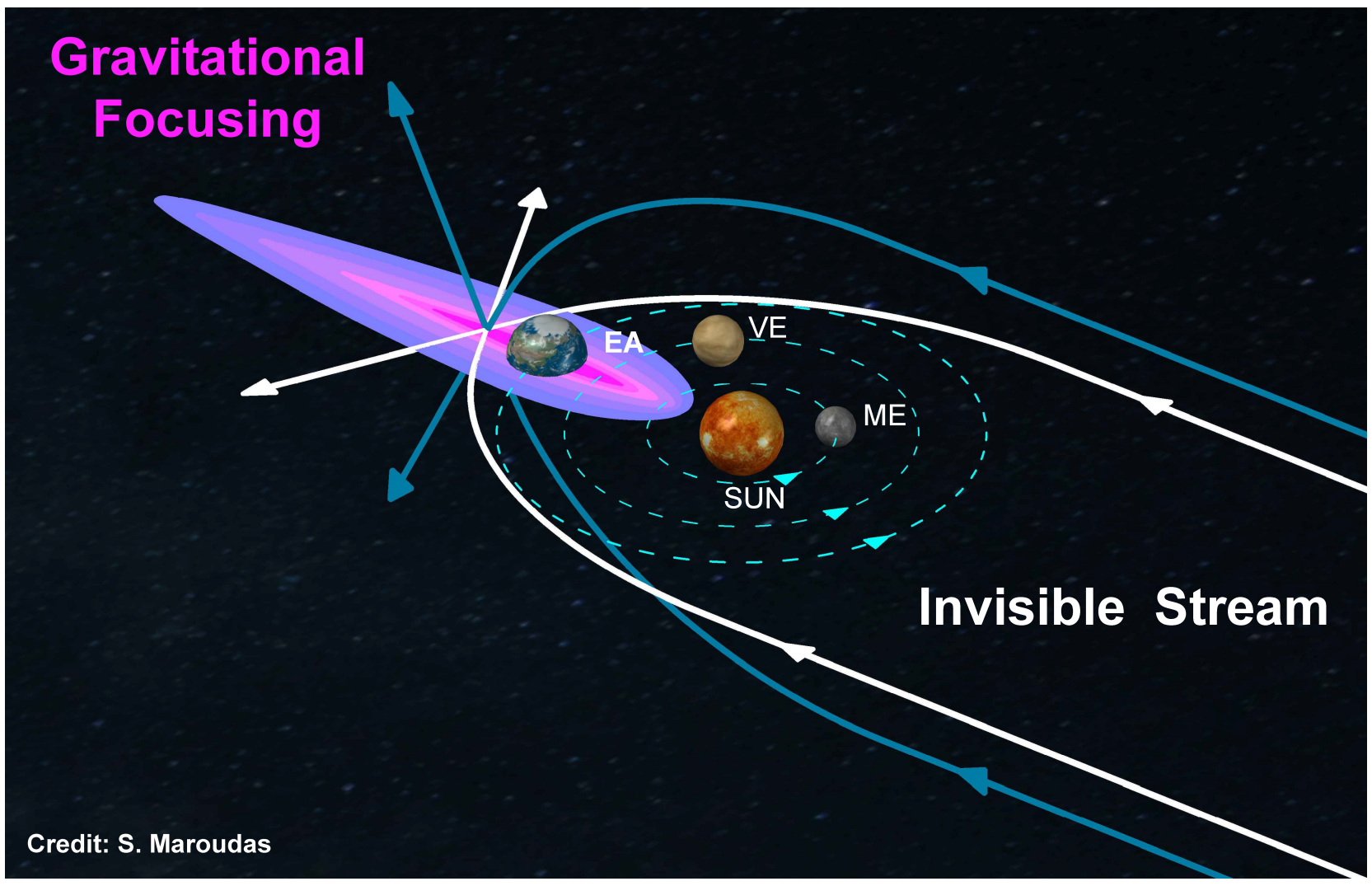

**Figure 14.1:** Schematic illustration of gravitational focusing effects of invisible massive streams by the Sun and its planets towards the Earth (same as Fig. 4.1a). In this configuration the GC is located on the lower right corner and in the opposite direction of the incident invisible streams.

As an example, there is a common feature between the Earth's and Sun's atmosphere and that is their ionised outer layers which are permeated with a magnetic field. Magnetised plasma or some other atmospheric properties could favour the interaction of the assumed invisible streaming matter incident on the Earth. As we will see in Chap. 15 the Earth's ionosphere shows an anomalously higher degree of ionisation in December known for several decades. Similarly, in the Earth's stratosphere (Chap. 16), there are observed annual temperature anomalies that can not be explained with conventional arguments. In addition, a variety of previous atmospheric observations have shown that the Earth's atmosphere senses somehow the 11 y solar cycle [338–341] however, with an underlying explanation missing [342, 343]. On the other hand, as outlined in Part III, even though the 11-year cycle itself remains otherwise as one of the biggest unsolved solar mysteries, it might be associated with the planetary motions [237] based on the reasoning of this work.

Therefore, here four diverse phenomena will be examined ranging from the Earth's upper atmosphere to living matter on the ground, and the Earth's sub-surface. A statistically significant planetary relationship of these observations would then be the signature for invisible streams as their contributing cause. Lastly, a comparison with solar activity will pinpoint possible similarities but also dissimilarities suggesting an exo-solar influence on the specific terrestrial observations.

15

# IONOSPHERIC ELECTRON CONTENT

## 15.1 Introduction

A similar characteristic between Earth's and Sun's atmosphere is that they both have common ionised outer layers permeated with a magnetic field. Additionally, magnetised plasma or some other atmospheric properties could favour an interaction with invisible streaming matter incident on the Earth or the Sun. At the same time, since we observe it continuously, the Earth's atmosphere can also act parasitically as a huge windowless and lowest threshold detector for the dark universe with built-in spatiotemporal resolution. Therefore, after solar observations, a search for planetary correlations in the Earth's atmosphere is a natural step.

Interestingly, several unexpected phenomena take place in the ionosphere of the Earth, which lies from about 70 km to 900 km altitude. The width of this dynamic region depends mainly on the solar conditions and is divided into the sub-regions D, E and F based on the wavelength of the solar radiation that is absorbed there (see Fig. 15.1). The first observation on the existence of anomalies in the Earth's ionosphere dates back to 1937 [344]. It was found that around December there is an anomalously high degree of ionisation compared to June of about 22% (see Fig. 15.2) which cannot be explained by the 7% annual solar irradiance modulation due to the varying distance between Earth and Sun [345].

In general, the global electron content of the ionosphere depends on the variable solar EUV irradiance. Though, the variations of the former contain unexplained anomalies [346,347]

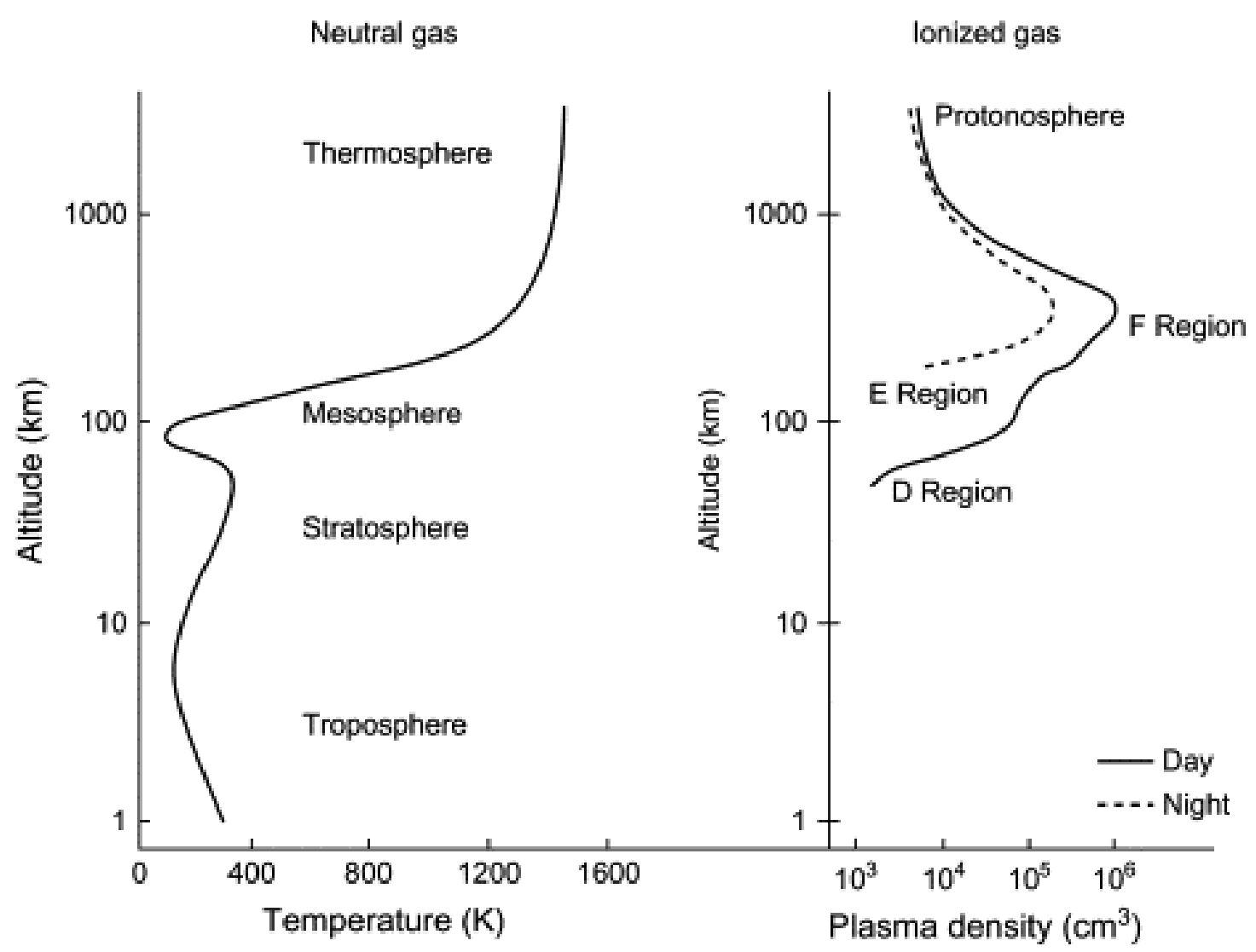

**Figure 15.1:** Typical profiles of neutral atmospheric temperature and ionospheric plasma density as a function of height (38).

(see Fig. 15.2). Therefore, given also the fact that EUV solar irradiance showed already a planetary relationship, it is worth investigating if the ionospheric anomalies are also connected to planetary lensing of invisible massive matter (see also Publications E.1, E.14, and E.15).

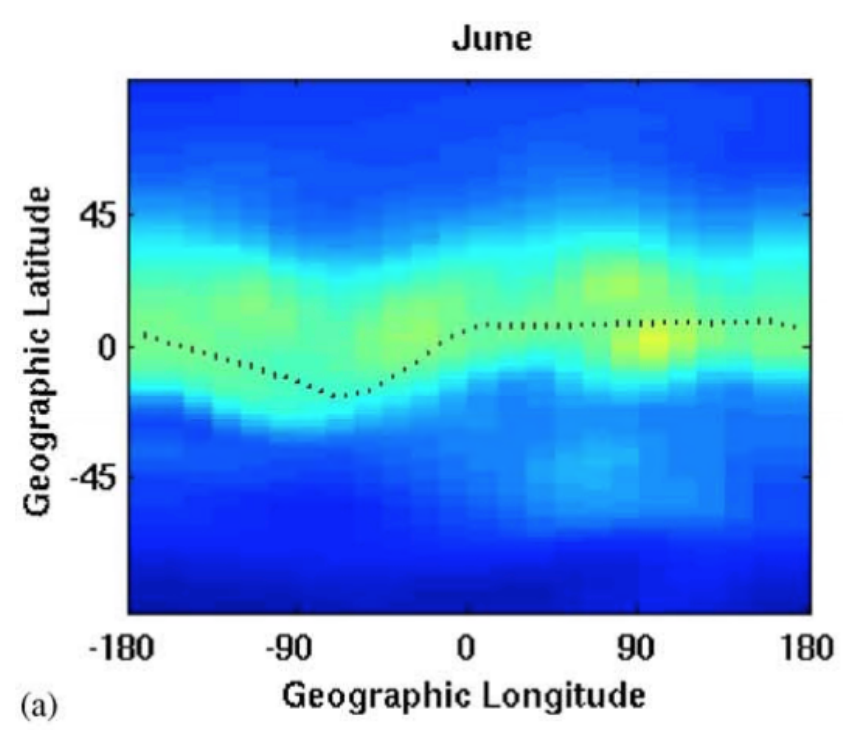

**(a)** Center on 21/06/2002.

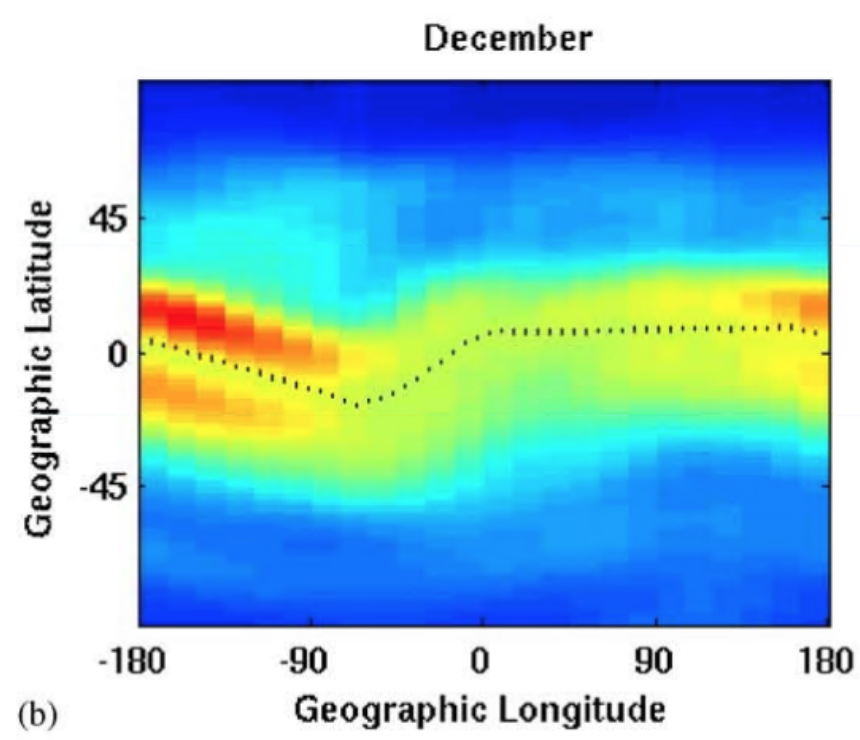

**(b)** Center of 21/12/2002.

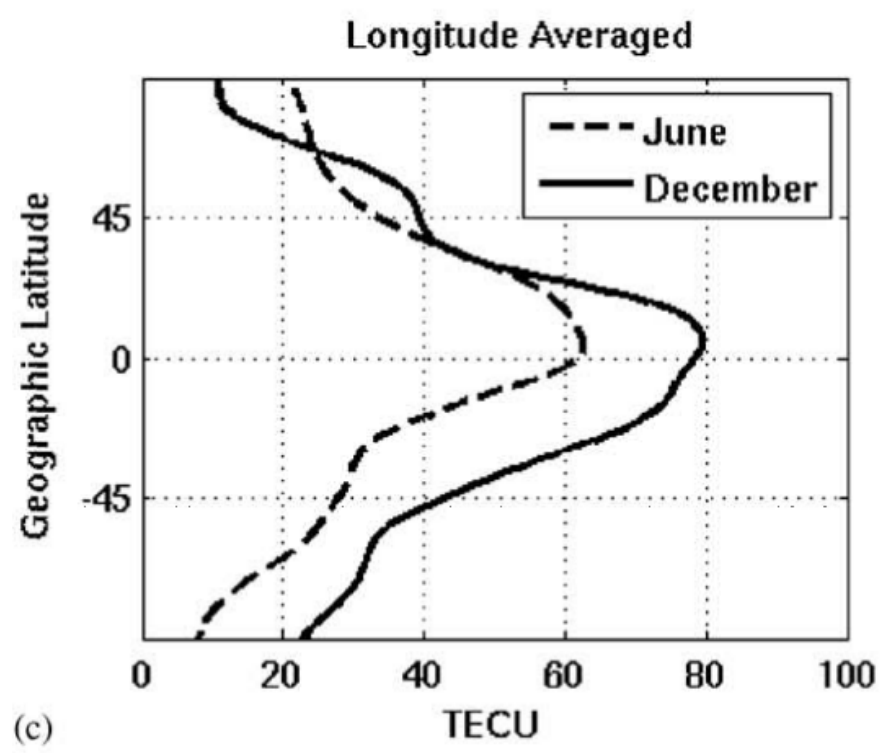

**(c)** TEC vs. latitude from Fig. 15.2a and 15.2b.

**Figure 15.2:** TEC at noon plotted in geographic latitude and longitude for 31 d average (39).

## 15.2 Data and methods

### 15.2.1 Data origin

The data for the global electron content of the ionosphere are derived from Global Positioning System (GPS) satellites [348]. The Total Electron Content (TEC) values are measured in Total Electron Content Units (TECUs), where 1 TECU= $10^{16}$ electrons/m$^2$. The original data used in this analysis are derived from Center for Orbit Determination in Europe (CODE) and contain 6574 daily-averaged measurements of TEC from 01/01/1995 to 30/12/2012 [349, 350]. The sum of the measurements is 126586.2 TECUs with an average of 19.3 TECUs per day. The minimum value is 6.4 and the maximum 59.8 TECUs, with their time series shown in Fig. 15.3. During the selected period we have the extreme deep solar minimum between 2008 − 2009, which induced a less pronounced behaviour of the TEC in the ionosphere. However, as seen in both Fig. 15.3 and 15.4a, the observed difference compared with the minimum of 1996 is not large [350].

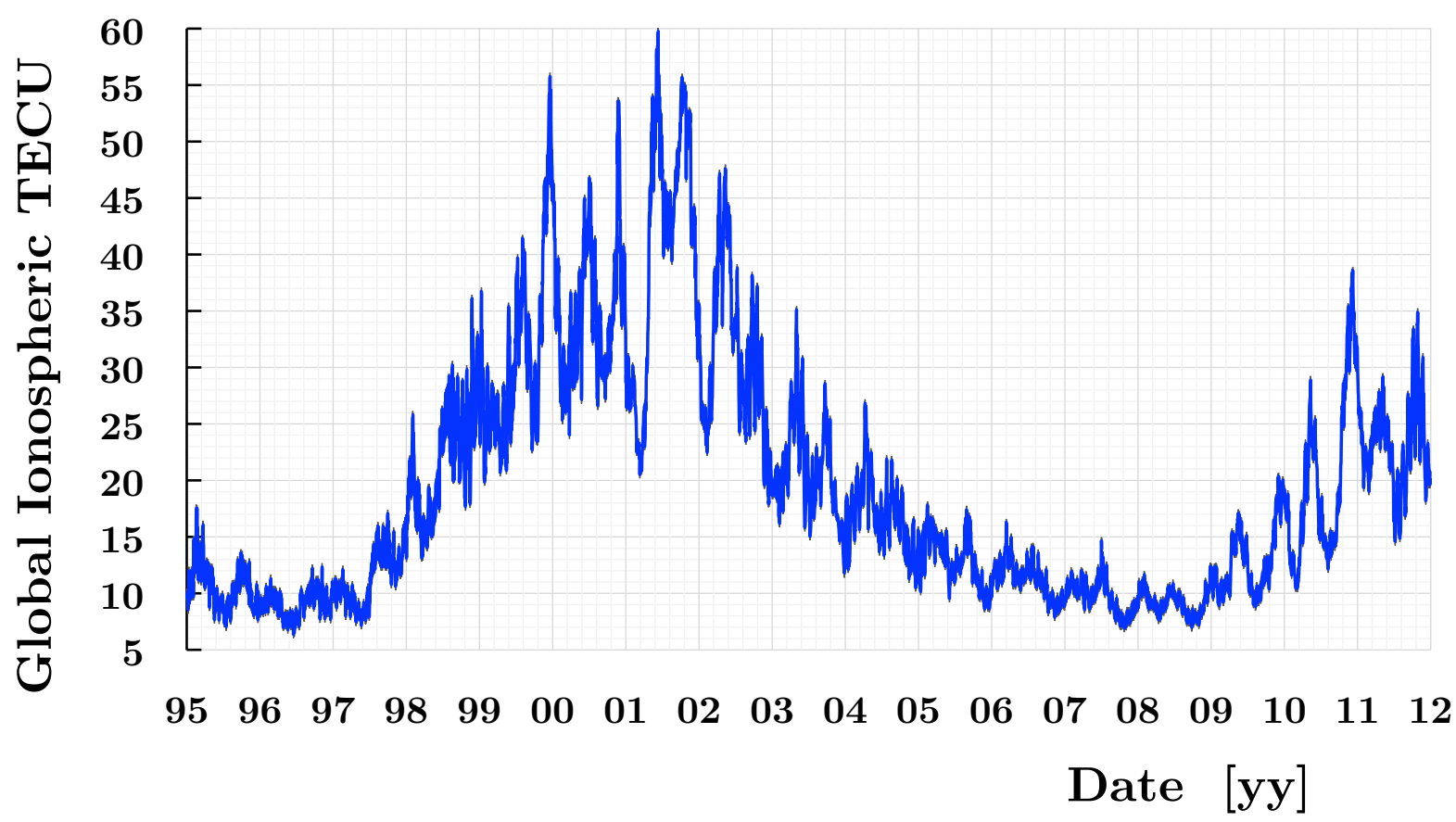

**Figure 15.3:** The raw daily TECs data from 01/01/1995 - 30/12/2012.

### 15.2.2 Data statistics

From the data used for Fig. 15.3, the various corresponding histograms of Fig. 15.4 have been created. They show the frequency distribution of the average daily TEC for the period 01/01/1995 - 30/12/2012. The ∼ 20% unexplained electron content variation between winter and summer solstices that was already visible in Fig. 15.4b, it appears even more clear in Fig. 15.4e.

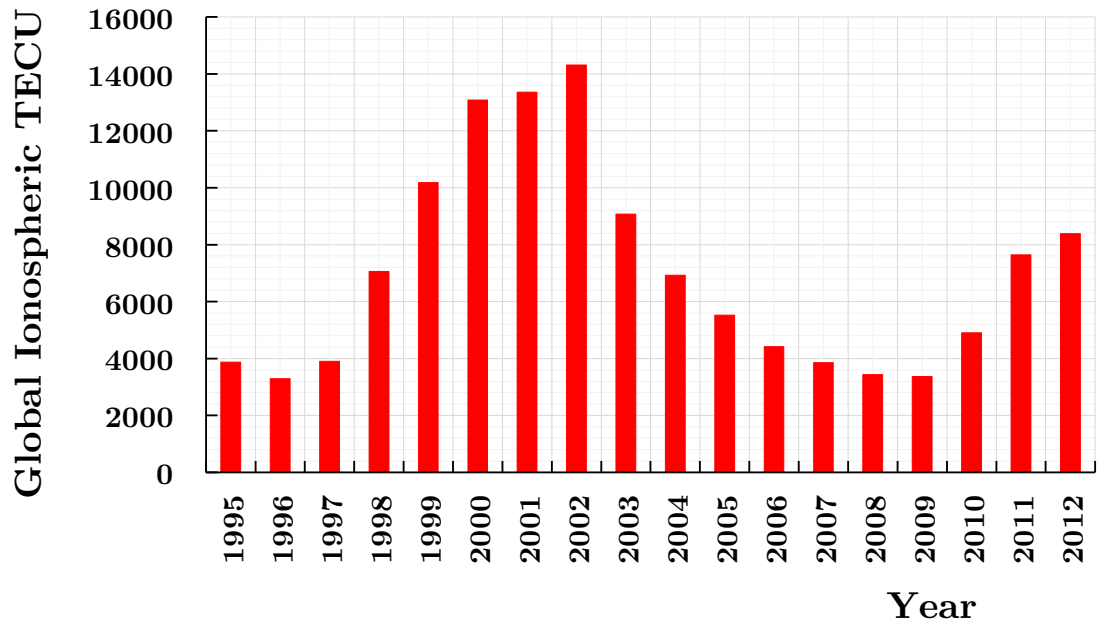

**(a)** TECUs per year.

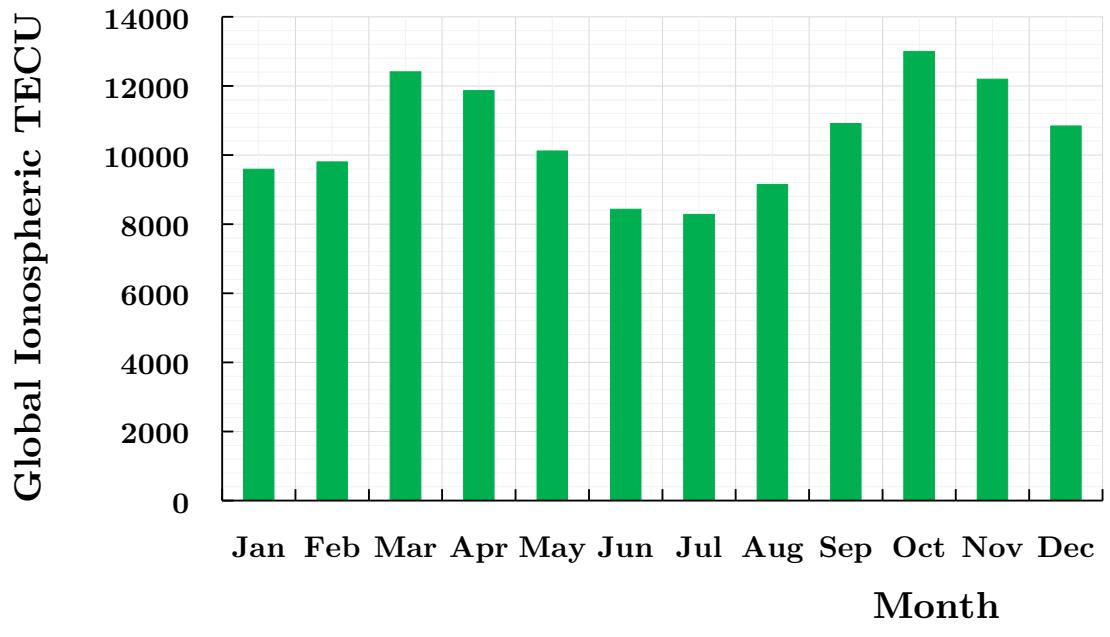

**(b)** TECUs per month.

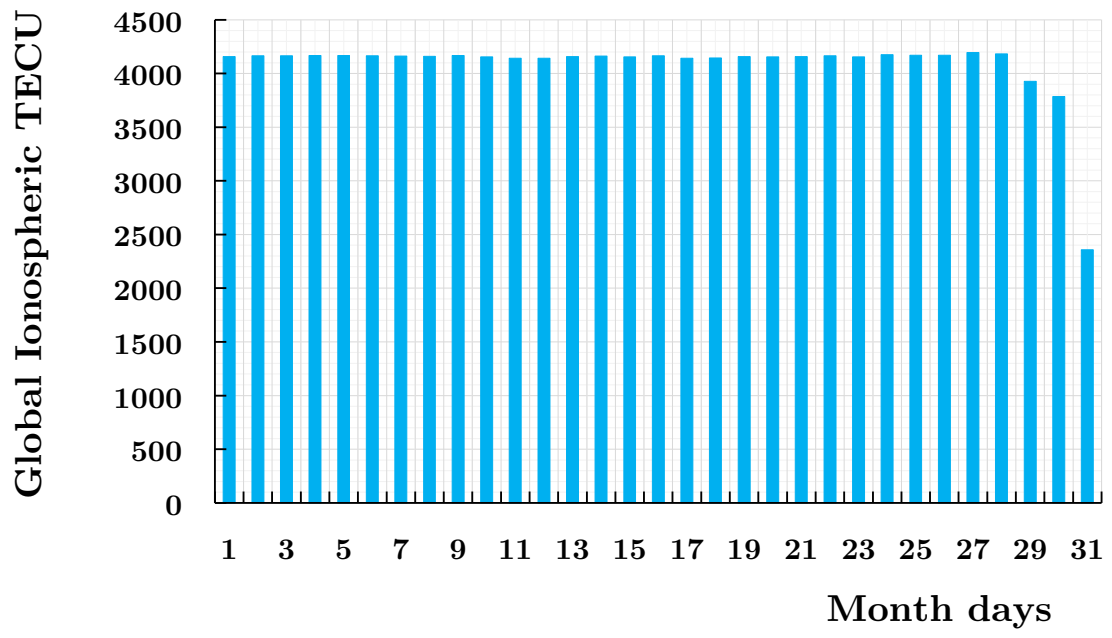

**(c)** TECUs per day of month.

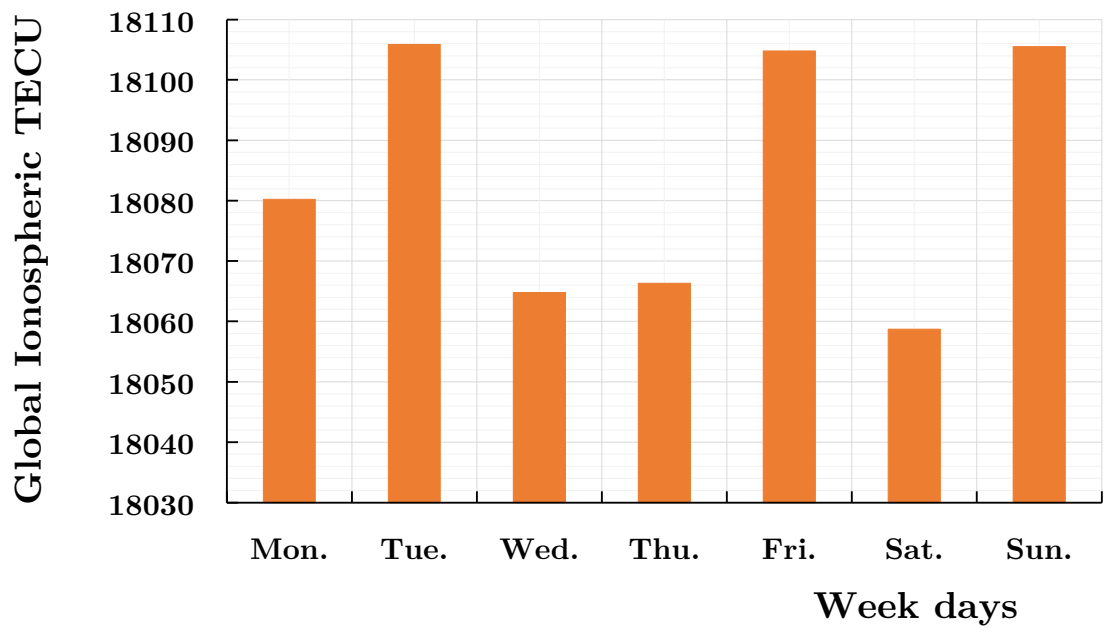

**(d)** TECUs per day of week.

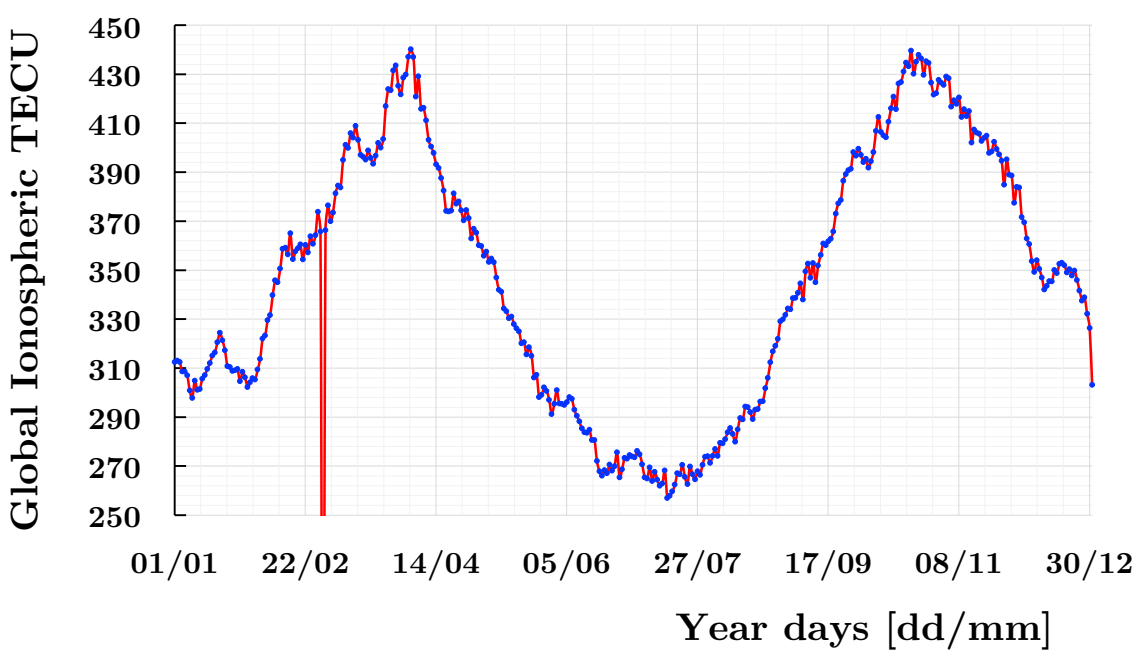

**(e)** TECUs per day of year. The observed point with the reduced amplitude corresponds to 29/02 which is a leap day added only to leap years and thus appears less frequently in the total period.

**Figure 15.4:** Histograms for the average ionospheric daily TECUs for the period 01/01/1995 - 30/12/2012.

## 15.3 Data analysis and results

### 15.3.1 Planetary longitudinal distributions

#### 15.3.1.1 Single planets

For a search of a planetary relationship, the daily TEC data of the ionosphere are projected on the corresponding planetary heliocentric longitudinal coordinates. The distribution of the TEC is then derived for the various planets in the solar system. The first step, shown in Fig. 15.5, is to use single planets without any constraint on the position of the rest of the planets. The inner planets Mercury, Venus and Earth as well as Mars, are used since they perform multiple orbits during the given period. More specifically we have about 75, 29, 18, and 10 revolutions for Mercury, Venus, Earth and Mars respectively with the observed differences between the maximum - minimum points being 5.4%, 8.9%, 38.9% and 19.6% respectively. In all four distributions we have 126586.2 TECUs in 6574 d.

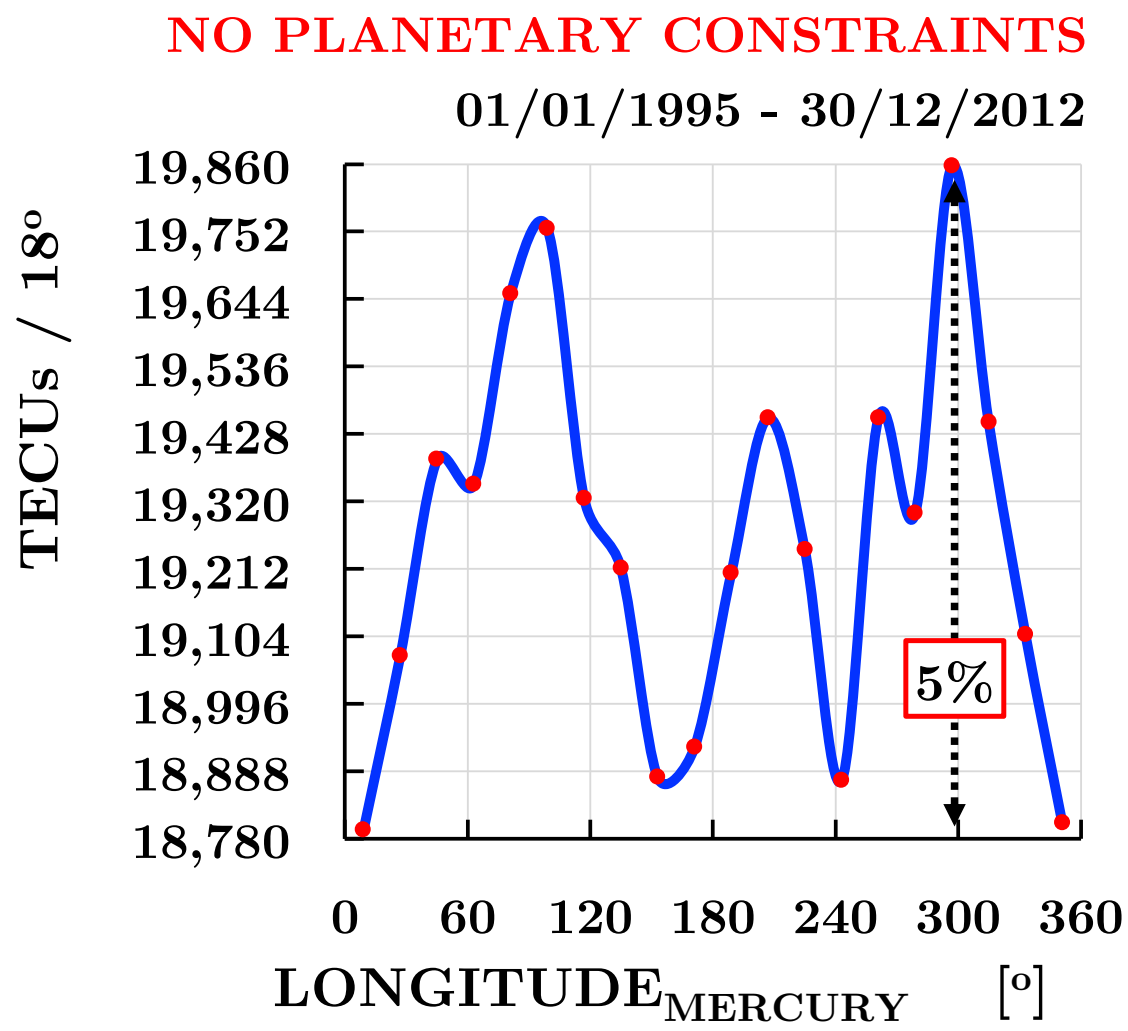

**(a)** Mercury for 18° bin.

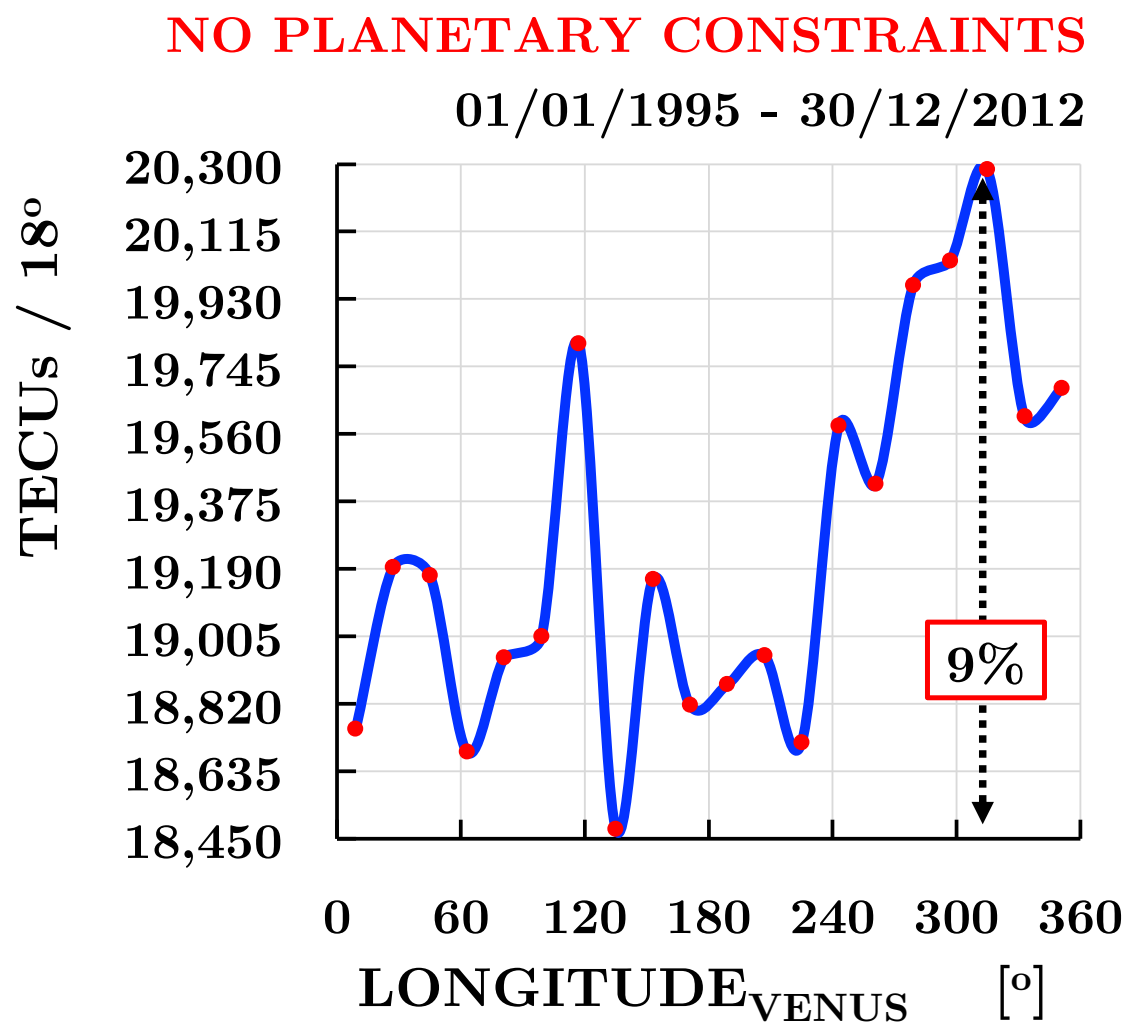

**(b)** Venus for 18° bin.

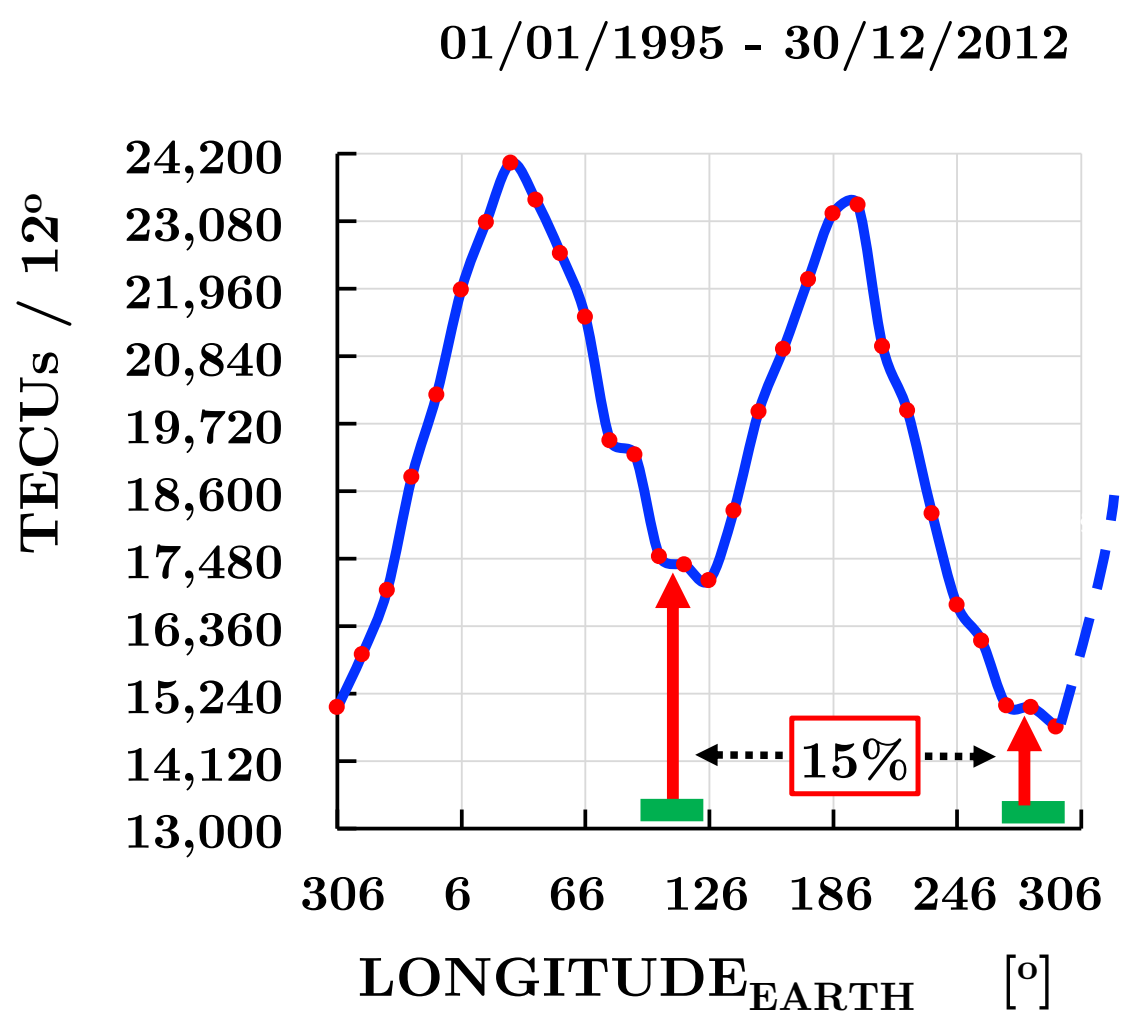

**(c)** Earth for 12° bin. The two green bars correspond to the two 30°-wide winter and summer solstices (90° to 120° and 270° to 300° respectively)

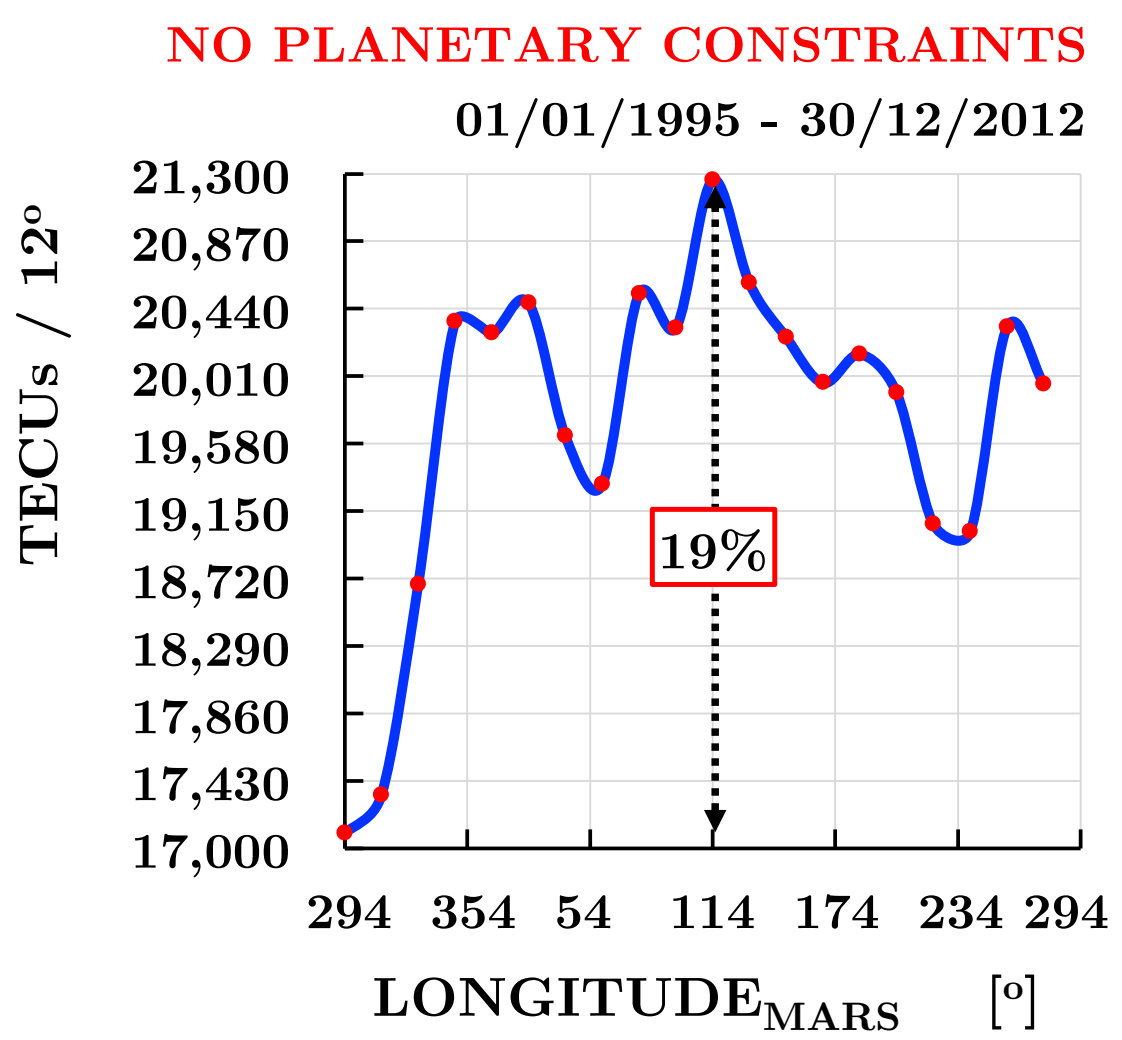

**(d)** Mars for 12° bin.

**Figure 15.5:** Planetary heliocentric longitude distributions of TEC of ionosphere for the period 01/01/1995 - 30/12/2012.

We note that in Fig. 15.5c we have reconstructed once more the shape observed in Fig. 15.4e and 15.4b. We clearly see the $\sim 15 - 20\%$ electron content variation between winter and summer solstices around the two minima in 90° to 120° and 270° to 300° which reflects the annual anomaly of the ionosphere [345] (see also Fig. 15.2).

#### 15.3.1.2 Combining planets

The next step is to search for possible correlations with the longitudinal positions of Mercury and Venus while requiring a second planet to propagate in some selected orbital arcs so as to see if the observed effects from Fig. 15.5 are enhanced with the combination of more planets. First, in Fig. 15.6, we use the reference frame of Mercury while Venus is propagating in two 180° opposite orbital arcs of 120° longitude. When we compare the two spectra in Fig. 15.6c we clearly see an otherwise unexpected difference between them (see also Fig. 15.6d). It is noted that within conventional physics, considering the long observation period of 18 y both distributions not only should be similar but should also have been rather isotropic without a significant peaking behaviour. In the case of Venus being in the 200° to 320° orbital arc (Fig. 15.6a) the number of days fulfilling this condition are 2207 with the difference between the maximum point to the minimum being 10.9%, while for the arc of 20° to 140° (Fig. 15.6b) we have 2161 d and a 14.3% difference. The selected orbital arcs were based on the same selections from previously analysed datasets where in all cases the biggest amplitude was observed in the direction of the GC around 266.5°.

On Fig. 15.7, Earth is constrained to propagate on two symmetrically 180° apart orbital arcs while Venus orbital position is observed. From Fig. 15.7c and 15.7d we clearly observe an outmatching of the 20° to 140° orbital arc over the 200° to 320° arc. When Earth is constrained to be between 200° to 320° we have in total 2246 d out of 6574 d, fulfilling this condition, and 21.5% amplitude. On the other hand when Earth is constrained to be between 20° to 140° case we have a 32% amplitude in 2134 d out of 6574 d.

Finally, in Fig. 15.8 we use the phase of Moon as the reference frame, while selecting only the periods around the two solstices, in a sector of 30°-wide around the minima seen in Fig. 15.5c. The 30°-width was selected based on the fact that during the Earth's propagation in these 30° arcs, the Moon completes one geocentric orbit. Furthermore, during the whole period of 18 y, Moon performs about 223 orbits around the Earth which should average out any randomly occurring TEC excursions. Moreover, Fig. 15.8a and 15.8b should have a similar shape, which as seen more clearly in Fig. 15.8c is not the case. When we subtract one spectrum from the other in Fig. 15.8d, we get a variation of a factor more than 6 between maximum and minimum. We also note that the position of the maximum coincides with the "new Moon" around 0°. This observation is consistent with the assumption that a massive stream of invisible matter comes from the direction of the GC and is gravitationally focused by the Sun and, assisted by the interposed Moon, towards the Earth (see also Fig. 14.1).

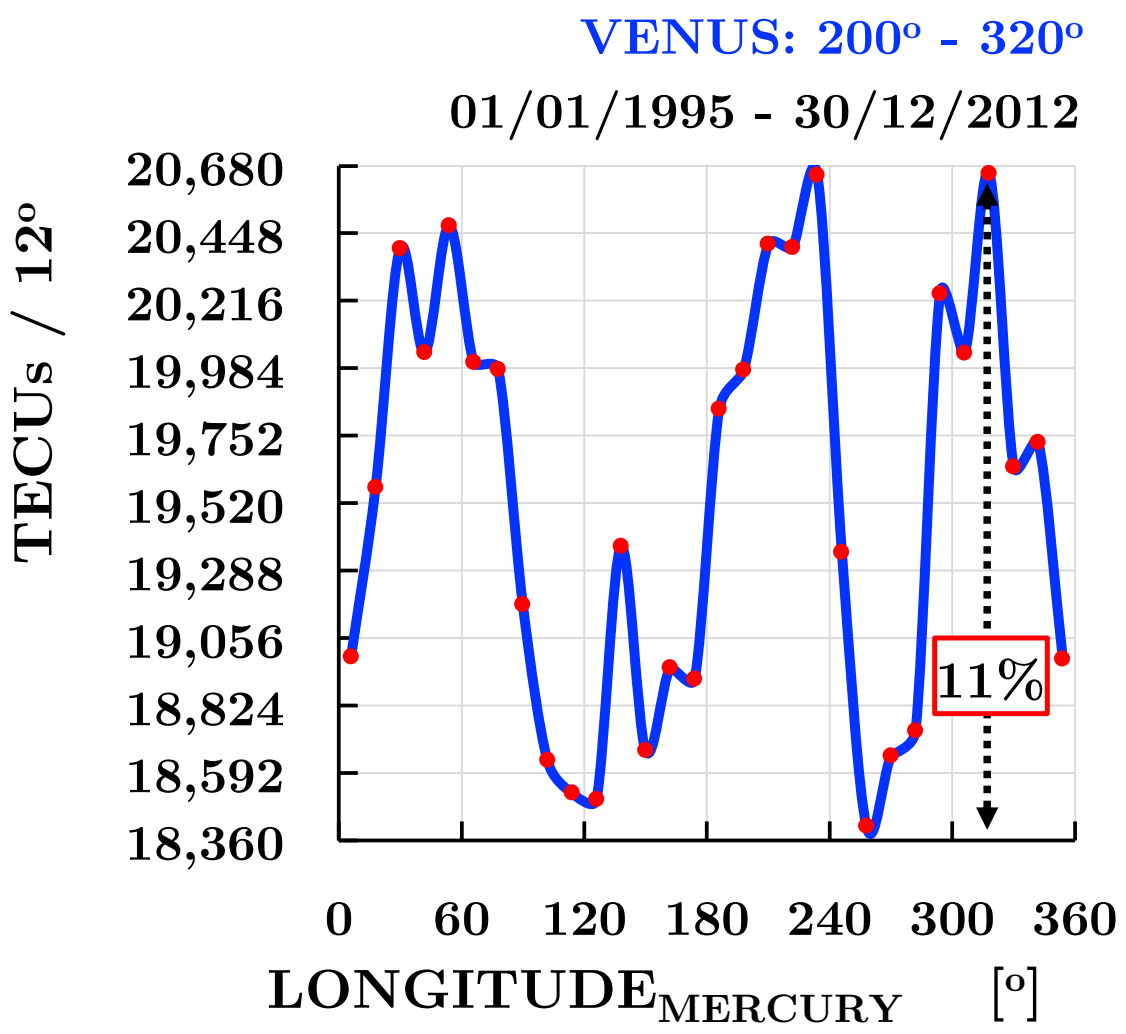

(a) Mercury when Venus is between 200° to 320°.

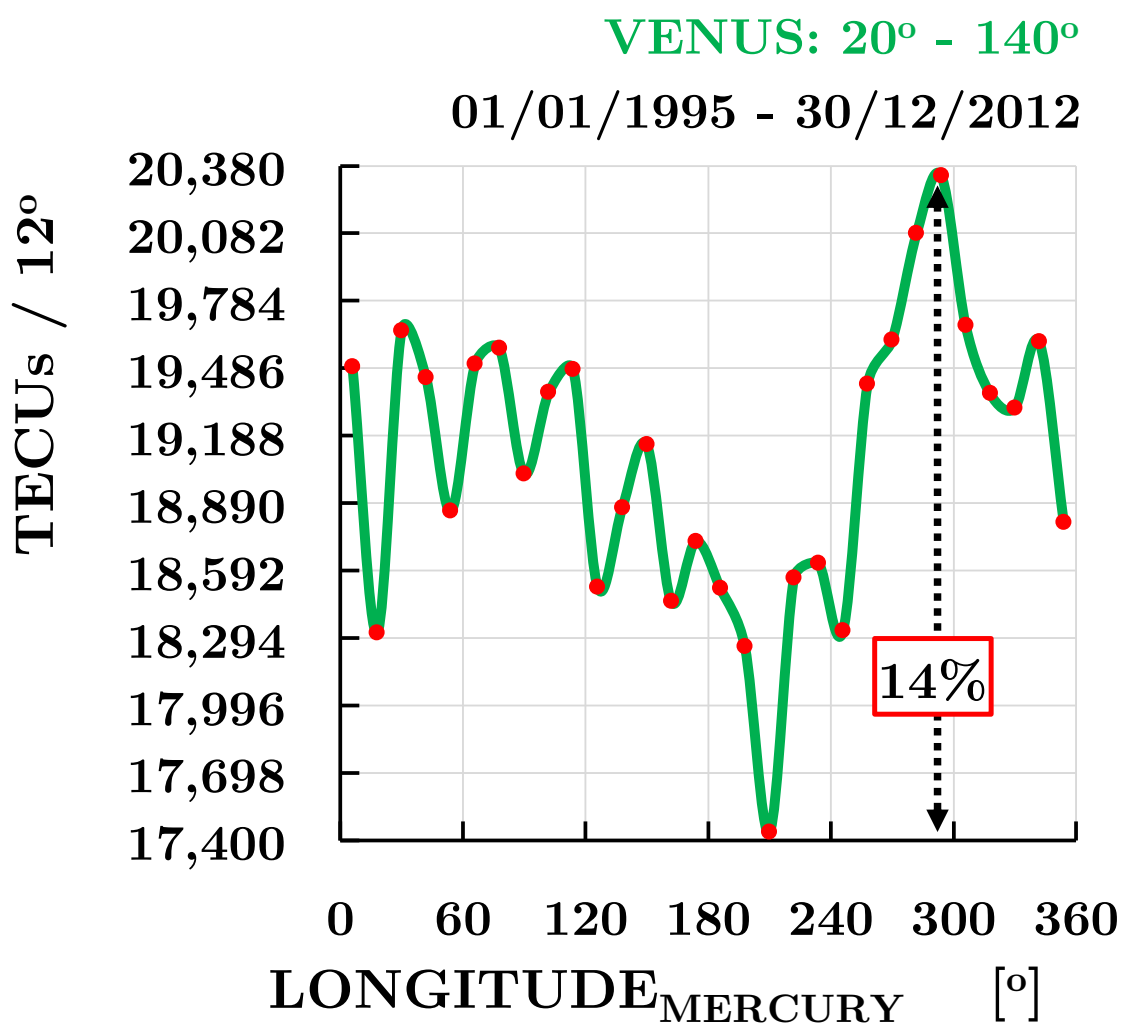

(b) Mercury when Venus is between 20° to 140°.

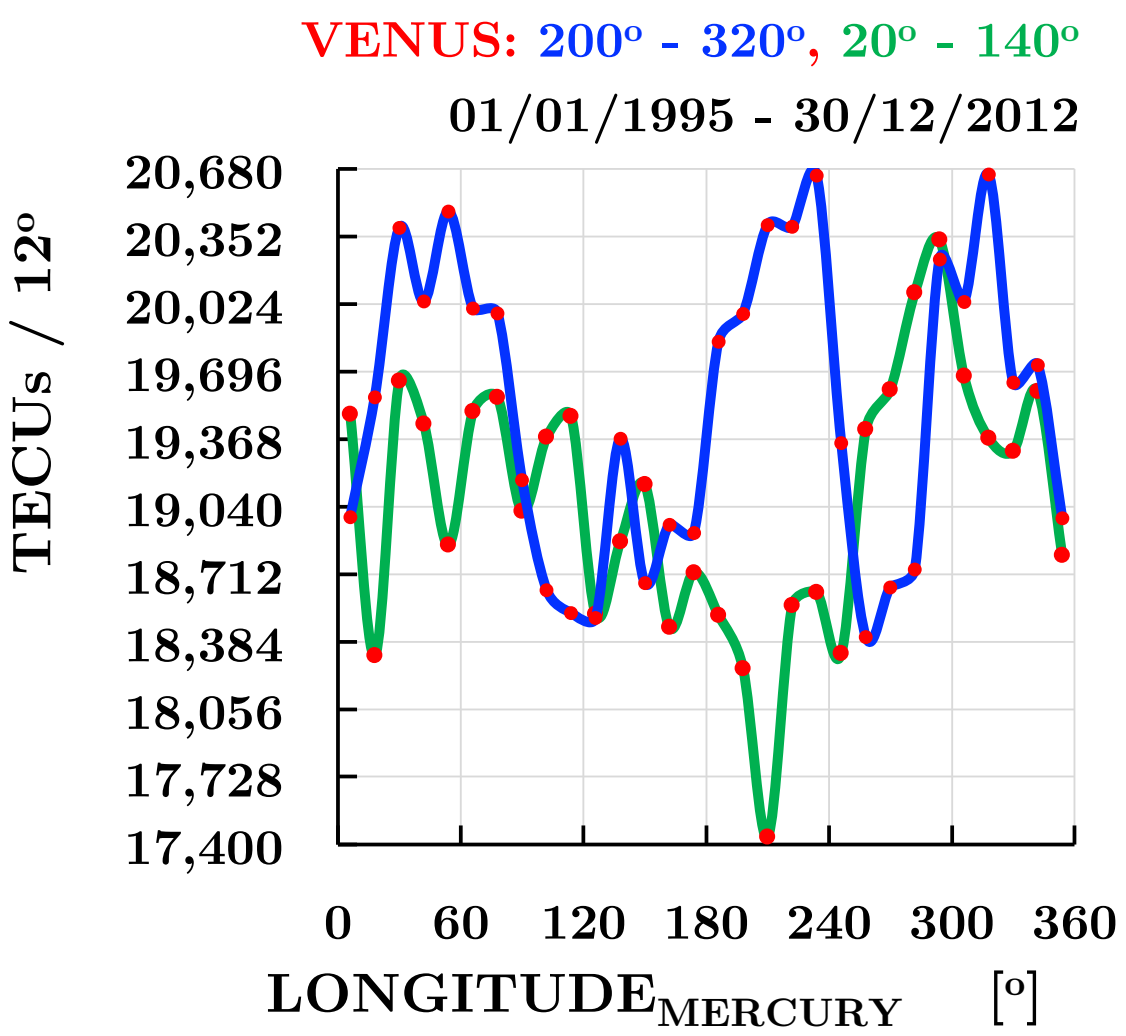

(c) Comparison between Fig. 15.6a and Fig. 15.6b.

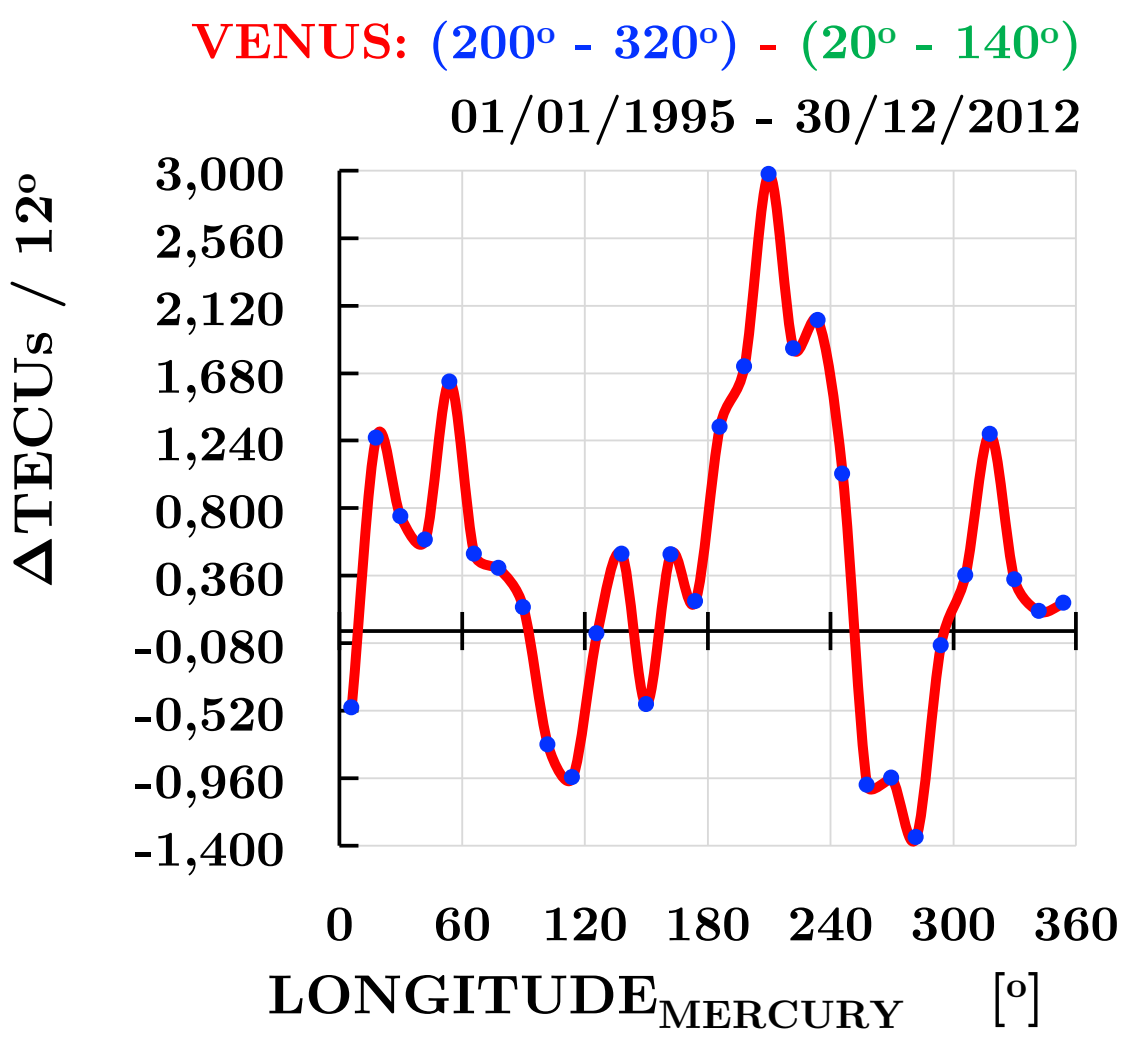

(d) Fig. 15.6a minus Fig. 15.6b.

**Figure 15.6:** Ionospheric TEC distribution for the reference frame of Mercury when Venus propagates between 180° opposite orbital arcs with 120° width, for 12° bin.

## 15.3.2 Fourier analysis

A Fourier analysis has been performed with all data from the period 01/01/1995 - 30/12/2012 aiming to provide an additional independent confirmation of the derived periodicities. The acquired results are then compared with the various planetary revolution periods and planetary synods.

The most interesting derived periodicity, seen in Fig. 15.9, corresponds to the $10^{th}$ biggest peak at $27.33\,d \pm 0.04\,d$ which has an amplitude of $\sim 4\,dB$. This coincides with Moon's sidereal month of $27.32\,d$ which is fixed to remote stars. On its own this observation points to an exo-solar impact on the ionisation of the ionosphere. We stress that such a narrow peak can

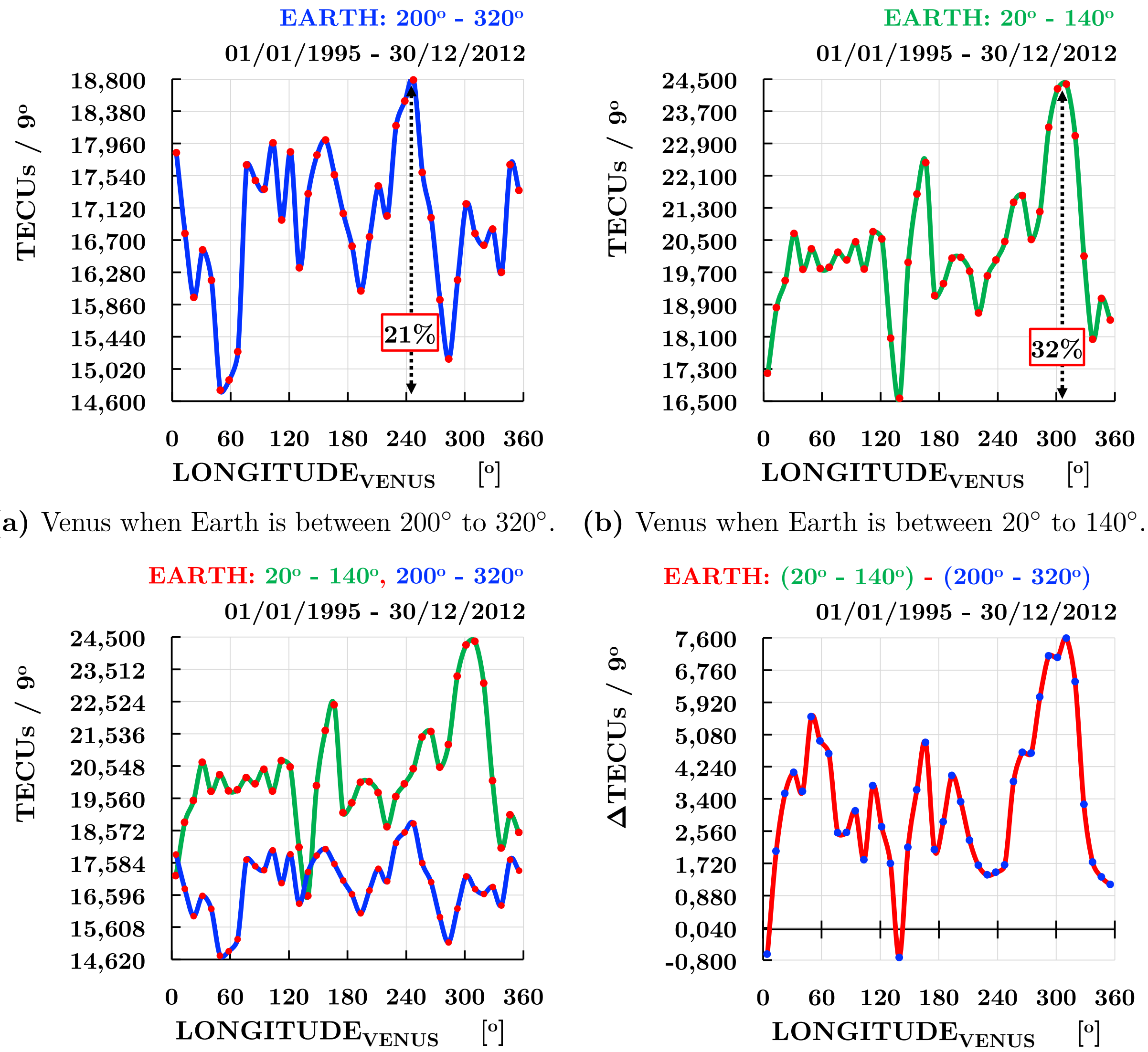

**(a)** Venus when Earth is between 200° to 320°. **(b)** Venus when Earth is between 20° to 140°.

**(c)** Comparison between Fig. 15.7a and Fig. 15.7b. **(d)** Fig. 15.7b minus Fig. 15.7a.

**Figure 15.7:** Ionospheric TEC distribution for the reference frame of Venus when Earth propagates between 180° opposite orbital arcs with 120° width, for 9° bin.

not originate from the differential rotation of the Sun which spans from about 25 d to 36 d between the equator and poles accordingly [269, 270].

In Fig. 15.9 we observe that there is also a tiny peak around 29.5 d ± 0.1 d which is the 39$^{th}$ biggest peak with amplitude of 1.1 dB. However, this is about 4 times less pronounced than the one around 27.33 d. We remind that 29.53 d, is Moon's synodic period fixed to the Sun. Also, this indicates a much stronger exo-solar impact on the ionosphere than the assumed dominating solar activity. This is an unexpected and highly significant result, since the ionosphere is known to be "strongly" influenced by the variable solar EUV irradiance.

Finally, the 14$^{th}$ biggest peak with 3.1 dB amplitude is around 223.8 d ± 2.9 d. This coincides within $\leq 1\sigma$ with the revolution period of Venus (224.7 d). Thus, the derived Fourier

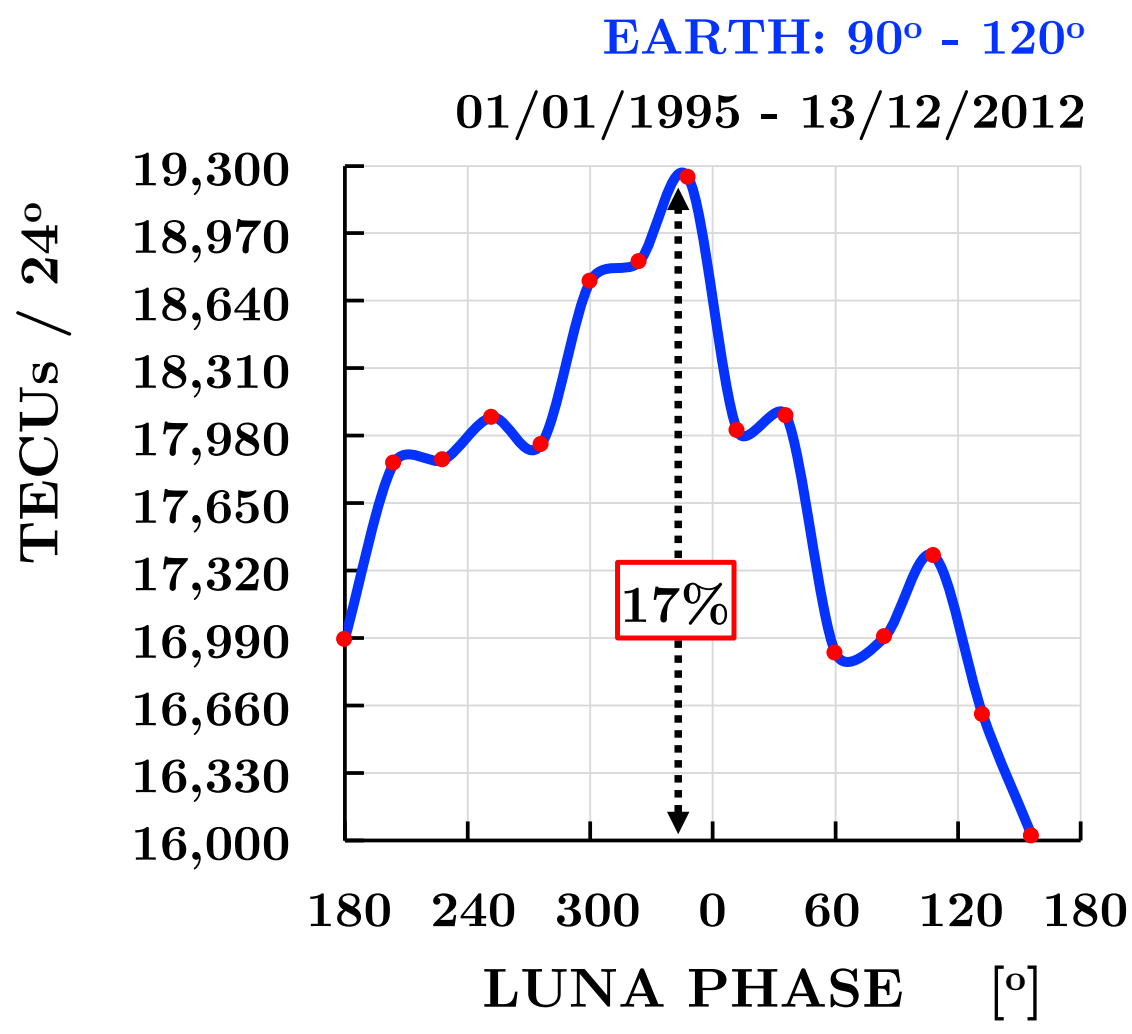

**(a)** Moon when Earth is between 90° to 120°.

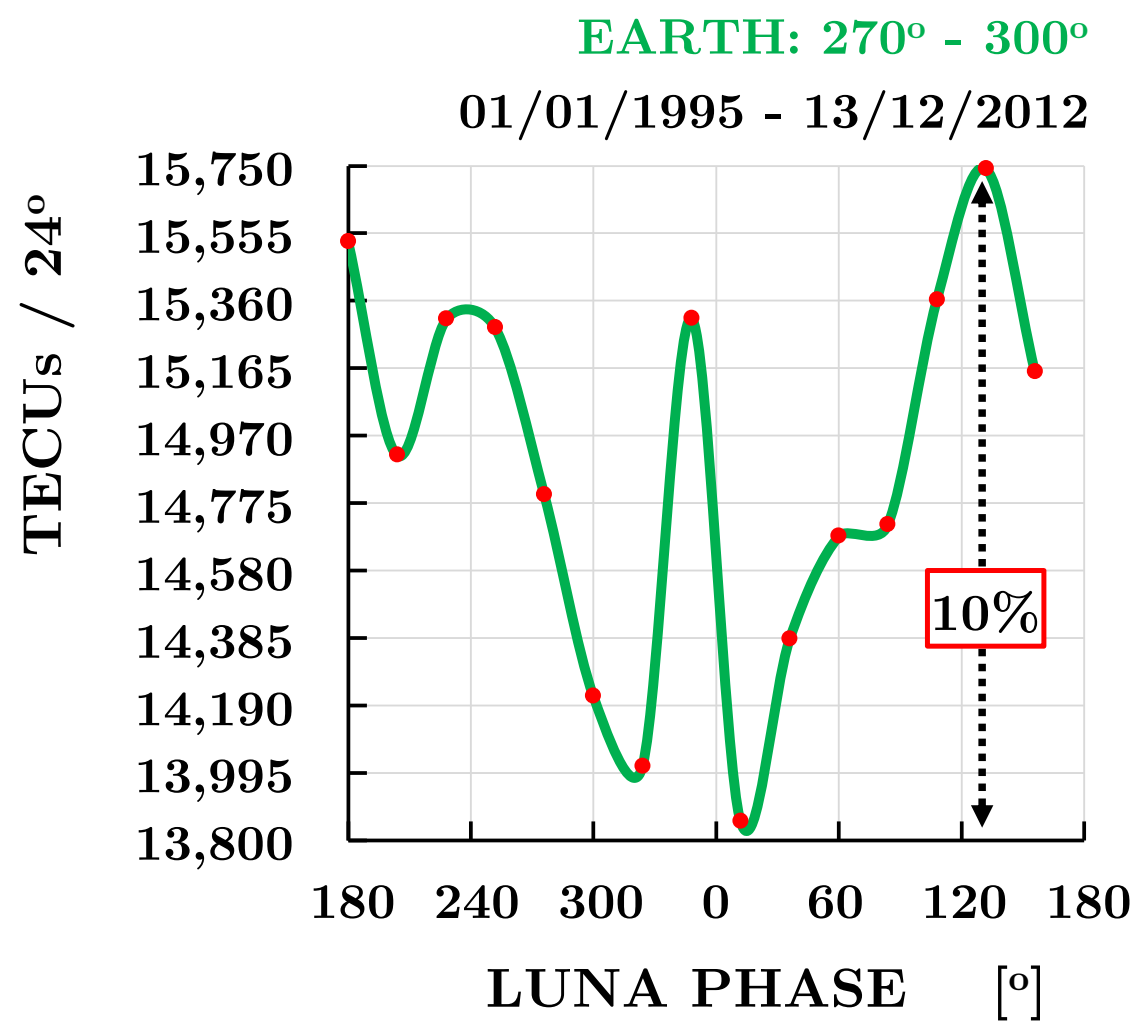

**(b)** Moon when Earth is between 270° to 300°.

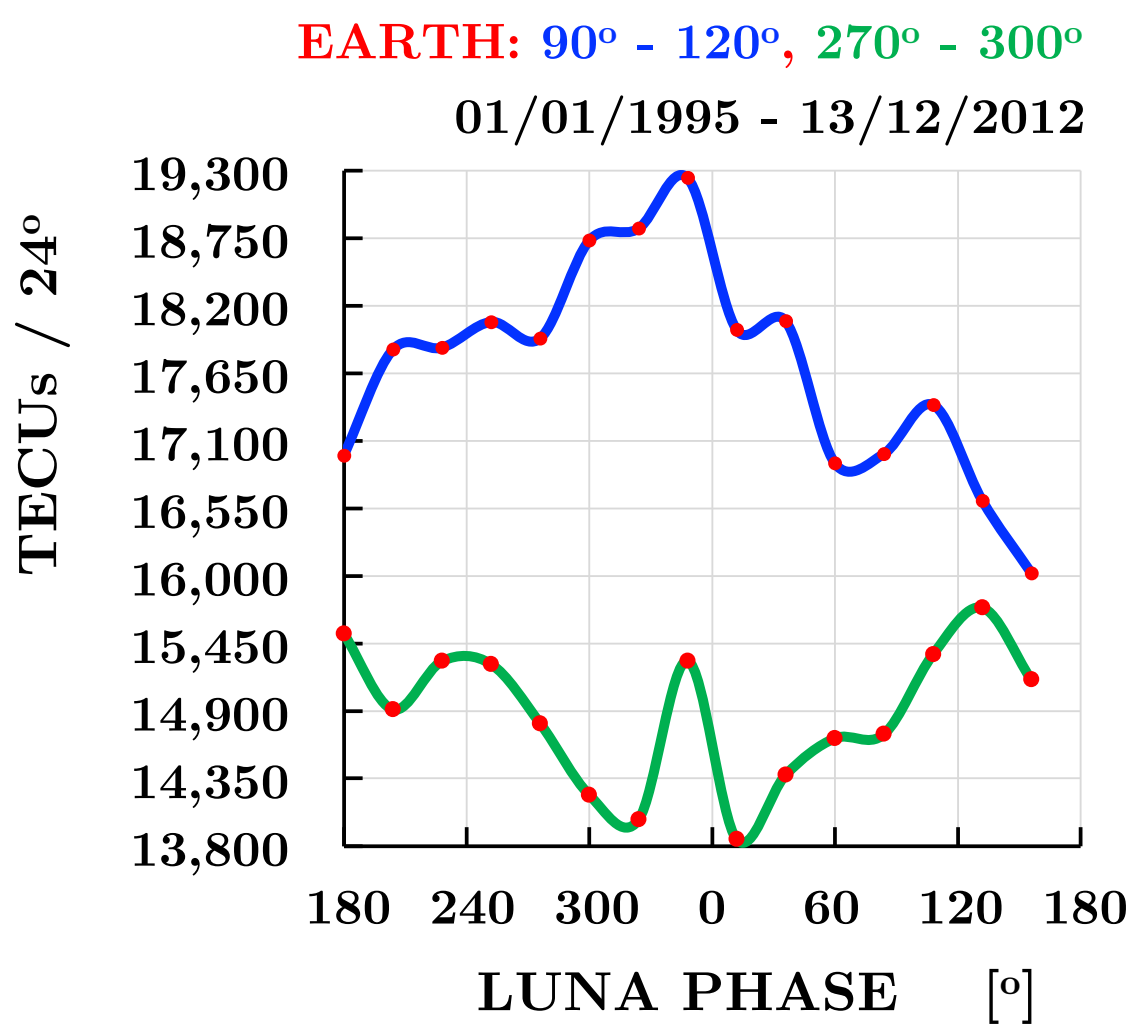

**(c)** Comparison between Fig. 15.8a and Fig. 15.8b.

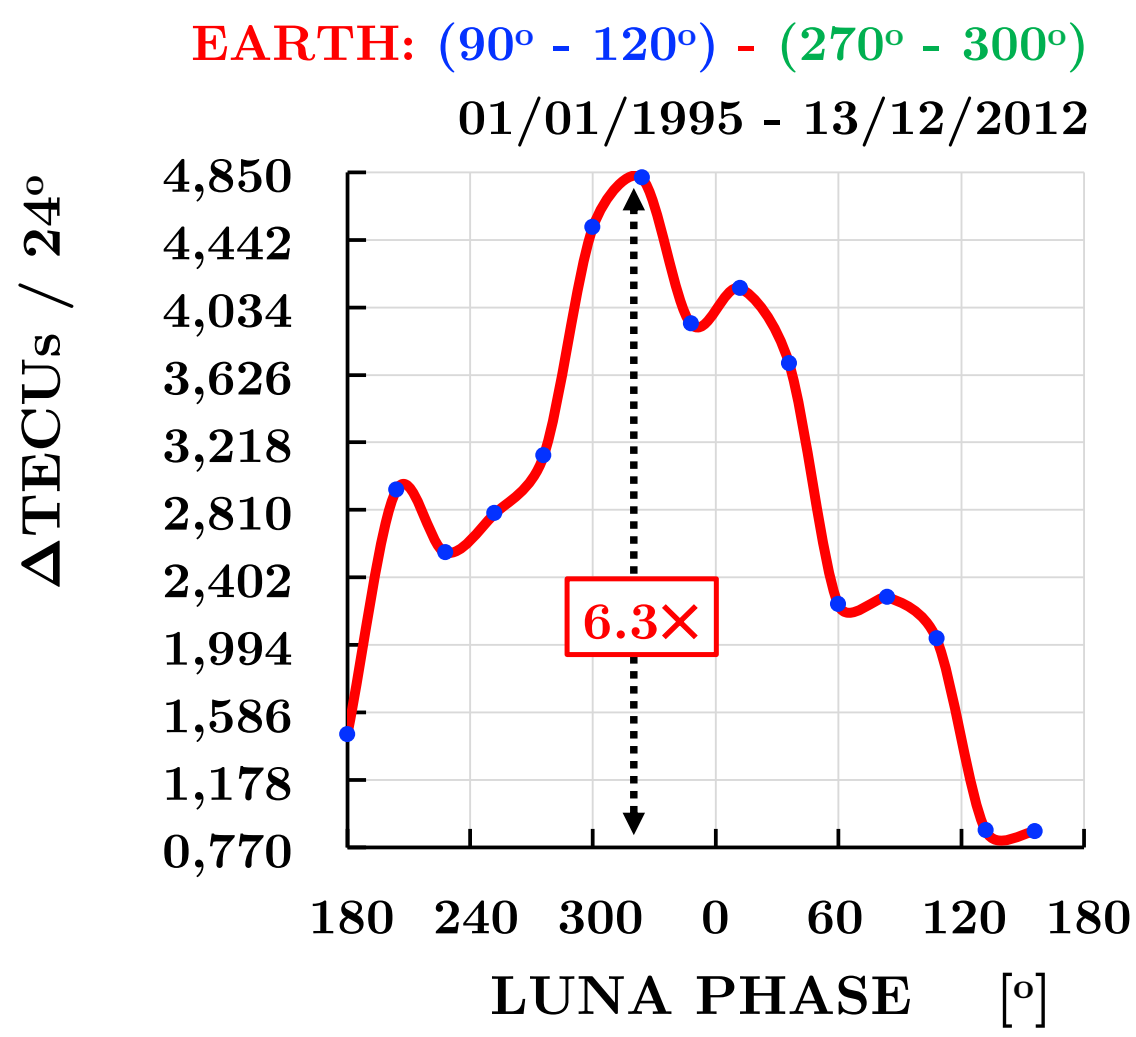

**(d)** Fig. 15.8a minus Fig. 15.8b.

**Figure 15.8:** Ionospheric TEC distribution as a function of the Moon's phase when Earth propagates between 180° opposite orbital arcs with 30° width, for 24° bin. The two orbital arcs were selected based on the observed difference in rate between the two minima seen in Fig. 15.5c.

periodicities verify further the unexpected results from the heliocentric longitude distributions presented before such as the the one in Fig. 15.8d which points to a considerable impact of the position of the Moon in modulating the TEC of the ionosphere.

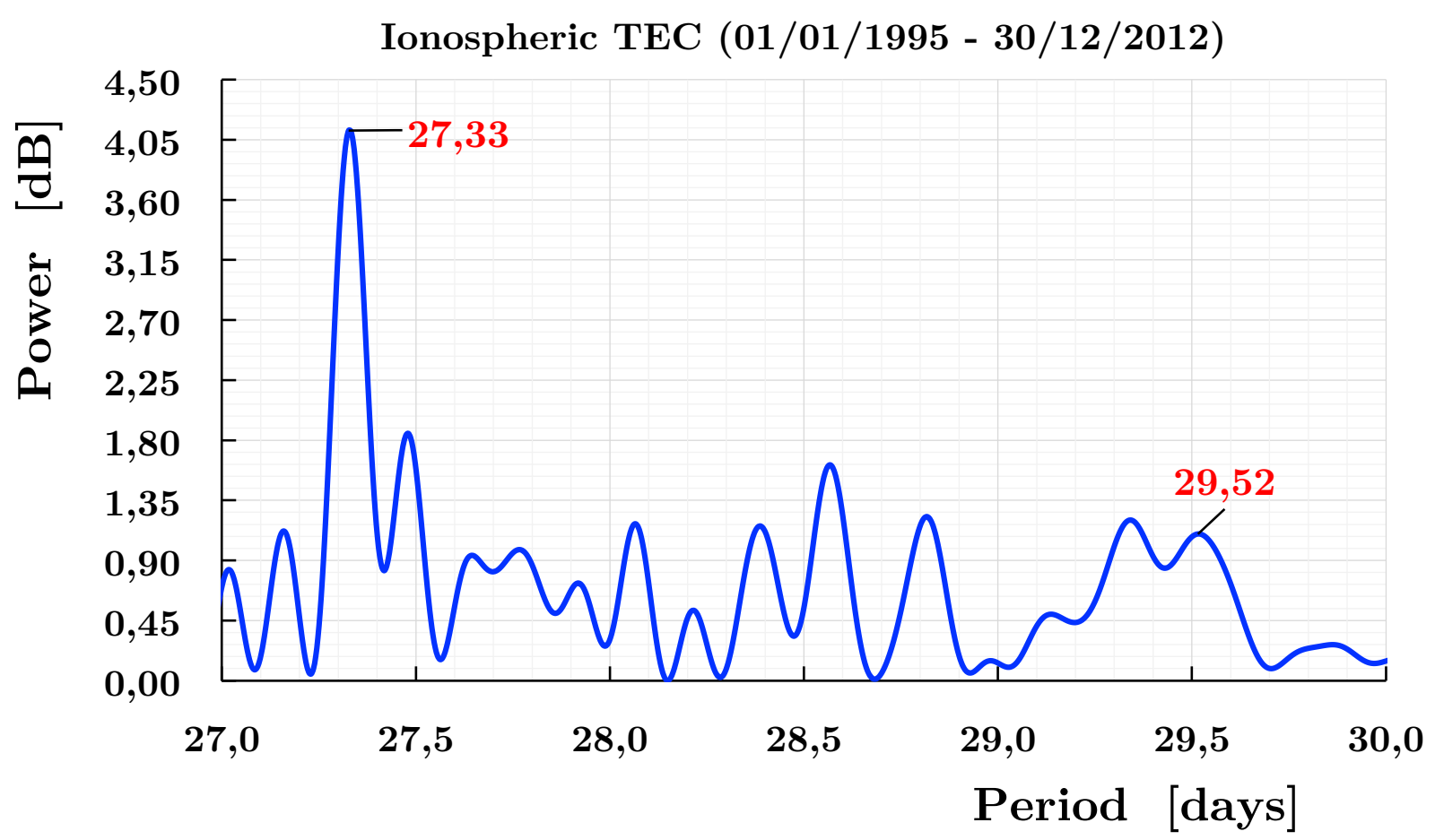

**Figure 15.9:** Fourier periodogram of the daily ionospheric TEC data for the period 01/01/1995 - 30/12/2012 zoomed in around 28.5 d.

## 15.4 Summary

The anomalously high electron content of the ionosphere shows an unexpected planetary relationship. The strongest observed correlation appears with the phase of Moon when Earth is around the two solstices. In fact, the position of the maximum difference coincides with the new Moon which is when the Moon propagates between Earth and Sun. This configuration goes along with the assumptions of this work and point to stream(s) of invisible matter coming from the direction of the Sun around December which is gravitationally focused at the Earth's position from the combined effect of the Sun and the interposed Moon. It is also noted that for an Earth observer, each year around middle of December ($\sim 86°$) we have the alignment GC $\rightarrow$ Sun $\rightarrow$ Earth within $\sim 5.5°$. This is shown in Fig. 15.10, where Mercury, Venus as well as the Moon have been placed in their preferred orbital ranges around 80° and 260° respectively as derived by their corresponding longitudinal distributions.

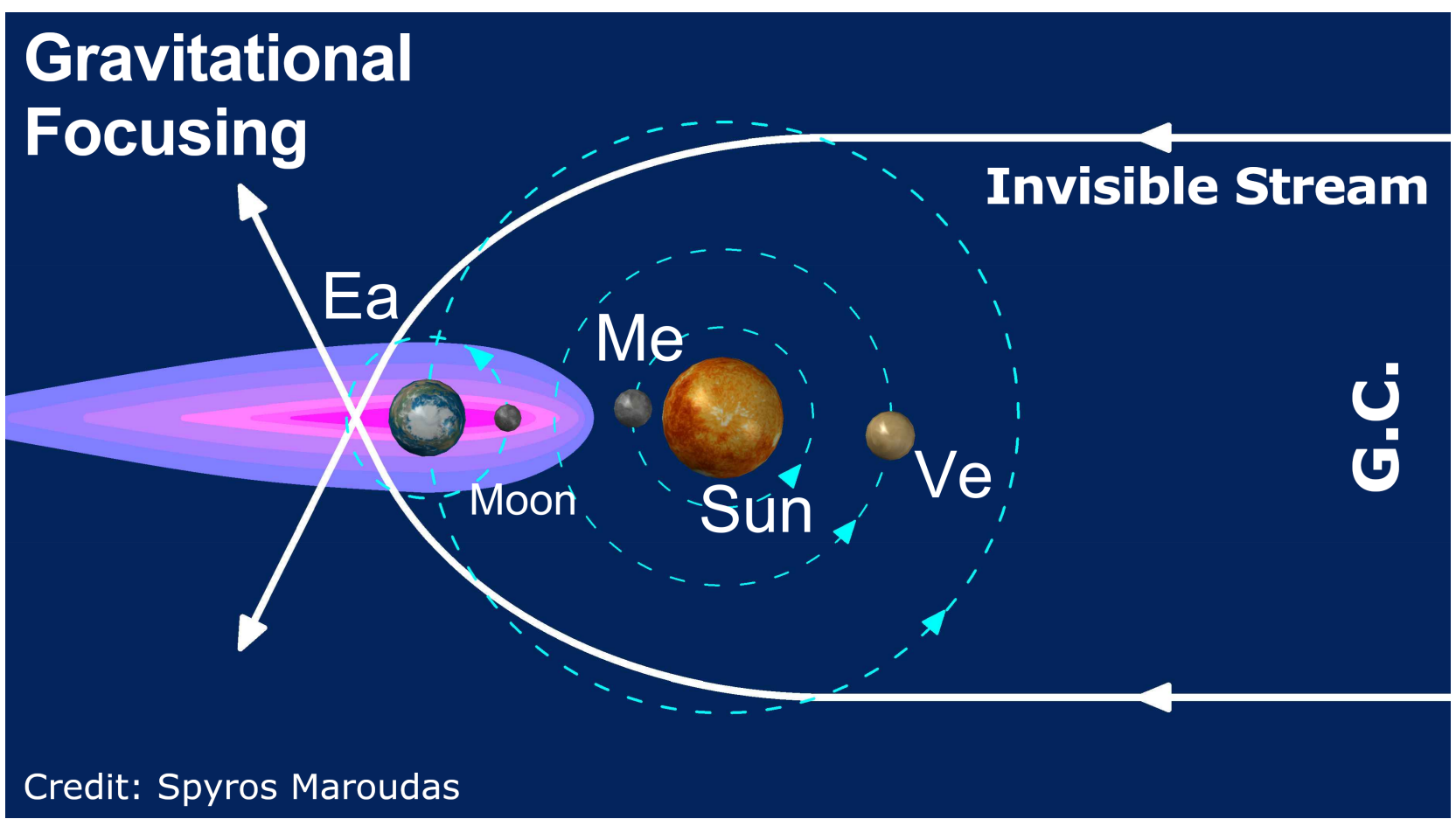

**Figure 15.10:** Schematic illustration of the GC - Sun - new Moon - Earth configuration focusing an invisible massive stream towards the Earth. Orbits and planet sizes are not to scale.

The significant Moon correlation is also supported by the Fourier analysis results, with the 27.33 d periodicity associated Moon's sidereal month referenced to remote stars being much stronger than the 29.53 d one, associated with Moon's synodic period referenced to the Sun. This observations points surprisingly to a strong exo-solar impact on the global TEC of the ionosphere. Finally, it is noted that a comparison of the various planetary distributions with EUV and F10.7 shows a significant positive linear correlation in most cases, which is expected based on their direct impact on the ionospheric electron density (see Appendix Sect. B.9).

Future analysis can include even more extended periods with even smaller than 1 d cadence time as well data from different mapping techniques. In addition, spatiotemporal variations of TEC (see [351] and references therein) can also be analysed to localise further the direction and duration of the envisaged streams. Finally, it would be interesting to search for planetary-related effects on the ionisation degree of other planetary atmospheres.

# 16 STRATOSPHERIC TEMPERATURE

## 16.1 Introduction

The stratosphere extends at an altitude of about 10 km to 50 km with the height of its lower boundary, called *tropopause*, varying from 7 km to 20 km depending on the latitude and seasons. The pressure in this region is between 200 hPa to 1 hPa (see Fig. 16.1a). With the mass of the atmosphere being proportional to pressure, the stratosphere contains about $10\% - 20\%$ of the total mass of the atmosphere with everything above the stratopause being just $\sim 0.1\%$ [352]. Interestingly, ozone ($O_3$) heats this layer of the atmosphere by absorbing UV radiation from the Sun, with the temperature increasing with altitude.

Temperature anomalies are known to exist in the stratosphere, at an altitude of $\approx 40$ km [353, 354]. Notably, some observations [354–356] are consistent with a "top-down" propagating signal from the stratosphere, thus raising the question of whether such anomalies are intrinsic to the atmosphere itself or are triggered by some external source. At the same time there is a variety of observations correlating the upper atmosphere with the 11 y solar cycle [338–341].

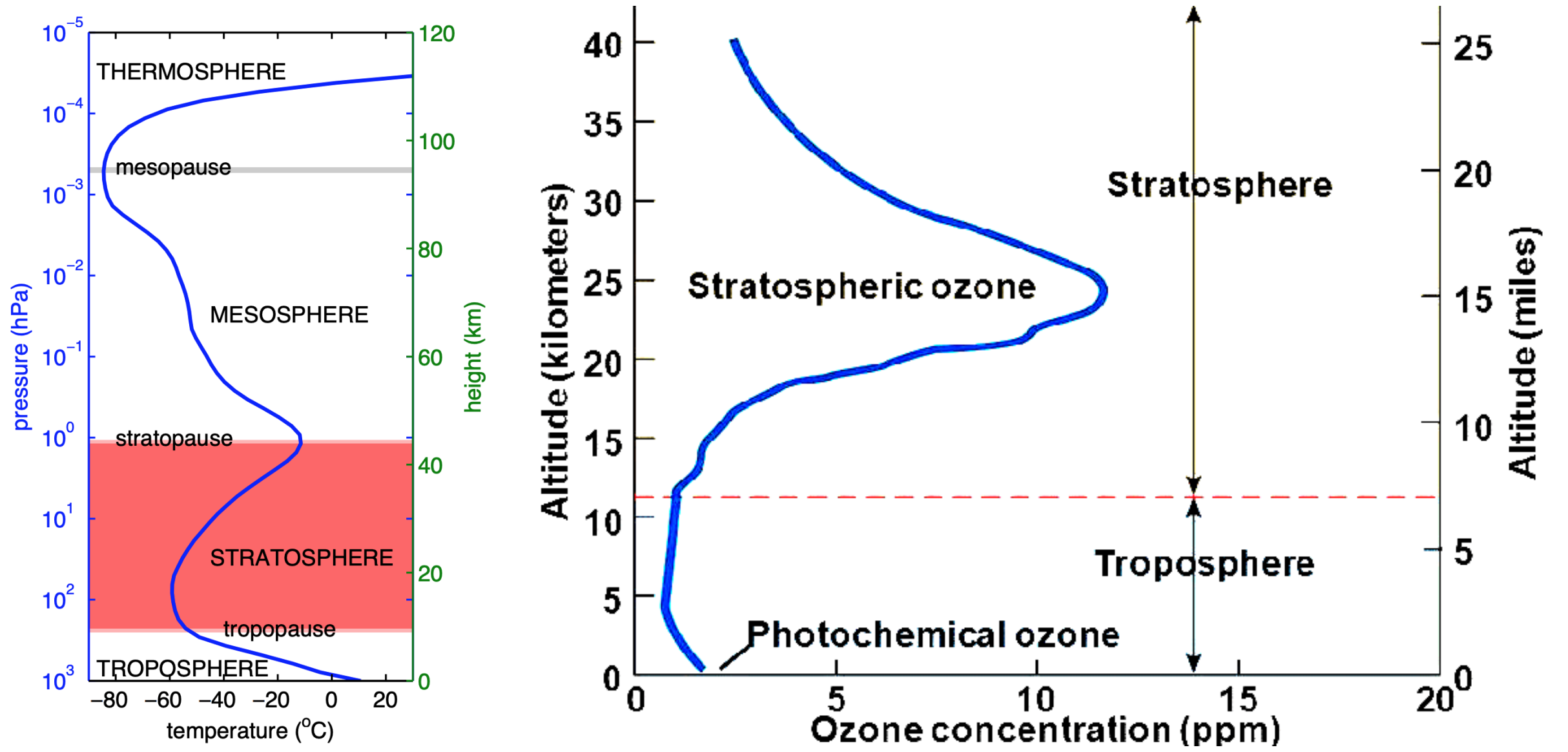

**(a)** Typical vertical atmospheric temperature profile based on January zonal mean temperature at 40° N (70).

**(b)** Ozone concentration in parts per million at different altitudes (71).

**Figure 16.1:** Typical characteristics of stratosphere.

These observations include changes in global and regional climate, the ozone production, zonal winds etc. On the absence of a conventional explanation [342, 343], the external planetary influence turns out to be the most probable scenario based on the assumptions of this work.

Notably, an observed peaking planetary relationship is the key signature for the involvement of the dark universe. The driving idea is based on gravitational focusing by the Sun and its planets towards the Earth of slow-moving invisible streaming matter. Whatever its ultimate properties are, it should interact somehow with the upper atmosphere in order to be able to cause the observed anomalous stratospheric behaviour. Additionally, if these stratospheric temperature anomalies, which can not be explained as being intrinsic to the atmosphere, also do not follow the simultaneously measured variable solar activity as this is manifested by EUV or the F10.7 solar proxy, which strongly affect the Earth's atmosphere, then, the only viable explanation seems to be that of an exo-solar impact from the dark universe (see also Publications E.3, and E.18).

## 16.2 Data and methods

### 16.2.1 Data origin

The stratospheric temperature data used in this analysis have been acquired from European Centre for Medium-range Weather Forecast (ECMWF) [357], and more specifically, from the ERA-Interim global reanalysis dataset [358]. The re-analysis data are created via a data assimilation scheme and models which ingest all the available observations every 6 h to 12 h over the period being analysed. In particular, about $7-9$ million observations are ingested at each time step, with the spatial resolution of the dataset being about $0.125° \times 0.125°$ on 60 vertical levels from the surface up to 0.1 hPa. The measurements which are used for the assimilation come mostly from satellite measurements but also from radiosondes, pilot balloons, aircraft and wind profilers. The number of available data from these sources remains constant during the specific period used except from the aircraft reports where the numbers increased significantly after 1998 [358].

The specific period used in this analysis is between 01/01/1986 and 31/08/2018. The grid points used were at 42.5°N, 13.5°E with a pixel size of about 10.7 km (at West-East direction) by 13.8 km (at North-South direction). This is close to Istituto Nazionale di Fisica Nucleare (INFN) Laboratori Nazionali del Gran Sasso (LNGS) in Italy, and was chosen due to the recently published results showing an annual long-term modulation of the cosmic muon flux during 17/5/2007 - 17/05/2017, with a measured amplitude of $(1.36 \pm 0.04)$ % [359]. The isobaric levels from the upper stratosphere that are used lie at 3 hPa, 2 hPa and 1 hPa corresponding to an approximate altitude of 38.5 km, 42.5 km and 47.5 km respectively. For each isobaric level two datasets were retrieved, one at 00:00 UTC and one at 12:00 UTC, as seen in Fig. 16.2. Finally, the overall derived uncertainty of each initial single temperature measurement is about $\pm 0.5$ K.

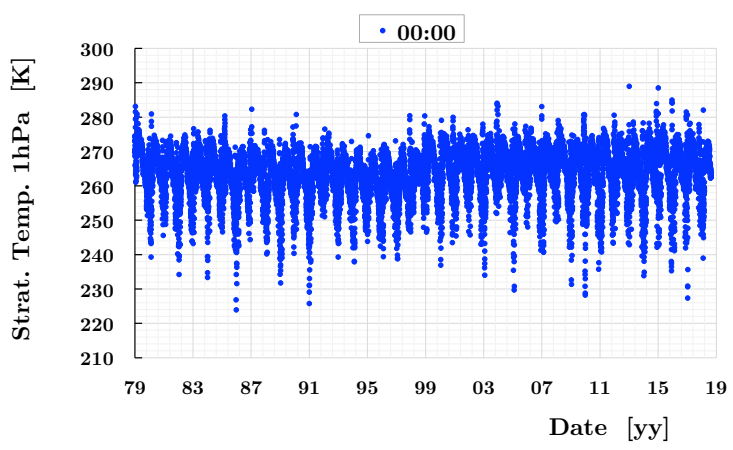

**(a)** 1 hPa at 00:00 UTC.

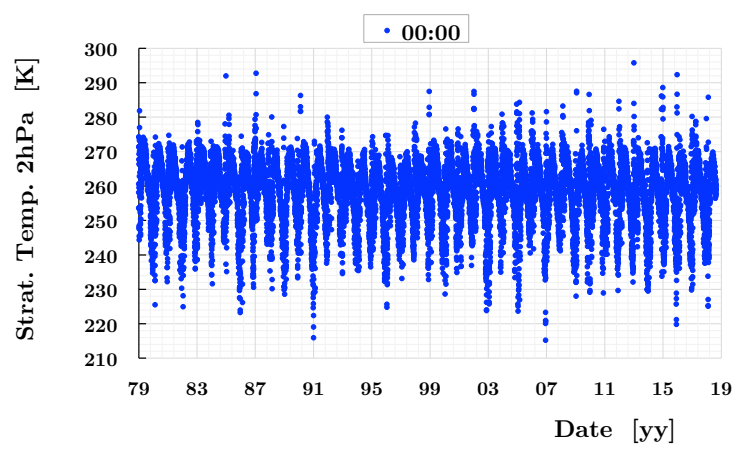

**(b)** 2 hPa at 00:00 UTC.

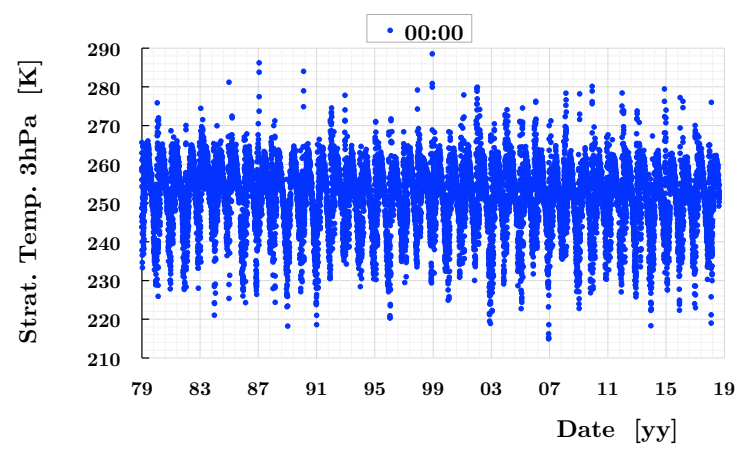

**(c)** 3 hPa at 00:00 UTC.

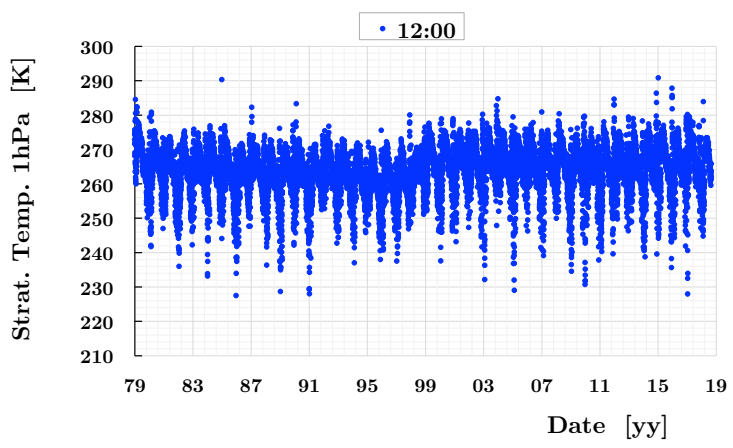

**(d)** 1 hPa at 12:00 UTC.

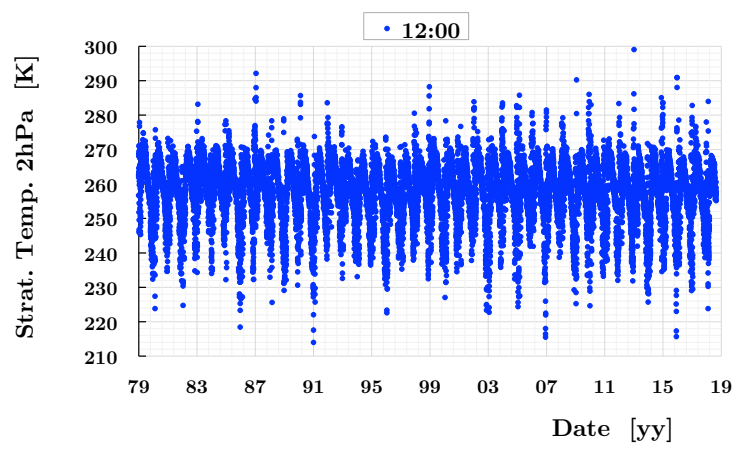

**(e)** 2 hPa at 12:00 UTC.

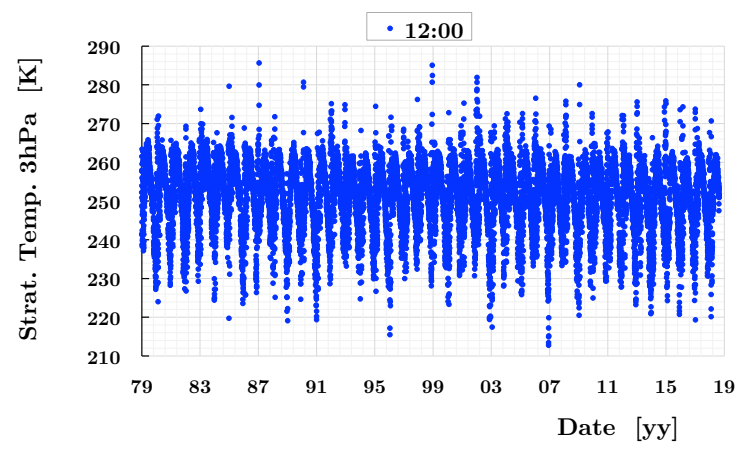

**(f)** 3 hPa at 12:00 UTC.

**Figure 16.2:** 1,2, and 3 hPa stratospheric temperature data for 00:00 UTC and 12:00 UTC for the period 01/01/1986 and 31/08/2018.

### 16.2.2 Data treatment

First of all, from the three isobaric levels at 1 hPa, 2 hPa, and 3 hPa seen in Fig. 16.2, an average temperature between the three layers has been derived: (1 hPa + 2 hPa + 3 hPa) /3. This procedure was performed for both of the two acquired datasets at 00:00 UTC and 12:00 UTC. As next, the arithmetic average between the two has been calculated [(00 : 00 + 12 : 00) /2] which corresponds to the average temperature at 18:00 UTC. Then, in each derived daily measurement the error of ±0.5 K has been designated. Finally, in each daily measurement the corresponding planetary heliocentric longitude position for 18:00 UTC was assigned in order for the analysis mentioned in Sect. 4.5.1 to take place.

As mentioned, only the data from the period 17/5/2007 - 17/05/2017 have been used for the current analysis because of possible comparison with the Borexino detector results [359]. However, the results from the entire data period are similar.

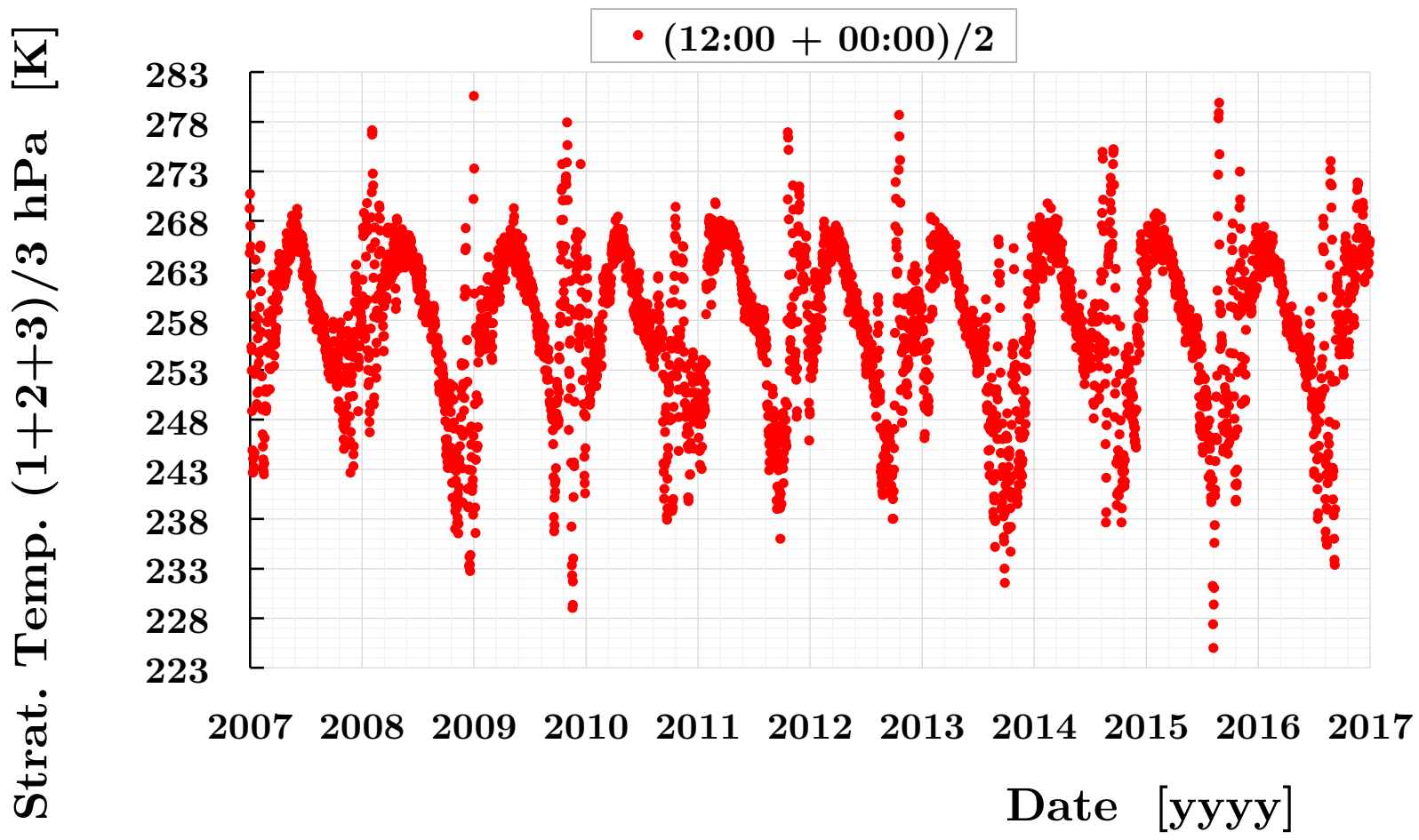

**Figure 16.3:** Data used in the main part of this analysis for the period 17/5/2007-17/05/2017. Each major tick mark in x-axis corresponds to 01/01 of each year.

Notably, from Fig. 16.3, we already see that around the months of December-January each year, significant temperature excursions on top of the rather smooth seasonal variations appear. Therefore, these specific Stratospheric Temperature Anomalies (STAs) will be followed further in the following analysis.

## 16.3 Data analysis and results

### 16.3.1 Planetary longitudinal distributions

#### 16.3.1.1 Single planets

To search for a planetary effect, the daily mean stratospheric temperature is projected on the associated planetary heliocentric longitudinal coordinates of the various planets for each given day. This is shown in Fig. 16.4 where only single planets are used without any constraint on the longitudinal position of the rest of the planets.

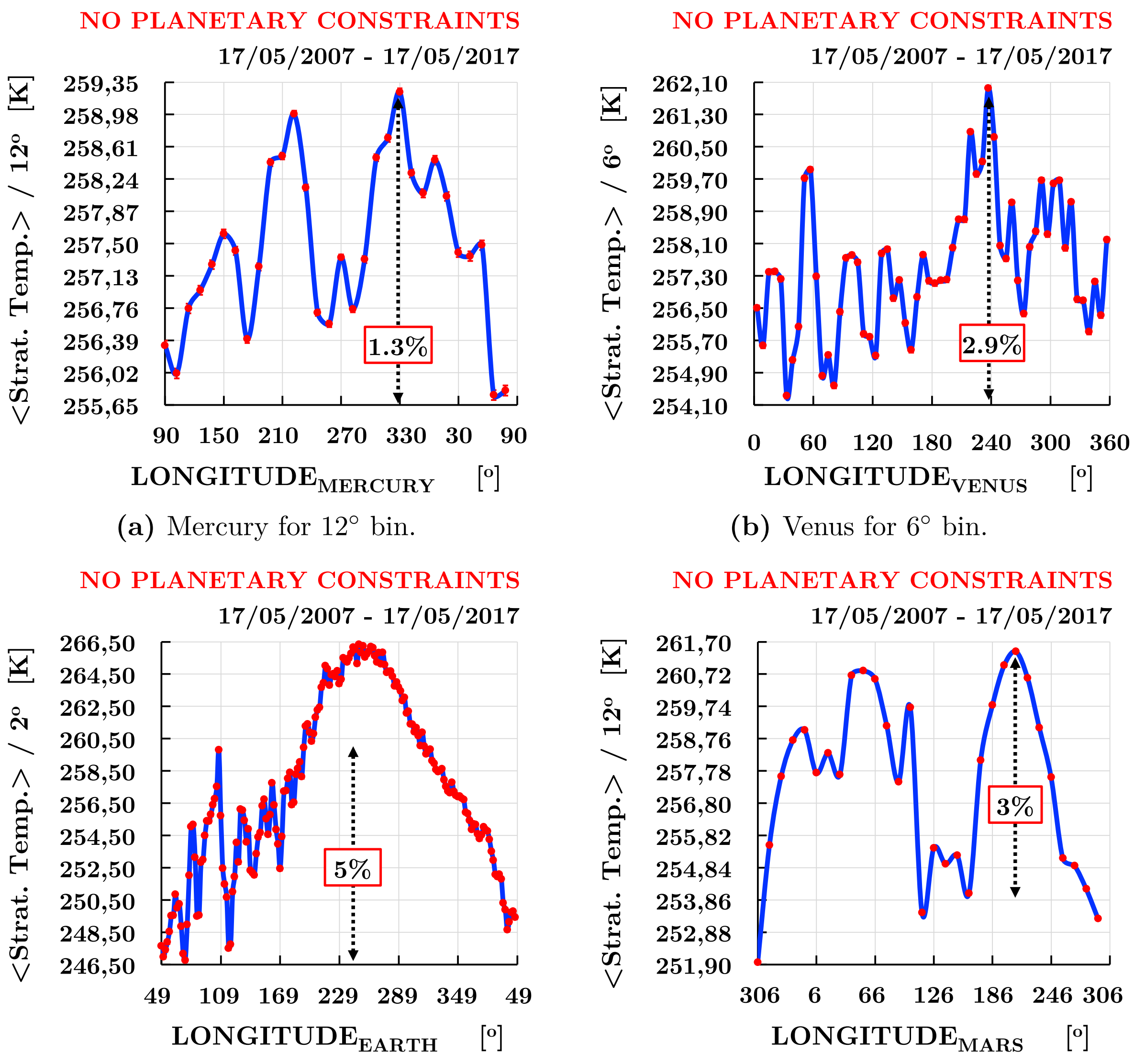

**(a)** Mercury for 12° bin.

**(b)** Venus for 6° bin.

**(c)** Earth for 2° bin.

**(d)** Mars for 12° bin.

**Figure 16.4:** Planetary heliocentric longitude distributions of stratospheric temperature for the period 17/5/2007 - 17/05/2017.

In Fig. 16.4c the 10 y averaged temperature distribution is projected to one year as a function of Earth's position with a bin of 12°. The annual temperature excursions observed in Fig. 16.3 around December to January give rise to a peak in Earth's spectrum around 100° which has a statistical significance of $54\sigma$ compared to the single point around 119°. It is also noted that the direction of the GC → Sun → Earth alignment is around 86.5°, which coincides with the onset of the 100° peak. Moreover, the observed wide smooth peak of the Earth around 260° with an amplitude of 7.4% corresponds to the summer period (around June) and therefore this behaviour is expected. Subsequently, the peaking first half of the Earth's spectrum is the one that we will focus on, to provide more information on this temporal heating of the upper stratosphere.

More reference frames are also used, besides, Earth like the ones of Mercury (Fig. 16.4a), Venus (Fig. 16.4b) and Mars (Fig. 16.4d) which also show a significant planetary relationship of the stratospheric temperature and have amplitudes of 1.3%, 2.9% and 3.6% respectively. It is noted that all the points in the plots contain error bars, though not directly visible since they are very small. This means that all the observed peaks are statistically significant and far above $5\sigma$. As an example, the single point in the peak around 330° compared to the point at 282° gives a difference of $43\sigma$. Similarly, for Venus the point around 237° compared to the single point on the left around 201° gives $45\sigma$. Interestingly, a performed simulation for the the two wide peaks of Mars shows that they should derive from the single wide peak of the Earth around 260° and therefore should not be considered as physical at this point (see Appendix Sect. A.2.3).

#### 16.3.1.2 Double planetary combinations

In order to search for enhanced effects double planetary combinations are also used by constraining an additional planet to propagate in a fixed heliocentric orbital arc during the selected time period. For example, in Fig. 16.5 the mean upper stratospheric temperature is plotted in the reference frame of Mercury while requiring Earth to propagate in two opposite orbital arcs with 130° width. As a result, in Fig. 16.5a, the maximum - minimum difference is 3.5% with the number of days fulfilling this constraint being 1284 d out of 3654 d. In the opposite position of Earth in Fig. 16.5b the amplitude is 0.4%, with the comparison of the two cases shown in Fig. 16.5c. This shows that the $\sim 44$ Mercury orbits smear out the original Earth's decreasing distribution shown Fig. 16.4c, which as we see in Fig. 16.5b it becomes totally flat showing no planetary connection. On the other hand the amplitude of Fig. 16.5a is comparable to the peak around 100° from Fig. 16.4c and in this case, no smearing out effect from the multiple orbits of Mercury is observed. This indicates a preferred position on the Earth's longitude around 50° to 180°. This also serves as a kind of simulation which shows that any unforeseen random fluctuations should appear similar in Fig. 16.5a and 16.5b. Instead, for the case of Fig. 16.5a we have statistically significant peaks far above $5\sigma$. The mean statistical standard deviation per bin in this plot is $\sim 0.08$ K. As an example, the wide peak around 21° in Fig. 16.5a has a statistical significance of $\sim 38\sigma$.

Similarly, in Fig. 16.6 the reference frame of Venus is used with the same constraints in Earth's longitudinal position. A comparable phenomenon is observed here, with the maximum - minimum difference on Fig. 16.6a being about 5.1% whereas in Fig. 16.6b, which basically represents the null effect, the difference is 0.8%. This confirms the observation made in Fig. 16.5.

Moving to the reference frame of Earth once more, in Fig. 16.7 we require Mercury

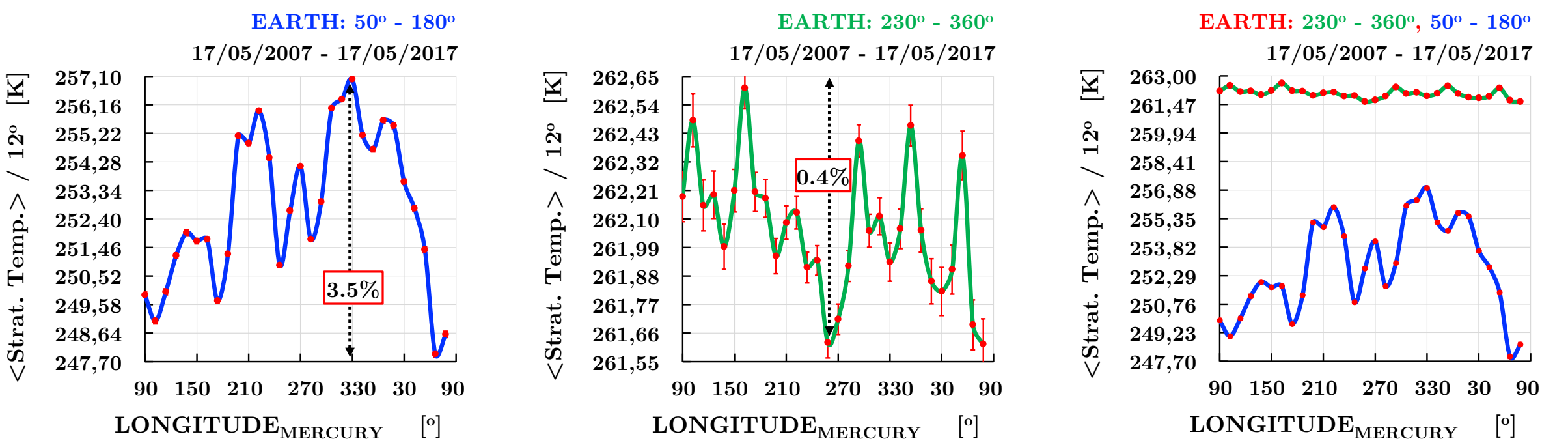

**(a)** Mercury when Earth is between 50° to 180°.
**(b)** Mercury when Earth is between 230° to 360°.
**(c)** Comparison between Fig. 16.5a and Fig. 16.5b.

**Figure 16.5:** Stratospheric temperature distribution for the reference frame of Mercury when Earth propagates between 180° opposite orbital arcs, for 12° bin.

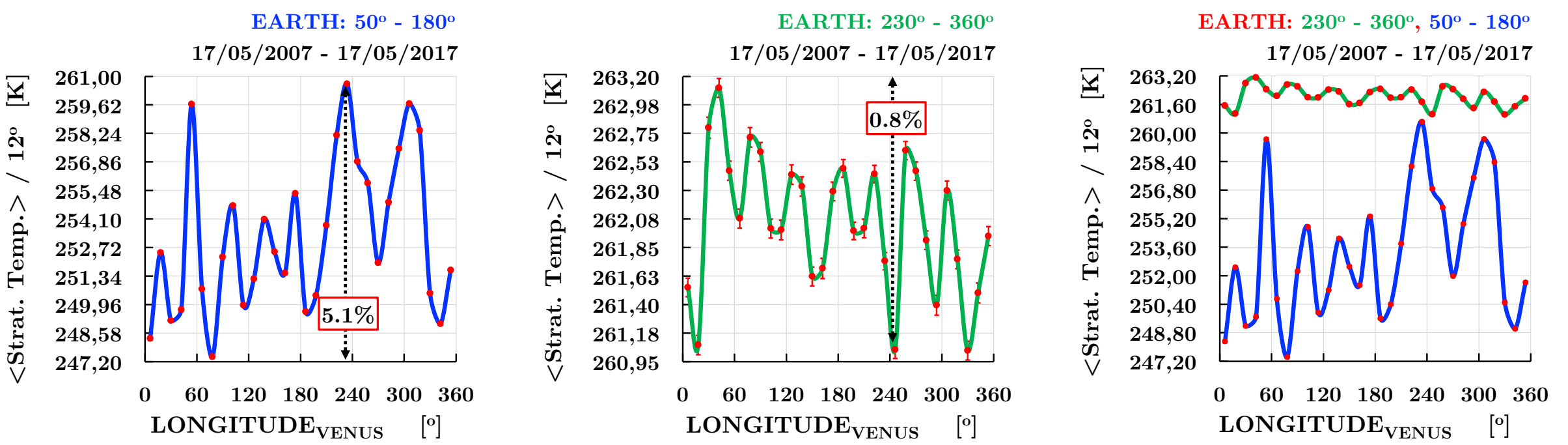

**(a)** Venus when Earth is between 50° to 180°.
**(b)** Venus when Earth is between 230° to 360°.
**(c)** Comparison between Fig. 16.6a and Fig. 16.6b.

**Figure 16.6:** Stratospheric temperature distribution for the reference frame of Venus when Earth propagates between 180° opposite orbital arcs, for 12° bin.

this time to propagate in two wide opposite orbital arcs with 180° width. The observed peak around 100° is much more pronounced in Fig. 16.7a, when Mercury is between 90° to 270°, rather than in the opposite case of 270° to 90° in Fig. 16.7b.

As next, assuming that no additional stratospheric planetary effects do not occur, the Earth spectrum in Fig. 16.4c should be replicated when requiring for example Jupiter in Fig. 16.8a to orbit between 61° to 180°. This orbital arc corresponds to four uninterrupted Earth orbits, which excludes any possible systematic effects. Once more we see that the peak around 100° appears better resolved with a better amplitude relative to the height of the wide peak around 250° pointing once more to planetary involvement. In addition, in Fig. 16.8b when Jupiter is 180° on the opposite side around 241° to 360° the peak around 100° disappears indicating the former case as the prominent location for Jupiter.

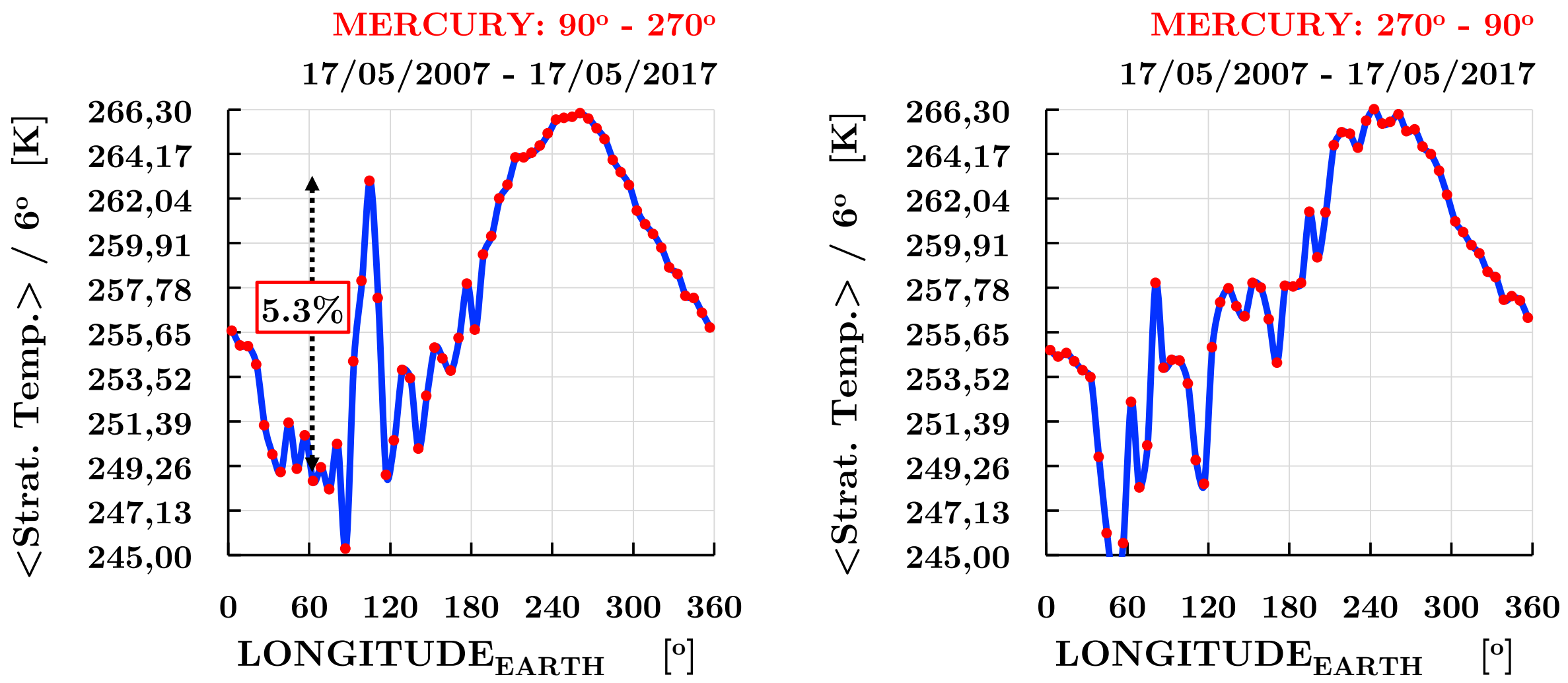

**(a)** Earth when Mercury is between 90° to 270°. **(b)** Earth when Mercury between 270° to 90°.

**Figure 16.7:** Stratospheric temperature distribution for the reference frame of Earth when Mercury propagates between 180° opposite orbital arcs, for 6° bin.

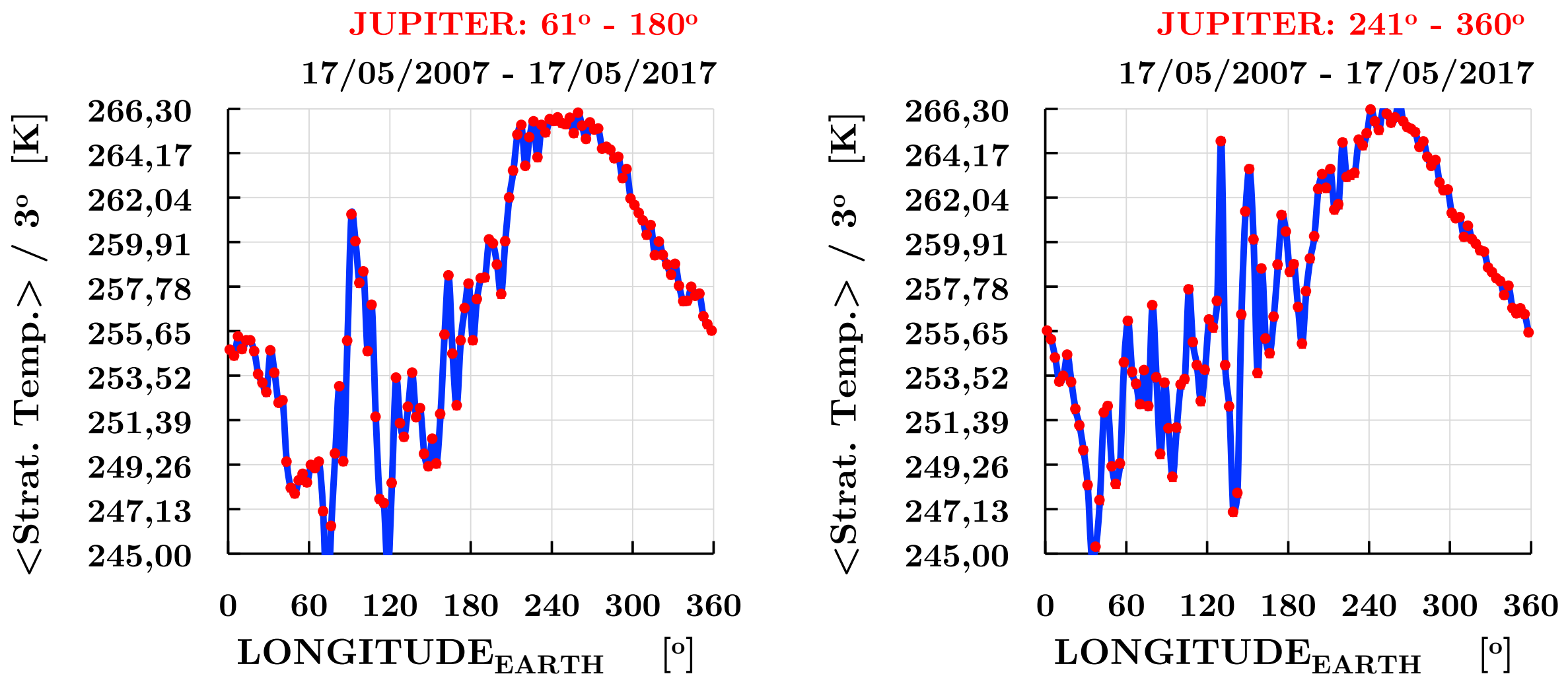

**(a)** Earth when Jupiter is between 61° to 180°. **(b)** Earth when Jupiter is between 241° to 360°.

**Figure 16.8:** Stratospheric temperature distribution as a function of Earth's position when Jupiter propagates between 180° opposite orbital arcs, for 3° bin.

#### 16.3.1.3 Multiple planetary combinations

In Fig. 16.9, multiple planetary combinations can be performed in order to get improved peaking distributions, with the peak around 100° having an even better SNR. In these multiple planetary constraints when observing the mean stratospheric temperature distribution as a function of the Earth's longitudinal position we can better pinpoint the optimal planetary positions and thus the direction and the possible track of the invisible stream(s).

As a next step, the whole available period of the temperature dataset is used 01/01/1979 - 31/08/2018 to improve also the confidence of the whole analysis procedure, as the total number

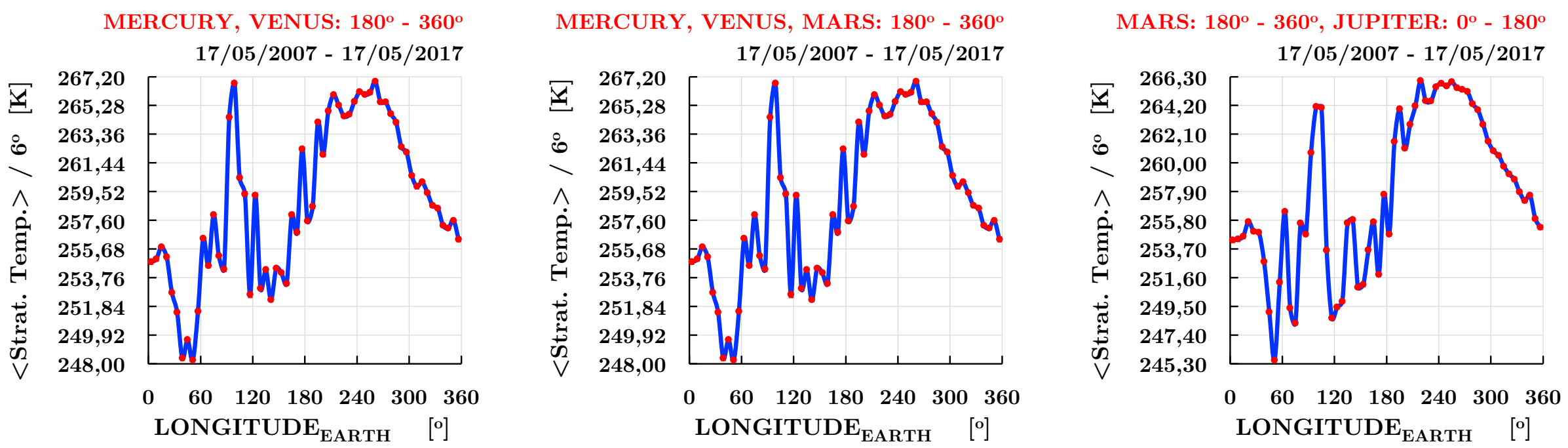

**(a)** Earth when Mercury and Venus are both between 180° to 360°.

**(b)** Earth when Mercury, Venus and Mars between 180° to 360°.

**(c)** Earth when Mars between 180° to 360° and Jupiter between 0° to 180°.

**Figure 16.9:** Stratospheric temperature distribution as a function of Earth's position combined with multiple planetary constraints, for 6° bin.

of days used is 14488, compared to the 3654 d for the period 17/5/2007-17/05/2017 which means that we can use about four times longer period. For the extended period, Earth has performed about 40 orbits while Mercury has performed 165 orbits, facts that should result in a flattening of the corresponding spectra if these temperature excursions are random fluctuations. In Fig. 16.10, Fig. 16.7 is selected to be used for the whole period adding also a constraint on Venus to be on the same heliocentric range as Mercury. The result shown in Fig. 16.10a is actually one of the strongest observed planetary relationship of the peaking upper stratospheric temperature covering the 39 y period. The observed difference on the peak around 100° with the opposite 180° orbital arc, at 270° to 90° is striking as seen in Fig. 16.10c.

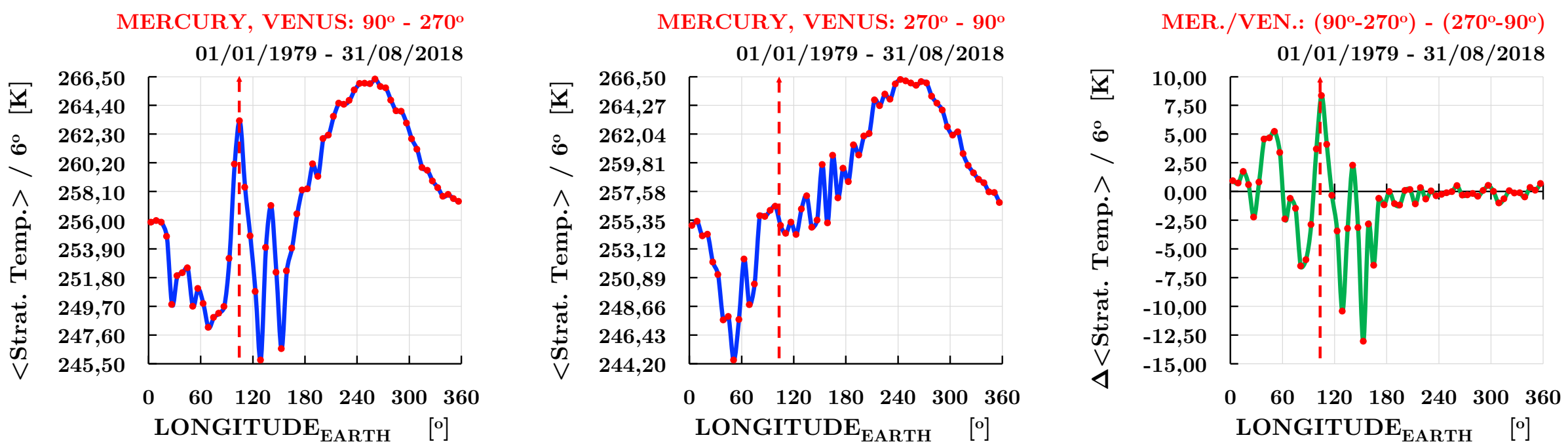

**(a)** Earth when Mercury and Venus are both between 90° to 270°.

**(b)** Earth when Mercury and Venus are both between 270° to 90°.

**(c)** Difference between Fig. 16.10a and Fig. 16.10b.

**Figure 16.10:** Earth's longitudinal distribution for the mean stratospheric temperature when Mercury and Venus propagate between 180° opposite orbital arcs, with bin = 6° for the whole period 01/01/1979 - 31/08/2018. The red dashed lines indicate the position of the peak at ∼ 105 deg.

The same constraints can be applied when we use the reference frame of Moon's phase while this is orbiting around the Earth. The results are presented in Fig. 16.11 where Earth's position is also fixed around the location of the peak at 100°. Again, the full period has been used and both Mercury and Venus have been required to propagate around 90° to 270° in Fig. 16.11a and on the opposite side 270° to 90° in Fig. 16.11b. Their notable difference is shown in Fig. 16.11c where in the location of the peak around 68° seen in Fig. 16.11a, we have about 12 K difference between the two cases. This result also proves the unexpected relationship of the stratospheric temperature distribution on Moon's position.

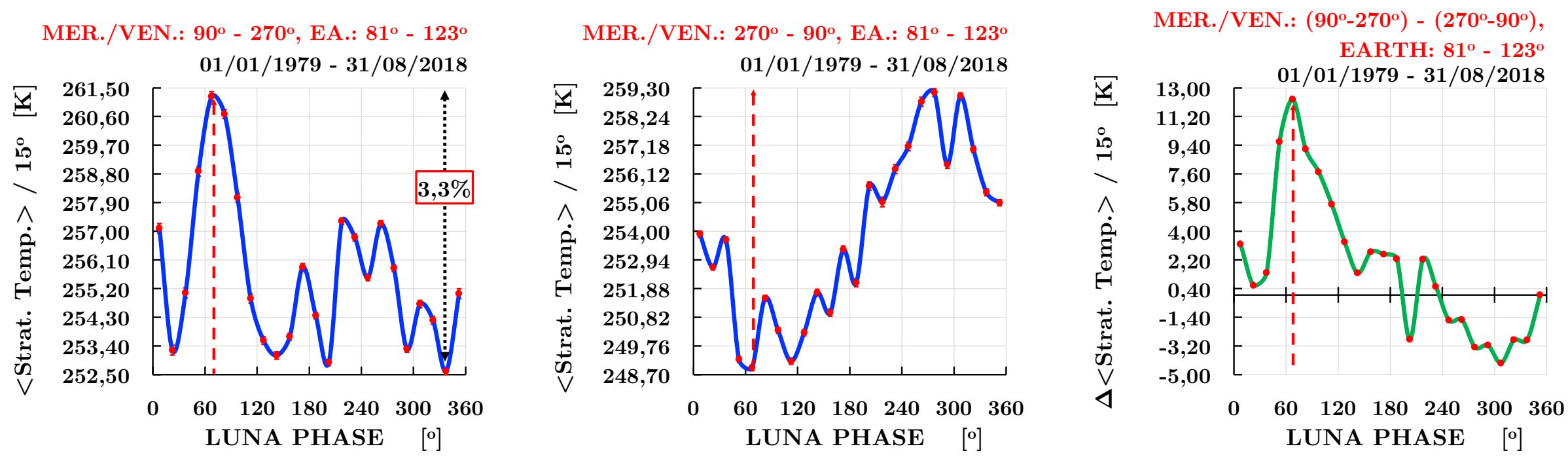

**(a)** Moon when Mercury and Venus are both between 90° to 270°.

**(b)** Moon when Mercury and Venus are both between 270° to 90°.

**(c)** Difference between Fig. 16.11a and Fig. 16.11b.

**Figure 16.11:** Stratospheric temperature distribution for Moon's phase when Earth is fixed at the location of the peak around 100° (see for example Fig. 16.7a) and at the same time Mercury and Venus propagate in a 180° window and its complimentary one, with bin = 15° for the whole period 01/01/1979 - 31/08/2018. The red dashed lines indicate the position of the peak at $\sim 70°$.

### 16.3.2 Fourier analysis

A Fourier analysis has been performed in the extended dataset of 01/01/1979 to 31/08/2018 for the 1 hPa to 3 hPa isobaric levels. From the derived periods it is worth noting that the 26$^{th}$ biggest peak seen in Fig. 16.12, which has an amplitude of about 16.3 dB and a frequency of $(27.31 \pm 0.02)$d, overlaps with Moon's sidereal month of 27.32 d, i.e. fixed to remote starts. At the same time the fact that the peak associated with Moon's synodic period, i.e. fixed to the Sun, around 29.52 d is two times smaller than the one around 27.31 d. This suggests an additional strong exo-solar impact on the Earth's stratosphere. Such an observation was also seen for the ionosphere in Sect. 15.3.2.

Other interesting peaks appearing to the periodogram that are close to synodic or orbital periods are the 8$^{th}$ biggest peak at 335.8 d $\pm$ 1.8 d with an amplitude of 52 dB which coincides within $\leq 1\sigma$ with the Venus - Mars synodic period of 334 d, and the 28$^{th}$ biggest peak with an amplitude of 15.8 dB and a frequency of 226.8 d $\pm$ 1.1 d which is very close to Venus revolution period of 224.7 d.

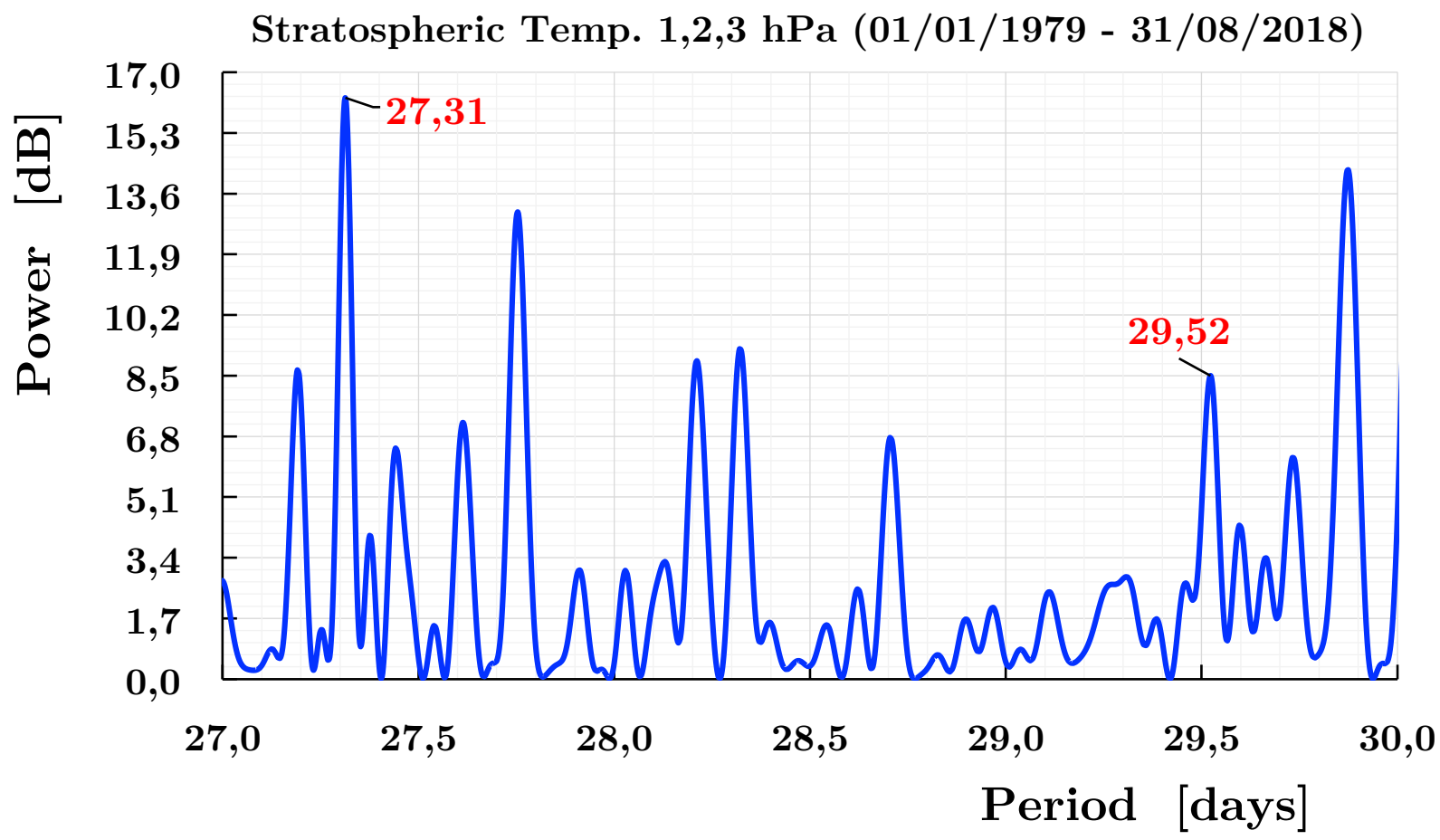

**Figure 16.12:** Fourier periodogram of the stratospheric temperature data from isobaric levels 1,2, 3 hPa for the period 01/01/1979 to 31/08/2018 zoomed around 28.5 d.

### 16.3.3 Energy deposition estimate

The atmosphere absorbs about 15 W/m$^2$ from the solar UV (∼200 nm to 300 nm), which in turn generates and interacts with the stratospheric ozone layer, and has a direct influence on the stratosphere [360, 361]. However, the impact of the solar cycle on the Earth's atmosphere is still a challenging question [340]. Having these in mind, it has been calculated that the temperature response and the UV flux modulation during the 11 y cycle is about 0.7 K to 1.1 K and only 0.2 W/m$^2$ respectively [341, 361].

In Fig. 16.13 the Earth's stratospheric temperature distribution has been created, but the data used for this distribution correspond to 12 y of solar maximum minus 9 y of solar minimum. The resulting maximum observed seasonal variation between solar maxima minus solar minima is about 13.4 K. Therefore, by scaling these values with the aforementioned ones about UV flux modulation, we get an estimated energy deposition of the exo-solar source on the order of 3 W/m$^2$. Even if this derived number is overestimated by a factor of 10, it still reflects a large unexpected impact from exo-solar origin.

The above estimation on the energy deposition could be due to some kind of invisible matter being gravitationally focused towards the Earth. This does not conflict the null results of underground experiments searching for DM. Some possible reasons could be (a) energy threshold effects and (b) the fact that the stratosphere is quasi not shielded to outer space in contrast with underground DM experiments, with its column density being $\rho_{\text{overhead}} \approx 1\,\text{g/cm}^2$.

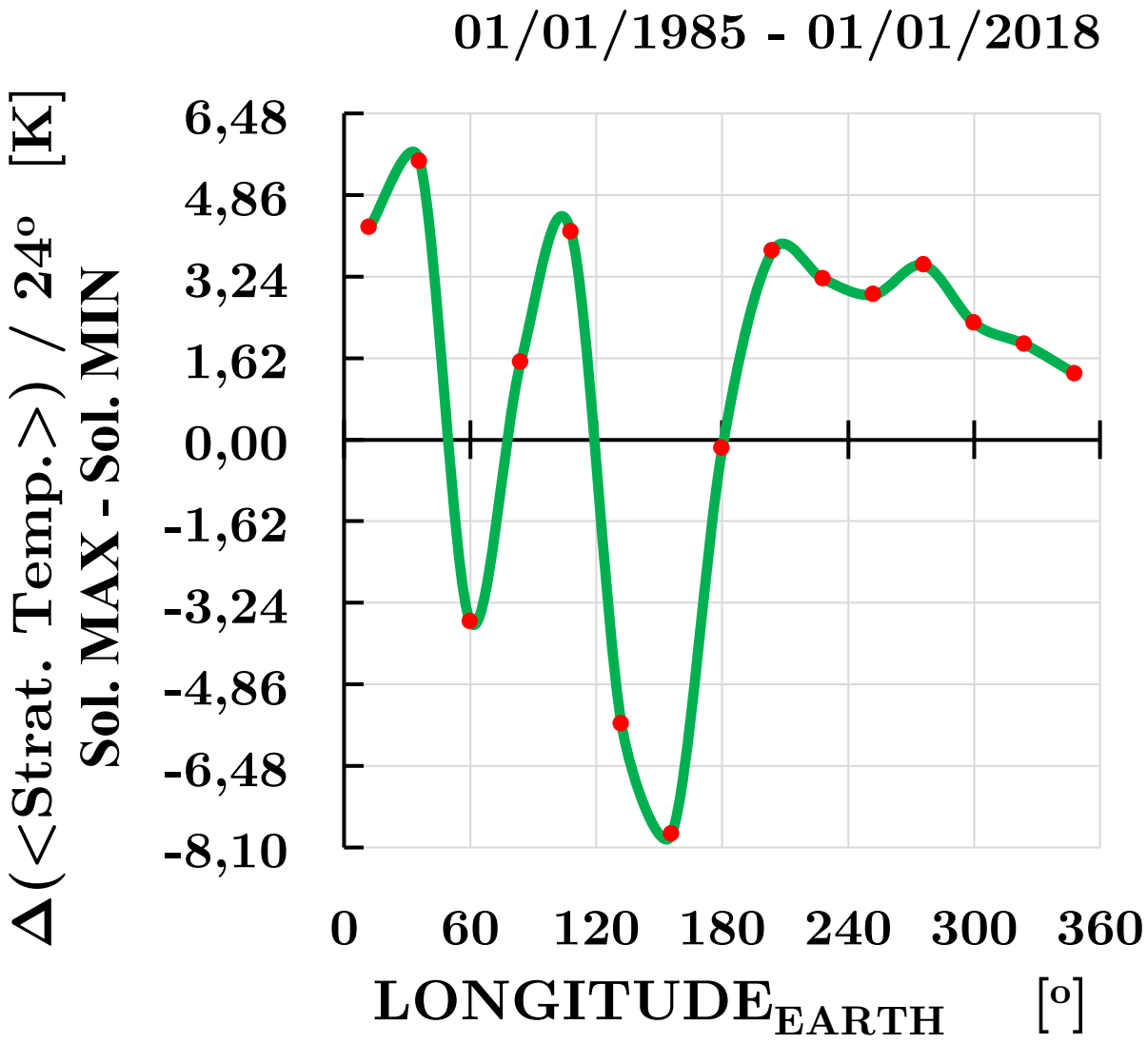

**Figure 16.13:** Seasonal mean upper stratospheric temperature difference between solar maximum and solar minimum. The time period used is 15 y for solar maximum (1989-1993, 1999-2003, 2011-2015) and 9 y for solar minimum (1985-1987, 1995-1997, 2007-2009).

## 16.4 Other considerations

### 16.4.1 Atmospheric temperatures at lower altitudes

An important comparison of the acquired results should be made with the lower atmospheric layers. For example, in Fig. 16.14 the distribution of Earth when Mercury and Venus are both between 90° to 270° seen in Fig. 16.10a, is plotted for four different altitudes. A clear observations is that the lower the altitude the more reduced the effect on the peak around 100° appears to be, which is consistent with the downwards moving invisible streaming matter scenario. The same applies to the rest of the distributions that have been presented above.

Finally, it is worth stressing, that the distribution in Fig. 16.14b represents the main ozone layer in the lower stratosphere at an altitude of ≈16 km to 31 km which is strongly affected by solar UV. However, the striking difference with Fig. 16.14a implies that the peaking distribution of the upper stratosphere at ≈ 38.5 km to 47.5 km is marginally or even not at all affected by the solar UV.

### 16.4.2 Other atmospheric phenomena

Earth's atmosphere is known to be a complex non-linear dynamic system where several phenomena take place at different spatiotemporal scales. As an example, the same energy deposition in the visible, UV or X-rays has different impacts on the atmosphere due to threshold effects. The same concept applies also to the unknown exo-solar source in the estimation of the energy deposition.

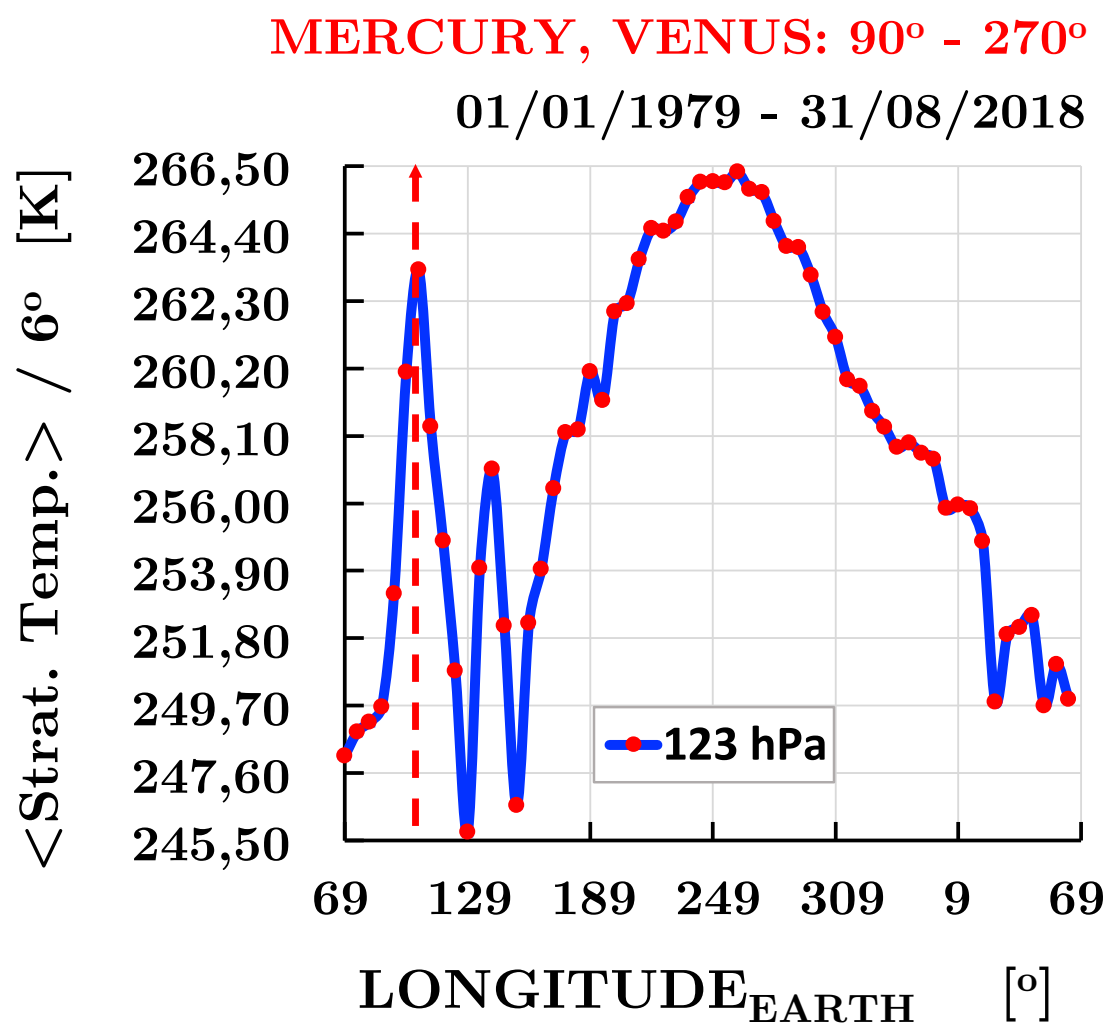

**(a)** Average between 1, 2 and 3 hPa.

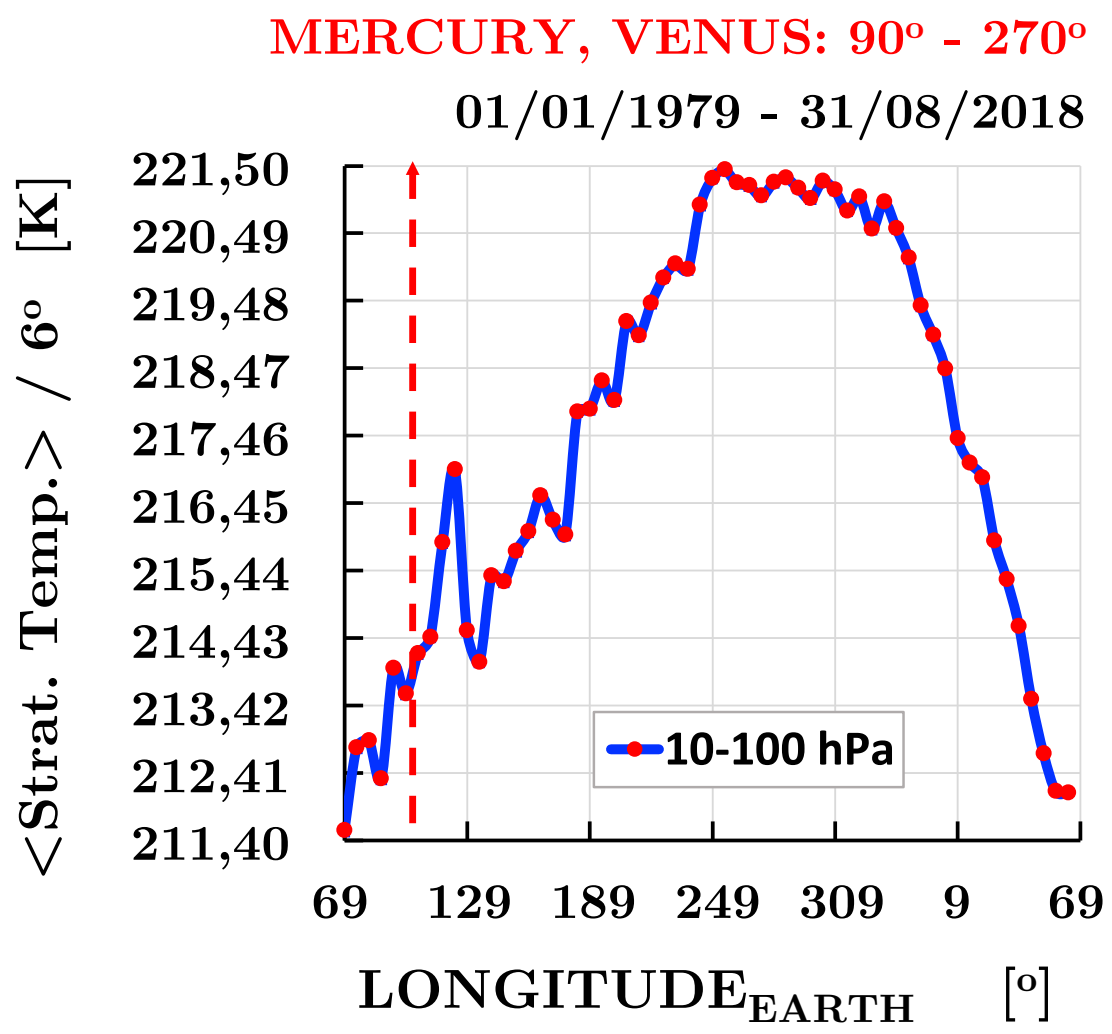

**(b)** Average between 10, 20, 30, 50, 70 and 100 hPa.

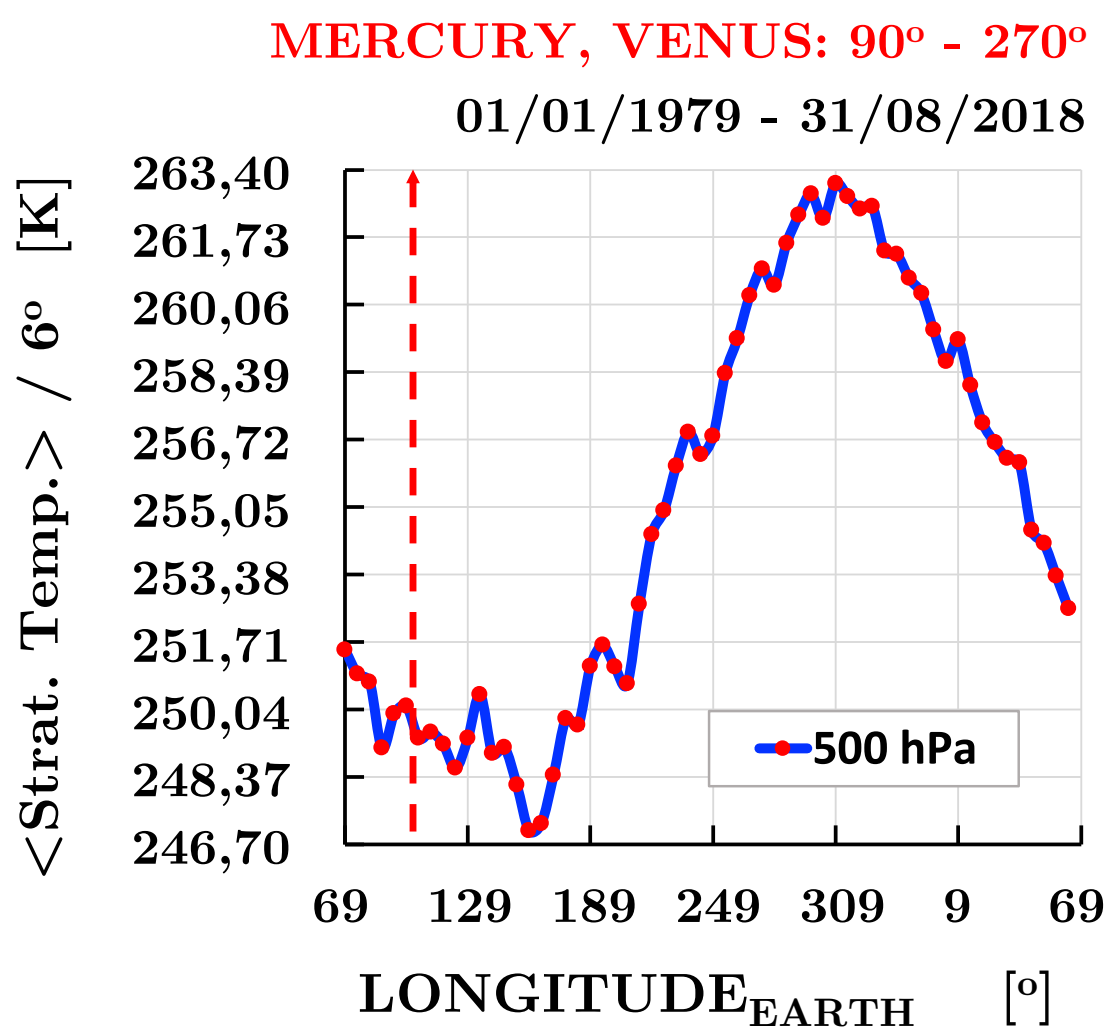

**(c)** 500 hPa.

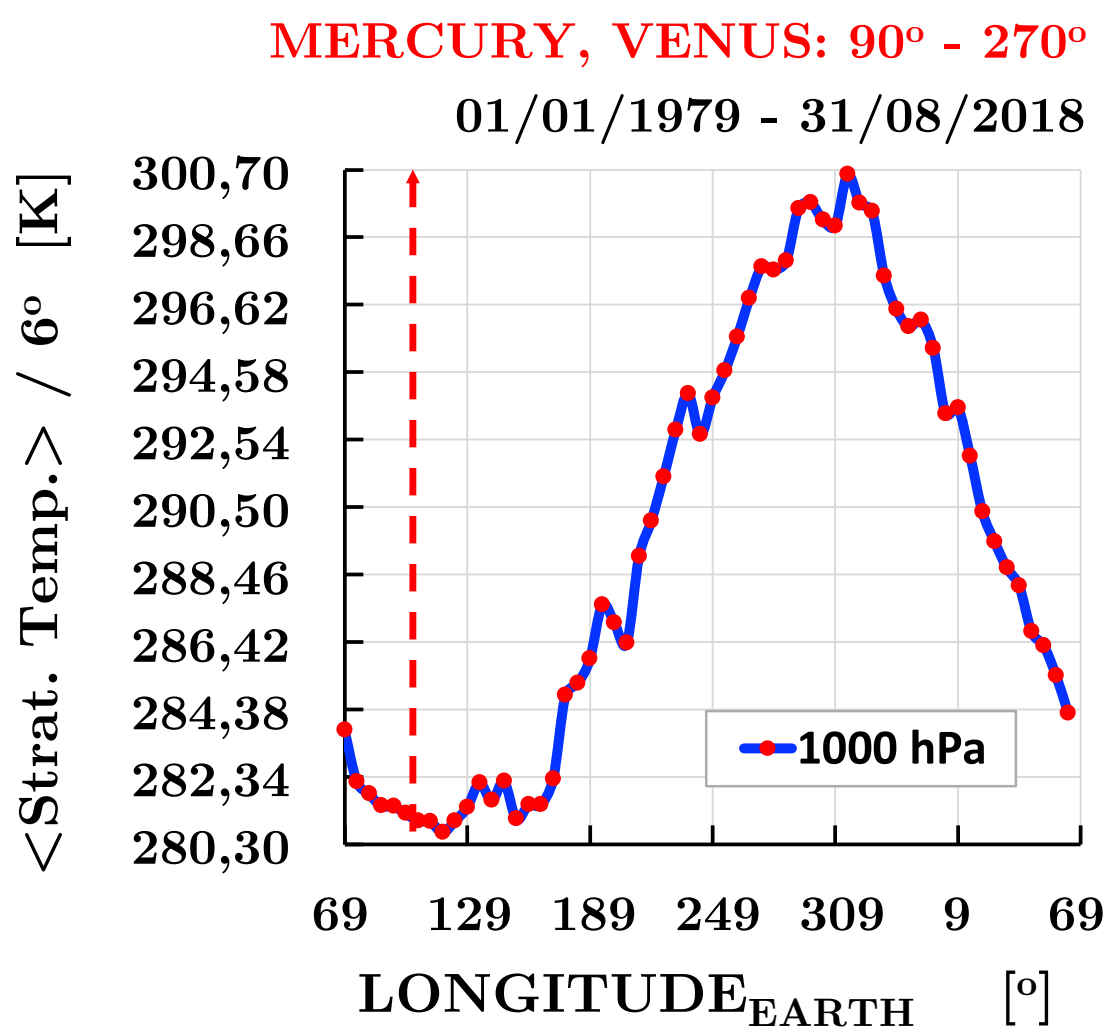

**(d)** 1000 hPa

**Figure 16.14:** Earth's longitudinal distribution for the mean stratospheric temperature when Mercury and Venus propagate between 90° to 270° for different isobaric levels for the whole period 01/01/1979 - 31/08/2018. The red dashed lines indicate the position of the peak at $\sim$ 105 deg.

An interesting atmospheric phenomenon that could have a contributing role to the heating of the upper atmosphere is that of *Sudden Stratospheric Warmings (SSWs)*. However, these events are rare and occur rather randomly, and therefore do not interfere with the annually occurring STAs. In the literature, we find also major SSWs provoking large temperature increases of the order of 50 K within a few days. These events, which require further investigation, occur mainly over the polar zone of the Northern Hemisphere and have a frequency of about 6 events per decade [362].

An additional STA is the so-called *Quasi Biennial Oscillation (QBO)* which is manifested as downward propagating easterly and westerly wind regimes and has a variable period with an average of about 28 months. The induced temperature variance is about 0.3 K to 1 K [363]. In the lower stratosphere the equatorial temperature modification due to QBO is on the order of ±4 K with a maximum of ±7 K near the isobaric levels 30 hPa to 50 hPa [364]. Once again, QBO cannot explain the anomalies seen here since its occurrence period is much larger than the yearly effect seen here.

Moreover, the general circulation of the atmosphere can be affected by major surface pressure differences like the *North Atlantic Oscillation (NAO)* or the *El Nino/Souhern Oscillation (ENSO)* variations in the Pacific. These phenomena which vary inter-annually can generate sea-surface temperature anomalies which can trigger stationary Rossby waves in the Extratropics [365]. NAO has no specific periodicity whereas ENSO has an irregular periodicity between 2 y to 7 y. Therefore, since these phenomena have different periodicities, and occur outside the geographic area selected here, they do not interfere with the presented results.

Finally, it is worth mentioning that in general the temperature of the stratosphere varies at different spatiotemporal scales due to the non-homogeneous distribution of the continents and the oceans around the globe, but also due to the variation of the impinging solar radiation.

## 16.5 Summary

A statistically significant planetary relationship of the mean upper stratospheric temperature is found. This behaviour is conventionally unexpected since any remote planetary interaction is smooth over a planetary orbit, but also no known phenomenon intrinsic to the atmosphere exists that can produce such an effect. At the same time, based on the dissimilarity with the solar activity (see Appendix Sect. B.10) an exo-solar impact on the Earth's stratosphere is suggested. This is strongly supported also by the Fourier analysis which gives a significant peak around $(27.31 \pm 0.02)$d associated with Moon's sidereal month (fixed to remote stars) and a weaker one at the Moon's synodic period around 29.52 d (fixed to the Sun).

Therefore, this can be the as yet overlooked signature of invisible "strongly" interacting massive streams which are gravitationally focused towards the Earth by the Sun and the planets including the self-focusing effect of the Earth itself. The calculated order of magnitude of the energy deposition of the exo-solar source in the atmosphere is quite large, on the order of $3\,\mathrm{W/m^2}$ (see Publications E.3 and E.18). This could eventually help explore new sources for clean energy (see Publication E.23). At the same time, a downwards propagating signature is consistent with the observed reduction of the overall effect when moving towards lower altitudes (Fig. 16.14).

The obtained results are also in agreement with the ones from Chap. 15, where the Earth's ionosphere (above $\sim 100$ km) shows an anomalously high degree of ionisation around December.

Both results coincide within a few days with the annual alignment between Earth, (Moon), Sun and GC at 86.5°. However, the daily ionospheric TEC data refer to the entire ionosphere and therefore, a direct comparison is not possible. Once spatially accurate ionospheric data become available, a comparison for a possible correlation between the stratosphere and the ionosphere above is expected to yield interesting results.

Future analyses can extend these first results to the whole Earth's atmosphere including all altitude levels (both lower and higher). Additionally, smaller cadences than 1 d and different times than 18:00 UTC, as well as wider time periods could be scrutinised in future analyses, such as specific periods corresponding only to solar maxima and solar minima. This way, the nature of the assumed invisible matter and its interaction with the Earth's atmosphere as well as the velocity distribution of the invisible matter could be probed.

# 17

# EARTHQUAKES

## 17.1 Introduction

Chap. 15 and 16 have provided evidence that unexpected planetary relationships exist for various phenomena occurring in the Earth's atmosphere. The most feasible scenario seems to be that of the planetary gravitational focusing and self-focusing of streaming low-speed invisible matter whereas long-range planetary forces such as tidal forces are actually excluded to be behind such observations as they are far too feeble [229]. More specifically, there has been an extensive search for the possibility of tidal forces inducing seismicity. However, the results have been mixed and no direct evidence with statistical significance has been found [366–368].

Interestingly, EQs and especially the biggest ones have already shown a non-random distribution [369] including a correlation with the solar activity [370] which we have already seen that it exhibits a significant planetary relationship. In addition, several EQ precursor anomalies have been found within the atmosphere, and more specifically within the ionosphere and even in the magnetosphere. This initiated a multidisciplinary approach aiming to investigate pre-EQ processes to understand the underlying physical mechanisms as well as for better forecasting [371]. In the ionosphere there have been observed strong TEC anomalies taking place a few hours / days before the occurrence of large EQs (Fig. 17.1) [372–376]. These excursions in the global electron density of the ionosphere before large EQs support a downwards propagating motion and are consistent with the hypotheses of this work. It is noted that the TEC of the ionosphere has already shown significant planetary relationships. More EQ precursors have also been found on disturbances on the Earth's magnetic field [377], as well as on bursts of charged particles from the Van Allen belts which seem to precede especially big EQs by 4 h to 5 h [378–380]. Recently, a correlation of the global seismic activity was also made with the variation of cosmic rays exhibiting a time lag of approximately two weeks [381].

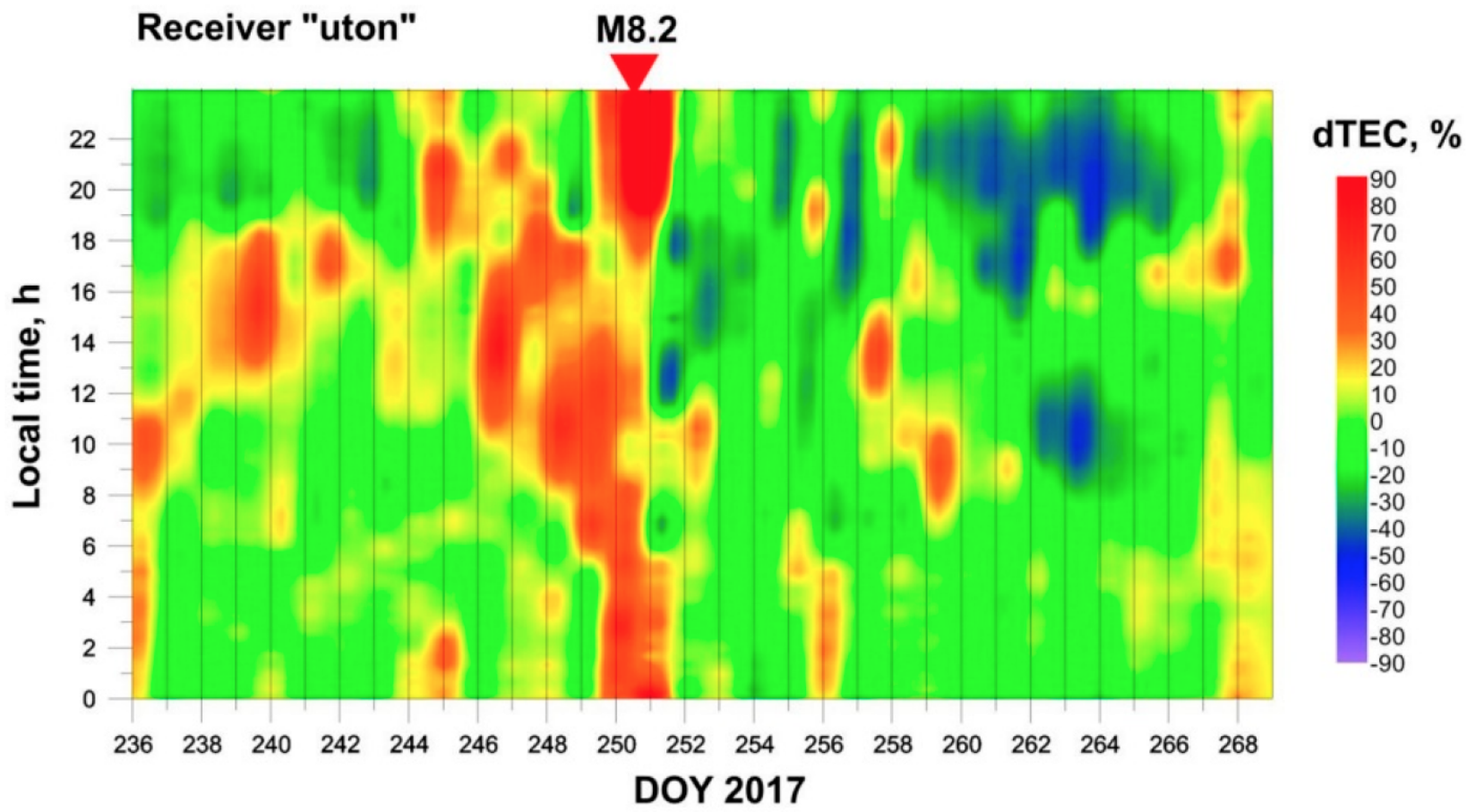

**Figure 17.1:** An increase in the TEC prior to a $M = 8.2$ EQ in Mexico. In x-axis the days of year 2017 starting from 24/08 (236) to 25/09 (268) are shown (40).

Therefore, it makes sense to extend the investigation for a planetary relationship to the occurrence of the violent EQs since gravitational focusing of highly interacting streams or clusters of invisible massive matter by the Sun and/or the planets including the Moon towards the Earth could provide a triggering mechanism. As a result, a first investigation can take place in the most extreme EQs all around the globe, which can provide the required statistics since they are monitored uninterruptedly and in great detail since decades.

## 17.2 Global earthquake data

### 17.2.1 Data origin

The list of EQs used in this work has been acquired from the U.S. Geological Survey (USGS) Earthquake Hazards Program, which is part of the National Earthquake Hazards Reduction Program (NEHRP) led by the National Institute of Standards and Technology (NIST). More specifically, the data have been downloaded from the publicly available Advanced National Seismic System (ANSS) Comprehensive Earthquake Catalog provided by the National Earthquake Information Center (NEIC) and contain data from all around the world from 28/12/1566 - today [382].

For a first analysis, the selected 15 y period that was selected was 01/01/2001 - 31/12/2015,

with EQs having a magnitude ($M$) above 5.2. This dataset contains 16396 EQs in a span of 5478 d with an average of $\sim 3$ EQs per day. The magnitude ranges from $M = 5.2$ to $M = 9.0$ with an average of $M \sim 5.6$.

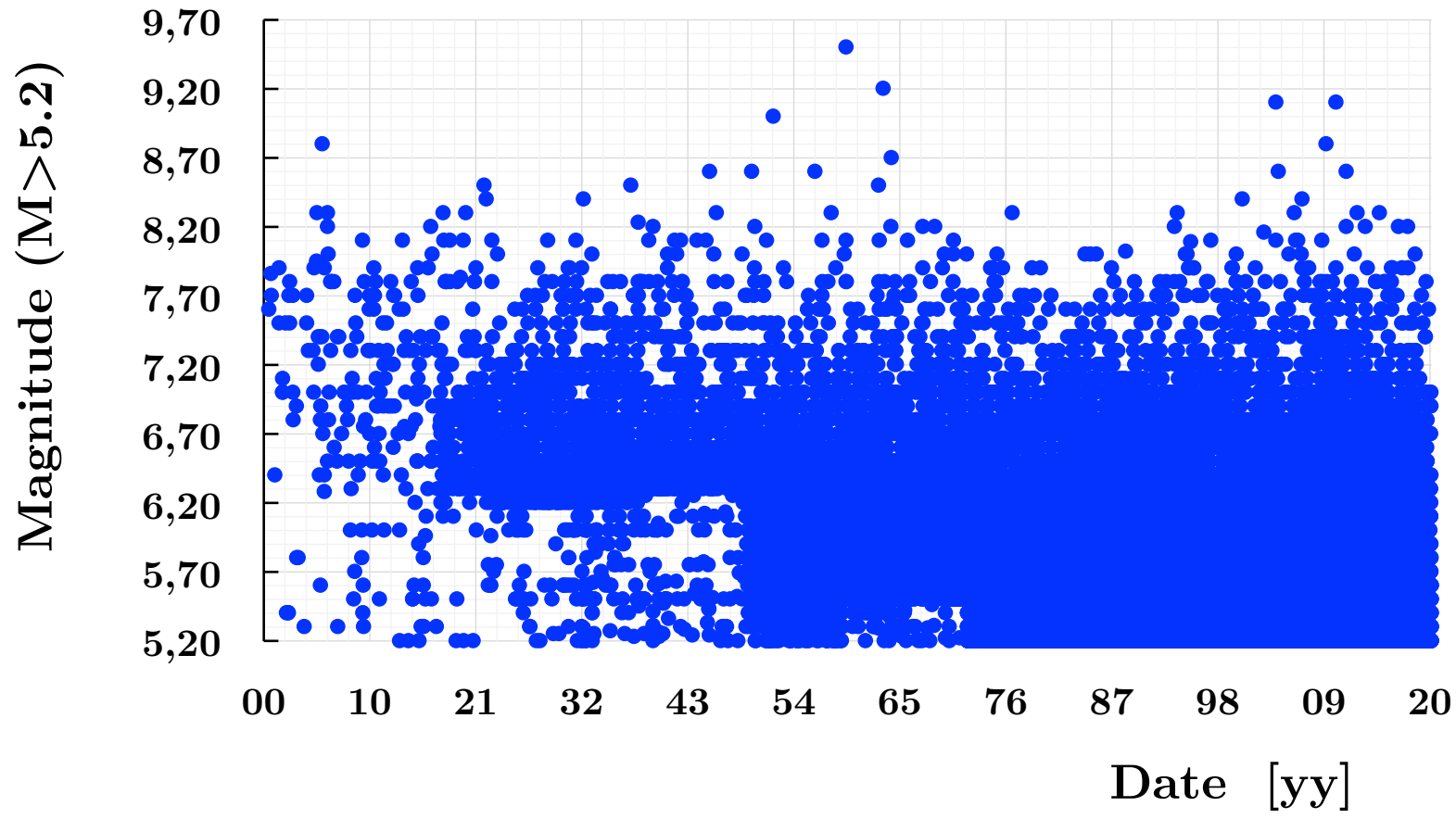

**Figure 17.2:** The raw global EQ data with $M > 5.2$ for the period 01/01/1900 - 31/01/2021.

However, a more extended period ranging from 01/01/1900 to 31/01/2021 has also been acquired containing in total 57033 EQs with $M > 5.2$. The corresponding number of days fulfilling this constraint is 21853 with the average number of EQs per day, including only the dates with these EQs, is $\sim 2.6$ with a maximum of 199 EQs. The magnitude of these EQs shown in Fig. 17.2 ranges from $M = 5.2$ to $M = 9.5$ with an average of $M \sim 5.6$. The various distribution statistics for the magnitude of these EQs are shown via various histograms in Fig. 17.3.

### 17.2.2 Data treatment

The downloaded raw data contain a variety of information such as:

- Timestamp when the event occurred (in UTC).
- Longitude ($-180.0 - 180.0$ deg).
- Latitude ($-90.0 - 90.0$ deg).
- Depth of the event ($0 - 1000$ km).
- Uncertainty of reported depth of the event ($0 - 100$ km).
- Magnitude of the event ($-1 - 10$). The magnitude is a logarithmic measure.
- Uncertainty of reported magnitude of the event.
- Total number of seismic stations used to determine the EQ location.
- The largest azimuthal gap between azimuthally adjacent stations (in deg).
- Uncertainty of reported location of the event (in km).
- Horizontal distance from the epicentre to the nearest station ($0.4 - 7.1$ deg where 1 deg corresponds to $\sim 111.2$ km).
- Root Mean Square (RMS) travel time residual using all weights (in s).

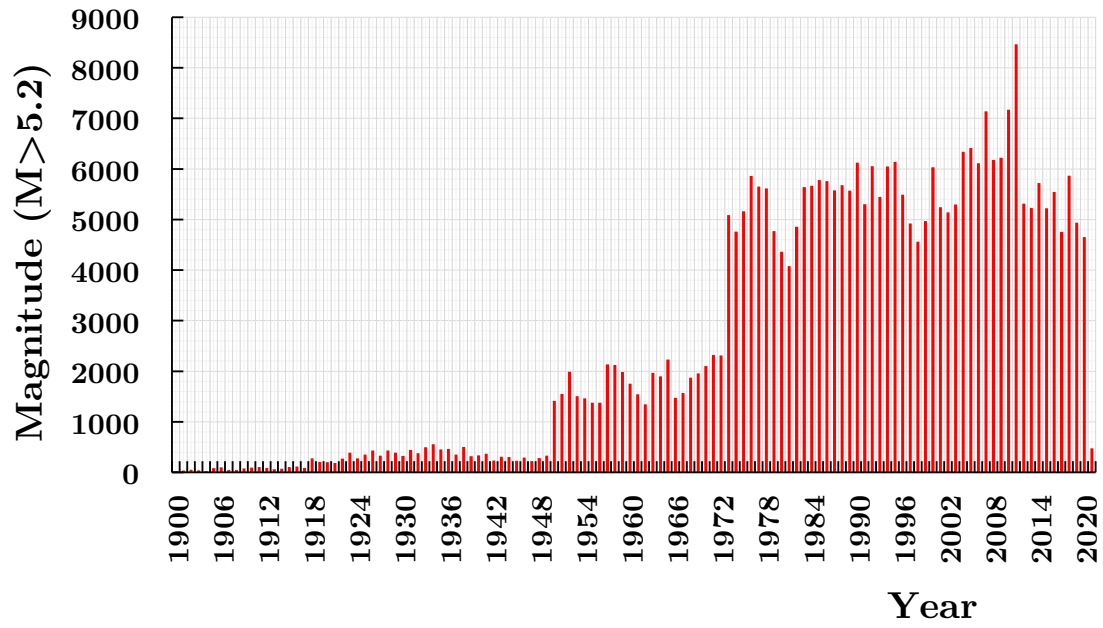

**(a)** EQs magnitude per year.

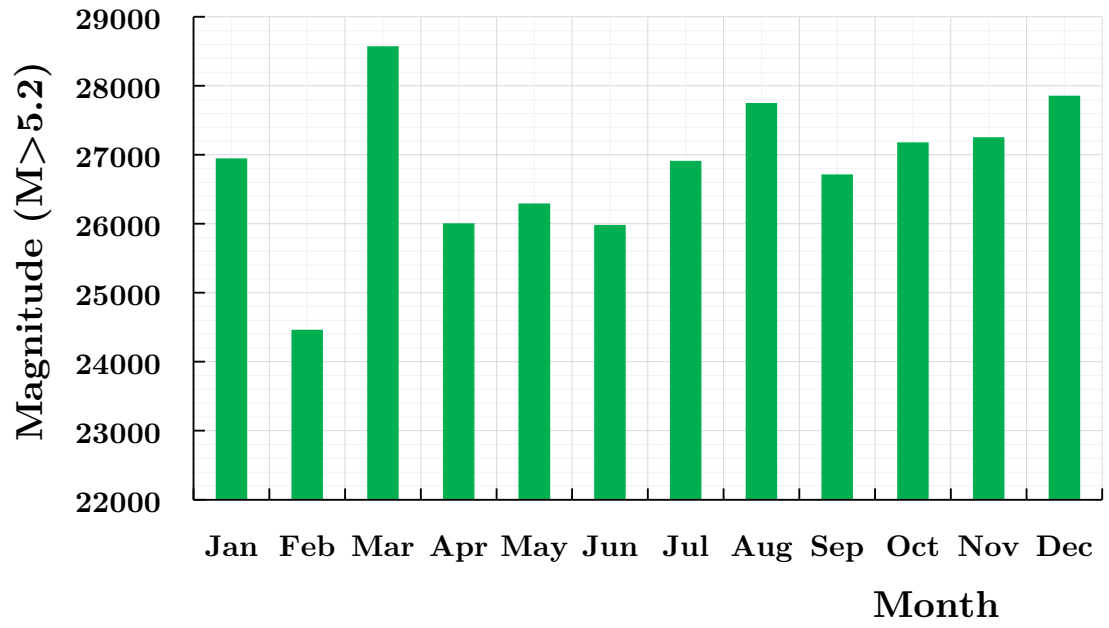

**(b)** EQs magnitude per month.

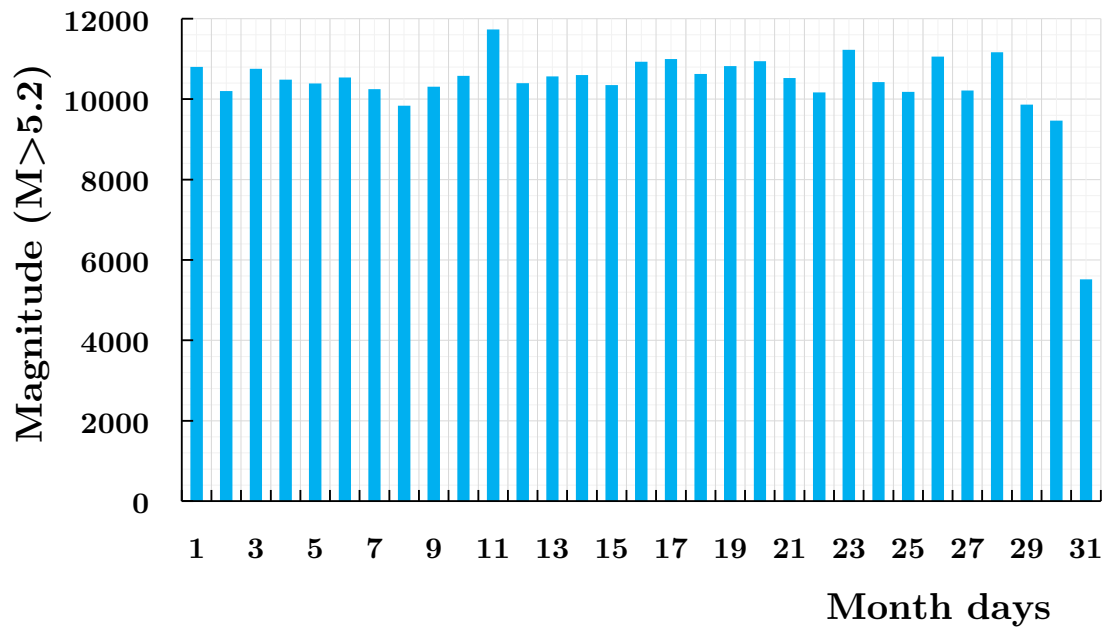

**(c)** EQs magnitude per day of month.

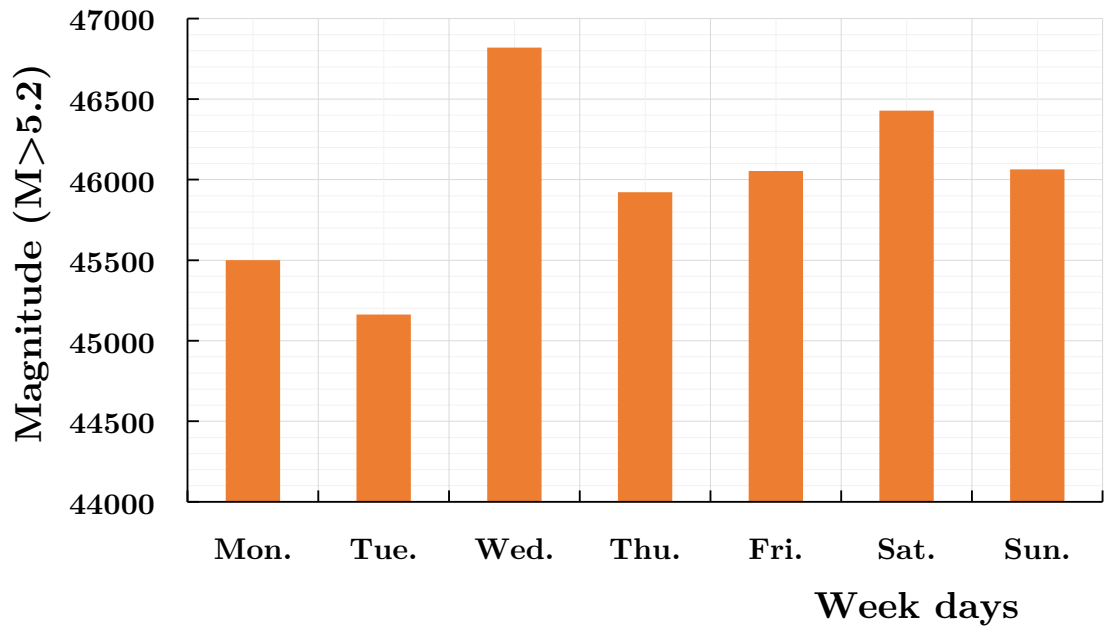

**(d)** EQs magnitude per day of week.

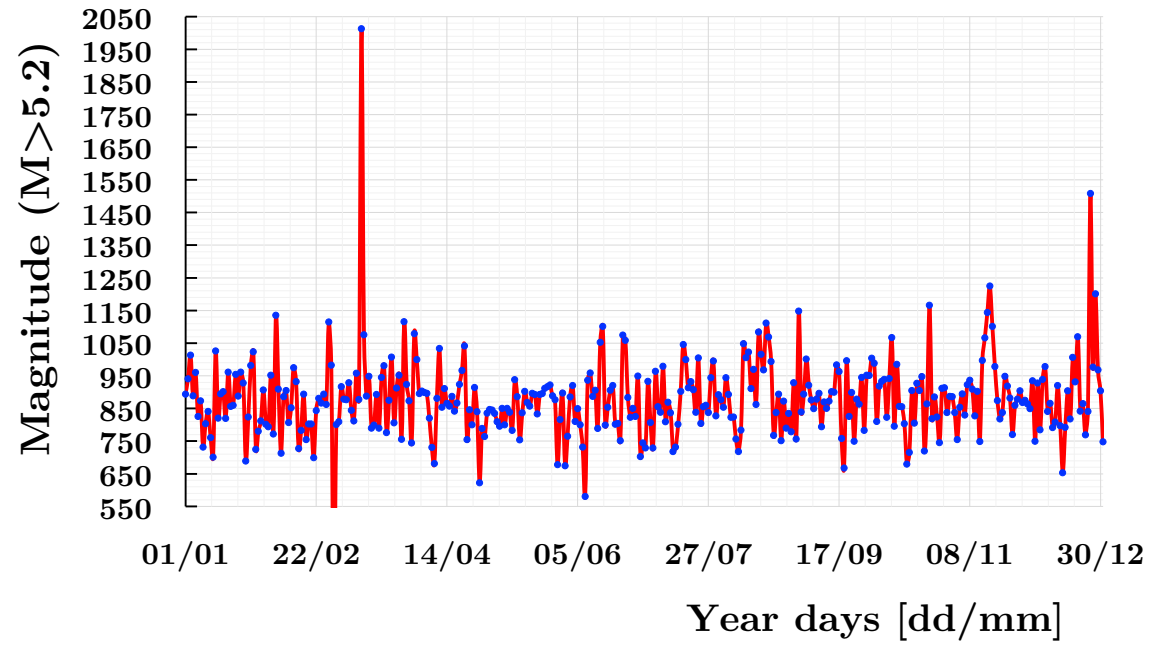

**(e)** EQs magnitude per day of year. The observed point with the reduced amplitude corresponds to 29/02 which is a leap day added only to leap years and thus appears less frequently in the total period. The point with the increased amplitude as we will see corresponds to 11/03/2011 where a large number of EQs were recorded.

**Figure 17.3:** Histograms for the magnitude of EQs with $M > 5.2$ for the period 01/01/1900 - 31/01/2021.

From these data, according to the timestamp of each even, a daily list has been created with the sum of the number of global EQs with $M > 5.2$ per day which are plotted in Fig. 17.4a. In the dates without any recorded EQ the number 0 was assigned. Furthermore, in each date, at 00:00 UTC, the corresponding planetary heliocentric longitude has been assigned.

As seen in Fig. 17.4a there is a number of dates with with an increased number of EQs per day comparing to the mean value which are attributed mainly to aftershocks. Such effects in single-days will result in high-amplitude single-bin excursions in the distribution spectra. Therefore, these dates have been removed from the analysis. More specifically, for the main analysis and the creation of the planetary longitude distribution spectra, the daily number of earthquakes has been constrained to be less than 25 (see Fig. 17.4b). This results to the exclusion of 10 d out of 5478 d ($\sim 0.2\%$) and leaves a total of 15703 EQs, from the total of 16396 EQs ($\sim 4.2\%$) of magnitude $M > 5.2$ for the 15 y-time interval. Such a small reduction in the total amount of selected days does not make any observable difference for the calculation of the number of stay days per bin of heliocentric longitude for each planet, and therefore no additional calculation or correction has to be made.

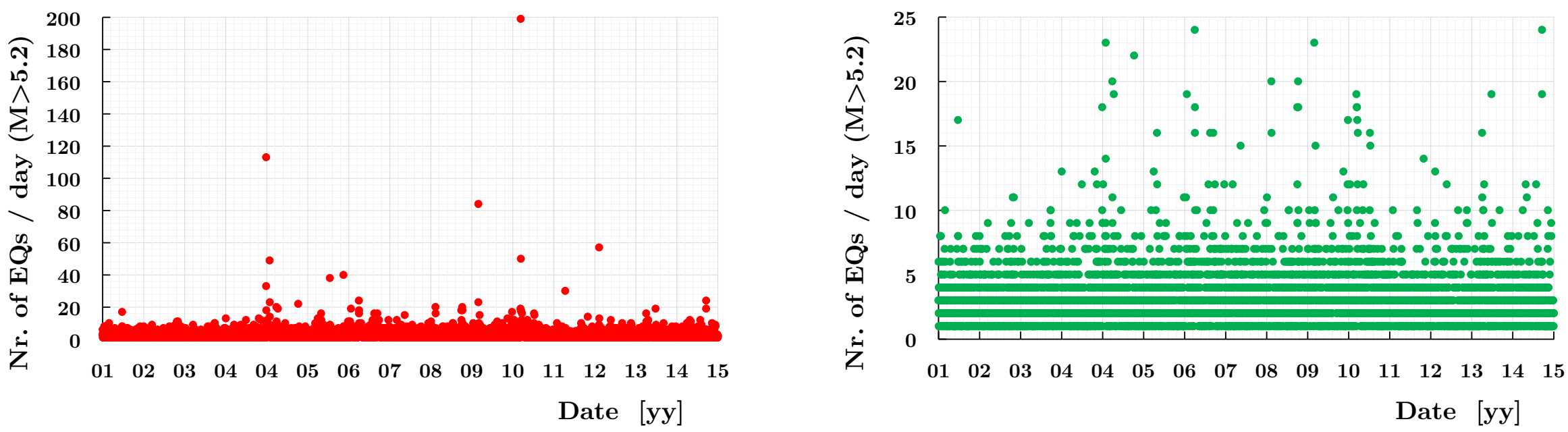

**(a)** All EQs. **(b)** $EQs < 25$/d.

**Figure 17.4:** Summed Nr. of EQs with $M > 5.2$ per day from 01/01/2001 - 31/12/2015.

An additional analysis has also been made for the period 01/01/2001 - 31/12/2015 considering the biggest EQs with a magnitude $M > 5.9$. In that case the constraint in the number of $EQs$/d is $0 - 10$ which in total excludes only 3 d:

- 26/12/2004 with 12 EQs.
- 27/2/2010 with 13 EQs.
- 11/3/2011 with 42 EQs.

This results to a total of 2320 EQs with $M > 5.9$. In these plots even though the observed effect is bigger, the overall significance is smaller due to the reduced number of total EQs. Therefore, these plots due to space limitations are omitted.

## 17.3 Data analysis and results

### 17.3.1 Planetary longitudinal distributions

#### 17.3.1.1 Single planets

After assigning the daily values of the heliocentric longitude of each planet to the daily number of EQs, the various planetary distributions are produced. The first part involves the selection of single planets without any longitudinal constraints on the positions of the rest of the planets. For the calculation of the relative standard deviation in each plot we keep in mind that for the period of 01/01/2001 - 31/12/2015 (= 5468 d) and in all the plots without any planetary constraints, we have in total 15703 EQs with $M > 5.2$, since we have applied the constrain of $0 - 24$ EQs per day. In Fig. 17.5 the various spectra with the distribution of EQs as a function of the planetary positions of the inner planets are presented. More specifically, in the spectrum of Mercury in Fig. 17.5a, we have a total maximum - minimum difference of 24.9% with the mean relative statistical error per point, due to the 18° bin, being $\sigma \sim 3.6\%$. This means that for the peak around 100° compared with the minimum point on the left around 10° we have a statistical significance of $\frac{3.4-2.4}{\sqrt{(0.04 \cdot 3.2)^2+(0.04 \cdot 2.4)^2}} \sim 5.5\sigma$. Similarly, for Venus in Fig. 17.5b we have the same mean error and the total observed difference between the minimum and maximum point is 23.6%. As an example the single point around 9° compared to a single minimum point on the base of the peak around 63° gives $4.3\sigma$. Finally, for Earth in Fig. 17.5c there is an observed amplitude of 20.5%. In this case, the calculated mean statistical error per point is $\sim 3.1\%$ giving a $5.1\sigma$ difference for the single point around 180° compared to the point around 252°.

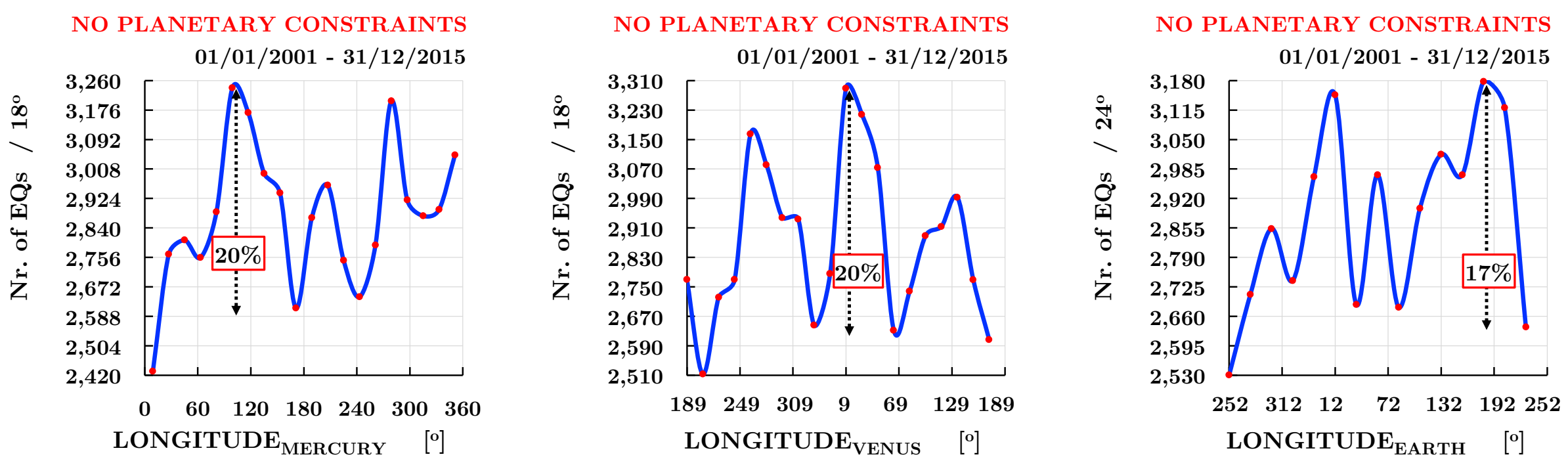

**(a)** Mercury for 18° bin. $\sigma \sim 3.6\%$.

**(b)** Venus for 18° bin. $\sigma \sim 3.6\%$.

**(c)** Earth for 24° bin. $\sigma \sim 3.1\%$.

**Figure 17.5:** Planetary heliocentric longitude distributions of the Nr. of EQs for the period 01/01/2001 - 31/12/2015. The mean relative standard deviation per bin for each case is also given.

As next, we can constrain the selected period used in Fig. 17.5 even further, and select,

for example, only the period of the deep solar minimum of 01/01/2008-31/12/2009 (= 731 d) where we end up with 2231 EQs. This period corresponds to perhaps the deeper solar minimum in the last century. In this case, in all plots in Fig. 17.6 the observed significance is improved. For example, in Fig. 17.6a the total observed amplitude is 62.1% with the statistical error per point being 9.5%. Similarly, in Fig. 17.6b Venus the maximum - minimum difference is 59.3% and the calculated Poisson standard deviation per bin is 11.6%. Finally, in the spectrum with Moon's phase in Fig. 17.6c we have a 41.4% amplitude. The calculated statistical accuracy per point for this case is $\sigma \sim 7.4\%$ (see also Publication E.19).

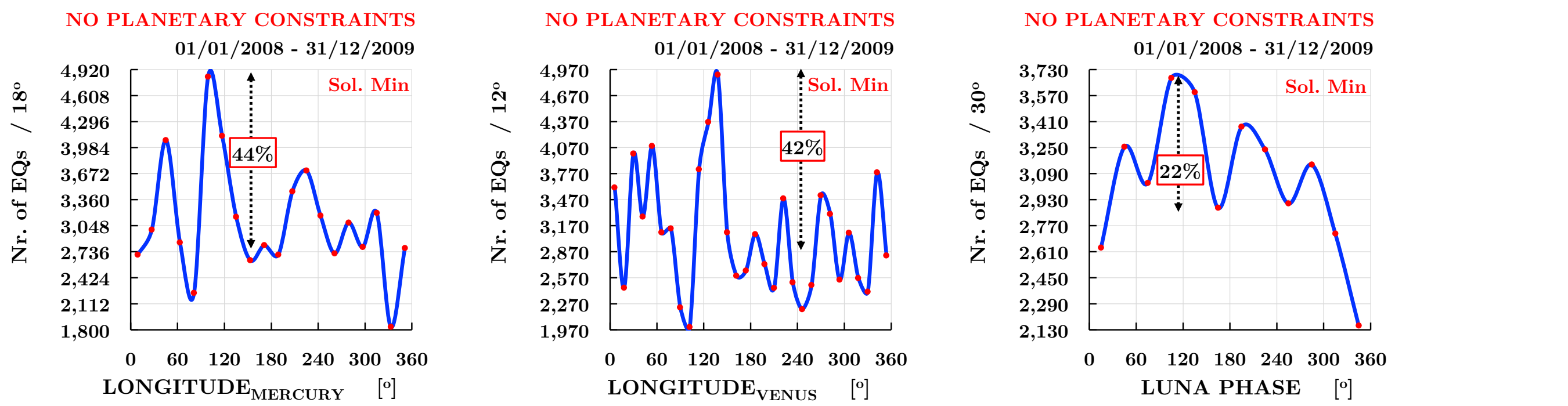

**(a)** Mercury for 18° bin. $\sigma \sim 9.5\%$.

**(b)** Venus for 12° bin. $\sigma \sim 11.6\%$.

**(c)** Moon's phase for 30° bin. $\sigma \sim 7.4\%$.

**Figure 17.6:** Planetary heliocentric longitude distributions of the Nr. of EQs for the solar minimum of 2008 – 2009. The mean relative standard deviation per bin for each case is also given.

#### 17.3.1.2 Combining planets

The next part of the analysis involves the usage of constraints in the heliocentric longitudinal position of other planets. This way the distribution of the number of EQs is plotted for a double planetary configuration which can provide hints on the direction(s) of the possible stream(s) and their path towards the Earth. In Fig. 17.7, Mercury's longitudinal position has been used as a reference frame while other planets were constrained on various heliocentric locations. For example the overall amplitude is maximised when Jupiter is constrained to be between 310° and 50° for a bin of 12° as seen in Fig. 17.7a. In that case the difference between the minimum and maximum point is 58.3%, whereas the number of EQs corresponding to these 1103 d, which fulfil the above conditions, is 3584. The standard error per bin is calculated at about 9.2%. This gives for the peak around 114° a statistical significance of $\sim 5.5\sigma$ compared to the minimum point around 186°. Similarly, in Fig. 17.7b we get a 67.1% amplitude when we select a bin of 9° and Moon's phase is between 80° and 180°. The number of EQs for this case is 4373 while the number of days is 1513 giving a statistical accuracy per point of 9.6%. Therefore, the big peak observed here around 113° has a high statistical significance of $\sim 6.5\sigma$.

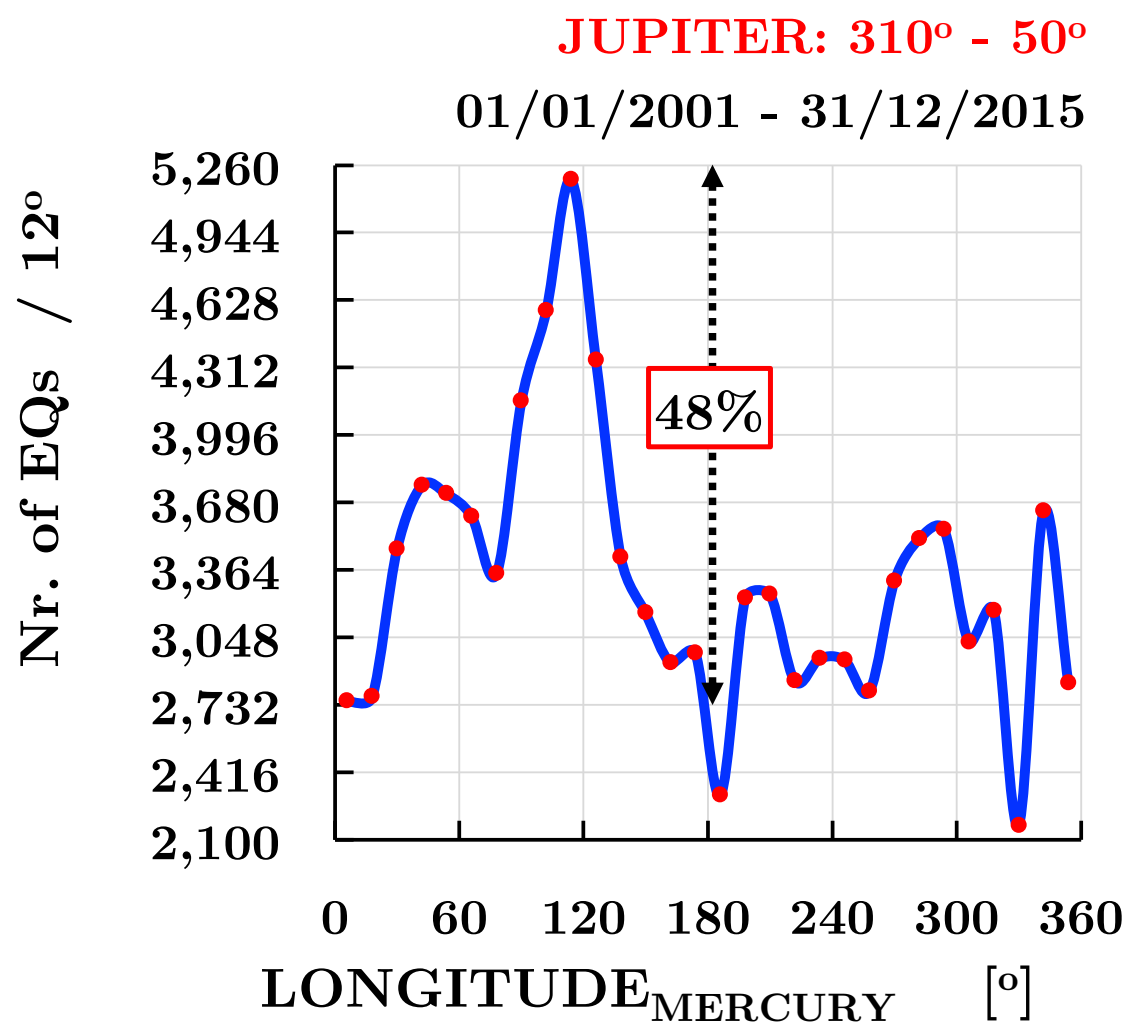

(a) Jupiter's heliocentric longitude allowed to be between 310° to 50°, for bin = 12°. $\sigma \sim$ 9.2%.

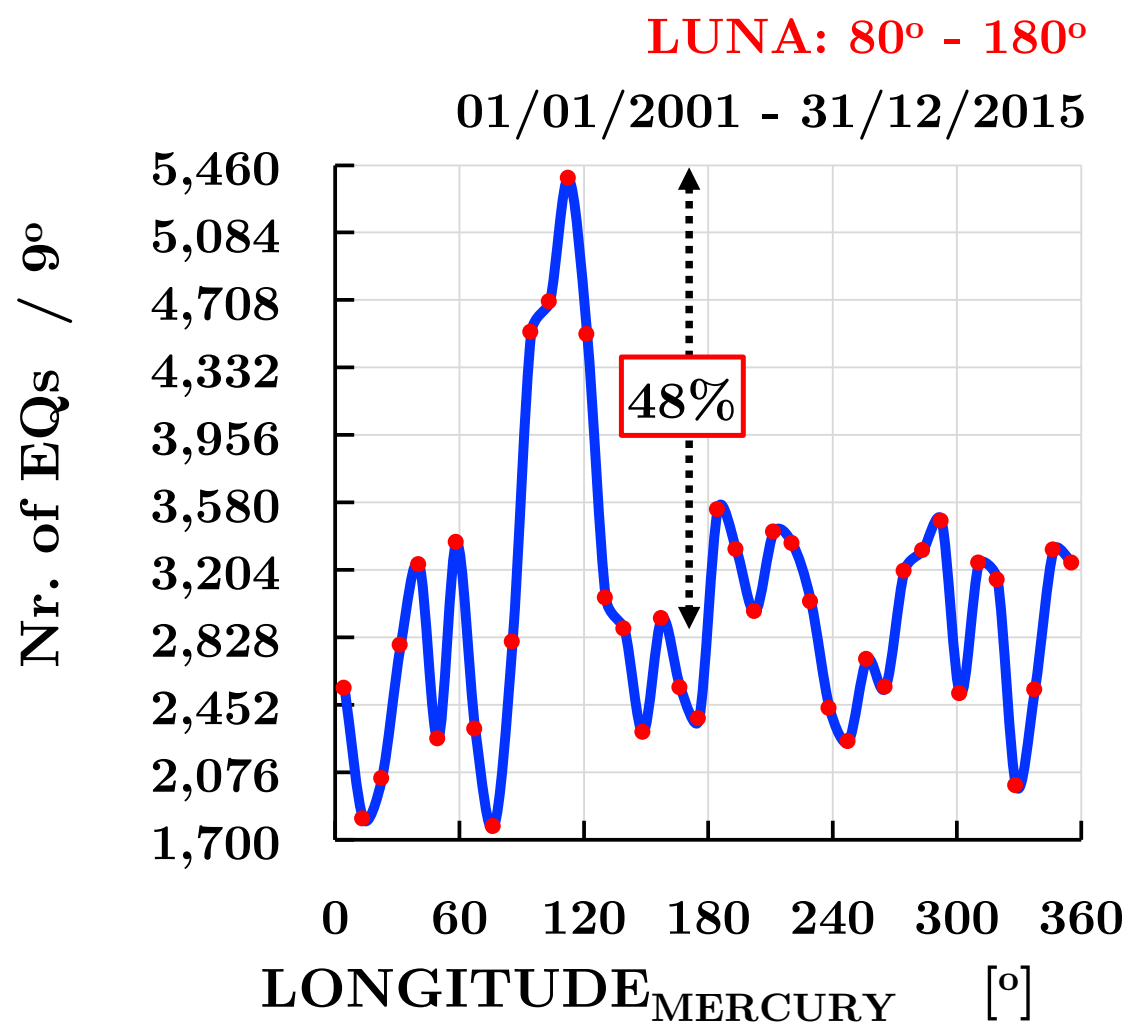

(b) Moon's phase allowed to be between 80° to 180°, for bin = 9°. $\sigma \sim$ 9.6%.

**Figure 17.7:** Nr. of EQs as a function of Mercury longitude while other planets are constrained, for the period 01/01/2001 - 31/12/2015. The mean relative standard deviation per bin for each case is also given.

In Fig. 17.8, the reference frame is this time set on Venus, while Mercury and Earth are constrained on various heliocentric locations to derive the biggest amplitude. In the case of Mercury in Fig. 17.8a being vetoed between 50° to 150° for a bin of 9°, we get a 64.6% amplitude with the total days being 1044 and the number of EQs for these days being 3084. Thus, the relative error per bin point is 11.4% which for the peak around 140° and the point on the base around 167° gives a maximum to minimum difference of $\sim 5.2\sigma$. Then, in Fig. 17.8b when Earth is on the window between 100° to 200°, we get 56.7% amplitude for the 8° bin. The number of EQs is 4472 for 1488 d. The statistical calculated error per point becomes 10%. In this case between the point at the peak around 260° and the point at 324° we have a statistical significance of $\sim 5.1\sigma$.

As next, in Fig. 17.9 the distribution of the number of EQs is observed as a function of Earth while Mercury, Venus and Jupiter are independently constrained to be in a 100° orbital arc. More specifically, in Fig. 17.9a Mercury's orbit is constrained to 50° to 150°. In this case, for 6° bin, the modulation observed is 75.9%. The corresponding total number of EQs is 3084 while the total days fulfilling the aforementioned parameters are 1044. Therefore, the Poisson relative error per point based on the sample size is 14% which for the peak around 93° compared to the minimum point around 75° gives us a difference of $\sim 5.3\sigma$. In Fig. 17.9b Venus' heliocentric longitude is between 220° to 320° range resulting to a 45.1% observed amplitude on Earth's position. For this case the number of days is 1506 while the number of EQs is 4434, giving a standard error per point of 5.8%. The wide peak in this case has a significance of $\sim 6.2\sigma$. In addition, when Jupiter in Fig. 17.9c is located between 300° to

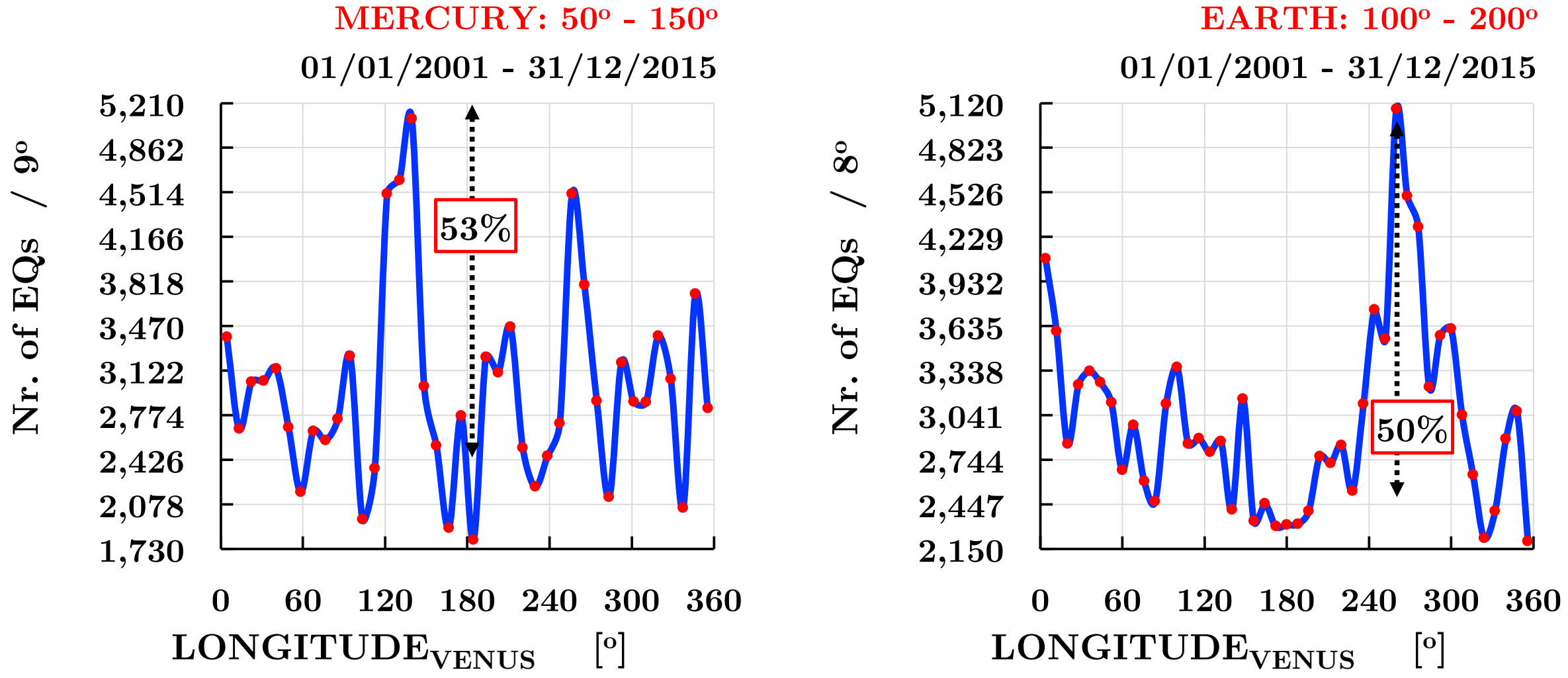

**(a)** Venus when Mercury's longitude is allowed to be between 50° to 150°, for bin = 9°. $\sigma \sim 11.4\%$.

**(b)** Venus when Earth's longitude is allowed to be between 100° to 200°, for bin = 8°. $\sigma \sim 10\%$.

**Figure 17.8:** Nr. of EQs as a function of Venus longitude while other planets are constrained, for the period 01/01/2001 - 31/12/2015. The mean relative standard deviation per bin for each case is also given.

40°, the overall amplitude for 6° bin is 75.3%. The number of EQs for this case is 3670 in 1108 d. The relative standard error per bin is 12.8%. This gives us for the peak around 171° a statistical significance of $\sim 5.1\sigma$.

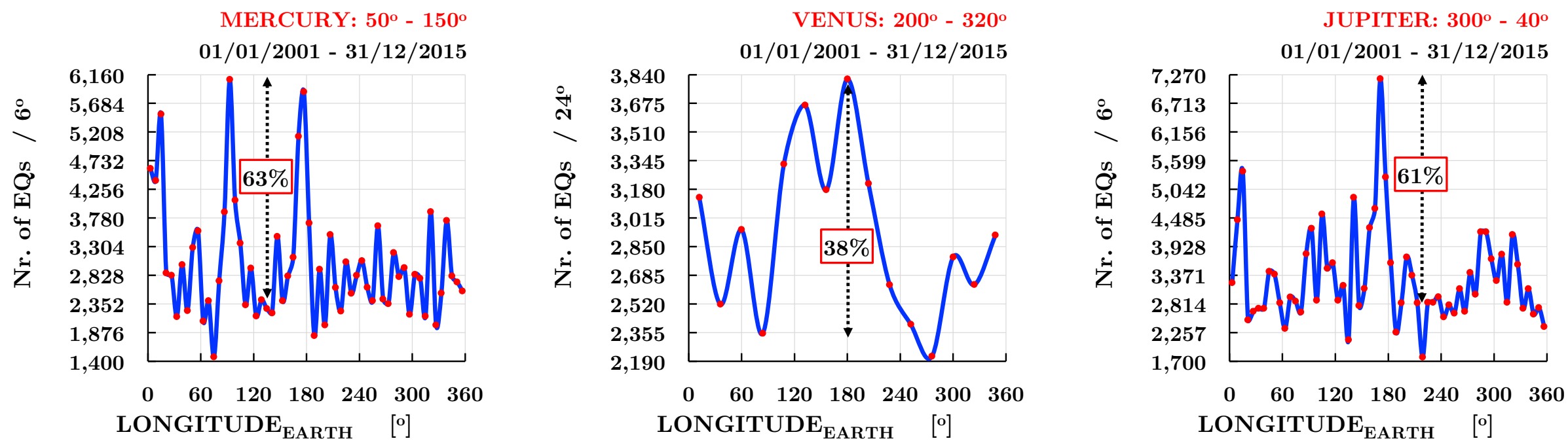

**(a)** Earth when Mercury's longitude is allowed to be between 50° to 150°, for bin = 6°. $\sigma \sim 14\%$.

**(b)** Earth when Venus' longitude is allowed to be between 220° to 320°, for bin = 24°. $\sigma \sim 5.8\%$.

**(c)** Earth when Jupiter's longitude is allowed to be between 300° to 40°, for bin = 6°. $\sigma \sim 12.8\%$.

**Figure 17.9:** Nr. of EQs as a function of Earth longitude while other planets are constrained, for the period 01/01/2001 - 31/12/2015. The mean relative standard deviation per bin for each case is also given.

Finally, in Fig. 17.10, the reference frame is set on Moon, while the Earth is between 90° to 190° heliocentric longitude around the Sun. For the bin of 18° we get a 41.7% amplitude. The number of EQs for this configuration is 4411 while the corresponding number of days is 1478. The relative statistical standard deviation per bin is calculated for this case to be 6.7%. As an example, in this case the peak around 153° has a statistical significance of $\sim 5.6\sigma$ between the two points in the maximum and the minimum of the peak.

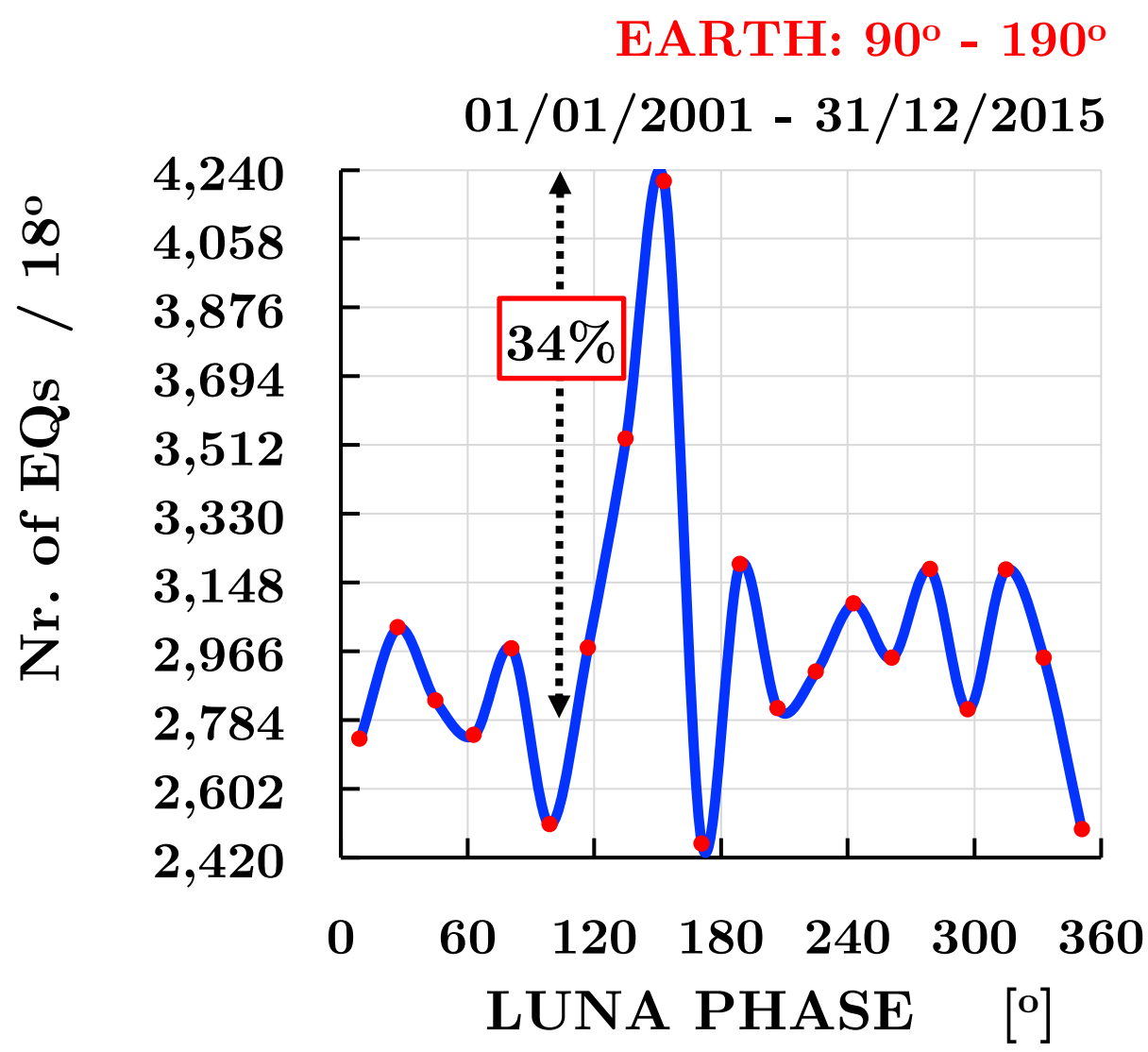

**Figure 17.10:** Nr. of EQs as a function of Moon's phase while Earth's longitude is vetoed between 90° to 190°, for bin = 18°. The mean relative standard deviation per bin is $\sigma \sim 6.7\%$.

The above set of plots and peaks in Fig. 17.7 through 17.10 can work as a probe for optimal planetary positions while validating the observed statistically significant unexpected planetary relationship of the number of EQs. We should note once more here that the exact longitudinal positions of the constrained planets were derived automatically by the algorithm as the best cases for the parameters chosen. Therefore, we can conclude that the optimum longitudinal range for the number of EQs for the specific period and for the specific constraints on the number of EQ per day $(0 - 25)$ seems to be for Mercury around 50° to 150°, for Venus around 220° to 320°, for Earth about 100° to 200° and for Jupiter around 310° to 50°.

### 17.3.2 Fourier analysis

A Fourier analysis has been performed in both the extended dataset of 01/01/1900 - 31/01/2021 and the 15 y dataset of 01/01/2001 - 31/12/2015. The periodicities derived from the Lomb periodogram are compared with the usual planetary revolution periods as well as the known synodical periods between all planets.

For the extended dataset of 01/01/1900 - 31/01/2021 some interesting periodicities appear. The 2$^{\text{nd}}$ biggest peak around 4048.4 d with 54.7 dB amplitude, is close to 11 y solar cycle (= 4015 d). This peak has a FWHM defined by a gaussian fit of $\sim 276.4$ d. Therefore, the average 11 y solar cycle period is within half the FWHM. Also, the 4$^{\text{th}}$ biggest peak, which is around 380.76 d with 14.4 dB and a FWHM of about 7.3 d, overlaps with 378 d which is the synodic period of Earth - Saturn. Similarly, the 14$^{\text{th}}$ biggest peak is around 102 d with 8.5 dB amplitude and 0.4 d FWHM, which is very close to 101 d synodic period of Mercury - Mars. Then, there is a peak at 115.8 d with 3.9 dB amplitude and about 0.4 d FWHM, which is near the synod of Mercury - Earth which is defined at 116 d. Moreover, the 41$^{st}$ biggest peak with 582.9 d and 5.7 dB amplitude with about 7.2 d FWHM is within the error of ±3.1 d with the 584 d corresponding to the Earth - Venus synod. All these peaks seem to further establish the significance of the effects in the double planetary configurations, and thus strengthening the planetary relationship of EQs. Last but not least, the 46$^{th}$ biggest peak is located at 27.32 d ± 0.02 d with an amplitude of 5.5 dB. This periodicity is similar to what was found in the Earth's ionosphere and stratosphere, and coincides with Moon's orbital position fixed to remote starts of 27.32 d. Therefore, this is pointing to an unexpected significant exo-solar impact on the whole Earth.

As mentioned before, the same procedure has been performed in the 15 y dataset (01/01/2001 - 31/12/2015) both when considering all the EQs with $M > 5.2$ and when restricting them to $0 - 25 EQs$/d. It is worth mentioning that in both cases, the peak around 27.32 d becomes even better resolved with an even higher amplitude. More specifically, in the latter case, the peak shown in Fig. 17.11 which is the 17$^{th}$ largest peak, is located at 27.32 d ± 0.05 d and has an amplitude of 6.65 dB. In comparison, around the frequency of 29.53 d, which is associated with Moon's synodic period i.e. fixed to the Sun, no peak is observed which suggests an exo-solar triggering of the EQs.

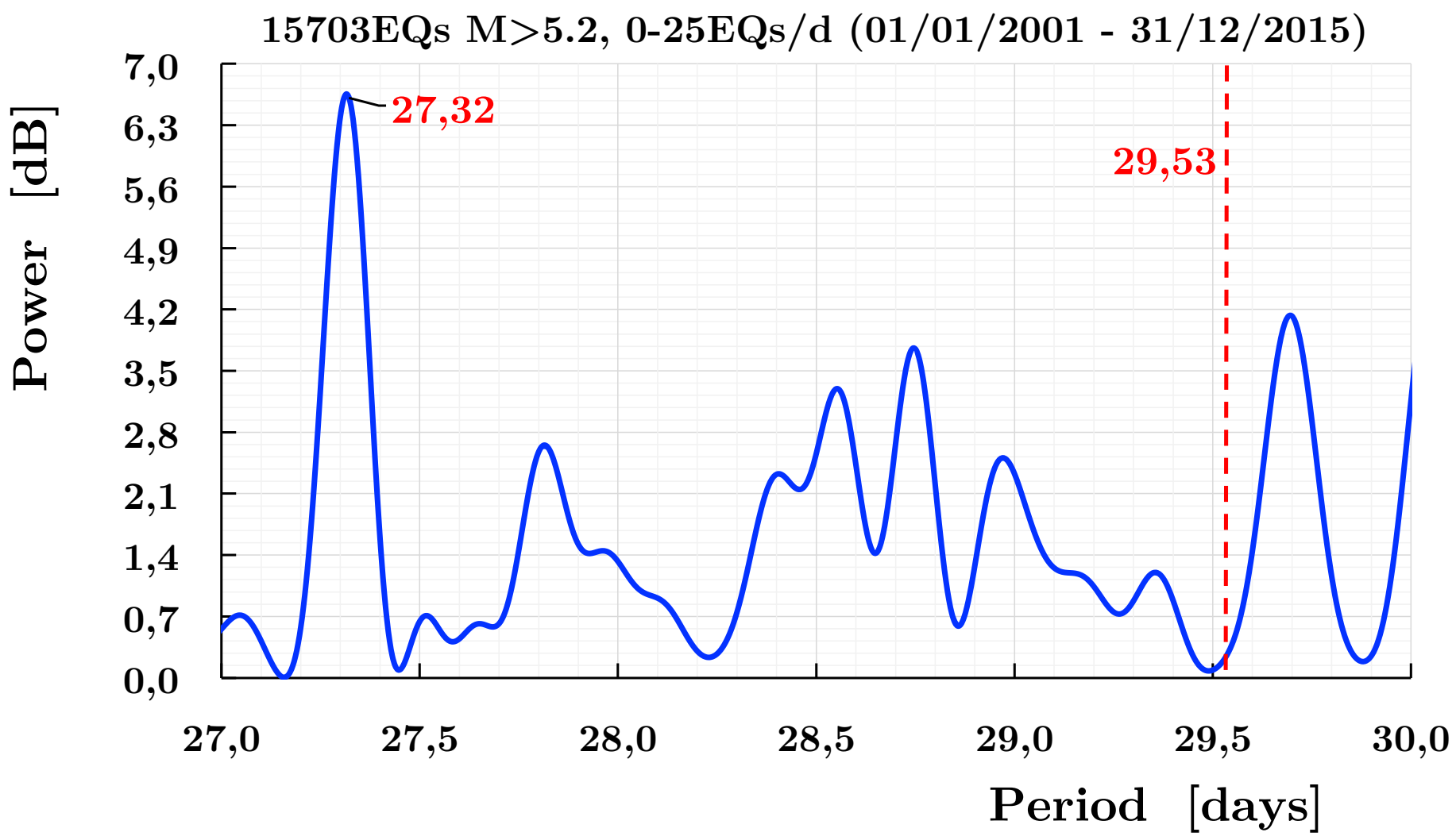

**Figure 17.11:** Fourier periodogram of Nr. of EQ with $M > 5.2$ and 0-25 EQ/day for 01/01/2001 - 31/12/2015 zoomed in around 28.5 d.

## 17.4 Summary

From the various planetary distributions both with single and combined planets, it is clear that the number of EQs shows a statistically significant ($> 5\sigma$) planetary relationship which does not coincide with solar activity (see Appendix Sect. B.11.1). In fact a very small negative correlation has been observed which is in agreement with earlier observations [383]. Interestingly, the biggest effects were observed for the EQs with the biggest magnitude even though the statistical significance is smaller due to smaller statistics. This is consistent with the assumptions of this work and in line with other searches (see for example [368, 384]) which indicated that the largest EQs showed the most significant results.

Furthermore, the derived through a Fourier analysis period of 27.32 d, which is associated with the Moon's sidereal period i.e. fixed to remote stars, and the absence of a peak around 29.53 d, which is associated with Moon's synodic period i.e. fixed to the Sun, indicate a significant exo-solar triggering on the EQs. Additionally, a correlation of the biggest EQs with the ionospheric TEC was found which depicts an increasing trend several days before these EQs occur. This long-term EQ precursor strengthens the claim for an external triggering of the EQs pointing to a downwards propagated motion (see Appendix Sect. B.11.2 and Publications E.18 and E.19).

For the explanation of these conventionally unexpected results, the only viable scenario is through gravitationally focused "strongly" interacting streaming invisible matter towards Earth's location. In addition, the relatively narrow-peaking planetary relationships exclude conventional explanations such as remote planetary interaction or intrinsic Earth-related physics. Such a violent phenomenon as the seismic activity requires a "strongly" interacting invisible matter from the dark sector which can not include the conventional DM approaches such as WIMPs or axions. On the other hand, alternative candidates, such as AQN could be the reason behind such phenomena, since it has already been argued that they can produce acoustic and seismic signals [385] when hitting the Earth through annihilation events [386].

Moreover, since the Earth's seismic activity is monitored uninterruptedly since long time, the streaming invisible matter scenario introduces the possibility for a novel detector through the occurrence of EQs. For a complete analysis to be performed, more EQ parameters have to be used like the depth and location of each EQ but also their specific magnitude. This has the potential to single out specific preferred parameters which could provide valuable information for forecasting but also on the interaction of the assumed invisible massive streams with normal matter. Finally, more time periods should be examined including complete solar cycles, or solar minima and maxima separately. This way, the identification of the direction(s) and duration of the interacting stream(s) might be defined, but could also provide new perspectives in seismological interpretations and forecasting.

# 18

# MELANOMA

## 18.1 Introduction

A possible reason that DM has eluded detection by all state-of-the-art experimental efforts could be due to energy threshold effects. However, living matter (e.g. bacterial, flora, humans) could still sense particles classified as invisible, due to its inherent sensitivity to external influences. Therefore, humans themselves could be the target and at the same time the detector of otherwise invisible irradiation. In addition, long-term exposure could resemble the effects of known radiation such as UV radiation or charged particles that are known to cause cancer. The scenario of gravitational (self-)focusing of streaming invisible matter by the Sun and the planets and the resulting flux enhancement at the site of the Earth could be at the origin of various as yet unexplained diseases such as cancer. If this is the case, then an otherwise unexpected planetary relationship of a biological observation would be a remarkable signature for invisible streams being a contributing cause.

An example of such a possible disease is melanoma, a particularly malignant skin cancer that has already been observed to have an 11 y periodicity [387, 388]. Cancer is a dataset with high statistics which constitutes one of the biggest mysteries in medicine with implications of increased significance for public health. In fact skin cancer is the most common form of cancer, accounting for about 40% − 50% of all the diagnosed cancers in United States of America (USA) [389]. Melanomas constitute about 4% of skin cancers but they account for almost half of all skin cancer deaths [390]. This investigation is triggered by a previous study showing a planetary relationship of melanoma occurrence, following monthly data obtained from Surveillance Epidemiology End Results database (SEER) for the period 01/01/1973 to 31/12/2010 (38 y) [391–393]. In the following analysis, the data on melanoma diagnosis that have been recovered, have a 30-times shorter cadence, i.e. daily registration was available instead of monthly mean values (see also Publications E.2 and E.4).

## 18.2 Data and methods

### 18.2.1 Data origin

The selected daily data have been acquired from the Australian Institute of Health and Welfare and more specifically from Australian Cancer Database (ACD) which contains a list of primary malignant cancers diagnosed in Australia, including Tasmania, since 1982 [394]. The acquired dataset contains exact dates of diagnosis of all cutaneous and ocular melanoma cases on the ACD for 01/01/1982 to 31/12/2014. More specifically the data include histology codes 8270 − 8790 (melanoma) and the topography codes: C44 (non-genital skin), C51 (skin of vulva), C60 (skin of penis), C63.2 (skin of scrotum), C69 (eye) and C80 (unknown primary site). The last site, C80, is included because the term "melanoma, site not specified" is recommended to be grouped with melanoma of the skin for reporting purposes. For this first analysis results, the sum of all codes has been used.

Additionally, each data, contain an accuracy indicator consisting of three letters. Although the first letter generally relates only to the accuracy of the day, the second to the month and the third to the year, there are some cases in which the indicator could mean that two or all three components were estimated. The following table Tab. 18.1 explains all the possibilities.

**Table 18.1:** The various accuracy indicators of the ACD data and their meaning.

| Value | Meaning |
|---|---|
| A (can occur in any position) | The component has been specifically recorded by the cancer registry as accurate. |
| B (can occur in any position) | The cancer registry does not (or did not at that time) have an accuracy indicator. Nevertheless, this component is likely to be accurate. |
| E (can occur in any position) | The true value of the component is unknown and has been estimated by the cancer registry. Note that if one or both of the other components also have the value E, these components may have been estimated jointly rather than independently. |
| FFF | One or more components were estimated by the cancer registry but it is not recorded which one(s). |
| $U__$ (U in 1$^{st}$ position; other positions can be A, B or E) | The true value of 'day' is unknown and was supplied by the registry as a missing value. AIHW has set it to the middle day (rounded down) of the range of possible days in which diagnosis could have occurred. This is usually the middle day of the month but can differ if the person was born and/or died in the same month as diagnosis, i.e. you can't be diagnosed before birth or after death. |
| $UU_$ (UU in 1$^{st}$ and 2$^{nd}$ positions; 3$^{rd}$ position can be A, B or E) | The true values of 'day' and 'month' are unknown and were supplied by the registry as missing values. AIHW has set the diagnosis date to the middle day (rounded down) of the range of possible days in which diagnosis could have occurred. This is usually the middle day of the year but can differ if the person was born and/or died in the same year as diagnosis, i.e. you can't be diagnosed before birth or after death. |
| UUU | Impossible because the year of diagnosis is never missing. |

Therefore, to restrict the analysis to records with an accurate day, month and year, then only the groups that consist of As and Bs are used (i.e. AAA, AAB, ABA, ..., BBB), since the codes that contain an E, F or U, mean that there is some level of uncertainty in the data. This results in eight usable codes altogether and a total of 248425 cases out of 278450 cases for all codes seen in Fig. 18.1 and for the period 01/01/1982 to 31/12/2014 (= 12 053 d). More specifically, we have 242416 cases for C44, 416 cases for C51, 31 cases for C60, 15 cases for C63, 4057 cases for C69 and 1505 cases for C80 seen in Fig. 18.1a. Finally, in the dates without a reported case, the value zero has been assigned.

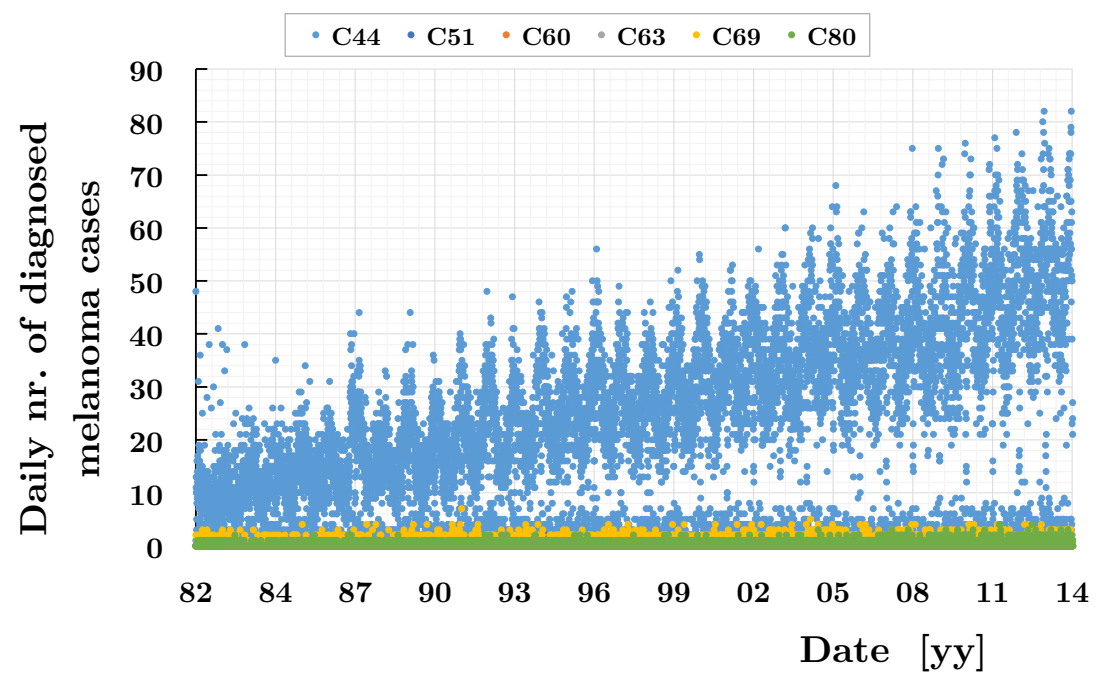

**(a)** The eight separate melanoma codes.

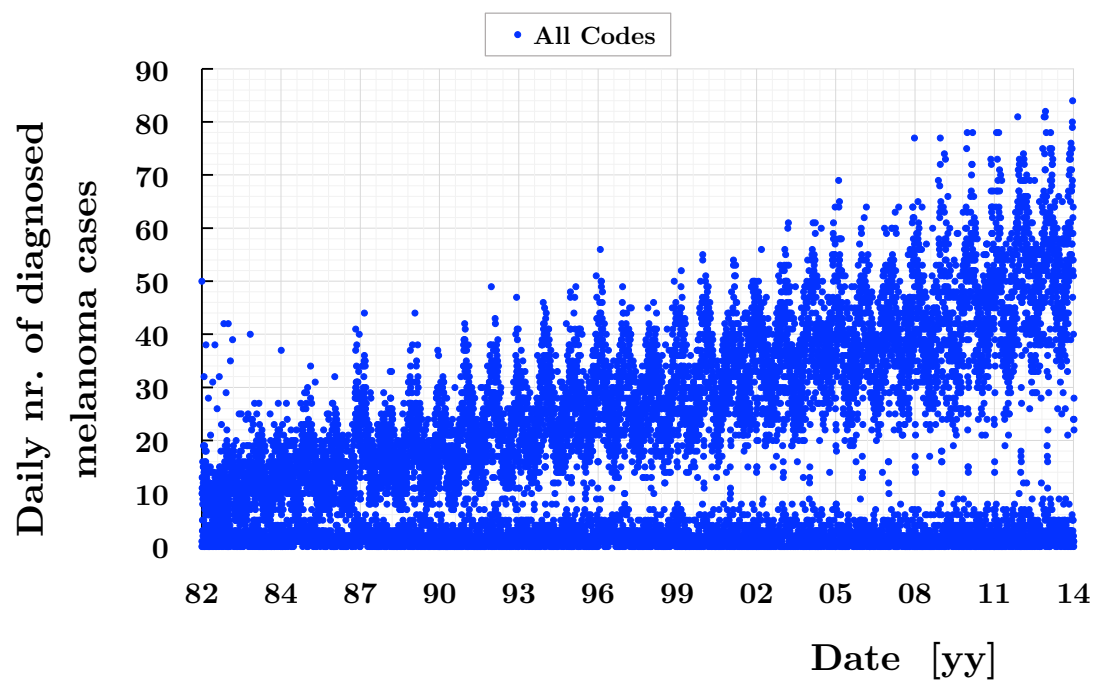

**(b)** All melanoma codes.

**Figure 18.1:** Daily Nr. of diagnosed melanoma cases with As and Bs accuracy indicators from 01/01/1982 - 31/12/2014.

In Fig. 18.2, the various frequency distributions for the number of diagnoses shown in Fig. 18.1b are created which include records with accuracy indicators A and B. It is worth noting Fig. 18.2a where the mean observed growth normalised to the population is about 60% per decade. A similar trend was also observed in the northern hemisphere in USA [391].

### 18.2.2 Data treatment

#### 18.2.2.1 Incomplete registration

The first correction that has to be made on the data is deducted from Fig. 18.2e. During each calendar year, there are six national holidays for which melanoma diagnosis and registration are unavailable, namely 26-28 January, 25 April and 25-26 December. For the analyses in this search, we have adjusted for these days with incomplete registration by interpolating the mean value from two days before and two days after each gap. The result is shown in Fig. 18.3. There is one more point at 29/02 which has fewer counts, but this is normal since this is a leap day which appears only once every four years, which means that this point does not have to be corrected. Furthermore, the rest of the region during the Christmas period has not been modified since it contains about 20 days. The total number of cases after this correction is increased to 250775 from the 248425 cases in the beginning.

In addition, due to the existence of a long-lasting gap around Christmas towards New

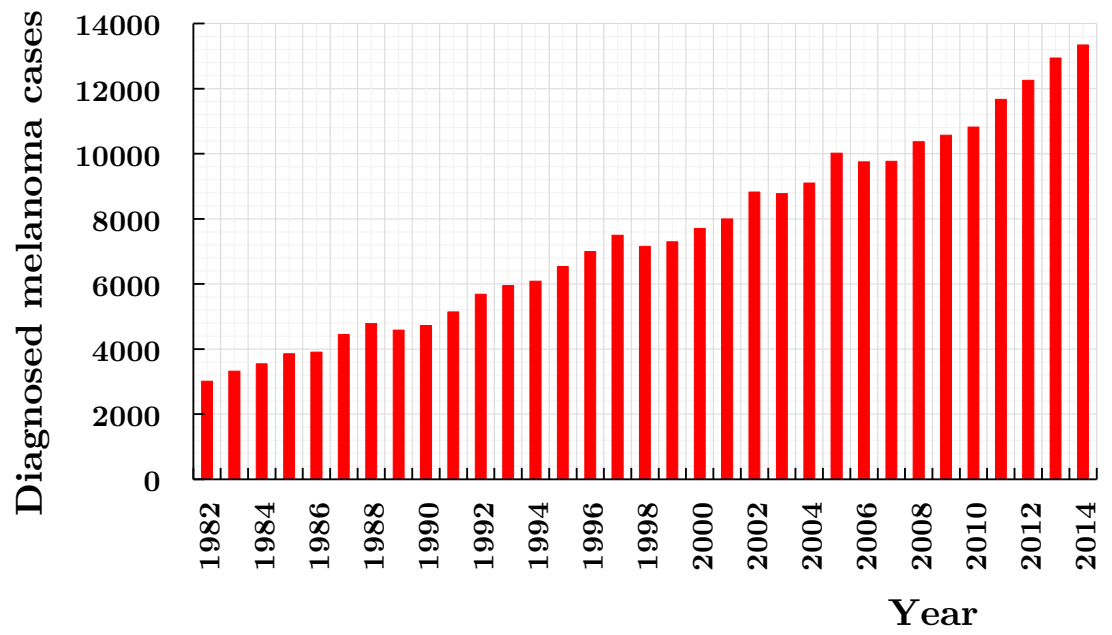

**(a)** Nr. of cases per year.

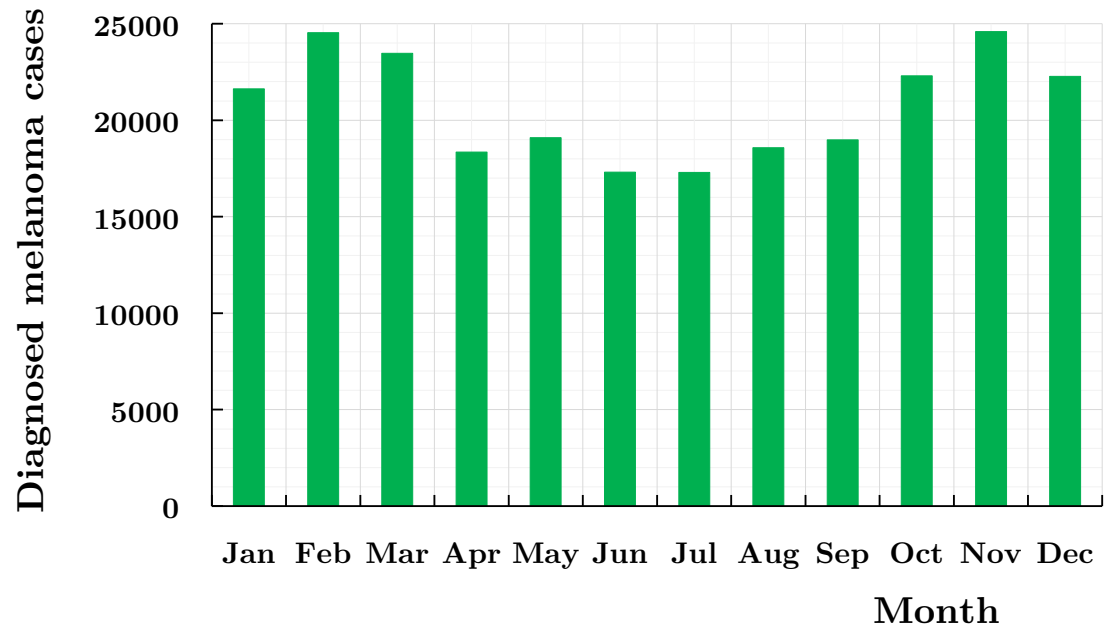

**(b)** Nr. of cases per month.

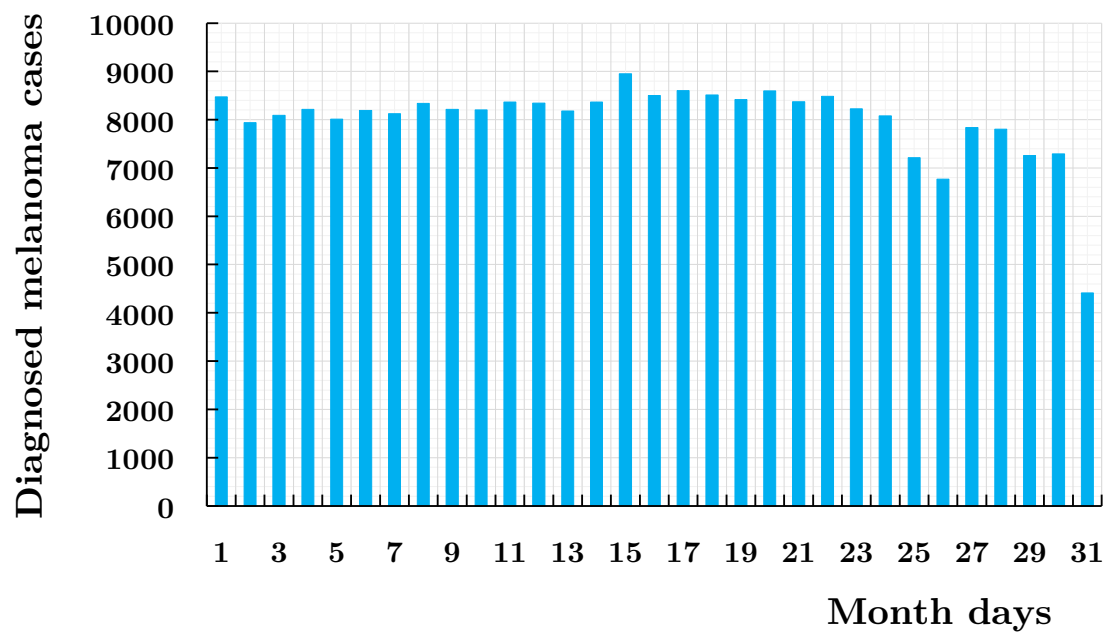

**(c)** Nr. of cases per day of month.

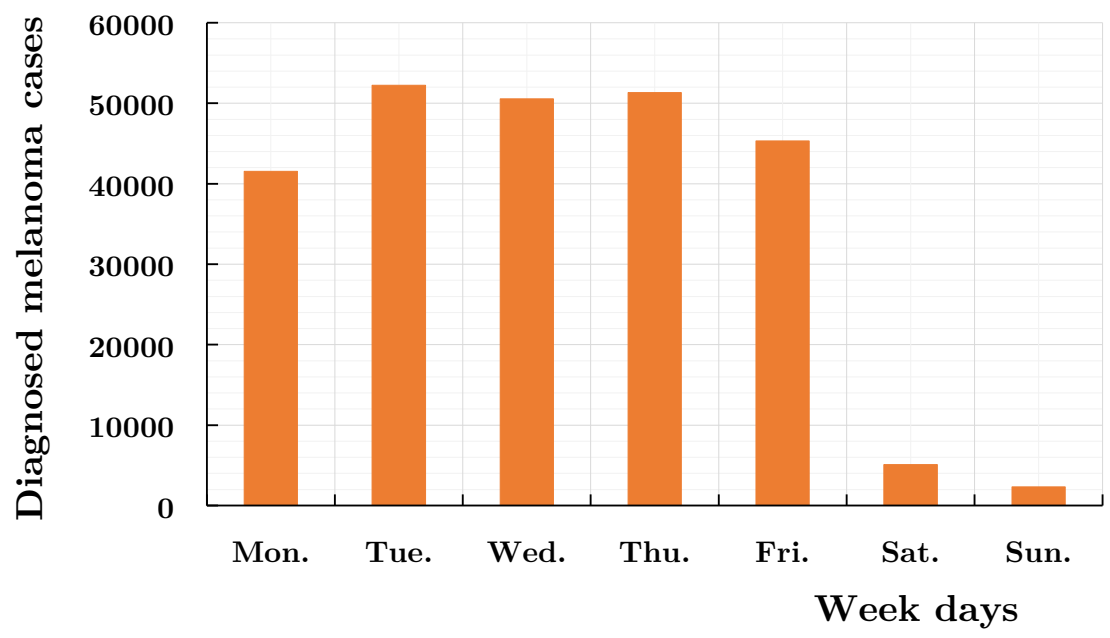

**(d)** Nr. of cases per day of week.

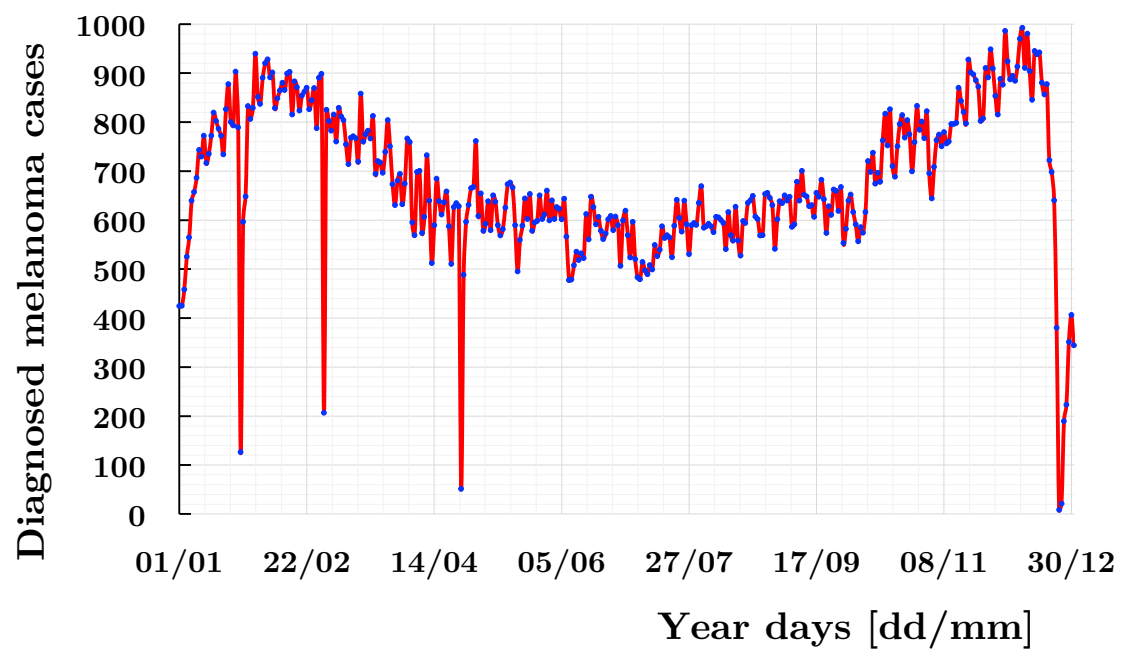

**(e)** Nr. of cases per day of year.

**Figure 18.2:** Histograms for the Nr. of diagnosed melanoma cases from all codes for 01/01/1982 - 31/12/2014.

Year, as seen also in Fig. 18.3, and to avoid any artefact, we have excluded from our analysis a period of about 1.5 month, from middle December to the end of January. For Earth's orbital position this corresponds to the range 87° to 127° heliocentric longitudes.

#### 18.2.2.2 Weekly correction

The next important correction is based on Fig. 18.2d, which shows that during the weekends there is a drop in the number of cases. This effect appears also on the Fourier result of the raw data with a peak equal to 7 d (not shown here). Therefore, to exclude that such weekly gaps could cause artefacts, we have performed the following adjustment. For each week, the total number of diagnosed melanoma cases are counted and divided by 7. This rate is used for each day of the corresponding week (see Fig. 18.4). This does not change the total number of diagnosed melanoma cases. As a result, the Fourier peak at 7 d was suppressed. We also note that the remaining difference observed in Monday, Tuesday and Wednesday from the rest of the weekdays is due to the existence of more days in leap years.

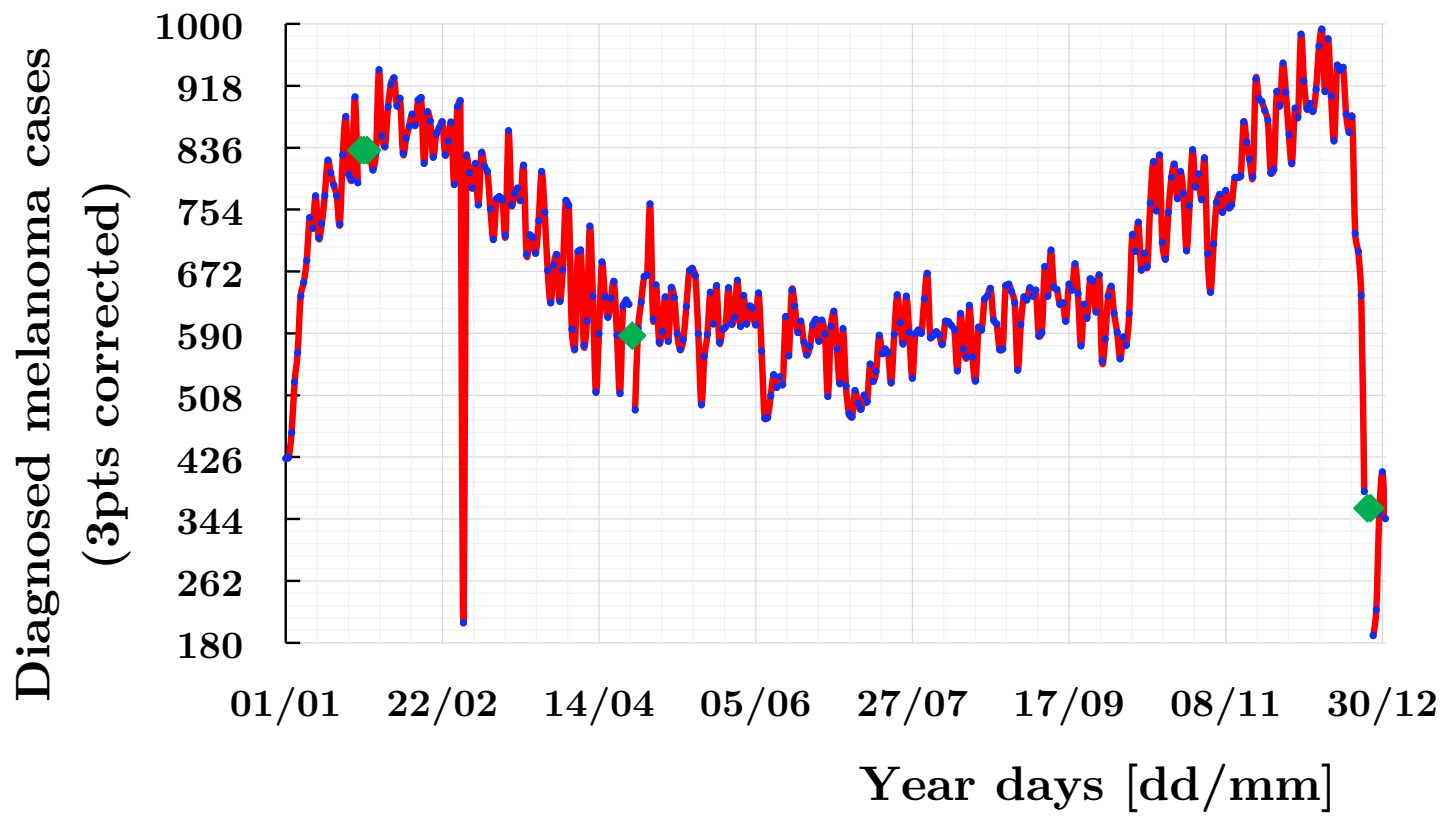

**Figure 18.3:** Nr. of diagnosed melanoma cases per day of year, with the three corrected cases with incomplete registration records due to national holidays shown in green (see also Fig. 18.2e).

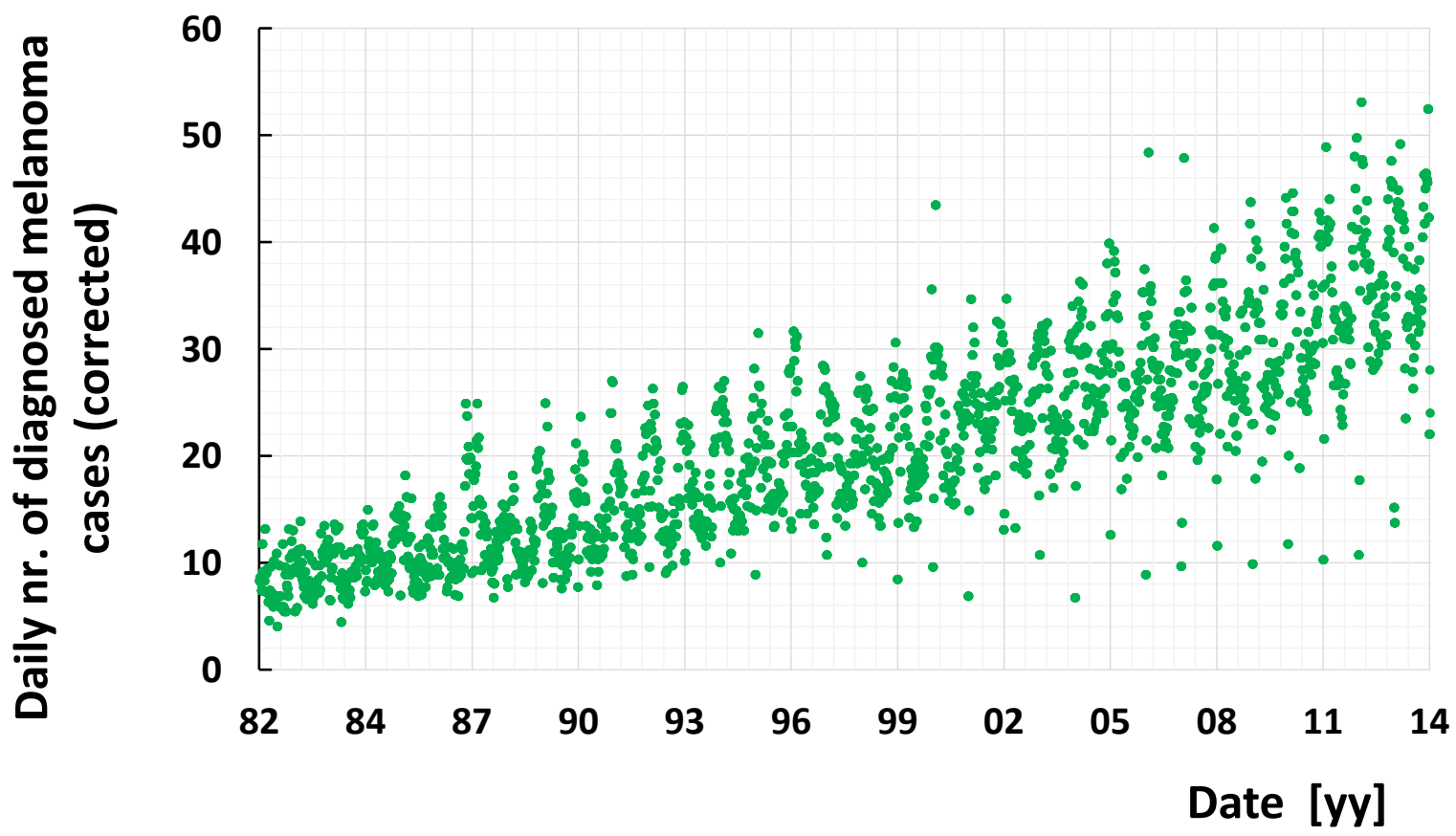

**(a)** Time distribution of corrected data.

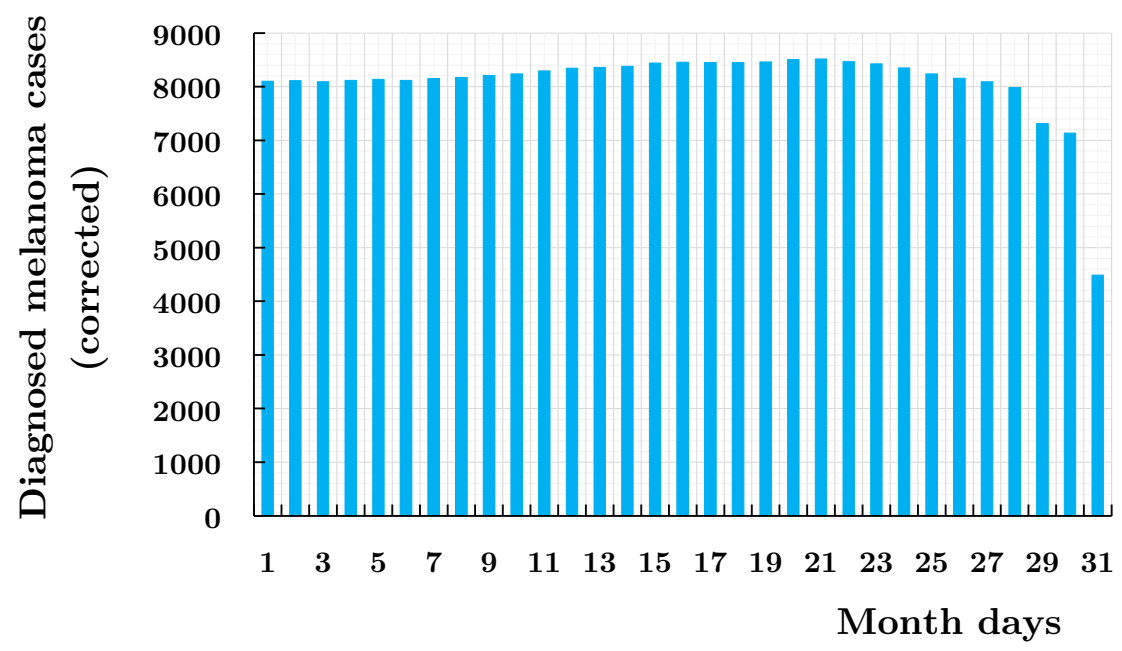

**(b)** Histogram of corrected data with the Nr. of cases per day of month.

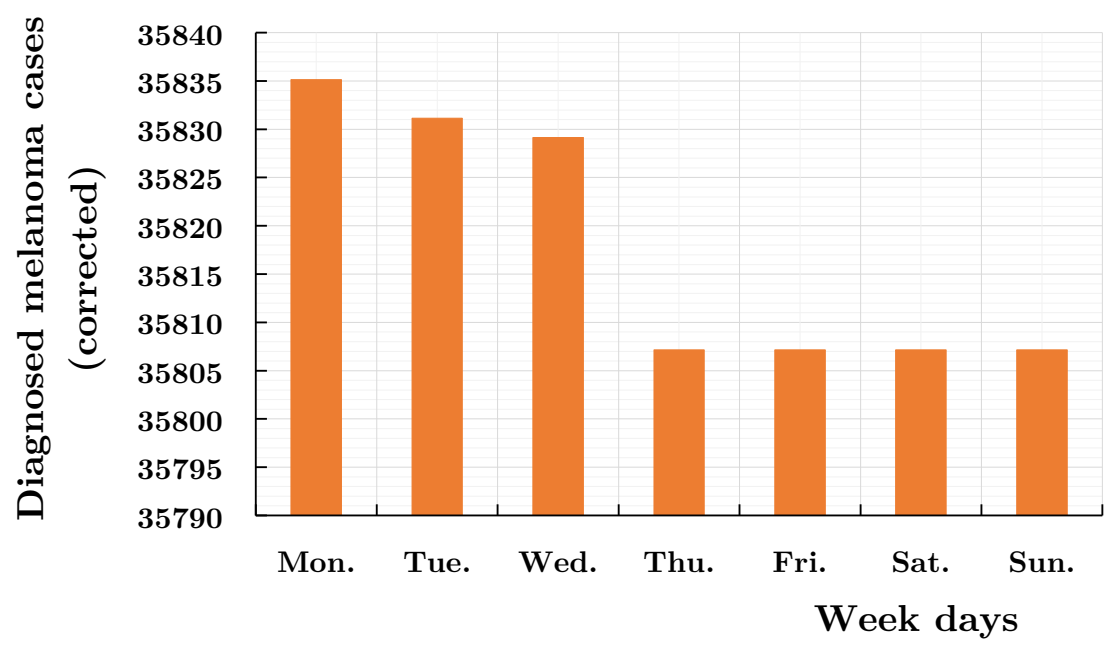

**(c)** Histogram of corrected data with the Nr. of cases per day of week.

**Figure 18.4:** Weekly corrected Nr. of diagnosed melanoma cases including also the 3-point correction from Fig. 18.3.

# 18.3 Data analysis and results

## 18.3.1 Planetary longitudinal distributions

### 18.3.1.1 Single planets

As usual, each day with a corresponding number of melanoma cases is assigned with a specific heliocentric longitude value for the position of each planet. More specifically, for the analysis in this dataset, the Earth's heliocentric position is mainly used as it contains the most significant result. As first, to verify once more the correction procedure described in Sect. 18.2.2, the daily diagnosed melanoma cases from both the raw and the corrected data are projected on the corresponding Earth's orbital position given by its heliocentric longitude in Fig. 18.5. As expected, the two short empty points, indicated by the green arrows in Fig. 18.5a, which correspond to the two public holidays dates of 26-28/01 (around 125° to 127°) and 25/04 (around 214°) do not change the overall spectral shape when the aforementioned corrections for the incomplete registrations are applied. Furthermore, the corrected data in Fig. 18.5b result in a smoother spectral shape which is to be expected due to the weekly correction. This is because, after one Earth orbit, there is a phase shift of about 1.2 d which every $\sim 3$ y the phase shift becomes half a week. Therefore, when the weekly correction is applied, when summing up 33 y the out-of-phase effects are minimised. This means that the short-term fluctuations that are observed in the raw data in Fig. 18.5a are averaged-out and the derived time dependence in Fig. 18.5b fluctuate less while retaining its key features.

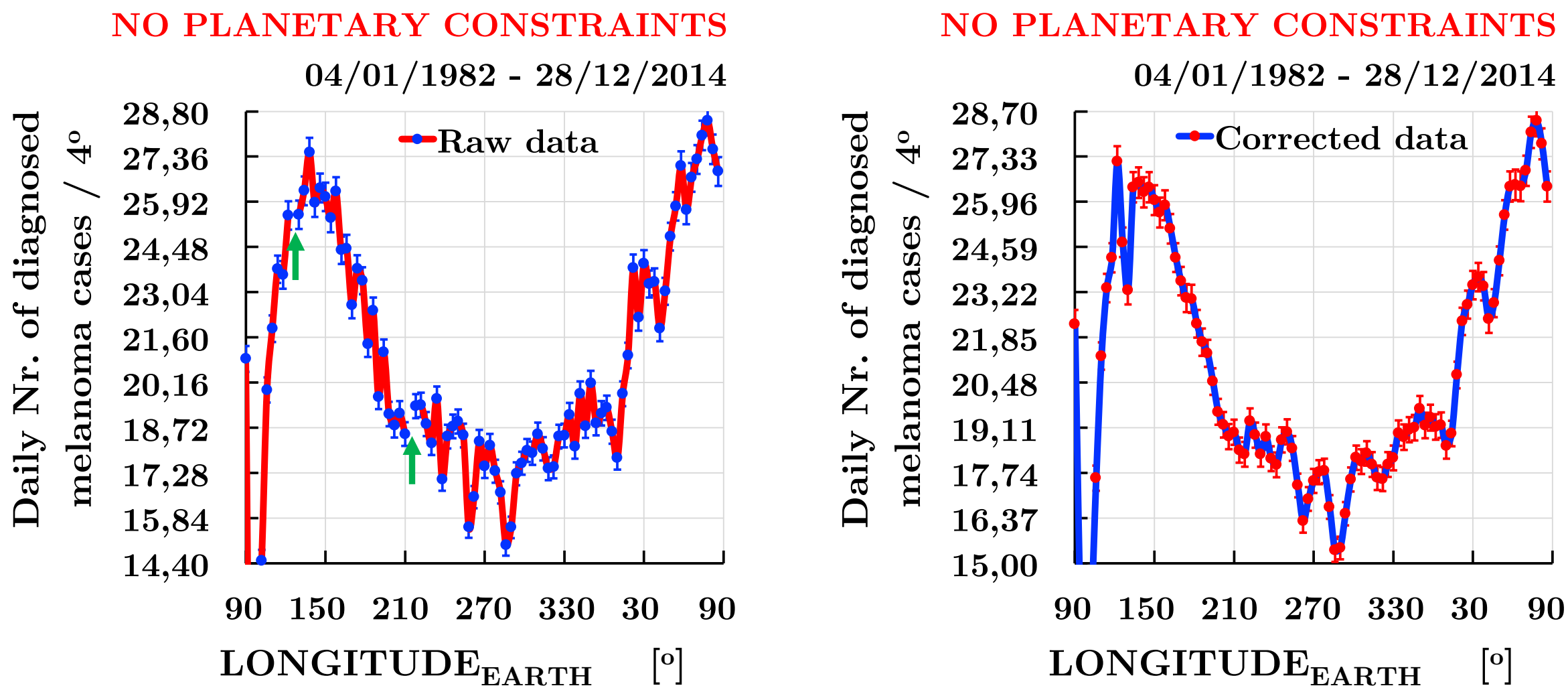

(a) Raw data. (b) Corrected data.

**Figure 18.5:** Daily Nr. of melanoma diagnoses as a function of Earth's heliocentric longitude for the period 04/01/1982 - 28/12/2014. As expected, the corrected data result in a smoother distribution compared to the raw data but the overall shape remains unchanged. In addition, a strong fluctuation is observed which is in contrast to the expected seasonal distribution following the solar UV exposure.

Assuming Poisson statistics the estimated statistical significance for example for the peak around 274° in Fig. 18.5b is far above $5\sigma$ which means that the apparent modulation could not be due to random fluctuations. It is noted that the bin of 4° in Earth corresponds to about 4 d and therefore in each point of the distribution we have added $33 \times 4$ d since the measurements were repeated 33 times over the time interval of 33 y. This leads to the conclusion, that assuming there was no planetary relationship, the summation of 33 y would cause a strong averaging on any fluctuation thus causing the appearance of a random distribution. More specifically, for the peak at 274°, taking into account the difference between the two values at the maximum and its next two values of the minimum on the left, we get a significance of $\geq 6\sigma$. The calculated mean statistical deviation per bin of 4° is about 1.9%. Such a large significance is possible because of the large amplitude of the modulation of the daily cases ($\sim 20\%$). This calculation can also serve as a rough estimate of the significance of the other modulation peaks in the next plots.

#### 18.3.1.2 Combining planets

The next step is the computation of the time dependence of melanoma diagnoses when requiring a second planet to be propagating in a specific region. In this specific case, Moon's phase is chosen and a comparison is made between 0° to 180° and 180° to 360° (Fig. 18.6). It is noted that 0° on Moon's phase corresponds to New Moon whereas 180° corresponds to full moon. As a result, the faint modulation seen in Fig. 18.5, it is much better resolved in Fig. 18.6a. Moreover, the dissimilarity of the two cases in Fig. 18.6 leads already to the conclusion that the location of the Moon has an impact on the melanoma incidence rate throughout one year.

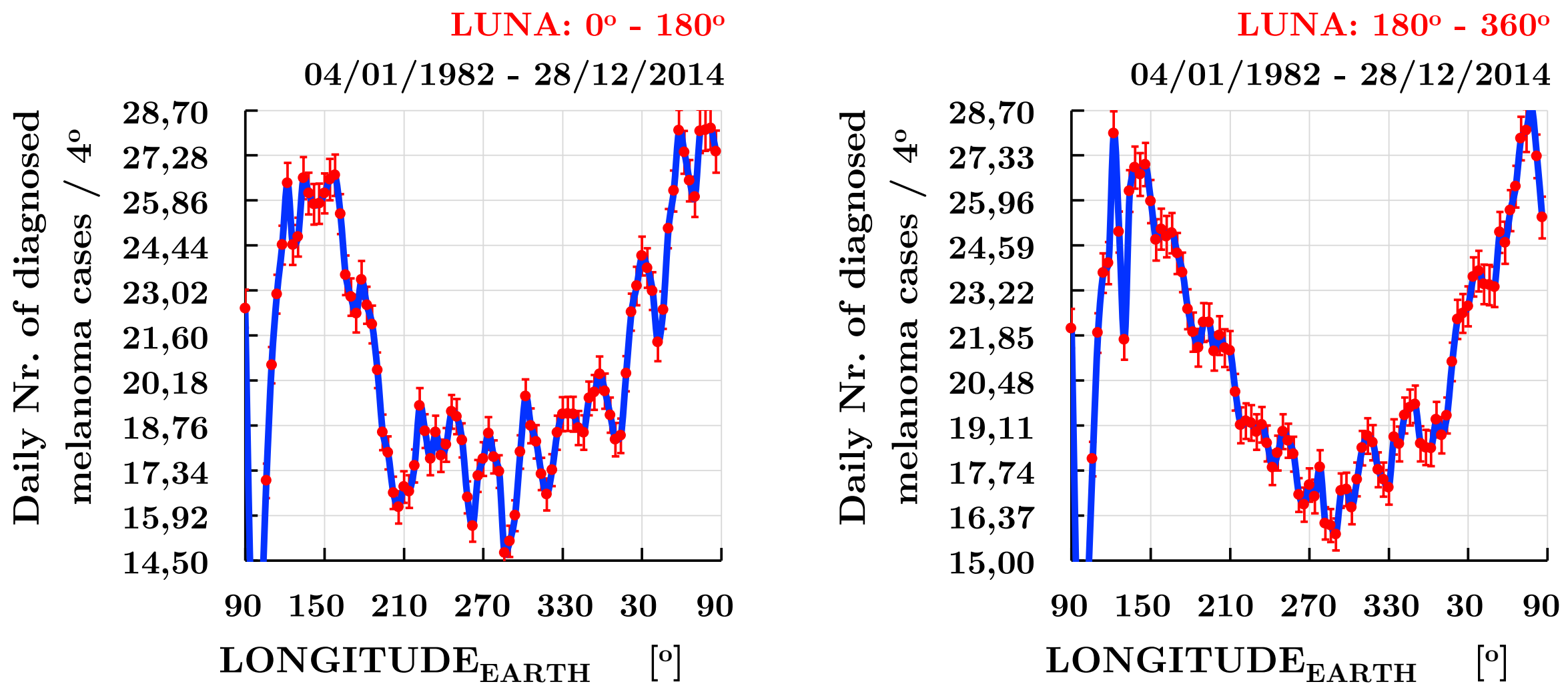

**(a)** Earth when Moon is between 0° to 180°. **(b)** Earth when Moon is between 180° to 360°.

**Figure 18.6:** Daily Nr. of melanoma diagnoses as a function of Earth's heliocentric longitude when Moon's phase is constrained to two opposite arcs.

In order to exclude systematics, Fig. 18.6a is plotted once more in Fig. 18.7 with the full period of 33 y being split into two parts. As it is obvious the modulation is indeed present in both sub-datasets which excludes a random modulation due to temporal excursions. Interestingly, the last 14 y (Fig. 18.7b), the time dependence seem to have a more pronounced modulation than the first 12 y (Fig. 18.7a). Such an effect reflects also the long-term increase in the melanoma rate that was observed in Fig. 18.2a. In fact, the comparison with the solar activity (see Sect. B.12.2) shows a clear dissimilarity between the two. This strengthens the evidence for the existence of an additional exo-solar long-term invisible stream(s) component(s) influencing the melanoma occurrence. From the two Fig. 18.7a and 18.7b, the relative mean melanoma incidence rate can be calculated. The total average increase observed in Fig. 18.7c is about 64.2%.

Finally, the plot from Fig. 18.6a is plotted once more with a smaller bin of 2° in Fig. 18.8, which corresponds to about 2 d. Fig. 18.8 reveals even more clearly the relatively

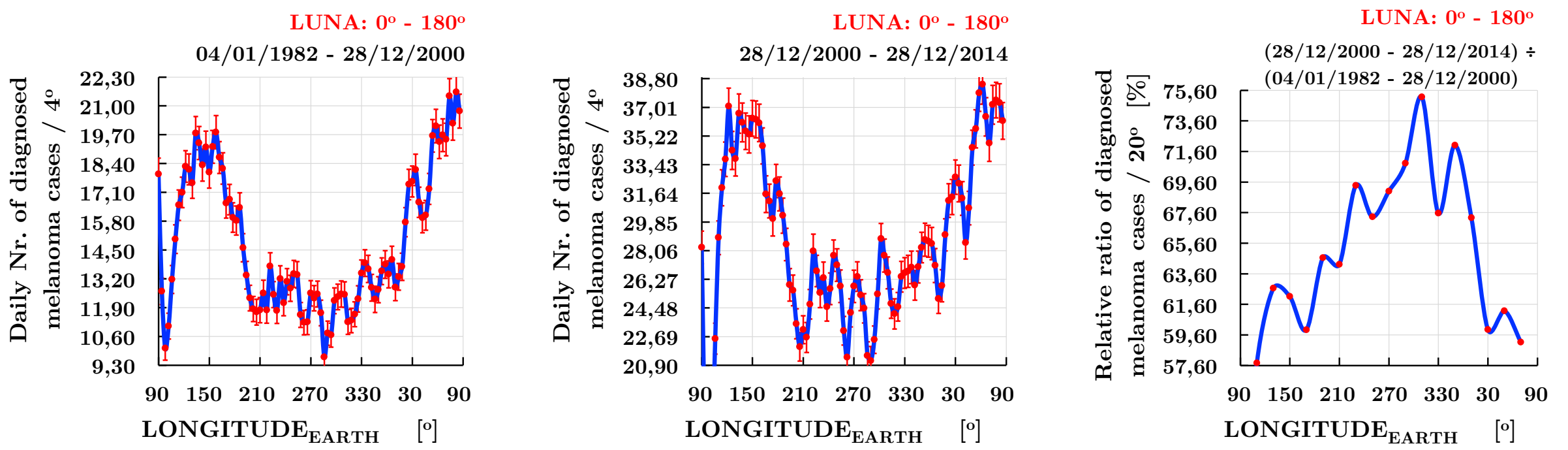

**(a)** 04/01/1982 - 28/12/2000. **(b)** 28/12/2000 - 28/12/2014. **(c)** Relative change of melanoma diagnosis between the two time intervals 2000-2014 (Fig. 18.7b) and 1982-2000 (Fig. 18.7a) for 20° bin.

**Figure 18.7:** Evolution over time of the Nr. of melanoma cases as a function of Earth's heliocentric longitude with Moon being between 0° to 180°, for two consecutive time intervals.

short modulating behaviour of melanoma diagnoses. The region from 127° to 89° corresponds to about 329 d. Therefore the 12 observable peaks correspond to a periodicity of about $329\,\mathrm{d}/12 = (27.4 \pm 0.2)\,\mathrm{d}$, assuming one bin ($\sim 2\,\mathrm{d}$) uncertainty from the 329 d time interval. This number is very close to Moon's sidereal month of 27.32 d as is obtained in Fig. 18.8.

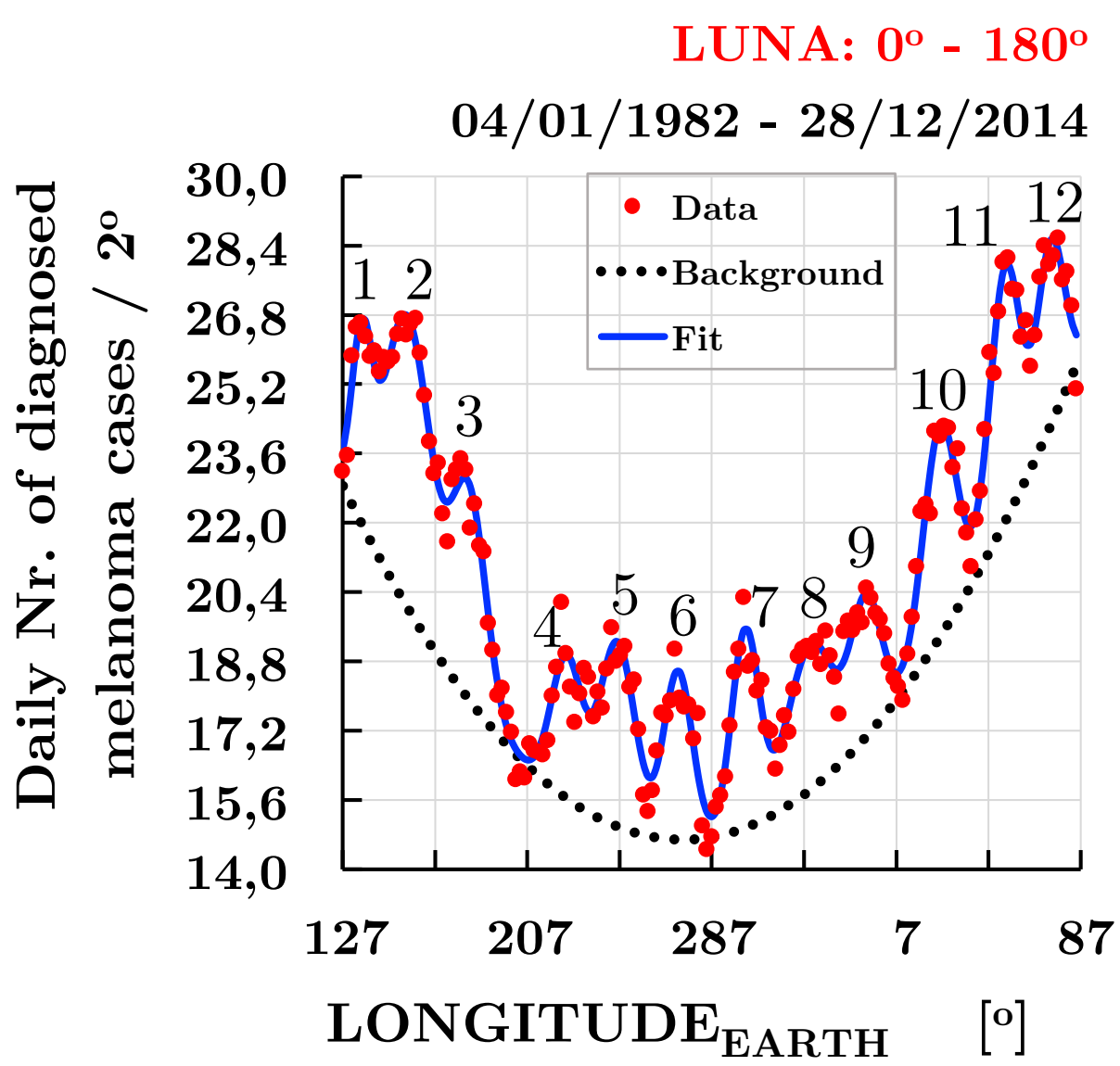

**Figure 18.8:** Daily diagnosed melanoma cases for 33 consecutive Earth orbits (1982-2014) as a function of Earth's heliocentric longitude, while the Moon is orbiting around Earth between 0° to 180°. A free parameter fit (in blue) of a second-degree polynomial (dotted line) is also shown.

To better quantify the 12 peaks in Fig. 18.8, a peak finding algorithm was applied which recovered them one by one. The data were then free-fitted arriving to 12 Gaussian peaks plus a polynomial of second degree as background (see blue solid line in Fig. 18.8). The free parameters were found through Python least square fitting technique. The corresponding average FWHM of these peaks is $(18 \pm 2)$ d, whereas the mean distance between neighbouring peaks is $(27.5 \pm 0.8)$ d. The large uncertainty is due to the limited number in evaluating the mean distance and the relatively large width of each peak. It is noted that the 27.32 d periodicity which corresponds to Moon's sidereal period, i.e. fixed to remote stars, is within $1\sigma$ from the observed periodicity, and thus points to an exo-solar influence.

### 18.3.2 Fourier analysis

In order to find the precise periodicity observed in Fig. 18.8, a Fourier analysis is performed. The Lomb periodogram result, shown in Fig. 18.9a, shows a significant peak around $(27.32 \pm 0.03)$ d, which strongly favours the Moon's sidereal month, i.e. fixed to the stars, of 27.32 d. The amplitude of the peak is 4.5 dB and is the 34$^{st}$ biggest peak in the whole Fourier spectrum. In the plot it is also observable, but with a smaller amplitude of $\sim 2.2$ dB and significance, a peak around 29.53 d which can be attributed to Moon's synodic period of 29.53 d, i.e. fixed to the Sun. This peak has a FWHM of 0.07 d which gives an error of $\pm 0.03$ d. This observation on its own indicates an increased exo-solar impact in the number of diagnosed melanoma cases. Finally, it is mentioned that the strong peak appearing at around 28.1 d is the 13$^{th}$ sub-harmonic of the strong annual peak at 365.2 d and serves as a calibration peak in the specific range.

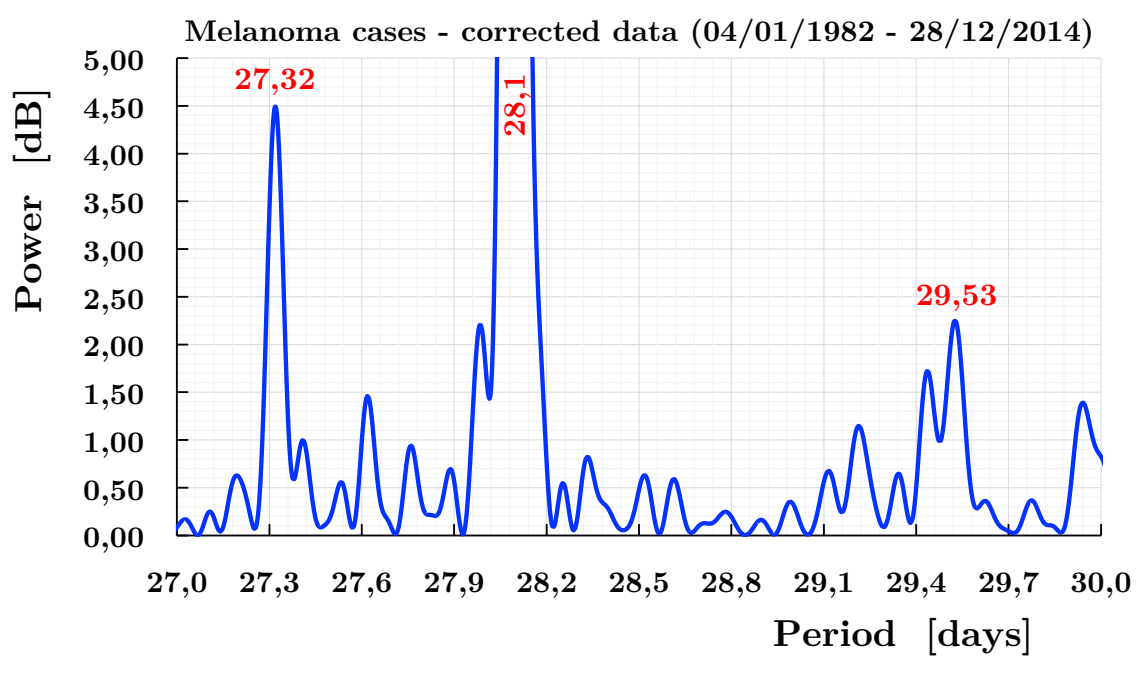

(a) Corrected melanoma data from 04/01/1982 to 28/12/2014.

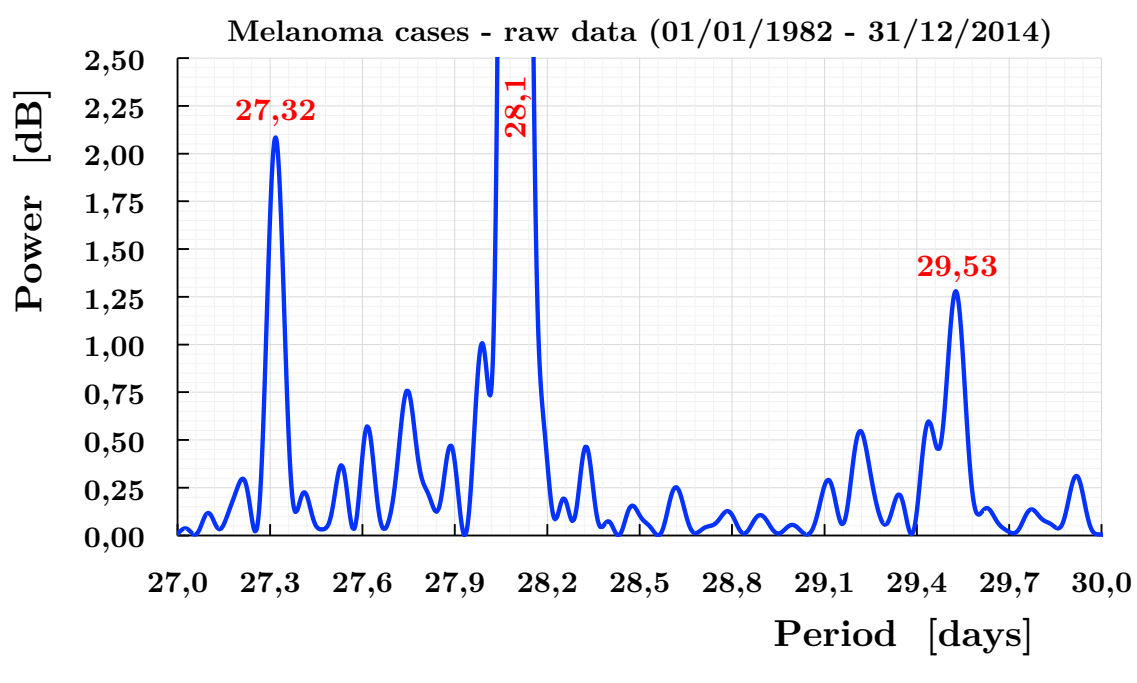

(b) Raw melanoma data from 01/01/1982 to 31/12/2014.

**Figure 18.9:** Fourier periodogram of the melanoma data in Australia.

The same peak around 27.32 d $\pm$ 0.03 d overlapping with Moon's sidereal month of 27.32 d is derived also in the very raw data without any correction, but with a smaller amplitude of 2.1 dB as seen in Fig. 18.9b. In this case also the peak around 29.53 d $\pm$ 0.03 d appears once more, but again with a smaller amplitude than the 27.32 d one and smaller significance.

## 18.4 Summary

In the absence of any external influence, the seasonal rate of melanoma should have been steady with a broad maximum during local summertime due to the increase in the solar UV exposure. However, the daily number of diagnosed melanoma cases in Australia show a significant modulation repeating every $\sim 4$ weeks with the exact periodicity coinciding with Moon's sidereal month of 27.32 d. This is also verified by a Fourier analysis with the 29.53 d period corresponding to Moon's synodic period, i.e. fixed to the Sun, being much weaker than the 27.32 d peak. This behaviour does not fit neither the UV exposure in the area nor the solar activity as it is manifested by the solar proxy F10.7 (see Appendix Sect. B.12). As a result, these findings point to an additional important exo-solar impact on the melanoma incident rate.

Moreover, the UV increase around Australia from $1996 - 2017$, shows an increase with a mean value of $+0.25\%$ per decade, whereas the net increase in melanoma cases is about 60% per decade (see Appendix Sect. B.12.1). In addition, this increase in the diagnosed melanoma diagnoses also can not be explained by the increase of the population in Australia from 15289000 inhabitants in 1982 to 23985000 in 2015 [395].

The underlying physics concept of invisible streaming massive matter can provide a possible explanation both for the surprising short oscillatory behaviour of $\sim 27$ d and the normalised long term increase of the diagnosed melanoma cases. The gravitational focusing by the Moon towards the Earth can modulate strongly the influx of slow moving (up to about 50 km/s to 400 km/s) stream(s) during its geocentric orbit. This is supported also by the derived exo-solar impact from the Fourier analysis. Within the same scenario of the dark universe, the observed unexpected increase on the melanoma diagnoses can be reflected to the corresponding concurrent increase of incoming invisible streaming matter. This is also supported by a similar long-term increase in melanoma incidence which was observed in the northern hemisphere in USA [391]. This poses a new physics-inspired interpretation for a widely known and long-overlooked medical observation on the number of diagnosed melanoma cases (see Publications E.2 and E.4).

With this kind of investigation, the manifestation of the dark universe has been extended from the upper atmosphere to the lowest atmosphere where humans live. These observations underline the need for more similar investigations to other diseases or biomedical phenomena (see also Publication E.8). Finally, different types of cancer, as well as different topography codes, can be analysed individually and compared which each other to provide more hints on the quest to decipher the nature of the assumed invisible streams and their form of interaction with the atmosphere as well as with living matter.

# 19

# DISCUSSION

The observationally driven results of this Part IV point to an external influencing factor on the manifestation of a wide range of terrestrial phenomena. The derived planetary relationships of observations such as the ionisation of the upper atmosphere, the stratospheric temperature excursions, the occurrence of large EQs, as well as the recorded melanoma cases, similarly to Part III, can not be explained conventionally. The only viable explanation suggested through this work is through some kind of low-speed, highly-interacting invisible massive streams from the dark sector. It is noted that even without the unexpected planetary observations made in this work, these phenomena could not be explained otherwise.

The reasoning of this work is based on the gravitational focusing by the solar system bodies including the intrinsic self-focusing by the Earth, of low-speed invisible streaming massive mater towards the Earth's location. As an example, the typical speeds of $\sim 250\,\mathrm{km/s}$ for dark sector constituents can be effectively gravitationally influenced by the Sun and its planets including the Moon since the gravitational deflection goes inverse with the velocity squared ($1/v^2$). Then, the assumed "strong" interaction of this invisible matter with the Earth's atmosphere as well as with the "living matter" could produce significant effects, with the planetary relationships being the key signature for the observations in this work. The GC remains a promising place in space with the GC → Sun → Earth alignment within 5.5° taking place around 18$^{\text{th}}$ of December each year with the Earth being around 86.5° heliocentric longitude. It is stressed that the aforementioned anomalies in the ionosphere as well as in the stratosphere are also taking place around December - January.

The long-term observations shown in Tab. 19.1, along with the conventionally unexpected peaking longitudinal distributions of all observables, indicate a significant relationship with the orbits of the planet(s). In addition, a downwards propagating signature in stratosphere that was hinted from previous studies [354–356], has been derived also here (Sect. 16.4.1).

At the same time, the comparison with the solar activity shows a different behaviour in most cases which strengthens the claim of an additional external and exo-solar origin (see Appendix Sect. B). In Tab. 19.2 a statistical correlation analysis has been performed for the general trend of the terrestrial observations compared with the solar ones. These results verify the above claim with the significant positive linear correlation found only for the TEC which is expected considering the known strong influence of solar activity on the ionosphere.

**Table 19.1:** The various datasets that have been analysed in this Part IV with their corresponding chapter reference and the acquired time-period.

| Dataset | Chapter | Time period |
|---|---|---|
| TECU | 15 | 01/01/1995 - 30/12/2012 |
| Strato. Temp. | 16 | 01/01/1979 - 31/08/2018 |
| EQs | 17 | 01/01/1900 - 31/01/2021 |
| Melanoma | 18 | 04/01/1982 - 31/12/2014 |

**Table 19.2:** Pearson's correlation coefficients matrix for the four terrestrial observations compared with the various concurrent manifestations of solar activity. The coefficients showing a positive linear correlation are in blue while the ones showing negative correlation are in red. The statistically significant results on the 0.05 level are marked with *. The maximum available periods are used for these calculations.

| | TECU | Strat. Temp. | EQs | Melanoma |
|---|---|---|---|---|
| M-Flares | 0.33* | -0.02* | -0.03* | -0.10* |
| X-Flares | 0.10* | $4.78 \times 10^{-4}$ | -0.005 | -0.03* |
| EUV | 0.92* | 0.02 | -0.02 | -0.21* |
| Sunspots | 0.85* | 0.01 | -0.02* | -0.16* |
| F10.7 | 0.90* | $-8.12 \times 10^{-4}$ | -0.04* | -0.14* |
| $\Delta R$ | -0.64* | -0.14* | 0.05* | 0.22* |
| FIP | 0.55* | -0.11* | -0.04 | 0.21* |
| Ly-$\alpha$ | 0.92* | 0.01 | -0.08* | -0.12* |

Then, in Tab. 19.3 the correlation analysis has been performed for the various terrestrial observations with each other. In this analysis there was no high degree of correlation found for all comparisons, pointing to different invisible matter components affecting these diverse phenomena.

**Table 19.3:** Pearson's correlation matrix for the four terrestrial observations. The coefficients showing a positive linear correlation are in blue while the ones showing negative correlation are in red. The statistically significant results on the 0.05 level are marked with *. The maximum available periods are used for these calculations.

| | TECU | Strat. Temp. | EQs | Melanoma |
|---|---|---|---|---|
| TECU | 1* | -0.04* | -0.04* | -0.01 |
| Strat. Temp. | -0.04* | 1* | -0.02* | -0.18* |
| EQs | -0.04* | -0.02* | 1* | 0.03* |
| Melanoma | -0.01 | -0.18* | 0.03* | 1* |

Interestingly, similarly to what has been found from the Lomb periodograms in solar observations, in all the terrestrial observations (Fig. 19.1) we also observe some narrow peaks around 27.32 d. This period which matches Moon's sidereal month i.e. fixed to remote stars. These peaks are also stronger than the ones around 29.53 d which is related to Moon's synodic period i.e. fixed to the Sun. This observation with narrow peaks can not be explained by solar differential rotation, which ranges from 25 d to 36 d between the equator and the poles [269, 270]. Thus, it strengthens the aforementioned claim of an additional exo-solar influence. This can be explained only by the gravitational focusing scenario of this work, with Moon's position modulating the flux of the assumed incoming stream(s). After all the Moon can itself focus particles with $\sim 10^{-4}$ c onto the Earth with a flux amplification of up to $10^4$. The exact values of the periodicities observed from the periodogram analysis are 27.33 d ± 0.04 d, 27.31 d ± 0.02 d, 27.32 d ± 0.06 d and 27.32 d ± 0.02 d for TECU, stratospheric temperatures, EQs and melanoma cases respectively.

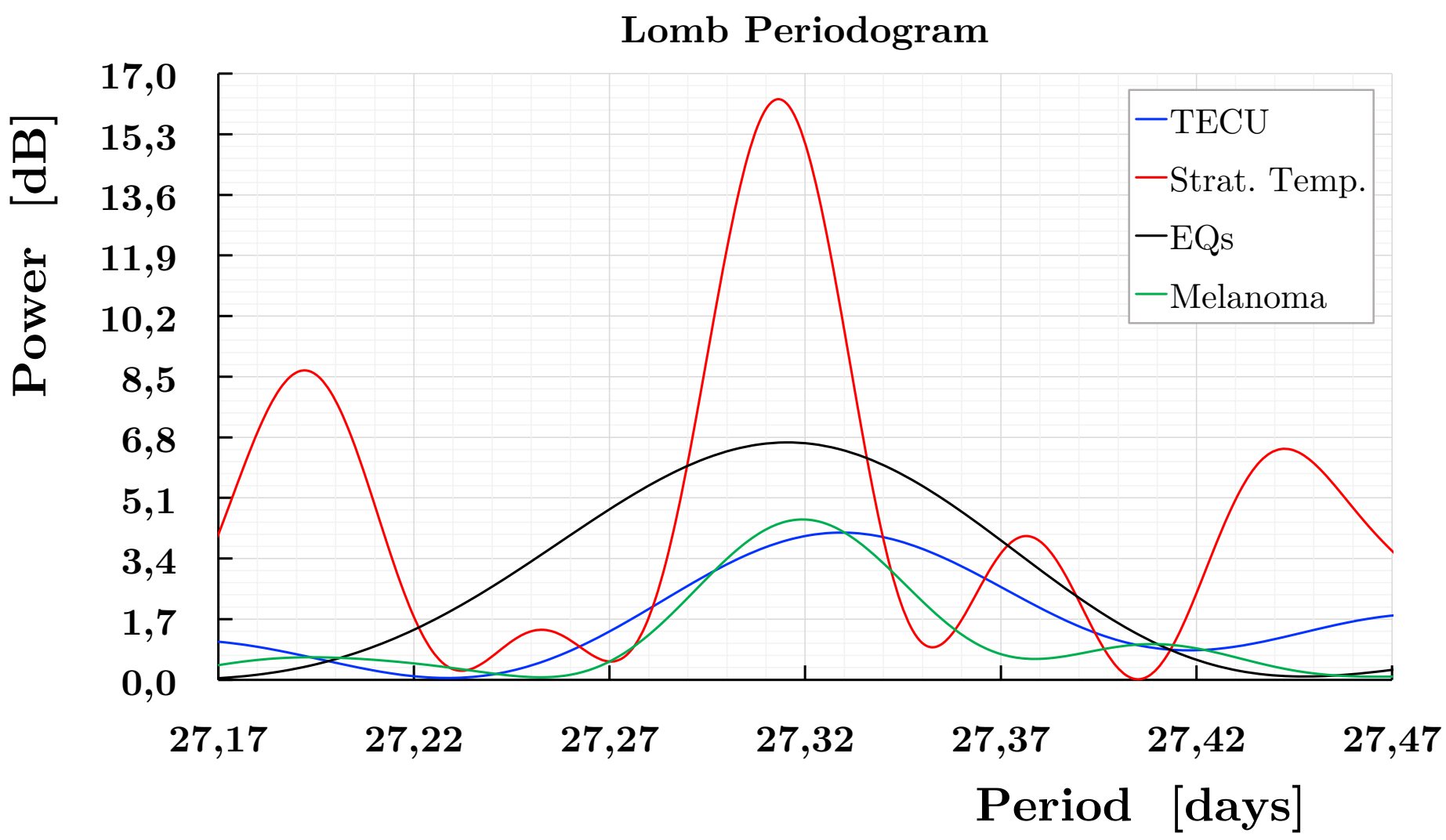

**Figure 19.1:** Fourier spectra for the four terrestrial observations around 27.32 d corresponding to Moon's sidereal month.

This work is suggestive for further investigations in the upper and lower Earth's atmosphere. This might provide additional information on the preferred direction(s) of the putative stream(s) or clusters and the strength of their interaction. Furthermore, based on the results of this work, an interesting analysis that could provide novel and valuable information can be performed on extraterrestrial seismology including Sunquakes but also Moonquakes [396, 397] and even Marsquakes [398]. Additionally, biomedical observations such as that of melanoma occurrence are also suggestive for new interdisciplinary approaches in physics and biomedical research alike and can provide an independent confirmation on the existence of invisible streams.

However, the ultimate demonstration of causality can be found depending upon actual direct detection of the assumed streams using for example micromegas-type detectors, along with an experimental demonstration of associated biomedical effects (see Publications E.4 and E.8). As an example, dark photons in the stratospheric altitude could be probed for direct detection since they can mix with real photos with the same total energy (see Fig. 19.2). This idea of direct detection, stemming from the planetary relationship of the stratospheric temperature excursions, could be extended to several candidates from the dark sector by using a suitable multi-purpose detector (see Publication E.7). This way, the potential experimental setup is transferred from underground or ground to the upper atmosphere. The reason is that a potential signature of such particles at the Earth's surface might be shielded by the atmosphere from resonant absorption of molecules or large cross-sections making the ground and underground experiments quasi-blind.

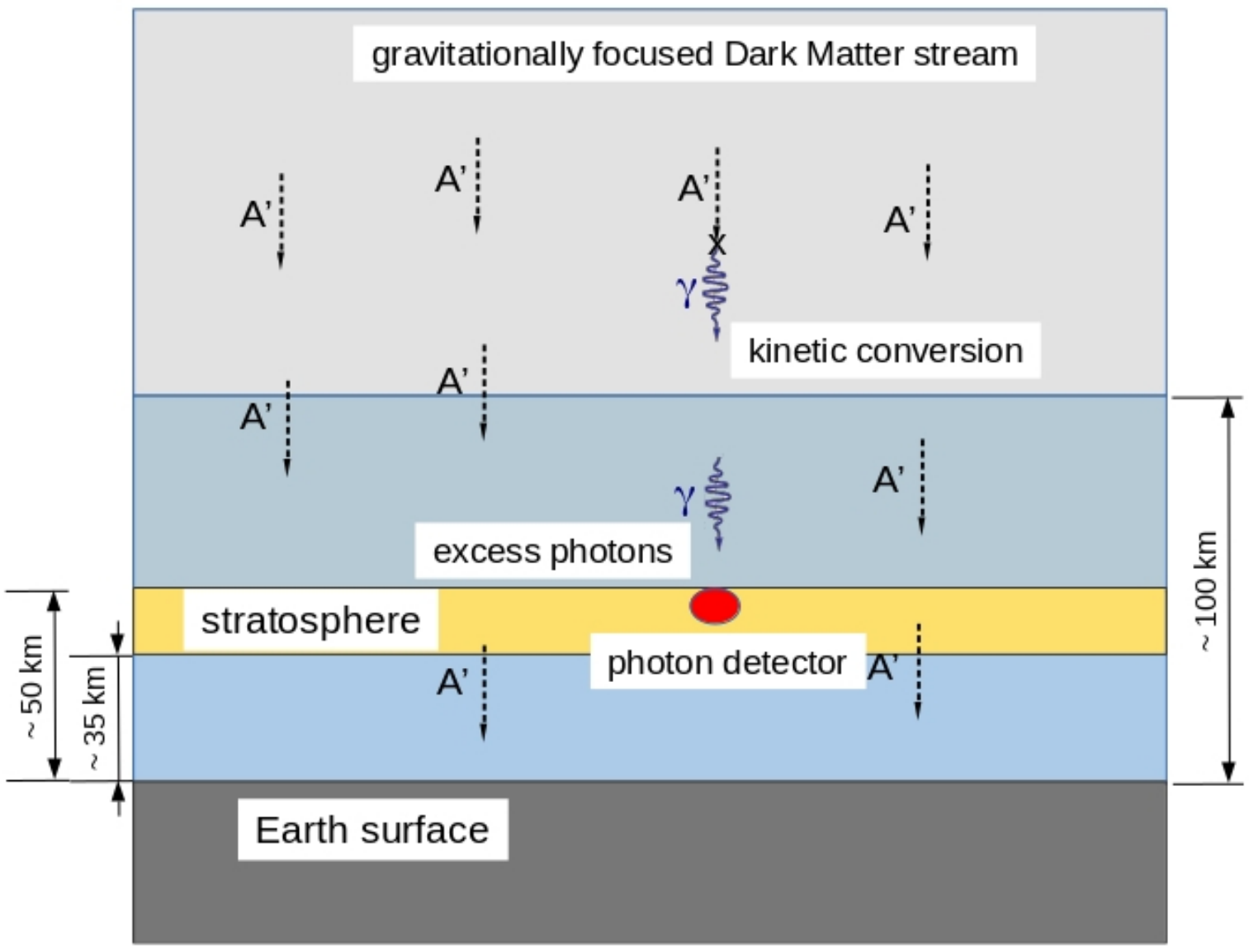

**Figure 19.2:** Schematic illustration (not to scale) of a direct detection scheme for incident dark photons in the stratosphere (see Publication E.7).

Such a novel detection strategy, being complementary to underground and ground experiments, together with the newly created analysis techniques described in this work lay the foundations for the exploration of unconventional signatures from the dark universe while setting the framework also for direct detection, which has not yet provided any signature for all existing experiments. Finally, it is worth stressing that current underground experiments searching for DM such as DArk MAtter (DAMA) [399], XENON1T [400] and Borexino [359], should re-analyse their data searching for possible planetary relationship (see also Publication E.12). Such a streaming-oriented scrutiny has the potential to unravel overlooked signatures from the dark sector.

# Part V:

# Direct Dark Matter Search

*"In theory there is no difference between theory and practice. In practice there is."*

—Yogi Berra (1925–2015),
*American Athlete*

# 20

# INTRODUCTION

The theoretically and cosmologically well-motivated DM has not been discovered yet by direct DM searches despite the extensive experimental efforts. Therefore, based on the theoretical framework presented on Part II, but also on derived potential signatures for the existence of streaming invisible matter through the solar and terrestrial observations, a new experimental approach for DM detection can be formulated. In this part of the thesis, we will focus on DM axions and ground-based detectors. Although axions are weakly interacting, they could constitute a component of a low-speed invisible stream and therefore they can also experience similar gravitational lensing effects.

For the specific case of axions, the current search strategy adopted by all DM experiments focuses on an assumed isotropic halo distribution with a broad velocity distribution around 240 km/s and an average density of $\sim 0.45\,\mathrm{GeV/cm^3}$ [401,402]. This choice could have been the reason for the so far fruitless detection efforts (E.12). However, assuming DM axion streams propagating near the ecliptic plane of our solar system, the experimental parameters could be adjusted accordingly with the detectors being sensitive to both halo and streaming DM axions. Low speed ($0.01\,c$ to $0.2\,c$) streams aligned in the Sun $\rightarrow$ Earth direction can be gravitationally focused by the Sun towards the Earth with the resulting flux enhancement being up to the level of $\sim 10^{8\pm3}$. Except the Sun, the Earth's intrinsic mass distribution can also result to a flux amplification of up to $10^9$ via self-focusing due to the Earth's intrinsic mass distribution (E.9). Finally, as seen in Fig. 20.1, gravitational effects by other solar system bodies such as the Moon can result to density enhancements on the level of $10^5$ for axions and 100 for WIMPs [225]. Therefore, even a temporally lensed small DM stream could still surpass by a lot the local DM density which would give a temporally large DM flux exposure on an axion haloscope such as the CERN Axion Solar Telescope (CAST)-Center for Axion and Precision Physics (CAPP) sub-detector (see Publications E.6, E.11, E.22, E.24, and E.26).

Since burst-like axion flux enhancements happen for as long as the stream alignment takes place, three experimental requirements are imposed for an axion haloscope such as CAST-CAPP to fully exploit them.

1. The covered frequency range has to be widened to the maximum, since the mass of the

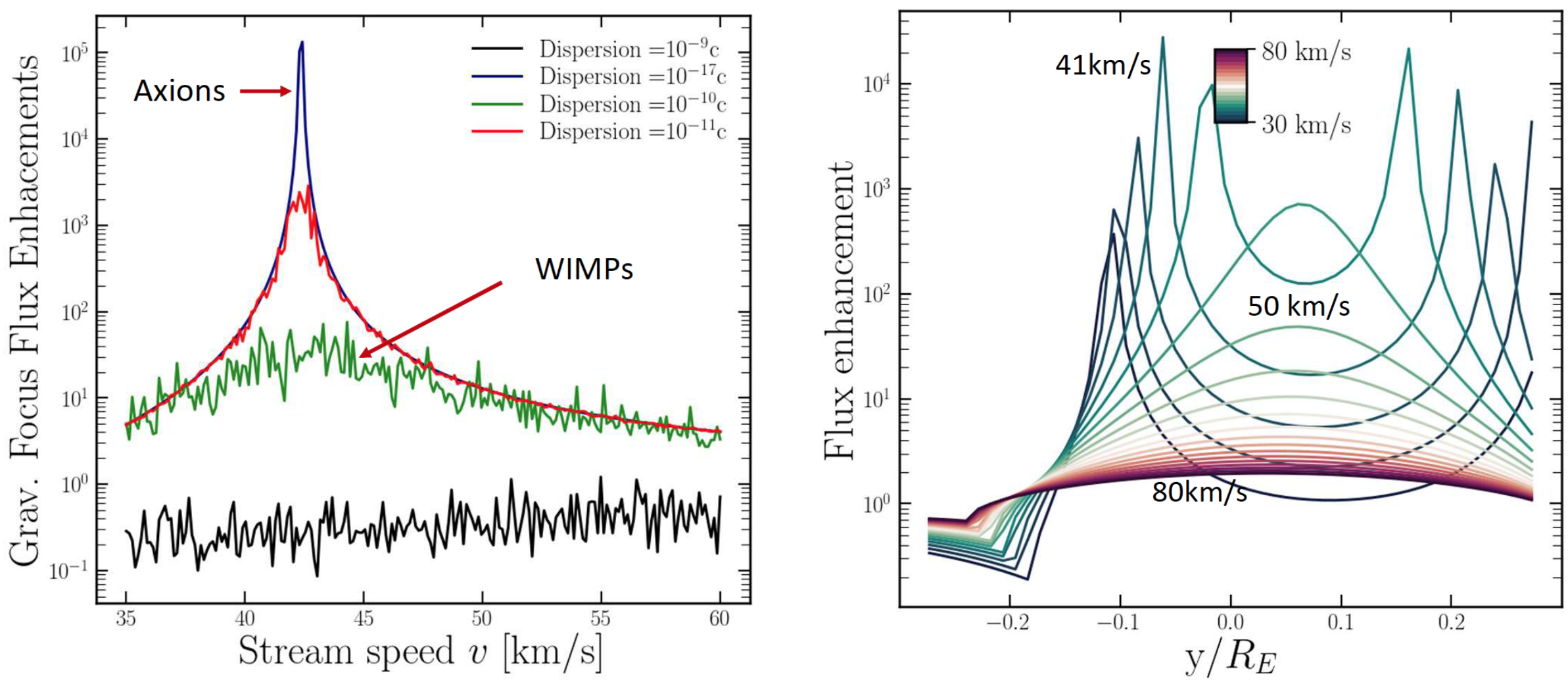

**(a)** Flux enhancement as a function of the velocity of the stream particles, including dispersion.

**(b)** Flux enhancement in various positions for a range of speeds between 30 km/s to 80 km/s.

**Figure 20.1:** Gravitational effects by the Moon towards the Earth on a fine-grained stream (41).

axions is unknown.

2. The scanning time should be shortened as much as possible.

3. The duty cycle of the fast scanning axion antenna should be maximised since the exact time of the alignment is still unknown.

These features can allow searching for gravitationally focused short axion transients, though with a decreased detection sensitivity since the integrated time per frequency step is smaller. This decrease is however mitigated by the large flux enhancement due to the putative gravitational lensing effect. As an example, in Fig. 20.2 the gravitational focusing by the Moon towards the Earth is shown with the different amplifications as a function of the transit time. Such a signal can manifest itself as a $\sim 20$ min transient in a DM axion detector.

It is noted that even though the timing of the alignment of streams with the Earth is unknown, interesting dates could be derived such as the known alignment of GC $\rightarrow$ Sun $\rightarrow$ Earth around the 18$^{\text{th}}$ of December, that is repeated annually. For more details see Publications E.10 and E.16.

This new detection scheme, when implemented by a network of haloscopes instead of a single haloscope could fulfil the above mentioned requirements and search also for transient events from streaming DM axions. In the next chapters, the axion theory as well as the implementation of the aforementioned technique in CAST-CAPP sub-detector will be presented, along with the various technical details and the analysis results (see also Publications E.6, E.22, E.24, and E.25).

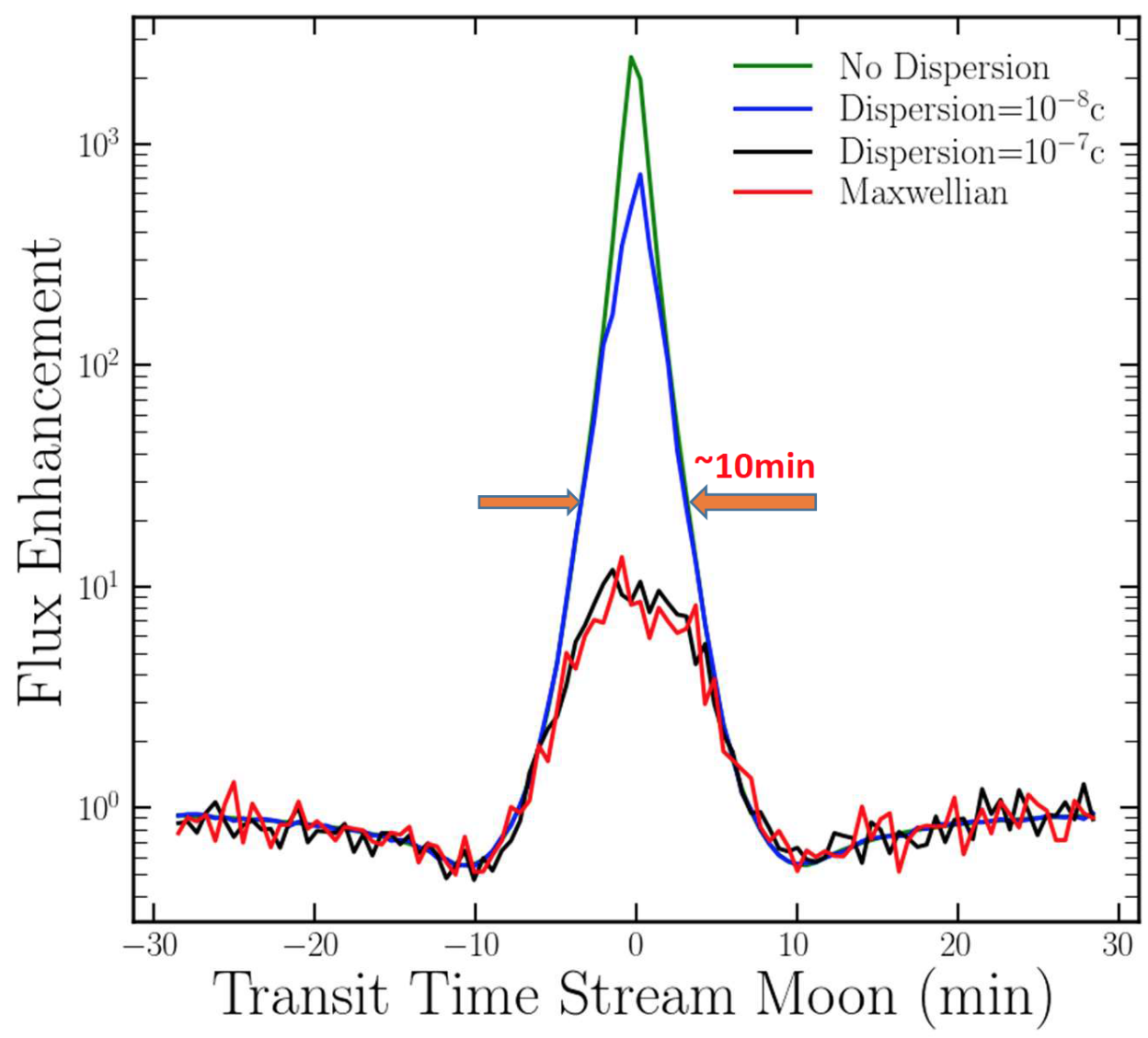

**Figure 20.2:** Flux enhancement by the Moon as function of the transit time for different dispersion values and for a stream with a selected $\upsilon = 41\,\text{km/s}$ where the flux amplification is maximised for axions (41).

# 21

# AXIONS

## 21.1 The strong CP problem

### 21.1.1 QCD Lagrantzian

In the QCD Lagrantzian where there is an effective contribution of the form [403]:

$$\mathcal{L}_{QCD,\bar{\theta}} = \frac{g_s^2}{32\pi^2}\bar{\theta}G^a_{\mu\nu}\tilde{G}^{\mu\nu,\alpha} \tag{21.1}$$

where: $g_s$ : the coupling constant of the strong interaction,

$G^a_{\mu\nu}$ : the gluon field strength tensor,

$G^{\mu\nu,\alpha}$ : its dual,

$\bar{\theta}$ : a phase which characterises the strength of the CP violation.

$\bar{\theta}$ has contributions from both the strong and the weak sectors:

$$\bar{\theta} = \theta + Arg \det M \tag{21.2}$$

where $M$ is the quark mass matrix and $\theta$ is a value that characterises the QCD vacuum. This term violates Parity and Time reversal invariance but conserves Charge conjugation invariance, so it violates CP [404]. This is called the *Strong CP problem*.

### 21.1.2 Electric Dipole Moment of the neutron

The neutron is a spin-1/2 particle with a non zero magnetic dipole moment $\mu$. From the Wigner-Eckart theorem in quantum mechanics, the expectation value of any vector operator, like Electric Dipole Moment (EDM) ($d$) points along the spin quantisation direction. Therefore, if the neutron EDM exists, it would be either parallel or antiparallel to the magnetic moment ($\mu$) [405]. As a consequence, we can expect that CP-violating effects in the strong interactions will give rise to baryon EDMs. The neutron EDM can be measured more precisely than other baryonic EDMs, since neutrons are electrically neutral and long-lived and therefore are the most promising candidates for such a search.

The dependence of the magnitude of the neutron EDM with the $\theta$ angle has been calculated to be [406, 407]:

$$d_n \simeq e\bar{\theta}\frac{m_q}{{m_n}^2} \sim \left(6 \times 10^{-17}\right) \cdot \bar{\theta}\ \mathrm{e\,cm} \tag{21.3}$$

with

$$m_q = \frac{m_u m_d}{m_u + m_d} \tag{21.4}$$

where: $m_u$ : the up quark mass,

$m_d$ : the down quark mass.

To date, no neutron EDM has been observed [408] with the limit being at $d_n = (0.0 \pm 1.1_{\mathrm{stat}} \pm 0.2_{\mathrm{sys}}) \times 10^{-26}\,\mathrm{e\,cm}$ [409]. This implies that the angle $\bar{\theta}$ is very small and unexpectedly close to zero:

$$\left|\bar{\theta}\right| < 1.8 \times 10^{-11} \tag{21.5}$$

This is known as the *Strong CP problem* and points to the question of why this $\bar{\theta}$ angle, which is a free parameter is so small. This is basically a fine-tuning problem since $\bar{\theta}$ is a phase that in principle could acquire any value between 0 and $2\pi$ as it is produced by adding up QCD contributions of different nature. Therefore, it is considered unnatural that the value is so small.

### 21.1.3 New chiral symmetry and axions

The most cogent solution to the strong CP problem is through the introduction of a new global chiral $U(1)$ symmetry, known as Peccei-Quinn Symmetry $U(1)_{\mathrm{PQ}}$, which is necessarily spontaneously broken. Its introduction to the theory replaces the static CP-violating angle $\theta$ with a dynamical CP- conserving field- the axion field [410, 411]. As a result, under a $U(1)_{\mathrm{PQ}}$ transformation, the axion field $a(x)$ translates:

$$a(x) \underset{U(1)_{\mathrm{PQ}}}{\rightarrow} a(x) + a f_a \tag{21.6}$$

where $f_a$ is the order parameter associated with the breaking of the $U(1)_{\mathrm{PQ}}$. This way the $\bar{\theta}$-term becomes a dynamic variable that affects the value of the QCD potential [410]. The potential is minimised for $\bar{\theta} = 0$ (see Fig. 21.1), which explains why CP violation is suppressed in the strong interactions.

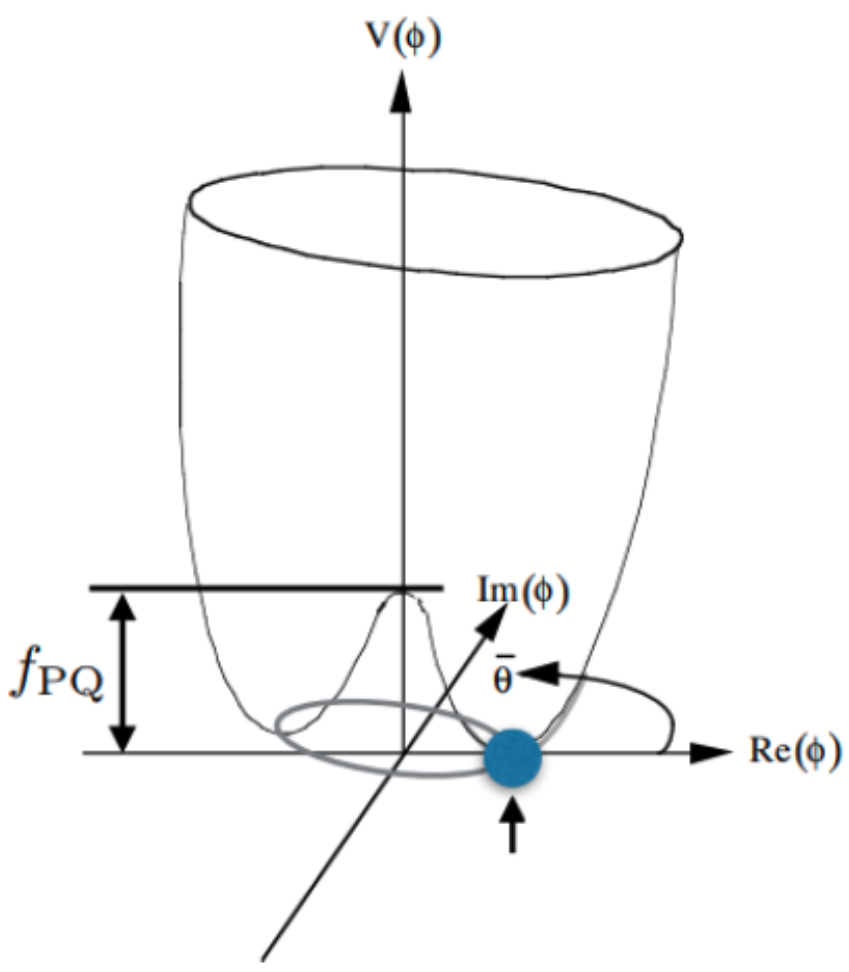

**Figure 21.1:** Effective potential for the axion field (42).

Soon after the Peccei-Quinn dynamical solution to the strong CP problem, Steven Weinberg [412] and Frank Wilczek [413] realised that as a result of the of the broken $U(1)_{\mathrm{PQ}}$ symmetry by instanton effects, a neutral pseudoscalar Nambu-Goldstone boson $a(x)$ is created which was named *axion*.

## 21.2 Axion theory and phenomenology

### 21.2.1 Axion properties

#### 21.2.1.1 Mass

If $U(1)_{\mathrm{PQ}}$ was an exact spontaneously broken symmetry the axion $a\,(x)$ would be a massless generic Goldstone boson. But the chiral anomaly generates an explicit $U(1)_{\mathrm{PQ}}$-breaking term:

$$\mathcal{L}_{agg} = \frac{{g_\gamma}^2}{32\pi^2 f_a} a F_\alpha^{\mu\nu} \tilde{F}_{\mu\nu,\alpha} \tag{21.7}$$

where: $g_\gamma$ : model dependent coefficient of order one,

$f_a$ : decay constant of the axion.

Therefore, a non vanishing axion potential $V_a\,(a/f_a) = \mathcal{L}_{agg}$ implies that the axion gets a nonzero mass which is given by:

$${m_a}^2 = \left.\frac{\partial^2 V_a}{\partial a^2}\right|_{a=-f_a\bar{\theta}} \tag{21.8}$$

The calculation of Eq. 21.8 is nontrivial as the low-temperature QCD is strongly coupled (a good approximation is calculated in [414]), but a dimensional analysis can give:

$${m_a}^2 \sim \frac{\Lambda_{\mathrm{QCD}}^4}{{f_a}^2} \tag{21.9}$$

where: $\Lambda_{\mathrm{QCD}}$ : the zero-temperature QCD topological susceptibility.

From the interactions of axions with the SM fields, we can derive an even more precise expression of the axion mass:

$${m_a}^2 = \frac{m_\pi^2 f_\pi^2}{{f_a}^2} \frac{m_u m_d}{(m_u + m_d)^2} \tag{21.10}$$

where: $m_\pi$ : mass of the pion,

$f_\pi$ : decay constant of the pion.

Eq. 21.10 is model-independent the same way as Eq. 21.7. What is important to notice is that the mass of the axion is inversely proportional to the coupling constant, which means that a larger coupling constant leads to a smaller mass. If $z = m_u/m_d$ is the ratio of the up and down quark masses, then Eq. 21.10 with the help of Eq. 21.9 can become:

$$m_a = \frac{m_\pi f_\pi}{f_a} \frac{\sqrt{z}}{1+z} = \frac{{\Lambda_{\mathrm{QCD}}}^2}{f_a} \tag{21.11}$$

#### 21.2.1.2 Couplings

The strength of the axion coupling to matter and radiation is characterised by the coupling constants $g_{a\gamma\gamma}$, $g_{aee}$, etc. The axion can couple to photons, electrons and nucleons for $m_a \leq 2m_e \approx 1\,\text{MeV}$. However, the coupling to photons dominates. The Lagrangian for this procedure is:

$$\mathcal{L}_{a\gamma\gamma} = g_{a\gamma\gamma} a\,(x)\,\vec{E} \cdot \vec{B} \tag{21.12}$$

with the coupling constant, using also Eq. 21.11, being:

$$g_{a\gamma\gamma} = \frac{g_\gamma a}{\pi f_a} = \frac{g_\gamma a}{\pi {\Lambda_{\text{QCD}}}^2} m_a \tag{21.13}$$

More specifically, the dimensionless model-dependent coefficient $g_\gamma$ is given by [67]:

$$g_\gamma = \frac{1}{2}\left[\frac{E}{N} - \frac{2\,(4+z)}{3\,(1+z)}\right] \tag{21.14}$$

where: $z$ : the ratio of the up and down quark masses,

$N$ : the axon colour anomaly,

$E$ : the axion electromagnetic anomaly.

From the latest direct measurements of $m_u$ and $m_d$ the $z$-value can be calculated to $z = m_u/m_d = 0.47^{+0.06}_{-0.07}$ [158]. However, historically axion papers assume $z = 0.56$ (obtained from chiral perturbation theory) which corresponds to $\Lambda_{\text{QCD}} = 77.6\,\text{MeV}$, and transforms Eq. 21.14 to:

$$g_\gamma = \frac{1}{2}\left[\frac{E}{N} - 1.95\right] \tag{21.15}$$

In Fig. 21.2 the Feynman diagram for $a \to \gamma\gamma$ can be seen. From Eq. 21.13 we infer that the requirement that axions solve the strong CP problem implies that $g_{a\gamma\gamma} \propto m_a$, where the precise value of the proportionality constant depends on the model dependent $g_\gamma$ which in turn depends on the $E/N$ value. Therefore, since a broad range of $E/N$ values is possible, this means that a *model band* for axions is available which is shown as a diagonal yellow band in exclusion plots [415–417]. However, this band still does not exhaust all the possibilities as there are also classes of QCD axion models whose photon couplings populate the entire allowed region above the yellow band thus motivating axion search efforts over a wide range of masses and couplings [418, 419].

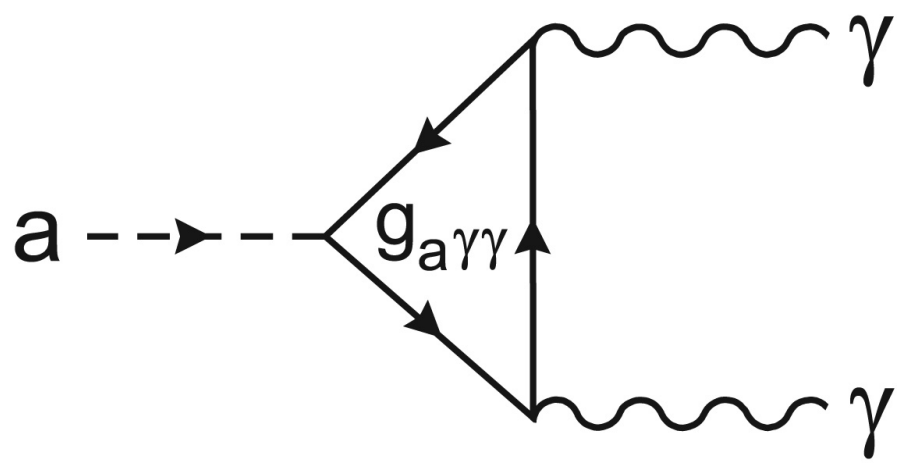

**Figure 21.2:** Fenynman diagram for the coupling of axions to two photons. Since the axion is a neutral particle it can not couple to photons at tree level. Instead this interaction must proceed through an electromagnetic or color anomaly (43).

### 21.2.2 The invisible axions

The original axion from the theory of Peccei-Quinn-Weinberg-Wilczek (PQWW) had a symmetry breaking scale $f_a$ around the weak scale:

$$\upsilon_{\text{weak}} = \left(\sqrt{2}G_F\right)^{-1/2} = 247\,\text{GeV} \tag{21.16}$$

this pointed to an axion mass $10\,\text{keV} \leq m_a \leq 1000\,\text{keV}$. Astrophysical observations from red giant evolution [420] as well as experiments looking for rare meson decays [421,422] rejected the PQWW axion. Then, new models with $f_a \gg \upsilon_{\text{weak}}$ arose in order to maintain the Peccei-Quin solution to the strong CP problem. From Eq. 21.9 it was clear that the new axion that was introduced had such a small mass that it could not be detected and therefore it was called the *invisible axion*.

#### 21.2.2.1 KSVZ model

The KSVZ model [423, 424] was the first of the invisible axion models. This model is called a *hadronic model* as the axions have bare couplings only to quarks and not to leptons. In the original KSVZ model $E/N = 0$ which from Eq. 21.15 corresponds to a value $g_\gamma \sim -0.97$. Using Eq. 21.13 and Eq. 21.15, we get a coupling limit [425]:

$$g_{a\gamma\gamma}{}^{\text{KSVZ}} \approx 0.38 \frac{m_a}{\text{GeV}^2} \tag{21.17}$$

#### 21.2.2.2 DFSZ model

The Dine, Fischler, Srednicki and Zhitnitski (DFSZ) model [426, 427] was developed several years after KSVZ. DFSZ is a leptonic model meaning that axions in this model have bare couplings to leptons as well as quarks. The axion-to-matter couplings are suppressed which results in a decreased value for the coupling constant comparing to the KSVZ model for a given $f_a$. In this model the $g_\gamma$ from Eq. 21.12 takes a value $g_\gamma \sim 0.36$ since $E/N = 8/3$ (suggested by GUTs). Using Eq. 21.13 and Eq. 21.15, we get a coupling limit which is weaker

than Eq. 21.17 [425]:

$$g_{a\gamma\gamma}{}^{\mathrm{DFSZ}} \approx 0.14 \frac{m_a}{\mathrm{GeV}^2} \tag{21.18}$$

### 21.2.3 Dark Matter axions

#### 21.2.3.1 Decay time

Despite having a very small mass [158]:

$$m_a \simeq 5.7\,\mathrm{meV} \left( \frac{10^9\,\mathrm{GeV}}{f_a} \right) \tag{21.19}$$

axions are effectively collisionless and can also be non-relativistic. Cold populations can be produced out of equilibrium and fill our universe. These, along with the fact that axions are supposed to be "invisible", makes them the perfect particle candidates for DM. More precisely since $g_{a\gamma\gamma} \propto 1/f_a$ and $m_a \propto 1/f_a$ this means that $g_{a\gamma\gamma} \propto m_a$ i.e. a very light axion has a very weak coupling to matter. This fact makes the axions difficult to detect but rather stable as the lifetime of the $a \to \gamma\gamma$ process, which is the dominating one, is [428]:

$$\tau_{a\gamma\gamma} = \frac{64\pi^3}{a^2 {g_\gamma}^2} \frac{{f_a}^2}{{m_a}^3} \tag{21.20}$$

Using Eq. 21.15 we get:

$$\tau_{a\gamma\gamma} \simeq 10^{24}\,\mathrm{s} \left[ \frac{1}{(E/N - 1.95)} \right]^2 \left( \frac{\mathrm{eV}}{m_a} \right)^5 \tag{21.21}$$

which for $E/N = 0$ (from original KSVZ model), becomes [158]:

$$\tau_{a\gamma\gamma} \sim 10^{24}\,\mathrm{s} \left( \frac{\mathrm{eV}}{m_a} \right)^5 \tag{21.22}$$

This means that the axion decay for $m_a \sim 1\,\mathrm{eV}$ would be $\tau_{a\gamma\gamma} \sim 10^{24}\,\mathrm{s}$, even bigger than the lifetime of the universe ($\sim 10^{17}\,\mathrm{s}$), and therefore primordial axions can still be in the universe. This property makes axions very promising DM candidates.

#### 21.2.3.2 Local density and wavelength

Assuming a local density of $\rho_a = \rho_{\mathrm{DM}} = 0.45\,\mathrm{GeV/cm}^3$ [401, 402] and a rest frame viral velocity $\langle v^2 \rangle^{1/2} = 270\,\mathrm{km/sec}$, then the axion number density in "empty" space within our galaxy is quite large $n_a = \rho_a / m_a \sim 2 \times 10^{13}\,\mathrm{cm}^{-3}$ for $m_a \sim 20\,\mu\mathrm{eV}$ [405]. This also means that the CDM relic axion has a big de Broglie wavelength which depends on the axion mass:

$$\lambda_a = \frac{2\pi h}{p_a} = \frac{\hbar}{m_a v_a} \sim 3 \times 10^8\,\mathrm{eV}^{-1} \sim 70\,\mathrm{m} \tag{21.23}$$

## 21.3 Axion searches

As pseudo-scalars, axions can be produced by the interaction of two photons one of which can be virtual, $\gamma + \gamma^* \to a$. This process is known as the *Primakoff effect* or *Sikivie effect* [429]. This means that photons and axions may mix in the presence of an external electromagnetic field through the Lagrangian of Eq. 21.12. This also means that the Lagrangian permits the conversion of an axion into a single real photon in the presence of an extra electromagnetic field. This is called the *inverse Primakoff effect*. In all axion searches based on the Primakoff effect, $\vec{E}$ represents the electric field of the real photon and $\vec{B}$ the external magnetic field, as it is much easier to produce and support a static magnetic field than the equivalent electric field (see Fig. 21.3). Finally, an important characteristic is that, in the case of relativistic axions, coherent axion-photon mixing in long magnetic field results in significant conversion probability even for weakly coupled axions [430].

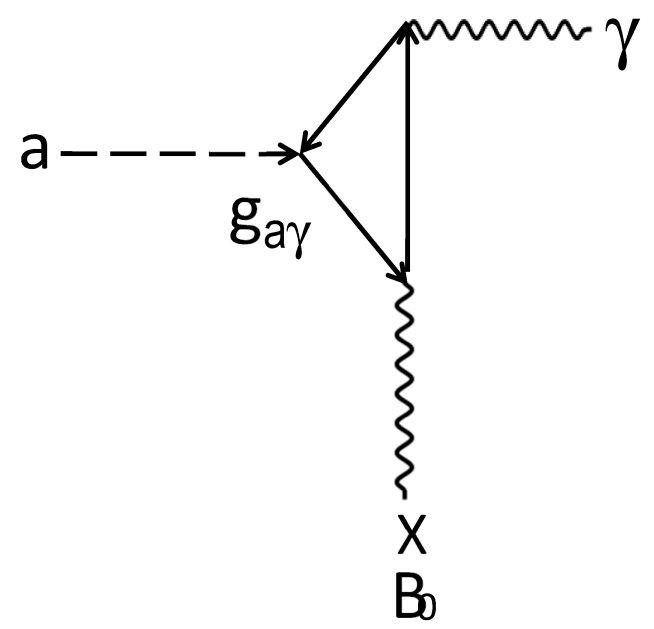

**Figure 21.3:** Inverse Primakoff effect in a static magnetic field ($B_0$) (44).

As next two of the most important approaches in axion search will be presented. For a more thorough and complete review the reader can refer to [158].

### 21.3.1 Solar searches

Axions can be produced in the Sun's nuclear burning core and they would be dominated by the Primakoff process $\gamma + Ze \to a + Ze$ [431, 432]. For KSVZ axions the integrated solar flux at the Earth would be [433]:

$$F_a = 7.4 \times 10^{11} {m_a}^2\,[\mathrm{eV}]\ \ \mathrm{cm}^{-2}\,\mathrm{s}^{-1} \tag{21.24}$$

with a thermal spectrum of mean energy $\sim 4.2\,\mathrm{keV}$.

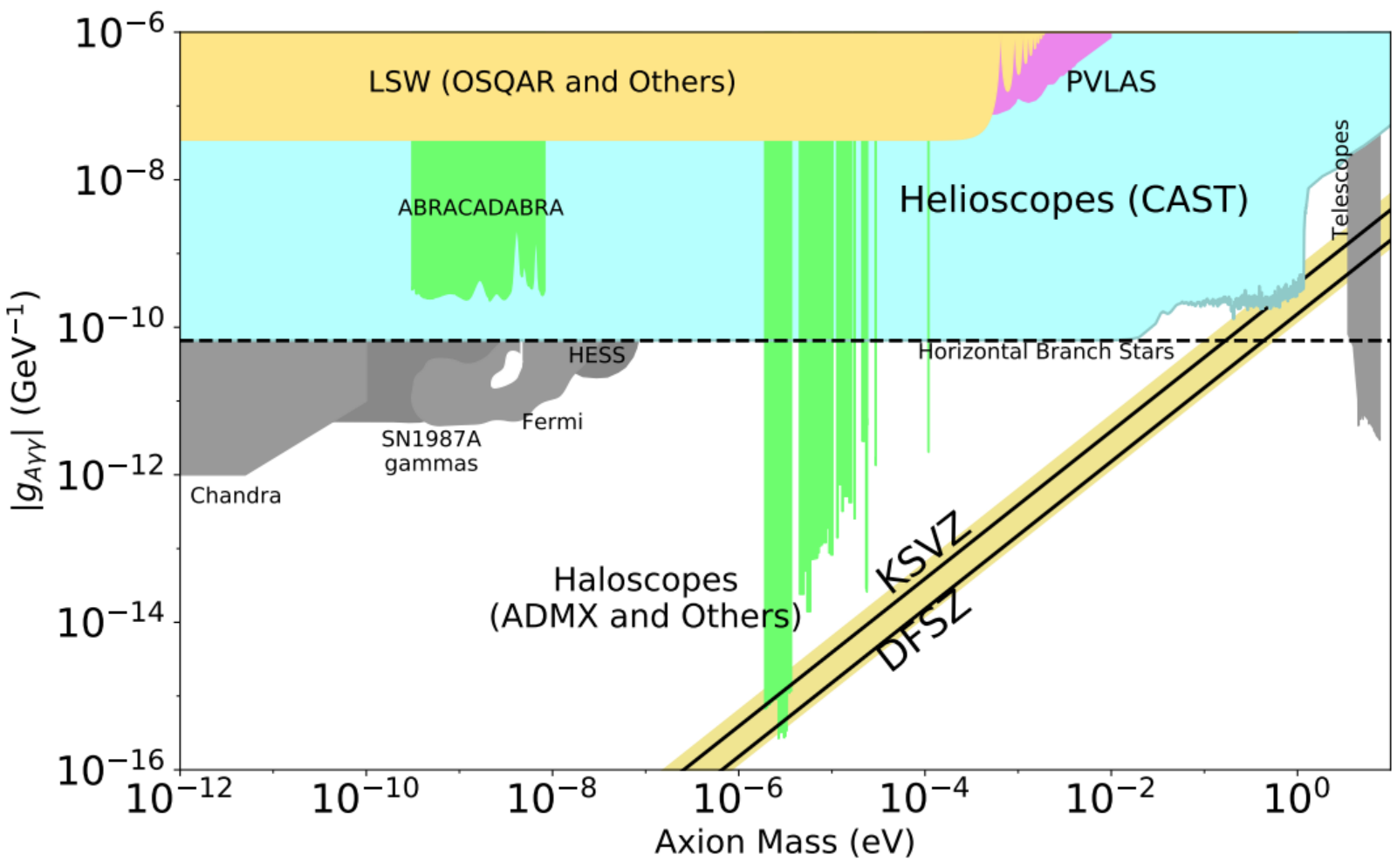

**Figure 21.4:** Exclusion plot for ALPs from various techniques and experiments (45).

For relativistic axions the conversion probability back to photons of the same energy in a uniform magnetic field is given by [433]:

$$P_{a\to\gamma} = \frac{1}{4}\left(g_{a\gamma\gamma}\vec{B}L\right)^2 |F(q)|^2 \tag{21.25}$$

where: $\vec{B}$ : the strength of the magnetic field,

$L$ : the length of the magnetic field,

$F(q)$ : form factor of the magnetic field with respect to the momentum mismatch between the massive axion and the massless photon of the same energy.

with

$$F(q) \equiv \int dx e^{iqx} \frac{B(x)}{B_0 L} \tag{21.26}$$

and

$$q = k_a - k_\gamma = \left(\omega^2 - {m_a}^2\right)^{1/2} - \omega \sim \frac{{m_a}^2}{2\omega} \tag{21.27}$$

In the limit $ql \ll 2\pi$, $|F(q)|$ is unity, whereas for $ql > 2\pi$ where the axions are no longer sufficiently relativistic to stay in phase with the photo for maximum mixing, it oscillates and falls off rapidly.

The basic layout of an axion *helioscope* requires a powerful magnet coupled to one or more X-ray detectors. When the magnet is aligned with the Sun, an excess of X-rays at the exit of the magnet is expected over the background measured at non-alignment periods (see Fig. 21.5). Therefore helioscopes focus on axions produced from blackbody photons in the solar core via the Primakoff effect. Since the interaction of these axions with ordinary matter is very weak, they can escape the solar interior, and stream to Earth.

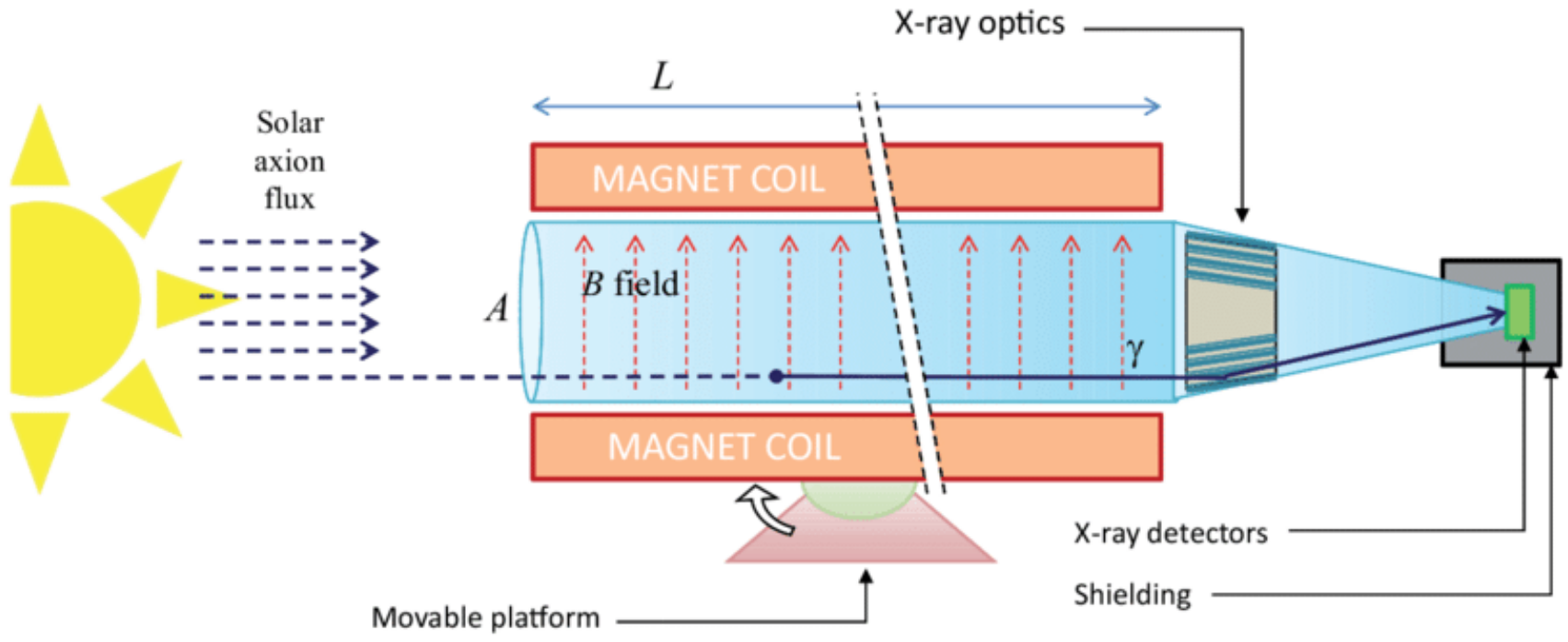

**Figure 21.5:** Basic setup of an axion helioscope converting solar axions in a strong laboratory magnetic field of cross-sectional area $A$ and length $L$ via the inverse Primakoff effect. The putative axion signal is then focused on the detector plane by X-ray optics (46).

#### 21.3.1.1 CAST experiment

The best limit so far comes from CAST experiment (see Fig. 21.6) which uses an LHC dipole magnet as the basis for an axion helioscope and follows the Sun for 1.5 h during sunrise and sunset during the whole year. The latest limit yields $g_{a\gamma\gamma} < 0.66 \times 10^{-10}\,\text{GeV}^{-1}$, valid for masses $m_a \leq 0.02\,\text{eV}$ [3], and is in on the same order as the most restrictive astrophysical bounds seen in Fig. 21.4.

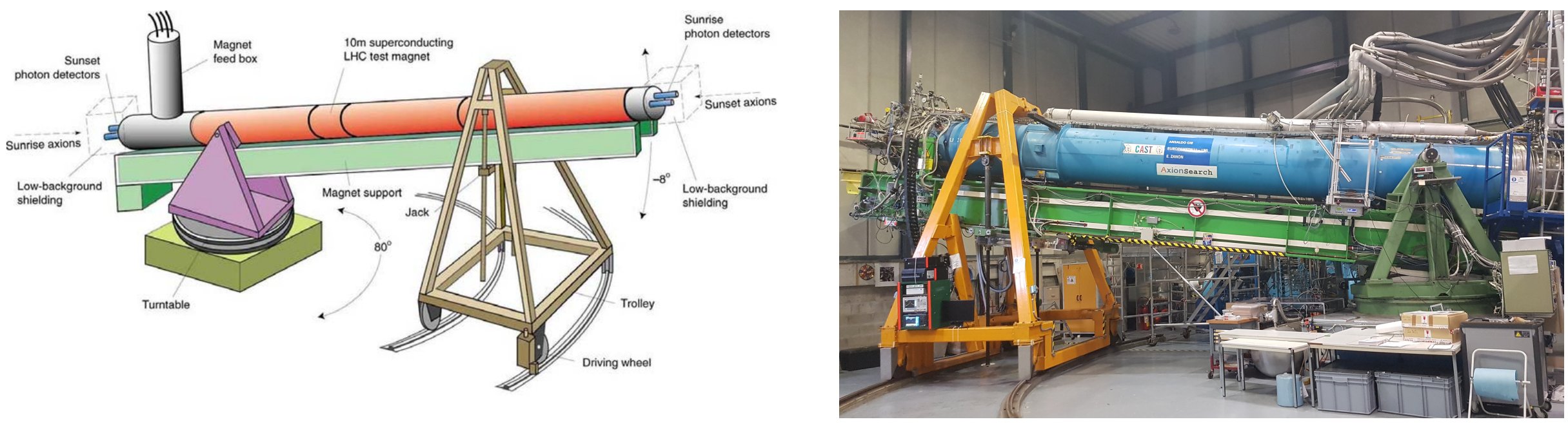

**(a)** Simplified layout of CAST experiment at CERN (72).

**(b)** Picture of CAST.

**Figure 21.6:** The CAST experiment at CERN.

CAST experiment also made use of the fact that the transition rate is the largest when the axion and photon amplitudes are coherent throughout the detection volume. To avoid destructive axion-photon interference, the length $L$ of the magnet should be:

$$L < \frac{2\pi\omega\hbar^2}{{m_a}^2 c^3} \tag{21.28}$$

where $\omega$ the angular frequency of the photon. For $L = 10\,\text{m}$ the coherence condition yields $m_a \leq 0.02\,\text{eV}$. Therefore, for massive axions of higher masses, the transition can be enhanced by filling the conversion region with an appropriate gas to give the photons an effective mass $m_\gamma$ which can be made to match $m_a$. In CAST, $^3He$ and $^4He$ were used. In general, photons do not follow the dispersion relation:

$$k^2 = \omega^2 - m_\gamma^2 \tag{21.29}$$

relevant for massive particles. However, for X-rays, when the energy of the photons is above all resonances, the $k^2 = \omega^2 - \omega_{pl}^2$ is a good approximation. In this case, the plasma frequency

$$\omega_{pl} = (4\pi a N_e/m_e)^{1/2} \equiv m_\gamma \tag{21.30}$$

where: $N_e$ : the number density of bound and free electrons,

endows the X-ray photon with an effective mass and therefore the full coherence of the axion and photon states can be restored and achieve the maximum theoretical conversion probability [434].

An important aspect of CAST experiment is that consists of four different detectors each one placed at the end points of the two magnet bores providing CAST some powerful tools to search for the dark sector. One of them is an X-ray photon detector searching for solar axions [435, 436]. Then, there are two microwave cavity detectors (see Sect. 21.3.2) CAST-CAPP which we will focus on Chap. 22 and 23 (see also Publications E.10, and E.16) and CAST-Relic Axion Detector Experimental Setup (RADES) [437, 438] searching for DM axions. Finally, there is the Kinetic Weakly Interacting Slim Particles (KWISP) detector [439, 440] which is an opto-mechanical force sensor searching for chameleons which are particle candidates for DE [441].

### 21.3.2 Microwave cavity experiments

The most important part of DM axion searches involves microwave cavity experiments through the axion haloscope technique which we will exploit in detail in Sect. 22.1. This technique assumes that axions constitute a significant fraction of the DM halo of our galaxy. These axions, as we will see can be detected through their resonant conversion into a quasi-monochromatic microwave signal in a high-Q cavity permeated by a strong magnetic field.

The first experiments which tested this technique were the Rochester-Brookhaven-Fermilab (RBF) and University of Florida (UF) experiments [2, 442, 443] using Heterojunction Field Effect Transistor (HFET) amplifiers, which managed to set limits in the range $4.5\,\mu\text{eV} < m_a < 16.3\,\mu\text{eV}$, although being $2-3$ orders of magnitude above the sensitivity for KSVZ and DFSZ axions (see Fig. 21.7).

Later, the Axion Dark Matter eXperiment (ADMX) experiment achieved sensitivity to KSVZ axions assuming they saturate the local DM density $\rho_a$ and are virialised over the mass range $1.9\,\mu\text{eV} - 3.3\,\mu\text{eV}$ [444]. The latest results from ADMX experiment, which replaced

HFET amplifiers with SQUID amplifiers, exclude DFSZ axion-photon couplings for axion masses 2.66 μeV − 3.31 μeV [445, 446] (see Fig. 21.7).

Since 2019, the CAST experiment has been converted from axion and chameleon helioscope also to an axion haloscope searching for DM axions with two detectors consisting of microwave cavities (see also Publications E.10, and E.16). These are the CAST-RADES [438, 447] and the CAST-CAPP which we will focus on Chap. 22. CAST-RADES has excluded axion-photon couplings with a constant of $g_{a\gamma\gamma} \geq 4 \times 10^{-13}\,\mathrm{GeV}^{-1}$ for axion masses 34.6738 μeV $< m_a <$ 34.6771 μeV [438]. On the other hand, as we will see in Chap. 23, CAST-CAPP recently excluded axion-photon couplings down to $g_{a\gamma\gamma} = 7 \times 10^{-14}\,\mathrm{GeV}^{-1}$ for the axion mass range 19.7 μeV $< m_a <$ 22.4 μeV. The two limits are shown in Fig. 21.7 (see also Publication E.6).

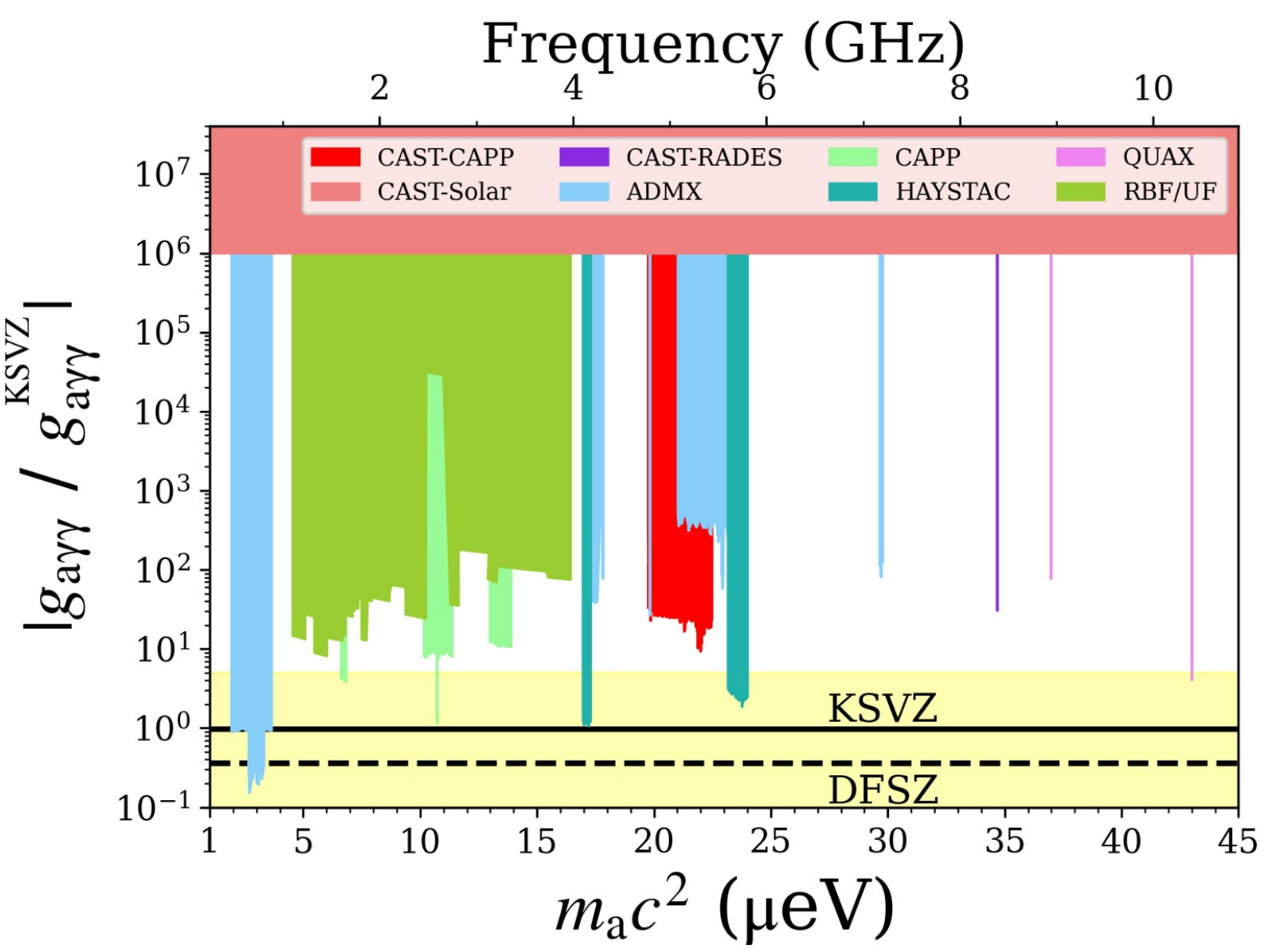

**Figure 21.7:** CAST-CAPP latest exclusion limit on the axion-photon coupling as a function of the axion mass compared to other axion searches [1–11] in the 1 μeV to 45 μeV axion mass range.

## 21.4 Streaming Dark Matter axions

The search for relic DM axions has been based so far on the assumed isotropic halo distribution of our galaxy with a broad velocity distribution around 240 km/s and an average density of $\rho_a \sim 0.45\,\mathrm{GeV/cm^3}$ [401, 402]. This choice could have been the reason why the axion has not been detected so far.

A new approach [221, 448, 449] could be established if we consider axion DM streams which propagate near the ecliptic plane of the solar system, and in particular streams which get occasionally aligned with the Sun → Earth direction. In such a configuration low speed incident particles, of $v < 10^{-2}c$, can be focused by the Sun downstream at the position of the Earth.

In the ideal case of a perfect alignment Stream → Sun → Earth the axion flux enhancement can be very large $\sim 10^6$ or even up to $\sim 10^{11}$. This temporal axion signal amplification can be utilised by axion DM experiments. Streaming DM could have a density which is $\sim 0.3$ to 23% of the local DM density. Therefore, even a temporally lensed tiny DM stream propagating along the Sun → Earth direction could still surpass the local DM density which would give rise to an unexpectedly large DM flux exposure of an axion haloscope like CAST-CAPP.

Furthermore, as we have already seen planets, including the Moon, can also work as gravitational lenses for slow speed particles with the flux enhancement at the site of the Earth being on the order of $10^6$. Specifically for the case of Moon, the flux of particles with speeds of $\sim 10^{-4}c$ can be magnified by about $10^4$ (see Fig. 20.1). This opens up additional windows of opportunity if an axion stream undertakes a temporal alignment with the direction of a Planet → Earth (see also Publications E.10, and E.16).

### 21.4.1 Theoretical background

As we have already seen the co-existence of streaming DM [159, 450] or the galactic dark disk hypothesis [194, 198] have already been discussed in the literature. An interesting case is for example the widely accepted phenomenon that the Milky Way disrupts the near Sgr dwarf elliptical galaxy during its multiple passages through the galactic disk. The tidally stripped stars give rise to the Sgr stream with the location of the solar system being probably close to the Sgr orbits. Therefore, this is expected to result also to streams of DM in our galaxy and its halo (see Sect. 2.2.2).

Another example of stream axion DM are dense small-scale structures, the axion mini-clusters [165, 166]. With time a fraction of them is disrupted and forms tidal streams, with the axion density being just an order of magnitude larger than the average. Stream-crossing events in our neighbourhood can occur about one in 20 y lasting for a few days. However, the axion mini-cluster(s) with mass $\sim 10^{-12}M_\odot$ can be trapped by the solar system during its formation. The estimated maximum axion density enhancement is about $10^5$ the DM average, with the Earth-crossing time of the dense axionic clumps being a few days per year. Such a trapped cluster mass is below a conservative bound on possible DM density in the solar system as it is derived from many high precision positional observations of planets and spacecraft [168, 169].

Finally, other streams of DM including caustics [167] may propagate along a gravitationally favourable direction like one from the Sun or a planet towards the Earth, which may further enhance the local flux as already mentioned. These would also produce a distinguishable narrow peak in the spectrum of microwave photons from axion conversion [167].

### 21.4.2 Experimental approach

In order to take advantage of burst-like axion flux enhancements due to temporally occurring axion stream alignments two criteria must be fulfilled:

- The covered frequency range must be as wide as possible as the exact mass of the axion $m_a$ is unknown.
- The scanning time must be as short as possible in order to take advantage of such transients events.

With these two features, gravitationally-focused short axion bursts can be explored, but with a decreased detection sensitivity due to the smaller integration time in each frequency step. However, if the aforementioned axion signal enhancement due to Sun's or any other planet's gravitational lensing effect can surpass the deceased detection sensitivity then such a detection scheme would be advantageous over the conventional one. Since an exact alignment date is difficult to be predicted, an axion haloscope should cover a full calendar year. Finally, taking into consideration that a single haloscope cannot fulfil on its own all the aforementioned requirements, together with the fact that the available axion mass range is very wide, a network of haloscopes distributed around the globe would be the most promising approach [448, 449].

### 21.4.3 Alignment with the Galactic Centre

An example direction of potential interest for DM experiments like CAST is the alignment *GC* → Sun → Earth within 5.5° which repeats every year around 18th of December (see Fig. 21.8a). Furthermore, once every few years this alignment also include the "New Moon", so that the alignment becomes GC → Sun → Moon → Earth (see Fig. 21.8b). Other alignments including different planets are also possible.

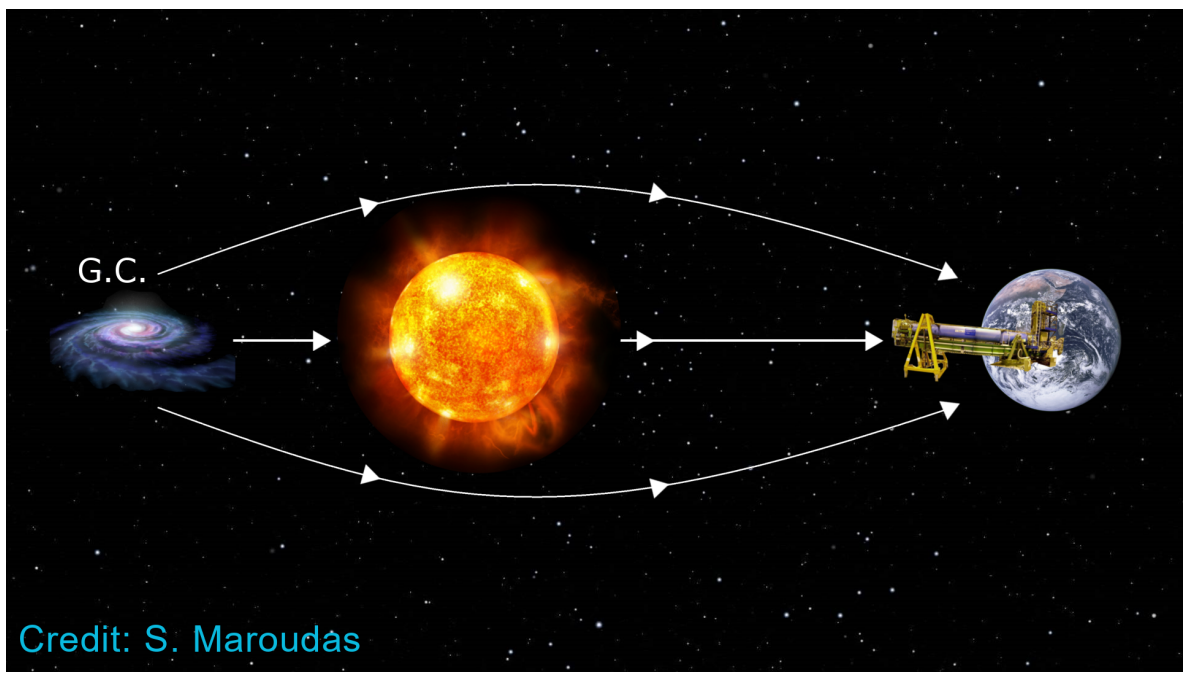

**(a)** Alignment of the GC with the Sun-Earth configuration every December.

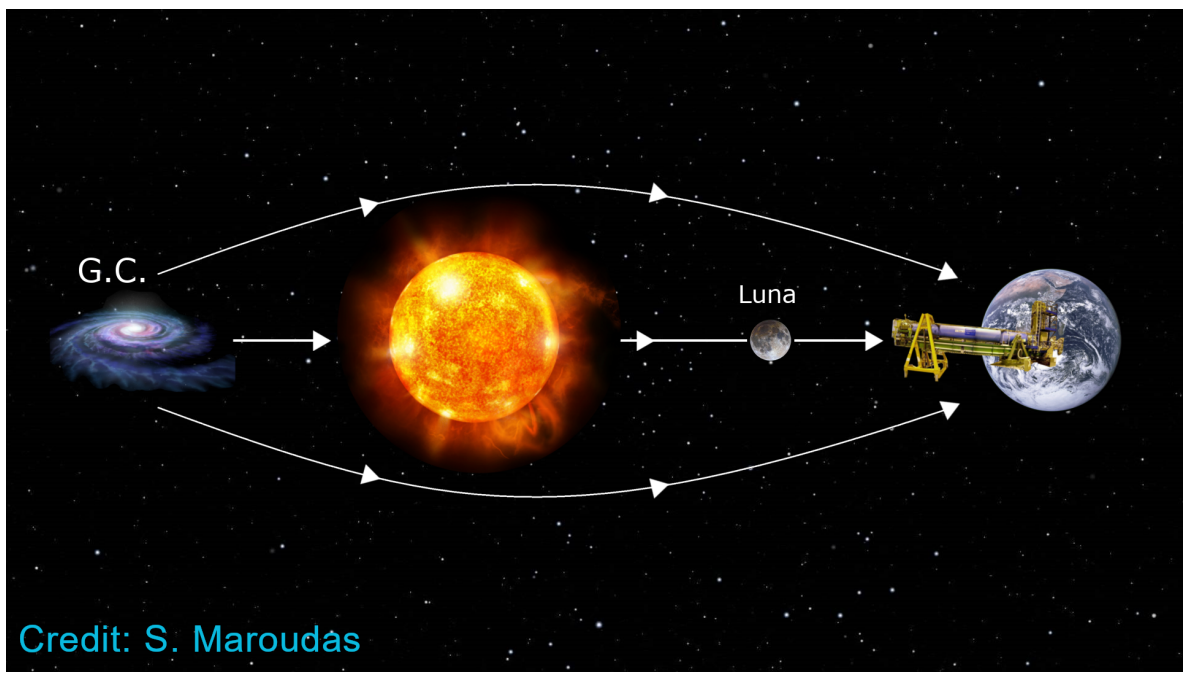

**(b)** Alignment of the GC with the Sun-Earth configuration including the New Moon.

**Figure 21.8:** Simplified illustration of a possible stream alignment coming from the GC with the Sun → Earth direction could be exploited by an appropriate axion haloscope like CAST-CAPP.

Such alignments happen within a different degree of accuracy which changes depending on the specific locations of the solar system bodies. For example, Fig. 21.9 gives the degree of alignment of Earth-GC with the intervening Sun in 2018 for the location of the CAST experiment. Such a configuration could enhance the flux of low-speed particles from this place at our site enormously. This holds also for any possible streams once they get aligned with the Sun → Earth reference frame. From Fig. 21.9 we can also deduce that while each time there is a minimum angle for the accuracy of the alignment, the actual alignment holds, for a few days before and after the minimum angle but with a smaller efficiency.

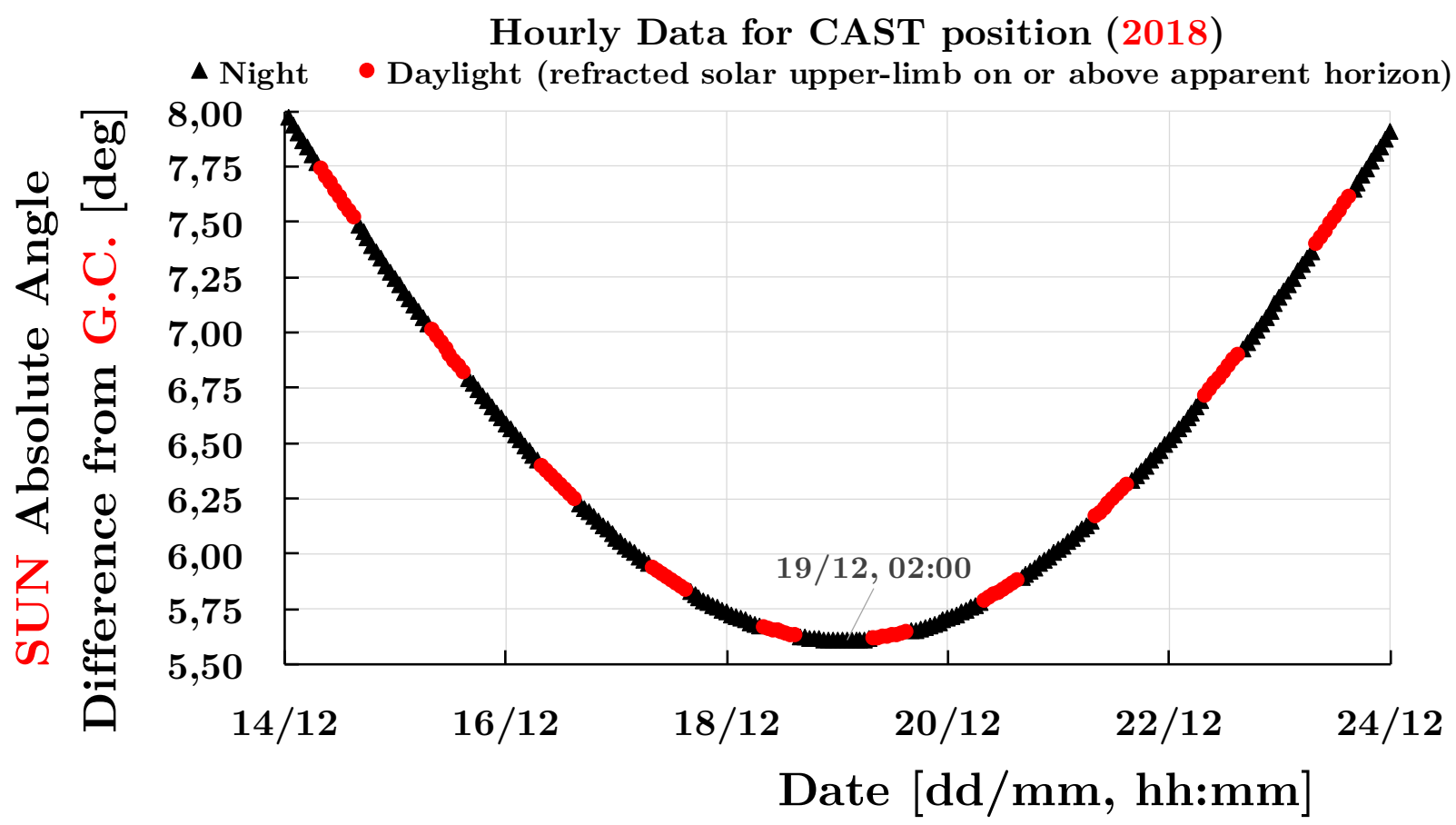

**Figure 21.9:** Absolute angle difference between Sun and GC as seen from the specific position of the CAST experiment around 18/12/2018.

To create Fig. 21.9, the specific geodetic coordinates of the CAST experiment are used, which are seen in Tab. 21.1.

**Table 21.1:** Geodetic coordinates describing the location of the CAST experiment.

| | |
|---|---|
| Longitude | 6.097 043 028° E |
| Latitude | 46.241 960 67° N |
| Altitude | 479.511 m |

For the coordinates of the Sun, the celestial coordinate system from Fig. 21.10 is used, which is defined by the astrometric right ascension (RA) and declination (DEC). The location of the Sgr A∗, a.k.a. the GC, in degrees is defined as RA = 266.416 837 08° and DEC = −29.007 810 56°. Then, the absolute angle difference of the Sun from the GC as seen from CAST, is calculated by:

$$\theta_{\text{abs}} = \sqrt{(\text{RA}_{\odot} - \text{RA}_{GC})^2 + (\text{DEC}_{\odot} - \text{DEC}_{GC})^2} \tag{21.31}$$

The same procedure can be performed for the calculation of the absolute alignment angle with other planets. As an example it is noted than on 18/12/2017 there was an alignment *GC* → Sun → New Moon → Earth within about 8.7° for the location of the CAST experiment.

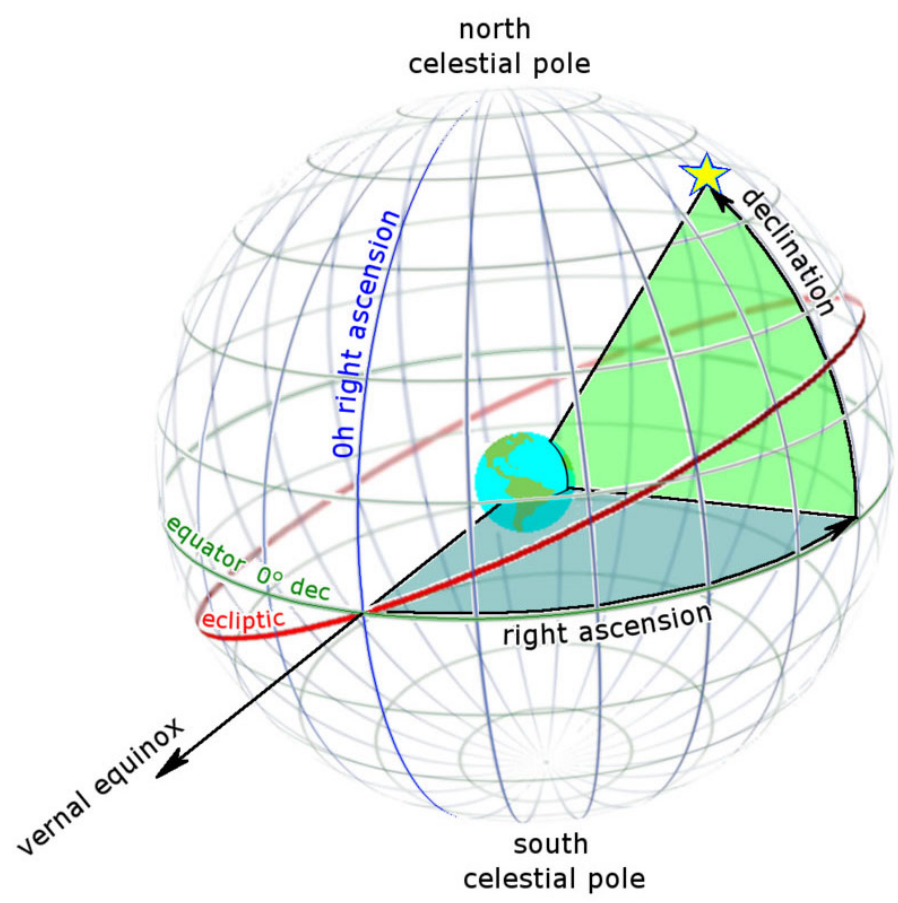

**Figure 21.10:** Celestial coordinate system where declination (green) is measured in degrees north and south of the celestial equator, whereas right ascension is measured east from the equinox. The red circle defines the sun's apparent path around the sky, which defines the ecliptic (47).

22

# CAST-CAPP DETECTOR

## 22.1 The axion haloscope

### 22.1.1 Detection principle

The most promising experiments for detecting DM axions are the so-called *axion haloscopes*, "à la Sikivie" [431, 432] where axions from the galactic halo can convert resonantly to microwave photons when they penetrate a cavity immersed in a strong magnetic field (Fig.

22.1). The Lagrangian for this interaction with the coupling of axions to photons was given in Eq. 21.12.

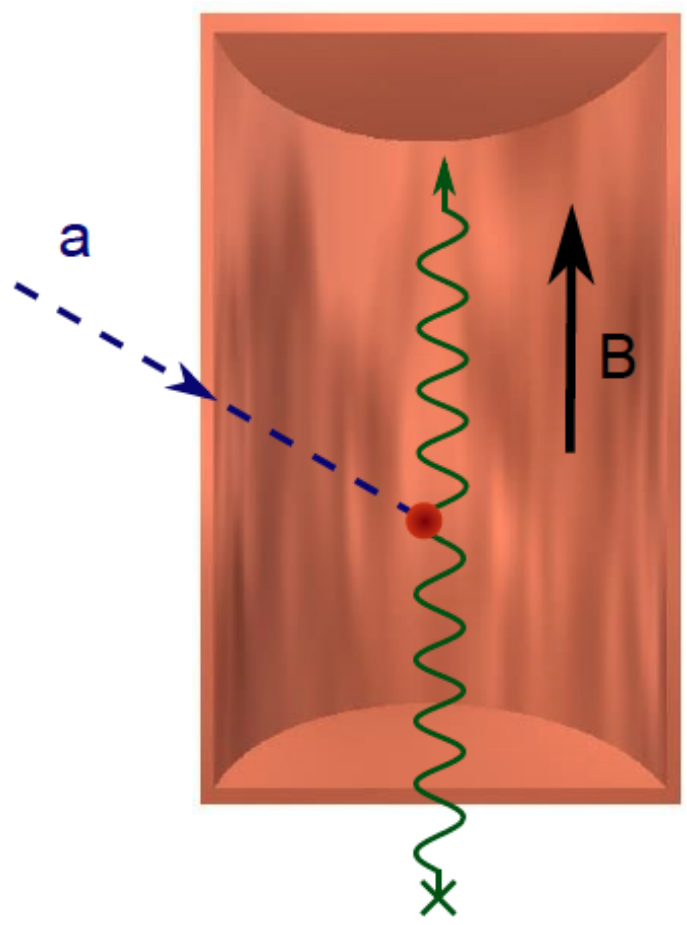

**Figure 22.1:** Axion detection principle in a haloscope. Axions $a$ scatter off of the virtual photons of the magnetic field $\vec{B}$ and convert to microwave photons $\gamma$ inside a resonant cavity (48).

More specifically, similarly to the Primakoff effect, the axion-to-photon conversion rate in a region of space where a strong magnetic field $\vec{B}$ is present, is further enhanced if the outgoing photon (represented by the electric field $\vec{E}$ in Eq. 21.12), is detected in a microwave cavity resonating to the corresponding frequency $\nu$ given by the axion mass:

$$m_a \sim h\nu/c^2 \tag{22.1}$$

Therefore, haloscopes (such as CAST), consist of cryogenic high-$Q$ microwave cavities and can detect axions in the µeV mass range. The microwave signal is then extracted through an antenna which is critically coupled to the cavity. The operation of such Radio Frequency (RF) cavities involves the appropriate choice of available electromagnetic modes, and, as the axion mass $m_a$ is unknown, a tuning mechanism for adjusting the resonant frequencies and thus covering a wide range of axion masses is indispensable. Finally, another condition that must be satisfied is that the characteristic size of the cavity $d$ should be smaller than the typical de Broglie wavelength of relic axions $\lambda_a$ as defined in Eq. 21.23:

$$d \leq \lambda_a \tag{22.2}$$

### 22.1.2 Conversion power

The on-resonance axion conversion power in a microwave cavity [431, 432] is proportional to the square of the magnetic field $\vec{B}^2$, the quality factor $Q_L$ of the cavity (the ratio of the cavity stored-energy to its losses), its volume $V$ and geometry factor $C_{lmn}$ [405]:

$$P_{\text{axion}} = \left(g_\gamma{}^2 \frac{a^2 \hbar^3 c^3 \rho_a}{\pi^2 \Lambda_{\text{QCD}}^4}\right) \left(\omega_0 \frac{1}{Q_L} \frac{\beta}{1+\beta}\right) \frac{\vec{B}^2}{\mu_0} V C_{lmn} \frac{Q_L{}^2}{1 + \left[2Q_L \left(\nu - \nu_0\right)/\nu_0\right]^2} \tag{22.3}$$

where: $\omega_0 = 2\pi\nu_0$ : resonant frequency of the unloaded cavity,

$\beta$ : normalised coupling parameter of the cavity,

$C_{lmn}$ : normalised coupling form factor or *geometry factor*,

$Q_L$ : loaded quality factor of the cavity,

$\delta\nu = \nu - \nu_0$ : distance to mode frequency of a point on the spectrum,

$\nu_0$ : mode frequency.

The geometry factor $C_{lmn}$ is a dimensionless quantity which is a measure of the coupling between the resonance electric field $\vec{E}_{lmn}$ in a cavity with volume $V$ and the static magnetic field $\vec{B}$ and is defined as [425]:

$$C_{lmn} = \frac{\left|\int_V d^3\vec{x} \vec{E}_{lmn}\left(\vec{x}\right) \cdot \vec{B}\right|^2}{\vec{B}^2 V \int_V d^3\vec{x} \varepsilon\left(\vec{x}\right) \left|\vec{E}_{lmn}\left(\vec{x}\right)\right|^2} \tag{22.4}$$

where: $\varepsilon\left(\vec{x}\right)$ : permittivity of the cavity.

Physically, the geometry factor parametrises the overlap between the cavity mode and the external magnetic field and is derived directly from the interaction Lagrangian in Eq. 21.12.

Furthermore, for any resonator with multiple loss channels, the total linewidth is the sum of partial linewidths, which means that the $Q$-factors associated with internal and external losses are:

$$\frac{1}{Q_L} = \frac{1}{Q_0} + \frac{1}{Q_r} \tag{22.5}$$

where: $Q_0$ : unloaded quality factor (limited by the nonzero resistance of the cavity),

$Q_r$ : parametrises the coupling to the receiver.

The dimensionless *cavity coupling parameter* can therefore be defined as:

$$\beta = \frac{Q_0}{Q_r} \tag{22.6}$$

in terms of which

$$Q_L = \frac{Q_0}{1+\beta} \tag{22.7}$$

From Eq. 22.3 we see that $P_{\text{axion}} \propto \beta/(1+\beta)^2$, which means that $P_{\text{axion}}$ is maximized for $\beta = 1$. Therefore, we can define the cases where:

- $\beta = 1$: The cavity is *critically coupled* to the receiver chain, meaning that half of the signal power is dissipated by the resistance of the cavity walls and half is coupled to the receiver.
- $\beta < 1$: The cavity is *under-coupled*, meaning that most of the power is absorbed internally. In this case $Q_L$ is close to $Q_0$.
- $\beta > 1$: The cavity is *over-coupled*, meaning that most of the conversion power is measurable, but the resonant enhancement of the signal is reduced.

### 22.1.3 Rectangular cavities in dipole magnets

A suitable experimental setup in a dipole field consists of tunable rectangular cavities submerged inside the magnet bore, with the magnetic field $\vec{B}$ parallel to the resonator lateral sides. In this case, the sensitivity is maximised if the Transverse Electric field (TE) modes are used. The mode frequency depends on the cavity width ($w$), height ($h$), length ($L$) and mode indexes $l$, $m$, $n$ as [451]:

$$f_{lmn} \propto \sqrt{\left(\frac{l}{w}\right)^2 + \left(\frac{m}{h}\right)^2 + \left(\frac{n}{L}\right)^2} \tag{22.8}$$

If $L$ is along the $z$ direction, and the external $\vec{B}$ field along the $y$ direction as seen in Fig. 22.2, it is convenient to choose modes in which $m = 0$ so that the resonant electric field $\vec{E}$ is parallel to the magnetic field $\vec{B}$ [452]:

$$\begin{aligned} \vec{E}_y &\propto \sin\left(\frac{l\pi}{w}x\right)\sin\left(\frac{n\pi}{L}z\right) \\ \vec{E}_x &= \vec{E}_z \equiv 0 \end{aligned} \tag{22.9}$$

Therefore, the fundamental $TE_{101}$ mode is the most favourable which gives a geometry factor of $C_{lmn} = 0.66$ for an empty cavity [452].

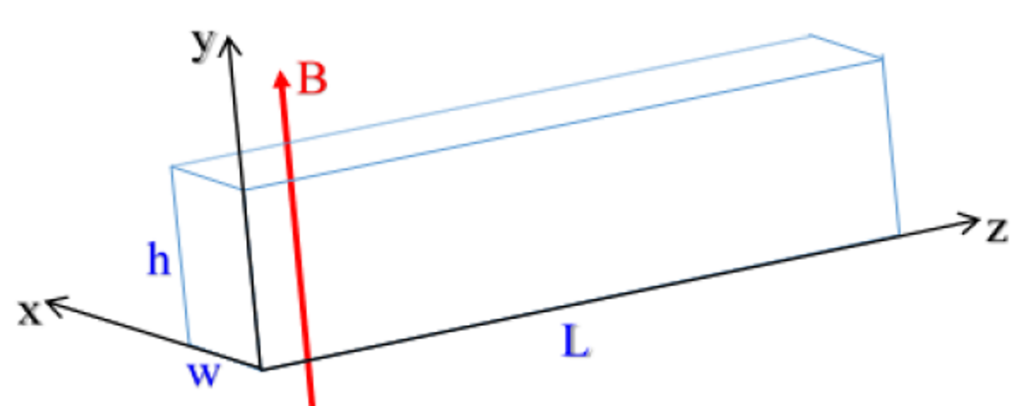

**Figure 22.2:** Cavity orientation with respect to the external static magnetic field $\vec{B}$.

### 22.1.4 Sensitivity

From the often used Sikivie method [431], the cavity on-resonance output power from axion to microwave photon conversion, for the specific CAST-CAPP parameters that will be presented in Sect. 22.2, is estimated to be:

$$\begin{aligned} P_{\text{axion}} = 2.1 \times 10^{-24}\,\text{W} \times \left(\frac{\rho_a}{0.45\,\text{GeV/cm}^3}\right)\left(\frac{g_\gamma}{0.97}\right)^2\left(\frac{\nu}{5\,\text{GHz}}\right)\left(\frac{\beta}{1+\beta}\right) \\ \times \left(\frac{\vec{B}}{8.8\,\text{T}}\right)^2\left(\frac{V}{224\,\text{cm}^3}\right)\left(\frac{Q_L}{2\times 10^4}\right)\left(\frac{C_{lmn}}{0.53}\right) \\ \times \left(\frac{1}{1+\left[2Q_L\left(\nu-\nu_0\right)/\nu_0\right]^2}\right) \end{aligned} \tag{22.10}$$

From Eq. 22.10 (see also Eq. 22.3) we see that the axion detection power depends on the frequency $\nu$ as well as the frequency distance from the resonant mode of the peak $|\nu - \nu_0|$. We also see that for CAST-CAPP detector, the expected axion power is $\sim 10^{-24}\,\text{W}$.

From Dicke's radiometer equation [453] the SNR is defined as the ratio of the signal power to uncertainty in noise power within the signal bandwidth:

$$SNR = \frac{P_{axion}}{\sigma_{\text{noise}}} = \frac{P_{\text{axion}}}{k_B T_S}\sqrt{\frac{t}{\delta\nu_a}} \tag{22.11}$$

where: $T_S$ : the sum of the physical temperature of the cavity $T_{\text{cav}}$ and the noise temperature of the receiver.

$t$ : the time spent integrating at a particular frequency.

$\delta\nu_a$ : the frequency bandwidth of the axion signal set by the local axion velocity distribution.

The signal bandwidth $\delta\nu_a$ for an axion detector is defined by the velocity dispersion of the axion [454]:

$$\delta\nu_a = \frac{m_a}{Q_a} \tag{22.12}$$

where: $Q_a \approx \frac{c^2}{\upsilon_a \Delta\upsilon_a} \sim 10^6$ : the *quality factor* of the axions,

such that $\upsilon_a$ and $\Delta\upsilon_a$ are the velocity and velocity dispersion respectively of the axions on Earth.

Since the axion mass $m_a$ is unknown, from Eq. 22.11, and by using Eq. 22.3, we can look at the scanning rate of the axion frequency, which indicates how fast a cavity experiment can scan a target frequency range with a given experimental sensitivity [455, 456]:

$$\begin{aligned}\frac{d\nu}{dt} &= \frac{\nu\delta\nu_a}{Q_L}\left(\frac{1}{SNR_{\text{target}}}\right)\left(\frac{P_a^{a\gamma\gamma}}{P_{\text{noise}}}\right)^2 \\ &= g_{a\gamma\gamma}{}^4\left[\frac{4\eta}{5SNR^2_{\text{target}}}\right]\left[\frac{\hbar c^3 a^2 \rho_a{}^2}{\pi^2\Lambda^4_{\text{QCD}}}\right]\left[\frac{\omega_0\vec{B}_0^2 V}{\mu_0 k_B T_S}\right]^2\left[\frac{\beta C_{lmn}}{1+\beta}\right]^2 Q_L Q_a N^2_{PM}\end{aligned} \tag{22.13}$$

where: $P_a^{a\gamma\gamma}$ : the axion signal power as mentioned in Eq. 22.10 for an axion-photon coupling strength,

$\delta\nu_a$ : the bandwidth of the axion signal,

$N_{PM}$ : the number of PM cavities described in Sect. 22.1.5.

and $P_{\text{noise}}$ is the noise power proportional to the noise temperature $T_S$:

$$P_{\text{noise}} = k_B T_S \delta\nu \tag{22.14}$$

Using Eq. 22.10, with the specific parameters of CAST-CAPP from Sect. 22.2, Eq. 22.13 becomes:

$$\begin{aligned}\frac{d\nu}{dt} \approx \frac{980\,\text{MHz}}{\text{y}} &\times \left[\frac{g_\gamma}{0.97\times 20}\right]^4\left[\frac{5\,\text{GHz}}{\nu_a}\right]^2\left[\frac{5}{SNR}\right]^2\left[\frac{10\,\text{K}}{T_S}\right]^2\times \\ \times\left[\frac{\vec{B}}{8.8\,\text{T}}\right]^4&\left[\frac{C_{lmn}}{0.53}\right]^2\left[\frac{\rho_a}{0.45\,\text{GeV}/\text{cm}^3}\right]^2\left[\frac{V}{224\,\text{cm}^3}\right]^2\left[\frac{Q_L}{2\times 10^4}\right]\left[\frac{Q_a}{10^6}\right]\left[\frac{\beta}{1+\beta}\right]^2\end{aligned} \tag{22.15}$$

Eq. 22.15 defines the performance of the experiment based on the given parameter values, and is the most useful figure of merit for an axion search with a microwave cavity. It is clear once more that large volume ($V$), large magnetic fields ($\vec{B}$), large quality factors ($Q_L$), and large geometrical coefficients ($C_{lmn}$) with low temperature ($T_S$) are very critical. For CAST-CAPP using a single cavity, the scanning rate required for a sensitivity of $20 \times KSVZ$ is 980 MHz/year whereas for four PM cavities the scanning rate becomes 1.3 GHz/month.

### 22.1.5 Phase-Matching concept

Since the sensitivity increases with the cavity volume as seen in Eq. 22.10, in order to increase the effective volume, multiple cavities can be used instead of enlarging a single cavity. The effect of using $N$ cavities instead of a single cavity on the SNR, due to volume expansion alone, is:

$$SNR_N = \sqrt{N}\cdot SNR_{\text{single}} \tag{22.16}$$

An additional enhancement of the SNR is possible through the *Phase-Matching* technique.

The concept is to get simultaneous read-outs from several frequency-matched cavities and then combine them coherently after pre-amplification. This idea of multiple-cavity design was first introduced in [457] with an experimental design using a quadrupole-cavity detector attempted in [458]. If the first stage amplification takes place before the signal combination, then the signal-to-noise ratio is improved linearly with the number of cavities [459]:

$$SNR_N = N \cdot SNR_{\text{single}} \tag{22.17}$$

More specifically, for uncorrelated noise, the total power becomes:

$$P_{\text{total}} = \sum_i P_i \tag{22.18}$$

which for equal $N$ powers $P_i \approx P$ gives $P_{total} \approx NP$. On the other hand for correlated noise the total power scales with $N^2$:

$$P_{\text{total}} = N^2 \sum_i P_i \tag{22.19}$$

Therefore, the enhancement of SNR for PM which coincides with Eq. 22.17 is:

$$\frac{SNR_N}{SNR_{\text{single}}} \sim \frac{P_{\text{correlated}}}{P_{\text{uncorrelated}}} = \frac{N^2}{N} = N \tag{22.20}$$

Due to the large wavelength $\lambda$ of the coherent axion field, the most challenging part on the PM of multiple cavities, is the frequency matching of all the cavities to the same resonant frequency. However, due to hardware limitations, there is a frequency mismatch that has to be permitted up to a certain level where the combined power is sufficiently high so that the resulting sensitivity is not degraded. This Frequency Matching Tolerance (FMT) has a dependence on the cavity quality factor $Q_0$ and target frequency [459]:

$$FMT\,(Q_0, \nu) = \frac{0.42\,\text{GHz}}{Q_0} \times \nu\,[\text{GHz}] \tag{22.21}$$

## 22.2 CAST-CAPP cavities

### 22.2.1 Experimental design

The CAST-CAPP detector (see Publications E.24, E.25, and E.26) consists of four tunable $23 \times 25 \times 390$ mm rectangular cavities with the total volume of each cavity being $V = 224\,\text{cm}^3$ [452]. The cavities are longitudinally split into two identical halves, and are made of stainless steel electroplated with $\sim 30\,\mu\text{m}$ of copper (see Fig. 22.3). They have been installed inside one of the two bores of CAST superconducting dipole magnet at European Organization for Nuclear Research (CERN), each of which has a ∼43 mm diameter. The direction of the split plane is placed parallel to the static magnetic field $\vec{B}$, along the cavity's small face. CAST dipole magnet, which was a former LHC prototype, has a 8.8 T magnetic field and 9.25 m magnetic length.

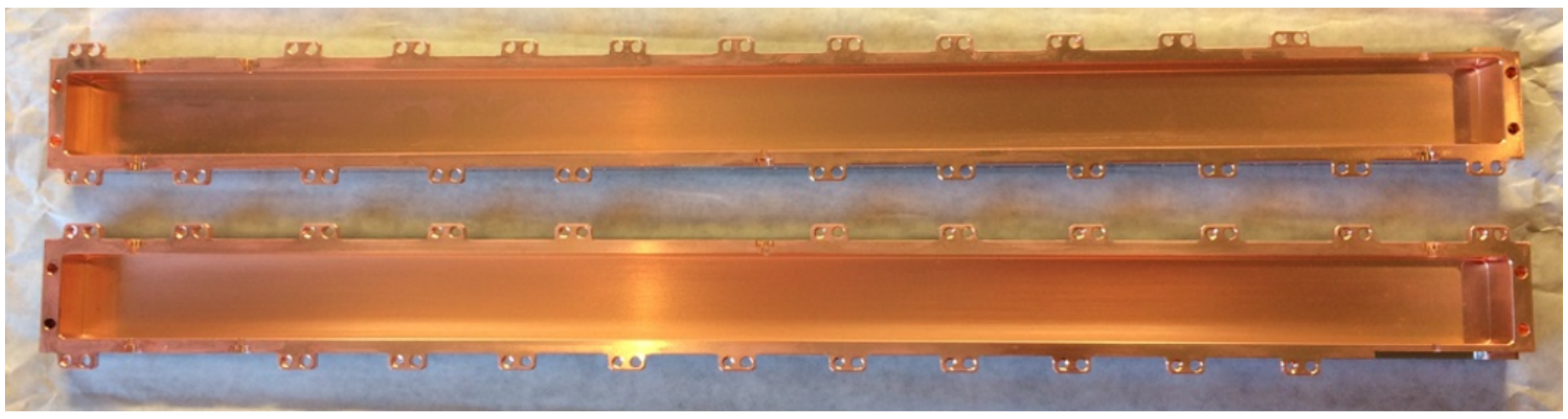

**Figure 22.3:** An open empty cavity showing the copper coating.

### 22.2.2 Tuning mechanism

The tuning mechanism consists of two dielectric strips made out of sapphire with permittivity $\varepsilon = 9$ and sizes $2.56 \times 12 \times 360$ mm which are symmetrically placed parallel to the longitudinal sides, and are moving simultaneously towards the centre as seen in Fig. 22.4a. The cavity tuning system including the electromagnetic impact from the locomotive mechanism situated inside the cavity was designed to have no mode crossings for the axion mode over the entire tuning range.

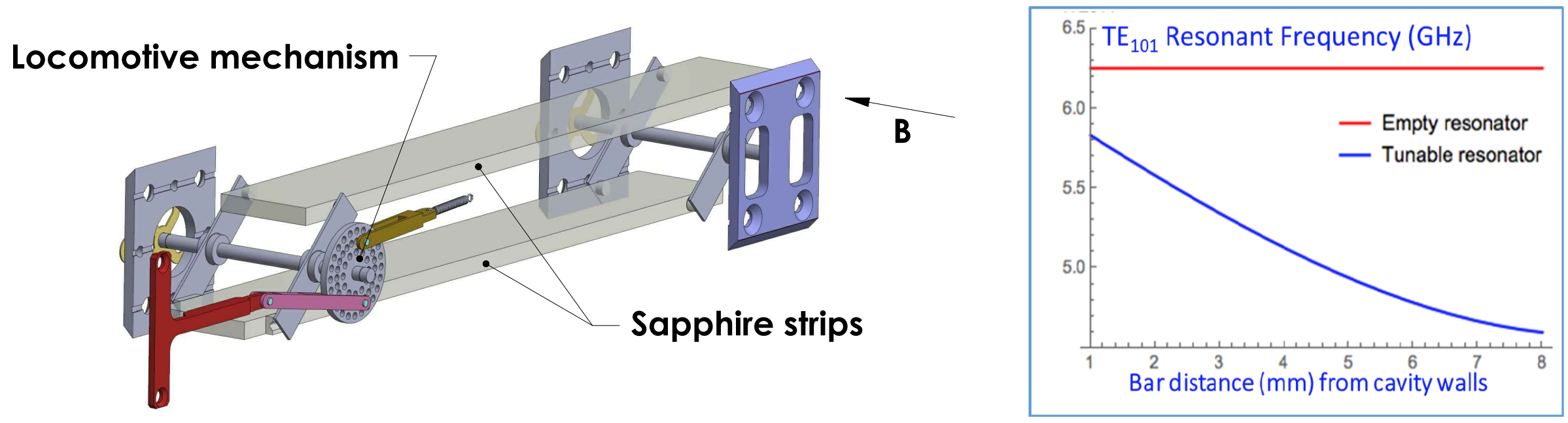

**(a)** Technical drawing of the tuning mechanism of CAST-CAPP. The static $\vec{B}$-field is shown by the arrow and is parallel to the two axes of the tuning mechanism.

**(b)** Simulation of the resonant frequency as a function of the position of the tuning strips [452].

**Figure 22.4:** CAST-CAPP tuning schematic with the two sapphire strips and frequency range.

The two parallel sapphire strips are activated by a piezoelectric motor seen in Fig. 22.6 through a locomotive mechanism delivering a tuning resolution of better than 100 Hz in stable conditions. The cavity frequency is changed by moving the strips towards or away from the split plane while keeping them parallel to each other. In Fig. 22.5a the details of the locomotive transmission cranks and wheel can be seen, connected on the right-hand side to the piezoelectric motor (called *Cryo Linear Actuators (CLAs)*) and on the left-hand side to one of the two sapphire shafts. The motion is transmitted to the tuning strips through eight spherical joints realised with 0.5 mm radius sapphire hemispheres (see Fig. 22.5). The current maximum tuning range for each cavity is on the order of 400 MHz corresponding to an axion mass range of ∼21–23 µeV, but an adjustment of the tuning gears can result to a full tuning range of ∼ 1 GHz (see Fig. 22.4b) [452].

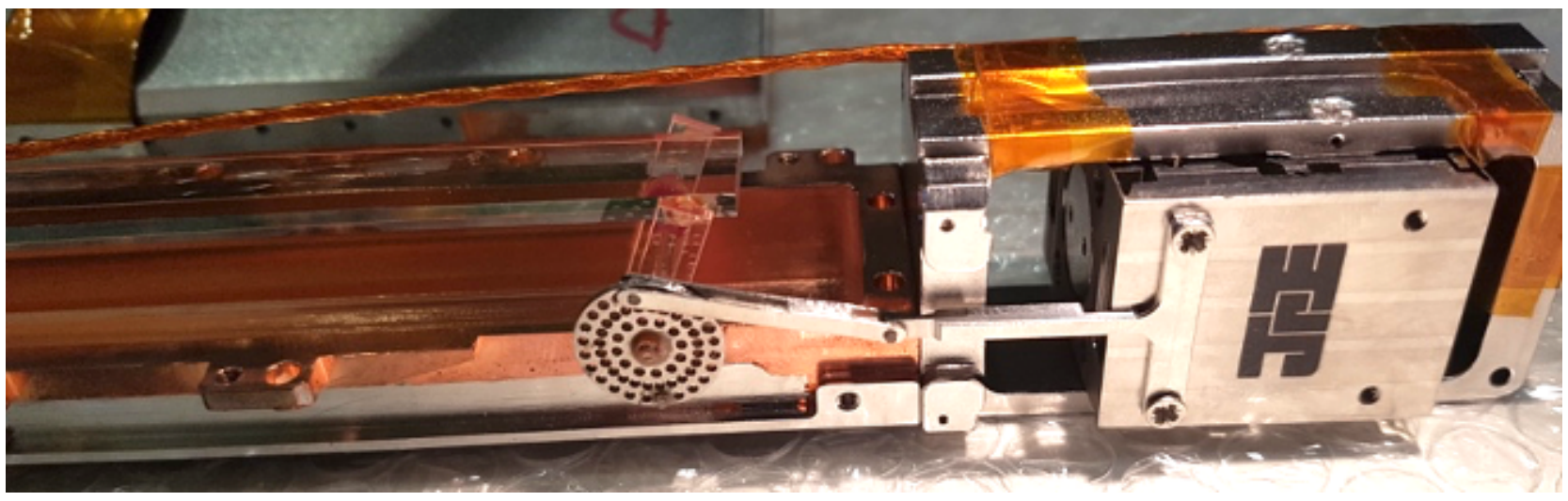

**(a)** The stepper motor providing the movement through the locomotive mechanism.

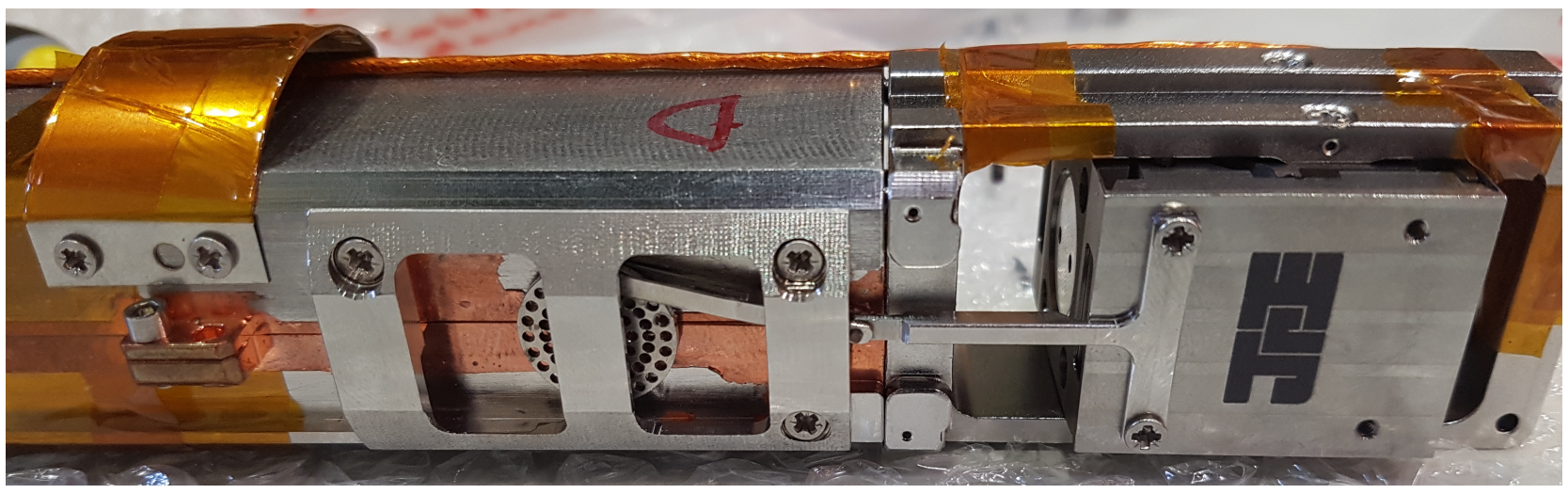

**(b)** Tuning mechanism as seen with the cavity closed.

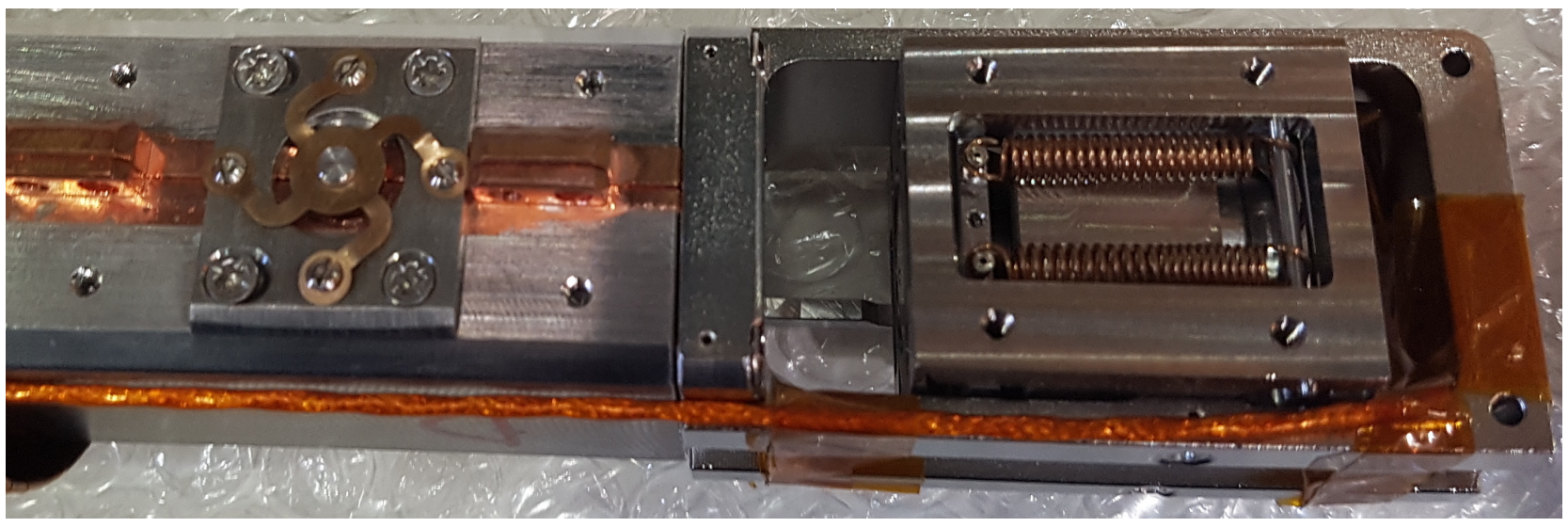

**(c)** Two springs on the rear side of the CTS which are used to preload the setup.

**Figure 22.5:** CAST-CAPP tuning mechanism.

The CLAs used are piezo-ceramic based actuators developed for accurate positioning in an ambient, vacuum and cryogenic environment. The model CLA1801 (see Fig. 22.6a) is currently used for all four cavities which has a diameter of 18 mm and is able to deliver 15 N peak force. Each CLA is a spindle/nut drive concept for which the nut is attached to the frame and the spindle is rotated by the piezo-based actuator. The electrical wiring is attached to the rotating part but decoupled for rotation utilising sliding contacts. With the use of the controller, it is possible to realise torque pulses in both directions on the spindle which enables the spindle to rotate with very small steps resulting in nm adjustability in a cryogenic environment. Each CLA is mounted on a Cryo Translation Stage (CTS) (see Fig. 22.6b) which is basically a positioning device driven by the CLA. The movement of the CTS is then transferred through the locomotive mechanism to the sapphire strips. Two BeCu 200 µm thick springs are also used to preload the setup (see Fig. 22.5c). Provisions to compensate for the differential thermal expansion between sapphire and stainless steel, while preserving smooth and precise operation, make these cavities able to sustain any number of cooling and magnetic field cycles. The detailed description of the controller is on Appendix Sect. C.1.6.

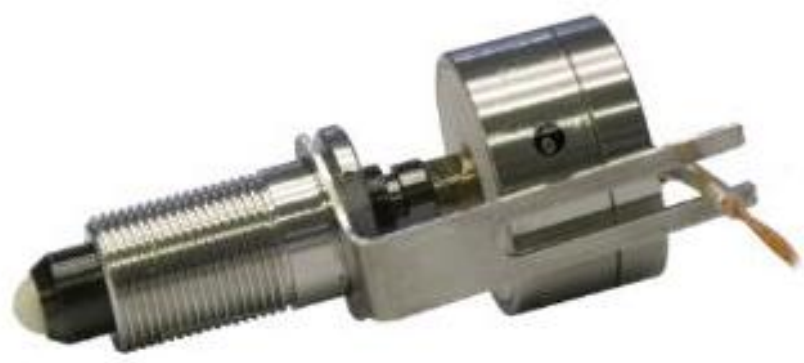

**(a)** CLA1801 actuator (73)

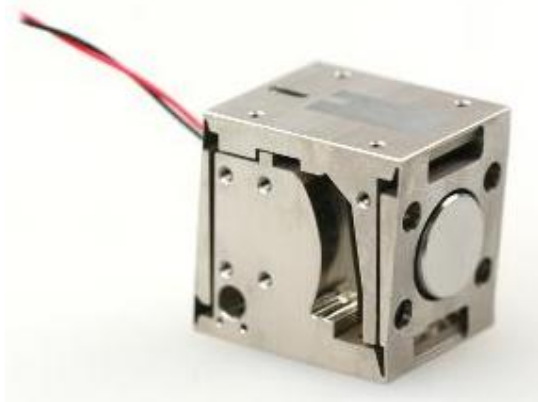

**(b)** CTS1 device (74)

**Figure 22.6:** JPE piezoelectric actuators used in CAST-CAPP detector.

#### 22.2.2.1 Mechanical vibrations

Inside the cavities, the two tuning sapphire strips have a very high mechanical $Q$ and therefore can accidentally act as a mechanical tuning fork, leading to strong frequency modulation of the mode of interest. To suppress these modulations a vibration-damping mechanism was introduced, consisting of two quartz glass tubes filled with teflon foils and glued on the middle of the sapphire tuning strips (see Fig. 22.7). The underlying idea was that the introduced friction from the teflon foils touching the sidewalls of the cavity will reduce the vibration. These dampers were selected based on the following properties: they are placed inside the cavities attached on the sapphire strips, not hampering significantly the RF performance, being compatible in cryogenic conditions and a strong magnetic field, but also not affect the overall tuning range of the cavities. Indeed, the vibration induced from CAST $He$ pumps has been reduced significantly increasing the measured effective $Q_L$ by a few %.

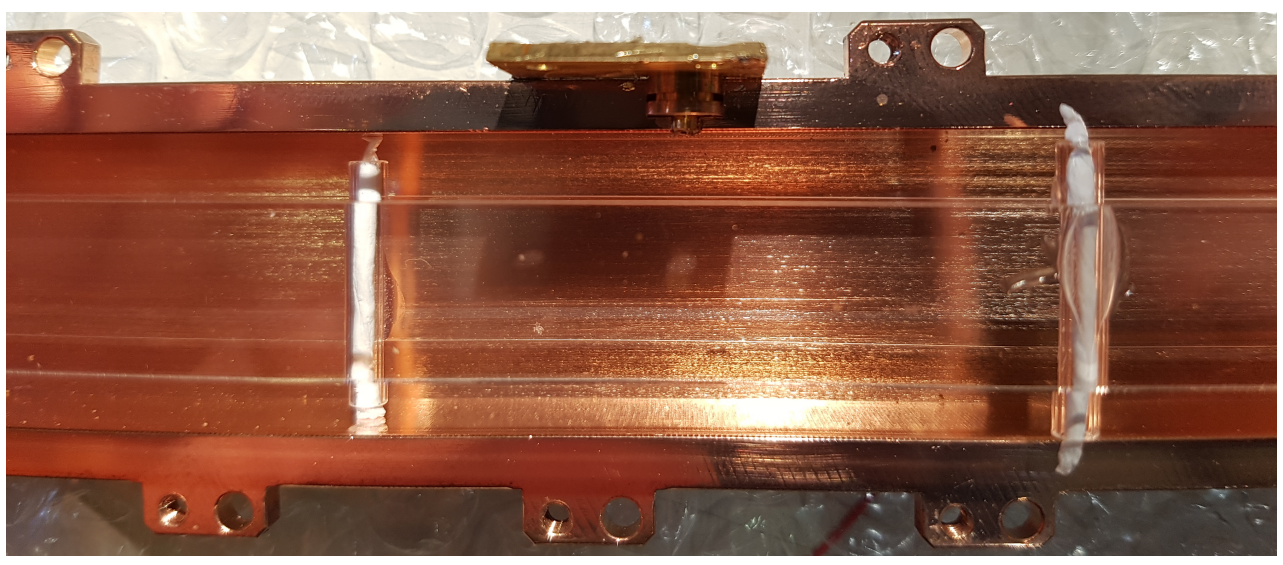

**Figure 22.7:** Vibration damping elements on one of the cavities made of quartz glass tubes filled with teflon.

### 22.2.3 Experimental setup

#### 22.2.3.1 Magnet bore setup

Each cavity assembly includes the cavity, a sapphire tuner, two RF couplers, one piezo-actuator and one cryogenic Low Noise Amplifier (LNA) placed at a few cm distance from the edge of its cavity and with an optimum orientation to the static magnetic field $\vec{B}$ (see Fig. 22.8). Furthermore, three of the cavities carry temperature sensors. As mentioned before, each cavity has been provided with two couplers, a weak coupler called *injection port* and a strong coupler called *main port*. The main port is coupled at $\sim 12\,\Omega$, which brings the cavities to near-critical coupling, within the available tuning range at cryogenic temperatures and magnetic field. More specifically, in CAST-CAPP, $\beta \sim 1$ was selected for all cavities in order to have optimum power transfer. The strong coupling criterion was based on the experimentally-done observation that the coupling factor increases by a factor of $\sim 4$ for a tunable cavity going from room to low temperature. The strong port is connected to a LNA via a 1.19 mm thick, $\sim$ 21 cm-long, RF cable. Finally, each LNA is mounted on a device, made in part of aluminium and part of 316-L stainless steel, which is able to lock the cavity inside the bore of CAST allowing its retrieval and heat exchanging with the bore while locked.

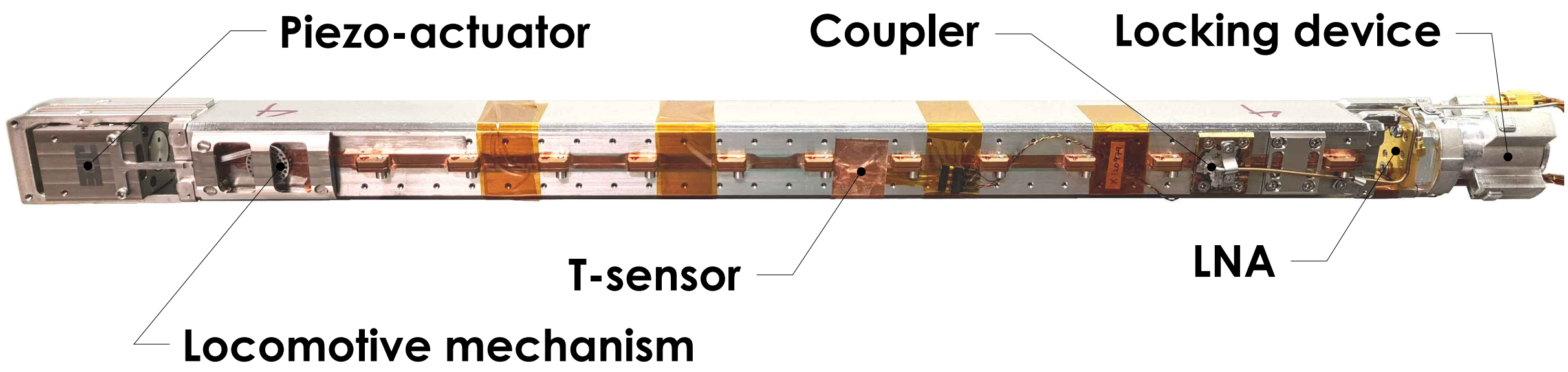

**Figure 22.8:** Typical CAST-CAPP cavity assembly.

However, to allow for an easier and safer cavity installation, the four cavities were banded together using customised inox non-magnetic metal sheet strips and were inserted all together inside the magnet bore of CAST (see Fig. 22.9). This way, the various RF and DC cables are fixed on the sides of the cavities and do not suffer any added friction due to the differential movement of the cables as happens in an independent installation of each cavity. Furthermore, this configuration facilitates the installation and extraction of the cavities and allows us to have much higher accuracy for the alignment of all cavities with the magnetic field, as the cavities cannot rotate differentially but can bend in tandem to account for the curvature of the magnet bore. Finally, an 8 mm endoscopic HD camera positioned on the extraction tool enables close monitoring of the whole installation procedure to ensure the orientation of the cavities as well as the positioning of the cables inside the magnet bore.

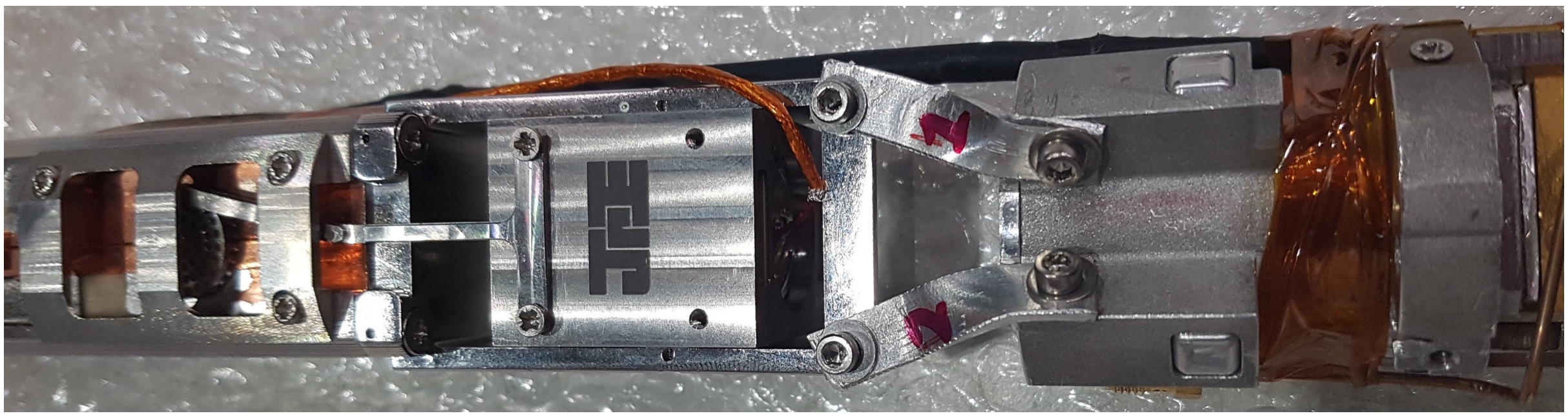

**(a)** The metal strips used to connect the cavities together

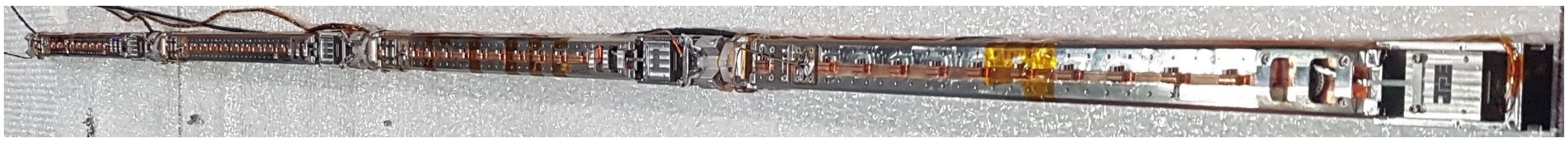

**(b)** All four CAST-CAPP cavities before insertion to the magnet bore.)

**Figure 22.9:** Connection of the four cavities together with wagon-like couplers made out of aluminium.

#### 22.2.3.2 Cryogenic cabling

For the RF connection of the input and output port of each cavity, ULT-10 RF cables from Keycom were used with 1.19 mm outer diameter and insertion loss 2.45 dB/m at 10 GHz made out of silver-plated Cu. So in total there are eight RF cables for the four cavities (two per cavity). As for the DC wiring, there are two DC wires per piezo actuator, three DC wires per LNA, four DC wires per temperature sensor, and tewenty DC wires for the $\vec{B}$-sensors (which were not installed). These are resulting in a total of 56 DC wires rooted from the magnet bore to room temperature. All these wires are connected to three sets of multi-pin DC connectors.

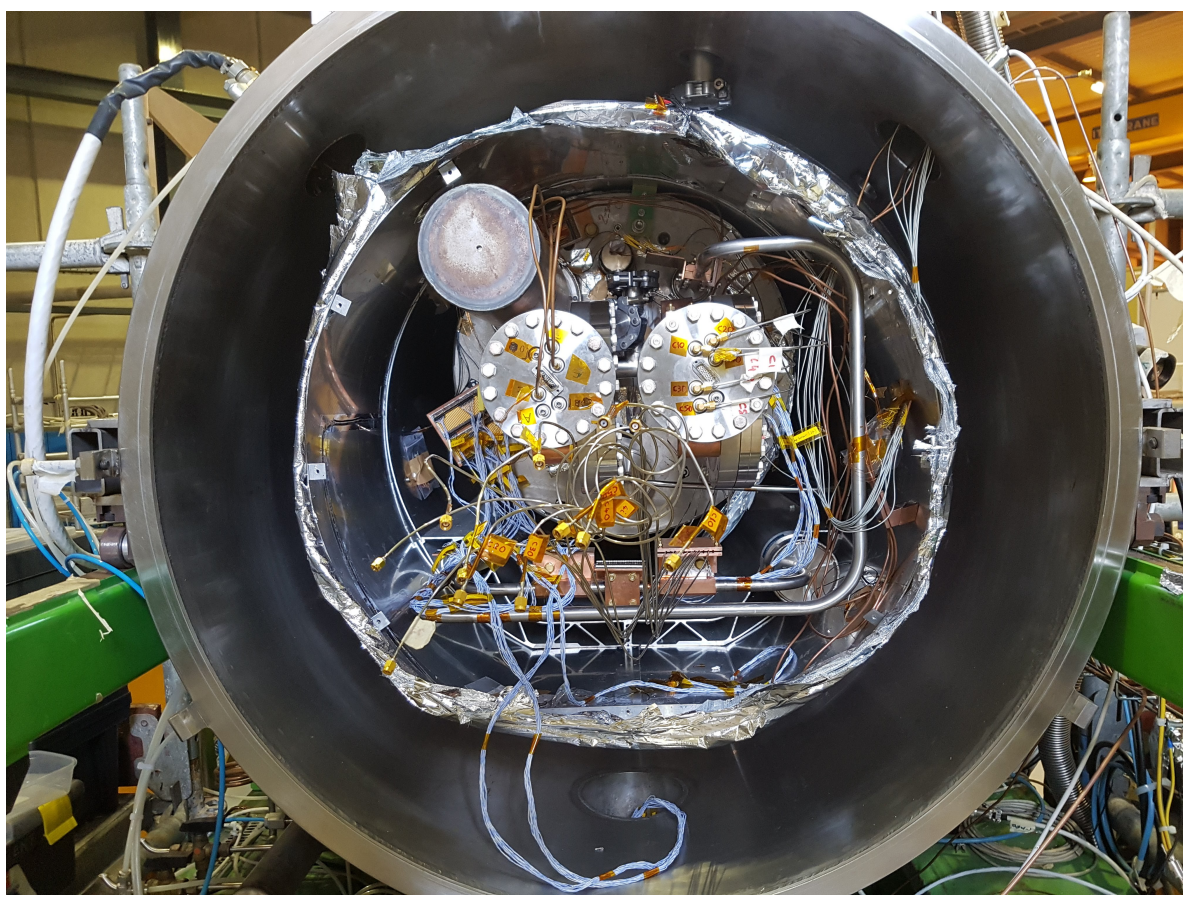

**(a)** Front view. The left flange belongs to the CAST-RADES detector.

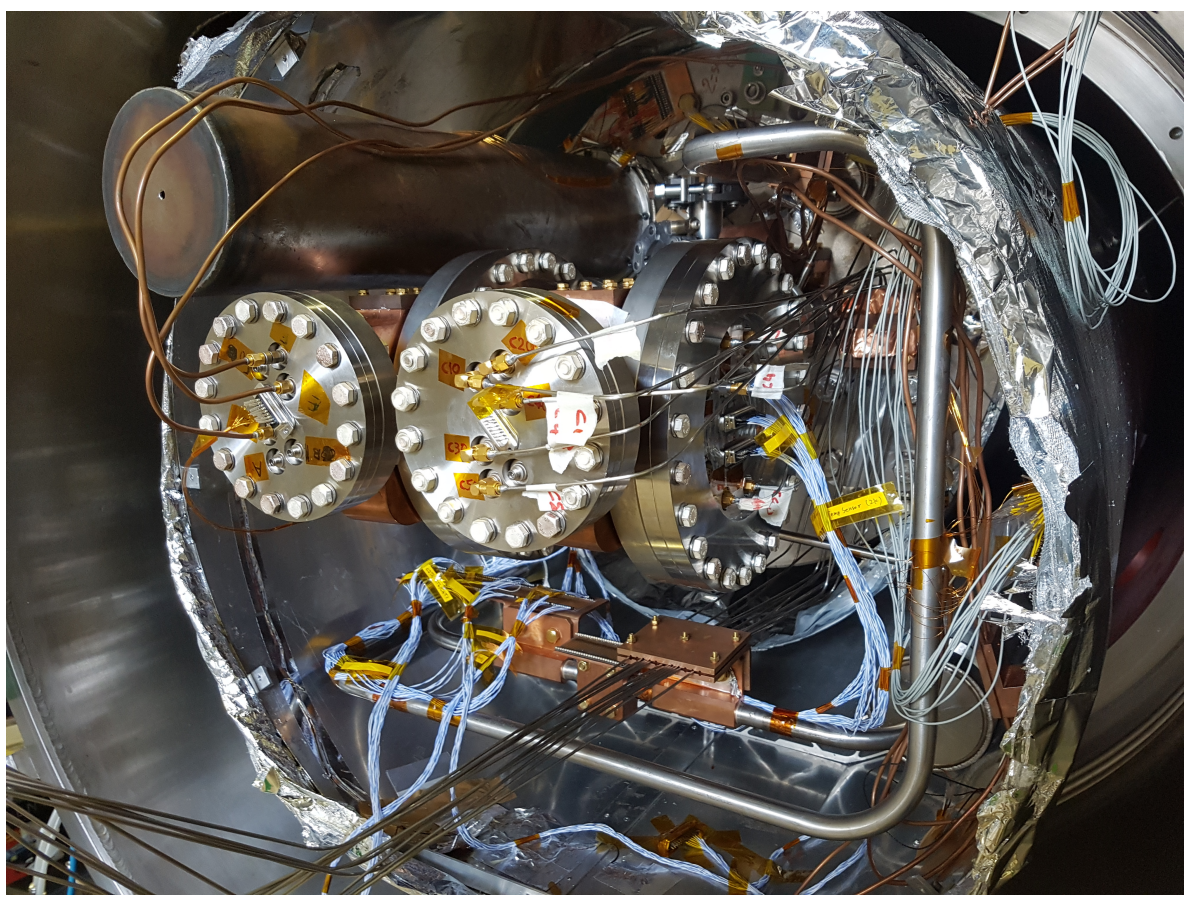

**(b)** Side view of the two CAST-CAPP flanges on the copper vessel.

**Figure 22.10:** The cold 70 K flanges connected on the copper vessel kept around 70 K transitioning the RF and DC cables from the magnet bore to an intermediate transition region.

These connectors along with the eight RF cables are then connected to two-sets of flanges each provided with electrical feedthroughs. The first set of flanges interfaces the magnet bore at around $\sim 2\,\mathrm{K}$, through a copper vessel to the cryostat in a transition region between 2 K and room temperature (see Fig. 22.10). The temperature of the copper vessel is kept around 70 K. The second set of flanges interfaces the cryostat to room temperature and are seen in Fig. 22.11). These are called *input* and *output* 300 K (or “warm”) flanges. The RF connections between the cold (70 K) and the warm flanges inside the cryostat are made through Storm Flex 0.047 in RF wires which have solid silver-plated copper-clad steel center conductor. These cables along with the DC cables are rooted through 80 K thermal plates for thermalisation.

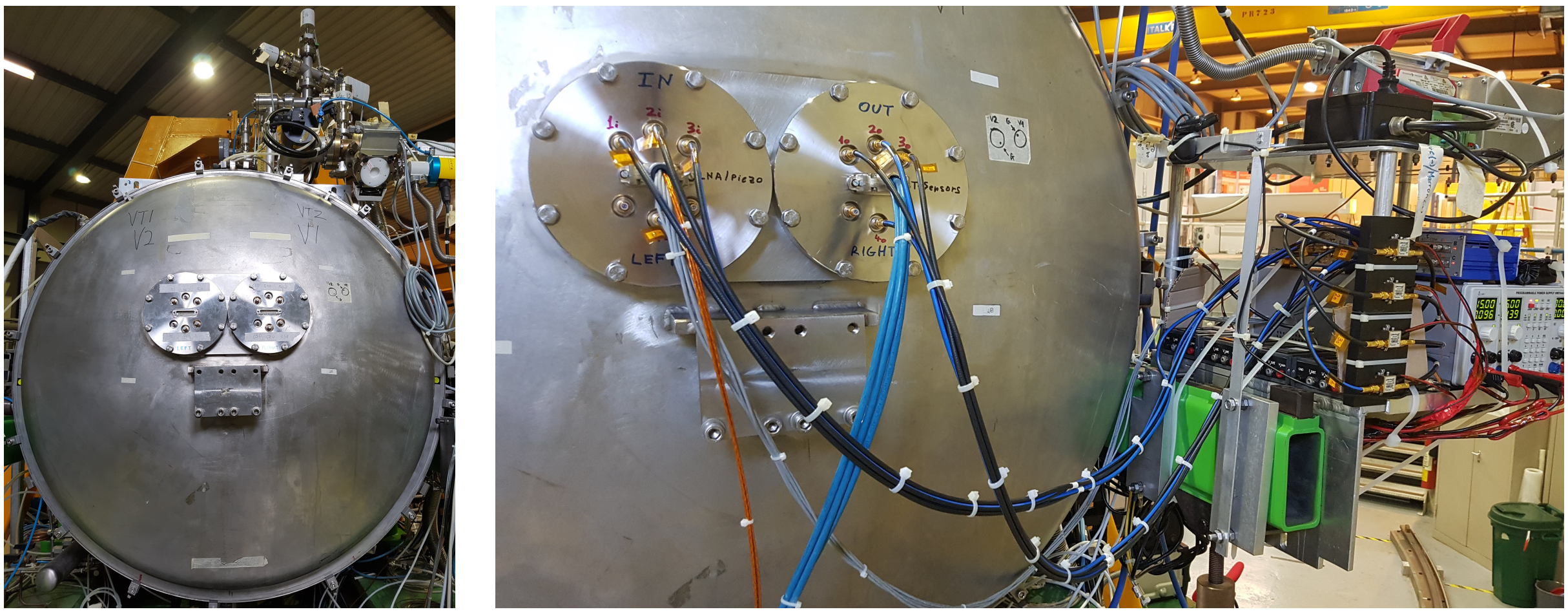

**(a)** Warm flanges in the MRB side of the CAST magnet.

**(b)** Connections from warm flanges to the room-temperature instruments.

**Figure 22.11:** The warm 300 K flanges transitioning the RF and DC cables from the cryostat to the outside of the magnet

#### 22.2.3.3 Setup outline

In Fig. 22.12 the schematic layout of the setup of CAST-CAPP detector outside the cryogenic bore of the CAST magnet is demonstrated. The various cables, from the outside instruments are connected with the cavities and the sensors inside the magnet bore through the electrical feedthroughs on input and output 300 K flanges as seen in Fig. 22.11b.

The Data Acquisition (DAQ) software system which was created in python programming language (see Appendix Sect. C) allows to connect all programmable instruments from Fig. 22.12 in the setup with the workstation Personal Computer (PC) via Universal Serial Bus (USB) or Local Area Network (LAN) connections, and performs automatic operations which enables a “hands-free” data-taking. The main procedure for data-taking is outlined in Appendix Sect. C.2. At the same time, in the DAQ procedure the storage, processing (Sect. 23.2.1) and analysis (Sect. 23.3) have been implemented in the routine and are done automatically after each daily data-taking run.

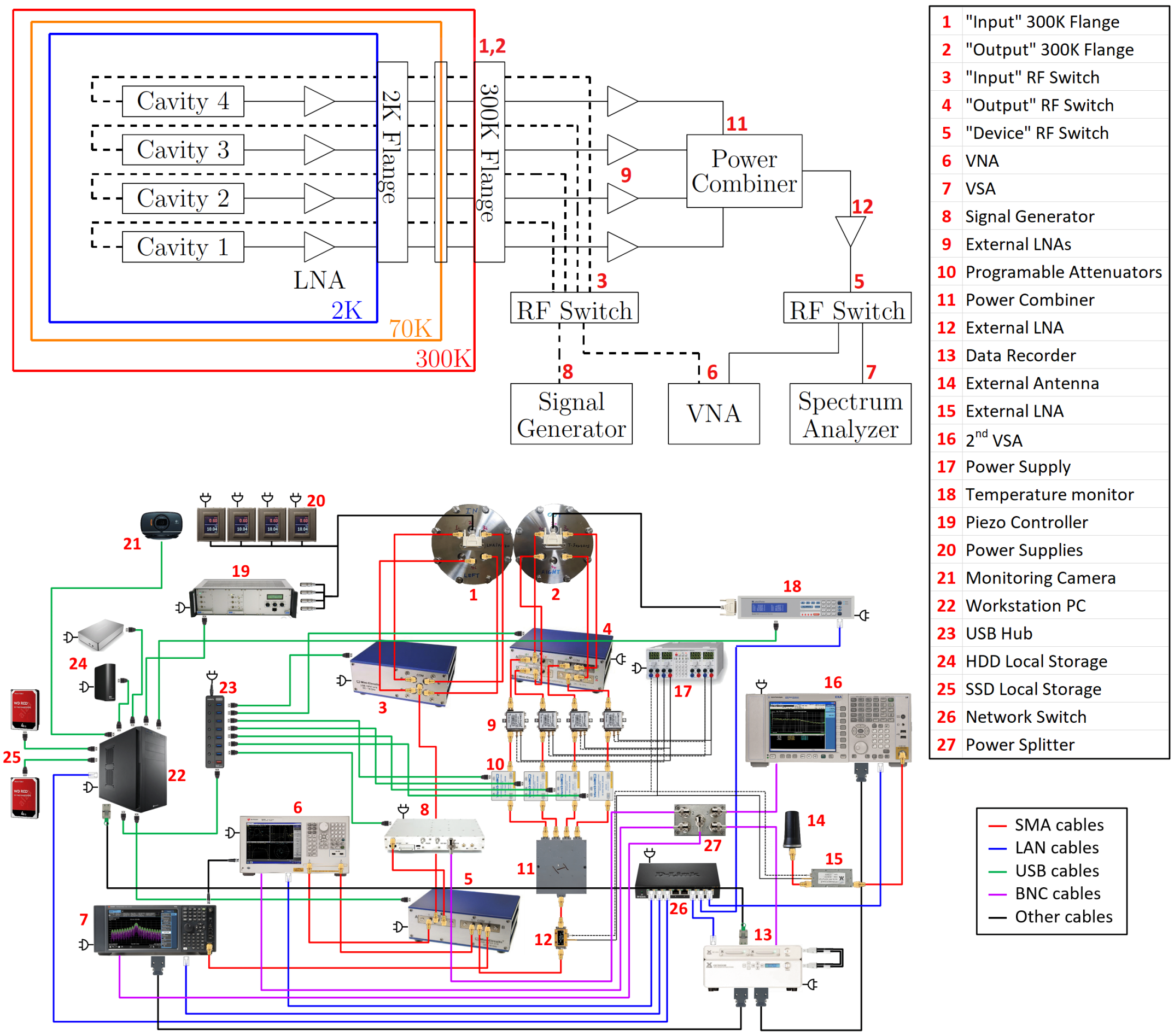

**Figure 22.12:** Top: Simplified outline of the CAST-CAPP setup (49). The dashed lines correspond to connections in each cavity input port whereas solid lines show connections in each cavity output port. The cavities are shown to be one on top of the other for visualisation purposes. They are actually aligned on the axis of their length as one after the other. Bottom: Detailed schematic for the room temperature hardware connections. The various instruments are explained in detail in Appendix Sect. C.1.

## 22.3 Measurements

### 22.3.1 Pilot tone

In several measurements, and as a test, a calibration peak from the Tone Synthesizer (TS) or the Vector Network Analyser (VNA) was used in order to verify that the whole data-taking system and the cavities themselves are responsive to an actual axion signal.

Additionally, the blind insertion of a well-defined pilot tone into the cavities in several random frequencies validates the analysis procedure, by making sure that it is sensitive to such simulated axion signals. Multiple narrowband Continuous Waveform (CW) signals as well as rectangular pulses with various widths and amplitudes were injected in a few specific frequencies so as not to contaminate all the bins of the resulting spectra. The nominal signal power was selected based on the expected action conversion power of an axion in the CAST-CAPP detector and the estimated signal power from several hours of data. Some example signal injections with a 1 Hz CW pulse and a frequency modulated 5 kHz signal within 10 ms and about $20\sigma$ significance, are shown in Fig. 22.13.

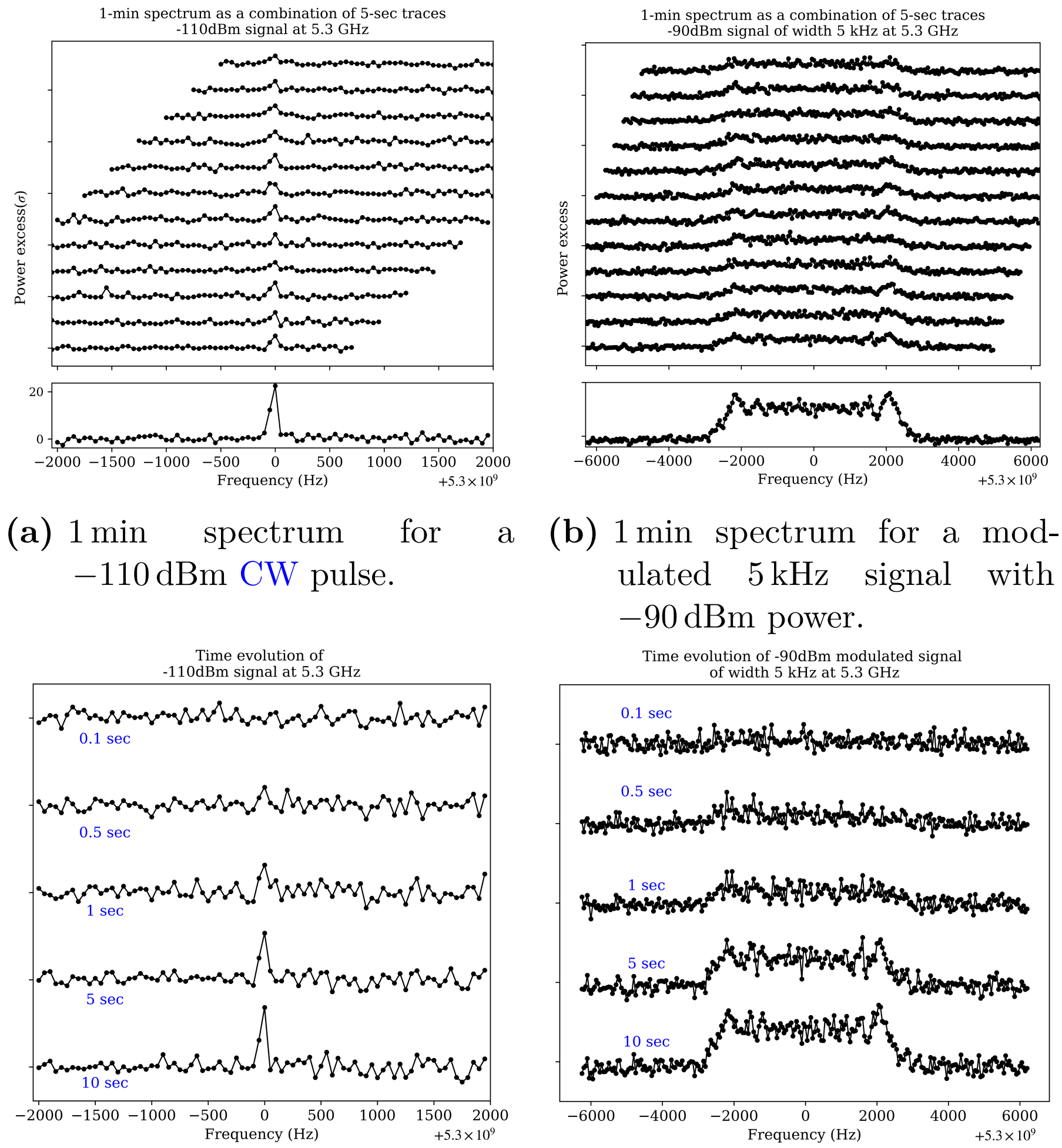

**(a)** 1 min spectrum for a $-110$ dBm CW pulse.

**(b)** 1 min spectrum for a modulated 5 kHz signal with $-90$ dBm power.

**(c)** Time evolution of the CW pulse.

**(d)** Time evolution of the modulated signal.

**Figure 22.13:** Hardware injected signals of a $-110$ dBm CW and a $-90$ dBm 5 kHz-wide sweeping signal. The combination of 12 5 min traces in a 1 min trace (upper row) as well as the time evolution (lower row) is showing the amplification of the signal as more and more data are combined, thus confirming the analysis procedure.

### 22.3.2 EMI/EMC parasites

As we will see also in Chap. 23, the data analysis is based upon power excess in the search statistics. To this end, and on the cost of having a considerable number of candidates, the threshold is set relatively low in order not to miss a potentially faint signal in the power spectra. The most time consuming and most difficult task of every RF-axion experiment around the world is the investigation of these excess candidates.

For this purpose, the real-time environmental background measurements in parallel with cavity measurements are an important part of the characterisation of potential candidates as ElectroMagnetic Interference (EMI)/ElectroMagnetic Compatibility (EMC) parasites. Since various intended and unintended ambient parasitic signals, are present in the CAST hall, a second simultaneous measurement channel has been installed with an independent Vector Signal Analyser (VSA) and an external omnidirectional antenna connected directly on the VSA operating at the same frequency band. This works like a veto counter. If a significant candidate signal is observed with both the VSA connected to the cavities which are inside the CAST magnet as well as with the second parasite-diagnostic channel with the external antenna at the same frequency, then this signal can easily be excluded as an EMI/EMC parasite. An example is shown in Fig. 23.11.

#### 22.3.2.1 Shielding efficiency

An important observation is that most of the outside parasite signals do not enter the cavities from the inside the magnet, as the magnetic body of CAST acts as a Faraday cage that shields the cavities from outside leakages. An example frequency range of 20 MHz around 5.19 GHz is shown in Fig. 22.14.

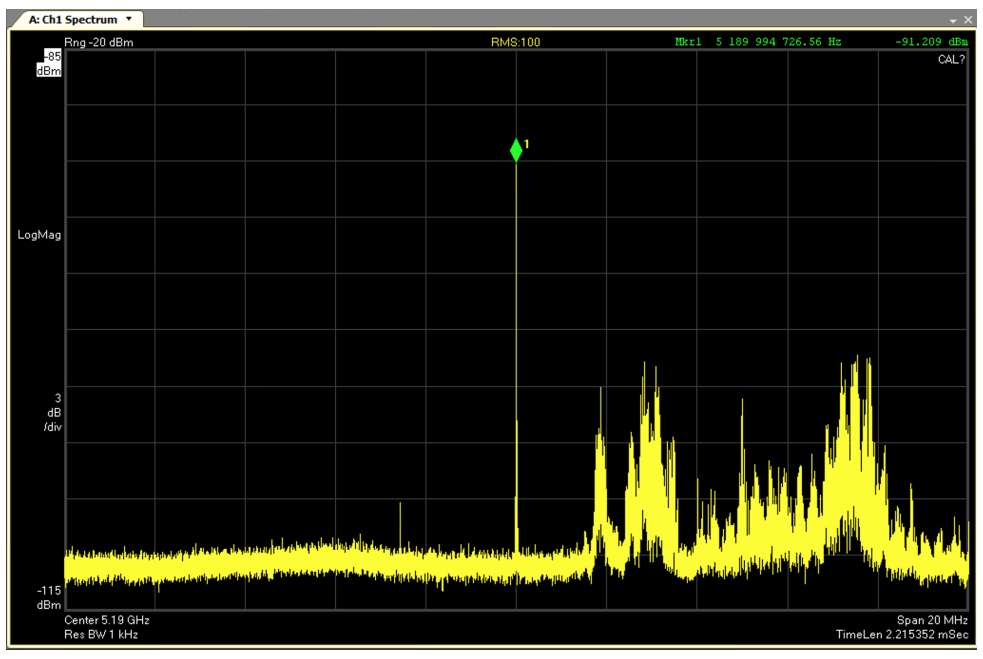

**(a)** Ambient EMI/EMC parasites around CAST hall.

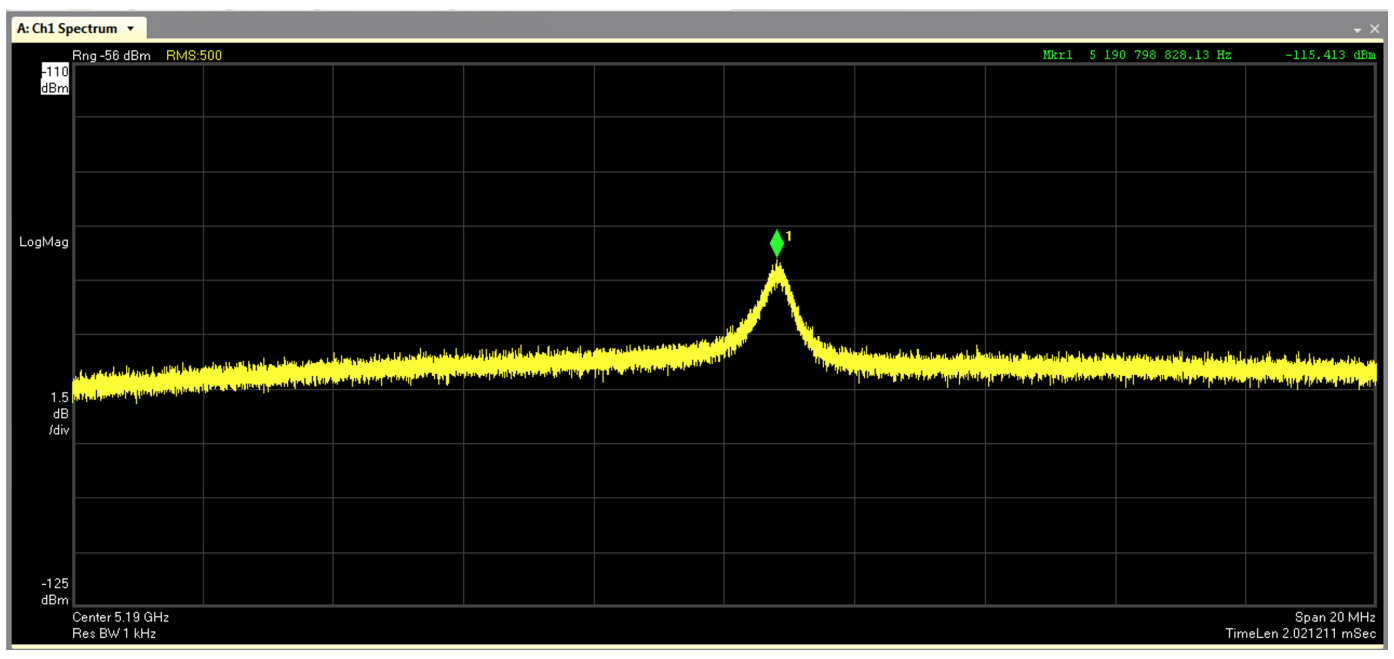

**(b)** Cavity #3 with only noise bump visible on the screen.

**Figure 22.14:** An example measurement around 5.18 GHz ± 10 MHz with both recording channels with the 89600 VSA mode of the VSA and a RBW of 1 kHz.

However, a few strong signals that make their way to the cavity measurements enter the DAQ chain through outside-the-cryostat connections, feedthroughs, cables and instruments. This is why the aforementioned simultaneous measurements between cavities and electromagnetic background noise are important for the proper characterisation of such candidates. As an example, using the VNA as a signal generator and radiating various well-defined signals through the external antenna in the CAST hall, it was verified that only strong signals beyond 30 dB amplitude can make their way int the cavity DAQ system. Therefore, signals which are less than 30 dB above the noise floor outside the magnet vessel can be excluded.

#### 22.3.2.2 Reducing EMI

As discovered with the use of the external antenna, an intended significant emitter in the CAST hall is the 5 GHz band from Wireless Local Area Network (WLAN) coming from the access points located around the area. To mitigate this issue, the corresponding channels 52-56 and 116-120, which were interfering with our measuring frequency band, have been disabled. The achieved huge suppression of these parasitic signals around the related frequencies is shown in Fig. 22.15.

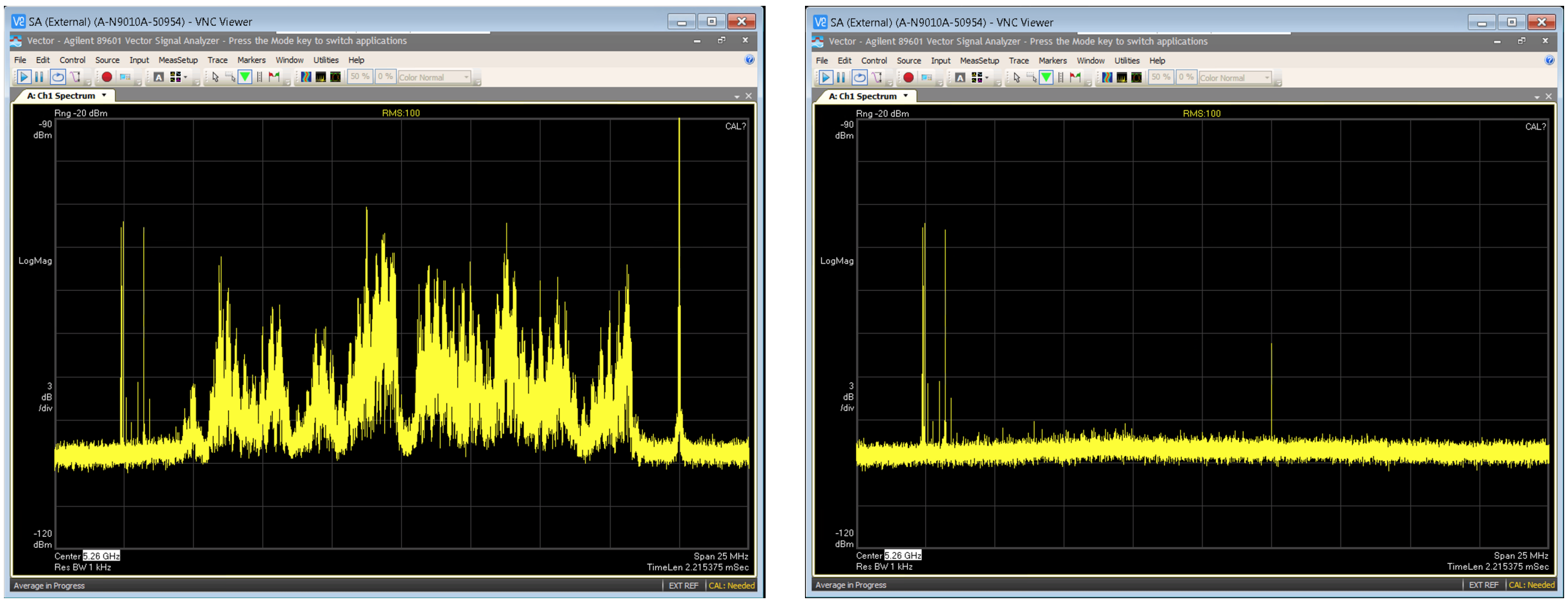

**(a)** Spectrum before disabling the channels. **(b)** Spectrum after disabling the channels..

**Figure 22.15:** Measurement by the second VSA showing power as a function of frequency for the 52-56 WLAN 5 GHz channels emitting in the CAST area.

### 22.3.3 Noise temperature

The usual method for the measurement of the noise figure of the cavities is the *Y-Factor technique* where the receiver input is connected to two loads at different known temperatures $T_C$ and $T_H > T_C$ and then the noise power in each configuration is measured. An easy way is using a Excess Noise Ratio (ENR) source which is turned on/of while measuring the change in the output noise power with a VSA. The Y factor is defined as the hot/cold noise power ratio where:

$$Y = \frac{P_H}{P_C} = \frac{T_H + T_A}{T_C + T_A} \tag{22.22}$$

which is solved for $T_A$.

In Fig. 22.16 the differences between noise source on/off for two different noise diodes of 8000 K and 200 000 K is presented. As clearly seen, even with the more powerful noise diode of 20 000 K in Fig. 22.16b, the difference is small at $\sim 0.72$ dB.

Therefore, an alternative method which was regularly used for the measurement of the noise figure of each cavity is through the 3 dB *signal generator method* (or *twice power method*).

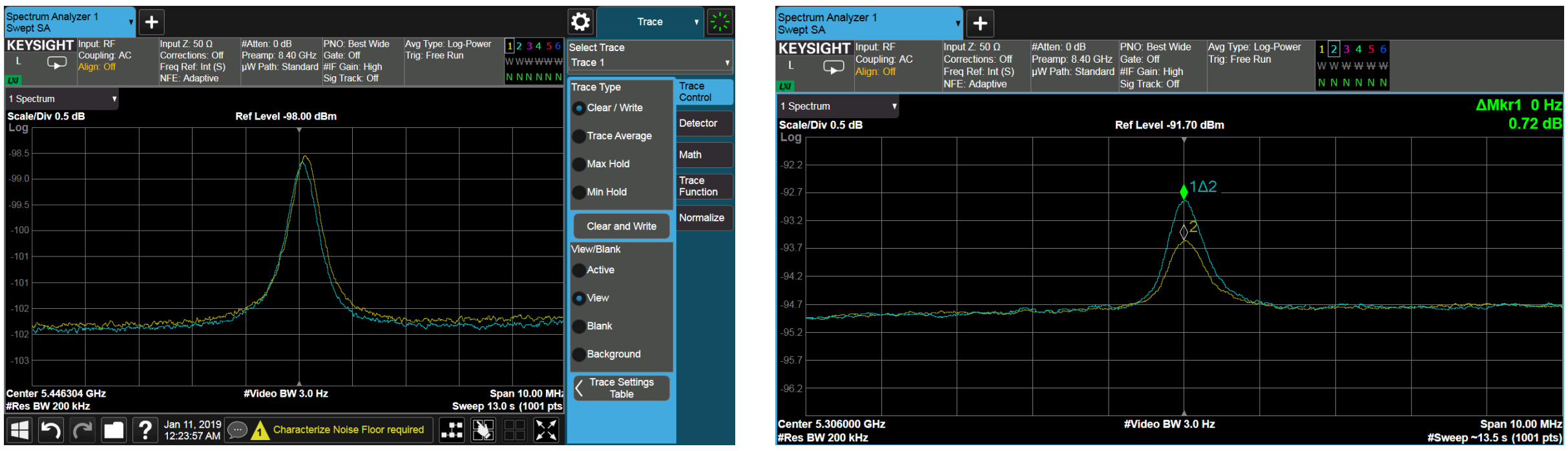

**(a)** 8000 K noise source turned on (yellow line) / off (blue line).

**(b)** 20 000 K noise source turned on (blue line) / off (yellow line).

**Figure 22.16:** Examples of noise temperature measurements with the Y-Factor method for cavity #2 in data-taking conditions.

As a TS, both the VNA and the signal generator were used depending on the availability of the instruments. The motivation for using the 3 dB method, is that much more signal strength can be provided than the usual 14 dB to 15 dB ENR available from a noise diode, such that we have big losses to the input of the LNA. An example of such a measurement is shown in Fig. 22.17. For the optimal visualisation of the desired 3 dB difference between signal injection on/off, the specific parameters of the swept SA mode of the VSA that are used include a Resolution Band Width (RBW) of 100 kHz, a Video Band Width (VBW) of 1 Hz and a span of 0 Hz.

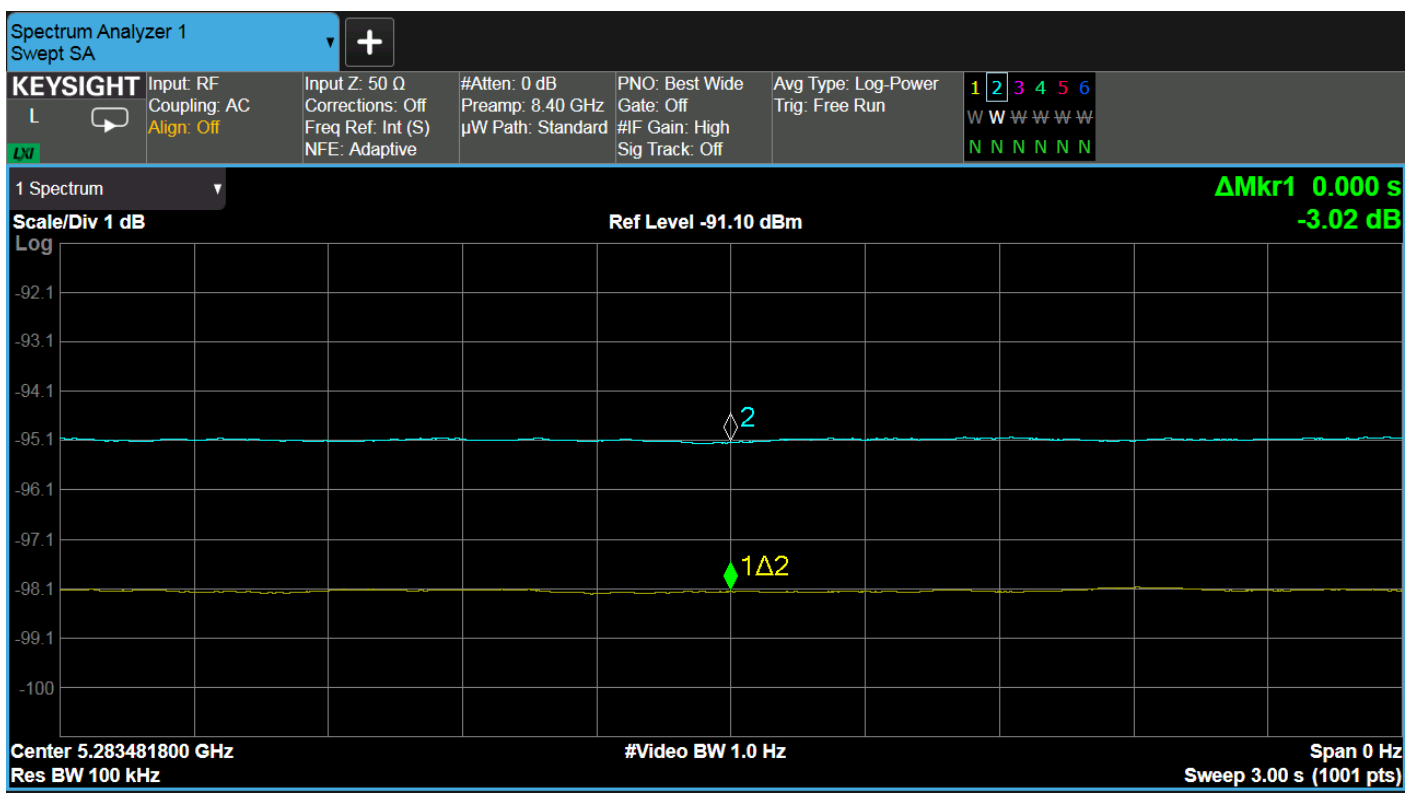

**Figure 22.17:** An example noise temperature measurement with the 3 dB method in data-taking conditions for cavity #2 without (yellow line) and with an input signal of −93.2 dBm (blue line).

From the noise figure measurement, and by considering the total attenuation to the input of the cavity as well as its many uncertainties, the noise temperature $T_S$ is calculated to have an upper limit of 10 K with an estimated uncertainty on the order of ±1 K. An additional calculation of the order of magnitude of $T_S$ is derived from the sum of the ambient temperatures of the cavities $T_{\text{cav}}$ plus the noise from the cryogenic LNA $T_{LNA}$, resulting also to an upper limit in $T_S$ of 10 K. Additionally, the loss of the connecting cable between the input of the amplifier and the output of the cavity at cryo was evaluated to be below 0.1 dB, and therefore its contribution to the system noise temperature $T_S$ is estimated to be safely below 0.2 K.

#### 22.3.3.1 Attenuation estimation

Due to the presence of the LNA on the output cable of each cavities, it is impossible to make a direct reflection measurement of the strongly coupled port of each cavity using the VNA, as there is no direct access. However, for the input cables which go all the way to the weakly coupled input port of each cavity, we can make a reflection measurement with the following procedure. The first step is to use the *lowpass impulse* (see Fig. 22.18a) and *lowpass step* (see Fig. 22.18b) mode of the VNA in time-domain option. This way the reflection from the cavity can be located on the spectrum (marker #6 in Fig. 22.18b). Then by using the *bandpass impulse* mode (see Fig. 22.18c), the attenuation, as well as the electrical delay of all the cables up to the input port of each cavity, can be measured. In this example the two-way attenuation measured is −50.12 dB which means −25.06 dB one-way attenuation. Similarly, the one-way electrical delay measured here is about 78 ns.

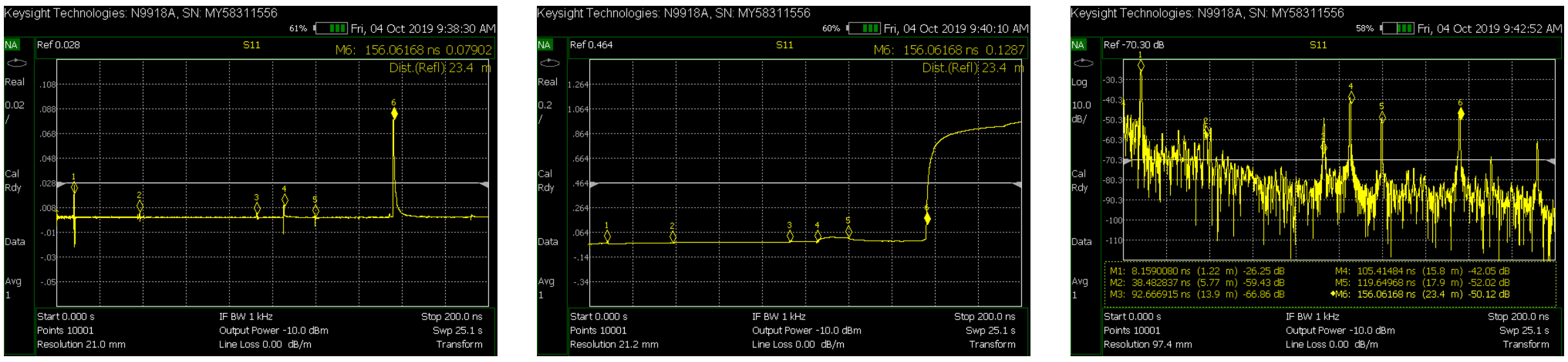

**(a)** Time lowpass impulse mode. **(b)** Time lowpass step mode. **(c)** Time bandpass impulse mode.

**Figure 22.18:** Cavity #1 S11 reflection measurements at the very end of the input cable for data-taking conditions. No calibration kit has been used in these specific measurements.

The same measurements have been performed for the rest of the cavities with the calculated one-way input attenuation for cavities #2, #3 and #4 being −24.96 dB, −23.85 dB and −21.43 dB respectively. These measurements were performed at the end of the input SubMiniature version A (SMA) cable in (6) of Fig. 22.12. The same ones have been done by connecting the measuring VSA directly to the feedthroughs of the input 300 K flange. This measurement indicated an attenuation of about −10 dB for the cables and connections in room temperature and about −15 dB for all connections and cables inside the magnet. An additional information that can be obtained from Fig. 22.18a and Fig. 22.18b is about he losses of each cable connection all the way to the cavity. Finally, all these measurements were also performed as a function of the temperature of the cavities during cool-down and as a function of the magnetic field $\vec{B}$ during ramping-up of the magnet.

### 22.3.4 Phase-Matching

To apply the PM technique we first need to perform a frequency matching of the selected cavities, which has to be within the allowed FMT, and then an amplitude matching of the resonant peaks. Finally, PM is achieved by adjusting the output signals of each cavity aiming for phase matching of the axion signal. Obviously the temperatures of the selected cavities also have to be similar within a few K.

The combination of the output signals from the four cavities through the power combiner is done after the individual amplification of each signal which means that the derived sensitivity of CAST-CAPP when using all four PM cavities is four times bigger comparing with the usage of single cavities. Finally, as explained in relation 22.2 the last condition for axion haloscopes is that their size should be smaller than the corresponding de Broglie wavelength $\lambda$ of the non-relativistic axions. Therefore, for CAST-CAPP which is searching for axions with mass of approximately $20\,\mu\text{eV}/c^2$, and considering $\beta = v/c \sim 10^{-3}$, we have from Eq. 21.23:

$$\lambda_a \sim 62\,\text{m} \tag{22.23}$$

When combining all four cavities of CAST-CAPP the total length of the detector is approximately $4 \times 0.5\,\text{m} = 2\,\text{m}$, since the length of each individual cavity is approximately 0.5 m including the locking mechanism and the tuning motors. Therefore, this length is well bellow the limit of Eq. 22.23, and thus the condition 22.2 is still fulfilled.

#### 22.3.4.1 Frequency-matching

For the case of CAST-CAPP detector we have $Q_L \simeq 2 \times 10^4$, and $\nu \simeq 5\,\text{GHz}$. Therefore, from Eq. 22.21 we get a $FMT \simeq 105\,\text{kHz}$. However the precise piezoelectric motors allow for a smooth tuning with an accuracy of less than 10 kHz which is well bellow the allowed FMT of around 100 kHz. As a result, this technically challenging concept of frequency-matching has been achieved with all four coherent cavities within $\sim \pm 10\,\text{kHz}$ in data-taking conditions (see Fig. 22.19).

#### 22.3.4.2 Amplitude-matching

As mentioned, an additional condition for PM is the matching of the amplitudes of the resonance peak of the cavities that are used, in order to get an equal weight on the SNR from the individual cavities. From Eq. 22.11 we have seen that for the four cavities:

$$SNR \sim \frac{\sum_{i=1}^{4} A_i S_i}{\sum_{i=1}^{4} A_i n_i} \tag{22.24}$$

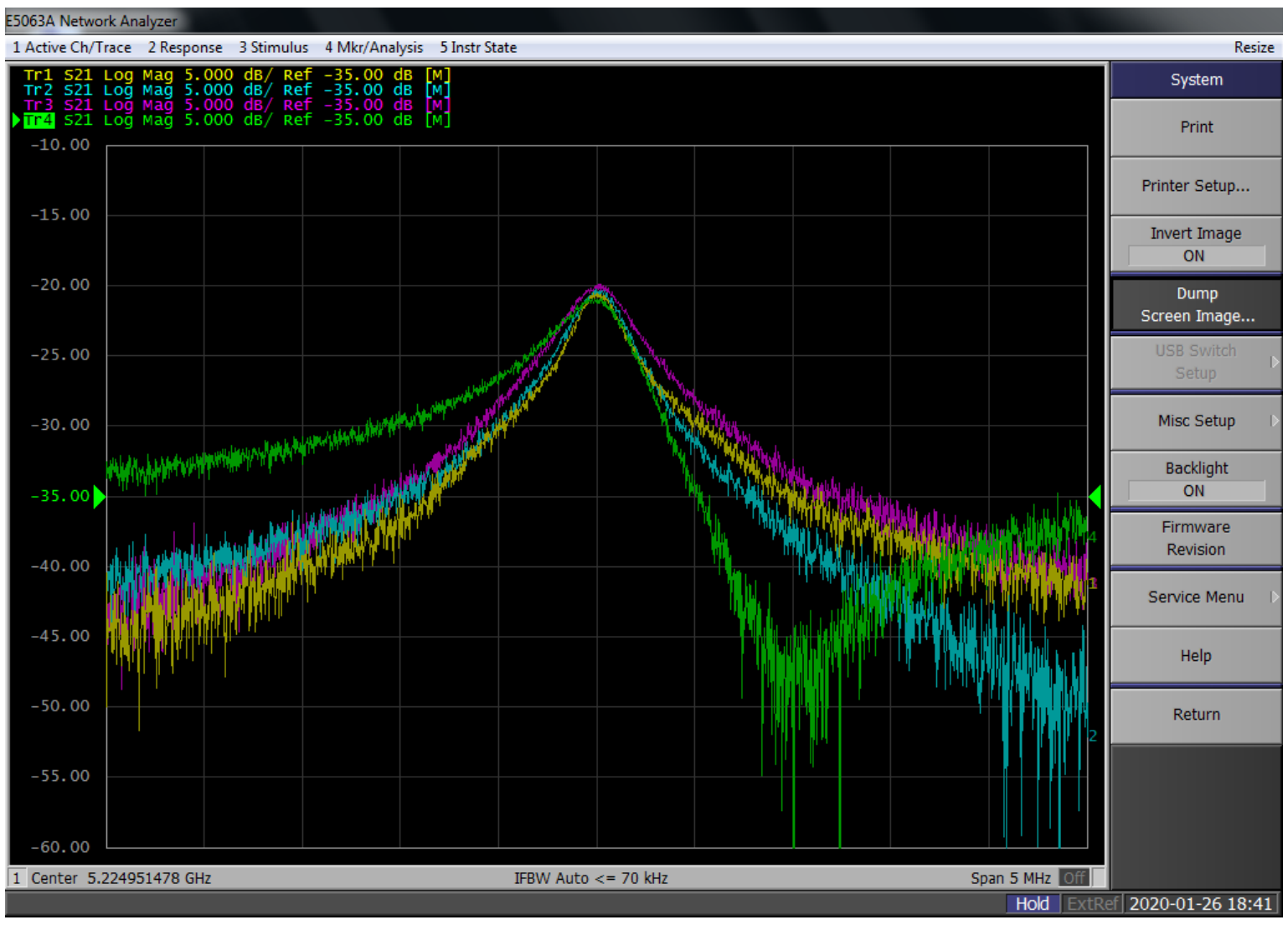

**Figure 22.19:** All four cavities of CAST-CAPP phase-matched within ±10 kHz and ±0.5 dB in data-taking conditions.

where: $A_i$ : the amplitudes of the resonant peaks of each cavity,

$S_i$ : the signal powers of individual cavities,

$n_i$ : noise of individual cavities.

Therefore, to get the $A_i$ terms to cancel out, so as to have a factor four enhancement of the SNR, the four amplitudes have to be equal: $A_1 = A_2 = A_3 = A_4$. If, for example, at least one of the amplitude terms is considerably higher than the rest, then all the rest amplitude terms would be negligible and the resulting SNR would be similar to that of a single cavity.

The amplitude matching of the cavities, is done by four phase-independent programmable attenuators, one for each cavity ((10) in Fig. 22.12). The weakest signal is used as a reference and then the rest of the signals are adjusted via the phase-independent programmable step attenuators for each cavity path. The tolerance of these attenuators is 0.25 dB, and the amplitude-matching is done when the resonant peaks are observed by the VNA from transmission measurement right before each data taking step (see Fig. 22.19). The reason this is done independently before each measurement is because the amplitude of the resonant peak of each cavity changes while the cavity is tuned to different frequencies.

#### 22.3.4.3 Phase-matching of output signals

The last condition for the PM technique is the matching of the phases of sine waves, in case there is a signal. This was done by adjusting the electrical delays of each cavity line before the signals arrive at the power combiner. The adjustment is done by inserting low-loss delay lines, between the external amplifiers ((9) in Fig. 22.12) and the programmable attenuators ((10) in Fig. 22.12). The delay line is a frequency dependent phase shifter with a linear relation of phase-shift vs. frequency. This is what we are aiming at in order to achieve tracking of the phase-matching conditions over our frequency range without re-tuning the phase-shifting elements at each frequency step. More specifically the phases were adjusted for coarse tuning by using short cable sections below 5 cm length and for the remaining difference, by continuously adjustable coaxial line stretchers. The result of the phase adjustment within 1° between the four cavities is shown in Fig. 22.20.

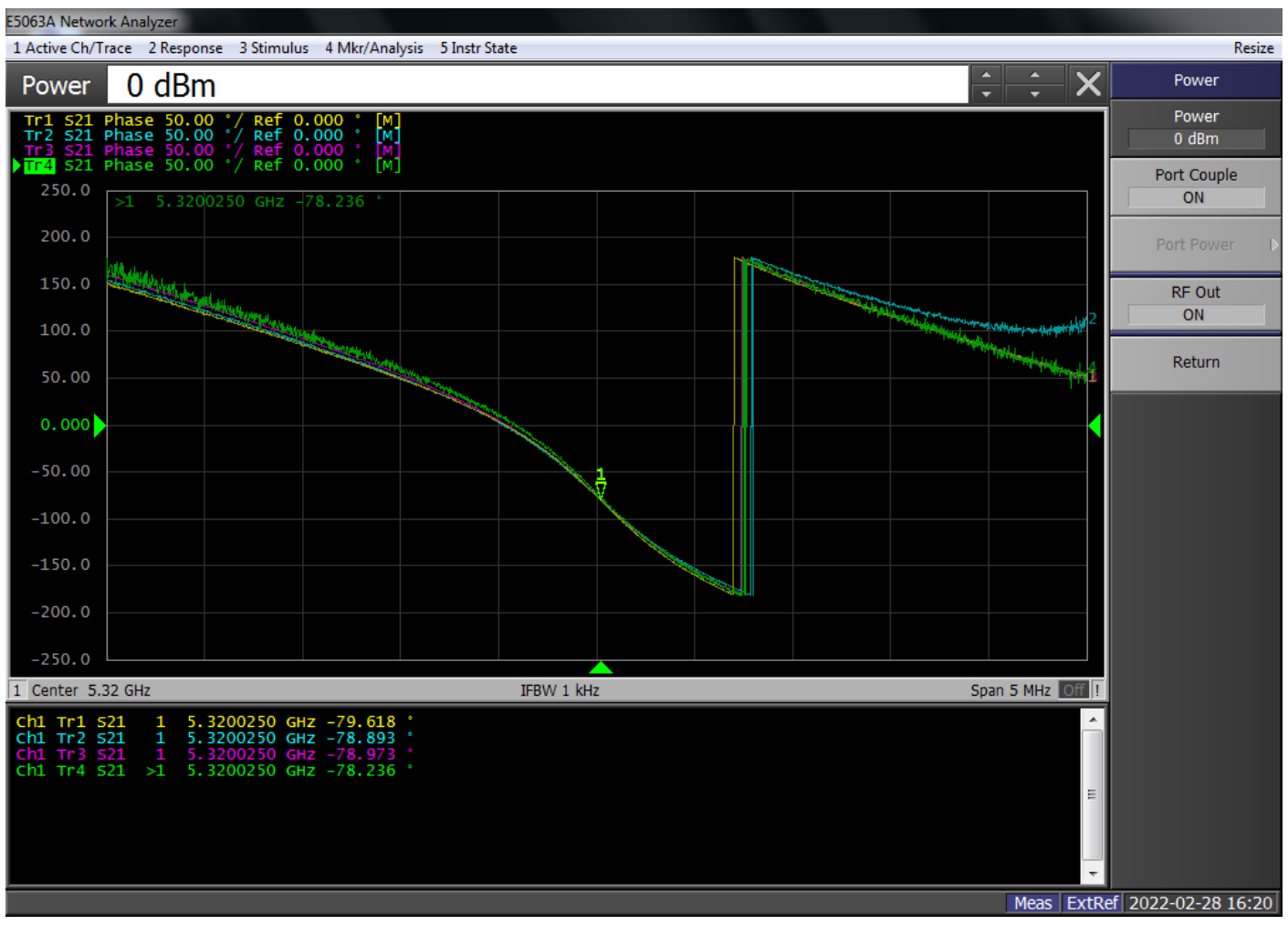

**Figure 22.20:** Transmission S21 measurement with the VNA in *Phase* format after the phases for each cavity have been adjusted within 1°. Each trace corresponds to a single cavity i.e. yellow trace is cavity #1 blue is cavity #2 and so on.

### 22.3.4.4 Procedure verification

In Fig. 22.21 the outputs of the three PM cavities (yellow line) are compared with the cases where one (blue line) or all three cavities (magenta) are detuned. It is noted that the noise for all the cases looks the same as it is expected. Here the combination of uncorrelated noise is in line with the theoretical expectations from Eq. 22.18.

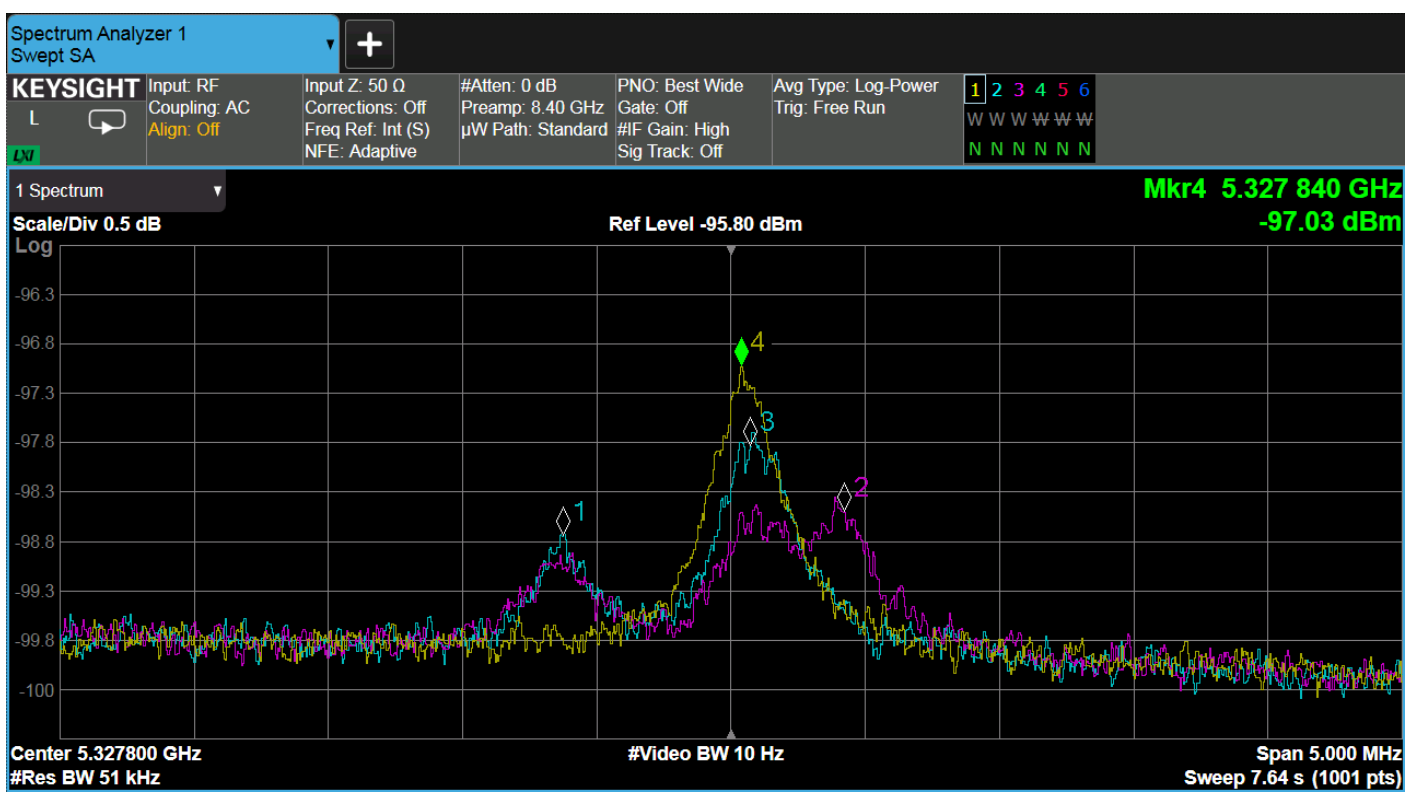

**Figure 22.21:** Swept SA mode of the VSA with the traces of three PM cavities combined (yellow line), two PM cavities and one detuned (blue line), and three single cavities in slightly different frequencies (magenta line) are shown. The RBW and VBW are chosen to be 51 kHz and 10 Hz respectively.

Finally, an additional crosscheck of the PM procedure is made by inserting into the cavities a well-defined coherent signal, using an appropriate power splitter and measure the cavity responses with the VSA. For this reason the procedure from Sect. 22.3.4.3 has to be repeated for the input lines so as a coherent signal to be injected from the TS to all four cavities simultaneously mimicking the axion field. The power gain in this case for the coherent signal is 16 times for the four cavities compared to a single cavity, based on Eq. 22.19. This way, the measured signal height of the four PM cavities is four times the height of a single cavity when the SNR in the individual cavities is the same. An example for three cavities is shown in Fig. 22.22 with the measured power of the coherent signal increasing by about a factor of 9 (9.54 dB). The minor deviation from the theoretical value are due to imperfections of the 4-to-1 power combiner and the reading accuracy of the marker. It is noted, that in this case the signal injected from the signal generator is far above the noise level and the SNR is very high.

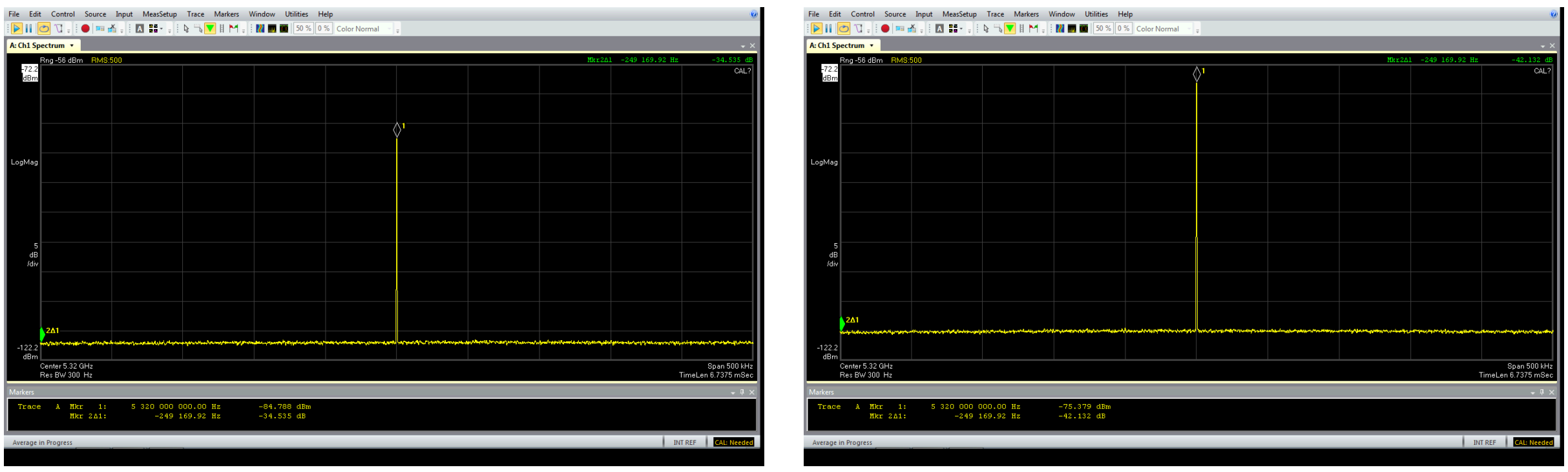

**(a)** Single cavity. **(b)** Three PM cavities.

**Figure 22.22:** Comparison of a coherent CW signal injected by the TS on the center of the resonance peak between a single and three PM cavities.

### 22.3.5 Data-taking strategy

The CAST-CAPP detector takes advantage of both the PM technique as well as the fast tuning system which allows the cavities to take data in a broad frequency range of about 660 GHz. To cover the maximum parameter space both single and PM cavity configurations are used. At the same time, the CAST-CAPP detector is tuned as fast as possible to take advantage of streaming DM without losing sensitivity in halo DM axions. An important part of this strategy is the processing and analysis of the raw data in a daily basis aiming for an accurate adjustment of the data-taking schedule for the following days. Following the predefined protocol, a given frequency is revisited multiple times for shorter periods rather than dwelling on it once for a long time interval. Thus, the axion mass range is scanned horizontally with the sensitivity improving progressively. This gives sensitivity not only to ALPs but also to axion mini-clusters, caustics and streams. In this work, the fast tuning mechanism allows primarily to quickly re-tune the cavities to a certain frequency in order to cross-check the nature of an outlier.

The data from CAST-CAPP are taken when there is no solar tracking with the CAST magnet since mechanical vibrations from the movement of the magnet are transmitted in the tuning strips of the cavities and contaminate the measurements (see Sect. 22.2.2.1). The time during the solar tracking of the CAST magnet which is required for the solar axion search (see Sect. 21.3.1), is used from CAST-CAPP for post-data-taking processes such as the offloading, storage, processing and analysis. After the offloading procedure has finished, the data-taking can restart in parallel with the running of the rest of the procedures (storage, processing and analysis).

After the analysis of the data, the first step is the comparison of the data taken from the cavities with the ones taken from the second independent channel measuring the background EMI/EMC noise. In the case of a presence of a possible significant excess only in the cavity channel, a rescanning of the corresponding frequency(ies) with the same and with different cavity combinations is immediately planned. The new data are then analysed both independently as well as combined with the original measurement(s). The reason for using also different cavities is the exclusion of any cavity-electronics-related noise. The different cavity combinations should also not affect the detection capabilities, except the sensitivity boost for axion candidates due to PM. If a candidate persists, then the next step is an extra rescanning by tuning the corresponding cavity(ies) to a higher-order resonant mode which does not couple to axion. The final step of the candidate elimination procedure involves rescanning with a varying magnetic field. This is to verify the expected behaviour following the $\vec{B}^2$ dependence as shown in Eq. 22.3. As an example, a rescanning of the candidate frequency with $\vec{B} = 0\,\mathrm{T}$ would eliminate an axion-generated signal. Lastly, some $\vec{B} = 0\,\mathrm{T}$ data are also required to help to rule out any electronics-related background which can not always be picked up by the external antenna as it is not part of the cavity setup.

## 22.4 Streaming DM axions

### 22.4.1 Fast resonant frequency tuning

In addition to searches for conventional axions, CAST-CAPP introduced for the first time in axion DM research, the *fast resonant scanning* technique. Thanks to it's fast scanning mechanism, CAST-CAPP detector is also sensitive to DM axion tidal or cosmological streams as well as to theoretically motivated axion mini-clusters (see Chap. 2 and Sect. 21.4). A quite wide axion mass range can be scanned within a time period of a few hours to eventually take advantage of streaming DM towards the Earth (see Sect. 21.4.2). We have already seen in Chap. 3 that the axion flux enhancement due to gravitational focusing by the Sun can be up to $\sim 10^8$ and in the ideal case as high as $\sim 10^{11}$. Apparently, the faster the scanning the shorter axion bursts can be detected. The maximum scanning speed of CAST-CAPP is 10 MHz/min, and therefore, its full tuning range of ∼400 MHz can be covered within ∼1 h. The maximum tuning speed is limited by the torque of the piezoelectric motors in cryogenic conditions.

The ideal step size of tuning that is used is on the order of the cavity frequency mode peak width $\sim \nu_0/Q_L$. For our setup with $\nu_0 \sim 5$ GHz and $Q_L \sim 20000$ this results to $\sim 250$ kHz but in order to account for slight changes in both variables the tuning step size was defined at 200 kHz. The integration time was chosen to be 1 min for each 200 kHz data taking step in every run.

Finally, it should be mentioned that although this technology is capable of detecting transient events from streaming DM, caustics or axion mini-clusters, an unambiguous detection of such events requires an independent detector at another location.

### 22.4.2 Wide-band scanning

An alternative to the fast tuning technique that has been tested and established for the first time is the *wide-band scanning* technique which is based on an out-of-resonance scanning, abandoning the resonance enhancement factor $Q$. This is the fastest technique that could continuously scan for DM axions but the sensitivity does not allow a simultaneous search for halo axions. For that reason, this technique was devised as "plan B". The idea is based on the fact that the power output of the axion to RF-photon conversion is proportional to the quality factor with a Lorentzian tail. The RF receiver's sensitivity is maximised around the cavity's resonant frequency, but the sensitivity away from the resonance is not zero. It is decreased by the response of the approximately critically coupled (towards the receiver) cavity in transmission according to its $\nu_0$ and $Q_L$. More specifically, the power of the cavity resonance mode is inversely proportional to the square of the frequency distance from the centre of the resonance mode.

However, this lack of sensitivity could be more than compensated by a temporally large axion flux enhancement. Thus, by changing the centre frequency of the VSA while keeping the resonant frequency of the cavity $\nu_0$ at a fixed position CAST-CAPP detector can detect an enhanced streaming axion signal. The minimum scanning period for this technique is ∼5 min for the maximum ∼160 MHz useable range and can be done automatically for several hours. Using all four cavities with their resonant frequencies fixed in a position where the distance between them is maximum, a strong streaming axion signal can be detected in a total range of about 1 GHz. It is worth noting that the electronics readout of the microwave cavities are kept broad-band, limited only by the span of the VSA that is used to read out the data (see also Publication E.26).

The transmissivity of the RF-chain system, has been tested as shown in Fig. 22.23. In this example, a variety of signal powers have been injected to cavity #1 with a TS and a measurement of the signal amplitudes as a function of frequency distance from the centre of the resonance peak has been made. More specifically, signals ranging from −120 dBm to −50 dBm with 10 dBm steps have been injected, and the amplitude with respect to noise floor has been measured from 0 (center of resonance peak) to −300 MHz with 10 MHz step, as the span of the VSA was also 10 MHz. As seen in Fig. 22.23, signals smaller than −110 dBm, as well as all the signals in frequencies below −160 MHz were not detected. For the signal amplitudes, it is noted that the attenuation from the position of the injected signals, to the input of the cavity, is about 25 dB. Therefore, the actual signals arriving inside the cavity were smaller from the ones ejected by the TS by about 25 dB. Finally, the left side of the resonance peak was chosen as there are no more resonances and the mismatch of the input frequency with the cavity resonance means that most of the power is reflected due to impedance mismatch. On the other hand, on the right side of the resonance peak, there are additional resonant modes with a negligible coupling strength that can still make it through.

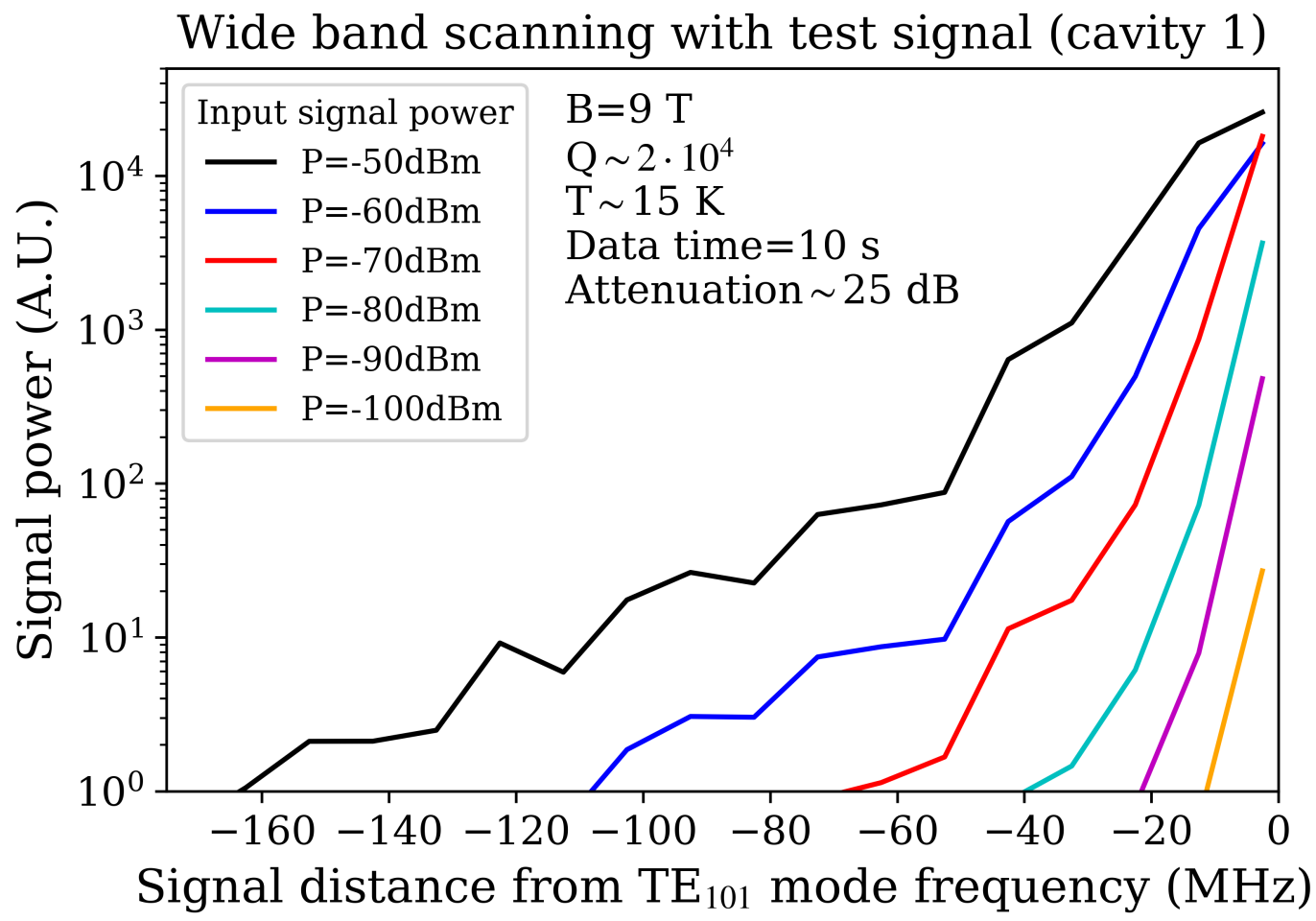

**Figure 22.23:** The measured power transmission vs. off the cavity resonance frequency for wide-band scanning measurement mode in data-taking conditions. Output measured power is plotted as a function of the frequency distance from the $TE_{101}$ resonant mode which couples to axions.

# 23

# CAST-CAPP RESULTS

## 23.1 Data-taking results

### 23.1.1 Commissioning runs

During the first installation of the CAST-CAPP cavities in CAST magnet in 2018, the first characterisation of the performance of the cavities along with the tuning mechanism, the amplifiers as well as the clarification of the quality of the recorded data, took place. However, due to malfunctioning of several systems, including two piezoelectric tuners, and a cryogenic LNAs, only 134 h of data with cavity #4 on a fixed frequency, with a 5 MHz span were taken. These data correspond to a total size of 13.3 TB.

Similarly, during the first half of 2019, an upgraded was performed on the whole DAQ chain both in hardware and software. An extra small run with only cavity #3 took place, with 27 h at a fixed frequency with 5 MHz span, and 260 min while tuning the cavity in a total frequency range of 240 MHz. All these data correspond to a total of 10.8 TB in CERN's tape system. These data however are not included in the results that will be presented here due to uncertainties in several experimental parameters.

Moreover, after the final installation of all four cavities in the CAST magnet in August 2019, the wide-band scanning technique, presented in Sect. 22.4.2, was first established and tested. With this technique measurements with all four single cavities were taken, resulting in a total of 10 h for 300 MHz frequency range. The analysis of these data is currently postponed to a later stage.

### 23.1.2 Data acquisition statistics

The main data-taking of CAST-CAPP took place from 12/09/2019 to 21/06/2021 with the total data size of these data in CERN Tape Archive (CTA) being 387 TB. During this period both the fast frequency tuning technique presented in Sect. 22.4.1 was used as well as the PM technique described in Sect. 22.1.5. The total scanned frequency range is $\sim$ 660 MHz, and spans from 4.774 GHz to 5.434 GHz which corresponds to axion masses between 19.74 µeV to 22.47 µeV. The tuning was also performed in steps of 200 kHz for a total acquisition time of 4123.8 h.

Additionally, the actual data-taking campaign including holidays and intervention periods lasted 342 d. The detailed periods are shown in Tab. 23.1. This means that the efficiency of aforementioned the data-taking campaign was about 12 h/d with the upper limit due to hardware and software limitations being at about 20 h/d. This difference between the theoretical and the actual number is explained by data-taking time assigned to $\vec{B} = 0\,\mathrm{T}$ measurements, wide-band scanning measurements (see Sect. 22.4.2), as well us magnet quenches, and various software and hardware interventions.

**Table 23.1:** CAST-CAPP data acquisition periods and efficiency.

| Dates [dd/mm/yyy] | Number of days | Data-taken [h] | Efficiency [h/d] |
|---|---|---|---|
| 12/09/2019 - 01/12/2019 | 81 | 637.9 | 7.88 |
| 15/01/2020 - 28/01/2020 | 14 | 187.8 | 13.41 |
| 28/07/2020 - 13/10/2020 | 78 | 635.3 | 8.15 |
| 18/11/2020 - 16/12/2020 | 29 | 576.5 | 19.88 |
| 02/02/2021 - 21/06/2021 | 140 | 2086.3 | 14.90 |
| Total | 342 | 4123.8 | 12.06 |

In Fig. 23.1 we can see the amount of data-taking time for each cavity including both single and PM cavity configurations as a function of the covered frequency range. This is in line with Tab. C.1 as well as with Tab. 23.2 and 23.3.

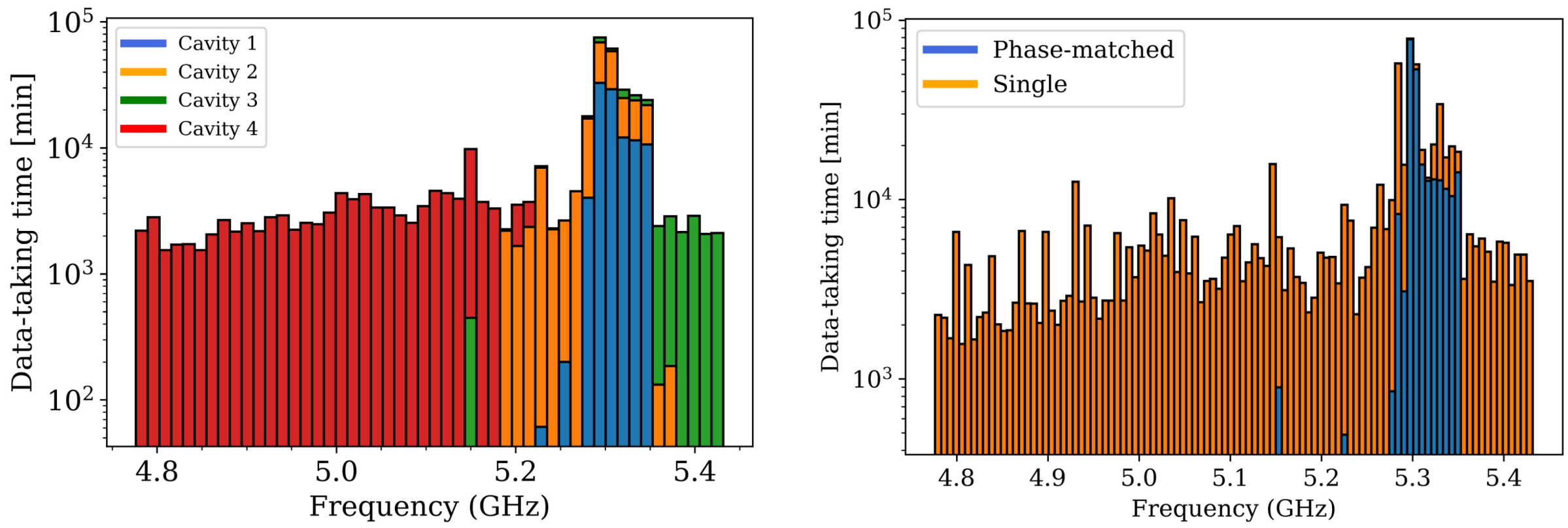

**(a)** Individual cavities including both single and phase-matched configurations (49).

**(b)** Single and phase-matched cavity configurations.

**Figure 23.1:** Histograms of data-taking time for the various cavity configurations as a function of the covered frequency range.

#### 23.1.2.1 $\vec{B} = 8.8\,\mathrm{T}$ data

The specific measurements performed with $\vec{B} = 8.8\,\mathrm{T}$ with all four CAST-CAPP cavities while using them as single cavities is shown in Tab. 23.2. On the other hand, when the signals of several cavities where combined coherently through PM in various configurations, resulted in the data presented in Tab. 23.3.

**Table 23.2:** CAST-CAPP data acquisition statistics for single cavities with $\vec{B} = 8.8\,\mathrm{T}$ .

| Cavity configurations | Time [d] | Frequency range [GHz - GHz] | Total frequency range [MHz] |
|---|---|---|---|
| Cavity #1 | 1.5 | 5.255 - 5.354 | 98.4 |
| Cavity #2 | 27.2 | 5.182 - 5.381 | 199.0 |
| Cavity #3 | 12.7 | 5.293 - 5.434 | 140.7 |
| Cavity #4 | 66.1 | 4.774 - 5.246 | 471.9 |
| Total | 107.5 | 4.774 - 5.434 | 660.2 |

The total data-taking time for both single and PM cavities can be derived by summing the results from Tab. 23.2 and Tab. 23.3, which gives 171.8 d with the total frequency range being 4.774 GHz - 5.434 GHz giving a total of 660.2 MHz of covered parameter space. An additional information from these numbers is that 62.57% of the data were taken with single cavities while 37.43% with PM cavities.

**Table 23.3:** CAST-CAPP data acquisition statistics for PM cavities with $\vec{B} = 8.8\,\mathrm{T}$.

| Cavity configurations | Time [d] | Frequency range [GHz - GHz] | Total frequency range [MHz] |
|---|---|---|---|
| PM Cavities 1-2 | 54.7 | 5.277 - 5.354 | 76.0 |
| PM Cavities 1-3 | 1 | 5.284 - 5.299 | 15.3 |
| PM Cavities 2-3 | 0.1 | 5.293 - 5.298 | 5.5 |
| PM Cavities 3-4 | 0.3 | 5.148 - 5.153 | 5.1 |
| PM Cavities 1-2-3 | 8.2 | 5.292 - 5.330 | 37.7 |
| PM Cavities 1-2-3-4 | 0.1 | 5.222 - 5.227 | 5.0 |
| Total | 64.3 | 5.148 - 5.354 | 206 |

#### 23.1.2.2 $\vec{B} = 0\,\mathrm{T}$ data

In order to exclude possible axion candidates, a considerable amount of background data, i.e. with $\vec{B} = 0\,\mathrm{T}$ have also been recorded. In Tab. 23.4 the specific frequency ranges and data-acquisition time for each cavity when used as a single entity is presented, while in Tab. 23.5 the case for PM cavities is shown.

**Table 23.4:** CAST-CAPP data acquisition statistics for single cavities with $\vec{B} = 0\,\mathrm{T}$.

| Cavity configurations | Time [h] | Frequency range [GHz - GHz] | Total frequency range [MHz] |
|---|---|---|---|
| Cavity #1 | 1.1 | 5.285 - 5.290 | 5.1 |
| Cavity #2 | 307.6 | 5.190 - 5.342 | 151.4 |
| Cavity #3 | 55.7 | 5.323 - 5.433 | 110.0 |
| Cavity #4 | 25.1 | 4.789 - 5.187 | 398.6 |
| Total | 389.4 | 4.789 - 5.433 | 644.7 |

**Table 23.5:** CAST-CAPP data acquisition statistics for PM cavities with $\vec{B} = 0\,\mathrm{T}$.

| Cavity configurations | Time [h] | Frequency range [GHz - GHz] | Total frequency range [MHz] |
|---|---|---|---|
| PM Cavities 1-2 | 5.2 | 5.285 - 5.290 | 5.2 |

The total data-taking time for both single and PM cavities when $\vec{B} = 0\,\mathrm{T}$ as seen from Tab. 23.4 and 23.5 is 394.6 h or 16.4 d with the maximum frequency range being 4.789 GHz - 5.433 GHz giving a total of 644 MHz.

#### 23.1.2.3 External antenna data

The first identification of the parasitic environmental background electromagnetic signals was done for the extended frequency range of 100 MHz - 6 GHz with 1 min measurements per 5 MHz steps. The reason for scanning lower frequencies than the minimum achieved by CAST-CAPP is to search for possible harmonics appearing in higher frequencies due to intermodulation. This procedure was performed both with $\vec{B} = 8.8\,\mathrm{T}$ and $\vec{B} = 0\,\mathrm{T}$ in order to allow a search for correlations and/or differences between the two cases, resulting to a total of 1180 measurements. The total data-taking time for each set of measurements was about 19.7 h resulting in a total data-taking time of about 39.4 h.

As presented in Sect. 22.3.2, from 18/11/2020 until 21/06/2021 a simultaneous data-taking campaign with a second measuring channel for EMI/EMC ambient parasites was performed. Since then, this second VSA channel, that was directly connected to an external omnidirectional antenna, operated for 89.2 d, covering a frequency range between 4.797 GHz - 5.402 GHz corresponding to a total range of 605.0 MHz.

## 23.2 Data processing and quality checks

### 23.2.1 Data processing

The data which are recorded with the IQC recorder from the two measuring VSAs channels i.e. the cavities and external antenna, consist of in-phase (I) and quadrature (Q) components of Time Domain (TD) voltage samples and are saved in binary file format (see also Appendix Sect. C.1.5).

These binary I/Q data are first converted into complex TD samples. To detect a potential axion signal with a constant frequency of the conversion photon the TD data are converted into Frequency Domain (FD), where a peak is expected as a manifestation of an axion signal. For this purpose, Discrete Fourier Transformation (DFT) is applied on the TD data using an Fast Fourier Transform (FFT) algorithm in order to convert them into FD.

Specifically, the TD data are split into chunks and then are converted to FD by applying a FFT to each individual chunk. This way I/Q samples in FD are obtained, which are related to the RMS power:

$$P_{RMS} = \frac{I^2 + Q^2}{2Z_0} \tag{23.1}$$

where: $Z_0 = 50\,\Omega$ : the VSA input impedance.

All the chunks which belong to a single datafile are then combined to obtain the FFT processed spectrum (see Fig. 23.2a). As next, the average of all FD chunks is calculated and they are converted into power units to form a power spectrum. The RMS power values can also be converted to dBm units using Eq. 23.2.

$$P\,(\mathrm{dBm}) = 10\log\left(\frac{P_{\mathrm{RMS}}\,[\mathrm{W}]}{1\,\mathrm{mW}}\right) \tag{23.2}$$

The number of samples $N$ in each chunk is determined by the RBW which is adjusted at $\delta\nu = 50\,\text{Hz}$. This means that time length of each chunk is $1/\delta\nu = 0.02\,\text{s}$. At the same time, the data-taking rate is 5 Mega samples/s which means that the number of samples per FFT chunk is $N = 0.02\,\text{s} \times 5$ Mega samples/s $= 10^5$. Based on the data-taking rate, in each 1 min measurement there are 300Mega samples. As a result, there are $300\text{M}/10^5 = 3000$ FFT chunks of samples in each 1 min file which are averaged to obtain the processed spectra. When these 3000 spectra are averaged, the distribution becomes Gaussian due to the central limit theorem [460] (see Fig. 23.2b).

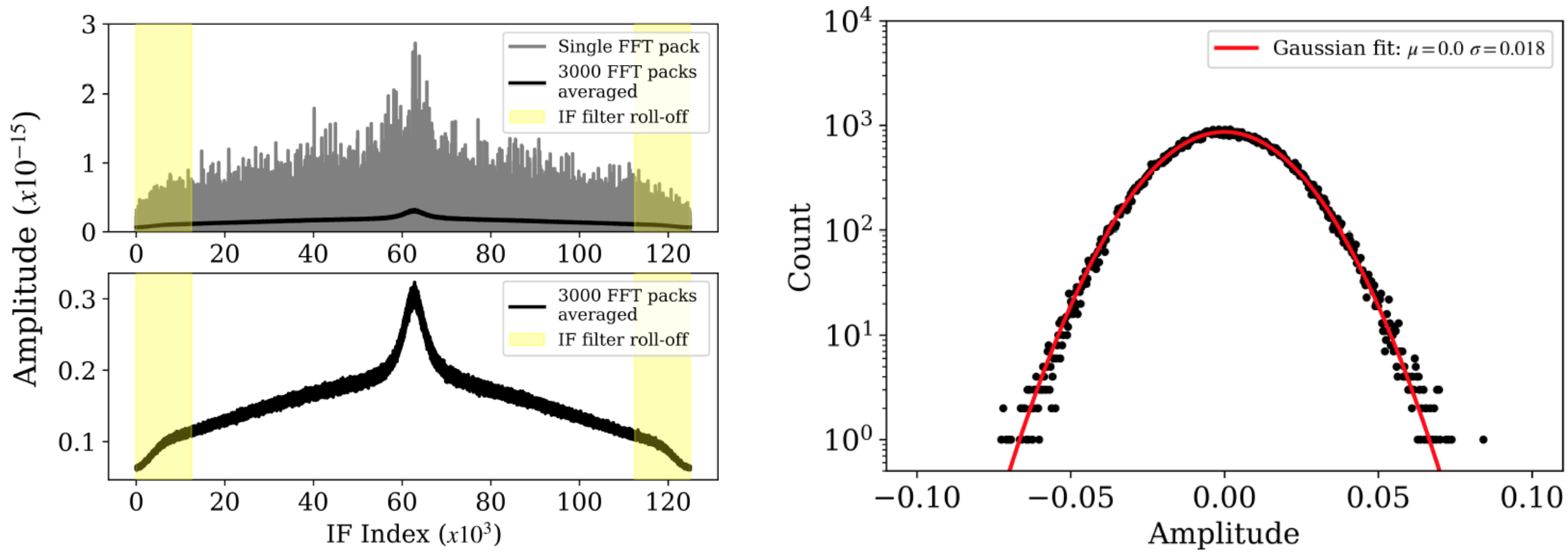

**(a)** Single (grey) and 3000 (black) averaged chunks of a 1 min file.

**(b)** Distribution of 3000 average chunks consisting a 1 min file.

**Figure 23.2:** CAST-CAPP example of FFT spectrum (49)

Finally, the actual bandwidth of each one of the recorded raw spectra is 6.25 MHz, with the $\sim 12.5\%$ of the left and right edges of each spectrum representing the roll-off of the Intermediate Frequency (IF) filter. As a result, both ends are cropped in the analysis with the number of samples in each spectrum becoming $N = 10^5$. This way about 25% of the samples in each spectrum are removed, to have refined spectra of 5 MHz bandwidth as expected (see Fig. 23.2a).

### 23.2.2 Quality checks

Several selection criteria are applied to each processed 1 min file before the analysis procedure takes place. This way, the data which are not qualified, are discarded from further consideration but kept in storage in case a revision of the selection criteria takes place in future analyses. The main quality checks focus on the removal of undesired non-systematic effects coming from mechanical vibrations (see Sect. 22.2.2.1) or any electronic-related issues. The criteria, which are presented in Tab. 23.6, are defined based on the real-time observations of the various parameters as well as on the analysis of the distribution of the samples, which in turn indicate the various thresholds required to eliminate extreme candidates.

More specifically, the various criteria applied to each specific 1 min measurement are the

**Table 23.6:** Data qualification criteria. "Before" and "after" labels indicate the measurements before and after each measurement block, respectively. Criteria 1-7 apply to all measurements while 8-10 refer only to PM data-taking.

| Nr. | Parameters | Criteria |
|---|---|---|
| 1 | Frequency stability | $\Delta\nu_0 < 100\,\text{kHz}$ |
| 2 | Amplitude variation | $\Delta A_0 < 3\,\text{dB}$ |
| 3 | Quality factor | $10^3 < Q_L < 4 \times 10^4$ |
| 4 | Quality factor shift | $\Delta Q_L < 7 \times 10^3$ |
| 5 | Temperature variation | $\Delta T_{\text{cav}} < 3\,\text{K}$ |
| 6 | Temperature | $1\,\text{K} < T_{\text{cav}} < 273\,\text{K}$ |
| 7 | Magnetic field variation | $\Delta\vec{B} < 0.1\,\text{T}$ |
| 8 | Frequency mismatch | $< 20\,\text{kHz}$ (before) & $< 80\,\text{kHz}$ (after) |
| 9 | Amplitude mismatch | $< 1\,\text{dB}$ |
| 10 | Temperature mismatch | $< 3\,\text{K}$ |

following:

1. The centre frequency of the resonant peak of each cavity should not shift more than 100 kHz during each measurement. Before and after each transmission measurement with the VNA, the center frequency $\nu_0$ of the $\text{TE}_{101}$ mode is measured. The absolute difference between these two values should not be bigger than the FWHM of the cavity which is defined by the Q-factor:

$$\left|{\nu_0}^{\text{initial}} - {\nu_0}^{\text{final}}\right| < 10^5\,\text{Hz} \approx \frac{1}{5}\frac{\nu_0}{Q_L} \tag{23.3}$$

2. Due to mechanical vibrations the amplitude of the resonant peak could also be altered (see Sect. 22.2.2.1). Another reason could be due to amplifier gain fluctuations. Therefore, a limit of 3 dB has been set on the allowed amplitude shift between the two measurements by VNA.
3. The $Q_L$ of each cavity when measured by VNA in transmission should always be between $10^3 - 4 \times 10^4$. The actual values, by taking into account also tuning, as well as simulation are between $10^4 - 2 \times 10^4$. Therefore, the limits that are set here are conservative.
4. Similarly, the resonant peak of the cavities can be distorted, and thus change the observed $Q_L$. Therefore, the maximum limit for the allowed shift of the $Q_L$ value is $7 \times 10^3$. This limit has been verified by observations.
5. The temperature of the cavities could change due to different LNA bias or due to heating from tuning (see Appendix Sect. C.1.7). Therefore, for these cases, the limit imposed on the temperature shift, before and after each 1 min measurement is 3 K.
6. For the case of a wrong temperature reading or an electronically related spike from the temperature sensors, the temperature values for each cavity should be in the range $1 - 273\,\text{K}$. There are no further restrictions for intermediate temperatures since the analysis quality is not affected by higher-temperature data since their weight in the procedure is much lower than the usual low-temperature data.

7. Finally, the $\vec{B}$-field of the cavities should be the same within 0.1 T between each initial and final measurement. This quality-check has been set for future use with the $\vec{B}$-field sensors. For the moment, it is only used for the case of the user-imposed value of $\vec{B} = 0\,\mathrm{T}$ to separate these measurements from the $\vec{B} = 8.8\,\mathrm{T}$ measurements or any other measurements in intermediate magnetic fields.

There are three more criteria for the case of PM cavities where, as explained in Sect. 22.3.4 the mismatch of the various parameters has to be minimised:

8. The center frequency mismatch between the selected PM cavities should not exceed 20 kHz for the initial and 80 kHz for the final measurements. By initial and final, once more, it is meant the measurements performed by the VNA before and after each 1 min recording. As mentioned in Sect. 22.3.4.1 the theoretical FMT is around 100 kHz, therefore the limits that are set here are conservative. The initial limit of 20 kHz is set due to the 10 kHz accuracy of the tuning mechanism, whereas the final limit accounts for possible vibrations.
9. As mentioned in Sect. 22.3.4.2, the amplitude mismatch between the PM cavities should not exceed 1 dB for both initial and final measurements. This limit is once more set to account for vibrations, as the accuracy of our programming attenuators allows for an amplitude-matching within 0.25 dB.
10. The temperature mismatch among the PM cavities should not exceed 3 K in both initial and final measurements. This limit is set to exclude cases with heterogeneous temperature distribution among the PM cavities.

#### 23.2.2.1 Statistics of discarded files

For $\vec{B} = 8.8\,\mathrm{T}$ the total number of 1 min files that were processed was 262491 with the amount of files not passing the quality checks being 11593, i.e. $\sim 4.4\%$ disqualification. In Tab. 23.7 the specific number of data-files per specific disqualified reason is shown.

For $\vec{B} = 0\,\mathrm{T}$ data the total amount of discarded files was 461/24138 translating to $\sim 1.9\%$. In Tab. 23.8 the specific number of data-files per specific disqualified reason is shown.

Finally, for the data taken simultaneously with the independent VSA connected to the external antenna, there is in principle no reason to discard any data from the quality checks. However, in order to compare the exact same amount of data in both channels for direct comparisons, the exact files which are discarded in the $\vec{B} = 8.8\,\mathrm{T}$ and $\vec{B} = 0\,\mathrm{T}$ measurements with the cavities, their corresponding ones in the EMI/EMC monitoring channel, are also discarded from analysis (see Tab. 23.7 and 23.8).

**Table 23.7:** Statistics of disqualified $\vec{B} = 8.8\,\mathrm{T}$ data due to the applied quality checks from Tab. 23.6 and percentage with respect to the total number of measurements. Some files fail in more than one criteria but in the total disqualified file number they are counted only once.

| Criterion Nr. | Disqualification reason | Nr. of data-files | Percentage [%] |
|---|---|---|---|
| 1 | $\nu_0$ shift | 176 | 0.067 |
| 2 | Amplitude shift | 353 | 0.13 |
| 3 | $Q_L$ initial value | 479 | 0.18 |
| 3 | $Q_L$ final value | 587 | 0.22 |
| 4 | $Q_L$ shift during measurement | 7256 | 2.76 |
| 8 | PM frequency mismatch initial | 561 | 0.21 |
| 8 | PM frequency mismatch final | 556 | 0.21 |
| 9 | PM amplitude mismatch initial | 3096 | 1.18 |
| 9 | PM amplitude mismatch final | 2864 | 1.09 |
| Total | All | 11593 | 4.42 |

**Table 23.8:** Statistics of disqualified $\vec{B} = 0\,\mathrm{T}$ data due to the applied quality checks from Tab. 23.6 and percentage with respect to the total number of measurements. Some files fail in more than one criteria but in the total disqualified file number they are counted only once.

| Criterion Nr. | Disqualification reason | Nr. of data-files | Percentage [%] |
|---|---|---|---|
| 1 | $\nu_0$ shift | 4 | 0.017 |
| 3 | $Q_L$ initial value | 1 | 0.0041 |
| 4 | $Q_L$ shift | 435 | 1.80 |
| 8 | PM frequency mismatch initial | 6 | 0.025 |
| 8 | PM frequency mismatch final | 1 | 0.0041 |
| 9 | PM amplitude mismatch initial | 12 | 0.050 |
| 9 | PM amplitude mismatch final | 7 | 0.029 |
| Total | All | 461 | 1.91 |

## 23.3 Data analysis

After the disqualification of the low-quality files, the analysis procedure that is followed for halo DM axions is similar to the widely accepted procedures from other haloscopes [4, 405, 445, 461], yet with slight modifications based on the specific experimental conditions of CAST-CAPP. The various related parameters that are used in the analysis together with their respective uncertainties are shown in Tab. 23.9.

### 23.3.1 Intermediate Frequency interferences

First, the processed spectra that pass the quality checks, are scanned for parasite signals. The resulting contaminated bins acquired from this procedure are deleted. A usual contamination comes from IF interferences which appear in the same single bins in all processed spectra. Such interferences originate mainly from receiver electronics or numerical artefacts.

**Table 23.9:** CAST-CAPP parameters and related uncertainties used for the analysis

| Parameters | Explanation | Values | Uncertainty |
|---|---|---|---|
| $\vec{B}$ | Static dipole magnetic field | 8.8 T | $10^{-3}$ |
| $V$ | Cavity volume | 224 cm$^3$ | 0.1 cm$^3$ |
| $C_{lmn}$ | Form factor | 0.53 | 10% |
| $\beta$ | Main port coupling factor | 1 | 0.3 |
| $Q_L$ | Loaded quality factor | 20000 | 3% |
| $T_S$ | System noise temperature | 9 K | 1 K |
| $\eta$ | Signal attenuation coefficient | 0.717 | 0.01 |

These IF bins can not be a candidate signal since they appear as a narrow line on a fixed index of processed spectra in contrast with RF interferences which occur in fixed frequencies.

The whole dataset is initially split into three groups and averaged according to the bin indexes. Then, the bins in each group exceeding the $5\sigma$ threshold are flagged, and if they appear in at least two out of the three groups, they are removed. It is noted that in the processed spectra there are $1.25 \times 10^5$ bins with only 0.04 bins being expected to exceed the $5\sigma$ threshold due to statistical fluctuations.

The total number of these flagged bins are 6, originating from 2 IF interferences (see Fig. 23.3) [460]. Finally, it is worth mentioning, that due to tuning of the cavities, all the bins in the combined spectrum have contributions from multiple processed spectra, and therefore in the final combined spectrum there is no single empty bin due to the removal of these IF interferences.

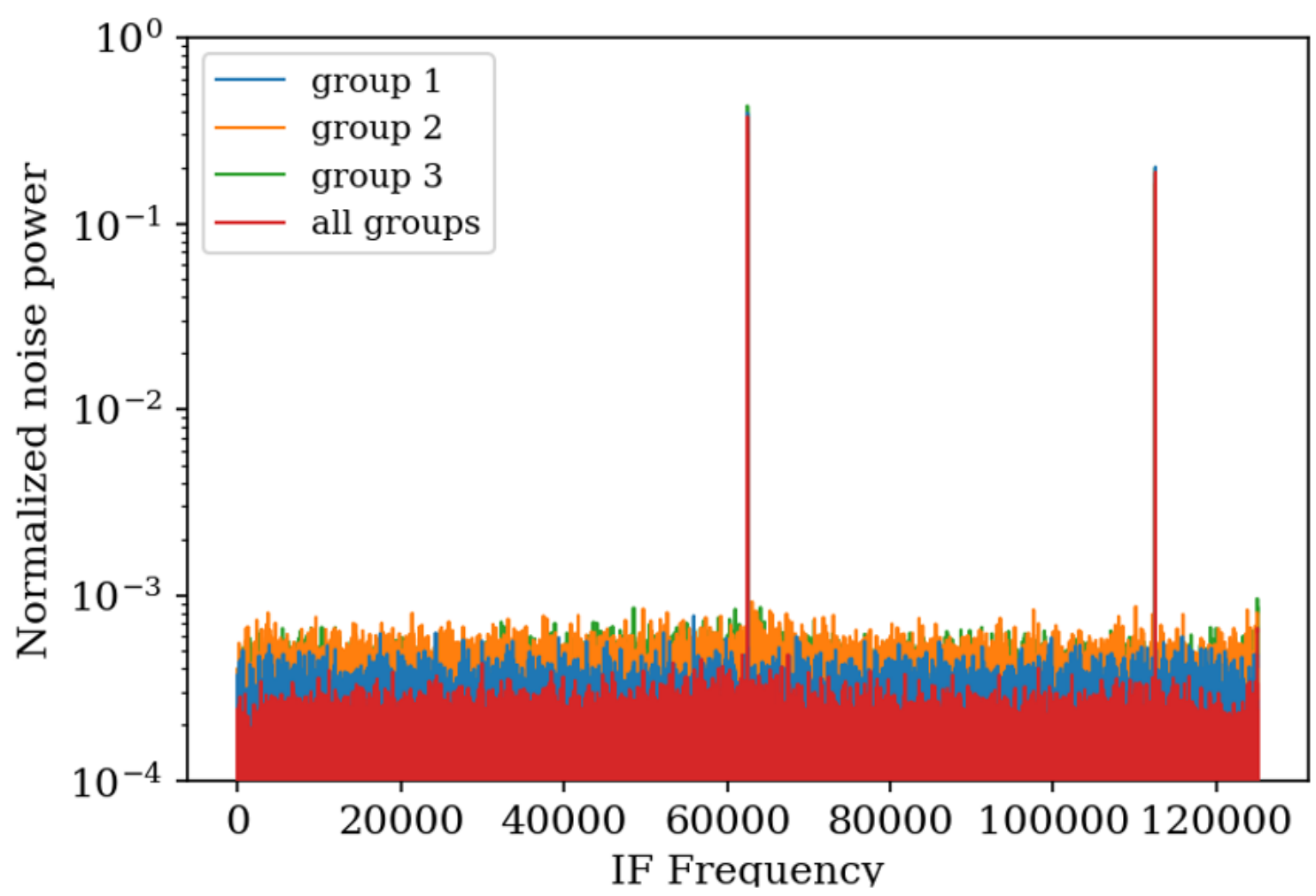

**Figure 23.3:** Two IF interferences appearing in two out of the three groups. (49).

## 23.3.2 Spectrum flattening

In order for the power spectrum to be analysed searching for a potential axion signal, the noise shape on the spectrum caused by the receiver chain has to be removed (see Fig. 23.2a).

This shape corresponds partially to the noise bump of the cavity plus a background added by the LNAs. This shape is also visible directly from the VSA (when in FFT mode) as seen in Fig. C.6.

The first step is to use a Savitzky-Golay (SG) filter which models the noise shape in the processed spectrum. This digital low-pass filter, which is widely used for data-smoothing, contains least-square fits of low order polynomials ($4^{\text{th}}$ in our case) on a sample window (with a length of $W = 1001$ bins in our case) that moves by one bin at each step of evaluation. In each step, the centre bin of the window is equated to the corresponding value of the polynomial fit [460].

Then, the data are divided with the SG filter and the spectrum becomes flattened as seen in Fig. 23.4.

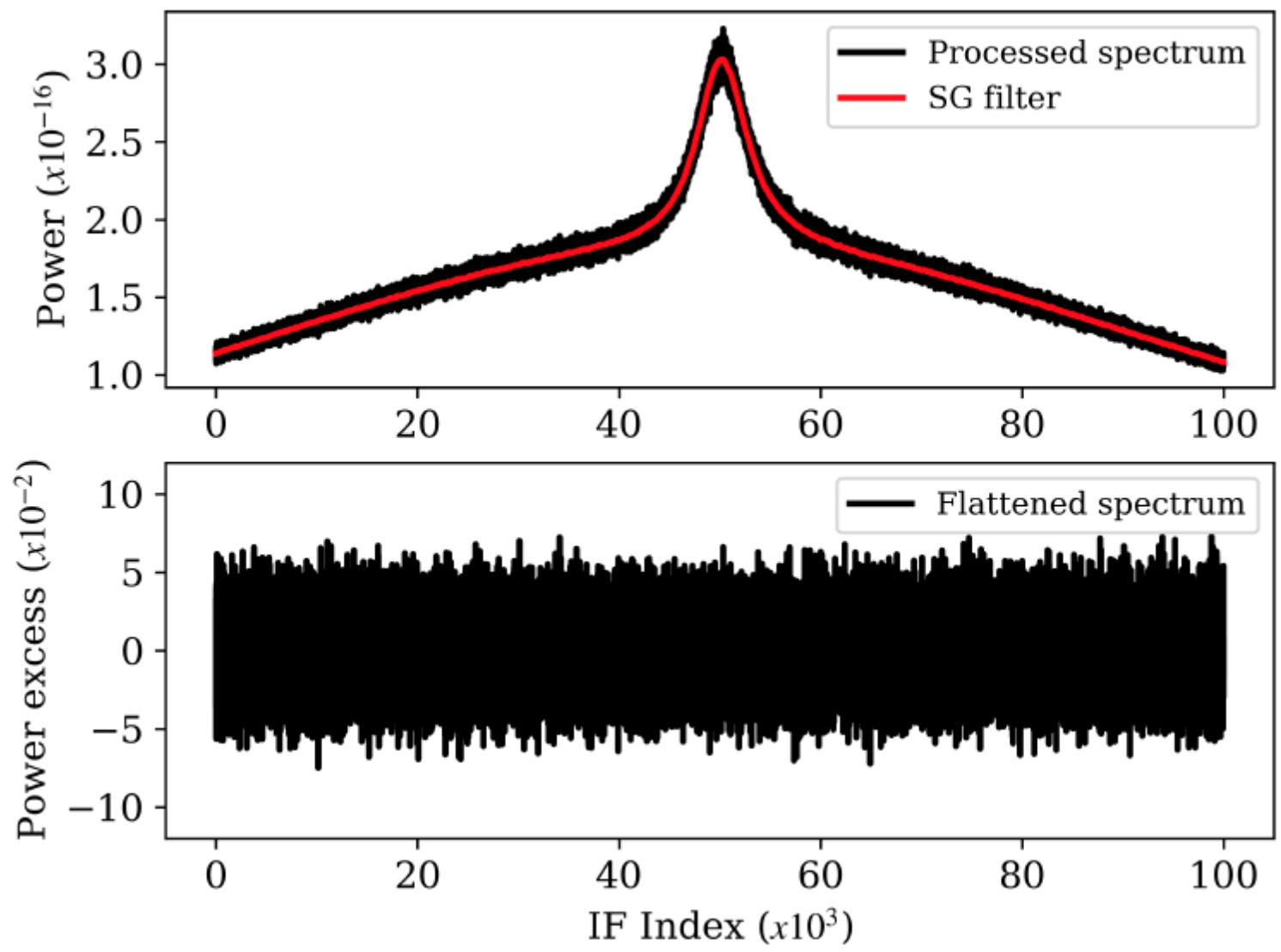

**Figure 23.4:** Example of a 1 min processed spectrum which is divided with the SG filter to become flattened (49).

Subsequently, we have to subtract the numeric value 1 from the flattened spectra in order to obtain a mean value of $\mu = 0$ with a standard deviation of $\sigma = (\delta\nu \cdot t)^{-1/2} \approx 0.018$ of a Gaussian distribution. This way the distribution represents Johnson noise (thermal).

### 23.3.3 Combination of multiple spectra

#### 23.3.3.1 Scaling

The first step of the combination of multiple spectra, is the scaling of each flattened spectrum by $P_{\text{noise}}/P_{\text{axion}}$ in to account for the fact that an axion signal would not have the same amplitude across the flattened spectrum. $P_{\text{axion}}$ is the axion signal power curve across the bandwidth of each spectrum as defined by Eq. 22.10, whereas $P_{\text{noise}}$ is the mean noise power from Eq. 22.14 [461].

More specifically, from the last term in Eq. 22.10, we get a Lorentzian profile on the axion conversion power as expected due to the shape of the resonance mode of the cavities. Therefore, the processed spectra are scaled according to the centre frequency $\nu_0$ of each measurement and the quality factor $Q_L$ by dividing them with a Lorentzian function (see the top plot in Fig. 23.5). In every step of the analysis, it is assumed that the axion power is all deposited in a single bin of the spectrum. So far it has been assumed that it fits in a bin of $\delta\nu = 50$ Hz. With the scaling of the flattened spectra, each bin will have a $\mu = 0$ except from the one bin with the potential axion signal which will have $\mu = 1$.

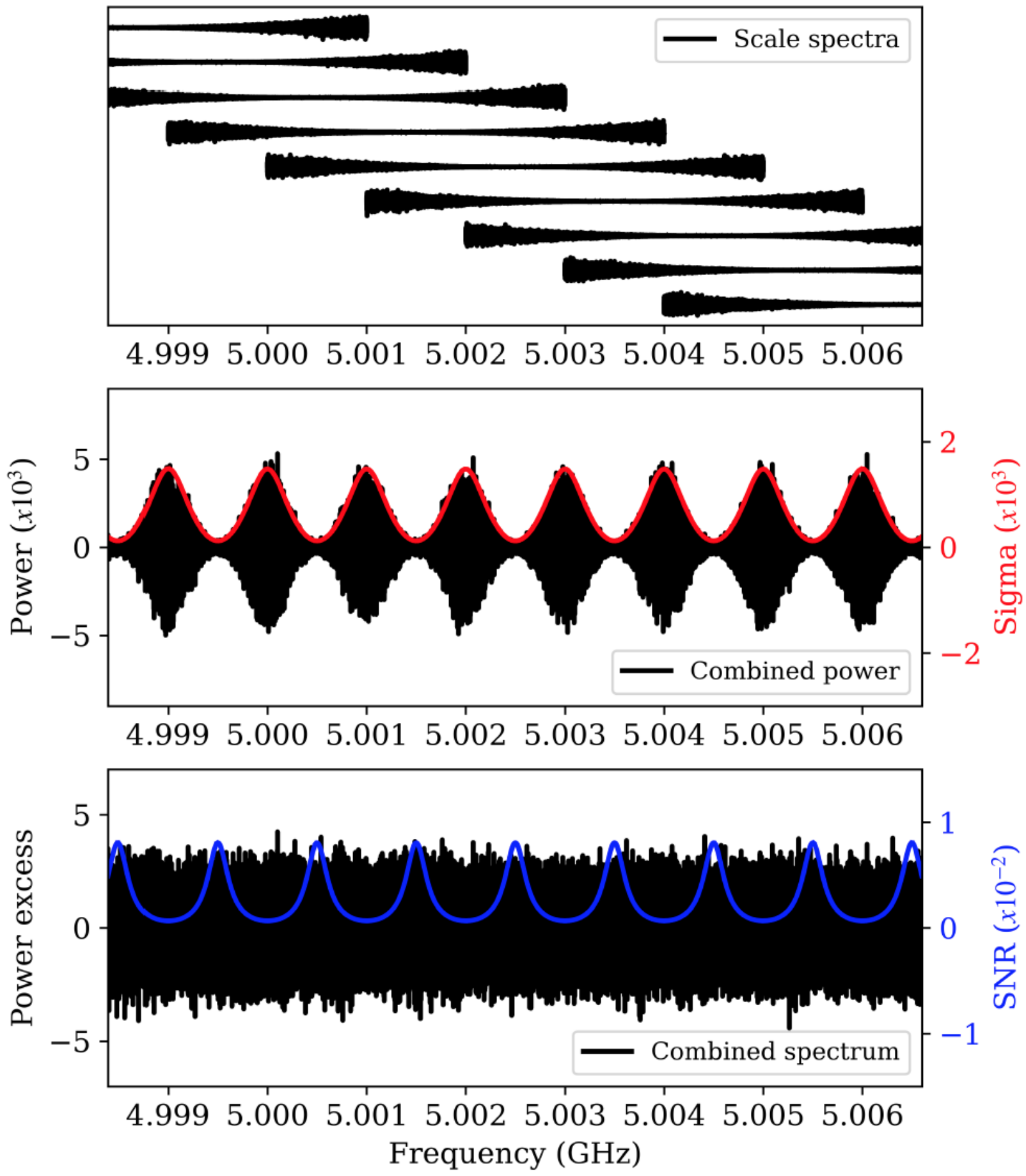

**Figure 23.5:** Spectrum combination procedure on 9 sample spectra. The scaled spectra are aligned according to the RF index (top). The weighted averaging of the scaled spectra is then taken (middle). The normalised combined spectrum (bottom) (49).

#### 23.3.3.2 Vertical averaging

To generate the so-called *combined spectrum*, the scaled spectra should be aligned according to an RF index instead of the IF index. This procedure is called *vertical averaging* of the scaled spectra bins where IF bins from different spectra corresponding to the same RF bin are combined based on an optimally weighted vertical sum (see middle plot in Fig. 23.5). The weights are equal to inverse bin variances, as determined by the Maximum Likelihood (ML) estimation which gives the minimum standard deviation and the maximum SNR.

#### 23.3.3.3 Normalisation

The final step involves the normalisation of the combined spectrum in order to have a standard sample distribution. To do this, the amplitude $p_k$ of each bin $k$ of the combined spectrum is divided by its standard deviation $\sigma_k$. This results to normalised samples with a normal distribution of $\mu = 0$ and $\sigma = 1$, except for the perspective bin containing the axion power which should have a mean of $(\sigma_k)^{-1}$ and $\sigma = 1$ (see bottom plot in Fig. 23.5).

It is noted that the RBW of the combined spectrum remains at $\delta\nu = 50\,\text{Hz}$. The resulting combined spectrum along with its noise distribution for the full dataset of CAST-CAPP is shown in Fig. 23.6.

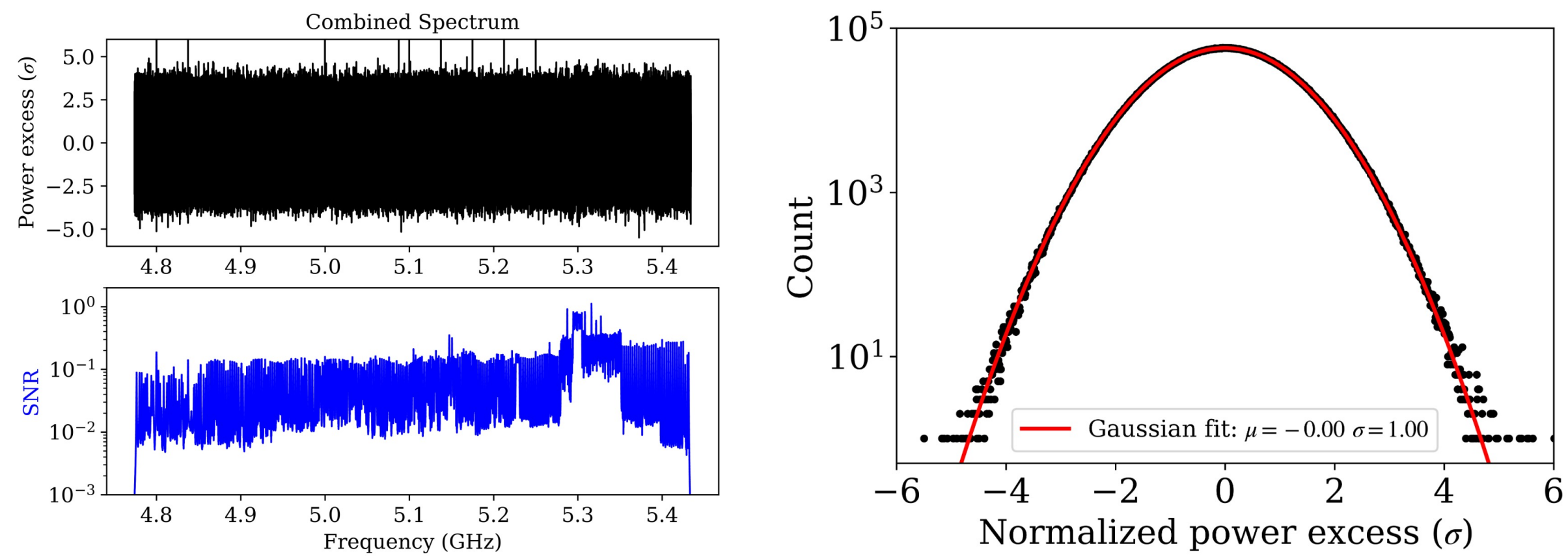

**(a)** Combined spectrum (black) and its SNR (blue) for an axion signal. The narrow lines which are above $5\sigma$ have been cropped here for visual purposes. Their origin is explained in Sect. 23.3.5.3.

**(b)** Noise distribution of the combined spectrum.

**Figure 23.6:** The combined spectrum and its projected noise distribution for $\vec{B} = 8.8\,\text{T}$ for CAST-CAPP detector.

### 23.3.4 Grand spectrum

#### 23.3.4.1 Rebinned spectrum

For the SNR of a halo axion in the combined spectrum to be maximised, a *rebinning* of all the bins in the combined spectrum has to be performed in order for their width to match the axion signal width of 7 kHz. This way $\sim 90\%$ of the axion signal curve is expected to fit in a single bin. In the previous considerations, when the power spectra were flattened, it was assumed that the whole axion conversion power is deposited in only one bin of those spectra.

The first important step for this rebinning, is the *horizontal averaging*, i.e. take a weighted average of 28 adjacent combined spectrum bins using ML weights [460]. The resulting rebinned spectrum has a RBW of 1.4 kHz with the number of bins being decreased by a factor 1/28.

#### 23.3.4.2 Convolution

The last step involves the application of a window function that has the shape of the expected axion power curve in the lab frame (see Fig. 23.7). It is noted that the shape of the galactic halo axions depends on the energy distribution over the frequency space. Therefore, the rebinned spectrum is scaled and transformed into the *grand spectrum* by considering the expected axion signal shape. The normalised grand spectrum is then generated by dividing the power of each bin by its standard deviation.

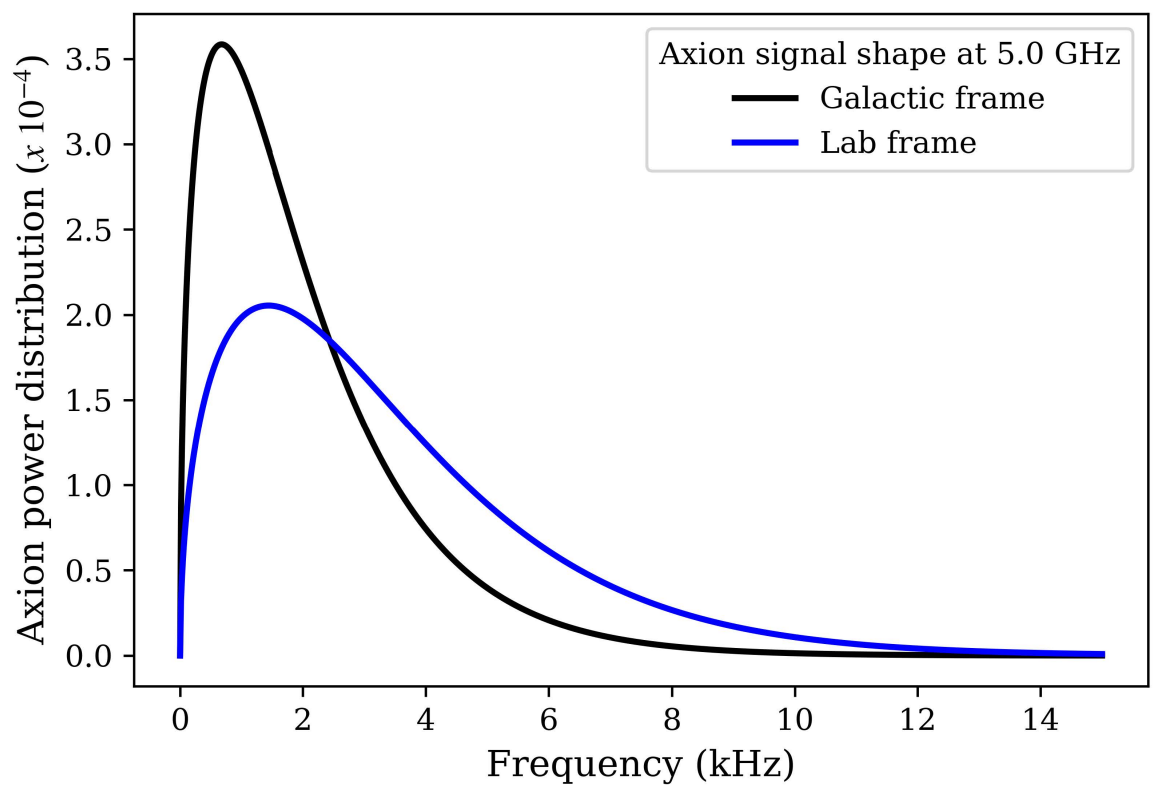

**(a)** Galactic (in black) vs. lab (in blue) frame. The signal width covers 90% of the signal power at 4.3 kHz for the galactic frame and 7 kHz for the lab frame.

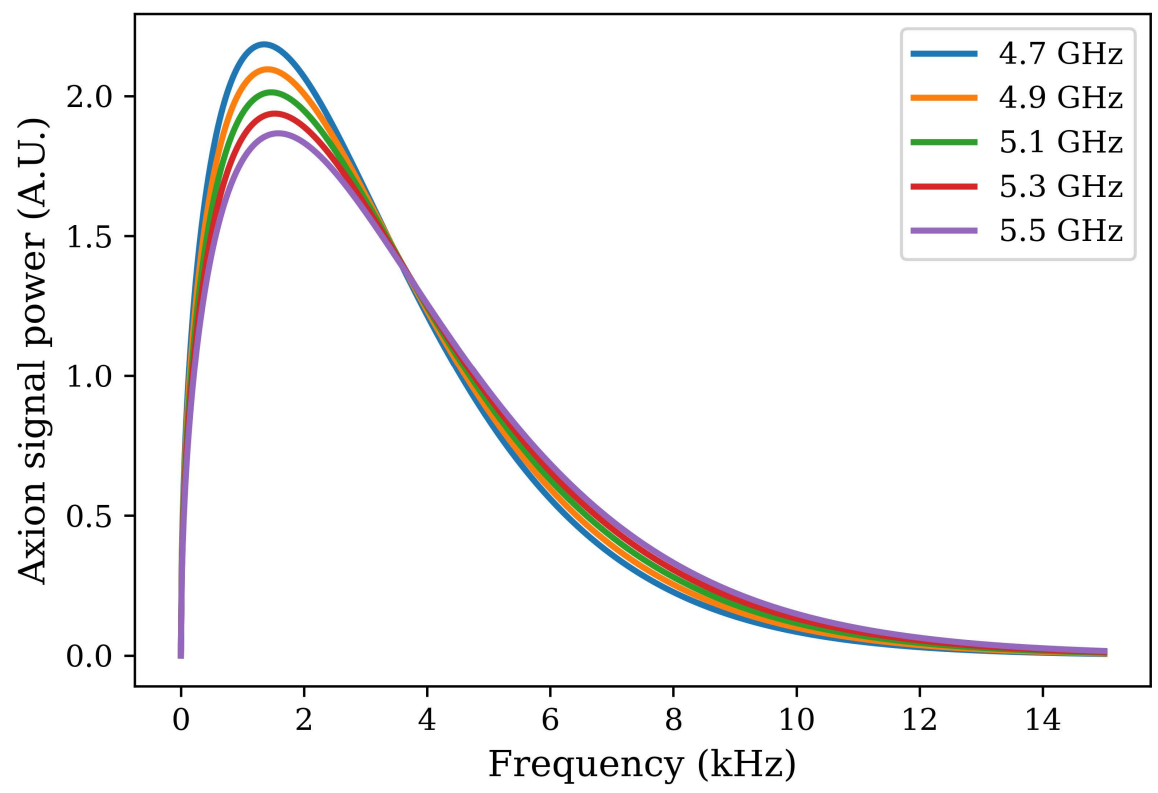

**(b)** Variation of the signal shape across the CAST-CAPP frequency range. The signal amplitude and width vary both by $\sim 8\%$.

**Figure 23.7:** Expected halo axion signal shape for CAST-CAPP detector. The x-axis corresponds to $\nu - \nu_a$ i.e. the frequency distance from $\nu_a = 5\,\text{GHz}$ (49).

#### 23.3.4.3 SG negative correlations

It is noted that the SG filter that is used for the flattening of the spectra, causes negative correlations among the consecutive bins of the combined spectrum in a range smaller than that of the filter window. This causes the standard deviation of the grand spectrum to be reduced from $\sigma = 1$ to $\sigma = 0.74$.

For this reason, a simulation of $10^4$ processed spectra has been constructed using sample noise baselines from the real data. The elements of the covariance matrix which are defined by this simulation are used to calculate the ML estimate of the standard deviation of the grand spectrum bins [460]. Therefore, the results of the simulation are applied to the real data to correct the changes from the SG filter induced bin height correlations to the rebinned and grand spectrum standard deviations.

By applying the correction, the sigma of the rebinned spectrum becomes $\sigma = 0.99$, while the "corrected" grand spectrum which uses the aforementioned corrected rebinned spectrum is described by $\sigma = 0.96$. The final grand spectrum along with its noise distribution for the whole CAST-CAPP dataset is depicted in Fig. 23.8a. It is mentioned that this grand spectrum is the very final one after the rescan and candidate identification procedure that will be outlined later. The grand spectrum before that procedure is shown in Fig. 23.12.

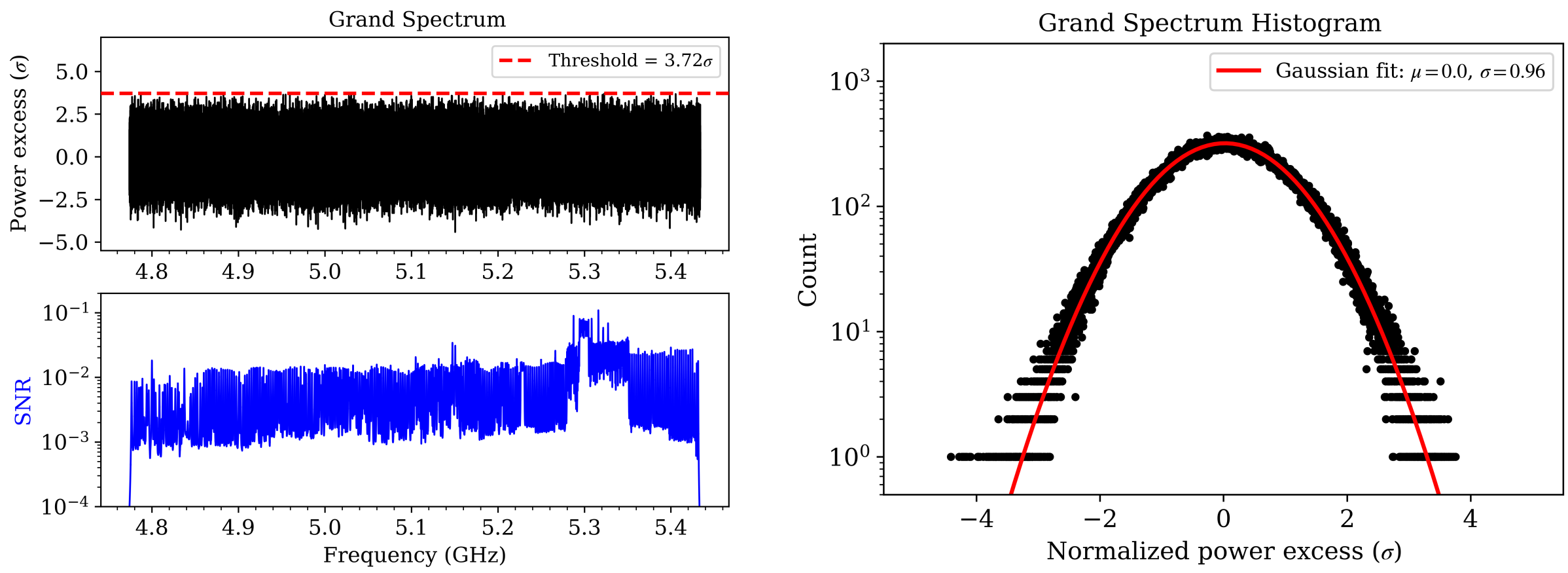

**(a)** Grand spectrum (black) with SNR (blue) for a KSVZ axion.

**(b)** Noise distribution of the grand spectrum.

**Figure 23.8:** The grand spectrum and its projected noise distribution for $\vec{B} = 8.8\,\mathrm{T}$ for CAST-CAPP detector.

## 23.3.5 Axion candidate search

### 23.3.5.1 Simulated signal

In addition to hardware injected signals outlined in Sect. 22.3.1, a software-generated axion signal has also been injected directly into the raw data. With this simulation, the retrieval of such a signal and its exact characteristics via the analysis procedure is also cross-checked. Based on the specific CAST-CAPP characteristic numbers, a $20 \times$ KSVZ axion signal with 7 kHz width should be visible after 90 min of averaging. The simulation that was performed containing 110 min of data is shown in Fig. 23.9 and verified the predictions and the analysis procedure.

### 23.3.5.2 Threshold definition

To check whether there is an axion signal with positive power excess in the grand spectrum or not, the *hypothesis testing steps* have to be followed. Following our prediction we set a target $SNR_{\text{target}}$ of $5\sigma$. This is the minimum condition for a KSVZ axion in the grand spectrum to fulfil the hypothesis testing procedure. Then the threshold value $\Theta$ for candidate bin selection

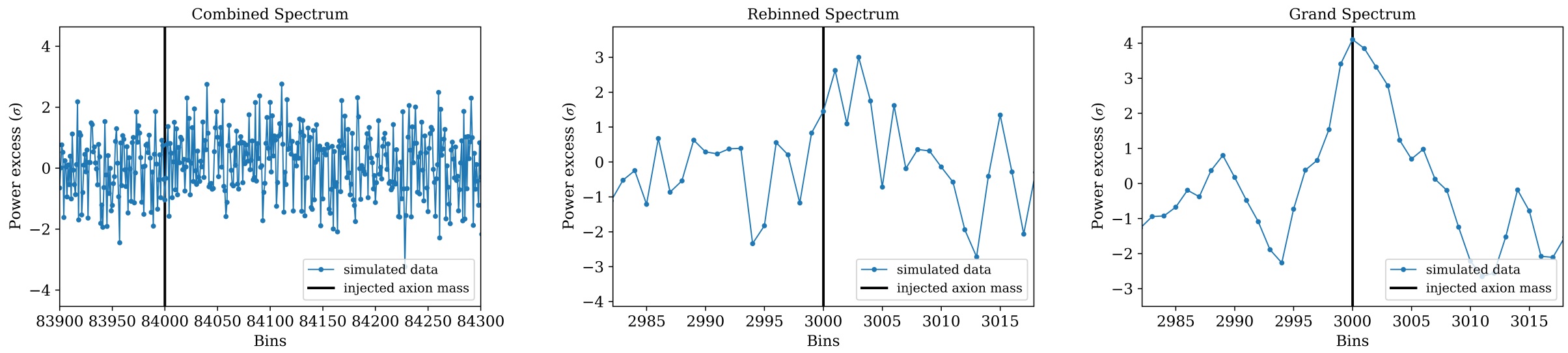

**(a)** Combined spectrum. **(b)** Rebinned spectrum. **(c)** Grand spectrum.

**Figure 23.9:** Simulation of 110 min simulated raw data including a $20 \times$ KSVZ axion. The effect of the analysis procedure is observed as a signal with higher SNR. The derived statistical significance is about $4\sigma$.

can be derived as a function of the $SNR_\text{T}$ and the Confidence Level (CL) (see Eq. 23.4). The axions should appear as spectrum bin(s) with positive excess power. As a result, any bin with a height exceeding this threshold is marked as an axion candidate. In other words, the threshold value represents the minimum bin height in the power spectrum to accept it as an external signal with $5\sigma$ confidence.

In Fig. 23.10 the bins containing an axion signal of the target SNR ($\mu = SNR_\text{target}$) are compared with the distribution of the bins containing only noise ($\mu = 0$). The standard deviation $\sigma^G = 1$ is the same in both cases. X-axis is the bin amplitude in the grand spectrum. The CL for axion detection is defined by the area under the Gaussian Probability Density Function (PDF) of the bin with the axion signal (blue) appearing on the right side of the threshold, i.e. the sum of blue and green coloured areas.

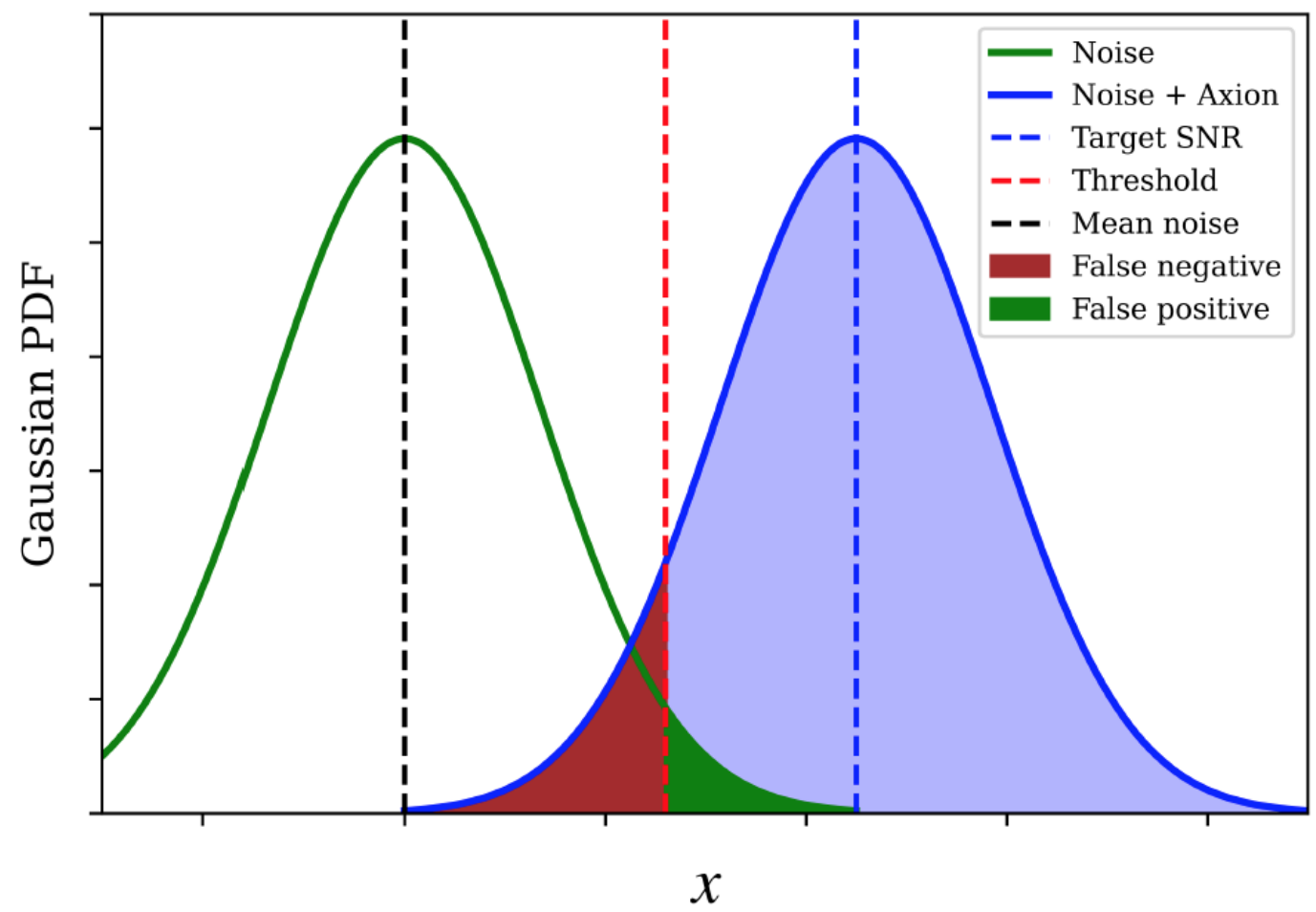

**Figure 23.10:** Hypothesis testing of an axion signal in the grand spectrum (49).

More specifically the threshold $\Theta$ is defined by the $SNR_\text{target}$ and the Cumulative Distribution Function (CDF) of the bins with axion power:

$$\Theta = 2SNR_\text{target} - CDF_S^{-1}(CL) \tag{23.4}$$

whereas CL is equal to $1 - CDF_S(\Theta)$. As a result, for $SNR_{\text{target}} = 5$ and $CL = 90\%$ we get $\Theta = 3.72$. This is the threshold on the grand spectrum bins for them to be flagged as candidates (see red dashed line in Fig. 23.8a).

The green area in Fig. 23.10 represents the false-positive candidate bins in the grand spectrum that will have to be scrutinised by rescanning them. For $\sim 472k$ number of bins in the grand spectrum and $\Theta = 3.72$, and for a normal distribution of $\mu = 0$ and $\sigma = 1$, we get 47 statistically expected candidates. However, due to the SG filter distortions we have $\sigma = 0.96$ which reduces the number of expected axion candidates to 25. As we will see next, from the analysis of the acquired data, the actual number that was derived for the statistical candidates was 40, with the difference from the theoretical expectation being due to the double-counting of several close-frequency candidates as well as the theoretical uncertainty induced by the aforementioned distortions in the distribution.

#### 23.3.5.3 Rescan strategy

When a possible candidate is found with an excess above the pre-defined threshold, a narrow range around it is rescanned to verify its consistency. The same is done for all the listed candidates including multiple cavity configurations (either single or PM). If a candidate is persistent enough, then this may still originate from the interference by the electronics in the receiver chain, therefore, the candidate spectrum is compared with the independent spectrum from the external antenna data scanning for EMI/EMC parasites. In case of the absence of a power excess of similar or larger amplitude in the second EMI/EMC monitoring channel, an extra rescanning is performed without the magnetic field ($\vec{B} = 0\,\text{T}$) for the case that the excess originates from an interference by the electronics in the receiver chain.

For the case of halo DM axions, a significant reduction of the number of candidates can also take place with the examination of the expected linewidth (around 10 kHz) and line shape of the halo axion signal which is expected to be a Maxwell–Boltzmann distribution (see Fig. 23.7). Of course, in addition to the two aforementioned properties (linewidth and shape), an axion signal will be distinguished from interference lines by its consistent occurrence at a fixed frequency.

More specifically, the grand spectrum bins that were flagged as rescan candidates were 60 out of the $\sim 472k$ (see Fig. 23.12). By comparing with the simultaneous measuring EMI/EMC channel (see Fig. 23.11 for an example), 9 out of 60 candidates were discarded from the remaining analysis with three of them being rejected following background measurements without magnetic field. In addition, 11 out of the remaining 51 candidates were identified as blind signal injections. These have been removed both from the combined and the grand spectrum in Fig. 23.6 and 23.8. The remaining 40 statistical candidates disappeared during the first rescanning step and therefore were excluded as halo DM axions.

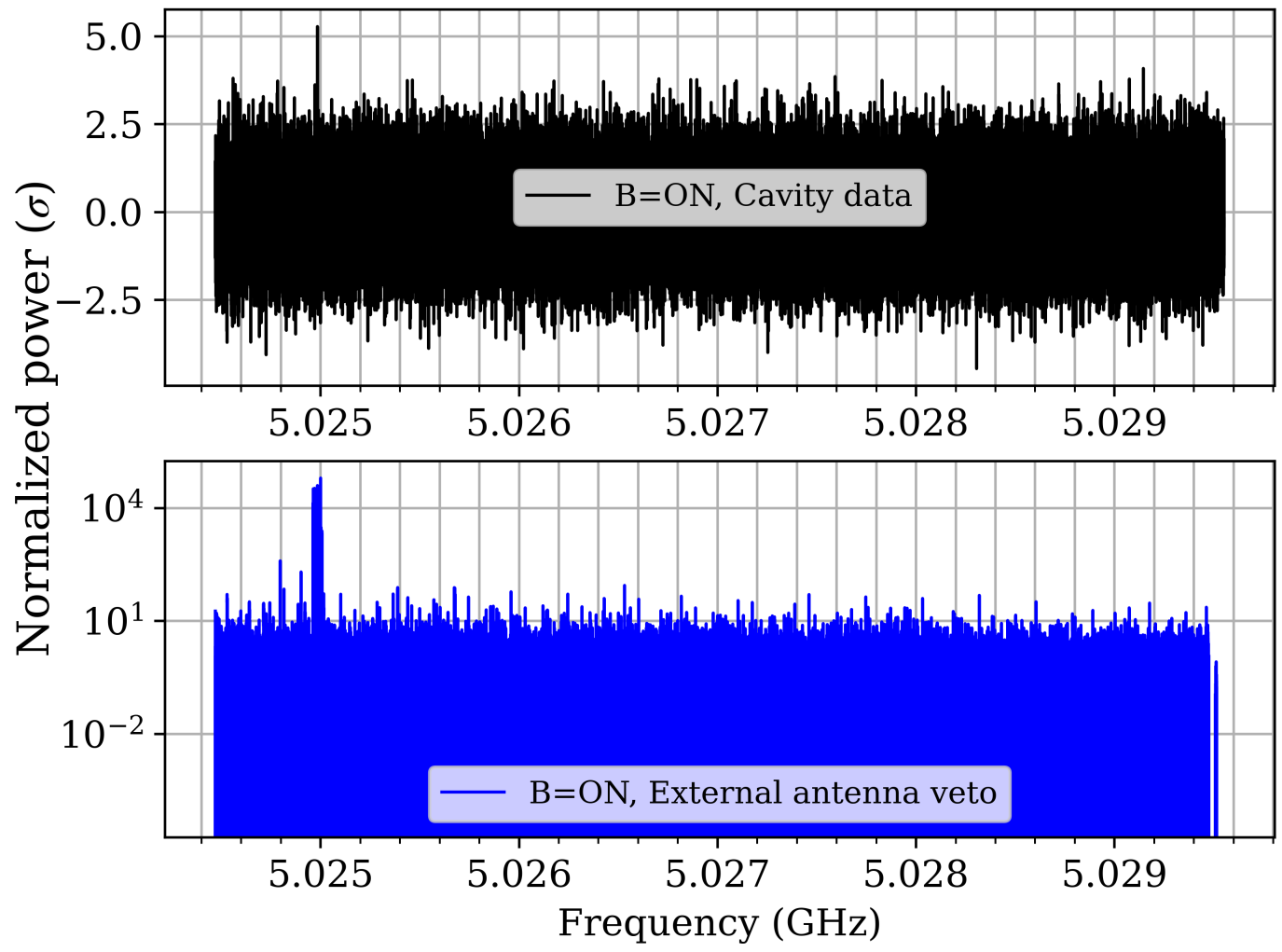

**Figure 23.11:** Example of a comparison of daily combined spectra for cavity data (upper figure) and the external antenna measuring the ambient EMI/EMC parasites. If a parasitic signal appears in both channels it is excluded from further consideration as an axion candidate.

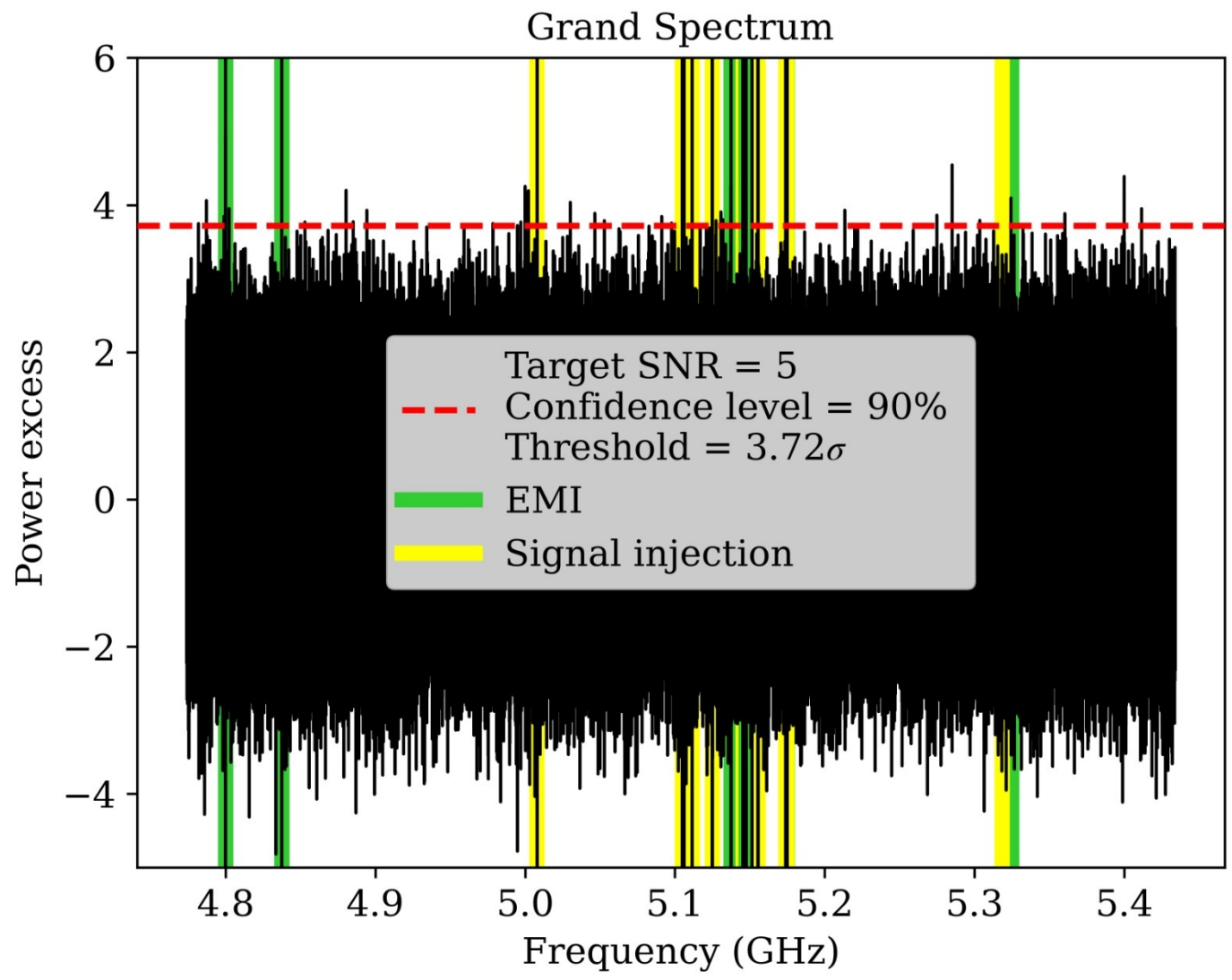

**Figure 23.12:** Grand spectrum before the rescan procedure took place. The green lines were verified from the second channel EMI/EMC parasites while the yellow lines indicate the blind signal injections. The black lines which are above the threshold were identified as statistical candidates and were therefore rescanned resulting to Fig. 23.8.

### 23.3.6 Exclusion plot

Since all the power excesses above the threshold turned out to be simple statistical fluctuations and electromagnetic parasites, an exclusion limit in the axion-photon coupling can be calculated. The expected SNR for a KSVZ axion is $SNR_{\text{target}} \propto \left(g_{\gamma}{}^{KSVZ}\right)^2$. In fact, each bin of the grand spectrum has a $SNR_g \propto \left(\sigma_g^G\right)^{-1}$ where $g$ denotes the RF index for the grand spectrum. Therefore, based on the various SNRs for each bin of the grand spectrum, an axion coupling higher that a minimum ${g_\gamma}^{\min}$ can be excluded:

$$\left|{g_\gamma}^{\min}\right|_g = \left|{g_\gamma}^{KSVZ}\right| \sqrt{\frac{SNR_{\text{target}}}{\left(\sigma_g^G\right)^{-1}}} \tag{23.5}$$

where $\sigma_g^G$ the standard deviation of bins that are corrected for the correlation induced by the SG filter.

In Fig. 23.13 the exclusion of axion-photon coupling at 90% CL as a function of the axion mass $m_a$ for virialised DM axions is shown. The yellow band around the theoretical KSVZ line indicates the expected axion-photon couplings predicted by the various theoretical models. A previously unexplored parameter space for axion masses between 19.74 µeV to 22.47 µeV has been excluded with axion-photon couplings down to $g_{a\gamma\gamma} = 8 \cdot 10^{-14}\,\text{GeV}^{-1}$. The estimated overall uncertainty calculated from the individual uncertainties shown in Tab. 23.9 is 10% (see also Publication E.6).

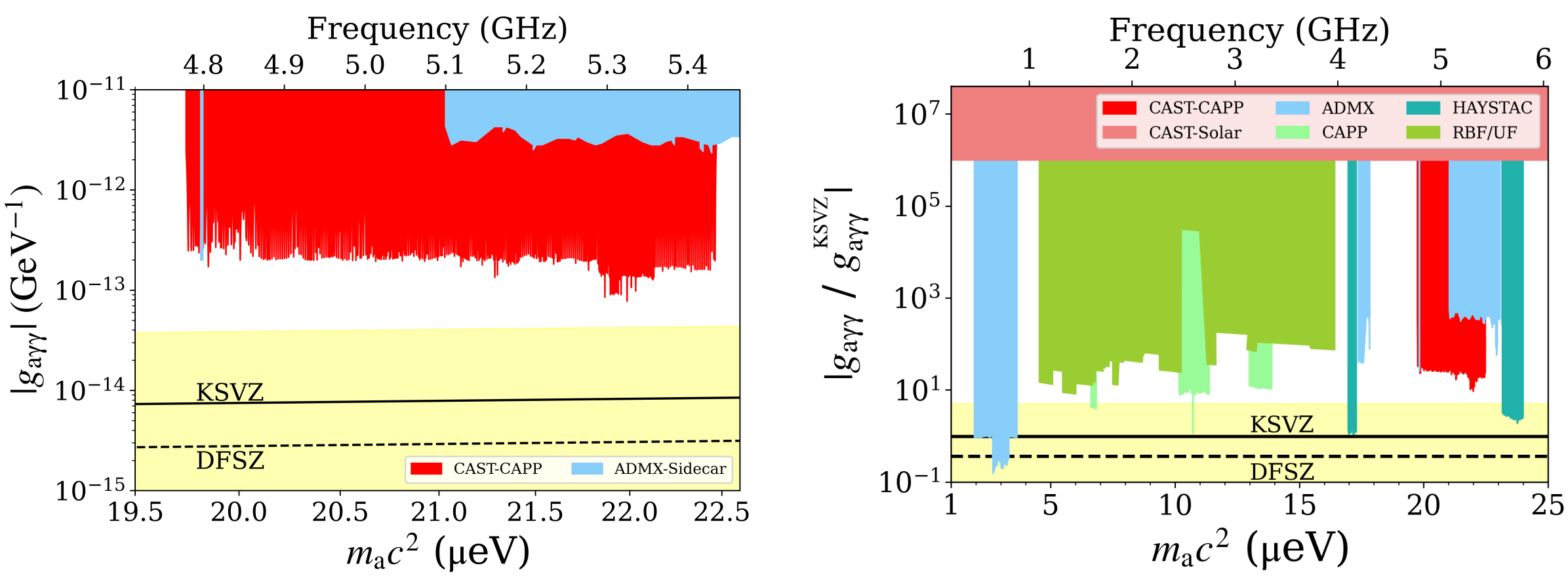

**(a)** Frequency range between 19.5 µeV to 22.5 µeV.

**(b)** Frequency range between 1 µeV to 25 µeV.

**Figure 23.13:** CAST-CAPP exclusion limit on the axion-photon coupling as a function of the axion mass $m_a$ for galactic halo DM axions at 90% confidence level compared to other axion searches [1–11].

### 23.3.7 Streaming DM axions

The same procedure for the exclusion plot can be performed for the cases of transient events such as streams or axion mini-clusters. As shown in Sect. 21.4 and 22.4 the fast resonant scanning technique which allows for a scanning speed of ∼ 10 MHz/min as well as the wide-band electronics of CAST-CAPP, could be decisive for the discovery of such short-lasting events. In these cases, the axion flux enhancement due to gravitational focusing effects by the Sun or even the intrinsic Earth's mass distribution can be up to $\sim 10^8$. As a result, strong daily modulations with eventually transient excursions can be observed.

For this reason, an independent analysis procedure has been created which allows the combination of daily processed spectra in order for CAST-CAPP to be sensitive to transient daily signals. Combining such alternative spectra, narrow transient events can be scrutinised. The second measurement channel which records simultaneously the environmental background for possible EMI/EMC parasites is decisive for such searches.

As an example, Fig. 23.14a shows the combined spectrum of a measurement performed uninterruptedly for $\sim 4.5$ hours on 24/11/2020 (from 19:19 to 23:53 local time), while using the fast resonant scanning method over a range of 42 MHz from 5.3547 GHz to 5.3967 GHz. This spectrum is a "first" following the above reasoning. No significant lines were observed in this specific measurement, and therefore it has not been further considered. The corresponding exclusion plot, for the aforementioned specific measuring time interval, assuming streaming DM ALPs with a modest flux enhancement by a factor of $10^2 - 10^4$ is shown in Fig. 23.14b. In this case, the width of the streaming axions has been assumed to be much smaller than that of the halo axions, at 50 Hz. An even smaller width can be exploited by changing the RBW while re-processing the raw data files. This is why the storage of the data-files in the time domain, even though they are big in size, are of great importance for such searches. It is noted, that Fig. 23.14b, serves only as an example to highlight the increased sensitivity for the detection of gravitationally-focused streaming axions and therefore can not be used as a "regular" exclusion plot like Fig. 23.13.

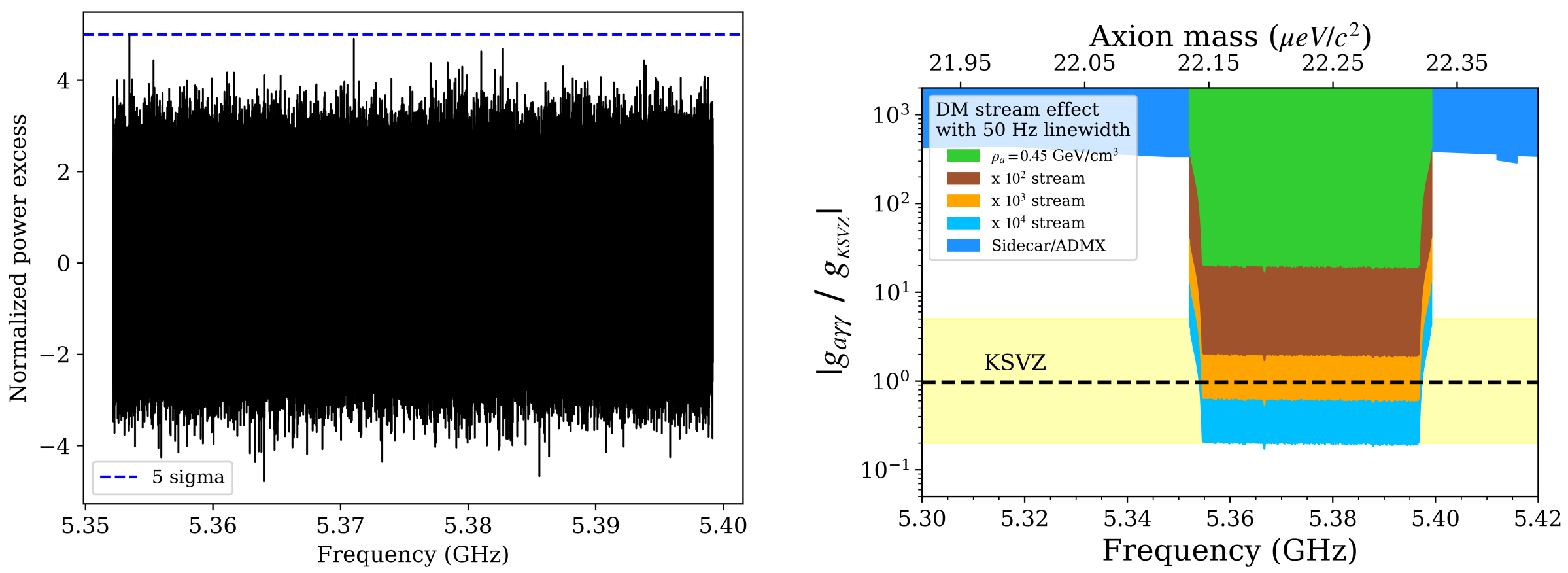

**(a)** Combined spectrum during a continuous series of measurements with the fast resonant scanning technique over 42 MHz on 24/11/2020. A possible transient axion enhancement would show up as a line above the threshold. Its amplitude above the noise level would have allowed to reconstruct the actual flux enhancement.

**(b)** "Local" exclusion plot for the specific period of measurements assuming streaming DM axions with flux enhancements $10^2$ (brown colour), $10^3$ (orange colour), and $10^4$ (blue colour).

**Figure 23.14:** CAST-CAPP example of a measurement for transient axion events at 24/11/2020 (from 19:19 to 23:53 local time).

## 23.4 Future prospects

### 23.4.1 Sensitivity prospects

In order to calculate the dependance of $g_{a\gamma\gamma}$ with the data-taking time $t$ as well as with the local DM density we use Eq. 23.5 and 22.11 and solve as a function of $t$ and $\rho_a$ accordingly:

$$g_\gamma{}^{\text{min}} \approx g_\gamma{}^{\text{KSVZ}} \left( \frac{SNR_{\text{target}} \cdot k_B \cdot T_S \cdot \sqrt{\frac{\delta\nu_a}{t}}}{P_{\text{axion}}} \right)^{\frac{1}{2}} \tag{23.6}$$

where $P_{\text{axion}}$ is given by Eq. 22.10.

This means that the exclusion plot changes accordingly with the $\sqrt{\frac{1}{\rho_a}}$ as well as with the fourth root of time $\sqrt[4]{t}$. Assuming halo DM axions, with a local density of $\rho_a \sim 0.45\,\text{GeV/cm}^3$ [401, 402], then the perspectives for CAST-CAPP detector with respect to KSVZ model as a function of the data acquisition time for three different scanning bandwidths (5 MHz, 50 MHz and 500 MHz) are shown in Fig. 23.15a. Similarly, in Fig. 23.15b the time to reach the theoretical lines of KSVZ and DFSZ model as a function of the covered frequency range for four and three PM cavities is shown.

Then, assuming streaming DM axions with a modest flux enhancement of $10^2$, the corresponding data-taking prospects are shown in Fig. 23.16. The major improvement of Fig. 23.16 comparing to Fig. 23.15 is evident. These calculations are based on Eq. 22.15. In this case, the reach to the theoretical KSVZ line would actually be a matter of hours if not minutes, depending on the actual flux enhancement.

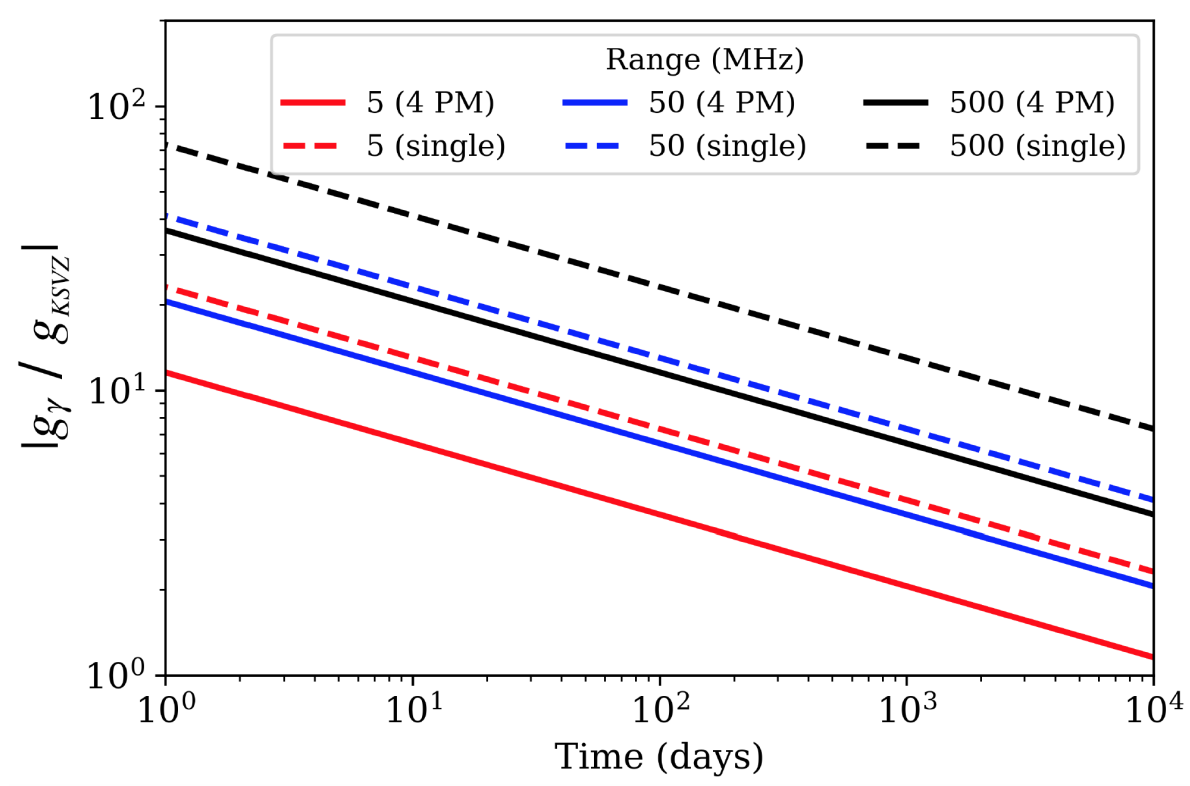

(a) Axion-photon coupling exclusion prospects as function of time for three different tuning ranges. The solid lines correspond to four PM cavities whereas the dashed lines to the usage of a single cavity (49).

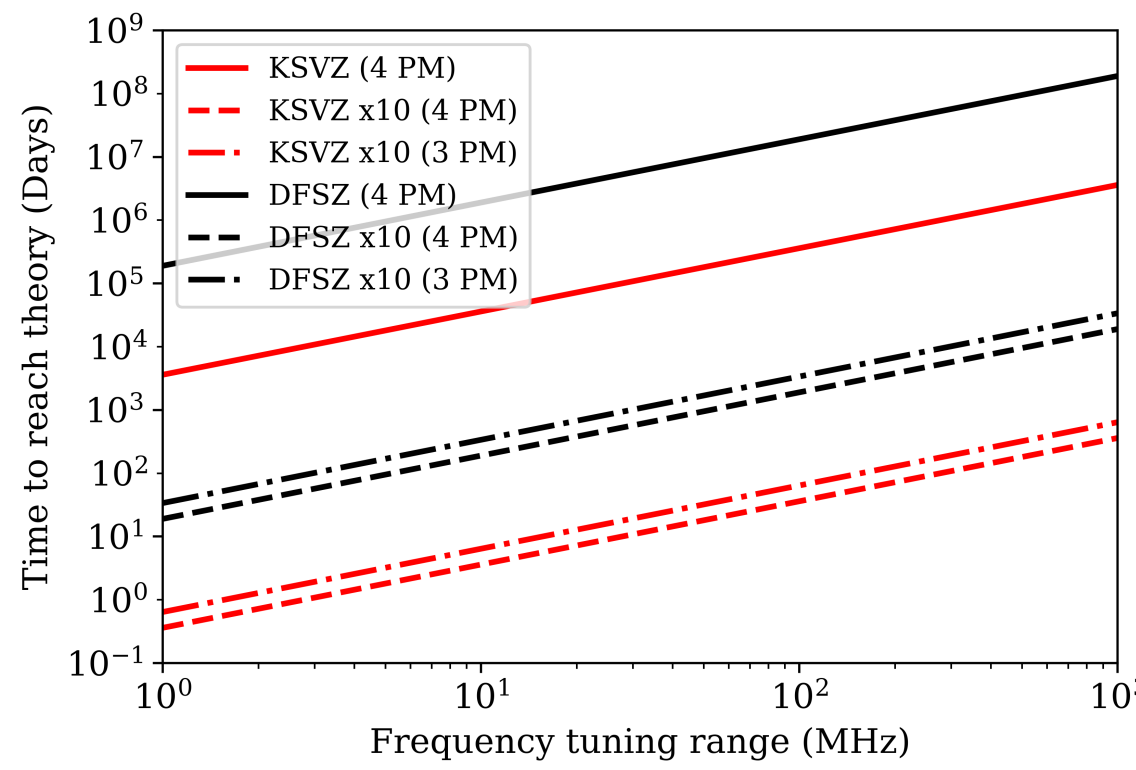

(b) Data-taking time required to reach the KSVZ and DFSZ limits as a function of the covered frequency range with four and three PM cavities.

**Figure 23.15:** Exclusion prospects of CAST-CAPP detector for halo DM axions with the usage of either a single or all four PM cavities.

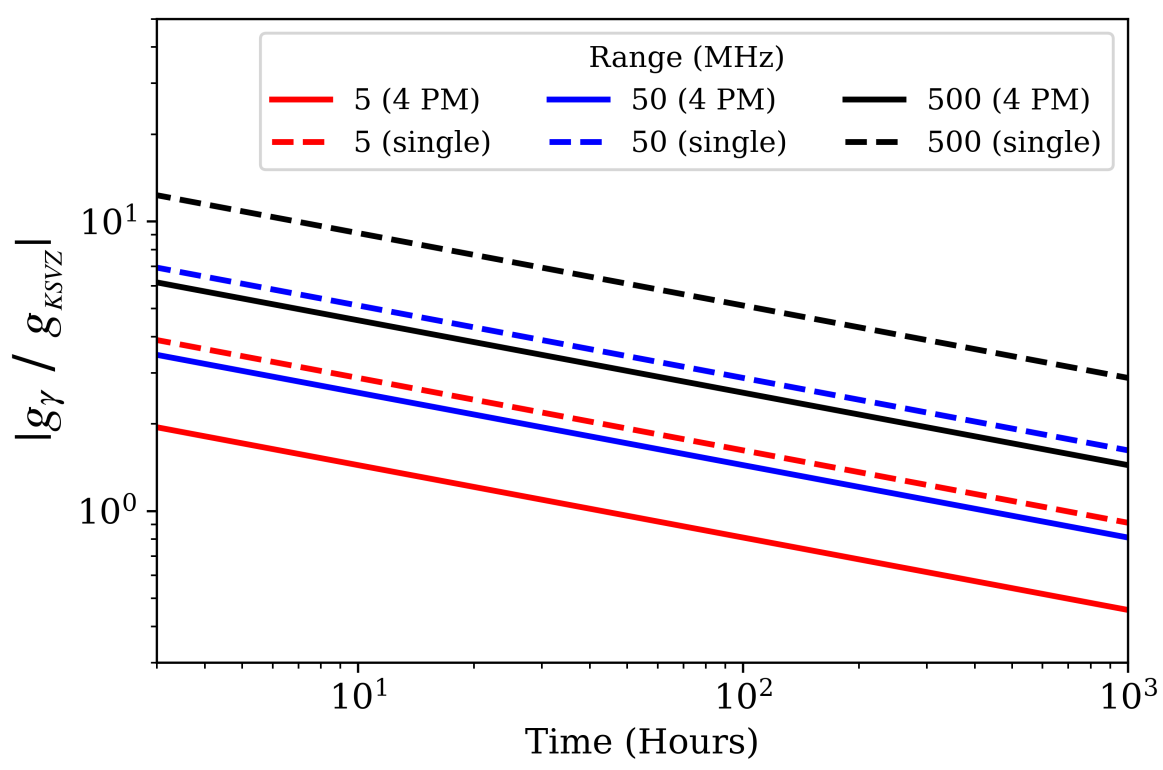

**(a)** Exclusion prospects as function of time for three different tuning ranges. The solid lines correspond to four PM cavities whereas the dashed lines to the usage of a single cavity. The x-axis is in hour.

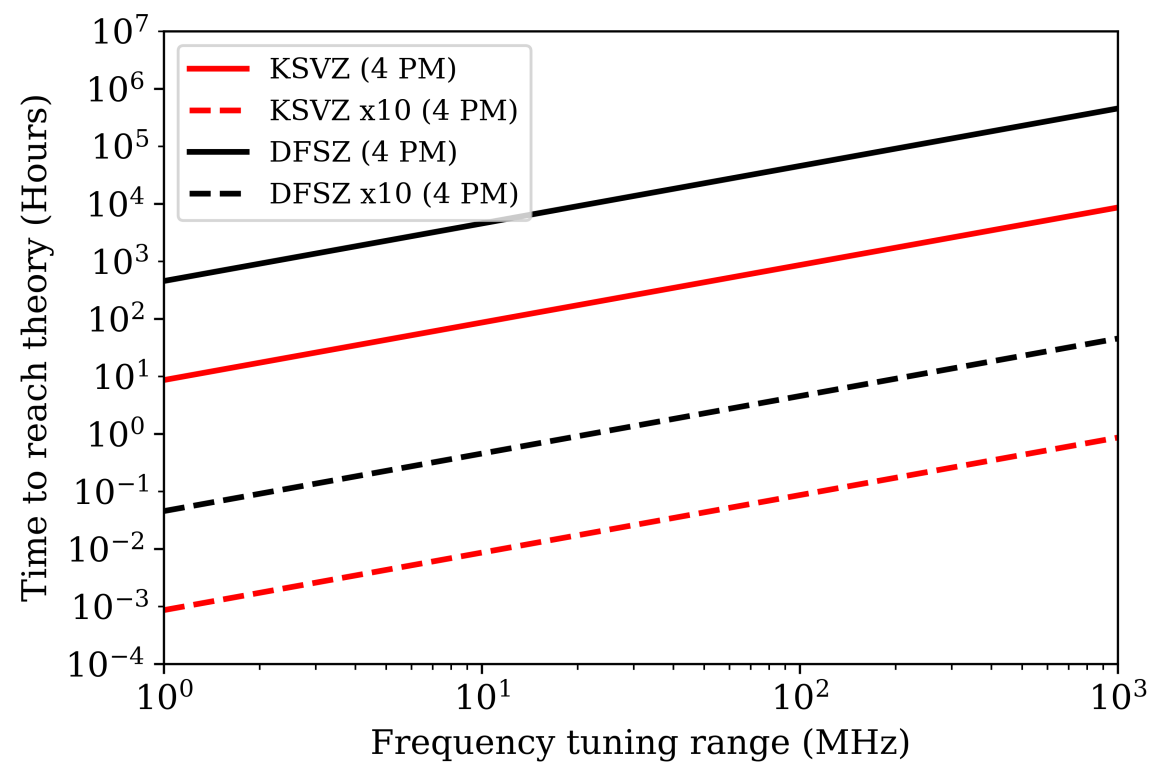

**(b)** Data-taking time required to reach the KSVZ and DFSZ limits (solid lines), as well as 10× these limits (dashed lines) as function of the covered frequency range (in MHz) using four PM cavities.

**Figure 23.16:** Exclusion prospects of CAST-CAPP detector for streaming DM axions with a modest enhancement of the local density by $10^2$.

## 23.5 Summary

CAST-CAPP detector has scanned an unexplored parameter space extending the axion search towards higher rest mass values from 19.74 µeV to 22.47 µeV excluding axion-photon couplings for virialised galactic axions down to $g_{a\gamma\gamma} = 8 \cdot 10^{-14}$ GeV$^{-1}$. This brings CAST-CAPP in the forefront of the worldwide landscape of DM axion search with the reached sensitivity being about one order of magnitude better than other experiments in its frequency range.

A 10 m-long dipole magnet with a high degree of field homogeneity was used and operated for the first time as a DM axion haloscope. In addition, four identical cavities have been coherently combined through the phase-matching technique which significantly increases the SNR. This technique is unique in DM research and can be extended in future large-scale high-frequency haloscopes in order to compensate the decreased sensitivity due to smaller volume (see Eq. 22.8). The successful scan of a wide mass range placed CAST-CAPP at the cutting-edge of cavity tunability in axion research. This was achieved thanks to the unique design of the locomotive mechanism and the cavity geometry which led to the absence of mode-crossings for the resonant mode over the entire frequency range.

Finally, additional searches for transient events can be performed parasitically, even though the protocol for signal identification is different, allowing to establish new avenues in DM research. The discovery potential increases with the probability to witness transients. For this reason, the novel fast resonant scanning technique together with the wide-band electronics are important to take full advantage of such events (see also Publications E.6, E.22, E.24, and E.25). In the future, the raw time domain data will be utilised for transient event searches as well as daily modulations.

# 24

# DISCUSSION

The CAST-CAPP sub-detector has demonstrated for the very first time the successful integration of the fast resonant scanning technique, and has pioneered the new technology of PM multiple cavities in a dipole magnet which maximises the detection sensitivity. The PM capability is still unique in axion searches and it has been successfully performed for the first time by CAST-CAPP. On the other hand, the fast resonant scanning technique enabled the capability of detecting short-lasting transient evens resulting from streaming DM or axion mini-clusters in addition to conventional searches for halo DM axions.

Regarding galactic axion searches, CAST-CAPP detector has successfully scanned with four cavities a quite wide axion mass range of about 660 MHz from 19.74 µeV to 22.47 µeV in a total data-taking time of $\sim 172$ d with the world-class results exceeding by almost two orders of magnitude the previous exclusion limits (Fig. 23.13).

At the same time, the fast resonant scanning technique which allows a maximum scanning speed of about 10 MHz/1 min was utilised searching for axion transients, while setting local limits for the period of scanning. Even while searching for galactic axions, the detector capabilities were kept wide-band to allow for the detection of shortly-lasting axion bursts. An important addition for the latter search-method, was the integration of a second measuring channel in the DAQ chain, with a simultaneous measuring VSA operating at the same frequency band as the cavities, and connected to an external omnidirectional antenna. This allowed the mapping of the ambient EMI/EMC parasitic signals at the area around the CAST experiment and was able to provide a quick characterisation of potential outliers. A specifically designed independent analysis procedure allowed the combination of processed spectra daily to be sensitive to transient signals on a daily basis. These combined spectra, which were compared with the corresponding ones from the second channel were studied for transient narrow lines. The results for the whole dataset following the aforementioned procedure is currently underway and may have a surprise. In the future, the analysis strategy for streaming axions can be modified further to allow a search for hourly-lasting signals. Lastly, an alternative wide-band scanning technique has been established for the first time, which is based on an out of resonance scanning, abandoning the resonance enhancement factor $Q$. This could be more than compensated by a temporally large axion flux enhancement (see Publications E.24, E.26).

It is noted that for an unambiguous detection of transient events, an independent detector is required at another location. Therefore, more DM experiments should follow this streaming strategy with the optimum scenario being the creation of a network of coordinated wide-band axion antennae haloscopes distributed worldwide which would cover a large parameter space as fast as possible. The duty cycle of the experiments should also be kept to a maximum to take advantage of such short-lasting signals. The reasoning for axions applies to some degree also to other DM candidates such as WIMPs (see Publications E.9, E.10). This constitutes an important addition to the direct search strategies for DM which have yet proven unsuccessful (see also Publication E.7).

# Part VI: Conclusions

*"Facts are more mundane than fantasies, but a better basis for conclusions."*

—Amory Lovins (1947–),
*American Writer*

One of the main challenges in modern physics is the detection of the constituents of DM with the strongest evidence coming from large-scale cosmic observations. However, so far a large number of direct DM searches have failed to provide a convincing evidence. The large scale observations suggest that the ordinary DM is relatively isotropic distributed at least in our solar neighbourhood. Nonetheless, the co-existence of dark streams, debris flows and the dark disk hypothesis have been widely discussed in the literature. At the same time, due to the non-relativistic effects of some candidates, solar and planetary gravitational lensing becomes efficient, resulting to a significant flux amplification by up to several orders of magnitude.

Following this work, the existence of such invisible streams could trigger the puzzling behaviour of the active Sun and unsolved mysteries in the Earth's atmosphere in a rather viable way. Such phenomena include the triggering mechanism of solar flares, the unnaturally hot solar corona, the sunspot cycle as well as the Earth's ionospheric ionisation excess around December, the stratospheric temperature anomalies around January and EQs. At the same time, the deficiency of the current solar and atmospheric models to explain and predict the origin of such phenomena also hints on the existence of an additional external influence. The generic dark candidate constituents are referred in this work as "invisible massive matter" to distinguish them from ordinary DM candidates such as axions and WIMPs which have been experimentally excluded. These invisible constituents from the dark sector should have a velocity distribution which would allow planetary and solar gravitational focusing to occur and interact "strongly" with normal matter in order to produce macroscopically observable effects. Fitting invisible massive matter candidates include AQNs, monopoles and dark photons. The key signature for the streaming scenario is a statistically significant planetary relationship which should be derived based on the optimum alignment of the planets with the incoming stream(s). Such a planetary relationship can not be explained with conventional approaches like remote planetary tidal forces since they are too feeble to cause such significant and diverse effects, while their impact extends smoothly over an orbital period.

In fact, by performing a statistical analysis of the time distribution for a variety of solar and terrestrial long-term observables projected on the heliocentric longitudinal position of the various planets, several conventionally unexpected statistically significant peaking distributions have been derived. Interestingly, in some cases an optimum longitudinal direction was found coinciding with the direction of the GC. Such small-scale anomalous observations showed planetary relationships providing thus accumulating evidence for the working hypothesis of slow streaming invisible "strongly"-interacting massive matter being gravitationally focused by the solar system bodies towards the Sun or the Earth, enhancing enormously their impinging flux. Additionally, by performing a Fourier analysis, narrow peaks are observed around 27.32 d, a periodicity which matches surprisingly Moon's sidereal month fixed to remote starts. These peaks are also often stronger than the corresponding ones around 29.53 d associated with the Moon's synodic period fixed to the Sun. These observations point on their own to an additional significant exo-solar impact modulated also by the Moon's orbital position. The 27 d periodicity is conventionally attributed to the differential solar rotation, which however can not explain such narrow periodicities since it spans from about 25 d to 36 d between the solar equator and poles respectively. In the future, more refined analyses including more or even all the solar system bodies with more constraints applied in even more long-term datasets might provide further insights on the direction(s) of the putative stream(s), their velocity distribution and the assumed interaction with normal matter.

The streaming scenario can also have considerable implications to direct DM searches and in detectors such as CAST-CAPP consisting of microwave cavities searching for DM axions. Low-speed particles can be focused by the solar system bodies downstream at the position of the Earth resulting to transitory marked flux enhancement of several orders of magnitude. This would give rise to an unexpectedly large transient DM flux exposure to an axion haloscope like CAST-CAPP. Therefore, a novel experimental approach has been implemented to take advantage of such burst-like axion flux enhancements due to temporally occurring streaming axion alignments. A unique fast-tuning mechanism was employed which together with the wide-band hardware of the detector and the introduced phase-matching technique which enhances significantly the SNR, allows CAST-CAPP to exploit such transient events (work in progress). At the same time CAST-CAPP excluded axion-photon couplings for virialised axions in a wide mass range with the sensitivity being about one order of magnitude better than other experiments in the related frequency range.

Concluding, the conventionally unexpected statistically significant planetary correlations that were found in this work on a wide and diverse range of solar and terrestrial observations, as well as the proposed novel search strategy for direct DM searches, allow to establish new insights in astroparticle physics indicating that the dark universe may indeed not be invisible as it is widely assumed!

# APPENDICES

*"I would have written a shorter letter, but I didn't have the time."*

—Blaise Pascal (1623–1662),
*French Mathematician, Physicist, Philosopher*

# A

# SIMULATIONS

## A.1 Basic interdependent kinematical effects

Due to the different revolution periods of the various planets, the appearance of one or more peaks in the longitudinal distribution of a single planet could be derived due to kinetic effects from other planets. This results from kinematical effects based on interdependent planetary revolution periods. For this reason a generic simulation is made that highlights the basic distribution characteristics for the longitudinal reference frames of Mercury, Venus, Earth, Mars, Jupiter and Saturn, and demonstrates the interdependent behaviour of different planets. These kind of simulations are very important as they can eventually point to the real planetary dominating peak(s) of a dataset.

On the other hand, if the available time resolution of a dataset is large e.g. monthly data, this means the distributions of the inner planets will not be conclusive since these planets either can not be used at all due to binning effects, or due to the absence of narrow peaks. However, as we will see, the appearance of certain peaks in the distributions of outer planets can point indirectly to the existence of a dominating peak in an inner planet. Even more details can be acquired from the identification of the specific amplitude of the peaks, their width as well as their exact heliocentric longitude position. From these details, the exact location, amplitude and shape of the underlying dominating peak can be extracted.

In the next sections some examples of the impact of a single $\pi$-distributed peak in the distributions of the various planets are depicted. Naturally, the appearance of different shaped peaks as well as in different locations and with different amplitudes as well as the appearance of more than one peak and the combination of more planets can result in very different distributions. Therefore the following cases are indicative, and for different and/or more complex distributions, a case-specific simulation has to be performed. Finally, it is noted that the appearance of some peaking distributions that will be observed in the next simulations is clearly due to orbital kinematical effects and the different phases between the planets.

### A.1.1 Mercury

The first test is performed on Mercury, where a generic $\pi$-distribution is introduced in each orbit of Mercury. Then the heliocentric longitudinal distributions of the other planets are created. The peak in Mercury is placed around 180° and has a width of 55 d and an amplitude of 60%. It is noted that in the plots in Fig. A.1 any eccentricity effect is already removed.

As we see for example in Fig. A.1c, the single peak in Mercury can produce four peaks in Earth's spectrum.

### A.1.2 Venus

As next, in Fig. A.2 the same procedure is performed for the distribution of Venus where a $\pi$-shaped peak with a 60% amplitude and 65 d width is added every 224.701 d. In this case we see three clear peaks in the distribution of Mars. This is expected since the ration of their orbital periods is $\frac{T_{Mars}}{T_{Venus}} \simeq 3.057$.

### A.1.3 Earth

In Fig. A.3 the introduced peak around 180° in the longitudinal distribution of Earth has a width of 94 d and an amplitude of 50%. In this case we see two peaks in the distribution of Mars (see Fig. A.3d), eight peaks in Venus (see Fig. A.3c) and an oscillatory peaking behaviour in Jupiter with twelve peaks (see Fig. A.3e).

### A.1.4 Mars

In Fig. A.4 the same procedure is performed for the reference frame of Mars. In this example the introduced $\pi$-shaped peak around 180° has a width of 100 d and an amplitude of

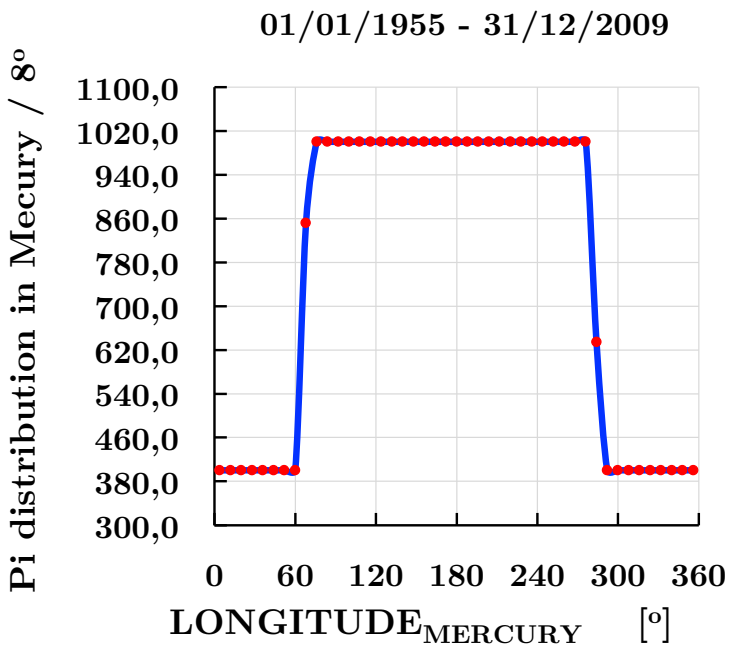

**(a)** An artificial pi-distribution in Mercury around 180° with 55 d width and 60% amplitude.

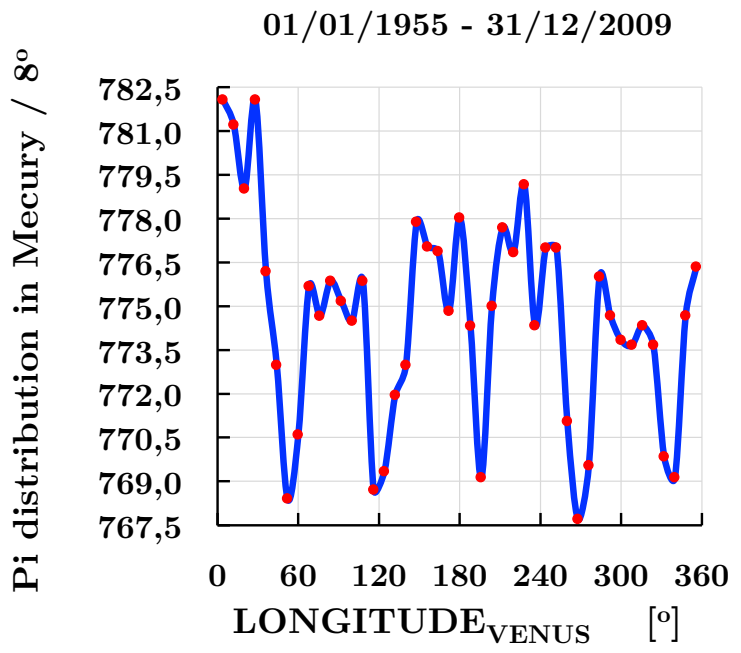

**(b)** Distribution of Venus with 8° bin resulting in a 1.84% modulation.

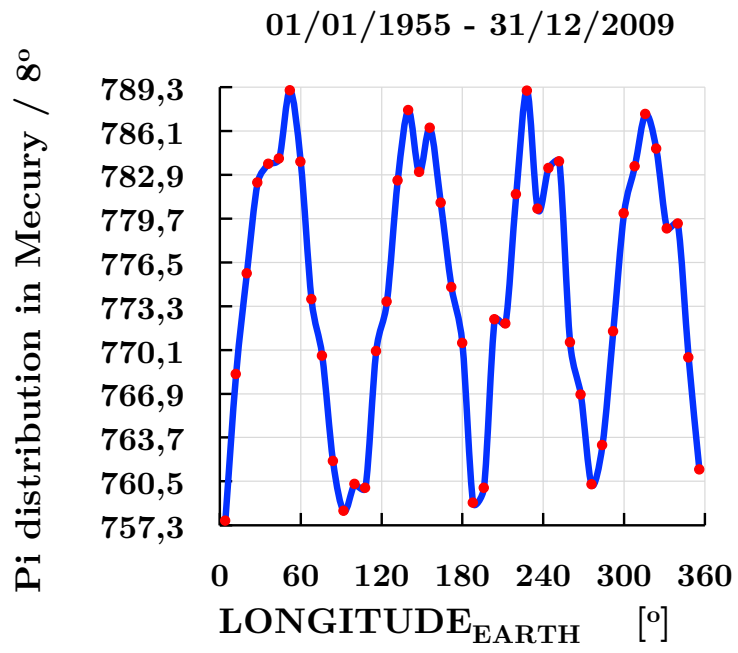

**(c)** Distribution of Earth with 8° bin resulting in a 3.99% modulation.

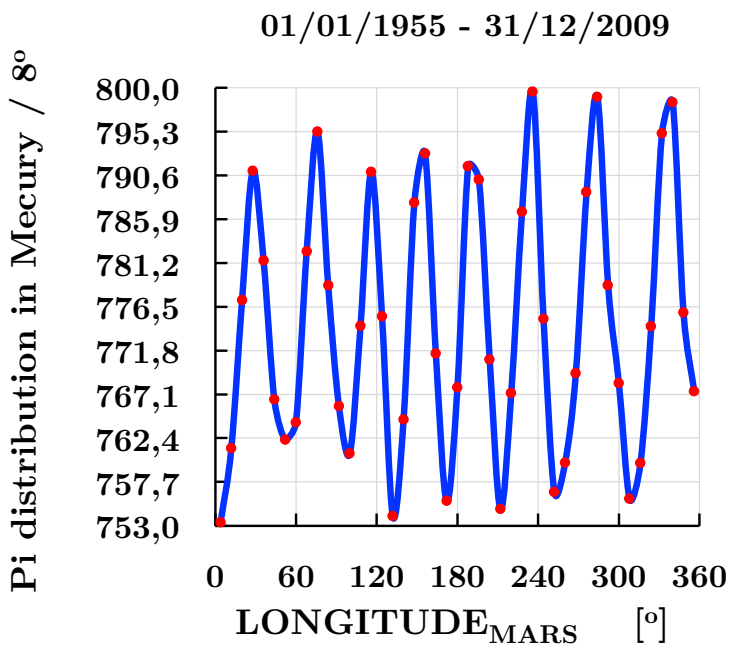

**(d)** Distribution of Mars with 8° bin resulting in a 5.78% modulation.

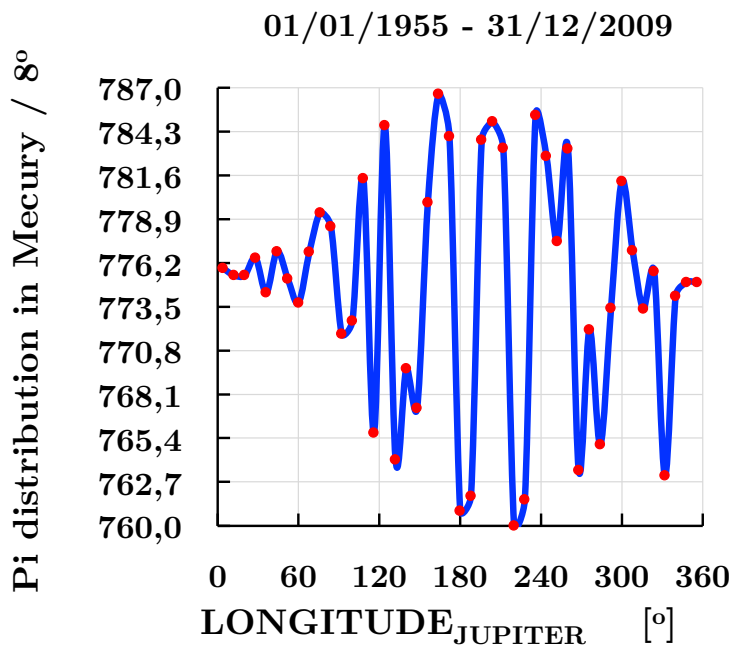

**(e)** Distribution of Jupiter with 8° bin resulting in a 3.38% modulation.

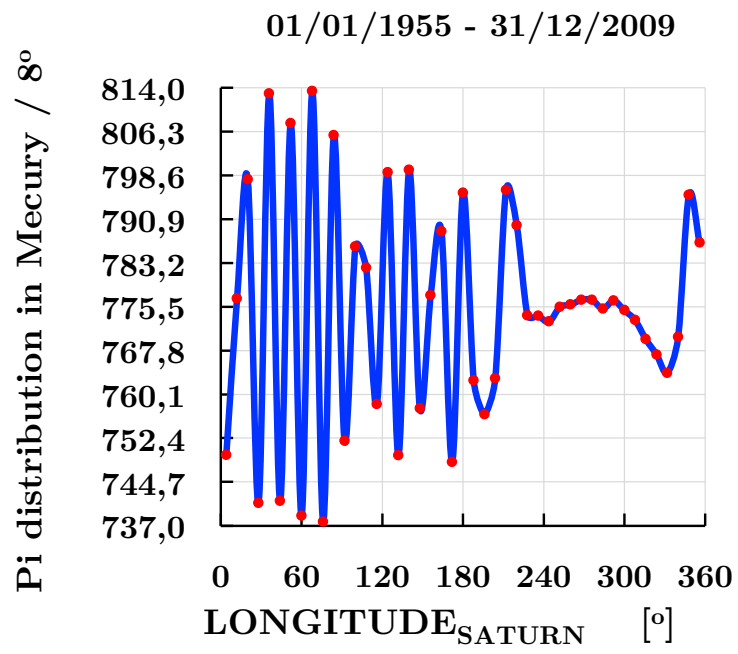

**(f)** Distribution of Saturn with 8° bin resulting in a 9.31% modulation.

**Figure A.1:** An artificial $\pi$-distribution in Mercury around 180° and its influence on the longitudinal distributions of the rest of the planets.

60%. As a result, in Venus (Fig. A.4c) and Earth (Fig. A.4d) a wide peak is observed around 0°.

### A.1.5 Jupiter

In Fig. A.5 the reference frame is set this time on Jupiter and an artificial simulated peak is injected in each of Jupiter's orbit. In this specific example the introduced $\pi$-shaped peak around 180° heliocentric longitude has a width of 400 d and an amplitude of 60%. These results to five peaks in the longitudinal distribution of Saturn (Fig. A.5f)

### A.1.6 Saturn

Finally, a simulated peak is injected in each of Saturn's orbit with its effect on the reference frames of the rest of the planets shown in Fig. A.6. The introduced $\pi$-shaped peak around 180° heliocentric longitude has a width of 600 d and an amplitude of 60%. This results to two small peaks in the distribution of Earth (Fig. A.6d) and two in the distribution of

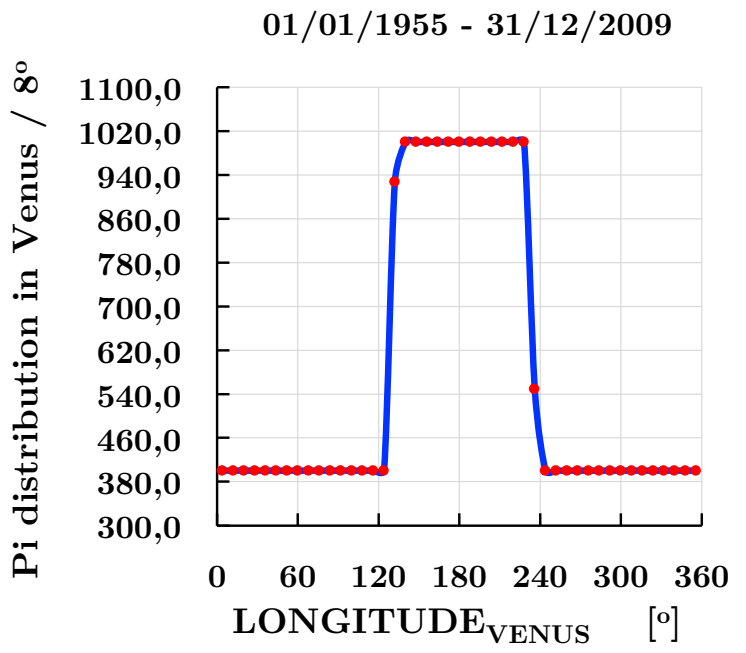

(a) An artificial $\pi$-distribution in Venus around 180° with 65 d width and 60% amplitude.

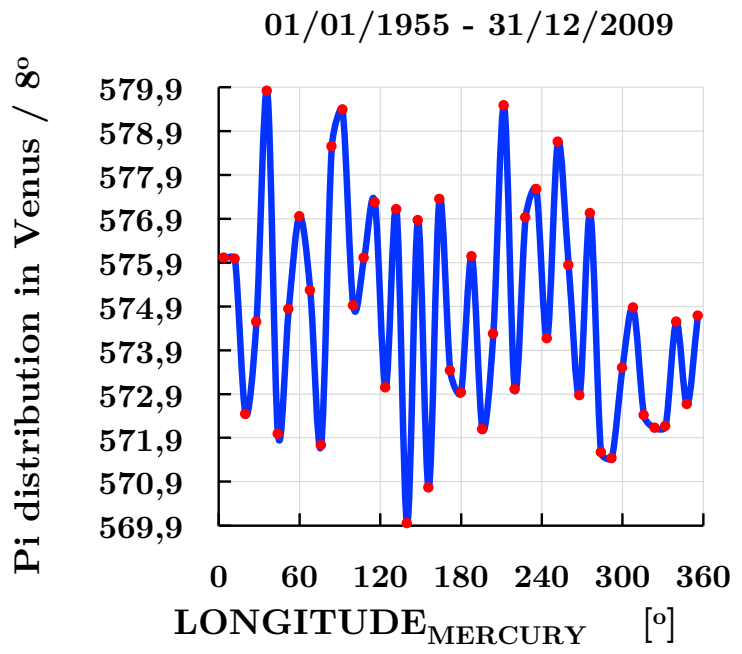

(b) Distribution of Mercury with 8° bin resulting in a 1.70% modulation.

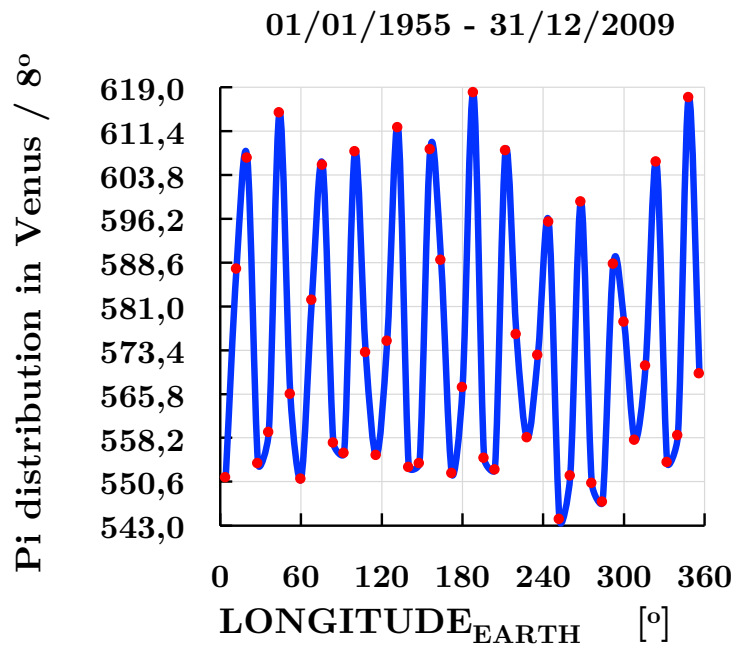

(c) Distribution of Earth with 8° bin resulting in a 11.97% modulation.

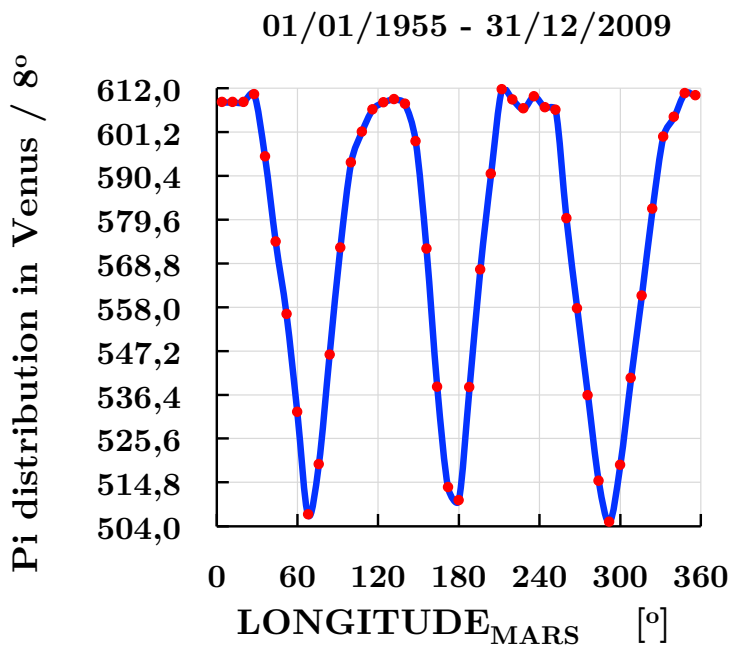

(d) Distribution of Mars with 8° bin resulting in a 17.43% modulation.

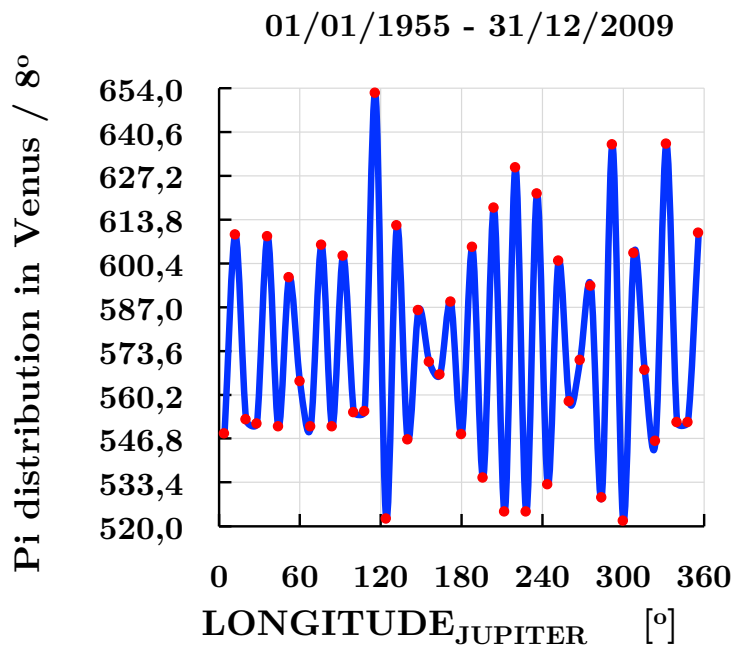

(e) Distribution of Jupiter with 8° bin resulting in a 20.07% modulation.

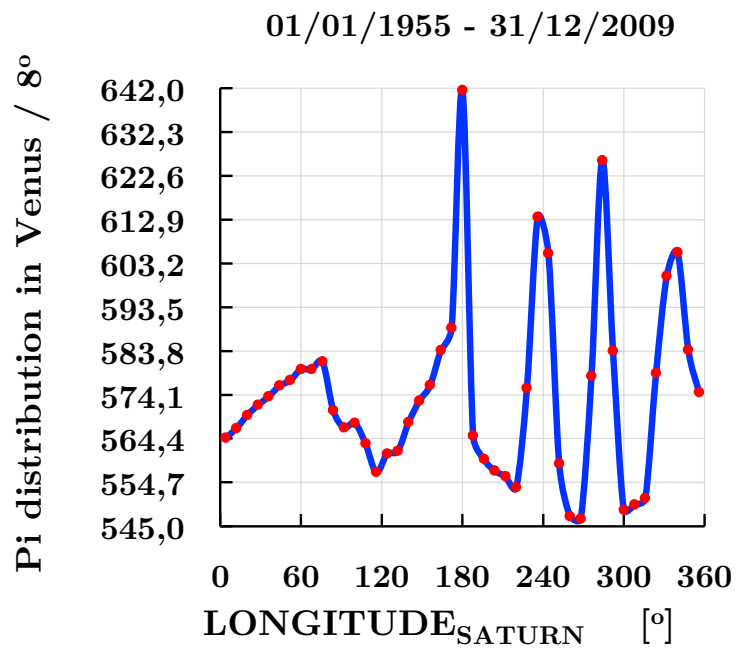

(f) Distribution of Saturn with 8° bin resulting in a 14.80% modulation.

**Figure A.2:** An artificial $\pi$-distribution in Venus around 180° and its influence on the longitudinal distributions of the rest of the planets.

Jupiter (Fig. A.6f).

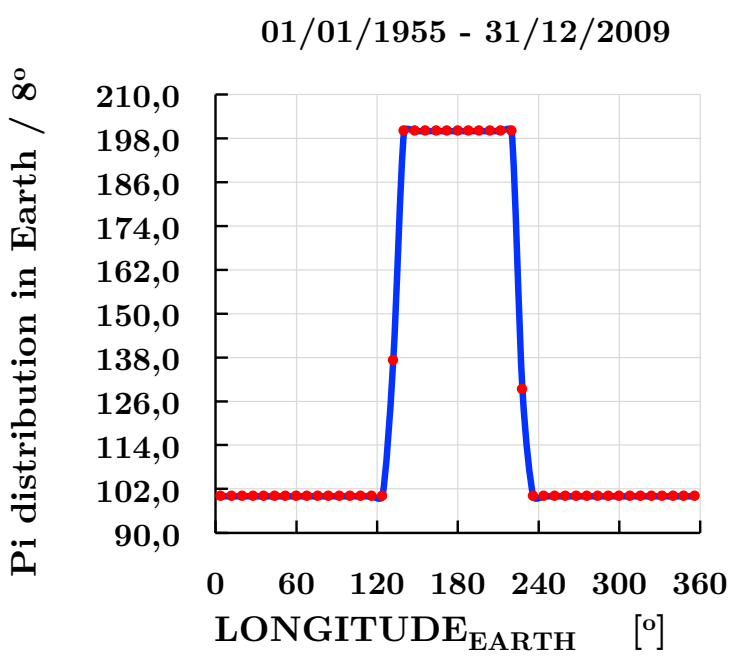

**(a)** An artificial $\pi$-distribution in Earth around 180° with 94 d width and 50% amplitude.

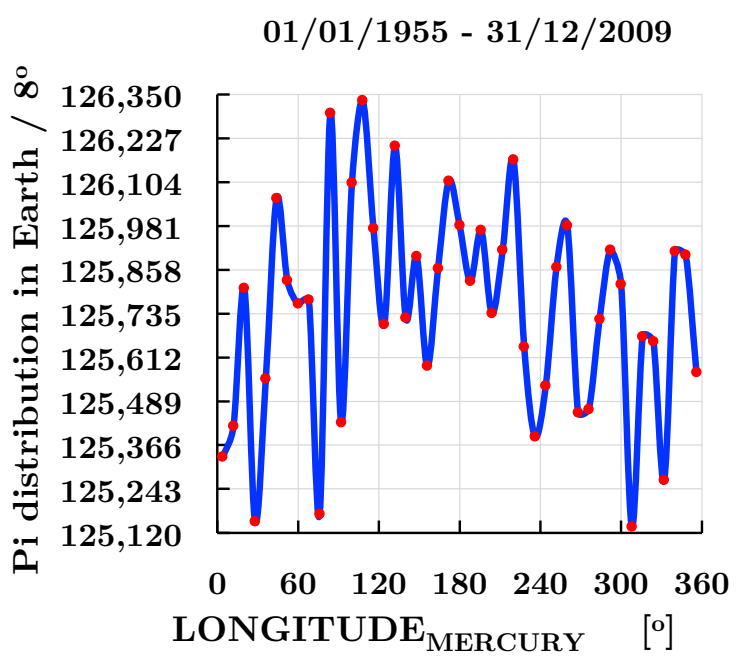

**(b)** Distribution of Mercury with 8° bin resulting in a 0.95% modulation.

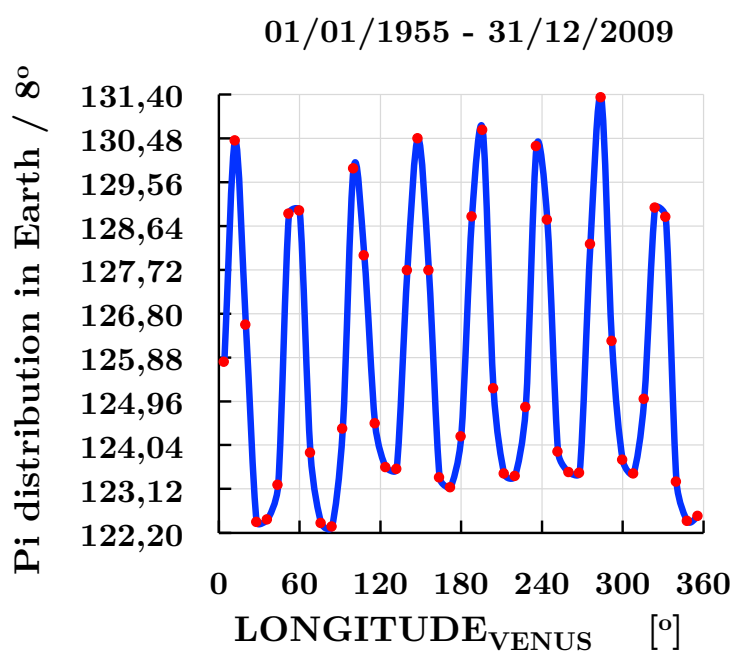

**(c)** Distribution of Venus with 8° bin resulting in a 6.86% modulation.

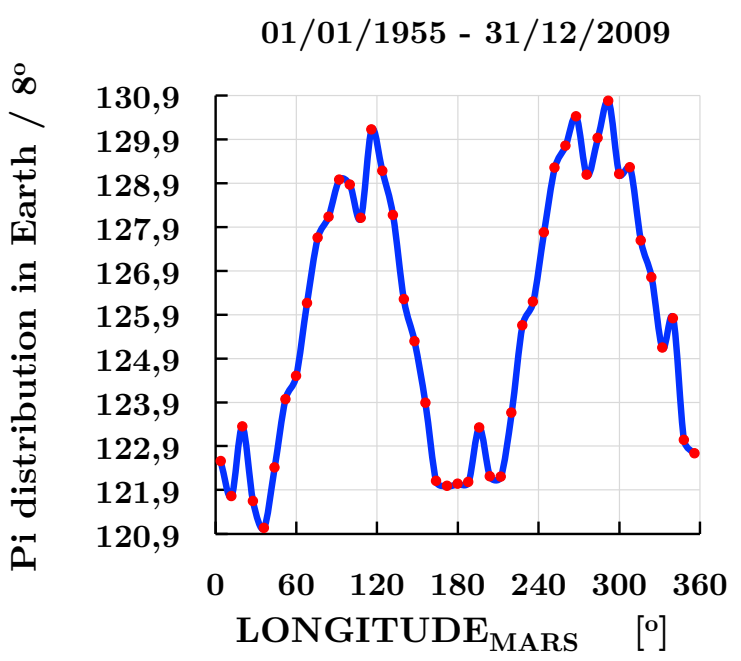

**(d)** Distribution of Mars with 8° bin resulting in a 7.45% modulation.

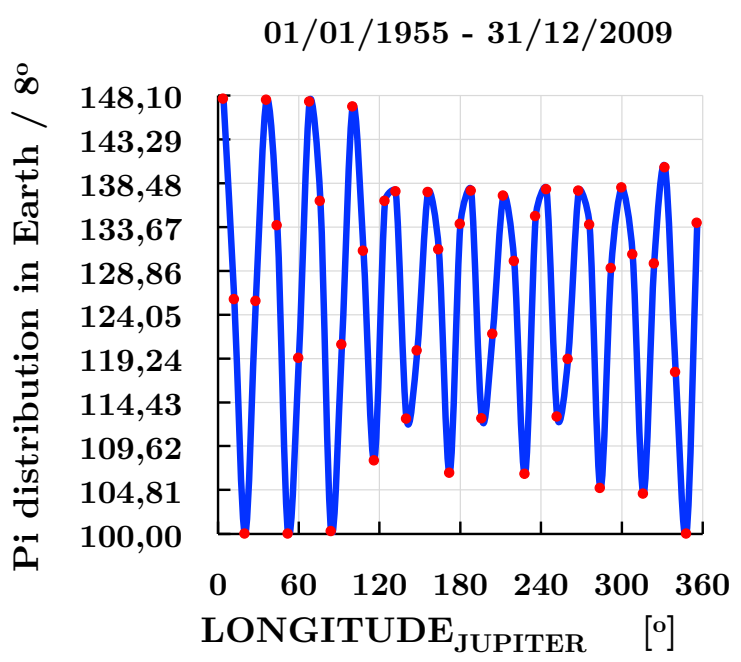

**(e)** Distribution of Jupiter with 8° bin resulting in a 32.30% effect.

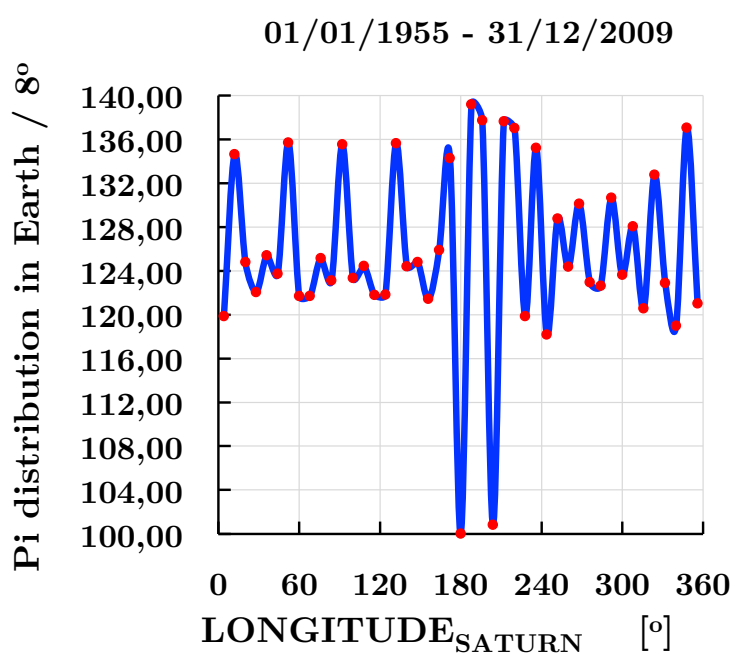

**(f)** Distribution of Saturn with 8° bin resulting in a 28.14% modulation.

**Figure A.3:** An artificial $\pi$-distribution in Earth around 180° and its influence on the longitudinal distributions of the rest of the planets.

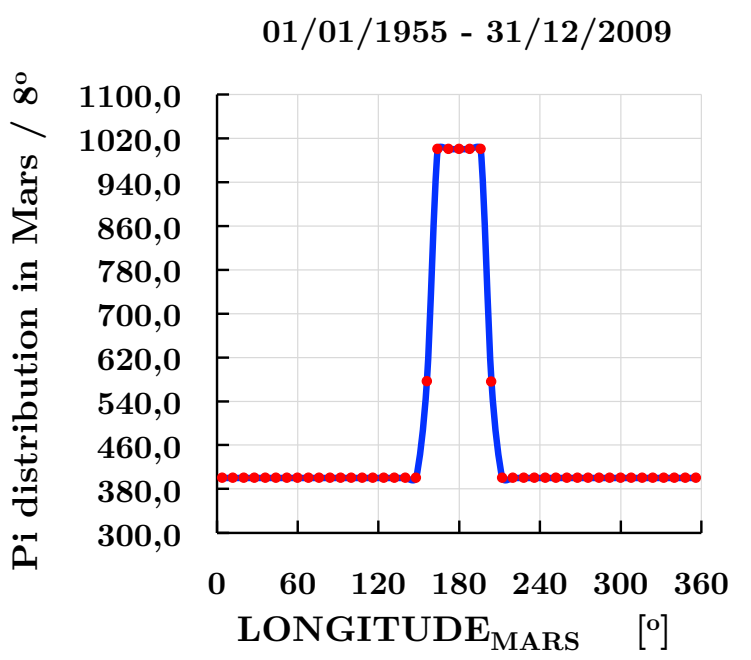

**(a)** An artificial $\pi$-distribution in Mars around 180° with 100 d width and 60% amplitude.

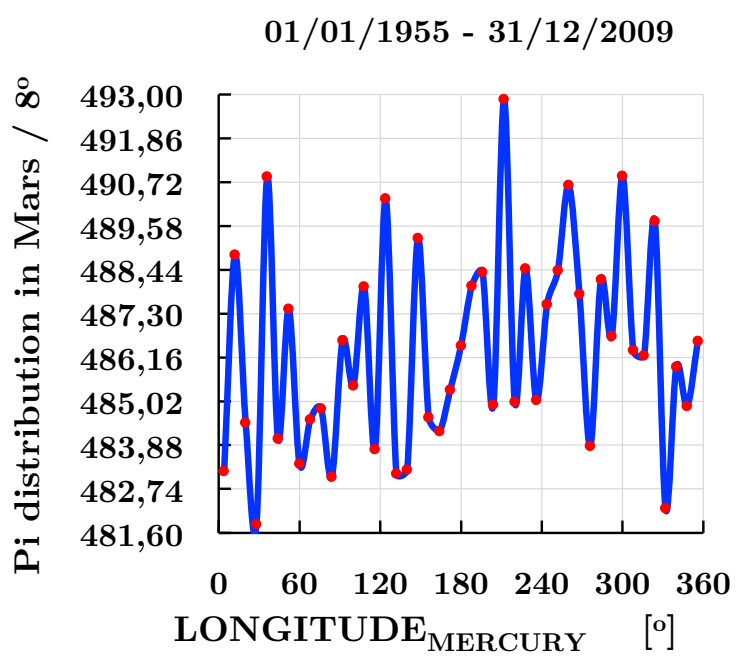

**(b)** Distribution of Mercury with 8° bin resulting in a 2.24% modulation.

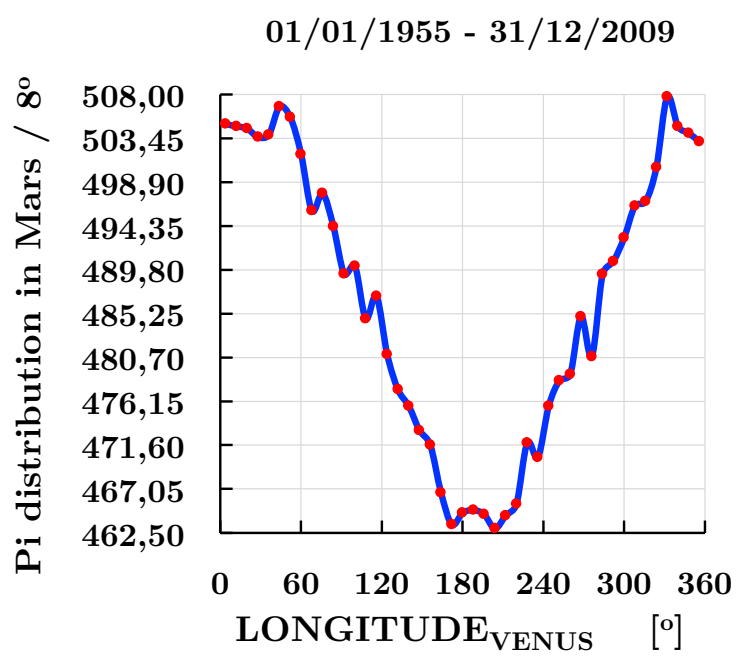

**(c)** Distribution of Venus with 8° bin resulting in a 8.83% modulation.

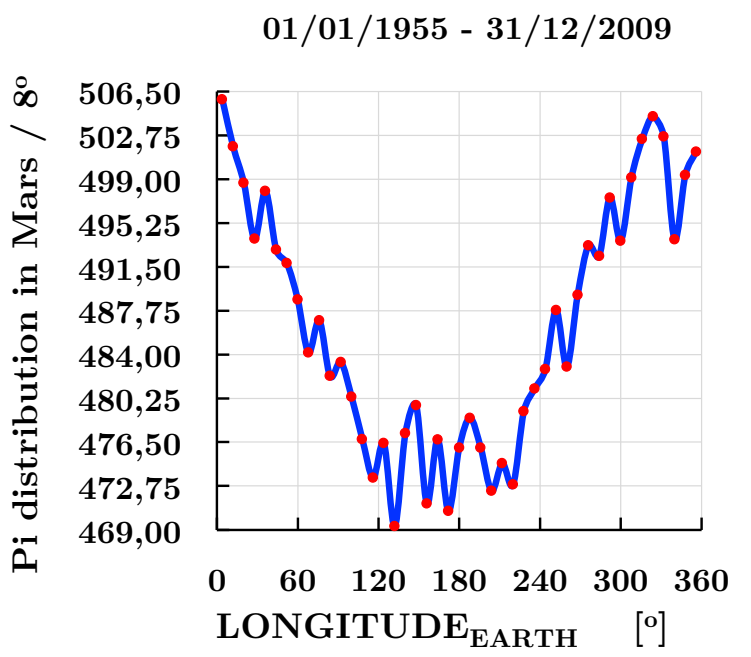

**(d)** Distribution of Earth with 8° bin resulting in a 7.22% modulation.

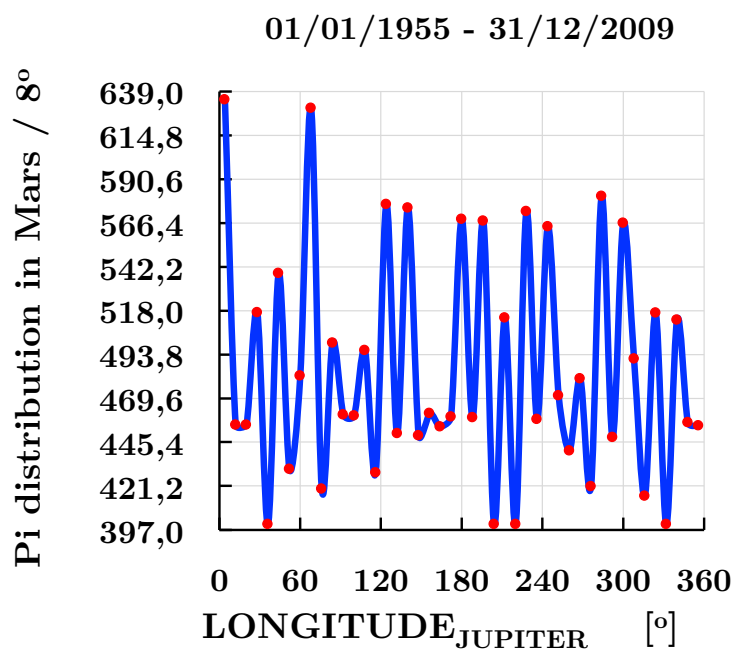

**(e)** Distribution of Jupiter with 8° bin resulting in a 36.96% modulation.

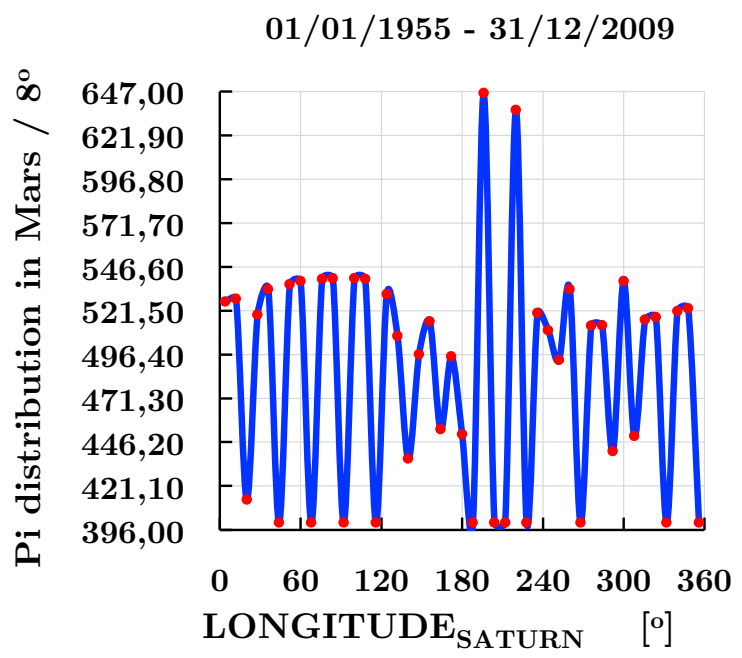

**(f)** Distribution of Saturn with 8° bin resulting in a 38.07% modulation.

**Figure A.4:** An artificial $\pi$-distribution in Mars around 180° and its influence on the longitudinal distributions of the rest of the planets.

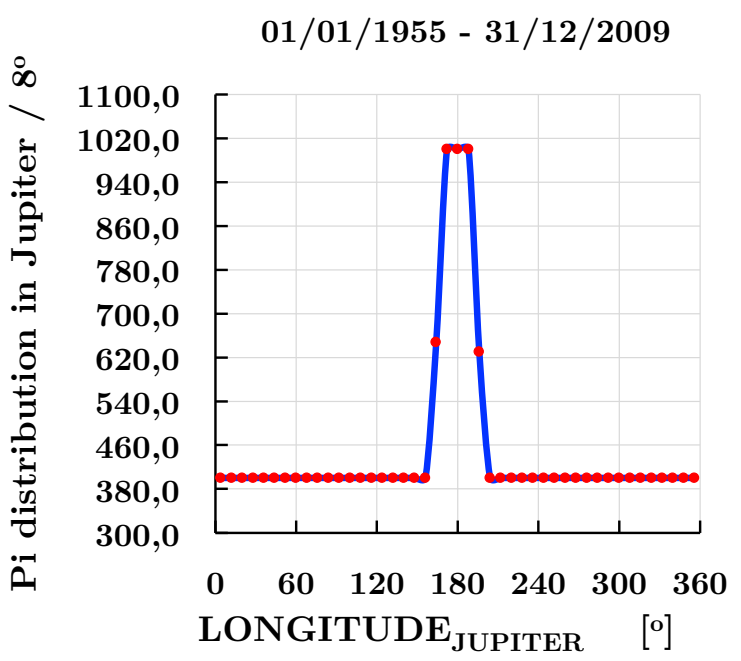

**(a)** An artificial $\pi$-distribution in Jupiter around 180° with 400 d width and 60% amplitude.

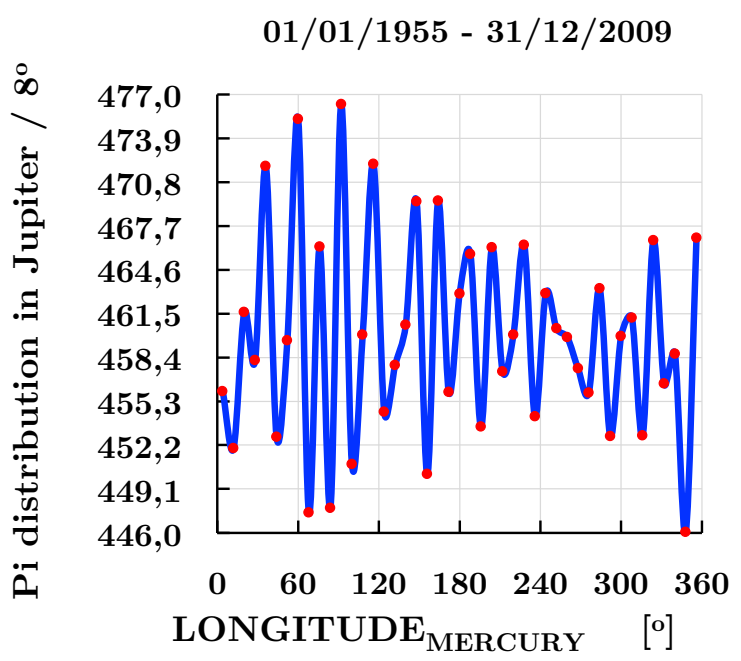

**(b)** Distribution of Mercury with 8° bin resulting in a 6.35% modulation.

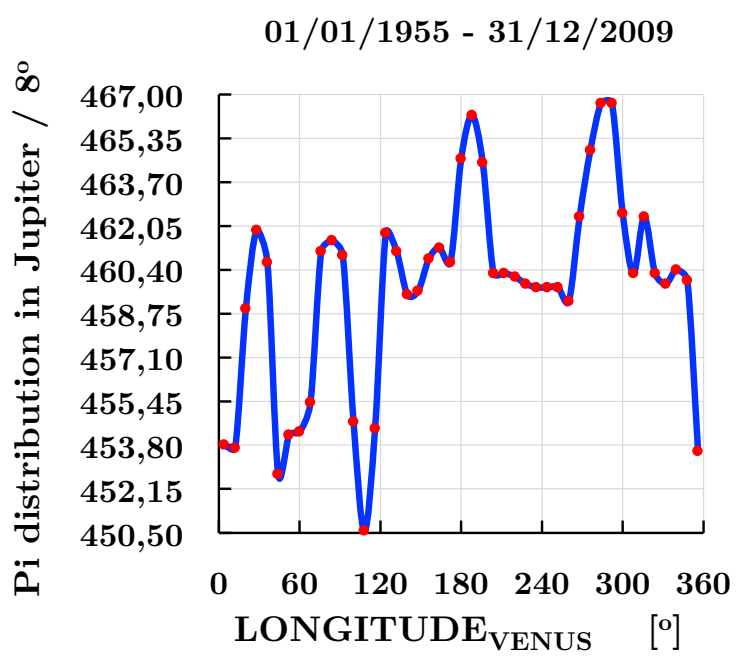

**(c)** Distribution of Venus with 8° bin resulting in a 3.45% modulation.

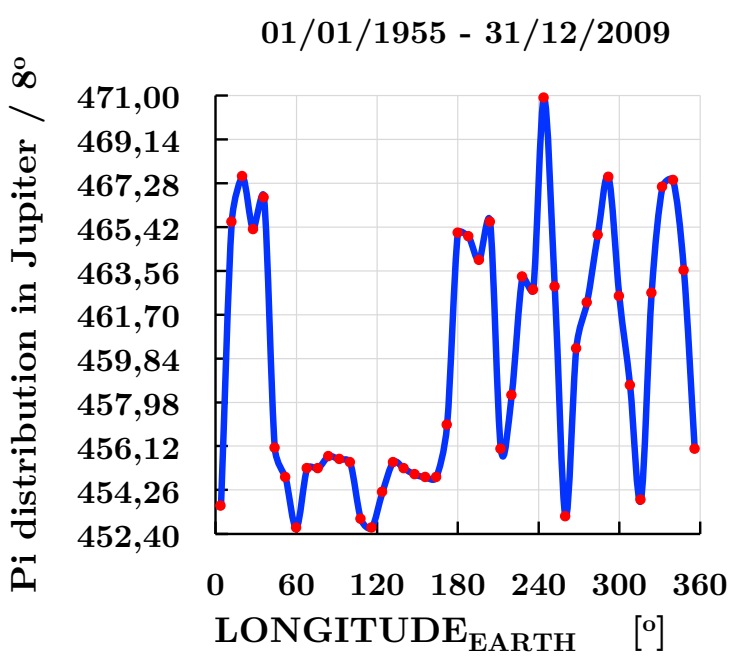

**(d)** Distribution of Earth with 8° bin resulting in a 3.87% modulation.

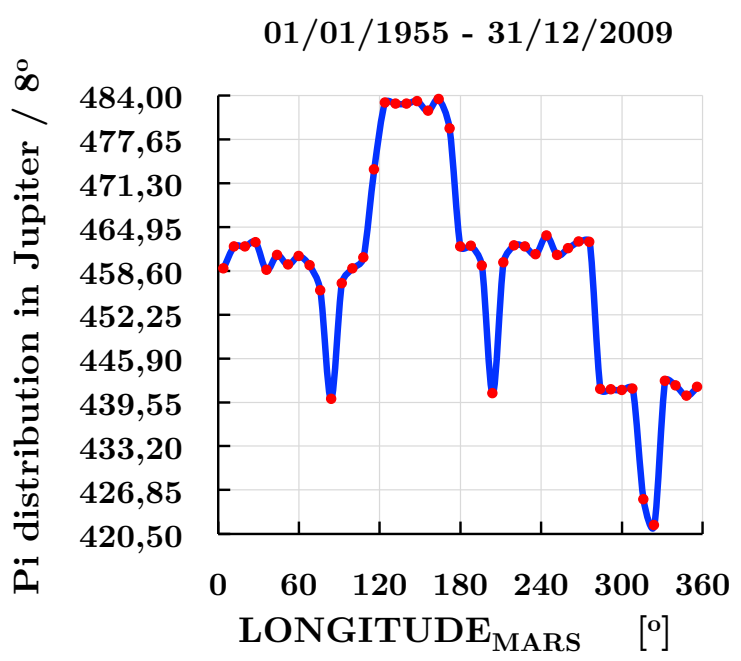

**(e)** Distribution of Mars with 8° bin resulting in a 12.76% modulation.

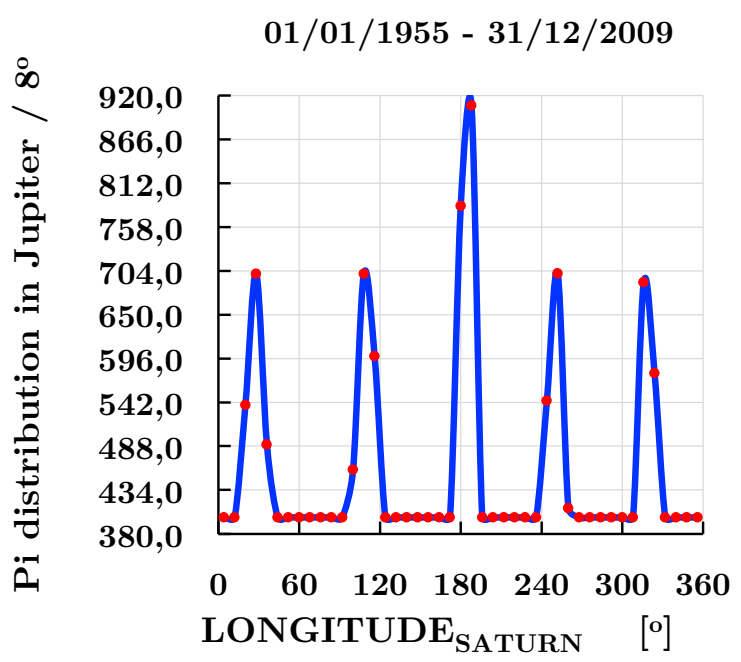

**(f)** Distribution of Saturn with 8° bin resulting in a 55.92% modulation.

**Figure A.5:** An artificial $\pi$-distribution in Jupiter around 180° and its influence on the longitudinal distributions of the rest of the planets.

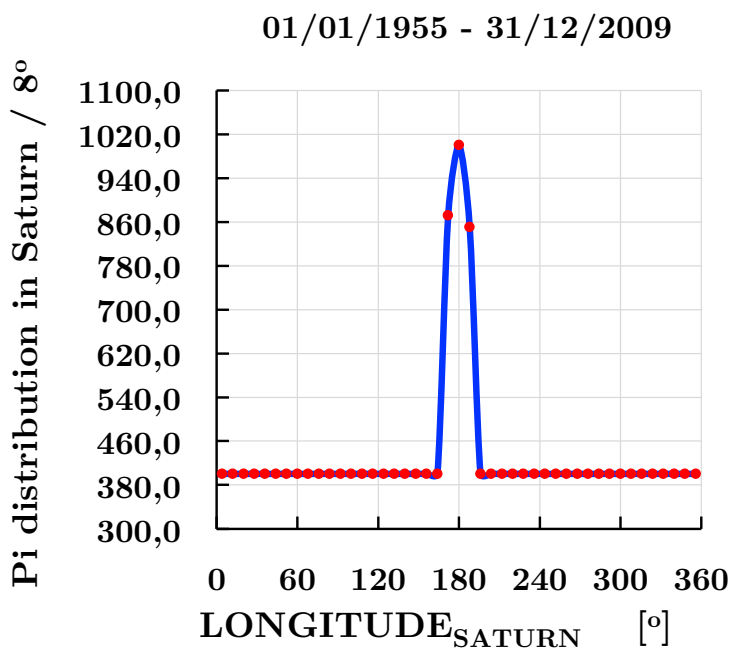

(a) An artificial $\pi$-distribution in Saturn around 180° with 600 d width and 60% amplitude.

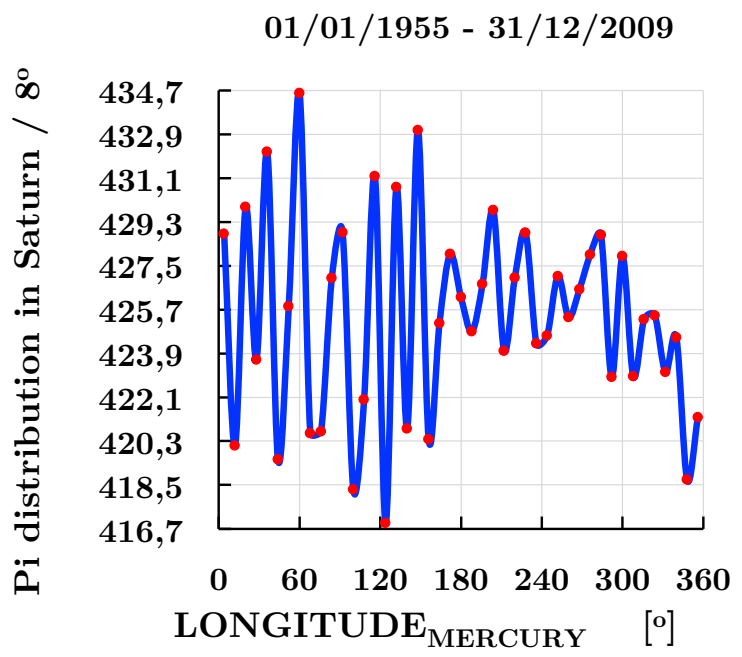

(b) Distribution of Mercury with 8° bin resulting in a 4.06% modulation.

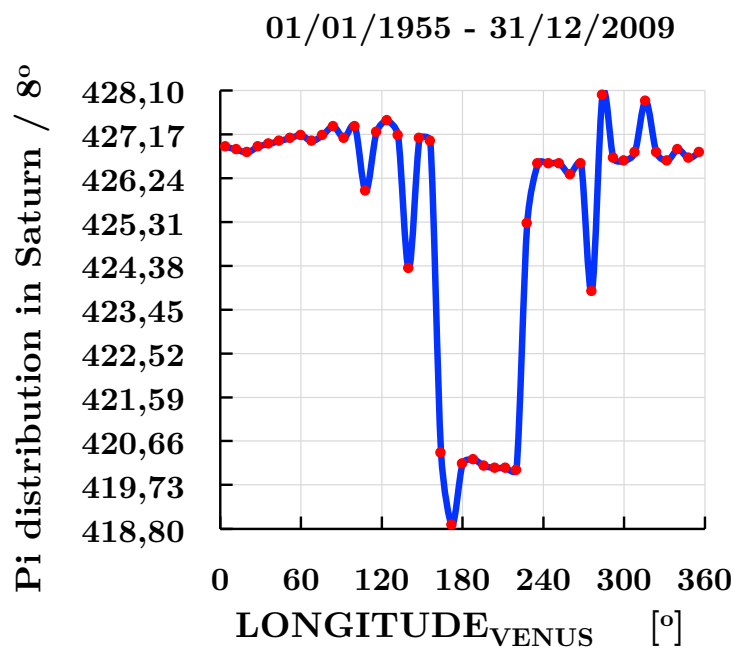

(c) Distribution of Venus with 8° bin resulting in a 2.13% modulation.

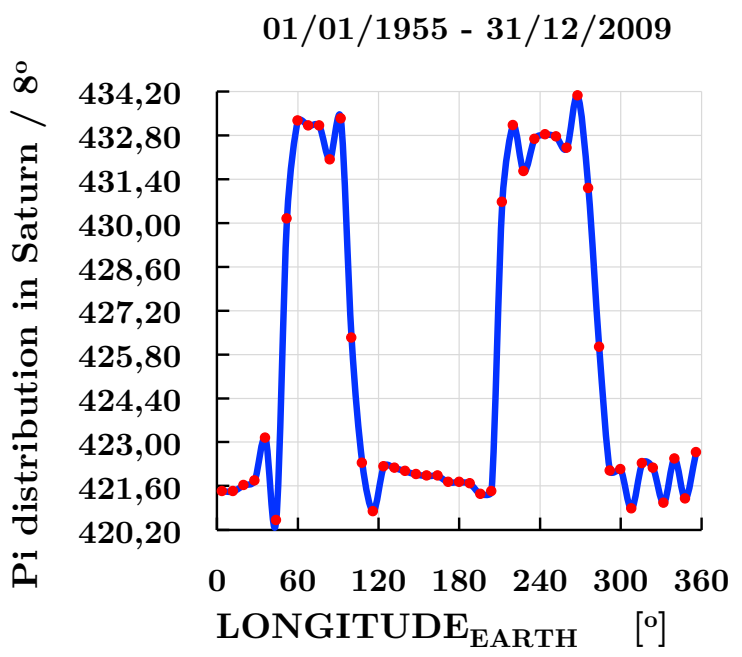

(d) Distribution of Earth with 8° bin resulting in a 3.12% modulation.

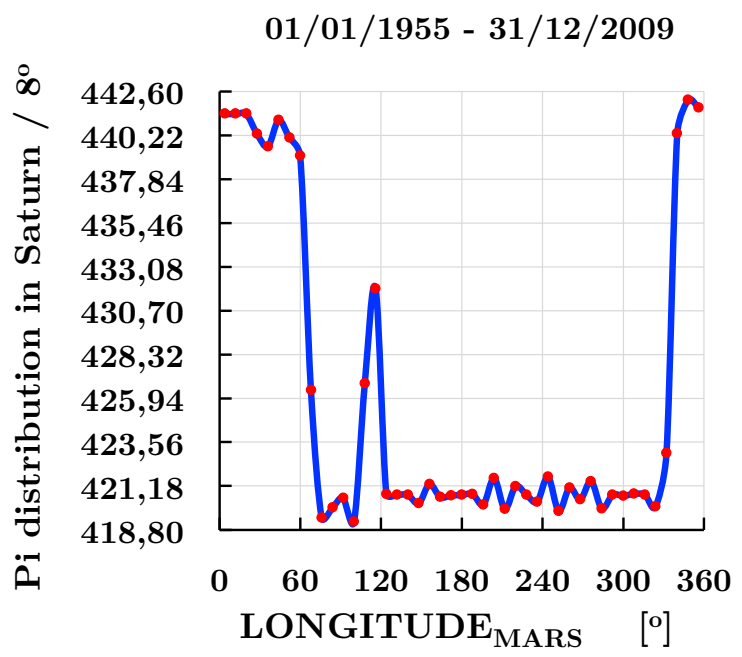

(e) Distribution of Mars with 8° bin resulting in a 5.18% modulation.

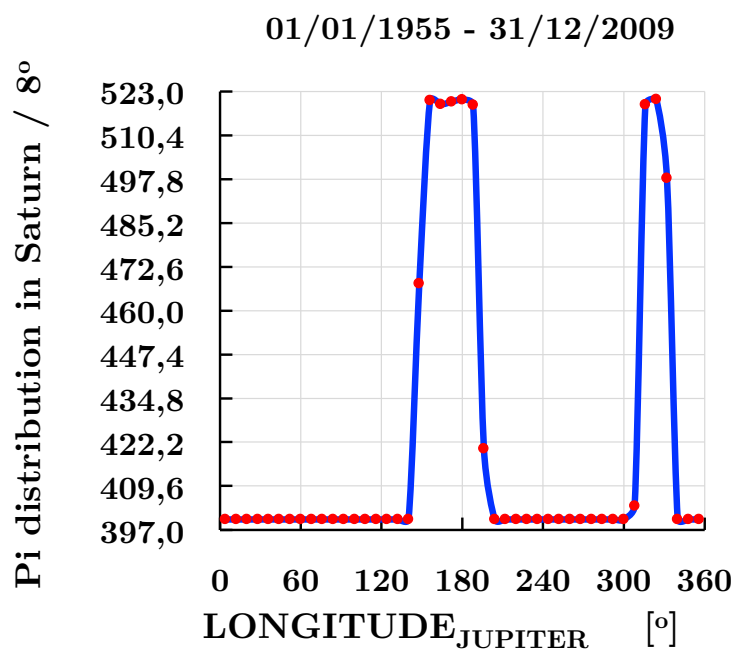

(f) Distribution of Jupiter with 8° bin resulting in a 23.19% modulation.

**Figure A.6:** An artificial $\pi$-distribution in Saturn around 180° and its influence on the longitudinal distributions of the rest of the planets.

# A.2 Case-specific simulations

## A.2.1 F10.7

### A.2.1.1 Jupiter simulation

To verify that the single wide peak observed in Jupiter in Fig. 9.3e does not create any artificial kinematic peaks in the distributions of the rest of the planets, a simulation is performed, where a peak resembling the one of Jupiter is created.

More specifically, to make a direct comparison a numeric value of 170 has been given to every 130° heliocentric angle of Jupiter and a value of 76 to every 318°. Then, a linear interpolation has been performed between these values to create a triangle-like peak similar to the one observed in Fig. 9.3e. The resulting distribution is shown in Fig. A.7e. As it is observed, in all distributions in Fig. A.7 when compared with Fig. 9.3, no F10.7 distribution can be derived from the single wide peak of Jupiter, since both the shapes and the observed effects of the real F10.7 data are different than the simulation. This fact strengthens the credibility of the results and the various peaks observed in the distributions of single planets for F10.7.

It is noted that the same simulation and its results hold for the case of sunspots where a similar peak in Jupiter's longitudinal distribution is observed (Fig. 8.6e).

## A.2.2 Solar radius

As we have already seen, a single peak appearing in the distribution of Venus results in three peaks in the distribution of Mars (Fig. A.2d), whereas a single peak in the distribution of Earth results in an oscillatory peaking behaviour of Jupiter with twelve peaks (Fig. A.3e). Therefore, to verify this more precisely in the case of the peaks observed in Sect. 10.3.1, and to verify the accuracy of the interpretation, a simulation with the exact characteristics of the individual peaks observed in the solar radius distributions of Venus and Earth is performed.

### A.2.2.1 Artificial peak in Venus

The first step is the creation of a $\pi$-shaped distribution in the daily data of Venus' heliocentric position. In order to simulate the exact characteristics of the peak seen in Fig. 10.4a we used a $\pi$ - distribution with 188 d width around 155° and an amplitude of about 18.70%. More specifically, the "background" days where given a numeric value of 100 while the "peak" days around 155° a numeric value of 123. This artificial peak was inserted in each orbit of Venus and for the same time period of $1996 - 2017$.

The distribution of these data as a function of Venus' heliocentric longitudinal position is

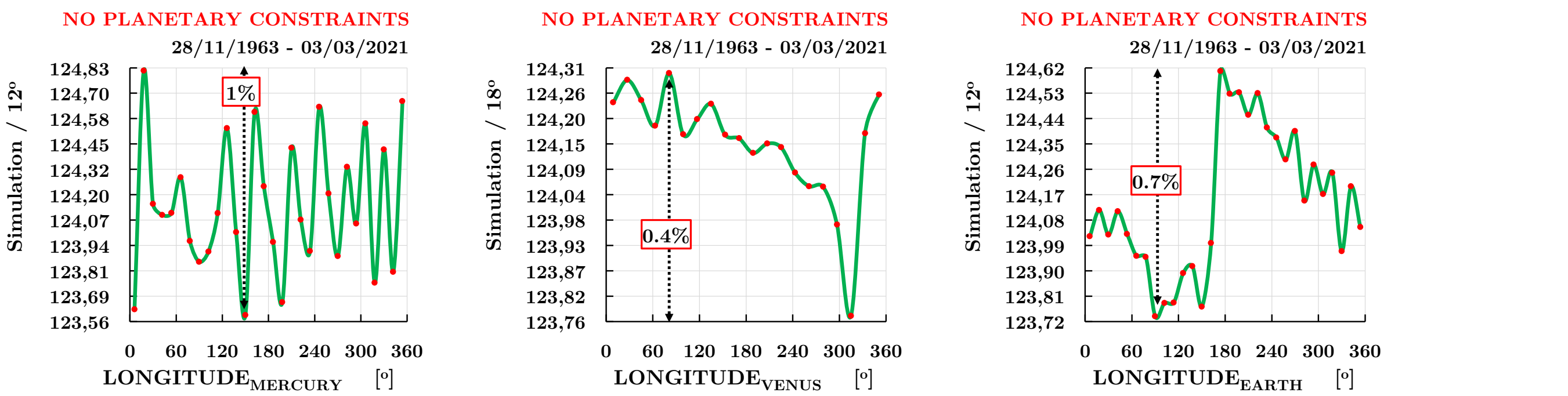

**(a)** Mercury for 12° bin. **(b)** Venus for 18° bin. **(c)** Earth for 12° bin.

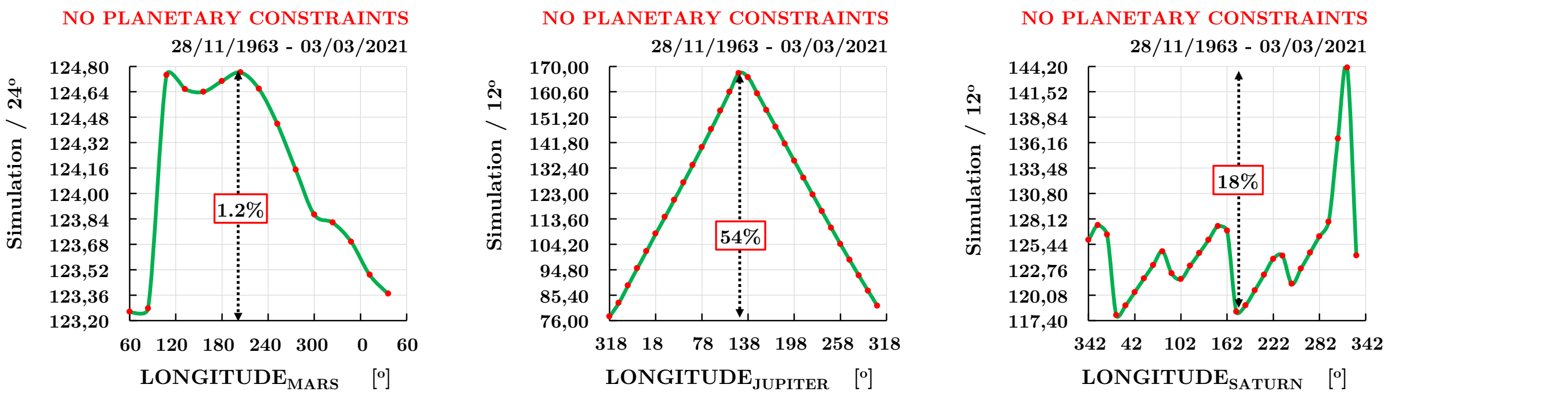

**(d)** Mars for 24° bin. **(e)** Jupiter for 12° bin. **(f)** Saturn for 12° bin.

**Figure A.7:** Planetary heliocentric longitude distributions with simulated Jupiter data for the period 28/11/1963 - 03/03/2021.

shown in Fig. A.8a. As next, these daily data, from Fig. A.8a were grouped into 72 d bins and then a linear interpolation between these values has been performed the same way as with the solar radius data to derive the daily values. This procedure was described in Sect. 10.2.2.1 and is the same handling between both the real and the simulated data. In fact, this procedure has been performed multiple times, choosing various widths and amplitudes of the simulated distribution in order to eventually reach the closest values of the distribution with the real data. The resulted distribution of Venus is shown in Fig. A.8b. The amplitude of the peak located around $99.19° \pm 1.15°$ is 6.77% and the fitted Gaussian gives a $FWHM \simeq 251.16° \pm 15.69°$ which corresponds to about $156\,d \pm 9\,d$. These numbers are very close to the ones of the peak observed in Fig. 10.4a.

Using the linearly interpolated data from Fig. A.8b, the heliocentric distribution of Mars has been plotted in Fig. A.8c. This shows that indeed, as expected, a peak in the Venus' spectrum results in three resolved peaks in the Mars' heliocentric longitudinal distribution, though each sub-peak with a smaller amplitude. The amplitude of the biggest peak (the 2$^{nd}$ one) in Mars is about 2.94%. As derived from a multiple Gaussian fit, the three peaks are located around $177.46° \pm 0.82°$, $296.18° \pm 0.77°$ and $71.34° \pm 0.77°$ and have a FWHM of $51.61° \pm 3.82°$, $60.59° \pm 4.18°$ and $56.33° \pm 4.50°$ respectively. This kinematical relationship is due to the fact that during a Mars orbit, Venus itself completes about 3.05 orbits. As a result, the relationship between solar radius and Venus is better established though the shape of Mars'

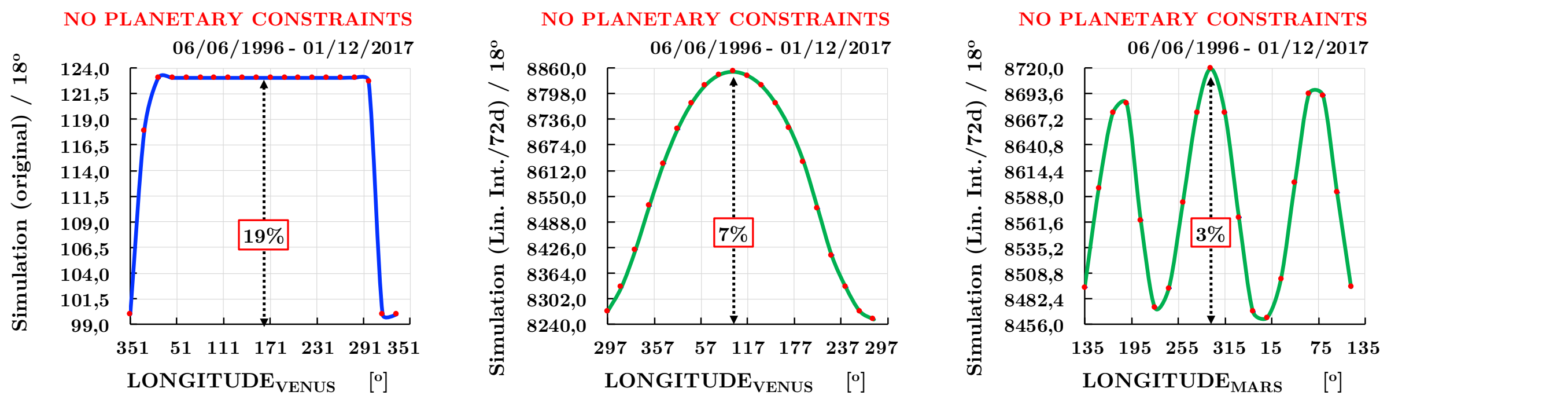

**(a)** Raw $\pi$ distribution in Venus.

**(b)** Linearly interpolated data from 72-binned days as a function of Venus' heliocentric longitude.

**(c)** Linearly interpolated data as a function of Mars' heliocentric longitude.

**Figure A.8:** Simulated $\pi$ distribution in Venus for bin = 18°.

spectrum in Fig. 10.4c which shows three peaks similarly to the simulation (Fig. A.8c).

#### A.2.2.2 Artificial peak in Earth

The exact same procedure has been performed for Earth, instead of Venus. The $\pi$-distribution used here has a 176 d width around 160° and an amplitude of about 28.06% (see Fig. A.9a). More specifically, the "background" days where given a numeric value of 100 while the "peak" days around 160° a value of 139. These values were chosen in order to eventually have similar characteristics with the solar radius data results shown in Fig. 10.4b. The next step was to group the daily data into 72 d bins and then perform a linear interpolation between these values in order to derive again the daily values as done with the real data of solar radius. The resulted distribution of Earth is shown in Fig. A.9b. The amplitude of the peak located around $122.82° \pm 1.08°$, for a bin = 18°, is 27.71% and the fitted Gaussian function gives a $FWHM \simeq 193.17° \pm 7.70°$ (Fig. A.9b).

Using these simulated data (from Fig. A.9b) the heliocentric distribution of Mars has been plotted in Fig. A.9c, which proves that a single peak in the spectrum of Earth results in two resolved peaks in the Mars' heliocentric longitudinal distribution. This is in contrast with the previous simulation seen in Fig. A.8c, where we had three peaks in the spectrum of Mars resulting from a single peak in the spectrum of Venus. The amplitude of the biggest peak in Mars in Fig. A.9c is about 6.75%. The peaks are located around 9° and 171° and have a FWHM of about $103.49° \pm 8.72°$ and $76.45° \pm 6.50°$ respectively. Their distance is about 326 d. This behaviour, is due to the ratio of the orbital periods of Mars and Earth being close to 2 within 6%.

This result strengthens further the planetary correlation of solar radius with Venus' heliocentric position (Fig. 10.4a) due the fact that an artificial peak in Earth, like the one seen in Fig. 10.4b, can not cause the appearance of the three peaks in Mars' spectrum observed in

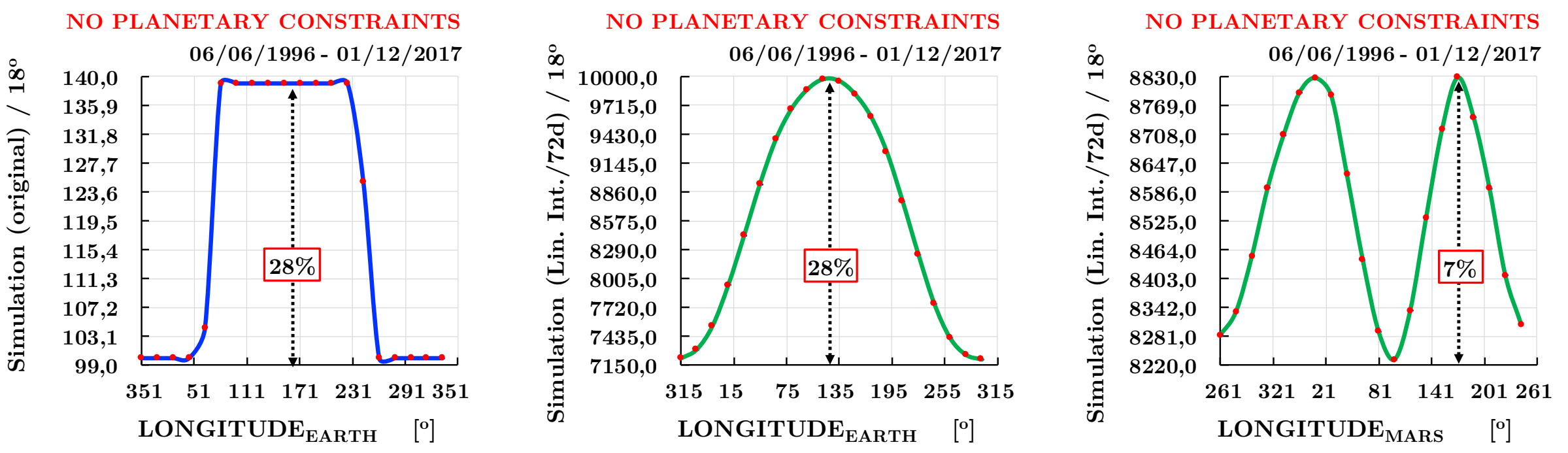

**(a)** Raw $\pi$ distribution.

**(b)** Linearly interpolated data from 72-binned days as a function of Earth's heliocentric longitude.

**(c)** Linearly interpolated data as a function of Mars' heliocentric longitude.

**Figure A.9:** Simulated $\pi$ distribution in Earth for bin = 18°.

Fig. 10.4c.

### A.2.3 Stratospheric temperature

In order to crosscheck the existence of interdependent planetary effects such as the appearance of the two peaks in Mars spectrum in Fig. 16.4d due to the existence of a wide peak in Earth from Fig. 16.4c a case-specific simulation is performed.

Therefore, a $\pi$-distribution with 18 d width has been introduced in the daily positional data of Earth around 260°. This artificial peak was inserted in each orbit of the Earth along the time period 2007 – 2017 with the "background" days having the numeric value 1000 while the "peak" days around 260° a numeric value of 1075 resulting to a $\sim 7\%$ effect in Earth similar to the one seen in Fig. A.10c.

From Fig. A.10d it is confirmed that the two peaks observed in Mars in the real stratospheric data in Fig. 16.4d, result from the wide peak of Earth around 260°(Fig. 16.4c). Similarly, it is also verified that the peaks as well as the amplitudes in the remaining planetary distributions in Fig. A.10 are different from the distributions observed in Fig. 16.4 and therefore the latter plots are not, at least directly, affected by the wide smooth peak of the Earth around 260° which corresponds to summer period (around June).

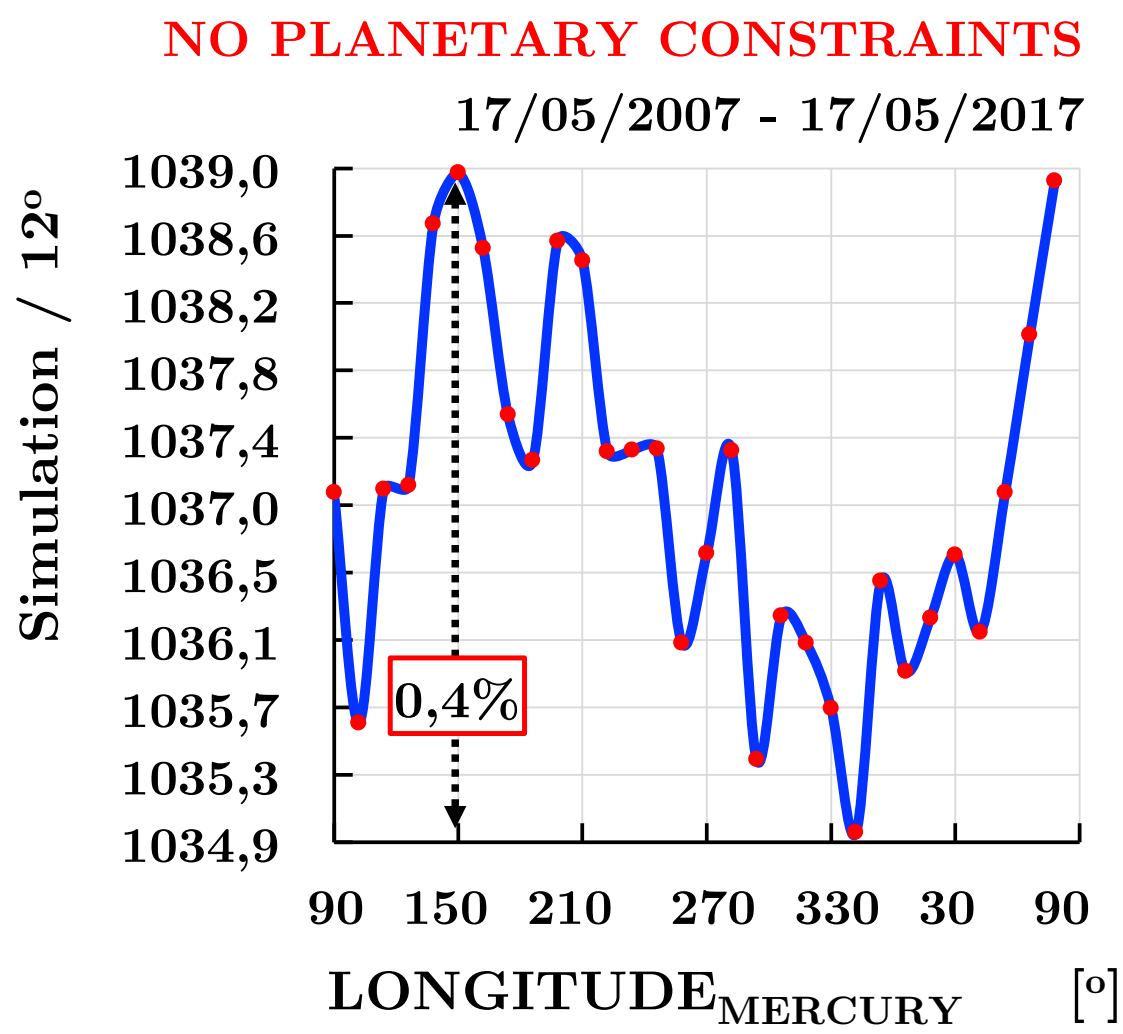

**(a)** Mercury's distribution for 12° bin.

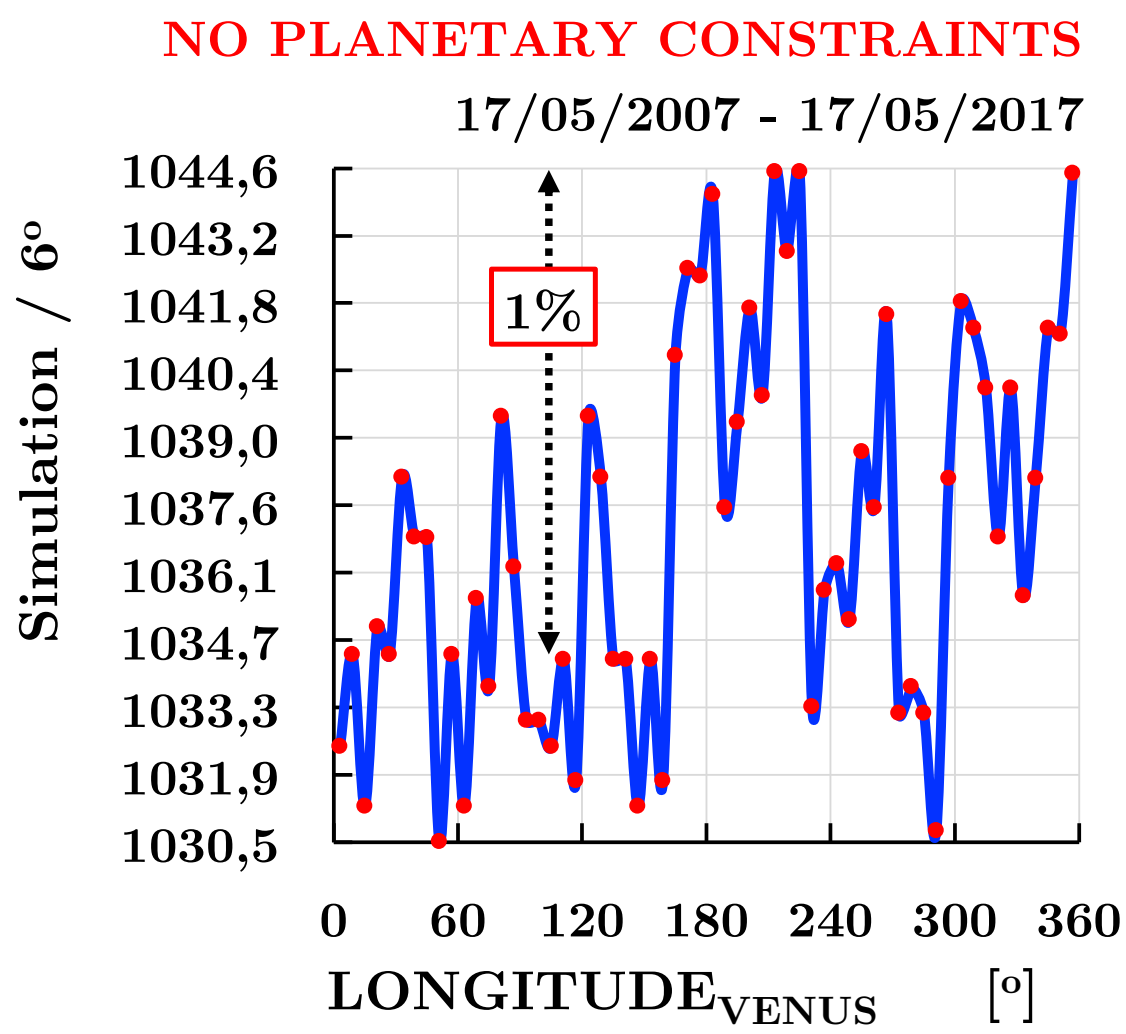

**(b)** Venus' distribution for 6° bin.

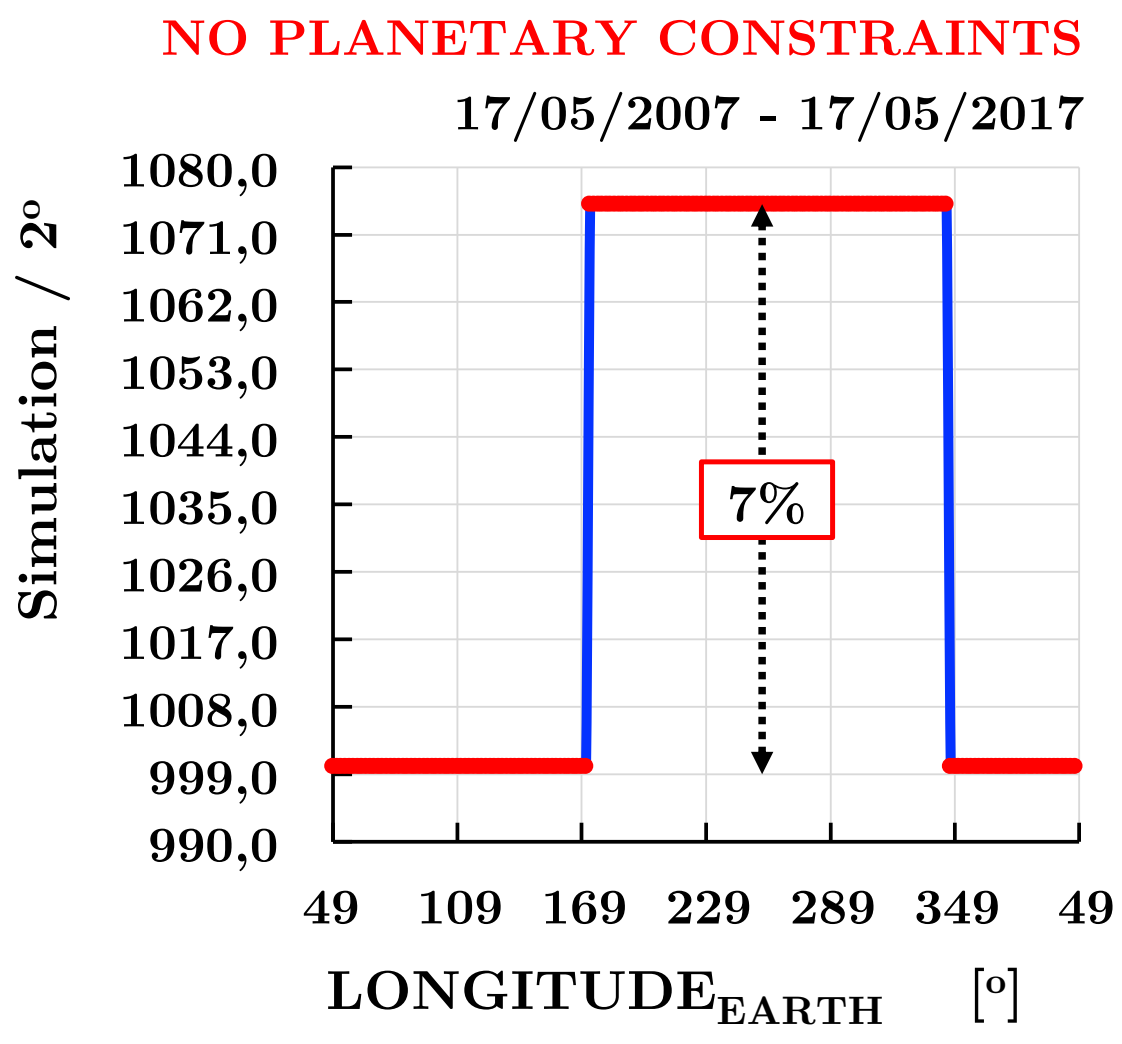

**(c)** Earth's distribution for 2° bin.

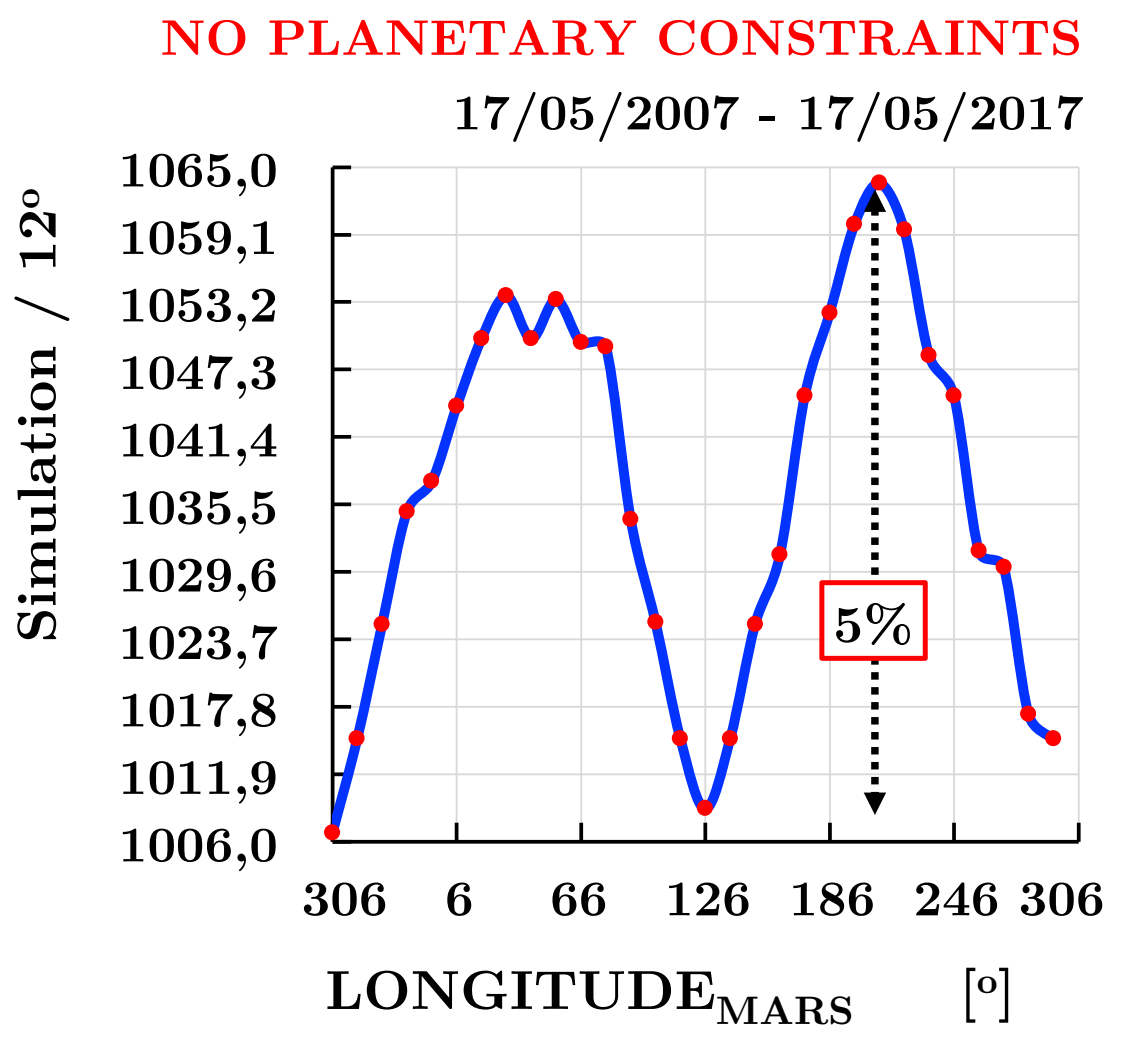

**(d)** Mars' distribution for 12° bin.

**Figure A.10:** Simulation of a $\pi$ distribution in Earth with a width of 180 d around 260° with $\sim 7\%$ amplitude.

B

# DATASET COMPARISONS

## B.1 Introduction

An important part of this work is the comparison between various observations to search for possible similarities or dissimilarities which can provide a hint on the underlying triggering mechanism of each case. The same constraints, as well as the same period, has to be used for these comparisons to be valid. To evaluate the strength of the relationship between the variables a correlation analysis is performed on the various planetary longitudinal distributions of the datasets under comparison.

The procedure involves the calculation of the *Pearson product-moment correlation coefficient*, $r$, which is a measure of the strength of a linear association between the two compared variables [462]. Basically, it is the covariance of the two variables divided by the product of their standard deviations. The Pearson correlation attempts to draw a line of best fit through the data of the two variables, and the parameter $r$ indicates how far away are the data points to the line of best fit. $r$ can take values ranging from $+1$ to $-1$, where $r = 0$ indicates no association between the two variables, $r > 0$ indicates a positive association meaning as the value of one variable increases so does the value of the other variable, and $r < 0$ indicates a negative association. The usual guideline for $|r|$ suggests that for $0.1 - 0.3$ we have a small association, for $0.3 - 0.5$ we have a medium correlation whereas for $r$ between $0.5 - 1$ we have a strong correlation between the two variables.

To determine the significance of the correlation between the two variables, the *p-value* is calculated. This gives the probability to get the current result if the correlation coefficient was zero. In other words, the p-value quantifies how significantly different is the correlation coefficient from zero. In practice, if this probability is lower than the conventional 5%, meaning $p < 0.05$, then the correlation coefficient is statistically significant, meaning that there is strong evidence against the null hypothesis. On the other hand, for $p > 0.05$ the null hypothesis has merit. In all the comparisons made in this work, a 2-tailed significance test is used and the correlations that are significant at the 0.05 level are marked in green colour.

The basic assumptions for the data that are cross-checked for the Pearson's correlation to be valid include the plotting of the two variables to verify a linear relationship between them, a check that both continuous variables follow a bivariate normal distribution, a check of the homoscedasticity of the two variables along the line of best fit and an investigation of possible univariate or multivariate outliers [463]. When one or more of these assumptions are violated different non-parametric techniques are followed including the calculation of *Spearman's rank correlation coefficient* and *Kendall's rank coefficient*. As an example when the data are not normally distributed then Spearman's correlation is used. In practice, all coefficients are calculated using Origin or SPSS Statistics and compared in case of disagreement [464].

# B.2 Solar flares

## B.2.1 X-flares

An important comparison of the derived results from M-flares is with the corresponding number of the X-flares for the same period since they X-flares are the biggest and strongest flares. Due to the small number of the X-flares, the statistics are lower but they can still provide an additional confirmation of the planetary relationship seen with the M-flares.

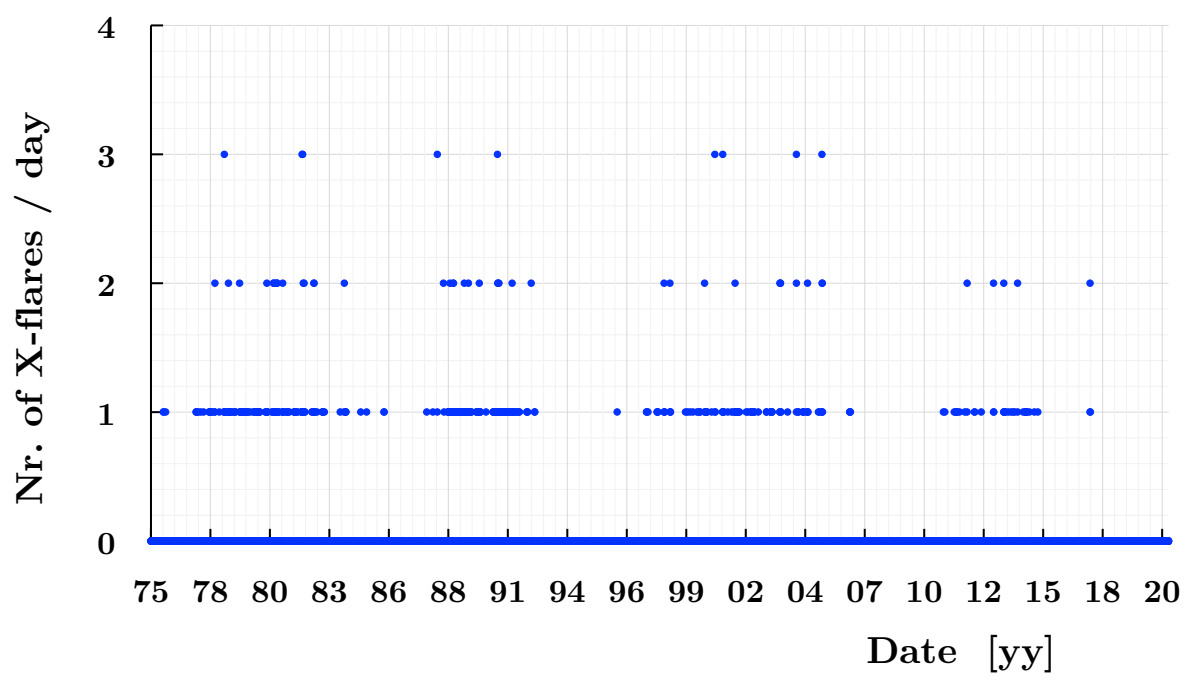

**Figure B.1:** The daily Nr. of X-class solar flares for the period 01/09/1975 - 12/03/2021..

More specifically, the total number of X-flares for the selected period is 495 with an average of 0.0298 X-flares per day and a maximum of 3 X-flares per day. The number of days without X-flares (= 0) is 16196 out of 16 630 d i.e. 97.39%. This means that between 01/09/1975 and 12/03/2021 434 d contained X-flares. Based on the data from Fig. B.1, the various histograms are created in Fig. B.2 and show the frequency distribution of the number of X-flares.

#### B.2.1.1 Single planets

The first step, shown in Fig. B.3, involves the comparison of the distributions of X-flares with M-flares as a function of the heliocentric longitude for the used planets without any further constraints as done in Fig. 6.4. The derived standard error per point for a bin of 6° (like Fig. B.3g) is $\sigma \sim 34.82\%$. Accordingly, for a bin of 12° (like Fig. B.3e and B.3f) the mean error per point is $\sigma \sim 24.62\%$, while for bin = 24° it is $\sigma \sim 17.41\%$. The statistical correlation analysis

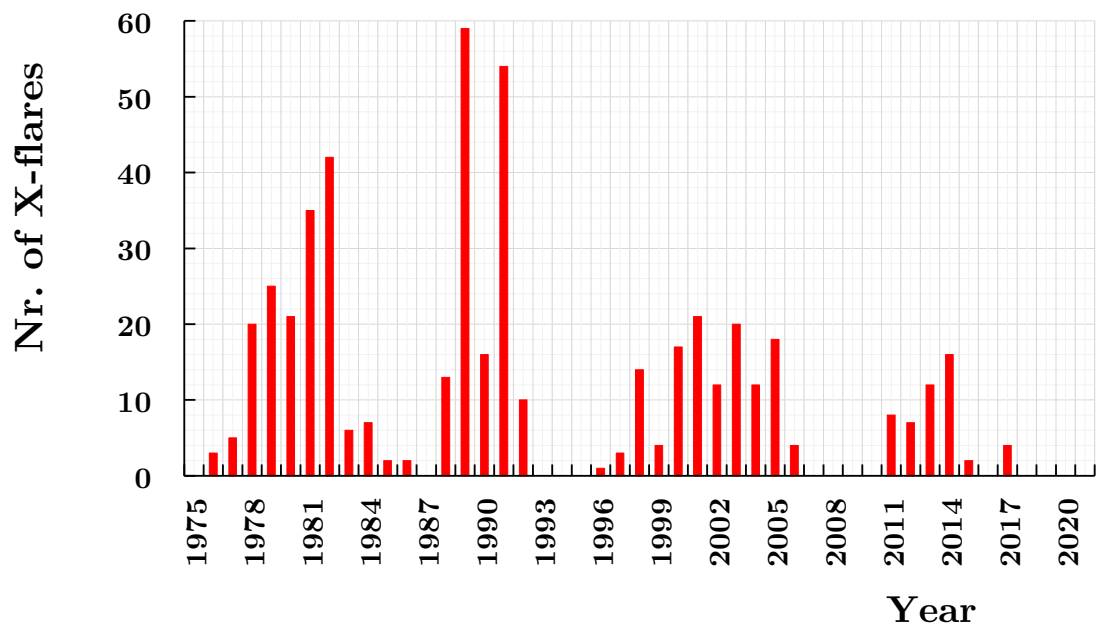

**(a)** Nr. of X-flares per year.

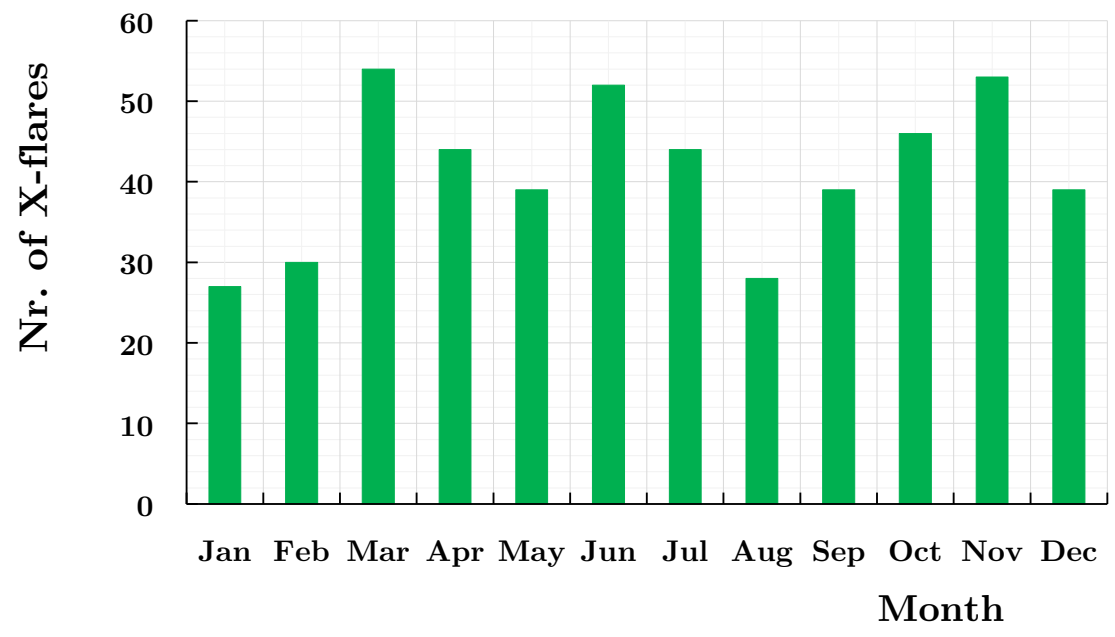

**(b)** Nr. of X-flares per month.

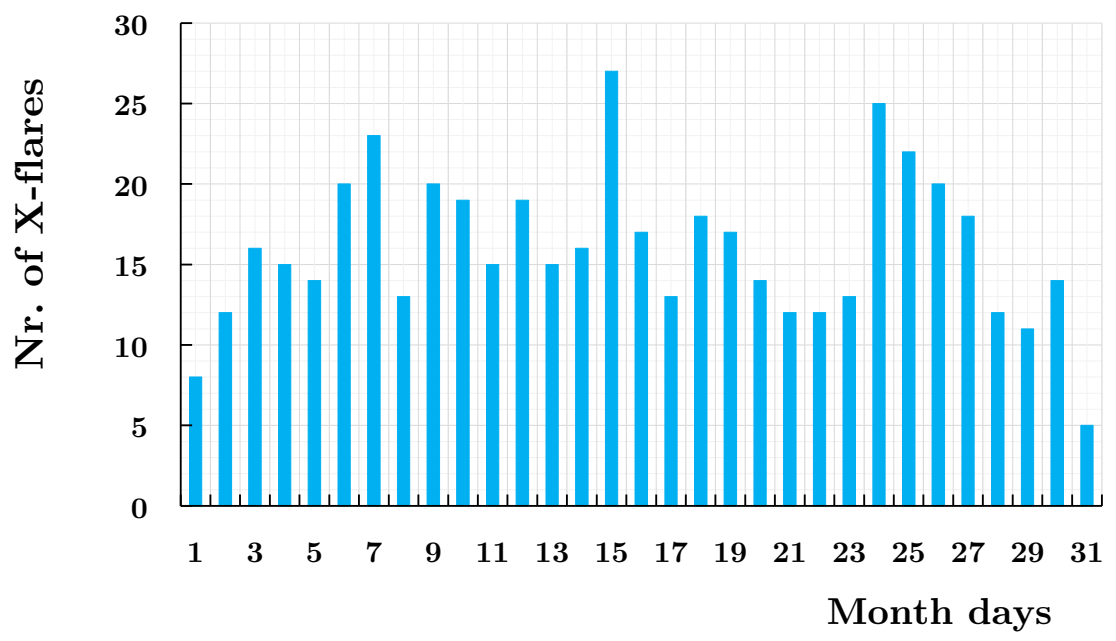

**(c)** Nr. of X-flares per day of month.

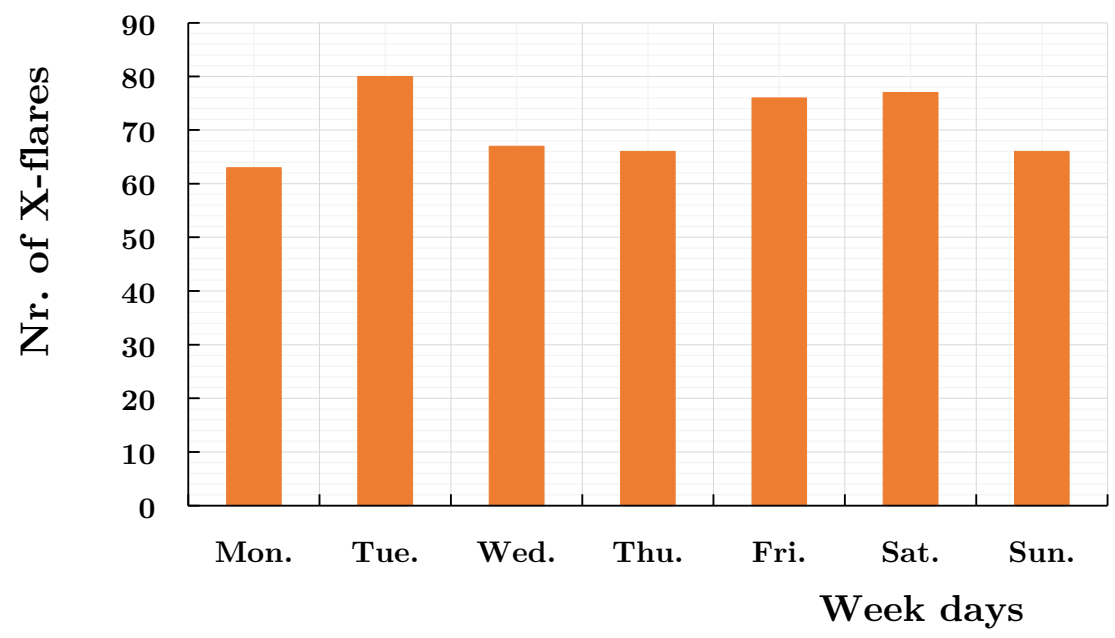

**(d)** Nr. of X-flares per day of week.

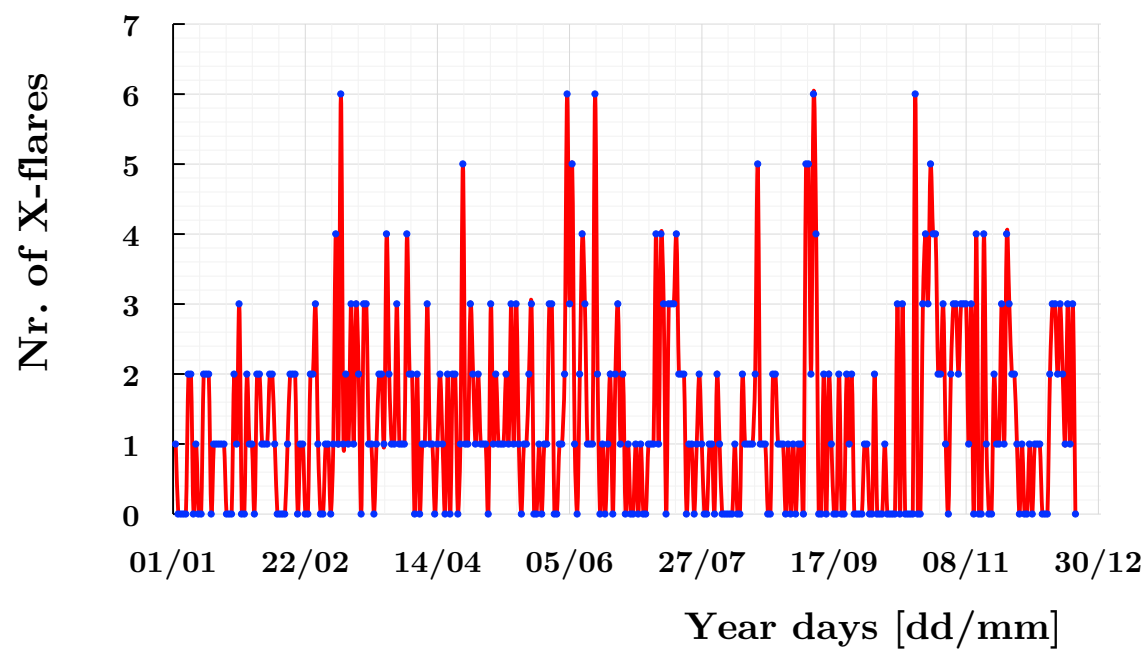

**(e)** Nr. of X-flares per day of year.

**Figure B.2:** Histograms for the Nr. of X-flares for the period 01/09/1975 - 12/03/2021.

between the single planetary distributions of M- and X-flares in Fig. B.3 gives ($r_{a,e} = 0.57$, $p_{a,e} = 9.74 \times 10^{-4}$), ($r_{b,f} = 0.52$, $p_{b,f} = 0.0035$), ($r_{c,g} = 0.52$, $p_{c,g} = 1.85 \times 10^{-5}$) and ($r_{d,h} = 0.48$, $p_{d,h} = 0.07$) for the corresponding subfigures a to h. This indicates a statistically significant strong positive linear correlation for the reference frames of Mercury, Venus and no statistically significant correlation for Mars.

#### B.2.1.2 Combining planets

A comparison of the planetary distribution spectra when a second planet is constrained to propagate in a specific longitude range, is shown in Fig. B.4. For the case of Venus being in the range 200° to 320° we have in total 203 flares, and 5584 d with the total amplitude being 79.24%, while for the case of Venus being in the range 20° to 140° we have a total 149 flares

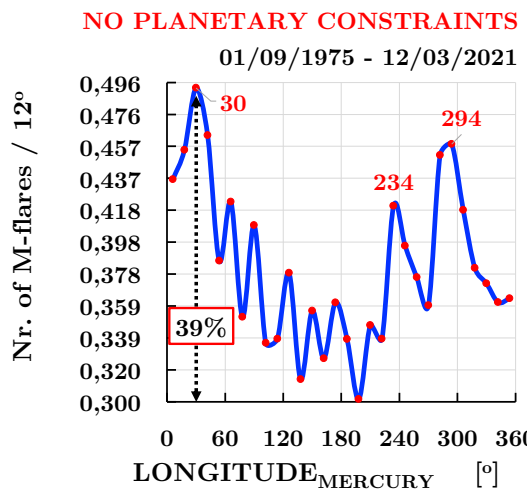

(a) Nr. of M-flares distribution for Mercury with 12° bin.

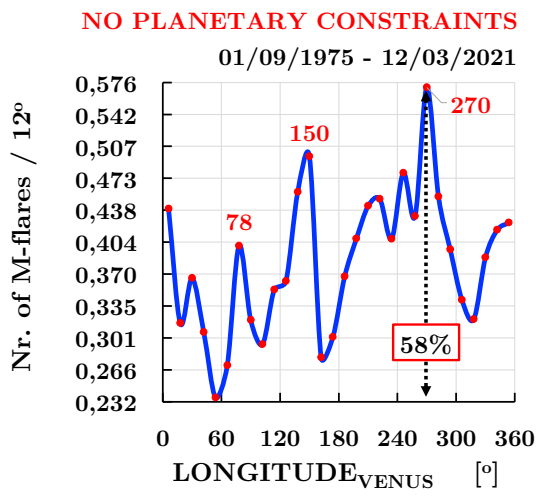

(b) Nr. of M-flares distribution for Venus with 12° bin.

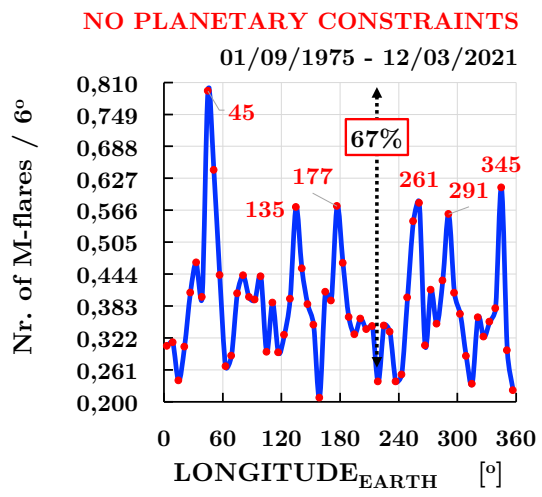

(c) Nr. of M-flares distribution for Earth with 6° bin.

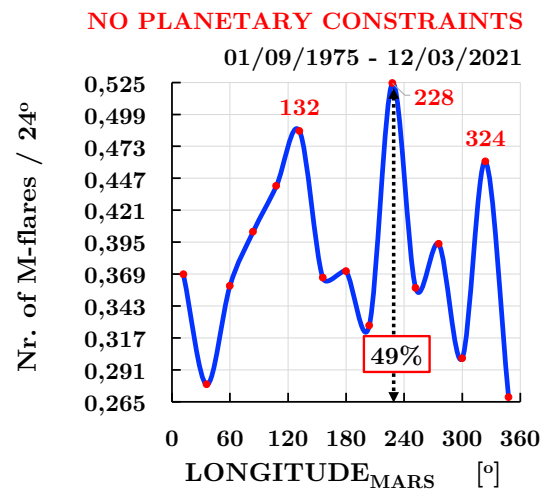

(d) Nr. of M-flares distribution for Mars with 24° bin.

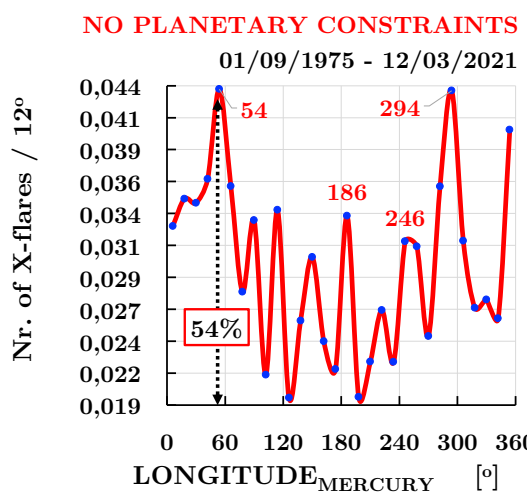

(e) Nr. of X-flares distribution for Mercury with 12° bin.

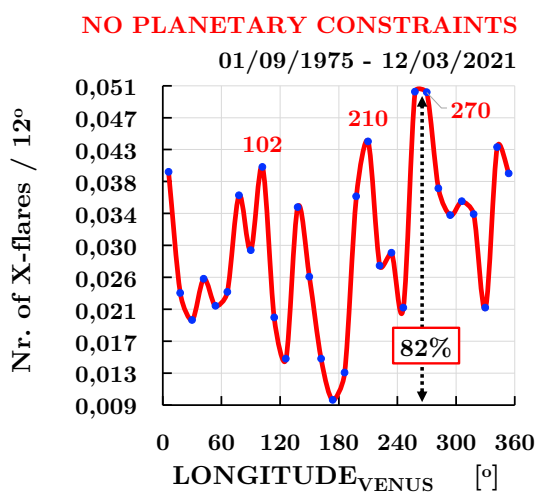

(f) Nr. of X-flares distribution for Venus with 12° bin.

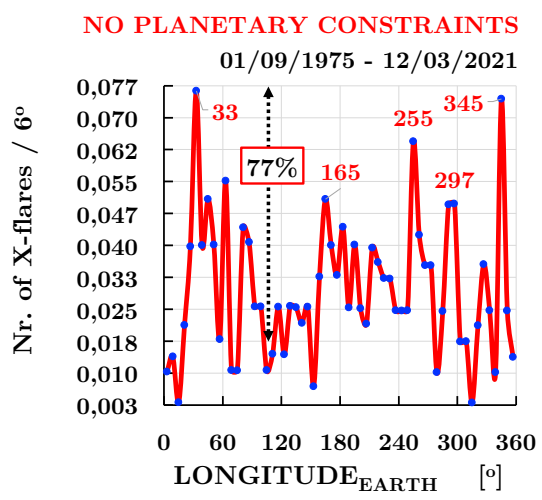

(g) Nr. of X-flares distribution for Earth with 6° bin.

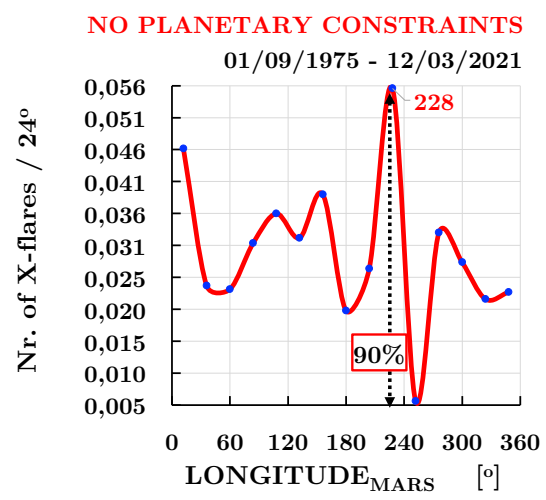

(h) Nr. of X-flares distribution for Mars with 24° bin.

**Figure B.3:** Comparison of the longitudinal distributions for the Nr. of M-flares vs. X-flares for the period 01/09/1975 - 12/03/2021.

for 5510 d and an amplitude of 90.17%. In this case, again (as in Fig. 6.5) the rate is bigger on the 200° to 320° orbital arc, with the difference between the two spectra regarding the total number of X-flares being (203 - 149) 2.88$\sigma$.

The correlation coefficients for the two positions of Venus in Fig. B.4a and B.4c are ($r_{a,c} = 0.42,\ p_{a,c} = 0.048$) and ($r_{a,c} = 0.79,\ p_{a,c} = 8.82 \times 10^{-6}$) while for their difference in Fig. B.4b and B.4d we get ($r_{b,d} = 0.69,\ p_{b,d} = 2.47 \times 10^{-4}$). These numbers indicate a statistically significant association in all distributions with a medium and a strong correlation between the two datasets for the case of Venus being between 200° to 320° and 20°-140° respectively, and a high degree of correlation for their difference $(200° - 320°) - (20° - 140°)$.

It is noted that the case of Fig. 6.6 is not reevaluated for X-flares due to the many zeroes appearing in the spectrum as a consequence of the small number of X-flares (102) fulfilling the condition of Mercury's orbital position being between 200° to 260° range.

#### B.2.1.3 Multiplication spectra

For the comparison of the multiplication spectra of the X-flares with the corresponding ones of the M-flares presented in Sect. 6.3.2, the first step is the calculation of the number of X-flares corresponding to each solar cycle. In Tab. B.1 the various numbers are shown in comparison with Tab. 6.2. Due to the lower number of X-flares on the 24$^{th}$ solar cycle, and

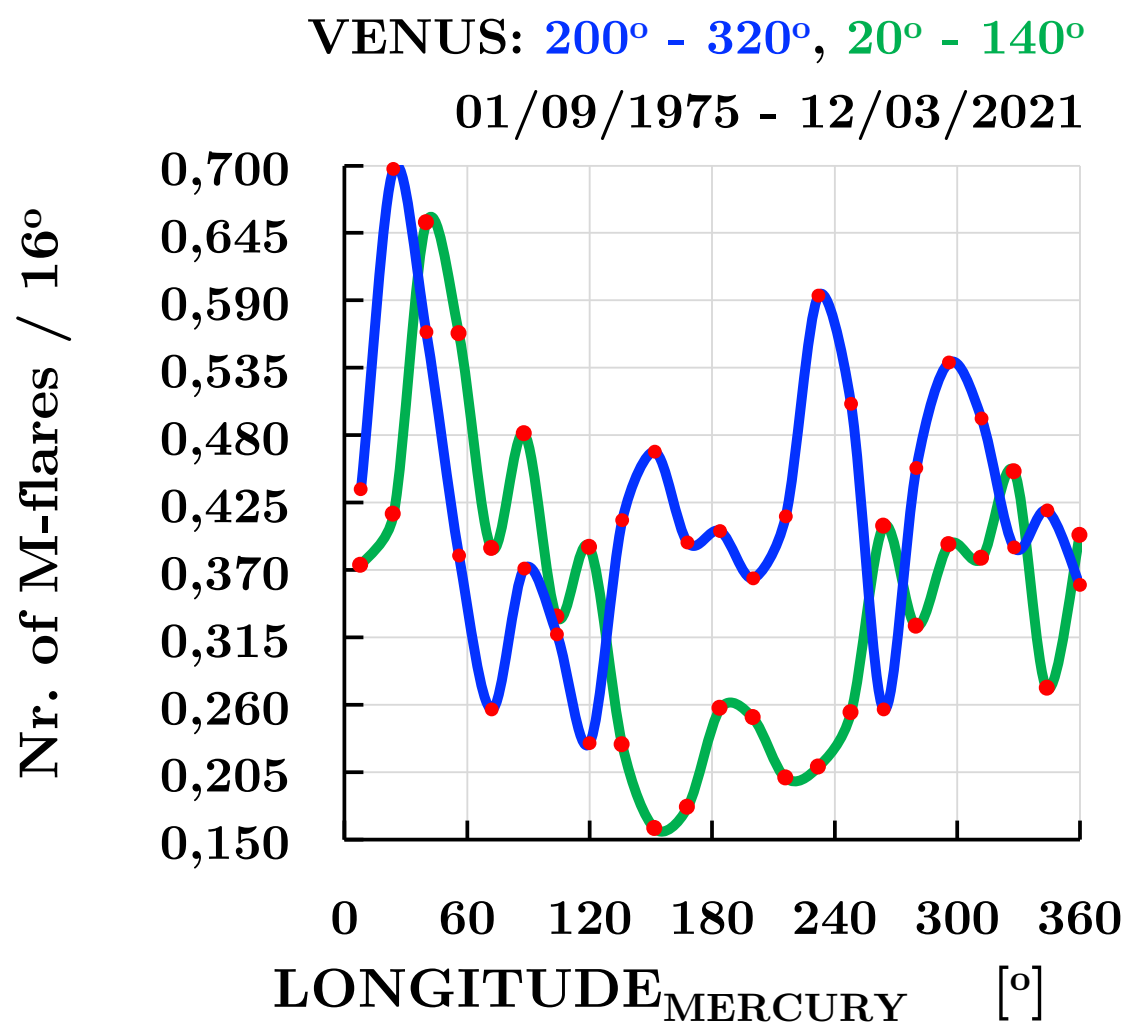

**(a)** M-flare distribution when comparing the two cases of Venus.

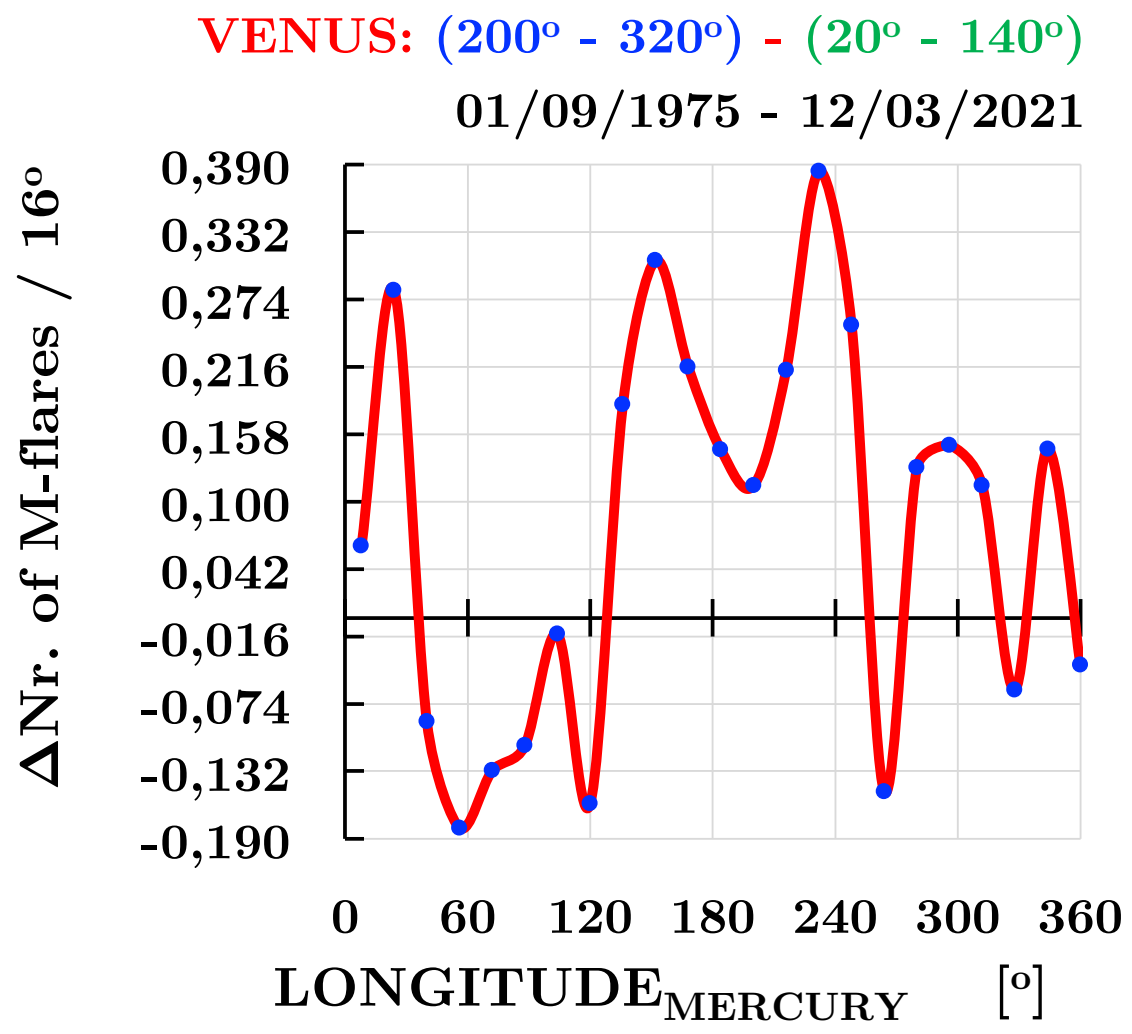

**(b)** M-flare distribution when subtracting the two cases of Venus.

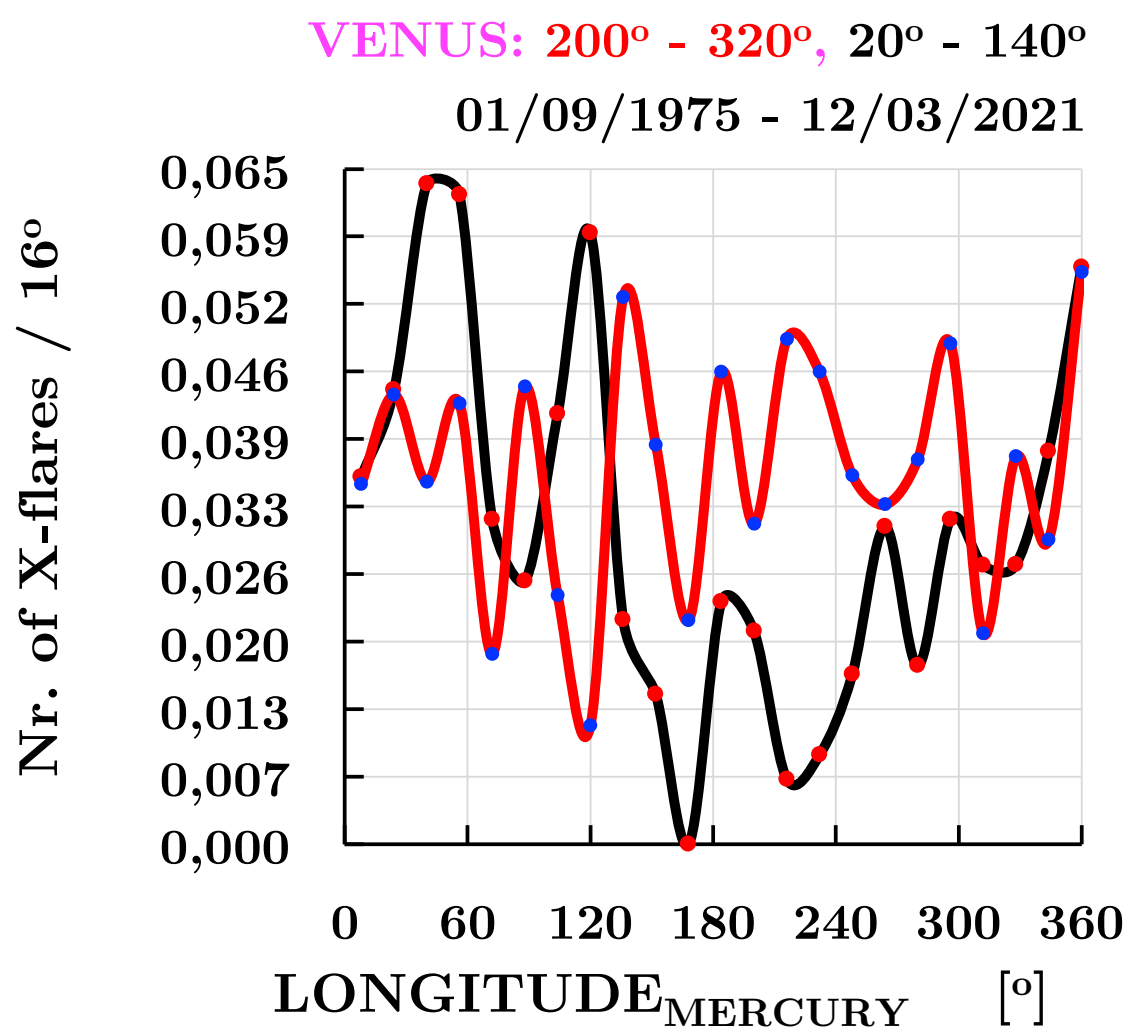

**(c)** X-flare distribution when comparing the two cases of Venus.

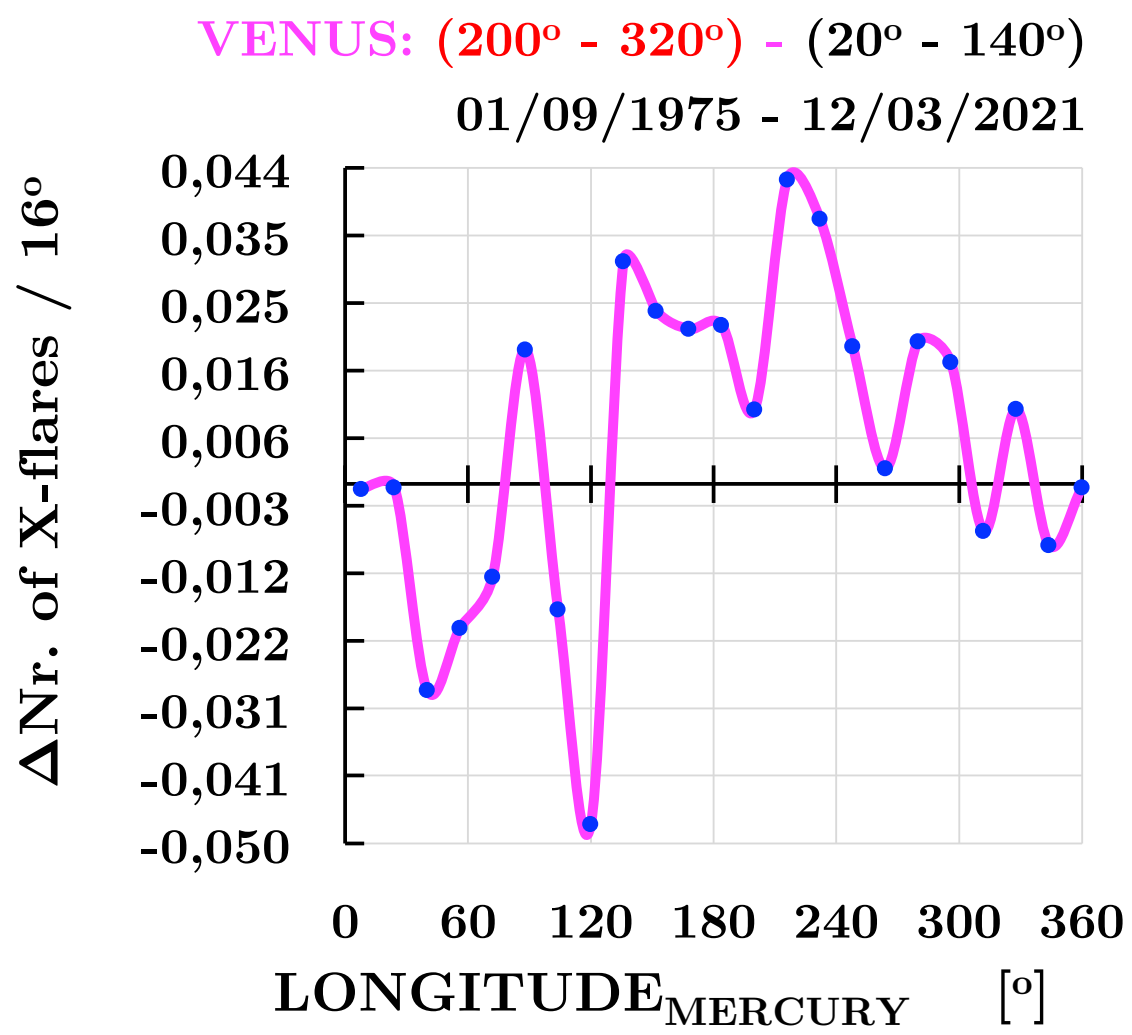

**(d)** X-flare distribution when subtracting the two cases of Venus.

**Figure B.4:** Comparison of the longitudinal distributions for the Nr. of M-flares vs. X-flares for the reference frame of Mercury when Venus is between 200° to 320° and 20° to 140° with bin = 16° for the period 01/09/1975 - 12/03/2021.

therefore weaker statistics, it has not been included in the calculation of the multiplication spectra for the X-flares. Therefore, only the three solar cycles (21$^{st}$ to 23$^{rd}$) have been used in this case and therefore no statistical correlation analysis is performed.

The results for the inner planets are presented in Fig. B.5. The first qualitative observation is that the peaks already appearing in Fig. B.3e through B.3g show up even better resolved in the multiplication spectra in Fig. B.5a through B.5c, and with a better SNR proving that also the distributions of X-flares are not random. Furthermore, when comparing Fig. B.5 with Fig. 6.7, we notice that most of the peaks overlap around the same heliocentric longitudes in both M- and X-flares even though the last solar cycle has not been used for X-flares due to lower statistics. As an example, the peak around 294° in the reference frame of Mercury appears in both cases (see Fig. 6.7a and B.5a), the same for the peak around 270° to 280° for Venus (see Fig. 6.7b and B.5b) and the peaks around 45° and 260° for Earth (see Fig. 6.7c and B.5c).

**Table B.1:** Exact dates of the latest four solar cycles, and the corresponding Nr. of days and X-flares.

| Solar cycle Nr. [-] | Start date [UTC] | End date [UTC] | Days [-] | Nr. of X-flares [-] |
|---|---|---|---|---|
| 21 | 01/03/1976 | 01/09/1986 | 3837 | 168 |
| 22 | 02/09/1986 | 01/08/1996 | 3622 | 153 |
| 23 | 02/08/1996 | 01/12/2008 | 4505 | 125 |
| 24 | 02/12/2008 | 01/05/2020 | 4169 | 49 |

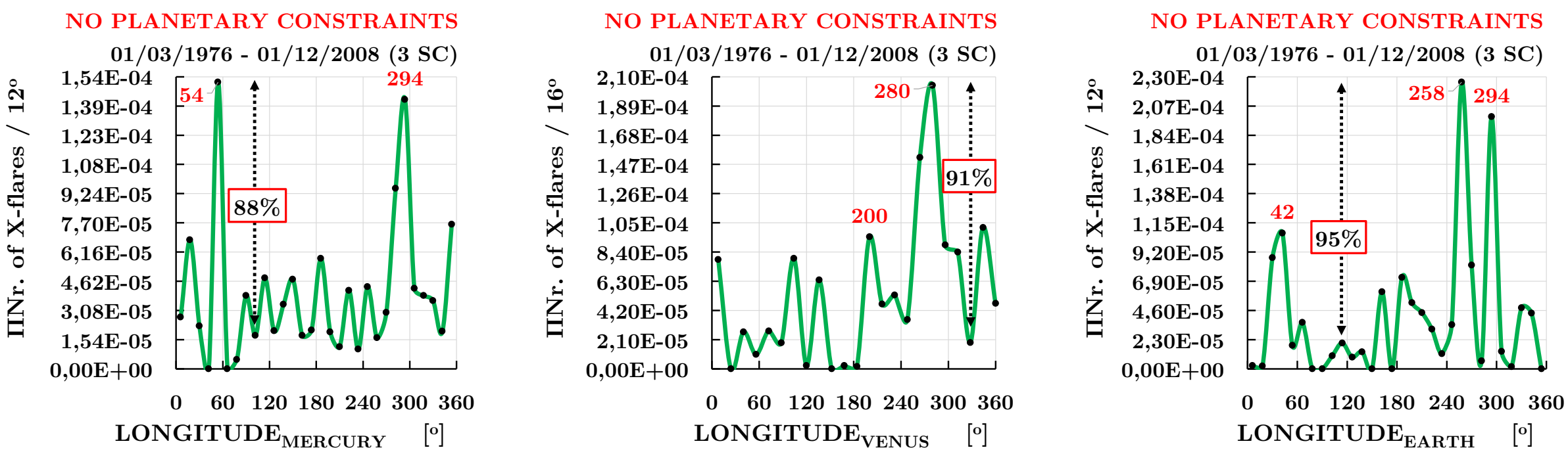

**(a)** Mercury for bin = 12°. **(b)** Venus for bin = 16°. **(c)** Earth for bin = 12°.

**Figure B.5:** Multiplication spectra for the Nr. of X-flares using the latest three solar cycles for Mercury, Venus and Earth.

#### B.2.1.4 Fourier analysis

Similarly to Sect. 6.3.3, a corresponding Lomb periodogram is produced for the number of X-flares for the acquired period of 1975 – 2021. The various observed significant peaks are in agreement with the case of the M-flares supporting further the observed planetary relationship of the number of flares. The 4$^{th}$ biggest peak in amplitude, has a period of about 144.24 d ± 0.531 d and an amplitude of 17.69 dB and overlaps with the same peak in M-flares and the synod of Mercury - Venus of 145 d (see Tab. 4.1). This specific peak is shown in Fig. B.6. Moreover, the 23$^{rd}$ biggest peak is located at 222.51 d ± 1.66 d with an amplitude of 10.35 dB which is close to the Venus orbital period of 224.7 d. Finally, at 371.59 d ± 3.35 d the peak with an amplitude of 8.34 dB overlaps with the Earth - Uranus synodical period of 370 d. There are some more peaks coinciding with planetary synods shown in Tab. 4.1, however with less significance and therefore are not presented here. It is noted that the provided errors in the various periods have been obtained through the FWHM which is derived from a Gaussian fit.

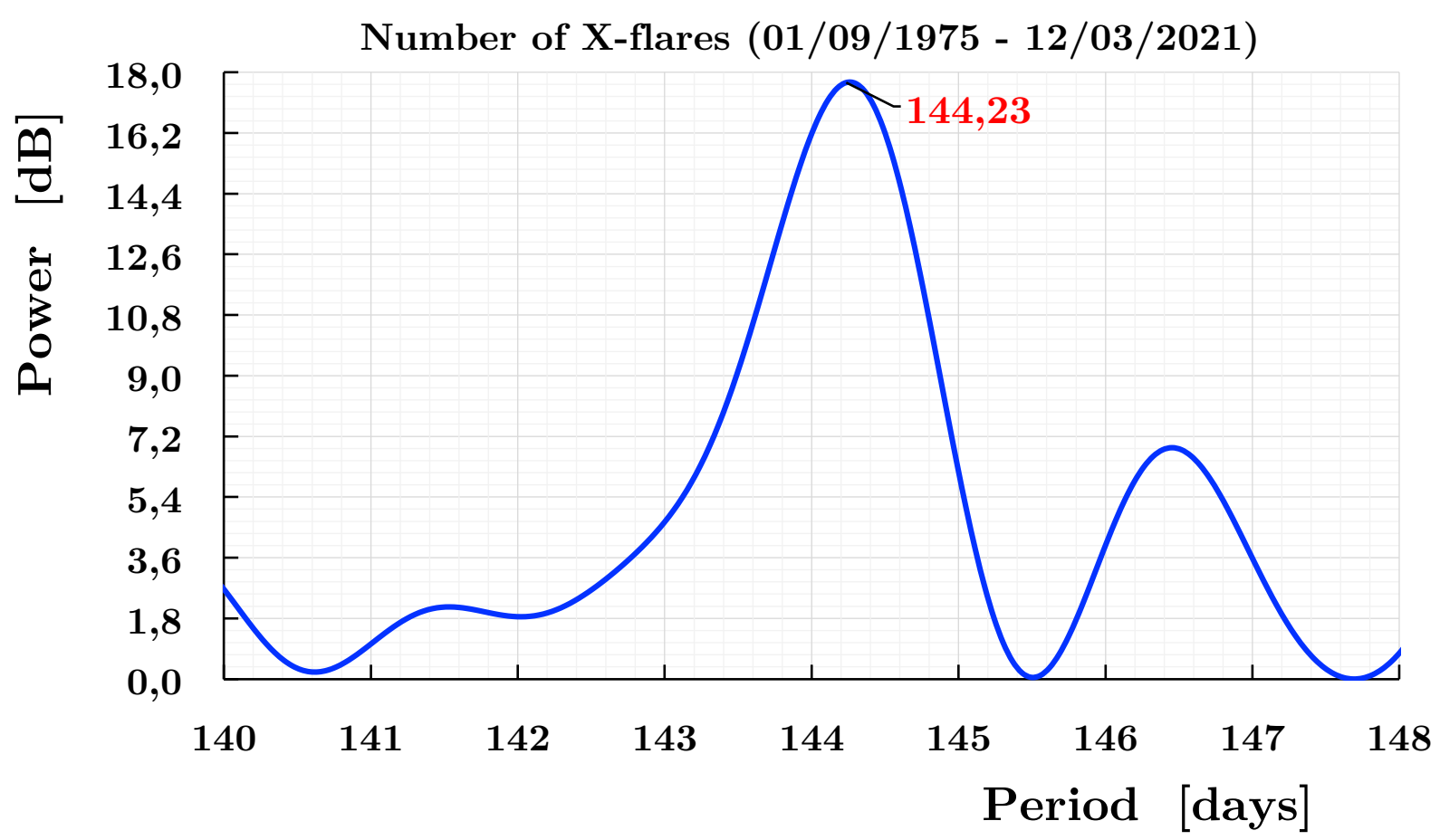

**Figure B.6:** Fourier periodogram of the Nr. of X-flares from 01/09/1975 to 12/03/2021 zoomed in to the period between 142 d to 147 d.

## B.3 EUV

### B.3.1 Comparison with M-flares

A comparison of the derived EUV results is made with other solar observations to look for similarities that could point to a common origin or dissimilarities that could point to a different planetary relationship. In this section, this is done with the number of M-flares.

In Fig. B.7 the time series of the two datasets from Fig. 7.4 and Fig. 6.2 are presented for the same overlapping period. By performing a statistical correlation analysis, we get the Pearson's correlation coefficient and its corresponding p-value ($r = 0.35$, $p = 0$) which points to a moderate degree of statistically significant correlation on the general trend of EUV and the number of M-flares.

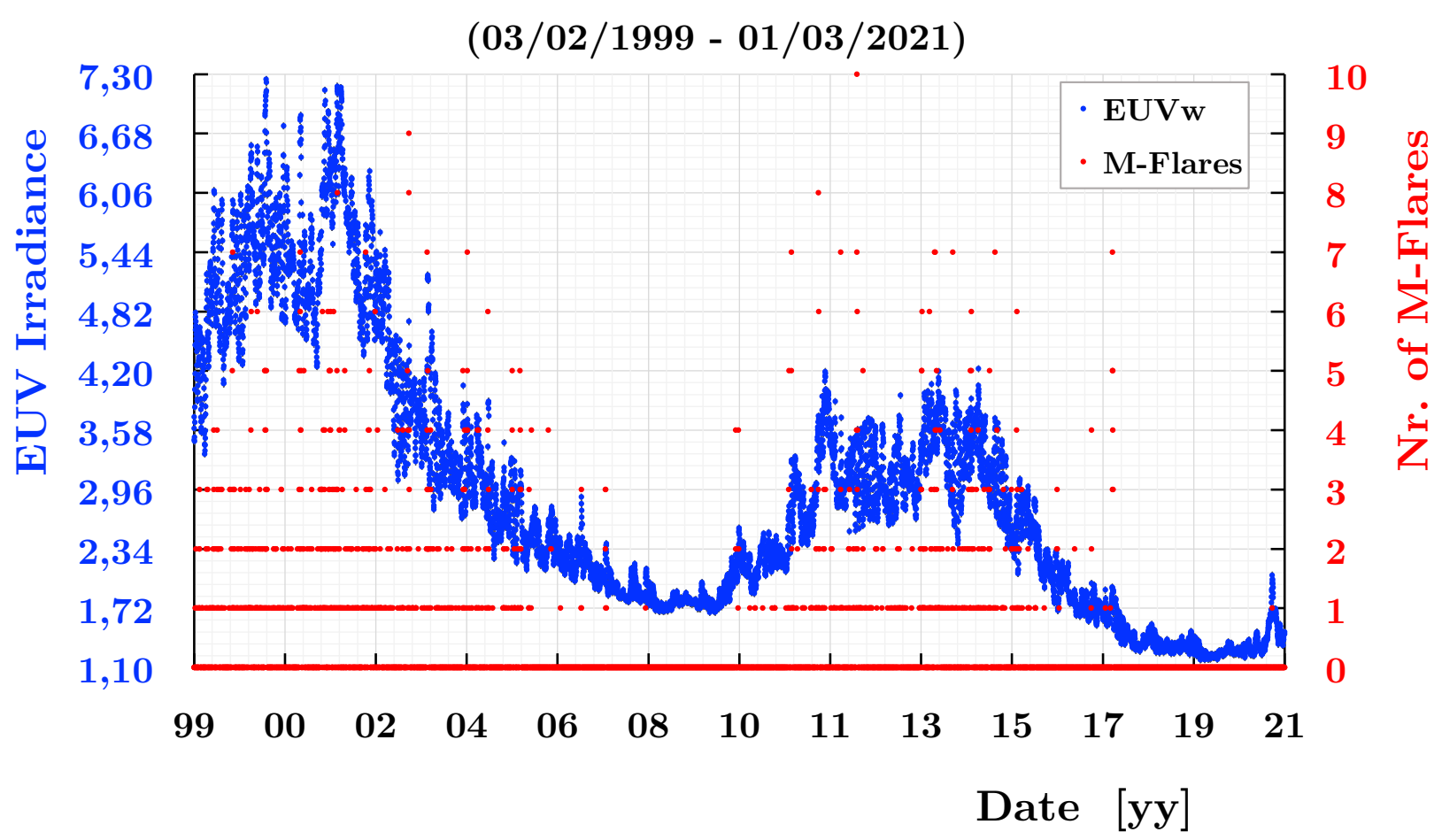

**Figure B.7:** Daily averaged EUV solar irradiance and Nr. of M-flares for the same period 03/02/1999 - 01/03/2021.

#### B.3.1.1 Single planets

In Fig. B.8, the various EUV planetary distributions from Fig. 7.6 are compared with the corresponding ones for the number of M-flares. The same overlapping period has been selected in order to have a direct comparison between the two datasets. The same procedure has been performed for Moon's phase around the Earth from Fig. 7.6f with the result being shown in Fig. B.9. The total corresponding number of M-flares for the period of 03/02/1999 - 01/03/2021 (in total 8063 d) is 2044. This means that for a bin = 18° the standard deviation per point is $\sigma \sim 9.9\%$, whereas for a bin = 24° it is $\sigma \sim 8.6\%$.

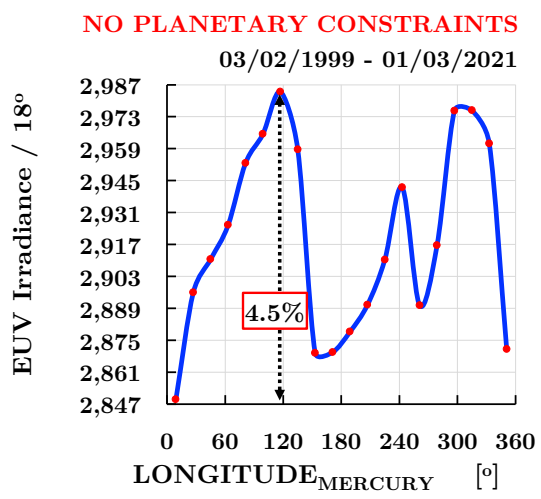

**(a)** EUV distribution for Mercury with 18° bin.

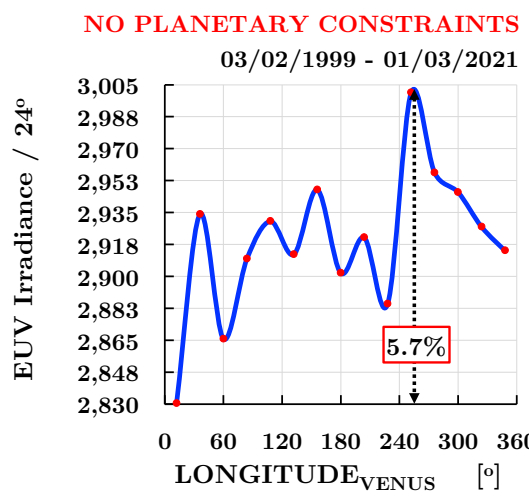

**(b)** EUV distribution for Venus with 24° bin.

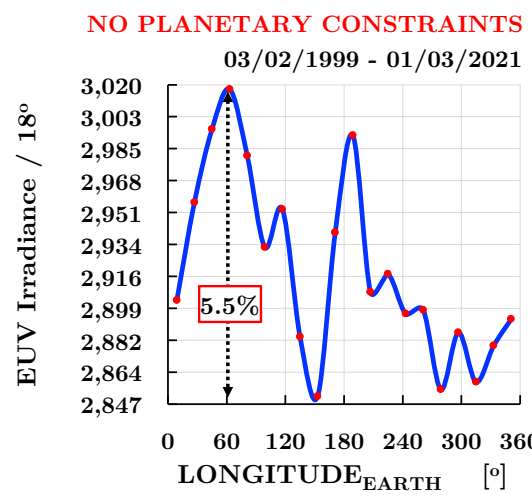

**(c)** EUV distribution for Earth with 18° bin.

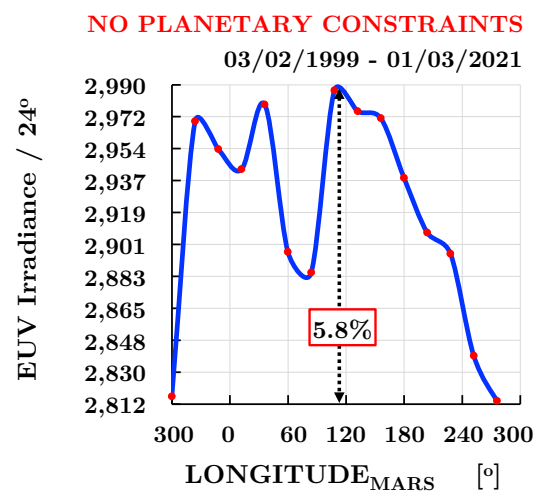

**(d)** EUV distribution for Mars with 18° bin.

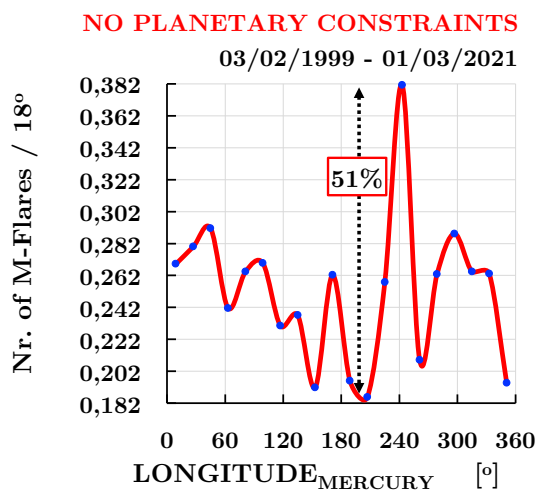

**(e)** Nr. of M-flares distribution for Mercury with 18° bin.

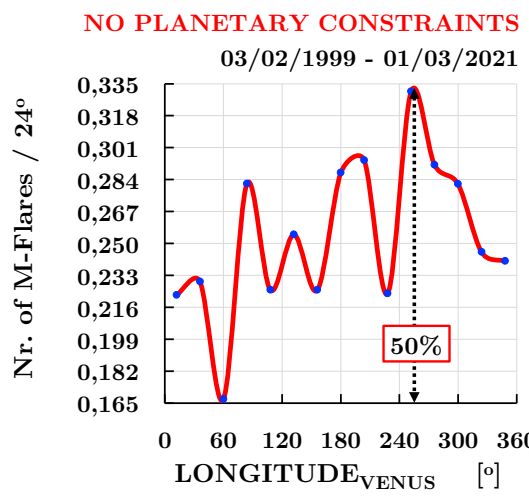

**(f)** Nr. of M-flares distribution for Venus with 24° bin.

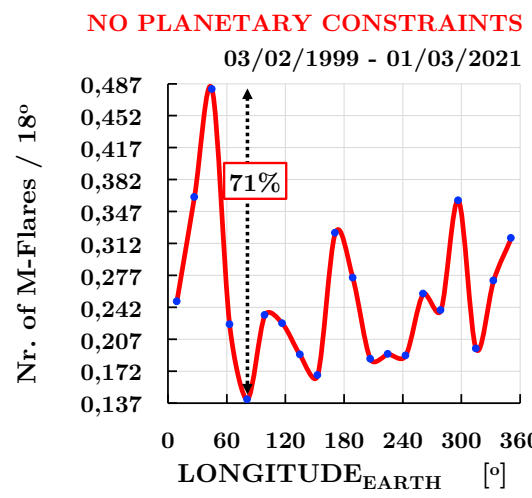

**(g)** Nr. of M-flares distribution for Earth with 18° bin.

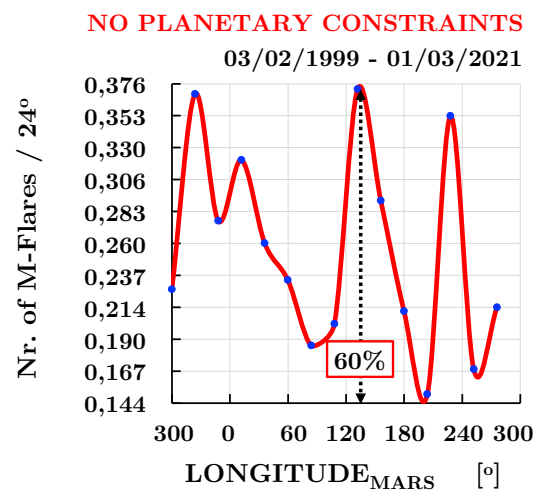

**(h)** Nr. of M-flares distribution for Mars with 18° bin.

**Figure B.8:** Comparison of distributions of EUV solar irradiance vs. Nr. of M-flares for the period 03/02/1999 - 01/03/2021.

From Fig. B.8 we see that even though there is a degree of similarity between the distributions of EUV and solar flares pointing to a common planetary effect, there are also some major differences like in the distributions of Mercury in Fig. B.8a and B.8e where for example the peak around 117° in EUV is missing in the distribution of M-flares, whereas the peak around 305° seems to be much less significant. We also see some differences in the distributions of Earth and Mars both on the relative amplitudes of the various peaks but also on their shapes. and widths. To quantify these observations a statistical correlation analysis has been performed. More specifically, the corresponding Pearson correlation coefficients for the various subfigures are calculated to be ($r_{a,e} = 0.38$, $p_{a,e} = 0.099$), ($r_{b,f} = 0.64$, $p_{b,f} = 0.010$), ($r_{c,g} = 0.29$, $p_{c,g} = 0.216$) and ($r_{d,h} = 0.47$, $p_{d,h} = 0.076$). Therefore, the only statistically significant correlation is found for the case of Venus (Fig. B.8b and B.8f) which points to a positive linear high degree of correlation. On the other hand, in Fig. B.9, the two plots are totally different with the numbers ($r_{a,b} = 0.24$, $p_{a,b} = 0.396$) pointing to an alternative and/or additional external influencing factor on the EUV, correlating also with the longitudinal location of the Moon.

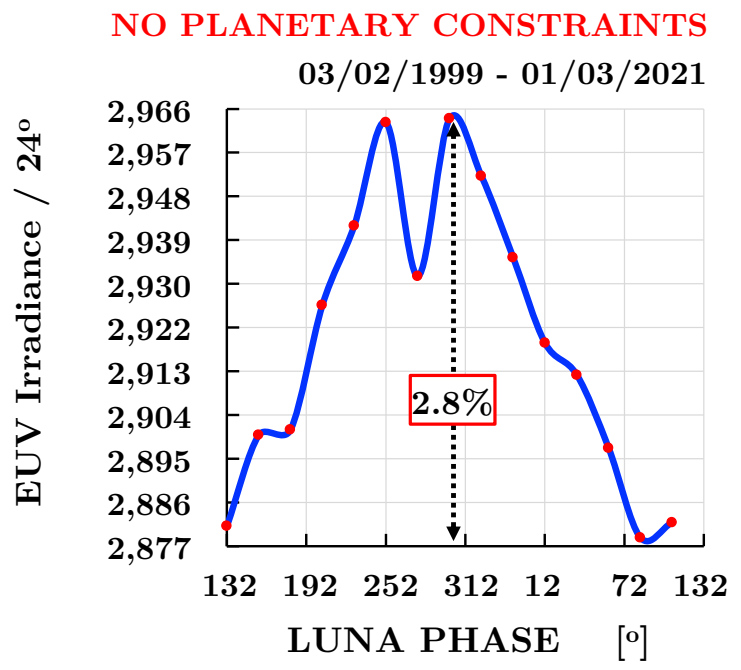

(a) EUV distribution.

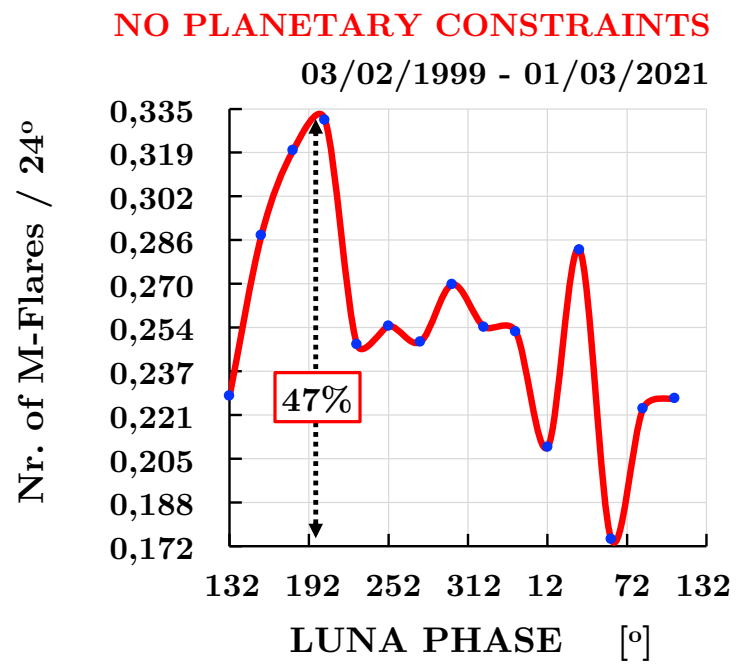

(b) Nr. of M-flares distribution.

**Figure B.9:** Comparison of Moon's distribution with bin = 24° for EUV irradiance vs. Nr. of M-flares for the period 03/02/1999 - 01/03/2021.

#### B.3.1.2 Combining planets

The next step for the comparison between EUV and M-flares is to replicate the plots that contain a planetary constrain like Fig. 7.7. The result is shown in Fig. B.10, where we see that the peak around 239° when Venus is between 200° to 320° appears in both datasets whereas the peak around 32° when Venus is between 20° to 140° appears more enhanced only in the case of EUV. The correlation coefficients for the two positions of Venus in Fig. B.10a and B.10c are ($r_{a,c} = 0.42,\ p_{a,c} = 0.0077$) and ($r_{a,c} = 0.26$, $p_{a,c} = 0.109$) while for their difference in Fig. B.10b and B.10d we get ($r_{b,d} = 0.38,\ p_{b,d} = 0.015$). These numbers indicate a medium strength of association between the two datasets for the case of Venus being between 200° to 320° and for their difference $(200° - 320°) - (20° - 140°)$, and no statistically significant association when Venus is between 20°-140°.

Finally, the same comparison between EUV and M-flares is performed for Fig. 7.9 with the result shown in Fig. B.11. In this case we see a different spectrum for the case of M-flares (Fig. B.11b) compared to that of EUV (Fig. B.11a) where only one significant peak is observed on the distribution of M-flares with the three remaining peaks from EUV not appearing on the M-flare distribution. The first large peak in the distribution of the M-flares is located around $31.37° \pm 11.82°$, whereas in EUV it is around $37.67° \pm 20.14°$. The Pearson coefficient for this case is ($r_{a,b} = 0.48,\ p_{a,b} = 0.0017$), which indicates a moderate degree of linear correlation.

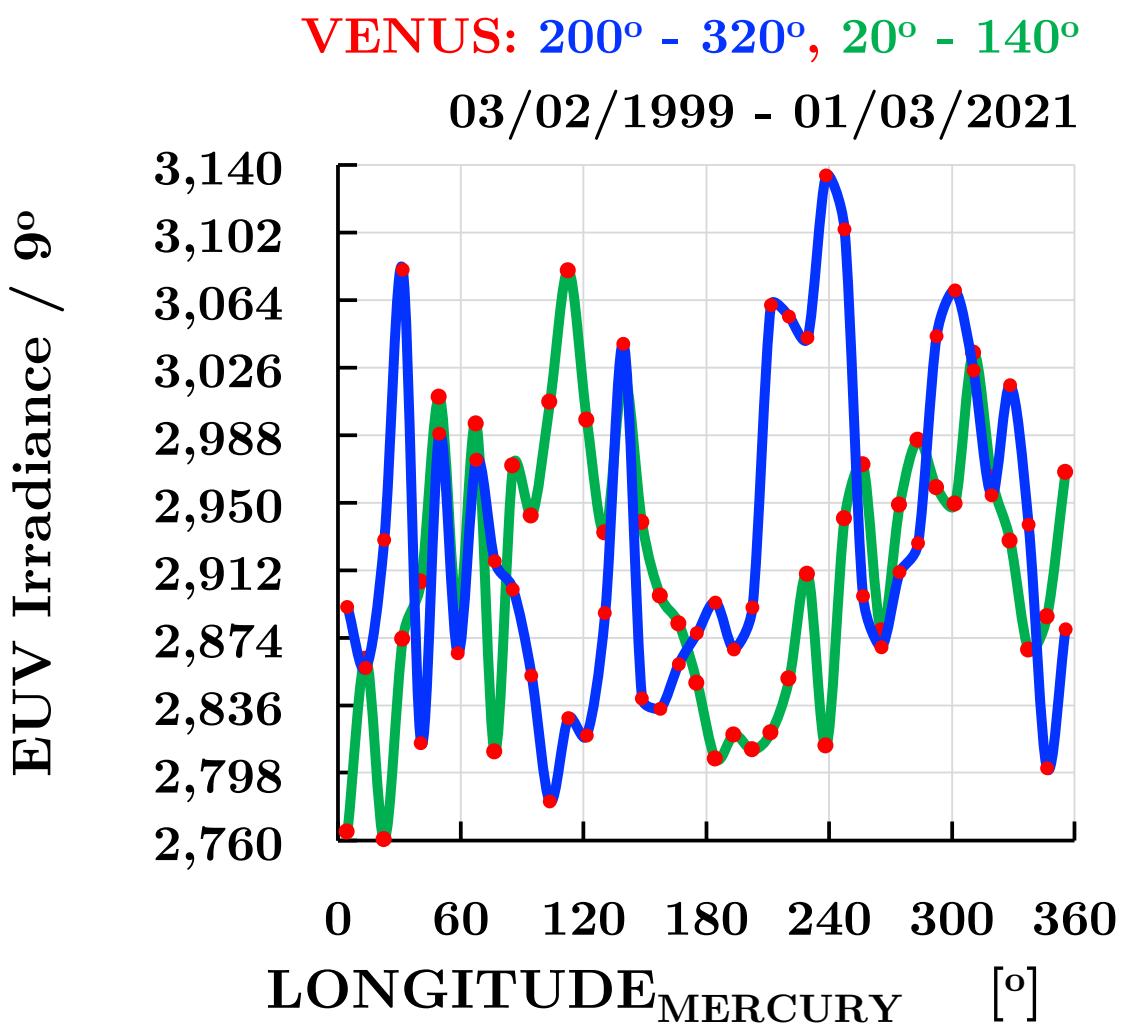

**(a)** EUV distribution when comparing the two cases of Venus.

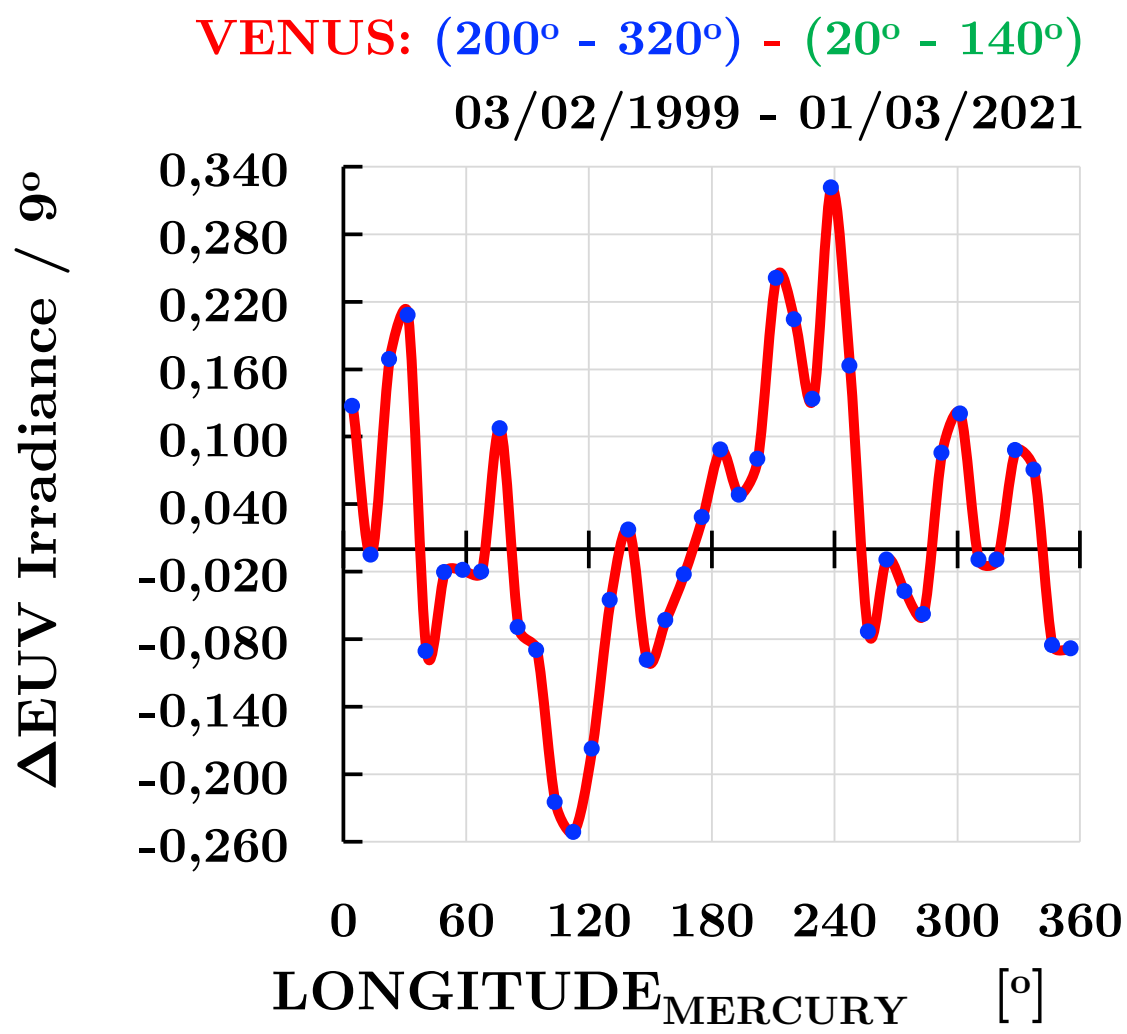

**(b)** EUV distribution when subtracting the two cases of Venus.

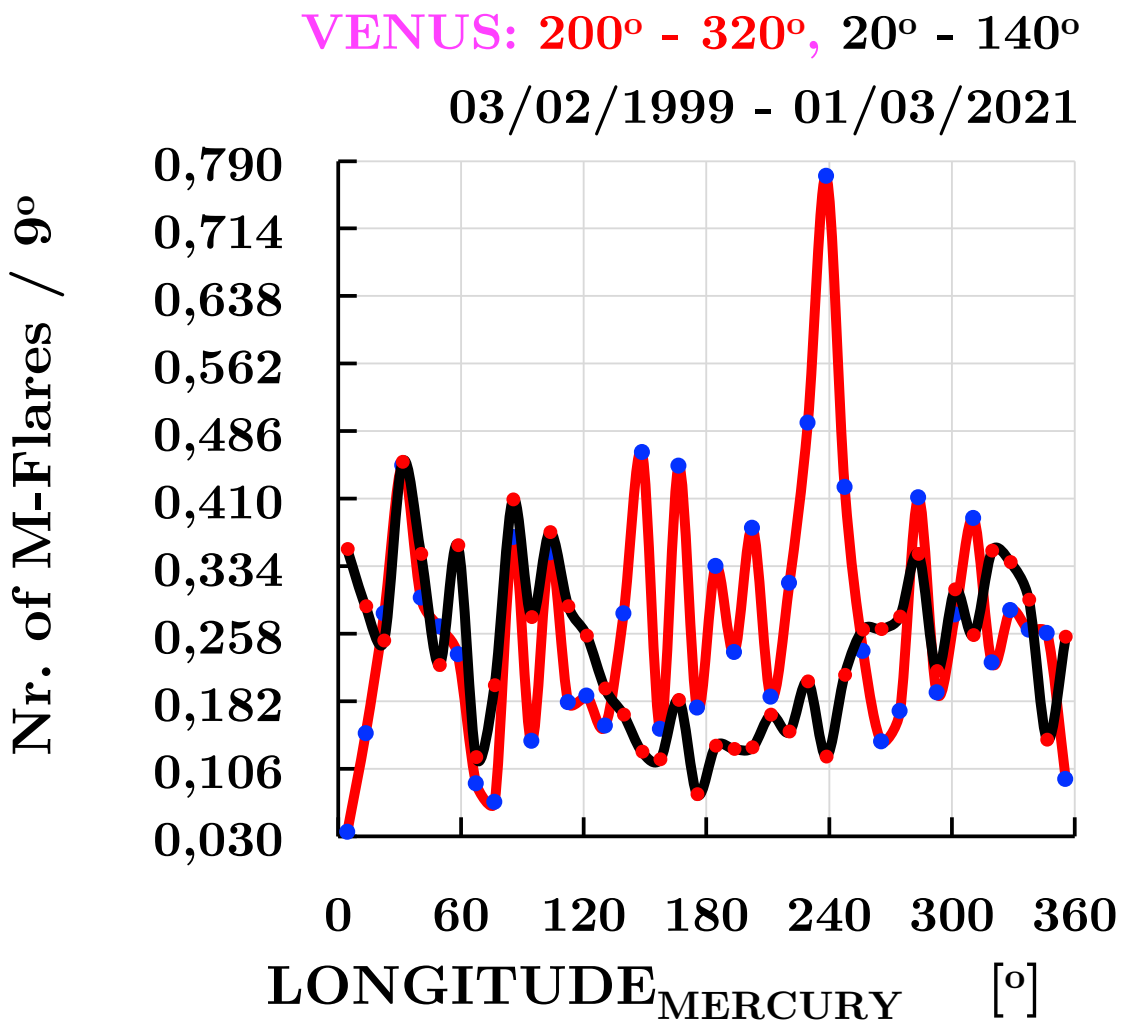

**(c)** Nr. of M-flares distribution when comparing the two cases of Venus.

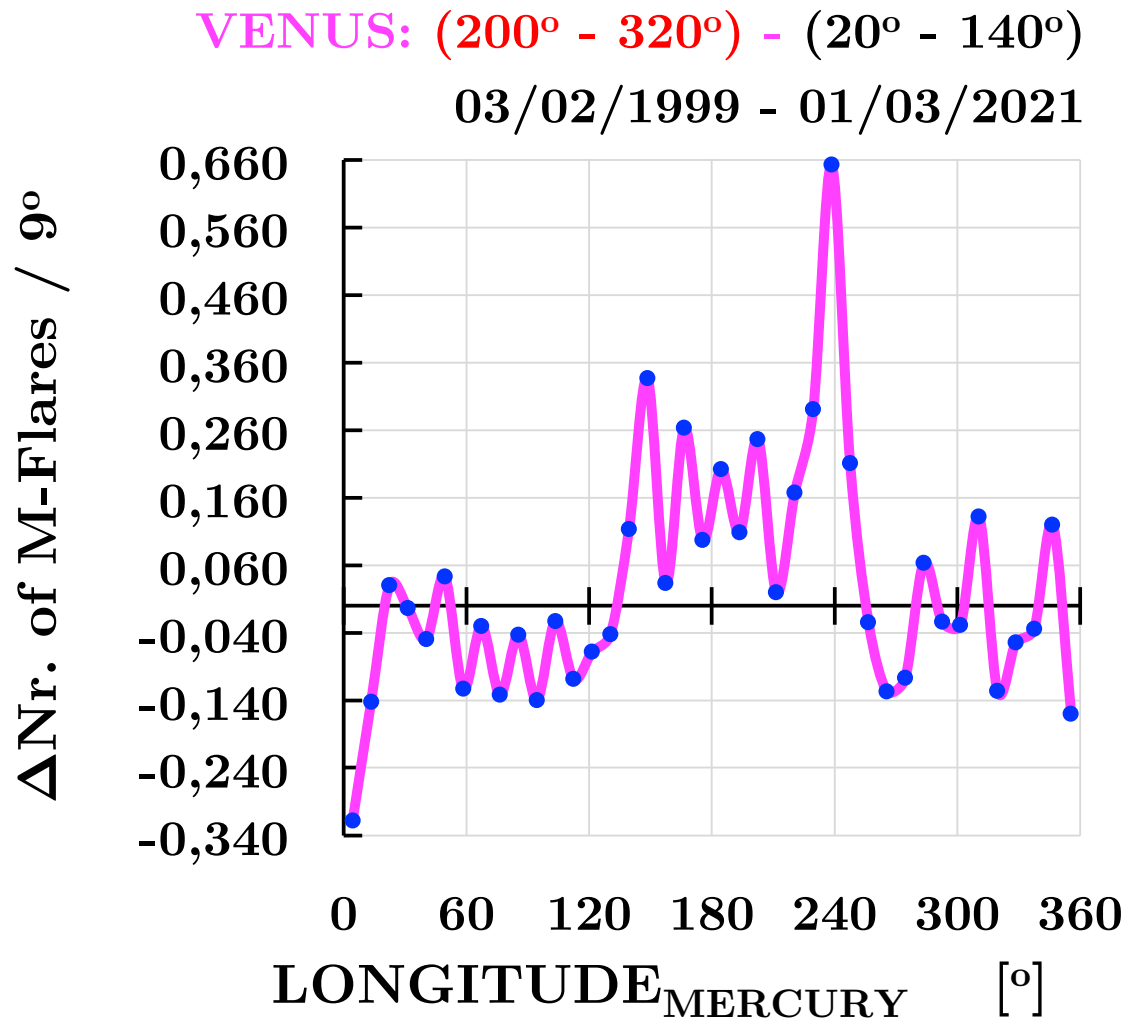

**(d)** Nr. of M-flares distribution when subtracting the two cases of Venus.

**Figure B.10:** Comparison of EUV vs. Nr. of M-flares for Mercury's distribution when Venus is between 200° to 320° and 20° to 140° with bin = 9° for the period 03/02/1999 - 01/03/2021.

## B.3.2 Correlation with X-flares

An alternative test has also been performed with the biggest X-flares, and more specifically with the ones having an integrated flux bigger than 1 J/m$^2$. For the period 01/09/1975 - 12/03/2021, there are 63 out of 495 X-Flares that fulfil this condition. However, considering the available EUV period (01/01/1996 - 01/03/2021) we are left with 7 flares. These flares are shown in Fig. B.12.

As next, a period of 120 d before and after each one of the 7 flares was defined, as seen in Tab. B.2. This was made in order to plot the summed EUV distribution during these periods and see if there is a correlation. The date of each flare is defined as a reference point (time = 0).

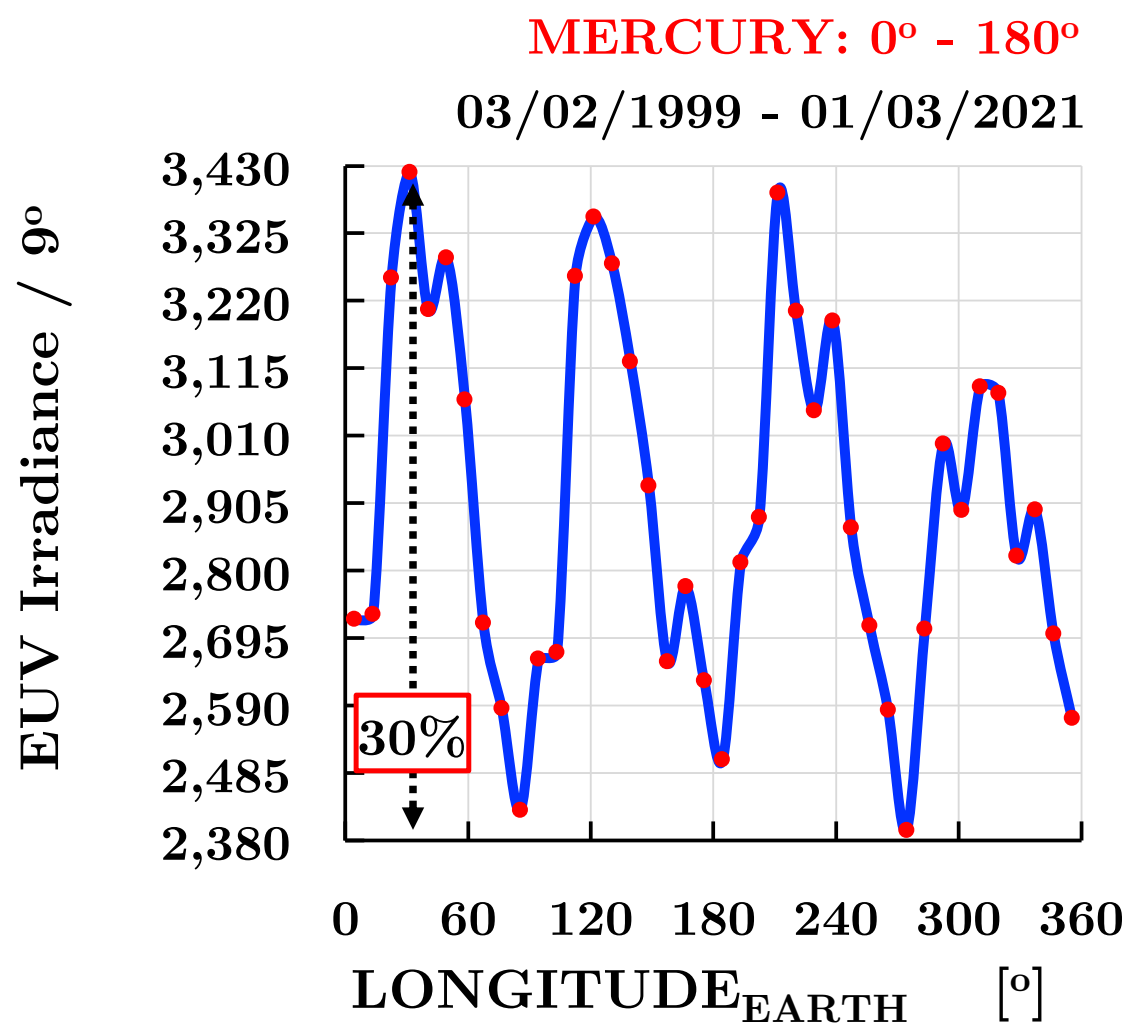

(a) EUV distribution.

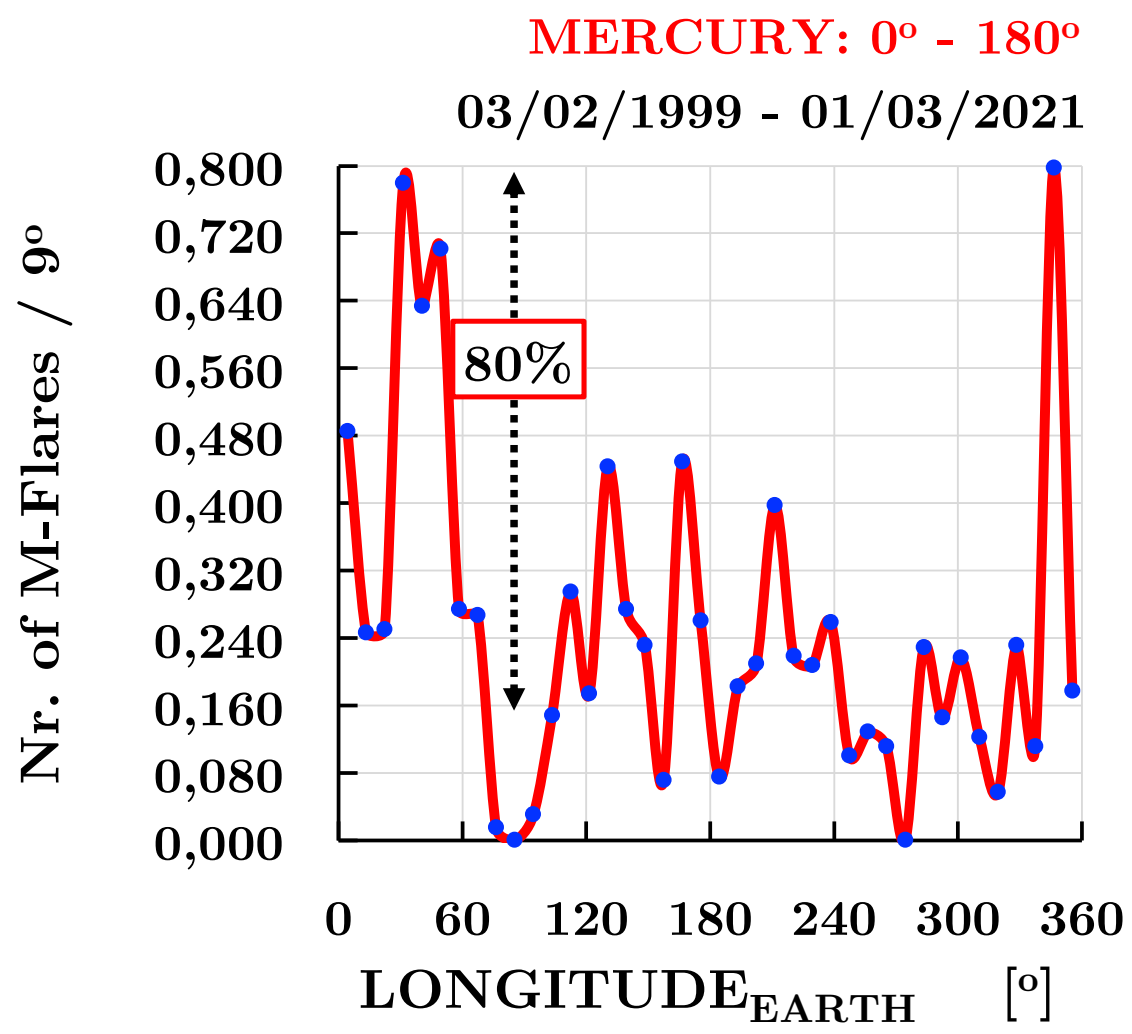

(b) Nr. of M-flares distribution.

**Figure B.11:** Comparison of Earth's distribution when Mercury is between 0° to 180° with bin = 9° for EUV irradiance vs. Nr. of M-flares for the period 03/02/1999 - 01/03/2021.

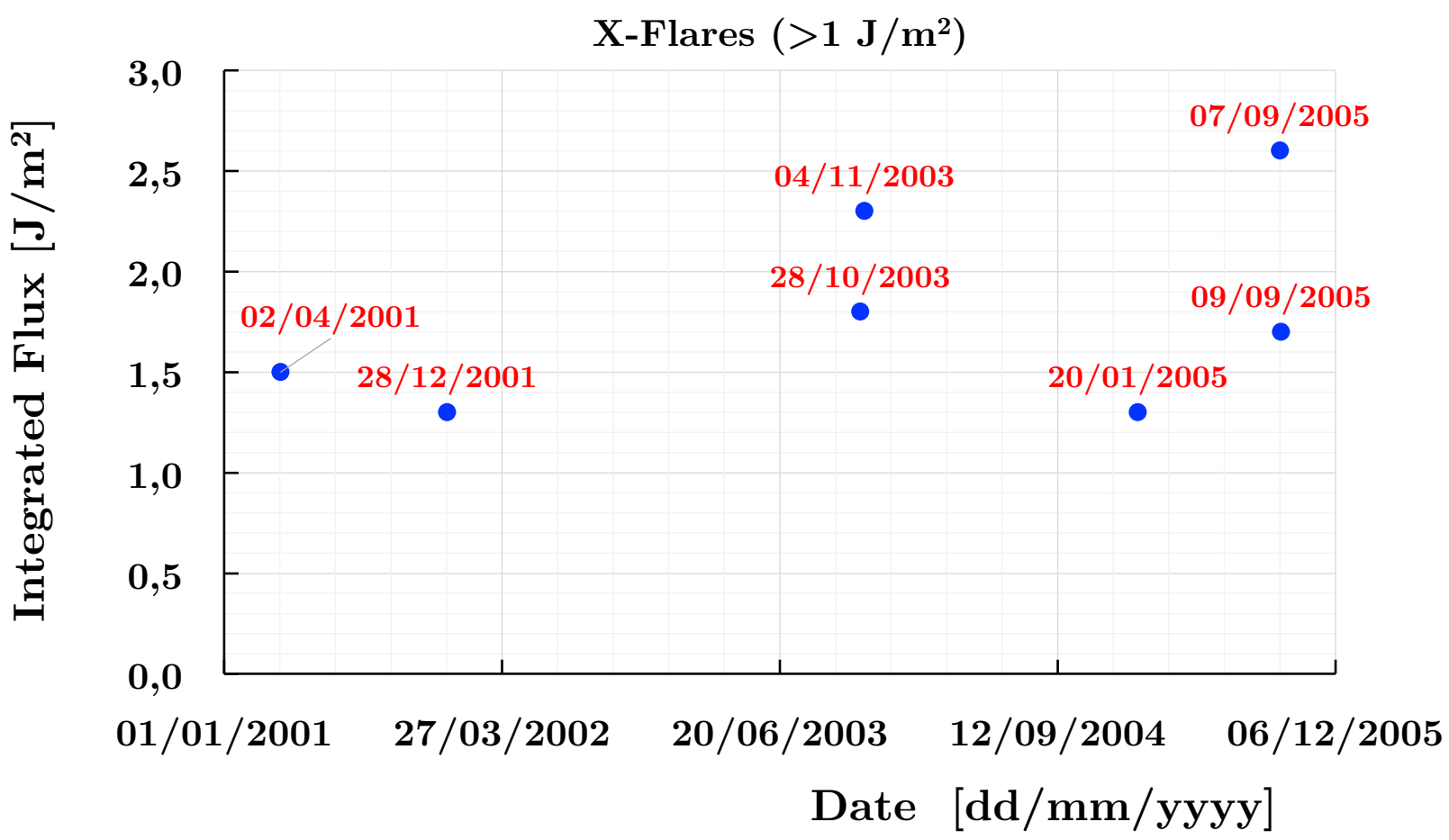

**Figure B.12:** X-flares with integrated flux > 1 J/m$^2$ after 01/01/1996.

Since some periods are overlapping, and to avoid double counting, the mean date between the two nearby dates, around 28/10/2003 and 07/09/2005, was used. The resulting new corrected five dates along with their corresponding ±120 d periods are shown in Tab. B.3.

In Fig. B.13, the result for the corresponding periods from both Tab. B.2 and B.3 is shown in red and blue accordingly, where the 7 and 5 EUV distributions were summed day by day. The centre of this plot at 0 d corresponds to the onset date of each flare. It is clear that the EUV distribution for both cases remains the same. On the left side, before the occurrence of the big X-flares, a gap appears in the EUV which lasts about $1-2$ months, indicating a long procedure happening in the Sun. Furthermore, even ∼1 week before each flare the EUV emission is increased, as if the Sun was aware that a flare was about to be generated. We recall that X-class solar flares last on average 24 min with a maximum of about 2 h [465]. It is noted that previously published results show a decrease of the solar EUV/UV emission 50 d to 10 d before and X-ray event, but also a depletion of the TSI i.e. the flux of solar light received at all wavelengths at the top of the Earth's atmosphere, of about 5 d before the X-ray burst [466, 467].

**Table B.2:** Dates with X-flares with integrated flux > 1 J/m$^2$ after 01/01/1996 and the corresponding periods used for the EUV. The overlapping periods are marked in red.

| Flare Date [UTC] | Date −120 d [UTC] | Date +120 d [UTC] |
|---|---|---|
| 02/04/2001 | 03/12/2000 | 31/07/2001 |
| 28/12/2001 | 30/08/2001 | 27/04/2002 |
| 28/10/2003 | 30/06/2003 | 25/02/2004 |
| 04/11/2003 | 07/07/2003 | 03/03/2004 |
| 20/01/2005 | 22/09/2004 | 20/05/2005 |
| 07/09/2005 | 10/05/2005 | 05/01/2006 |
| 09/09/2005 | 12/05/2005 | 07/01/2006 |

**Table B.3:** No overlapping corrected X-flare dates with integrated flux > 1 J/m$^2$ with the corresponding periods used for the EUV. The corrected periods are marked in red.

| Adjusted Flare Date [UTC] | Date −120 d [UTC] | Date +120 d [UTC] |
|---|---|---|
| 02/04/2001 | 03/12/2000 | 31/07/2001 |
| 28/12/2001 | 30/08/2001 | 27/04/2002 |
| 01/11/2003 | 04/07/2003 | 29/02/2004 |
| 20/01/2005 | 22/09/2004 | 20/05/2005 |
| 08/09/2005 | 11/05/2005 | 06/01/2006 |

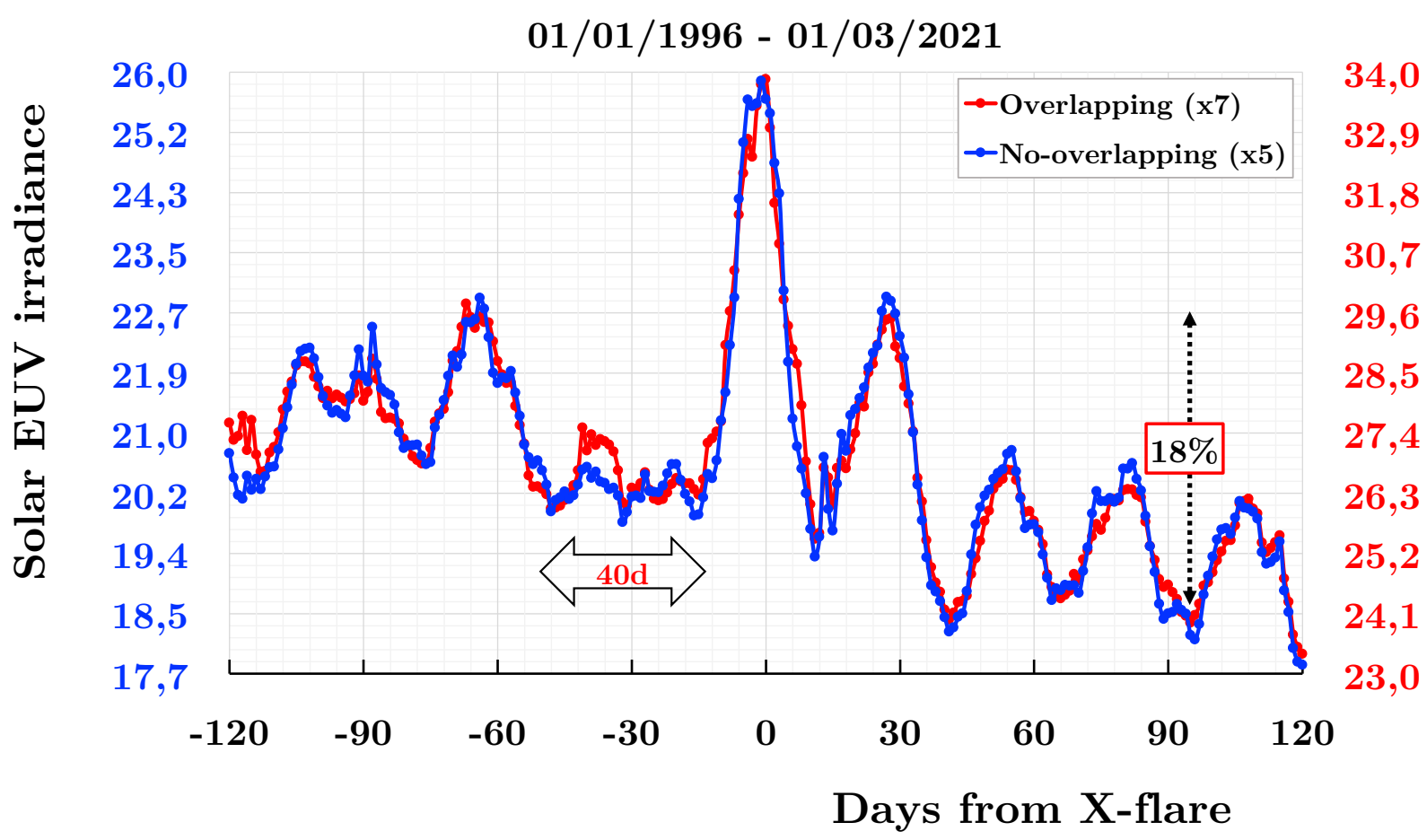

**Figure B.13:** EUV distribution before and after each selected big X-flare (> 1 J/m$^2$).

## B.4 Sunspots

### B.4.1 Comparison with solar EUV

The various planetary distributions of the sunspots are compared with the corresponding ones from EUV. Since the period of EUV spans from 01/01/1996 to 01/03/2021, the sunspot distributions are recreated in order for the overlapping period 01/01/1996 - 28/02/2021 to be the same. The rest of the conditions like the bin size as well as the starting point of the x-axis remain the same to allow for an additional comparison with the original plots from Sect. 8.3. A correlation analysis has also been performed on the various distributions.

In Fig. B.14 the time series of the two datasets are overlaid for the same overlapping period. As expected, the Pearson correlation coefficient and p-value ($r = 0.89$, $p = 0$) point to a strong, statistically significant linear correlation on the general trend of sunspots and EUV irradiance. As next specific comparisons also on the small-scale characteristics of the datasets are performed based on the derived planetary dependencies.

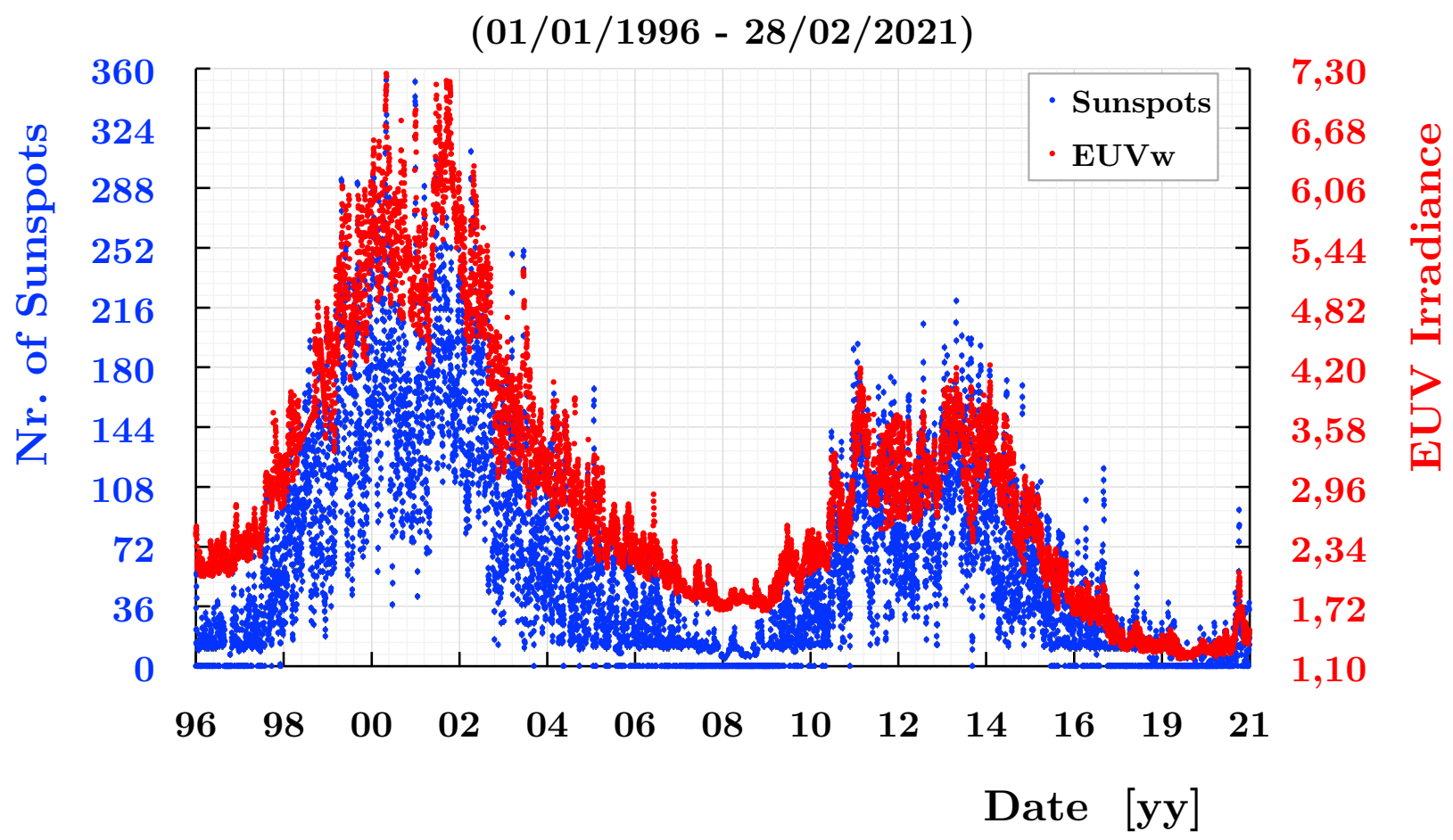

**Figure B.14:** Daily sunspot Nr. and EUV solar irradiance for the same period 01/01/1996 - 28/02/2021.

#### B.4.1.1 Single planets

In Fig. B.15 and B.16 the cases where single planets are used as reference frames are shown. The observed EUV distributions appear to be comparable with the corresponding cases for the number of sunspots. A notable difference is observed for the case of Earth in Fig. B.15c and B.15g as well as for the relative differences of the various peaks in Mars. The Pearson correlation coefficients for the various cases are ($r_{a,e} = 0.66$, $p_{a,e} = 7.73 \times 10^{-5}$), ($r_{b,f} = 0.82$, $p_{b,f} = 1.65 \times 10^{-4}$), ($r_{c,g} = 0.33$, $p_{c,g} = 0.224$) and ($r_{d,h} = 0.59$, $p_{d,h} = 5.72 \times 10^{-4}$) for the subfigures a to h in Fig. B.15, and ($r_{a,b} = 0.88$, $p_{a,b} = 0$) for Fig. B.16 showing a statistically significant correlation for all the cases except Earth, with Jupiter having an almost perfect linear association. We also notice that the amplitudes in Fig. B.15a through B.15d and Fig. B.16a are bigger than the corresponding ones from Fig. 8.6 where the extended period was used. That is to be expected since more revolutions corresponds to more averaging and therefore individual phenomena correlating with specific periods disappear. This is also explained by the fact that every solar cycle is different but also due to different planetary configurations taking place in different periods considering that outer planets such as Saturn, Uranus and Neptune rotate much slower and may not even complete a full revolution around the Sun during the selected period.

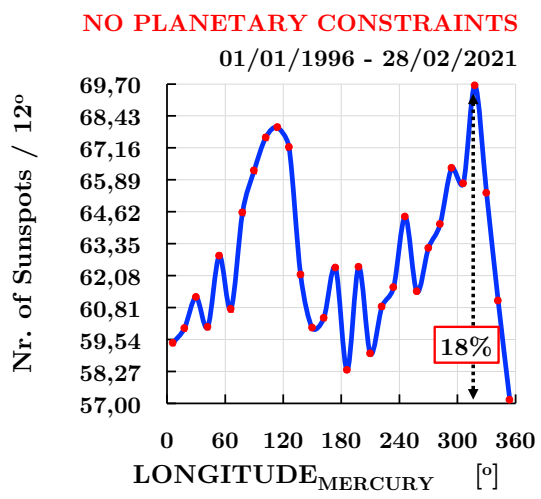

**(a)** Sunspot distribution for Mercury with 12° bin.

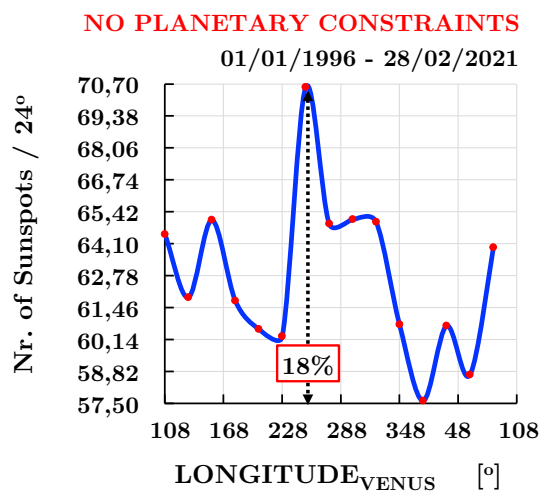

**(b)** Sunspot distribution for Venus with 24° bin.

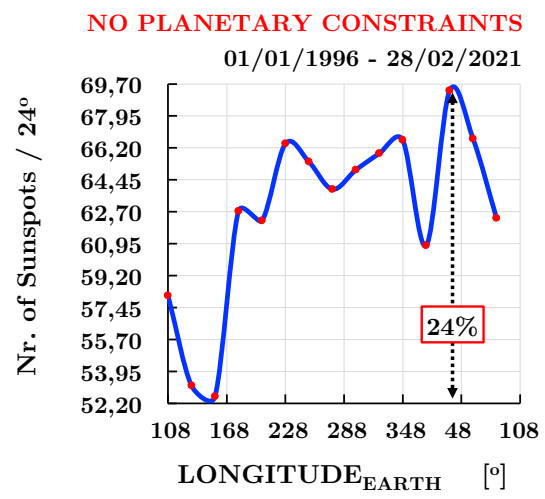

**(c)** Sunspot distribution for Earth with 24° bin.

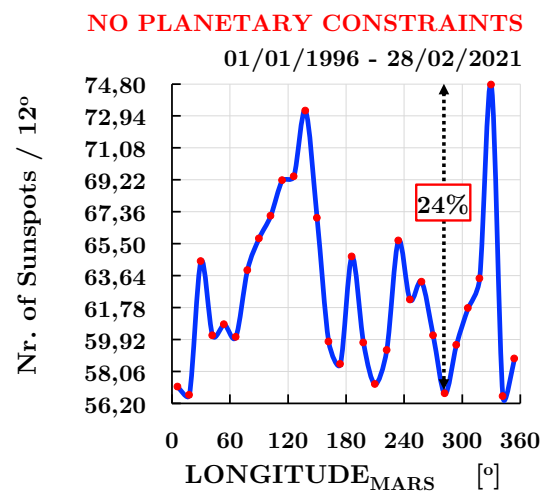

**(d)** Sunspot distribution for Mars with 12° bin.

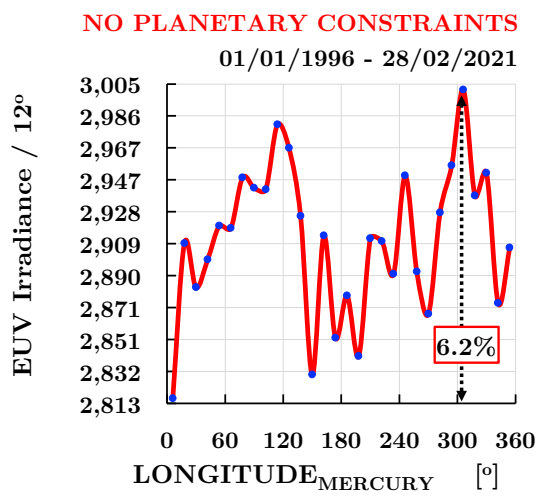

**(e)** EUV distribution for Mercury with 12° bin.

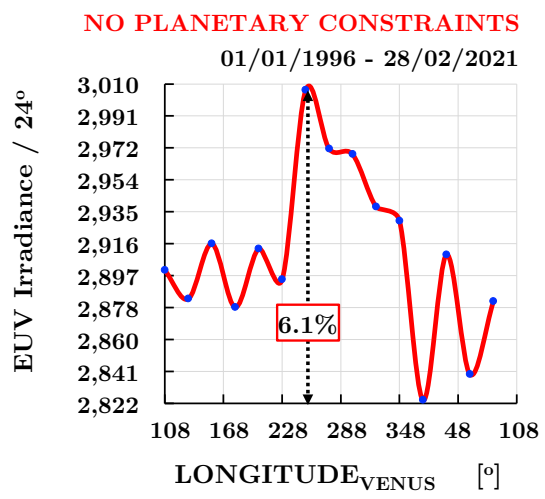

**(f)** EUV distribution for Venus with 24° bin.

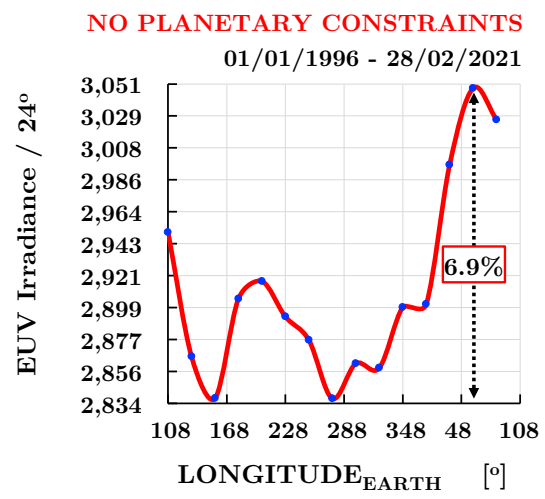

**(g)** EUV distribution for Earth with 24° bin.

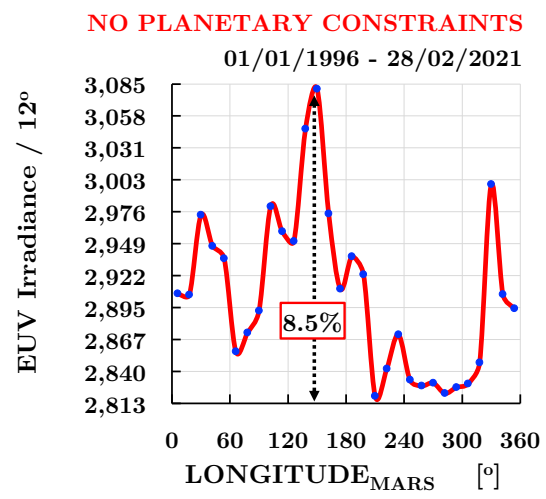

**(h)** EUV distribution for Mars with 12° bin.

**Figure B.15:** Comparison of distributions for the Nr. of sunspots vs. EUV solar irradiance for the period 01/01/1996 - 28/02/2021.

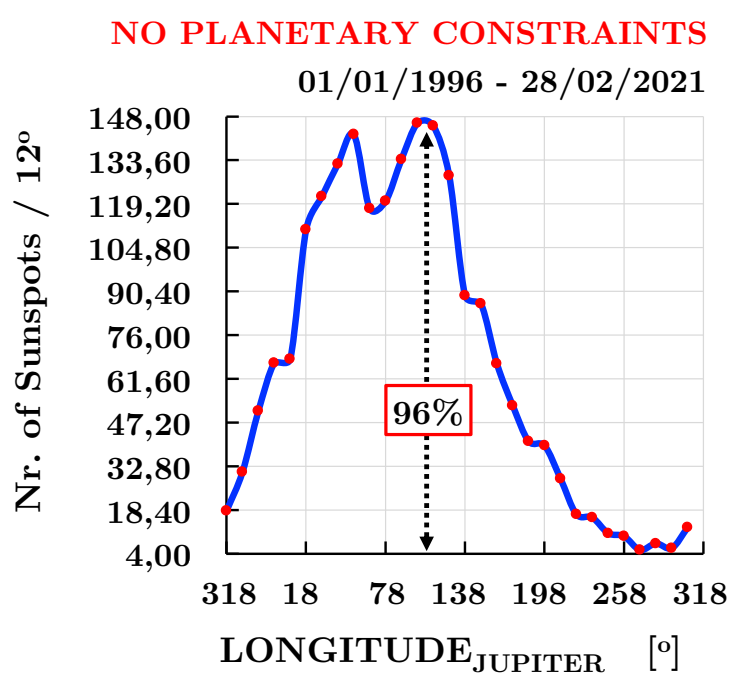

**(a)** Sunspot distribution.

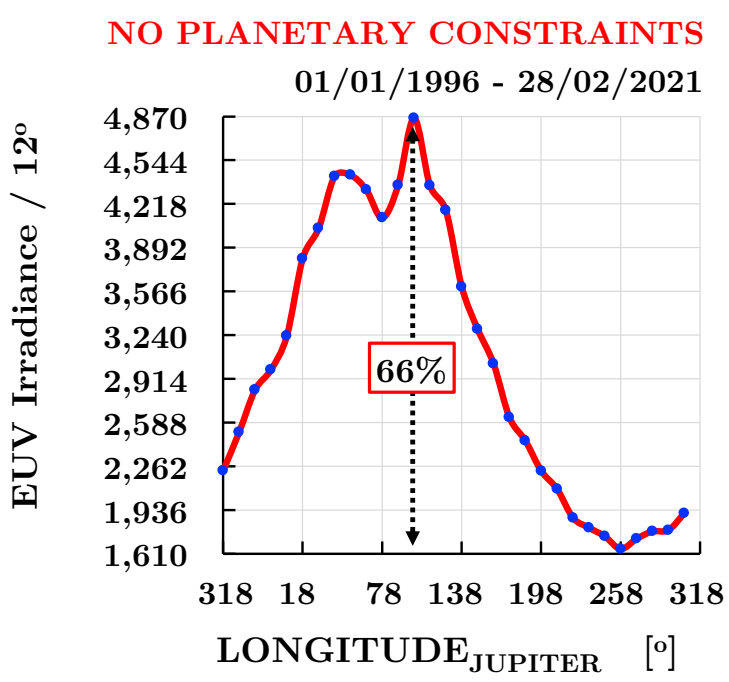

**(b)** EUV distribution.

**Figure B.16:** Comparison of distributions for the Nr. of sunspots vs. EUV solar irradiance for Jupiter with 12° bin for the period 01/01/1996 - 28/02/2021.

#### B.4.1.2 Combining planets

The comparison procedure is also performed for the plots with a planetary combination. In Fig. B.17, the distributions from Fig. 8.7c and 8.8 are compared with the corresponding ones from the EUV solar irradiance. Even though some differences are observed in the relative amplitudes of several peaks, in Fig. B.17a and B.17b when compared with Fig. B.17d and B.17e, the overall observed shape of the plots for the two datasets appear to be similar. This observation is supported by the calculated correlation coefficients; ($r_{a,d} = 0.76$, $p_{a,d} = 1.01 \times 10^{-6}$) and ($r_{a,d} = 0.69$, $p_{a,d} = 2.40 \times 10^{-5}$) for the two positions of Venus in Fig. B.17a and B.17d and ($r_{b,e} = 0.68$, $p_{b,e} = 3.98 \times 10^{-5}$), ($r_{c,f} = 0.99$, $p_{c,f} = 0$) for the remaining two cases where Mars' and Moon's position is constrained respectively.

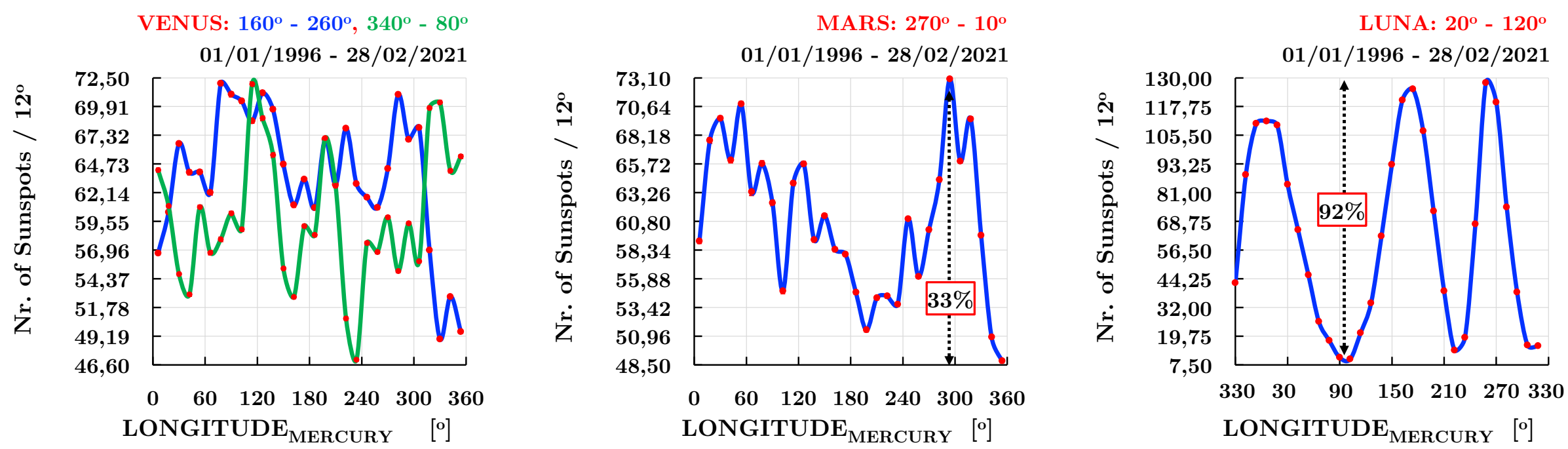

**(a)** Sunspot distribution for Mercury while Venus propagates between 160° to 260° and 340° to 80°.

**(b)** Sunspot distribution for Mercury while Mars propagates between 270° to 10°.

**(c)** Sunspot distribution for Mercury while Moon propagates between 20° to 120°.

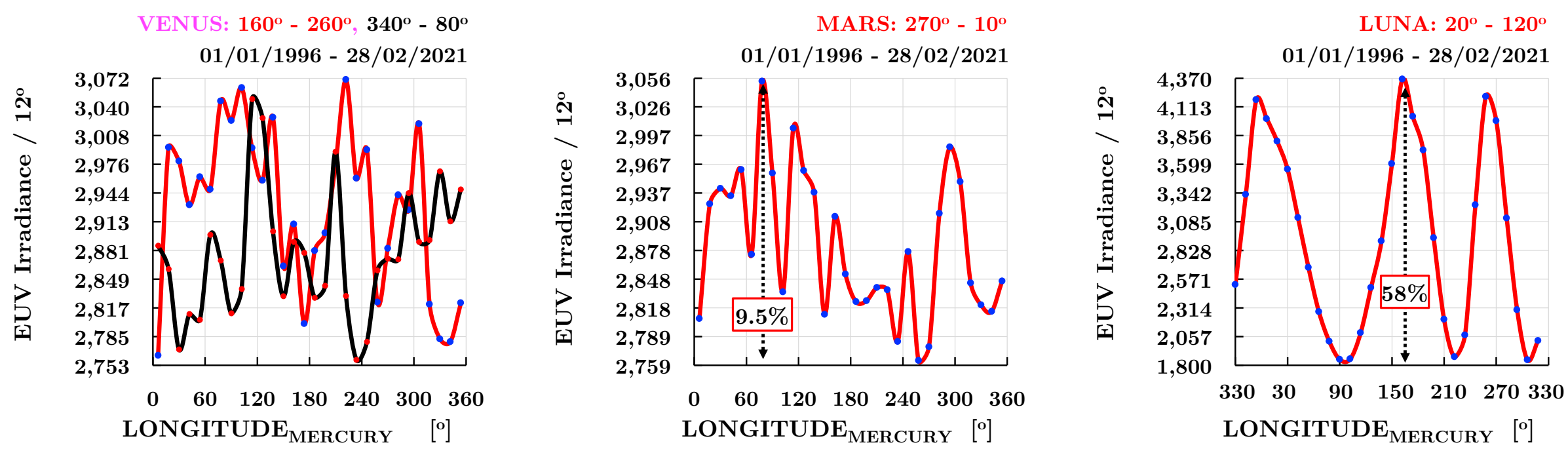

**(d)** EUV distribution for Mercury while Venus propagates between 160° to 260° and 340° to 80°.

**(e)** EUV distribution for Mercury while Mars propagates between 270° to 10°.

**(f)** EUV distribution for Mercury while Moon propagates between 20° to 120°.

**Figure B.17:** Comparison of distributions for the Nr. of sunspots vs. EUV solar irradiance for the reference frame of Mercury while other planets are constrained for the period 01/01/1996 - 28/02/2021 with 12° bin.

Moving on to the reference frame of Venus, in Fig. B.18 the plots from Fig. 8.9 are reevaluated for the same period of EUV and are set side by side with the corresponding EUV distributions. Once more they are found to have a significant and very high degree of linear correlation, with the Pearson's correlation coefficients for the various subfigures being ($r_{a,d} = 0.69$, $p_{a,d} = 2.14 \times 10^{-5}$), ($r_{b,e} = 0.94$, $p_{b,e} = 2.82 \times 10^{-14}$) and ($r_{c,f} = 0.97$, $p_{c,f} = 0$).

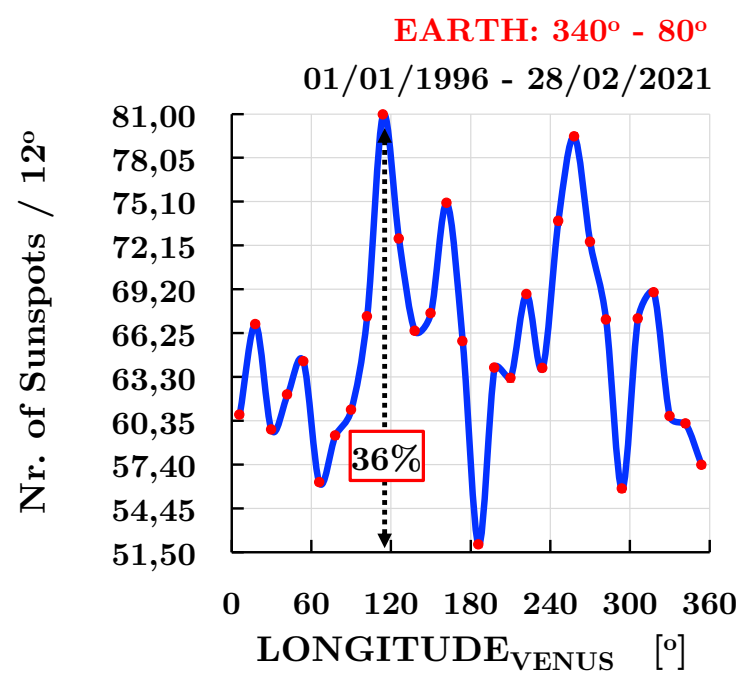

**(a)** Sunspot distribution for Venus while Earth propagates between 340° to 80°.

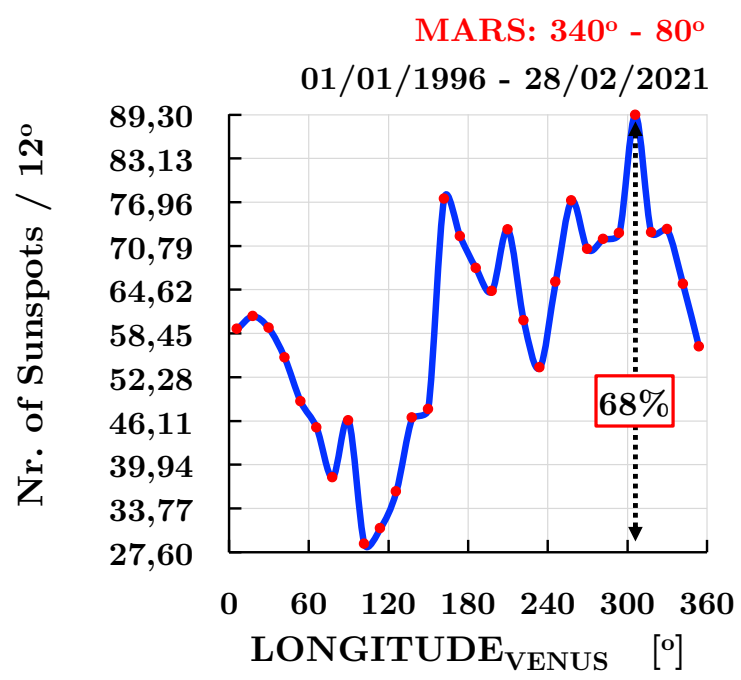

**(b)** Sunspot distribution for Venus while Mars propagates between 340° to 80°.

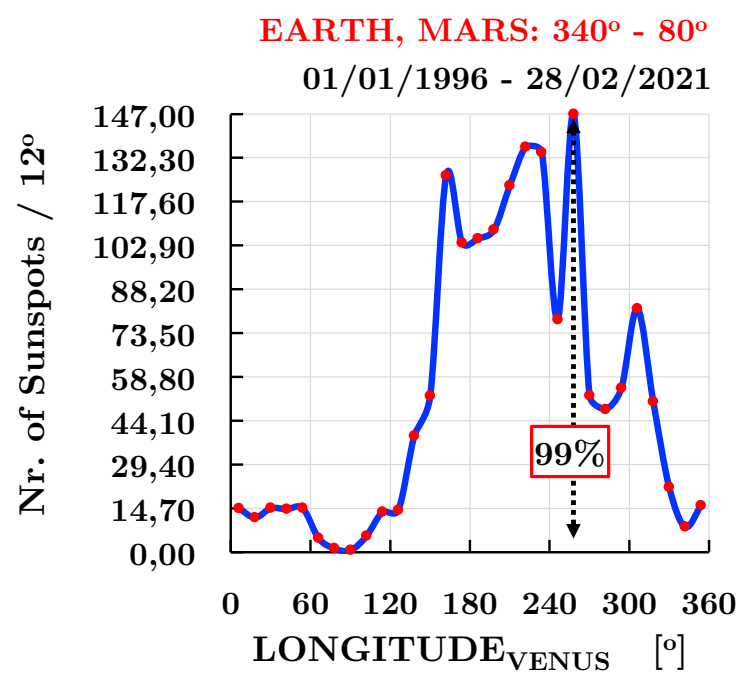

**(c)** Sunspot distribution for Venus while Earth and Mars propagate both between 340° to 80°.

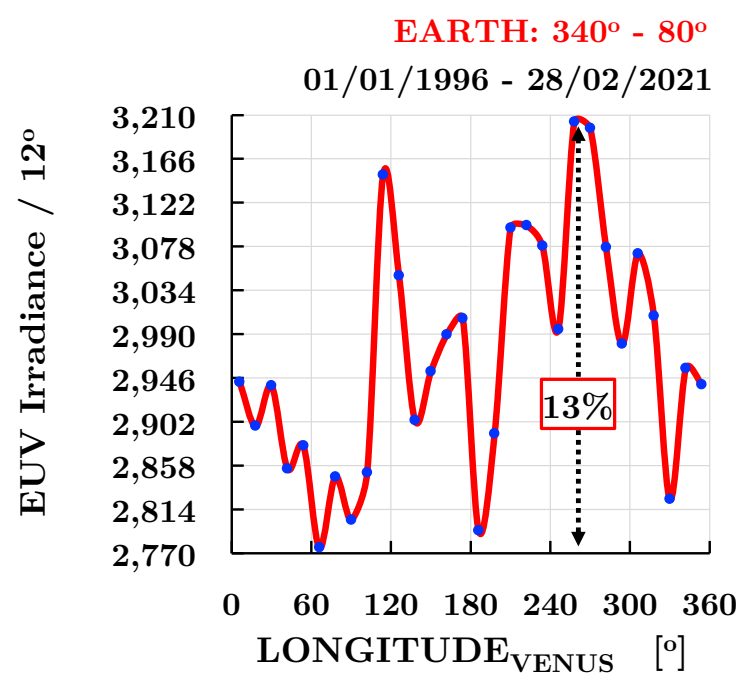

**(d)** EUV distribution for Venus while Earth propagates between 340° to 80°.

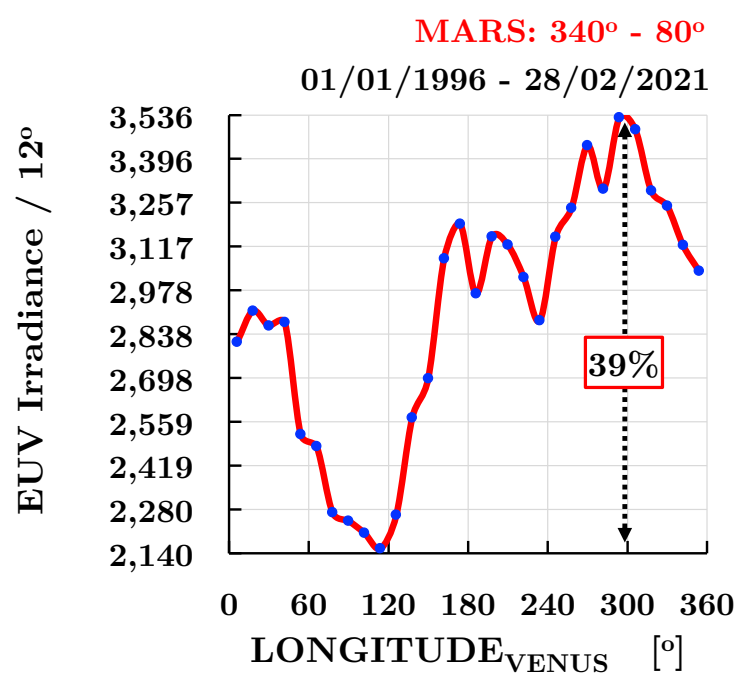

**(e)** EUV distribution for Venus while Mars propagates between 340° to 80°.

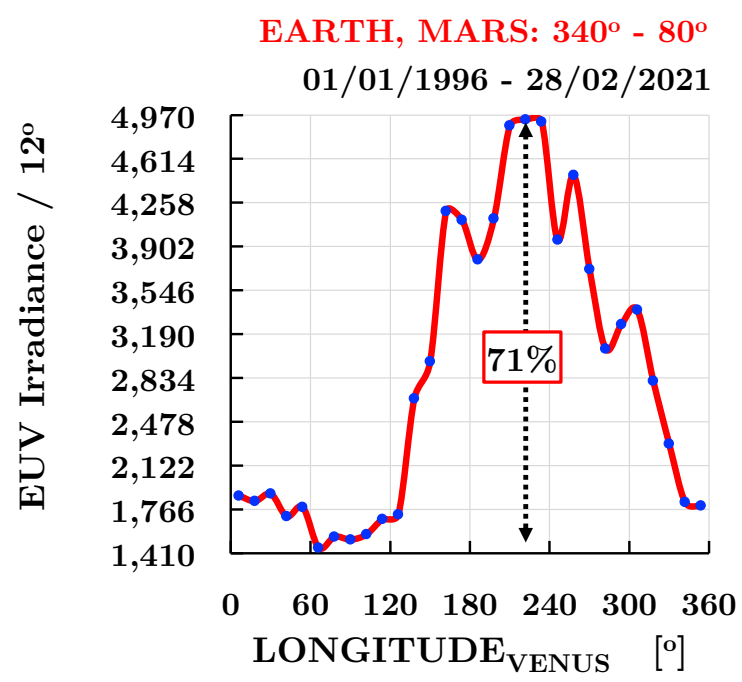

**(f)** EUV distribution for Venus while Earth and Mars propagate both between 340° to 80°.

**Figure B.18:** Comparison of distributions for the Nr. of sunspots vs. EUV solar irradiance for the reference frame of Venus while other planets are constrained for the period 01/01/1996 - 28/02/2021 with 12° bin.

In Fig. B.19, an identical behaviour with eight peaks at the same heliocentric longitudes of Mars is observed between the number of sunspots and the EUV irradiance with ($r_{a,b} = 0.97$, $p_{a,b} = 0$). Comparing also Fig. B.19a with Fig. 8.10a we see a small shift on the positions of the peaks by an average of about 19.60° towards higher heliocentric longitudes for the period 01/01/1996 - 28/02/2021.

Finally, in Fig. B.20 the case from Fig. 8.11c when Moon's phase is observed without constraints and when Mercury propagates around $90° \pm 50°$ is compared with the distributions of EUV. Again, a similar behaviour is observed for the two manifestations of solar activity with the correlation coefficient ($r_{a,b} = 0.99$, $p_{a,b} = 4.44 \times 10^{-16}$) indicating an almost perfect positive linear correlation. In addition, contrasting Fig. B.20a and 8.11c we see that for the case of Mercury propagating between 40° to 140° the wide peak for the period 01/01/1996

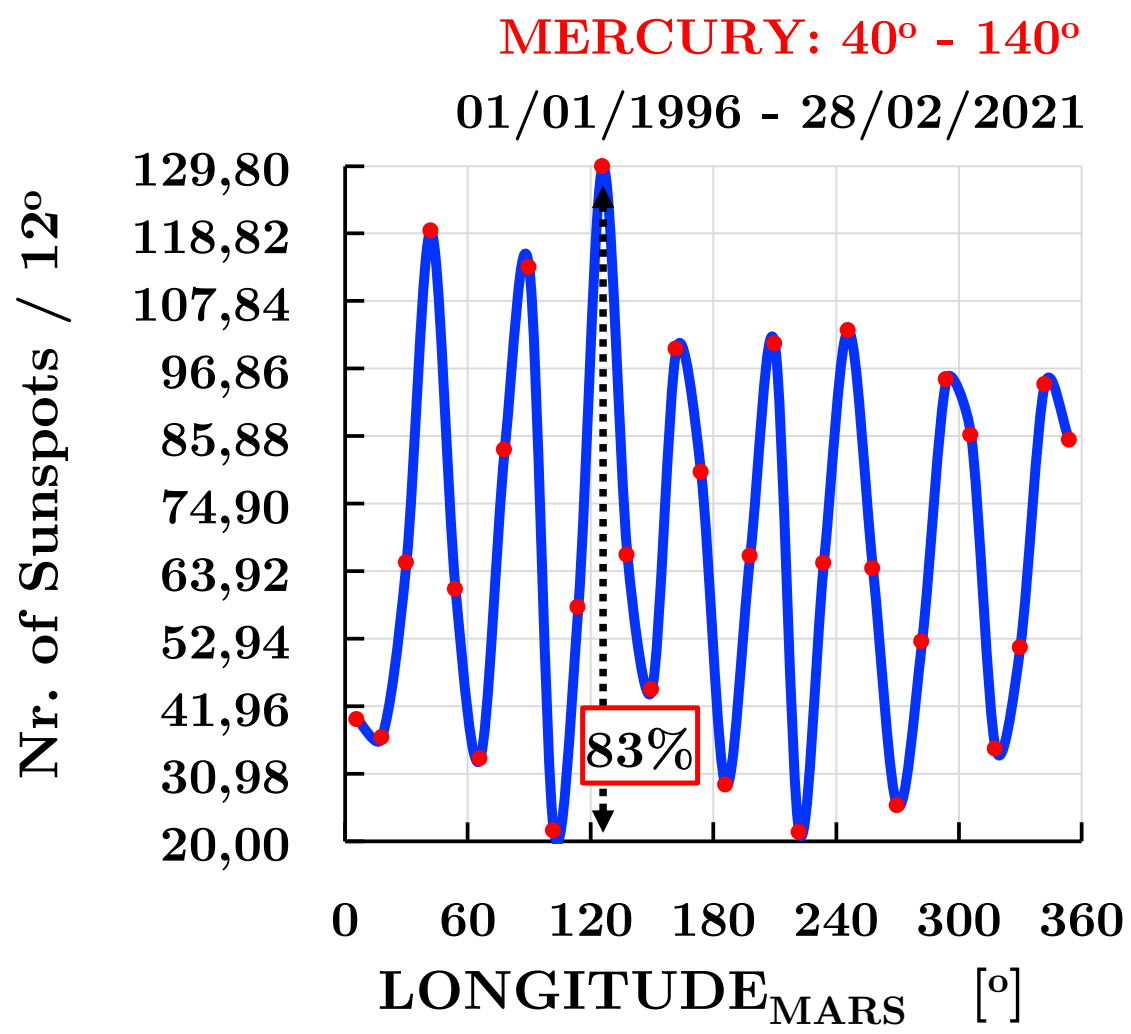

(a) Sunspot distribution.

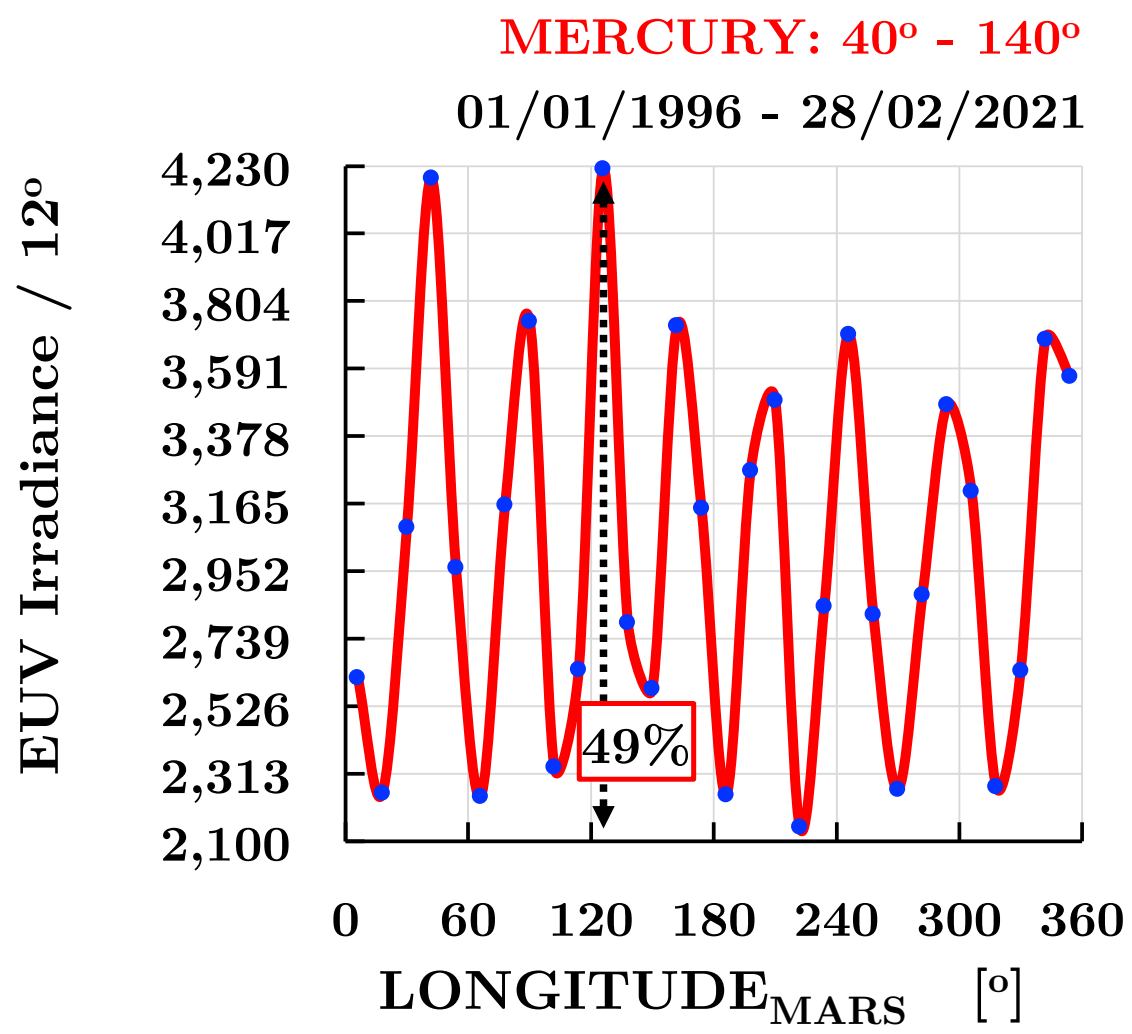

(b) EUV distribution.

**Figure B.19:** Comparison of distributions for the Nr. of sunspots vs. EUV solar irradiance for the reference frame of Mars while Mercury propagates between 40° to 140° for the period 01/01/1996 - 28/02/2021 with 12° bin.

- 28/02/2021 is centred around 254.91° with a FWHM = 213.20°, whereas for the period 01/03/1900 - 28/02/2021 it was centred around 314.92° with a FWHM = 188.76°. This indicates a phase shift for Moon's phase of $\sim 60.01°$ towards lower longitudes.

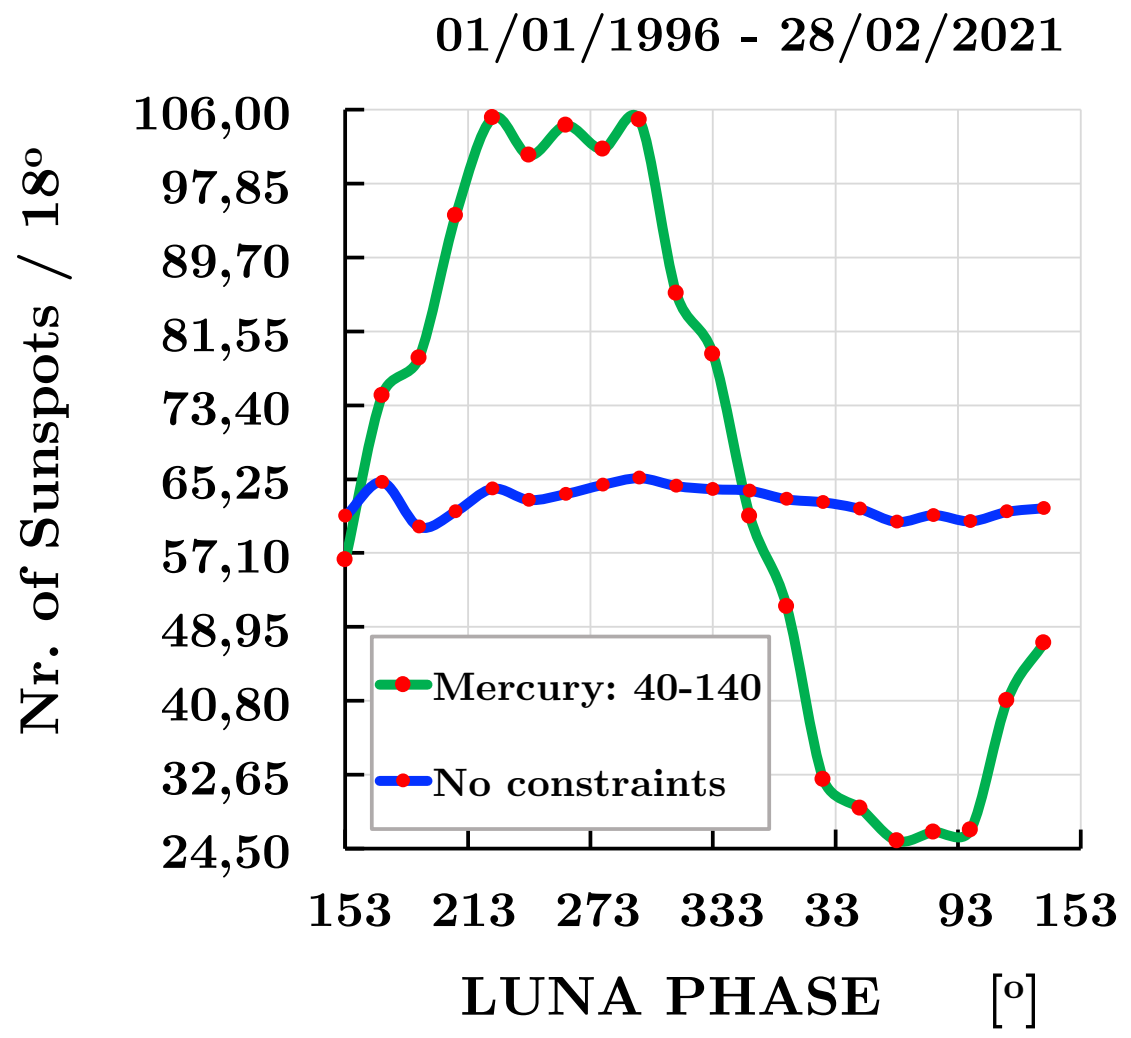

(a) Sunspot distribution.

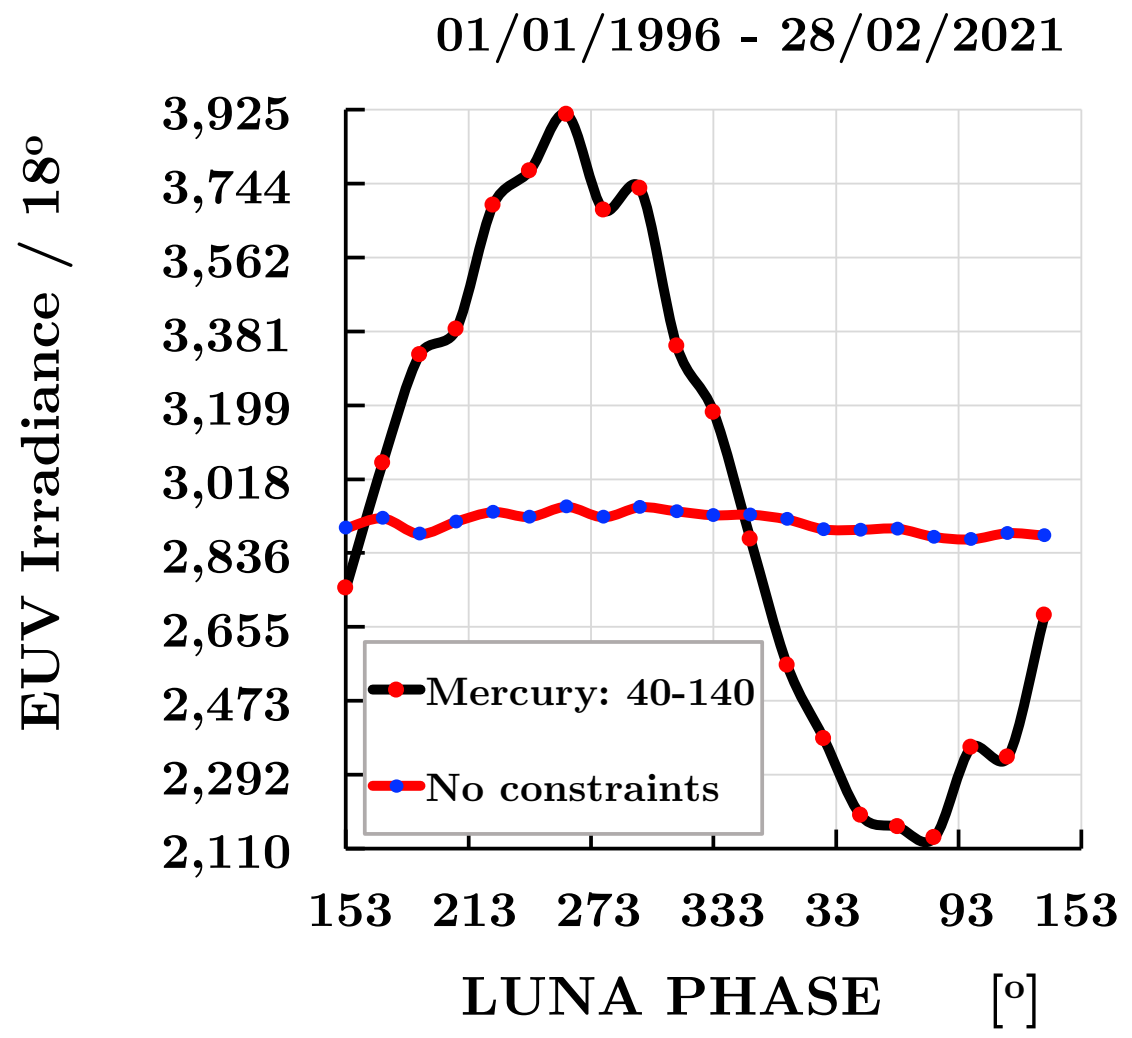

(b) EUV distribution.

**Figure B.20:** Comparison of distributions for the Nr. of sunspots vs. EUV solar irradiance for the reference frame of Moon's phase without any additional constraints and when Mercury propagates between 40° to 140° for the period 01/01/1996 - 28/02/2021. The bin is set at 18°.

# B.5 F10.7

## B.5.1 Comparison with solar EUV

A comparison of the F10.7 results is made with the EUV in order to look for similarities or differences pointing to a similar or alternative origin of the observed planetary relationships. Since the available time range is different between the datasets with 1999 – 2021 for EUV and 1963 – 2021 for F10.7, the same overlapping period is selected.

In Fig. B.21 the time series of the two solar proxies are shown for the same period. As expected also from the observed general trend, the calculated Pearson correlation coefficient and p-value ($r = 0.94$, $p = 0$) show a high degree of statistically significant correlation. Following the general trend similarity, the specific planetary dependencies are also compared individually.

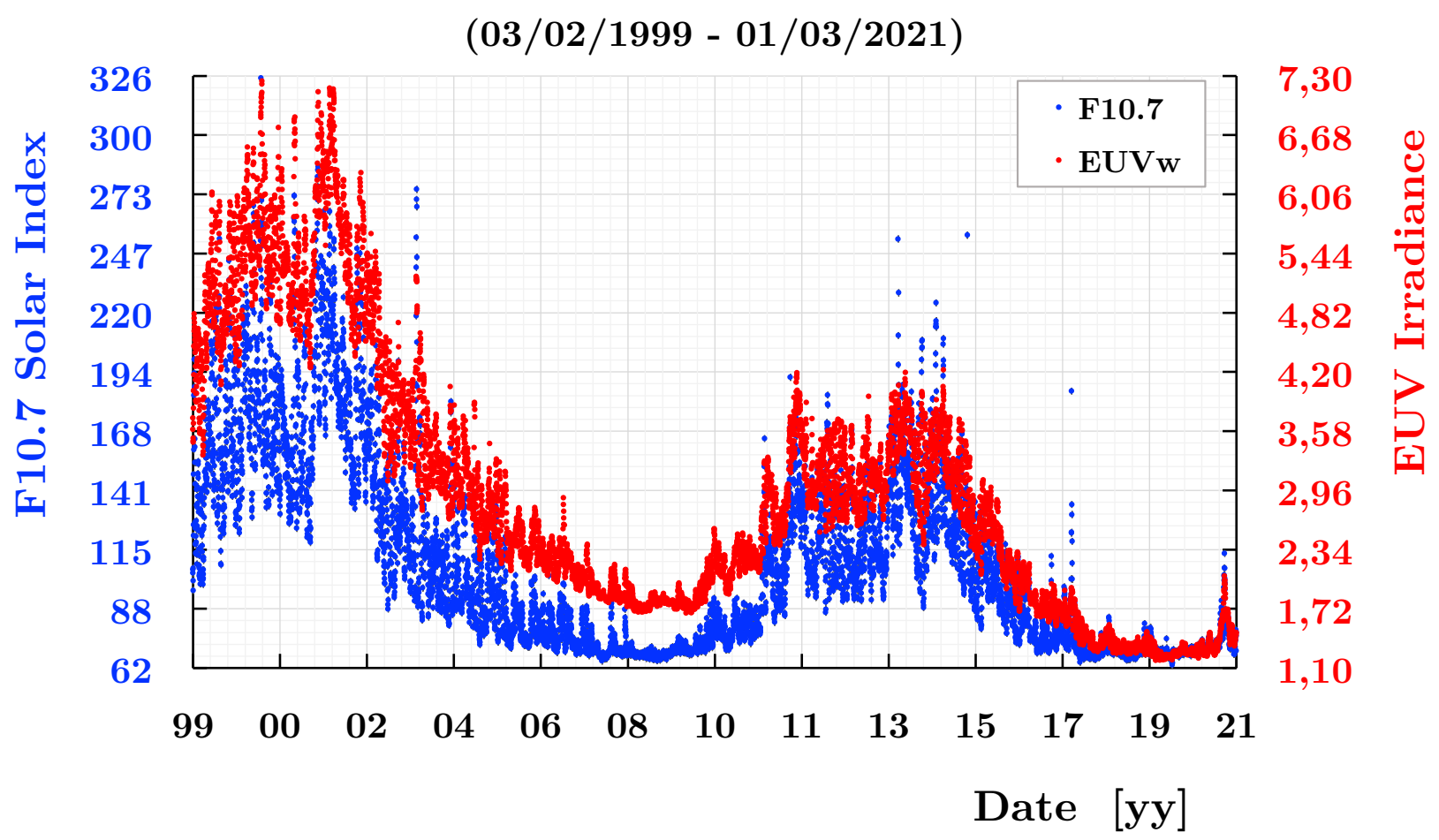

**Figure B.21:** Daily F10.7 solar index and EUV irradiance for the same period 03/02/1999 - 01/03/2021.

#### B.5.1.1 Single planets

In Fig. B.22 and B.23, the F10.7 single-planet distributions from Fig. 9.3 are calculated once more for the period 03/02/1999 - 01/03/2021, and then compared with the corresponding spectra of the EUV solar irradiance. Saturn is not compared in this case due to the limited available time range compared with Saturn's long orbital period. The various spectra in Fig. B.22 and B.23 show a high degree of similarity between the two solar datasets pointing to the same origin of planetary relationship. This is also verified by the statistical correlation analysis. The calculated Pearson's correlation coefficients, all being statistically significant, are ($r_{a,e} = 0.92$, $p_{a,e} = 5.46\times10^{-13}$), ($r_{b,f} = 0.90$, $p_{b,f} = 7.64\times10^{-8}$), ($r_{c,g} = 0.74$, $p_{c,g} = 2.79\times10^{-6}$) and ($r_{d,h} = 0.78$, $p_{d,h} = 5.79 \times 10^{-4}$) for the subfigures a to h in Fig. B.22 and ($r_{a,b} = 0.99$, $p_{a,b} = 0$) for Fig. B.23.

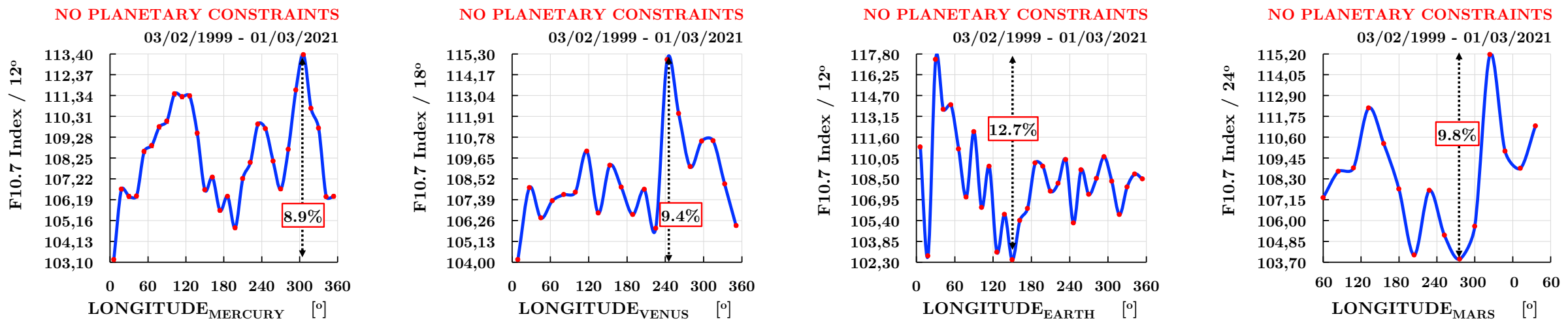

(a) F10.7 distribution for Mercury with 12° bin.
(b) F10.7 distribution for Venus with 18° bin.
(c) F10.7 distribution for Earth with12° bin.
(d) F10.7 distribution for Mars with 24° bin.

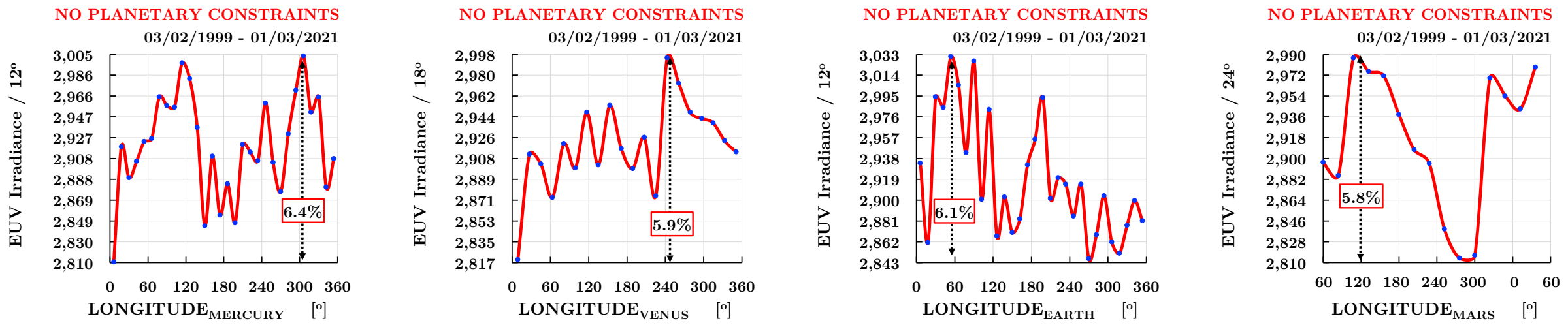

(e) EUV distribution for Mercury with 12° bin.
(f) EUV distribution for Venus with 18° bin.
(g) EUV distribution for Earth with 12° bin.
(h) EUV distribution for Mars with 24° bin.

**Figure B.22:** Comparison of distributions of F10.7 vs. EUV solar irradiance for the period 03/02/1999 - 01/03/2021.

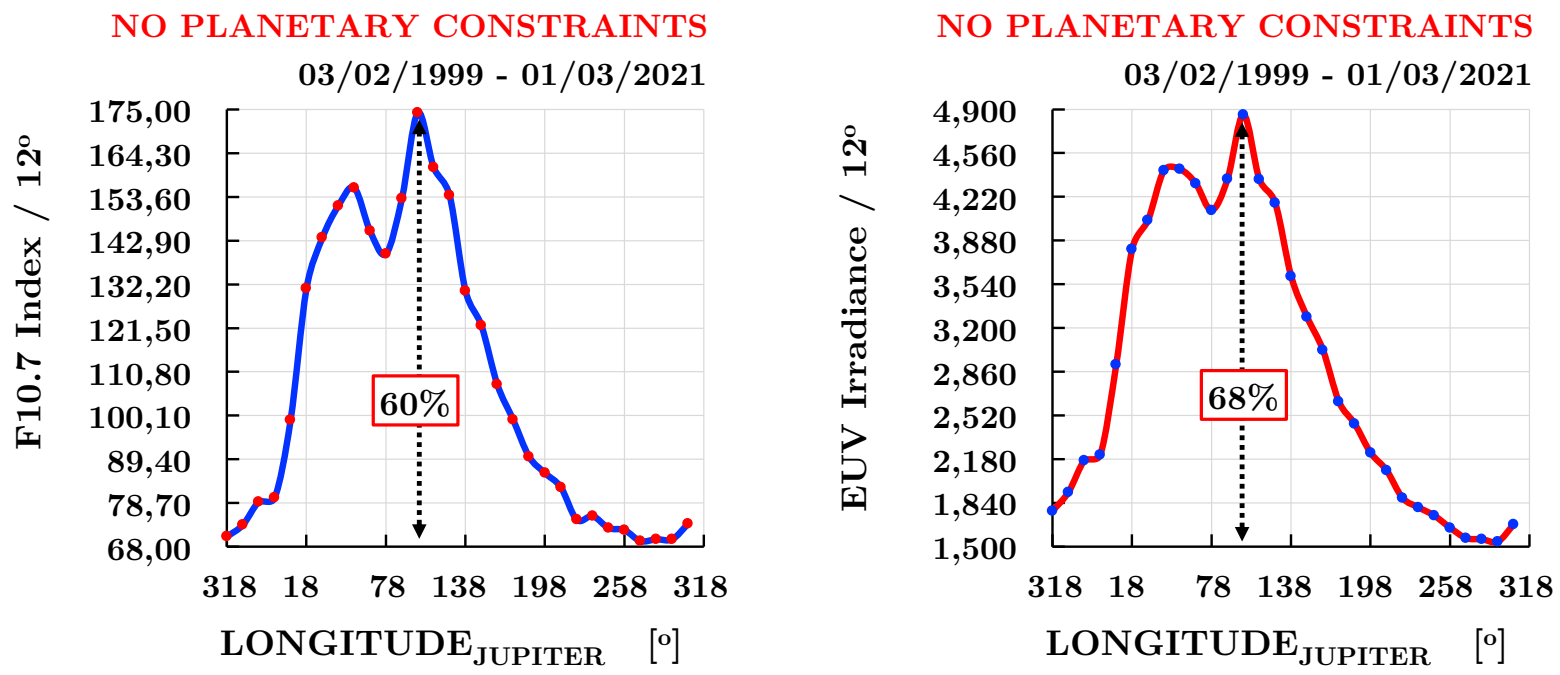

(a) F10.7 distribution.
(b) EUV distribution.

**Figure B.23:** Comparison of distributions of F10.7 vs. EUV solar irradiance for Jupiter with 12° bin for the period 03/02/1999 - 01/03/2021.

#### B.5.1.2 Combining planets

The same procedure is performed for the plots where a combination of the planetary longitudinal Fig. is made. As an example, in Fig. B.24 the plots from Fig. 9.4c, 9.5c and 9.8c are recalculated and compared with the EUV distributions. In all three cases, similar behaviour is observed for the two manifestations of solar activity. The correlation coefficients are ($r_{a,d} = 0.89$, $p_{a,d} = 6.11 \times 10^{-8}$), and ($r_{a,d} = 0.79$, $p_{a,d} = 2.10 \times 10^{-7}$) for the two constraints on Venus in Fig. B.24a and B.24d. Accordingly, for the two constraints on Mars in Fig. B.24b and B.24e we have ($r_{b,e} = 0.86$, $p_{b,e} = 1.26 \times 10^{-9}$) and ($r_{b,e} = 0.80$, $p_{b,e} = 1.29 \times 10^{-7}$) accordingly. And finally, for the constraint on Mercury's position in Fig. B.24c and B.24f we have again a perfect positive linear correlation with ($r_{c,f} = 0.99$, $p_{c,f} = 0$).

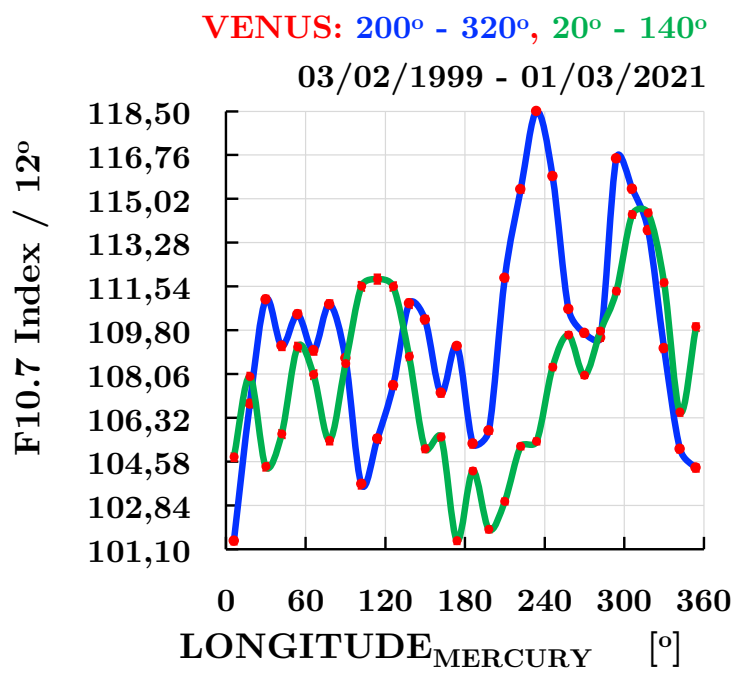

**(a)** F10.7 distribution for Mercury when Venus is between 200° to 320° and 20° to 140° for 12° bin.

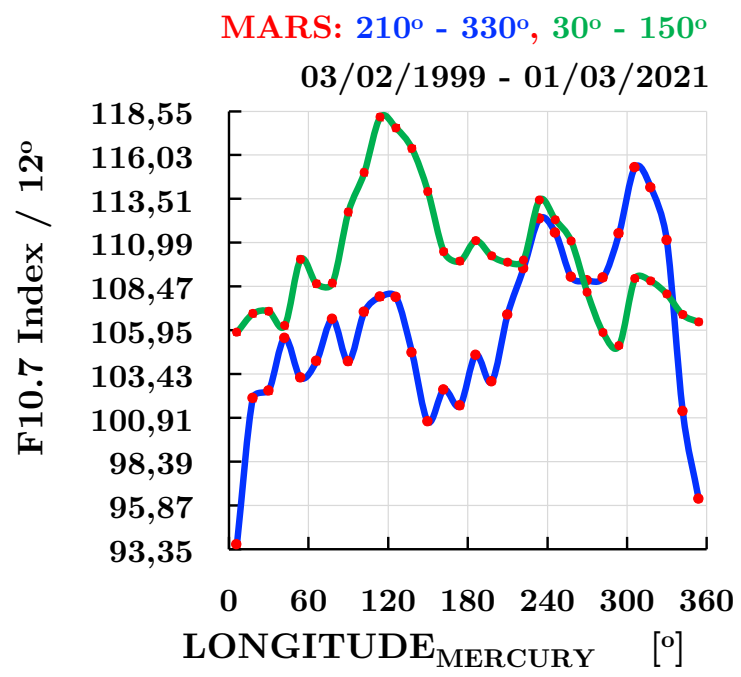

**(b)** F10.7 distribution for Mercury when Mars is between 210° to 330° and 40° to 150° for 12° bin.

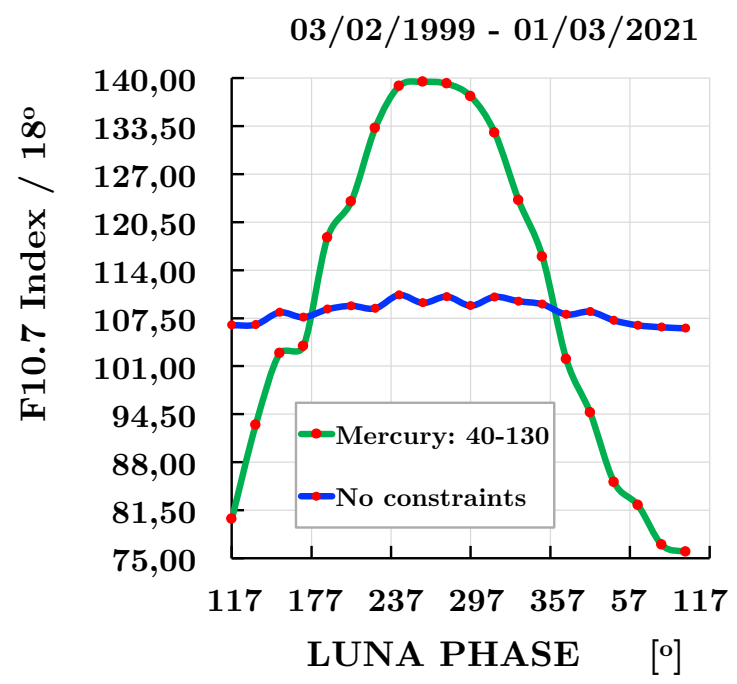

**(c)** F10.7 distribution for Moon's phase without constraints and when Mercury is between 40° to 130° for 18° bin.

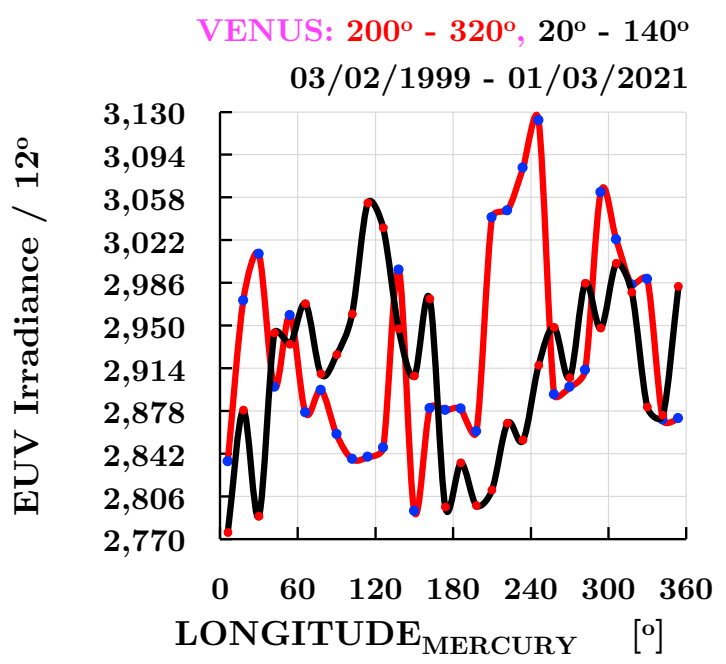

**(d)** EUV distribution for Mercury when Venus is between 200° to 320° and 20° to 140° for 12° bin.

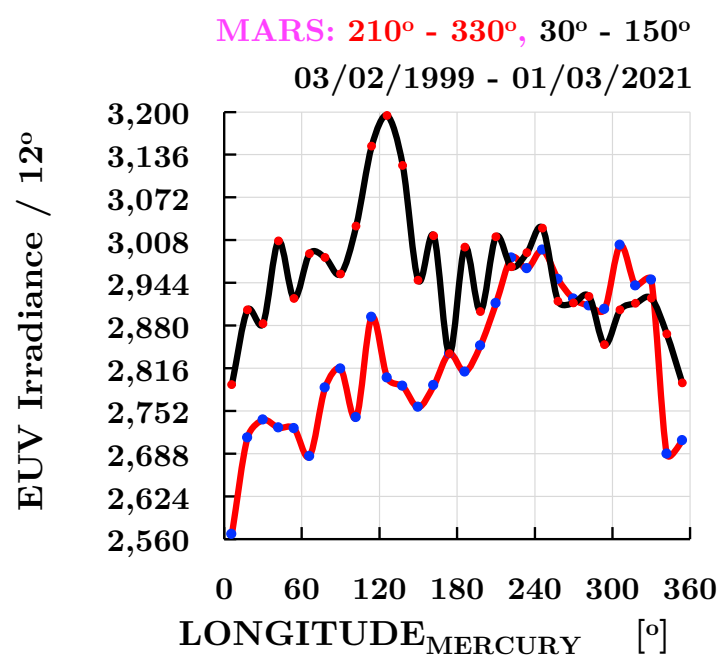

**(e)** EUV distribution for Mercury when Mars is between 210° to 330° and 40° to 150° for 12° bin.

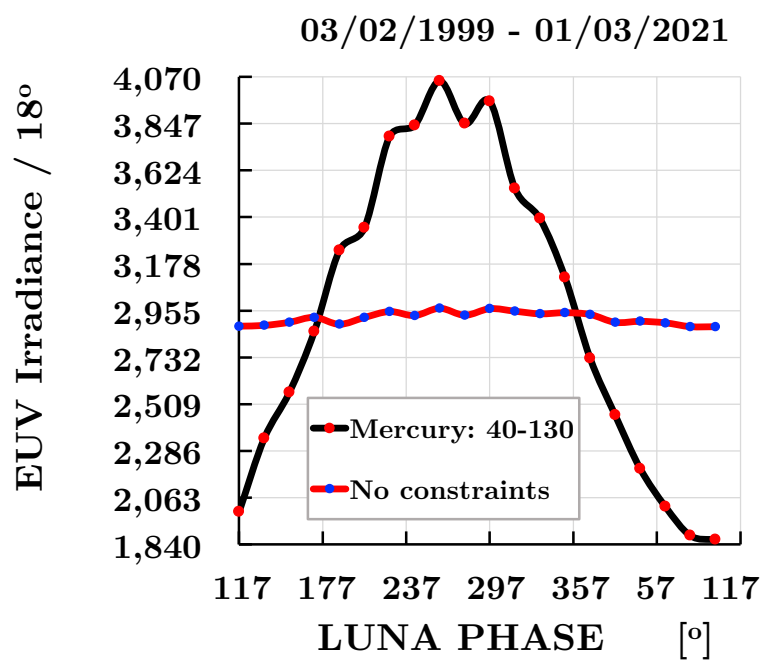

**(f)** EUV distribution for Moon's phase without constraints and when Mercury is between 40° to 130° for 18° bin.

**Figure B.24:** Comparison of distributions of F10.7 vs. EUV solar irradiance for the period 03/02/1999 - 01/03/2021.

The final test is performed in Fig. B.25 with the results of F10.7 solar index from Fig. 9.6 and 9.7 being compared for the same period with the EUV irradiance. Once more the similarity between the two cases is evident, with the correlation coefficients giving a significant high degree of correlation in all cases. For Fig. B.25a and B.25d we have ($r_{a,d} = 0.77$, $p_{a,d} = 6.33 \times 10^{-7}$), for Fig. B.25b and B.25e ($r_{b,e} = 0.91$, $p_{b,e} = 1.74 \times 10^{-12}$), whereas for Fig. B.25c and B.25f the Pearson correlation gives ($r_{c,f} = 0.96$, $p_{c,f} = 0$).

### B.5.2 Comparison with sunspots

An additional comparison is performed between F10.7 and the number of sunspots. The maximum available time range is selected for both datasets i.e. 28/11/1963 - 28/02/2021.

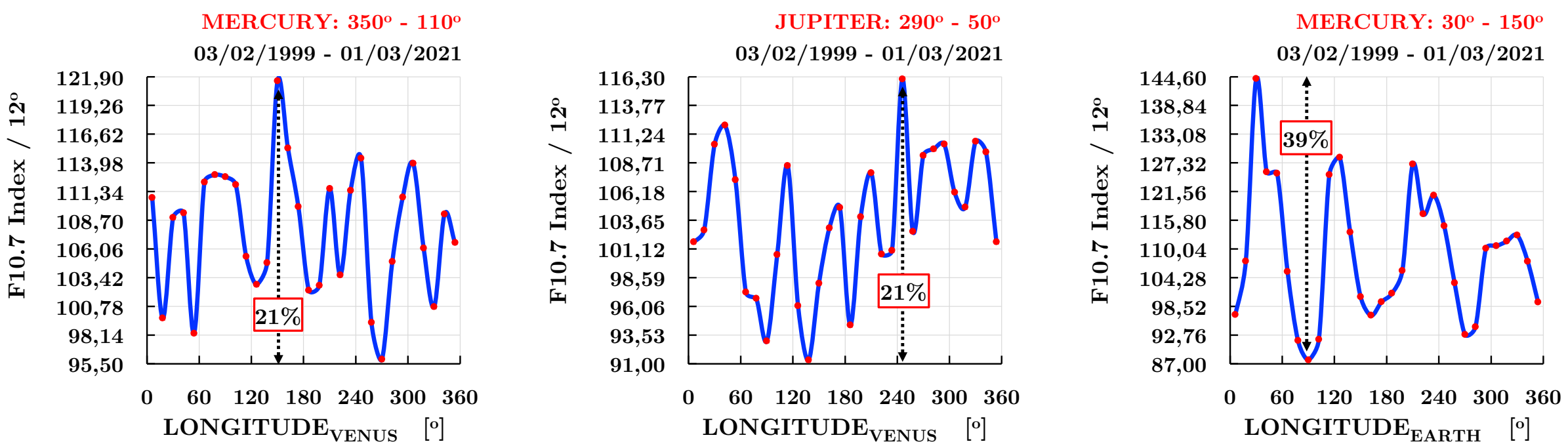

**(a)** F10.7 distribution for Venus when Mercury is between 350° to 110°.

**(b)** F10.7 distribution for Venus when Jupiter is between 290° to 50°.

**(c)** F10.7 distribution for Earth when Mercury is between 30° to 150°.

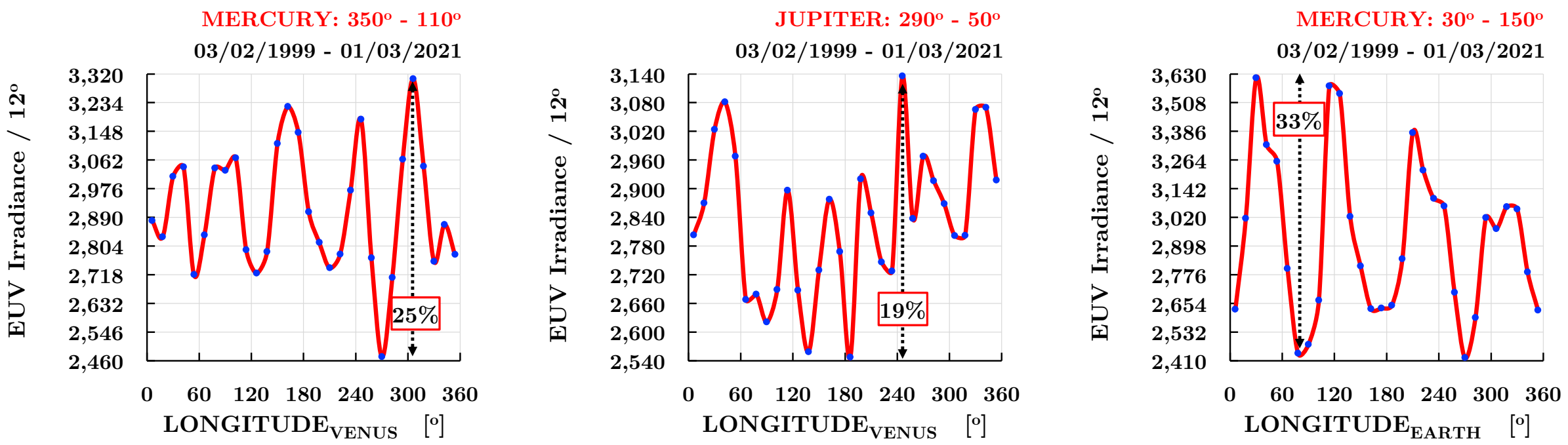

**(d)** EUV distribution for for Venus when Mercury is between 350° to 110°.

**(e)** EUV distribution for Venus when Jupiter is between 290° to 50°.

**(f)** EUV distribution for Earth when Mercury is between 30° to 150°.

**Figure B.25:** Comparison of distributions of F10.7 vs. EUV solar irradiance for the period 03/02/1999 - 01/03/2021 with 12° bin.

In Fig. B.26 the time series of the two datasets are overlapped. The calculated Pearson correlation coefficient and p-value for the observed general trend are ($r = 0.949,\ p = 0$). These show a high degree of statistically significant correlation as expected. As next the specific planetary dependencies are also individually compared.

### B.5.2.1 Single planets

In Fig. B.27 and B.28, the F10.7 single-planet distributions are compared with the corresponding spectra for the number of sunspots for 28/11/1963 - 28/02/2021. As with the general trend also all of the longitudinal distributions show a high degree of similarity between the two solar datasets pointing to the same origin of planetary relationship. This is once more verified by a statistical correlation analysis. The calculated statistically significant Pearson's correlation coefficients and p-values are ($r_{a,e} = 0.67,\ p_{a,e} = 4.59 \times 10^{-5}$), ($r_{b,f} = 0.93,\ p_{b,f} = 2.59 \times 10^{-9}$), ($r_{c,g} = 0.62,\ p_{c,g} = 2.33 \times 10^{-4}$) and ($r_{d,h} = 0.93,\ p_{d,h} = 6 \times 10^{-7}$) for the subfigures a to h in Fig. B.27 and ($r_{a,c} = 1,\ p_{a,c} = 0$), ($r_{b,d} = 0.99,\ p_{b,d} = 0$) for the subfigures of Jupiter and Saturn respectively in Fig. B.28.

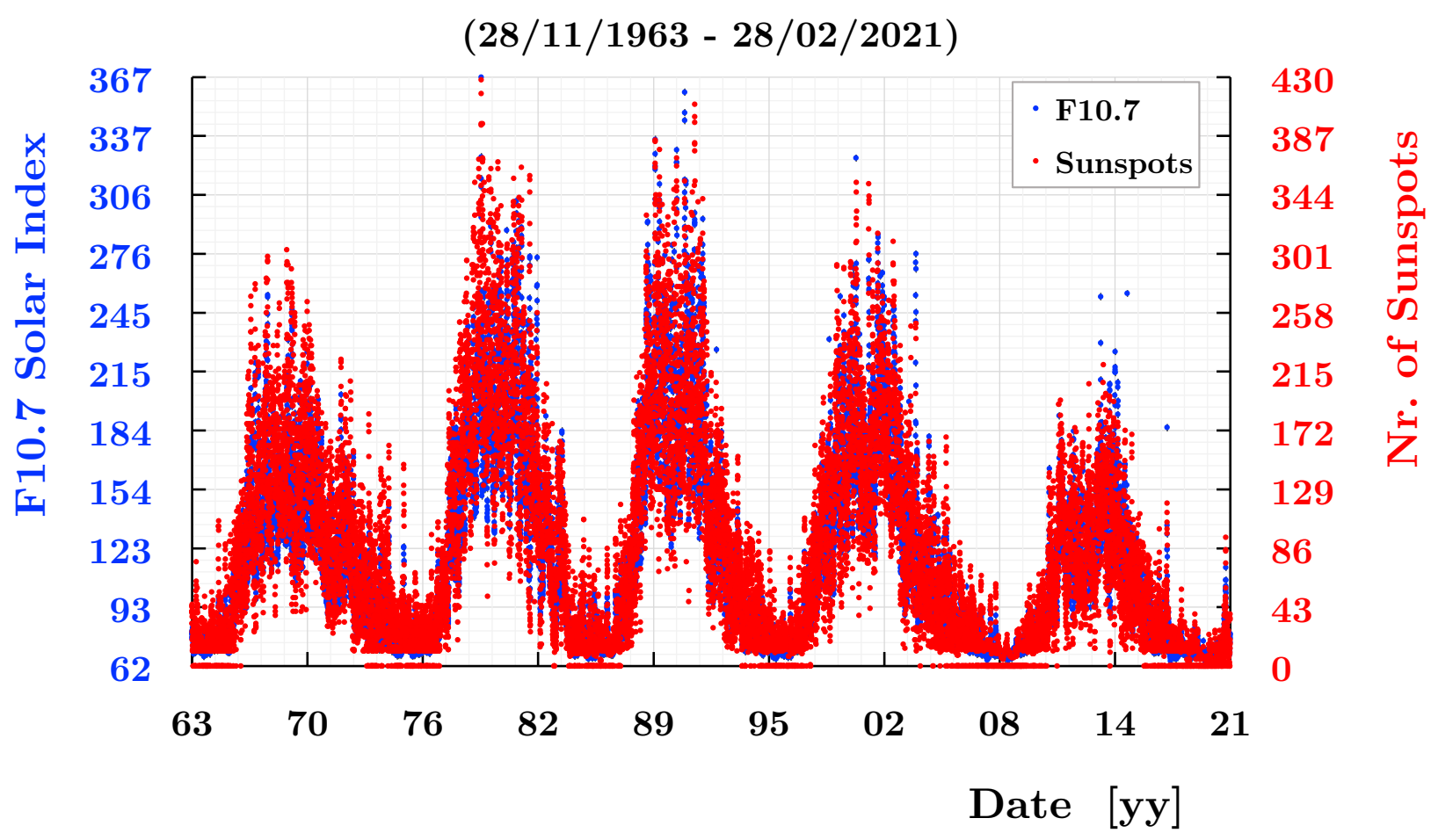

**Figure B.26:** Daily F10.7 solar index and Nr. of sunspots. The period used is 28/11/1963 - 28/02/2021.

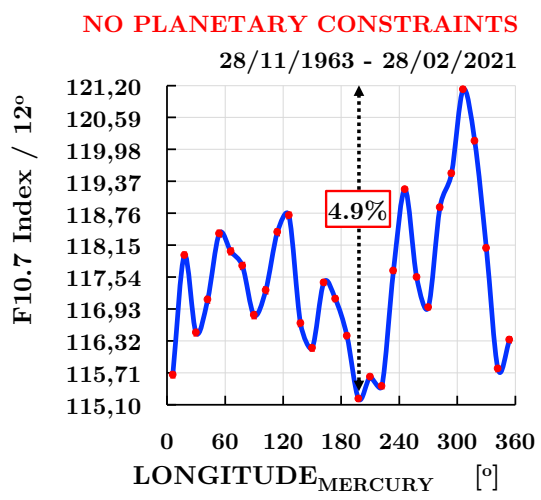

**(a)** F10.7 distribution for Mercury with 12° bin.

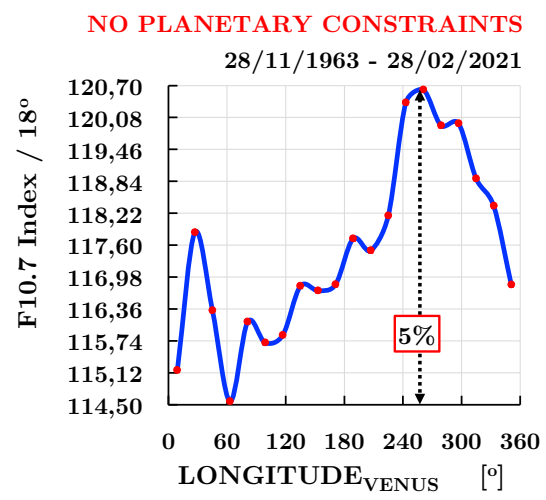

**(b)** F10.7 distribution for Venus with 18° bin.

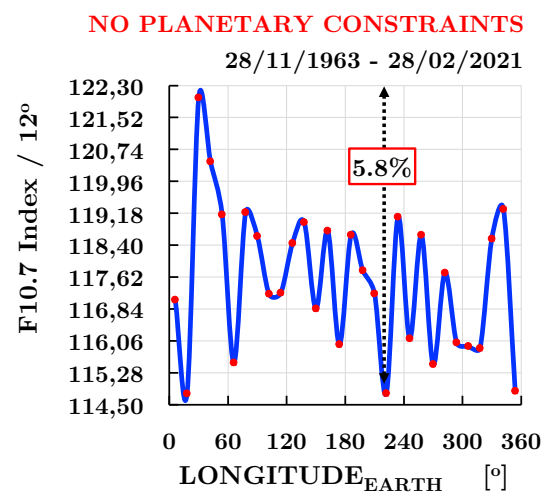

**(c)** F10.7 distribution for Earth with12° bin.

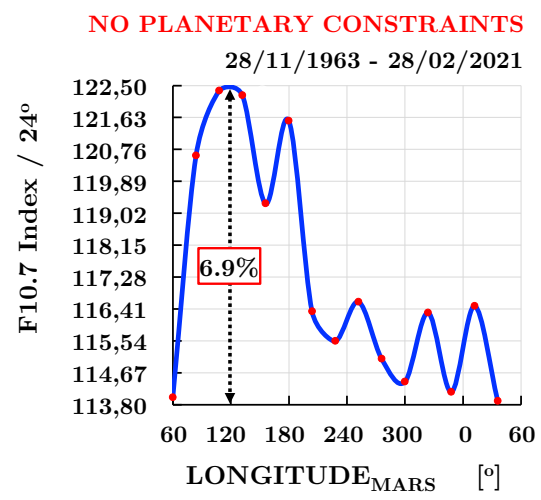

**(d)** F10.7 distribution for Mars with 24° bin.

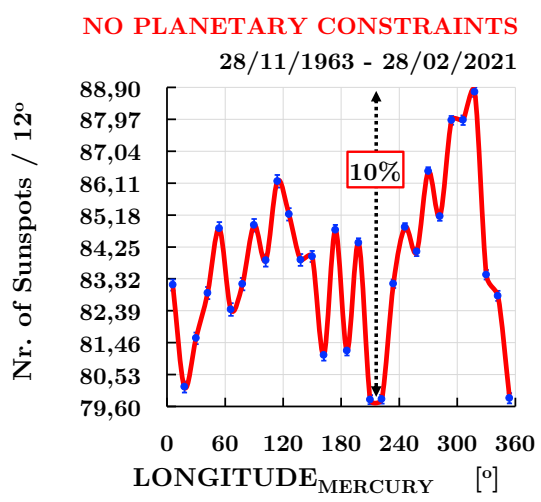

**(e)** Nr. of sunspots distribution for Mercury with 12° bin.

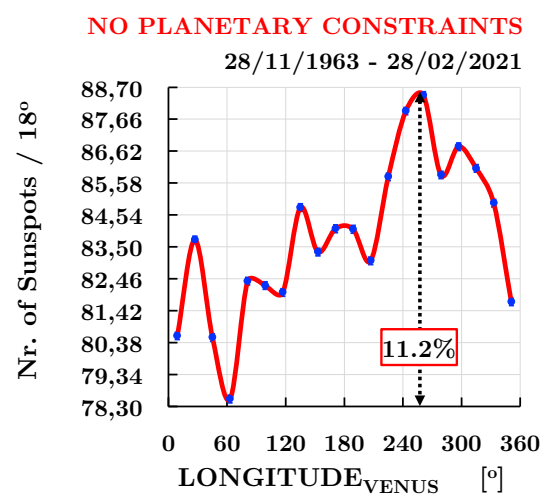

**(f)** Nr. of sunspots distribution for Venus with 18° bin.

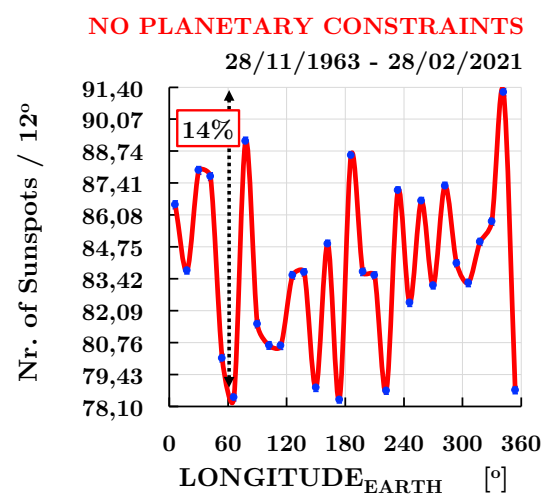

**(g)** Nr. of sunspots distribution for Earth with 12° bin.

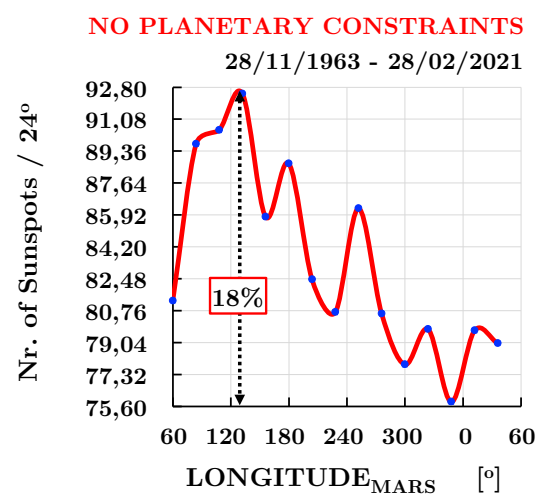

**(h)** Nr. of sunspots distribution for Mars with 24° bin.

**Figure B.27:** Comparison of distributions of F10.7 vs. Nr. of sunspots for the period 28/11/1963 - 28/02/2021.

### B.5.2.2 Combining planets

Similarly, a comparison is performed for the plots containing a combination of planets. In Fig. B.29 the distributions from Fig. 9.4c, 9.5c and 9.8c are reevaluated and compared with the corresponding distributions for the number of sunspots. In all three cases, we have a statistically significant positive linear correlation for the two manifestations of solar activity. The exact correlation coefficients are ($r_{a,d} = 0.90$, $p_{a,d} = 1.14 \times 10^{-11}$), and ($r_{a,d} = 0.77$,

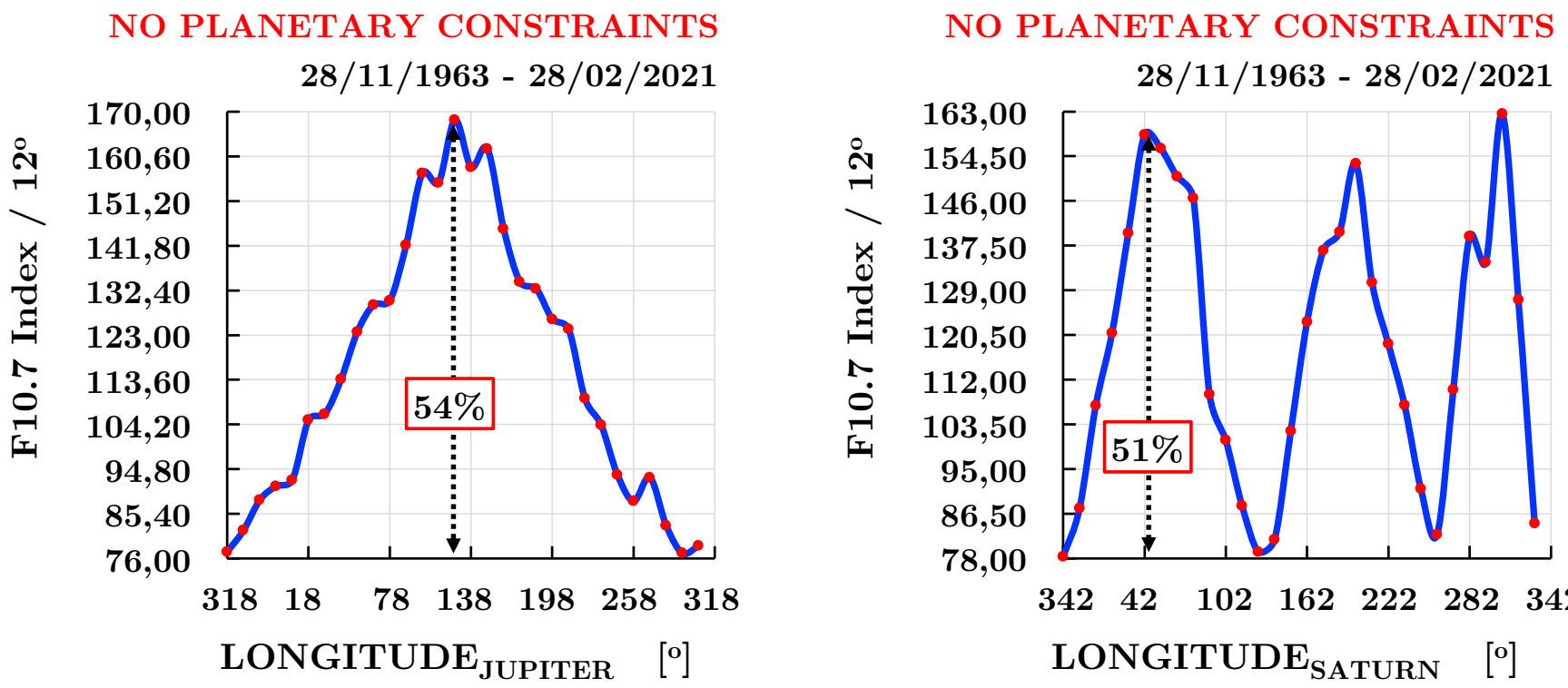

**(a)** F10.7 distribution for Jupiter. **(b)** F10.7 distribution for Saturn.

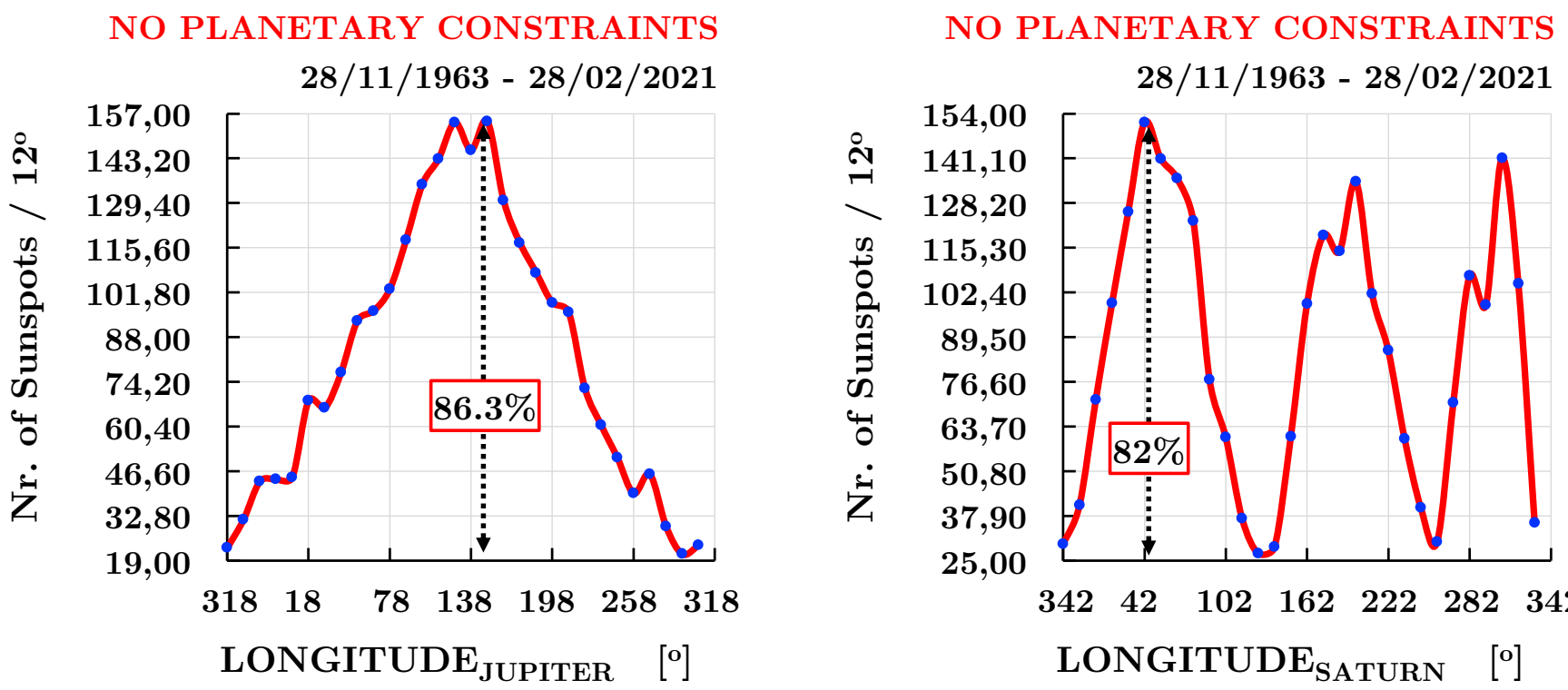

**(c)** Nr. of sunspots distribution for Jupiter. **(d)** Nr. of sunspots distribution for Saturn.

**Figure B.28:** Comparison of distributions of F10.7 vs. Nr. of sunspots for Jupiter and Saturn with 12° bin for the period 28/11/1963 - 28/02/2021.

$p_{a,d} = 7.85 \times 10^{-7}$) for the two constraints on Venus in Fig. B.29a and B.29d. Accordingly, for the two constraints on Mars in Fig. B.29b and B.29e we have ($r_{b,e} = 0.91$, $p_{b,e} = 2.41 \times 10^{-12}$) and ($r_{b,e} = 0.57$, $p_{b,e} = 8.98 \times 10^{-4}$) accordingly. Finally, for the constraint on Mercury's position in Fig. B.29c and B.29f we get ($r_{c,f} = 0.99$, $p_{c,f} = 4.44 \times 10^{-16}$).

The last test is performed in Fig. B.30 with the F10.7 longitudinal distributions from Fig. 9.6 and 9.7 being compared for the same period with the number of sunspots. The similarity between the two cases is apparent, with the correlation analysis indicating a statistically significant high degree of correlation in all cases. For Fig. B.30a and B.30d we have ($r_{a,d} = 0.94$, $p_{a,d} = 1.11 \times 10^{-14}$), for Fig. B.30b and B.30e ($r_{b,e} = 0.95$, $p_{b,e} = 6.66 \times 10^{-16}$), whereas for Fig. B.30c and B.30f we get ($r_{c,f} = 0.92$, $p_{c,f} = 8.19 \times 10^{-13}$). Therefore, for all these cases we have a statistically significant high degree of linear correlation.

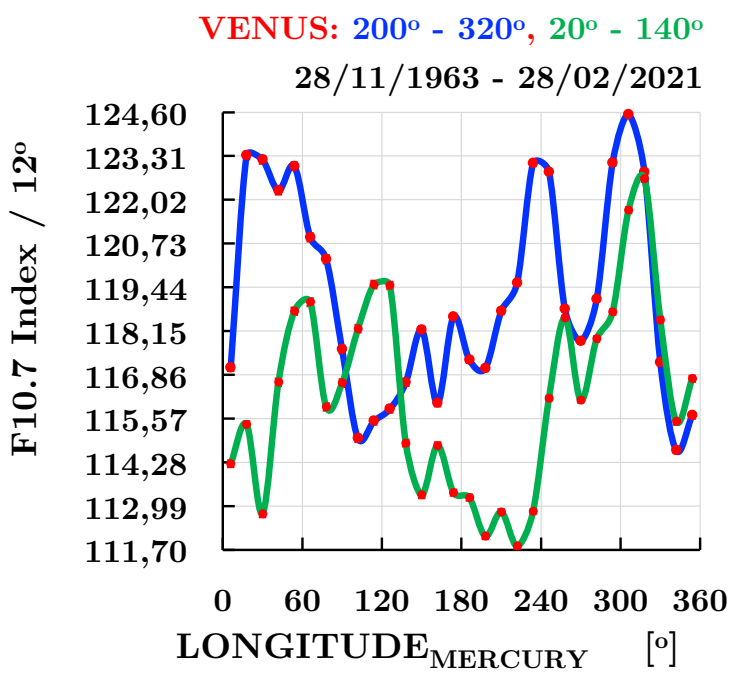

**(a)** F10.7 distribution for Mercury when Venus is between 200° to 320° and 20° to 140° for 12° bin.

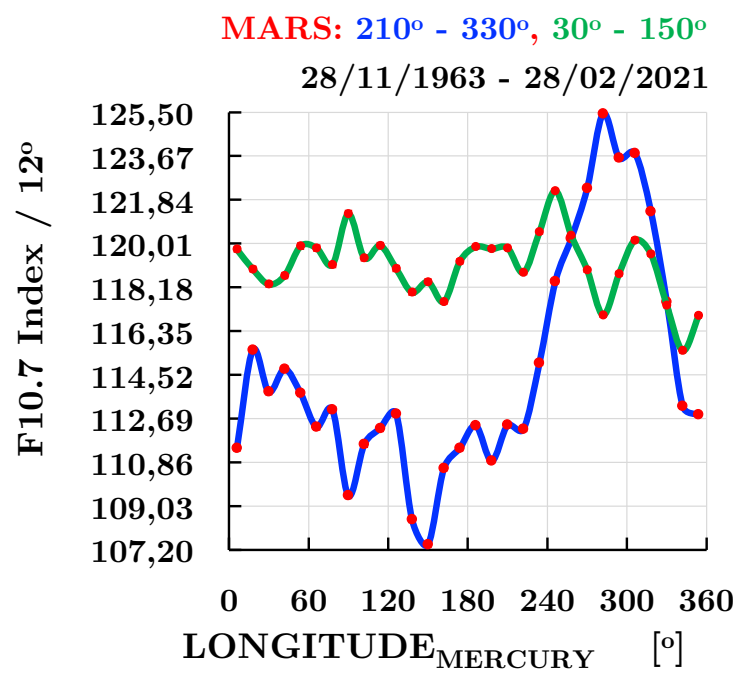

**(b)** F10.7 distribution for Mercury when Mars is between 210° to 330° and 40° to 150° for 12° bin.

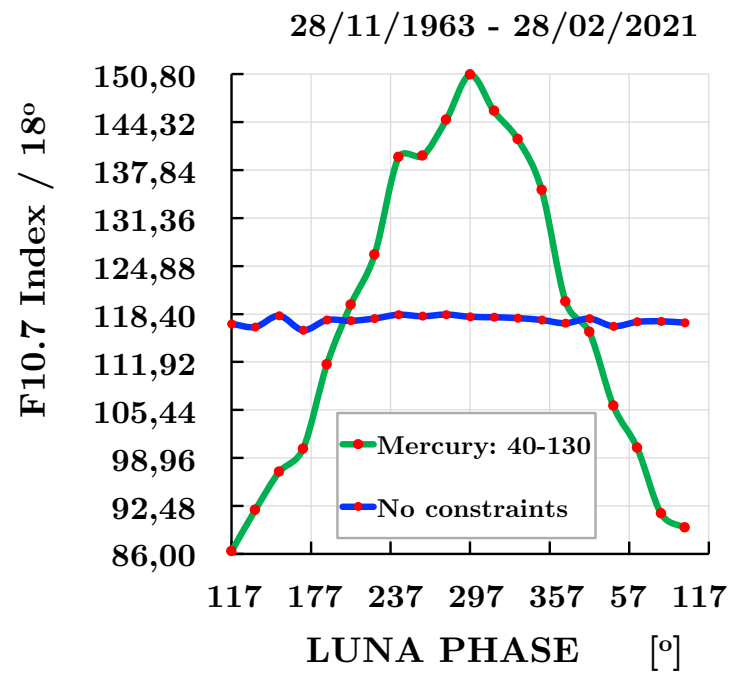

**(c)** F10.7 distribution for Moon's phase without constraints and when Mercury is between 40° to 130° for 18° bin.

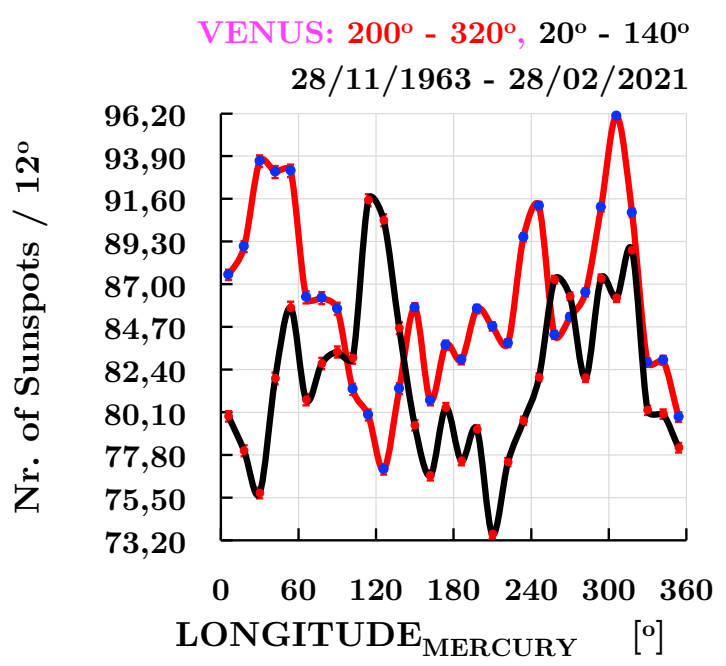

**(d)** Nr. of sunspots distribution for Mercury when Venus is between 200° to 320° and 20° to 140° for 12° bin.

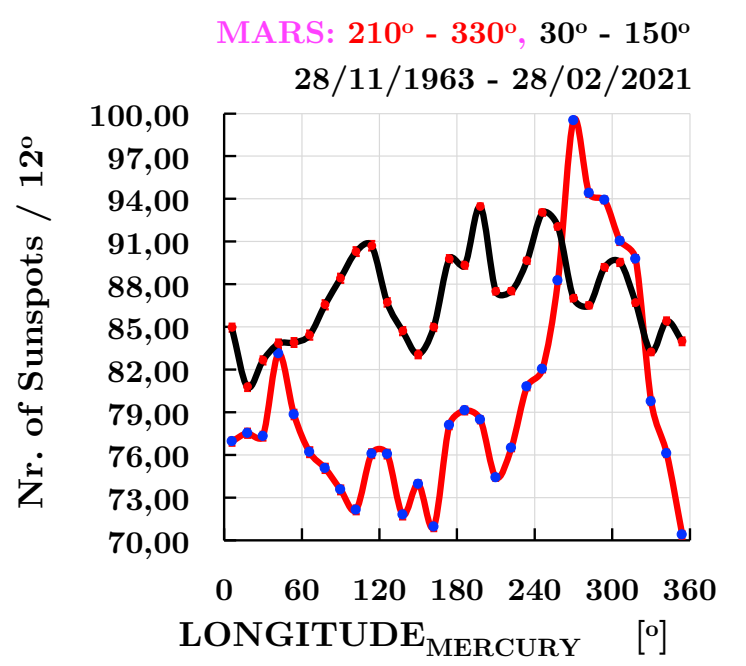

**(e)** Nr. of sunspots distribution for Mercury when Mars is between 210° to 330° and 40° to 150° for 12° bin.

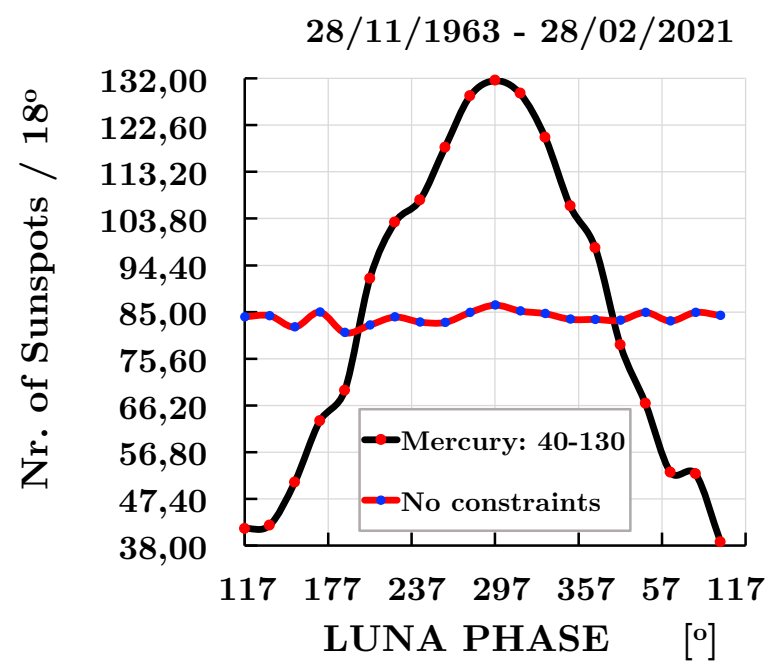

**(f)** Nr. of sunspots distribution for Moon's phase without constraints and when Mercury is between 40° to 130° for 18° bin.

**Figure B.29:** Comparison of distributions of F10.7 vs. Nr. of sunspots for the period 28/11/1963 - 28/02/2021.

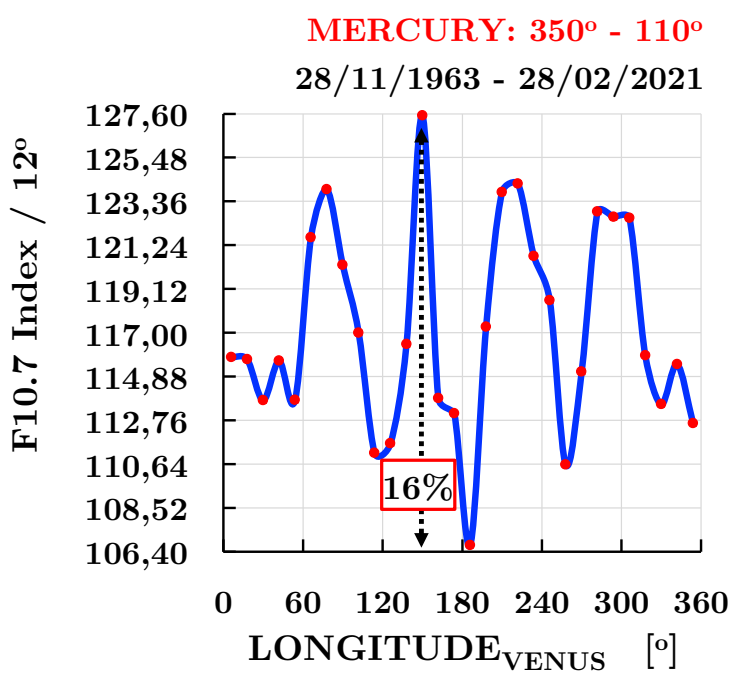

(a) F10.7 distribution for Venus when Mercury is between 350° to 110°.

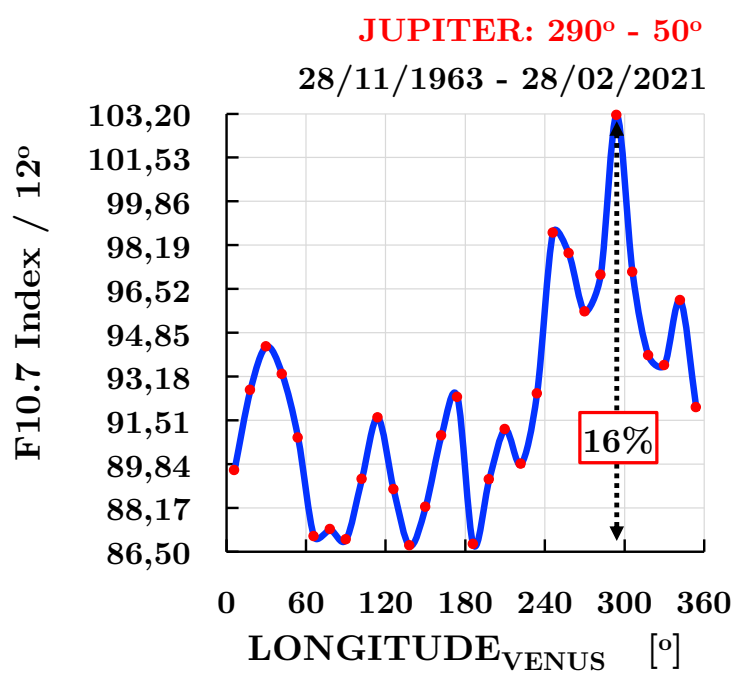

(b) F10.7 distribution for Venus when Jupiter is between 290° to 50°.

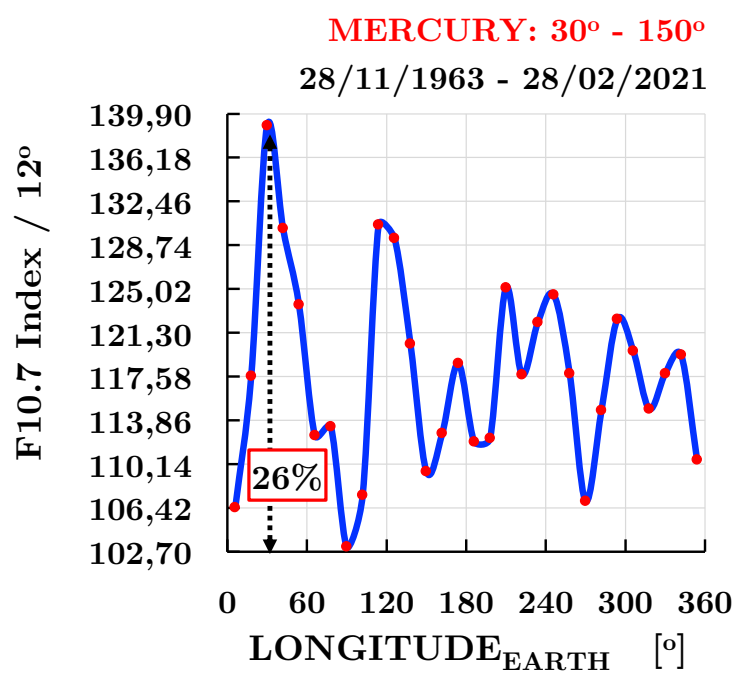

(c) F10.7 distribution for Earth when Mercury is between 30° to 150°.

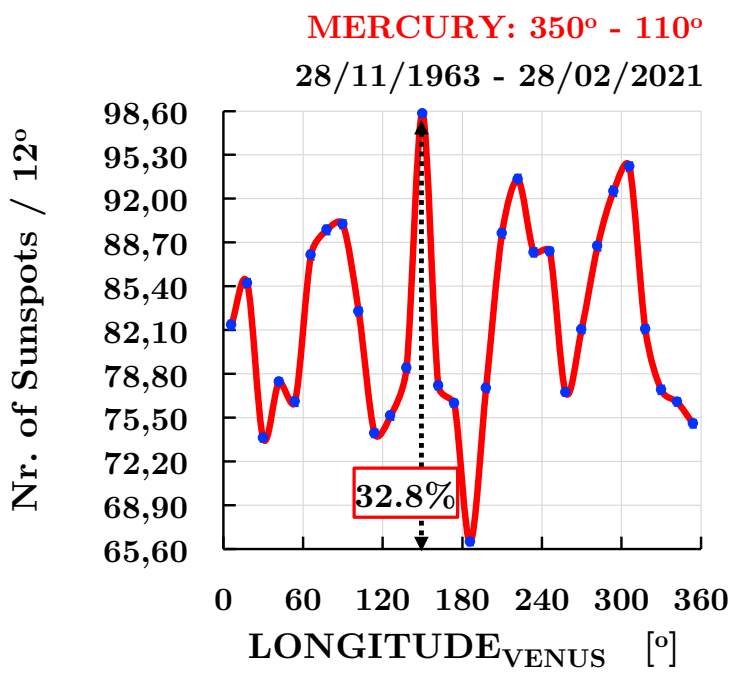

(d) Nr. of sunspots distribution for for Venus when Mercury is between 350° to 110°.

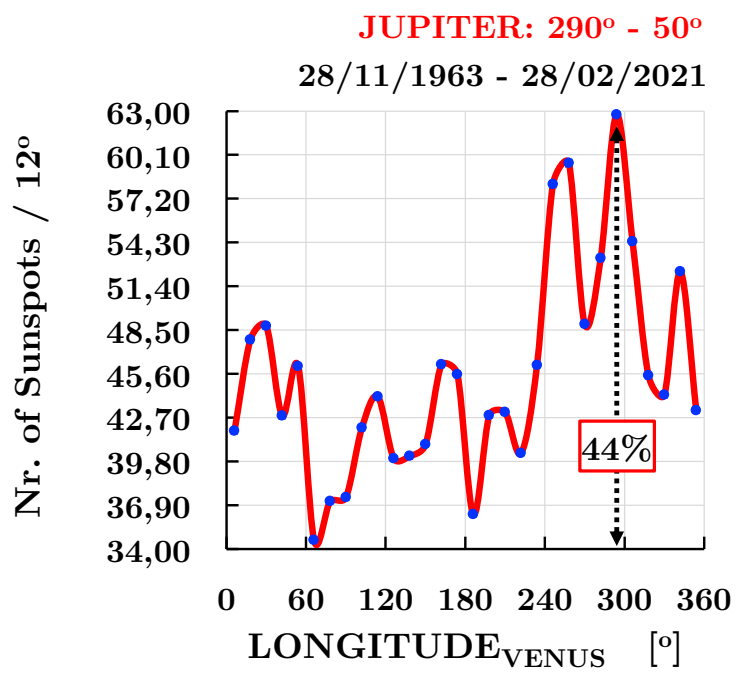

(e) Nr. of sunspots distribution for Venus when Jupiter is between 290° to 50°.

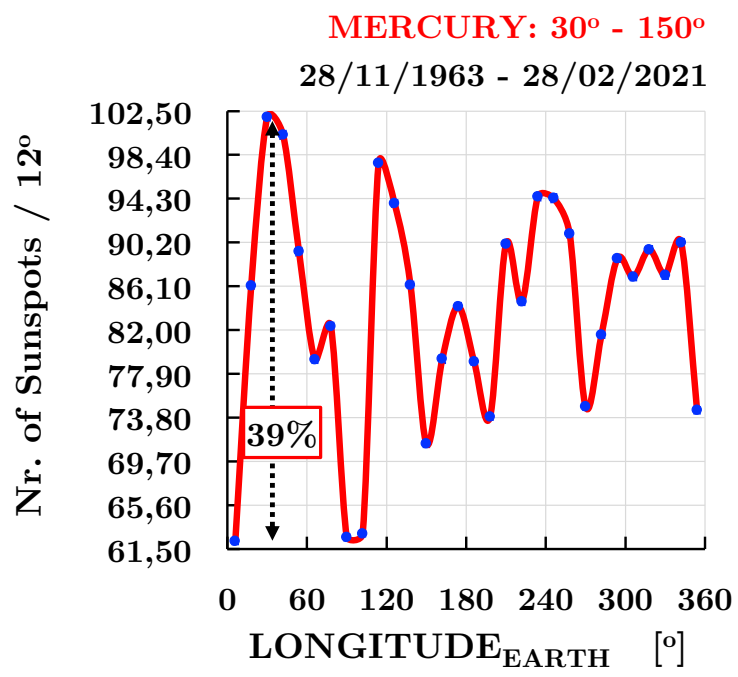

(f) Nr. of sunspots distribution for Earth when Mercury is between 30° to 150°.

**Figure B.30:** Comparison of distributions of F10.7 vs. Nr. of sunspots for the period 28/11/1963 - 28/02/2021 with 12° bin.

# B.6 Solar radius

## B.6.1 Data curation verification

An independent cross-check of the data treatment procedure performed in Sect. 10.2.2 is made with other available daily datasets which show planetary relationship. This can strengthen the credibility of the aforementioned simulation including the linear interpolation procedure in arriving from 72 d cadence to daily values. For this purpose, the initial daily data were grouped in 72 d bins and then a linear interpolation between these values has been performed the same way as with the solar radius data. The analysis with two independent datasets, as we will see, provided the expected behaviour, verifying the reliability of the applied simulation procedure.

### B.6.1.1 M-flares

The first check has been performed using the M-flares. The first step was to group the daily M-flare data ranging from 01/09/1975 to 12/03/2021 into bins of 72 d. This way 231 values were created out of the original 16624 daily values. The two datasets before and after this treatment are presented in Fig. B.31a and B.31b respectively. The next step is to perform a linear interpolation between the 72 d-binned values. This is the same procedure that was performed with the solar radius data. The result is shown in Fig. B.31c.

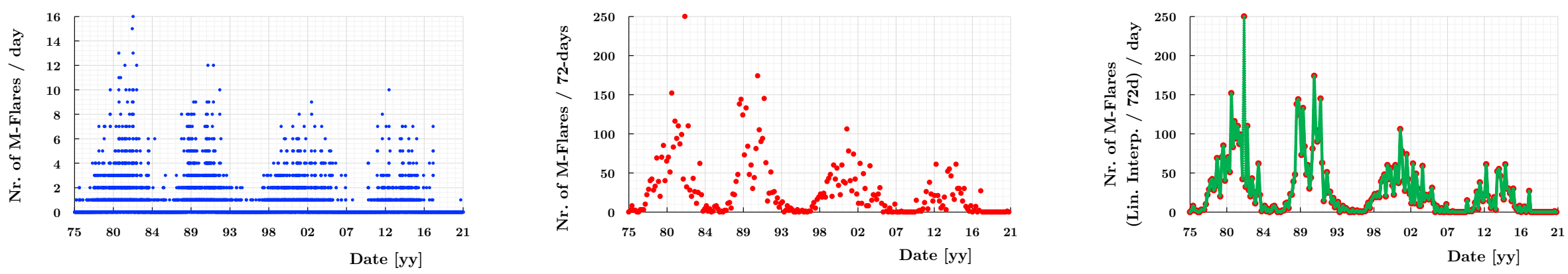

**(a)** Original daily values. **(b)** Grouped 72 d values. **(c)** Linearly interpolated values from Fig. B.31b.

**Figure B.31:** The time evolution of the M-flare data before and after treatment.

Next, the planetary distribution for Venus, without any constraints, is compared between the original daily M-flare data and the data derived from the aforementioned curation. In Fig. B.32a and B.32c the longitudinal distribution of Venus for the raw daily data of the number of M-flares is plotted for a bin of 18° and 30° accordingly. In these cases, we can see a peak at around 260° with an amplitude of 52.65% and 35.83% respectively. These spectra are compared with the spectra on the right in Fig. B.32b and B.32d which correspond to the curated data. We see that the two narrow peaks after the treatment show up as a wide peak in Venus with a much smaller amplitude of about 17.94% and 17.30% for the two bin sizes accordingly. This is exactly the behaviour expected from this kind of treatment.

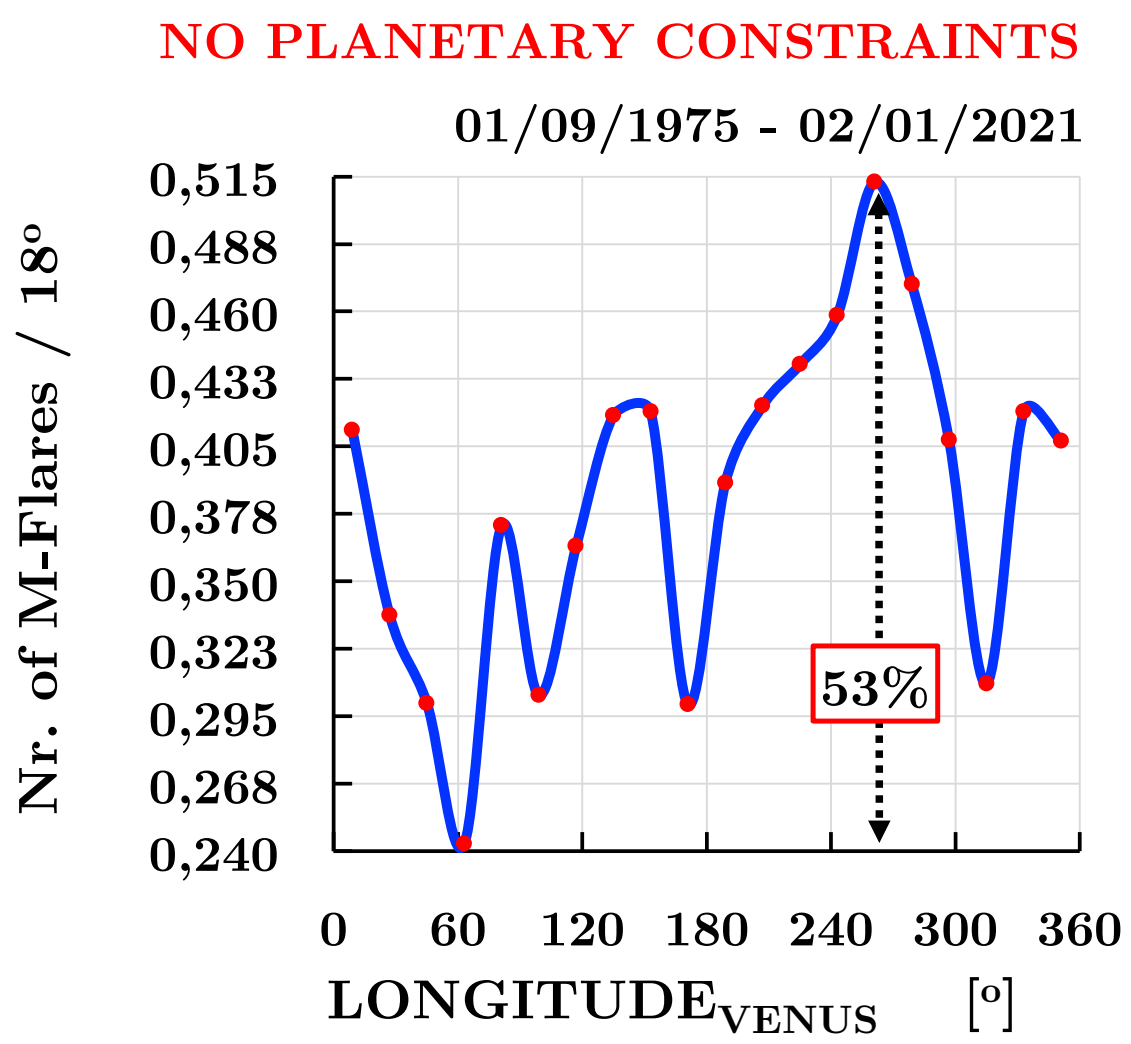

(a) Original data for a bin of 18°.

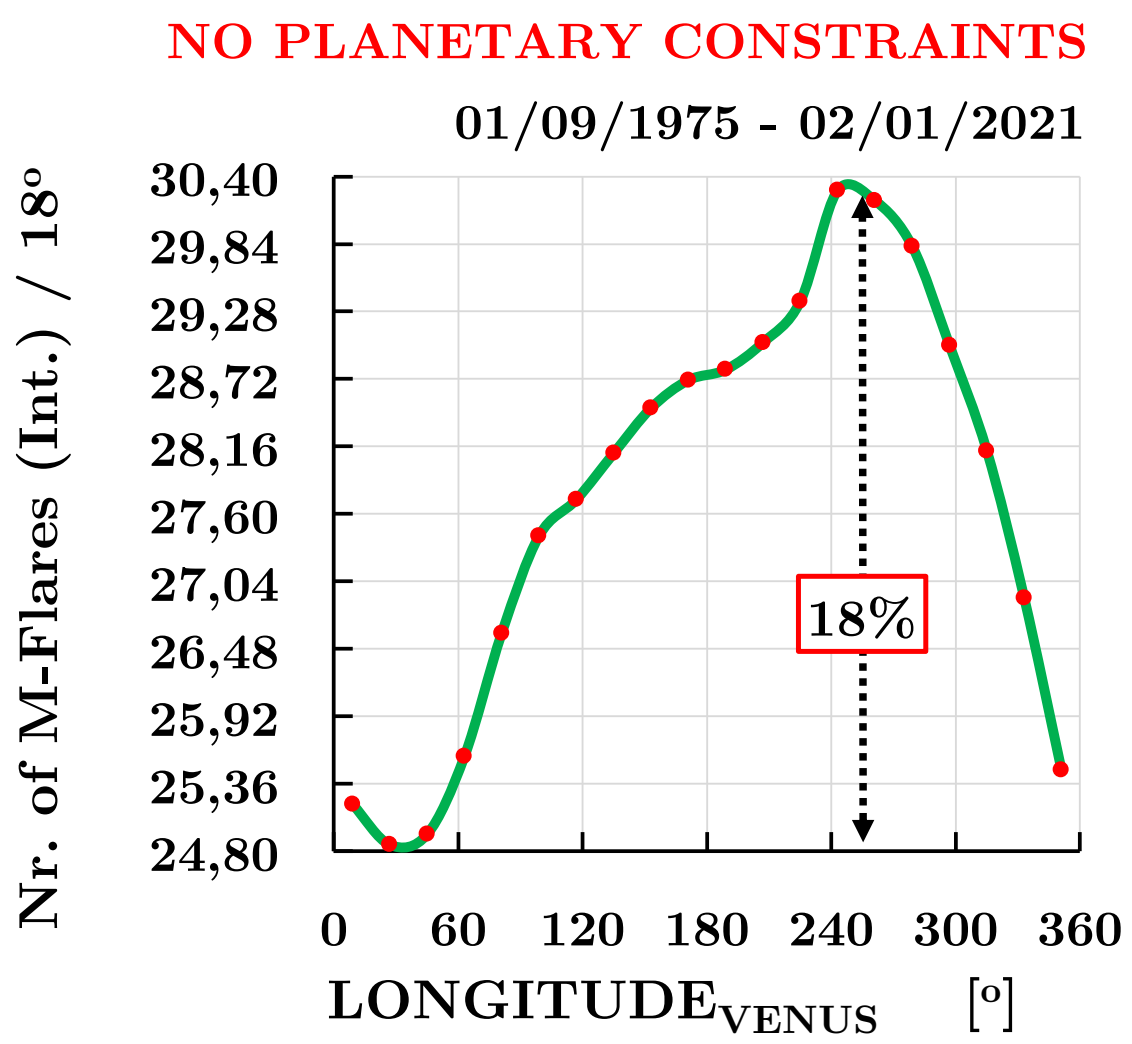

(b) Interpolated data for a bin of 18°.

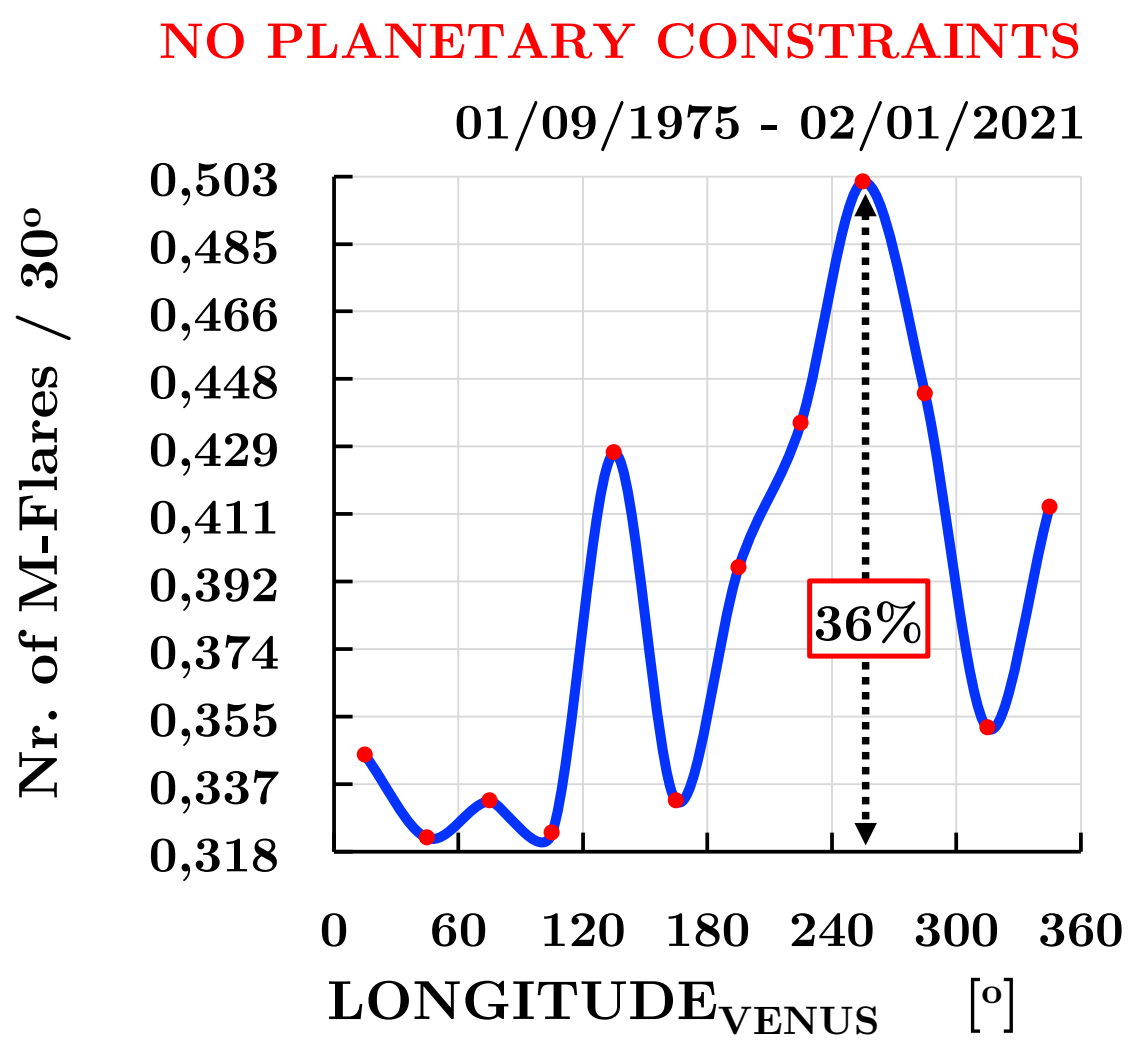

(c) Original data for a bin of 30°.

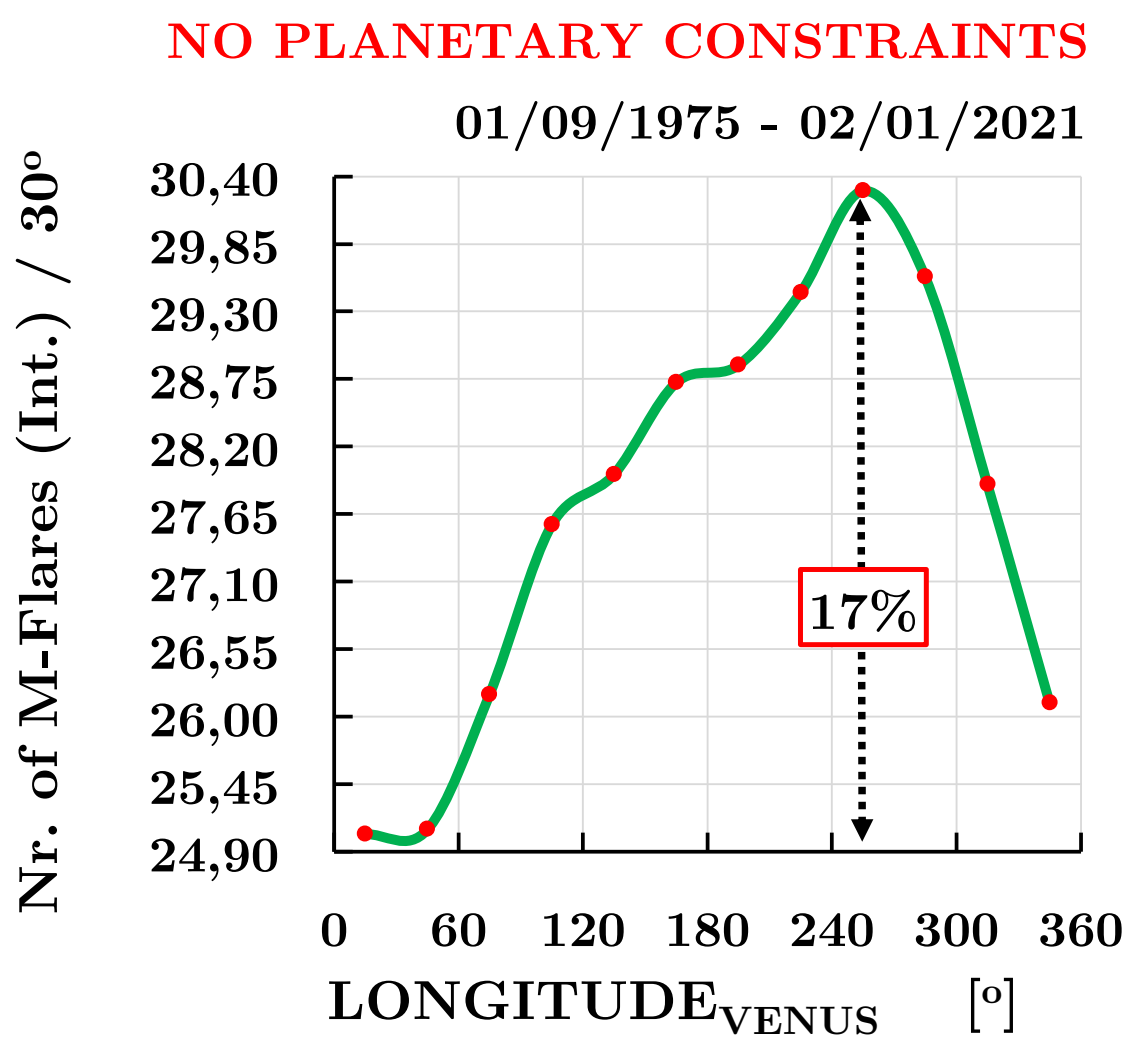

(d) Interpolated data for a bin of 30°.

**Figure B.32:** Venus heliocentric longitude distribution for original data of Nr. of M-flares from Fig. B.31a (left) and linearly interpolated data from 72 d-binned values from Fig. B.31c (right).

Moreover, the planetary distribution for Earth, without any constraints, is also compared for the original M-flare data and the curated data in Fig. B.33. The maximum - minimum difference in Fig. B.33a is 45.35% whereas in Fig. B.33b it reduces to 17.15%. We also note that the whole spectrum is shifted towards lower longitudes by about 36 d which is expected due to the grouping of the original data in 72 d.

#### B.6.1.2 TEC

The exact same procedure has been performed for the TEC data of the Earth's ionosphere. Therefore, in Fig. B.34a we see the original daily values spanning from 01/01/1995 to 09/12/2012, whereas in Fig. B.34b these values have been grouped in bins of 72 d. Then, in

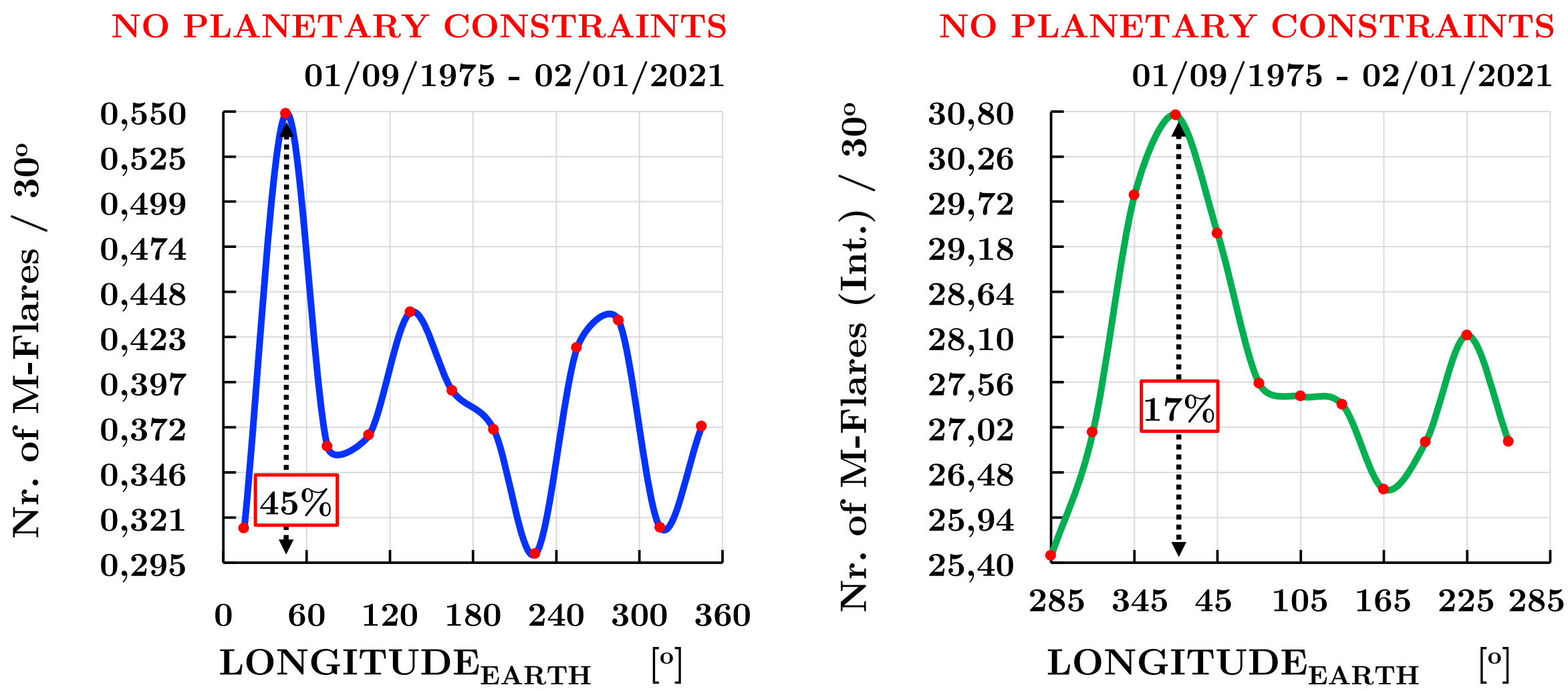

**(a)** Original daily data **(b)** Interpolated data.

**Figure B.33:** Earth's heliocentric longitude distribution for original data of M-flares from Fig. B.31a and linearly interpolated data from 72 d-binned values from Fig. B.31c for a bin of 30°.

Fig. B.34c the data are being transformed back into daily values by linearly interpolating between the grouped 72 d values.

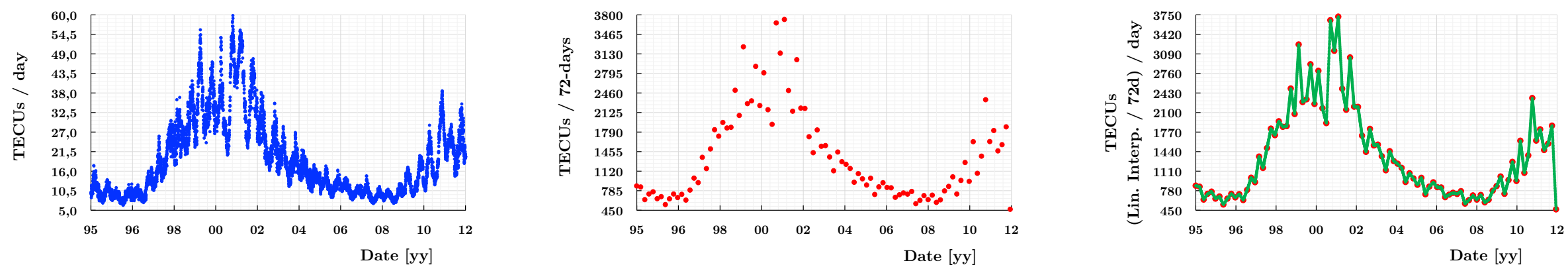

**(a)** Original daily values. **(b)** Grouped 72 d values. **(c)** Linearly interpolated values from Fig. B.34b

**Figure B.34:** The time evolution of the TEC data from Earth's atmosphere before and after treatment.

Fig. B.35 shows the daily total electron content as a function of the heliocentric longitude of Venus, with no planetary constraint applied, for the original daily TEC data and the linearly interpolated ones. Again, we see that the peak in Fig. B.35a around 290° has an amplitude of 7.17% for the original daily TEC data, which then in Fig. B.35b transforms into a wider peak with a smaller amplitude of about 4% for the treated data. Additionally, the peak is shifted to the left by about 36 d due to the 72 d grouping. Finally, as expected, the even narrower three small peaks on the left in Fig. B.35a become a single wide peak in Fig. B.35b.

Finally, in Fig. B.36 an additional TEC distribution is shown in the Earth's reference frame instead of Venus'. In this case, the shape of two peaks, in the original data in Fig. B.36a, with the biggest one having an amplitude of about 38.29%, is preserved after the data treatment in Fig. B.36b however, with the amplitude being reduced to 24.75%. Once more, there is a shift of the whole spectrum in the simulated data compared with the original daily data by about 36 d to the left due to the grouping of the former. This is also apparent from the x-axis values starting from 297° in the raw data and 261° in the treated data.

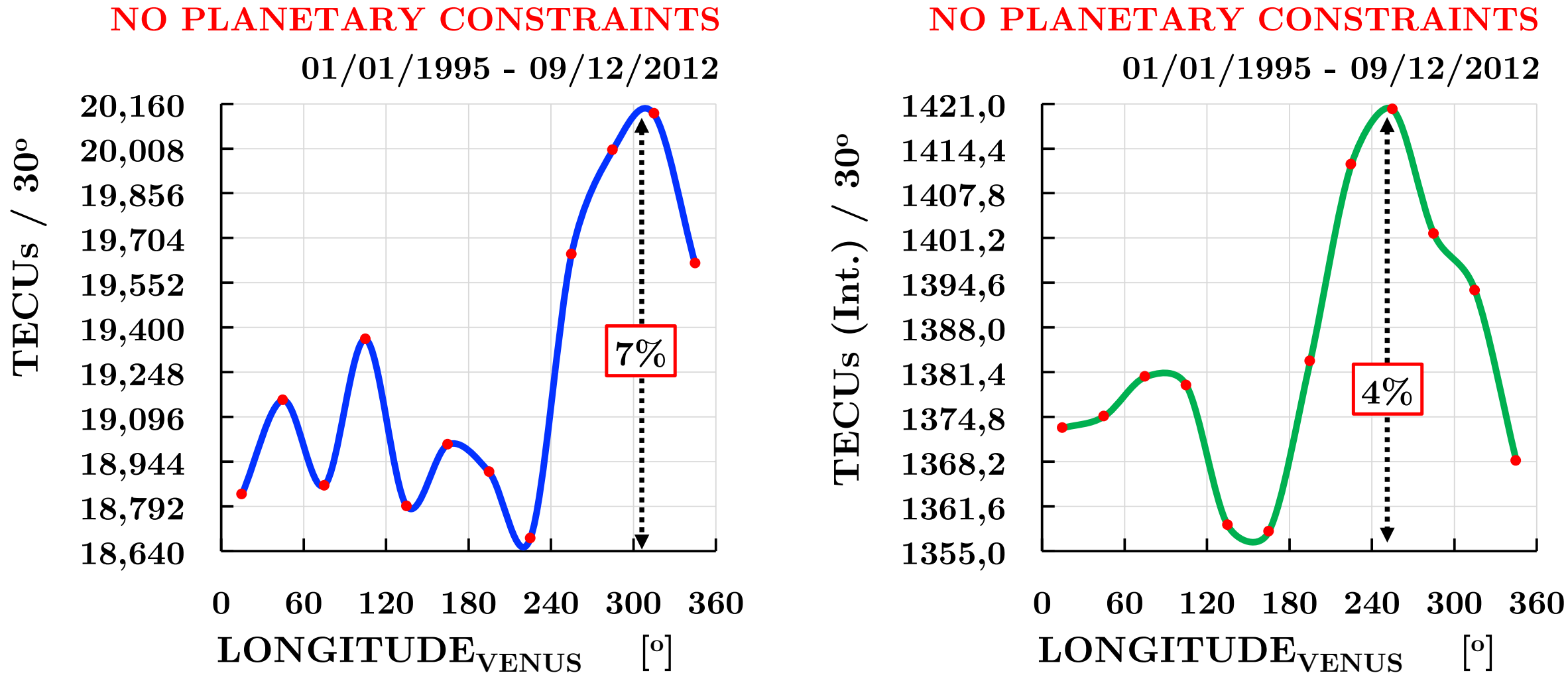

(a) Original daily data

(b) Interpolated data.

**Figure B.35:** Venus heliocentric longitude distribution for original data of TEC from Fig. B.34a and linearly interpolated data from 72 d-binned values from Fig. B.34c for a bin of 30°.

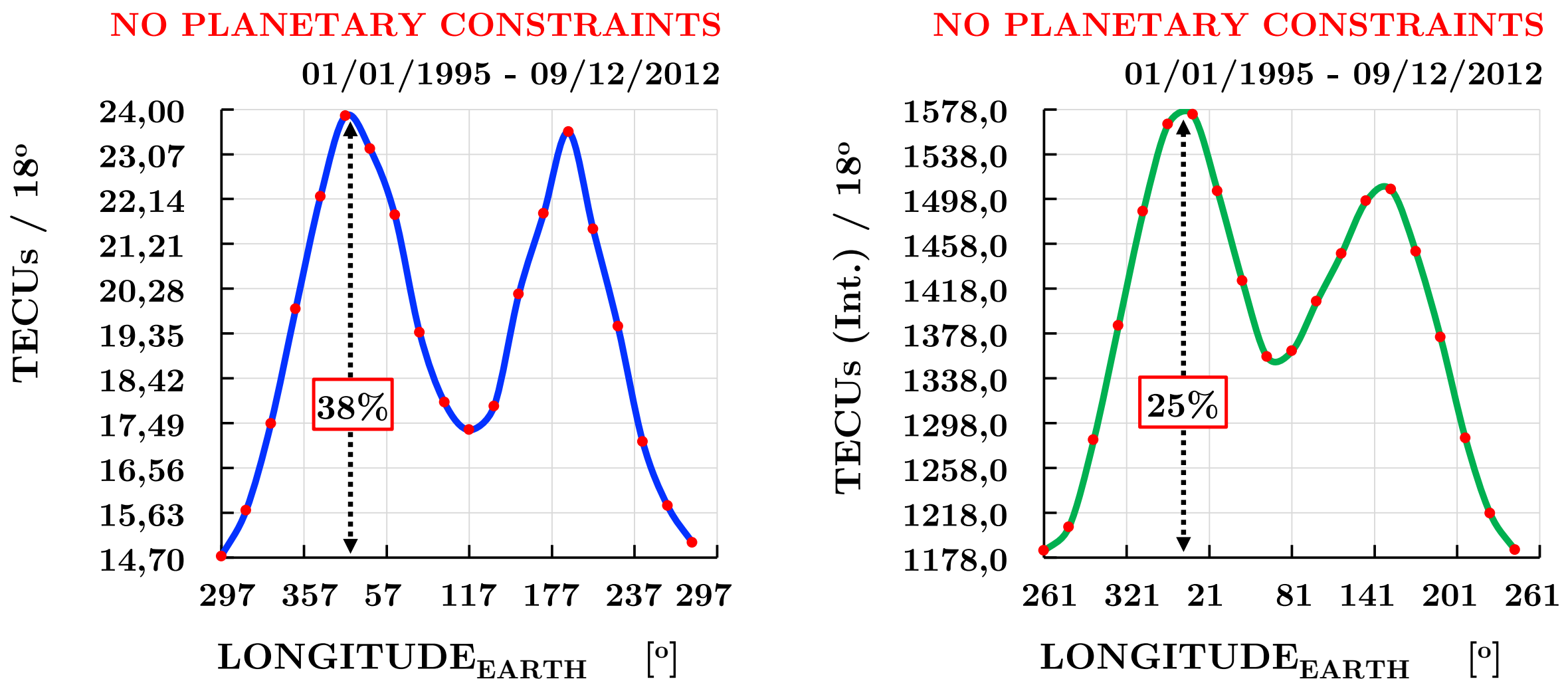

(a) Original daily data

(b) Interpolated data.

**Figure B.36:** Earth's heliocentric longitude distribution for original data of TEC from Fig. B.34a and linearly interpolated data from 72 d-binned values from Fig. B.34c for a bin of 18°.

## B.6.2 Comparison with F10.7

### B.6.2.1 Data treatment

An additional comparison of the solar radius data is performed with the F10.7 solar proxy. The matching time series of these datasets are overlaid in Fig. B.37a. In order of a direct comparison to take place, the F10.7 data are arranged with a cadence of 72 d and then linearly interpolated to derive again the daily values. The time series of the interpolated solar radius and F10.7 data are shown in Fig. B.37b.

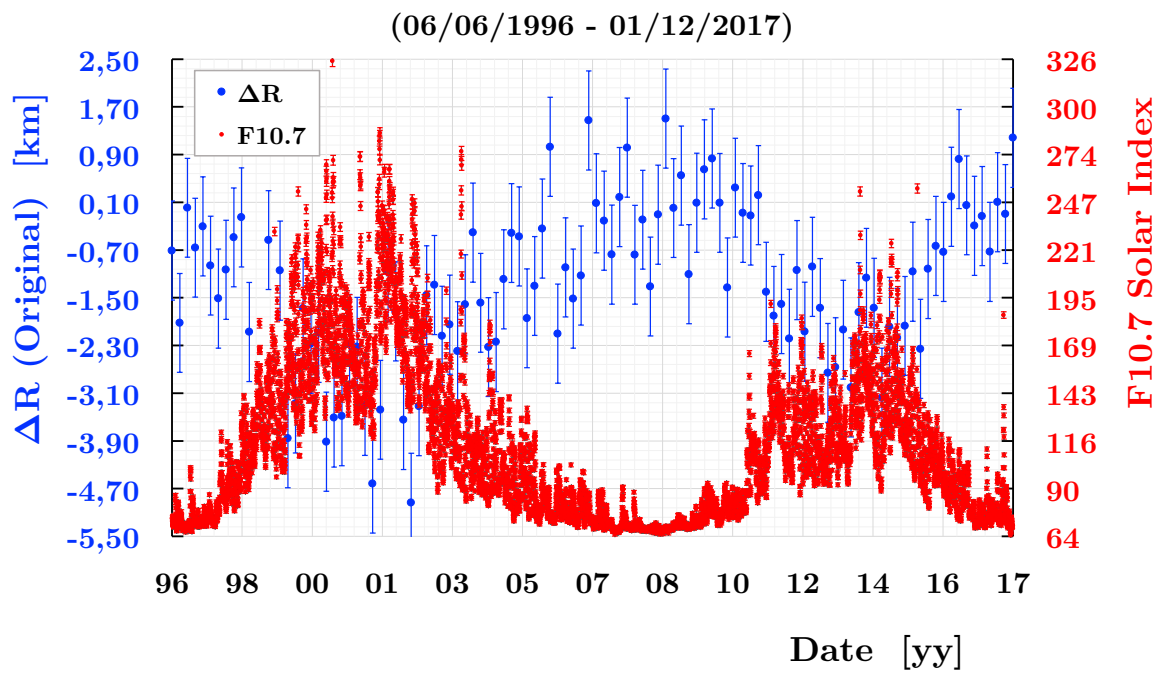

(a) Original data.

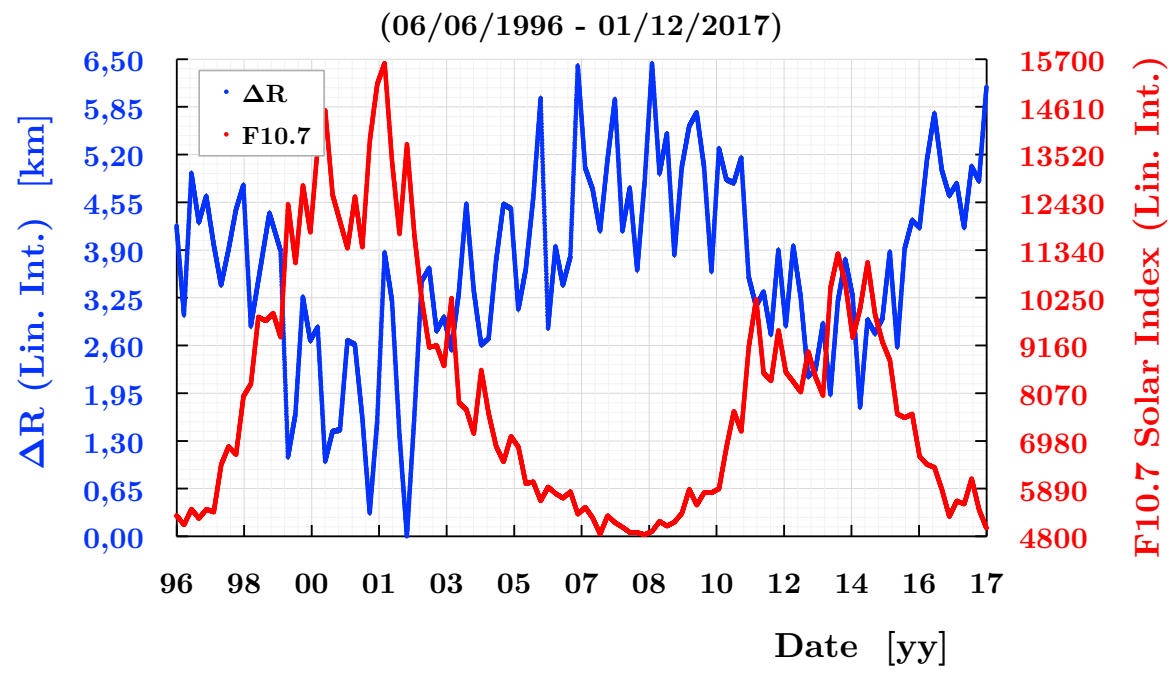

(b) Linearly interpolated data.

**Figure B.37:** Solar radius data and F10.7 solar index for the same period 06/06/1996 - 01/12/2017.

Be performing the statistical correlation analysis, on the two time series of Fig. B.37b, we get for the Pearson's correlation coefficient and p-value ($r = -0.79, p = 0$). This indicates a statistically significant high degree of negative linear correlation for the general trend of the interpolated solar radius and F10.7 data for the same overlapping period, which is in accordance with the anti-correlation found in [300,301]. Notably, the comparison between the interpolated solar radius data and the original daily values of the F10.7 show no difference with the numbers for the Pearson's coefficient and p-value being ($r = -0.70, p = 0$).

### B.6.2.2 Planetary distributions

In Fig. B.38 a direct comparison of Fig. 10.4a with the corresponding one from the F10.7 in Fig. B.38b is shown. For the comparison, the same period (06/06/1996 to 01/12/2017) and the same bin = 18° have been chosen. In the F10.7 spectrum of Venus in Fig. B.38b there is a single narrow significant peak around 243°. The total maximum - minimum amplitude observed for this case is 11.58%. This is totally different with the single wide peak observed in Fig. B.38a, which implies that the Venus related relationship for the solar radius data (as seen from Fig. B.38a) is different from that of the solar activity. In other words the cause of the solar activity seems to be different from then one behind the 11 y rhythmically varying solar radius at the level of about $\sim 10^{-5}$.

In Fig. B.38c the distribution of Venus is shown but for the interpolated F10.7 data. In the ideal case the plots Fig. B.38b and Fig. B.38c would be identical. The achieved similarity shows the quality degree of the simulation procedure, since a peaking shape with the original data in Fig. B.38b reappears in the simulated spectrum in Fig. B.38c. Finally, as expected the smaller amplitude observed in Fig. B.38c ($\sim 3.06\%$), as well as the larger values in the y-axis of Fig. B.38c, compared to Fig. B.38b are due to the grouping of 72 d and should be seen as relative values. Furthermore, a comparison between Fig. B.38a and B.38c, via the calculation of the Pearson's correlation coefficient, yields ($r_{a,c} = -0.29$, $p_{a,c} = 0.212$), which means a non-significant low degree of negative correlation.

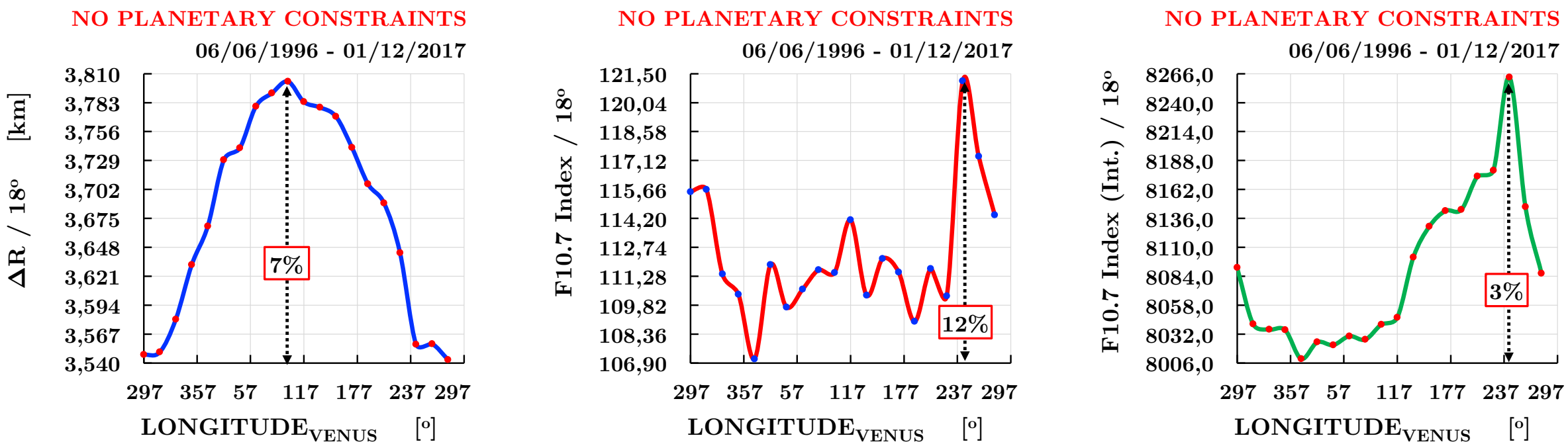

**(a)** Solar radius data. **(b)** Original daily F10.7 data. **(c)** F10.7 grouped in 72 d and linearly interpolated data.

**Figure B.38:** Comparison of the heliocentric longitude distributions of Venus for the solar radius and the F10.7 data for a bin of 18°.

The same comparison between solar radius and F10.7 data can be done with the distribution of Mars in Fig. B.39. We see that the two plots in Fig. B.39a and B.39b are different. In the solar radius data in Fig. B.39a we have three clear peaks whereas in the heliocentric distribution of Mars for F10.7 data for the same period and bin in Fig. B.39b we have a single peak and also in a different location, with the total amplitude on the latter plot being 13.78%.

Once more, as done in Fig. B.38, the F10.7 daily data have been binned every consecutive 72 d, and then, the same linear interpolation between two neighbouring 72 d mean values has been applied in order to derive the daily values. Again, in the ideal case the plots Fig. B.39b and Fig. B.39c would be identical. In our case the archived similarity between both spectral shapes confirms once more the quality of the applied simulation. And as expected, the reconstructed plot Fig. B.39c is a little wider and smaller in amplitude than the original one in Fig. B.39b. The amplitude of the peak in Fig. B.39c is about 5.5%, with the total maximum - minimum difference being 7.25%. Lastly, the Pearson's correlation in this case between Fig. B.39a and B.39c gives ($r_{a,c} = 0.14$, $p_{a,c} = 0.554$), thus indicating no significant correlation between the two plots.

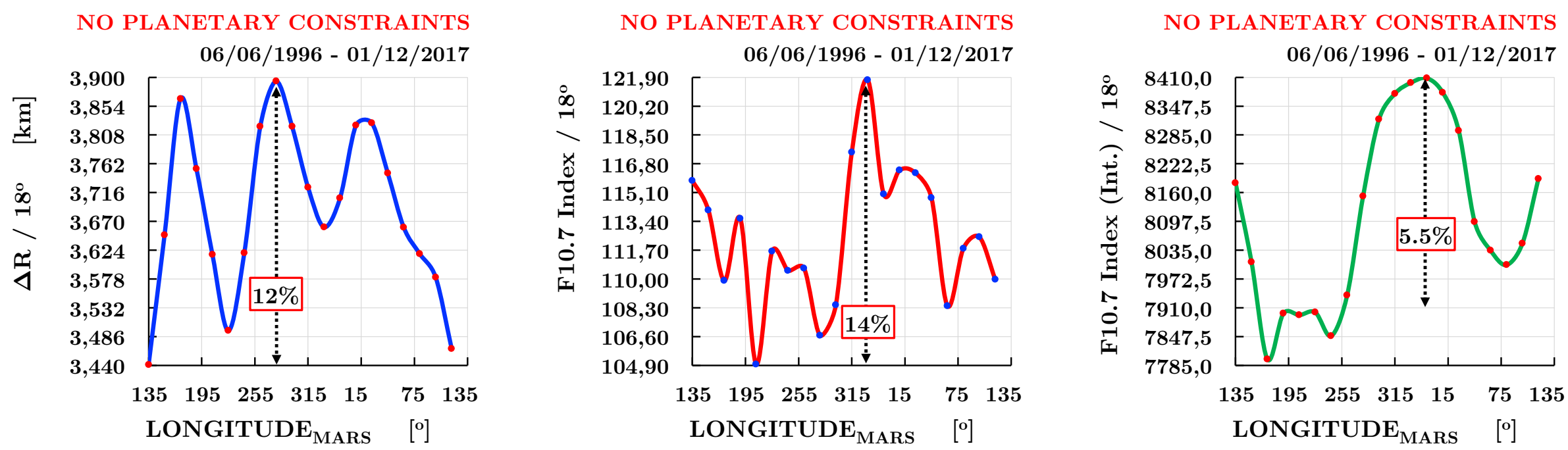

**(a)** Solar radius data. **(b)** Original daily F10.7 data. **(c)** F10.7 grouped in 72 d and linearly interpolated data.

**Figure B.39:** Comparison of the heliocentric longitude distributions of Mars for the solar radius and the F10.7 data for a bin of 18°.

## B.6.3 Comparison with tidal forces

### B.6.3.1 Data treatment

Since the distribution of Venus in Fig. 10.4a exhibits a single wide peak, instead of narrow ones as in other datasets, an additional comparison is performed with the tidal forcing scenario to exclude definitely every conventional explanation. The daily data containing the estimate of the combined effects of the planetary tides on the Sun by the eight planets have been acquired from [242] (red curve from Figure 10 in [242]). The daily tidal data spanned from 01/01/1950 to 31/12/2030, but for a direct comparison, the same dates as with the solar radius data have been chosen i.e. 06/06/1996 - 01/12/2017. The time series of the two datasets are shown in Fig. B.40a. Then, in Fig. B.40b the linearly interpolated data are shown. For the tidal data, a grouping in bins of 72 d has been performed before the linear interpolation. For the treated data, the statistical correlation analysis gives a Pearson's correlation coefficient and p-value of ($r = -0.056$, $p = 7.01 \times 10^{-7}$), which shows no association between the two datasets with the result being statistically significant. This result on its own excludes the correlation of the solar radius with the tidal forces, however a direct comparison of the small-scale characteristics through the planetary distributions is also made.

### B.6.3.2 Planetary distributions

In Fig. B.41 a comparison between the Fig. 10.4a distribution with the corresponding one from the total tidal induced irradiance in Fig. B.41b is shown. For this comparison, the same period (06/06/1996 to 01/12/2017) and the same bin size have been chosen. The tidal spectrum of Venus in Fig. B.41b shows five distinct peaks with a similar amplitude and with the total maximum - minimum difference being 25.68% and the total irradiance being 2630.07 W/m$^2$. Then, in Fig. B.41c the distribution of Venus is shown for the grouped and interpolated tidal

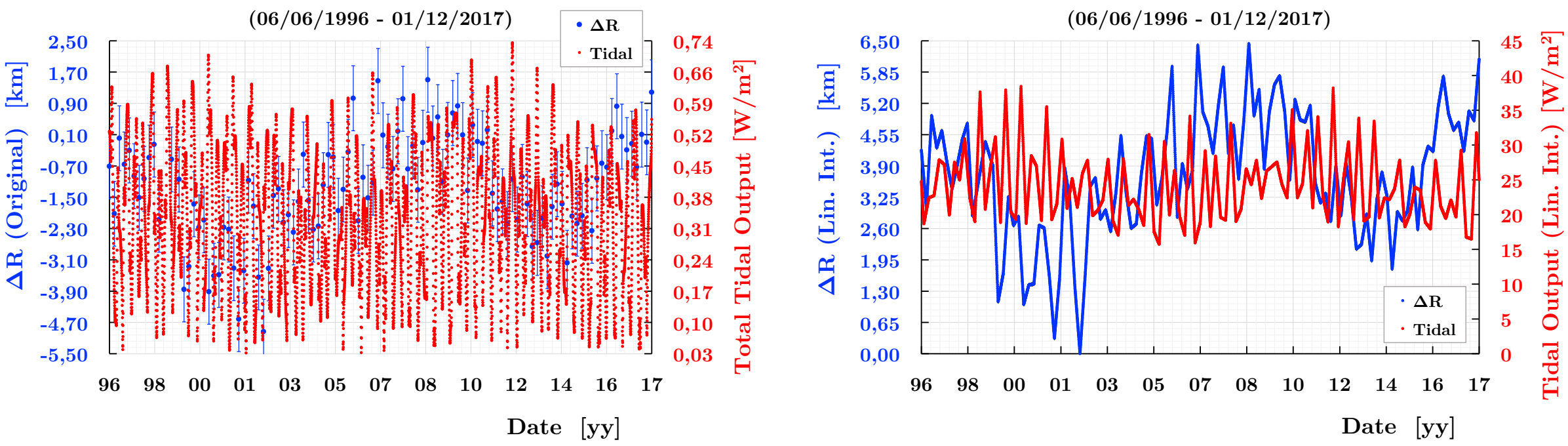

**(a)** Original data. **(b)** Linearly interpolated data.

**Figure B.40:** Solar radius and total tidal induced irradiance for the same period 06/06/1996 - 01/12/2017.

data. In that case we have two prominent peaks with a total amplitude of 6.14%. The reduced amplitude from Fig. B.41b to Fig. B.41c is explained from the grouping and interpolation procedure, as shown also in Appendix Sect. B.6.2.2. However, the significant difference between Fig. B.41a and Fig. B.41c is evident. The calculation of a Pearson's correlation coefficient and p-value also gives ($r_{a,c} = 0.23$, $p_{a,c} = 0.334$), validating a non-significant correlation between the two. Though, because the assumption of linearity between the two variables is not perfectly achieved, we also calculate the non-parametric tests of Spearman's and Kendall's rank correlation which give ($r_{S_{a,c}} = 0.17$, $p_{S_{a,c}} = 0.482$) and ($\tau_{a,c} = 0.10$, $p_{a,c} = 0.516$), which show again a statistically not significant small correlation between solar radius and tidal irradiance.

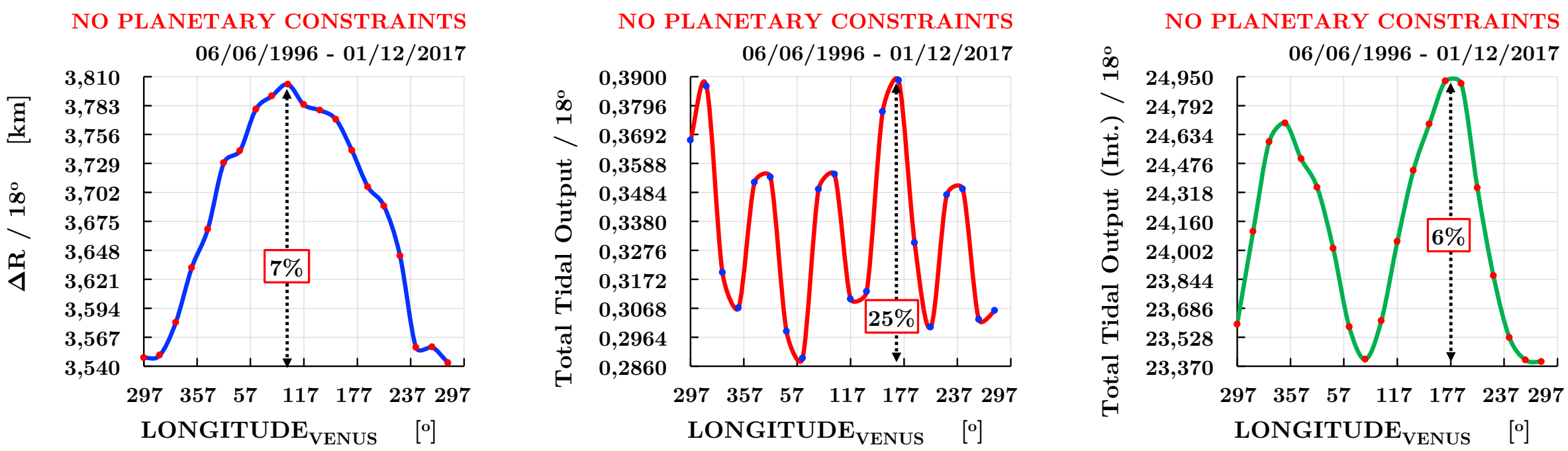

**(a)** Solar radius data. **(b)** Original daily tidal data. **(c)** Tidal data grouped in 72 d and linearly interpolated.

**Figure B.41:** Comparison of the heliocentric longitude distributions of Venus for the solar radius and the tidal data for a bin of 18°.

Finally, in Fig. B.42 the same comparison is performed for Mars. Fig. B.42b shows the distribution of the total tidal induced irradiance exhibiting a maximum - minimum difference of 11.78% and six peaks. For a more accurate comparison of the tidal effects on the Sun with the solar radius data, the same interpolation procedure is performed in the tidal data as before. This is exhibited in Fig. B.42c which gives an overall modulation of 3.92%. For the case of Venus in Fig. B.41, the statistical correlation analysis for Fig. B.42a and B.42c yields $(r_{a,c} = 0.32,\ p_{a,c} = 0.175)$, therefore a non-statistically significant positive correlation.

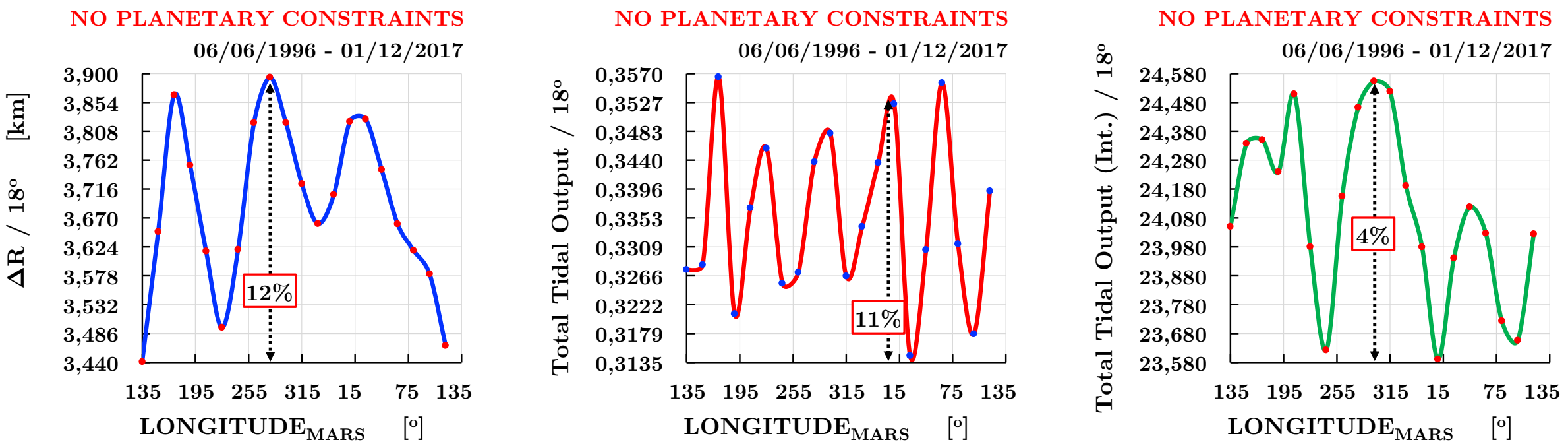

**(a)** Solar radius data. **(b)** Original daily tidal data. **(c)** Tidal data grouped in 72 d and linearly interpolated.

**Figure B.42:** Comparison of the heliocentric longitude distributions of Mars for the solar radius and the tidal data for a bin of 18°.

# B.7 Coronal composition

## B.7.1 Comparison with F10.7

Based on [316] the variation of coronal composition is highly correlated with the F10.7 solar proxy not only in the general trend but also in some of the small-scale details. Therefore, here we will try to replicate this result and get an insight into the degree of similarity between the two datasets. In Fig. B.43 the time series of FIP bias and F10.7 are overlapped for the same period of 30/04/2010 to 11/05/2014. By performing a statistical correlation analysis, we get a Pearson's correlation coefficient and p-value for these two datasets of $(r = 0.48,\ p = 0)$ which show a statistically significant moderate degree of linear correlation.

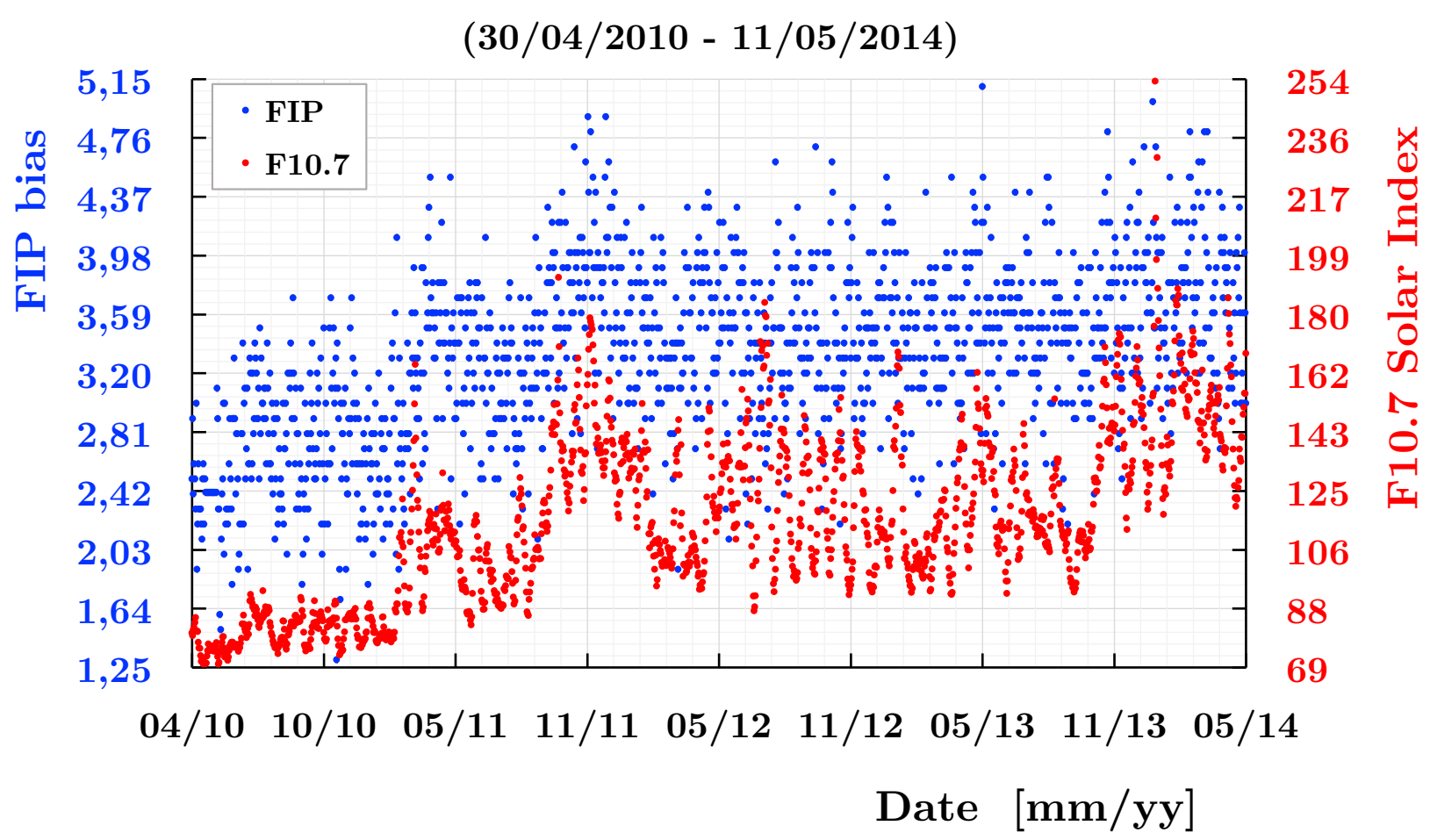

**Figure B.43:** Daily FIP data and F10.7 solar index for the same period 30/04/2010 - 11/05/2014.

#### B.7.1.1 Single planets

Moving on from the general trend to the derived planetary distributions, the first comparison of the two datasets is performed for the distributions of Fig. 11.3. In Fig. B.44 the same time period has been chosen for the F10.7 solar radio flux data. We see that the distribution of Mercury in Fig. B.44a and B.44e, and Venus in Fig. B.44b and B.44f differ, while Earth and Mars look more similar. However, in order to quantify the observation, the linear Pearson correlation coefficient has been calculated for these plots to be ($r_{a,e} = -0.28$, $p_{a,e} = 0.31$), ($r_{b,f} = 0.35$, $p_{b,f} = 0.05$), ($r_{c,g} = 0.80$, $p_{c,g} = 2.38 \times 10^{-5}$) and ($r_{d,h} = 0.85$, $p_{d,h} = 6.56 \times 10^{-5}$) accordingly. This means that only the spectra of Earth and Mars seem to have a statistically significant linear positive correlation.

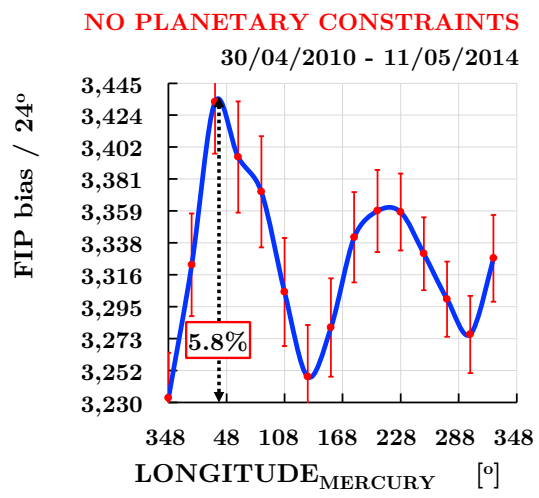

**(a)** FIP bias distribution for Mercury with 24° bin.

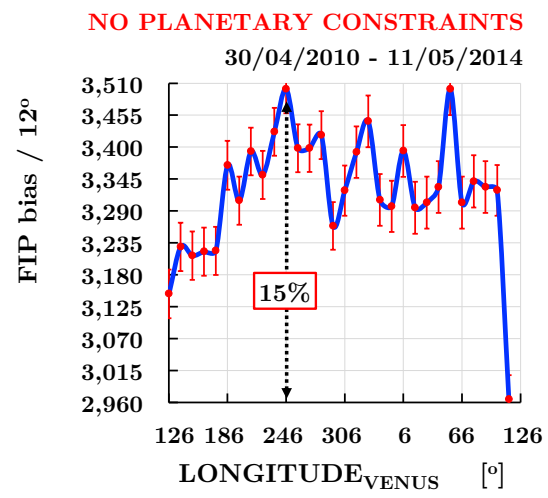

**(b)** FIP distribution for Venus with 12° bin.

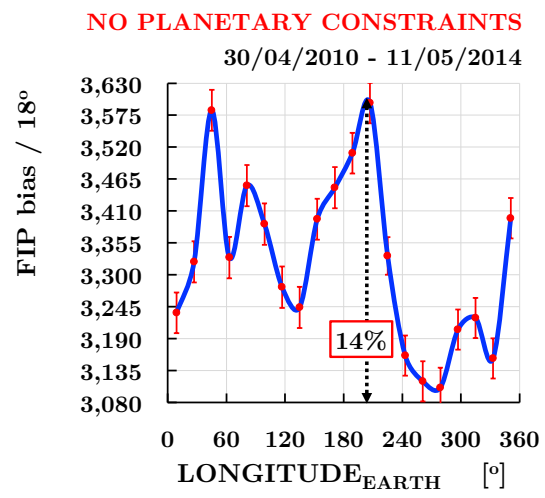

**(c)** FIP distribution for Earth with 18° bin.

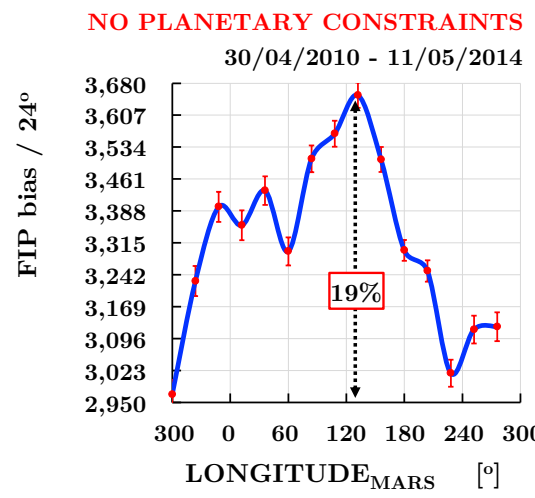

**(d)** FIP distribution for Mars with 24° bin.

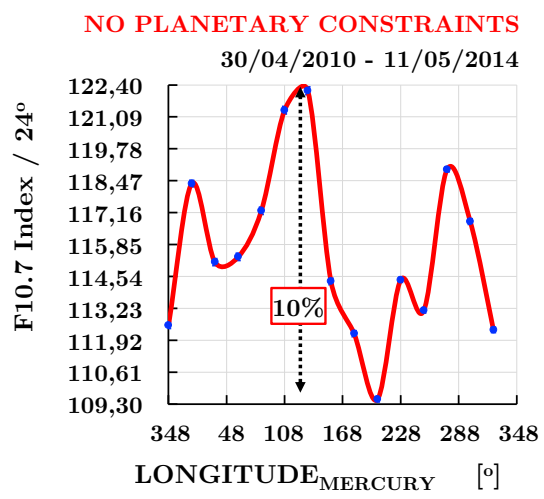

**(e)** F10.7 distribution for Mercury with 24° bin.

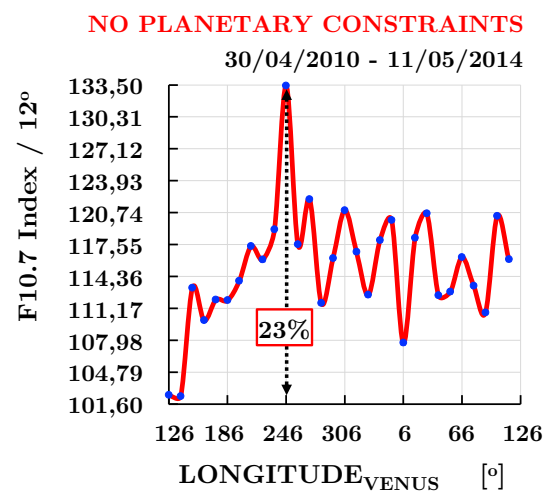

**(f)** F10.7 distribution for Venus with 12° bin.

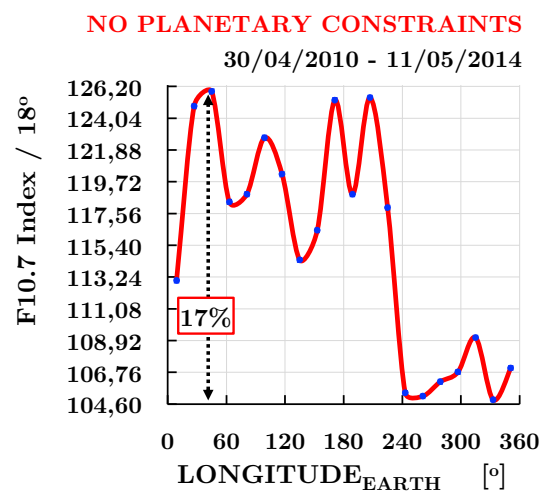

**(g)** F10.7 distribution for Earth with 18° bin.

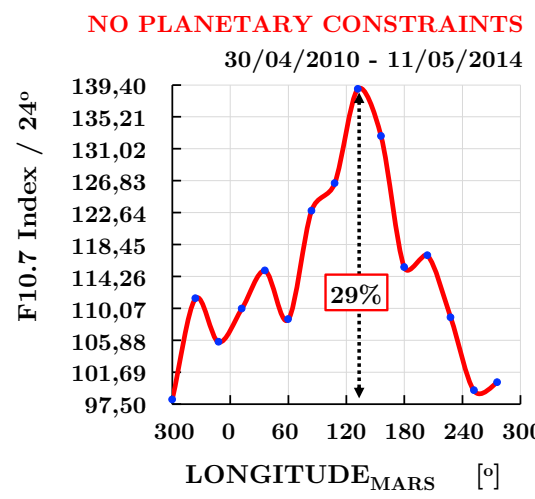

**(h)** F10.7 distribution for Mars with 24° bin.

**Figure B.44:** Comparison of distributions of FIP bias vs. F10.7 solar index for the period 30/04/2010 - 11/05/2014.

#### B.7.1.2 Combining planets

An additional comparison is made with the plots containing a combination of planets since in these cases we have the most statistically significant effects. In Fig. B.45 the plots from Fig. 11.4 are compared with the corresponding ones from F10.7 solar proxy. We see that in these cases there is an even more clear dissimilarity. For example in the case of Venus being constrained between 340° to 80° in Fig. B.45a and B.45d, there is a peak around 130° in F10.7 whereas in the distribution of FIP there is a minimum. Similarly in the case of Mars being around 180° to 280° in Fig. B.45c and B.45f the peak around 60° in FIP bias corresponds to a minimum in F10.7 whereas the peak around 130 degree in F10.7 corresponds to a minimum in FIP bias. Indeed, the Pearson correlation for these three cases gives ($r_{a,d} = 0.15$, $p_{a,d} = 0.59$), ($r_{b,e} = 0.55$, $p_{b,e} = 0.033$) and ($r_{c,f} = -0.13$, $p_{c,f} = 0.64$) respectively. This confirms the above claim and shows a significant positive correlation of $r = 0.55$ only for the Fig. B.45b and B.45e.

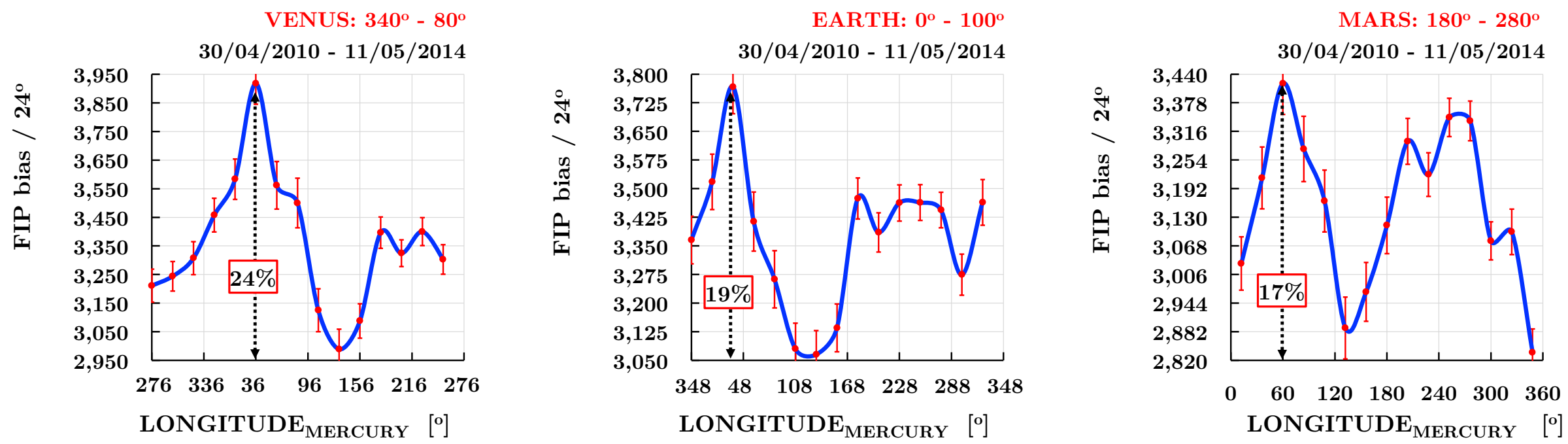

**(a)** FIP bias distribution for Mercury when Venus is between 340° to 80°.

**(b)** FIP bias distribution for Mercury when Earth is between 0° to 100°.

**(c)** FIP bias distribution for Mercury when Mars is between 180° to 280°.

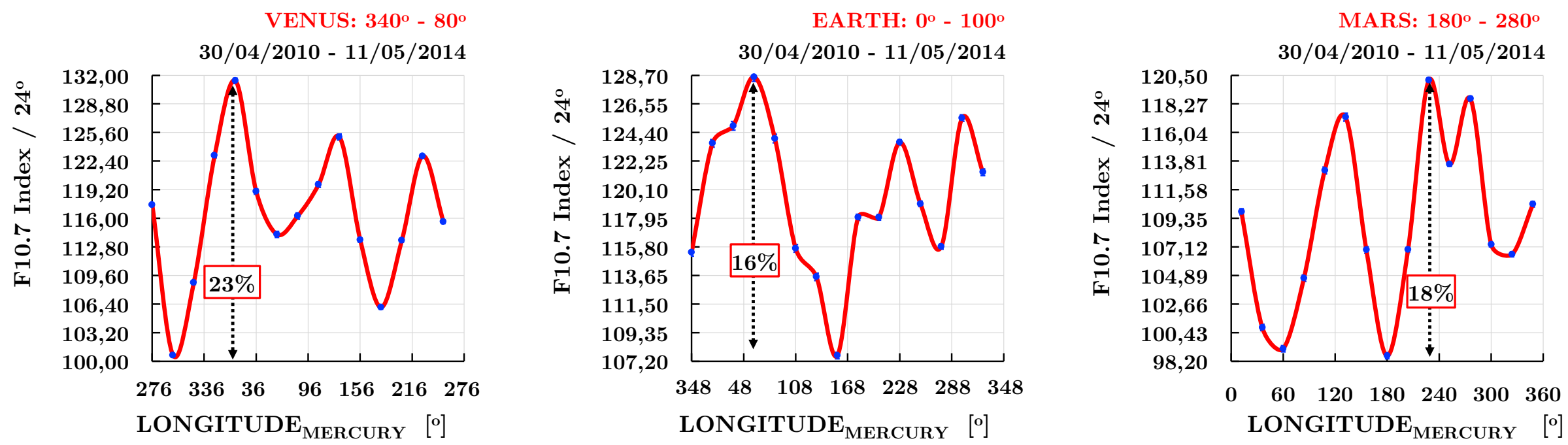

**(d)** F10.7 distribution for Mercury when Venus is between 340° to 80°.

**(e)** F10.7 distribution for Mercury when Earth is between 0° to 100°.

**(f)** F10.7 distribution for Mercury when Mars is between 180° to 280°

**Figure B.45:** Comparison of distributions of FIP bias vs. F10.7 solar index for the reference frame of Mercury when other planets are constrained to propagate in a specific longitude region for bin = 24° for the period 30/04/2010 - 11/05/2014.

The same procedure has been performed in Fig. B.46 where the plots from Fig. 11.5 and 11.6 have been used. The correlation has been calculated to ($r_{a,c} = 0.23$, $p_{a,c} = 0.41$) for Fig. B.46a-B.46c and ($r_{b,d} = 0.76$, $p_{b,d} = 1.17 \times 10^{-4}$) for Fig. B.46b-B.46d showing a significant positive correlation only for the latter case.

## B.7.2 Comparison with solar EUV

Since the selected FIP bias measurements are obtained by spectroscopic measurements on the EUV [316], a comparison of the derived planetary relationship with the EUV irradiance would provide interesting results. In Fig. B.47 the time series of FIP bias are compared with

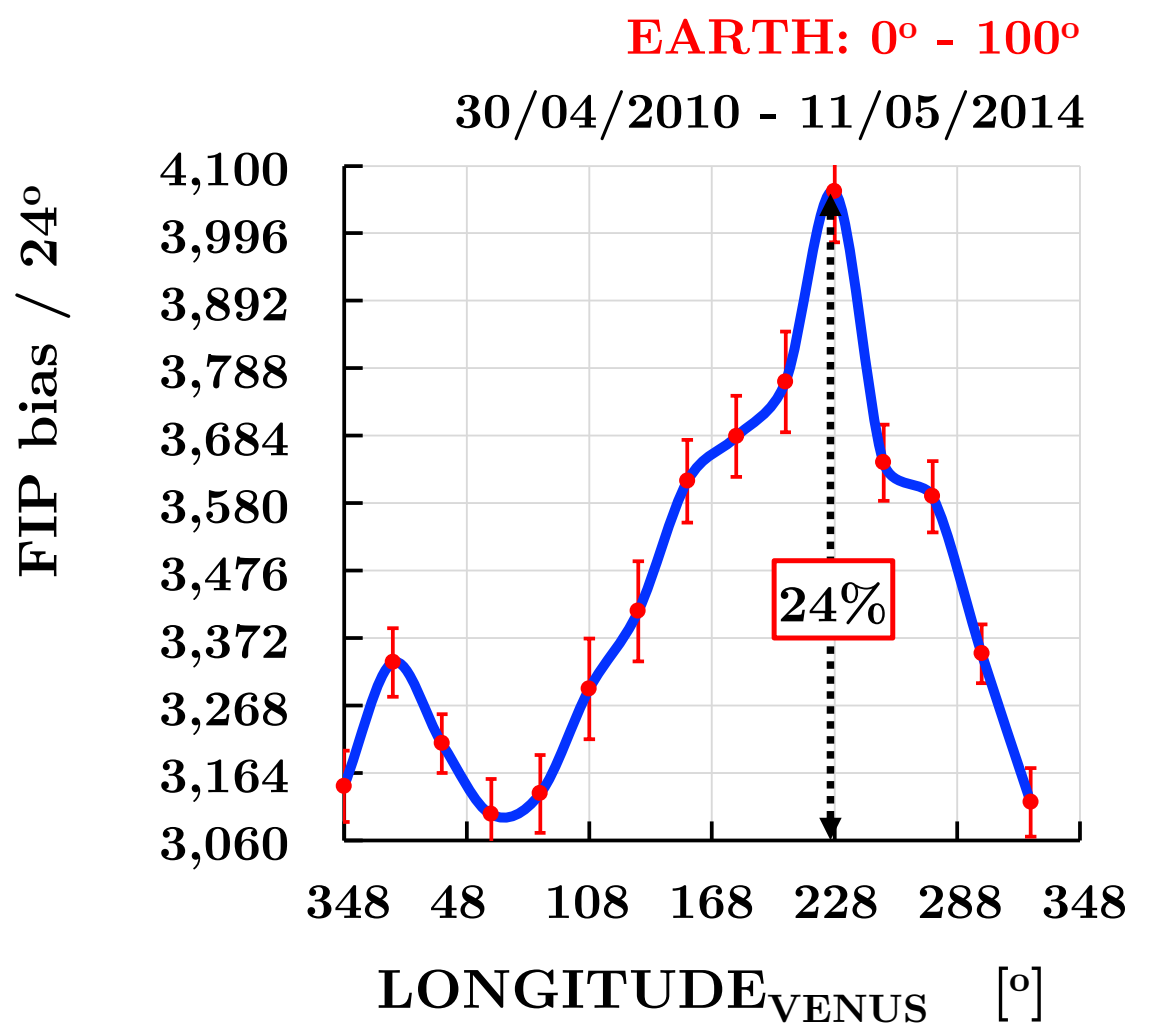

(a) FIP bias distribution for Venus when Earth is between 0° to 100° for bin = 24°.

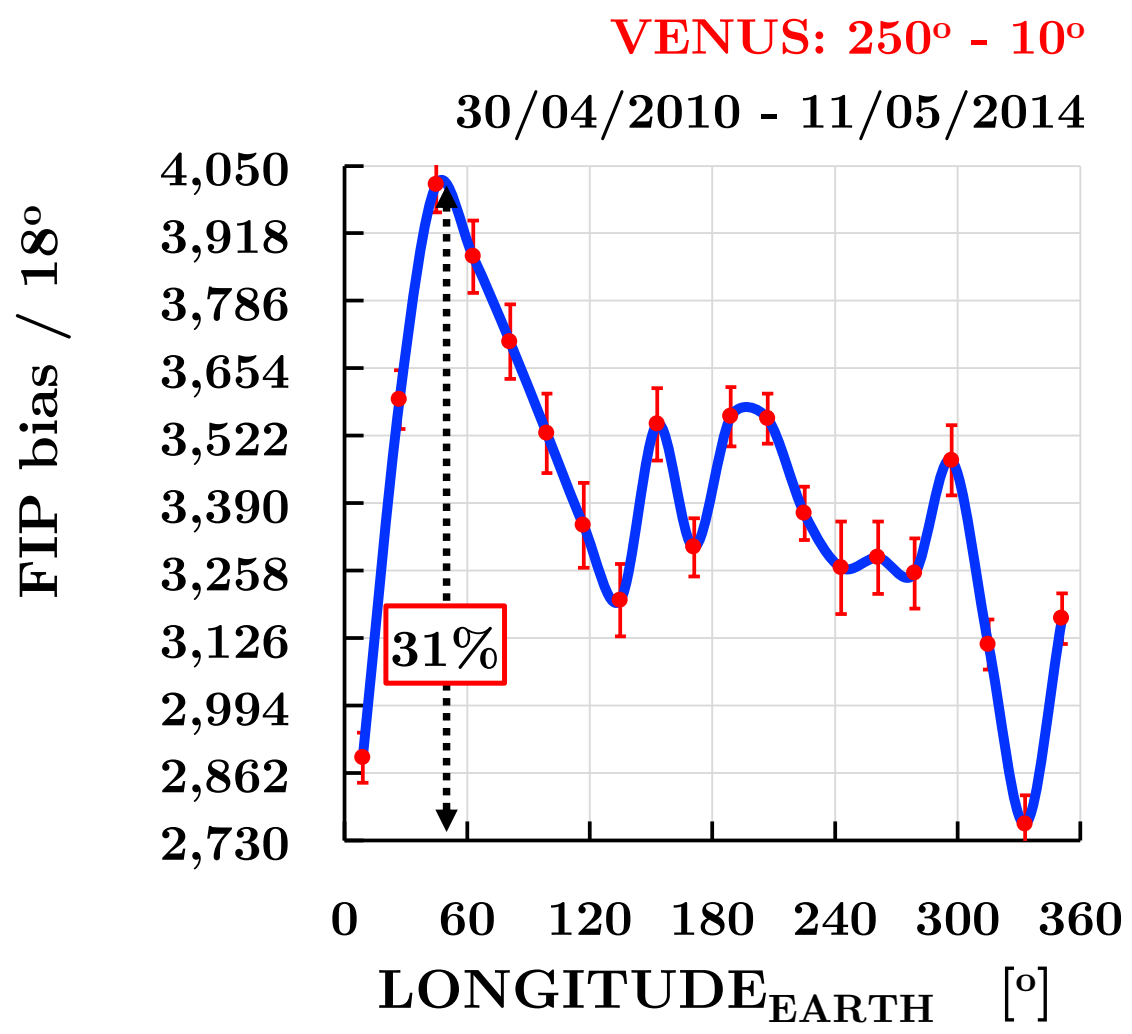

(b) FIP bias distribution for Earth when Venus is between 250° to 10° for bin = 18°.

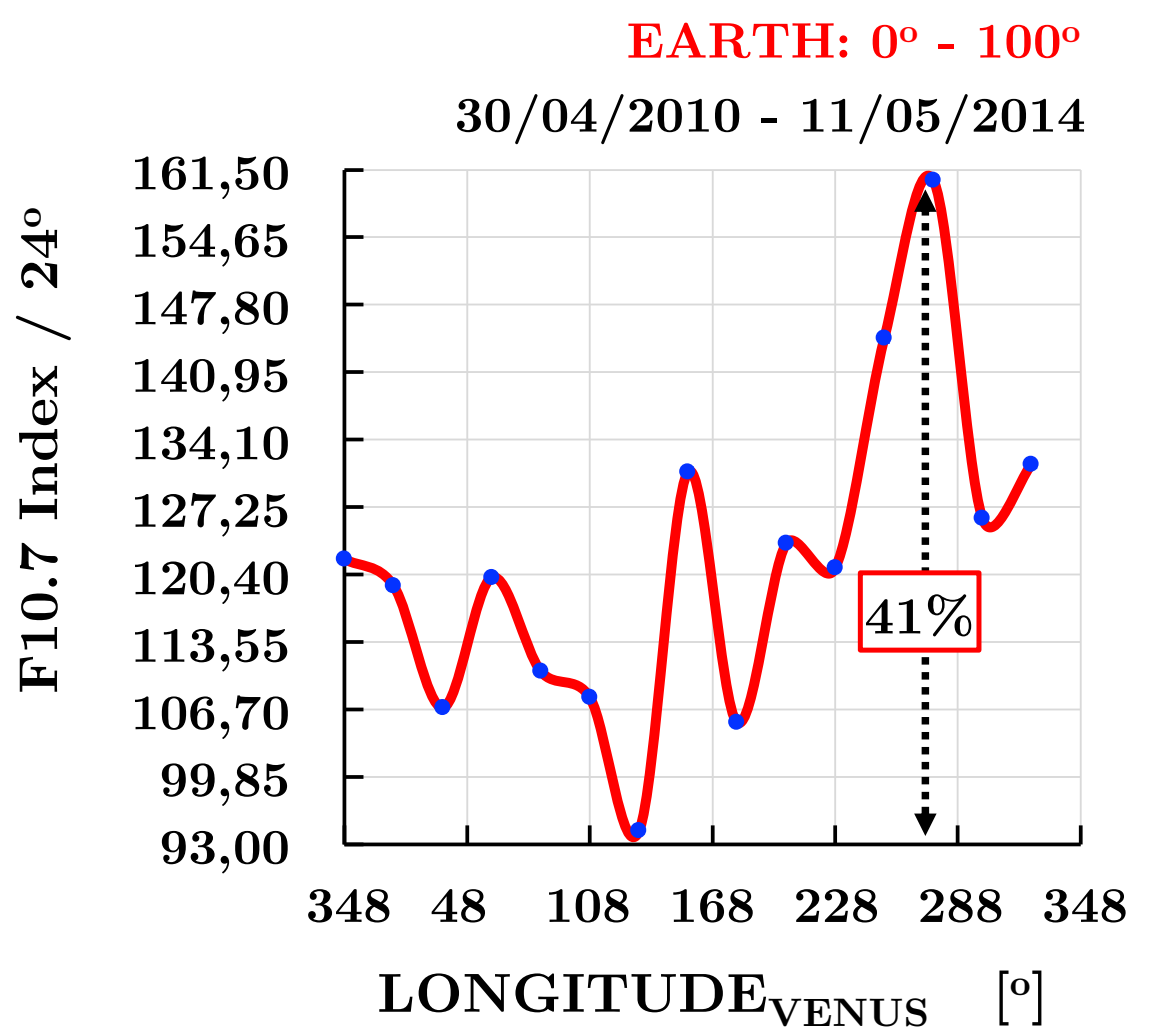

(c) F10.7 distribution for Venus when Earth is between 0° to 100° for bin = 24°.

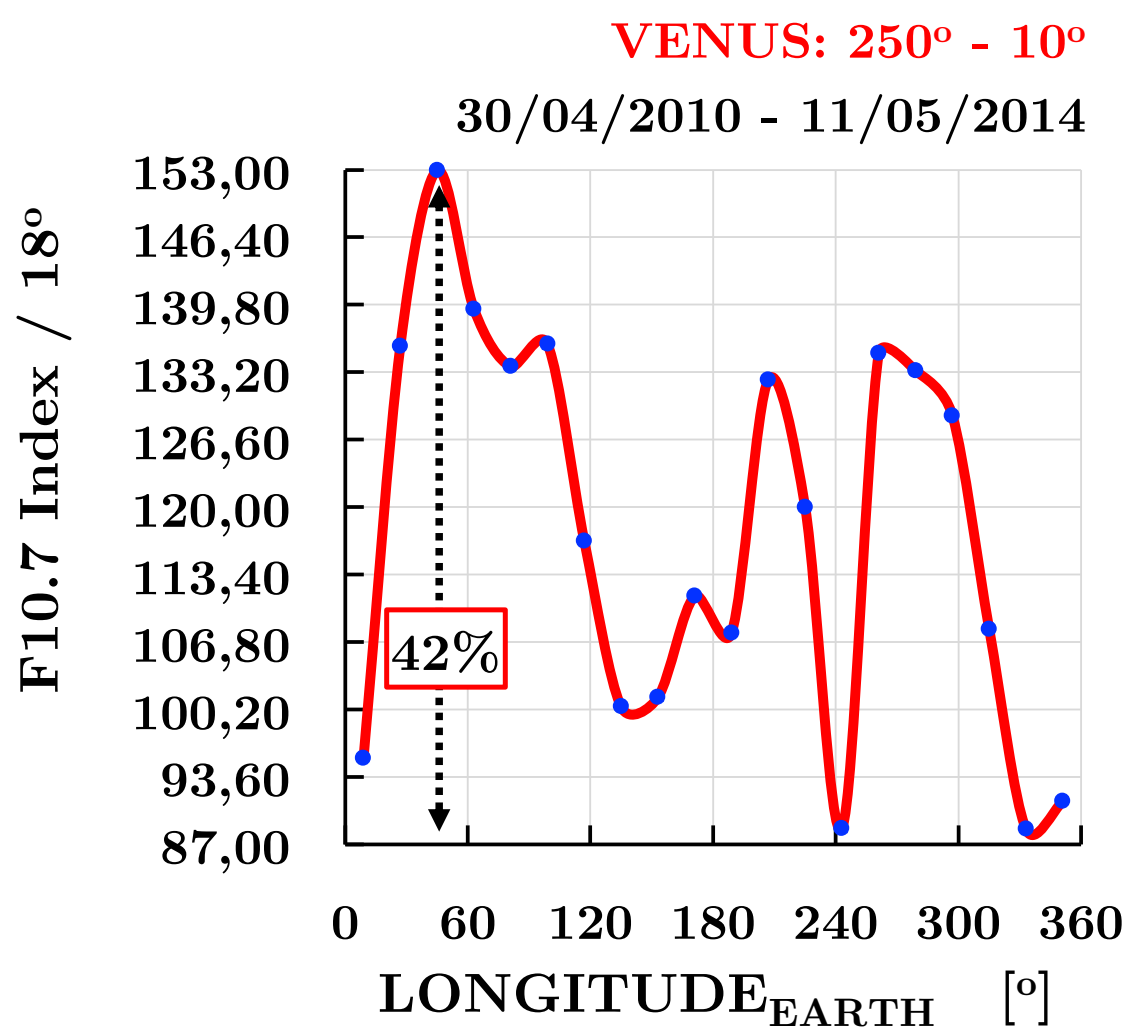

(d) F10.7 distribution for Earth when Venus is between 250° to 10° for bin = 18°.

**Figure B.46:** Comparison of distributions of FIP bias vs. F10.7 solar index for the reference frames of Venus and Earth while one of them is allowed, in each case, to propagate in a specific longitudinal region for the period 30/04/2010 - 11/05/2014.

the EUV solar irradiance. In this case the Pearson's correlation coefficient and p-value are calculated to be ($r = 0.54$, $p = 0$) which show a marginally high degree of positive correlation which is statistically significant. We remind that in the comparison of the general trend of FIP bias with F10.7 we had $r = 0.48$.

#### B.7.2.1 Single planets

In Fig. B.48, we compare the distributions from Fig. 11.3 with the corresponding ones in EUV. The results are similar with the F10.7 case where there seems to be some degree of similarity between the spectra together with a few discrepancies. For example the wide peak of Mercury in Fig. B.48a around 210° shows an opposite behaviour in Fig. B.48e. More specifically, the quantitative correlations for each case give ($r_{a,e} = -0.031$, $p_{a,e} = 0.91$), ($r_{b,f} = 0.47$, $p_{b,f} = 0.009$), ($r_{c,g} = 0.82$, $p_{c,g} = 8.65 \times 10^{-6}$) and ($r_{d,h} = 0.91$, $p_{d,h} = 2.69 \times 10^{-6}$), with all the results except Mercury where no correlation was found, being statistically significant. More specifically, for the case of Venus a moderate degree of correlation was found while for Earth and Mars we have a high degree of correlation.

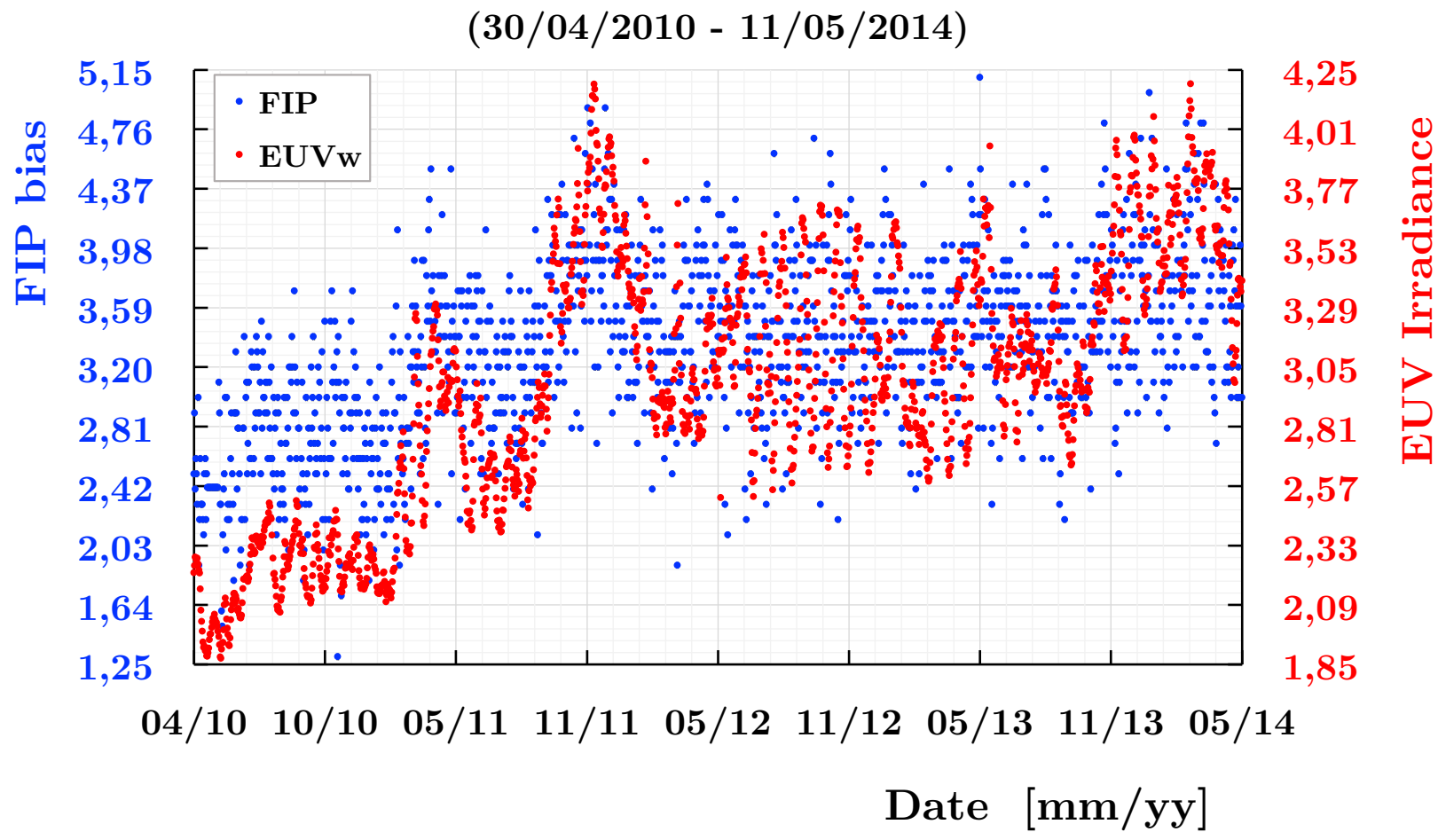

**Figure B.47:** Daily FIP data and EUV solar irradiance for the same period 30/04/2010 - 11/05/2014.

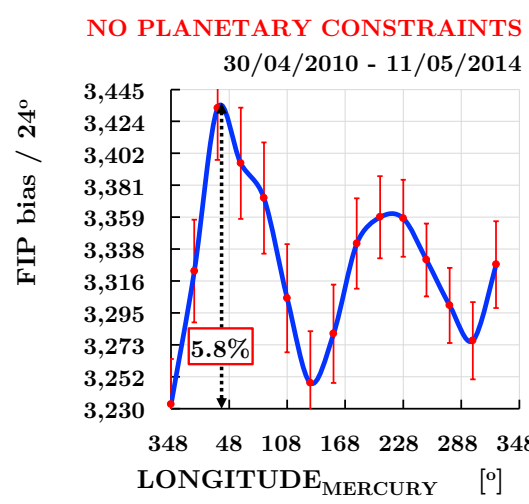

**(a)** FIP distribution for Mercury with 24° bin.

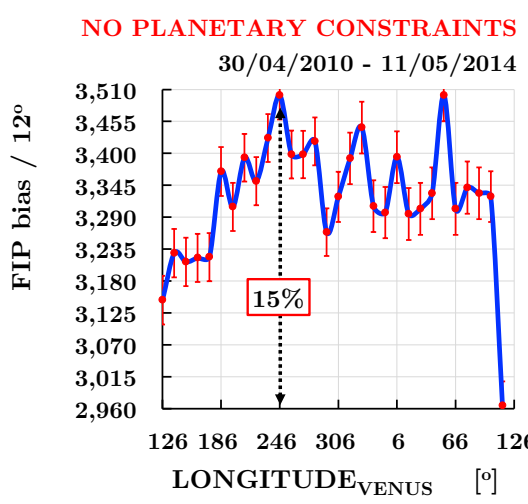

**(b)** FIP distribution for Venus with 12° bin.

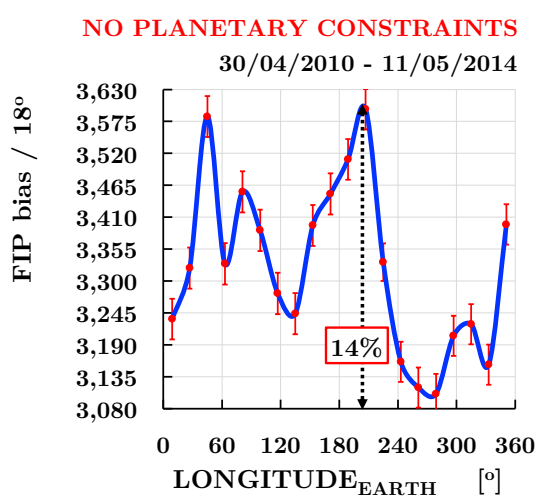

**(c)** FIP distribution for Earth with 18° bin.

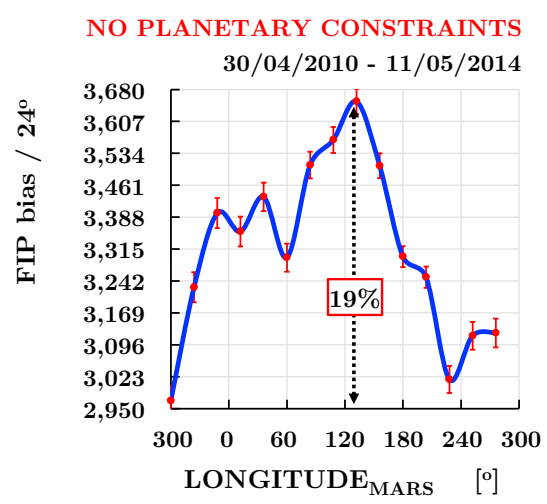

**(d)** FIP distribution for Mars with 24° bin.

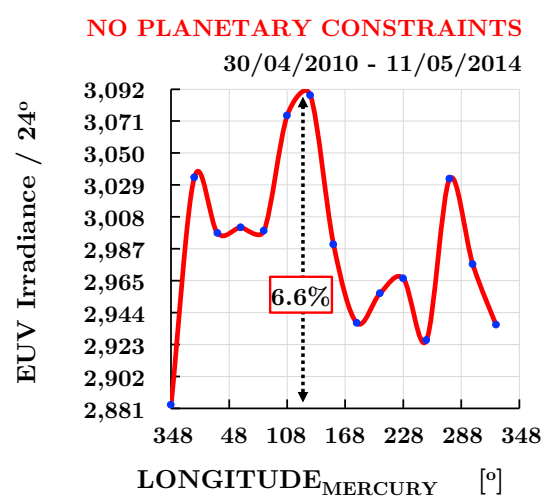

**(e)** EUV distribution for Mercury with 24° bin.

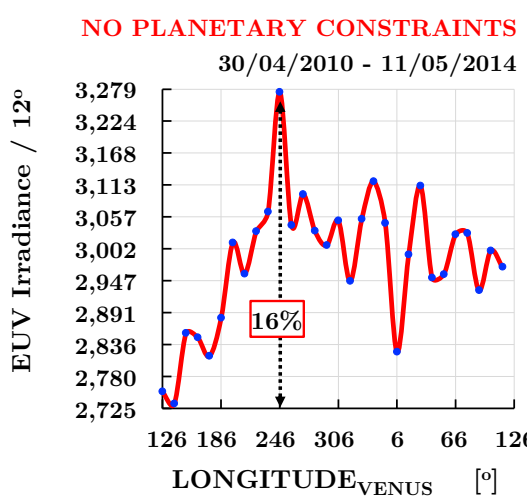

**(f)** EUV distribution for Venus with 12° bin.

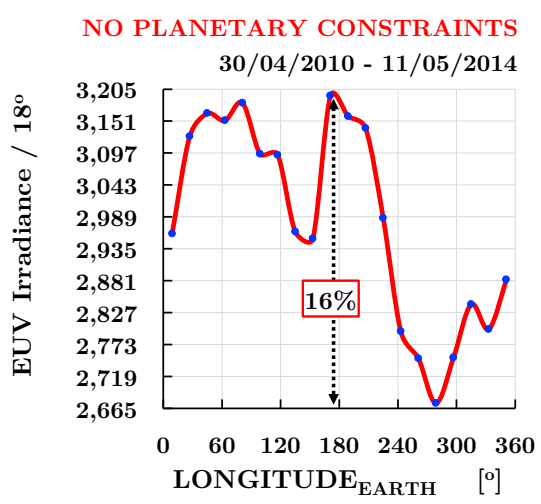

**(g)** EUV distribution for Earth with 18° bin.

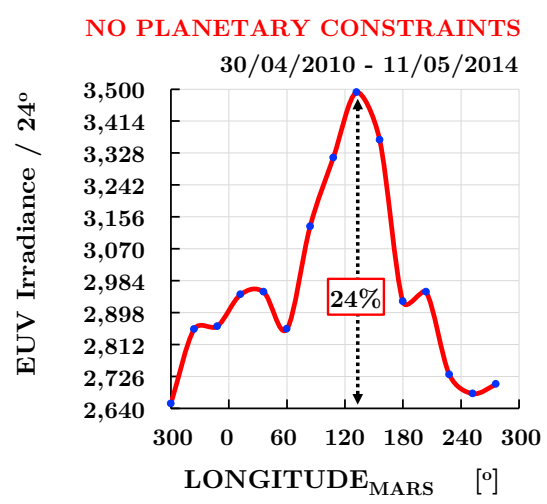

**(h)** EUV distribution for Mars with 24° bin.

**Figure B.48:** Comparison of distributions of FIP bias vs. EUV solar irradiance for the period 30/04/2010 - 11/05/2014.

### B.7.2.2 Combining planets

The next step is the comparison of the plots including additional planetary constraints. The reference frame of Mercury is initially selected in Fig. B.49 with Venus, Earth and Mars being constrained in a 100°-wide windows with a bin of 24°. Again, similarly to Fig. B.45 there are significant differences in the cases where Venus and Mars being constrained in Fig. B.49a-B.49d and Fig. B.49c-B.49f, where we have ($r_{a,d} = 0.084$, $p_{a,d} = 0.77$) and ($r_{c,f} = 0.20$, $p_{c,f} = 0.46$) respectively. In contrast, for the case of Earth being constrained in Fig. B.49b-B.49e we have a similar shape with a derived significant correlation ($r_{b,e} = 0.53$, $p_{b,e} = 0.042$).

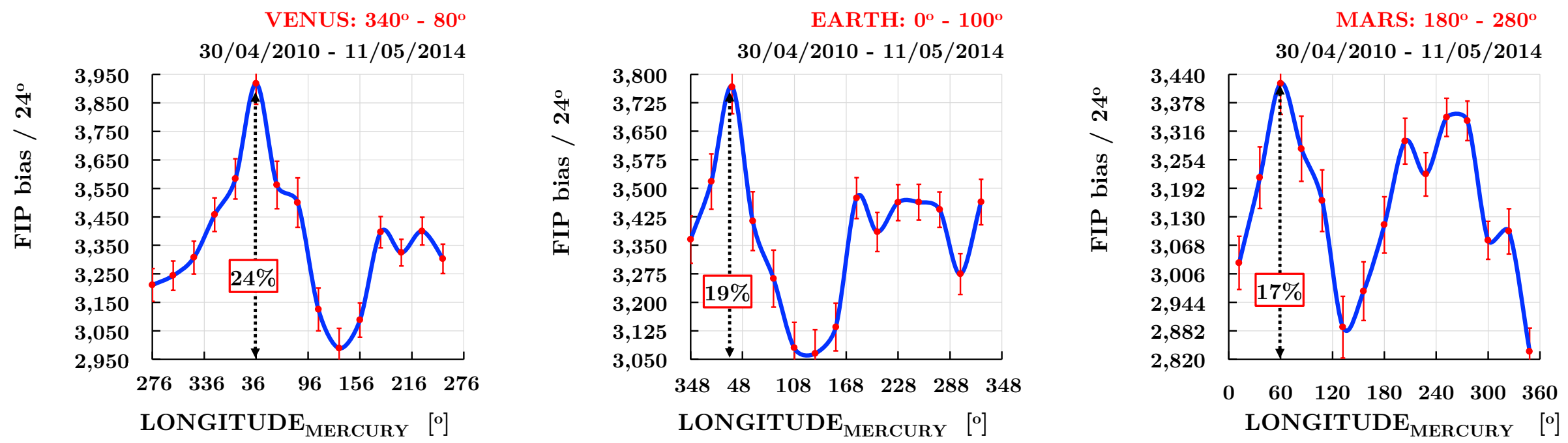

**(a)** FIP bias distribution for Mercury when Venus is between 340° to 80°.

**(b)** FIP bias distribution for Mercury when Earth is between 0° to 100°.

**(c)** FIP bias distribution for Mercury when Mars is between 180° to 280°.

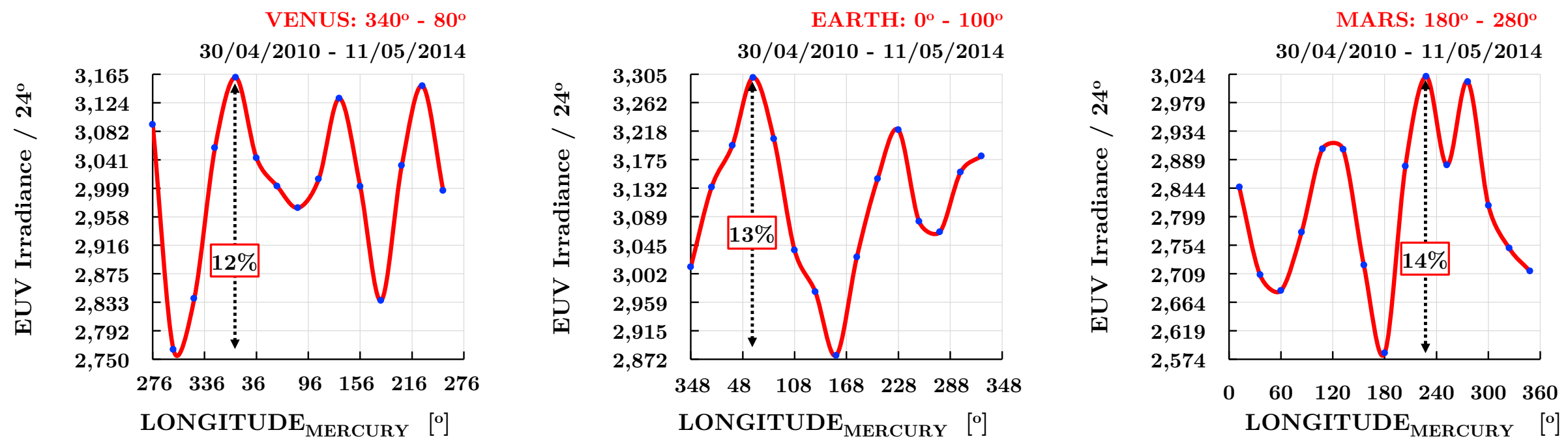

**(d)** EUV distribution for Mercury when Venus is between 340° to 80°.

**(e)** EUV distribution for Mercury when Earth is between 0° to 100°.

**(f)** EUV distribution for Mercury when Mars is between 180° to 280°

**Figure B.49:** Comparison of distributions of FIP bias vs. EUV solar irradiance for the reference frame of Mercury when other planets are constrained to propagate in a specific longitudinal region for bin = 24° for the period 30/04/2010 - 11/05/2014.

The last comparison is shown in Fig. B.50 where Venus and Earth are plotted when the orbital position of the second planet is constrained on a 100° and 120°-wide longitudinal range respectively. The correlations between the two sets of spectra were calculated at ($r_{a,c} = 0.25$, $p_{a,c} = 0.36$) and ($r_{b,d} = 0.83$, $p_{b,d} = 7.33 \times 10^{-6}$) with no relationship being found on the former case and a high degree of positive correlation for the latter case.

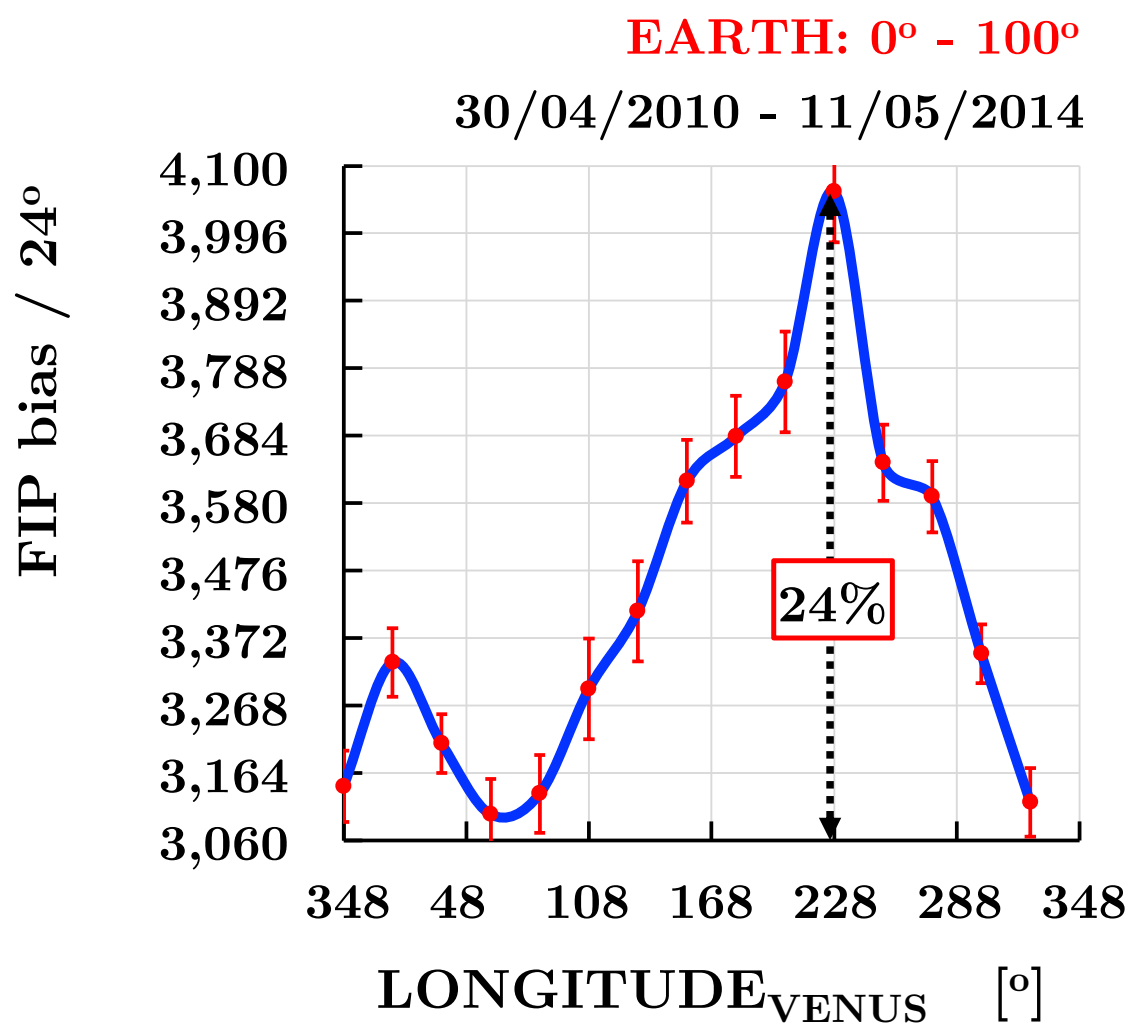

**(a)** FIP bias distribution for Venus when Earth is between 0° to 100° for bin = 24°.

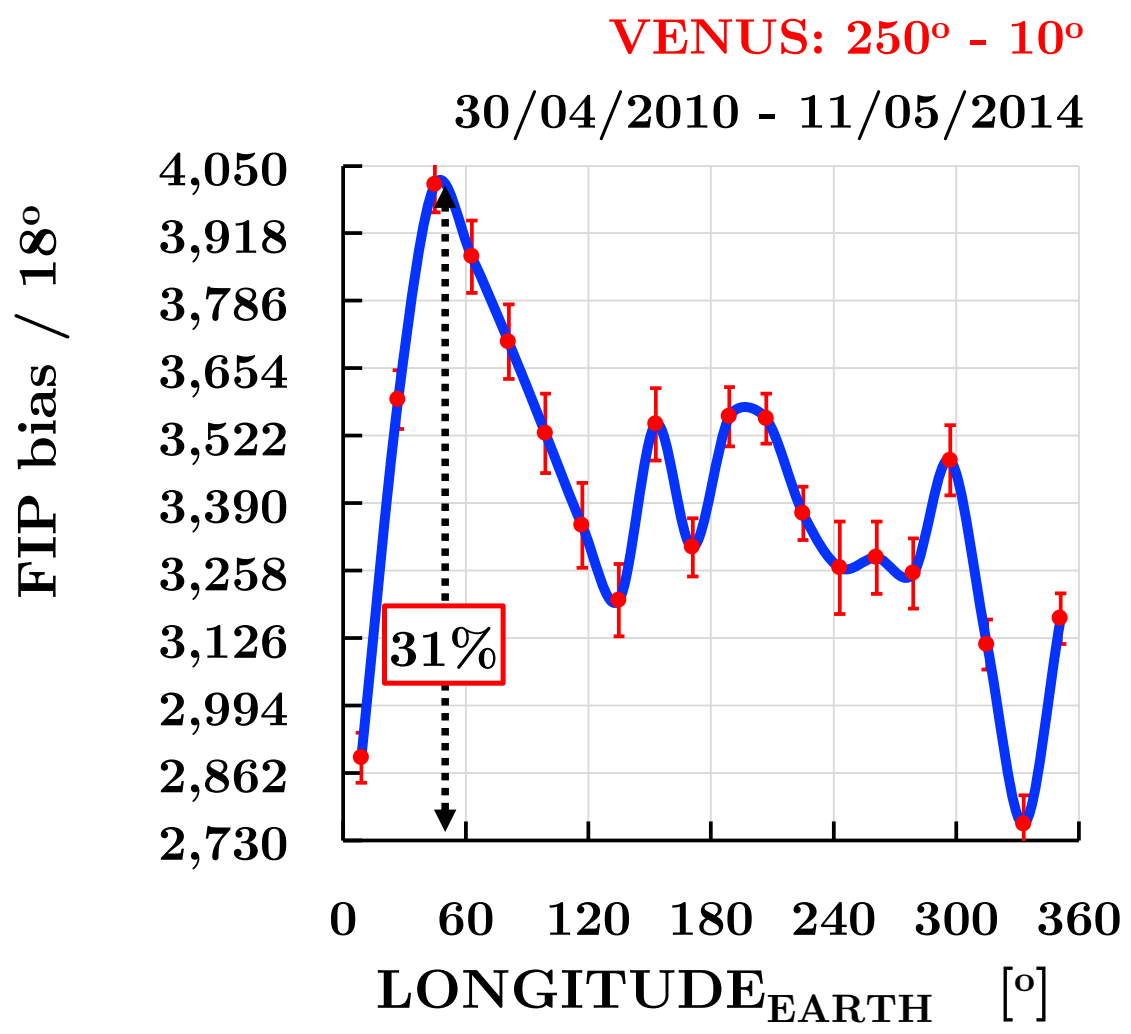

**(b)** FIP bias distribution for Earth when Venus is between 250° to 10° for bin = 18°.

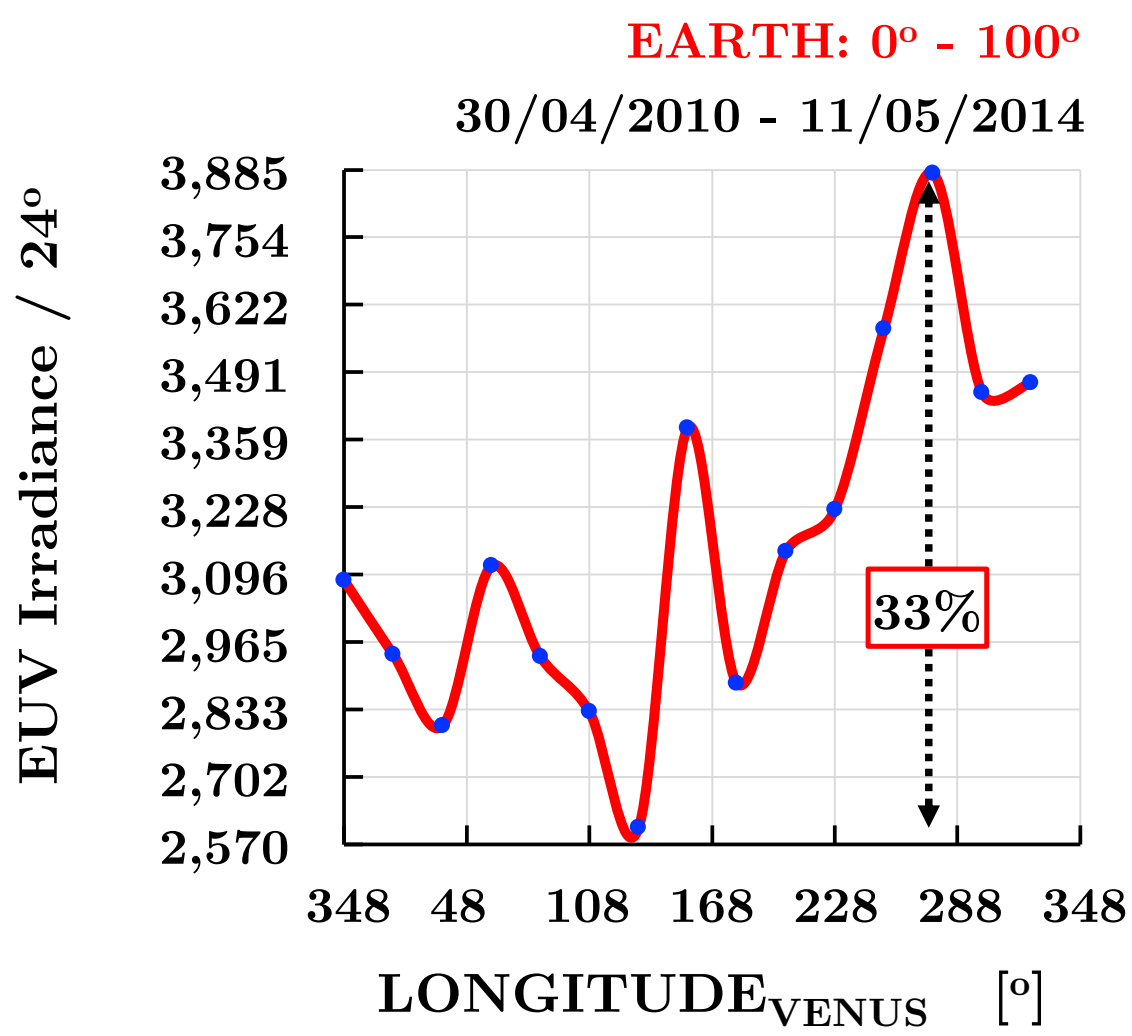

**(c)** EUV distribution distribution for Venus when Earth is between 0° to 100° for bin = 24°.

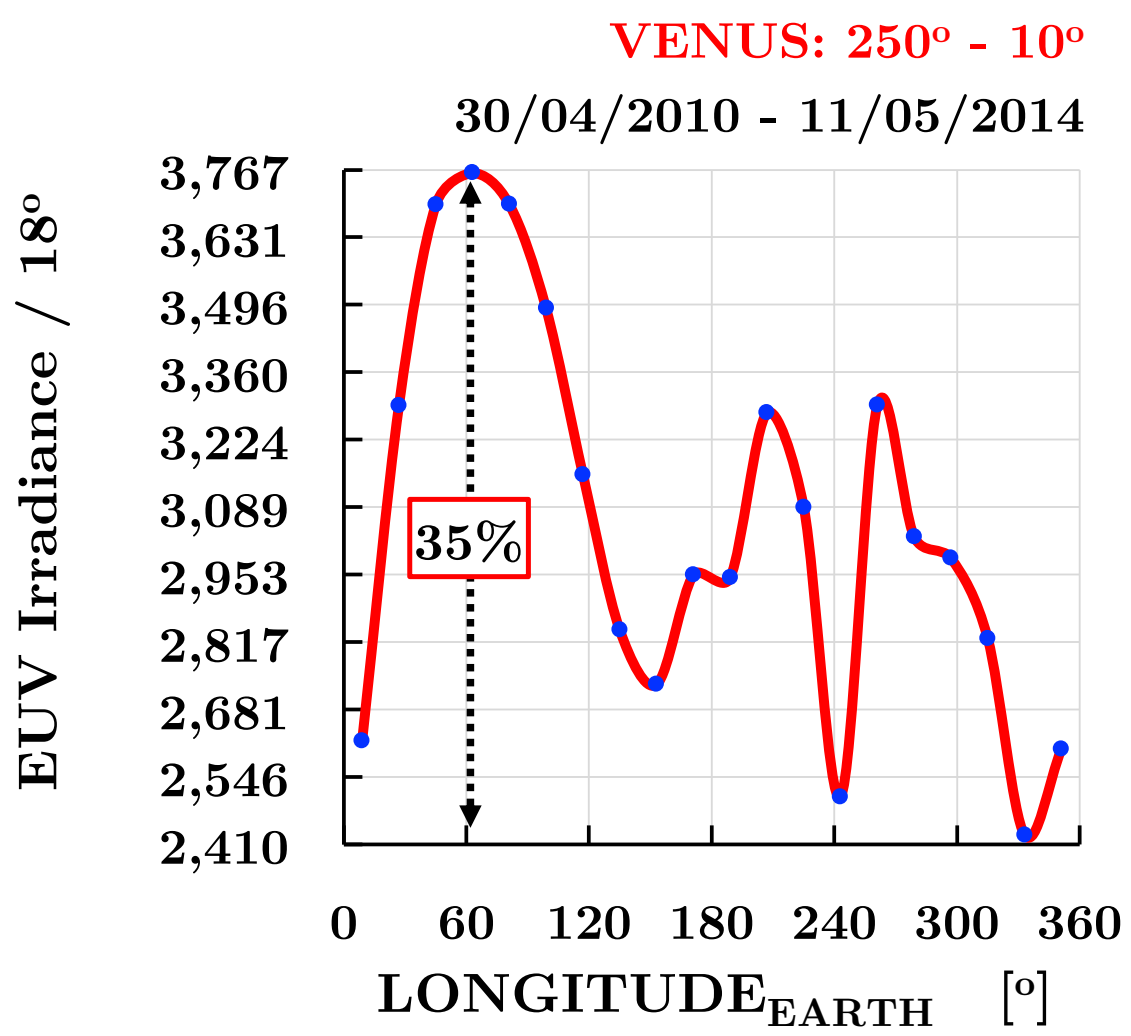

**(d)** EUV distribution distribution for Earth when Venus is between 250° to 10° for bin = 18°.

**Figure B.50:** Comparison of distributions of FIP bias vs. EUV solar irradiance for the reference frames of Venus and Earth while one of them is allowed, in each case, to propagate in a specific longitude region for the period 30/04/2010 - 11/05/2014.

## B.8 Lyman-alpha

### B.8.1 Comparison with solar EUV

The most obvious comparison of the Ly-$\alpha$ irradiance planetary distributions is with the corresponding ones for the EUV irradiance. EUV has already exhibited significant planetary relationship, thus a direct comparison between the two cases would point to a common or a different origin of the two solar proxies. Since the currently available EUV dataset spans from 01/01/1996 to 01/03/2021 the various Ly-$\alpha$ plots are recalculated to be compared under the same conditions. In Fig. B.51 the time series of the daily data of the two datasets are shown. By performing a statistical correlation analysis we get a Pearson's correlation coefficient and p-value of ($r = 0.94,\ p = 0$). As expected, this points to a very high degree of association between the general trends of two datasets with the results being statistically significant as indicated by the p-value.

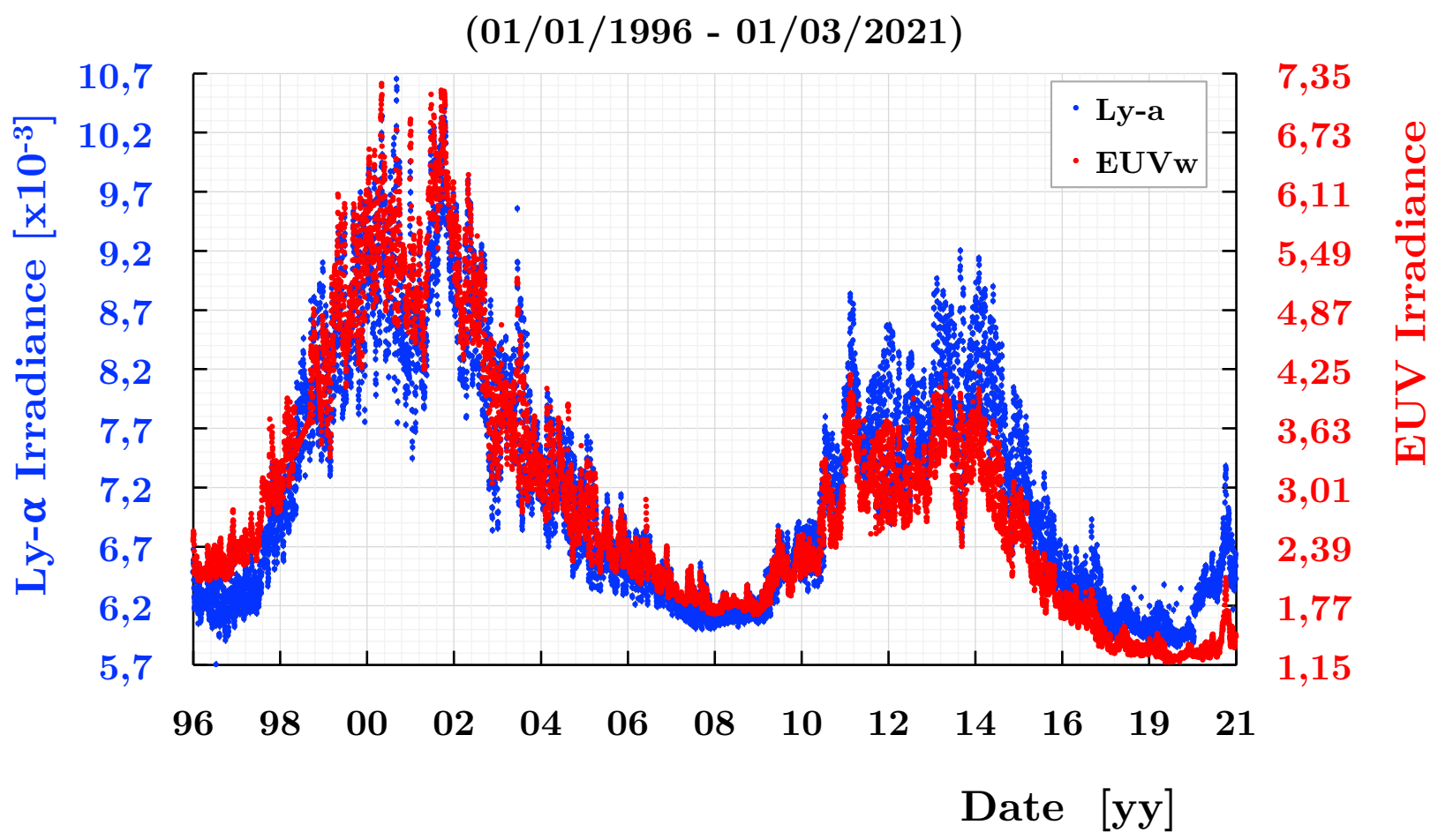

**Figure B.51:** Daily Ly-$\alpha$ and EUV solar irradiances for the same period 01/01/1996 - 01/03/2021.

### B.8.1.1 Single planets

The next step is to compare the longitudinal distributions resulting from single planets for the two datasets. This comparison is shown in Fig. B.52 and B.53 where we see similar distributions for the two cases. Indeed, a high degree of positive correlation in all distributions is found, with the results being all statistically significant at the 0.05 level. More specifically, the various Pearson correlation coefficients with the corresponding p-values are ($r_{a,d} = 0.70,\ p_{a,d} = 3.5 \times 10^{-3}$), ($r_{b,e} = 0.92,\ p_{b,e} = 1.27 \times 10^{-6}$), ($r_{c,f} = 0.95,\ p_{c,f} = 4.43 \times 10^{-8}$), for Mercury, Venus and Earth in Fig. B.52 and ($r_{a,c} = 0.93,\ p_{a,c} = 5.10 \times 10^{-7}$), ($r_{b,d} = 0.99,\ p_{b,d} = 0$) for Mars and Jupiter in Fig. B.53 respectively. It is noted that Saturn is not compared as it has not completed a full revolution around the Sun during the period of 25 y used here.

An additional important observation comes from the comparison of Fig. B.52a, B.52b, B.52c, B.53a and B.53b with Fig. 12.3a through 12.3e which correspond to the full period of 74 y. We observe that for the case of Mercury the peak around 300° becomes more pronounced in the last 25 y, whereas for Earth only the peak around 90°, from Fig. B.52c survives but it shifts towards lower heliocentric longitudes by about 30°. In the case of Jupiter, the wide peak from Fig. 12.3e shifts by about 90° from $\sim 160°$ to $\sim 70°$. Lastly, in all cases of the period 01/01/1996 - 01/03/2021 the observed amplitudes are slightly higher. This behaviour is expected since different planetary configurations take place in different years especially

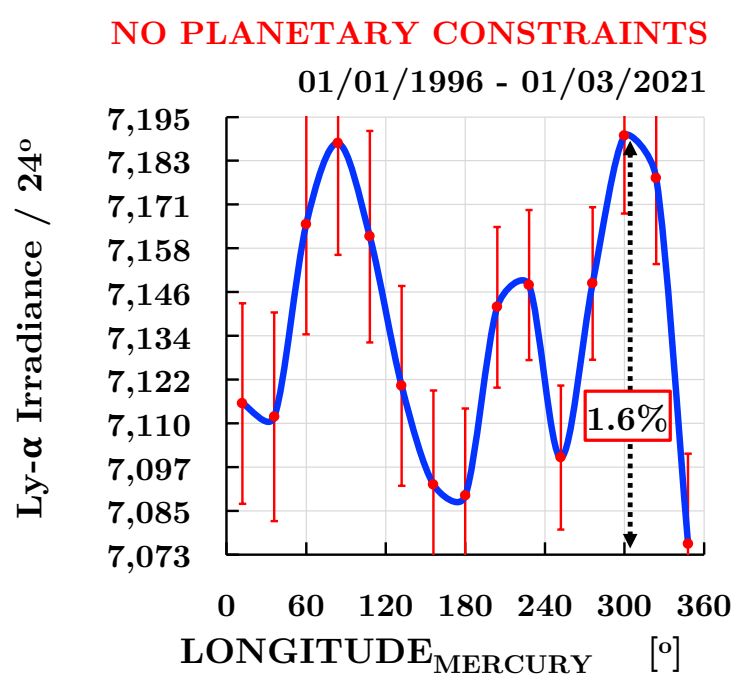

**(a)** Ly-$\alpha$ distribution for Mercury.

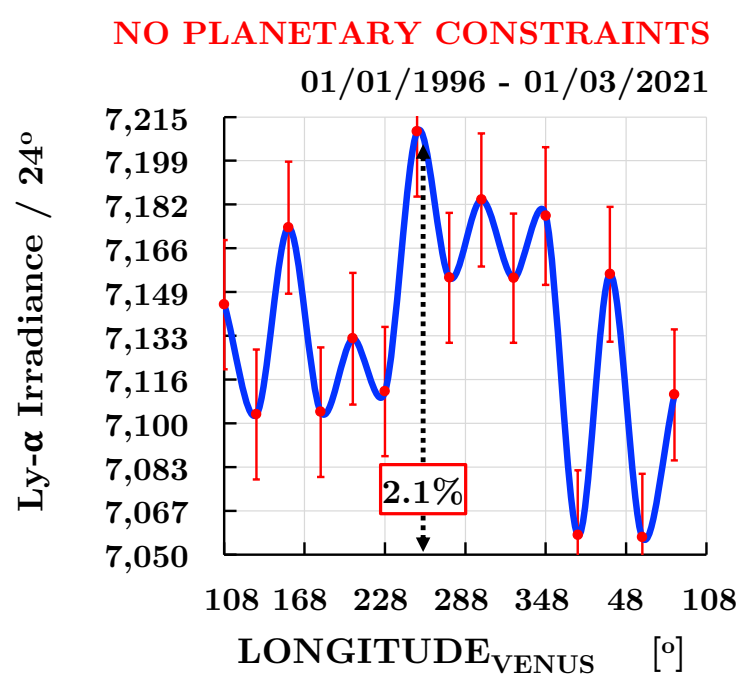

**(b)** Ly-$\alpha$ distribution for Venus.

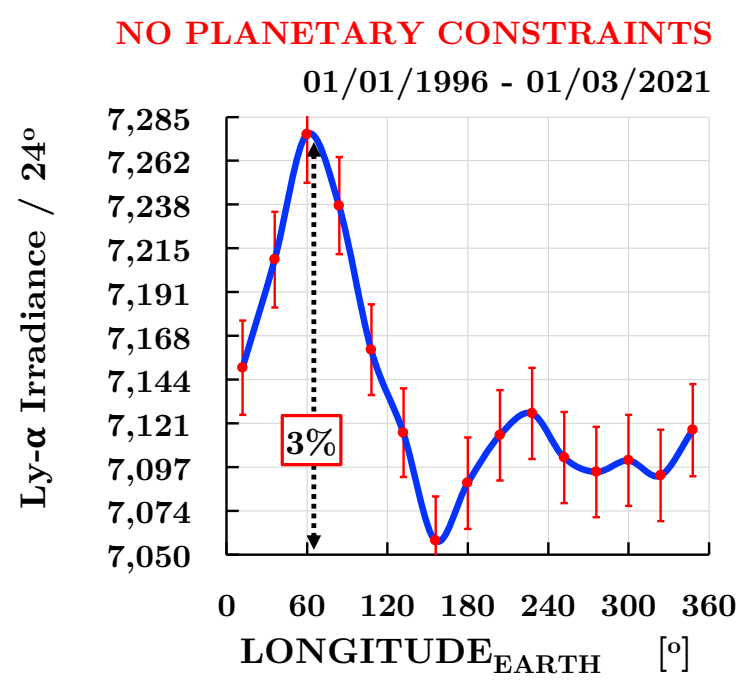

**(c)** Ly-$\alpha$ distribution for Earth.

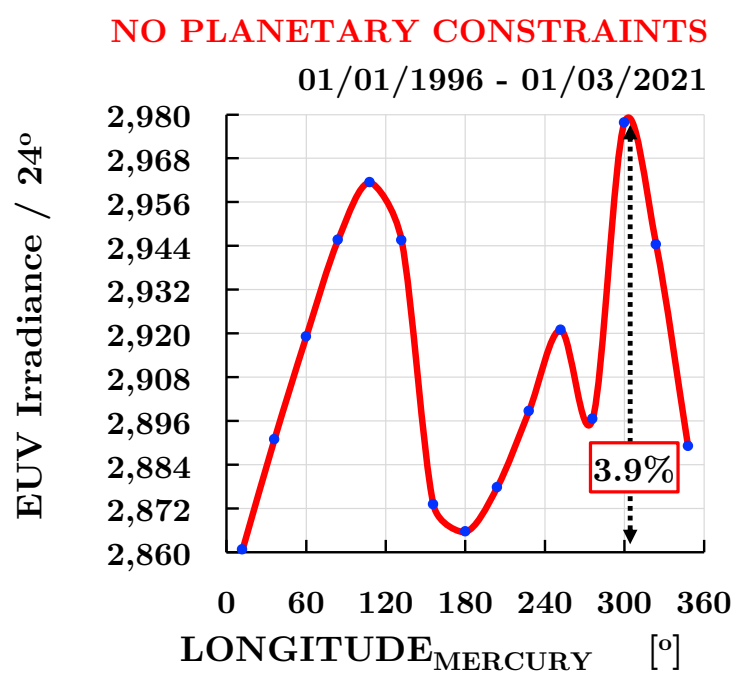

**(d)** EUV distribution for Mercury.

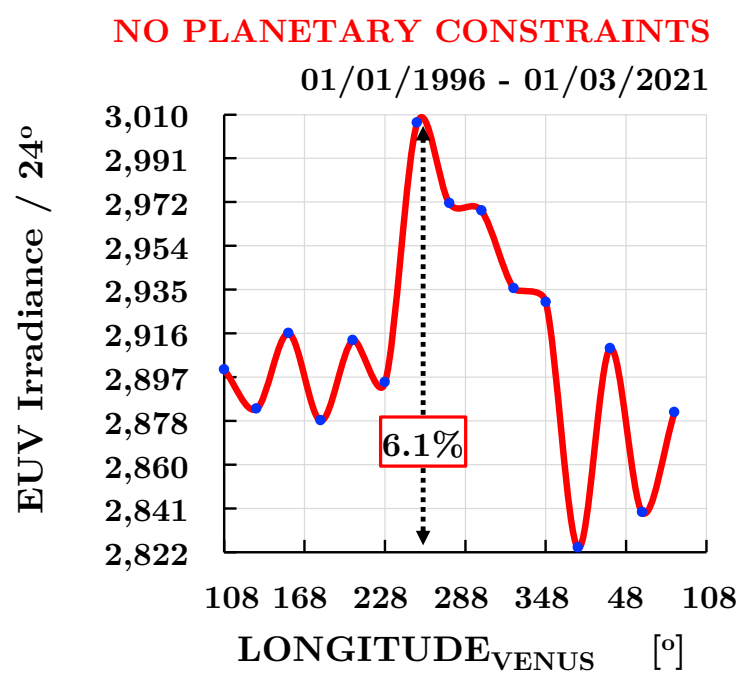

**(e)** EUV distribution for Venus.

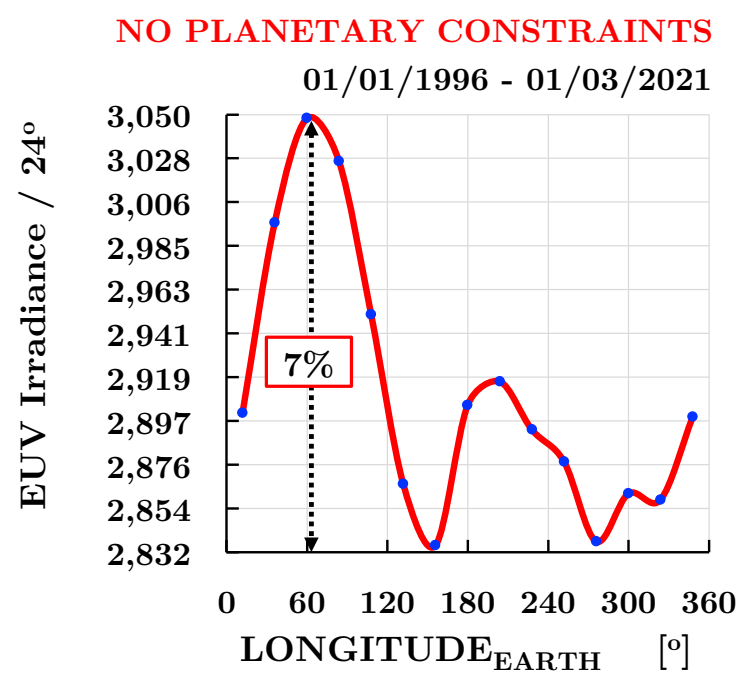

**(f)** EUV distribution for Earth.

**Figure B.52:** Comparison of inner planet's distributions of Ly-$\alpha$ vs. EUV irradiance for the period 01/01/1996 - 01/03/2021 with 24° bin.

regarding the outer planets which can influence strongly the resulting distributions.

### B.8.1.2 Combining planets

The following step involves the comparison of the distributions resulting from planetary combinations. In Fig. B.54 the case from Fig. 12.4 is shown. Firstly, we observe that the distributions are similar for both Ly-$\alpha$ and EUV with the correlation coefficients pointing to a strong positive linear correlation ($r_{a,c} = 0.86$, $p_{a,c} = 9.01 \times 10^{-7}$) for the case of Mars being constrained between 250° to 10° and a moderate but significant correlation ($r_{a,c} = 0.47$, $p_{a,c} = 3.90 \times 10^{-2}$) for the case of Mars being constrained between 70° to 190°.

Furthermore, for the 25 y period the distributions no longer show the same planetary relationship as in full 74 y presented in Fig. 12.4c and 12.4d. This time the distribution without constraints and with constraints, in Fig. B.54b, does not show a significant difference but also in the comparison between the two opposite orbital arcs of Mars in Fig. B.54a we observe the opposite behaviour to Fig. 12.4c. As mentioned, this observation could be explained the differences induced by different solar cycles to the various planetary distributions and subsequently due to the different locations of the outer planets such as Saturn, Uranus and Neptune which rotate much slower than the inner ones and are in different locations during

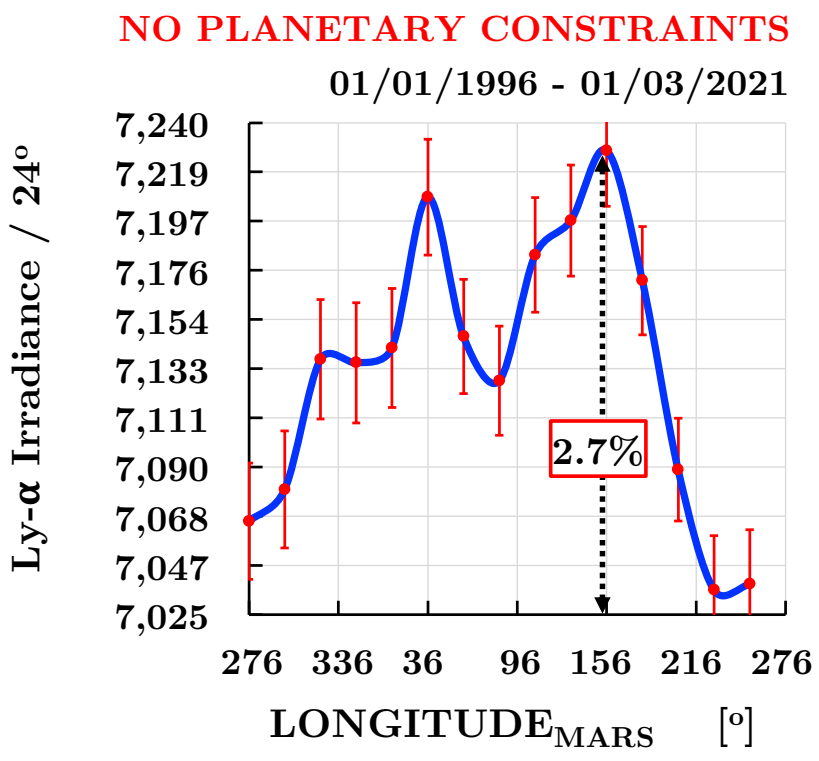

**(a)** Ly-$\alpha$ distribution for Mars with 24° bin.

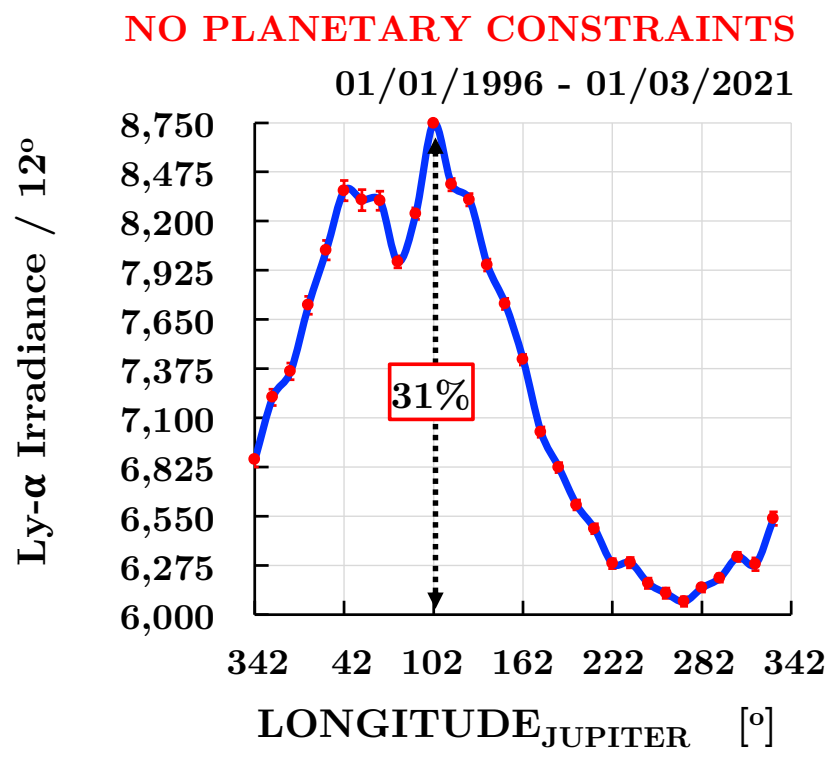

**(b)** Ly-$\alpha$ distribution for Jupiter with 12° bin.

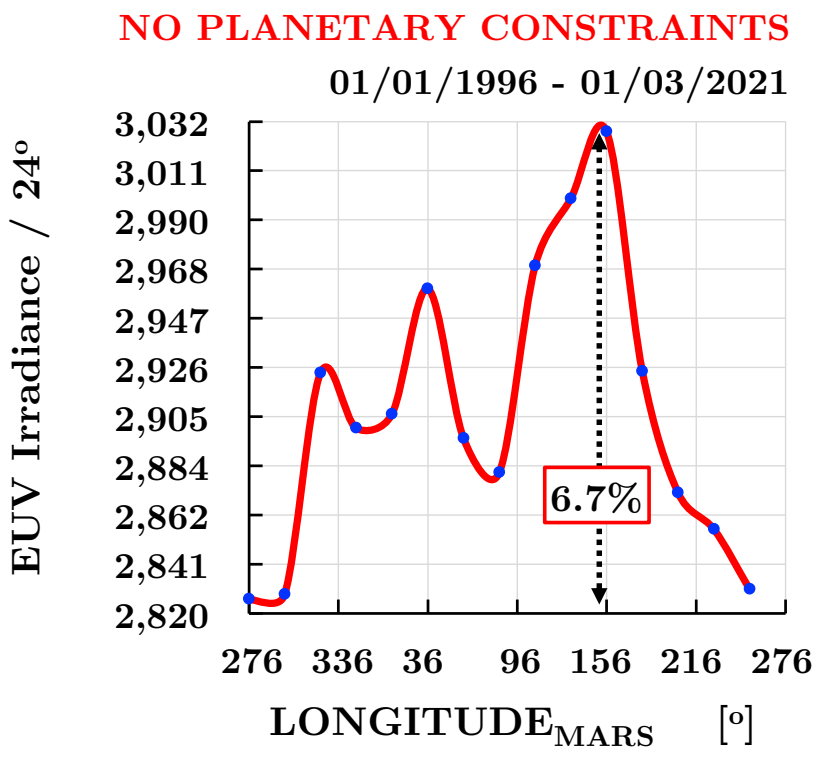

**(c)** EUV distribution for Mars with 24° bin.

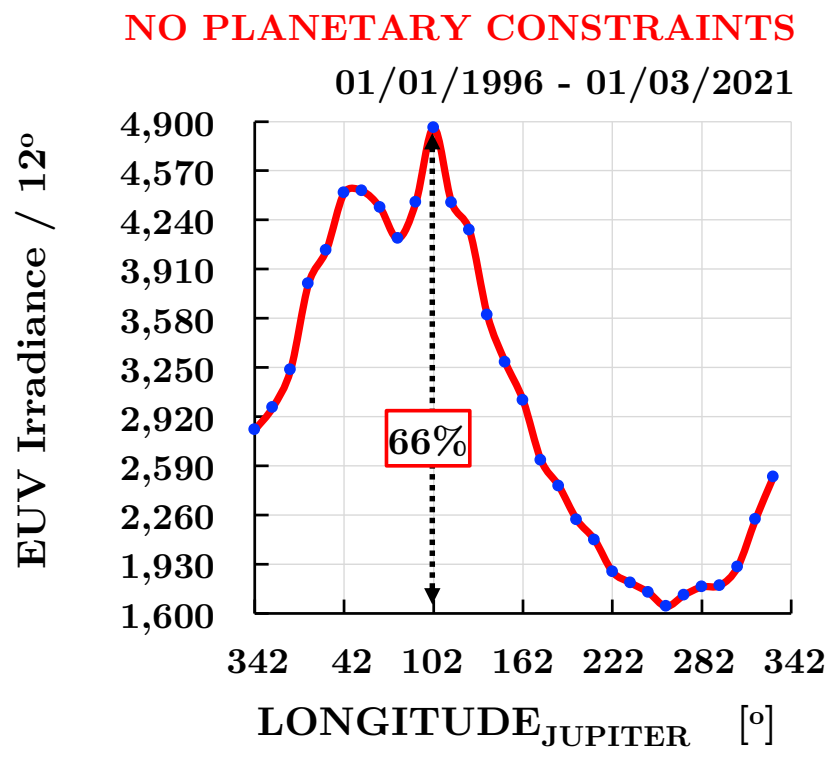

**(d)** EUV distribution for Jupiter with 12° bin.

**Figure B.53:** Comparison of Mars' and Jupiter's distributions of Ly-$\alpha$ vs. EUV irradiance for the period 01/01/1996 - 01/03/2021.

each solar cycle influencing in different manners the incoming streams of invisible matter.

In Fig. B.55 we compare the two solar irradiances for the reference frame of Venus when Earth and Mars are constrained in a 100°-wide window. A strong correlation is found for all the cases with the values of the correlation coefficients and p-values being ($r_{a,d} = 0.88$, $p_{a,d} = 2.24 \times 10^{-10}$), ($r_{b,e} = 0.91$, $p_{b,e} = 5.52 \times 10^{-12}$) and ($r_{c,f} = 0.98$, $p_{c,f} = 0$) for each compared column of plots respectively.

The same procedure is performed for Fig. 12.6 and is presented on Fig. B.56. In this case the correlation between Ly-$\alpha$ and EUV for the case of Mars propagating around $310° \pm 60°$ is perfectly linear with ($r_{a,b} = 0.99$, $p_{a,b} = 0$).

Finally, in Fig. B.57 the distribution from Fig. 12.7 is compared for the two datasets. Once more, the statistical output shows a statistically significant strong linear correlation, with the Pearson correlation giving ($r_{a,b} = 0.99$, $p_{a,b} = 6.66 \times 10^{-16}$). It is noted that we also observe a shift of the center of the peak of about 47.43°, from 316.96° in Fig. 12.7 to 266.53° in Fig. B.57.

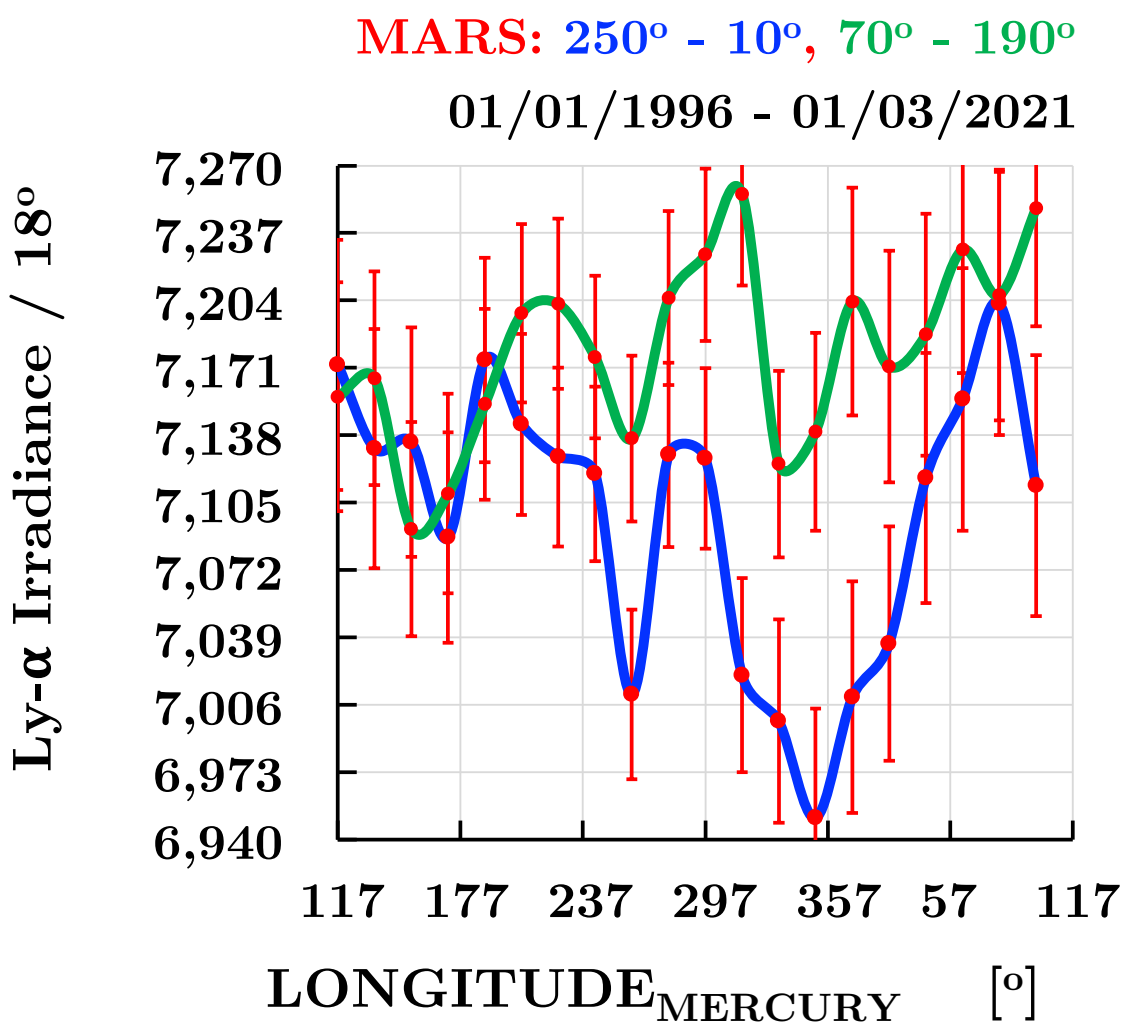

(a) Ly-$\alpha$ distribution for Mercury when Mars is between 250° to 10° and between 70° to 190°.

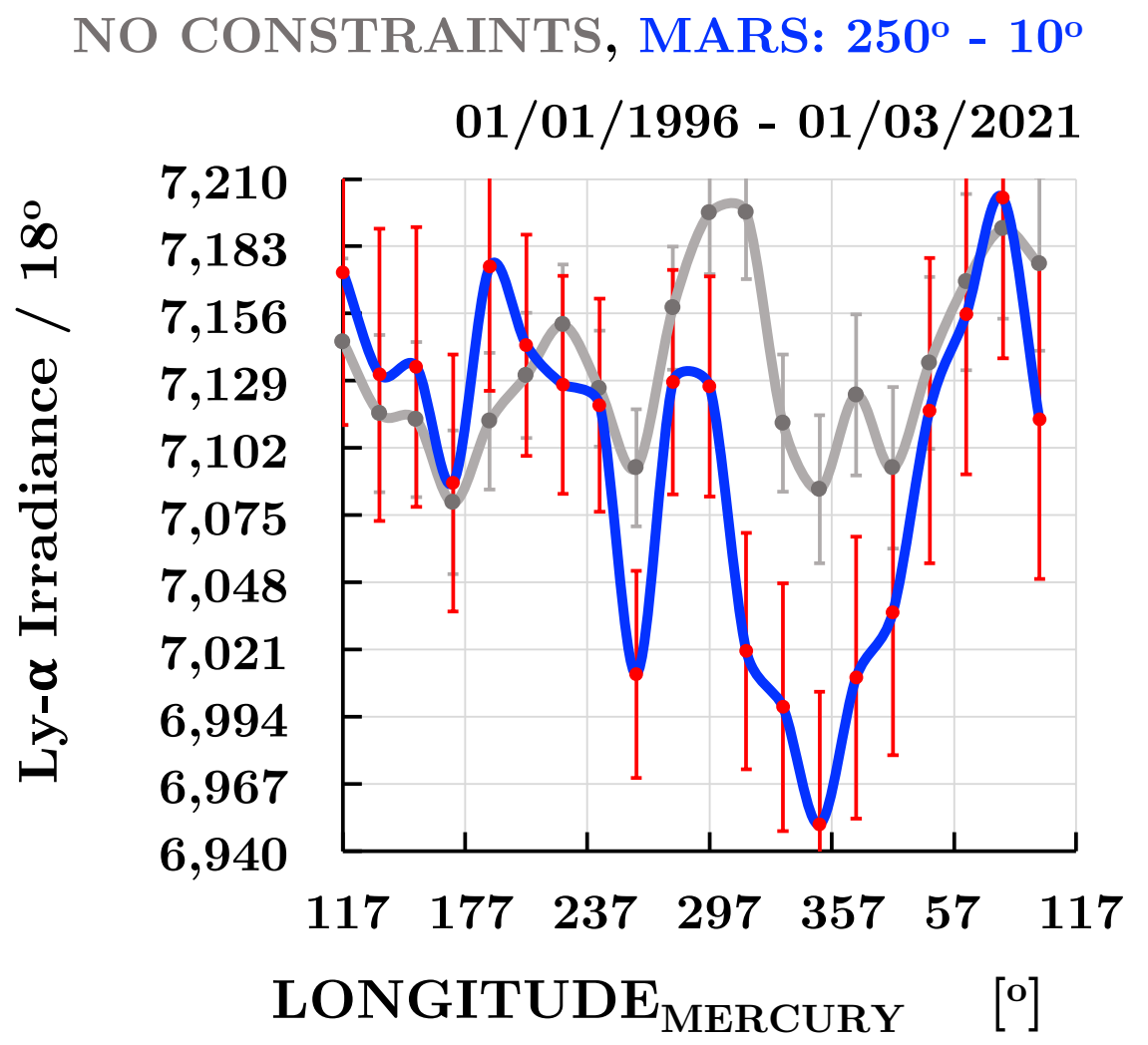

(b) Ly-$\alpha$ distribution for Mercury without constraints and when Mars is between 250° to 10°.

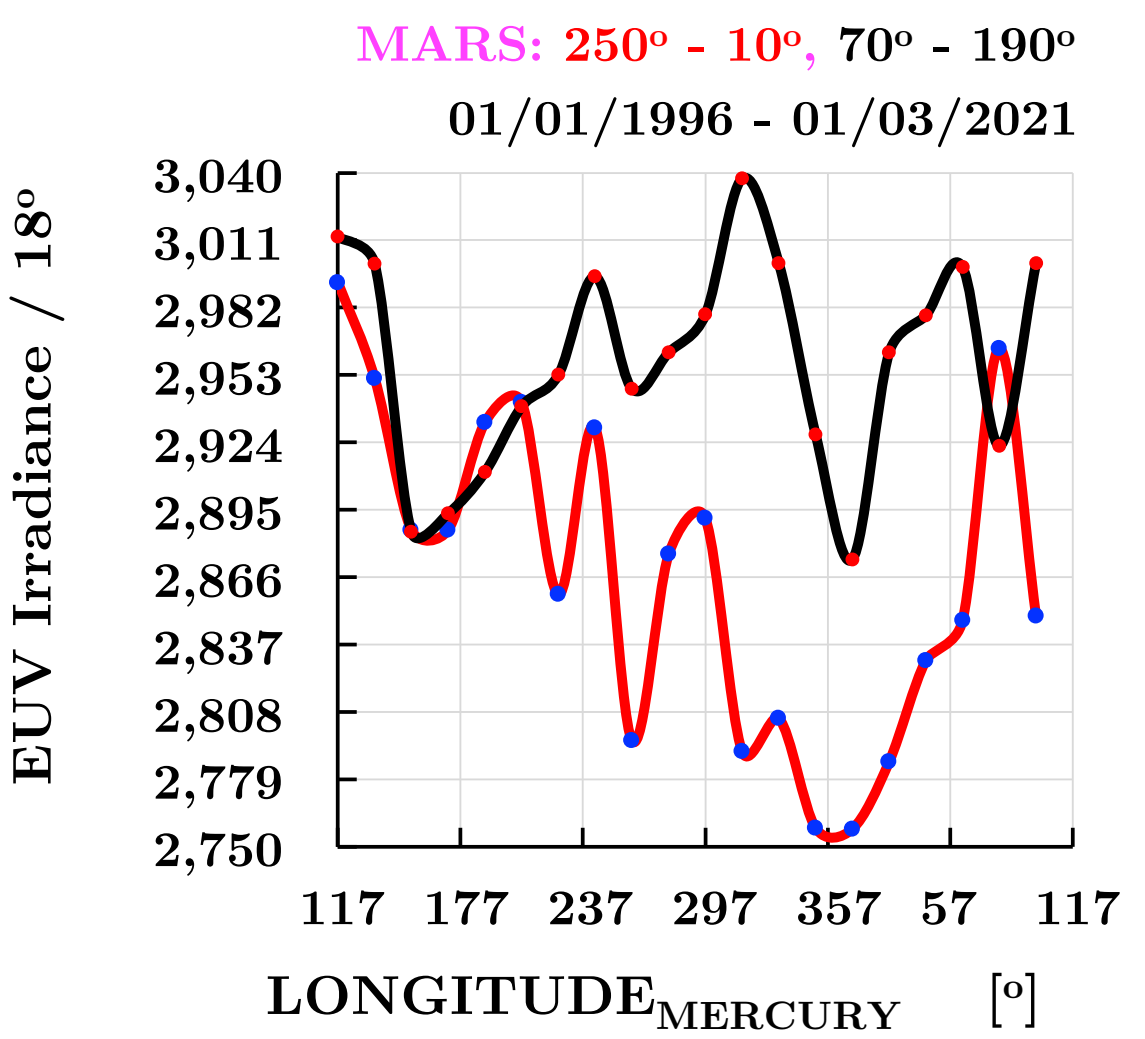

(c) EUV distribution for Mercury when Mars is between 250° to 10° and between 70° to 190°.

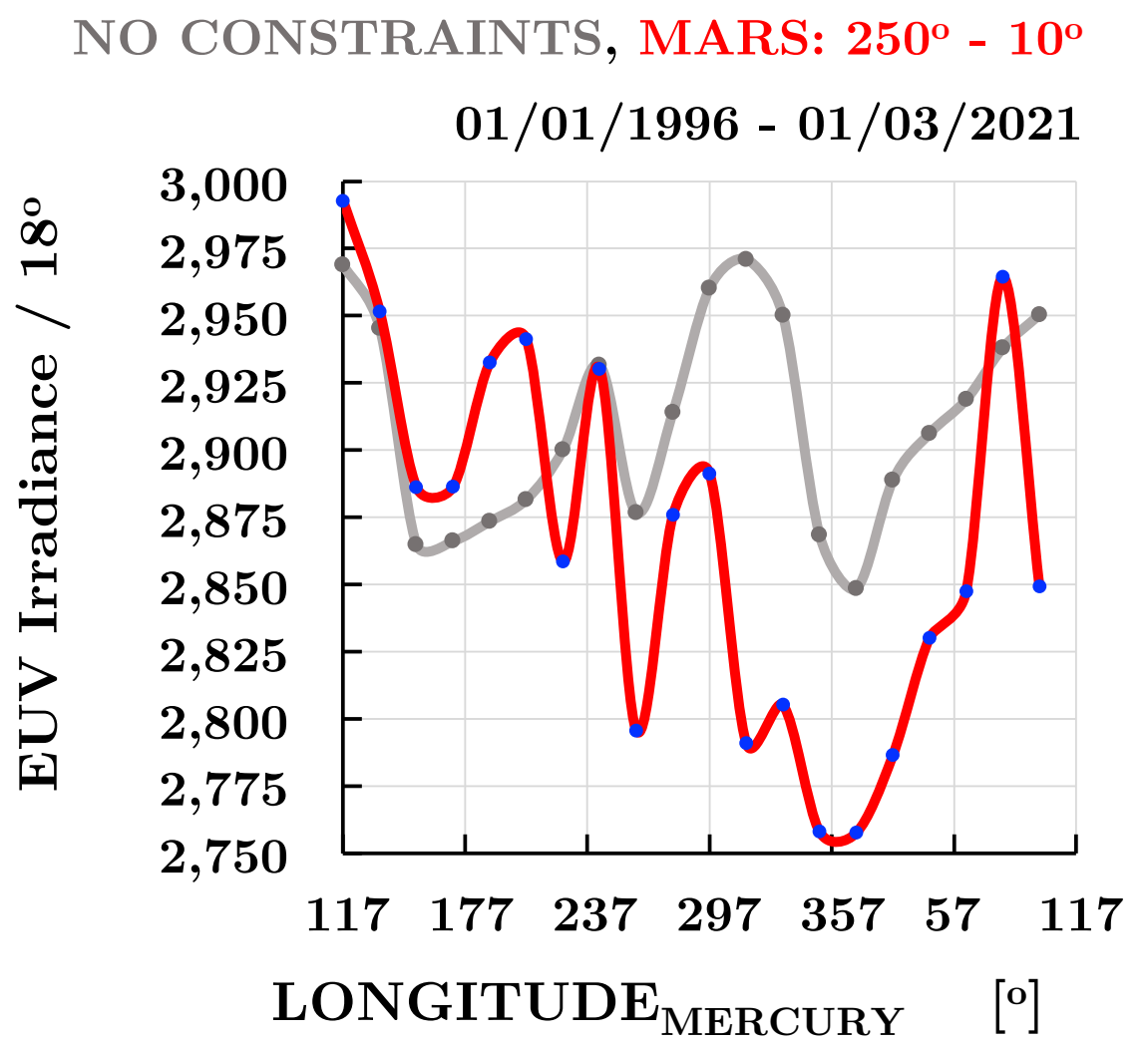

(d) EUV distribution for Mercury without constraints and when Mars is between 250° to 10°.

**Figure B.54:** Comparison of distributions of Ly-$\alpha$ vs. EUV irradiance for the reference frame of Mercury without constraints and when Mars is constrained to propagate in two 180° opposite regions with bin = 18° for the period 01/01/1996 - 01/03/2021.

## B.8.2 Comparison with F10.7

The next comparison is made with an alternative proxy of the solar activity, namely the F10.7 solar radio flux which has already been shown in the literature to exhibit a long-term correlation and a smaller short-term correlation for a variation of 27 d [323]. F10.7 index spans from 28/11/1963 to 03/03/2021. Thus, since this period was overlapping in both datasets it was also used for the comparison with Ly-$\alpha$. The two time series are presented in Fig. B.58 where we see an almost perfect overlapping. The statistical correlation analysis yields ($r = 0.94$,

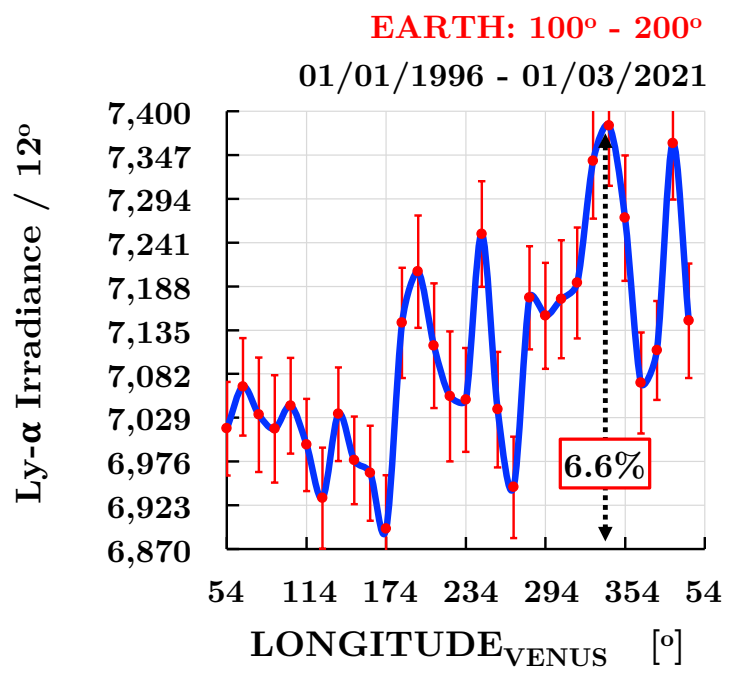

**(a)** Ly-$\alpha$ distribution for Venus when Earth is between 100° to 200°.

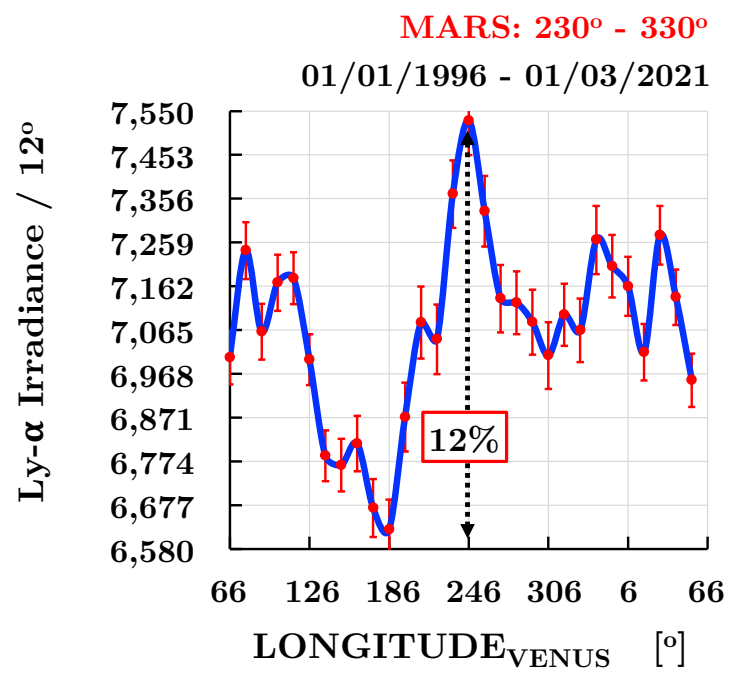

**(b)** Ly-$\alpha$ distribution for Venus when Mars is between 230° to 330°.

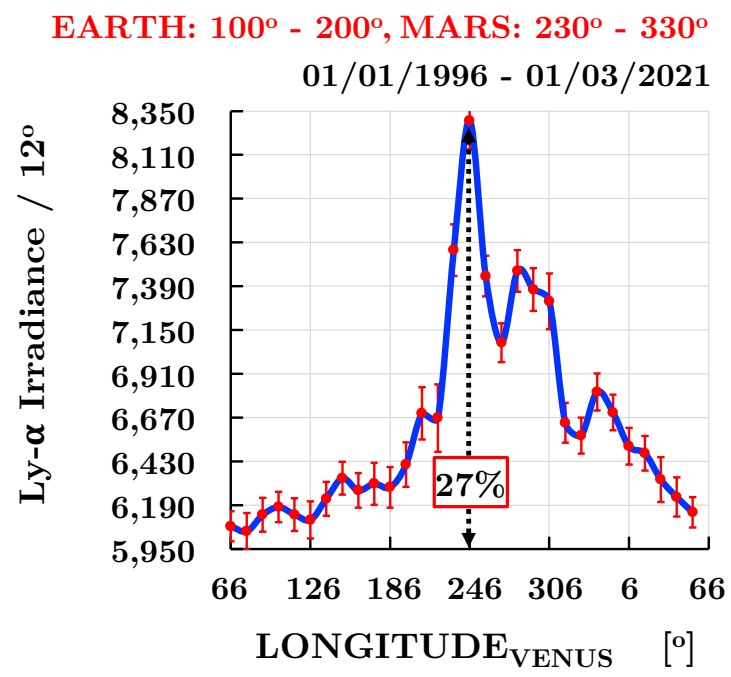

**(c)** Ly-$\alpha$ distribution for Venus when Earth is between 100° to 200° and at the same time Mars is between 230° to 330°.

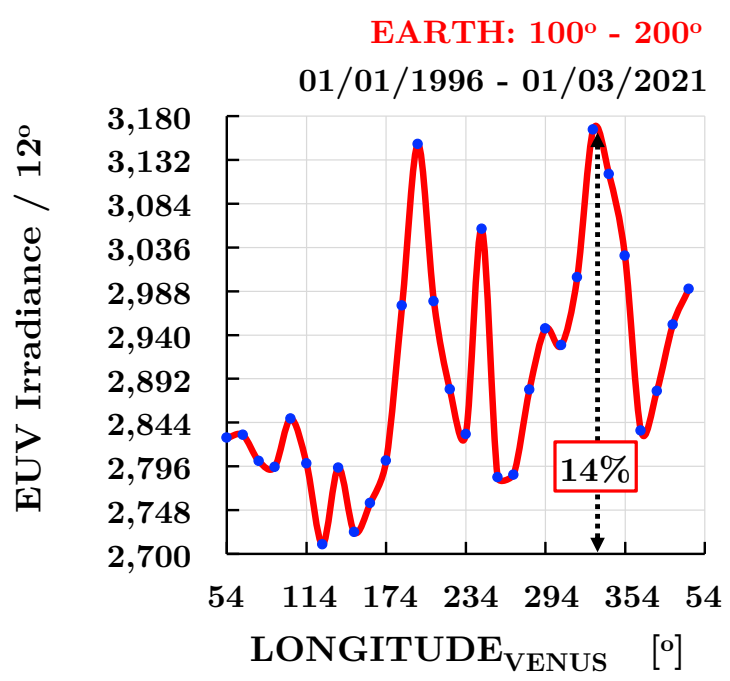

**(d)** EUV distribution for Venus while Earth is between 100° to 200°.

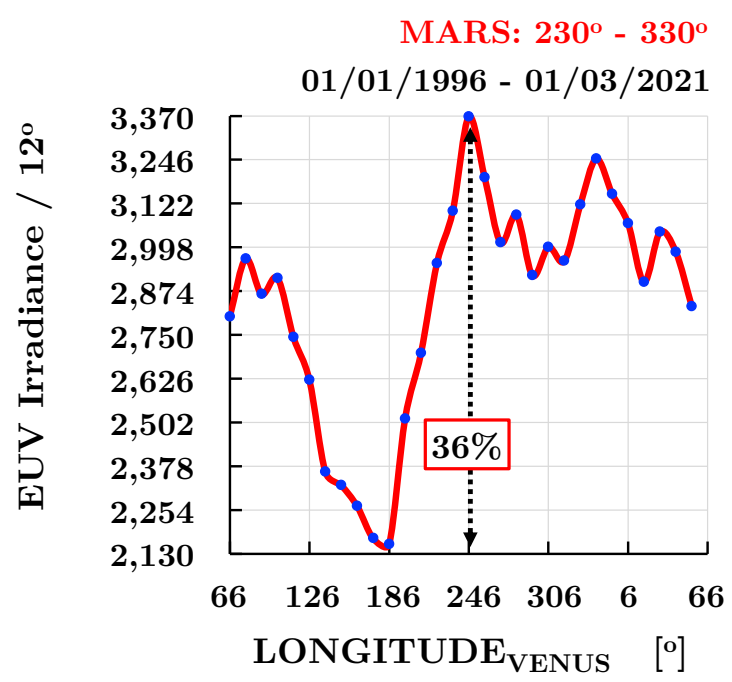

**(e)** EUV distribution for Venus when Mars is between 230° to 330°.

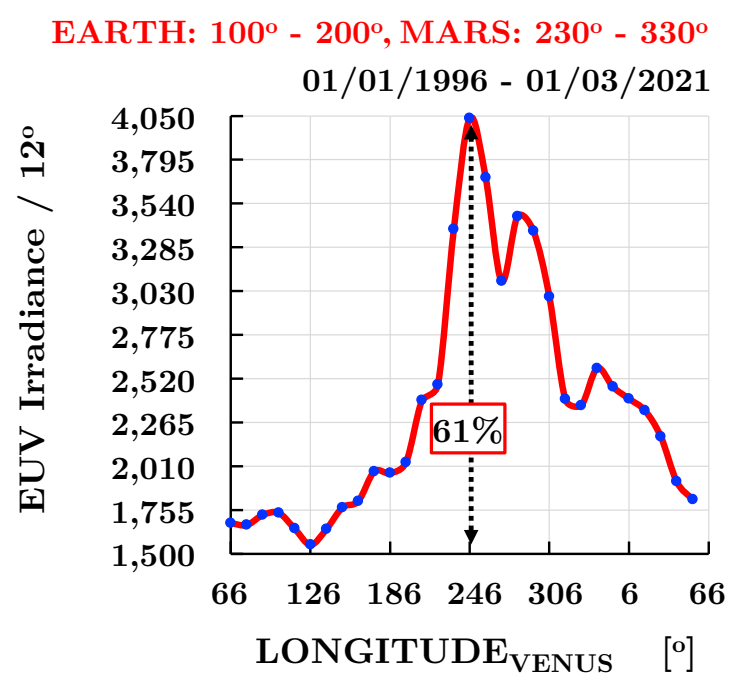

**(f)** EUV distribution for Venus when Earth is between 100° to 200° and at the same time Mars is between 230° to 330°.

**Figure B.55:** Comparison of distributions of Ly-$\alpha$ vs. EUV irradiance for the reference frame of Venus when applying various constraints on the positions of other planets with bin = 12° for the period 01/01/1996 - 01/03/2021.

$p = 0$) which, as with the case of EUV, shows a strong positive linear correlation on the general trend of Ly-$\alpha$ and F10.7.

#### B.8.2.1 Single planets

The single planet distributions from Fig. 12.3 are the first ones to be compared with F10.7. The results are shown in Fig. B.59 and B.60. A statistical correlation analysis between the two datasets show a moderate correlation for Mercury with ($r_{a,e} = 0.53,\ p_{a,e} = 40.20 \times 10^{-2}$) and a strong one for the rest of the cases. More specifically the various Pearson correlation coefficients and p-values give ($r_{b,f} = 0.89,\ p_{b,f} = 7.03 \times 10^{-6}$), ($r_{c,g} = 0.58,\ p_{c,g} = 2.40 \times 10^{-2}$), ($r_{d,h} = 0.88,\ p_{d,h} = 1.83 \times 10^{-5}$) for Venus, Earth, and Mars and an almost perfect linear correlation ($r_{a,c} = 0.99,\ p_{a,c} = 0$) and ($r_{b,d} = 0.98,\ p_{b,d} = 0$) for Jupiter and Saturn respectively. We also note that in Fig. B.59a through B.59d and Fig. B.60a and B.60b the distributions

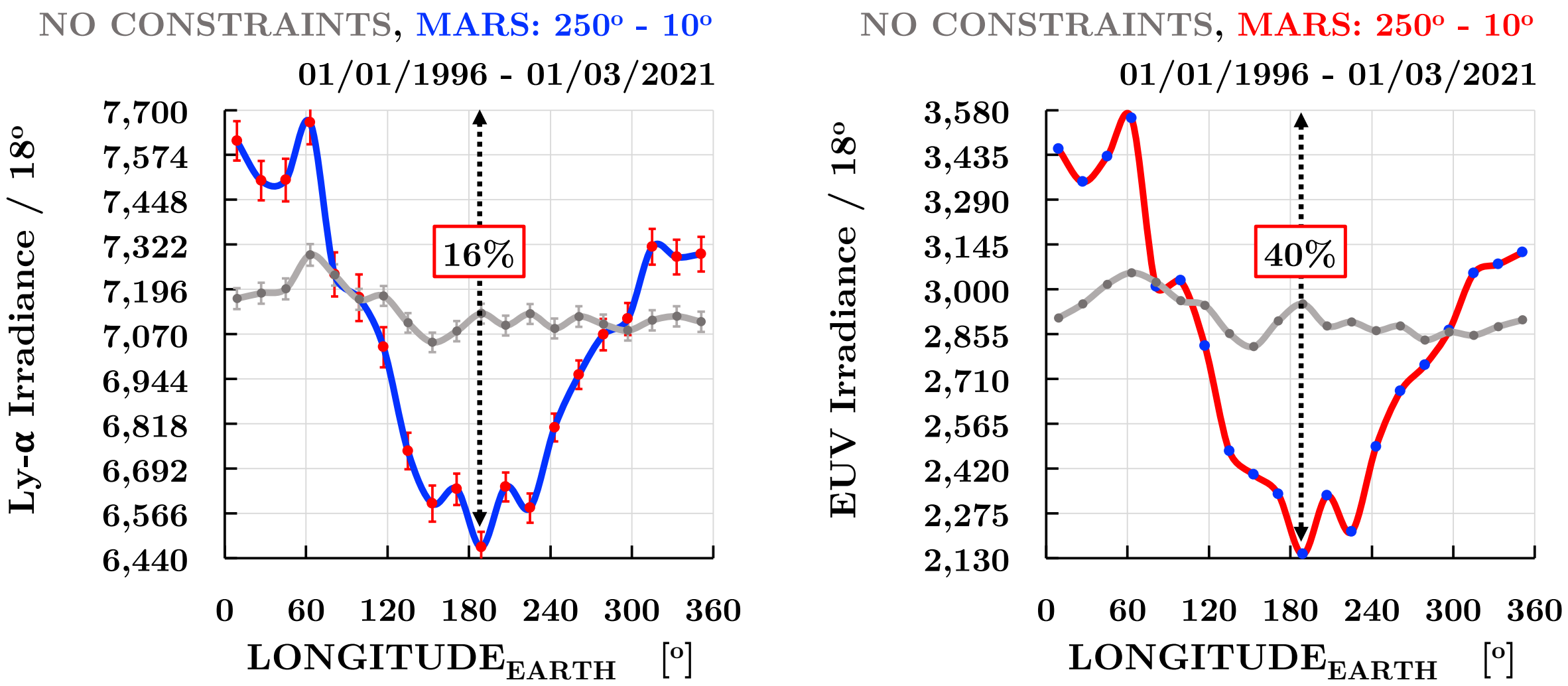

**(a)** Ly-$\alpha$ distribution. **(b)** EUV distribution.

**Figure B.56:** Comparison of distributions of Ly-$\alpha$ vs. EUV irradiance for the reference frame of Earth without constraints and when Mars is constrained to move between 250° to 10° with bin = 18° for the period 01/01/1996 - 01/03/2021.

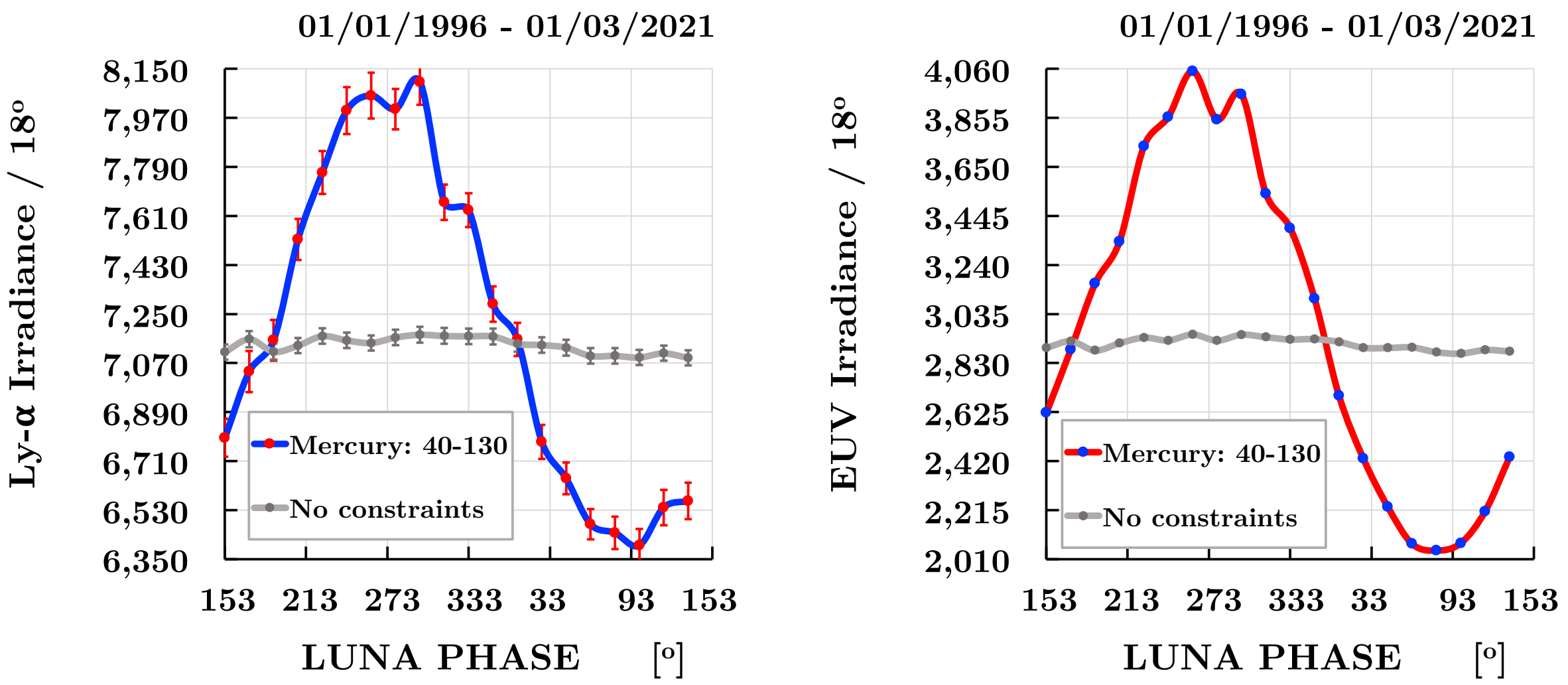

**(a)** Ly-$\alpha$ distribution. **(b)** EUV distribution.

**Figure B.57:** Comparison of distributions of Ly-$\alpha$ vs. EUV irradiance as a function of Moon's phase without constraints and when Mercury is constrained to move between 40° to 130° with bin = 18° for the period 01/01/1996 - 01/03/2021.

of Mercury, Venus, Jupiter and Saturn are similar to the ones in Fig. 12.3 where the whole period of 74 y was used, except for the cases of Earth and Mars.

#### B.8.2.2 Combining planets

Moving to combinations of planetary heliocentric longitude positions, in Fig. B.61 we observe a large positive linear relationship between Ly-$\alpha$ and EUV. The correlation coefficients are found to be ($r_{a,c} = 0.89$, $p_{a,c} = 1.13 \times 10^{-7}$) for the case of Mars being between 250° to

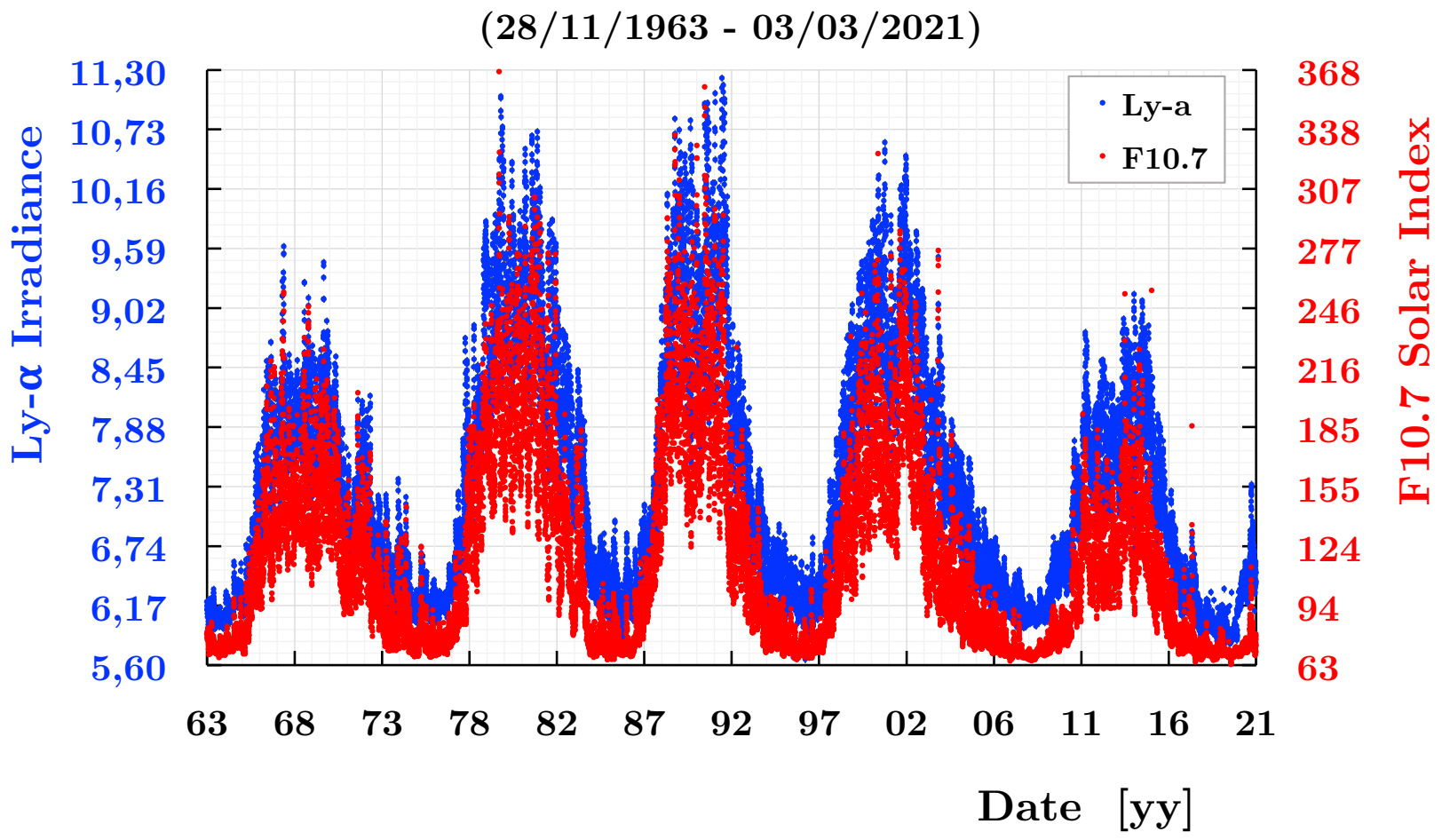

**Figure B.58:** Daily Ly-$\alpha$ irradiance and F10.7 solar index for the same period 28/11/1963 - 03/03/2021.

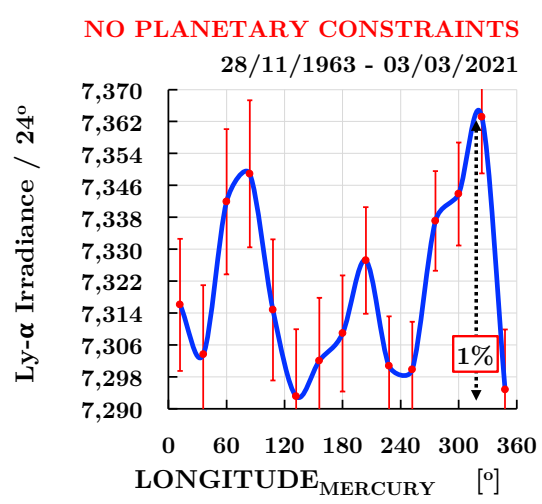

**(a)** Ly-$\alpha$ distribution for Mercury.

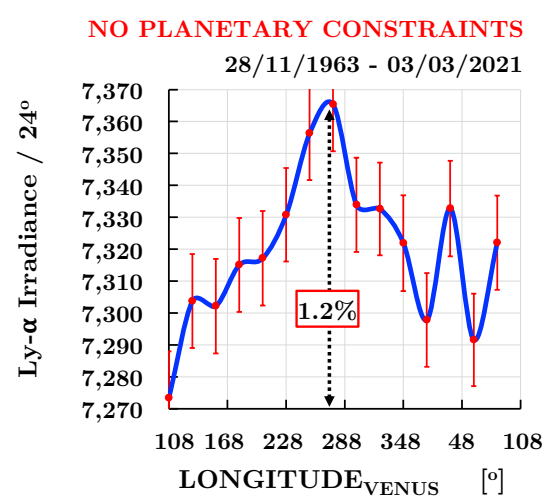

**(b)** Ly-$\alpha$ distribution for Venus.

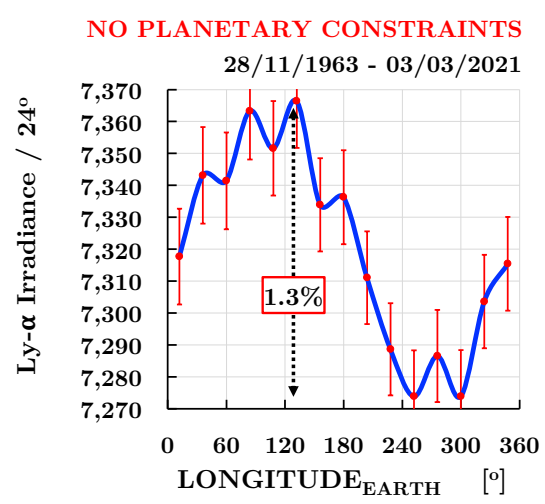

**(c)** Ly-$\alpha$ distribution for Earth.

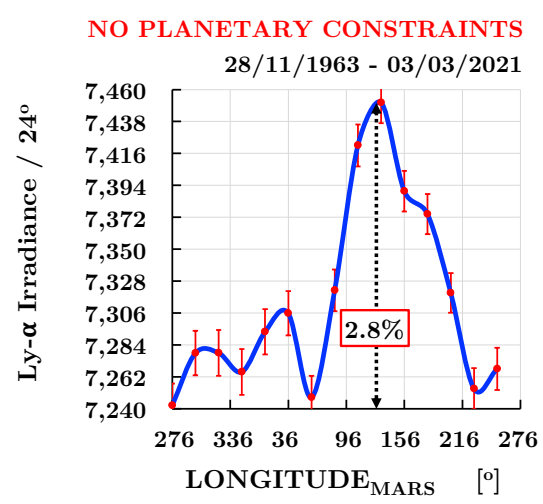

**(d)** Ly-$\alpha$ distribution for Mars.

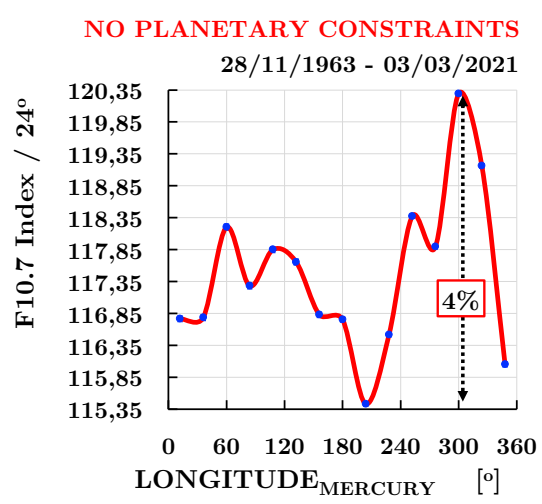

**(e)** F10.7 distribution for Mercury.

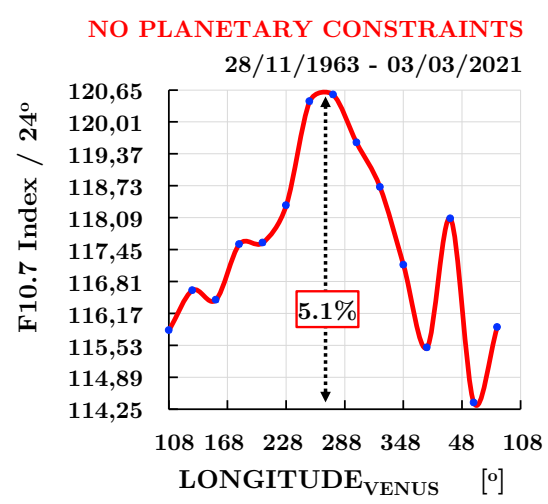

**(f)** F10.7 distribution for Venus.

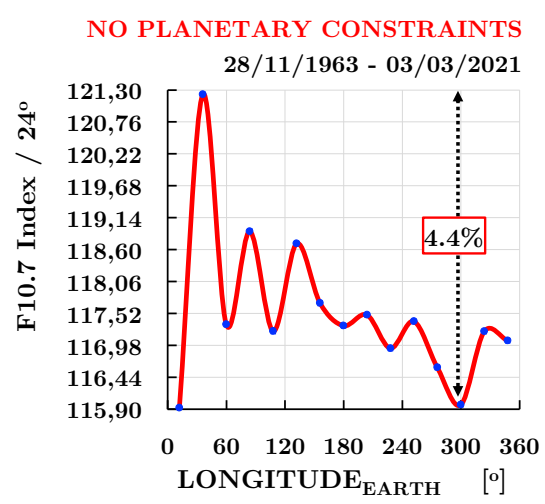

**(g)** F10.7 distribution for Earth.

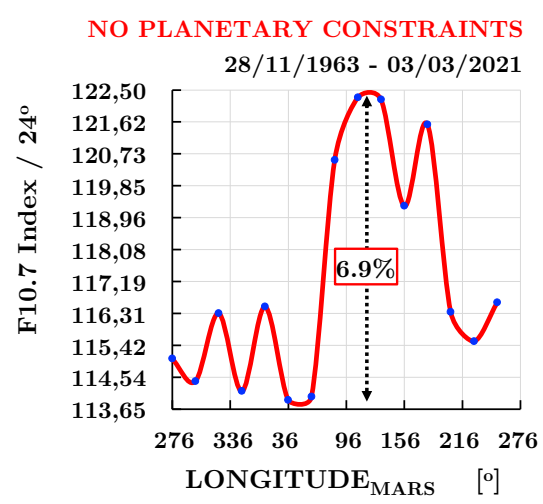

**(h)** F10.7 distribution for Mars.

**Figure B.59:** Comparison of planetary distributions of Ly-$\alpha$ vs. F10.7 solar radio flux irradiance for the period 28/11/1963 - 03/03/2021 with 24° bin.

10°, and ($r_{a,c} = 0.68,\ p_{a,c} = 8.79 \times 10^{-4}$) for the case of Mars being between 70° to 190° in Fig. B.61a and B.61c.

In Fig. B.62 the plots from Fig. 12.5 are reconstructed for Ly-$\alpha$ and F10.7 for the same period of 28/11/1963 - 03/03/2021. Likewise with the previous cases, we observe a strong linear relationship for all three distributions. The Pearson's correlation coefficients along with the corresponding p-values are ($r_{a,d} = 0.92,\ p_{a,d} = 1.01 \times 10^{-12}$), ($r_{b,e} = 0.87,\ p_{b,e} = 5.02 \times 10^{-10}$) and ($r_{c,f} = 0.99,\ p_{c,f} = 0$) for Fig. B.62a-B.62d, B.62b-B.62e and B.62c-B.62f respectively.

The spectrum of Earth from Fig. 12.6 is also compared with the F10.7 proxy in Fig. B.63. The calculated correlation when Mars propagates between 250° to 10° is ($r_{a,b} = 0.93,$

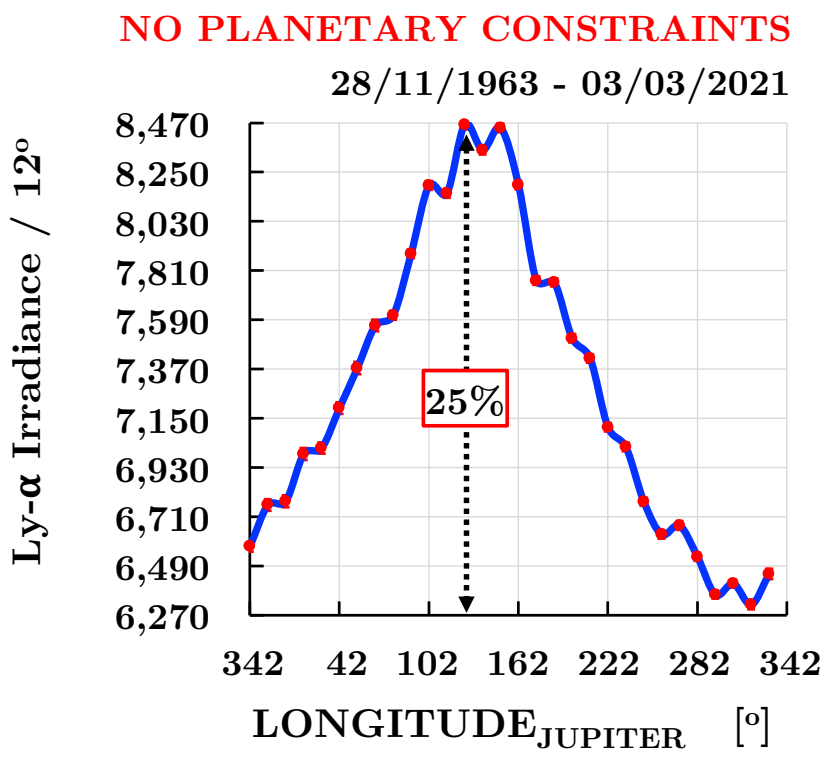

**(a)** Ly-$\alpha$ distribution for Jupiter.

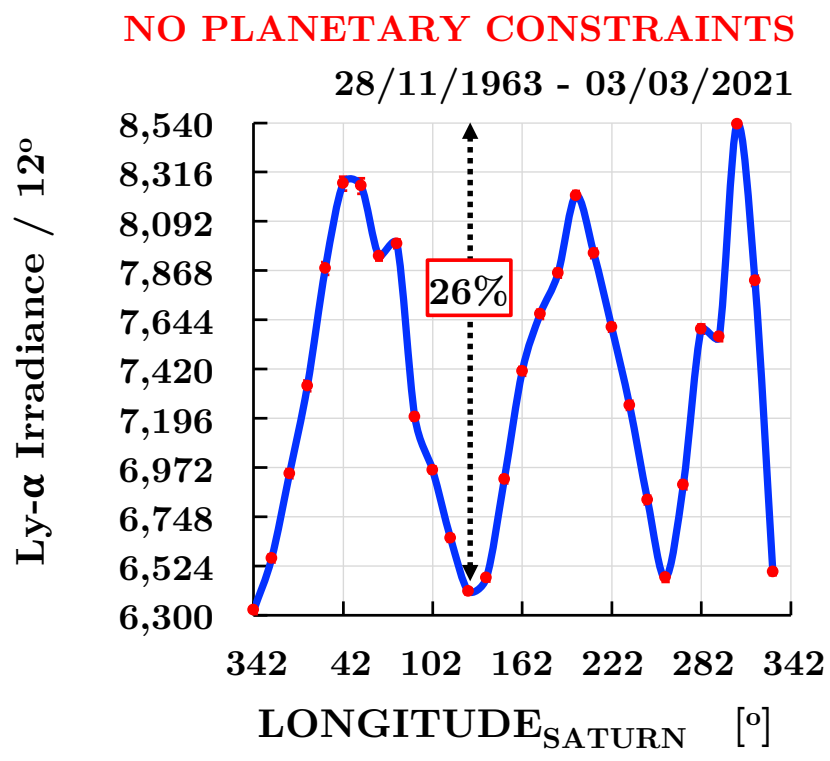

**(b)** Ly-$\alpha$ distribution for Saturn.

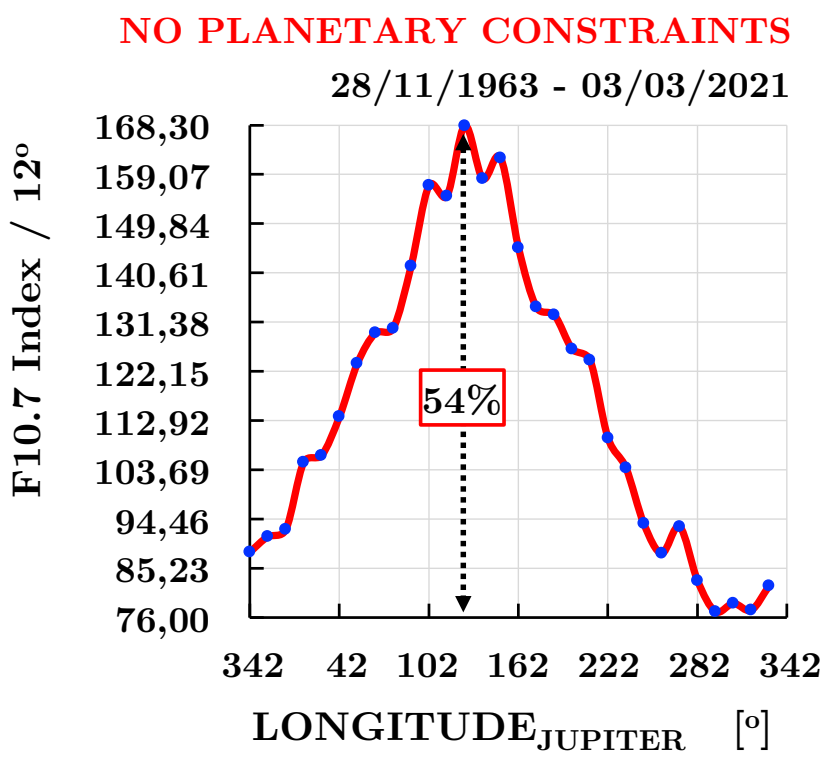

**(c)** F10.7 distribution for Jupiter.

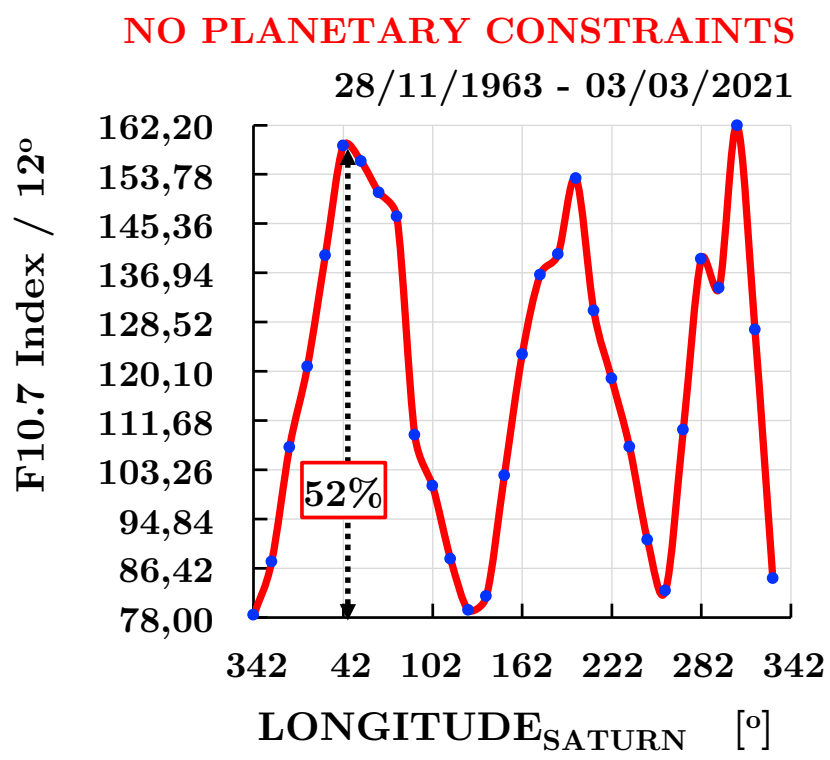

**(d)** F10.7 distribution for Saturn.

**Figure B.60:** Comparison of Jupiter and Saturn distributions of Ly-$\alpha$ vs. F10.7 solar radio flux for the period 28/11/1963 - 03/03/2021 with 12° bin.

$p_{a,b} = 2.34 \times 10^{-9}$). As in Fig. B.56, also in Fig. B.63 there is an observed difference with Fig. 12.6 where the blue spectrum seems to be shifted by about 180°.

Finally, Fig. 12.7c which pointed to a significant planetary relationship of Ly-$\alpha$ with the phase of the Moon when constrained by Mercury between 40° to 130°, is compared with the F10.7 solar proxy in Fig. B.64. The two datasets show a remarkable linear correlation of ($r_{a,b} = 0.99$, $p_{a,b} = 6.66 \times 10^{-16}$). The shift of the wide peak in this case is from 313.96° in Fig. 12.7c to 293.12° in Fig. B.64a as defined from a Gaussian fit function.

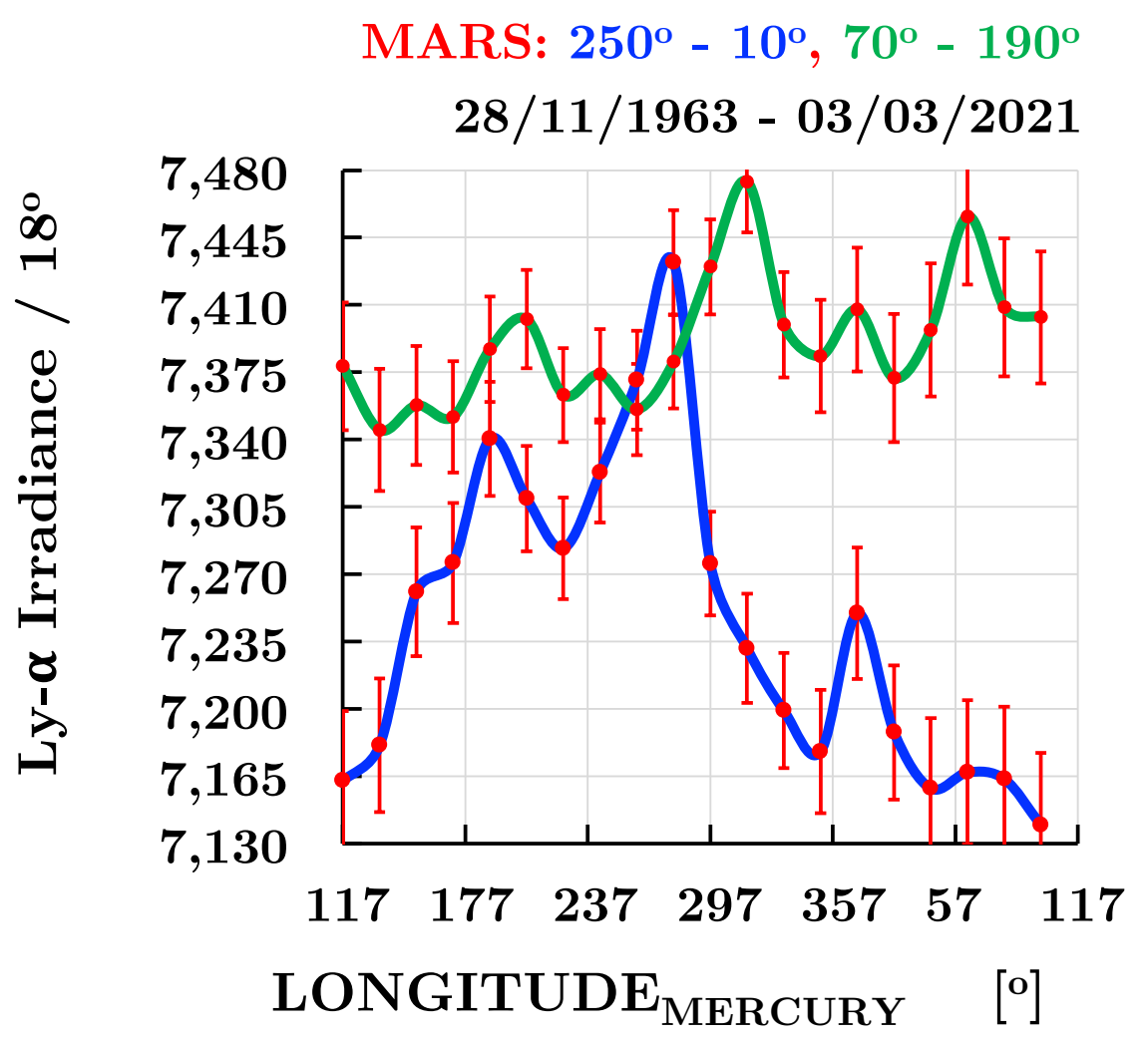

(a) Ly-$\alpha$ distribution for Mercury when Mars is between 250° to 10° and between 70° to 190°.

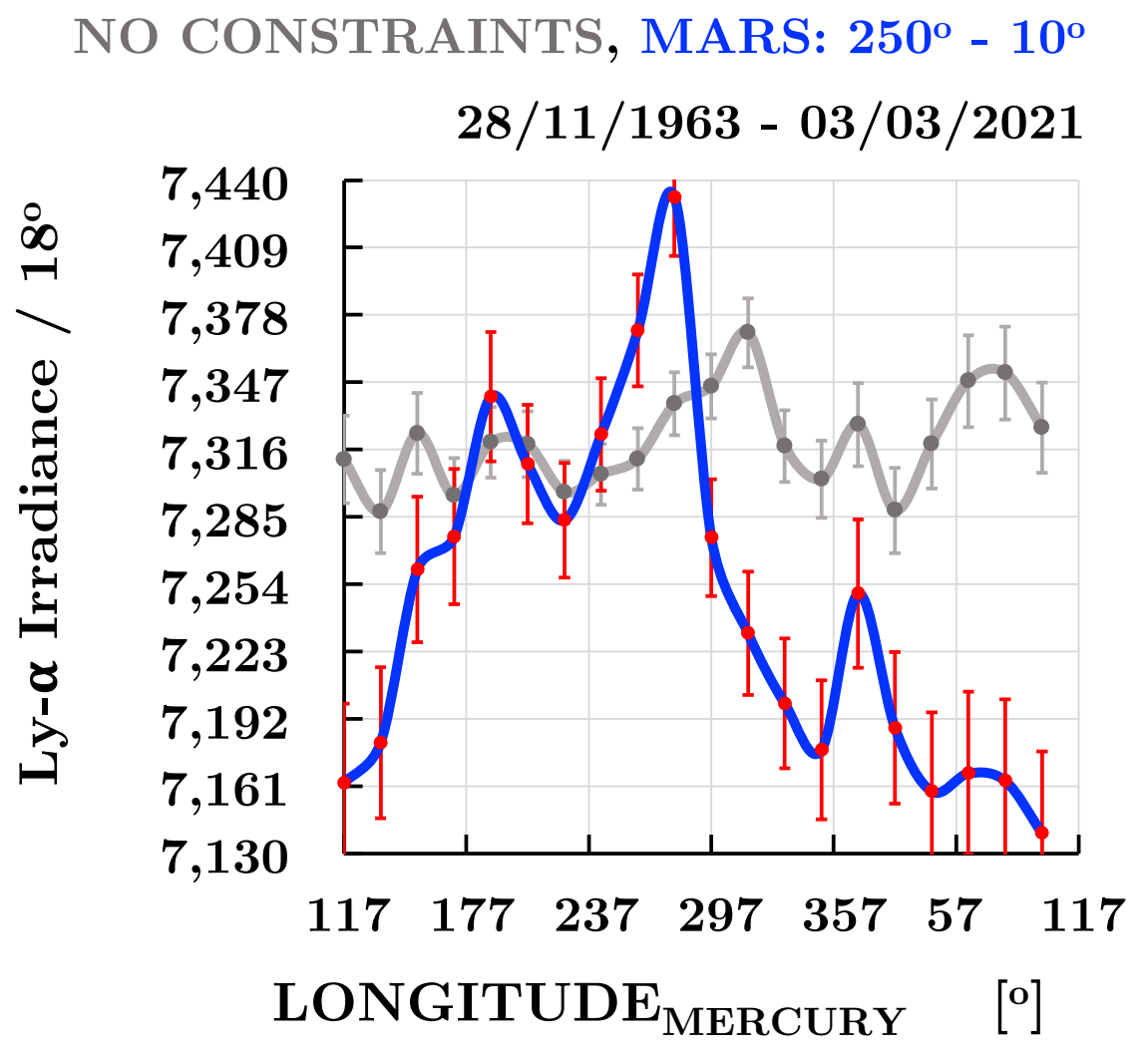

(b) Ly-$\alpha$ distribution for Mercury without constraints and when Mars is between 250° to 10°.

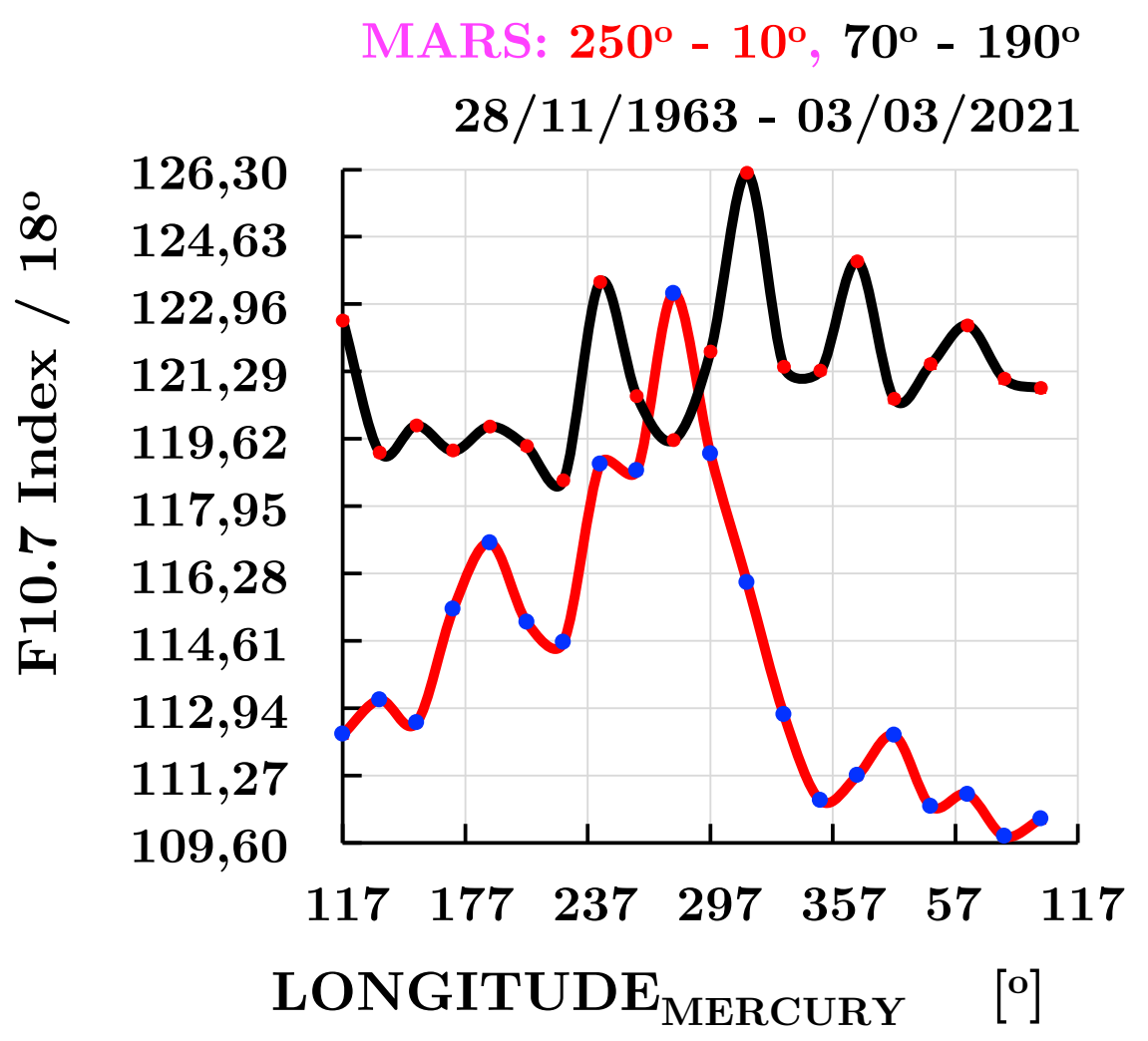

(c) F10.7 distribution for Mercury when Mars is between 250° to 10° and between 70° to 190°.

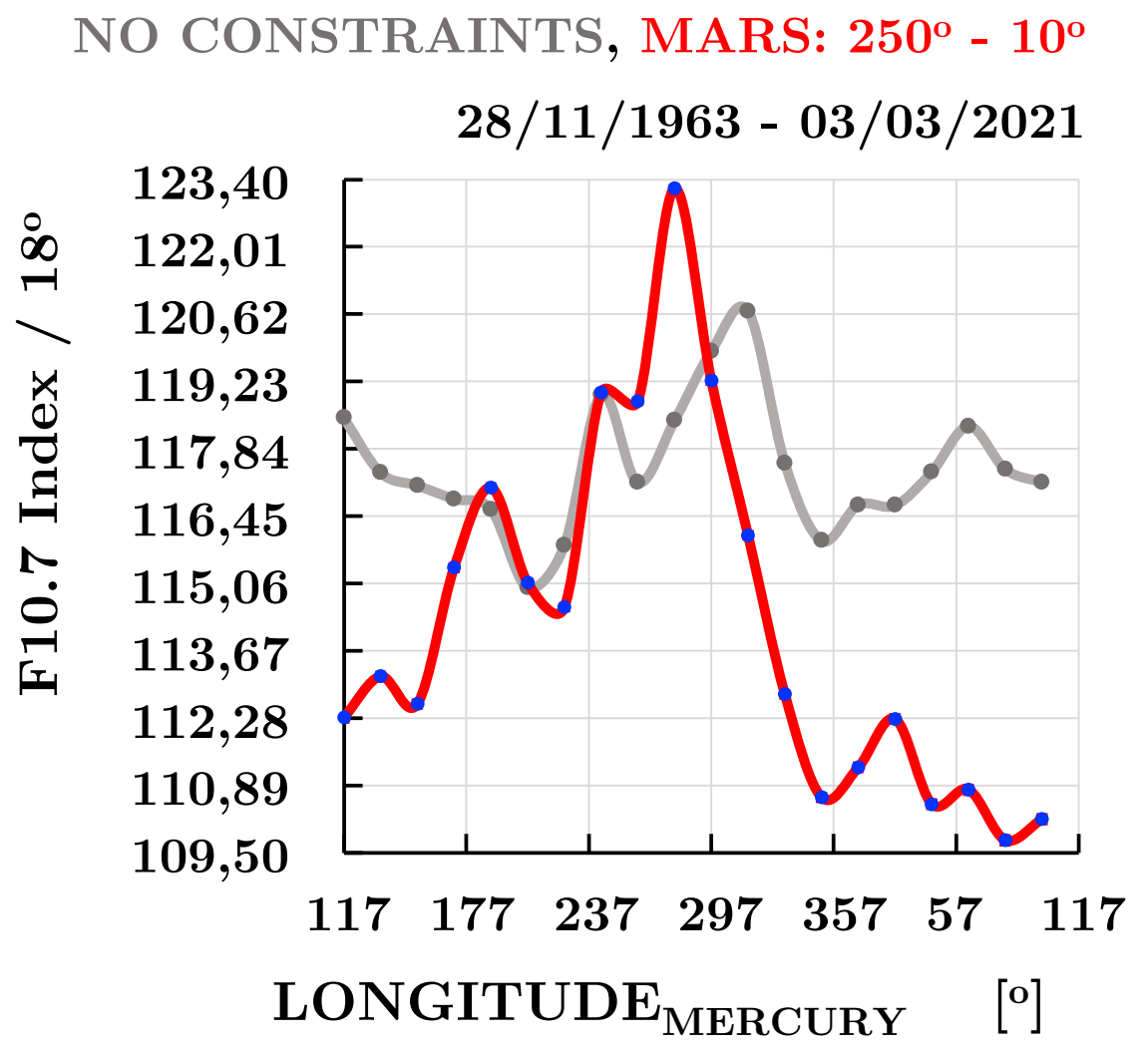

(d) F10.7 distribution for Mercury without constraints and when Mars is between 250° to 10°.

**Figure B.61:** Comparison of distributions of Ly-$\alpha$ vs. F10.7 solar radio flux for the reference frame of Mercury without constraints and when Mars is constrained to propagate in two 180° opposite regions wih bin = 18° for the period 28/11/1963 - 03/03/2021.

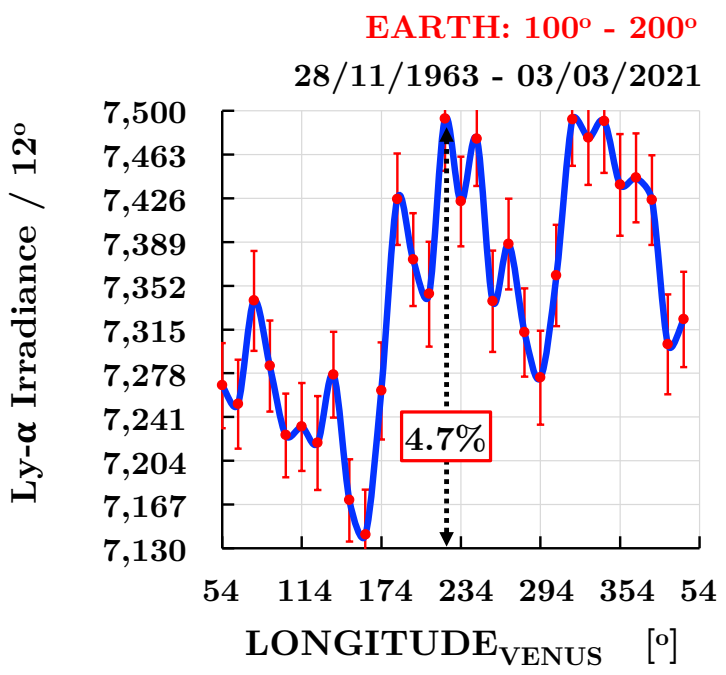

(a) Ly-$\alpha$ distribution for Venus when Earth is between 100° to 200°.

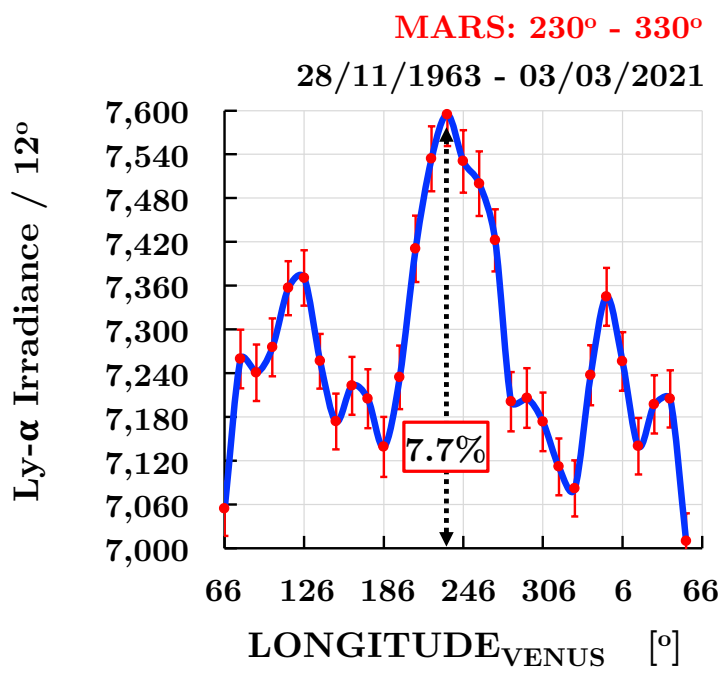

(b) Ly-$\alpha$ distribution for Venus when Mars is between 230° to 330°.

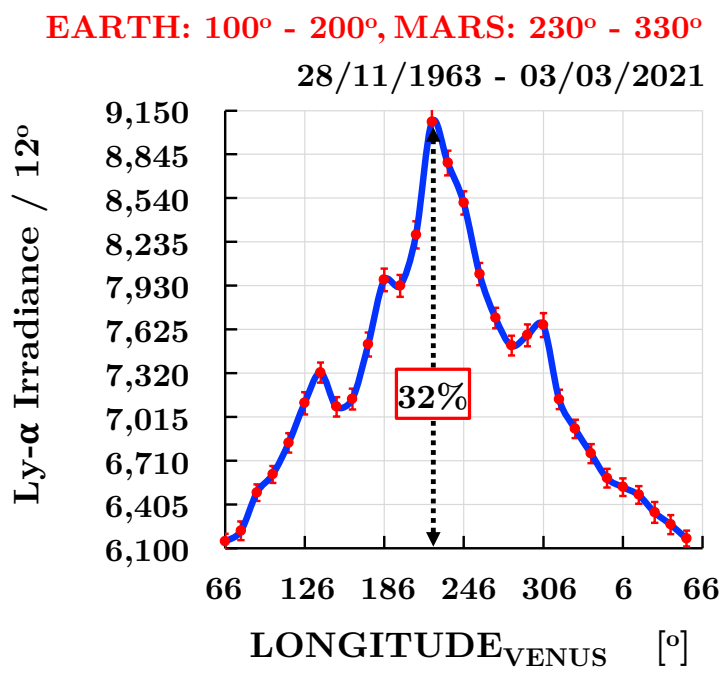

(c) Ly-$\alpha$ distribution for Venus when Earth is between 100° to 200° and at the same time Mars is between 230° to 330°.

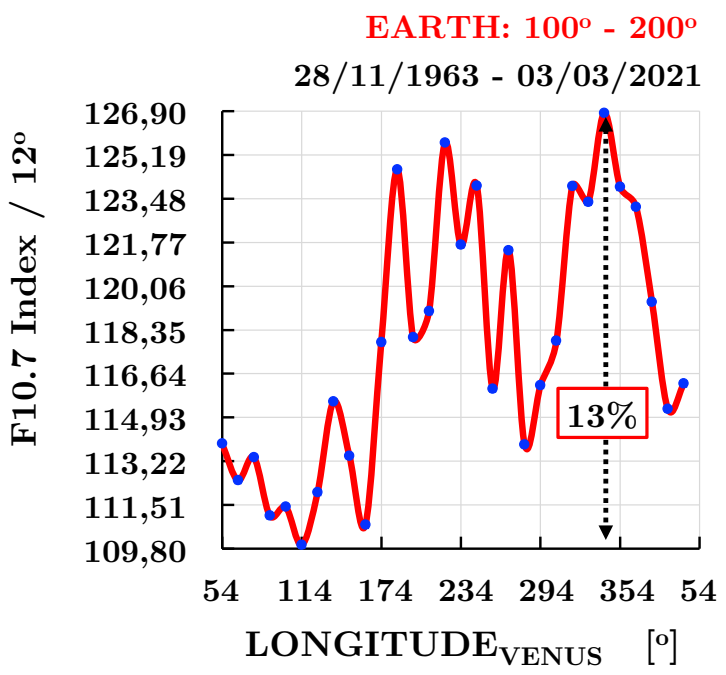

(d) F10.7 distribution for Venus while Earth is between 100° to 200°.

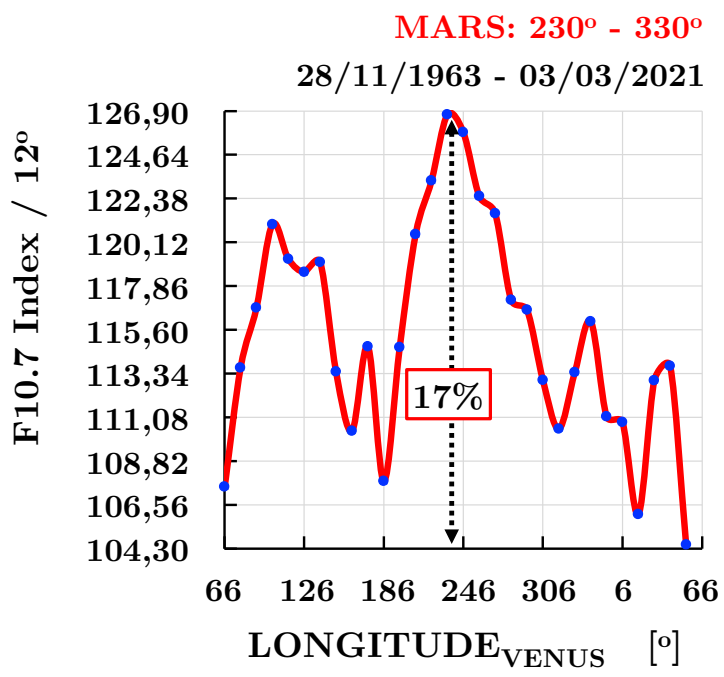

(e) F10.7 distribution for Venus when Mars is between 230° to 330°.

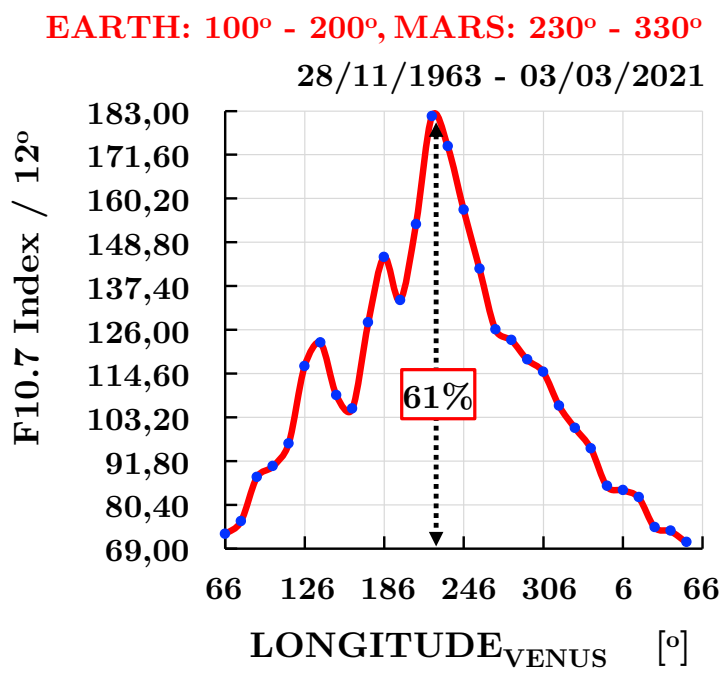

(f) F10.7 distribution for Venus when Earth is between 100° to 200° and at the same time Mars is between 230° to 330°.

**Figure B.62:** Comparison of distributions of Ly-$\alpha$ vs. F10.7 solar radio flux for the reference frame of Venus when applying various constraints on the positions of other planets with bin = 12° for the period 28/11/1963 - 03/03/2021.

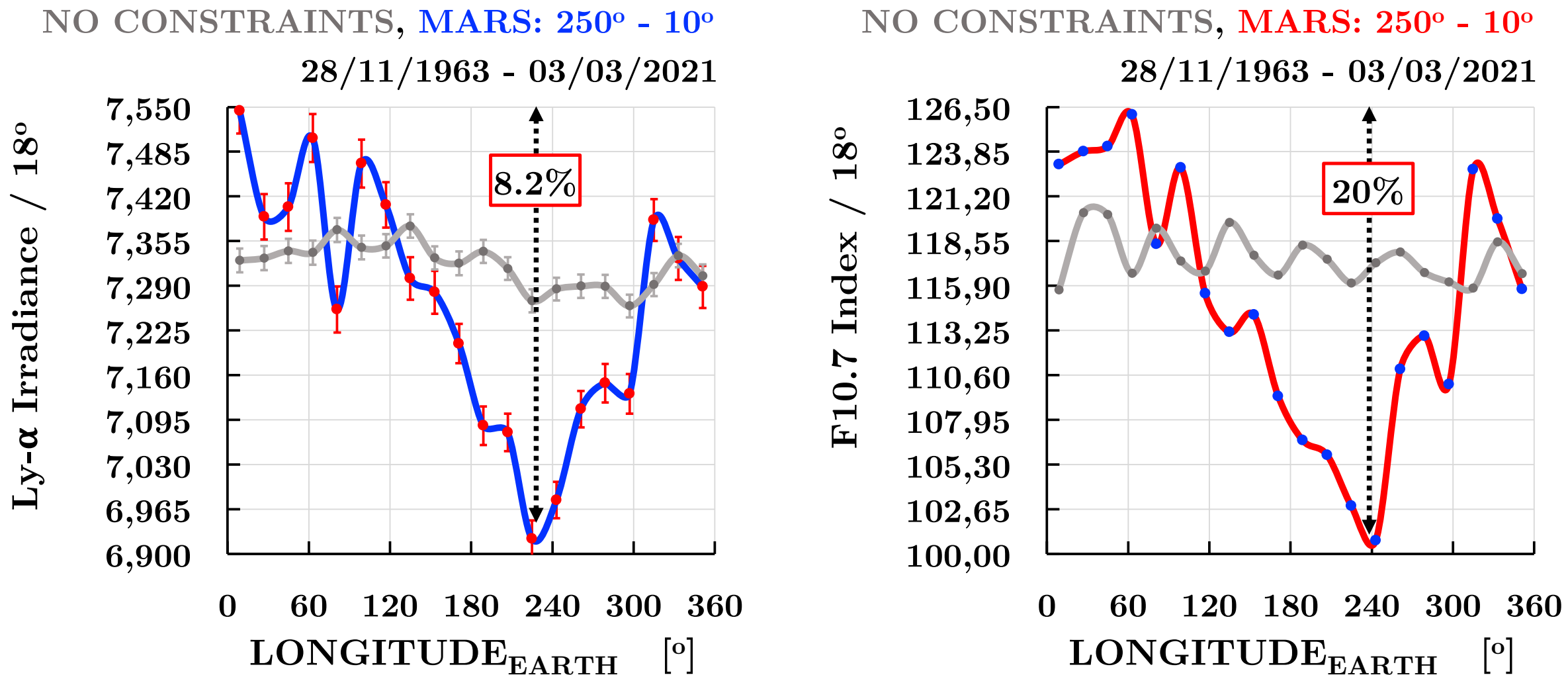

**(a)** Ly-$\alpha$ distribution. **(b)** F10.7 distribution.

**Figure B.63:** Comparison of distributions of Ly-$\alpha$ vs. F10.7 solar radio flux for the reference frame of Earth without constraints and when Mars is constrained to move between 250° to 10° with bin = 18° for the period 28/11/1963 - 03/03/2021.

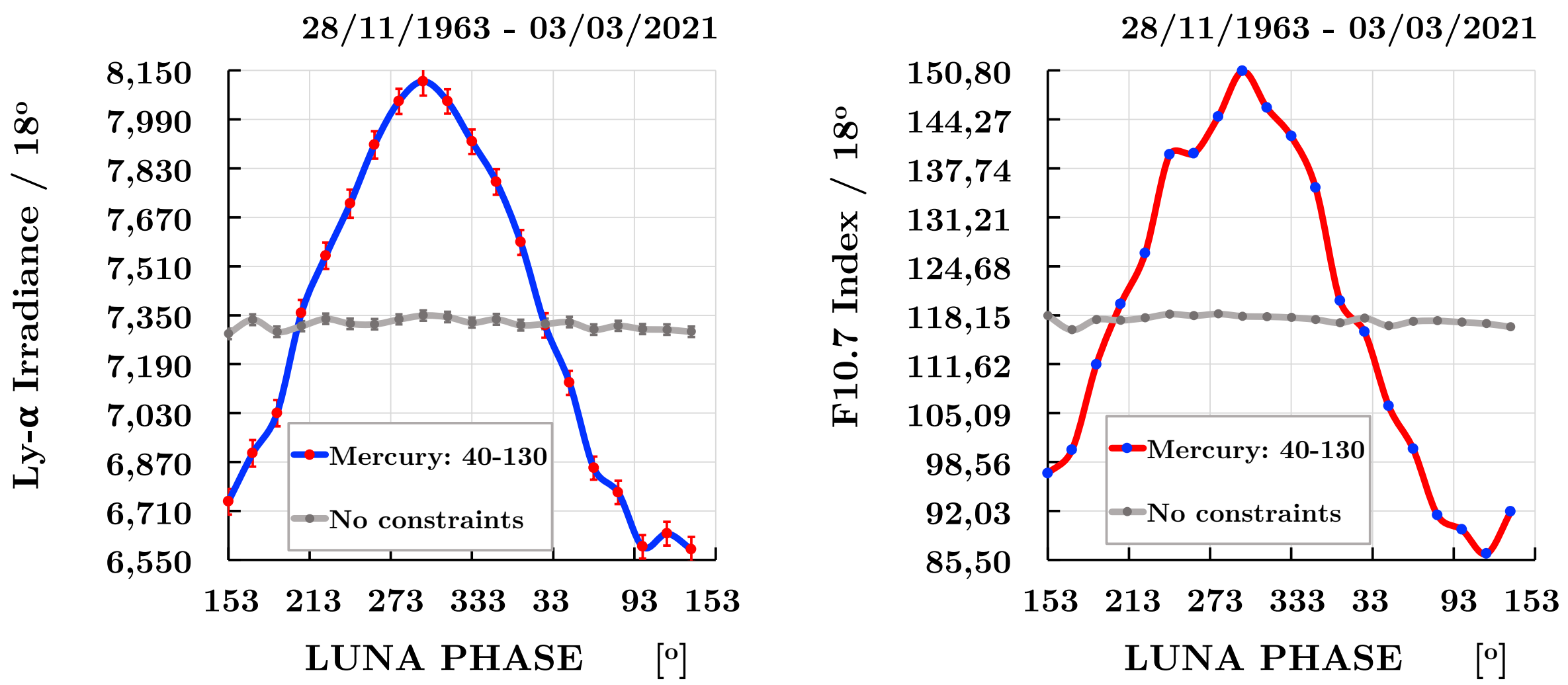

**(a)** Ly-$\alpha$ distribution. **(b)** F10.7 distribution.

**Figure B.64:** Comparison of distributions of Ly-$\alpha$ vs. F10.7 solar radio flux as a function of Moon's phase without constraints and when Mercury is constrained to move between 40° to 130° with bin = 18° for the period 01/01/1996 - 01/03/2021.

## B.9 Ionospheric electron content

### B.9.1 Comparison with solar EUV

The most obvious comparison of the results from Sect. 15.3.1 is made with EUV solar irradiance, since we know that it strongly affects the ionosphere. The reason is that the rays from the Sun meet the outer layers of the atmosphere first and the short wave-length radiation is absorbed there causing ionisation from which ionosphere gets its name [468, 469]. In Fig. B.65 the time-series two datasets are overlaid for the same overlapping period. From the statistical correlation analysis, we get ($r = 0.92$, $p = 0$) which, as expected, indicates a strong positive linear correlation for the general trend of TEC and EUV.

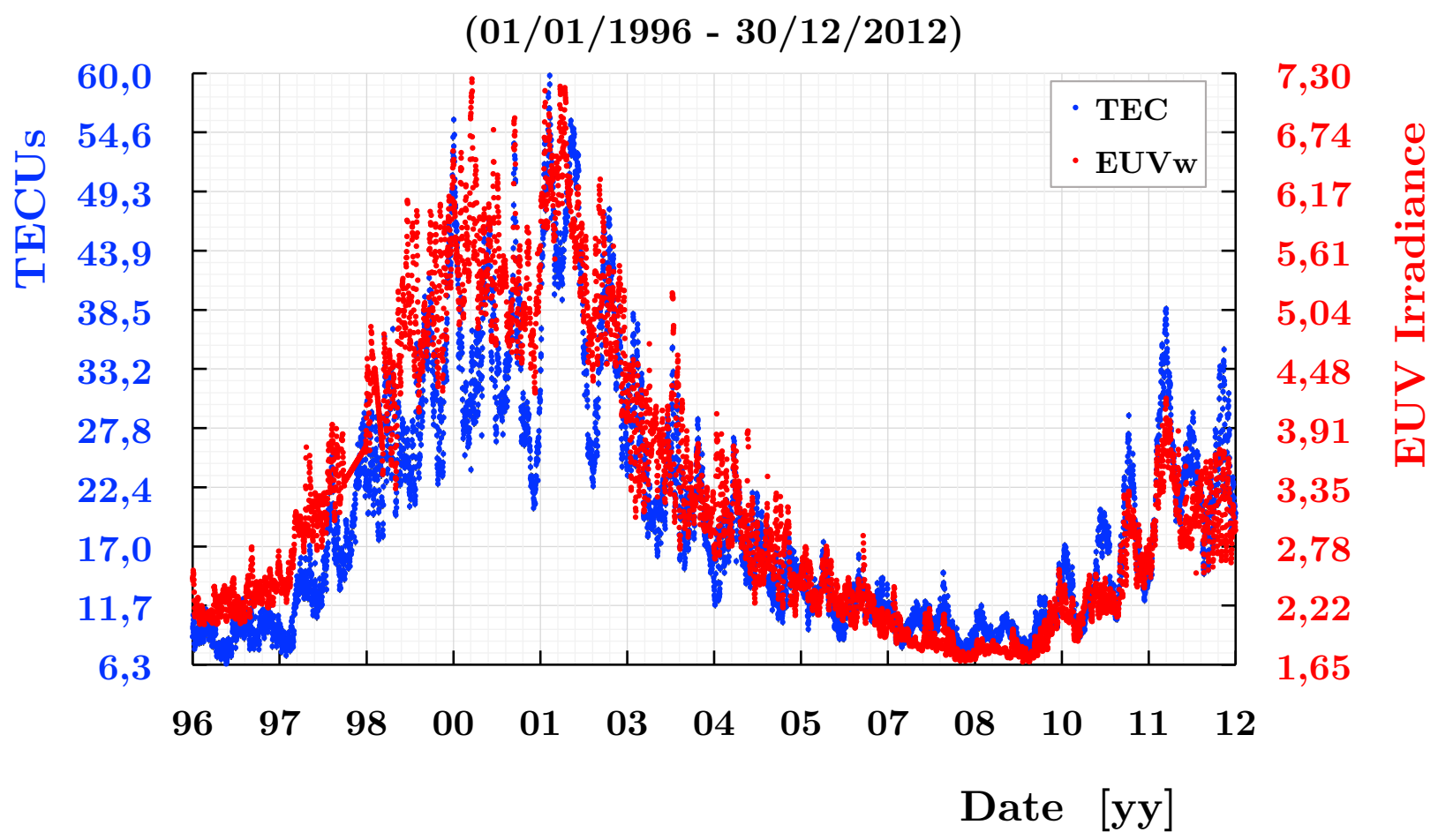

**Figure B.65:** TEC data and EUV solar irradiance for the same period 01/01/1996 - 30/12/2012.

#### B.9.1.1 Single planets

As next, in Fig. B.66 we reconstruct the distributions from single planets shown in Fig. 15.5 but for the case of solar EUV. We note that since EUV data start from 01/01/1996, the plots from Fig. 15.5 were recalculated for the period 01/01/1996 - 30/12/2012 (6209 d) in order to have an exact one-to-one comparison with the corresponding ones from EUV. These new plots for TEC are Fig. B.66a through B.66d. We see that there is a resemblance between the two datasets in Mercury, Venus and Mars spectra but the ones from Earth in Fig. B.66c and B.66g show some difference especially on the differential amplitude of the two peaks. For a quantitative assessment the Pearson's correlation coefficients for the subfigures a-e, b-f, c-g and d-h are calculated, with all of them being statistically significant. More specifically we have ($r_{a,e} = 0.73$, $p_{a,e} = 2.43 \times 10^{-4}$), ($r_{b,f} = 0.91$, $p_{b,f} = 3.77 \times 10^{-8}$), ($r_{c,g} = 0.45$, $p_{c,g} = 0.012$) and ($r_{d,h} = 0.78$, $p_{d,h} = 5.49 \times 10^{-5}$). Therefore all but Earth's distributions have a high degree of positive correlation whereas Earth shows a medium association.

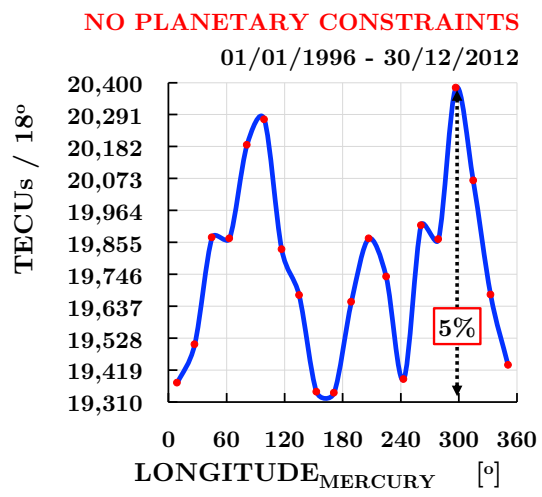

(a) TEC distribution for Mercury with 18° bin.

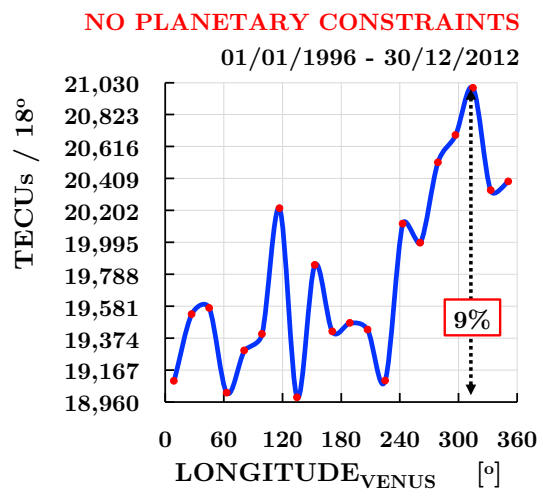

(b) TEC distribution for Venus with 18° bin.

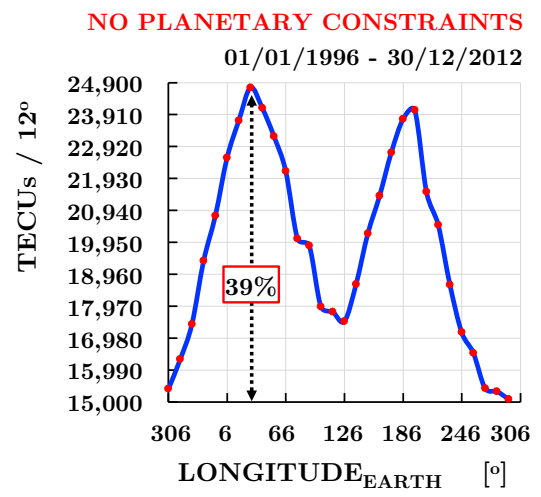

(c) TEC distribution for Earth with12° bin.

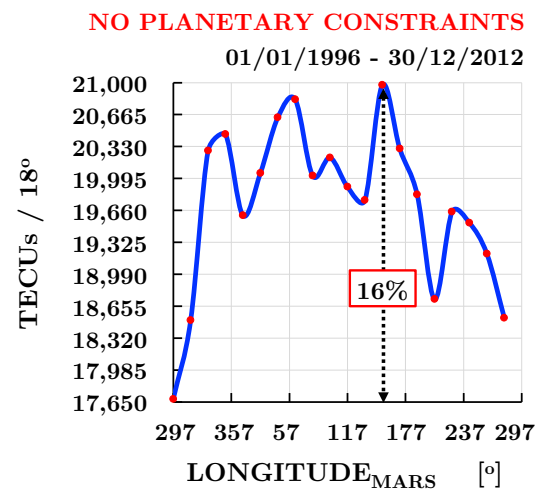

(d) TEC distribution for Mars with 18° bin.

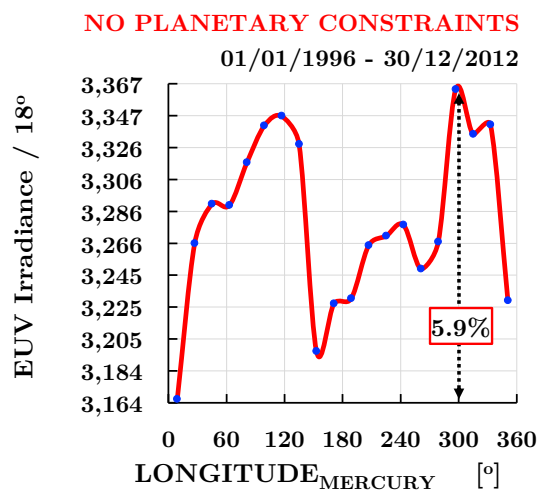

(e) EUV distribution for Mercury with 18° bin.

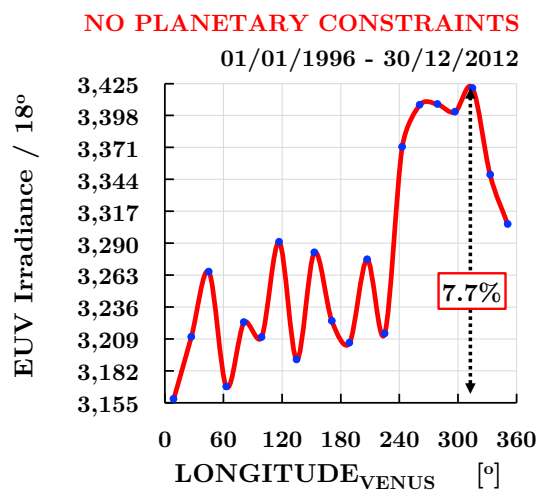

(f) EUV distribution for Venus with 18° bin.

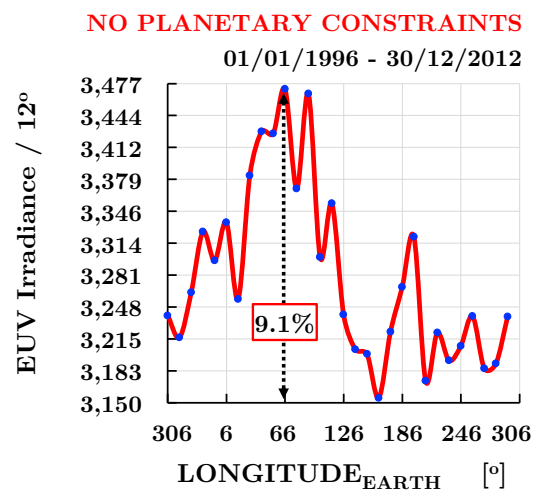

(g) EUV distribution for Earth with 12° bin.

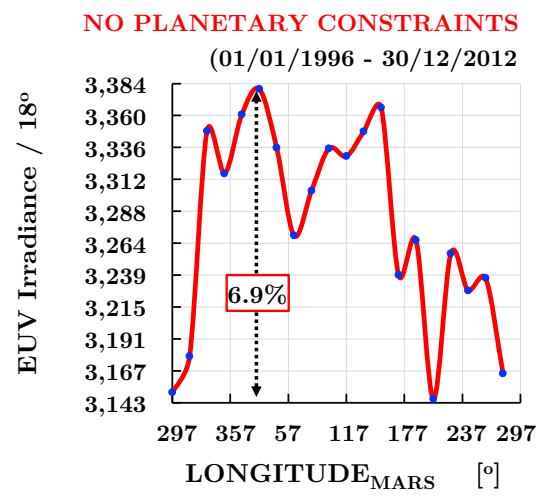

(h) EUV distribution for Mars with 18° bin.

**Figure B.66:** Comparison of distributions of ionospheric TEC vs. solar EUV intensity for the period 01/01/1996 - 30/12/2012.

#### B.9.1.2 Combining planets

We then compare the distribution with the biggest effect in ionospheric TEC, which is Fig. 15.8. However, we need, once more, to recalculate the cases for the TEC distribution starting from 01/01/1996 since this is when EUV data start, and go until 23/12/2012 when Moon completes 210 full orbits around the Earth. These new plots are shown in Fig. B.67a and B.67b. The corresponding spectra for EUV are shown in Fig. B.67c and B.67d where we see that even though the shape might seem similar, the amplitude is much smaller, pointing to an additional external factor on the ionisation of the ionosphere. The calculated correlation coefficients for the constraints on Earth's position in Fig. B.67a and B.67c are ($r_{a,c} = 0.90$, $p_{a,c} = 3.49 \times 10^{-6}$) and ($r_{a,c} = 0.29$, $p_{a,c} = 0.296$), whereas for their difference in Fig. B.67b and B.67d we have ($r_{a,c} = 0.84$, $p_{a,c} = 7.98 \times 10^{-5}$).

### B.9.2 Comparison with F10.7

The next comparison is made with the F10.7 solar radio flux line since it is assumed to have both a direct impact on ionospheric electron density through ionisation processes, and an indirect impact through thermospheric heating that operates on longer temporal scales [351]. Similarly to Fig. B.65, in Fig. B.68 the TEC time series are compared with F10.7 data. The Pearson's correlation coefficient for this case is ($r = 0.90$, $p = 0$) which, as with EUV, shows a

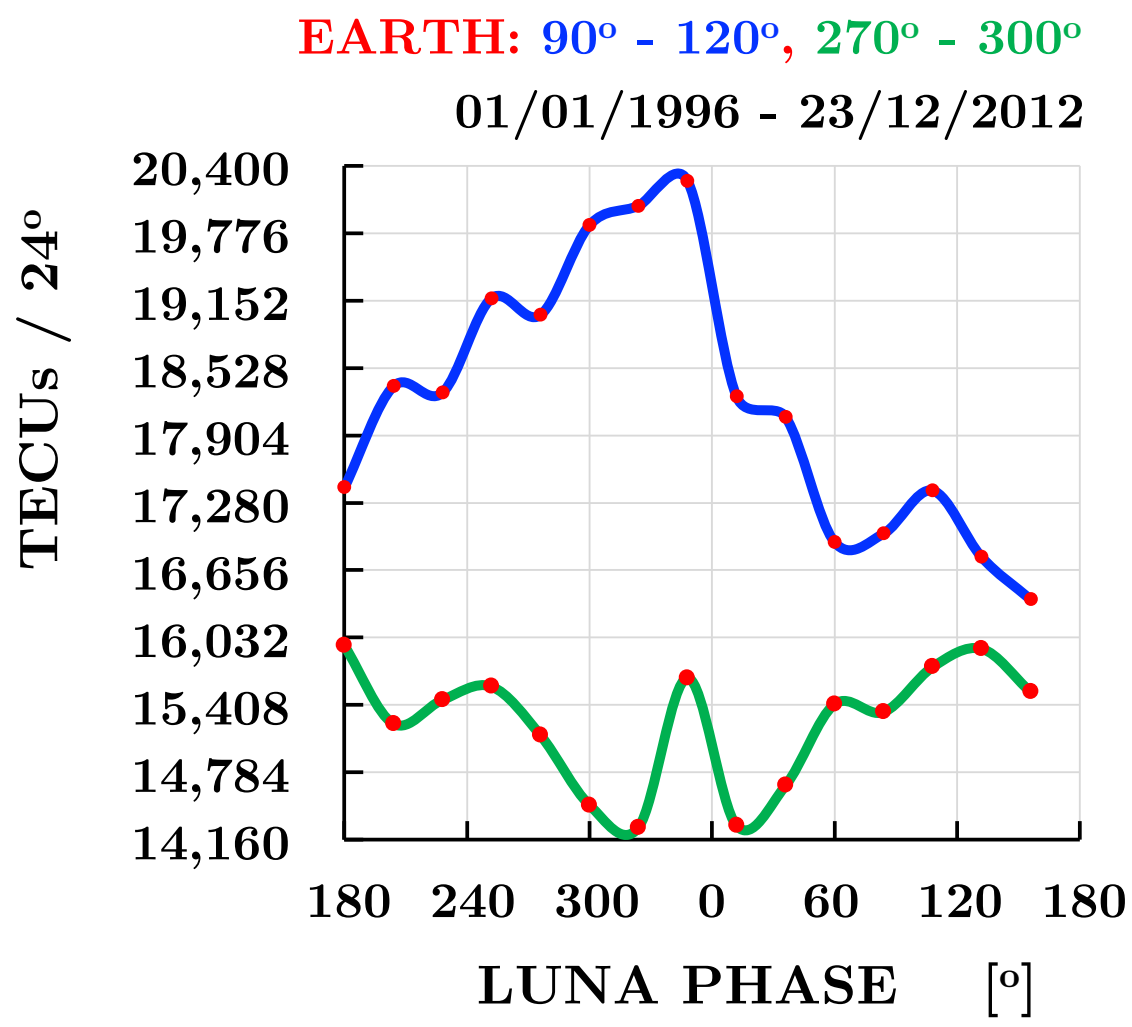

**(a)** TEC distribution when comparing the two cases of Earth.

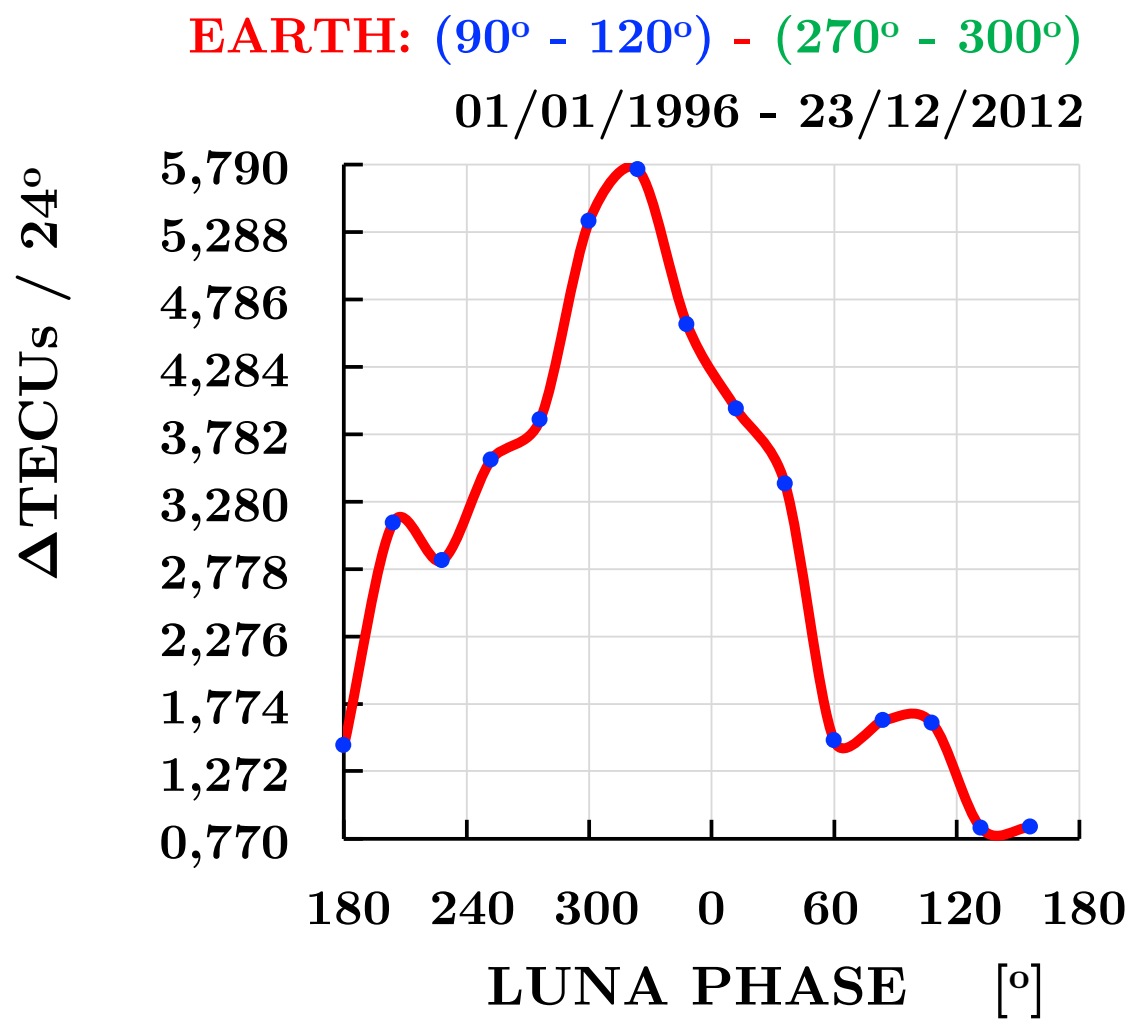

**(b)** TEC distribution when subtracting the two cases of Earth.

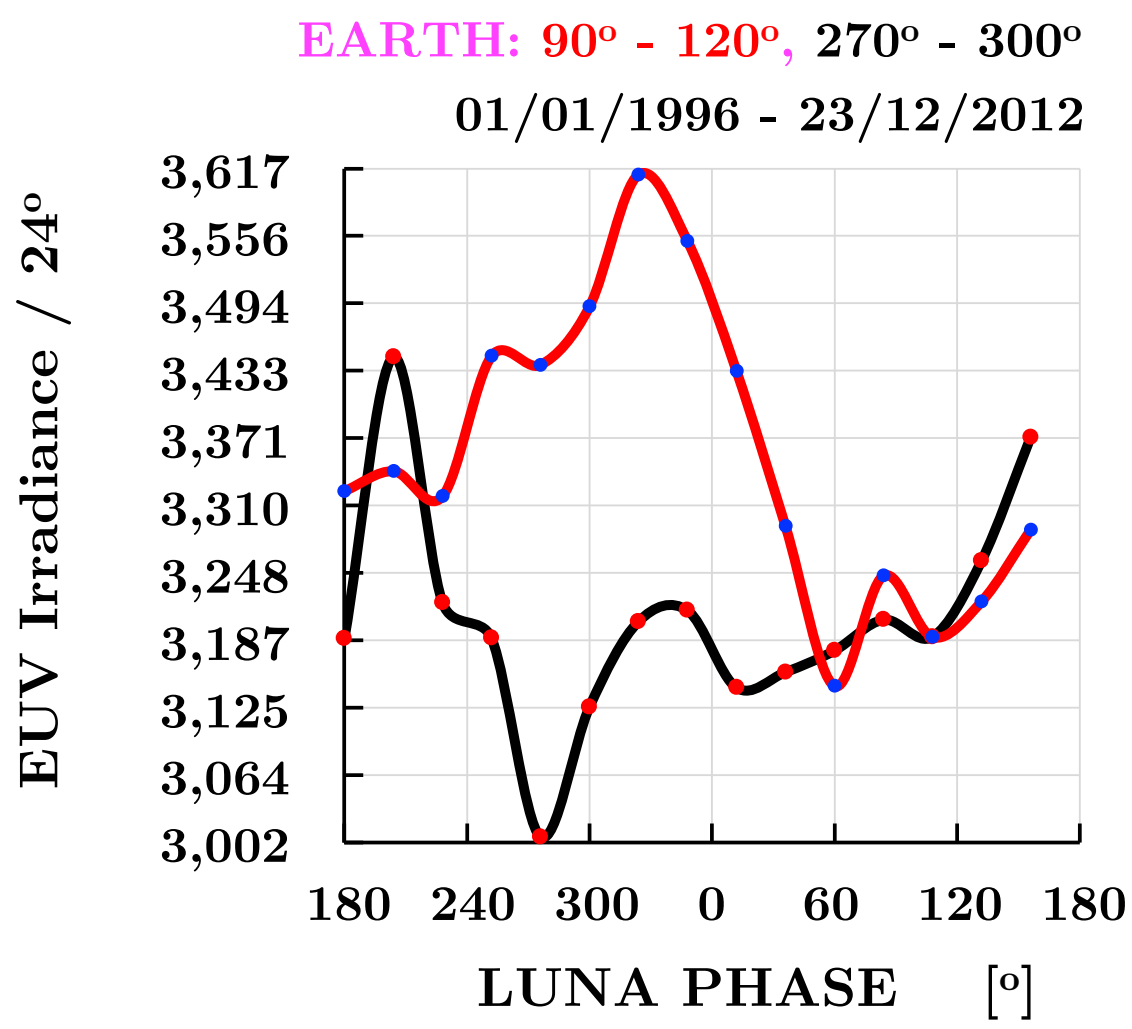

**(c)** EUV distribution when comparing the two cases of Earth.

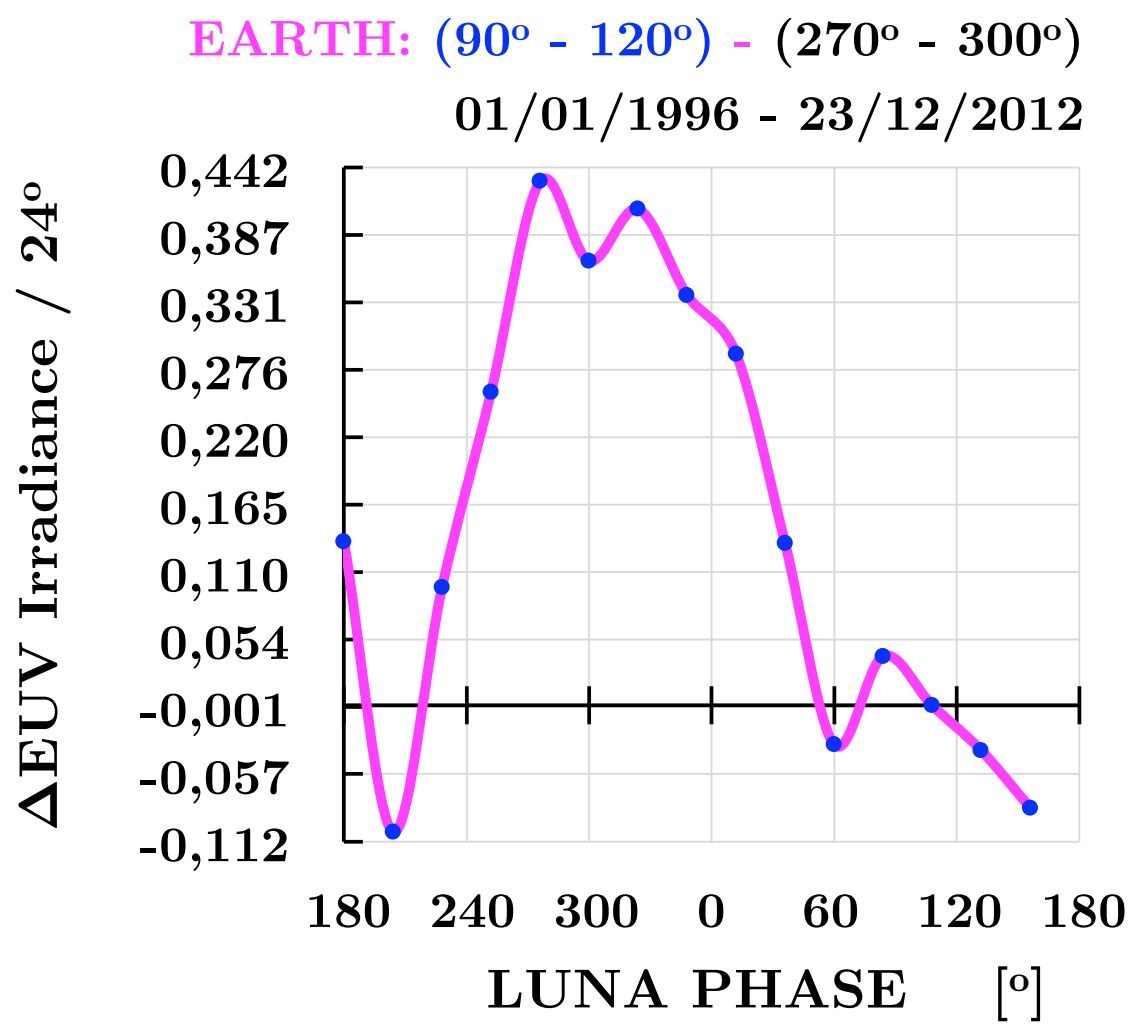

**(d)** EUV distribution when subtracting the two cases of Earth.

**Figure B.67:** Comparison of ionospheric TEC vs. solar EUV intensity for Moon's distributions when Earth is between 90° to 120° and 270° to 300° with bin = 24° for the period 01/01/1996 - 23/12/2012.

significant positive linear correlation of high degree.

#### B.9.2.1 Single planets

In Fig. B.69 we repeat he procedure as in Fig. B.66, but with the F10.7 distributions. Therefore, we compare directly the spectra from Fig. 15.5 using the same period 01/01/1995 - 30/12/2012. Again, we see a high degree of similarity between the two datasets with the exception of Earth distribution (Fig. B.69c and B.69g) which seems to have the most different shape between TEC and F10.7 in the compared distributions. This observation is verified by the Pearson's correlation coefficients and p-values which give ($r_{a,e} = 0.78$, $p_{a,e} = 5.22 \times 10^{-5}$),

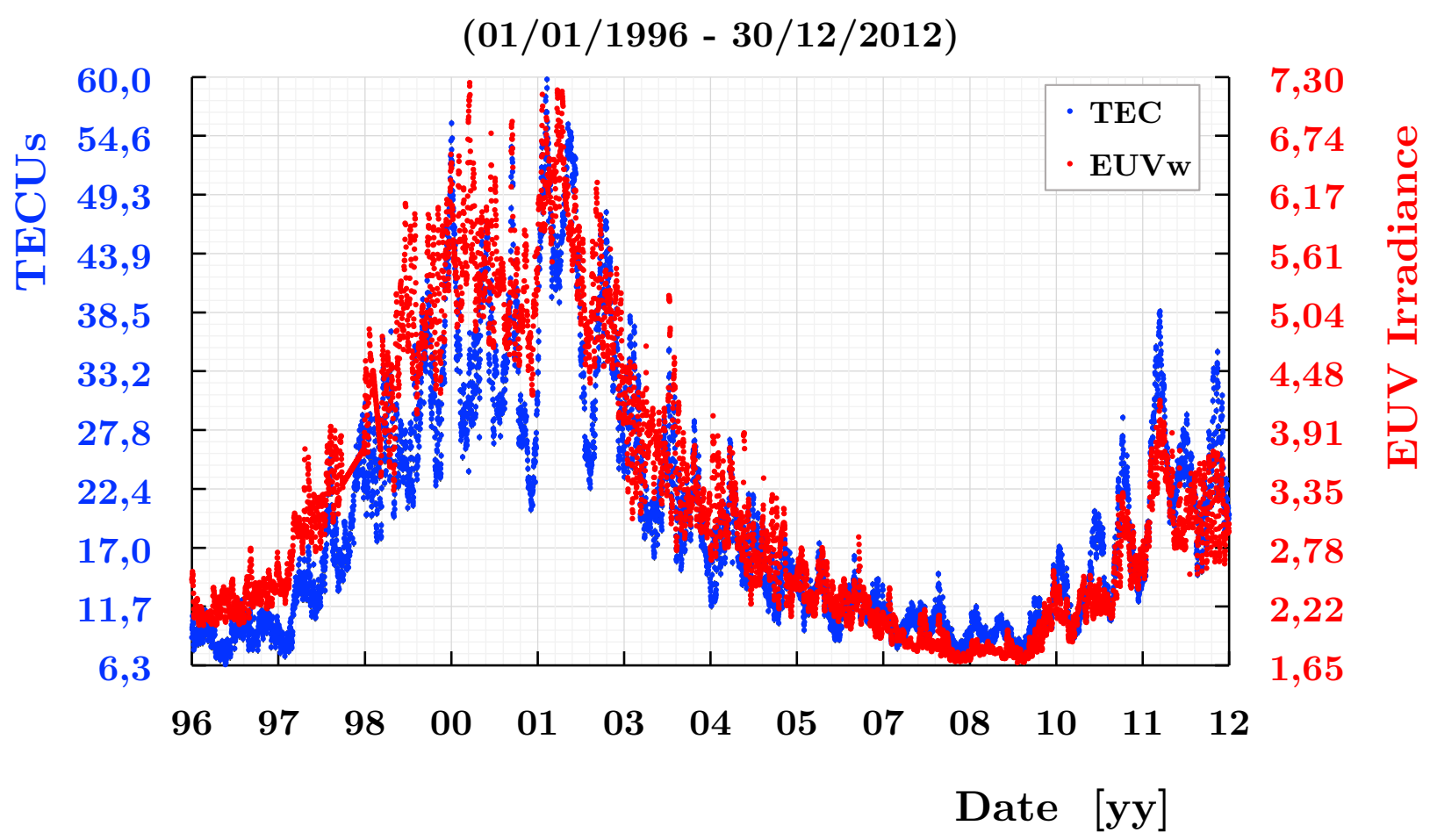

**Figure B.68:** TEC data and EUV solar irradiance for the same period 01/01/1996 - 30/12/2012.

($r_{b,f} = 0.84$, $p_{b,f} = 4.12 \times 10^{-6}$), ($r_{c,g} = 0.27$, $p_{c,g} = 0.15$) and ($r_{d,h} = 0.77$, $p_{d,h} = 7.48 \times 10^{-5}$).

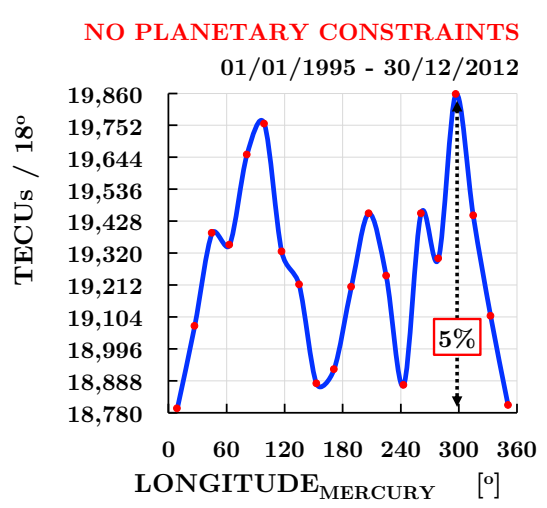

(a) TEC distribution for Mercury with 18° bin.

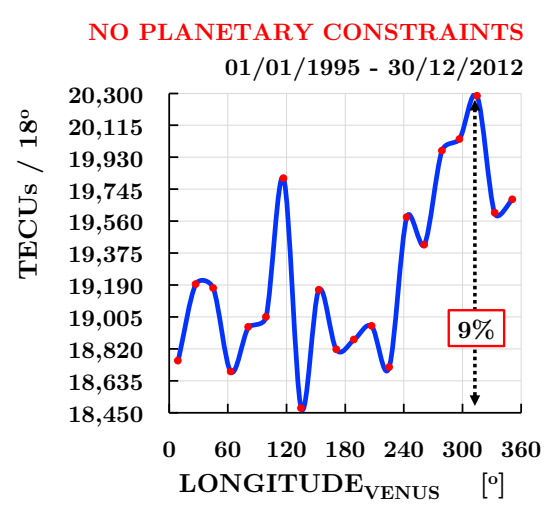

(b) TEC distribution for Venus with 18° bin.

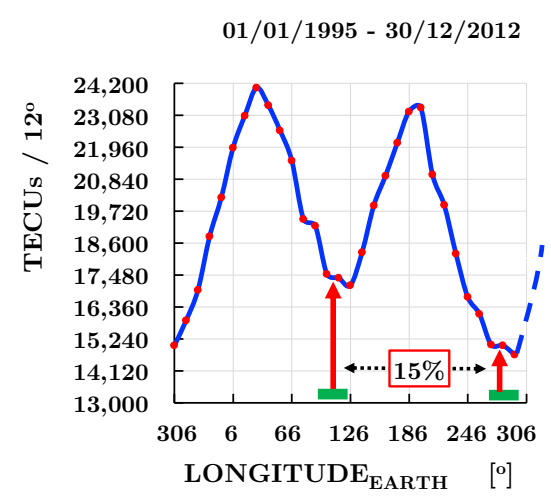

(c) TEC distribution for Earth with12° bin.

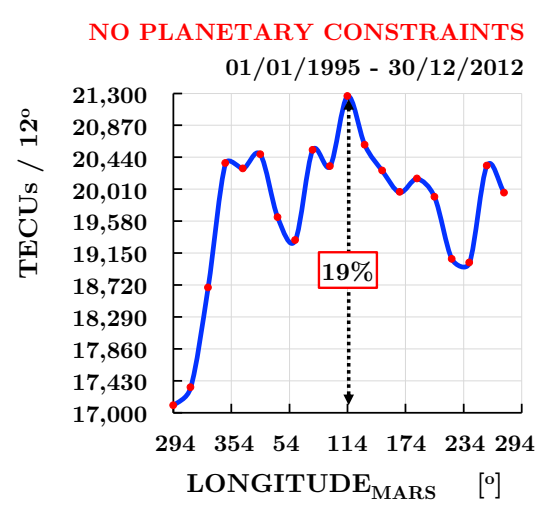

(d) TEC distribution for Mars with 18° bin.

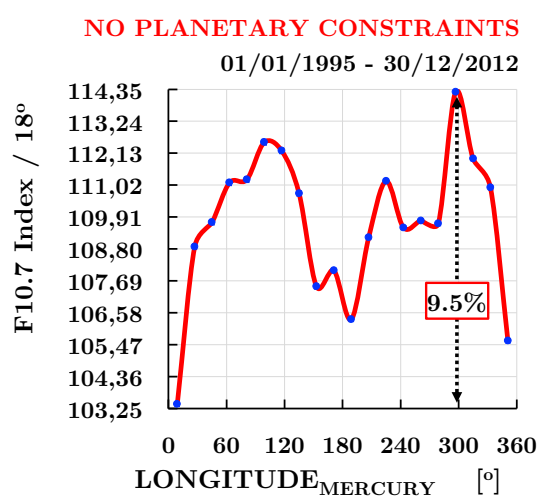

(e) F10.7 distribution for Mercury with 18° bin.

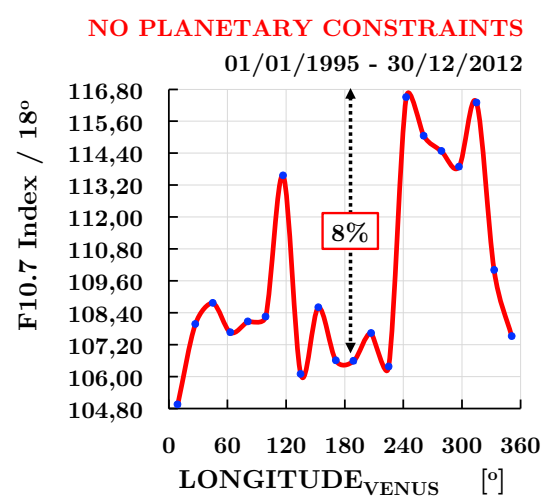

(f) F10.7 distribution for Venus with 18° bin.

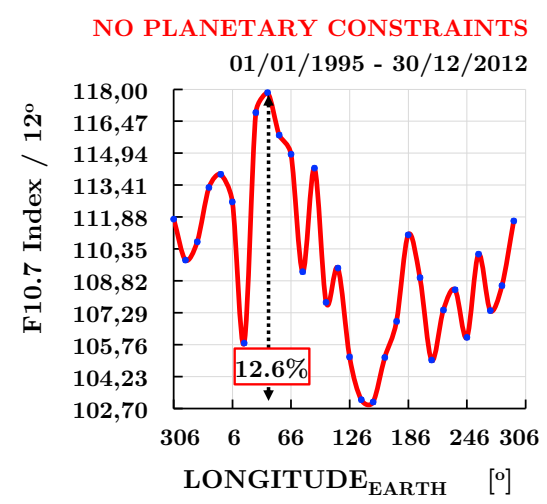

(g) F10.7 distribution for Earth with 12° bin.

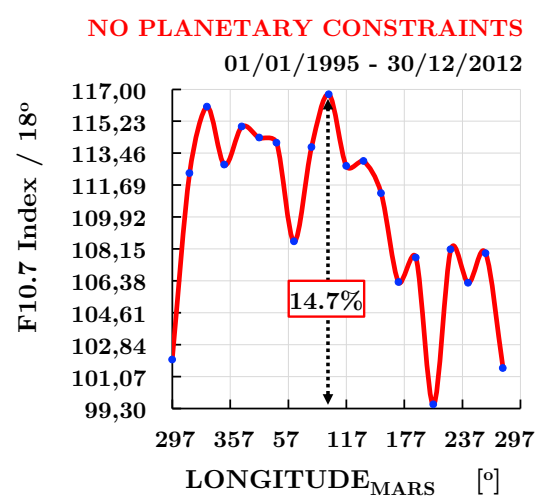

(h) F10.7 distribution for Mars with 18° bin.

**Figure B.69:** Comparison of distributions of ionospheric TEC vs. F10.7 solar proxy for the period 01/01/1995 - 30/12/2012.

#### B.9.2.2 Combining planets

Finally, Fig. 15.8 is recreated for F10.7 since it is the one with the biggest amplitude observed so far in all of the datasets analysed in this work. The comparison results are shown in Fig. B.70, where in contrast with Fig. B.67, in this case there seems to be a dissimilarity between F10.7 and TEC in shape but mostly on the overall observed amplitude, which as with the case of EUV hints on the presence of an additional exo-solar impact on the TEC of the Earth's ionosphere. More specifically, when constraining Earth between 90° to 120° and 270° to 300° in Fig. B.70a and B.70c the correlation coefficient is ($r_{a,c} = 0.63,\ p_{a,c} = 0.012$) and ($r_{a,c} = 0.31,\ p_{a,c} = 0.258$). However, the homoscedasticity is violated in the first case therefore non-parametric methods are used such as the Spearman and Kendall coefficients. These give ($r_{S_{a,c}} = 0.75,\ p_{S_{a,c}} = 0.0013$) and ($\tau_{a,c} = 0.56,\ p_{a,c} = 0.0035$). Therefore, there is a strong association for the case of Earth being in the 90° to 120° orbital arc, but a non-significant small degree of correlation for the 270° to 300° orbital arc. Then, when subtracting one case from the other in Fig. B.70b and B.70d we get ($r_{a,c} = 0.56,\ p_{a,c} = 0.029$).

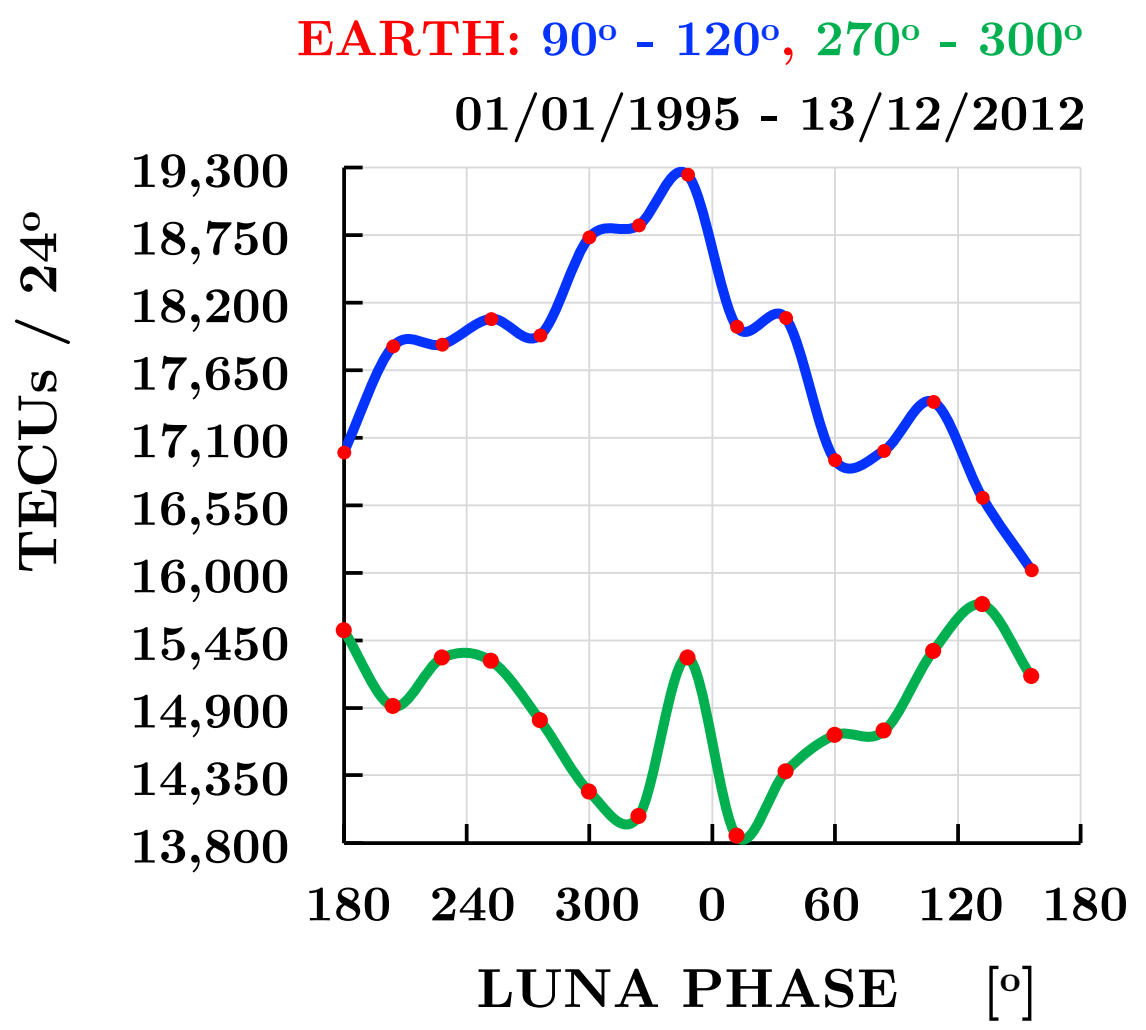

**(a)** TEC distribution when comparing the two cases of Earth.

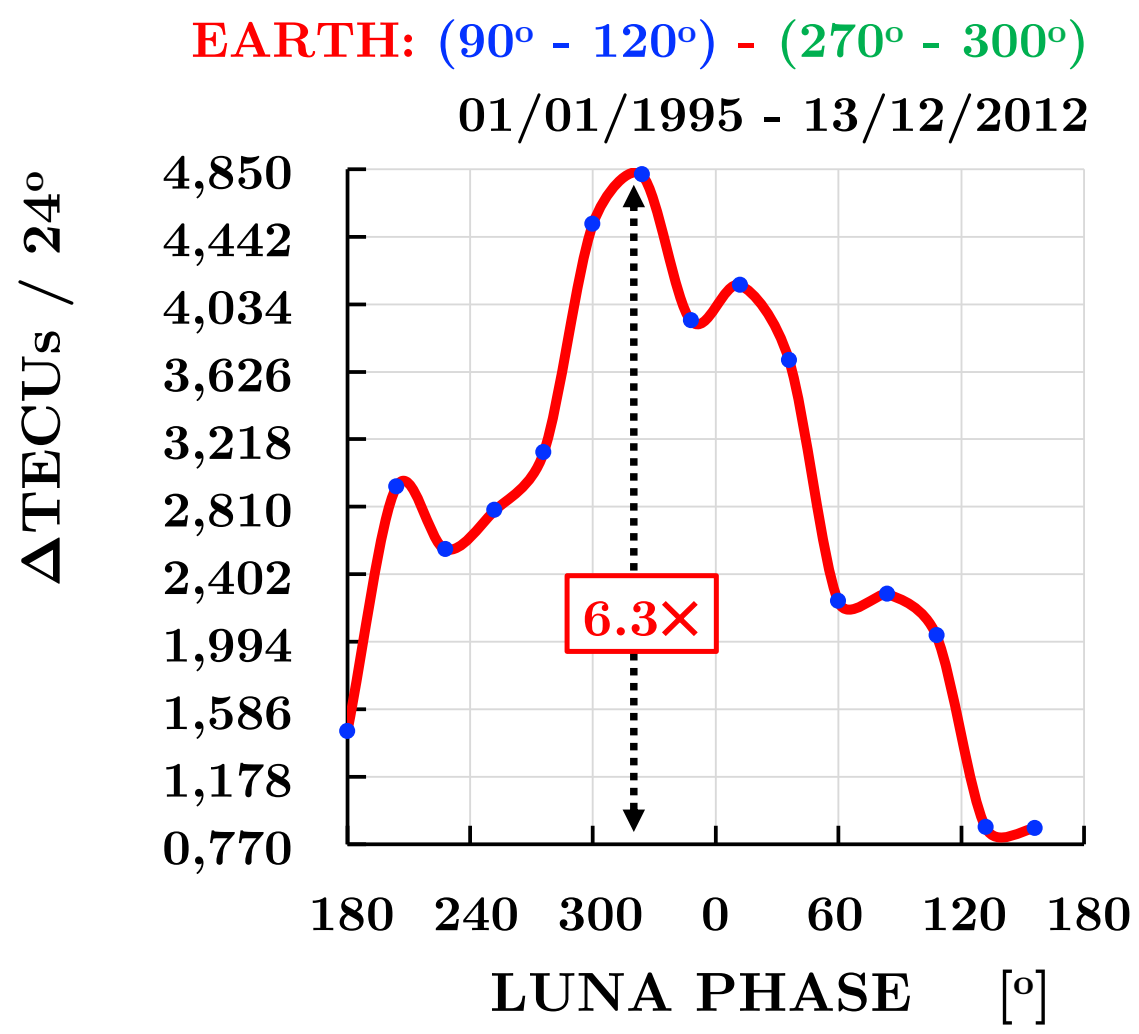

**(b)** TEC distribution when subtracting the two cases of Earth.

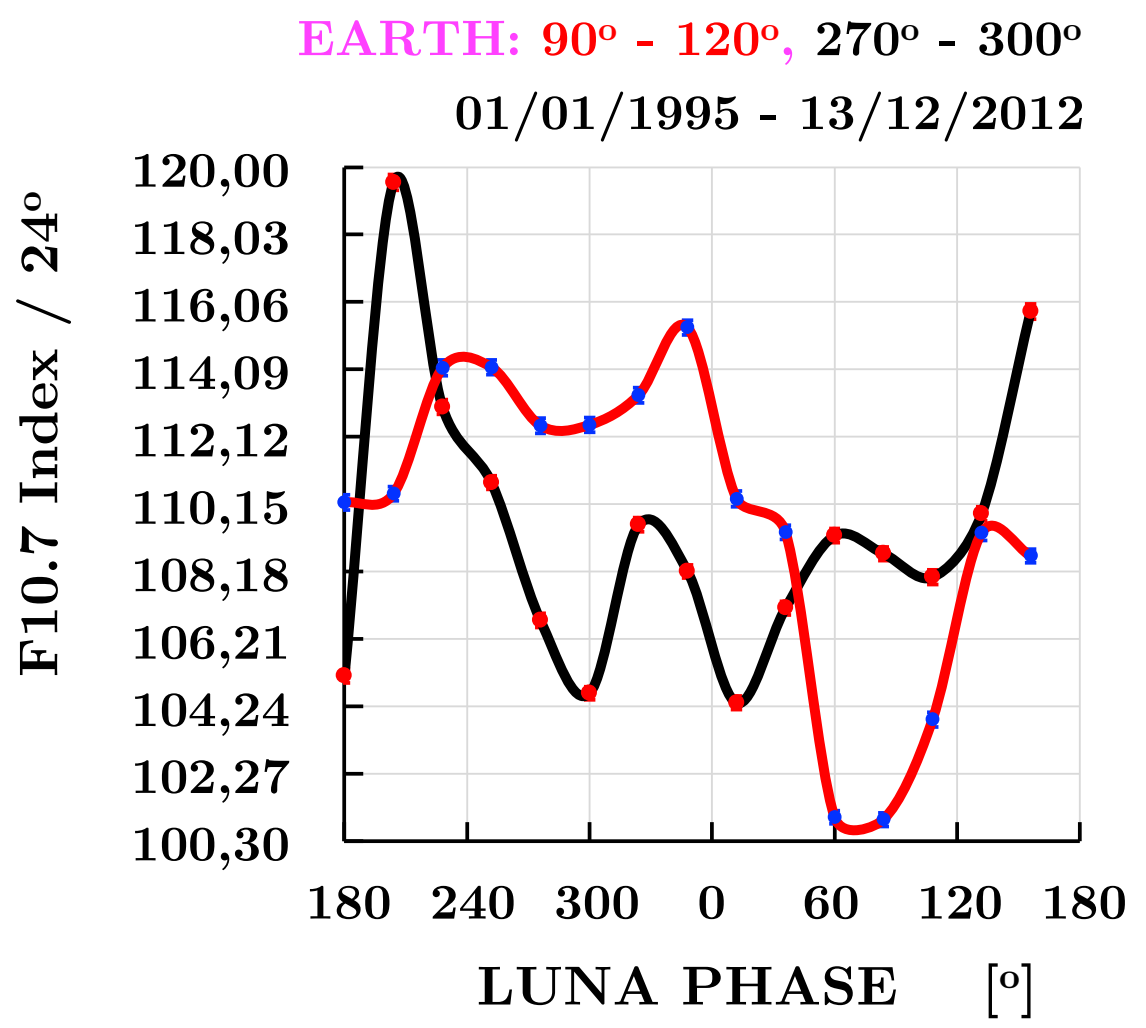

**(c)** F10.7 distribution when comparing the two cases of Earth.

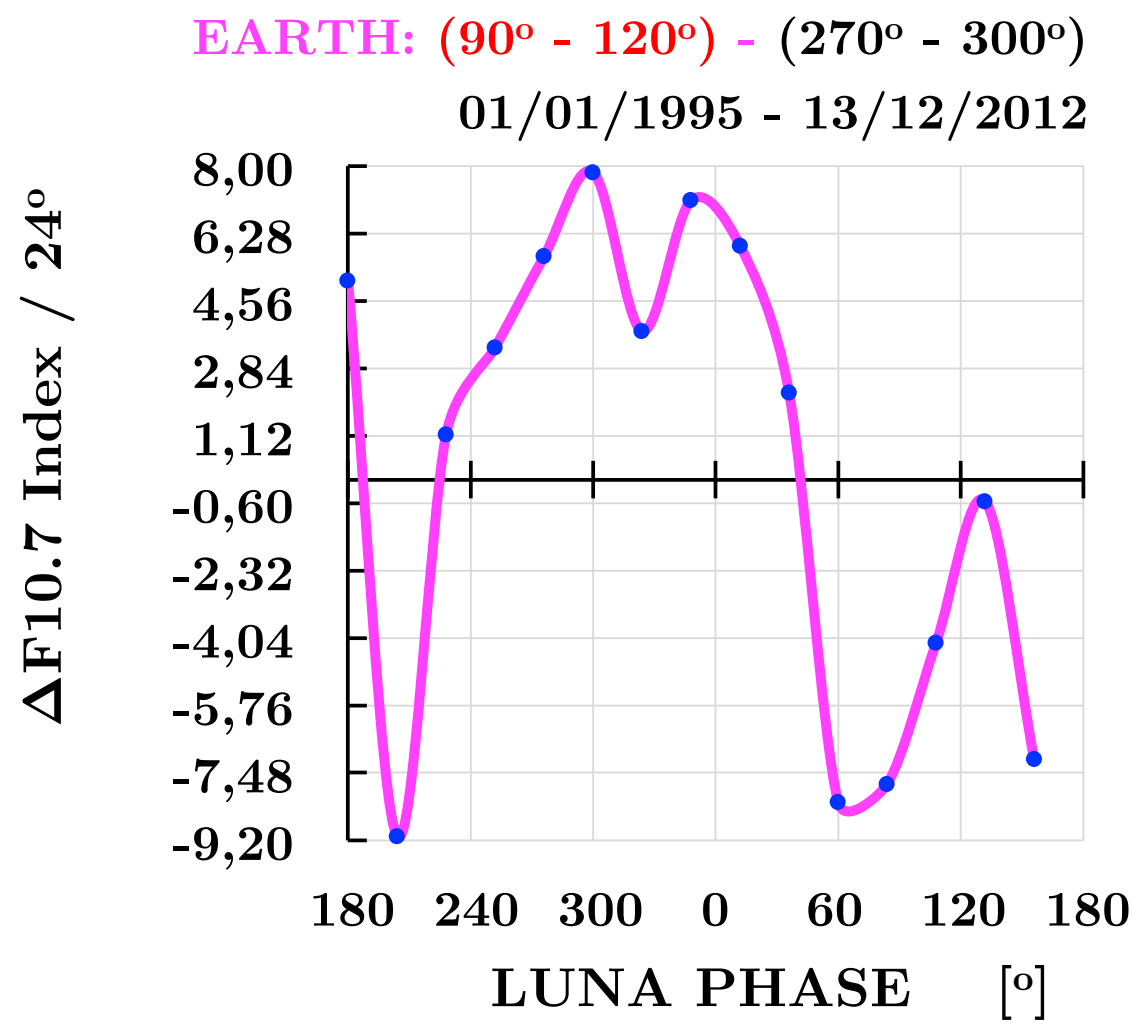

**(d)** F10.7 distribution when subtracting the two cases of Earth.

**Figure B.70:** Comparison of ionospheric TEC vs. F10.7 solar proxy for Moon's distributions when Earth is between 90° to 120° and 270° to 300° with bin = 24° for the period 01/01/1996 - 23/12/2012.

# B.10 Stratospheric temperature

## B.10.1 Comparison with solar EUV

An important question that arises is whether the observed planetary correlation of the upper stratosphere appears indirectly due to the planetary relationship of the solar activity itself. The first most reasonable comparison of the results of this analysis is with the solar EUV irradiance due to its known direct impact on the atmosphere above ∼ 10 km. A dissimilarity between the spectra of the stratospheric temperature, and the corresponding ones of EUV would provide an additional confirmation of the exo-solar impact on the Earth's atmosphere. A statistical correlation analysis is performed to quantify the degree of association on the shapes of the various distributions of the two datasets.

In Fig. B.71 the time-series of the average stratospheric temperatures are compared with the EUV solar irradiance for the same overlapping period. The Pearson's correlation coefficient and the corresponding p-value are calculated to be ($r = -0.08$, $p = 3.22 \times 10^{-7}$) which indicates a very small degree of negative correlation in the general trend.

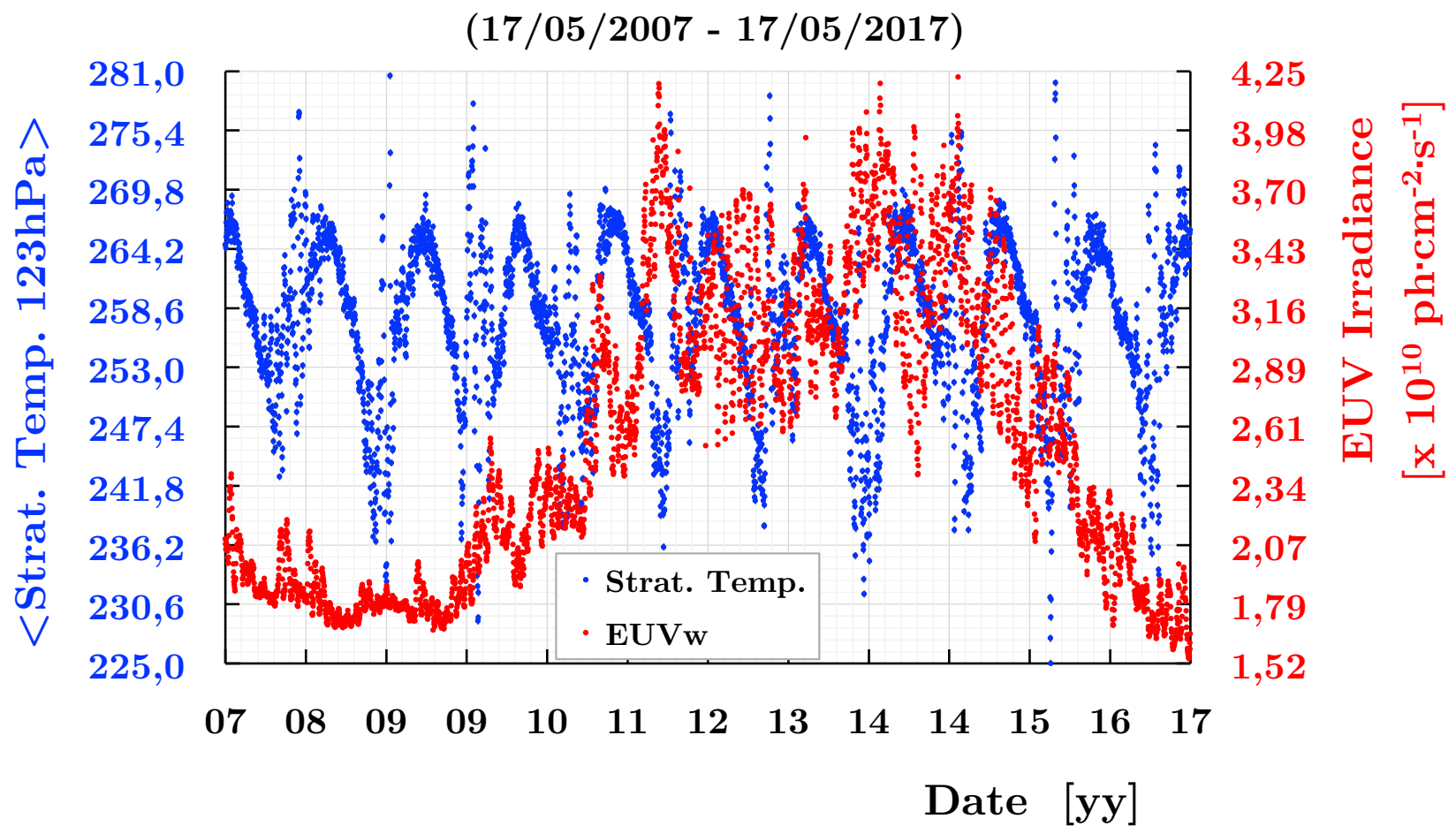

**Figure B.71:** Average stratospheric temperature data and EUV solar irradiance (17/05/2007 - 17/05/2017).

#### B.10.1.1 Single planets

In Fig. B.72 the single planetary distributions of the stratospheric temperature from Fig. 16.4 are compared directly with the corresponding spectra of EUV using the exact same time period. The results show that for all the spectral shapes are clearly different. In fact, the calculated Pearson's linear correlation coefficients are ($r_{a,e} = -0.14$, $p_{a,e} = 0.469$), ($r_{b,f} = 0.20$, $p_{b,f} = 0.123$), ($r_{c,g} = -0.49$, $p_{c,g} = 4.66 \times 10^{-12}$) and ($r_{d,h} = -0.36$, $p_{d,h} = 0.048$) for the subfigures a to h in Fig. B.72. This means that a statistically significant negative correlation exists for the distributions of Earth and Mars whereas no significant association is found for Mercury and Venus.

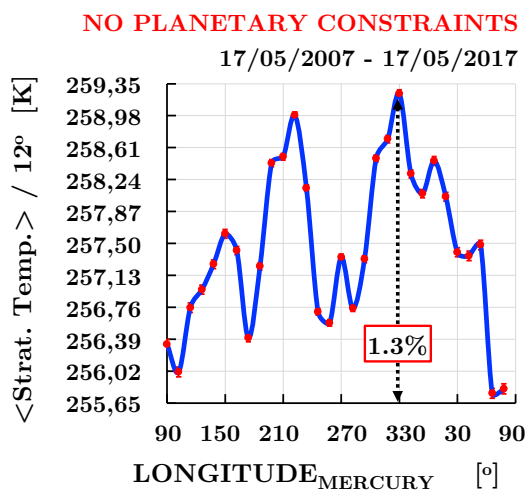

**(a)** Stratospheric temperature distribution for Mercury with 12° bin.

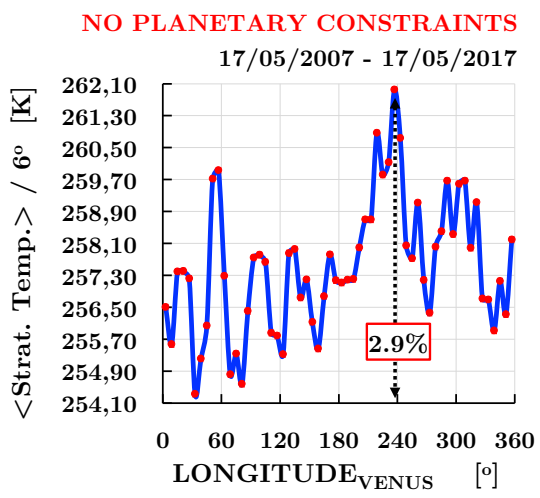

**(b)** Stratospheric temperature distribution for Venus with 6° bin.

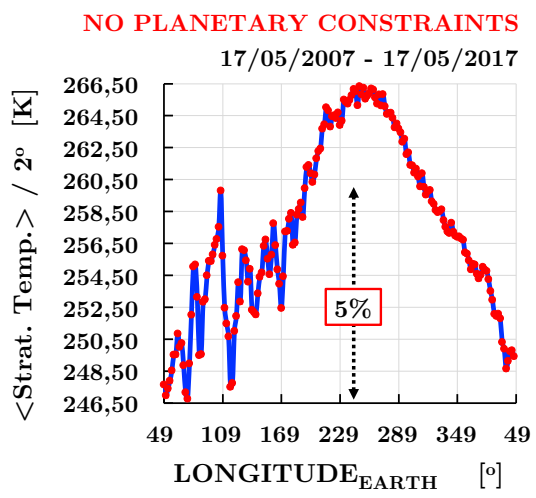

**(c)** Stratospheric temperature distribution for Earth with2° bin.

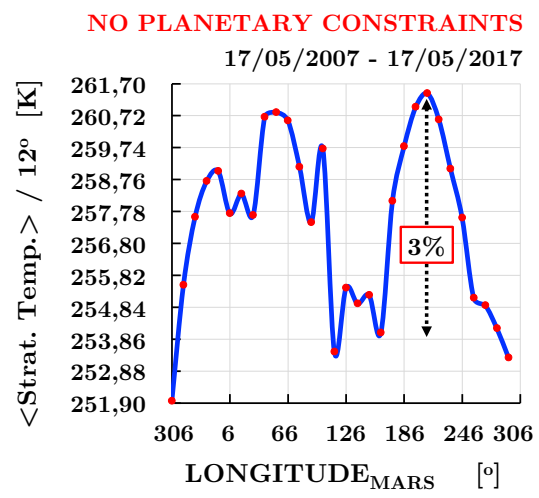

**(d)** Stratospheric temperature distribution for Mars with 12° bin.

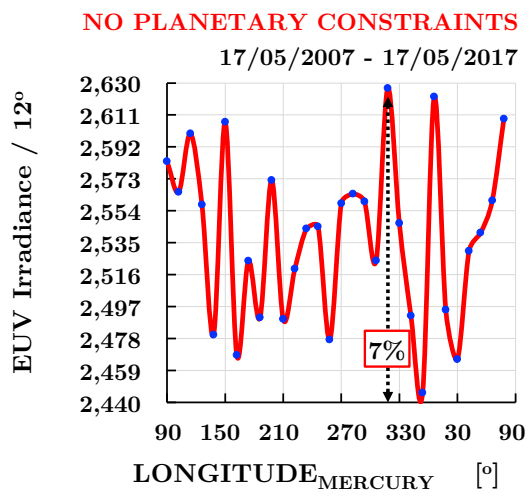

**(e)** EUV distribution for Mercury with 12° bin.

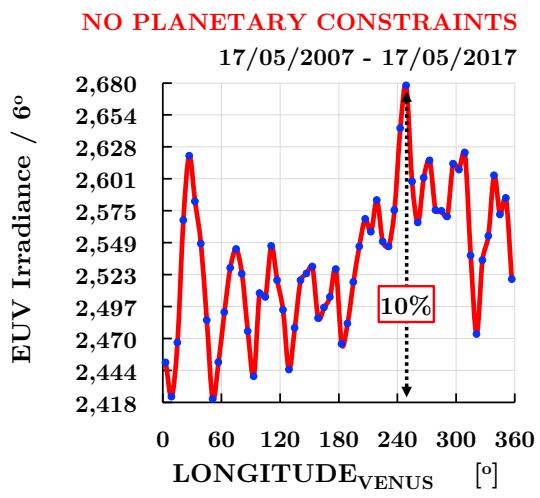

**(f)** EUV distribution for Venus with 6° bin.

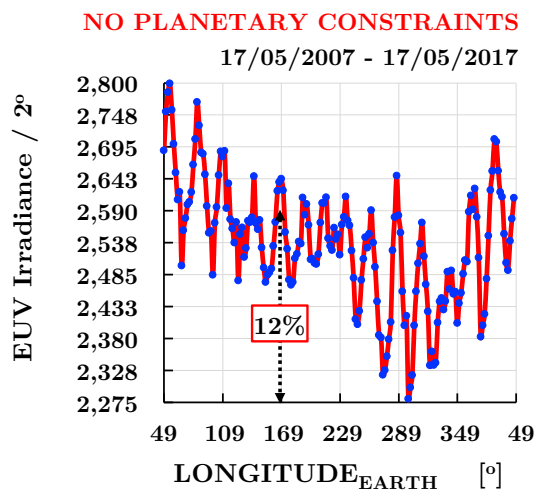

**(g)** EUV distribution for Earth with 2° bin.

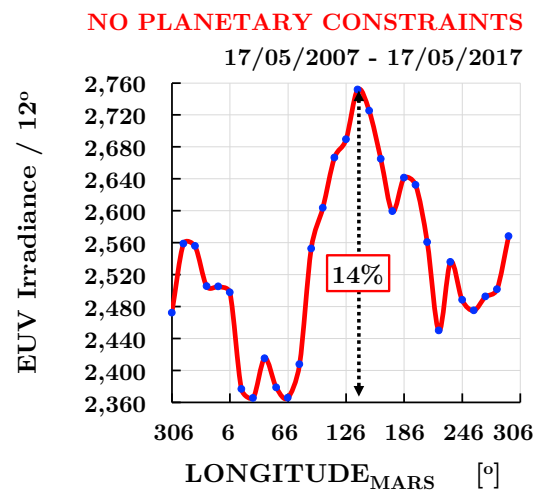

**(h)** EUV distribution for Mars with 12° bin.

**Figure B.72:** Comparison of distributions of stratospheric temperature vs. EUV intensity for the period 17/5/2007 - 17/05/2017.

#### B.10.1.2 Combining planets

The most interesting plots of stratospheric temperature containing also a planetary constraint are also compared with the same conditions with EUV. These results are shown in Fig. B.73. All four distributions show a clear dissimilarity which further excludes solar radiation from being the main and only cause of the upper atmospheric temperature distribution. Calculating the Pearson's correlation coefficients, we get ($r_{a,e} = 0.21$, $p_{a,e} = 0.266$) and ($r_{a,e} = -0.60$, $p_{a,e} = 4.94 \times 10^{-4}$) for the distributions with two constraints in Earth's position in Fig. B.73a and B.73e, ($r_{b,f} = 0.39$, $p_{b,f} = 0.033$) and ($r_{b,f} = 0.20$, $p_{b,f} = 0.293$) for the two distributions with constraints in Earth's position in Fig. B.73b and B.73f, ($r_{c,g} = -0.25$, $p_{c,g} = 0.055$) for Fig. B.73c and B.73g and ($r_{d,h} = -0.57$, $p_{d,h} = 7.75 \times 10^{-12}$) for subfigures d-h.

### B.10.2 Comparison with F10.7

To further exclude that the Sun's irradiation causes the STAs observed here, the daily measured solar activity proxy F10.7 radio line (at ≈ 2.8 GHz) can also be used. In Fig. B.74 the time-series of the stratospheric temperatures are overlaid with the F10.7 solar index for

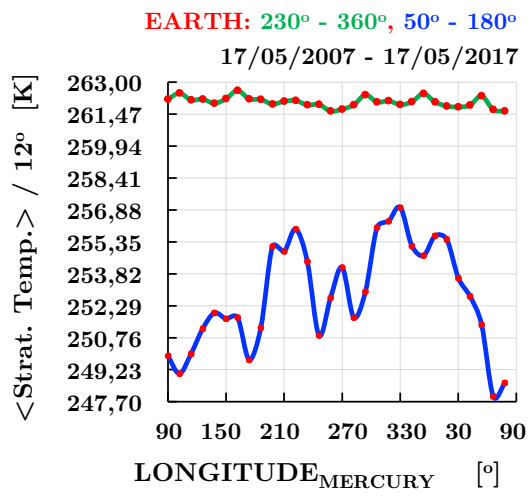

**(a)** Mercury's temperature distribution when Earth is between 50° to 180° and 230° to 60°.

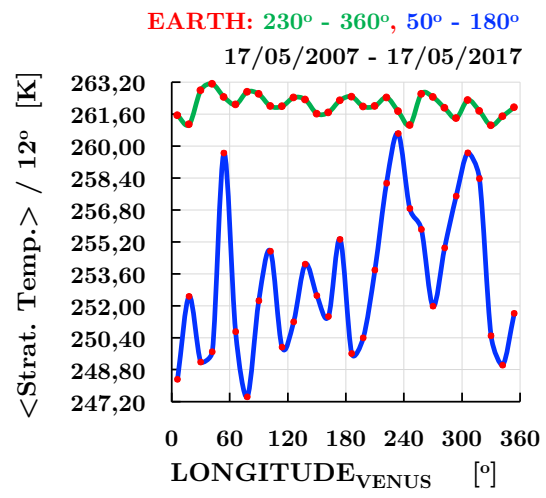

**(b)** Venus' temperature distribution when Earth is between 50° to 180° and 230° to 60°.

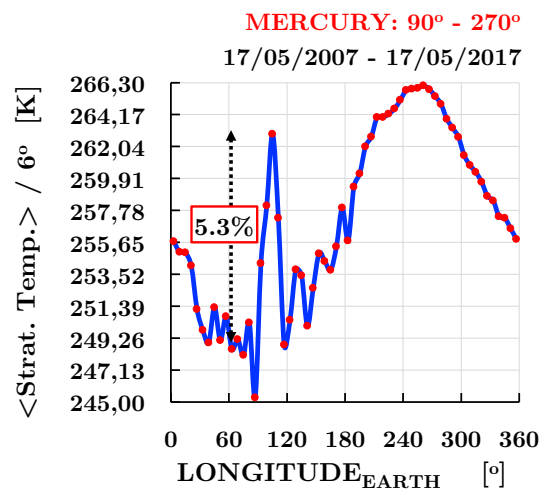

**(c)** Earth's temperature distribution when Mercury is between 90° to 270°.

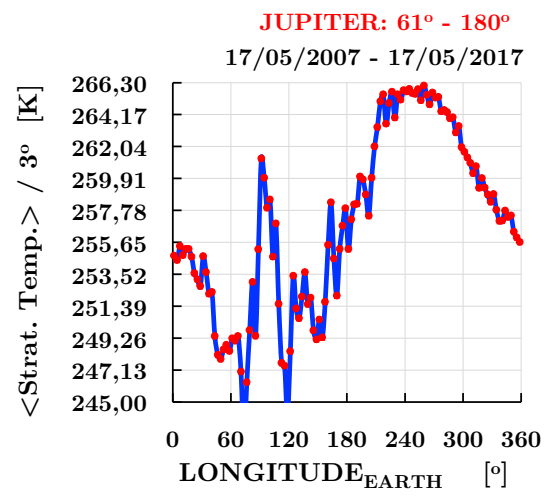

**(d)** Earth's temperature distribution when Jupiter is between 61° to 180°.

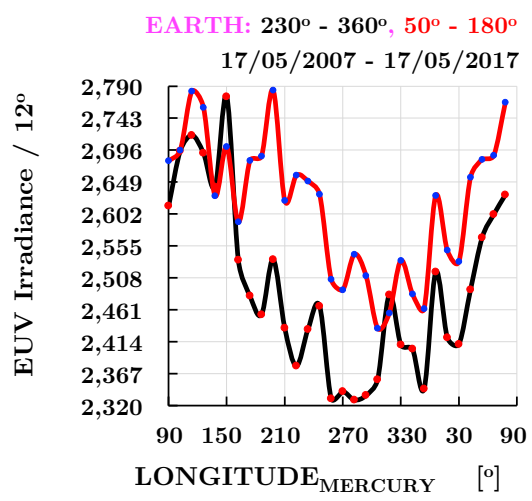

**(e)** Mercury's EUV distribution when Earth is between 50° to 180° and 230° to 60°.

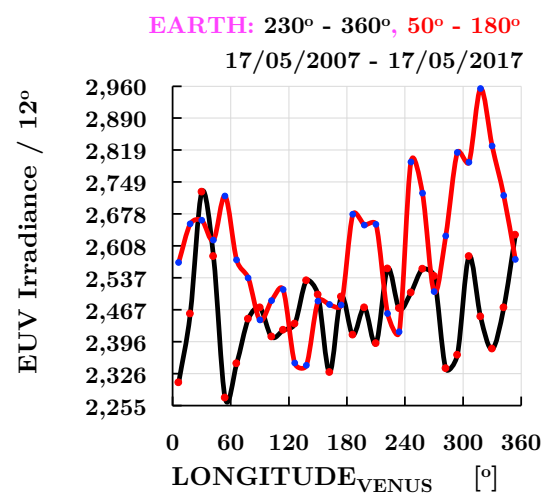

**(f)** Venus' EUV distribution when Earth is between 50° to 180° and 230° to 60°.

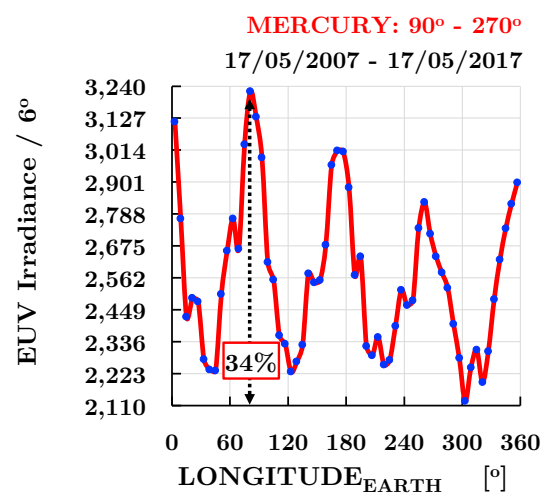

**(g)** Earth's EUV distribution when Mercury is between 90° to 270°.

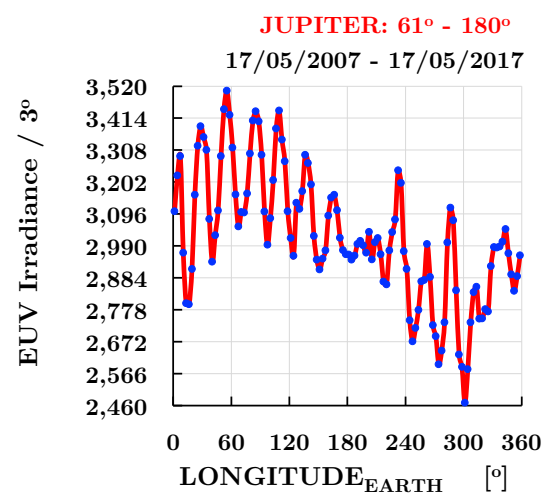

**(h)** Earth's EUV distribution when Jupiter is between 61° to 180°.

**Figure B.73:** Comparison of positionally-constrained longitudinal distributions of stratospheric temperature vs. EUV intensity for the period 17/5/2007 - 17/05/2017.

the same period. In this case the correlation analysis gives ($r = -0.05$, $p = 8.36 \times 10^{-4}$) which, as with EUV, indicates a very small degree of negative but statistically significant correlation.

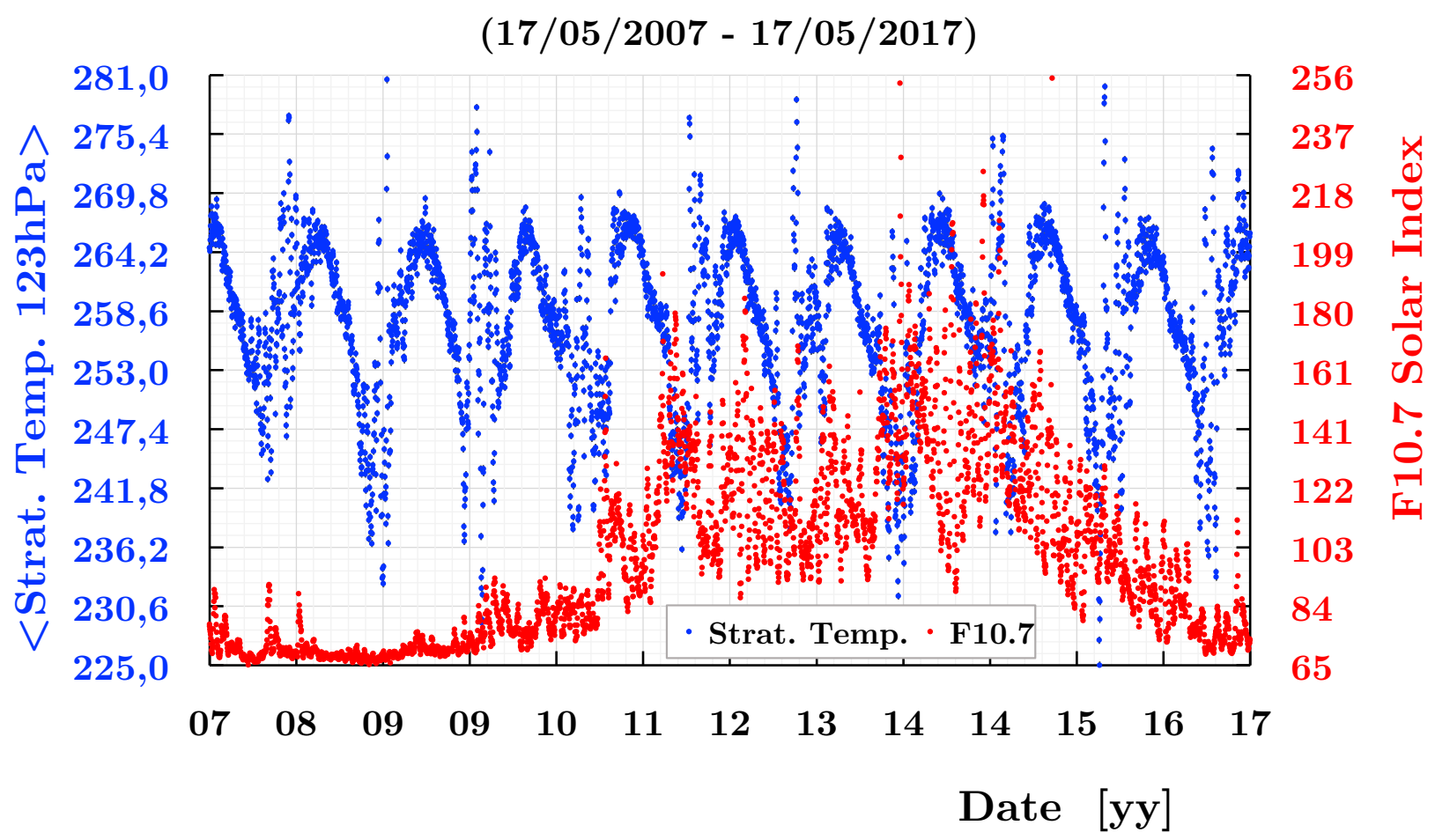

**Figure B.74:** Average stratospheric temperature data and F10.7 solar index for the same period 17/05/2007 - 17/05/2017.

#### B.10.2.1 Single planets

In Fig. B.75, the longitudinal signle-planet distributions of the stratospheric temperature are compared with the corresponding ones from F10.7. The results indicate no similarity between all distributions. The correlation analysis for the compared subfigures a-h gives ($r_{a,e} = -0.24$, $p_{a,e} = 0.206$), ($r_{b,f} = 0.22$, $p_{b,f} = 0.090$), ($r_{c,g} = -0.26$, $p_{c,g} = 5.23 \times 10^{-4}$) and ($r_{d,h} = -0.36$, $p_{d,h} = 0.048$) showing an even negative linear correlation for Earth and Mars and no significant correlation for the spectra of Mercury and Venus.

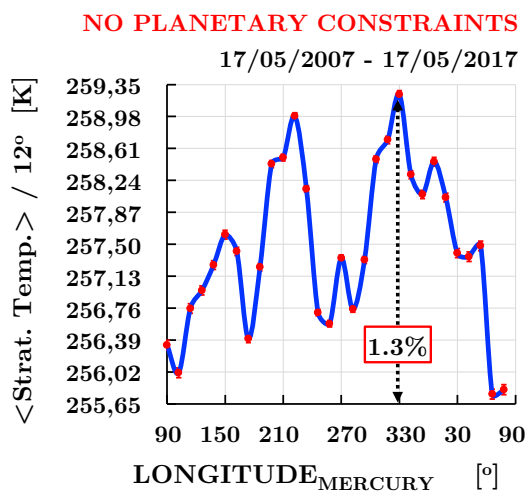

**(a)** Stratospheric temperature distribution for Mercury with 12° bin.

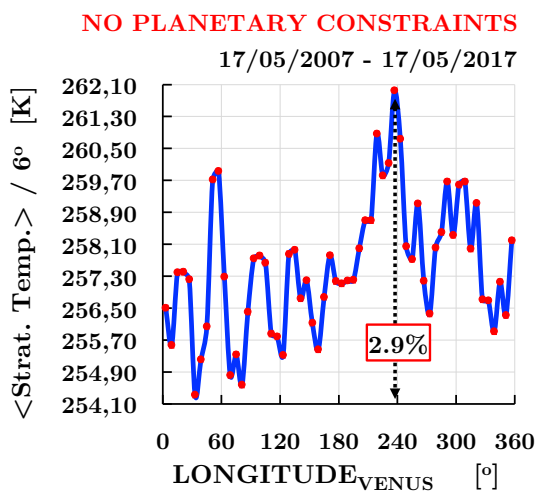

**(b)** Stratospheric temperature distribution for Venus with 6° bin.

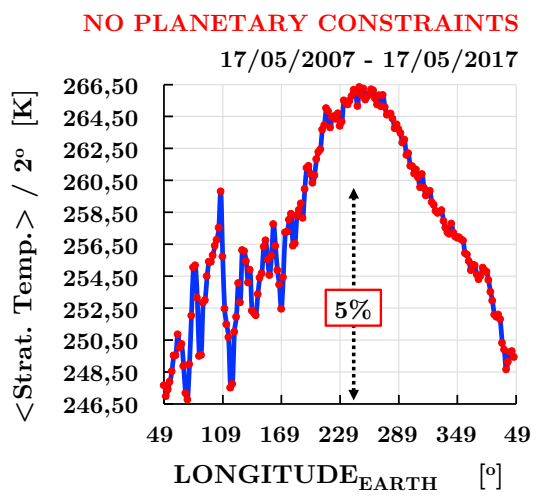

**(c)** Stratospheric temperature distribution for Earth with2° bin.

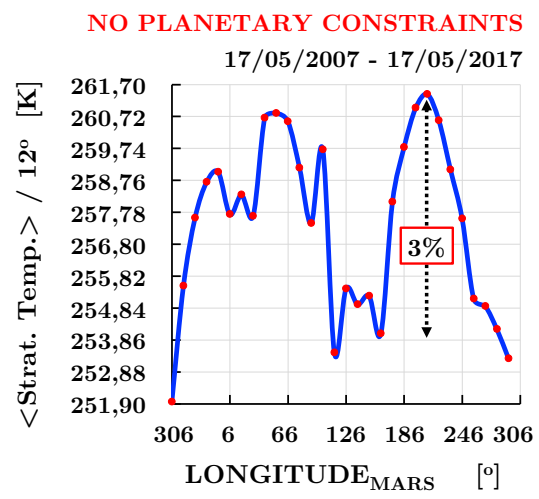

**(d)** Stratospheric temperature distribution for Mars with 12° bin.

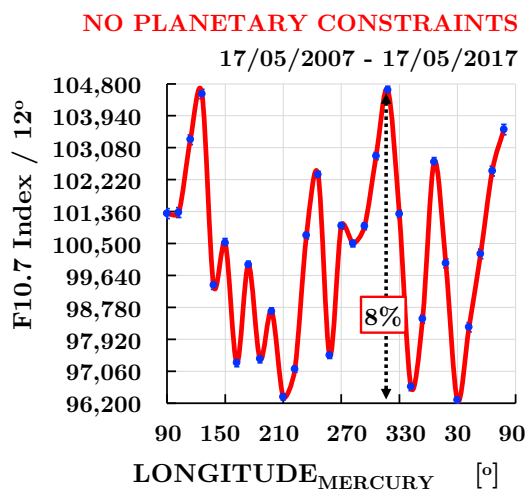

**(e)** F10.7 distribution for Mercury with 12° bin.

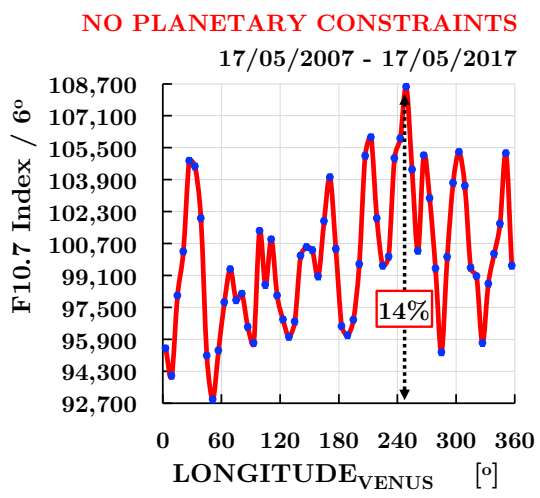

**(f)** F10.7 distribution for Venus with 6° bin.

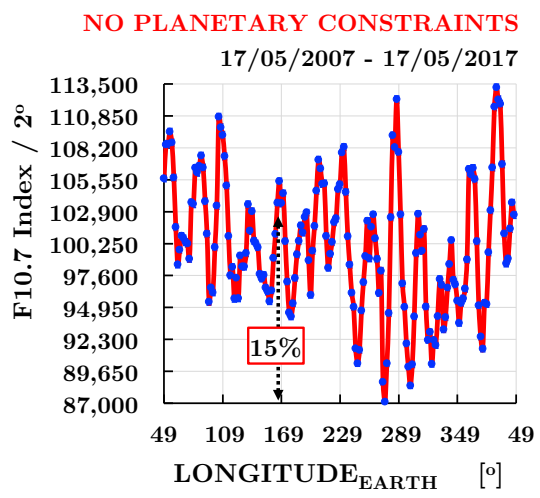

**(g)** F10.7 distribution for Earth with 2° bin.

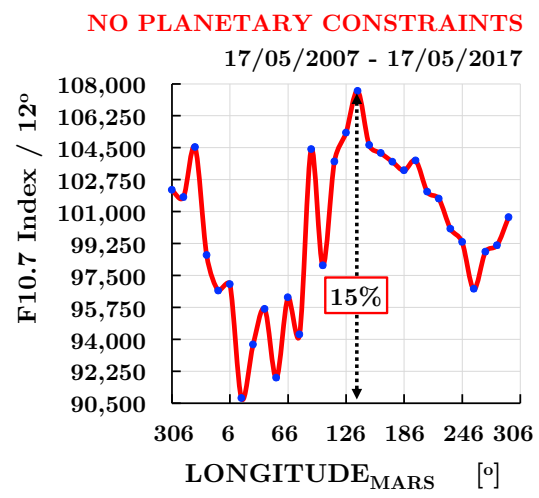

**(h)** F10.7 distribution for Mars with 12° bin.

**Figure B.75:** Comparison of stratospheric temperature vs. F10.7 solar index (17/5/2007 - 17/05/2017).

#### B.10.2.2 Combining planets

Finally, to further exclude the known solar radiation as being the only driving source behind the stratospheric temperature distributions, in Fig. B.76 the main observed planetary differences on Earth's reference frame while Mercury and Venus propagate in 180° opposite orbital arcs are compared with the corresponding ones from F10.7. It is obvious that the F10.7 spectra are strikingly different, thus confirming once more the aforementioned exo-solar claim. The Pearson's correlation coefficient for the two cases confirms this claim givint ($r_{a,c} = -0.15$, $p_{a,c} = 0.238$) for Fig. B.76a and B.76c and ($r_{b,d} = 0.36$, $p_{b,d} = 0.08$) for Fig. B.76b and B.76d. It is noted that the remaining spectra, like the ones shown in Fig. B.73, also demonstrate similar results but are not shown here due to space considerations.

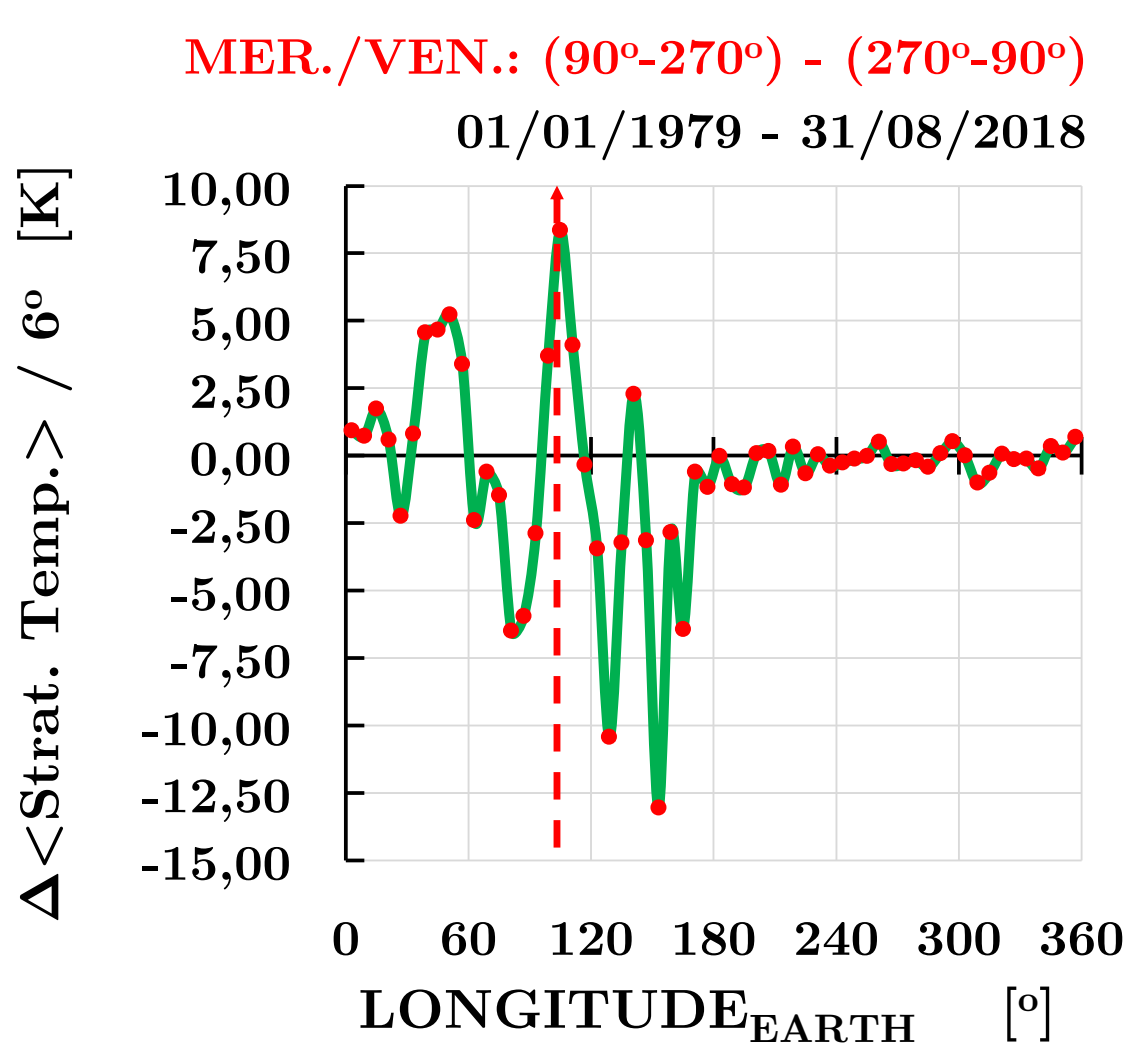

(a) Earth's temperature distribution difference when Mercury and Venus are both between 90° to 270° and 270° to 90°.

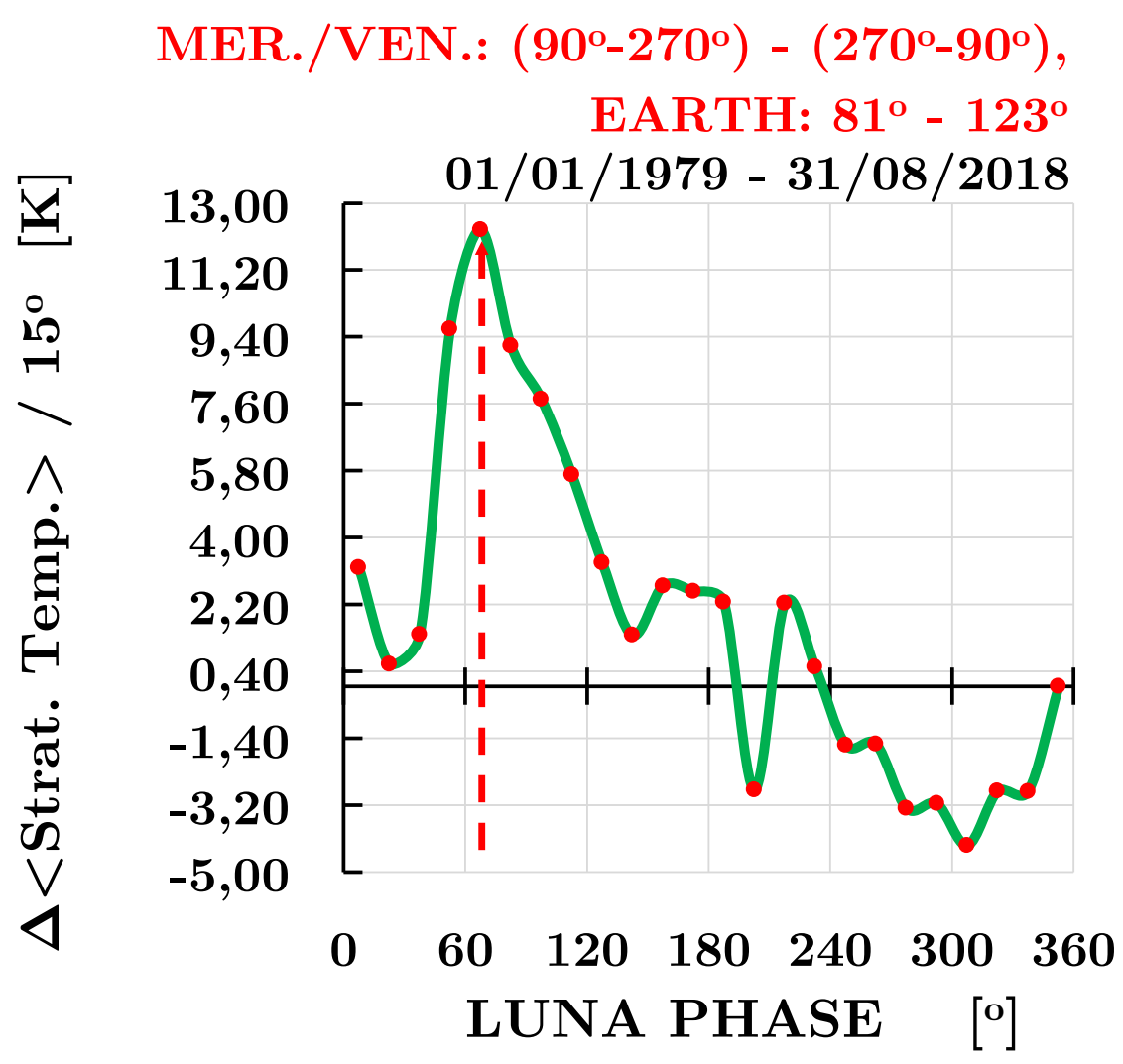

(b) Moon's temperature distribution difference when Mercury and Venus are both between 90° to 270° and 270° to 90° while Earth is fixed between 81° to 123°.

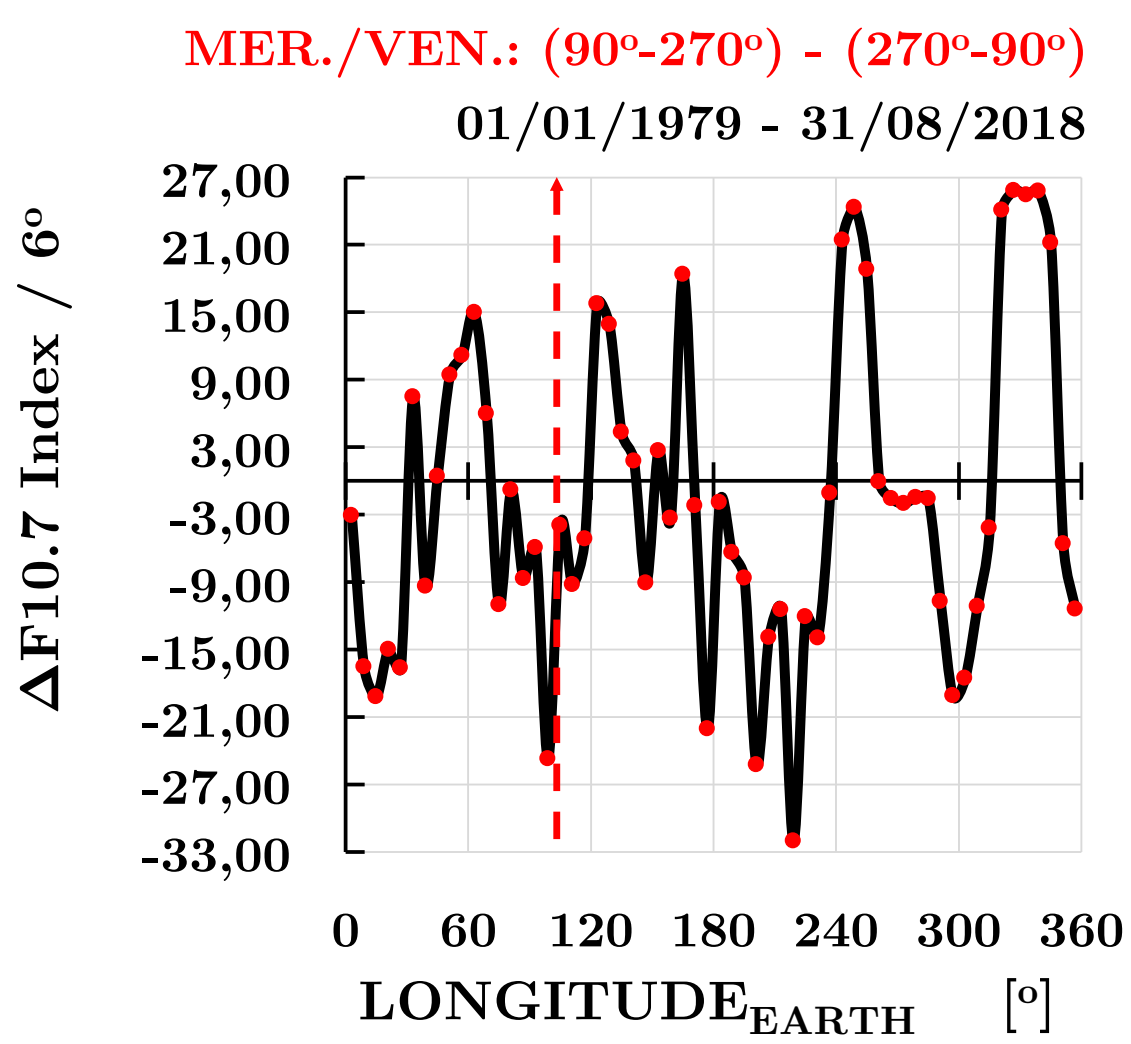

(c) Earth's F10.7 distribution difference when Mercury and Venus are both between 90° to 270° and 270° to 90°.

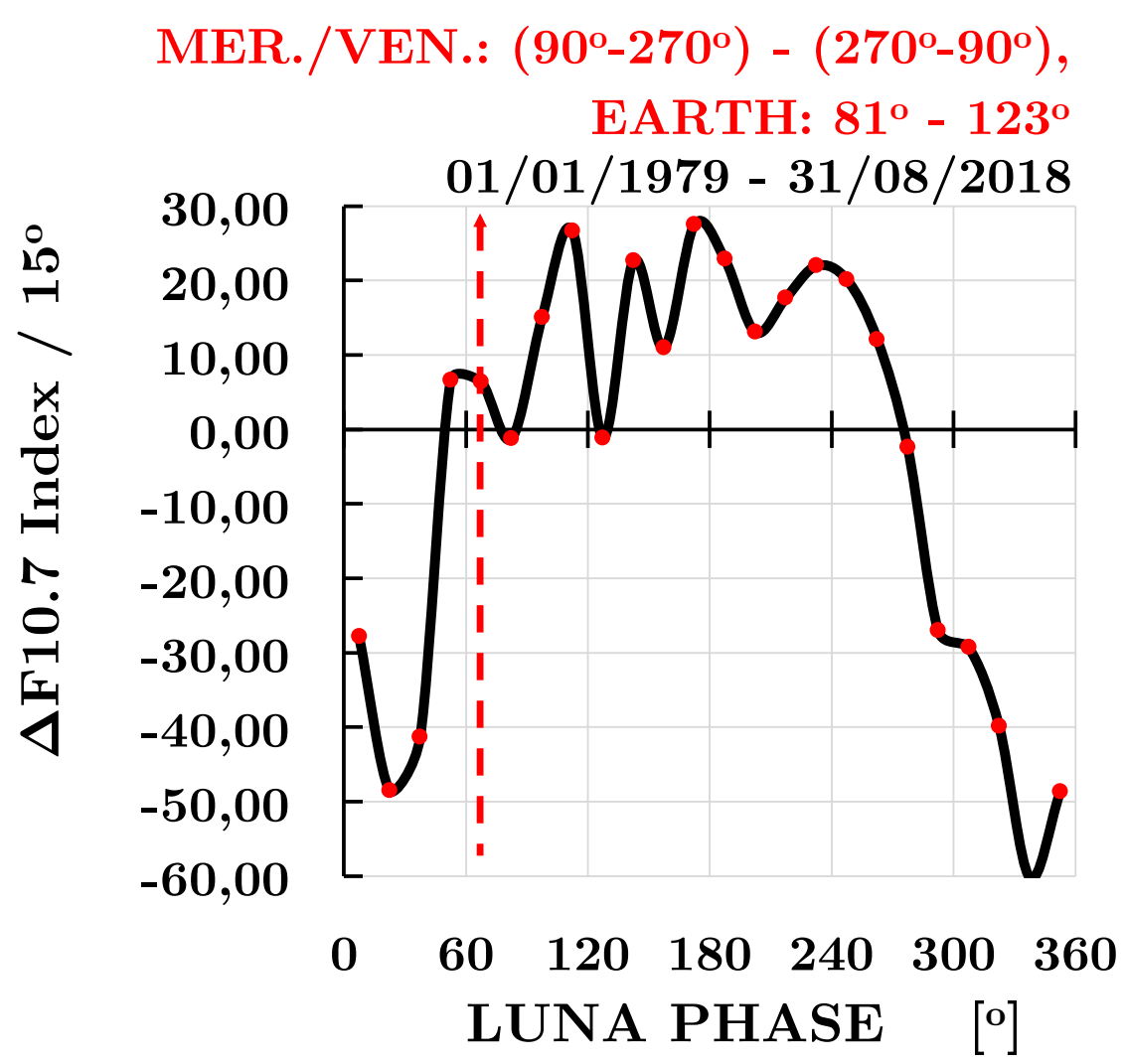

(d) Moon's F10.7 distribution difference when Mercury and Venus are both between 90° to 270° and 270° to 90° while Earth is fixed between 81° to 123°.

**Figure B.76:** Comparison of multiply positionally-constrained longitudinal distributions of stratospheric temperature vs. F10.7 solar index for the period 01/01/1979 - 31/08/2018. The red dashed lines indicate the position of the two peaks in the stratospheric temperature distributions.

# B.11 Earthquakes

## B.11.1 Comparison with F10.7

In [370] a correlation of the occurrence of large EQs with $M > 5.6$ with the solar activity has been found. At the same time, in Sect. 17.3.2 an additional correlation of the number of EQs with the 11 y solar cycle was also inferred based on the derived $\sim$ 11 y peak from the Fourier periodogram. Therefore, as next, a direct comparison is performed with the F10.7 solar proxy line. In Fig. B.77 the time series of the data from these two observations are depicted for the same overlapping period. The statistical correlation analysis yields ($r = -0.086$, $p = 1.50 \times 10^{-10}$) which points to a very small negative linear correlation for the general trend of the two datasets agreeing with other observations [383]. As next the various planetary distributions are also compared.

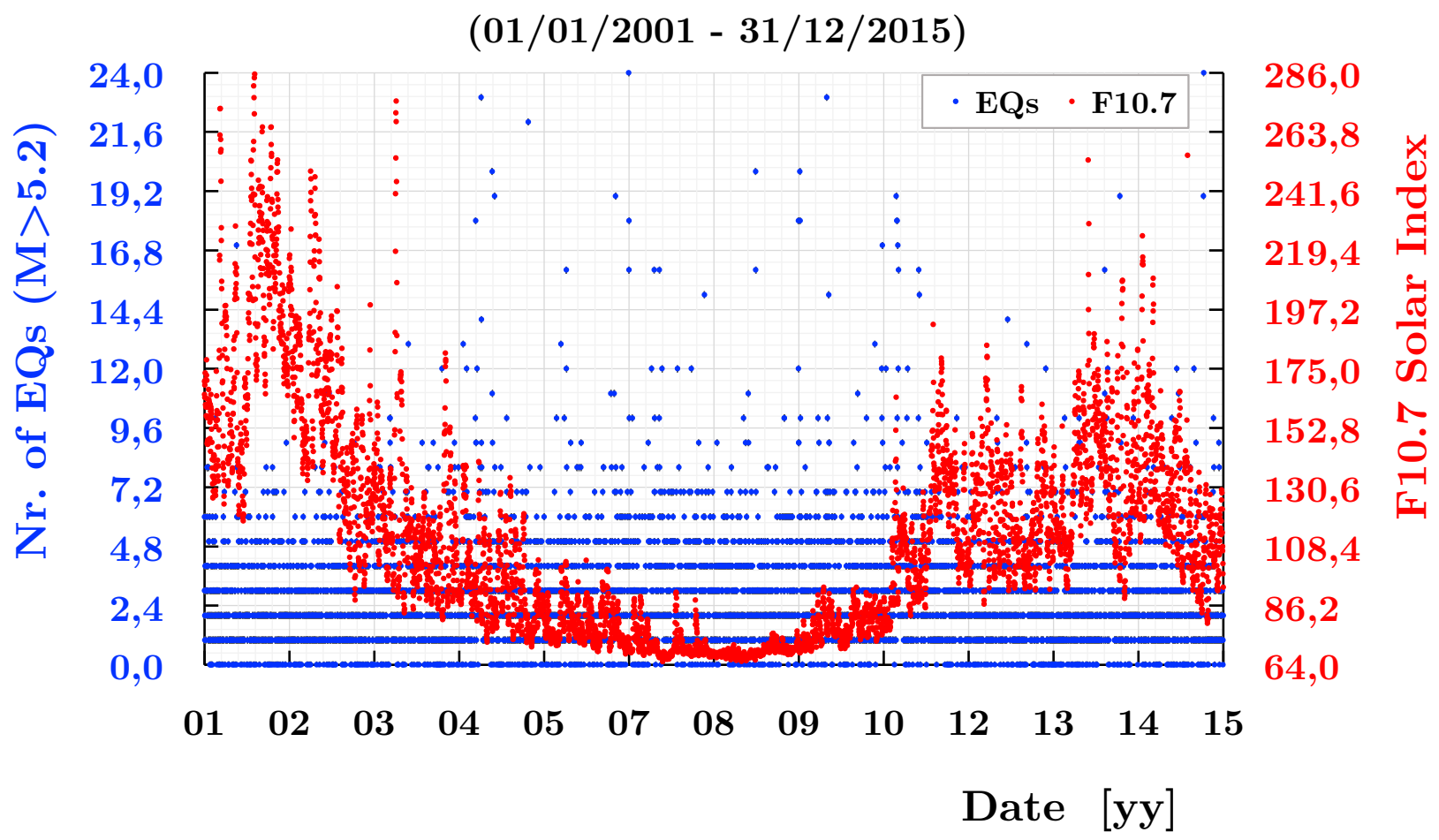

**Figure B.77:** Nr. of EQs and F10.7 solar index for the same period 01/01/2001 - 31/12/2015.

### B.11.1.1 Single planets

In Fig. B.78, the corresponding spectra of F10.7 for single planetary positions, are compared with the ones from Fig. 17.5 for the same period of 01/01/2001 to 31/12/2015. The three plots of F10.7 Fig. B.78d through B.78f contain also error bars which are less that 1% and therefore not directly visible in the plots. Therefore, a first approximation shows that there is a dissimilarity between the EQs and solar activity, which means that the origin of the planetary relationships seem to be exo-solar in origin. The statistical correlation analysis confirms the above observation as it finds no statistically significant correlation. More specifically, the calculated Pearson's correlation coefficients and p-values are ($r_{a,d} = 0.35$, $p_{a,d} = 0.128$) for Fig. B.78a and B.78d, ($r_{b,e} = -0.05$, $p_{b,e} = 0.828$) for Fig. B.78b and B.78e and ($r_{c,f} = -0.09$, $p_{c,f} = 0.752$) for Fig. B.78c and B.78f.

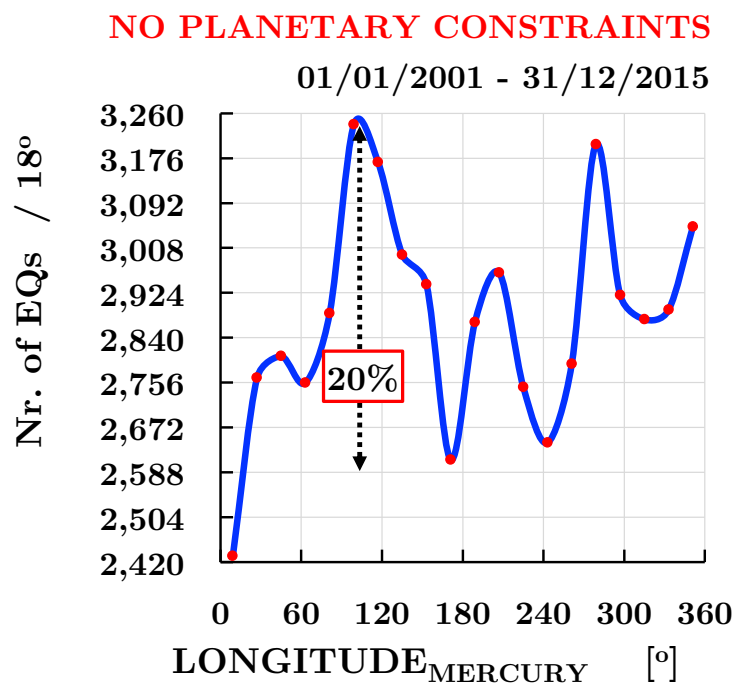

(a) Nr. of EQs as a function of the position of Mercury for 18° bin.

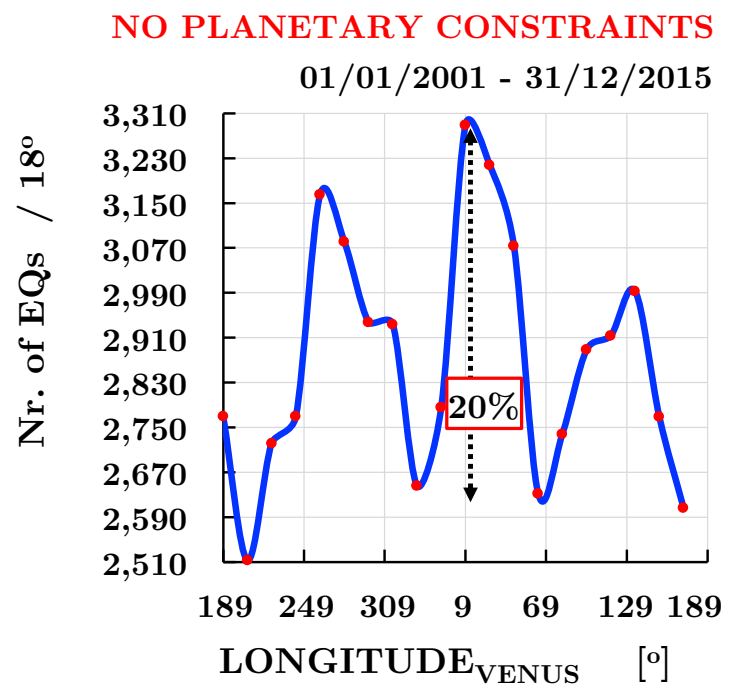

(b) Nr. of EQs as a function of the position of Venus for 18° bin.

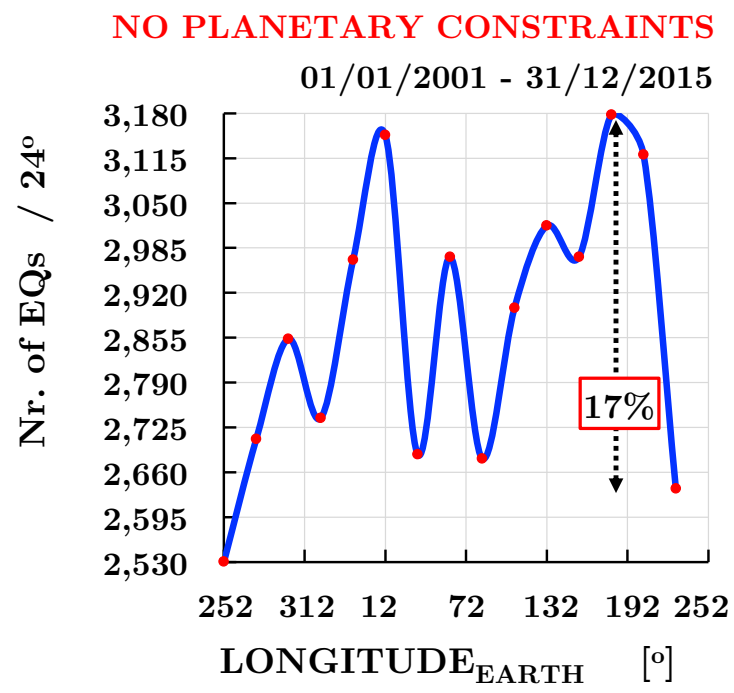

(c) Nr. of EQs as a function of the position of Earth for 24° bin.

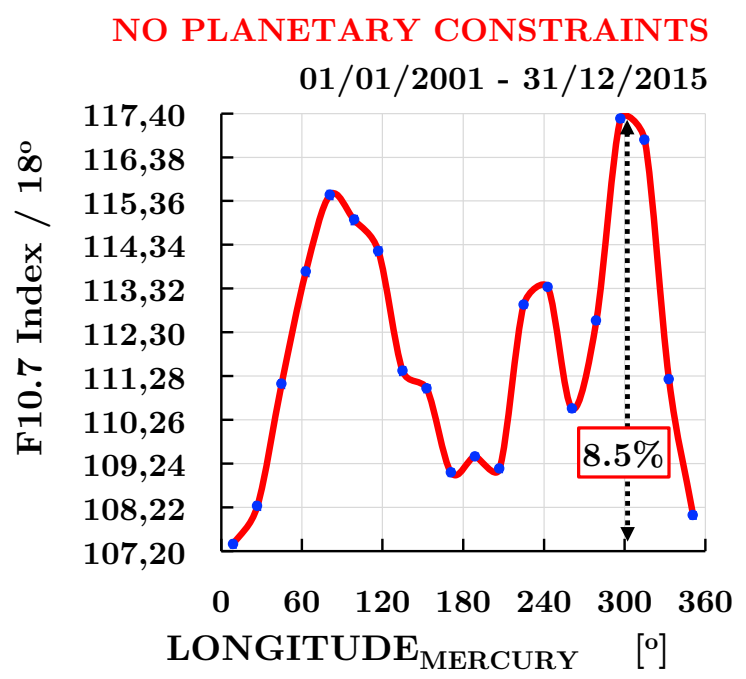

(d) F10.7 as a function of the position of Mercury for 18° bin.

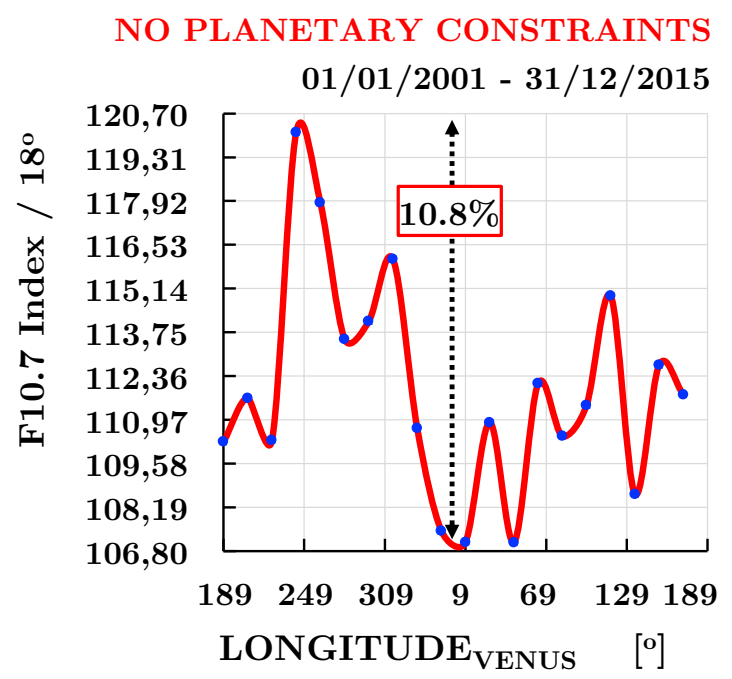

(e) F10.7 as a function of the position of Venus for 18° bin.

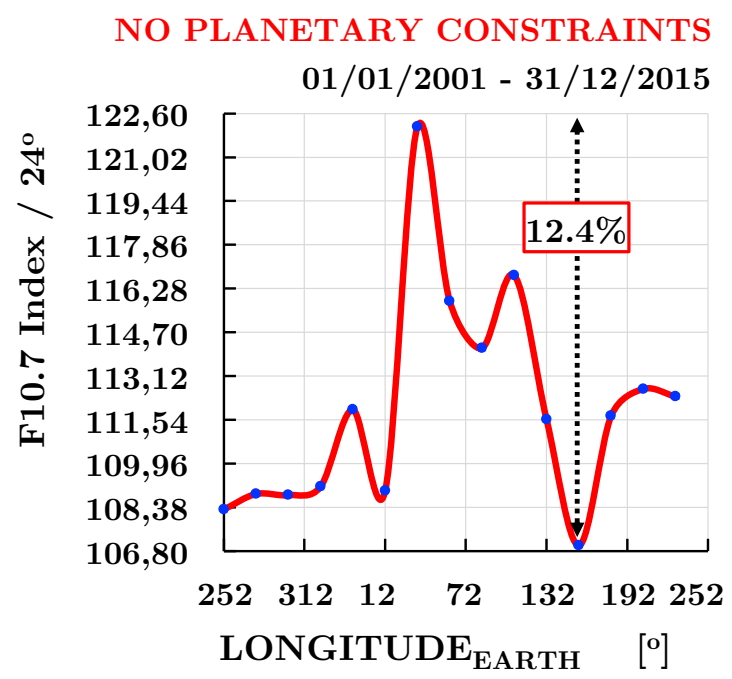

(f) F10.7 as a function of the position of Earth for 24° bin.

**Figure B.78:** Comparison of F10.7 solar proxy with the Nr. or EQs for the same conditions without any constraints.

#### B.11.1.2 Combining planets

As a next step, also the plots with a planetary constraint are compared for the number of EQs and F10.7. First of all, the distributions from Fig. 17.7 are compared with the F10.7 spectra in Fig. B.79. The derived Pearson's correlation coefficients for the two cases of Mercury are ($r_{a,c} = 0.13$, $p_{a,c} = 0.508$) for Fig. B.79a and B.79c and ($r_{b,d} = -0.30$, $p_{b,d} = 0.056$) for Fig. B.79b and B.79d. This means that neither of the two cases shows a significant linear correlation.

The same procedure is performed in Fig. B.80 for the plots from Fig. 17.8 where Venus' reference frame is selected. In this case the correlation analysis gives ($r_{a,c} = -0.23$, $p_{a,c} = 0.162$) for Fig. B.80a and B.80c and ($r_{b,d} = -0.063$, $p_{b,d} = 0.682$) for Fig. B.80b and B.80d. However, due to the appearance of two outliers in the latter case, the Spearman and Kendall coefficients are also calculated giving ($r_{S_{b,d}} = 0.15$, $p_{S_{b,d}} = 0.331$) and ($\tau_{b,d} = 0.11$, $p_{b,d} = 0.304$) respectively, therefore again we have non-significant correlations.

In Fig. B.81 the various distributions from Fig. 17.9 are compared with F10.7 solar radio line for the Earth's reference frame when other planet's position are vetoed in a specific

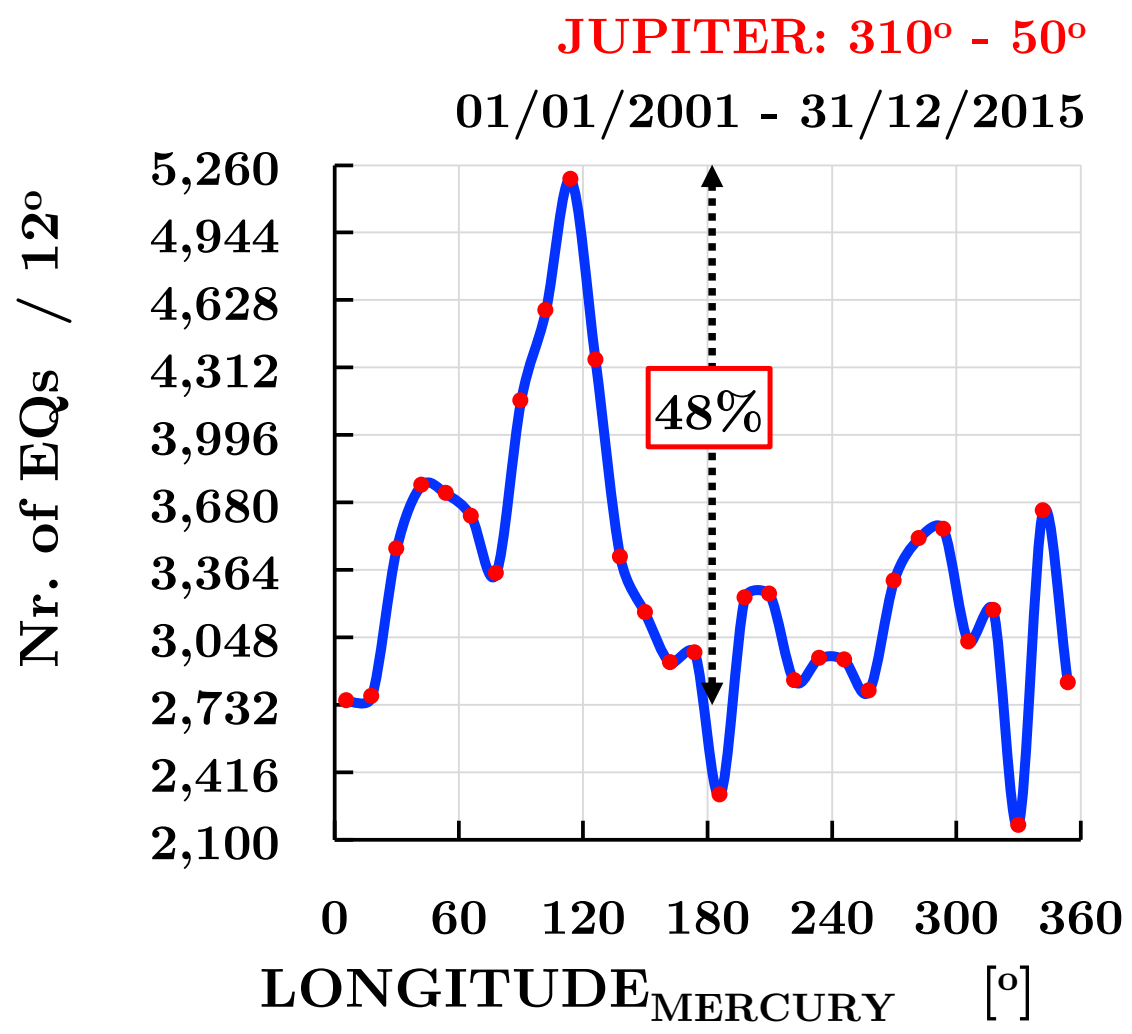

**(a)** Nr. of EQs distribution for Mercury when Jupiter propagates between 310° to 50°.

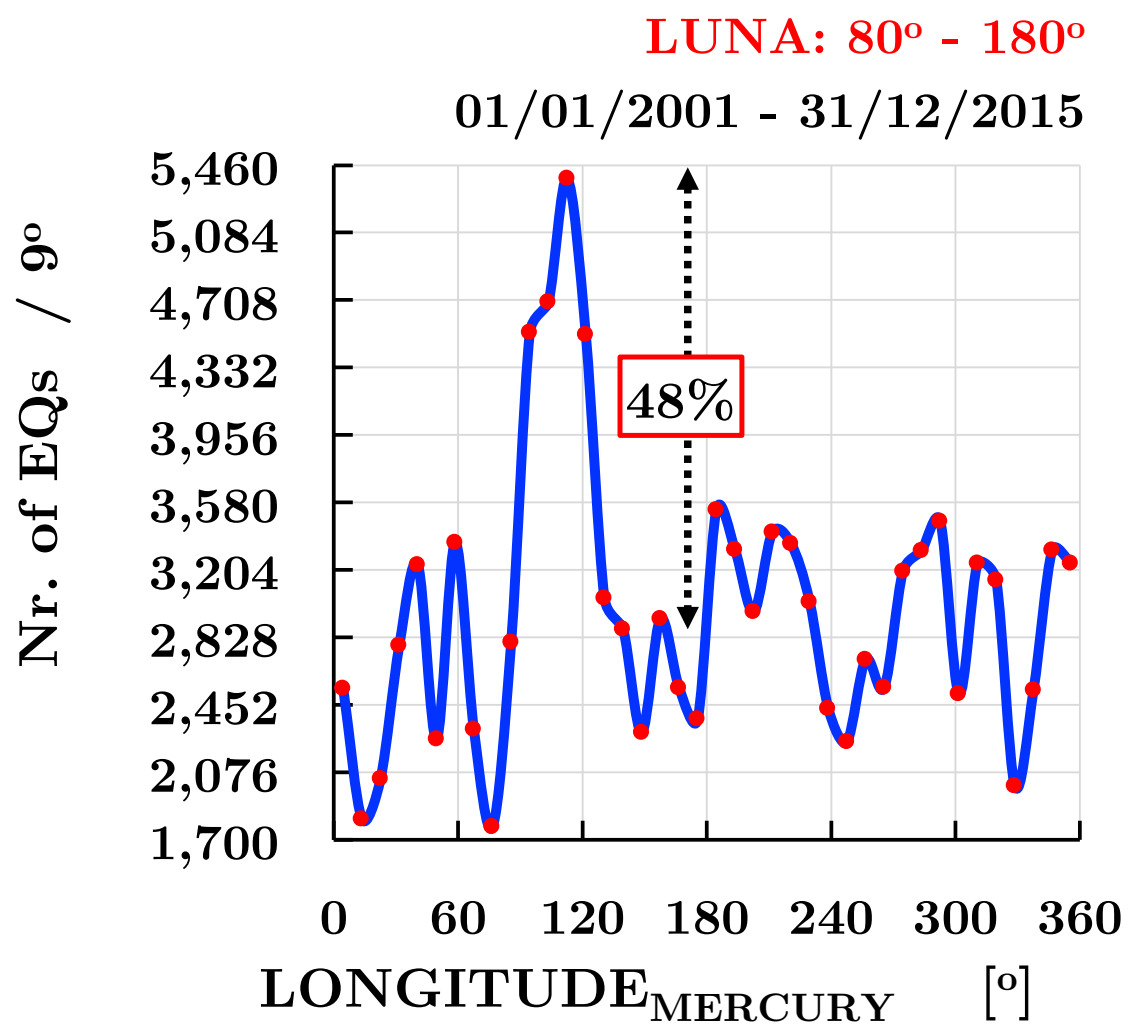

**(b)** Nr. of EQs distribution for Mercury when Moon's phase is between 80° to 180°.

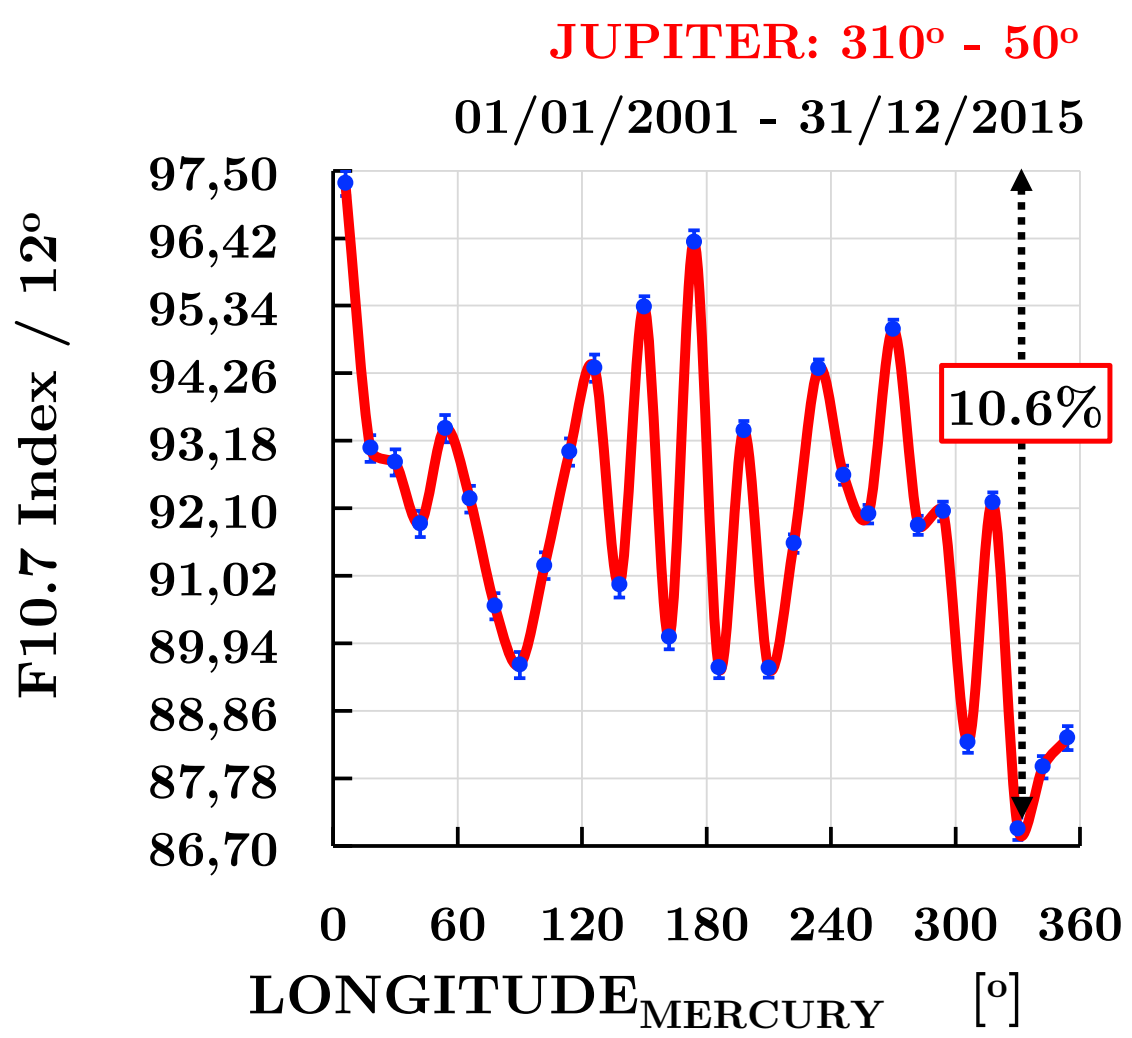

**(c)** F10.7 distribution for Mercury when Jupiter propagates between 310° to 50°.

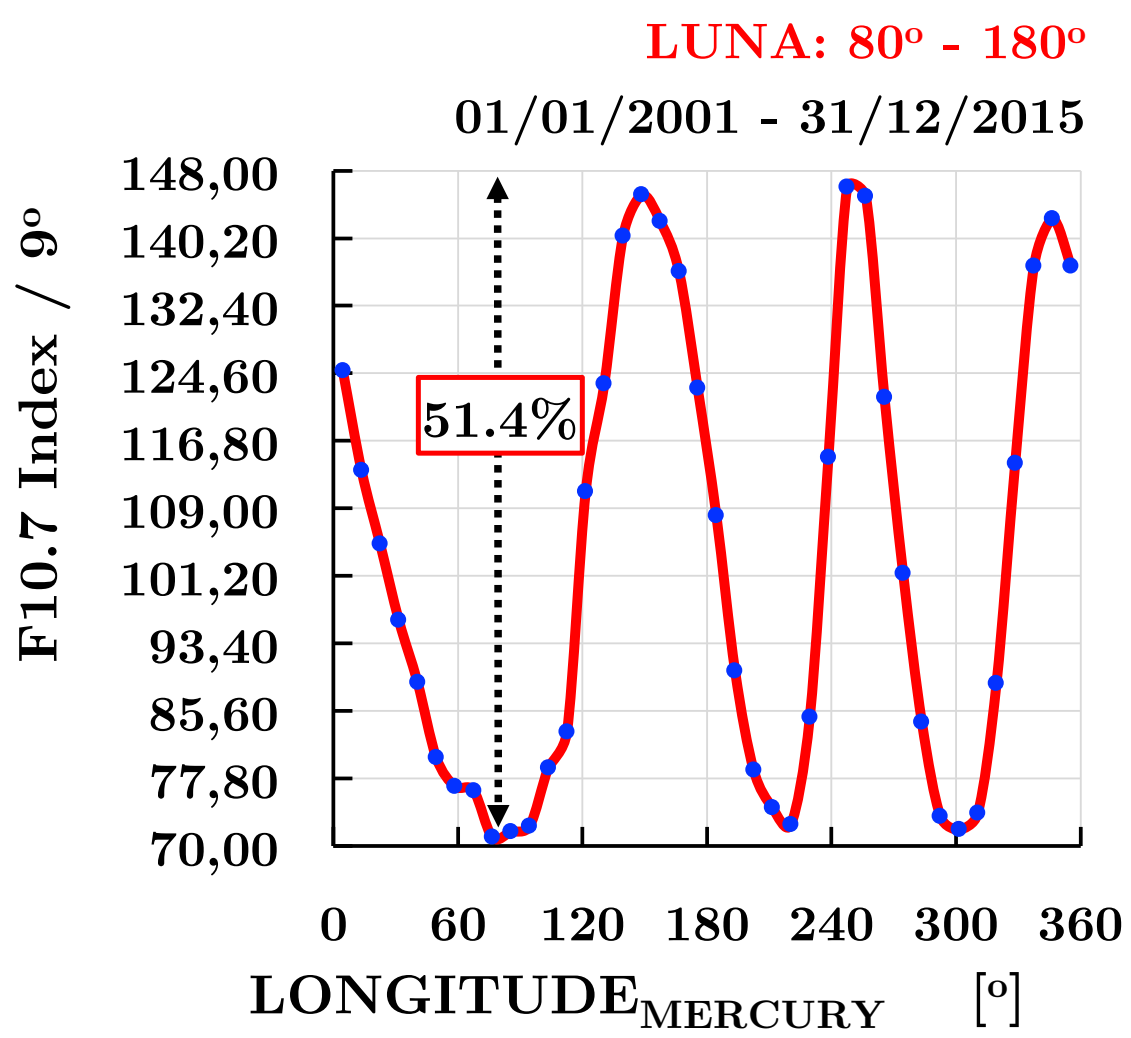

**(d)** F10.7 distribution for Mercury when Moon's phase is between 80° to 180°.

**Figure B.79:** Nr. or EQs vs. F10.7 solar proxy for the reference frame of Mercury while Jupiter's and Moon's positions are constrained.

heliocentric longitudinal range. For Fig. B.81a through B.81f the correlation coefficients are accordingly ($r_{a,d} = -0.24$, $p_{a,d} = 0.069$), ($r_{b,e} = -0.35$, $p_{b,e} = 0.202$) and ($r_{c,f} = -0.09$, $p_{c,f} = 0.495$). In all cases we get a negative linear correlation but not significant in the 0.05 level.

Lastly, Moon's phase when Earth's orbital position is constrained from Fig. 17.10 is produced for the same period for the F10.7 solar proxy. The comparison with the EQ number is shown in Fig. B.82. The correlation coefficient along with the p-value is calculated to ($r_{a,b} = 0.096$, $p_{a,b} = 0.688$) which means that there is no significant linear correlation between the two distributions.

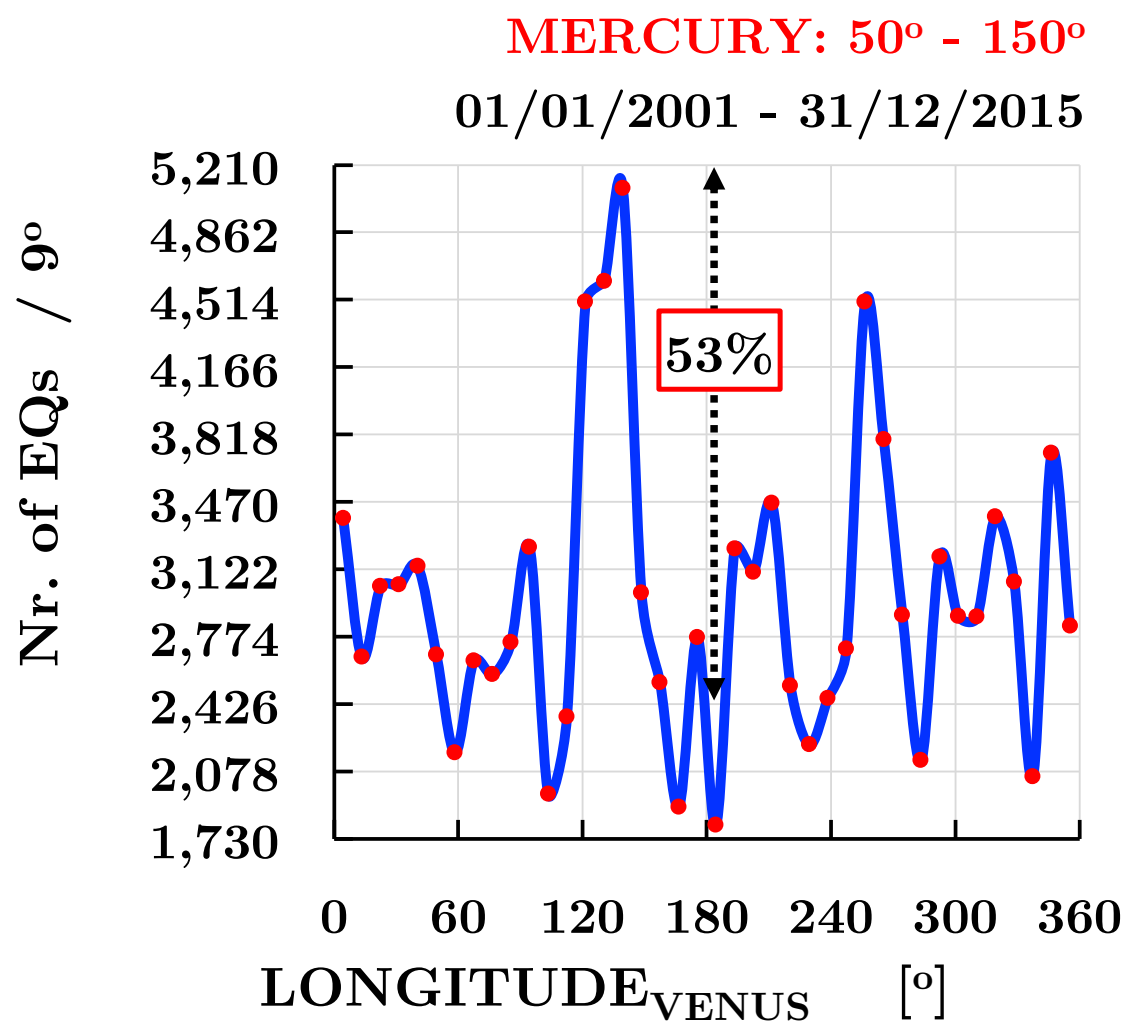

**(a)** Nr. of EQs distribution for Venus when Mercury propagates between 50° to 150°.

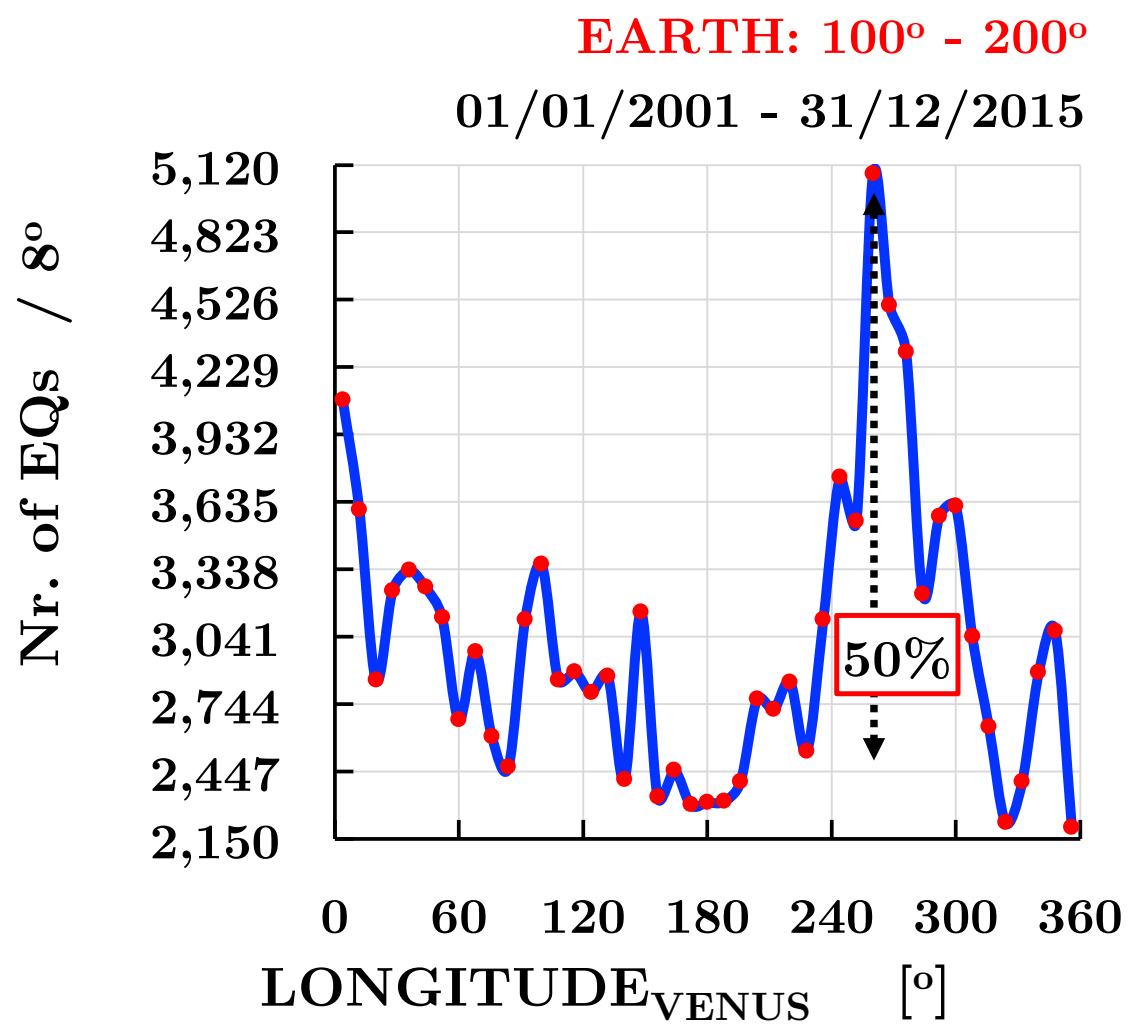

**(b)** Nr. of EQs distribution for Venus when Earth propagates between 100° to 200°.

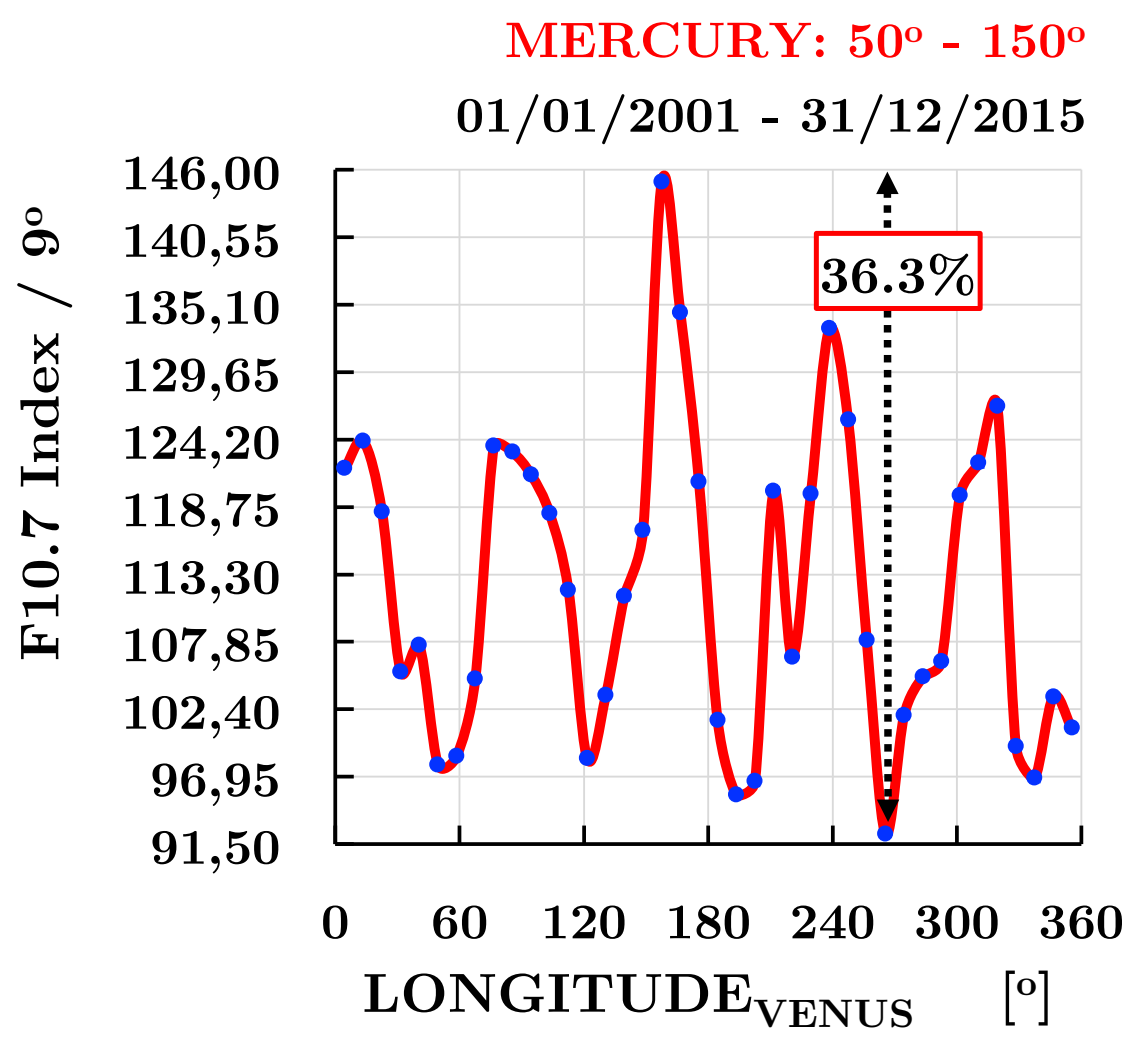

**(c)** F10.7 distribution for Venus when Mercury propagates between 50° to 150°.

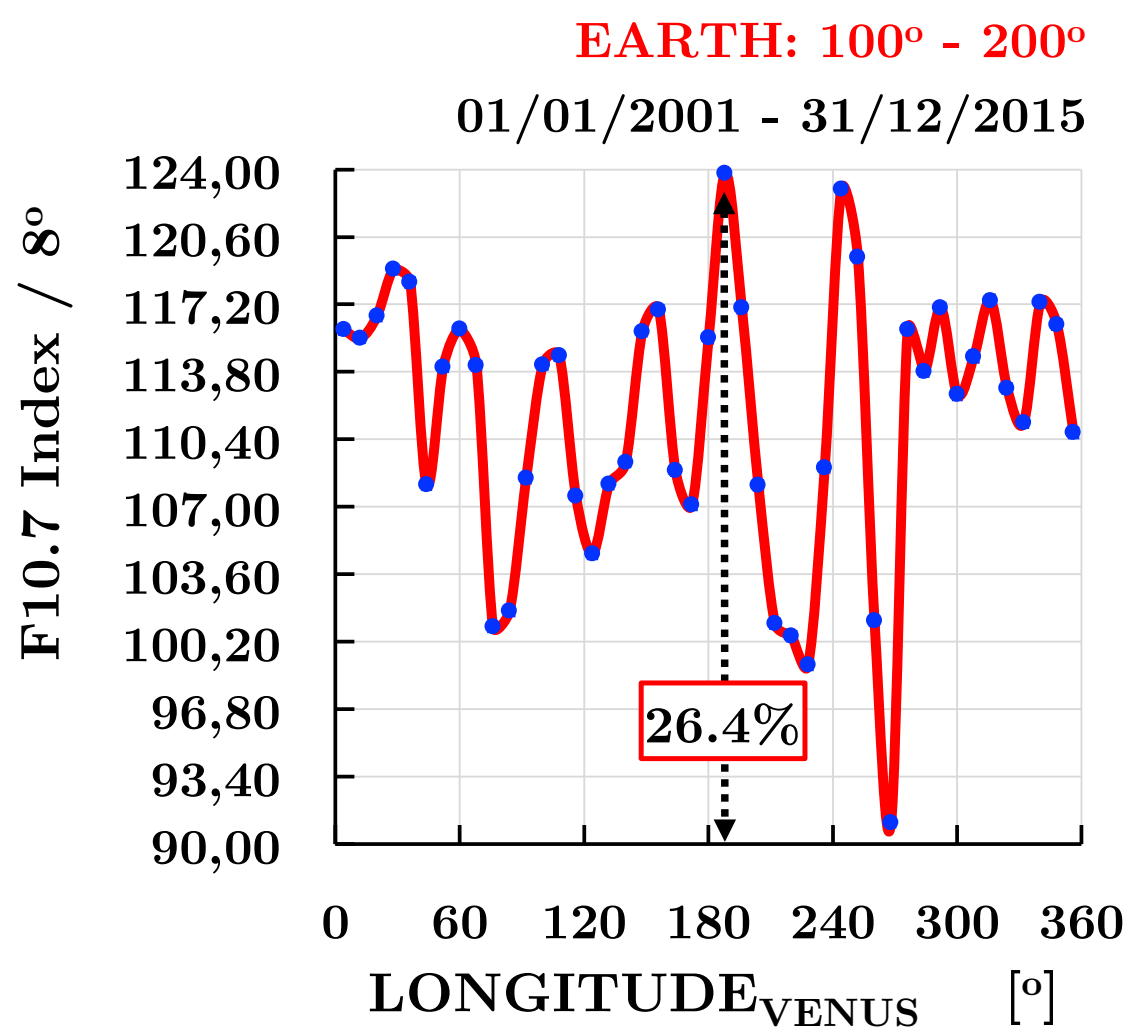

**(d)** F10.7 distribution for Venus when Earth propagates between 100° to 200°.

**Figure B.80:** Nr. of EQs vs. F10.7 solar proxy for the reference frame of Venus while Mercury's and Earth's positions are constrained.

## B.11.2 Correlation with TEC

### B.11.2.1 Earthquakes with $M \geq 8.0$

For the TEC dataset an alternative evaluation procedure is performed. First, the biggest EQs with a magnitude equal to or higher than 8.0 are chosen for the period that the TEC is available, i.e. 01/01/1995 to 31/12/2012. During that period there were 20 EQs fulfilling these conditions, which are shown in Fig. B.83.

As next, a period of 90 d before and after each one of these 20 EQs was defined, as seen in Tab. B.4, in order to plot the total TEC distribution during these periods. The date of each

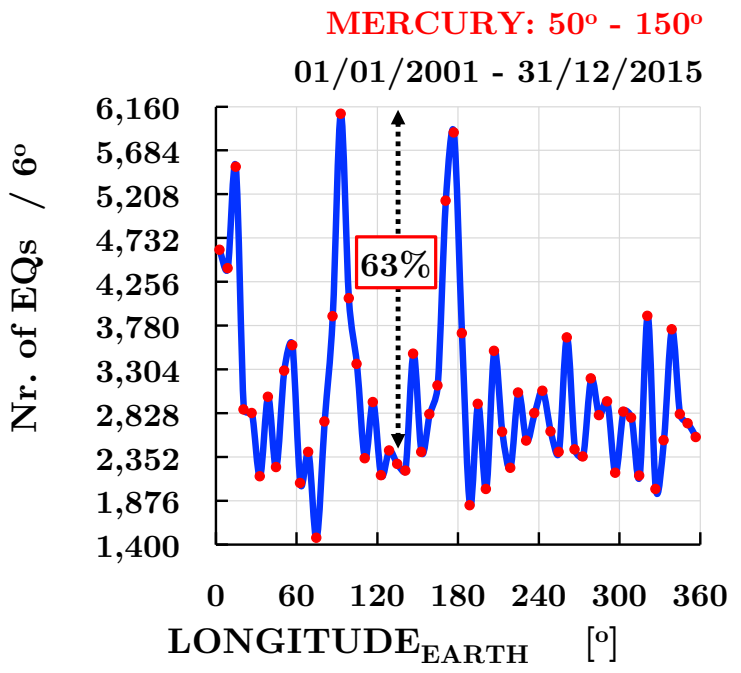

**(a)** Nr. of EQs distribution for Earth when Mercury propagates between 50° to 150°.

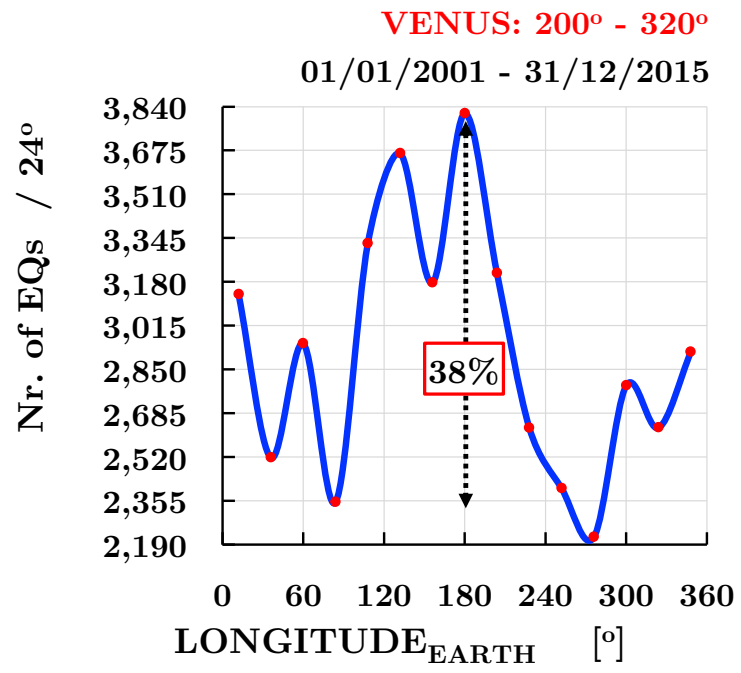

**(b)** Nr. of EQs distribution for Earth when Venus propagates between 220° to 320°.

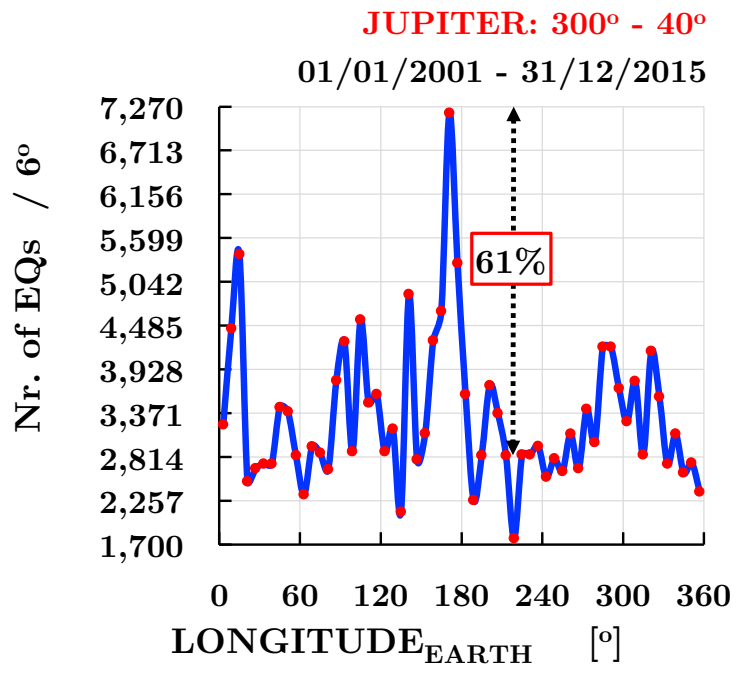

**(c)** Nr. of EQs distribution for Earth when Jupiter propagates between 300° to 40°.

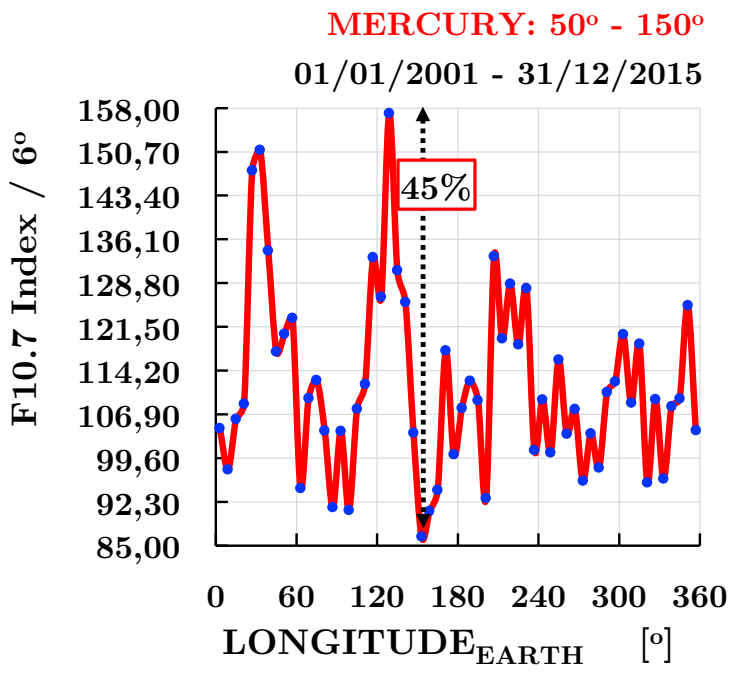

**(d)** F10.7 distribution for Earth when Mercury propagates between 50° to 150°.

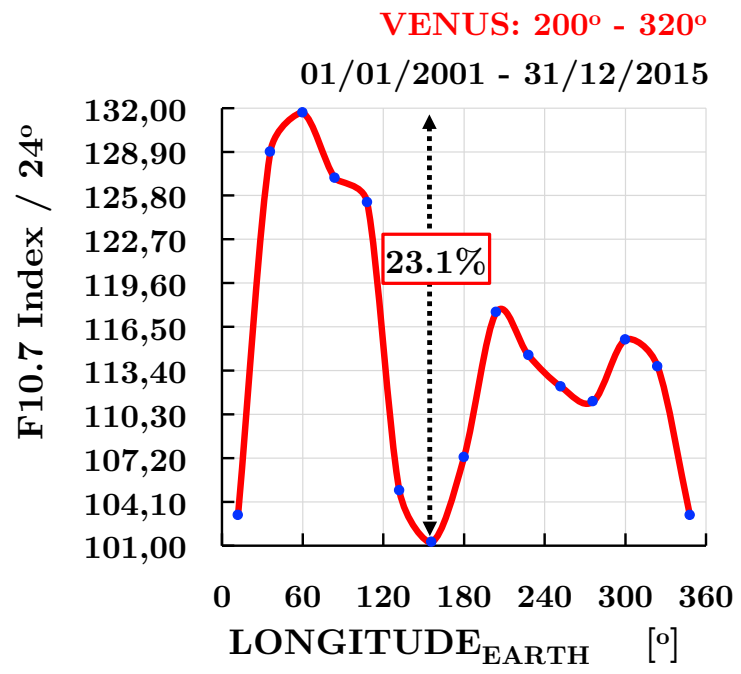

**(e)** F10.7 distribution for Earth when Venus propagates between220° to 320°.

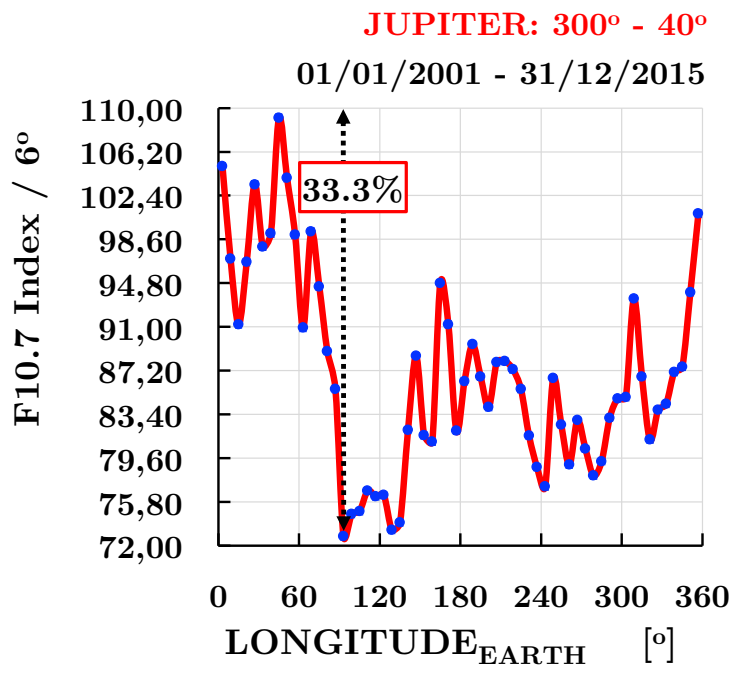

**(f)** F10.7 distribution for Earth when Jupiter propagates between 300° to 40°.

**Figure B.81:** Nr. of EQs vs. F10.7 solar proxy for the reference frame of Earth while the rest planets are constrained.

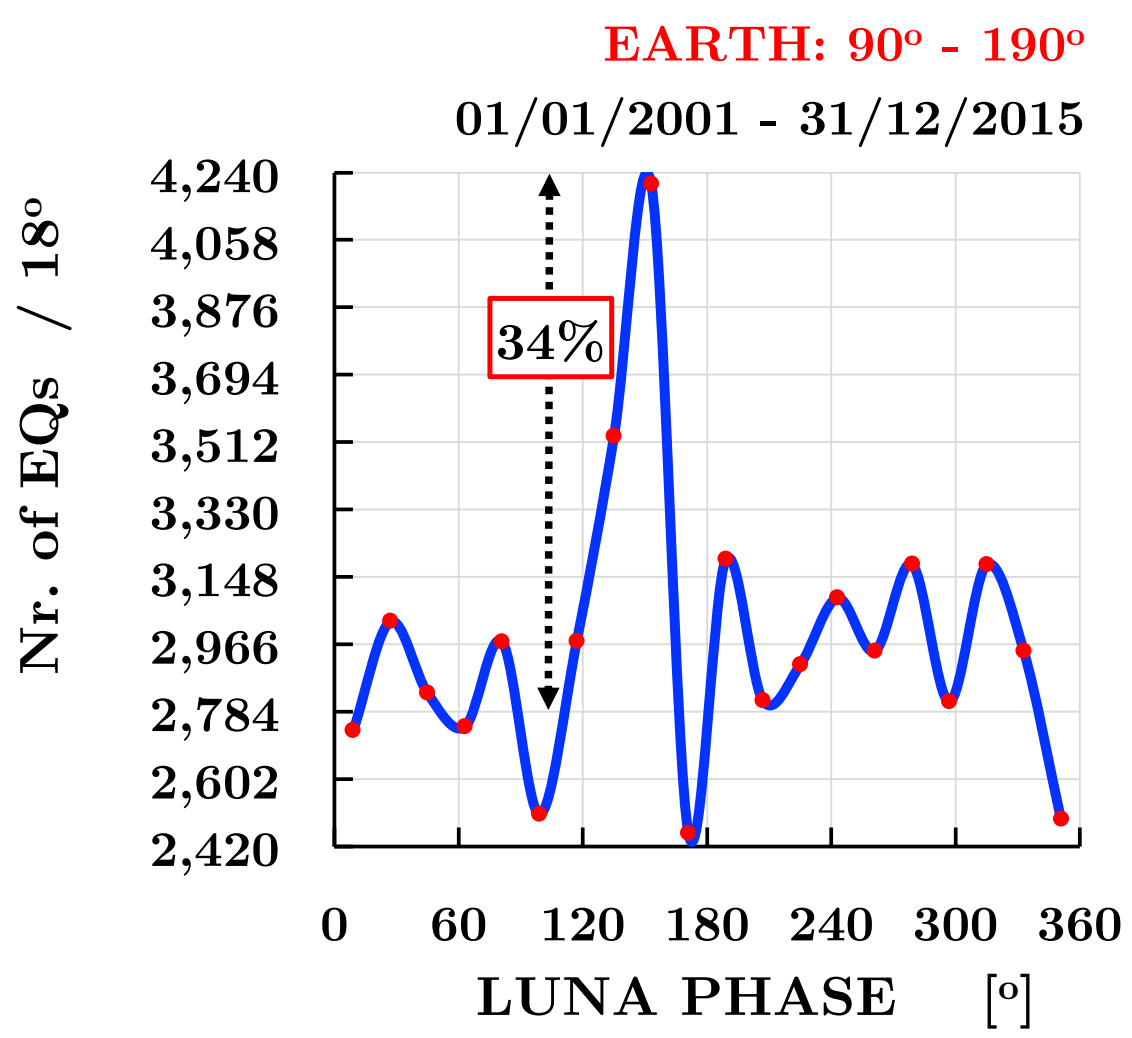

**(a)** Nr. of EQs distribution.

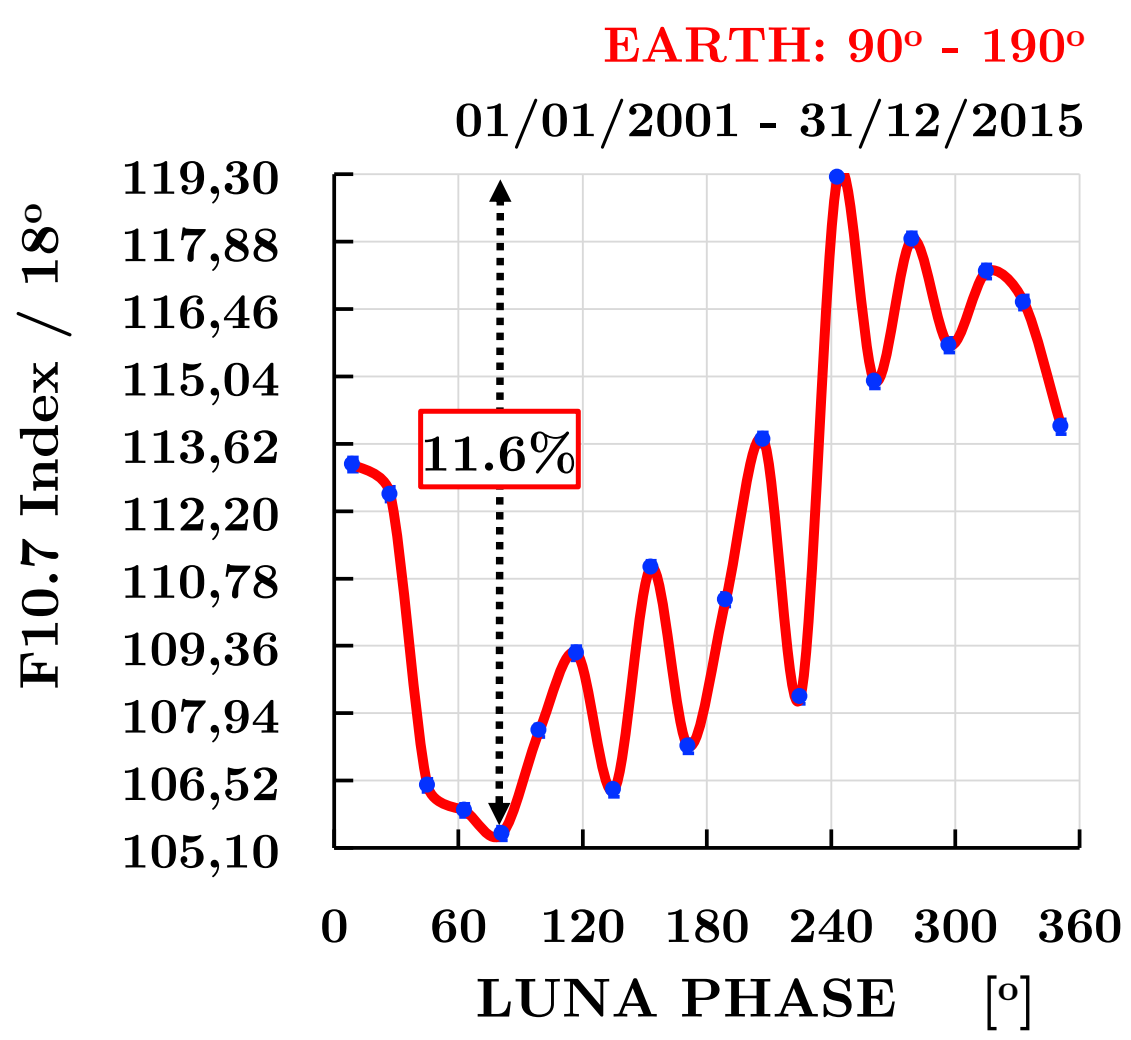

**(b)** F10.7 distribution.

**Figure B.82:** Nr. or EQs vs. F10.7 solar proxy as a function of Moon's phase while Earth's orbital position is constrained between 90° to 190°.

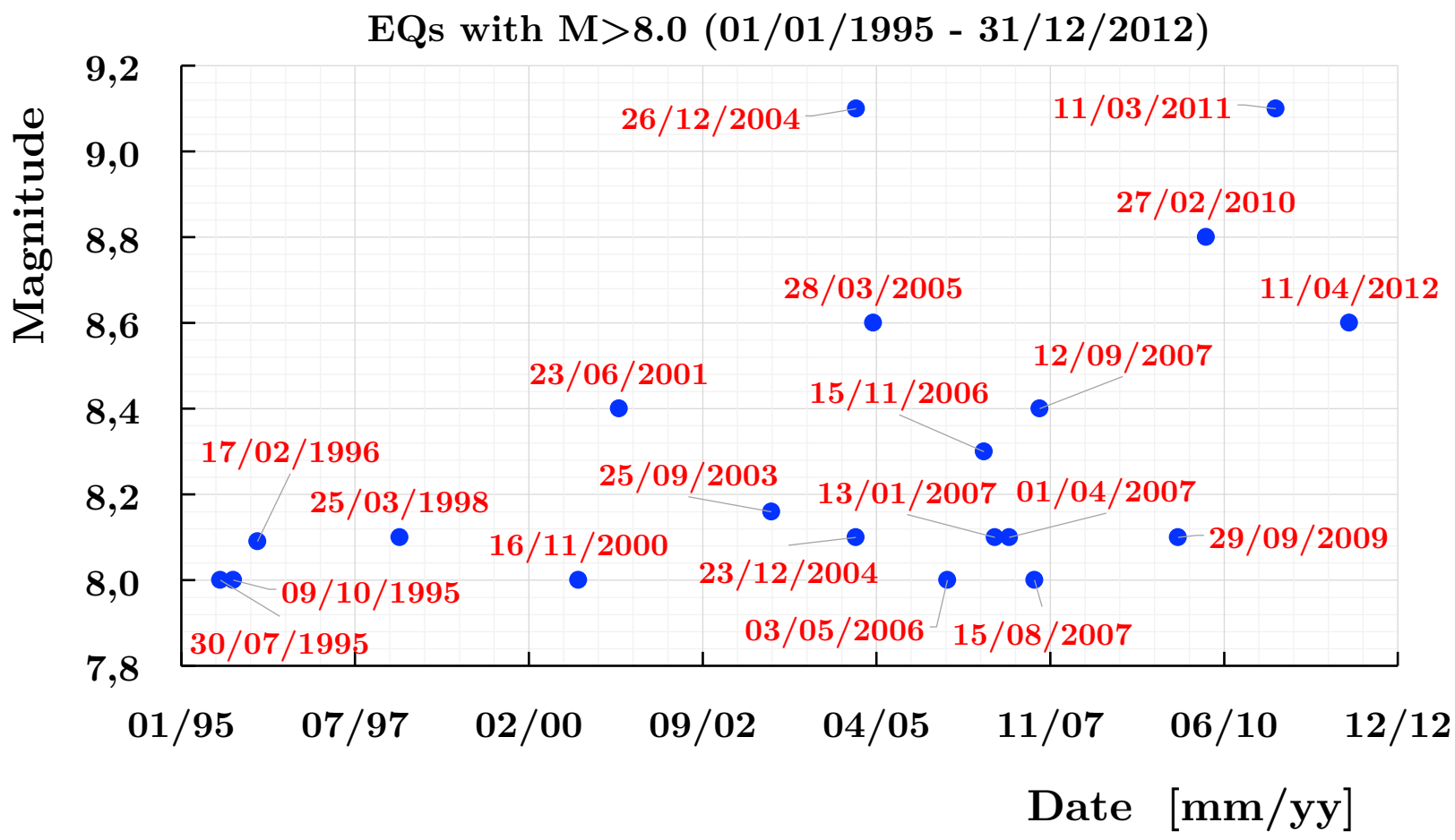

**Figure B.83:** EQs with $M \geq 8.0$ for the period 01/01/1995 - 31/12/2012.

EQ is defined as a reference point (time = 0).

**Table B.4:** Dates of EQs with $M \geq 8.0$ between 01/01/1995 - 31/12/20126 and the corresponding periods used for TEC. The overlapping periods are marked in red.

| EQ Date [UTC] | Date −90 d [UTC] | Date +90 d [UTC] | Magnitude |
|---|---|---|---|
| 30/07/1995 | 01/05/1995 | 28/10/1995 | 8.00 |
| 09/10/1995 | 11/07/1995 | 07/01/1996 | 8.00 |
| 17/02/1996 | 19/11/1995 | 17/05/1996 | 8.09 |
| 25/03/1998 | 25/12/1997 | 23/06/1998 | 8.10 |
| 16/11/2000 | 18/08/2000 | 14/02/2001 | 8.00 |
| 23/06/2001 | 25/03/2001 | 21/09/2001 | 8.40 |
| 25/09/2003 | 27/06/2003 | 24/12/2003 | 8.16 |
| 23/12/2004 | 24/09/2004 | 23/03/2005 | 8.10 |
| 26/12/2004 | 27/09/2004 | 26/03/2005 | 9.10 |
| 28/03/2005 | 28/12/2004 | 26/06/2005 | 8.60 |
| 03/05/2006 | 02/02/2006 | 01/08/2006 | 8.00 |
| 15/11/2006 | 17/08/2006 | 13/02/2007 | 8.30 |
| 13/01/2007 | 15/10/2006 | 13/04/2007 | 8.10 |
| 01/04/2007 | 01/01/2007 | 30/06/2007 | 8.10 |
| 15/08/2007 | 17/05/2007 | 13/11/2007 | 8.00 |
| 12/09/2007 | 14/06/2007 | 11/12/2007 | 8.40 |
| 29/09/2009 | 01/07/2009 | 28/12/2009 | 8.10 |
| 27/02/2010 | 29/11/2009 | 28/05/2010 | 8.80 |
| 11/03/2011 | 11/12/2010 | 09/06/2011 | 9.10 |
| 11/04/2012 | 12/01/2012 | 10/07/2012 | 8.60 |

Since some of the 180 d periods marked in red in Tab. B.4 seem to be overlapping, in order to avoid double counting, the mean value between each two periods with overlapping dates was used. The resulting 16 dates along with the used TEC period ±90 d are listed in Tab. B.5. This procedure is exactly the same as the one performed in Appendix Sect. B.3.2 correlating the EUV distribution with X-flares.

The result from the sum of the global TEC for the corresponding time periods from both Tab. B.4 and B.5 is depicted in Fig. B.84 with red and blue colour respectively. In this presentation, the center of x-axis in 0 d corresponds to the onset date of each EQ whereas the

**Table B.5:** No-overlapping corrected dates for EQs with $M \geq 8.0$ between 01/01/1995 - 31/12/20126 and the corresponding periods used for TEC. The corrected periods are marked in red.

| EQ Date [UTC] | Date −90 d [UTC] | Date +90 d [UTC] |
|---|---|---|
| 03/09/1995 | 05/06/1995 | 02/12/1995 |
| 17/02/1996 | 19/11/1995 | 17/05/1996 |
| 25/03/1998 | 25/12/1997 | 23/06/1998 |
| 16/11/2000 | 18/08/2000 | 14/02/2001 |
| 23/06/2001 | 25/03/2001 | 21/09/2001 |
| 25/09/2003 | 27/06/2003 | 24/12/2003 |
| 24/12/2004 | 25/09/2004 | 24/03/2005 |
| 28/03/2005 | 28/12/2004 | 26/06/2005 |
| 03/05/2006 | 02/02/2006 | 01/08/2006 |
| 14/12/2006 | 15/09/2006 | 14/03/2007 |
| 01/04/2007 | 01/01/2007 | 30/06/2007 |
| 29/08/2007 | 31/05/2007 | 27/11/2007 |
| 29/09/2009 | 01/07/2009 | 28/12/2009 |
| 27/02/2010 | 29/11/2009 | 28/05/2010 |
| 11/03/2011 | 11/12/2010 | 09/06/2011 |
| 11/04/2012 | 12/01/2012 | 10/07/2012 |

TEC is summed day by day. The result for both cases with overlapping and non-overlapping dates remains the same and shows a slow increase in the global electron content of the ionosphere starting about 45 d before the big EQs (see also Publication E.18). This long-term EQ precursor is different than the short-term ones observed in the literature and strengthens the claim for the external triggering of the EQs pointing to a downwards propagated motion, i.e. an in-falling low-speed invisible stream.

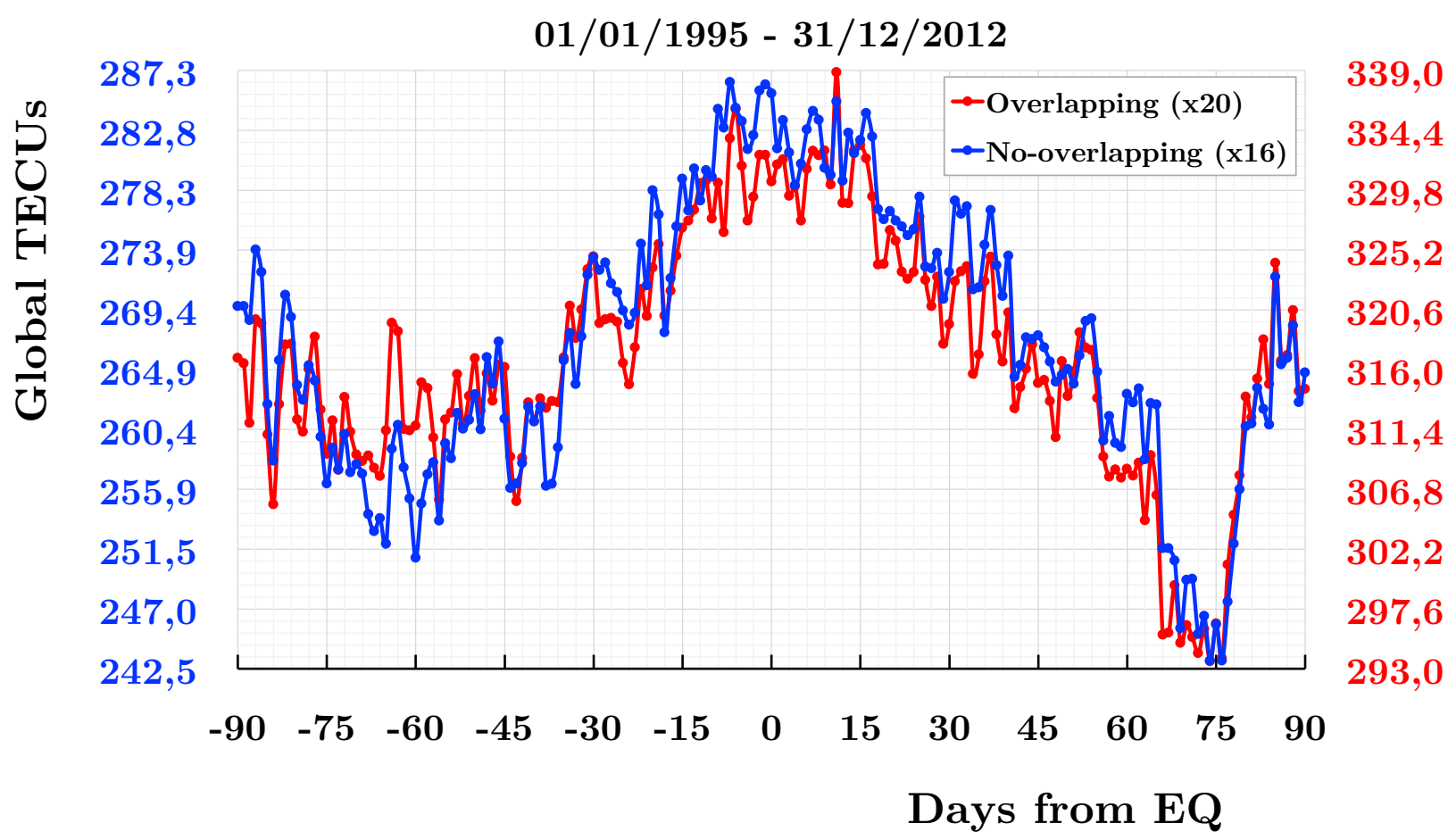

**Figure B.84:** TEC distribution 90 d before and after each EQ with $M \geq 8.0$ for the period 01/01/1995 - 31/12/2012.

### B.11.2.2 Earthquakes with $M \geq 8.6$

The same procedure has been performed for the biggest EQs with a magnitude $M \geq 8.6$. For the period 01/01/1995 to 31/12/2012 there were 5 EQs fulfilling these conditions, which are shown in Tab. B.6. Based also on Fig. B.84 the small overlapping observed on the first two dates can not cause any difference on the resulting correlation and therefore it is ignored.

**Table B.6:** Dates of EQs with $M \geq 8.6$ between 01/01/1995 - 31/12/20126 and the corresponding periods used for TEC.

| EQ Date [UTC] | Date −90 d [UTC] | Date +90 d [UTC] | Magnitude |
|---|---|---|---|
| 26/12/2004 | 27/09/2004 | 26/03/2005 | 9.10 |
| 28/03/2005 | 28/12/2004 | 26/06/2005 | 8.60 |
| 27/02/2010 | 29/11/2009 | 28/05/2010 | 8.80 |
| 11/03/2011 | 11/12/2010 | 09/06/2011 | 9.10 |
| 11/04/2012 | 12/01/2012 | 10/07/2012 | 8.60 |

The summation of the global TEC for the periods of Tab. B.6 is shown in Fig. B.85 where the time dependence of the global degree of ionisation of the ionosphere within ±90 d is plotted with reference to the occurrence of each of the 5 biggest EQs with $M \geq 8.6$. Like in Fig. B.84, the relationship between the occurrence of EQs and TEC is apparent, showing an increase in the global TEC of the ionosphere for several days before the occurrence of the big EQs (see also Publication E.19).

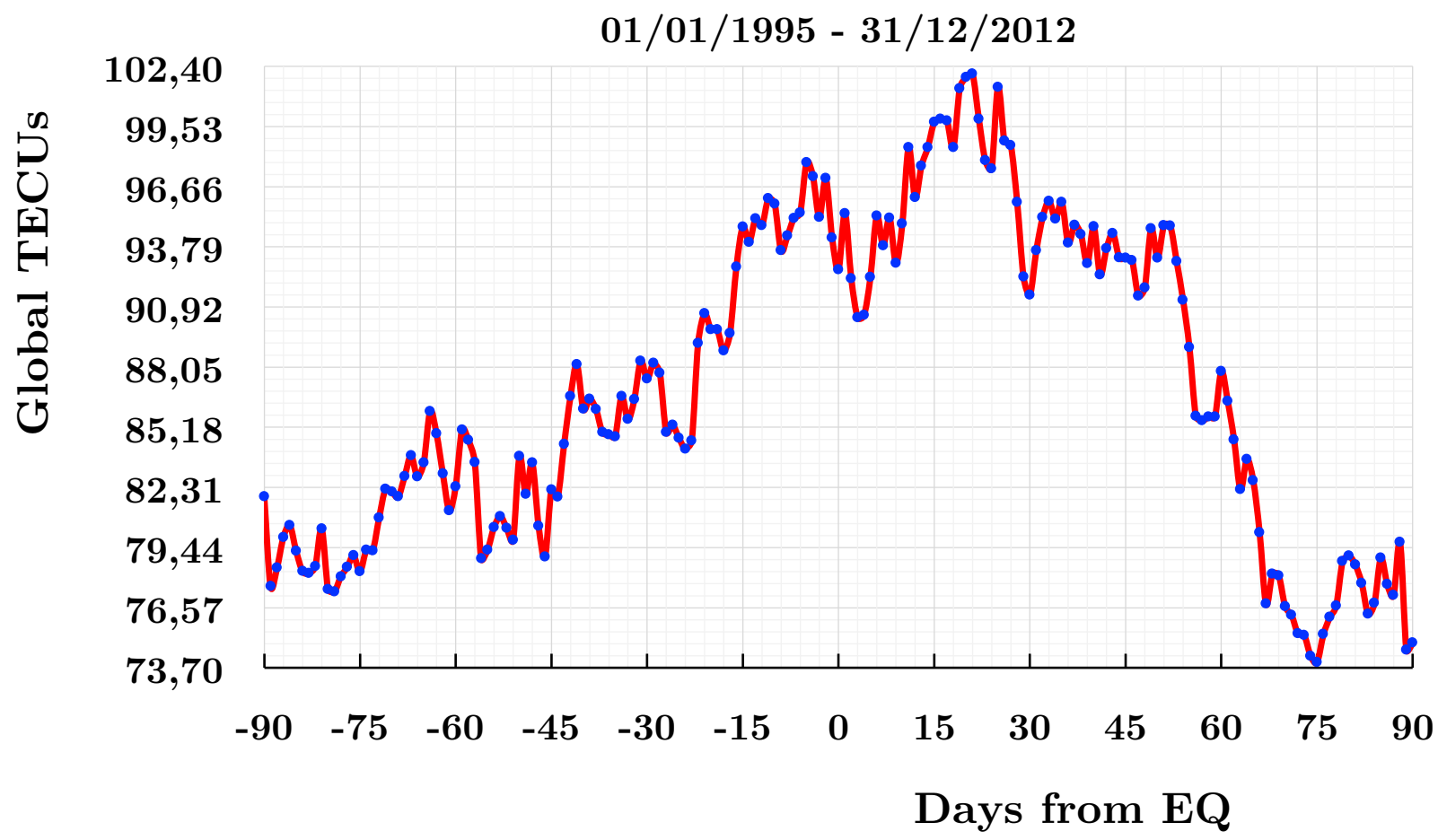

**Figure B.85:** TEC distribution 90 d before and after each EQ with $M \geq 8.6$ for the period 01/01/1995 - 31/12/2012.

# B.12 Melanoma

## B.12.1 Comparison with solar UV

The first and most obvious comparison is made with the UV irradiance since it is widely accepted that the solar UV is linked to melanoma risk [470, 471]. If no other external cause for melanoma exists, then its seasonal rate would be steady with a broad maximum during local summertime due to increased solar UV exposure. For a direct crosscheck, the data corresponding to the local UV intensity on the Earth's surface have been acquired from ERA5 database [472]. More specifically, a large grid point of about 4325 km × 3325 km across Australia with outer coordinates $-10.40°S/11.75°E$ and $-43.7°S/155.0°E$, has been chosen, as seen in Fig. B.86. Then the mean value of the assimilated UV intensity from several daily reanalysis values across this surface were used.

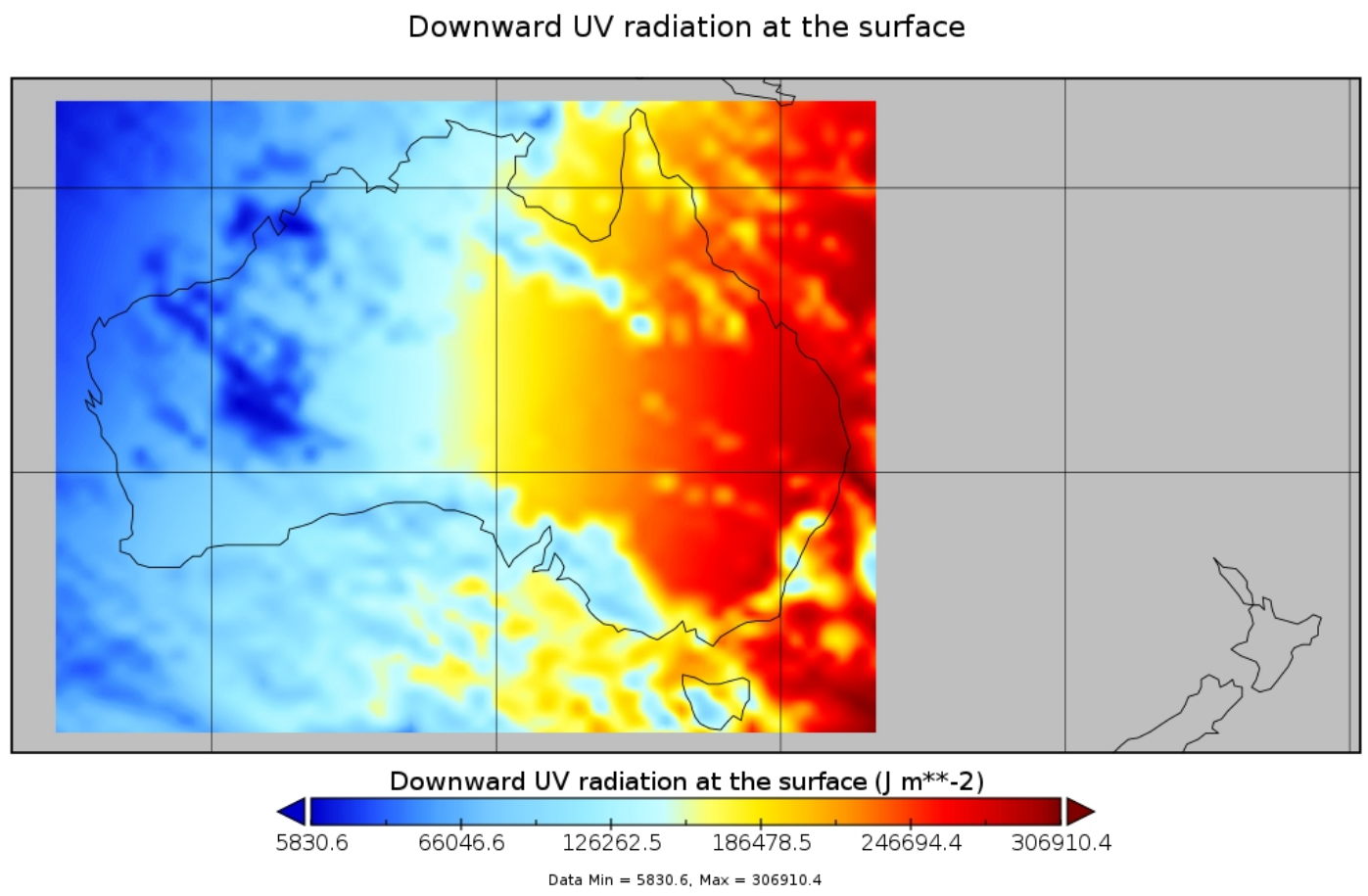

**Figure B.86:** Grid point of UVB data downloaded for the comparison with the daily cases of melanoma in Australia.

In Fig. B.87 the time series of melanoma cases and UV for the same location and time period are shown. From the statistical correlation analysis we get a Pearson correlation coefficient and p-value of ($r = 0.24$, $p = 0$) showing a statistically significant medium positive linear correlation between the time series of the two datasets.

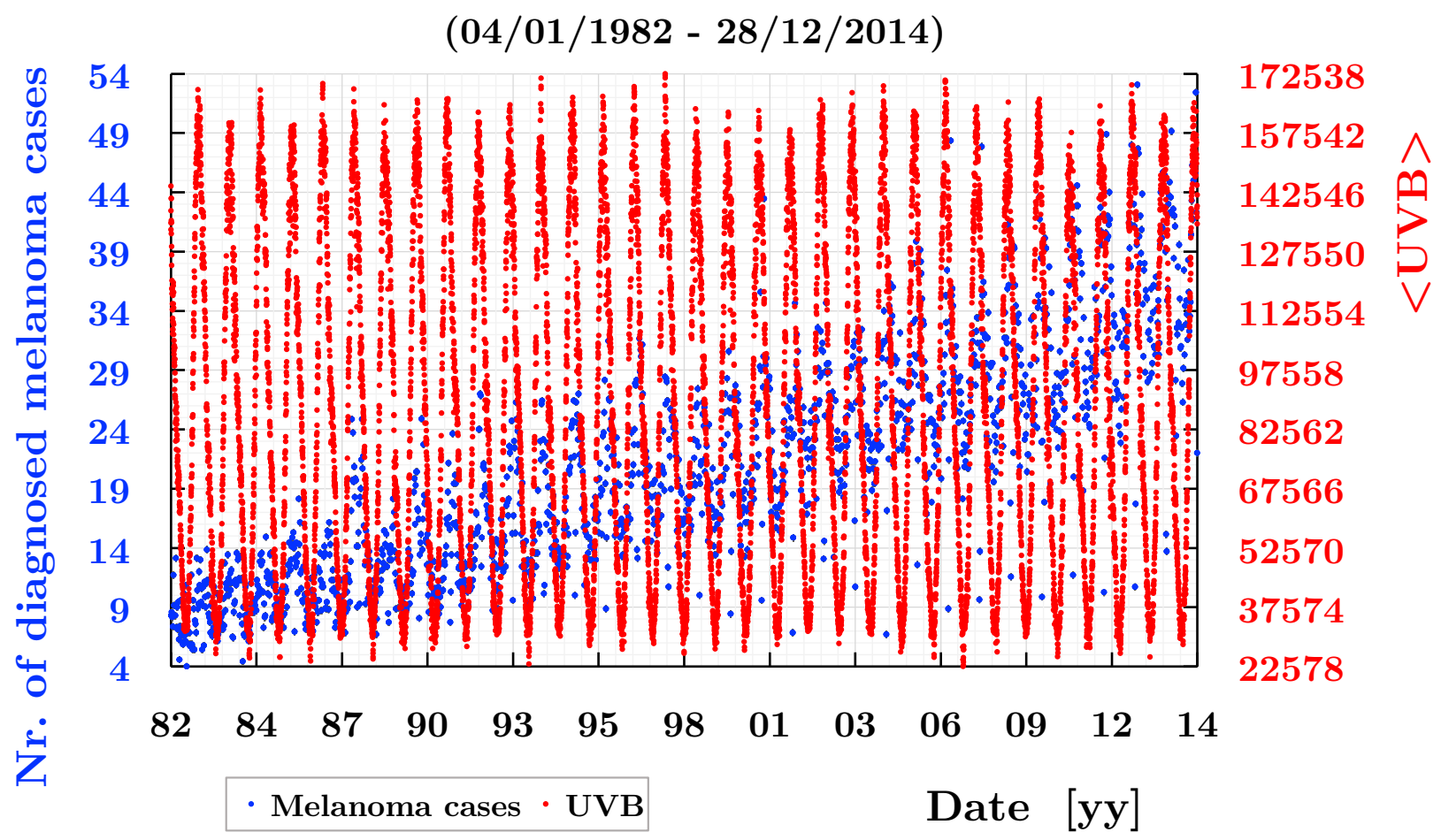

**Figure B.87:** Nr. of daily diagnosed melanoma cases and UV radiation for the same period 04/01/1982 - 28/12/2014.

As next, we compare the distribution from Fig. 18.8 which had the most significant effect, with that of the UV for the same location and period, and with the same conditions. The black dashed line in Fig. B.88a gives the trend for a steady time-dependent melanoma rate as expected if solar UV was the sole cause of melanoma incidence, whereas the blue line is simply connecting through the points. A typical statistical error ($\pm 1\sigma$) is also provided. The appearance of the 12 peaks in Fig. B.88a is in contrast with the smooth distribution of UV in Fig. B.88b. The overall trend of Fig. B.88a follows that of UV in Fig. B.88b but the modulating shape of melanoma stands out. Therefore, solar UV can not account for 100% of melanoma incidence as it can not explain the measured 12 large melanoma modulation peaks in Fig. B.88a with an amplitude of $\sim 20\%$. By calculating the Pearson's correlation coefficient we get ($r_{a,b} = 0.81,\ p_{a,b} = 0$) which shows a statistically significant high degree of linear correlation between UV and melanoma distributions which is expected based on the similarity of the large-scale structure observed in Fig. B.88.

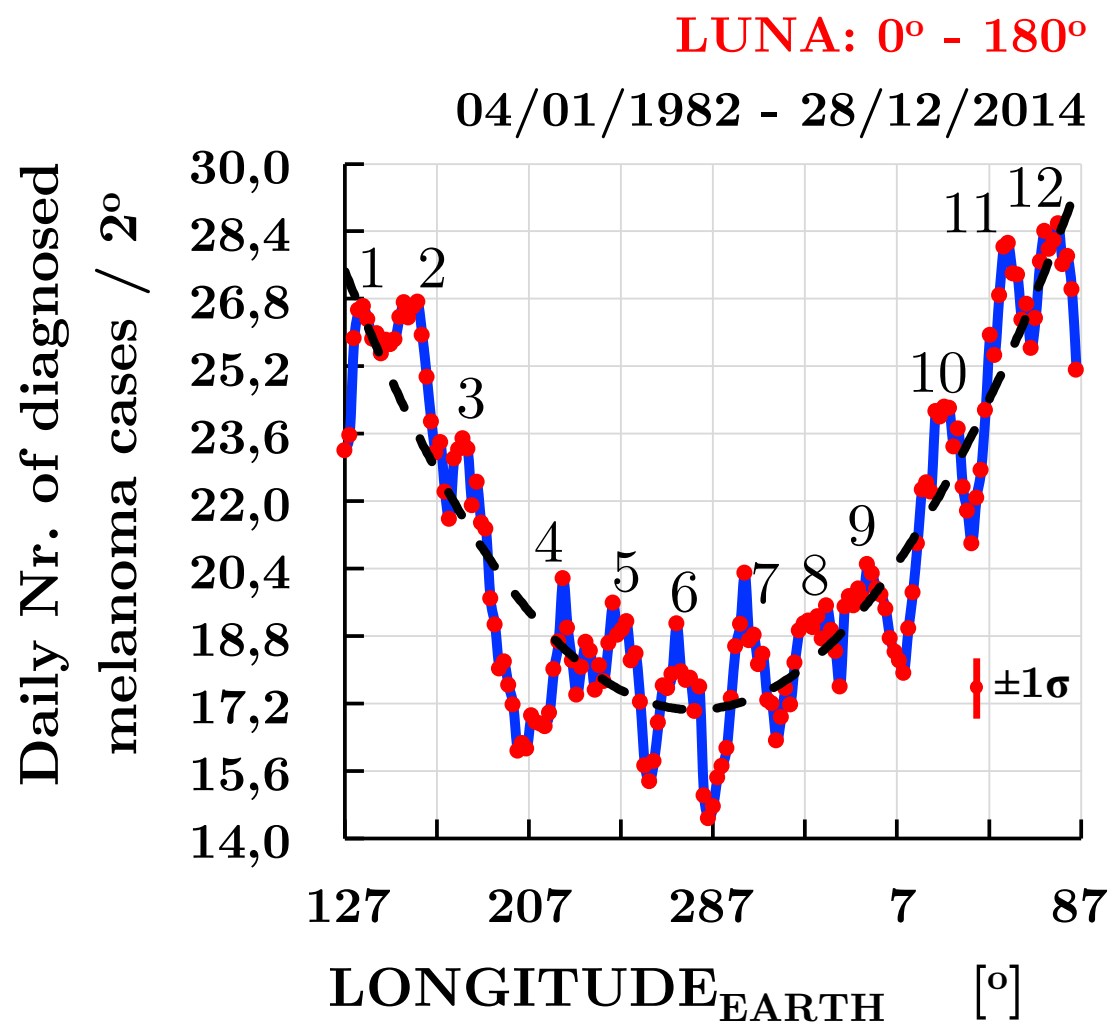

(a) Daily Nr. of diagnosed melanoma cases. The black dashed line corresponds to a second degree polynomial fit with $y = 0.0018x^2 - 0.3366x + 33.204$.

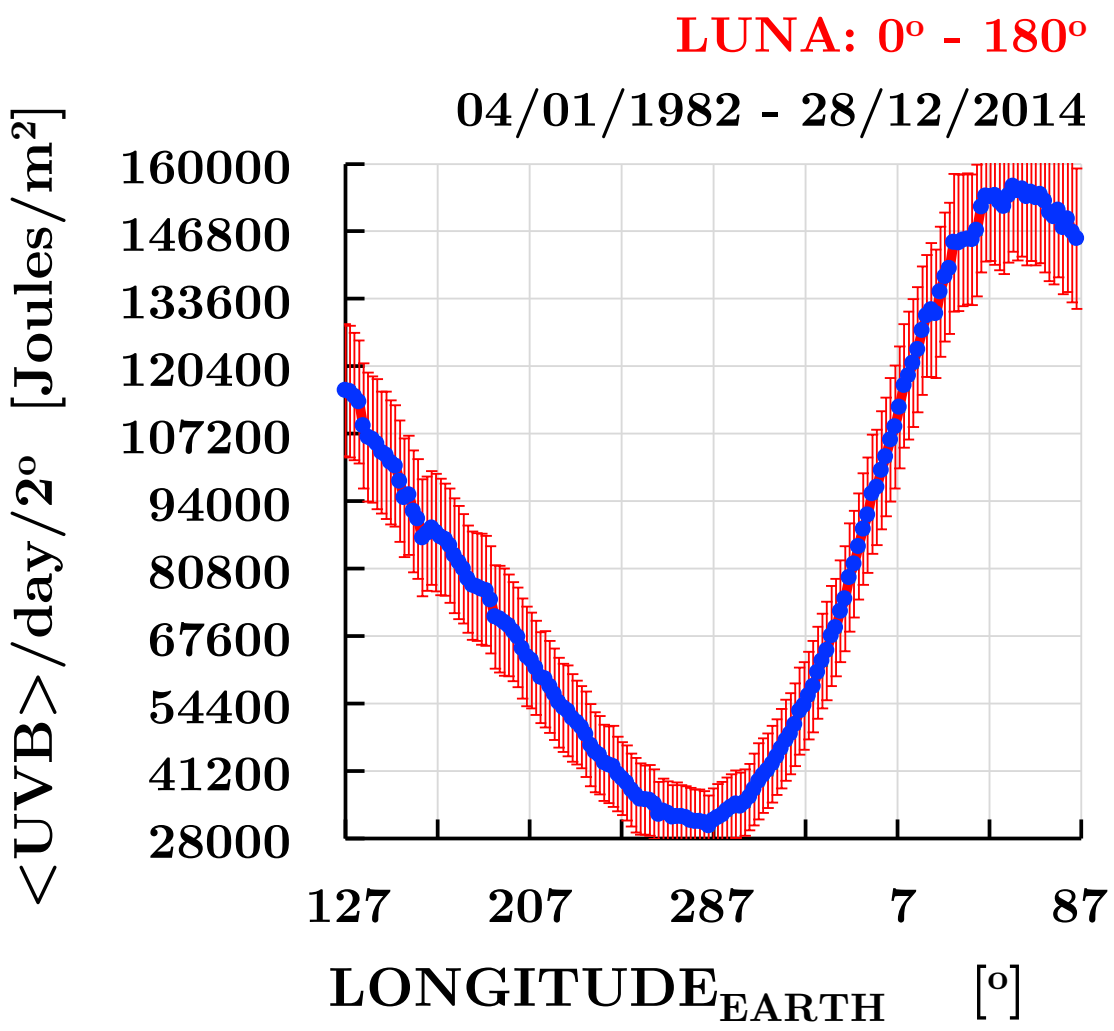

(b) Corresponding UVB distribution in Australia.

**Figure B.88:** Comparison of the time dependence of melanoma cases vs. UV intensity as a function of Earth's heliocentric longitude when Moon's phase is vetoed to propagate between 0° to 180° for bin = 2°.

Furthermore, the UV increase around Australia from $1996 - 2017$, shows an increase with a mean value of +0.25% per decade, as measured from the Australian Radiation Protection and Nuclear Safety Agency (ARPANSA) network of broadband sensors [473]. On the other hand, the net increase in melanoma cases is about 60% per decade following the population normalised $\sim 3.62$ times steady increase in 33 y seen in Fig. 18.2a. For comparison, in Indonesia [474] and New Zealand [475] which neighbour Australia, the UV indexes are reported to have, as expected, seasonal modulation but remain constant over the last 10 y to 20 y. It is also noted that in the southern hemisphere the UV may have even decreased slightly although in New Zealand has remained constant in all seasons [474, 476].

## B.12.2 Comparison with F10.7

The next most obvious comparison is with the F10.7 solar radio flux as it is widely used as an indice of solar activity. The resulting Pearson's correlation coefficient and corresponding p-value for the two overlapping time series in Fig. B.89 are ($r = -0.14,\ p = 0$) which indicates a very small degree of negative linear correlation.

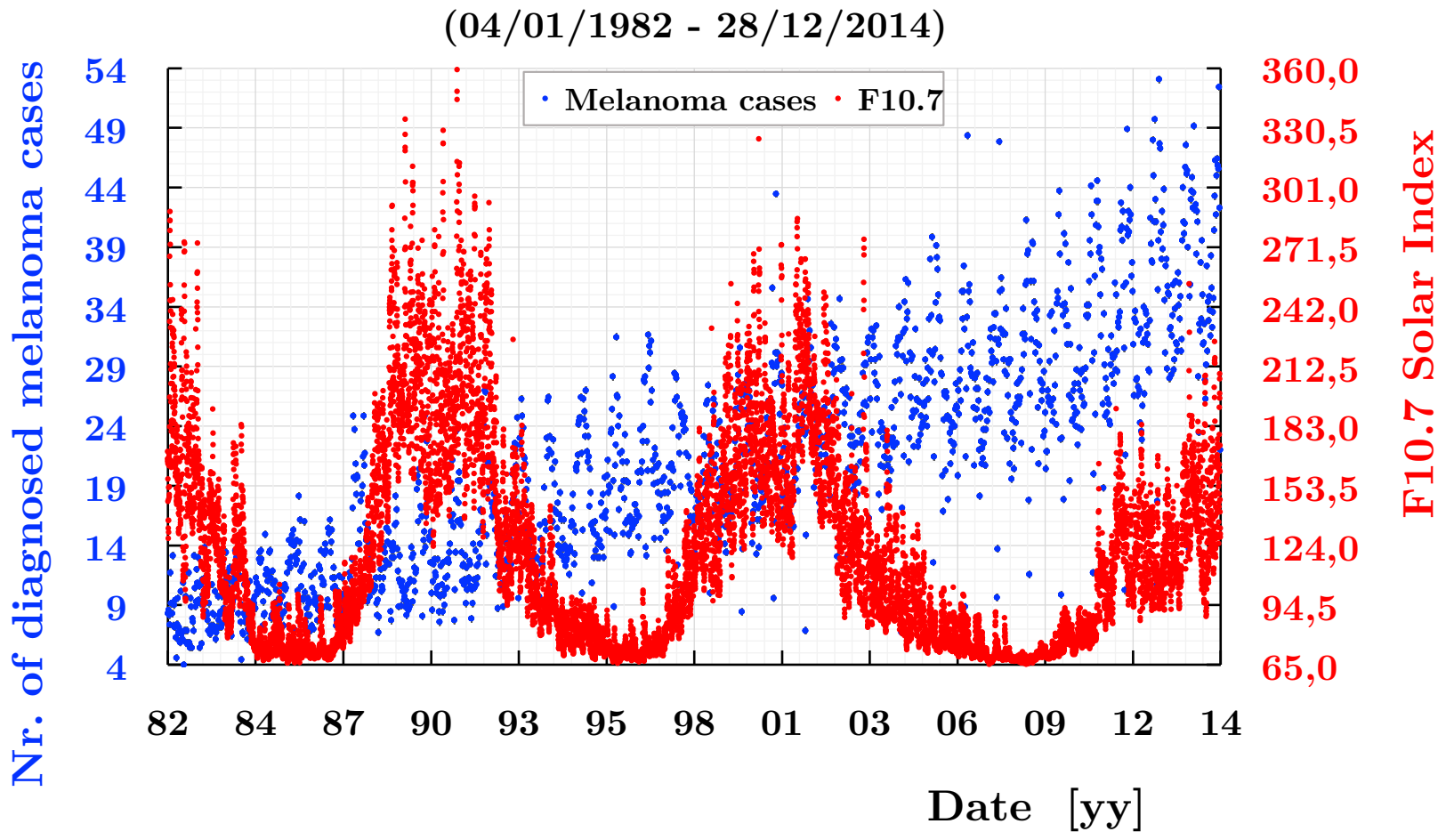

**Figure B.89:** Nr. of daily diagnosed melanoma cases and F10.7 solar index for the same period 04/01/1982 - 28/12/2014.

In Fig. B.90 the heliocentric longitude distributions for the various planets between the two datasets are compared. The same period 1982−2015, with the same bin and constraints have been used. It is clear that the distributions are quite different, with the only peak overlapping in Fig. B.90a and B.90c being the one at around 33°. However, in F10.7 distribution this peak, which is the strongest, and relative narrow ($\sim 12\,\mathrm{d}$), appears less pronounced when constraining the Moon to be between 0° to 180° in Fig. B.90d. In contrast, the corresponding melanoma peak in Fig. B.90a appears even better resolved with slightly larger amplitude in Fig. B.90b. Therefore, this correlation between melanoma incidence and F10.7 in the specific peak of 33° is not strong. As a result, once more, at least part of the melanoma distribution does not seem to be solar driven. The calculated linear Pearson's correlation coefficients verify this claim giving ($r_{a,c} = 0.26,\ p_{a,c} = 0.014$) for Fig. B.90a and B.90c and ($r_{b,d} = 0.0056$, $p_{b,d} = 0.958$) for Fig. B.90b and B.90d. This means that only a small positive association is found between Fig. B.90a and B.90c.

An additional comparison is made for the rate observed in Fig. 18.7c. The exact same distribution has been created for the F10.7 data in Fig. B.91b. The conclusion from Fig. B.91 is a clear dissimilarity between the two datasets verified by the calculated correlation coefficient of ($r_{a,b} = 0.022$, $p_{a,b} = 0.932$). Further, the melanoma rate indicates an average increase of 64.20% between the two periods, while the F10.7 decreases by about 14.59% thus strengthening once more exo-solar effects. The same result is derived for the raw spectrum of Earth i.e. without

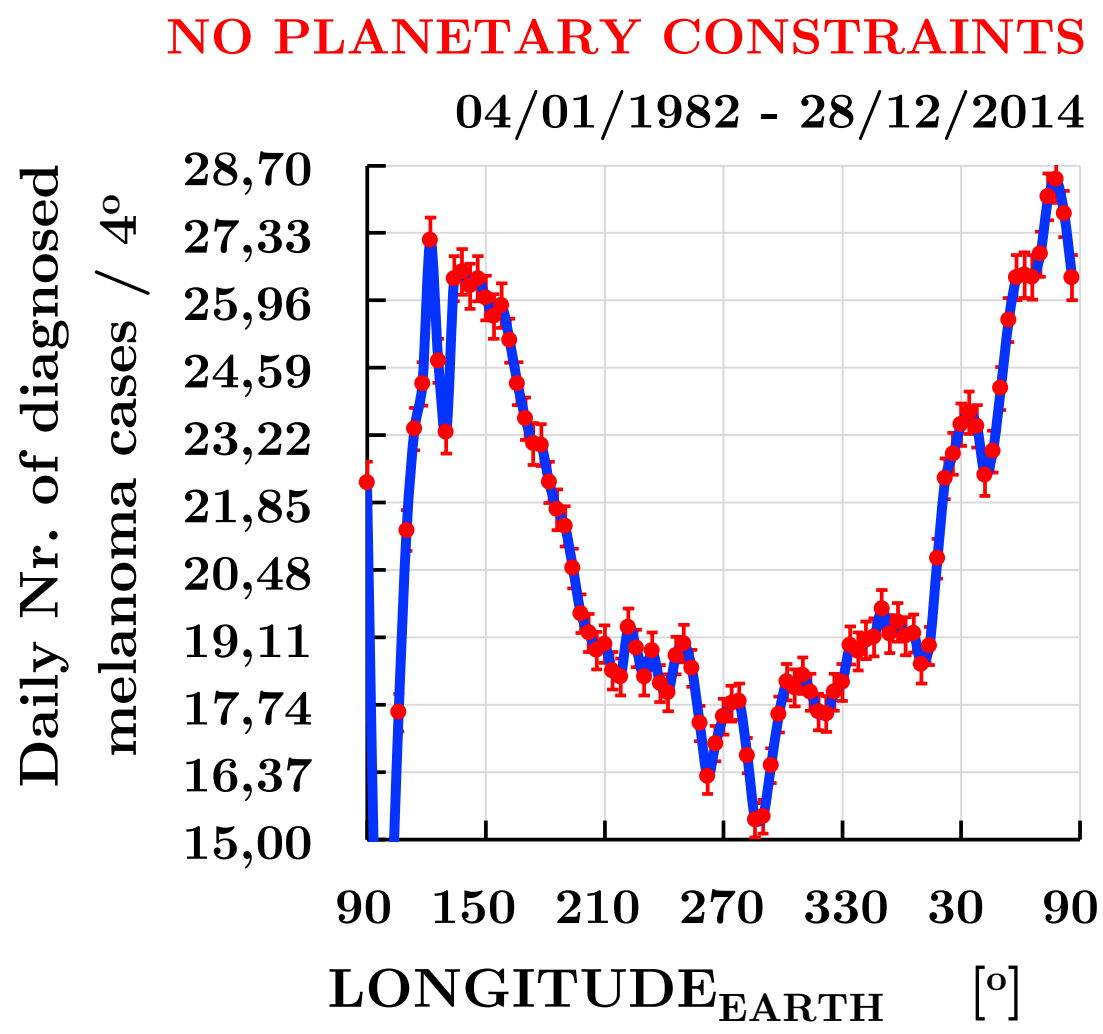

(a) Melanoma cases without additional planetary constraints.

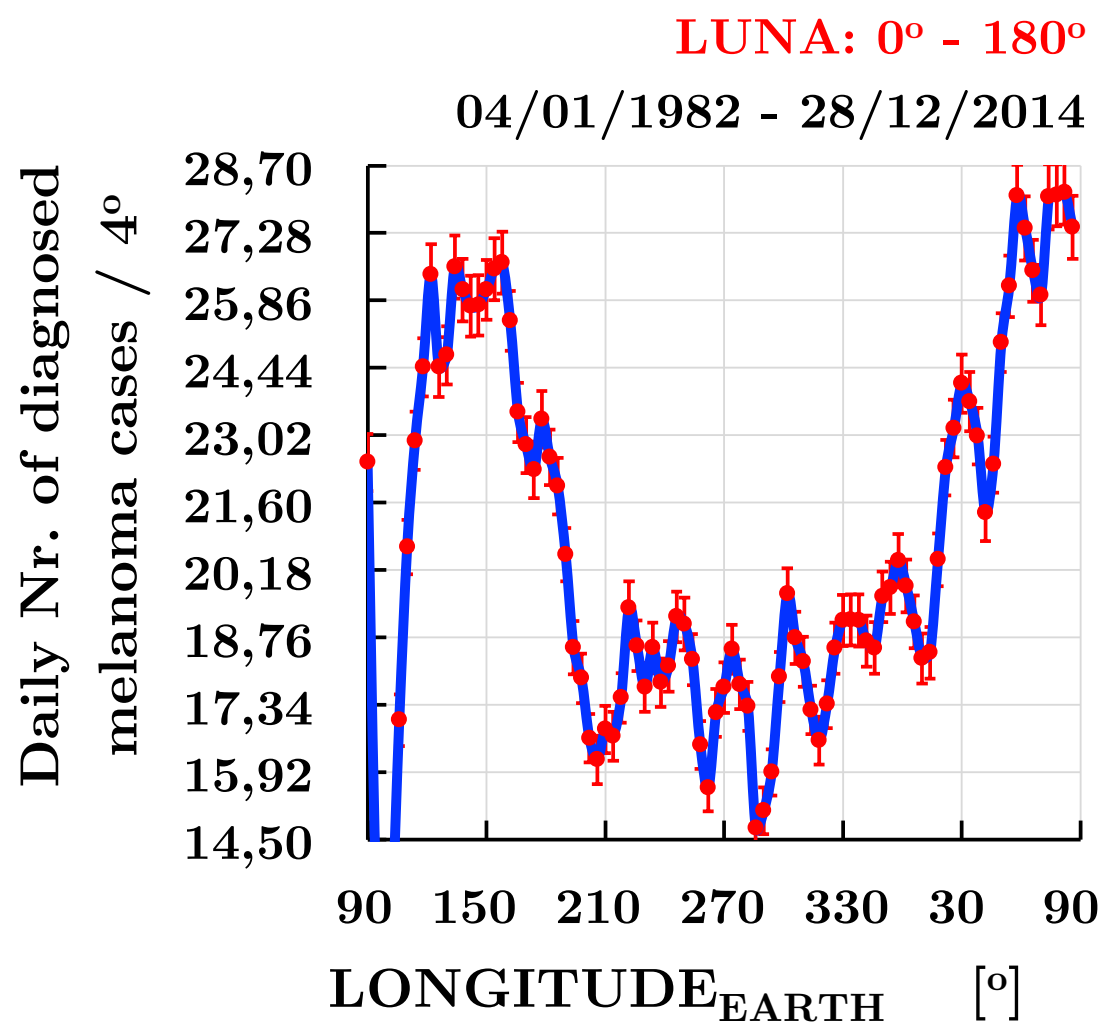

(b) Melanoma cases when Moon is between 0° to 180°.

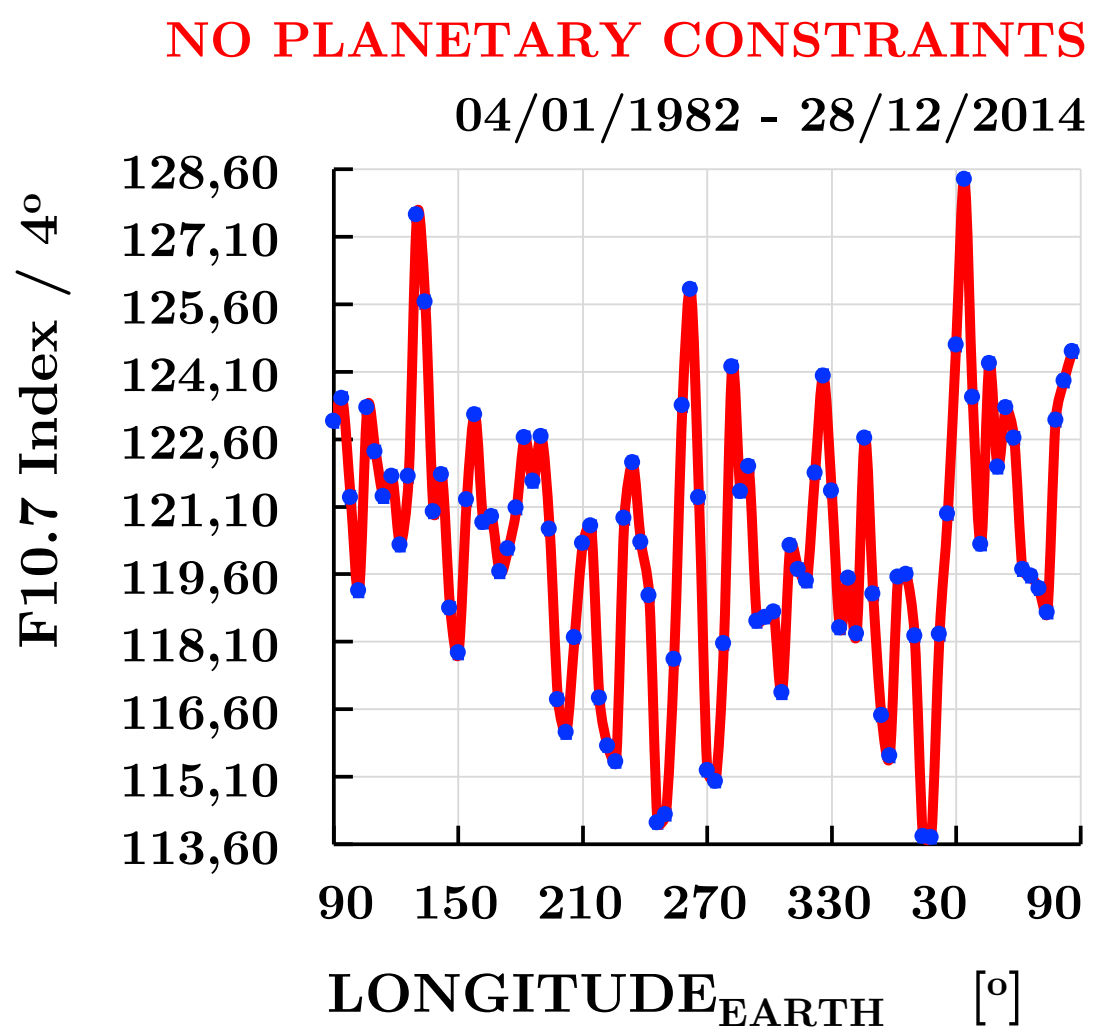

(c) F10.7 without additional planetary constraints.

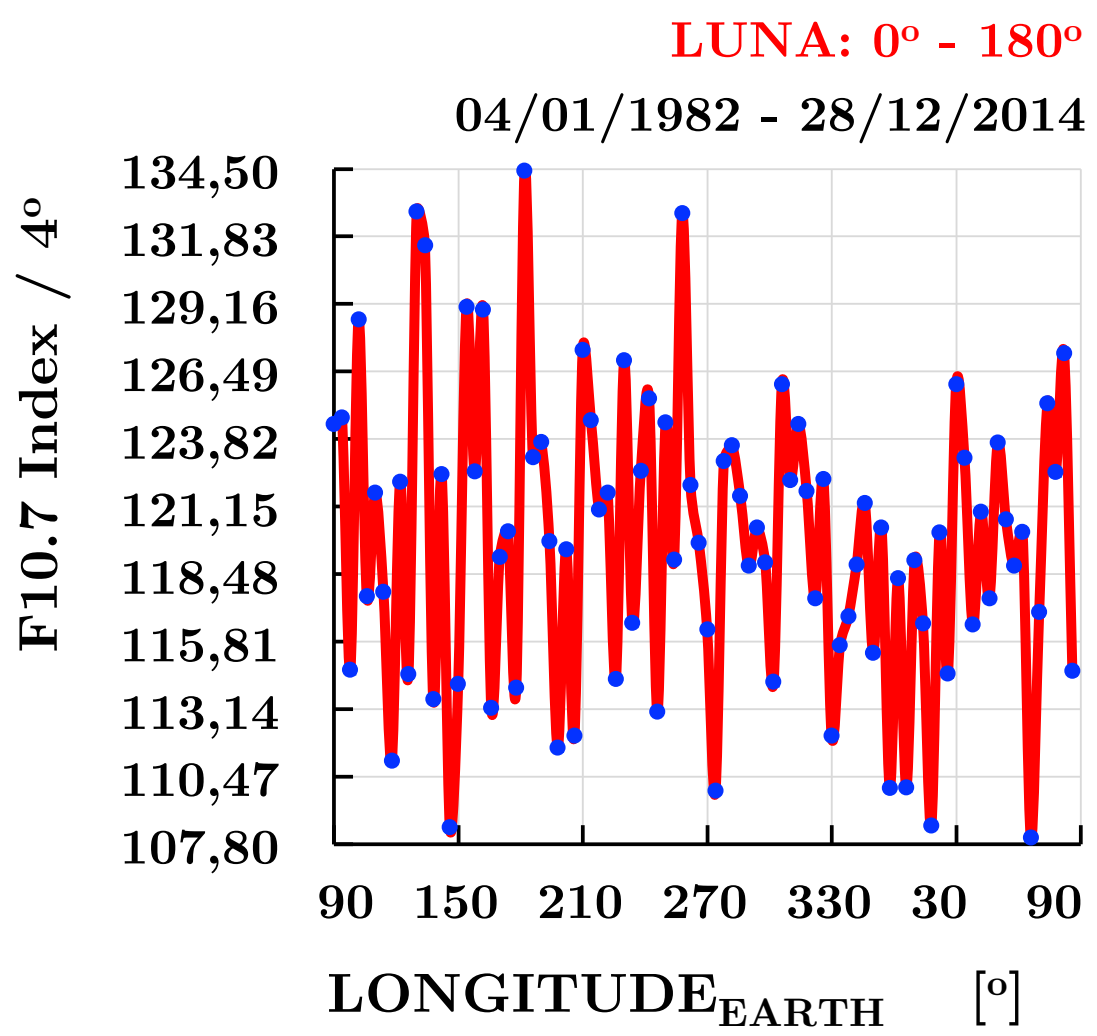

(d) F10.7 when the Moon is between 0° to 180°.

**Figure B.90:** Comparison of the distribution of melanoma cases vs. F10.7 intensity as a function of Earth's heliocentric longitude for bin = 4°.

constraining Moon's orbital position. More specifically, for the melanoma daily incidence rate in this case, we have an average increase between the two periods 2000 − 2014 and 1982 − 2000 of 64.49%, whereas in the case of F10.7 we have an average decrease of 12.76%.

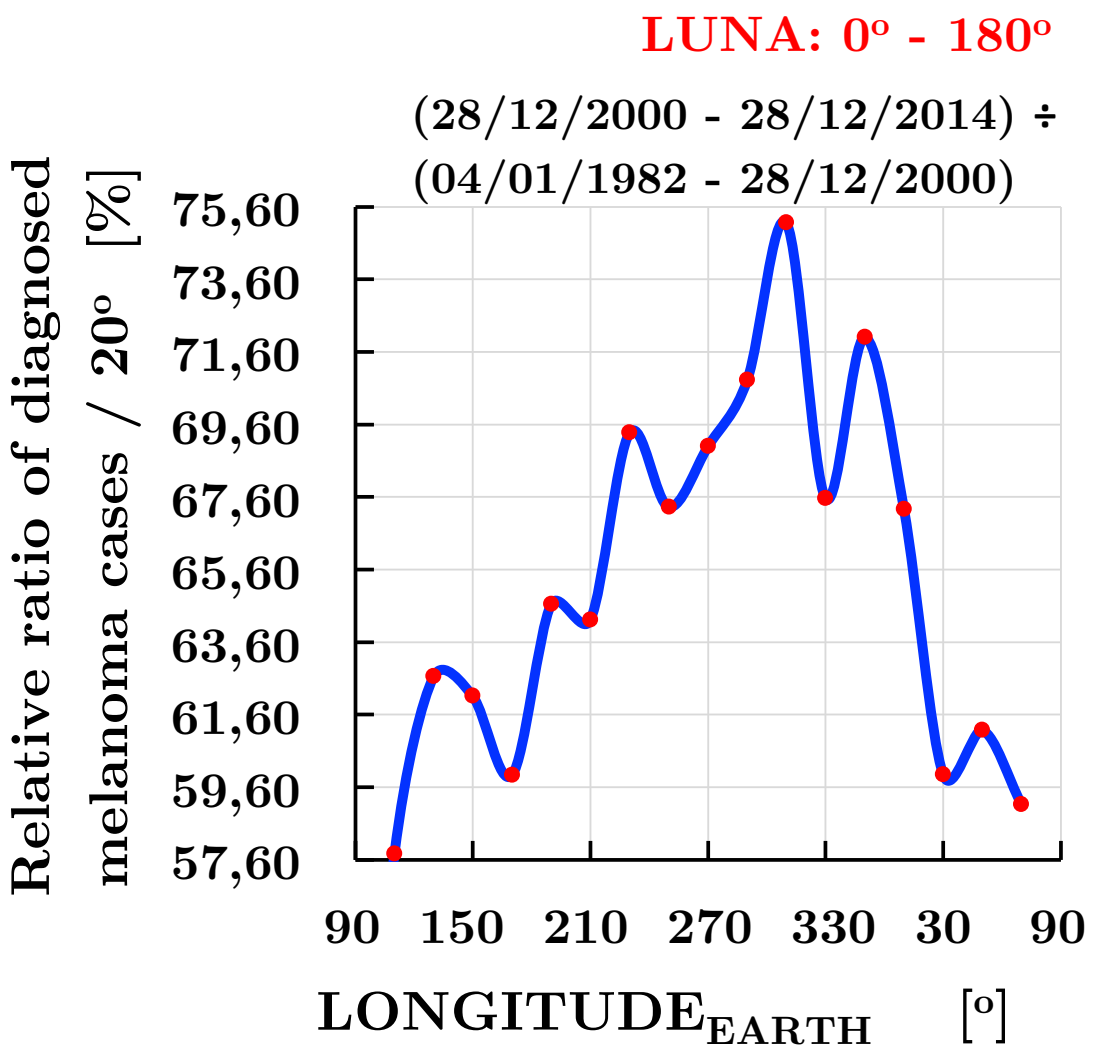

(a) Melanoma relative ratio.

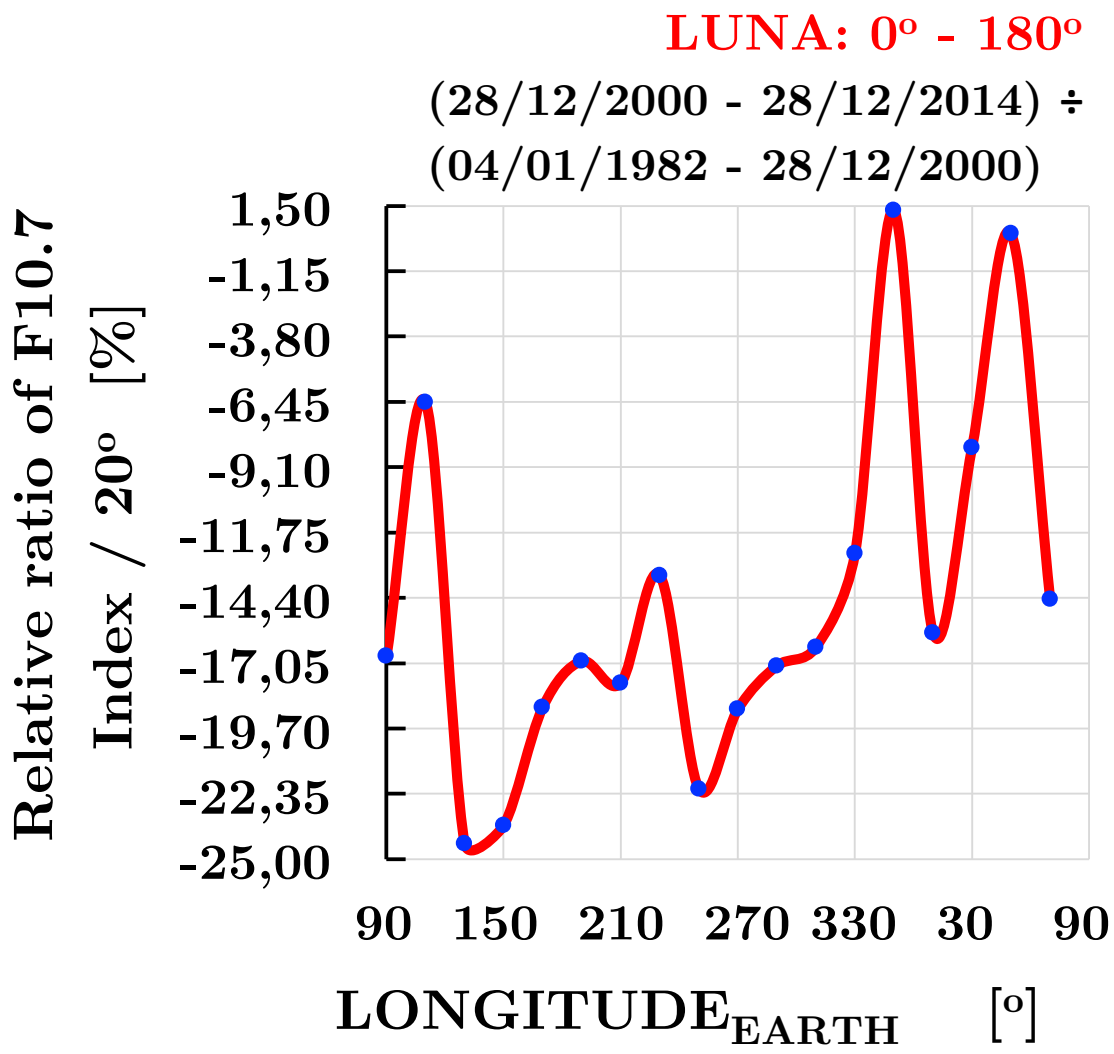

(b) F10.7 relative ratio.

**Figure B.91:** The relative change of melanoma diagnosis vs. F10.7 for the reference frame of Earth, while the Moon is allowed in the range 0° to 180°, between the time interval 2000-2014 and 1982-2000, for bin = 20°.

# C

# CAST-CAPP DATA ACQUISITION SYSTEM

## C.1 Basic instruments

The various instruments that are used in the CAST-CAPP setup outside the CAST magnet, based on Fig. 22.12, are the following:

(1) 300 K input flange on the Magnet Return Box (MRB) side of the CAST magnet (see Fig. 22.11).
(2) 300 K output flange on the MRB side of the CAST magnet (see Fig. 22.11).
(3) Mini-Circuits "1SP4T" RF switch, denoted as *input switch* (see Fig. C.3a), is connected through SMA connectors to the input RF cables of each cavity and directs the signal sent from the VNA (6), or from the signal generator (8) to either one of the cavities.
(4) Mini-Circuits "4SPDT" RF switch, denoted as *output switch* (see Fig. C.3b), regulates the connection between output RF cables and the data processing instruments (VSA (7) or VNA (6)).
(5) Mini-Circuits "2SPDT-16A RF" switch, denoted as *device switch* (see Fig. C.3c), connects accordingly the cavity input and/or output cables to the desired instruments.
(6) Keysight "E5063A ENA Series" VNA (see Fig. C.1a) is used to make reflection S11 and transmission S12 measurements to the cavities.
(7) Keysight "N9030B PXA" VSA (see Fig. C.1b) is used to measure the output signal from the cavities.

(8) National Instruments "FSW-0020" signal generator with $0.65 - 20$ GHz range and a fixed 13 dBm output power (see Fig. C.1g).

(9) Four external Mini-Circuits wide-band LNAs "ZX60-83LN-S+" (see Fig. C.2b) operating in the range $0.5 - 8$ GHz one for each cavity (see Fig. C.2b) providing a gain of about 22 dB.

(10) Four Mini-Circuits "RCDAT-8000-30" programable attenuators with 0.25 dB resolution and $0 - 30$ dB attenuation range. These attenuators are adjusting the attenuation for each cavity to match the corresponding resonance amplitudes for PM.

(11) KDI / Aeroflex "D406M 8624" power combiner which combines the signals from the four cavities for PM.

(12) Miteq "AFD3-0208-40-ST" LNA providing about 30 dB gain.

(13) X-COM "IQC 5000B" data recorder (see Fig. C.1d) is connected directly to the two VSAs ((7) and (16)) from which it records the data to two Redundant Array of Inexpensive Disks (RAIDs) connected to the "IQC-MEM" module.

(14) Omnidirectional SMA WiFi antenna working at $1.7 - 6$ GHz range used to record the ambient EMI/EMC parasites in the CAST area.

(15) Aydin Ga As FET "4080A38" external LNA providing an amplification of the signal from the WiFi antenna (14).

(16) Keysight "EXA N9010A" VSA is used to measure the parasitic signals in the CAST environment as seen from the external antenna (14) (see Fig. C.1c).

(17) Hameg HM7044 power supply providing power to LNAs (12) and (15) with 15 V and to LNAs (9) with 6 V (see Fig. C.2d).

(18) Lakeshore 224 Temperature monitor (see Fig. C.1f) monitors and records all four temperature sensors from the cavities.

(19) Janssen Precision Engineering (JPE) piezoelectric controller with two Cryo Actuator Base Cabinets (CABs) which are connected to the 4 CLAs of the four cavities (see Fig. C.1e). It provides manual and automatic left/right movement with varying speed and steps.

(20) Low Noise Factory power supplies "LNF-PS_3" (see Fig. C.2c) of the four cryogenic LNAs "LNF-LNC4_8D" (see Fig. C.2a) providing a constant current power supply.

(21) Logitech webcam to monitor continuously the displays of (20) for possible changes in voltage or current due to spurious events.

(22) Workstation PC with 8-core Intel i7 processor at 3.4 GHz, 32 GB Random Access Memory (RAM) and 10 GBps network card.

(23) Turbo-X 8-port USB 3.0 hub for the connection of all USB cables to (22).

(24) Local storage consisting of two external Hard Disk Drive (HDD) hard drives with total storage capacity 16 TB.

(25) Local storage consisting of two internal Solid-State Drives (SSDs) with total storage capacity 4 TB.

(26) TP-Link 8port network switch which connects all related instruments ((6), (7), (13), (16), (18) and (22)) to a LAN.

(27) Mini-Circuits "ZSC-4-1+" 4-ways power splitter operating at $0.1 - 200$ MHz connecting a 10 MHz reference through Bayonet Neill–Concelman (BNC) cables to all available instruments. More specifically, (7) has been selected as master, providing the source of the reference signal and (6), (8), (13) and (16) as slaves acquiring the reference signal.

Finally, all instruments mentioned above including all the power supplies of the various amplifiers, are connected to an APC 1.98 kW 2200 V A line interactive smart-Uninterruptible

Power Supply (UPS) to protect them from possible voltage spikes, voltage dips, fluctuations and complete power failures. The main controllable instruments that are used for the data acquisition are shown in Fig. C.1.

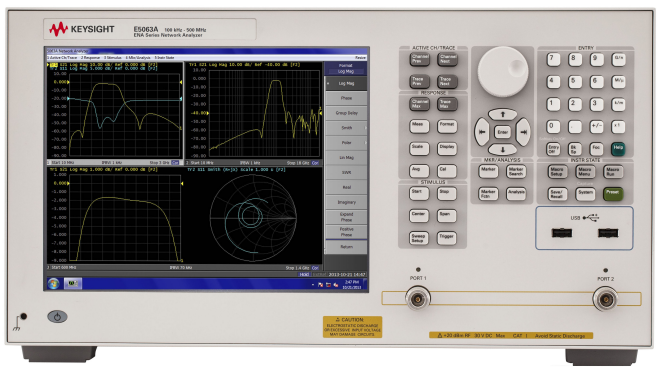

**(a)** Keysight's E5063A ENA Vector Network Analyser (75).

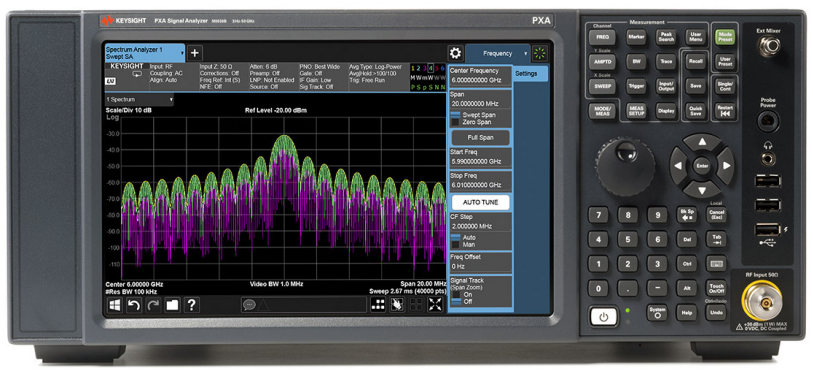

**(b)** Keysight's N9030B PXA Vector Signal Analyser (76).

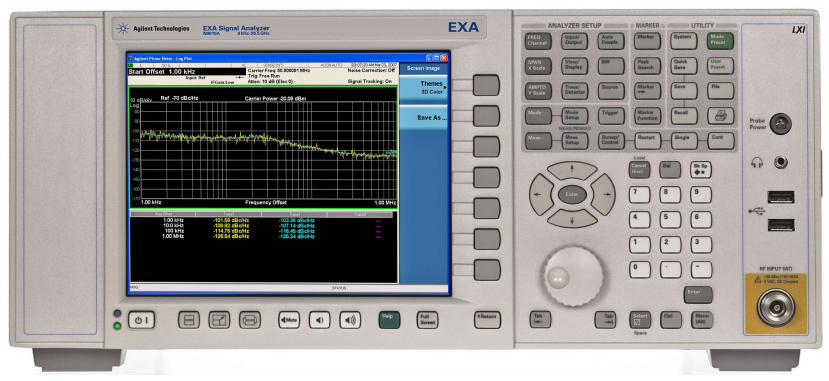

**(c)** Keysight's N9010A EXA Vector Signal Analyser (77).

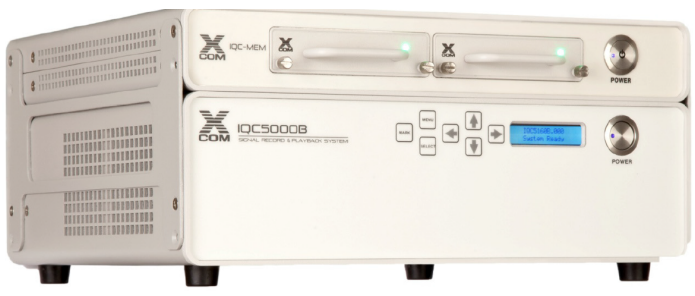

**(d)** X-COM's IQC5000B RF Recorder with the IQC5000B-MEM module on top (78).

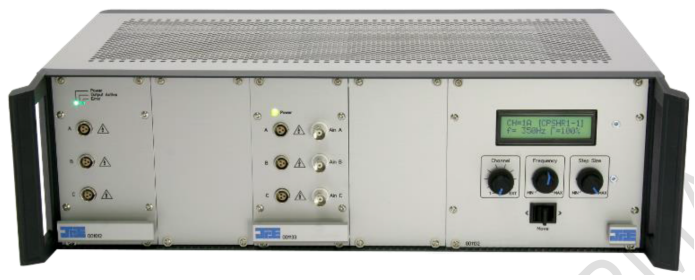

**(e)** JPE piezoelectric controller (73).

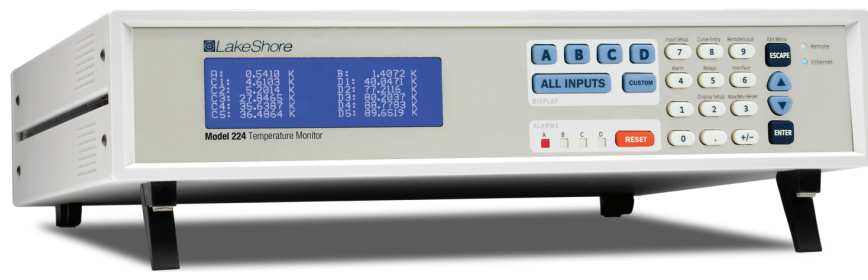

**(f)** Lakeshore 224 Temperature monitor (79)

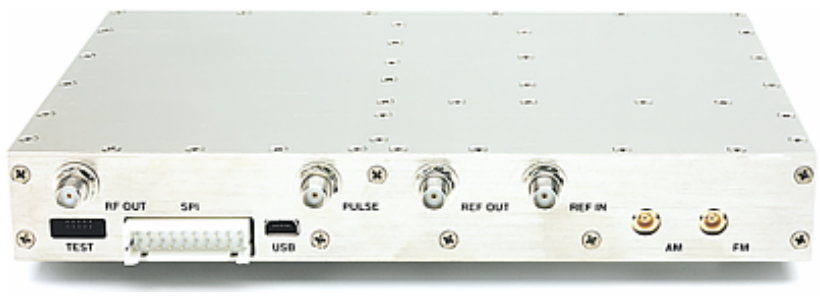

**(g)** National Instruments FSW-0020 signal generator (80)

**Figure C.1:** Main instruments controlled by CAST-CAPP DAQ chain

### C.1.1 Amplifiers

The amplification of the signal from the strongly-coupled port of each cavity is performed by a LNAs from Low Noise Factory providing a 39 dB amplification with a 2 K noise for the frequency range of CAST-CAPP in cryogenic conditions and with $\vec{B} = 8.8\,\mathrm{T}$ (Fig. C.2a). Then, the signal of each cavity is transitioned through RF cables from the cryogenic environment to room temperature by means of thermal plates. These cables have certain frequency-dependent losses which have to be taken into account for the overall gain. Then, an external signal amplification of about 22 dB is made for each one of the four cavities with external room-temperature amplifiers (Fig. C.2b) before the connection with the RF switches which send the signal towards/from the VSA or VNA depending on the type of measurement. Finally, a third stage amplification on the output signal of about 30 dB is performed after the power combiner with a Miteq “AFD3-0208-40-ST” LNA.

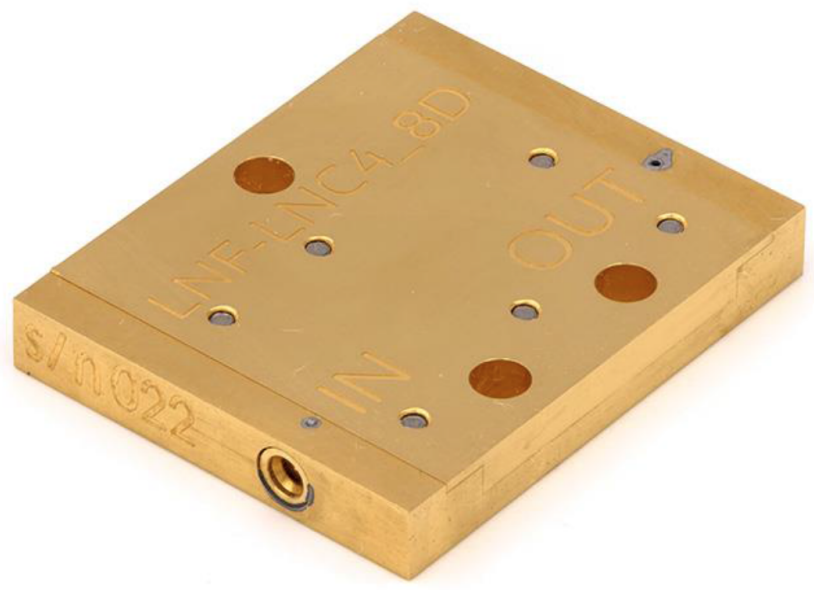

**(a)** LNF-LNC4_8D LNA from Low Noise Factory (81).

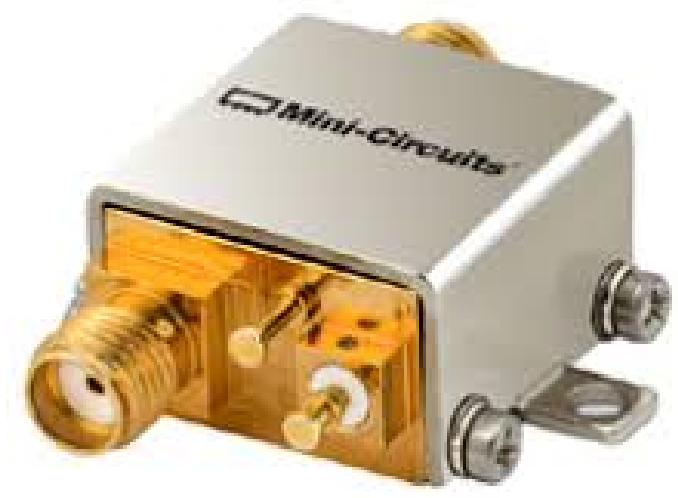

**(b)** ZX60-83LN-S+ LNA from Mini-Circuits (82).

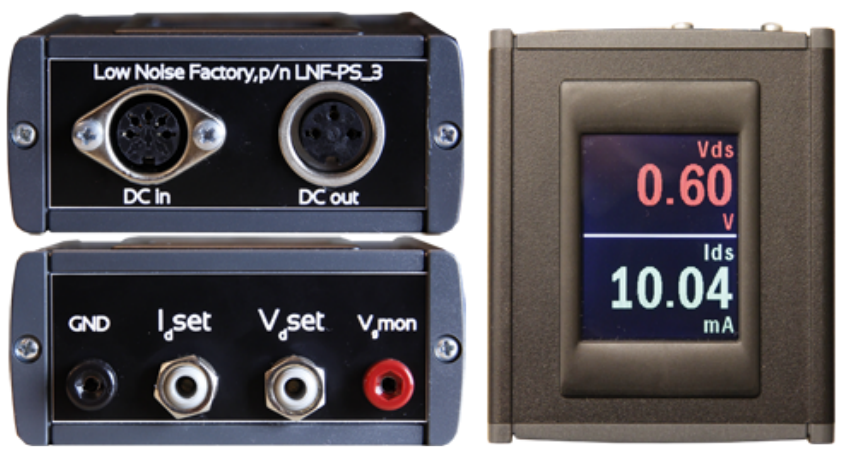

**(c)** LNF-PS_3 power supply for Low Noise Factory's LNAs (Fig. C.2a) (83).

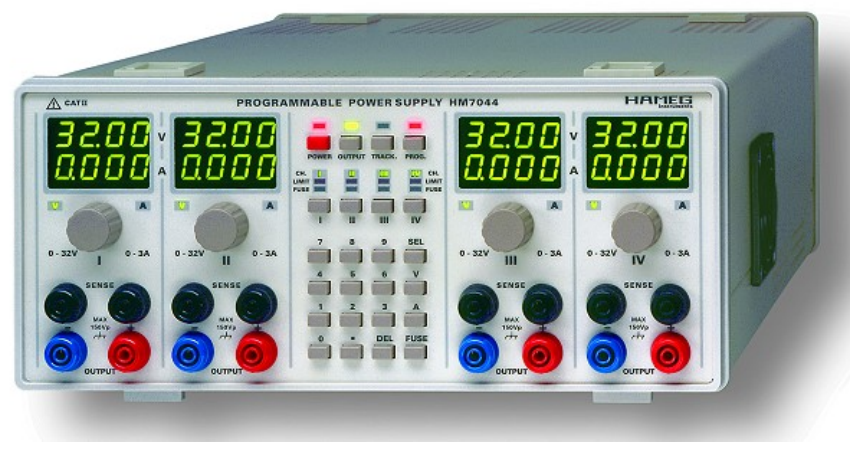

**(d)** Hameg HM7044 power supply which powers up the external LNAs in Fig. C.2b (84).

**Figure C.2:** Main LNAs used in CAST-CAPP detector and their power supplies.

### C.1.2 RF switches

All the active connections to the various data-taking instruments are mainly performed by the three RF switches shown in Fig. C.3 which are connected via USB cables with the workstation PC.

The input switch (Fig. C.3a) handles the input signal to the cavities and selects each time one cavity to be used either for measurements with the VNA or for a signal injection from the TS. In the COM port, the input cable coming from the device switch is connected, whereas in the rest of the four ports the four input SMA cables are connected going to the input warm flange and eventually to the four cavities.

The output switch (Fig. C.3b) selects the output signal coming from one or more cavities. For that purpose, in each of the four COM ports of the four sub-switches, the output signal coming through SMA cables from the output warm flange from each of the four cavities is connected (see also Fig. C.3e). Then, in each port-2 of all four sub-switches a 50 Ω load has been connected, whereas in each port-1 of all four switches an RF cable is connected which end up in the external amplifiers ((9) in Fig. 22.12). This way depending on which cavity the user wants to select, the automated DAQ system controls each of the four switches to send the signals either to the port of the 50 Ω load or to the external amplifiers. For example, if single cavity #1 is to be used, then the sub-switches B, C and D are switching to port-2 with the 50 Ω load whereas sub-switch A to port-1. If for example PM cavities 1-2-3 are to be used, then the switching configuration is A1, B1, C1, D2.

The device switch (Fig. C.3c) has two different sub-switches A and B shown in Fig. C.3f. The sub-switch A handles the input connections whereas sub-switch B handles the output connections. So, in port-1 of sub-switch A the input port of the VNA is connected and in port-2 the TS is connected, so that both instruments can send signals inside the cavities through the input port. Then, in sub-switch B in port-1, the output port of the VNA is connected whereas in port-2 the VSA is connected. Therefore, as an example, if a transmission measurement with the VNA is to be made, then the switching configuration is A1, B1.

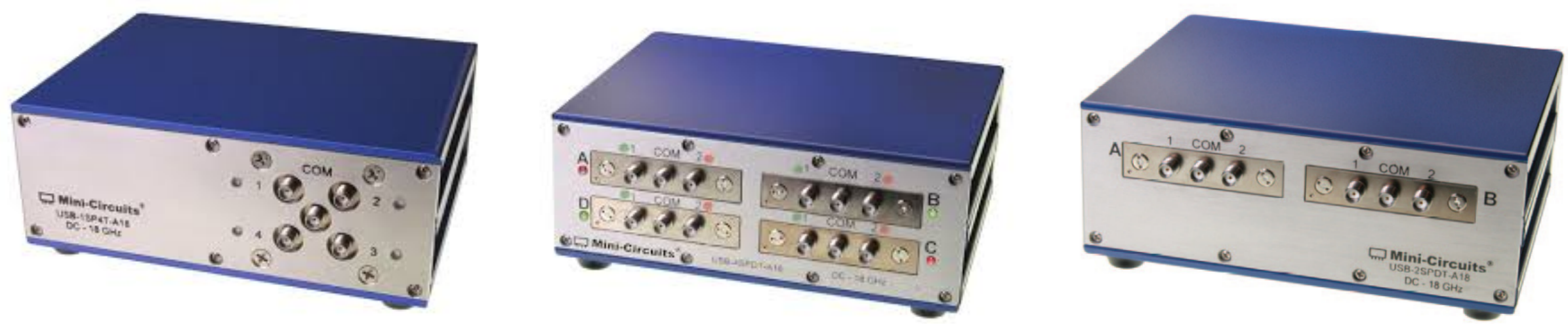

**(a)** Inpu switch (85). **(b)** Output switch (86). **(c)** Device switch (87).

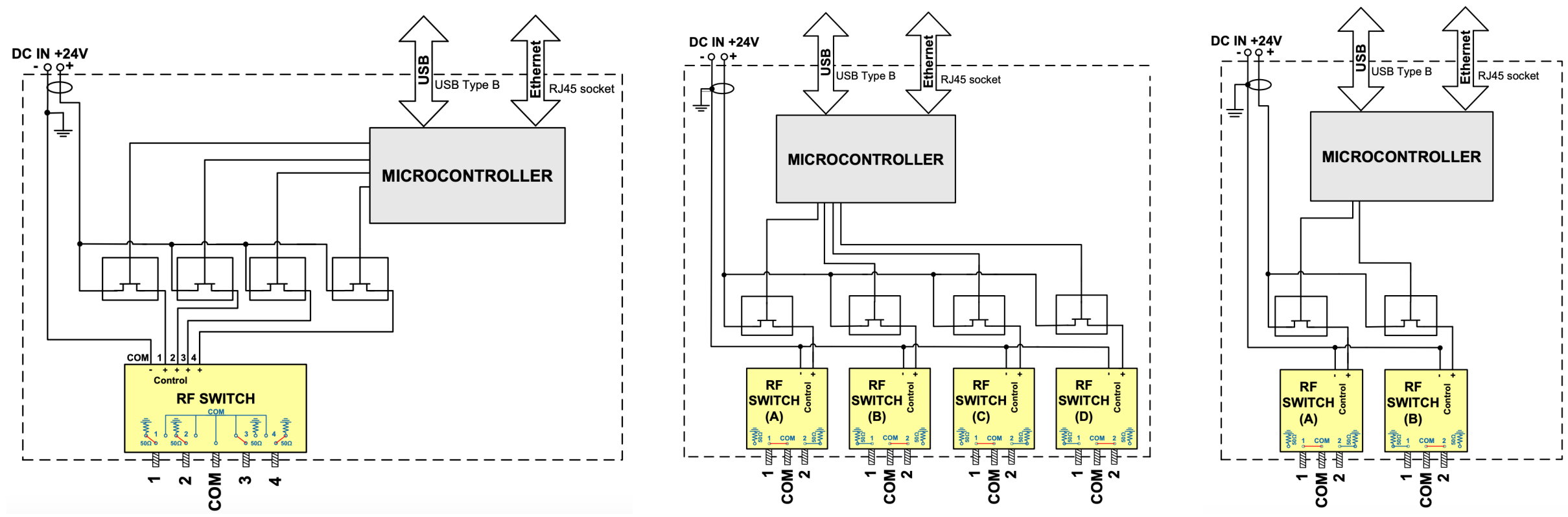

**(d)** Block diagram of the input switch(85). **(e)** Block diagram of the output switch (86). **(f)** Block diagram of the device switch (87).

**Figure C.3:** RF switches and their block diagrams used in CAST-CAPP detector.

### C.1.3 Vector Network Analyser

The VNA shown in Fig. C.1a is mainly used for transmission S21 measurements where a strong signal, from port-2 of VNA, is sent to the cavity through the weakly coupled input port and is sweeping the frequency range that the user has selected. The signal then comes out of the cavity from the near-critically coupled output port and is detected in port-1 of the VNA. This way the resonance mode frequency of the cavity can be retrieved and the quality factor ($Q_L$) can be calculated as seen in Fig. C.4. As mentioned in Sect. 22.1.3, the $TE_{101}$ mode which is used for axion search, corresponds to the lowest resonance frequency and the largest geometry factor. This is also verified from simulation. More specifically, the geometry factor for CAST-CAPP cavities is defined by simulations to be about $C_{lmn} = 0.53$ with a 10% uncertainty at 5.4 GHz.

Then the $Q_L$ calculation is based on the ratio of the centre frequency $\nu_0$ to bandwidth, where the bandwidth is selected at −3 dB. The bandwidth of the band-pass filter is determined by placing a marker on the maximum level, which corresponds to the centre frequency $\nu_0$. Therefore, the bandwidth is determined by measuring the values of the higher and lower frequencies at −3 dB below the peak.

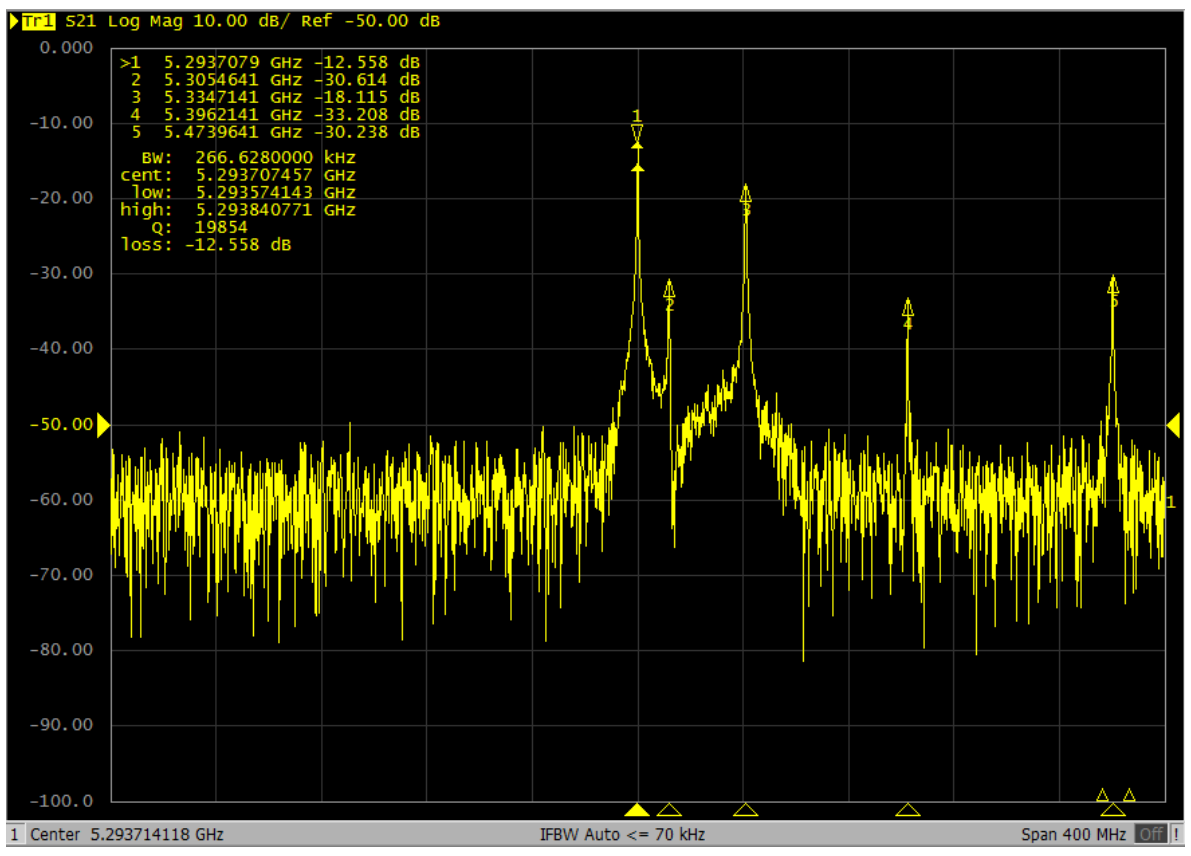

**(a)** Cavity #1.

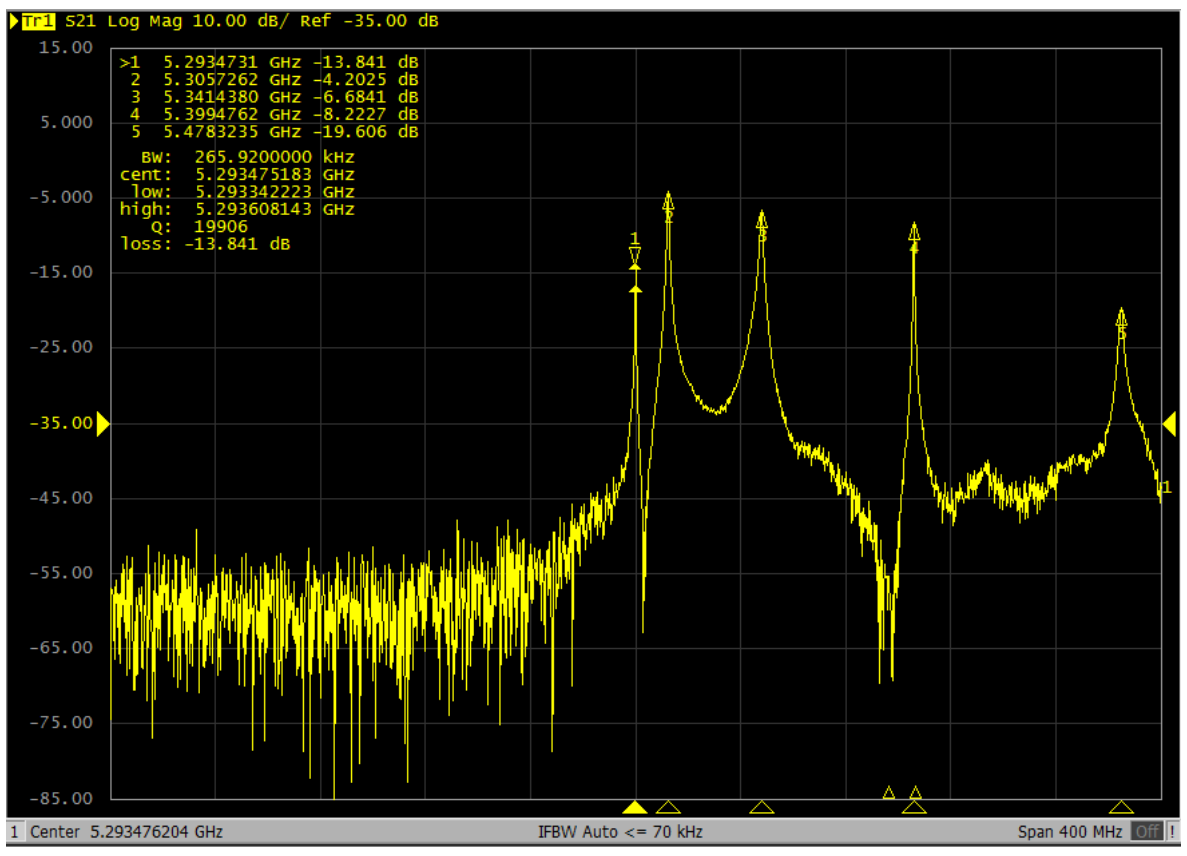

**(b)** Cavity #2.

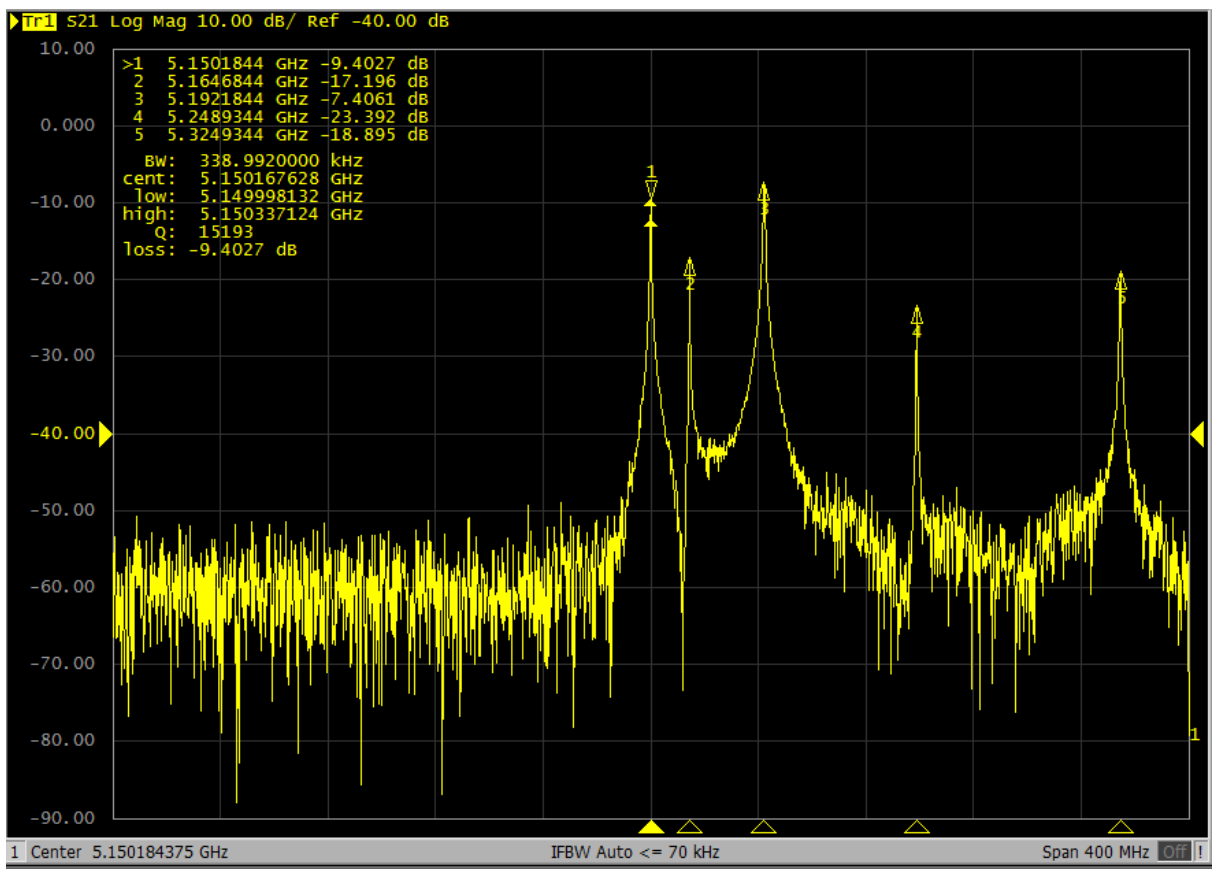

**(c)** Cavity #3.

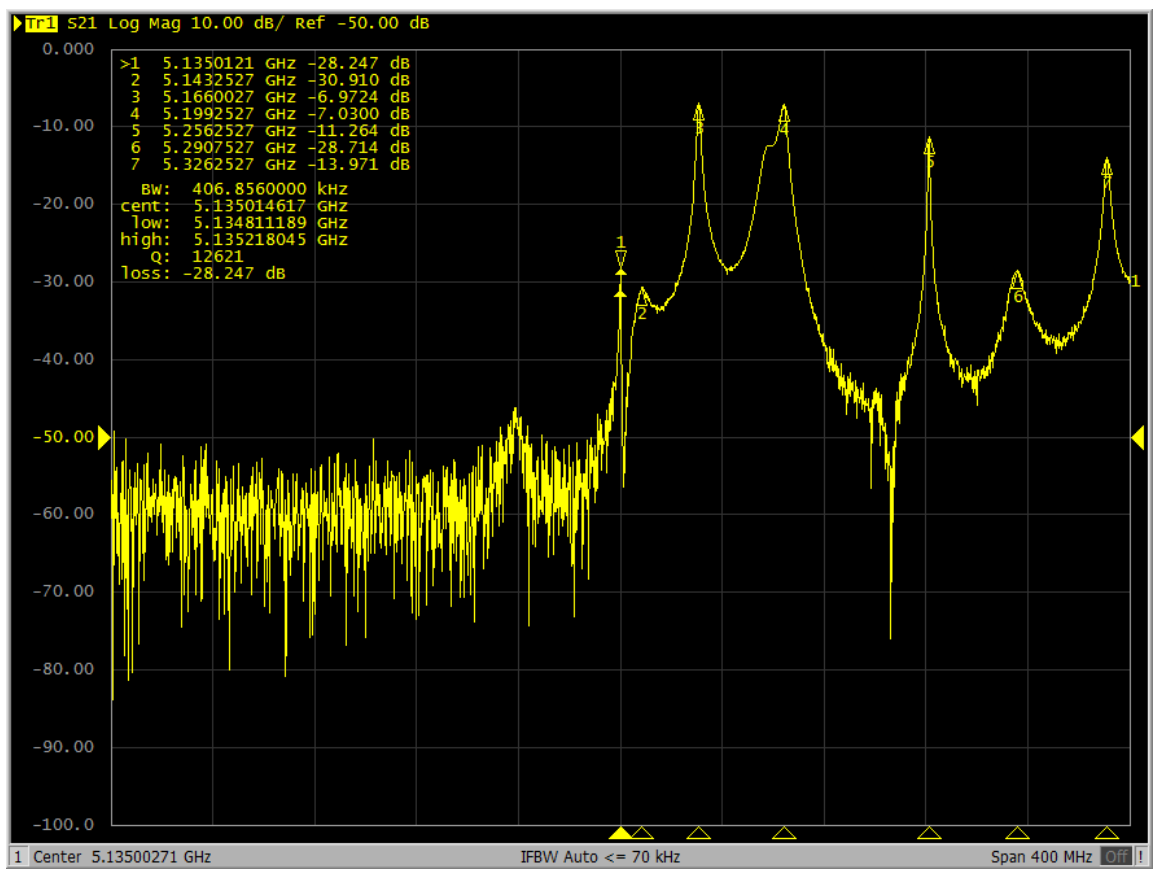

**(d)** Cavity #4.

**Figure C.4:** Transmission S21 measurements from the VNA for all four cavities in data-taking conditions with 400 MHz span. Marker #1 corresponds to the resonant peak which is used for the coupling to axions, while the rest of the markers correspond to higher order modes. The cavities are tuned to random positions. In $Q_L$ measurements a much more narrow span is used at around 5 MHz.

## C.1.4 Vector Signal Analyser

In CAST-CAPP detector two VSAs are used. One for taking data from the critically-coupled output port of the cavities and one for taking data from the external omnidirectional antenna measuring the electromagnetic background in the CAST hall. Both, are used with the same settings in “IQ Analyser”, “Complex Spectrum mode” which is FFT based. A window/span of 5 MHz is used for all measurements as seen in Fig. C.5.

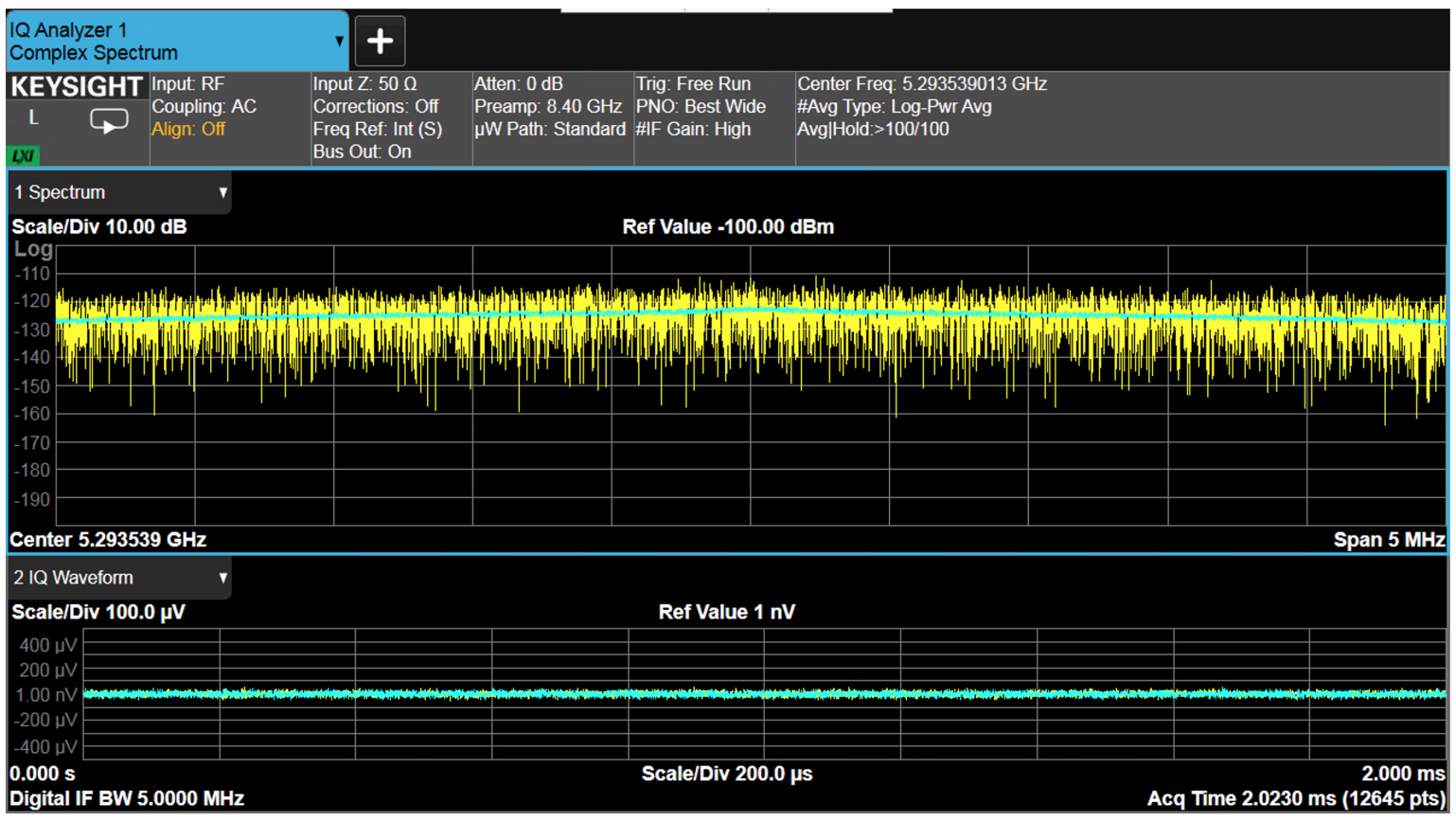

**Figure C.5:** VSA screen and the various settings in data-taking mode. The upper window corresponds to the frequency-domain with the yellow line corresponding to the current trace whereas the blue one to the averageddata. The lower window corresponds to time-domain from which the I/Q values are recorded. The latter screen is basically an I/Q waveform of voltage versus time.

However, to manually check the quality of the noise bump of the cavities, the VSA is also used in the “89600 VSA mode” between the data-taking runs. The noise bump corresponding to the resonance peak is expected to have an amplitude of at least 1 dB above the noise floor. In Fig. C.6 we can see that the noise bump of cavity #3 in the resonance peak is $\sim 2$ dB above noise level which indicates the good quality of the receiver chain and the near-critical coupling of the output port of the cavity. Finally, the difference between signal on/off has to be at least 10 dB. This is why an additional external amplifier ((12) in Fig. 22.12) was used, which gives in total about 35 dB difference between signal on/off.

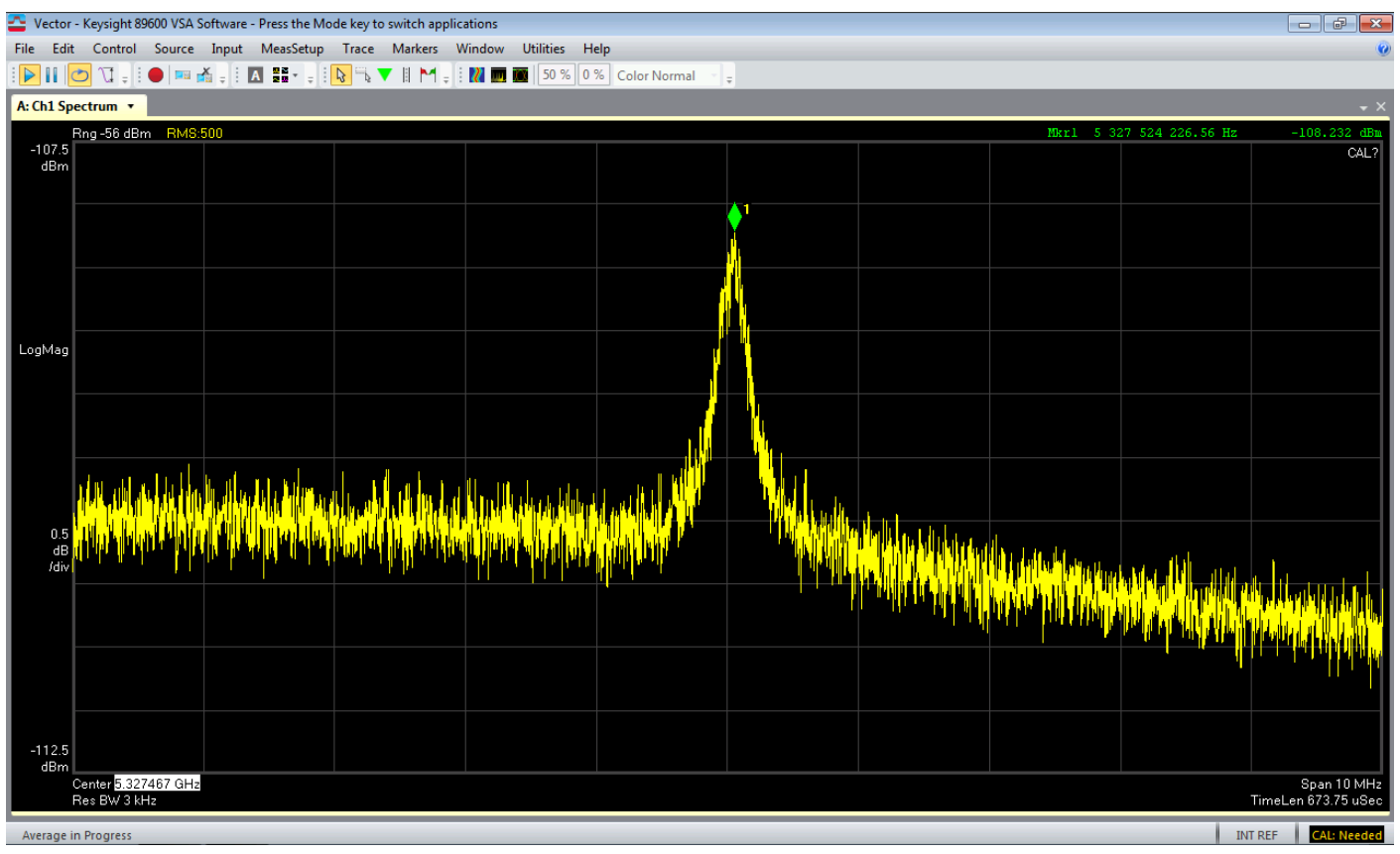

**Figure C.6:** Noise bump of a cavity in cryogenic conditions as seen through the “89600 VSA mode”.

## C.1.5 Recorder

The IQC data recorder (Fig. C.1d) is connected directly to the I/Q channels of the two VSAs via two Low Voltage Differential Signaling (LVDS) serial communication data cables designed for very high speed data transfer. The data are saved in the two integrated SSD drives of the IQC-MEM module, which are merged into a single RAID, and are time series of I/Q samples appearing as 16-bit binary data. These I/Q voltage readings carry the information of amplitude and phase. In each data file (of xdat) format, a human-readable header file (xhdr) is also recorded which contains data-taking parameters such as centre frequency, acquisition bandwidth (span), sample rate and start/end timestamps.

The data are recorded in 1 min measurements and correspond to a file size of 1.5 GB per channel. The two channels correspond to the two VSAs connected to the cavities and the external antenna. The selected bandwidth is also above the minimum requirements dictated by the width of the resonance peak $\nu_0/Q_L \sim 300\,\text{kHz}$. Furthermore, the sampling rate is 6.25 MHz per channel which correspond to $6.25 \times 10^6$ I/Q samples per second. This in turn corresponds to $375 \times 10^6$ data samples in each 1 min file.

Finally, the transfer of the files from the internal RAID SSD disks to the local SSDs of the workstation PC is performed through a Peripheral Component Interconnect Express (PCIe) connection which allows for fast offload speeds up to 450 MB/s.

### C.1.6 Piezo controller

The controller for the piezoelectric motors (Fig. C.1e) has the capability to control the movement of each one of the four CLAs independently either manually or through an Application Programming Interface (API). Each command is based on five parameters. The controller ID, which is selecting between the four controllers (CLAs), the temperature $T\,[\text{K}]$ of the environment in which the actuator is used, the frequency $F\,[\text{Hz}]$ of operation input $(0-600\,\text{Hz})$, the relative piezo step size parameter input $R\,[\%]$ (0-100), and the number of actuation steps $N\,[-]$ limited from 1 to 50000.

When the cavities are moved automatically via the DAQ software, the controller is switched to external control mode, and a continuous movement is performed until the tuner reaches the desired frequency as it is read by the VNA. For that purpose, the temperature of the corresponding cavity is measured before the movement and is given as an input to the controller. Also, the frequency of operation is chosen depending on the speed of movement. For a more accurate, smooth and eventually faster tuning, the frequency of operation is reduced to half when the frequency of the cavity is close to the target frequency. For the continuous movement, it is given $N = 0$ on the number of steps, and the piezo movement is stopped when the target frequency is reached. Finally, for maximum torque, when the piezo has difficulties in moving to certain positions, the maximum frequency $F = 600\,\text{Hz}$ and minimum temperature $T = 0\,\text{K}$ is used to enhance and/or recover the piezo movement.

The maximum, as well as the optimum frequency range of each cavity, is defined by external conditions such as temperature, magnetic field strength, mechanical vibrations as well as by the CTS conditions such as spring deformations, minimal axes misalignments etc. In Tab. C.1 these two ranges for each cavity are defined. During tuning within the automatic DAQ software safety measures are taken to avoid movement of the CLAs outside their optimal range. The shifted frequency range of cavity #4 compared with the rest of the three cavities was due to a repair on the sapphire axis holding the two sapphire strips together. However, the repair and upgrade of this cavity, has also resulted to an increase its achieved total frequency range. Notably, with an upgrade of the piezoelectric motors, the tuning range can even further be extended to 1 GHz and slightly beyond (4.6 GHz to 5.8 GHz) as shown by simulations [452] and demonstrated also on the bench at room temperatures with a slight adjustment of the tuning gears.

**Table C.1:** Maximum and optimal frequency range for the four cavities of CAST-CAPP detector using a single position in the tuning gears.

| | Maximum frequency range [GHz] | Optimal frequency range [GHz] |
|---|---|---|
| Cavity #1 | 5.13 - 5.50 | 5.20 - 5.34 |
| Cavity #2 | 5.07 - 5.49 | 5.20 - 5.35 |
| Cavity #3 | 5.03 - 5.50 | 5.25 - 5.45 |
| Cavity #4 | 4.74 - 5.40 | 4.75 - 5.23 |

### C.1.7 Temperature monitor

The temperature monitor (Fig. C.1f) is placed outside the cryostat and connects to four different Cernox temperature sensors installed in the cavities inside the magnet. More specifically, there is one temperature sensor installed in the middle of cavity #1 and one in the middle of cavity #2. There are also two sensors mounted on the front and the back of cavity #2. For cavity #2 the average temperature between the two sensors (front and back) is used for the analysis. Finally, for the calculation of the temperature of cavity #3, the average temperature readings from cavities #2 and #4 are used.

Two important factors for the temperatures of the cavities are the cryogenic LNAs, seen in Fig. C.2a, which are attached to the front side of each cavity and the piezoelectric motors which are placed on the rear side of the cavities. The ambient temperature of the magnet is around 1.7 K, whereas the temperature difference of the cavities between LNAs on/off is about $8-9$ K. This results to cavity temperatures between $T_{\text{cav}} \sim 10-14$ K for nominal bias of the LNAs and around 5 K with the LNAs turned off. However, the nominal heat dissipation of 7.15 mW from the LNAs could be reduced to 1.5 mW by fine-tuning the voltage/current bias on the power supplies of these LNAs (Fig. C.2c), without a sacrifice in their noise ($T_{LNA} \approx 2$ K). As a result, the average final physical temperatures of the cavities are $T_{\text{cav}} \sim 8$ K or less. This difference is shown in Fig. C.7.

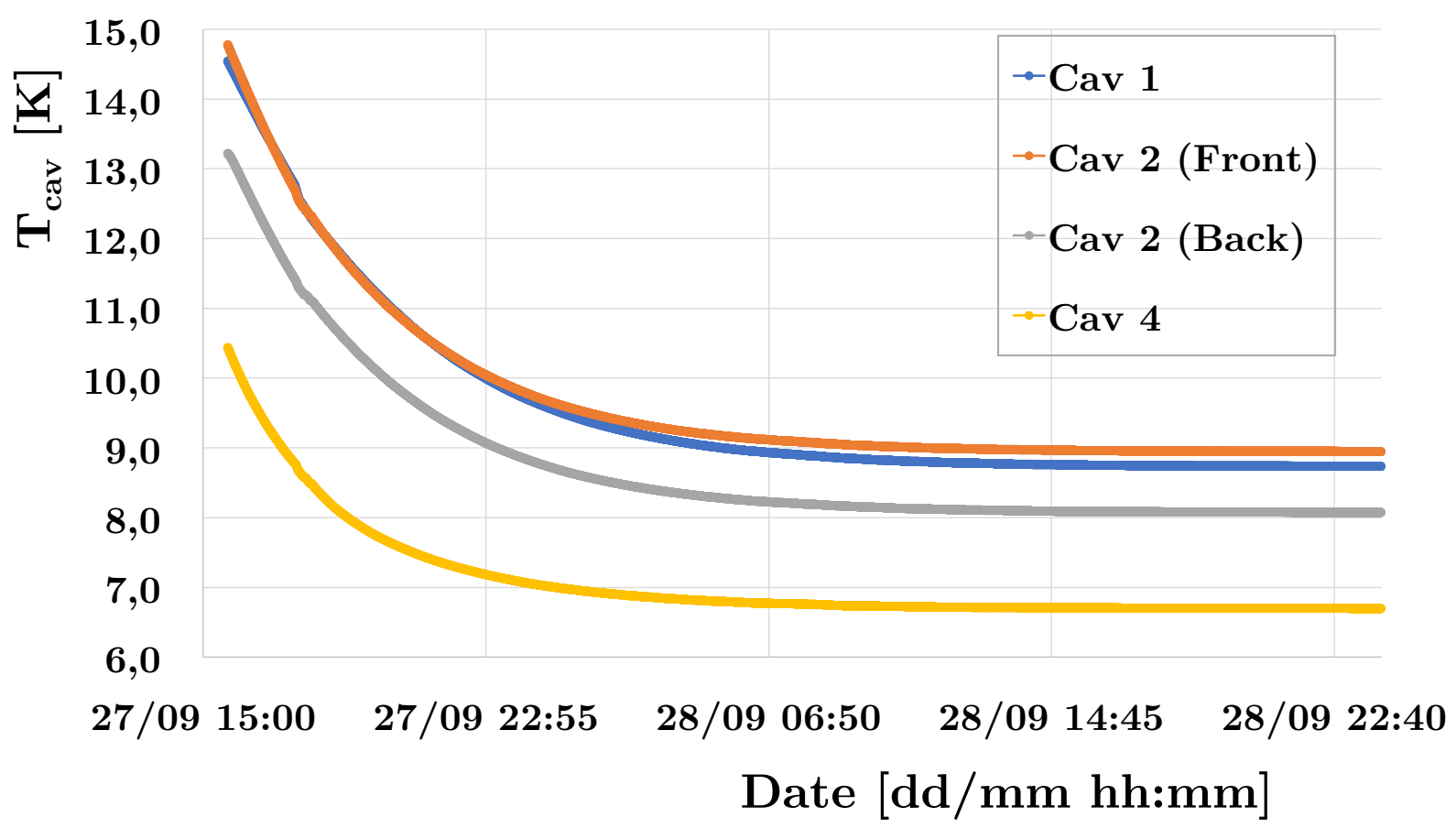

**Figure C.7:** Changes on the temperatures of the cavities when adjusting the bias on the power supplies of the four LNAs for minimal heat dissipation.

An additional small heat dissipation (on the order of $\sim 0.3\,\text{K}$) comes from the movement of the piezoelectric motors, but without any big effect on the temperatures recorded while taking data. This is because the tuning of the cavities takes place before data-recording is started. This heat dissipation can increase or decrease by small amounts depending on the size of the frequency steps while tuning.

Finally, the last contributing factor comes from the thermal conductivity of the cavities. The locking mechanism of the cavities with the magnet bore provides the thermal connection needed (see Fig. 22.8). However, the connection of all the cavities into a single system, reduced the anchoring points of the locking system from four per cavity to four for all the cavities. This results in a temperature difference between the cavities. For example, cavity #4 which is on the front side of the magnet and has its four anchoring points locked on the magnet, has a temperature around 7 K whereas the rest of the cavities around 8 K.

In Fig. C.8 the evolution of the temperature in the four sensors during the cool-down of CAST magnet with liquid He is presented. It is worth mentioning that during the cool-down of the cavities the resonant frequency is shifting towards higher frequencies due to slight deformation of the main body of each cavity and the various parts of the locomotive mechanism. The cool-down from room temperature to 8 K takes approximately 1.5 weeks as seen in Fig. C.8. During warm-up the opposite behaviour in the resonance peak of the cavities is observed. Finally, as expected, it is noted that a shift of the resonant peak of each cavity takes place also during ramping-up or ramping down of the CAST $\vec{B} = 8.8\,\text{T}$ magnet by about $0.3\,\text{MHz}$.

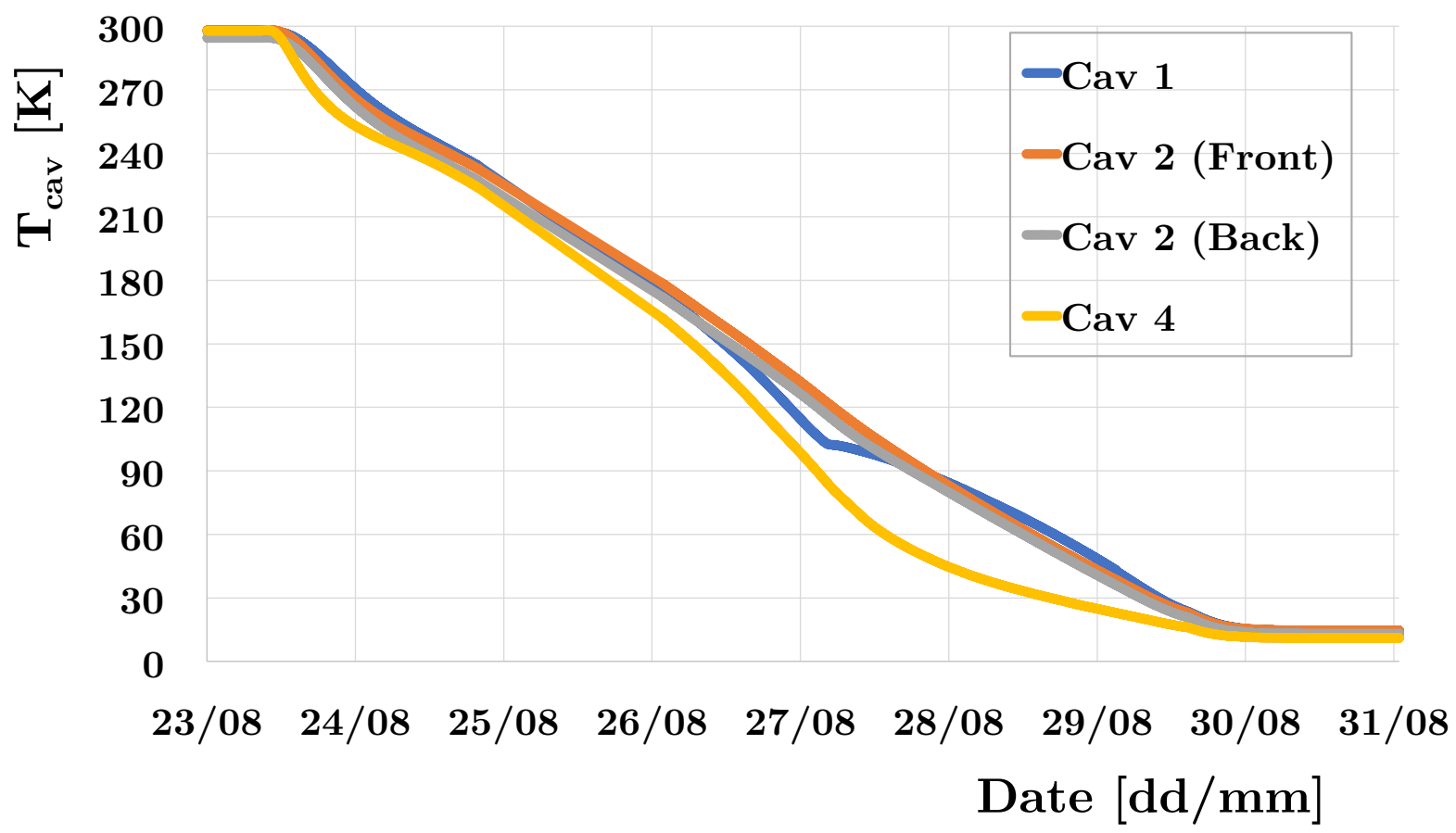

**Figure C.8:** Changes of the readings of the four temperature sensors placed on top of the cavities during the cool-down of CAST magnet on August 2019.

## C.2 Main procedure

Using the programming language python a variety of scripts have been created to control the various data-taking instruments in order to automatically take data with the minimum user involvement. At the same time the processing (Sect. 23.2.1) and analysis procedures (Sect. 23.3) have also been integrated in the general data-taking procedure. These python scripts together with the additional library scripts to control the various instruments will not be shown here due to space considerations.

### C.2.1 Single cavity

For a measurement with a single cavity, the first part involves the connection through LAN or USB of the workstation PC to all the DAQ related instruments and a verification of their normal operation and temperature (Fig. C.9c). The user can then select the appropriate measurement he wishes to perform. To perform an automatic combination of single measurements, the function "*auto*" is used. Then the user selects the cavity he wishes to use and the data-taking time (usually 1 min). Next, the user can choose to tune or use a fixed frequency for the cavity selected. In either case, he selects the starting frequency. For the tuning case, he also adds the number of steps, the size of the tuning steps (200 kHz) and the tuning direction. For fixed-frequency measurements, he adds the number of repeating measurements. Finally, the user has to decide whether at the end of the measurement(s) he wishes to offload the data from IQC-Mem to local storage, and perform post-data-taking operations such as upload of the raw data files to CERN tape system, processing of the raw files, and analysis of the processed files.

The second part involves the various operations that are performed from the script automatically based on the aforementioned details selected by the user. In the beginning, the input and output switches are connected to the desired cavity. The script then connects through the device switch to the VNA and through an S21 transmission measurement and measures the exact location of $\nu_0$ and the $Q_L$ of the selected cavity. At the same time the physical temperature of the selected cavity is measured through the temperature monitor. Then, the DAQ software through the device switch connects the input signal to the TS and the output signal to the VSA, where it switches to the "Complex Spectrum mode". As next, through the IQC Recorder it performs a measurement from both VSAs for the desired time (usually 1 min) and span (usually 5 MHz). When the measurement is finished, it connects back to the VNA to measure again the resonant frequency $\nu_0$ and $Q_L$ in case there is an unexpected change during the data-taking attributed to external reasons such as mechanical oscillations. Furthermore, the temperature is measured again also for crosschecking reasons. Finally, if the user has selected to perform multiple measurements the script goes into a loop to take all the measurements. If the user has selected tuning, then the script through the piezo controller tunes the selected cavity to the desired frequency step and then a new temperature measurement is performed to calculate the temperature change due to heating from the movement of the piezoelectric motor. Moreover, the number of steps that each piezo performs to reach the target frequency are recorded for classification reasons and to mark possibly safe/problematic frequency positions. Finally, after all the measurements are over, the data files which are saved in the two SSD RAIDs, of total size 4 TB of the IQC-Mem, depending on the user selection, are offloaded to the locally mounted SSD drives, uploaded to CERN tape system (Fig. C.9e), processed (Fig. C.9f) and analysed (Fig. C.9g). At the end of processing and analysis, the resulting files are also uploaded to CERN Advanced STORage manager (CASTOR) with dedicated scripts.

### C.2.2 Phase-matched cavities

In the case of PM, the first step is the tuning of all the selected cavities to the same desired frequency within a tolerance of 10 kHz. Next, based on the individual measurements of the amplitude of the resonance from the VNA, the programmable attenuators adjust, if needed, the attenuation that is added for each cavity in order to match the different amplitudes. This is done based on the configuration which has the minimum inserted attenuation. The next addition to the above-mentioned procedure for single cavities is the checking of the conditions for frequency-matching, amplitude-matching and temperature-matching before and after every 1 min measurement. In the case of multiple measurements, if one of the two conditions (frequency-matching or amplitude-matching) is not fulfilled, then the script automatically adjusts the frequencies (by tuning) and/or the amplitudes (from the attenuators) to correct the discrepancy and be within the allowed values. If there is an ambient temperature-mismatching and/or a frequency/amplitude mismatch at any point which can not be corrected,

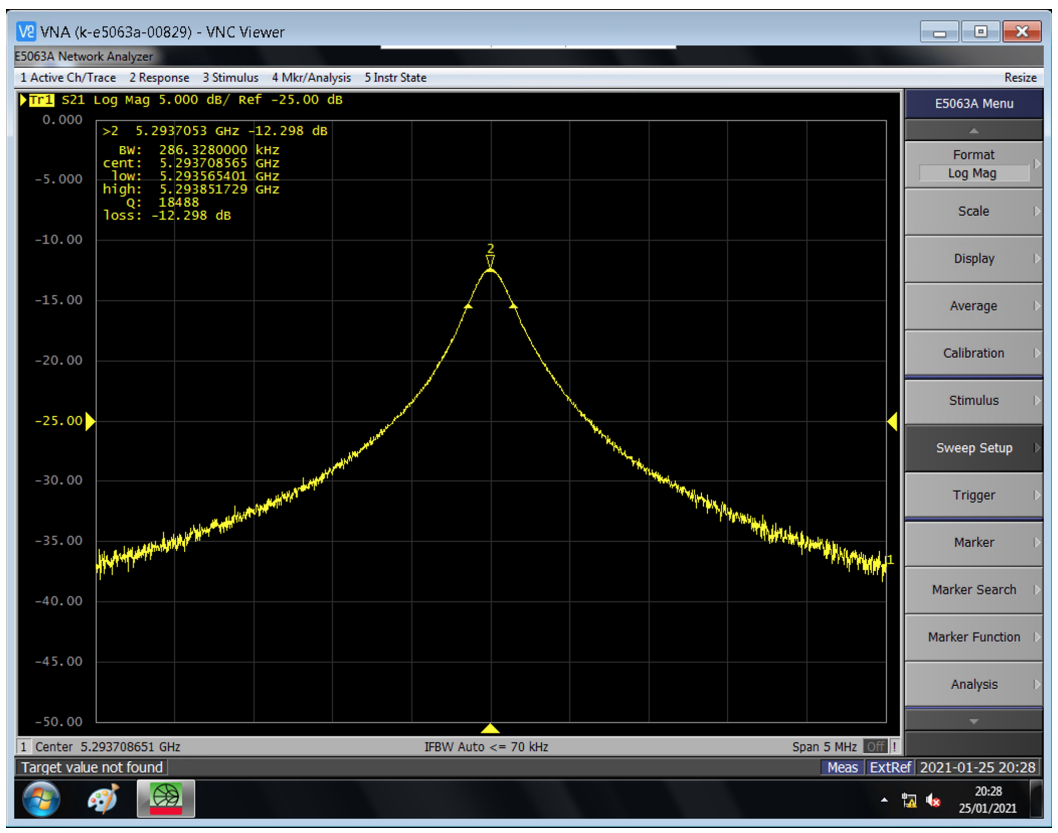

(a) VNA screen.

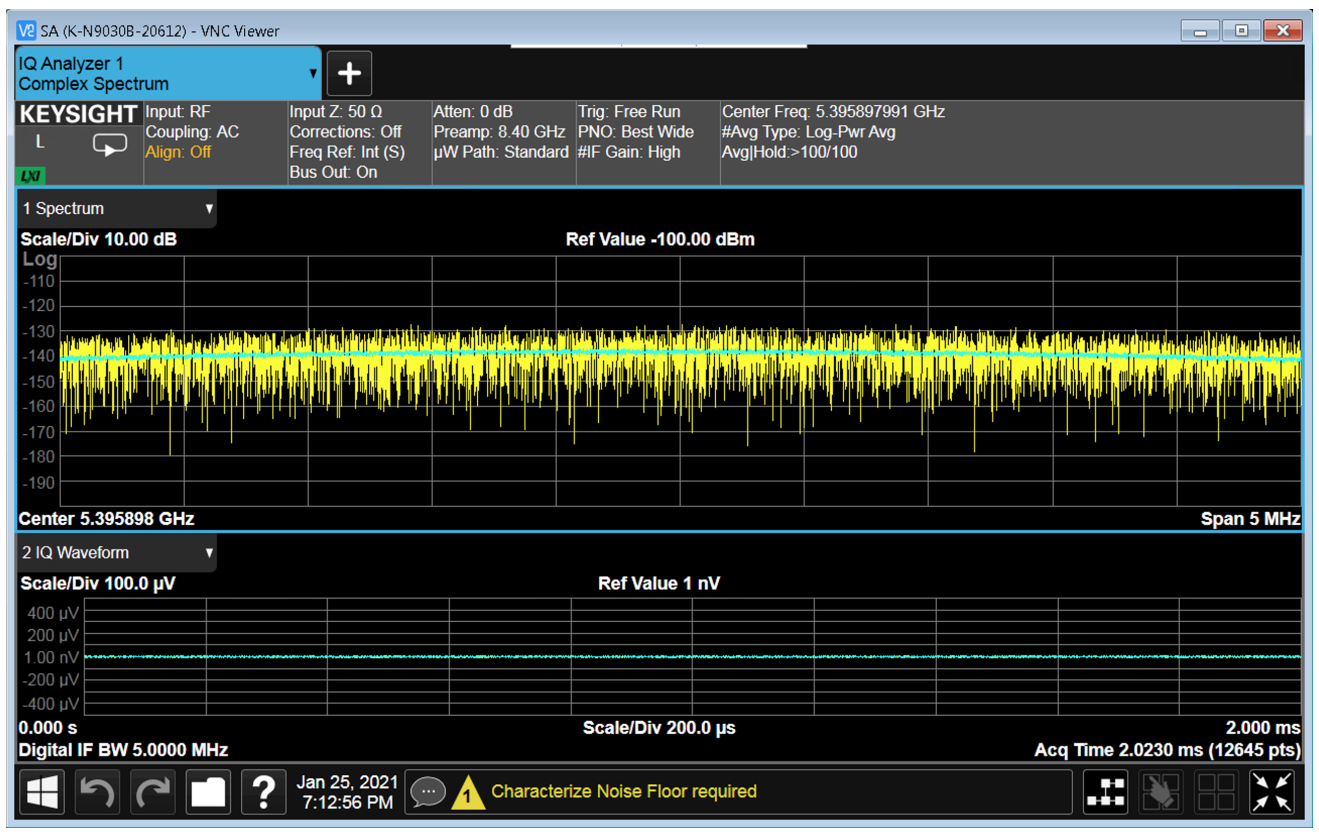

(b) VSA screen.

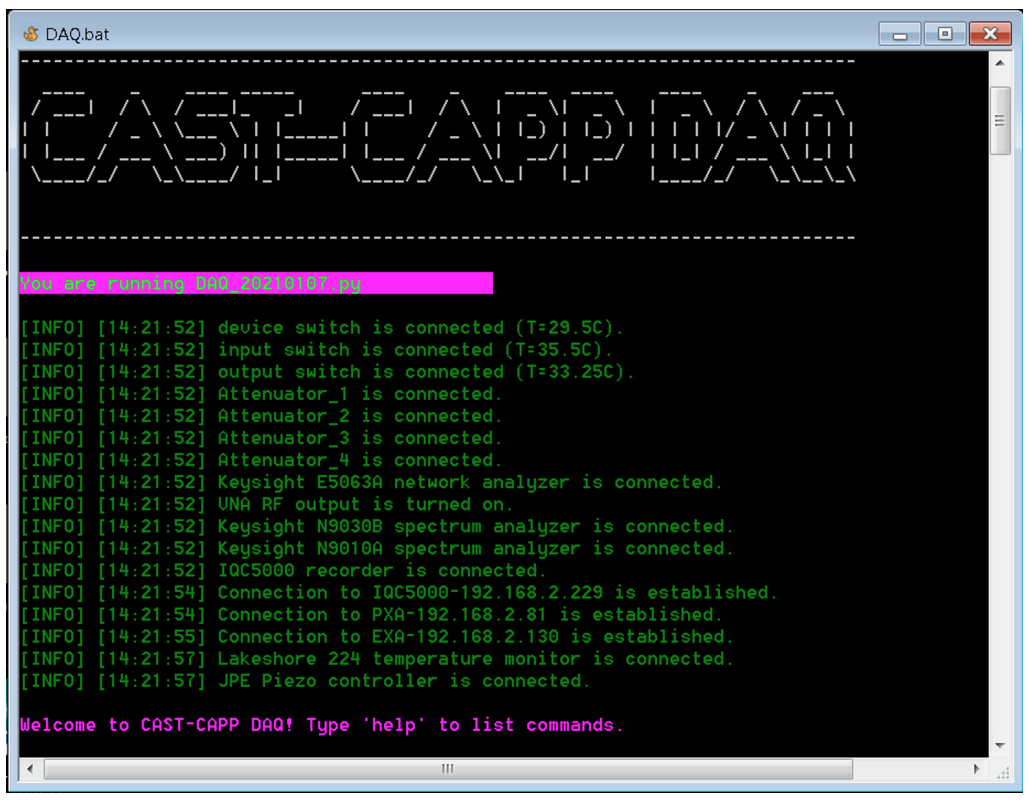

(c) DAQ script.

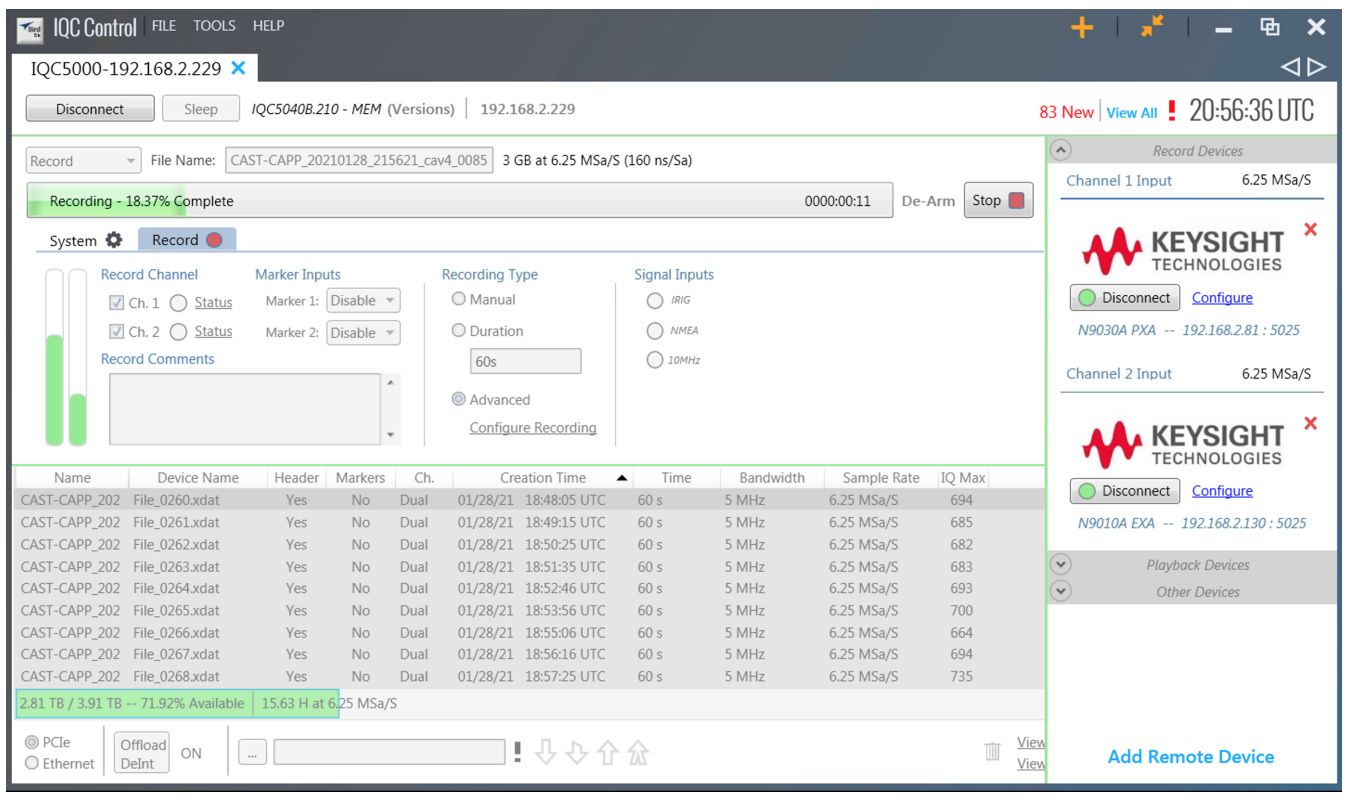

(d) IQC recorder screen.

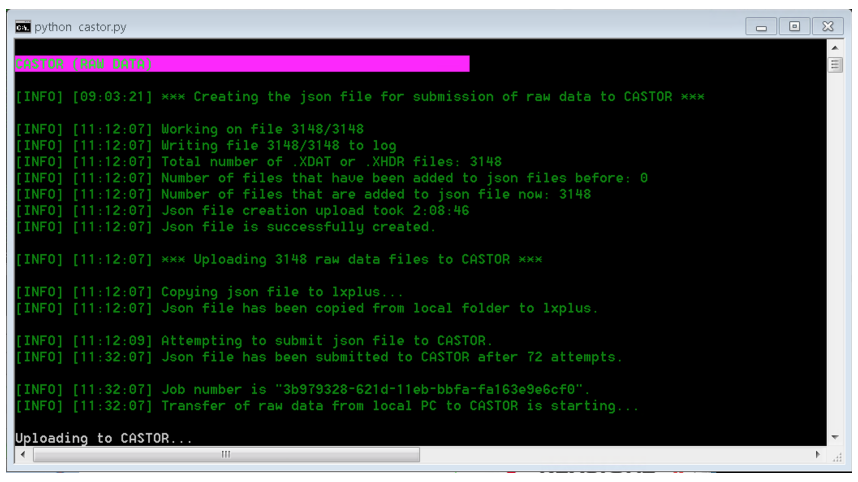

(e) CASTOR script.

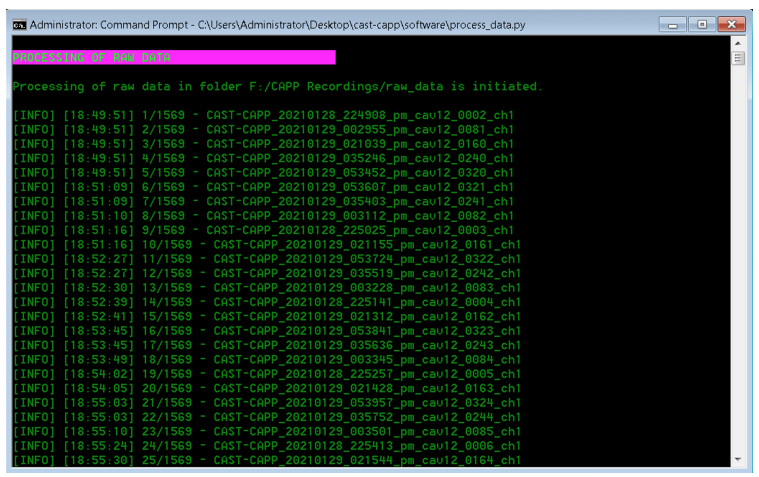

(f) Processing script.

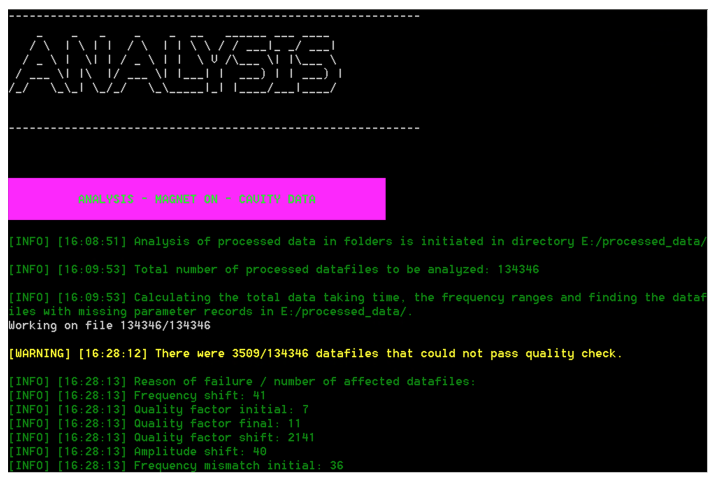

(g) Analysis script.

**Figure C.9:** Screen of the workstation PC with the main python script of DAQ, processing and analysis running at the same time, along with the screens of the VNA and VSA and IQC recorder.

the measurements are aborted. Finally, it is worth noticing that whenever there is any critical or serious error within the whole procedure, the user is notified also by email.

### C.2.3 Storage

Each 1 min file is about 1.5 GB per channel. This means that the total amount of data taken every day (about 20 h) is about 4 TB and therefore no local storage capacity can work for several days in a row. To mitigate this issue, after each data-taking day all the raw files are uploaded to CERN's tape archiving system CASTOR, processed and then deleted. This way the data are safely stored at CERN's tape system and can be recovered at any time if an additional processing is required. This saves space on the local drives and allows for a faster re-start of data-taking for the next day without interruptions and dead-times.

For the uploading procedure, a designated script has been created. Furthermore, the local SSD hard drives where the data are initially offloaded from the IQC-Mem, have been transformed into an File transfer Protocol (FTP) server. The read/write speed of the local SSD drives is about 940 MB/s. The first part of the uploading script is the inclusion of the filenames with exact directories in a json file. For each file apart from the source (the location on the local FTP server), the destination in CASTOR is also defined. Moreover, each specific checksum is also calculated, for the procedure to verify that the uploaded files in CASTOR are the same as the ones in the local FTP server. Finally, at the end of the json file various parameters such as retries, timer and overwrite option are defined. The next step in the procedure involves copying the json file to a designated Linux Public Login User Service (LXPLUS) directory. Lastly, the json file is submitted through File Transfer Service (FTS) to CASTOR. The whole uploading procedure with a 10 GB/s network card takes about 1.5 h with a throughput of around 500 MB/s per file. This procedure is repeated at the end of every data-taking day when the CAST solar tracking starts.

The processed and analysis files are saved in a different external hard drive for safety reasons, and backed-up regularly. The processed files are significantly smaller than the raw data files, i.e. 4 MB per 1 min file. For the daily upload also of these processed files at the end of each data-taking day, a new FTP server has been created in a different local HDD. Summing all the files from CAST-CAPP detector, including the processed and the analysed data from the various data-taking runs, results in a current total size of data in CASTOR of about 646.1 TB.

During the last data-taking run CERN's tape system has been replaced by a new high-performance system called *CTA*. As a result, all the data were migrated to the new system and the submission scripts have also been updated accordingly to match the latest protocols.

D

# DATA ANALYSIS ALGORITHMS

For the realisation of the various graphs of this thesis there were two main data analysis algorithms that were created and used. They were created using the programming language VBA but several versions of them existed also in Fortran and C++. The various commands inside these two algorithms are linked to spreadsheets in Microsoft Office Excel (via a creation of a module in the developer tab) where they read/write data on corresponding “Worksheets”, “Cells”, “Columns”, “Rows” etc.

## D.1 Main analysis algorithm

The “main analysis algorithm” was created in order to be able to calculate and plot the planetary longitudinal distribution of a given dataset. More precisely, the user provides the inquiry by typing on the corresponding cell of the worksheet “Results” (see Fig. D.1a) the needed calculation date period, data-range, bin size and the minimum and maximum planetary range of a planet that is to be constrained (0° to 360° is given if no longitudinal constraint is needed). Next, the user selects the planet target as a function of which he wishes to plot the given data (provided in the worksheet “Data”, as seen is Fig. D.1b), and runs the script by clicking on the grey “RUN” box. The algorithm then performs the corresponding calculations and automatically plots the various results.

```
Sub Main_Analysis_Script()
'Author: Marios R. Maroudas

Dim StartTime As Double
Dim MinutesElapsed As String

StartTime = Timer 'Remember the time when the macro starts.

```

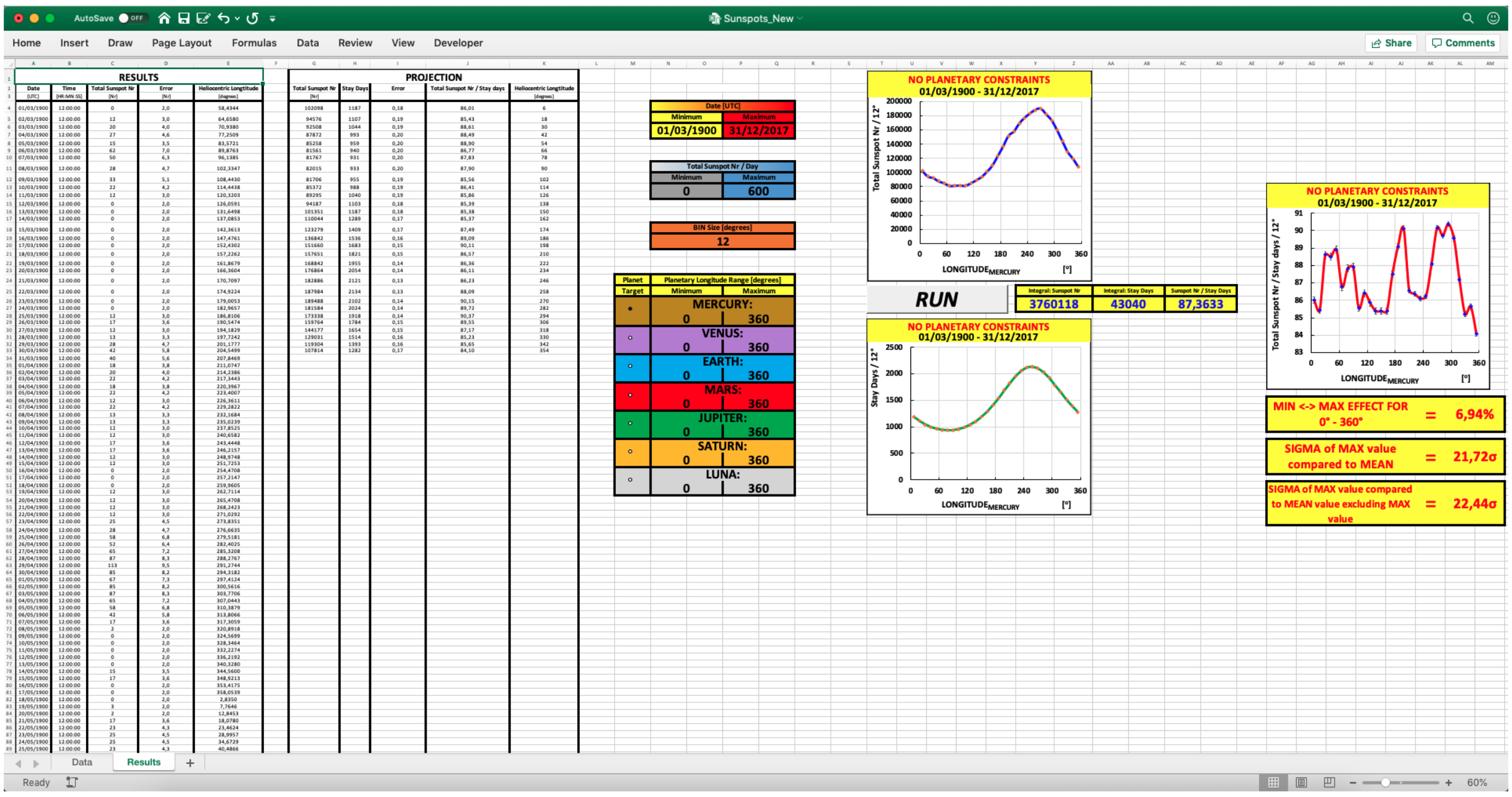

(a) "Results" worksheet.

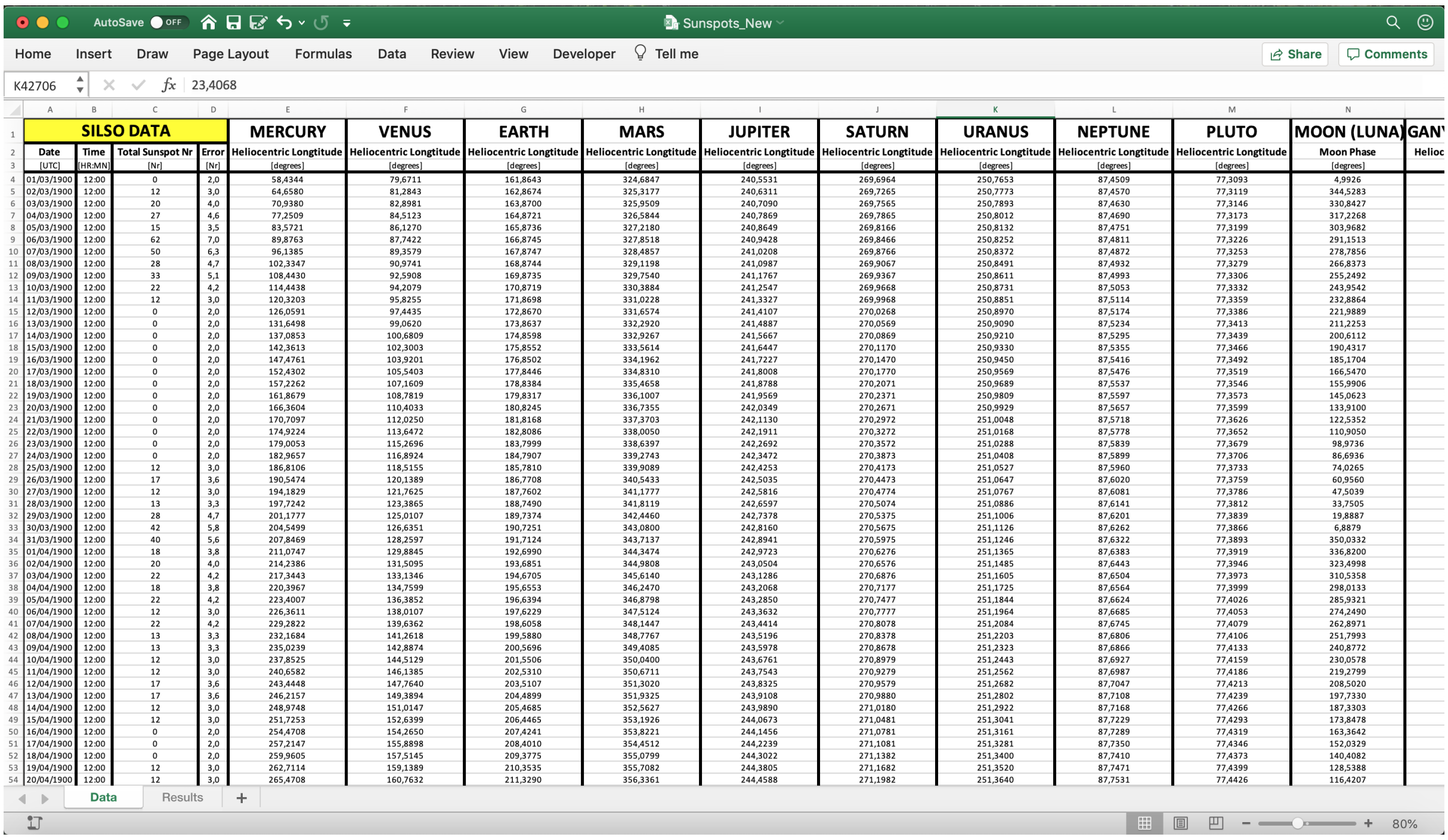

(b) "Data" worksheet.

**Figure D.1:** Example of the Excel Spreadsheet corresponding to the "Main Analysis" algorithm for the Sunspots dataset.

```
FinalRow = Worksheets("Data").Cells(Rows.Count, 1).End(xlUp).Row 'Find the last row of data from the first
    column.
Worksheets("Results").Range("A4:K1000000").ClearContents 'Delete the previous results from previous runs.
r = 4 ' "r" row goes through the data row to row.
y = 4 ' In "y" row we write the results.

'Message when wrong dates are given:
If Worksheets("Results").Cells(6, 14).Value < Worksheets("Data").Cells(4, 1).Value Or
    Worksheets("Results").Cells(6, 16).Value > Worksheets("Data").Cells(FinalRow, 1).Value Or
```

```
Worksheets("Results").Cells(6, 14).Value >= Worksheets("Results").Cells(6, 16).Value Then
    MsgBox "The DATES you have provided are wrong!" & vbNewLine & vbNewLine & "Please give correct dates.", vbCritical
    Exit Sub
End If

'Message when wrong data-range is given:
If Worksheets("Results").Cells(13, 14).Value >= Worksheets("Results").Cells(13, 16).Value Or Worksheets("Results").Cells(13, 14).Value < 0 Or Worksheets("Results").Cells(13, 16).Value < 0 Then
    MsgBox "The DATA range you have provided is wrong!" & vbNewLine & vbNewLine & "Please give correct minimum and maximum numbers.", vbCritical
    Exit Sub
End If

'Message when wrong bin size is given:
If Worksheets("Results").Cells(19, 14).Value <= 0 Or Worksheets("Results").Cells(19, 14).Value > 360 Then
    MsgBox "The BIN size you have provided is wrong!" & vbNewLine & vbNewLine & "Please give a correct BIN size number.", vbCritical
    Exit Sub
End If

'Message when wrong longitude range is given:
If Worksheets("Results").Cells(28, 14).Value = Worksheets("Results").Cells(28, 16).Value Or Worksheets("Results").Cells(28, 14).Value < 0 Or Worksheets("Results").Cells(28, 14).Value > 360 Or Worksheets("Results").Cells(28, 16).Value > 360 Or Worksheets("Results").Cells(28, 16).Value < 0 Then
    MsgBox "The MERCURY longitude range you have provided is wrong!" & vbNewLine & vbNewLine & "Please give correct numbers.", vbCritical
    Exit Sub
ElseIf Worksheets("Results").Cells(32, 14).Value = Worksheets("Results").Cells(32, 16).Value Or Worksheets("Results").Cells(32, 14).Value < 0 Or Worksheets("Results").Cells(32, 14).Value > 360 Or Worksheets("Results").Cells(32, 16).Value > 360 Or Worksheets("Results").Cells(32, 16).Value < 0 Then
    MsgBox "The VENUS longitude range you have provided is wrong!" & vbNewLine & vbNewLine & "Please give correct numbers.", vbCritical
    Exit Sub
ElseIf Worksheets("Results").Cells(36, 14).Value = Worksheets("Results").Cells(36, 16).Value Or Worksheets("Results").Cells(36, 14).Value < 0 Or Worksheets("Results").Cells(36, 14).Value > 360 Or Worksheets("Results").Cells(36, 16).Value > 360 Or Worksheets("Results").Cells(36, 16).Value < 0 Then
    MsgBox "The EARTH longitude range you have provided is wrong!" & vbNewLine & vbNewLine & "Please give correct numbers.", vbCritical
    Exit Sub
ElseIf Worksheets("Results").Cells(40, 14).Value = Worksheets("Results").Cells(40, 16).Value Or Worksheets("Results").Cells(40, 14).Value < 0 Or Worksheets("Results").Cells(40, 14).Value > 360 Or Worksheets("Results").Cells(40, 16).Value > 360 Or Worksheets("Results").Cells(40, 16).Value < 0 Then
    MsgBox "The MARS longitude range you have provided is wrong!" & vbNewLine & vbNewLine & "Please give correct numbers.", vbCritical
    Exit Sub
ElseIf Worksheets("Results").Cells(44, 14).Value = Worksheets("Results").Cells(44, 16).Value Or Worksheets("Results").Cells(44, 14).Value < 0 Or Worksheets("Results").Cells(44, 14).Value > 360 Or Worksheets("Results").Cells(44, 16).Value > 360 Or Worksheets("Results").Cells(44, 16).Value < 0 Then
    MsgBox "The JUPITER longitude range you have provided is wrong!" & vbNewLine & vbNewLine & "Please give correct numbers.", vbCritical
    Exit Sub
ElseIf Worksheets("Results").Cells(48, 14).Value = Worksheets("Results").Cells(48, 16).Value Or Worksheets("Results").Cells(48, 14).Value < 0 Or Worksheets("Results").Cells(48, 14).Value > 360 Or Worksheets("Results").Cells(48, 16).Value > 360 Or Worksheets("Results").Cells(48, 16).Value < 0 Then
    MsgBox "The SATURN longitude range you have provided is wrong!" & vbNewLine & vbNewLine & "Please give correct numbers.", vbCritical
```

```
    Exit Sub
ElseIf Worksheets("Results").Cells(52, 14).Value = Worksheets("Results").Cells(52, 16).Value Or Worksheets("Results").Cells(52, 14).Value < 0 Or Worksheets("Results").Cells(52, 14).Value > 360 Or Worksheets("Results").Cells(52, 16).Value > 360 Or Worksheets("Results").Cells(52, 16).Value < 0 Then
    MsgBox "The LUNA longitude range you have provided is wrong!" & vbNewLine & vbNewLine & "Please give correct numbers.", vbCritical
    Exit Sub
End If

'Running starts for each line "r":
For r = 4 To FinalRow
    'Splitting cases for Mercury:
    If Worksheets("Results").Cells(28, 14).Value < Worksheets("Results").Cells(28, 16).Value Then 'For the normal case when we provide "Min < Max". i.e. Mercury: 200 - 320 degrees
        Case_Mercury = Worksheets("Data").Cells(r, 5).Value >= Worksheets("Results").Cells(28, 14).Value And Worksheets("Data").Cells(r, 5).Value <= Worksheets("Results").Cells(28, 16).Value
    Else 'For the case when we provide "Max < Min". i.e. Mercury: 300 - 100 degrees
        Case_Mercury = (Worksheets("Data").Cells(r, 5).Value >= Worksheets("Results").Cells(28, 14).Value And Worksheets("Data").Cells(r, 5).Value <= 360) Or (Worksheets("Data").Cells(r, 5).Value >= 0 And Worksheets("Data").Cells(r, 5).Value <= Worksheets("Results").Cells(28, 16).Value)
    End If

    'Splitting cases for Venus:
    If Worksheets("Results").Cells(32, 14).Value < Worksheets("Results").Cells(32, 16).Value Then
        Case_Venus = Worksheets("Data").Cells(r, 6).Value >= Worksheets("Results").Cells(32, 14).Value And Worksheets("Data").Cells(r, 6).Value <= Worksheets("Results").Cells(32, 16).Value
    Else
        Case_Venus = (Worksheets("Data").Cells(r, 6).Value >= Worksheets("Results").Cells(32, 14).Value And Worksheets("Data").Cells(r, 6).Value <= 360) Or (Worksheets("Data").Cells(r, 6).Value >= 0 And Worksheets("Data").Cells(r, 6).Value <= Worksheets("Results").Cells(32, 16).Value)
    End If

    'Splitting cases for Earth:
    If Worksheets("Results").Cells(36, 14).Value < Worksheets("Results").Cells(36, 16).Value Then
        Case_Earth = Worksheets("Data").Cells(r, 7).Value >= Worksheets("Results").Cells(36, 14).Value And Worksheets("Data").Cells(r, 7).Value <= Worksheets("Results").Cells(36, 16).Value
    Else
        Case_Earth = (Worksheets("Data").Cells(r, 7).Value >= Worksheets("Results").Cells(36, 14).Value And Worksheets("Data").Cells(r, 7).Value <= 360) Or (Worksheets("Data").Cells(r, 7) >= 0 And Worksheets("Data").Cells(r, 7).Value <= Worksheets("Results").Cells(36, 16).Value)
    End If

    'Splitting cases for Mars:
    If Worksheets("Results").Cells(40, 14).Value < Worksheets("Results").Cells(40, 16).Value Then
        Case_Mars = Worksheets("Data").Cells(r, 8).Value >= Worksheets("Results").Cells(40, 14).Value And Worksheets("Data").Cells(r, 8).Value <= Worksheets("Results").Cells(40, 16).Value
    Else
        Case_Mars = (Worksheets("Data").Cells(r, 8).Value >= Worksheets("Results").Cells(40, 14).Value And Worksheets("Data").Cells(r, 8).Value <= 360) Or (Worksheets("Data").Cells(r, 8).Value >= 0 And Worksheets("Data").Cells(r, 8).Value <= Worksheets("Results").Cells(40, 16).Value)
    End If

    'Splitting cases for Jupiter:
    If Worksheets("Results").Cells(44, 14).Value < Worksheets("Results").Cells(44, 16).Value Then
        Case_Jupiter = Worksheets("Data").Cells(r, 9).Value >= Worksheets("Results").Cells(44, 14).Value And Worksheets("Data").Cells(r, 9).Value <= Worksheets("Results").Cells(44, 16).Value
    Else
```

```
        Case_Jupiter = (Worksheets("Data").Cells(r, 9).Value >= Worksheets("Results").Cells(44, 14).Value And Worksheets("Data").Cells(r, 9).Value <= 360) Or (Worksheets("Data").Cells(r, 9).Value >= 0 And Worksheets("Data").Cells(r, 9).Value <= Worksheets("Results").Cells(44, 16).Value)
    End If

    'Splitting cases for Saturn:
    If Worksheets("Results").Cells(48, 14).Value < Worksheets("Results").Cells(48, 16).Value Then
        Case_Saturn = Worksheets("Data").Cells(r, 10).Value >= Worksheets("Results").Cells(48, 14).Value And Worksheets("Data").Cells(r, 10).Value <= Worksheets("Results").Cells(48, 16).Value
    Else
        Case_Saturn = (Worksheets("Data").Cells(r, 10).Value >= Worksheets("Results").Cells(48, 14).Value And Worksheets("Data").Cells(r, 10).Value <= 360) Or (Worksheets("Data").Cells(r, 10).Value >= 0 And Worksheets("Data").Cells(r, 10).Value <= Worksheets("Results").Cells(48, 16).Value)
    End If

    'Splitting cases for Luna:
    If Worksheets("Results").Cells(52, 14).Value < Worksheets("Results").Cells(52, 16).Value Then
        Case_Luna = Worksheets("Data").Cells(r, 14).Value >= Worksheets("Results").Cells(52, 14).Value And Worksheets("Data").Cells(r, 14).Value <= Worksheets("Results").Cells(52, 16).Value
    Else
        Case_Luna = (Worksheets("Data").Cells(r, 14).Value >= Worksheets("Results").Cells(52, 14).Value And Worksheets("Data").Cells(r, 14).Value <= 360) Or (Worksheets("Data").Cells(r, 14).Value >= 0 And Worksheets("Data").Cells(r, 14).Value <= Worksheets("Results").Cells(52, 16).Value)
    End If

   ' Asking the relevant conditions:
        If Worksheets("Data").Cells(r, 1).Value >= Worksheets("Results").Cells(6, 14).Value And Worksheets("Data").Cells(r, 1).Value <= Worksheets("Results").Cells(6, 16).Value Then 'Condition for dates.
            If Worksheets("Data").Cells(r, 3).Value >= Worksheets("Results").Cells(13, 14).Value And Worksheets("Data").Cells(r, 3).Value <= Worksheets("Results").Cells(13, 16).Value Then 'Condition for each data range.
                If Case_Mercury And Case_Venus And Case_Earth And Case_Mars And Case_Jupiter And Case_Saturn And Case_Luna Then 'Conditions for longitude ranges for all the planets.
                    Worksheets("Results").Cells(y, 1).Value = Worksheets("Data").Cells(r, 1).Value
                    Worksheets("Results").Cells(y, 1).Font.Color = Worksheets("Data").Cells(r, 1).Font.Color
                    Worksheets("Results").Cells(y, 2).Value = Worksheets("Data").Cells(r, 2).Value
                    Worksheets("Results").Cells(y, 2).Font.Color = Worksheets("Data").Cells(r, 2).Font.Color
                    Worksheets("Results").Cells(y, 3).Value = Worksheets("Data").Cells(r, 3).Value
                    Worksheets("Results").Cells(y, 3).Font.Color = Worksheets("Data").Cells(r, 3).Font.Color
                    Worksheets("Results").Cells(y, 4).Value = Worksheets("Data").Cells(r, 4).Value
                    Worksheets("Results").Cells(y, 4).Font.Color = Worksheets("Data").Cells(r, 4).Font.Color
                        If Worksheets("Results").Shapes("Option Button 1").OLEFormat.Object.Value = 1 Then
                            Worksheets("Results").Cells(y, 5).Value = Worksheets("Data").Cells(r, 5).Value
                            Worksheets("Results").Cells(y, 5).Font.Color = Worksheets("Data").Cells(r, 5).Font.Color
                        ElseIf Worksheets("Results").Shapes("Option Button 2").OLEFormat.Object.Value = 1 Then
                            Worksheets("Results").Cells(y, 5).Value = Worksheets("Data").Cells(r, 6).Value
                            Worksheets("Results").Cells(y, 5).Font.Color = Worksheets("Data").Cells(r, 6).Font.Color
                        ElseIf Worksheets("Results").Shapes("Option Button 3").OLEFormat.Object.Value = 1 Then
                            Worksheets("Results").Cells(y, 5).Value = Worksheets("Data").Cells(r, 7).Value
                            Worksheets("Results").Cells(y, 5).Font.Color = Worksheets("Data").Cells(r, 7).Font.Color
                        ElseIf Worksheets("Results").Shapes("Option Button 4").OLEFormat.Object.Value = 1 Then
                            Worksheets("Results").Cells(y, 5).Value = Worksheets("Data").Cells(r, 8).Value
                            Worksheets("Results").Cells(y, 5).Font.Color = Worksheets("Data").Cells(r,
```

```
                        8).Font.Color
                    ElseIf Worksheets("Results").Shapes("Option Button 5").OLEFormat.Object.Value = 1 Then
                        Worksheets("Results").Cells(y, 5).Value = Worksheets("Data").Cells(r, 9).Value
                        Worksheets("Results").Cells(y, 5).Font.Color = Worksheets("Data").Cells(r, 9).Font.Color
                    ElseIf Worksheets("Results").Shapes("Option Button 6").OLEFormat.Object.Value = 1 Then
                        Worksheets("Results").Cells(y, 5).Value = Worksheets("Data").Cells(r, 10).Value
                        Worksheets("Results").Cells(y, 5).Font.Color = Worksheets("Data").Cells(r, 10).Font.Color
                    ElseIf Worksheets("Results").Shapes("Option Button 7").OLEFormat.Object.Value = 1 Then
                        Worksheets("Results").Cells(y, 5).Value = Worksheets("Data").Cells(r, 14).Value
                        Worksheets("Results").Cells(y, 5).Font.Color = Worksheets("Data").Cells(r, 14).Font.Color
                    End If
                y = y + 1
            End If
        End If
    End If
Next r

'Depending on the bin size we create the projections:

FinalRow2 = Worksheets("Results").Cells(Rows.Count, 1).End(xlUp).Row 'Find the last row which has data from the first column of results (to minimise running time).

y = 4
r = 4
c = 0
b = 0
d = 0
e = 0

For b = 0 To 359
    For r = 4 To FinalRow2
        If Worksheets("Results").Cells(r, 5).Value >= b And Worksheets("Results").Cells(r, 5).Value < (b + Worksheets("Results").Cells(19, 14).Value) And Not IsEmpty(Worksheets("Results").Cells(r, 5).Value) Then
        c = c + ActiveSheet.Cells(r, 3).Value 'Calculates the relevant data (i.e. EUV) for the bin that has been selected.
        d = d + 1 'Calculates the "stay days" for all the dates that are constrained also from the amplitude of the data for the relevant bin.
        e = e + (Worksheets("Results").Cells(r, 4).Value * Worksheets("Results").Cells(r, 4).Value) 'Calculates the standard deviation.
        End If
    Next r
Worksheets("Results").Cells(y, 11).Value = (b + ((b + Worksheets("Results").Cells(19, 14).Value))) / 2 'Here we place the data in the middle of each bin. For example if we have bin=6deg then the first number will be placed in long=3deg.
Worksheets("Results").Cells(y, 7).Value = c
Worksheets("Results").Cells(y, 8).Value = d
    If Worksheets("Results").Cells(y, 8).Value = 0 Then 'When the division gives zero as a result.
       Worksheets("Results").Cells(y, 9).Value = 0
       Worksheets("Results").Cells(y, 10).Value = 0
    Else
       Worksheets("Results").Cells(y, 9).Value = Sqr(e) / d
       Worksheets("Results").Cells(y, 10).Value = (Worksheets("Results").Cells(y, 7).Value) / (Worksheets("Results").Cells(y, 8).Value)
```

```
    End If
d = 0
c = 0
e = 0
b = (b + Worksheets("Results").Cells(19, 14).Value - 1) 'With "Next b" it will move one unit, so this has to be
     reduced 1 unit.
y = y + 1
Next b

'Here we configure the relevant chart labels, fonts etc. for every planet case that has been selected:

If Worksheets("Results").Shapes("Option Button 1").OLEFormat.Object.Value = 1 Then
    Worksheets("Results").ChartObjects("Chart1").Activate
        With ActiveChart.Axes(xlCategory)
            .HasTitle = True
            .AxisTitle.Characters.Text = "LONGITUDEMERCURY [o]"
            .AxisTitle.Font.ColorIndex = 1
            .AxisTitle.Font.Size = 20
            .AxisTitle.Font.Bold = msoTrue
            .AxisTitle.Characters(Start:=10, Length:=7).Font.Subscript = True
            .AxisTitle.Characters(Start:=10, Length:=7).Font.Size = 24
            .AxisTitle.Characters(Start:=38, Length:=1).Font.Superscript = True
            .AxisTitle.Left = 155
        End With
    Worksheets("Results").ChartObjects("Chart2").Activate
        With ActiveChart.Axes(xlCategory)
            .HasTitle = True
            .AxisTitle.Characters.Text = "LONGITUDEMERCURY [o]"
            .AxisTitle.Font.ColorIndex = 1
            .AxisTitle.Font.Size = 20
            .AxisTitle.Font.Bold = msoTrue
            .AxisTitle.Characters(Start:=10, Length:=7).Font.Subscript = True
            .AxisTitle.Characters(Start:=10, Length:=7).Font.Size = 24
            .AxisTitle.Characters(Start:=38, Length:=1).Font.Superscript = True
            .AxisTitle.Left = 155
        End With
    Worksheets("Results").ChartObjects("Chart3").Activate
        With ActiveChart.Axes(xlCategory)
            .HasTitle = True
            .AxisTitle.Characters.Text = "LONGITUDEMERCURY [o]"
            .AxisTitle.Font.ColorIndex = 1
            .AxisTitle.Font.Size = 20
            .AxisTitle.Font.Bold = msoTrue
            .AxisTitle.Characters(Start:=10, Length:=7).Font.Subscript = True
            .AxisTitle.Characters(Start:=10, Length:=7).Font.Size = 24
            .AxisTitle.Characters(Start:=38, Length:=1).Font.Superscript = True
            .AxisTitle.Left = 155
        End With
ElseIf Worksheets("Results").Shapes("Option Button 2").OLEFormat.Object.Value = 1 Then
    Worksheets("Results").ChartObjects("Chart1").Activate
        With ActiveChart.Axes(xlCategory)
            .HasTitle = True
            .AxisTitle.Characters.Text = "LONGITUDEVENUS [o]"
            .AxisTitle.Font.ColorIndex = 1
            .AxisTitle.Font.Size = 20
            .AxisTitle.Font.Bold = msoTrue
            .AxisTitle.Characters(Start:=10, Length:=7).Font.Subscript = True
```

```
            .AxisTitle.Characters(Start:=10, Length:=5).Font.Size = 24
            .AxisTitle.Characters(Start:=36, Length:=1).Font.Superscript = True
            .AxisTitle.Left = 180
        End With
    Worksheets("Results").ChartObjects("Chart2").Activate
        With ActiveChart.Axes(xlCategory)
            .HasTitle = True
            .AxisTitle.Characters.Text = "LONGITUDEVENUS [o]"
            .AxisTitle.Font.ColorIndex = 1
            .AxisTitle.Font.Size = 20
            .AxisTitle.Font.Bold = msoTrue
            .AxisTitle.Characters(Start:=10, Length:=7).Font.Subscript = True
            .AxisTitle.Characters(Start:=10, Length:=5).Font.Size = 24
            .AxisTitle.Characters(Start:=36, Length:=1).Font.Superscript = True
            .AxisTitle.Left = 180
        End With
    Worksheets("Results").ChartObjects("Chart3").Activate
        With ActiveChart.Axes(xlCategory)
            .HasTitle = True
            .AxisTitle.Characters.Text = "LONGITUDEVENUS [o]"
            .AxisTitle.Font.ColorIndex = 1
            .AxisTitle.Font.Size = 20
            .AxisTitle.Font.Bold = msoTrue
            .AxisTitle.Characters(Start:=10, Length:=7).Font.Subscript = True
            .AxisTitle.Characters(Start:=10, Length:=5).Font.Size = 24
            .AxisTitle.Characters(Start:=36, Length:=1).Font.Superscript = True
            .AxisTitle.Left = 180
        End With
ElseIf Worksheets("Results").Shapes("Option Button 3").OLEFormat.Object.Value = 1 Then
    Worksheets("Results").ChartObjects("Chart1").Activate
        With ActiveChart.Axes(xlCategory)
            .HasTitle = True
            .AxisTitle.Characters.Text = "LONGITUDEEARTH [o]"
            .AxisTitle.Font.ColorIndex = 1
            .AxisTitle.Font.Size = 20
            .AxisTitle.Font.Bold = msoTrue
            .AxisTitle.Characters(Start:=10, Length:=7).Font.Subscript = True
            .AxisTitle.Characters(Start:=10, Length:=5).Font.Size = 24
            .AxisTitle.Characters(Start:=36, Length:=1).Font.Superscript = True
            .AxisTitle.Left = 180
        End With
    Worksheets("Results").ChartObjects("Chart2").Activate
        With ActiveChart.Axes(xlCategory)
            .HasTitle = True
            .AxisTitle.Characters.Text = "LONGITUDEEARTH [o]"
            .AxisTitle.Font.ColorIndex = 1
            .AxisTitle.Font.Size = 20
            .AxisTitle.Font.Bold = msoTrue
            .AxisTitle.Characters(Start:=10, Length:=7).Font.Subscript = True
            .AxisTitle.Characters(Start:=10, Length:=5).Font.Size = 24
            .AxisTitle.Characters(Start:=36, Length:=1).Font.Superscript = True
            .AxisTitle.Left = 180
        End With
    Worksheets("Results").ChartObjects("Chart3").Activate
        With ActiveChart.Axes(xlCategory)
            .HasTitle = True
            .AxisTitle.Characters.Text = "LONGITUDEEARTH [o]"
```

```
            .AxisTitle.Font.ColorIndex = 1
            .AxisTitle.Font.Size = 20
            .AxisTitle.Font.Bold = msoTrue
            .AxisTitle.Characters(Start:=10, Length:=7).Font.Subscript = True
            .AxisTitle.Characters(Start:=10, Length:=5).Font.Size = 24
            .AxisTitle.Characters(Start:=36, Length:=1).Font.Superscript = True
            .AxisTitle.Left = 180
        End With
ElseIf Worksheets("Results").Shapes("Option Button 4").OLEFormat.Object.Value = 1 Then
    Worksheets("Results").ChartObjects("Chart1").Activate
        With ActiveChart.Axes(xlCategory)
            .HasTitle = True
            .AxisTitle.Characters.Text = "LONGITUDEMARS [o]"
            .AxisTitle.Font.ColorIndex = 1
            .AxisTitle.Font.Size = 20
            .AxisTitle.Font.Bold = msoTrue
            .AxisTitle.Characters(Start:=10, Length:=7).Font.Subscript = True
            .AxisTitle.Characters(Start:=10, Length:=4).Font.Size = 24
            .AxisTitle.Characters(Start:=35, Length:=1).Font.Superscript = True
            .AxisTitle.Left = 186
        End With
    Worksheets("Results").ChartObjects("Chart2").Activate
        With ActiveChart.Axes(xlCategory)
            .HasTitle = True
            .AxisTitle.Characters.Text = "LONGITUDEMARS [o]"
            .AxisTitle.Font.ColorIndex = 1
            .AxisTitle.Font.Size = 20
            .AxisTitle.Font.Bold = msoTrue
            .AxisTitle.Characters(Start:=10, Length:=7).Font.Subscript = True
            .AxisTitle.Characters(Start:=10, Length:=4).Font.Size = 24
            .AxisTitle.Characters(Start:=35, Length:=1).Font.Superscript = True
            .AxisTitle.Left = 186
        End With
    Worksheets("Results").ChartObjects("Chart3").Activate
        With ActiveChart.Axes(xlCategory)
            .HasTitle = True
            .AxisTitle.Characters.Text = "LONGITUDEMARS [o]"
            .AxisTitle.Font.ColorIndex = 1
            .AxisTitle.Font.Size = 20
            .AxisTitle.Font.Bold = msoTrue
            .AxisTitle.Characters(Start:=10, Length:=7).Font.Subscript = True
            .AxisTitle.Characters(Start:=10, Length:=4).Font.Size = 24
            .AxisTitle.Characters(Start:=35, Length:=1).Font.Superscript = True
            .AxisTitle.Left = 186
        End With
ElseIf Worksheets("Results").Shapes("Option Button 5").OLEFormat.Object.Value = 1 Then
    Worksheets("Results").ChartObjects("Chart1").Activate
        With ActiveChart.Axes(xlCategory)
            .HasTitle = True
            .AxisTitle.Characters.Text = "LONGITUDEJUPITER [o]"
            .AxisTitle.Font.ColorIndex = 1
            .AxisTitle.Font.Size = 20
            .AxisTitle.Font.Bold = msoTrue
            .AxisTitle.Characters(Start:=10, Length:=7).Font.Subscript = True
            .AxisTitle.Characters(Start:=10, Length:=7).Font.Size = 24
            .AxisTitle.Characters(Start:=38, Length:=1).Font.Superscript = True
            .AxisTitle.Left = 168
```

```
        End With
    Worksheets("Results").ChartObjects("Chart2").Activate
        With ActiveChart.Axes(xlCategory)
            .HasTitle = True
            .AxisTitle.Characters.Text = "LONGITUDEJUPITER [o]"
            .AxisTitle.Font.ColorIndex = 1
            .AxisTitle.Font.Size = 20
            .AxisTitle.Font.Bold = msoTrue
            .AxisTitle.Characters(Start:=10, Length:=7).Font.Subscript = True
            .AxisTitle.Characters(Start:=10, Length:=7).Font.Size = 24
            .AxisTitle.Characters(Start:=38, Length:=1).Font.Superscript = True
            .AxisTitle.Left = 168
        End With
    Worksheets("Results").ChartObjects("Chart3").Activate
        With ActiveChart.Axes(xlCategory)
            .HasTitle = True
            .AxisTitle.Characters.Text = "LONGITUDEJUPITER [o]"
            .AxisTitle.Font.ColorIndex = 1
            .AxisTitle.Font.Size = 20
            .AxisTitle.Font.Bold = msoTrue
            .AxisTitle.Characters(Start:=10, Length:=7).Font.Subscript = True
            .AxisTitle.Characters(Start:=10, Length:=7).Font.Size = 24
            .AxisTitle.Characters(Start:=37, Length:=1).Font.Superscript = True
            .AxisTitle.Left = 168
        End With
ElseIf Worksheets("Results").Shapes("Option Button 6").OLEFormat.Object.Value = 1 Then
    Worksheets("Results").ChartObjects("Chart1").Activate
        With ActiveChart.Axes(xlCategory)
            .HasTitle = True
            .AxisTitle.Characters.Text = "LONGITUDESATURN [o]"
            .AxisTitle.Font.ColorIndex = 1
            .AxisTitle.Font.Size = 20
            .AxisTitle.Font.Bold = msoTrue
            .AxisTitle.Characters(Start:=10, Length:=7).Font.Subscript = True
            .AxisTitle.Characters(Start:=10, Length:=6).Font.Size = 24
            .AxisTitle.Characters(Start:=37, Length:=1).Font.Superscript = True
            .AxisTitle.Left = 168
        End With
    Worksheets("Results").ChartObjects("Chart2").Activate
        With ActiveChart.Axes(xlCategory)
            .HasTitle = True
            .AxisTitle.Characters.Text = "LONGITUDESATURN [o]"
            .AxisTitle.Font.ColorIndex = 1
            .AxisTitle.Font.Size = 20
            .AxisTitle.Font.Bold = msoTrue
            .AxisTitle.Characters(Start:=10, Length:=7).Font.Subscript = True
            .AxisTitle.Characters(Start:=10, Length:=6).Font.Size = 24
            .AxisTitle.Characters(Start:=37, Length:=1).Font.Superscript = True
            .AxisTitle.Left = 168
        End With
    Worksheets("Results").ChartObjects("Chart3").Activate
        With ActiveChart.Axes(xlCategory)
            .HasTitle = True
            .AxisTitle.Characters.Text = "LONGITUDESATURN [o]"
            .AxisTitle.Font.ColorIndex = 1
            .AxisTitle.Font.Size = 20
            .AxisTitle.Font.Bold = msoTrue
```

```
            .AxisTitle.Characters(Start:=10, Length:=7).Font.Subscript = True
            .AxisTitle.Characters(Start:=10, Length:=6).Font.Size = 24
            .AxisTitle.Characters(Start:=37, Length:=1).Font.Superscript = True
            .AxisTitle.Left = 168
        End With
ElseIf Worksheets("Results").Shapes("Option Button 7").OLEFormat.Object.Value = 1 Then
    Worksheets("Results").ChartObjects("Chart1").Activate
        With ActiveChart.Axes(xlCategory)
            .HasTitle = True
            .AxisTitle.Characters.Text = "MOON PHASE [o]"
            .AxisTitle.Font.ColorIndex = 1
            .AxisTitle.Font.Size = 20
            .AxisTitle.Font.Bold = msoTrue
            .AxisTitle.Characters(Start:=32, Length:=1).Font.Superscript = True
            .AxisTitle.Left = 204
        End With
    Worksheets("Results").ChartObjects("Chart2").Activate
        With ActiveChart.Axes(xlCategory)
            .HasTitle = True
            .AxisTitle.Characters.Text = "MOON PHASE [o]"
            .AxisTitle.Font.ColorIndex = 1
            .AxisTitle.Font.Size = 20
            .AxisTitle.Font.Bold = msoTrue
            .AxisTitle.Characters(Start:=32, Length:=1).Font.Superscript = True
            .AxisTitle.Left = 204
        End With
    Worksheets("Results").ChartObjects("Chart3").Activate
        With ActiveChart.Axes(xlCategory)
            .HasTitle = True
            .AxisTitle.Characters.Text = "MOON PHASE [o]"
            .AxisTitle.Font.ColorIndex = 1
            .AxisTitle.Font.Size = 20
            .AxisTitle.Font.Bold = msoTrue
            .AxisTitle.Characters(Start:=32, Length:=1).Font.Superscript = True
            .AxisTitle.Left = 204
        End With
End If

MinutesElapsed = Format((Timer - StartTime) / 86400, "hh:mm:ss") 'Determine how many seconds code took to run
MsgBox "The compilation has been successfully completed in " & MinutesElapsed & " [hh:mm:ss]" & vbNewLine &
     vbNewLine & "Yiii Haaaa!!!", vbInformation 'Informs the user

End Sub
```

## D.2 Optimal case algorithm

The "optimal case algorithm" was created in order to be able to find biggest min-max effects (peaks) around a specific longitude range for a specific planetary configuration. The algorithm takes as an input the step (in degrees) and the opening angle (in degrees) of a planet that is constrained and goes through all the cases one by one and finds the best case. There are also a couple of variations of this script in order to find the biggest sigma of a peak (including or not the maximum value) compared to the overall mean value of the result, instead of the biggest min-max effect. Of course in order for these variations to work an error bar in each value has to be provided in the original dataset.

Specifically, when each "optimal case" script is ran (see Fig. D.2), all the data from the "Data" worksheet are parsed, then all the data fulfilling the user-defined requirements (dates, data-range, step, opening read from "Best_Results" worksheet) are printed in worksheet "All_Results" (see Fig. D.2a). As next, the corresponding projections from "All_Results" worksheet are created in "All_Projections" worksheet (see Fig. D.2b) according to the provided bin size. Then, all the individual effects which are inside the provided longitude range set by the user (as read from "Best_Results" worksheet) are calculated according to the min and max values and are printed in "All_Effects" worksheet (see Fig. D.2c). Finally, the biggest effect (from "All_Effects" worksheet) is printed and plotted on the "Best_Results" worksheet (see Fig. D.2d), while the user is being notified if the algorithm has finished successfully (see Fig. D.3).

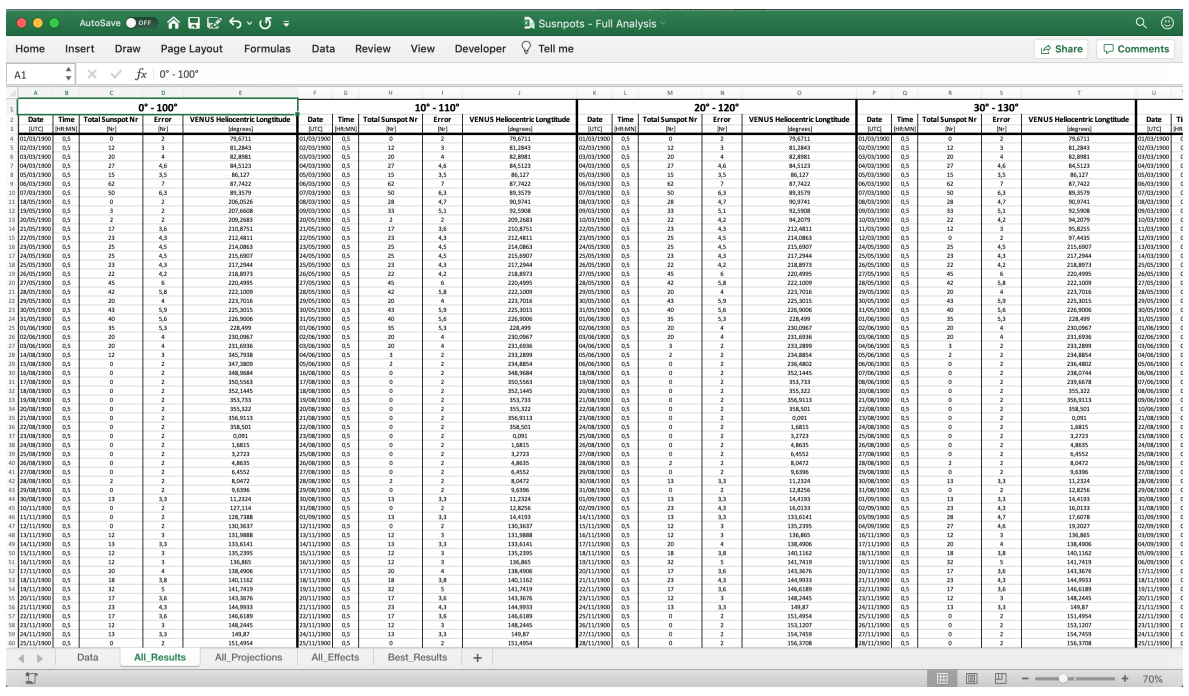

**(a)** "All_Results" worksheet.

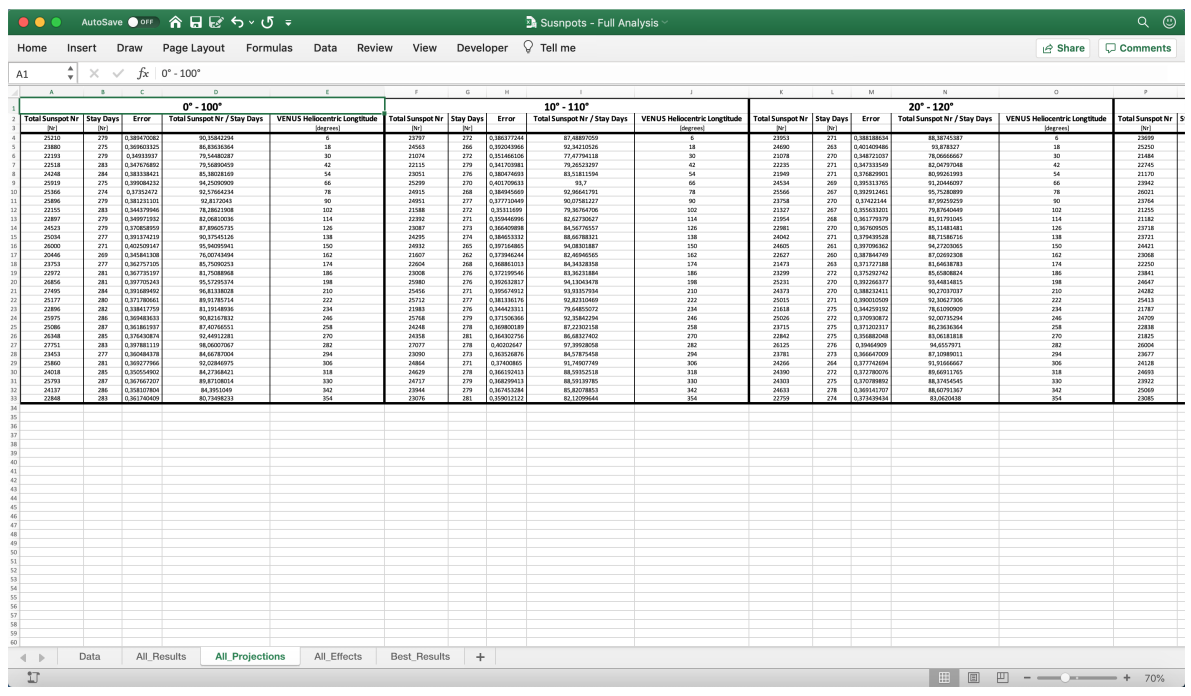

**(b)** "All_Projections" worksheet.

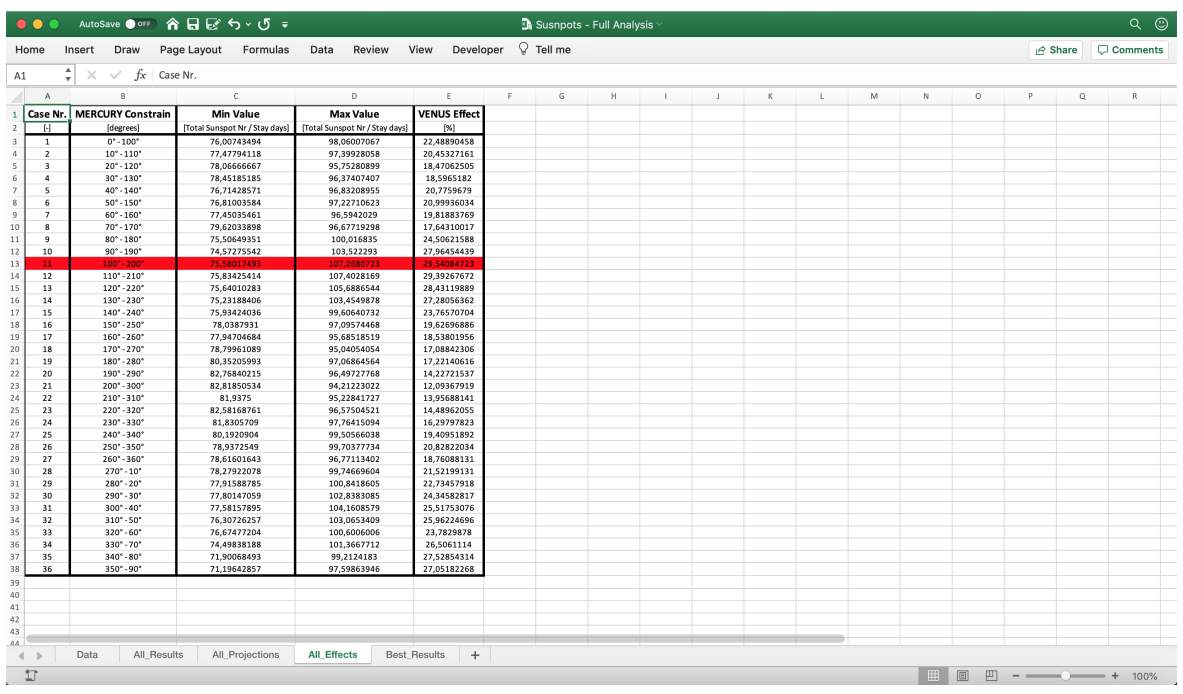

**(c)** "All_Effects" worksheet.

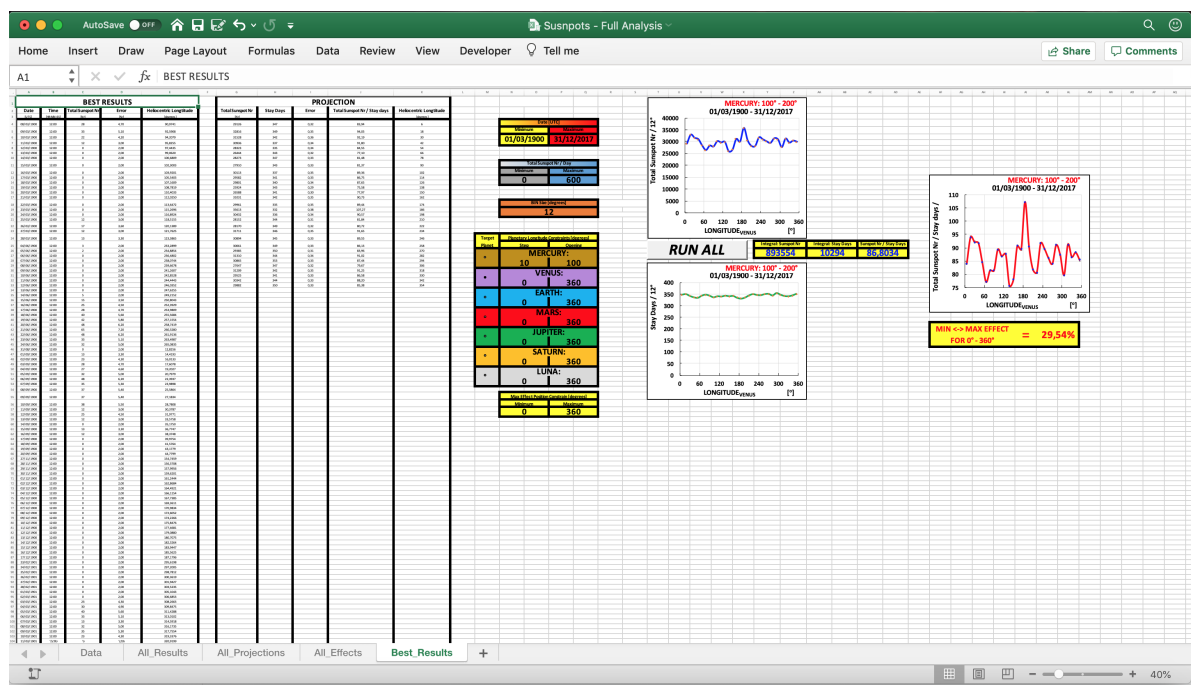

**(d)** "Best_Results" worksheet.

**Figure D.2:** Example of the Excel Spreadsheet corresponding to the "optimal case algorithm" for the sunspots dataset.

```
Sub Full_Analysis_Script()
'Author: Marios R. Maroudas

'We start running all the cases for our planets of choice (Only one planet can be constrained at a time).

```

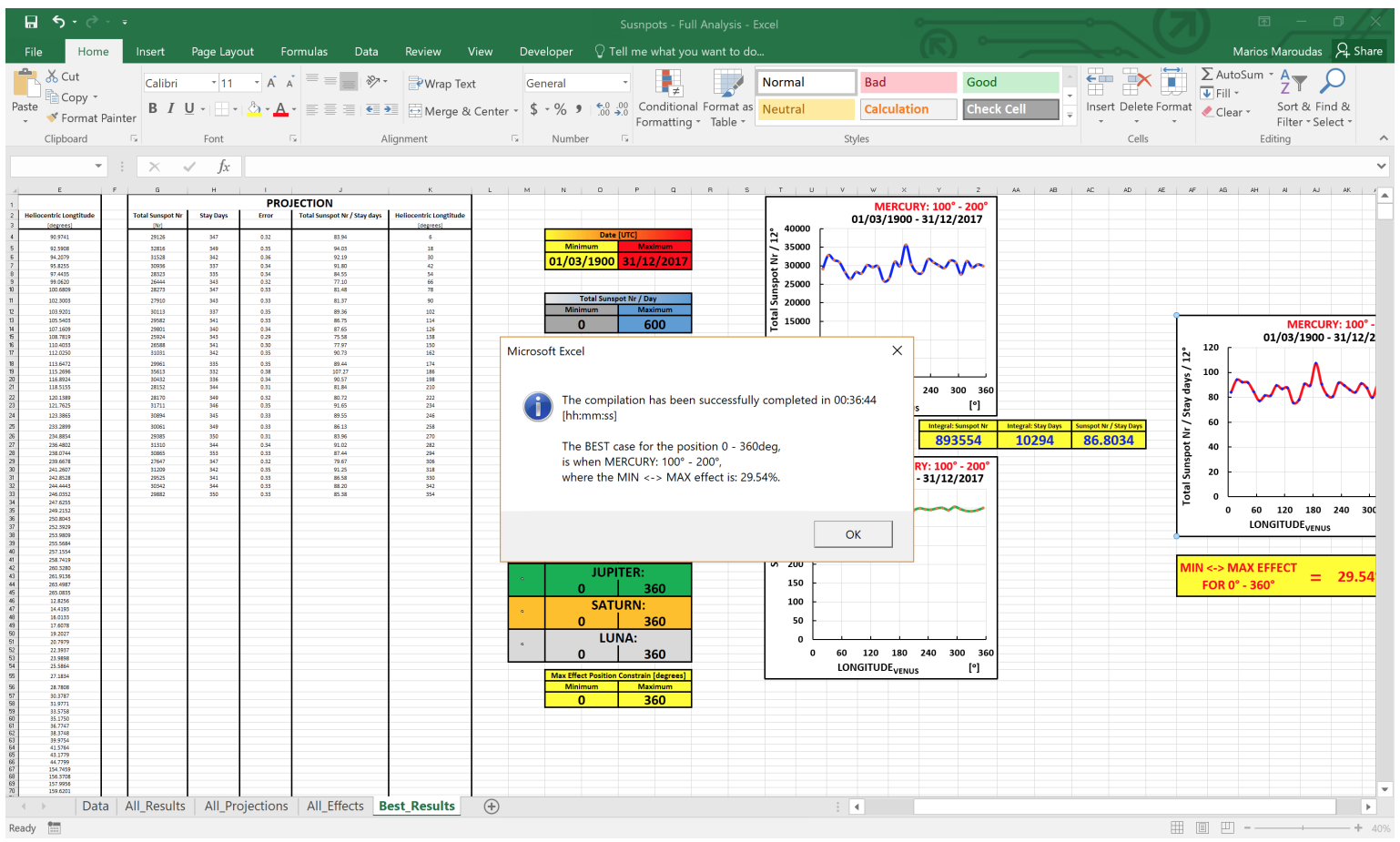

**Figure D.3:** Message printed to the user at the end of the run of “optimal case algorithm” informing him about the results and the data-taking time.

```
Dim StartTime As Double
Dim MinutesElapsed As String

StartTime = Timer 'Remember the time when our macro starts
FinalRow_Data = Worksheets("Data").Cells(Rows.Count, 1).End(xlUp).Row 'Find the last row of data from the first column.
Step = 0
Opening = 0

'Find the "step" and the "opening angle" that is to be used:
If Worksheets("Best_Results").Cells(28, 14).Value <> 0 Then
    Step = Worksheets("Best_Results").Cells(28, 14).Value
    Opening = Worksheets("Best_Results").Cells(28, 16).Value
    Constained_Planet = 5
    Worksheets("Best_Results").Cells(2, 16).Value = "MERCURY: "
ElseIf Worksheets("Best_Results").Cells(32, 14).Value <> 0 Then
    Step = Worksheets("Best_Results").Cells(32, 14).Value
    Opening = Worksheets("Best_Results").Cells(32, 16).Value
    Constained_Planet = 6
    Worksheets("Best_Results").Cells(2, 16).Value = "VENUS: "
ElseIf Worksheets("Best_Results").Cells(36, 14).Value <> 0 Then
    Step = Worksheets("Best_Results").Cells(36, 14).Value
    Opening = Worksheets("Best_Results").Cells(36, 16).Value
    Constained_Planet = 7
    Worksheets("Best_Results").Cells(2, 16).Value = "EARTH: "
ElseIf Worksheets("Best_Results").Cells(40, 14).Value <> 0 Then
    Step = Worksheets("Best_Results").Cells(40, 14).Value
    Opening = Worksheets("Best_Results").Cells(40, 16).Value
    Constained_Planet = 8
    Worksheets("Best_Results").Cells(2, 16).Value = "MARS: "
ElseIf Worksheets("Best_Results").Cells(44, 14).Value <> 0 Then
    Step = Worksheets("Best_Results").Cells(44, 14).Value
    Opening = Worksheets("Best_Results").Cells(44, 16).Value
    Constained_Planet = 9
    Worksheets("Best_Results").Cells(2, 16).Value = "JUPITER: "
ElseIf Worksheets("Best_Results").Cells(48, 14).Value <> 0 Then
    Step = Worksheets("Best_Results").Cells(48, 14).Value
    Opening = Worksheets("Best_Results").Cells(48, 16).Value
```

```
    Constained_Planet = 10
    Worksheets("Best_Results").Cells(2, 16).Value = "SATURN: "
ElseIf Worksheets("Best_Results").Cells(52, 14).Value <> 0 Then
    Step = Worksheets("Best_Results").Cells(52, 14).Value
    Opening = Worksheets("Best_Results").Cells(52, 16).Value
    Constained_Planet = 14
    Worksheets("Best_Results").Cells(2, 16).Value = "LUNA: "
End If

'Find the planet that is to be selected as a reference and its longitude will be printed:
If Worksheets("Best_Results").Shapes("Option Button 1").OLEFormat.Object.Value = 1 Then
    Target_Planet = 5
ElseIf Worksheets("Best_Results").Shapes("Option Button 2").OLEFormat.Object.Value = 1 Then
    Target_Planet = 6
ElseIf Worksheets("Best_Results").Shapes("Option Button 3").OLEFormat.Object.Value = 1 Then
    Target_Planet = 7
ElseIf Worksheets("Best_Results").Shapes("Option Button 4").OLEFormat.Object.Value = 1 Then
    Target_Planet = 8
ElseIf Worksheets("Best_Results").Shapes("Option Button 5").OLEFormat.Object.Value = 1 Then
    Target_Planet = 9
ElseIf Worksheets("Best_Results").Shapes("Option Button 6").OLEFormat.Object.Value = 1 Then
    Target_Planet = 10
ElseIf Worksheets("Best_Results").Shapes("Option Button 7").OLEFormat.Object.Value = 1 Then
    Target_Planet = 14
End If

'Message when wrong dates are given:
If Worksheets("Best_Results").Cells(6, 14).Value < Worksheets("Data").Cells(4, 1).Value Or Worksheets("Best_Results").Cells(6, 16).Value > Worksheets("Data").Cells(FinalRow_Data, 1).Value Or Worksheets("Best_Results").Cells(6, 14).Value >= Worksheets("Best_Results").Cells(6, 16).Value Then
    MsgBox "The DATES you have provided are wrong!" & vbNewLine & vbNewLine & "Please give correct dates.", vbCritical
    Exit Sub
End If

'Message when wrong data-range is given:
If Worksheets("Best_Results").Cells(13, 14).Value >= Worksheets("Best_Results").Cells(13, 16).Value Or Worksheets("Best_Results").Cells(13, 14).Value < 0 Or Worksheets("Best_Results").Cells(13, 16).Value < 0 Then
    MsgBox "The DATA range you have provided is wrong!" & vbNewLine & vbNewLine & "Please give correct minimum and maximum numbers.", vbCritical
    Exit Sub
End If

'Message when wrong BIN size is given:
If Worksheets("Best_Results").Cells(19, 14).Value <= 0 Or Worksheets("Best_Results").Cells(19, 14).Value > 360 Then
    MsgBox "The BIN size you have provided is wrong!" & vbNewLine & vbNewLine & "Please give a correct BIN size number.", vbCritical
    Exit Sub
End If

'Message when no planet has been selected to be constrained:
If Step = 0 And Opening = 0 Then
    MsgBox "You haven't chosen a planet to be constrained!" & vbNewLine & vbNewLine & "Please give different step and opening values.", vbCritical
    Exit Sub
```

```
End If

'Message for wrong "step" number:
If Step <= 0 Or Step >= 360 Then
    MsgBox "The Step you have provided is wrong!" & vbNewLine & vbNewLine & "Please give a correct number.",
        vbCritical
    Exit Sub
End If

'Message for wrong "opening" number:
If Opening <= 0 Or Opening >= 360 Then
    MsgBox "The Opening Angle you have provided is wrong!" & vbNewLine & vbNewLine & "Please give a correct
        number.", vbCritical
    Exit Sub
End If

'Message when "TargetPlanet" = "Constrained planet":
If Target_Planet = Constained_Planet Then
    MsgBox "The Target Planet is the same as the Constrained Planet!" & vbNewLine & vbNewLine & "Please give
        different values.", vbCritical
    Exit Sub
End If

'Message for wrong max effect position:
If Worksheets("Best_Results").Cells(57, 14).Value < 0 Or Worksheets("Best_Results").Cells(57, 14).Value > 360
    Or Worksheets("Best_Results").Cells(57, 14).Value >= Worksheets("Best_Results").Cells(57, 16).Value Or
    Worksheets("Best_Results").Cells(57, 16).Value < 0 Or Worksheets("Best_Results").Cells(57, 16).Value > 360
    Then
    MsgBox "The Max Effect Position you have provided is wrong!" & vbNewLine & vbNewLine & "Please give
        different minimum and maximum values.", vbCritical
    Exit Sub
End If

FinalColumn_Results = Worksheets("All_Results").Cells(4, Columns.Count).End(xlToLeft).Column
FinalColumn_Projections = Worksheets("All_Projections").Cells(4, Columns.Count).End(xlToLeft).Column
FinalRow_Projections = Worksheets("All_Projections").Cells(Rows.Count, FinalColumn_Projections).End(xlUp).Row
FinalRow_Effects = Worksheets("All_Effects").Cells(Rows.Count, 1).End(xlUp).Row
Worksheets("All_Results").Range(Worksheets("All_Results").Cells(1, 1),
    Worksheets("All_Results").Cells(FinalRow_Data, FinalColumn_Results)).ClearContents 'Delete the previous
    results from previous runs.
Worksheets("All_Results").Range(Worksheets("All_Results").Cells(1, 1),
    Worksheets("All_Results").Cells(FinalRow_Data, FinalColumn_Results)).ClearFormats 'Delete the previous
    formats from previous runs.
Worksheets("All_Projections").Range(Worksheets("All_Projections").Cells(1, 1),
    Worksheets("All_Projections").Cells(FinalRow_Projections, FinalColumn_Projections)).ClearContents
Worksheets("All_Projections").Range(Worksheets("All_Projections").Cells(1, 1),
    Worksheets("All_Projections").Cells(FinalRow_Projections, FinalColumn_Projections)).ClearFormats
If FinalRow_Effects >= 3 Then
    Worksheets("All_Effects").Range(Worksheets("All_Effects").Cells(3, 1),
        Worksheets("All_Effects").Cells(FinalRow_Effects, 5)).ClearContents
    Worksheets("All_Effects").Range(Worksheets("All_Effects").Cells(3, 1),
        Worksheets("All_Effects").Cells(FinalRow_Effects, 5)).ClearFormats
End If
Worksheets("Best_Results").Range(Worksheets("Best_Results").Cells(4, 1),
    Worksheets("Best_Results").Cells(FinalRow_Data, 11)).ClearContents
r = 4
Row = 4
```

```
Column = 1
Start_Value = 0
Stop_Value = Opening

For Start_Value = 0 To 359
    For r = 4 To FinalRow_Data
        If Stop_Value <= 360 Then
            Case_Stop = Worksheets("Data").Cells(r, Constained_Planet).Value >= Start_Value And
                 Worksheets("Data").Cells(r, Constained_Planet).Value <= Stop_Value
        Else
            Case_Stop = (Worksheets("Data").Cells(r, Constained_Planet).Value >= Start_Value And
                 Worksheets("Data").Cells(r, Constained_Planet).Value <= 360) Or (Worksheets("Data").Cells(r,
                 Constained_Planet).Value >= 0 And Worksheets("Data").Cells(r, Constained_Planet).Value <=
                 Stop_Value - 360)
        End If
        If Worksheets("Data").Cells(r, 1).Value >= Worksheets("Best_Results").Cells(6, 14).Value And
             Worksheets("Data").Cells(r, 1).Value <= Worksheets("Best_Results").Cells(6, 16).Value Then
             'Condition for dates.
            If Worksheets("Data").Cells(r, 3).Value >= Worksheets("Best_Results").Cells(13, 14).Value And
                 Worksheets("Data").Cells(r, 3).Value <= Worksheets("Best_Results").Cells(13, 16).Value Then
                 'Condition for each data range.
                If Case_Stop Then ' Condition for the data to be between "Start_Value" and "Stop_Value".
                    If Stop_Value <= 360 Then
                        Worksheets("All_Results").Cells(1, Column).Value = Start_Value & ChrW(176) & " - " &
                             Stop_Value & ChrW(176)
                    Else
                        Worksheets("All_Results").Cells(1, Column).Value = Start_Value & ChrW(176) & " - " &
                             Stop_Value - 360 & ChrW(176)
                    End If
                    Worksheets("All_Results").Cells(1, Column).Font.Size = 20
                    Worksheets("All_Results").Cells(1, Column).Font.Bold = True
                    Worksheets("All_Results").Range(Worksheets("All_Results").Cells(1, Column),
                         Worksheets("All_Results").Cells(1, Column + 4)).MergeCells = True
                    Worksheets("All_Results").Cells(2, Column).Value = Worksheets("Data").Cells(2, 1).Value
                    Worksheets("All_Results").Cells(2, Column).Font.Bold = True
                    Worksheets("All_Results").Cells(2, Column).Font.Size = 14
                    Worksheets("All_Results").Cells(3, Column).Value = Worksheets("Data").Cells(3, 1).Value
                    Worksheets("All_Results").Cells(2, Column + 1).Value = Worksheets("Data").Cells(2, 2).Value
                    Worksheets("All_Results").Cells(2, Column + 1).Font.Bold = True
                    Worksheets("All_Results").Cells(2, Column + 1).Font.Size = 14
                    Worksheets("All_Results").Cells(3, Column + 1).Value = Worksheets("Data").Cells(3, 2).Value
                    Worksheets("All_Results").Cells(2, Column + 2).Value = Worksheets("Data").Cells(2, 3).Value
                    Worksheets("All_Results").Cells(2, Column + 2).Font.Bold = True
                    Worksheets("All_Results").Cells(2, Column + 2).Font.Size = 14
                    Worksheets("All_Results").Cells(3, Column + 2).Value = Worksheets("Data").Cells(3, 3).Value
                    Worksheets("All_Results").Cells(2, Column + 3).Value = Worksheets("Data").Cells(2, 4).Value
                    Worksheets("All_Results").Cells(2, Column + 3).Font.Bold = True
                    Worksheets("All_Results").Cells(2, Column + 3).Font.Size = 14
                    Worksheets("All_Results").Cells(3, Column + 3).Value = Worksheets("Data").Cells(3, 4).Value
                    Worksheets("All_Results").Cells(2, Column + 4).Value = Worksheets("Data").Cells(1,
                         Target_Planet).Value & " " & Worksheets("Data").Cells(2, Target_Planet).Value
                    Worksheets("All_Results").Cells(2, Column + 4).Font.Bold = True
                    Worksheets("All_Results").Cells(2, Column + 4).Font.Size = 14
                    Worksheets("All_Results").Cells(3, Column + 4).Value = Worksheets("Data").Cells(3,
                         Target_Planet).Value
                    Worksheets("All_Results").Range(Worksheets("All_Results").Cells(1, Column),
                         Worksheets("All_Results").Cells(3, Column + 4)).BorderAround Weight:=xlThick
```

```
                    Worksheets("All_Results").Cells(Row, Column).Value = Worksheets("Data").Cells(r, 1).Value
                    Worksheets("All_Results").Cells(Row, Column).Font.Color = Worksheets("Data").Cells(r,
                        1).Font.Color
                    Worksheets("All_Results").Cells(Row, Column + 1).Value = Worksheets("Data").Cells(r, 2).Value
                    Worksheets("All_Results").Cells(Row, Column + 1).Font.Color = Worksheets("Data").Cells(r,
                        2).Font.Color
                    Worksheets("All_Results").Cells(Row, Column + 2).Value = Worksheets("Data").Cells(r, 3).Value
                    Worksheets("All_Results").Cells(Row, Column + 2).Font.Color = Worksheets("Data").Cells(r,
                        3).Font.Color
                    Worksheets("All_Results").Cells(Row, Column + 3).Value = Worksheets("Data").Cells(r, 4).Value
                    Worksheets("All_Results").Cells(Row, Column + 3).Font.Color = Worksheets("Data").Cells(r,
                        4).Font.Color
                    Worksheets("All_Results").Cells(Row, Column + 4).Value = Worksheets("Data").Cells(r,
                        Target_Planet).Value
                    Worksheets("All_Results").Cells(Row, Column + 4).Font.Color = Worksheets("Data").Cells(r,
                        Target_Planet).Font.Color
                    Row = Row + 1
                End If
            End If
        ElseIf Worksheets("Data").Cells(r, 3).Value > Worksheets("Best_Results").Cells(13, 16).Value Then
            GoTo EndLoop_Results
        End If
    Next r
EndLoop_Results:
If Row <> 4 Then
    Worksheets("All_Results").Range(Worksheets("All_Results").Cells(4, Column),
        Worksheets("All_Results").Cells(Row - 1, Column + 4)).Borders.Weight = xlThick
    Worksheets("All_Results").Range(Worksheets("All_Results").Cells(4, Column),
        Worksheets("All_Results").Cells(Row - 1, Column + 4)).Borders(xlInsideHorizontal).LineStyle = xlNone
    Worksheets("All_Results").Range(Worksheets("All_Results").Cells(2, Column),
        Worksheets("All_Results").Cells(Row - 1, Column + 4)).Borders(xlInsideVertical).Weight = xlThin
End If
r = 4
Row = 4
Column = Column + 5
Start_Value = (Start_Value + Step - 1)
Stop_Value = (Stop_Value + Step)
Next Start_Value

Worksheets("All_Results").Columns(1).Resize(, Column).EntireColumn.HorizontalAlignment = xlCenter
Worksheets("All_Results").Columns(1).Resize(, Column).EntireColumn.AutoFit

'Depending on the bin size we create the projections:

FinalColumn_Results = Worksheets("All_Results").Cells(4, Columns.Count).End(xlToLeft).Column
    If FinalColumn_Results = 1 Then
        MsgBox "No DATA have been found for these constraints." & vbNewLine & vbNewLine & "Please change the
            constraints and try again.", vbCritical
        Exit Sub
    End If
Row = 4
Column = 1
r = 4
Bin = 0
Data = 0
Stay_Days = 0
St_Dev = 0
```

```

For Column = 1 To FinalColumn_Results
   FinalRow_Results_N = Worksheets("All_Results").Cells(Rows.Count, Column).End(xlUp).Row
      If FinalRow_Results_N = 1 Then
         GoTo EndLoop_Projections
      End If
   Worksheets("All_Projections").Cells(1, Column).Value = Worksheets("All_Results").Cells(1, Column).Value
   Worksheets("All_Projections").Cells(1, Column).Font.Size = 20
   Worksheets("All_Projections").Range(Worksheets("All_Projections").Cells(1, Column), Worksheets("All_Projections").Cells(1, Column + 4)).MergeCells = True
   Worksheets("All_Projections").Cells(1, Column).Font.Bold = True
   Worksheets("All_Projections").Cells(2, Column).Value = Worksheets("All_Results").Cells(2, Column + 2).Value
   Worksheets("All_Projections").Cells(2, Column).Font.Bold = True
   Worksheets("All_Projections").Cells(2, Column).Font.Size = 14
   Worksheets("All_Projections").Cells(3, Column).Value = Worksheets("All_Results").Cells(3, Column + 2).Value
   Worksheets("All_Projections").Cells(2, Column + 1).Value = Worksheets("Best_Results").Cells(2, 8).Value
   Worksheets("All_Projections").Cells(2, Column + 1).Font.Bold = True
   Worksheets("All_Projections").Cells(2, Column + 1).Font.Size = 14
   Worksheets("All_Projections").Cells(3, Column + 1).Value = "[Nr]"
   Worksheets("All_Projections").Cells(2, Column + 2).Value = Worksheets("All_Results").Cells(2, Column + 3).Value
   Worksheets("All_Projections").Cells(2, Column + 2).Font.Bold = True
   Worksheets("All_Projections").Cells(2, Column + 2).Font.Size = 14
   Worksheets("All_Projections").Cells(2, Column + 3).Value = Worksheets("All_Results").Cells(2, Column + 2).Value & " / " & Worksheets("Best_Results").Cells(2, 8).Value
   Worksheets("All_Projections").Cells(2, Column + 3).Font.Bold = True
   Worksheets("All_Projections").Cells(2, Column + 3).Font.Size = 14
   Worksheets("All_Projections").Cells(2, Column + 4).Value = Worksheets("All_Results").Cells(2, Column + 4).Value
   Worksheets("All_Projections").Cells(2, Column + 4).Font.Bold = True
   Worksheets("All_Projections").Cells(2, Column + 4).Font.Size = 14
   Worksheets("All_Projections").Cells(3, Column + 4).Value = Worksheets("All_Results").Cells(3, Column + 4).Value
   Worksheets("All_Projections").Range(Worksheets("All_Projections").Cells(1, Column), Worksheets("All_Projections").Cells(3, Column + 4)).BorderAround Weight:=xlThick
   For Bin = 0 To 359
      For r = 4 To FinalRow_Results_N
         If Worksheets("All_Results").Cells(r, Column + 4).Value >= Bin And Worksheets("All_Results").Cells(r, Column + 4).Value < (Bin + Worksheets("Best_Results").Cells(19, 14).Value) And Not IsEmpty(Worksheets("All_Results").Cells(r, Column + 4).Value) Then
            Data = Data + Worksheets("All_Results").Cells(r, Column + 2).Value 'Calculates the relevant data (i.e. EUV) for the bin that has been selected.
            Stay_Days = Stay_Days + 1 'Calculates the "stay days" for all the dates that are constrained also from the amplitude of the data for the relevant bin.
            St_Dev = St_Dev + (Worksheets("All_Results").Cells(r, Column + 3).Value * Worksheets("All_Results").Cells(r, Column + 3).Value) 'Calculates the standard deviation.
         End If
      Next r
   Worksheets("All_Projections").Cells(Row, Column + 4).Value = (Bin + ((Bin + Worksheets("Best_Results").Cells(19, 14).Value))) / 2 'Here we place the data in the middle of each bin. For example if we have bin=6deg then the first number will be placed in long=3deg.

   Worksheets("All_Projections").Cells(Row, Column).Value = Data
   Worksheets("All_Projections").Cells(Row, Column + 1).Value = Stay_Days
      If Worksheets("All_Projections").Cells(Row, Column + 1).Value = 0 Then 'When the division gives zero as a result.
```

```
            Worksheets("All_Projections").Cells(Row, Column + 2).Value = 0
            Worksheets("All_Projections").Cells(Row, Column + 3).Value = 0
        Else
            Worksheets("All_Projections").Cells(Row, Column + 2).Value = Sqr(St_Dev) / Stay_Days
            Worksheets("All_Projections").Cells(Row, Column + 3).Value =
                 (Worksheets("All_Projections").Cells(Row, Column).Value) /
                 (Worksheets("All_Projections").Cells(Row, Column + 1).Value)
        End If
    Data = 0
    Stay_Days = 0
    St_Dev = 0
    Row = Row + 1
    Bin = (Bin + Worksheets("Best_Results").Cells(19, 14).Value - 1) 'With "Next b" it will move one unit, so
         this has to be reduced 1 unit.
    Next Bin
If Row <> 4 Then
    Worksheets("All_Projections").Range(Worksheets("All_Projections").Cells(4, Column),
         Worksheets("All_Projections").Cells(Row - 1, Column + 4)).Borders.Weight = xlThick
    Worksheets("All_Projections").Range(Worksheets("All_Projections").Cells(4, Column),
         Worksheets("All_Projections").Cells(Row - 1, Column + 4)).Borders(xlInsideHorizontal).LineStyle = xlNone
    Worksheets("All_Projections").Range(Worksheets("All_Projections").Cells(2, Column),
         Worksheets("All_Projections").Cells(Row - 1, Column + 4)).Borders(xlInsideVertical).Weight = xlThin
End If
Data = 0
Stay_Days = 0
St_Dev = 0
Row = 4
Bin = 0
EndLoop_Projections:
Column = Column + 4
Next Column

Worksheets("All_Projections").Columns(1).Resize(, Column).EntireColumn.HorizontalAlignment = xlCenter
Worksheets("All_Projections").Columns(1).Resize(, Column).EntireColumn.AutoFit

'We start finding all the effects for every case and identifying the biggest one:

FinalColumn_Projections = Worksheets("All_Projections").Cells(4, Columns.Count).End(xlToLeft).Column
FinalRow_Projections = Worksheets("All_Projections").Cells(Rows.Count, FinalColumn_Projections).End(xlUp).Row
Start_Position = 4
Stop_Position = FinalRow_Projections
For r = 4 To FinalRow_Projections
    If Worksheets("All_Projections").Cells(r, FinalColumn_Projections).Value >=
         Worksheets("Best_Results").Cells(57, 14).Value Then
        Start_Position = r
        Exit For
    End If
Next r
r = 4
For r = 4 To FinalRow_Projections
    If Worksheets("All_Projections").Cells(r, FinalColumn_Projections).Value >
         Worksheets("Best_Results").Cells(57, 16).Value Then
        Stop_Position = r - 1
        Exit For
    ElseIf Worksheets("All_Projections").Cells(r, FinalColumn_Projections).Value =
         Worksheets("Best_Results").Cells(57, 16).Value Then
        Stop_Position = r
```

```
        Exit For
    End If
Next r

Row = 3
Column = 4
Number = 1
Worksheets("All_Effects").Cells(1, 2).Value = Worksheets("Data").Cells(1, Constained_Planet).Value & "
    Constrain"
Worksheets("All_Effects").Cells(1, 5).Value = Worksheets("Data").Cells(1, Target_Planet).Value & " Effect"
Worksheets("All_Effects").Cells(2, 3).Value = "[" & Worksheets("Best_Results").Cells(2, 10).Value & "]"
Worksheets("All_Effects").Cells(2, 4).Value = "[" & Worksheets("Best_Results").Cells(2, 10).Value & "]"
For Column = 4 To FinalColumn_Projections
    Min_Value = 100000000 'We do not set the first value as minimum values for the case that the first value =
        zero.
    Max_Value = Worksheets("All_Projections").Cells(Start_Position, Column).Value
        For r = Start_Position To Stop_Position
            If Worksheets("All_Projections").Cells(r, Column).Value < Min_Value And
                Worksheets("All_Projections").Cells(r, Column).Value <> 0 Then
                Min_Value = Worksheets("All_Projections").Cells(r, Column).Value
            End If
            If Worksheets("All_Projections").Cells(r, Column).Value > Max_Value Then
                Max_Value = Worksheets("All_Projections").Cells(r, Column).Value
            End If
        Next r
        If Max_Value = 0 Then
            Min_Value = 0
            Max_Value = 0
            Effect = 0
        Else
            Effect = ((Max_Value - Min_Value) / Max_Value) * 100
        End If
    Worksheets("All_Effects").Cells(Row, 1).Value = Number
    Worksheets("All_Effects").Cells(Row, 2).Value = Worksheets("All_Projections").Cells(1, Column - 3).Value
    Worksheets("All_Effects").Cells(Row, 3).Value = Min_Value
    Worksheets("All_Effects").Cells(Row, 4).Value = Max_Value
    Worksheets("All_Effects").Cells(Row, 5).Value = Effect
    Row = Row + 1
    Number = Number + 1
    Column = Column + 4
Next Column

FinalRow_Effects = Worksheets("All_Effects").Cells(Rows.Count, 1).End(xlUp).Row
Worksheets("All_Effects").Range(Worksheets("All_Effects").Cells(3, 1),
    Worksheets("All_Effects").Cells(FinalRow_Effects, 5)).Borders.Weight = xlThick
Worksheets("All_Effects").Range(Worksheets("All_Effects").Cells(3, 1),
    Worksheets("All_Effects").Cells(FinalRow_Effects, 5)).Borders(xlInsideHorizontal).LineStyle = xlNone
Worksheets("All_Effects").Range(Worksheets("All_Effects").Cells(3, 1),
    Worksheets("All_Effects").Cells(FinalRow_Effects, 5)).Borders(xlInsideVertical).Weight = xlThick
Row = 3
Max_Effect = Worksheets("All_Effects").Cells(3, 5).Value
Number = 0
    For Row = 3 To FinalRow_Effects
        If Worksheets("All_Effects").Cells(Row, 5).Value >= Max_Effect Then
            Max_Effect = Worksheets("All_Effects").Cells(Row, 5).Value
            Number = Worksheets("All_Effects").Cells(Row, 1).Value
        End If
```

```
    Next Row
    Worksheets("Best_Results").Cells(41, 37).Value = Max_Effect / 100
    Worksheets("All_Effects").Cells(Number + 2, 1).Interior.ColorIndex = 3
    Worksheets("All_Effects").Cells(Number + 2, 2).Interior.ColorIndex = 3
    Worksheets("All_Effects").Cells(Number + 2, 3).Interior.ColorIndex = 3
    Worksheets("All_Effects").Cells(Number + 2, 4).Interior.ColorIndex = 3
    Worksheets("All_Effects").Cells(Number + 2, 5).Interior.ColorIndex = 3
    Worksheets("All_Effects").Columns("A:E").HorizontalAlignment = xlCenter
    Worksheets("All_Effects").Columns("A:E").EntireColumn.AutoFit
    Worksheets("Best_Results").Cells(2, 17).Value = Worksheets("All_Effects").Cells(Number + 2, 2).Value
    Column = ((Number - 1) * 5) + 1
    If Max_Effect = 0 Then
        MsgBox "There were NO effects above zero!" & vbNewLine & vbNewLine & "Please change the constraints and try again.", vbCritical
        Exit Sub
    End If

FinalRow_Results_N = Worksheets("All_results").Cells(Rows.Count, Column).End(xlUp).Row
    For Row = 4 To FinalRow_Results_N
        Worksheets("Best_Results").Cells(Row, 1).Value = Worksheets("All_Results").Cells(Row, Column).Value
        Worksheets("Best_Results").Cells(Row, 1).Font.Color = Worksheets("All_Results").Cells(Row, Column).Font.Color
        Worksheets("Best_Results").Cells(Row, 2).Value = Worksheets("All_Results").Cells(Row, Column + 1).Value
        Worksheets("Best_Results").Cells(Row, 2).Font.Color = Worksheets("All_Results").Cells(Row, Column + 1).Font.Color
        Worksheets("Best_Results").Cells(Row, 3).Value = Worksheets("All_Results").Cells(Row, Column + 2).Value
        Worksheets("Best_Results").Cells(Row, 3).Font.Color = Worksheets("All_Results").Cells(Row, Column + 2).Font.Color
        Worksheets("Best_Results").Cells(Row, 4).Value = Worksheets("All_Results").Cells(Row, Column + 3).Value
        Worksheets("Best_Results").Cells(Row, 4).Font.Color = Worksheets("All_Results").Cells(Row, Column + 3).Font.Color
        Worksheets("Best_Results").Cells(Row, 5).Value = Worksheets("All_Results").Cells(Row, Column + 4).Value
        Worksheets("Best_Results").Cells(Row, 5).Font.Color = Worksheets("All_Results").Cells(Row, Column + 4).Font.Color
    Next Row
    Row = 4
    For Row = 4 To FinalRow_Projections
        Worksheets("Best_Results").Cells(Row, 7).Value = Worksheets("All_Projections").Cells(Row, Column).Value
        Worksheets("Best_Results").Cells(Row, 8).Value = Worksheets("All_Projections").Cells(Row, Column + 1).Value
        Worksheets("Best_Results").Cells(Row, 9).Value = Worksheets("All_Projections").Cells(Row, Column + 2).Value
        Worksheets("Best_Results").Cells(Row, 10).Value = Worksheets("All_Projections").Cells(Row, Column + 3).Value
        Worksheets("Best_Results").Cells(Row, 11).Value = Worksheets("All_Projections").Cells(Row, Column + 4).Value
    Next Row

'Here we configure the relevant chart labels, fonts etc. for every planet case that has been selected:

If Worksheets("Best_Results").Shapes("Option Button 1").OLEFormat.Object.Value = 1 Then
    Worksheets("Best_Results").ChartObjects("Chart1").Activate
        With ActiveChart.Axes(xlCategory)
            .HasTitle = True
            .AxisTitle.Characters.Text = "LONGITUDEMERCURY [o]"
            .AxisTitle.Font.ColorIndex = 1
            .AxisTitle.Font.Size = 22
```

```
            .AxisTitle.Font.Bold = msoTrue
            .AxisTitle.Characters(Start:=10, Length:=7).Font.Subscript = True
            .AxisTitle.Characters(Start:=10, Length:=7).Font.Size = 26
            .AxisTitle.Characters(Start:=38, Length:=1).Font.Superscript = True
            .AxisTitle.Left = 130
        End With
    Worksheets("Best_Results").ChartObjects("Chart2").Activate
        With ActiveChart.Axes(xlCategory)
            .HasTitle = True
            .AxisTitle.Characters.Text = "LONGITUDEMERCURY [o]"
            .AxisTitle.Font.ColorIndex = 1
            .AxisTitle.Font.Size = 22
            .AxisTitle.Font.Bold = msoTrue
            .AxisTitle.Characters(Start:=10, Length:=7).Font.Subscript = True
            .AxisTitle.Characters(Start:=10, Length:=7).Font.Size = 26
            .AxisTitle.Characters(Start:=38, Length:=1).Font.Superscript = True
            .AxisTitle.Left = 130
        End With
    Worksheets("Best_Results").ChartObjects("Chart3").Activate
        With ActiveChart.Axes(xlCategory)
            .HasTitle = True
            .AxisTitle.Characters.Text = "LONGITUDEMERCURY [o]"
            .AxisTitle.Font.ColorIndex = 1
            .AxisTitle.Font.Size = 22
            .AxisTitle.Font.Bold = msoTrue
            .AxisTitle.Characters(Start:=10, Length:=7).Font.Subscript = True
            .AxisTitle.Characters(Start:=10, Length:=7).Font.Size = 26
            .AxisTitle.Characters(Start:=38, Length:=1).Font.Superscript = True
            .AxisTitle.Left = 130
        End With
ElseIf Worksheets("Best_Results").Shapes("Option Button 2").OLEFormat.Object.Value = 1 Then
    Worksheets("Best_Results").ChartObjects("Chart1").Activate
        With ActiveChart.Axes(xlCategory)
            .HasTitle = True
            .AxisTitle.Characters.Text = "LONGITUDEVENUS [o]"
            .AxisTitle.Font.ColorIndex = 1
            .AxisTitle.Font.Size = 22
            .AxisTitle.Font.Bold = msoTrue
            .AxisTitle.Characters(Start:=10, Length:=7).Font.Subscript = True
            .AxisTitle.Characters(Start:=10, Length:=5).Font.Size = 26
            .AxisTitle.Characters(Start:=36, Length:=1).Font.Superscript = True
            .AxisTitle.Left = 155
        End With
    Worksheets("Best_Results").ChartObjects("Chart2").Activate
        With ActiveChart.Axes(xlCategory)
            .HasTitle = True
            .AxisTitle.Characters.Text = "LONGITUDEVENUS [o]"
            .AxisTitle.Font.ColorIndex = 1
            .AxisTitle.Font.Size = 22
            .AxisTitle.Font.Bold = msoTrue
            .AxisTitle.Characters(Start:=10, Length:=7).Font.Subscript = True
            .AxisTitle.Characters(Start:=10, Length:=5).Font.Size = 26
            .AxisTitle.Characters(Start:=36, Length:=1).Font.Superscript = True
            .AxisTitle.Left = 155
        End With
    Worksheets("Best_Results").ChartObjects("Chart3").Activate
        With ActiveChart.Axes(xlCategory)
```

```
            .HasTitle = True
            .AxisTitle.Characters.Text = "LONGITUDEVENUS [o]"
            .AxisTitle.Font.ColorIndex = 1
            .AxisTitle.Font.Size = 22
            .AxisTitle.Font.Bold = msoTrue
            .AxisTitle.Characters(Start:=10, Length:=7).Font.Subscript = True
            .AxisTitle.Characters(Start:=10, Length:=5).Font.Size = 26
            .AxisTitle.Characters(Start:=36, Length:=1).Font.Superscript = True
            .AxisTitle.Left = 155
        End With
ElseIf Worksheets("Best_Results").Shapes("Option Button 3").OLEFormat.Object.Value = 1 Then
    Worksheets("Best_Results").ChartObjects("Chart1").Activate
        With ActiveChart.Axes(xlCategory)
            .HasTitle = True
            .AxisTitle.Characters.Text = "LONGITUDEEARTH [o]"
            .AxisTitle.Font.ColorIndex = 1
            .AxisTitle.Font.Size = 22
            .AxisTitle.Font.Bold = msoTrue
            .AxisTitle.Characters(Start:=10, Length:=7).Font.Subscript = True
            .AxisTitle.Characters(Start:=10, Length:=5).Font.Size = 26
            .AxisTitle.Characters(Start:=36, Length:=1).Font.Superscript = True
            .AxisTitle.Left = 155
        End With
    Worksheets("Best_Results").ChartObjects("Chart2").Activate
        With ActiveChart.Axes(xlCategory)
            .HasTitle = True
            .AxisTitle.Characters.Text = "LONGITUDEEARTH [o]"
            .AxisTitle.Font.ColorIndex = 1
            .AxisTitle.Font.Size = 22
            .AxisTitle.Font.Bold = msoTrue
            .AxisTitle.Characters(Start:=10, Length:=7).Font.Subscript = True
            .AxisTitle.Characters(Start:=10, Length:=5).Font.Size = 26
            .AxisTitle.Characters(Start:=36, Length:=1).Font.Superscript = True
            .AxisTitle.Left = 155
        End With
    Worksheets("Best_Results").ChartObjects("Chart3").Activate
        With ActiveChart.Axes(xlCategory)
            .HasTitle = True
            .AxisTitle.Characters.Text = "LONGITUDEEARTH [o]"
            .AxisTitle.Font.ColorIndex = 1
            .AxisTitle.Font.Size = 22
            .AxisTitle.Font.Bold = msoTrue
            .AxisTitle.Characters(Start:=10, Length:=7).Font.Subscript = True
            .AxisTitle.Characters(Start:=10, Length:=5).Font.Size = 26
            .AxisTitle.Characters(Start:=36, Length:=1).Font.Superscript = True
            .AxisTitle.Left = 155
        End With
ElseIf Worksheets("Best_Results").Shapes("Option Button 4").OLEFormat.Object.Value = 1 Then
    Worksheets("Best_Results").ChartObjects("Chart1").Activate
        With ActiveChart.Axes(xlCategory)
            .HasTitle = True
            .AxisTitle.Characters.Text = "LONGITUDEMARS [o]"
            .AxisTitle.Font.ColorIndex = 1
            .AxisTitle.Font.Size = 22
            .AxisTitle.Font.Bold = msoTrue
            .AxisTitle.Characters(Start:=10, Length:=7).Font.Subscript = True
            .AxisTitle.Characters(Start:=10, Length:=4).Font.Size = 26
```

```
            .AxisTitle.Characters(Start:=35, Length:=1).Font.Superscript = True
            .AxisTitle.Left = 165
        End With
    Worksheets("Best_Results").ChartObjects("Chart2").Activate
        With ActiveChart.Axes(xlCategory)
            .HasTitle = True
            .AxisTitle.Characters.Text = "LONGITUDEMARS [o]"
            .AxisTitle.Font.ColorIndex = 1
            .AxisTitle.Font.Size = 22
            .AxisTitle.Font.Bold = msoTrue
            .AxisTitle.Characters(Start:=10, Length:=7).Font.Subscript = True
            .AxisTitle.Characters(Start:=10, Length:=4).Font.Size = 26
            .AxisTitle.Characters(Start:=35, Length:=1).Font.Superscript = True
            .AxisTitle.Left = 165
        End With
    Worksheets("Best_Results").ChartObjects("Chart3").Activate
        With ActiveChart.Axes(xlCategory)
            .HasTitle = True
            .AxisTitle.Characters.Text = "LONGITUDEMARS [o]"
            .AxisTitle.Font.ColorIndex = 1
            .AxisTitle.Font.Size = 22
            .AxisTitle.Font.Bold = msoTrue
            .AxisTitle.Characters(Start:=10, Length:=7).Font.Subscript = True
            .AxisTitle.Characters(Start:=10, Length:=4).Font.Size = 26
            .AxisTitle.Characters(Start:=35, Length:=1).Font.Superscript = True
            .AxisTitle.Left = 166
        End With
ElseIf Worksheets("Best_Results").Shapes("Option Button 5").OLEFormat.Object.Value = 1 Then
    Worksheets("Best_Results").ChartObjects("Chart1").Activate
        With ActiveChart.Axes(xlCategory)
            .HasTitle = True
            .AxisTitle.Characters.Text = "LONGITUDEJUPITER [o]"
            .AxisTitle.Font.ColorIndex = 1
            .AxisTitle.Font.Size = 22
            .AxisTitle.Font.Bold = msoTrue
            .AxisTitle.Characters(Start:=10, Length:=7).Font.Subscript = True
            .AxisTitle.Characters(Start:=10, Length:=7).Font.Size = 26
            .AxisTitle.Characters(Start:=38, Length:=1).Font.Superscript = True
            .AxisTitle.Left = 145
        End With
    Worksheets("Best_Results").ChartObjects("Chart2").Activate
        With ActiveChart.Axes(xlCategory)
            .HasTitle = True
            .AxisTitle.Characters.Text = "LONGITUDEJUPITER [o]"
            .AxisTitle.Font.ColorIndex = 1
            .AxisTitle.Font.Size = 22
            .AxisTitle.Font.Bold = msoTrue
            .AxisTitle.Characters(Start:=10, Length:=7).Font.Subscript = True
            .AxisTitle.Characters(Start:=10, Length:=7).Font.Size = 26
            .AxisTitle.Characters(Start:=38, Length:=1).Font.Superscript = True
            .AxisTitle.Left = 145
        End With
    Worksheets("Best_Results").ChartObjects("Chart3").Activate
        With ActiveChart.Axes(xlCategory)
            .HasTitle = True
            .AxisTitle.Characters.Text = "LONGITUDEJUPITER [o]"
            .AxisTitle.Font.ColorIndex = 1
```

```
            .AxisTitle.Font.Size = 22
            .AxisTitle.Font.Bold = msoTrue
            .AxisTitle.Characters(Start:=10, Length:=7).Font.Subscript = True
            .AxisTitle.Characters(Start:=10, Length:=7).Font.Size = 26
            .AxisTitle.Characters(Start:=37, Length:=1).Font.Superscript = True
            .AxisTitle.Left = 145
        End With
ElseIf Worksheets("Best_Results").Shapes("Option Button 6").OLEFormat.Object.Value = 1 Then
    Worksheets("Best_Results").ChartObjects("Chart1").Activate
        With ActiveChart.Axes(xlCategory)
            .HasTitle = True
            .AxisTitle.Characters.Text = "LONGITUDESATURN [o]"
            .AxisTitle.Font.ColorIndex = 1
            .AxisTitle.Font.Size = 22
            .AxisTitle.Font.Bold = msoTrue
            .AxisTitle.Characters(Start:=10, Length:=7).Font.Subscript = True
            .AxisTitle.Characters(Start:=10, Length:=6).Font.Size = 26
            .AxisTitle.Characters(Start:=37, Length:=1).Font.Superscript = True
            .AxisTitle.Left = 145
        End With
    Worksheets("Best_Results").ChartObjects("Chart2").Activate
        With ActiveChart.Axes(xlCategory)
            .HasTitle = True
            .AxisTitle.Characters.Text = "LONGITUDESATURN [o]"
            .AxisTitle.Font.ColorIndex = 1
            .AxisTitle.Font.Size = 22
            .AxisTitle.Font.Bold = msoTrue
            .AxisTitle.Characters(Start:=10, Length:=7).Font.Subscript = True
            .AxisTitle.Characters(Start:=10, Length:=6).Font.Size = 26
            .AxisTitle.Characters(Start:=37, Length:=1).Font.Superscript = True
            .AxisTitle.Left = 145
        End With
    Worksheets("Best_Results").ChartObjects("Chart3").Activate
        With ActiveChart.Axes(xlCategory)
            .HasTitle = True
            .AxisTitle.Characters.Text = "LONGITUDESATURN [o]"
            .AxisTitle.Font.ColorIndex = 1
            .AxisTitle.Font.Size = 22
            .AxisTitle.Font.Bold = msoTrue
            .AxisTitle.Characters(Start:=10, Length:=7).Font.Subscript = True
            .AxisTitle.Characters(Start:=10, Length:=6).Font.Size = 26
            .AxisTitle.Characters(Start:=37, Length:=1).Font.Superscript = True
            .AxisTitle.Left = 145
        End With
ElseIf Worksheets("Best_Results").Shapes("Option Button 7").OLEFormat.Object.Value = 1 Then
    Worksheets("Best_Results").ChartObjects("Chart1").Activate
        With ActiveChart.Axes(xlCategory)
            .HasTitle = True
            .AxisTitle.Characters.Text = "MOON PHASE [o]"
            .AxisTitle.Font.ColorIndex = 1
            .AxisTitle.Font.Size = 22
            .AxisTitle.Font.Bold = msoTrue
            .AxisTitle.Characters(Start:=32, Length:=1).Font.Superscript = True
            .AxisTitle.Left = 183
        End With
    Worksheets("Best_Results").ChartObjects("Chart2").Activate
        With ActiveChart.Axes(xlCategory)
```

```
            .HasTitle = True
            .AxisTitle.Characters.Text = "MOON PHASE [o]"
            .AxisTitle.Font.ColorIndex = 1
            .AxisTitle.Font.Size = 22
            .AxisTitle.Font.Bold = msoTrue
            .AxisTitle.Characters(Start:=32, Length:=1).Font.Superscript = True
            .AxisTitle.Left = 183
        End With
    Worksheets("Best_Results").ChartObjects("Chart3").Activate
        With ActiveChart.Axes(xlCategory)
            .HasTitle = True
            .AxisTitle.Characters.Text = "MOON PHASE [o]"
            .AxisTitle.Font.ColorIndex = 1
            .AxisTitle.Font.Size = 22
            .AxisTitle.Font.Bold = msoTrue
            .AxisTitle.Characters(Start:=32, Length:=1).Font.Superscript = True
            .AxisTitle.Left = 183
        End With
End If

MinutesElapsed = Format((Timer - StartTime) / 86400, "hh:mm:ss") 'Determine how many seconds code took to run.
MsgBox "The compilation has been successfully completed in " & MinutesElapsed & " [hh:mm:ss]" & vbNewLine &
     vbNewLine & "The BEST case for the position " & Worksheets("Best_Results").Cells(57, 14).Value & " - " &
     Worksheets("Best_Results").Cells(57, 16).Value & "deg," & vbNewLine & "is when " &
     Worksheets("Best_Results").Cells(2, 15).Value & "," & vbNewLine & "where the MIN <-> MAX effect is: " &
     Format(Max_Effect, "0.00") & "%.", vbInformation 'Informs the user for the min-max effect.

End Sub
```

E

# RESULTING PUBLICATIONS

During the realisation of this thesis, several papers and articles were written, presented in international conferences and published in scientific journals and books. They are listed below by genre and chronological order based on their publication date [1].

Specifically, the following nine articles were published in **peer-reviewed journals**:

1. ***"The Sun and its Planets as detectors for invisible matter"*** was published in Physics of the Dark Universe on 27/06/2017, https://doi.org/10.1016/j.dark.2017.06.001.
2. ***"Response to Comment on Planetary Dependence of Melanoma"*** was published in Biophysical Reviews and Letters on 04/02/2019, https://doi.org/10.1142/S1793048019200029.
3. ***"Stratospheric temperature anomalies as imprints from the dark Universe"*** was published in Physics of the Dark Universe on 06/02/2020, https://doi.org/10.1016/j.dark.2020.100497.
4. ***"Observation of a 27 Days Periodicity in Melanoma Diagnosis"*** was published in Biophysical Reviews and Letters on 27/11/2020, https://doi.org/10.1142/S1793048020500083.
5. ***"On the Origin of the Rhythmic Sun's Radius Variation"*** was published in Symmetry on 05/02/2022, https://doi.org/10.3390/sym14020325.
6. ***"Search for Dark Matter Axions with CAST-CAPP"*** was published in Nature Communications on 19/10/2022, https://doi.org/10.1038/s41467-022-33913-6.
7. ***"Dark Matter Detection in the Stratosphere"*** was published in Symmetry on 29/05/2023, https://doi.org/10.3390/sym15061167.
8. ***"Planetary relationships to birth (imputed conception) rates in humans: A signature of cosmic origin?"*** was published in Biophysical Reviews and Letters on 14/08/2023, https://doi.org/10.1142/S1793048023500029.
9. ***"Gravitational focusing effects on streaming dark matter as a new detection concept"*** was published in Physical Review D on 27/12/2023, https://doi.org/10.1103/PhysRevD.108.123043.

Additionally, four more articles were published in the **arXiv** and **Qeios** repositories:

1. ***"Search for axions in streaming dark matter"*** was published in arXiv on 04/03/2017, https://doi.org/10.48550/arXiv.1703.01436.

[1]The provided list of publications is updated with a number of publications which were initiated during the writing of this dissertation but were published after the defence date of 30/06/2022.

2. ***"Response to Time Modulations and Amplifications in the Axion Search Experiments"*** was published in arXiv on 19/08/2019, https://doi.org/10.48550/arXiv.1908.07875.

3. ***"Response-suggestion to The XENON1T excess: an overlooked dark matter signature?"*** was published in arXiv on 26/06/2020, https://doi.org/10.48550/arXiv.2006.16907.

4. ***"Planetary relationship as a key signature from the dark sector"*** was published in Qeios on 07/09/2023, https://doi.org/10.32388/0XHYID.

Moreover, there were 17 invited and contributed talks made at international conferences some of which were later published as **proceedings papers**:

1. ***"Signals for invisible matter from solar - terrestrial observations"*** was presented by K. Zioutas at the "12$^{th}$ Patras Workshop on Axions, WIMPs and WISPs" International Conference held in Jeju island, Republic of Korea on 20–24 of June 2016, https://indico.desy.de/event/13889/contributions/14503, and was later published in the proceedings by "Verlag Deutsches Elektronen-Synchrotron Hamburg", http://dx.doi.org/10.3204/DESY-PROC-2009-03/Zioutas_Konstantin.

2. ***"Signals for invisible matter from solar - terrestrial observations"*** was presented by M. Maroudas at the "5$^{th}$ International Conference on New Frontiers in Physics" held in Kolymbari, Crete, Greece on 06–14 of July 2016, https://indico.cern.ch/event/442094/contributions/2228064, and was later published in the proceedings by "EPJ Web of Conferences", https://doi.org/10.1051/epjconf/201716407041.

3. ***"Search for streaming dark matter axions or other exotica"*** was presented by K. Zioutas at the "Axion Dark Matter" International Conference held in Nordita, Stockholm, Sweden on 05–09 of December 2016, https://indico.fysik.su.se/event/5859/contributions/7862/.

4. ***"Search for streaming dark matter axions or other exotica"*** was presented by A. Gardikiotis at the "13$^{th}$ Patras Workshop on Axions, WIMPs and WISPs" International Conference held in Thessaloniki, Greece on 15–19 of May 2017, https://indico.desy.de/event/16884/contributions/24932, and was later published in the proceedings by "Verlag Deutsches Elektronen-Synchrotron Hamburg", https://doi.org/10.3204/DESY-PROC-2017-02/gardikiotis_antonios, as well as in arXiv, https://doi.org/10.48550/arXiv.1803.08588.

5. ***"Invisible matter in the solar system: new observational evidence"*** was presented by K. Zioutas at the "IBS Conference on Dark World" held in Daejeon, Republic of Korea on 30 of October to 03 of November 2017, https://indico.ibs.re.kr/event/137/contributions/193.

6. ***"Everyday signatures for streaming invisible matter, dark matter, and missing antimatter mysteries"*** was presented by K. Zioutas at the "Discrete Symmetries in Particle, Nuclear and Atomic Physics and implications for our Universe" Workshop held in Trento, Italy on 08–12 of October 2018, https://indico.ectstar.eu/event/25/contributions/615.

7. ***"Search for Dark Mater Axions with CAST-CAPP"*** was presented by M. Maroudas at the "16th Patras Workshop on Axions, WIMPs and WISPs" International Conference held online in Trieste, Italy on 14–18 of June 2021, https://agenda.infn.it/event/20431/contributions/137685.

8. ***"The Dark Universe is not invisible"*** was presented by K. Zioutas at the "1st Electronic Conference on Universe" International Conference held online on 22–28 of February 2021, https://sciforum.net/paper/view/9313, and was later published in the proceedings by "Physical Sciences Forum", https://doi.org/10.3390/ECU2021-09313, as well as in arXiv, https://doi.org/10.48550/arXiv.2108.11647.

9. ***"Atmospheric temperature anomalies as manifestation of the dark Universe"*** was presented by K. Zioutas at the "15th International Conference on Meteorology, Climatology and Atmospheric Physics" held in Ioannina, Greece on 26–29 of September 2021, https://www.conferre.gr/allevents/comecap2020/files/abstactbookopt.pdf, and was later published in the proceedings, https://www.conferre.gr/allevents/comecap2020, as well as in arXiv, https://doi.org/10.48550/arXiv.2309.10779.

10. ***"Fingerprints of the dark Universe from geoscience"*** was presented by K. Zioutas at the "1st Mediterranean Geosciences Union" International Conference held in Istanbul, Turkey on 25–28 of November 2021 and was later published in the proceedings by "Springer", https://doi.org/10.1007/978-3-031-43218-7_96.

11. ***"Planetary relationship as the new signature from the dark Universe"*** was presented by K. Zioutas at the "25th Workshop on What Comes Beyond the Standard Models" International Conference held online in Bled, Slovenia on 04–10 of July 2022, http://bsm.fmf.uni-lj.si/bled2022bsm/talks/Bled_proceedings-2022Dec_03.pdf, and was later published in the proceedings by "Bled Workshops in Physics", http://bsm.fmf.uni-lj.si/bled2022bsm/talks/BledProceedings2022_Acceptable_06_12.pdf.

12. ***"Latest results on dark matter axions with CAST-CAPP"*** was presented by M. Maroudas at the "17$^{th}$ Patras Workshop on Axions, WIMPs and WISPs" International Conference held in Mainz, Germany on 08–12 of August 2022, https://indico.him.uni-mainz.de/event/109/contributions/770.

13. ***"Novel planetary signatures from the dark Universe"*** was presented by K. Zioutas at the "17$^{th}$ Patras Workshop on Axions, WIMPs and WISPs" International Conference held in Mainz, Germany on 08–12 of August 2022, https://indico.him.uni-mainz.de/event/109/contributions/850, and was later published in Astrophysics by "National Academy of Sciences of the Republic of Armenia", https://doi.org/10.54503/0002-3051-2023.76.4-591, as well as in Astrophysics by "Springer", https://doi.org/10.1007/s10511-024-09809-2.

14. ***"Unexpected planetary relationships of unexpected observations as the new signature in astro-particle physics"*** was presented by K. Zioutas at the "XI International Conference on New Frontiers in Physics" held in Crete, Greece on 30 of August to 12 of September 2022, https://indico.cern.ch/event/1133591/contributions/5016504.

15. ***"Dark matter search with CAST and beyond CAST"*** was presented by K. Zioutas at the "40$^{th}$ Conference on Recent Developments in High Energy Physics and Cosmology" held in Ioannina, Greece on 05–07 of April 2023, https://indico.cern.ch/event/1223490/contributions/5286010/, and was later accepted for publication in the proceedings by Springer-Nature.

16. ***"Searches for daily modulations with the CAST-CAPP detector"*** was presented by K. Özbozduman at the "18$^{th}$ Patras Workshop on Axions, WIMPs and WISPs" International Conference held in Rijeka, Croatia on 03–07 of July 2023, https://agenda.infn.it/event/34455/contributions/202562.

17. ***"Clean energy from the dark Universe?"*** was presented by K. Zioutas at the "26$^{th}$ International workshop: What comes beyond the standard models" held in Bled, Slovenia on 10–19 of July 2023, http://bsm.fmf.uni-lj.si/bled2023bsm/talks/AstriAbstractsBled2023.pdf, and was later published in the proceedings by "Bled Workshops in Physics", https://doi.org/10.51746/9789612972097.

There was also one article published in a **newsletter** journal:

1. ***"CAST: from Solar to Dark Matter Axions searches"*** was an article published in the Newsletter of the EP department of CERN on 12/12/2019, https://ep-news.web.cern.ch/content/cast-solar-dark-matter-axions-searches.

In addition, there was one contribution to a **book chapter**:

1. ***"Hunting Dark Matter Axions with CAST"*** was a chapter of the book "Advances in Cosmology" published by Springer Nature Switzerland AG on 05/12/2022, https://doi.org/10.1007/978-3-031-05625-3_8.

Finally, there was one **poster** presented at an international conference:

1. ***"CAST-CAPP detector Project"*** was a poster that was presented at the "15th Patras Workshop on Axions WIMPs and WISPs" International Conference held in Freiburg, Germany on 03–07 of June 2019, https://indico.desy.de/event/22598/contributions/47429.

## E.1 Article I

### The Sun and its planets as detectors for invisible matter

## E.2 Article II

### Response to comment on planetary dependence of melanoma”

K. Zioutas, E. Valachovic, **M. Maroudas**, “Response to Comment on Planetary Dependence of Melanoma””, Biophysical Reviews and Letters, Vol. 14, No. 01, pp. 11-15, 2019, https://doi.org/10.1142/S1793048019200029.

## E.3 Article III

### Stratospheric temperature anomalies as imprints from the dark Universe

K. Zioutas, A. Argiriou, H. Fischer, S.Hofmann, **M. Maroudas**, A. Pappa, Y. K. Semertzidis, "Stratospheric temperature anomalies as imprints from the dark Universe", Physics of the Dark Universe, Vol. 28, pp. 100497, 2020, https://doi.org/10.1016/j.dark.2020.100497.

## E.4 Article IV

### Observation of a 27 days periodicity in melanoma diagnosis

## E.5 Article V

### On the origin of the rhythmic Sun's radius variation

## E.6 Article VI

### Search for dark matter axions with CAST-CAPP

C. M. Adair, K. Altenmüller, V. Anastassopoulos, S. Arguedas Cuendis, J. Baier, K. Barth, A. Belov, D. Bozicevic, H. Bräuninger, G. Cantatore, F. Caspers, J. F. Castel, S. A. Çetin, W. Chung, H. Choi, J. Choi, T. Dafni, M. Davenport, A. Dermenev, K. Desch, B. Döbrich, H. Fischer, W. Funk, J. Galan, A. Gardikiotis, S. Gninenko, J. Golm, M. D. Hasinoff, D. H. H. Hoffmann, D. Díez Ibáñez, I. G. Irastorza, K. Jakovčić, J. Kaminski, M. Karuza, C. Krieger, Ç. Kutlu, B. Lakić, J. M. Laurent, J. Lee, S. Lee, G. Luzón, C. Malbrunot, C. Margalejo, **M. Maroudas**, L. Miceli, H. Mirallas, L. Obis, A. Özbey, K. Özbozduman, M. J. Pivovaroff, M. Rosu, J. Ruz, E. Ruiz-Chóliz, S. Schmidt, M. Schumann, Y. K. Semertzidis, S. K. Solanki, L. Stewart, I. Tsagris, T. Vafeiadis, J. K. Vogel, M. Vretenar, S. Youn, K. Zioutas, "Search for Dark Matter Axions with CAST-CAPP", Nature Communications, Vol. 13, No. 6180, pp. 1-9, 2022, https://doi.org/10.1038/s41467-022-33913-6.

## E.7 Article VII

### Dark matter detection in the stratosphere

G. Cantatore, S. A. Çetin, H. Fischer, W. Funk, M. Karuza, A. Kryemadhi, **M. Maroudas**, K. Özbozduman, Y. K. Semertzidis, K. Zioutas, "Dark Matter Detection in the Stratosphere", Symmetry, Vol. 15, No. 06, 1167, 2023, https://doi.org/10.3390/sym15061167.

## E.8 Article VIII

### Planetary Relationships to birth (imputed conception) rates in humans: A signature of cosmic origin?

## E.9 Article IX

### Gravitational focusing effects on streaming dark matter as a new detection concept

A. Kryemadhi, **M. Maroudas**, A. Mastronikolis, K. Zioutas, "Gravitational focusing effects on streaming dark matter as a new detection concept", Physical Review D, Vol. 108, No. 12, pp. 123043, 2023, https://doi.org/10.1103/PhysRevD.108.123043.

## E.10 arXiv Article I

### Search for axions in streaming dark matter

K. Zioutas, V. Anastassopoulos, S. Bertolucci, G. Cantatore, S. A. Çetin, H. Fischer, W. Funk, A. Gardikiotis, D.H.H. Hoffmann, S. Hofmann, M. Karuza, **M. Maroudas**, Y.K. Semertzidis, I. Tkatchev, "Search for axions in streaming dark matter", arXiv, 1703.0143, 2017, https://doi.org/10.48550/arXiv.1703.01436.

## E.11 arXiv Article II

### RESPONSE TO Time Modulations and Amplifications in the Axion Search Experiments

S. Bertolucci, H. Fischer, S. Hofmann, **M. Maroudas**, Y. K. Semertzidis, K. Zioutas, "RESPONSE TO Time Modulations and Amplifications in the Axion Search Experiments", arXiv, 1908.07875, 2019, https://doi.org/10.48550/arXiv.1908.07875.

## E.12 arXiv Article III

### Response-suggestion to the XENON1T excess: an overlooked dark matter signature?

K. Zioutas, G. Cantatore, M. Karuza, A. Kryemadhi, **M. Maroudas**, Y.K. Semertzidis, "Response-suggestion to The XENON1T excess: an overlooked dark matter signature?", arXiv, 2006.16907, 2020, https://doi.org/10.48550/arXiv.2006.16907.

## E.13 Qeios Article I

### Planetary relationship as a key signature from the dark sector

K. Zioutas, G. Cantatore, S. A. Çetin, A. Gardikiotis, E. Georgiopoulou, S. Hofmann, M. Karuza, A. Kryemadhi, **M. Maroudas**, A. Mastronikolis, K. Özbozduman, Y. K. Semertzidis, I. Tsagris, M. Tsagri, "Planetary relationship as a key signature from the dark sector", Qeios, 0XHYID, 2023, https://doi.org/10.32388/0XHYID.

## E.14 Proceedings Paper I

### Signals for invisible matter from solar - terrestrial observations

## E.15 Proceedings Paper II

### Signals for invisible matter from solar - terrestrial observations

S. Bertolucci, K. Zioutas, S. Hofmann, **M. Maroudas**, “Signals for invisible matter from solar - terrestrial observations”, 5$^{\text{th}}$ International Conference on New Frontiers in Physics ICNFP2016, EPJ Web of Conferences, Vol. 164, pp. 07041, 2017, https://doi.org/10.1051/epjconf/201716407041.

## E.16 Proceedings Paper III

### Search for streaming dark matter axions or other exotica

A. Gardikiotis, V. Anastassopoulos, S. Bertolucci, G. Cantatore, S. A. Çetin, H. Fischer, W. Funk, D.H.H. Hoffmann, S. Hofmann, M. Karuza, **M. Maroudas**, Y.K. Semertzidis, I. Tkachev, K. Zioutas, "Search for streaming dark matter axions or other exotica", Verlag Deutsches Elektronen-Synchrotron Hamburg, 2018, https://doi.org/10.3204/DESY-PROC-2017-02/gardikiotis_antonios; arXiv, 1803.08588, 2018, https://doi.org/10.48550/arXiv.1803.08588.

## E.17 Proceedings Paper IV

### The dark Universe is not invisible

K. Zioutas, V. Anastassopoulos, A. Argiriou, G. Cantatore, S. A. Çetin, A. Gardikiotis, D.H.H. Hoffmann, S. Hofmann, M. Karuza, A. Kryemadhi, **M. Maroudas**, E. L. Matteson, K. Özbozduman, T. Papaevangelou, M. Perryman, Y.K. Semertzidis, I. Tsagris, M. Tsagri, G. Tsiledakis, D. Utz, E. Valahovic, "The Dark Universe is not invisible", Physical Sciences Forum, Vol. 02, No. 01, 2021, https://doi.org/10.3390/ECU2021-09313; arXiv, 2108.11647, 2021, https://doi.org/10.48550/arXiv.2108.11647.

## E.18 Proceedings Paper V

### Atmospheric temperature anomalies as manifestation of the dark Universe

K. Zioutas, V. Anastassopoulos, A. Argiriou, G. Cantatore, S. A. Çetin, A. Gardikiotis, H. Haralambous, D.H.H. Hoffmann, S. Hofmann, M. Karuza, A. Kryemadhi, **M. Maroudas**, A. Mastronikolis, C. Oikonomou, K. Özbozduman, Y. K. Semertzidis, "Atmospheric temperature anomalies as manifestation of the dark Universe", 15th International Conference on Meteorology, Climatology and Atmospheric Physics, pp. 728-732, 2021, comecap2020/Proceedings_Final; arXiv, 2309.10779, 2023, https://doi.org/10.48550/arXiv.2309.10779.

## E.19 Proceedings Paper VI

### Fingerprints of the dark Universe from geoscience

K. Zioutas, V. Anastassopoulos, A. Argiriou, G. Cantatore, S. A. Çetin, A. Gardikiotis, J. Guo, H. Haralambous, D.H.H. Hoffmann, S. Hofmann, M. Karuza, A. Kryemadhi, **M. Maroudas**, A. Mastronikolis, C. Oikonomou, K. Özbozduman, Y. K. Semertzidis, "Fingerprints of the dark Universe from geoscience", Recent Research on Geotechnical Engineering, Remote Sensing, Geophysics and Earthquake Seismology, Springer, pp. 415-419, 2024, https://doi.org/10.1007/978-3-031-43218-7_96.

## E.20 Proceedings Paper VII

### Planetary relationship as the new signature from the dark Universe

K. Zioutas, V. Anastassopoulos, A. Argiriou, G. Cantatore, S. A. Çetin, A. Gardikiotis, M. Karuza, A. Kryemadhi, **M. Maroudas**, A. Mastronikolis, K. Özbozduman, Y. K. Semertzidis, M. Tsagri, I. Tsagris, "Planetary relationship as the new signature from the dark Universe", $25^{\text{th}}$ International Workshop on What Comes Beyond the Standard Models, Bled Workshops in Physics, Vol. 23, No. 01, pp. 256-261, 2022, BledProceedings2022_Acceptable_06_12.pdf.

## E.21 Proceedings Paper VIII

### Novel planetary signatures from the dark Universe

K. Zioutas, V. Anastassopoulos, A. Argiriou, G. Cantatore, S. Çetin, A. Gardikiotis, H. Haralambous, M. Karuza, A. Kryemadhi, **M. Maroudas**, A. Mastronikolis, C. Oikonomou, K. Özbozduman, Y. K. Semertzidis, M. Tsagri, I. Tsagris, "Novel Planetary Signatures from the Dark Universe", 17$^{th}$ Patras Workshop on Axions, WIMPs and WISPs, Astrophysics, National Academy of Sciences of the Republic of Armenia, Vol. 66, No. 4, pp. 591-601, 2023, https://doi.org/10.54503/0002-3051-2023.76.4-591; Astrophysics, Springer, Vol. 66, No. 4, pp. 550-558, 2024, https://doi.org/10.1007/s10511-024-09809-2.

## E.22 Proceedings Paper IX

### Dark matter search with CAST and beyond CAST

K. Zioutas, V. Anastassopoulos, A. Argiriou, G. Cantatore, S. A. Çetin, A. Gardikiotis, H. Haralambous, M. Karuza, A. Kryemadhi, **M. Maroudas**, A. Mastronikolis, C. Oikonomou, K. Özbozduman, Y. K. Semertzidis, M. Tsagri, I. Tsagris, "Dark matter search with CAST and beyond CAST", $40^{\text{th}}$ Conference on Recent Developments in High Energy Physics and Cosmology, Springer-Nature.

## E.23 Proceedings Paper X

### Clean energy from the dark Universe?

## E.24 Newsletter I

### CAST: from solar to dark matter axions searches

**M. Maroudas**, Y. K. Semertzidis, "CAST: from Solar to Dark Matter Axions searches", CERN EP Newsletter, 2019, cast-solar-dark-matter-axions-searches.

## E.25 Book Chapter I

### Hunting dark matter axions with CAST

## E.26 Poster I

### CAST-CAPP detector project

**M. Maroudas**, K. Özbozduman, "CAST-CAPP detector Project", Small Presentation and Poster on the 15$^{th}$ Patras Workshop on Axions WIMPs and WISPs, Freiburg, Germany, 2019, https://indico.desy.de/event/22598/contributions/47429/.

# LIST OF ACRONYMS

| Acronym | Full Name | Page(s) |
|---|---|---|
| **CDF** | Cumulative Distribution Function | 284, 285 |
| **CDM** | Cold Dark Matter | 19–23, 37, 44, 48, 52, 234 |
| **CERN** | European Organization for Nuclear Research | 249, 270, 399–402 |
| **CHAMP** | CHArged Massive Particle | 26 |
| **CL** | Confidence Level | 284, 285, 287 |
| **CLA** | Cryo Linear Actuator | 250, 251, 390, 396, 397 |
| **CMB** | Cosmic Microwave Background | 15, 19–22 |
| **CME** | Coronal Mass Ejection | 79, 81 |
| **CODE** | Center for Orbit Determination in Europe | 167 |
| **CTA** | CERN Tape Archive | 270, 402 |
| **CTS** | Cryo Translation Stage | 251, 397 |
| **CV** | Coefficient of Variation | 90 |
| **CW** | Continuous Waveform | 256, 257 |
| **DAMA** | DArk MAtter | 220 |
| **DAQ** | Data Acquisition | 255, 258, 270, 291, 392, 396, 397, 399, 400 |
| **DC** | Direct Current | xl, 253–255 |
| **DDDM** | Double-Disk Dark Matter | 44–47 |
| **DE** | Dark Energy | 9, 16, 26, 237 |
| **DFSZ** | Dine, Fischler, Srednicki and Zhitnitski | 232, 233, 237, 238, 289 |
| **DFT** | Discrete Fourier Transformation | 273 |
| **DM** | Dark Matter | 9, 10, 14–24, 26, 27, 29–33, 35–37, 39–49, 51, 52, 59–62, 64–68, 75, 80, 137, 159, 163, 187, 204, 205, 220, 223, 224, 233, 237–240, 243, 265–267, 277, 285, 287–292, 295, 296 |

| Acronym | Full Name | Page(s) |
|---|---|---|
| **ECMWF** | European Centre for Medium-range Weather Forecast | 178 |
| **EDM** | Electric Dipole Moment | 228 |
| **EMC** | ElectroMagnetic Compatibility | 258, 266, 273, 276, 285, 288, 291, 390 |
| **EMI** | ElectroMagnetic Interference | 258, 266, 273, 276, 285, 288, 291, 390 |
| **ENR** | Excess Noise Ratio | 259, 260 |
| **ENSO** | El Nino/Souhern Oscillation | 190 |
| **EQ** | Earthquake | 10, 193–204, 217, 219, 295, 375–378, 380–382 |
| **EROS** | Expérience pour la Recherche d'Objets Sombre | 30 |
| **ESA** | European Space Agency | 95 |
| **EUV** | Extreme UltraViolet | 66, 93–99, 102, 116–118, 126, 140, 146, 147, 155, 157, 158, 165, 166, 172, 175, 178, 320–326, 328, 329, 331–333, 349, 350, 353, 355, 356, 358, 359, 365, 366, 369–372, 380 |
| **FD** | Frequency Domain | 273 |
| **FFT** | Fast Fourier Transform | 273, 274, 279, 394 |
| **FIP** | First Ionisation Potential | 139–144, 159, 347–350 |
| **FMT** | Frequency Matching Tolerance | 249, 262, 276 |
| **FTP** | File transfer Protocol | 402 |
| **FTS** | File Transfer Service | 402 |
| **FWHM** | Full Width at Half Maximum | 91, 98, 99, 110–115, 121, 122, 124, 131, 135, 136, 142, 145, 150, 152–154, 202, 203, 214, 275, 308, 309, 319, 330 |

| Acronym | Full Name | Page(s) |
|---|---|---|
| **GAIA** | Global Astrometric Interferometer for Astrophysics | 37, 38, 43, 159, 160 |
| **GC** | Galactic Centre | 31, 38, 41, 45, 46, 68, 88, 92, 99, 102, 121, 122, 157, 170, 174, 181, 191, 217, 224, 241, 295 |
| **GIMP** | Gravitationally-Interacting Massive Particle | 26 |
| **GOES** | Geostationary Operational Environmental Satellites | 82, 83, 92 |
| **GPS** | Global Positioning System | 167 |
| **GUT** | Grand Unified Theory | 23, 25, 29, 232 |
| **HDD** | Hard Disk Drive | 390, 402 |
| **HDM** | Hot Dark Matter | 19–21, 24 |
| **HFET** | Heterojunction Field Effect Transistor | 237, 238 |
| **Hipparcos** | HIgh Precision PARallax COllecting Satellite | 37 |
| **HMI** | Helioseismic and Magnetic Imager | 129 |
| **IF** | Intermediate Frequency | 274, 277, 278, 280 |
| **INFN** | Istituto Nazionale di Fisica Nucleare | 179 |
| **JPE** | Janssen Precision Engineering | 390 |
| **JPL** | Jet Propulsion Laboratory | 71 |
| **KSVZ** | Kim, Shiftman, Vainshtein and Zakharov | xlii, 232–234, 237, 248, 283, 284, 286, 287, 289 |
| **KWISP** | Kinetic Weakly Interacting Slim Particles | 237 |
| **LAN** | Local Area Network | 255, 390, 399 |
| **LASP** | Laboratory for Atmospheric and Space Physics | 148 |

| Acronym | Full Name | Page(s) |
|---|---|---|
| **LAT** | Large Area Telescope | 160 |
| **LHC** | Large Hadron Collider | 33, 236, 250 |
| **LISIRD** | Lasp Interactive Solar Irradiance Datacenter | 148 |
| **LMSAL** | Lockheed Martin Solar and Astrophysics Laboratory | 92 |
| **LNA** | Low Noise Amplifier | 253, 254, 260, 261, 269, 275, 279, 390, 391, 397 |
| **LNGS** | Laboratori Nazionali del Gran Sasso | 179 |
| **LSP** | Lightest Supersymmetric Particle | 23 |
| **LVDS** | Low Voltage Differential Signaling | 395 |
| **LXPLUS** | Linux Public Login User Service | 402 |
| **MACHO** | MAssive Compact Halo Object | 19, 21, 22, 30 |
| **MACRO** | Monopole, Astrophysics and Cosmic Ray Observatory | 29 |
| **MDI** | Michelson Doppler Imager | 129 |
| **ML** | Maximum Likelihood | 281, 282 |
| **MoEDAL** | Monopole and Exotics Detector at the LHC | 33 |
| **MOND** | Modified Newton Dynamics | 27 |
| **MRB** | Magnet Return Box | 389 |
| **MSSM** | Minimal Supersymmetric Standard Model | 27 |
| **NAO** | North Atlantic Oscillation | 190 |
| **NASA** | National Aeronautics and Space Administration | 71, 74, 95, 109 |
| **NEHRP** | National Earthquake Hazards Reduction Program | 194 |
| **NEIC** | National Earthquake Information Center | 194 |
| **NGDC** | National Geophysical Data Center | 82 |

| Acronym | Full Name | Page(s) |
|---|---|---|
| **NIST** | National Institute of Standards and Technology | 194 |
| **NOOA** | National Oceanic and Atmospheric Administration | 82, 83, 92 |
| **NRC** | National Research Council Canada | 118, 126 |
| **NRCan** | Natural Resources Canada | 118 |
| **OGLE** | Optical Gravitational Lensing Experiment | 30 |
| **PBH** | Primordial Black Hole | 21, 22 |
| **PC** | Personal Computer | 255, 390, 392, 396, 399 |
| **PCIe** | Peripheral Component Interconnect Express | 396 |
| **PDF** | Probability Density Function | 284 |
| **PIDM** | Partially Interacting Dark Matter | 43, 44 |
| **PM** | Phase-Matching | 248, 249, 262–266, 270–272, 276, 277, 285, 289, 291, 390, 392, 400 |
| **PQWW** | Peccei-Quinn-Weinberg-Wilczek | 232 |
| **QBO** | Quasi Biennial Oscillation | 189, 190 |
| **QCD** | Quantum Chromodynamics | 20, 22, 227–231, 245, 248 |
| **RADES** | Relic Axion Detector Experimental Setup | 237, 238 |
| **RAID** | Redundant Array of Inexpensive Disk | 390, 396, 400 |
| **RAM** | Random Access Memory | 390 |
| **RBF** | Rochester-Brookhaven-Fermilab | 237 |
| **RBW** | Resolution Band Width | 260, 273, 281, 282, 288 |
| **RF** | Radio Frequency | 244, 252–255, 257, 267, 278, 280, 286, 389, 391, 392 |

| Acronym | Full Name | Page(s) |
|---|---|---|
| **RMS** | Root Mean Square | 195, 273 |
| **SDO** | Solar Dynamics Observatory | 129, 140 |
| **SDSS** | Sloan Digital Sky Survey | 17, 38, 52 |
| **SEER** | Surveillance Epidemiology End Results database | 206 |
| **SEM** | Solar EUV Monitor | 95 |
| **SEP** | Solar Energetic Particle | 139, 140 |
| **SG** | Savitzky-Golay | 279, 282, 285, 287 |
| **Sgr** | Sagittarius | 37–39, 239, 241 |
| **SHM** | Standard Halo Model | 35, 47, 49, 51, 52 |
| **SIDM** | Self-Interacting Dark Matter | 26, 43, 44 |
| **SILSO** | Sunspot Index and Long-term Solar Observations | 105 |
| **SIMP** | Strongly-Interacting Massive Particle | 26 |
| **SM** | Standard Model | 9, 23, 24, 27, 32, 47, 230 |
| **SMA** | SubMiniature version A | 261, 389, 390, 392 |
| **SNR** | Signal-to-Noise Ratio | 89, 90, 92, 184, 247–249, 262, 263, 265, 281, 283–287, 289, 290, 296, 318 |
| **SOHO** | Solar and Heliospheric Observatory | 95, 129 |
| **SQUID** | Superconducting Quantum Interference Device | 29, 238 |
| **SSD** | Solid-State Drive | 390, 396, 400, 402 |
| **SSW** | Sudden Stratospheric Warming | 189 |
| **STA** | Stratospheric Temperature Anomaly | 180, 189, 371 |
| **SUSY** | Supersymmetry | 21, 23, 25 |
| **SW** | Solar Wind | 139, 140 |
| **SWPC** | Space Weather Prediction Center | 83 |

| Acronym | Full Name | Page(s) |
|---|---|---|
| **TD** | Time Domain | 273 |
| **TE** | Transverse Electric field | 246, 275, 393 |
| **TEC** | Total Electron Content | 167–170, 173, 175, 191, 193, 194, 204, 217, 340, 341, 365–369, 378, 380–382 |
| **TECU** | Total Electron Content Unit | 167, 169, 219 |
| **TS** | Tone Synthesizer | 256, 260, 265, 268, 392, 393, 400 |
| **TSI** | Total Solar Irradiance | 159, 325 |
| **UF** | University of Florida | 237 |
| **UPS** | Uninterruptible Power Supply | 390 |
| **USA** | United States of America | 205, 207, 215 |
| **USB** | Universal Serial Bus | 255, 390, 392, 399 |
| **USGS** | U.S. Geological Survey | 194 |
| **UTC** | Coordinated Universal Time | 109, 118, 126, 142, 179, 180, 191, 195, 196 |
| **UV** | UltraViolet | 117, 147, 177, 187, 188, 205, 215, 325, 382–384 |
| **VBA** | Visual Basic for Applications | 72, 403 |
| **VBW** | Video Band Width | 260 |
| **VNA** | Vector Network Analyser | 256, 258, 260, 261, 263, 275, 276, 389, 391–393, 396, 400 |
| **VSA** | Vector Signal Analyser | 258–261, 265, 267, 268, 273, 276, 279, 291, 389–391, 393–396, 400 |
| **WDC** | World Data Center | 105 |
| **WDM** | Warm Dark Matter | 19, 21, 37, 43 |

| Acronym | Full Name | Page(s) |
|---|---|---|
| **WIMP** | Weakly Interacting Massive Particle | 19, 23, 26–28, 30, 31, 33, 39, 43, 46, 66–68, 163, 204, 223, 292, 295 |
| **WLAN** | Wireless Local Area Network | 259 |
| **WSO** | Wilcox Solar Observatory | 160 |
| **XRS** | X-Ray Sensor | 82 |

# LIST OF SYMBOLS

| Symbol | Units | Description |
|---|---|---|
| $a$ | - | Axion field |
| $\vec{B}$ | T | Magnetic field |
| $\beta$ | - | Coupling factor |
| $C_{lmn}$ | - | Geometry factor |
| $\Delta t$ | s | Sampling interval in time |
| $\delta\nu$ | Hz | Resolution bandwidth |
| $\delta\nu_a$ | Hz | Axion bandwidth |
| $\vec{E}$ | V/m | Electric field |
| $f_a$ | GeV | Axion decay constant |
| $g_{a\gamma\gamma}$ | $\mathrm{GeV}^{-1}$ | Axion-photon coupling |
| $g_\gamma$ | - | Unitless coupling constant |
| $\lambda$ | m | Wavelength |
| $m_a$ | $\mathrm{eV}/c^2$ | Axion mass |
| $\mu$ | - | Mean value |

| Symbol | Units | Description |
|---|---|---|
| $\omega$ | Hz | Angular frequency |
| $Q_0$ | - | Unloaded quality factor |
| $Q_a$ | - | Quality factor of the axion |
| $Q_L$ | - | Loaded quality factor |
| $\rho_a$ | GeV/cm$^3$ | Axion energy density |
| $\sigma$ | - | Standard deviation |
| $T_S$ | K | Noise temperature of the system |
| $\nu$ | Hz | Frequency |
| $\nu_0$ | Hz | Center frequency of cavity resonance mode |
| $\nu_a$ | Hz | Axion frequency |
| $\upsilon$ | km/sec | Velocity |
| $V$ | L | Volume |

# LIST OF CONSTANTS

| Symbol | Value | Units | Description |
|---|---|---|---|
| $a$ | $7.2974 \times 10^{-3}$ | - | Fine structure constant |
| $a_0$ | $5.292 \times 10^{-11}$ | m | Bohr radius |
| $AU$ | $1.4960 \times 10^{11}$ | m | Astronomical unit |
| $c$ | 299792458 | m/s | Speed of light |
| $e$ | $1.602 \times 10^{-19}$ | C | Electron charge magnitude |
| $\varepsilon_0$ | $8.854 \times 10^{-12}$ | $\mathrm{C^2/(N\,m^2)}$ | Vacuum electric permitivity |
| $G_F$ | $1.1664 \times 10^{-5}$ | $\mathrm{GeV^{-2}}$ | Fermi coupling constant |
| $G_N$ | $6.674 \times 10^{-11}$ | $\mathrm{N\,m^2/kg^2}$ | Gravitational constant |
| $g_N$ | 9.807 | $\mathrm{m/s^2}$ | Earth's surface gravity |
| $h$ | $6.626 \times 10^{-34}$ | $\mathrm{J\,Hz^{-1}}$ | Planck's constant |
| $H_0$ | $67.8 \pm 0.9$ | km/sec/Mpc | Present day Hubble expansion rate |
| $\hbar$ | $1.055 \times 10^{-34}$ | J s | Reduced Plank's constant |
| $h_0$ | 0.678 | - | Rescaled Hubble's constant |
| $k_B$ | $1.381 \times 10^{-23}$ | J/K | Boltzmann constant |
| $k_E$ | $8.988 \times 10^{9}$ | $\mathrm{N\,m^2/C^2}$ | Coulmb's constant |
| $L_\odot$ | $3.828 \times 10^{26}$ | W | Nominal Solar luminocity |

| Symbol | Value | Units | Description |
|---|---|---|---|
| $ly$ | $0.946 \times 10^{16}$ | m | Light year |
| $m_e$ | 0.5110 | MeV/$c^2$ | Electron mass |
| $M_\oplus$ | $5.972 \times 10^{24}$ | kg | Earth mass |
| $m_n$ | 939.6 | MeV/$c^2$ | Neutron mass |
| $m_p$ | 938.3 | MeV/$c^2$ | Proton mass |
| $M_\odot$ | $1.988 \times 10^{30}$ | kg | Solar mass |
| $\mu_0$ | $4\pi \times 10^{-7}$ | N/A$^2$ | Vacuum magnetic permeability |
| $N_A$ | $6.022 \times 10^{23}$ | mol$^{-1}$ | Avogadro constant |
| $pc$ | $3.0857 \times 10^{16}$ | m | Parsec |
| $R_\oplus$ | $6.3781 \times 10^{6}$ | m | Earth radius |
| $R_H$ | $1.374 \times 10^{26}$ | m | Hubble radius |
| $\rho_{\rm cr}$ | $1.8784 \times 10^{-29}$ | h g cm$^{-3}$ | Critical density |
| $R_\odot$ | $6.957 \times 10^{8}$ | m | Solar Radius |
| $sfu$ | $1 \times 10^{-22}$ | W/(m$^2$ Hz) | Solar flux units |
| $T_0$ | 2.7255 | K | Present day CMB temperature |

# TEXT REFERENCES

# IMAGE REFERENCES